%% file: main.tex
\documentclass[12pt,twoside]{mitthesis}
\usepackage{lgrind}
\usepackage{cmap}
\usepackage[T1]{fontenc}
\usepackage{subfig}
\usepackage{xspace}
\usepackage[countmax]{subfloat}
\usepackage{slashed}

\usepackage{amssymb}
\usepackage{amsmath}
\usepackage{mathtools}
\usepackage{cancel}
\usepackage{fullpage}
\usepackage{color}
\usepackage{braket}
\usepackage{subfig}
\usepackage{graphicx}

\usepackage{bbm}

\usepackage{color}
\definecolor{darkblue}{rgb}{0,0,0.5}
\definecolor{darkred}{rgb}{0.5,0,0}

\DeclareRobustCommand{\Sec}[1]{Sec.~\ref{#1}}
\DeclareRobustCommand{\Secs}[2]{Secs.~\ref{#1} and \ref{#2}}
\DeclareRobustCommand{\App}[1]{App.~\ref{#1}}
\DeclareRobustCommand{\Tab}[1]{Table~\ref{#1}}
\DeclareRobustCommand{\Tabs}[2]{Tables~\ref{#1} and \ref{#2}}
\DeclareRobustCommand{\Fig}[1]{Fig.~\ref{#1}}
\DeclareRobustCommand{\Figs}[2]{Figs.~\ref{#1} and \ref{#2}}
\DeclareRobustCommand{\Eq}[1]{Eq.~(\ref{#1})}
\DeclareRobustCommand{\Eqs}[2]{Eqs.~(\ref{#1}) and (\ref{#2})}
\DeclareRobustCommand{\Ref}[1]{Ref.~\cite{#1}}
\DeclareRobustCommand{\Refs}[1]{Refs.~\cite{#1}}

\DeclareRobustCommand{\Chap}[1]{Chap.~\ref{#1}}

\begin{document}

\include{cover}
\pagestyle{plain}
\include{contents}

\include{chap1}

\include{chap1a}

\pagestyle{plain}
\include{chap2}

\pagestyle{plain}
\include{chap8}
\include{chap4}

\include{chap5}

\include{chap_conc}

\appendix
\include{appa}

\include{appb}

\include{biblio}
\end{document}

%% file: cover.tex
%
%
%
%
%
%
%
%
%
%
%

\title{Effective Field Theories for the LHC}

\author{Ian James Moult}
\department{Department of Physics}

\degree{Doctor of Philosophy}

\degreemonth{June}
\degreeyear{2016}
\thesisdate{April 27, 2016}


\supervisor{Iain W. Stewart}{Professor of Physics}

\chairman{Nergis Mavalvala}{Associate Department Head of Physics}

\maketitle



\cleardoublepage
\setcounter{savepage}{\thepage}
\begin{abstractpage}
\input{abstract}
\end{abstractpage}


\cleardoublepage

\section*{Acknowledgments}

I am grateful to a large number of people who have made my time at MIT extremely enjoyable. In particular I have had the privilege of having a large number of excellent collaborators, and people from whom I have learned a great deal of physics. Of these, I must thank in particular, Frank Tackmann, Wouter Waalewijn, Markus Ebert, Lisa Zeune, Dan Kolodrubetz, Ilya Feige, Matt Schwartz, Jesse Thaler, Piotr Pietrulewicz, Matthew Low, Christopher Lee, Daekyung Kang and Hua Xing Zhu. I must also thank my office mates, in particular, Josephine Suh, and Lina Necib, with whom I spent a large amount of time in the office, and who made this experience very enjoyable. A large portion of the work presented in this thesis, and more generally the work that I did while at MIT, was in collaboration with Duff Neill and Andrew Larkoski. I enjoyed this collaboration immensely, and it was one of the major highlights of my time at MIT. I must thank them for teaching me a huge amount of physics, and for being fun to collaborate with.  Finally, I am particularly thankful to my advisor Iain Stewart for letting me transfer to the CTP after my first year, for being an amazing advisor, for teaching me a lot of physics, and for giving me many opportunities to pursue physics.


%% file: abstract.tex
%
%
%
In this thesis I study applications of effective field theories to understand aspects of QCD jets and their substructure at the Large Hadron Collider. In particular, I introduce an observable, $D_2$, which can be used for distinguishing boosted $W/Z/H$ bosons from the QCD background using information about the radiation pattern within the jet, and perform a precision calculation of this observable. To simplify calculations in the soft collinear effective theory, I also develop a helicity operator basis, which facilitates matching calculations to fixed order computations performed using spinor-helicity techniques, and demonstrate its utility by computing an observable relevant for studying the properties of the newly discovered Higgs boson.

%% file: contents.tex
\tableofcontents
\newpage
\listoffigures
\newpage
\listoftables

%% file: chap1.tex

\DeclareRobustCommand{\Sec}[1]{Sec.~\ref{#1}}
\DeclareRobustCommand{\Secs}[2]{Secs.~\ref{#1} and \ref{#2}}
\DeclareRobustCommand{\App}[1]{App.~\ref{#1}}
\DeclareRobustCommand{\Tab}[1]{Table~\ref{#1}}
\DeclareRobustCommand{\Tabs}[2]{Tables~\ref{#1} and \ref{#2}}
\DeclareRobustCommand{\Fig}[1]{Fig.~\ref{#1}}
\DeclareRobustCommand{\Figs}[2]{Figs.~\ref{#1} and \ref{#2}}
\DeclareRobustCommand{\Eq}[1]{Eq.~(\ref{#1})}
\DeclareRobustCommand{\Eqs}[2]{Eqs.~(\ref{#1}) and (\ref{#2})}
\DeclareRobustCommand{\Ref}[1]{Ref.~\cite{#1}}
\DeclareRobustCommand{\Refs}[1]{Refs.~\cite{#1}}

\newcommand\bn{{\bar n}}
\newcommand{\sdt}{\!\cdot\!}
\newcommand{\la}{\lambda}
\newcommand{\ord}[1]{\mathcal{O}(#1)}
\def\nn{{\nonumber}}
\newcommand{\lp}{\tilde p}        
\def\cP{\mathcal{P}}
\newcommand{\bnP}{\overline {\mathcal P}}
\def\cL{\mathcal{L}}
\newcommand{\hard}{\mathrm{hard}}
\newcommand{\dyn}{\mathrm{dyn}}
\newcommand{\tree}{\mathrm{tree}}
\newcommand{\BPS}{\mathrm{BPS}}

\newcommand{\SCETi}{\mbox{${\rm SCET}_{\rm I}$}\xspace}
\newcommand{\SCETii}{\mbox{${\rm SCET}_{\rm II}$}\xspace}
\def\cB{\mathcal{B}}
\def\cD{\mathcal{D}}
\def\cY{\mathcal{Y}}

\newcommand{\w}{\omega}
\def\tr{{\rm tr}}
\newcommand{\Sl}[1]{\slashed{#1}}

\newcommand{\df}{\mathrm{d}}

\newcommand{\hH}{\widehat{H}}
\newcommand{\hS}{\widehat{S}}
\newcommand{\hV}{\widehat{V}}

\chapter{Introduction}\label{chap:intro}

The study of fundamental physics has progressed over the past century by probing the structure of particles at smaller and smaller scales, or equivalently by colliding them at higher and higher energies. In 2010, the Large Hadron Collider (LHC) became the highest energy collider, with a center of mass energy of $7$ TeV (now $13$ TeV), allowing for a probe of nature at unexplored scales. Already in its first few years of running, this has enabled the discovery of the Higgs boson.

Essential to maximizing the impact of the LHC program is a detailed theoretical understanding of the processes expected at the LHC within the Standard Model (SM) of particle physics. Precise theoretical predictions allow for the study of particles in the SM, as well as the detection of deviations from the SM, which would indicate the presence of new physics. Precision predictions at the LHC are made difficult by the fact that the LHC collides protons, which are not fundamental particles, but are made up of partons which interact strongly via Quantum Chromodynamics.

\begin{figure}
\begin{center}
\subfloat[]{\label{fig:Peak}
\includegraphics[width=7cm]{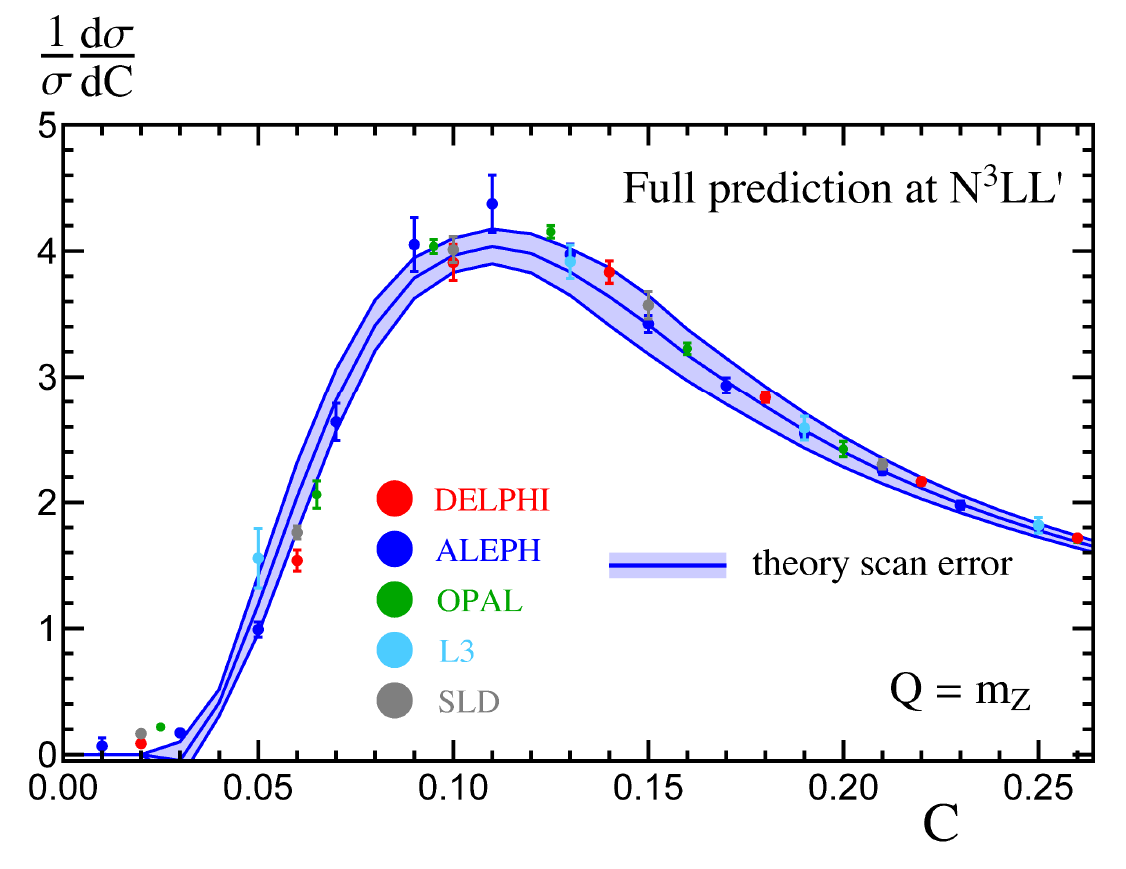}
}
$\qquad$
\subfloat[]{\label{fig:thrust} 
\includegraphics[width=7cm]{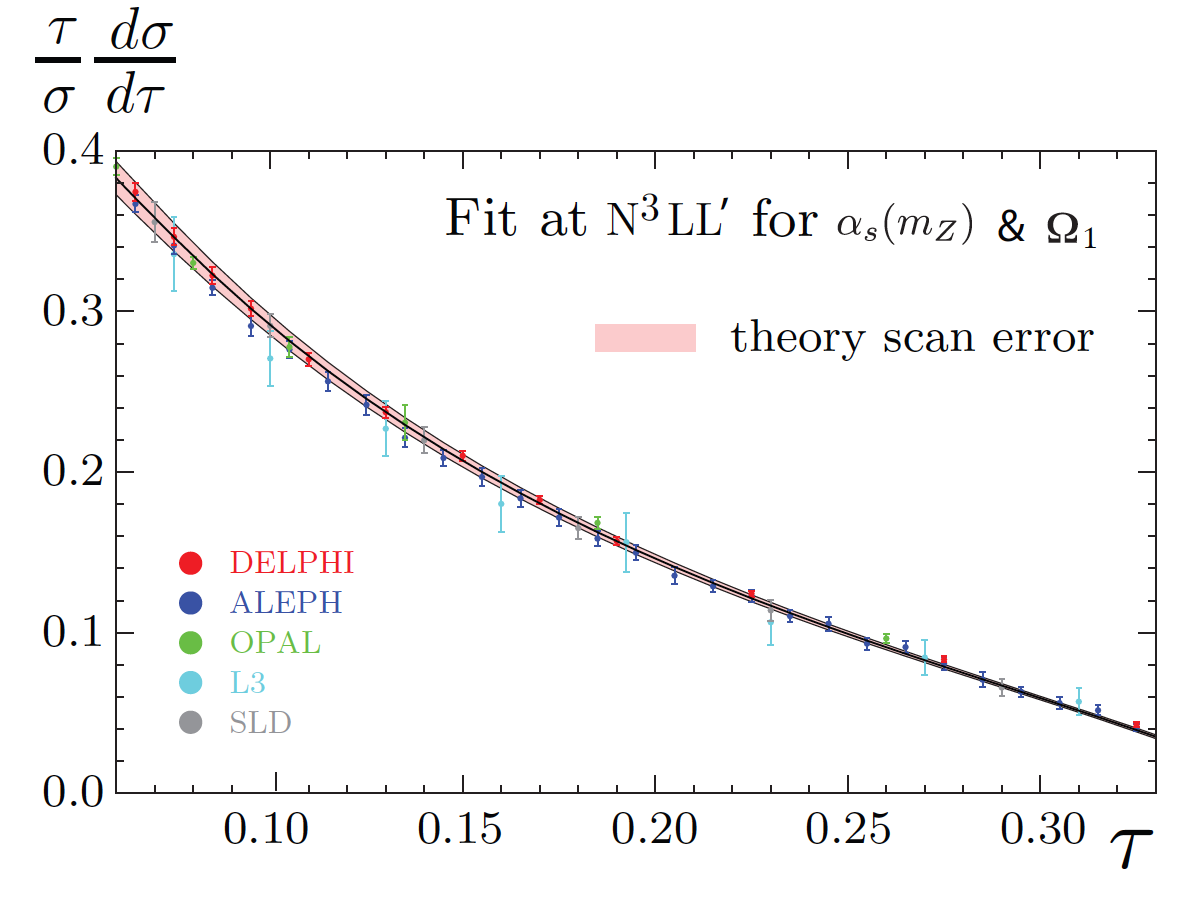}
}
\end{center}
\caption{Precision theoretical predictions for the $C$-parameter (left) (figure from \cite{Hoang:2014wka})  and Thrust (right) (figure from \cite{Abbate:2010xh}) compared with data from $e^+e^-$ colliders.
}
\label{fig:event_shapes}
\end{figure}

Quantum Chromodynamics is a non-abelian gauge theory. It is strongly coupled and confining at low energies, but becomes weakly coupled at high energies, allowing for a perturbative treatment of hard scattering processes. QCD has been extensively studied in hard scattering processes at $e^+e^-$ colliders by measuring the pattern of radiation produced in collisions using observables called event shapes. Event shapes are theoretically clean, and have been calculated to high precision. Two examples comparing theory predictions with experimental data are shown in \Fig{fig:event_shapes} \cite{Abbate:2010xh,Hoang:2014wka}.

The study of QCD at the extreme environments of the LHC is significantly more complicated, and requires a variety of new theoretical tools. This is due both to the strongly interacting nature of the colliding partons, as well as the level of sophistication of the measurements which can be performed on jets using the high granularity detectors at the LHC. 

\section{QCD at the LHC}

The study of QCD plays an important role in nearly all measurements made at the LHC. Due to the fact that the colliding partons, namely quarks and gluons, are strongly interacting, QCD radiation, in the form of collimated, highly energetic jets is produced copiously. The study of the dynamics of these jets is extremely interesting for understanding the dynamics of QCD. Furthermore, electroweak particles such as the $W/Z/H$, are produced in association with QCD jets. Therefore the theoretical foundation for understanding any process at the LHC is dominated by QCD.

Consider a process with $N$ final-state jets and $L$ leptons, photons, or other nonstrongly interacting particles, with the  underlying hard Born process
\begin{equation} \label{eq:interaction_intro}
\kappa_a (q_a)\, \kappa_b(q_b) \to \kappa_1(q_1) \dotsb \kappa_{N+L}(q_{N+L})
\,,\end{equation}
where $\kappa_{a,b}$ denote the colliding partons, and $\kappa_i$ denote the outgoing quarks, gluons, leptons, and other particles with momenta $q_i$. The incoming partons are essentially along the beam directions, $q_{a,b}^\mu = x_{a,b} P_{a,b}^\mu$, where $x_{a,b}$ are the momentum fractions and $P^\mu_{a,b}$ the (anti)proton momenta.

\begin{figure}
\begin{center}
\includegraphics[width=10cm]{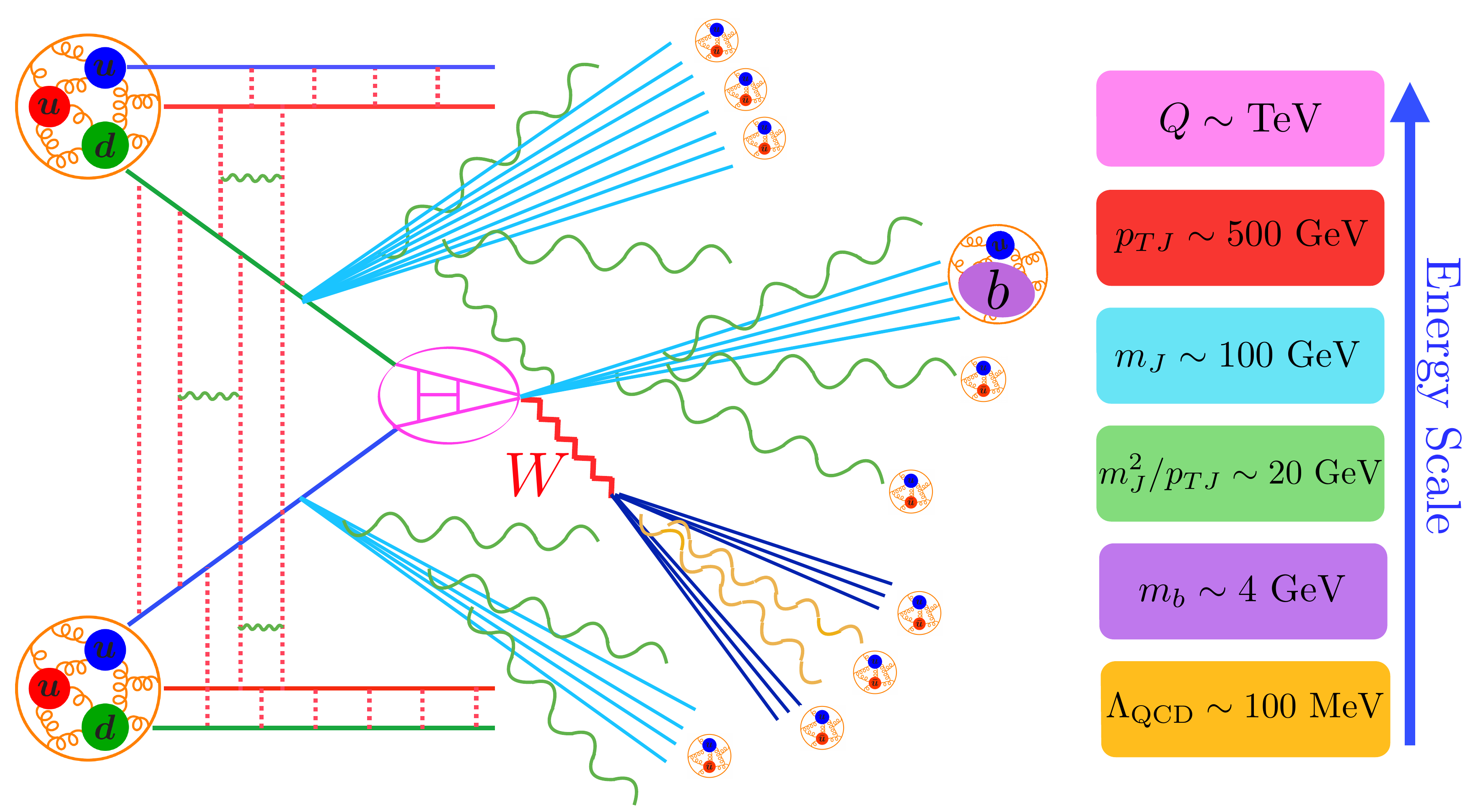}
\end{center}
\caption{A schematic picture of the factorized structure of an LHC collision and the relevant scales in the problem. Jets with both resolved and unresolved substructure are illustrated.
}
\label{fig:jet_collision}
\end{figure}

The active-parton exclusive jet cross section corresponding to \Eq{eq:interaction_intro} can be factorized for a variety of jet resolution variables.\footnote{Here active parton refers to initial-state quarks or gluons. Proofs of factorization with initial-state hadrons must also account for effects due to Glaubers~\cite{Collins:1988ig}, which may or may not cancel, and whose relevance depends on the observable in question~\cite{Gaunt:2014ska,Zeng:2015iba}. For a recent discussion in the context of SCET, see \cite{Rothstein:2016bsq}} The factorized expression for the exclusive jet cross section can be written schematically in the form
\begin{align} \label{eq:sigma_intro}
\df\sigma &=
\int\!\df x_a\, \df x_b\, \df \Phi_{N+L}(q_a \!+ q_b; q_1, \ldots)\, M(\{q_i\})
\\\nn &\quad \times
\sum_{\kappa} \tr\,\bigl[ \hH_{\kappa}(\{q_i\}) \hS_\kappa \bigr] \otimes
\Bigl[ B_{\kappa_a} B_{\kappa_b} \prod_J J_{\kappa_J} \Bigr]
+ \dotsb
\,,\end{align}
which is shown pictorially in \Fig{fig:jet_collision}. Here, $\df \Phi_{N+L}(q_a\!+ q_b; q_1, \ldots)$ denotes the Lorentz-invariant phase space for the Born process in \Eq{eq:interaction_intro}, and $M(\{q_i\})$ denotes the measurement made on the hard momenta of the jets (which in the factorization are approximated by the Born momenta $q_i$). The dependence on the underlying hard interaction is encoded in the hard function $\hH_{\kappa}(\{q_i\})$, where $\{q_i\} \equiv \{q_1, \ldots, q_{N+L}\}$, the sum over $\kappa \equiv \{\kappa_a, \kappa_b, \ldots \kappa_{N+L}\}$ is over all relevant partonic processes, and the trace is over color. Any dependence probing softer momenta, such as measuring jet masses or low $p_T$s, as well as the choice of jet algorithm, will affect the precise form of the factorization, but not the hard function $\hH_\kappa$. This dependence enters through the definition of the soft function $\hS_\kappa$ (describing soft radiation), jet functions $J_{\kappa_J}$ (describing energetic final-state radiation in the jets) and the beam functions $B_i$ (describing energetic initial-state radiation along the beam direction). More precisely, the beam function is given by $B_i = \sum_{i'} {\cal I}_{i i'} \otimes f_{i'}$ with $f_i$ the parton distributions of the incoming protons, and ${\cal I}_{i i'}$ a perturbatively calculable matching coefficient depending on the measurement definition~\cite{Stewart:2009yx}. The ellipses at the end of \Eq{eq:sigma_intro} denote power-suppressed corrections. All functions in the factorized cross section depend only on the physics at a single scale. This allows one to evaluate all functions at their own natural scale, and then evolve them to a common scale using their RGE. This procedure resums the large logarithms of scale ratios appearing in the cross section to all orders in perturbation theory. Furthermore, it enables the definition of universal non-perturbative quantities, such as the parton distribution functions, or fragmentation functions, which although they are not calculable from first principles, are universal, and thus predictivity is maintained. The description of even the simplest LHC observables therefore requires a detailed understanding of QCD, embodied in \Eq{eq:sigma_intro}. The complexity becomes far greater as realistic experimental measurements are imposed on the final state.

%
%
%

\begin{figure}
\begin{center}
\subfloat[]{\label{fig:diboson_ATLAS_1}
\includegraphics[width=7cm]{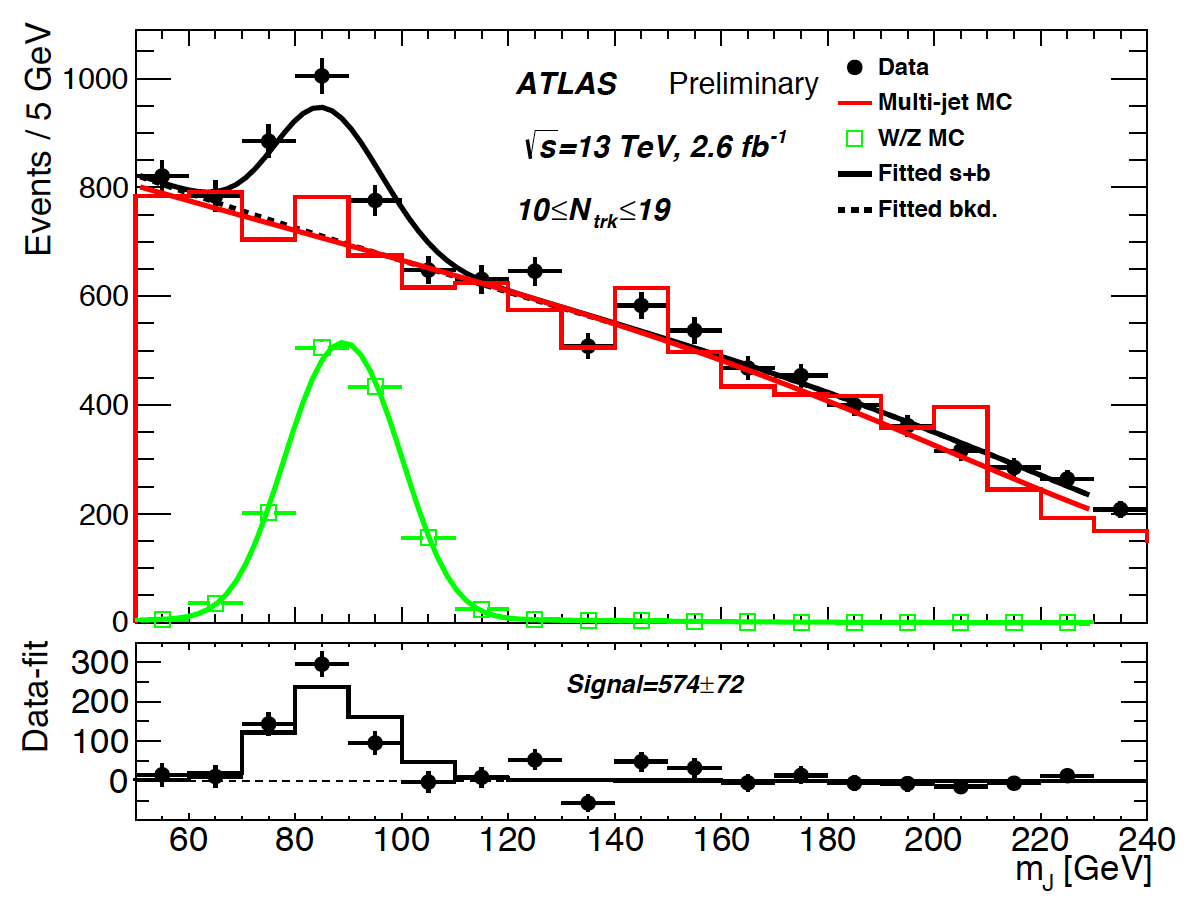}
}
$\qquad$
\subfloat[]{\label{fig:diboson_ATLAS_2} 
\includegraphics[width=7cm]{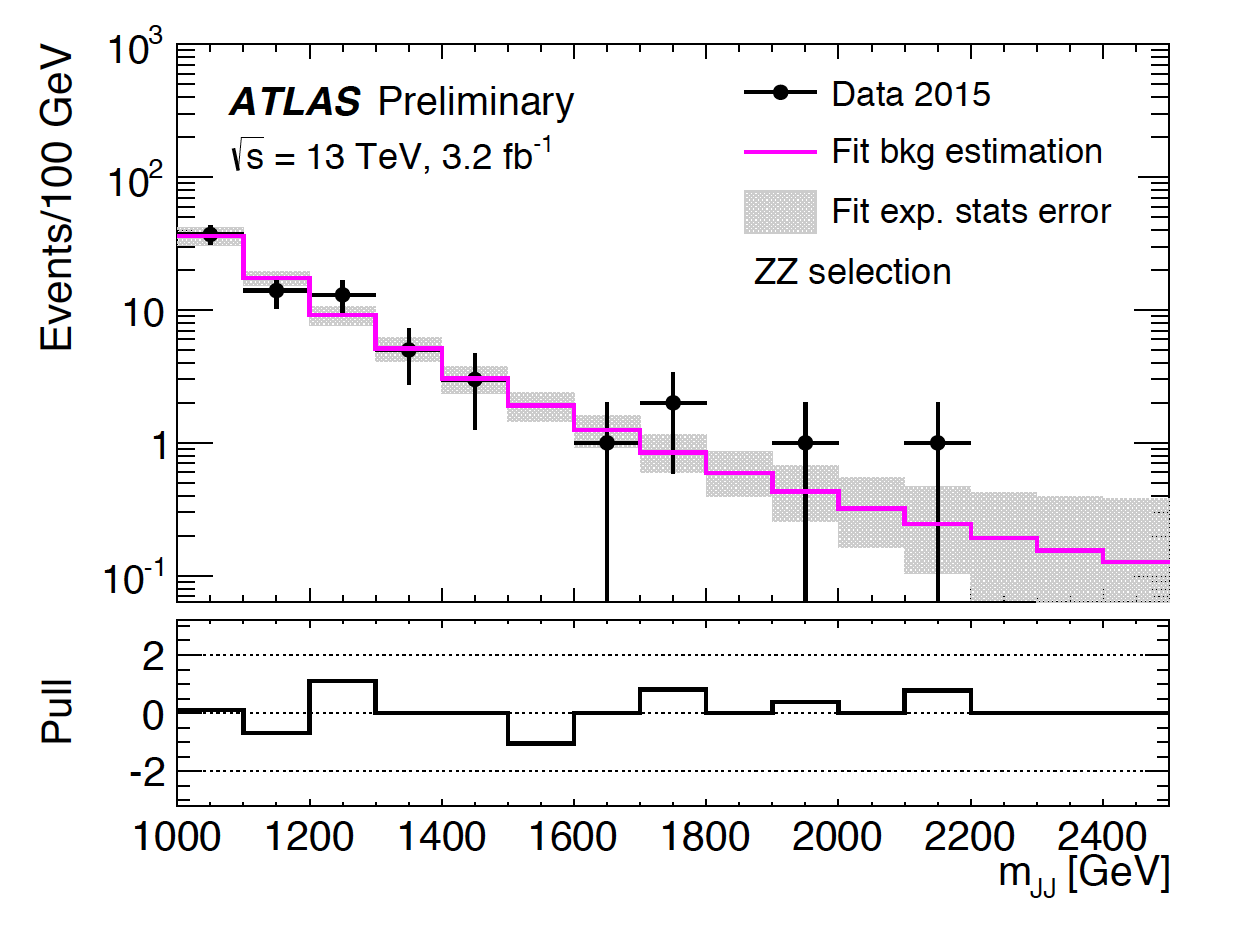}
}
\end{center}
\caption{A plot of the invariant mass spectrum of jets produced at the LHC, showing the contribution from hadronically decaying $W/Z$ bosons (left) and the dijet invariant mass spectrum used to search for resonances decaying to highly boosted $W/Z$ bosons (right) (figures from \cite{collaboration:2015aa}).
}
\label{fig:diboson_ATLAS}
\end{figure}

\begin{figure}
\begin{center}
\subfloat[]{\label{fig:btag_1}
\includegraphics[width=7cm]{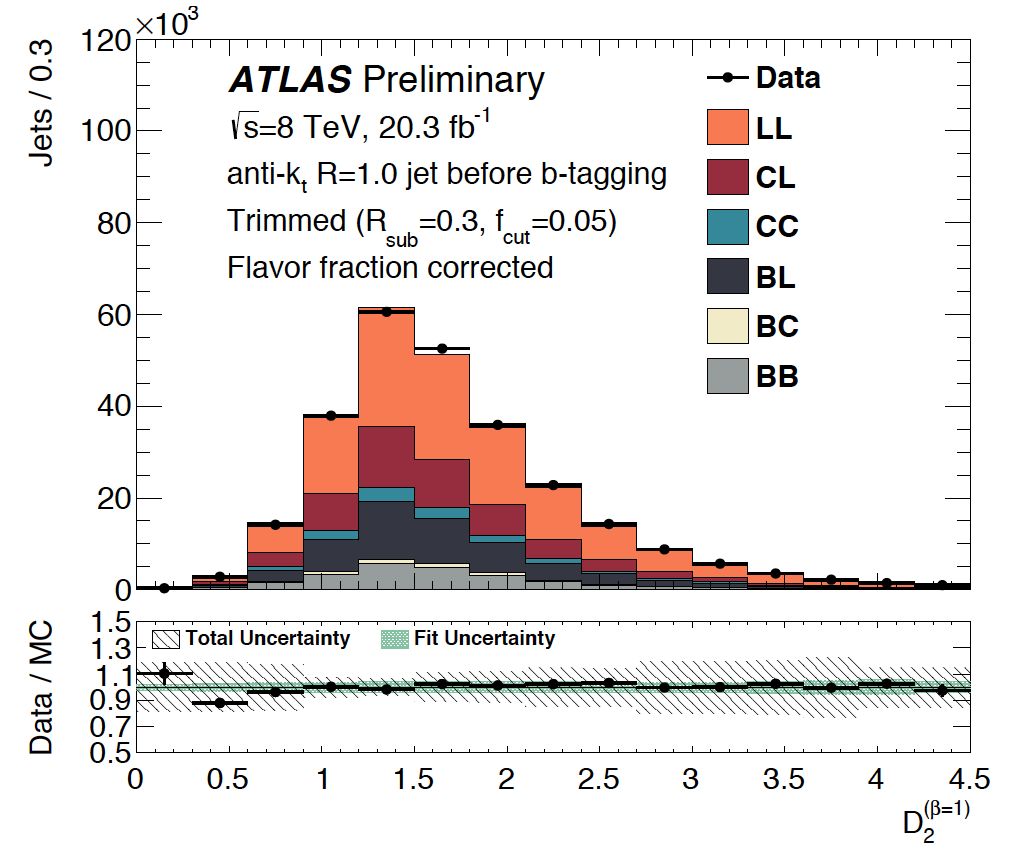}
}
$\qquad$
\subfloat[]{\label{fig:btag_1} 
\includegraphics[width=7cm]{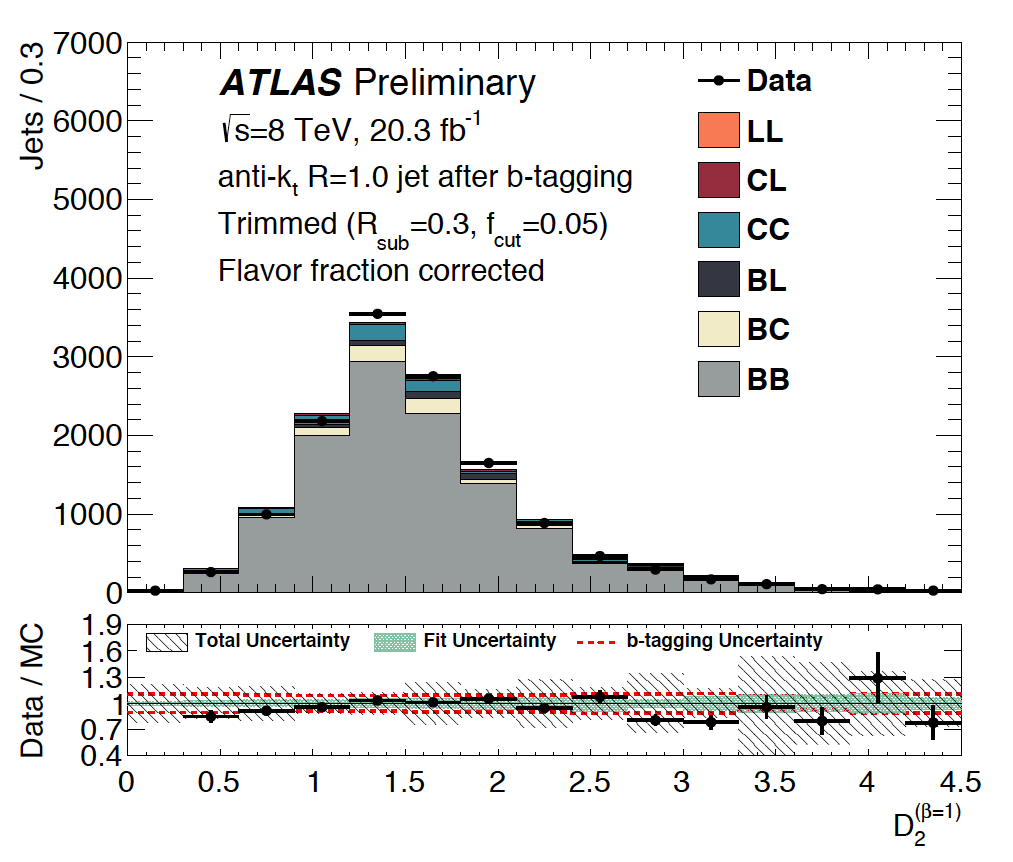}
}
\end{center}
\caption{The distribution of radiation within a jet, as measured using $D_2$ before and after a subjet $b$-tagging is applied, showing the sophistication of modern jet substructure measurements (figures from  \cite{collaboration:2016aa}).
}
\label{fig:btag}
\end{figure}

One of the major advances at the LHC in the precision study of jets is the study of jet substructure, which will allow for an unprecedented study of strong dynamics. The extremely fine granularity of the LHC detectors allows for the substructure within a jet to be accurately measured, as opposed to simply its global properties. Experimentally, this is extremely useful, as the radiation pattern within a jet provides clues to the nature of the particle that created the jet. For example, a hadronically decaying $W/Z/H$ particle will decay primarily into a jet consisting of two hard prongs, while an average QCD jet produced by a radiating quark or gluon will not have such a structure. This is shown schematically in \Fig{fig:jet_collision}, where a hadronically decaying $W$ jet is embedded into a complex LHC collision involving a number of other QCD jets. Similarly, jet substructure measurements can be used to determine the charge of a jet, to determine whether a jet originated from a quark or a gluon, or to identify jets coming from decaying top quarks. Jet substructure techniques have been used extensively in a variety of searches at the LHC. For example, they have been used to search for new heavy resonances decaying to boosted $W/Z$ bosons, which then decay hadronically. The boosted $W/Z$ bosons can be identified using jet substructure techniques, and the mass of the resonance can then be reconstructed. In \Fig{fig:diboson_ATLAS} we show on the left a plot of the jet mass spectrum, showing a clear peak at the $W/Z$ mass, and on the right, the dijet invariant mass spectrum obtained using jet substructure techniques \cite{collaboration:2015aa}. New resonances would appear as bumps in this distribution. As a further example of the level of sophistication of the experimental techniques available for the study of jet substructure, in \Fig{fig:btag}, we show a distribution of a jet shape observable $D_2$ \cite{Larkoski:2014gra}, to be discussed in detail in this thesis, for gluon jets decaying to $b$ quarks, $g\to b\bar b$ obtained using subjet $b$-tagging \cite{collaboration:2016aa}.

To be able to take advantage of these experimental measurements requires a detailed understanding of the substructure of QCD jets. One theoretically clean way that this can be achieved is to extend the ideas used in $e^+e^-$ event shapes to jet shapes, which probe the radiation pattern within a particular jet, and which can be specifically designed to identify particular features of the jet. The calculation of jet shapes relevant for the LHC, for instance for boosted $W/Z/H$ production are still in their infancy. They are significantly more complicated than $e^+e^-$ event shape calculations, as additional scales appear in the calculation associated with the substructure of the jet. To match the level of experimental sophistication in performing jet substructure measurements requires the development of new tools in QCD to accurately describe the substructure of QCD jets, and to extend the factorization formula of \Eq{eq:sigma_intro} to describe jet substructure observables.

\section{Soft Collinear Effective Theory}

Due to the complexity of QCD, effective field theories have played an important role, enabling calculations in specific limits of the theory. For example, effective field theories have been developed using power expansions in heavy quark masses, in the forward scattering limit, or more generally, in the presence of hierarchical kinematic scales. For studying QCD jets at the LHC, the appropriate effective field theory is the Soft Collinear Effective Field Theory (SCET).

SCET is an effective field theory of QCD describing the interactions of collinear and soft particles in the presence of a hard interaction, as relevant for jet production at the LHC \cite{Bauer:2000ew, Bauer:2000yr, Bauer:2001ct, Bauer:2001yt, Bauer:2002nz}. In this section we briefly describe notation commonly used in SCET,  give the Lagrangian of the theory, and show how field redefinitions can be used to factorize soft and collinear radiation, leading to an expression for a factorized cross section as in \Eq{eq:sigma_intro}.

SCET describes collinear particles (which are characterized by a large momentum along a particular light-like direction), as well as soft particles. It is therefore convenient to use light-cone coordinates. For each jet direction we define two light-like reference vectors $n_i^\mu$ and $\bn_i^\mu$ such that $n_i^2 = \bn_i^2 = 0$ and $n_i\sdt\bn_i = 2$. One typical choice for these quantities is
\begin{equation}
n_i^\mu = (1, \vec{n}_i)
\,,\qquad
\bn_i^\mu = (1, -\vec{n}_i)
\,,\end{equation}
where $\vec{n}_i$ is a unit three-vector. Given a choice for $n_i^\mu$ and $\bn_i^\mu$, any four-momentum $p$ can then be decomposed as
\begin{equation} \label{eq:lightcone_dec}
p^\mu = \bn_i\sdt p\,\frac{n_i^\mu}{2} + n_i\sdt p\,\frac{\bn_i^\mu}{2} + p^\mu_{n_i\perp}\
\,.\end{equation}
An ``$n_i$-collinear'' particle has momentum $p$ close to the $\vec{n}_i$ direction, so that the components of $p$ scale as $(n_i\!\cdot\! p, \bn_i \!\cdot\! p, p_{n_i\perp}) \sim \bn_i\!\cdot\! p$ $\,(\la^2,1,\la)$, where $\la \ll 1$ is a small parameter determined by the form of the measurement or kinematic restrictions. To ensure that $n_i$ and $n_j$ refer to distinct collinear directions, they have to be well separated, meaning
\begin{equation} \label{eq:nijsep}
  n_i\sdt n_j  \gg \la^2 \qquad\text{for}\qquad i\neq j
\,.\end{equation}
Two different reference vectors, $n_i$ and $n_i'$, with $n_i\cdot n_i' \sim \ord{\lambda^2}$ both describe the same jet and corresponding collinear physics. Thus, each collinear sector can be labelled by any member of a set of equivalent vectors, $\{n_i\}$. This freedom is manifest as a symmetry of the effective theory known as reparametrization invariance (RPI) \cite{Manohar:2002fd,Chay:2002vy}. Specifically, the three classes of RPI transformations are
\begin{alignat}{3}\label{eq:RPI_def}
&\text{RPI-I} &\qquad &  \text{RPI-II}   &\qquad &  \text{RPI-III} \nn \\
&n_{i \mu} \to n_{i \mu} +\Delta_\mu^\perp &\qquad &  n_{i \mu} \to n_{i \mu}   &\qquad & n_{i \mu} \to e^\alpha n_{i \mu} \nn \\
&\bar n_{i \mu} \to \bar n_{i \mu}  &\qquad &  \bar n_{i \mu} \to \bar n_{i \mu} +\epsilon_\mu^\perp  &\qquad & \bar n_{i \mu} \to e^{-\alpha} \bar n_{i \mu}\,.
\end{alignat}
Here, we have $\Delta^\perp \sim \lambda$, $\epsilon^\perp \sim \lambda^0$, and $\alpha\sim \lambda^0$. The parameters $\Delta^\perp$ and $\epsilon^\perp$ are infinitesimal, and satisfy $n_i\cdot \Delta^\perp=\bar n_i\cdot \Delta^\perp=n_i \cdot \epsilon^\perp=\bar n_i \cdot \epsilon^\perp=0$.

The effective theory is constructed by expanding momenta into label and residual components
\begin{equation} \label{eq:label_dec}
p^\mu = \lp^\mu + k^\mu = \bn_i \sdt\lp\, \frac{n_i^\mu}{2} + \lp_{n_i\perp}^\mu + k^\mu\,.
\,\end{equation}
Here, $\bn_i \cdot\lp \sim Q$ and $\lp_{n_i\perp} \sim \la Q$ are the large label momentum components, where $Q$ is the scale of the hard interaction, while $k\sim \la^2 Q$ is a small residual momentum. A multipole expansion is then performed to obtain fields with momenta of definite scaling, namely collinear quark and gluon fields for each collinear direction, as well as soft quark and gluon fields. Independent gauge symmetries are enforced for each set of fields.

The SCET fields for $n_i$-collinear quarks and gluons, $\xi_{n_i,\lp}(x)$ and $A_{n_i,\lp}(x)$, are labeled by their collinear direction $n_i$ and their large momentum $\lp$. They are written in position space with respect to the residual momentum and in momentum space with respect to the large momentum components. Derivatives acting on the fields pick out the residual momentum dependence, $i \partial^\mu \sim k \sim \la^2 Q$. The large label momentum is obtained from the label momentum operator $\cP_{n_i}^\mu$, e.g. $\cP_{n_i}^\mu\, \xi_{n_i,\lp} = \lp^\mu\, \xi_{n_i,\lp}$. When acting on a product of fields, $\cP_{n_i}$ returns the sum of the label momenta of all $n_i$-collinear fields. For convenience, we define $\bnP_{n_i} = \bn\sdt\cP_{n_i}$, which picks out the large momentum component.  Frequently, we will only keep the label ${n_i}$ denoting the collinear direction, while the momentum labels are summed over (subject to momentum conservation) and suppressed.

Soft degrees of freedom are described in the effective theory by separate quark and gluon fields. We will assume that we are working in the SCET$_\text{I}$ theory where these soft degrees of freedom are referred to as ultrasoft so as to distinguish them from the soft modes of SCET$_\text{II}$ \cite{Bauer:2002aj}.  In SCET$_\text{I}$, the ultrasoft modes do not carry label momenta, but have residual momentum dependence with $i \partial^\mu \sim \la^2Q$. They are therefore described by fields $q_{us}(x)$ and $A_{us}(x)$ without label momenta. The ultrasoft degrees of freedom are able to exchange residual momenta between the jets in different collinear sectors. Particles that exchange large momentum of $\ord{Q}$ between different jets are off-shell by $\ord{n_i\cdot n_j Q^2}$, and are integrated out by matching QCD onto SCET.  Before and after the hard interaction the jets described by the different collinear sectors evolve independently from each other, with only ultrasoft radiation between the jets.

SCET is formulated as an expansion in powers of $\la$, constructed so that manifest power counting is maintained at all stages of a calculation. As a consequence of the multipole expansion, all fields acquire a definite power counting \cite{Bauer:2001ct}, shown in \Tab{tab:PC}. The SCET Lagrangian is also expanded as a power series in $\lambda$
\begin{align} \label{eq:SCETLagExpand}
\cL_{\text{SCET}}=\cL_\hard+\cL_\dyn= \sum_{i\geq0} \cL_\hard^{(i)}+\sum_{i\geq0} \cL^{(i)} \,,
\end{align}
where $(i)$ denotes objects at ${\cal O}(\lambda^i)$ in the power counting. The Lagrangians $ \cL_\hard^{(i)}$ contain the hard scattering operators $O^{(i)}$, whose structure is determined by the matching process, as described in \Sec{sec:matching}. The $\cL^{(i)}$ describe the dynamics of ultrasoft and collinear modes in the effective theory, and their structure will be discussed shortly.

Factorization theorems used in jet physics are typically derived at leading power in $\lambda$. In this case, interactions involving hard processes in QCD are matched to a basis of leading power SCET hard scattering operators $O^{(0)}$, the dynamics in the effective theory are described by the leading power Lagrangian, $\cL^{(0)}$, and the measurement function, which defines the action of the observable, is expanded to leading power. Higher power terms in the $\lambda$ expansion, known as power corrections, arise from three sources: subleading power hard scattering operators $O^{(i)}$, subleading Lagrangian insertions, and subleading terms in the expansion of the measurement functions which act on soft and collinear radiation. The first two sources are independent of the form of the particular measurement, while the third depends on its precise definition.

\begin{table}
\begin{center}
\begin{tabular}{| l | c | c |c |c|c| r| }
  \hline                       
  Operator & $\cB_{n_i\perp}^\mu$ & $\chi_{n_i}$& $\cP_\perp^\mu$&$q_{us}$&$D_{us}^\mu$ \\
  Power Counting & $\lambda$ &  $\lambda$& $\lambda$& $\lambda^3$& $\lambda^2$ \\
  \hline  
\end{tabular}
\end{center}
\caption{
Power counting for building block operators in $\text{SCET}_\text{I}$.
}
\label{tab:PC}
\end{table}

In SCET, collinear operators are constructed out of products of fields and Wilson lines that are invariant under collinear gauge transformations~\cite{Bauer:2000yr,Bauer:2001ct}.  The smallest building blocks are collinearly gauge-invariant quark and gluon fields, defined as
\begin{align} \label{eq:chiB}
\chi_{{n_i},\w}(x) &= \Bigl[\delta(\w - \bnP_{n_i})\, W_{n_i}^\dagger(x)\, \xi_{n_i}(x) \Bigr]
\,,\\
\cB_{{n_i}\perp,\w}^\mu(x)
&= \frac{1}{g}\Bigl[\delta(\w + \bnP_{n_i})\, W_{n_i}^\dagger(x)\,i  D_{{n_i}\perp}^\mu W_{n_i}(x)\Bigr]
 \,. \nn
\end{align}
With this definition of $\chi_{{n_i},\w}$, when expanded to a single quark, we have $\w > 0$ for an incoming quark and $\w < 0$ for an outgoing antiquark. For $\cB_{{n_i},\w\perp}$, $\w > 0$ ($\w < 0$) corresponds to outgoing (incoming) gluons. In \Eq{eq:chiB},
\begin{equation}
i  D_{{n_i}\perp}^\mu = \cP^\mu_{{n_i}\perp} + g A^\mu_{{n_i}\perp}\,,
\end{equation}
is the collinear covariant derivative and
\begin{equation} \label{eq:Wn}
W_{n_i}(x) = \biggl[~\sum_\text{perms} \exp\Bigl(-\frac{g}{\bnP_{n_i}}\,\bn\sdt A_{n_i}(x)\Bigr)~\biggr]\,,
\end{equation}
is a Wilson line of ${n_i}$-collinear gluons in label momentum space. In general the structure of Wilson lines must be derived by a matching calculation from QCD. These Wilson lines sum up arbitrary emissions of ${n_i}$-collinear gluons off of particles from other sectors, which due to the power expansion always appear in the ${\bar{n}_i}$ direction. The emissions summed in the Wilson lines are $\ord{\lambda^0}$ in the power counting. The label operators in \Eqs{eq:chiB}{eq:Wn} only act inside the square brackets. Since $W_{n_i}(x)$ is localized with respect to the residual position $x$, we can treat
$\chi_{{n_i},\w}(x)$ and $\cB_{{n_i},\w}^\mu(x)$ as local quark and gluon fields from the perspective of ultrasoft derivatives $\partial^\mu$ that act on $x$. 

The complete set of collinear and ultrasoft building blocks for constructing hard scattering operators or subleading Lagrangians at any order in the power counting is given in \Tab{tab:PC}. All other field and derivative combinations can be reduced to this set by the use of equations of motion and operator relations~\cite{Marcantonini:2008qn}. Since these building blocks carry vector or spinor Lorentz indices they must be contracted to form scalar operators, which also involves the use of objects like $\{n_i^\mu, \bn_i^\mu, \gamma^\mu, g^{\mu\nu}, \epsilon^{\mu\nu\sigma\tau}\}$. 

As shown in \Tab{tab:PC}, both the collinear quark and collinear gluon building block fields scale as ${\cal O}(\lambda)$. For the majority of jet processes there is a single collinear field operator for each collinear sector at leading power.  (For fully exclusive processes that directly produce hadrons there will be multiple building blocks from the same sector in the leading power operators since they form color singlets in each sector.) Also, since $\cP_\perp\sim \lambda$, this operator will not typically be present at leading power (exceptions could occur, for example, in processes picking out P-wave quantum numbers). At subleading power, operators for all processes can involve multiple collinear fields in the same collinear sector, as well as $\cP_\perp$ operator insertions. The power counting for an operator is obtained by adding up the powers for the building blocks it contains. To ensure consistency under renormalization group evolution the operator basis in SCET must be complete, namely all operators consistent with the symmetries of the problem must be included.

Dependence on the ultrasoft degrees of freedom enters the operators through the ultrasoft quark field $q_{us}$, and the ultrasoft covariant derivative $D_{us}$, defined as 
\begin{equation}
i  D_{us}^\mu = i  \partial^\mu + g A_{us}^\mu\,,
\end{equation}
from which we can construct other operators including the ultrasoft gluon field strength. All operators in the theory must be invariant under ultrasoft gauge transformations. Collinear fields transform under ultrasoft gauge transformations as background fields of the appropriate representation. The power counting for these operators is shown in \Tab{tab:PC}. Since they are suppressed relative to collinear fields, ultrasoft fields typically do not enter factorization theorems in jet physics at leading power. An example where ultrasoft fields enter at leading power is $B \to X_s \gamma$ in the photon endpoint region, which is described at leading power by a single collinear sector, and an ultrasoft quark field for the b quark. 

The leading power SCET Lagrangian, $\cL^{(0)}$, describing the interactions of soft and collinear particles in the effective theory can be written\footnote{Here we assume that other possible modes, for example Glauber modes, do not contribute to the observable, and cancel out of the cross section.} 
\begin{align} \label{eq:leadingLag_2}
\cL^{(0)} &= \cL^{(0)}_{n \xi} + \cL^{(0)}_{n g} +  \cL^{(0)}_{us}\,, 
\end{align}
where
\begin{align}
\cL^{(0)}_{n \xi} &= \bar{\xi}_n\big(i n \cdot D_{ns} + i \slashed{D}_{n \perp} W_n \frac{1}{\overline{\cP}_n} W_n^\dagger i \slashed{D}_{n \perp} \big)  \frac{\slashed{\bar{n}}}{2} \xi_n\,, \\
\cL^{(0)}_{n g} &= \frac{1}{2 g^2} \tr \big\{ ([i D^\mu_{ns}, i D^\nu_{ns}])^2\big\} + \zeta \tr \big\{ ([i \partial^\mu_{ns},A_{n \mu}])^2\big\}+2 \tr \big\{\bar{c}_n [i \partial_\mu^{ns}, [i D^\mu_{ns},c_n]]\big\} \,, \nn
\end{align}
and the ultrasoft Lagrangian, $\cL^{(0)}_{us}$, is simply the QCD Lagrangian. Here we have used a covariant gauge with gauge fixing parameter $\zeta$ for the collinear gluons.
The various derivatives in \Eq{eq:leadingLag_2} are defined as
\begin{align}
iD^\mu_n &= i\partial^\mu_n +g A^\mu_n\,, 
& iD^\mu_{ns} &=i D^\mu_n +\frac{\bn^\mu}{2}gn \cdot A_{us}\,,\qquad
\\ 
i\partial^\mu_{ns}&=i \partial^\mu_n +\frac{\bn^\mu}{2} gn\cdot A_{us}\,,
& i \partial^\mu_n &= \frac{\bn^\mu}{2} n \cdot \partial + \frac{n^\mu}{2} \overline{\cP} + \cP_\perp^\mu
\,,\nn
\end{align}
and include ultrasoft-collinear interactions.

A particularly convenient aspect of SCET is that the factorization of soft and collinear degrees of freedom can be performed at the Lagrangian level using the BPS field redefinition. This decouples leading power interactions between soft and collinear particles, and moves them into Wilson lines appearing in the hard scattering operators. The BPS field redefinitions are given by
\begin{align} \label{eq:BPSfieldredefinition_2}
\cB^{a\mu}_{n\perp}\to \cY_n^{ab} \cB^{b\mu}_{n\perp} , \qquad \chi_n^\alpha \to Y_n^{\alpha \beta} \chi_n^\beta\,.
\end{align}
Additionally, ghost particles $c_n$, which are present after gauge fixing, transform under BPS field redefinitions according to
\begin{align}
c_n^a \to {\cal Y}_n^{ab} c_n^b.
\end{align}

After performing the BPS field redefinition, we have
\begin{align}
\cL^{(0)\text{BPS}} &= \cL^{(0)\text{BPS}}_{n \xi} + \cL^{(0)\text{BPS}}_{n g} +  \cL^{(0)}_{us}\,, 
\end{align}
where the ultrasoft Lagrangian is unchanged. The collinear quark Lagrangian is given by
\begin{align}
\cL^{(0)\text{BPS}}_{n \xi   }&=\bar \chi_n  \left(  i n \cdot \cD_{n}  + i \slashed{\cD}_{n \perp}  \frac{1}{\overline{\cP}_n}  i \slashed{\cD}_{n \perp}  \right)  \frac{\Sl \bn}{2} \chi_n\,,
\end{align}
where
\begin{align}
i \cD^\mu_n = W_n^\dagger  i D^\mu_{n} W_n\,,
\end{align}
and the collinear gluon Lagrangian is given by
\begin{align}
\cL^{(0)\text{BPS}}_{n g} &=  \frac{1}{2 g^2} \tr \big\{ ([i D^\mu_{n}, i D^\nu_{n}])^2\big\} + \zeta \tr \big\{ ([i \partial^\mu_{n},A_{n \mu}])^2\big\}+2 \tr \big\{\bar{c}_n [i \partial_\mu^{n}, [i D^\mu_{n},c_n]]\big\} \,,
\end{align}
explicitly showing that ultrasoft and collinear particles have been decoupled to leading power. After BPS field redefinition, since the leading power Lagrangian, the cross section can be written as a product of matrix elements, each involving only soft or collinear fields, as in \Eq{eq:sigma_intro}.

It is important to note that the BPS field redefinition does not, however, decouple ultrasoft and collinear interactions in the subleading Lagrangians. However, these contributions are power suppressed, and therefore one only needs to consider a fixed number of insertions of subleading Lagrangians, so that factorization is still manifest.

As presented here, SCET is appropriate for treating problems with a collinear scale, a soft scale, and a hard scale. In the context of jet substructure, where additional observables are measured on the jet, other scales may be present, and additional modes must be added to the effective field theory to treat such cases. This situation will be discussed in detail in \Chap{chap:D2_anal}. In such many scale problems, the power of the effective theory, which enables one to separate a problem into separate calculations at each scale, becomes even greater. The extension of SCET to treat various kinematic hierarchies is currently an active area of research.

%% file: chap1a.tex

%
%
\newcommand{\ecf}[2]{e_{#1}^{(#2)}} 
\newcommand{\ecfnobeta}[1]{e_{#1}}

\def\ea{e_\alpha}
\def\eb{e_\beta}
\def\eaba{e_\alpha^{\beta/\alpha}}
\def\ebab{e_\beta^{\alpha/\beta}}
\def\log{\text{log}}
\def\muss{\mu_{S\rightarrow S}}
\def\musj{\mu_{S\rightarrow J}}

\def \thetac {\theta_c}
\def \thetacs {\theta_{cs}}
\def \thetasj {\theta_{sj}}
\def \zs {z_{s}}
\def \zcs {z_{cs}}
\def \zsj {z_{sj}}

\def\tauo{{\tau_1}}
\def\taut{{\tau_2}}

\def\be{\begin{equation}}
\def\ee{\end{equation}}

\newcommand{\ptveto}{p_{T}^{veto}}

\newcommand{\Dobs}[2]{D_{#1}^{(#2)}} 
\newcommand{\Dobsnobeta}[1]{D_{#1}} 
\newcommand{\Xobs}[2]{X_{#1}^{(#2)}} 
\newcommand{\Cobs}[2]{C_{#1}^{(#2)}} 
\newcommand{\Cobsnobeta}[1]{C_{#1}} 

\def\nslash{n\hspace{-2mm}\slash}
\def\nbarslash{\bar n\hspace{-2mm}\slash}
\def\nslashinline{n\!\!\!\slash}
\def\nbarslashinline{\bar n\!\!\!\slash}
\def\nbar{\bar n}

\newcommand{\pythia}[1]{\textsc{Pythia\xspace #1}}
\newcommand{\madgraph}[1]{\textsc{MadGraph5\xspace #1}}
\newcommand{\fastjet}[1]{\textsc{FastJet\xspace #1}}
\newcommand{\herwig}[1]{\textsc{Herwig\xspace #1}}
\newcommand{\herwigpp}[1]{\textsc{Herwig++\xspace #1}}
\newcommand{\vincia}[1]{\textsc{Vincia\xspace #1}}

\DeclareRobustCommand{\Sec}[1]{Sec.~\ref{#1}}
\DeclareRobustCommand{\Secs}[2]{Secs.~\ref{#1} and \ref{#2}}
\DeclareRobustCommand{\App}[1]{App.~\ref{#1}}
\DeclareRobustCommand{\Tab}[1]{Table~\ref{#1}}
\DeclareRobustCommand{\Tabs}[2]{Tables~\ref{#1} and \ref{#2}}
\DeclareRobustCommand{\Fig}[1]{Fig.~\ref{#1}}
\DeclareRobustCommand{\Figs}[2]{Figs.~\ref{#1} and \ref{#2}}
\DeclareRobustCommand{\Eq}[1]{Eq.~(\ref{#1})}
\DeclareRobustCommand{\Eqs}[2]{Eqs.~(\ref{#1}) and (\ref{#2})}
\DeclareRobustCommand{\Ref}[1]{Ref.~\cite{#1}}
\DeclareRobustCommand{\Refs}[1]{Refs.~\cite{#1}}

\newcommand{\Nsub}[2]{\tau_{#1}^{(#2)}}
\newcommand{\Nsubnobeta}[1]{\tau_{#1}}

%

\chapter{Overview}\label{chap:overview}

In this section I provide a brief outline of the material presented in this thesis, and discuss the significance of the work in the context of other work performed during my thesis. Several key results and figures from different sections throughout this thesis are reproduced here. This section also contains comments on material related to the contents of the papers that appeared after the publication of the papers, for example measurements at the LHC, or related work by other groups.

The main focus of the work presented in this thesis is divided into two separate categories: the development of effective field theories for jet substructure, and the use of effective field theories for the calculation of precision cross sections involving jets at the LHC. Each of these will be discussed in turn.

\section{Effective Field Theories for Jet Substructure }

As discussed in \Chap{chap:intro}, one of the major advances in the study of jets at the LHC is the ability to study in detail the substructure of a jet. This necessitates new calculations in QCD. In this section I discuss my work in the area of jet substructure and put into context the work presented in \Chap{chap:PC} and \Chap{chap:D2_anal}.

\begin{figure}
\begin{center}
\subfloat[]{\label{fig:D2_summary_a}
\includegraphics[width=6.5cm]{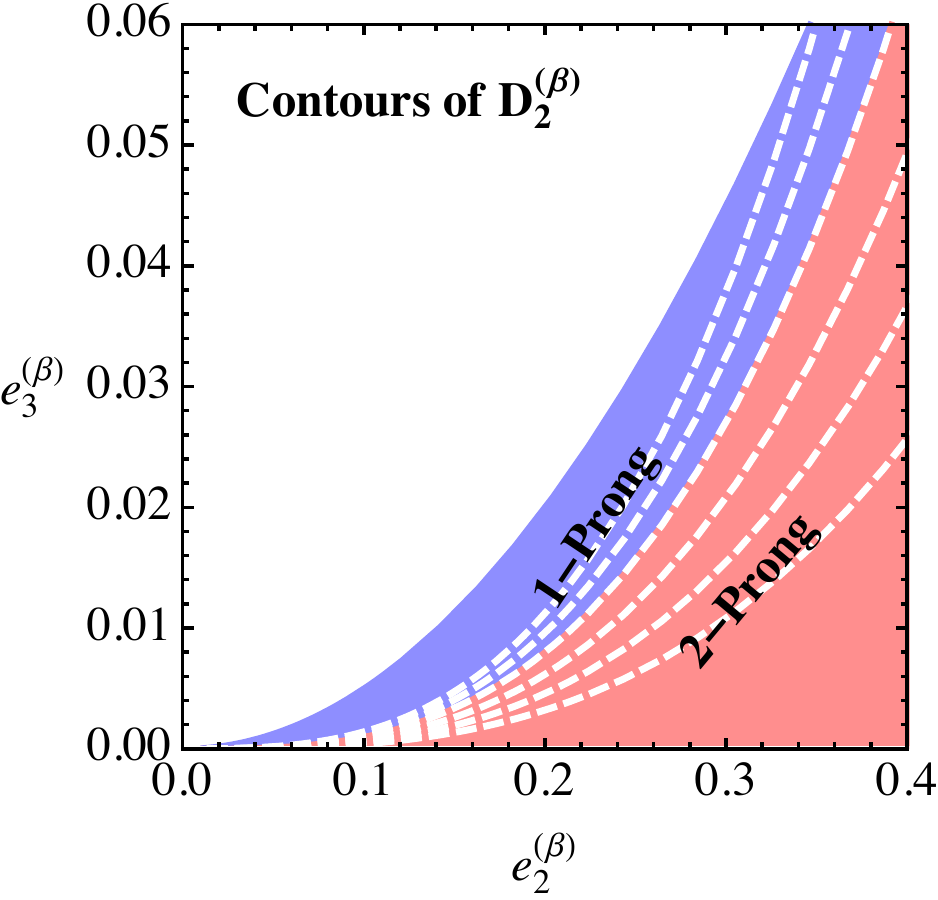}
}\qquad
\subfloat[]{\label{fig:D2_summary_b}
\includegraphics[width=7.25cm]{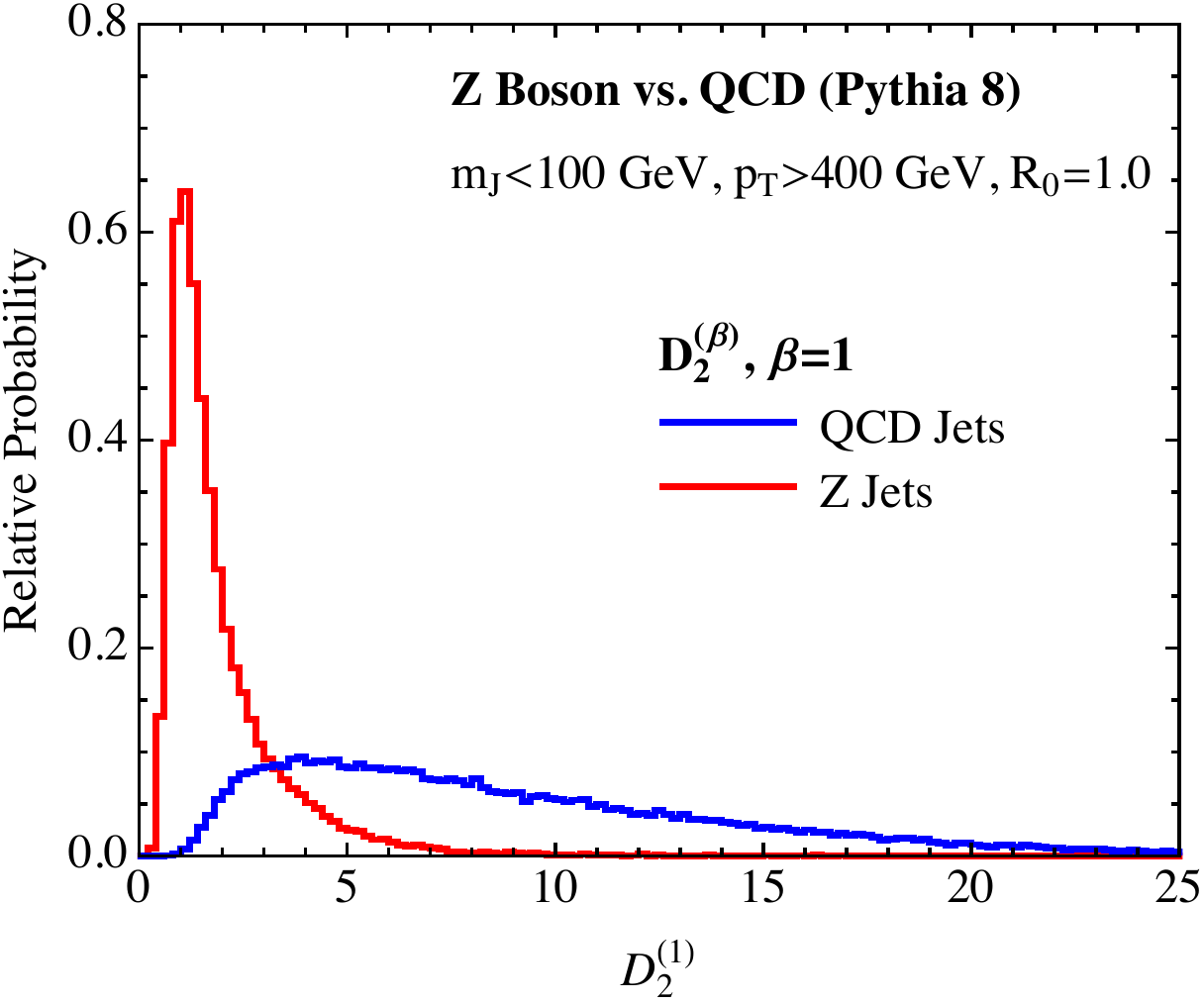}
}
\end{center}
\caption{ 
The phase space for $\ecf{2}{\beta}$ and $\ecf{3}{\beta}$ along with contours of constant $\Dobs{2}{\beta}$ separating the one- and two-prong regions of phase space (left). Monte Carlo distributions of the $D_2$ observable with $\beta=1$ (right) (figures from \cite{Larkoski:2014gra}).
}
\label{fig:D2_summary}
\end{figure}

Jet substructure observables play an important role at the LHC, as was reviewed in \Chap{chap:intro}. These observables are very complicated to calculate from first principles, and therefore are most often studied using Monte Carlo programs. When designing jet substructure observables, a guess and check attitude is most often taken. It is therefore an interesting question to understand the extent that basic scaling arguments based on the fact that QCD is nearly a scale invariant theory can be used to understand the parametric behavior of jet substructure observables, and to design new jet substructure observables. After which, dedicated calculations, which are quite time intensive, can be performed. In "Power Counting to Better Jet Observables" \cite{Larkoski:2014gra}, discussed in \Chap{chap:PC}, we introduced the power counting approach for studying jet substructure observables formed from combinations of Infrared and Collinear Safe (IRC) safe observables. This approach allows one to make parametric predictions about the behavior of jet substructure observables, which are robust to tunings of Monte Carlo generators, or models of hadronization. We then used this approach to derive the two-prong jet substructure discriminant $D_2$, a ratio observable formed from the energy correlation functions $\ecf{2}{\beta}$ and $\ecf{3}{\beta}$ \cite{Larkoski:2013eya}. The observable $D_2$ is designed to discriminate boosted two-prong substructure from background QCD jets, by identifying the scaling of contours separating the $\ecf{2}{\beta}$, $\ecf{3}{\beta}$ phase space, as is shown in \Fig{fig:D2_summary_a}. Its properties were studied using power counting arguments and then compared with Monte Carlo simulations. An example Monte Carlo plot showing the $D_2$ observable as measured on QCD jets, and boosted $Z$ jets, is shown in \Fig{fig:D2_summary_b}.

\begin{figure}
\begin{center}
\subfloat[]{\label{fig:D2_ATLAS_1}
\includegraphics[width=7cm]{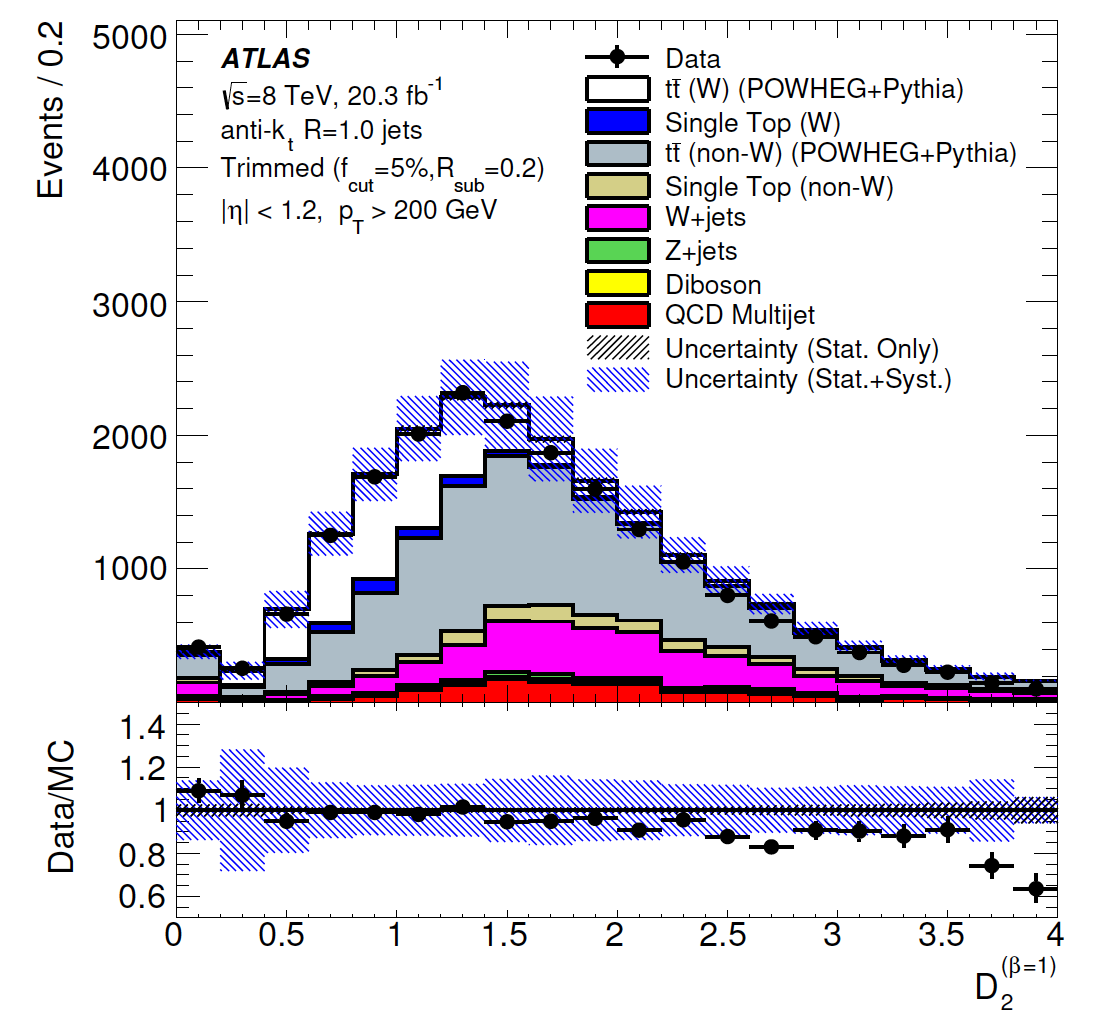}
}
$\qquad$
\subfloat[]{\label{fig:D2_ATLAS_2} 
\includegraphics[width=7.5cm]{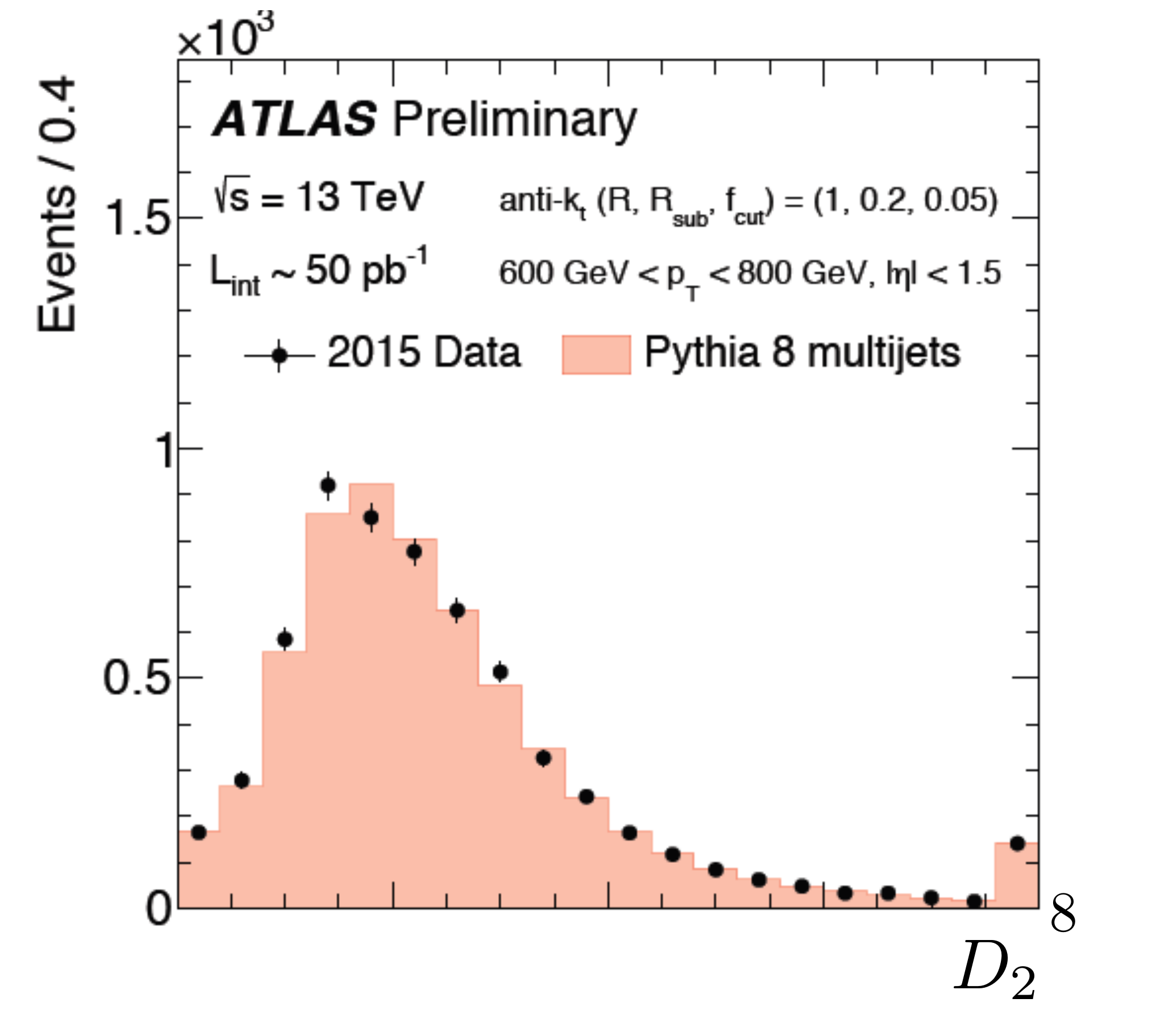}
}
\end{center}
\caption{ Measurements of the $D_2$ observable with angular exponent $\beta=1$ by the ATLAS collaboration at the LHC at $8$ TeV (left) (figure from \cite{Aad:2015rpa}) and $13$ TeV (right) (figure from  \cite{atlas_recent:2015}).
}
\label{fig:D2_ATLAS}
\end{figure}

The $D_2$ observable has been adopted by the ATLAS collaboration for tagging boosted two-prong substructure. Due to the importance of boosted two-prong tagging, for example for probing multi-TeV resonances, the $D_2$ observable was one of the first measurements at the $13$ TeV LHC. In \Fig{fig:D2_ATLAS}, measurements of the $D_2$ observable are shown at both $8$ and $13$ TeV as taken from \cite{Aad:2015rpa, atlas_recent:2015}. As mentioned in \Chap{chap:intro}, quite interestingly, the $D_2$ distribution has also been measured on double b-tagged $g\to bb$ jets \cite{collaboration:2016aa}. This demonstrates some of the many advances made in the area of jet substructure made at the LHC in the past several years, including the ability to $b$-tag subjets, which can play an important role, for example, in boosted top tagging.

\begin{figure*}[t]
\centering
\subfloat[]{
\includegraphics[width=7cm]{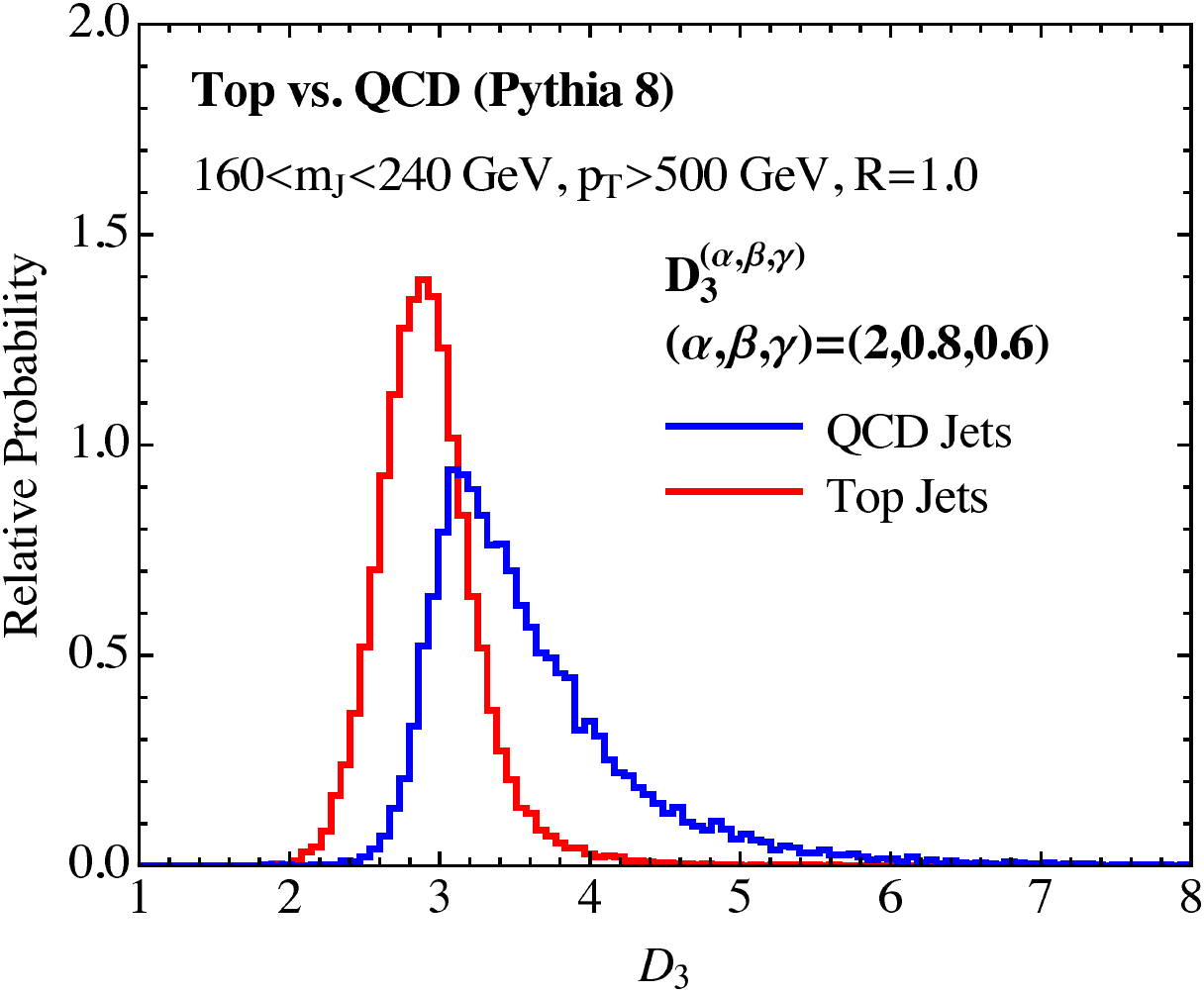}
}
\subfloat[]{
\includegraphics[width=7.2cm]{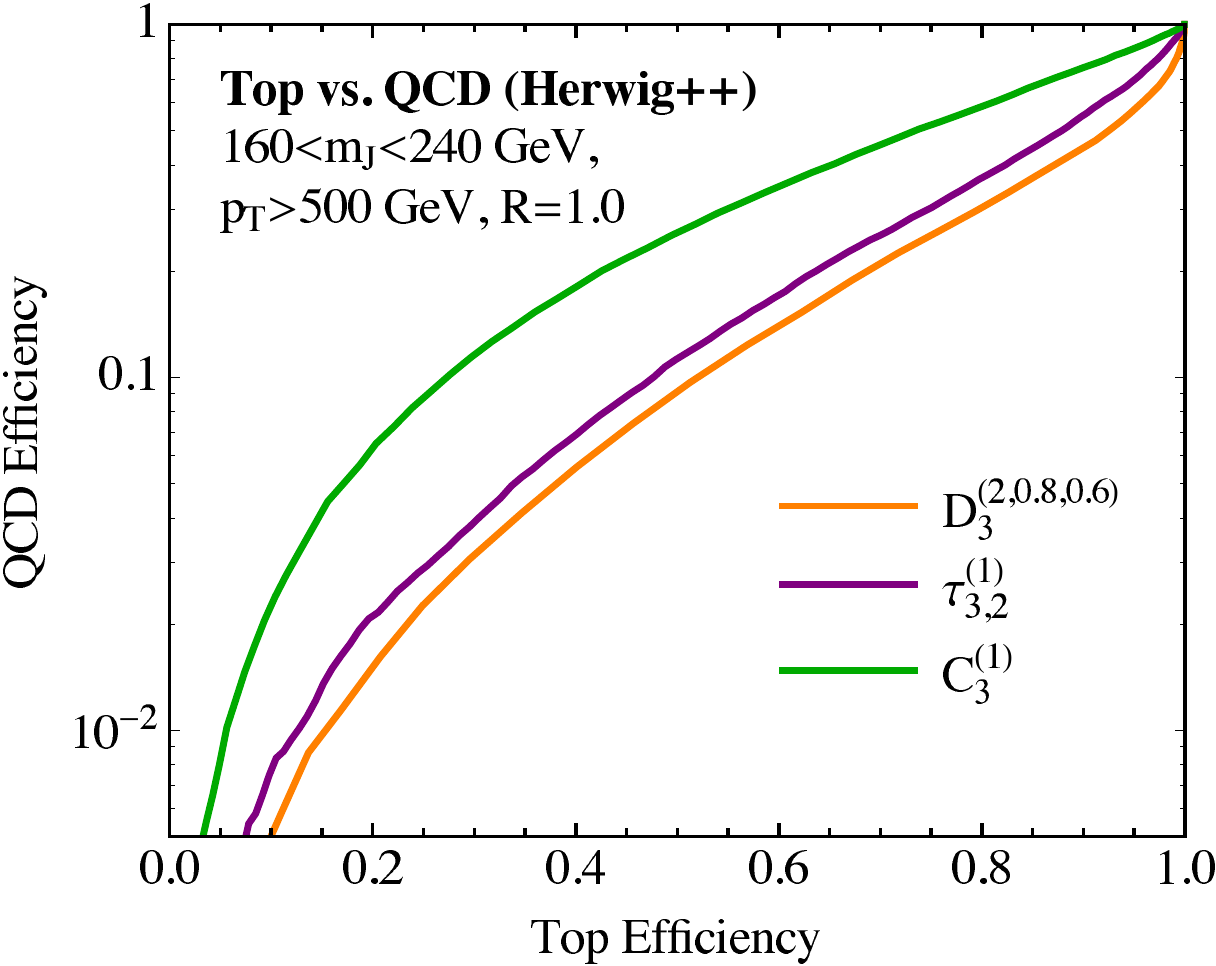}
}
\caption{A distribution of $D_3^{(2,0.8,0.6)}$ as measured on top quark and QCD jets in Monte Carlo (left). Signal vs.~Background efficiency curves comparing $C_3^{(1)}$, $D_3^{(2,0.8,0.6)}$, and $\tau_{3,2}^{(1)}$ (right) (figures from \cite{Larkoski:2014zma}).}
\label{fig:D3_summary}
\end{figure*}

Top quarks play an important role in the Standard Model as the heaviest quark, with a mass of $m_t \sim 172$ GeV. Because of this large mass they play an important role in many models of BSM physics, where they are expected to couple strongly to new physics responsible for mass generation.  Such models of new physics often predict resonances in the TeV scale which could decay to highly boosted top quarks. Furthermore, at LHC energies, highly boosted top quarks are also produced copiously by Standard Model processes, and provide a probe of the dynamics of the Standard Model in the TeV regime. The ability to study boosted top quarks at the LHC relies on the ability to distinguish them from the QCD background, similar to the case of boosted $W/Z/H$ jets. A number of jet substructure variables have therefore been proposed for this purpose. In "Building a Better Boosted Top Tagger" \cite{Larkoski:2014zma}, we extended the power counting arguments to identify the optimal boosted top tagger formed from the energy correlation functions. A plot of the proposed observable $D_3$ is shown in \Fig{fig:D3_summary}, and is compared with the standard N-Subjettiness observable $\tau_{3,2}$ \cite{Thaler:2011gf}, showing slightly improved performance. The $D_3$ observable is however, quite complicated, a reflection of the structure of the energy correlation function observables. It is currently being tested by the ATLAS collaboration. It would be interesting to identify a simpler boosted top quark observable formed from modified energy correlation functions.

The difficulty in performing analytic calculations of jet substructure observables, in particular those which are able to resolve an $n$-prong substructure within a jet, is that the calculation must be able to describe the jet not only when it has $n$ resolved subjets, but also when is has no resolved subjets, as well as for all intermediate options. Observables in QCD typically have singular behavior at edges of phase space regions, although it is also possible that observables exhibit singular behavior in the physical region. The location and structure of singular points for QCD observables plays an important role in their calculation, as naive perturbation theory breaks down in singular regions of phase space, and resummation is required. Observables typically also recieve large non-perturbative corrections in non-singular regions of phase space, which are difficult to calculate from first principles. It is therefore important to study the singular behavior of jet substructure observables.

\begin{figure*}[t]
\centering
\subfloat[]{\label{fig:Nsub_sum_a}
\includegraphics[width=8.5cm]{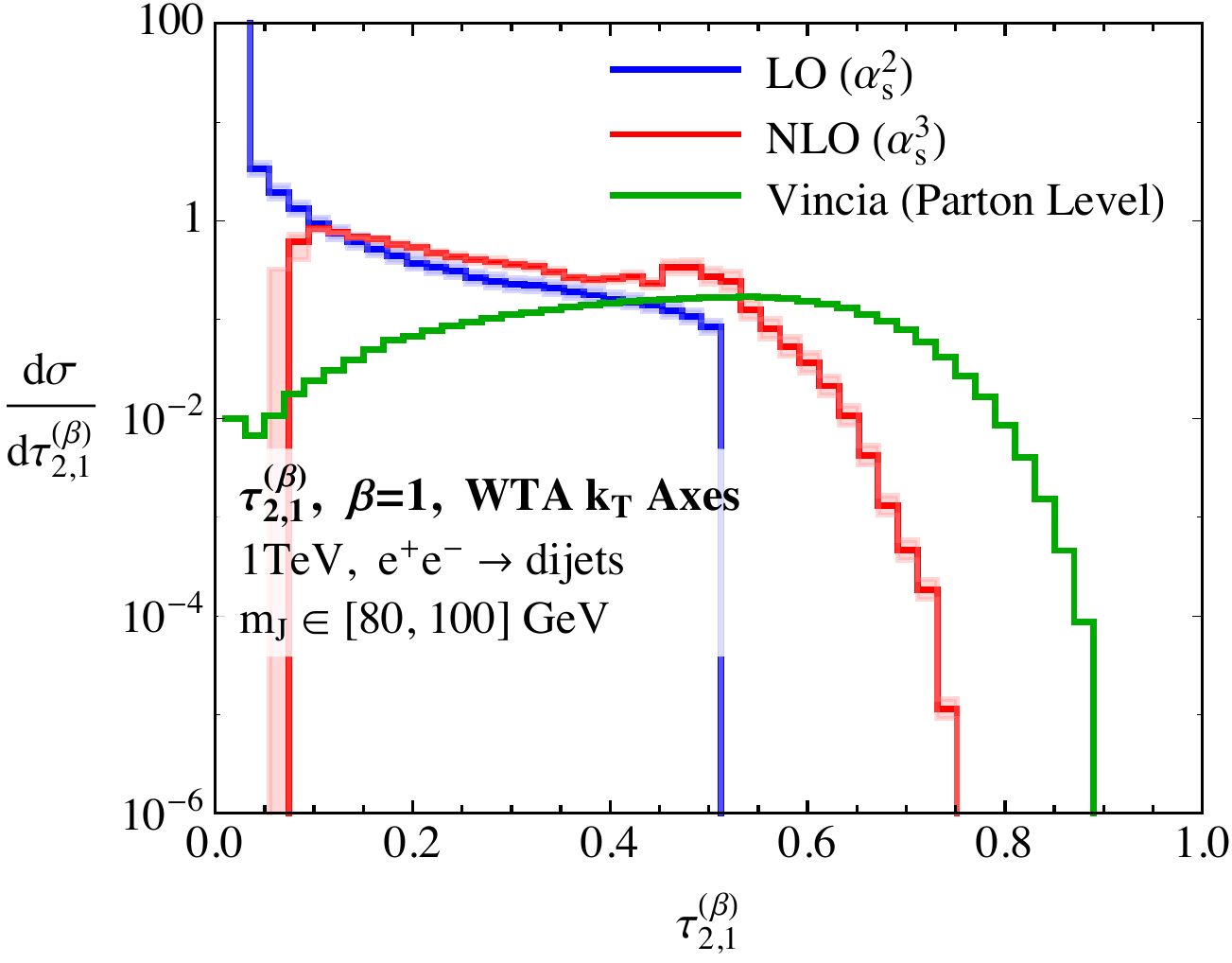}    
} 
\subfloat[]{\label{fig:Nsub_sum_b}
\includegraphics[width=8.4cm]{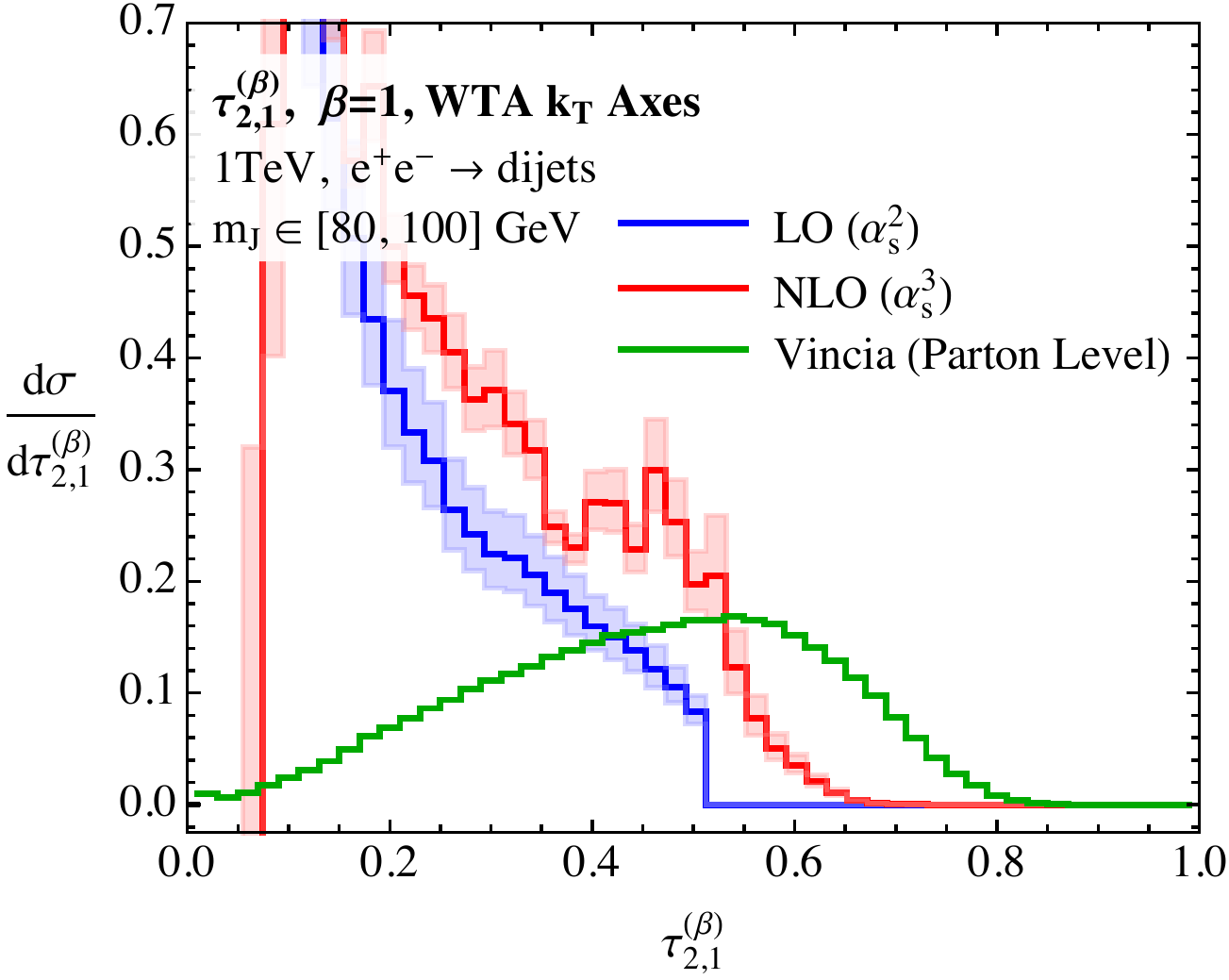} 
}
\caption{Distributions for $\Nsubnobeta{2,1}$ as defined using WTA $k_T$ axes with $\beta=1$ in both fixed order perturbation theory and parton level Monte Carlo. The distributions are shown on a logarithmic scale in (a), and on a linear scale in (b). The behavior observed in Monte Carlo as $\Nsubnobeta{2,1}\to1$ is not well reproduced in fixed order perturbation theory (figures from \cite{Larkoski:2015uaa} ).
}
\label{fig:Nsub_sum}
\end{figure*}

 In "The Singular Behavior of Jet Substructure Observables" \cite{Larkoski:2015uaa}, the singular behavior of two jet substructure observables $D_2$, and $N$-subjettiness \cite{Thaler:2010tr,Thaler:2011gf} were studied. The precise definition of these observables will be given in \Chap{chap:PC}, but importantly the definition of the $N$-subjettiness observable requires a definition of the $N$-axes used to define the observable. A variety of different definitions have been proposed. One of the most powerful is the winner take all axes definition \cite{Larkoski:2014uqa}. We studied the $N$-subjettiness ratio observable for a variety of different axes definitions.  This was performed using power counting arguments, as well as fixed order and parton shower Monte Carlo. In particular, it was shown that the singular behavior of the observable depends strongly on the definition of the $N$-subjettiness axes. In particular, depending on the definition of the axes, smooth distributions, singularities, or discontinuities can be observed in the distributions. Of particular phenomenological relevance, it was shown that with the winner take all axes definition, the $N$-jettiness ratio observable exhibits an infinite number of singularities in the physical region. This occurs because with a fixed number of partons, the winner take definition of $N$-subjettiness exhibits a maximal value, which can be surpassed by the addition of an extra parton. The first such shoulder is clearly seen in \Fig{fig:Nsub_sum}. This implies that the calculation in fixed order perturbation theory is extremely unstable. Furthermore, each of these singularities requires special treatment. It also implies the presence of large non-perturbative corrections throughout the entire distribution. This is in stark contrast to the behavior of the $D_2$ observable, which has a simple singular structure. In the presence of a mass cut, the $D_2$ observable isolates the singular region to $D_2 \to 0$, so that the effect of non-perturbative physics is isolated to this region of phase space, and is therefore controlled. 

Having understood, using power counting techniques, the properties of a variety of popular jet substructure observables, as well as introducing observables with appealing theoretical properties, the next goal is the analytic calculation of jet substructure observables from first principles QCD. To achieve this goal, a number of new techniques needed to be introduced, namely the study of multi differential measurements, and the introduction of effective field theories describing jet substructure.

Many jet substructure observables, particularly those designed to tag jets with hierarchical scales, for example boosted $W,Z,H$ jets, rely on the measurement of multiple observables on the same jet. Some common examples are the $N$-subjettiness ratio observables $\Nsub{2,1}{\beta}$ or the $\Dobs{2}{\beta}$. Theoretical calculations of these observables require calculations of multi-differential cross sections, that is cross sections differential in multiple jet measurements, for example, the angularities with two different angular exponents.

\begin{figure}[]
\begin{center}
\subfloat[]{\label{fig:anal_2015}
\includegraphics[width=7.0cm]{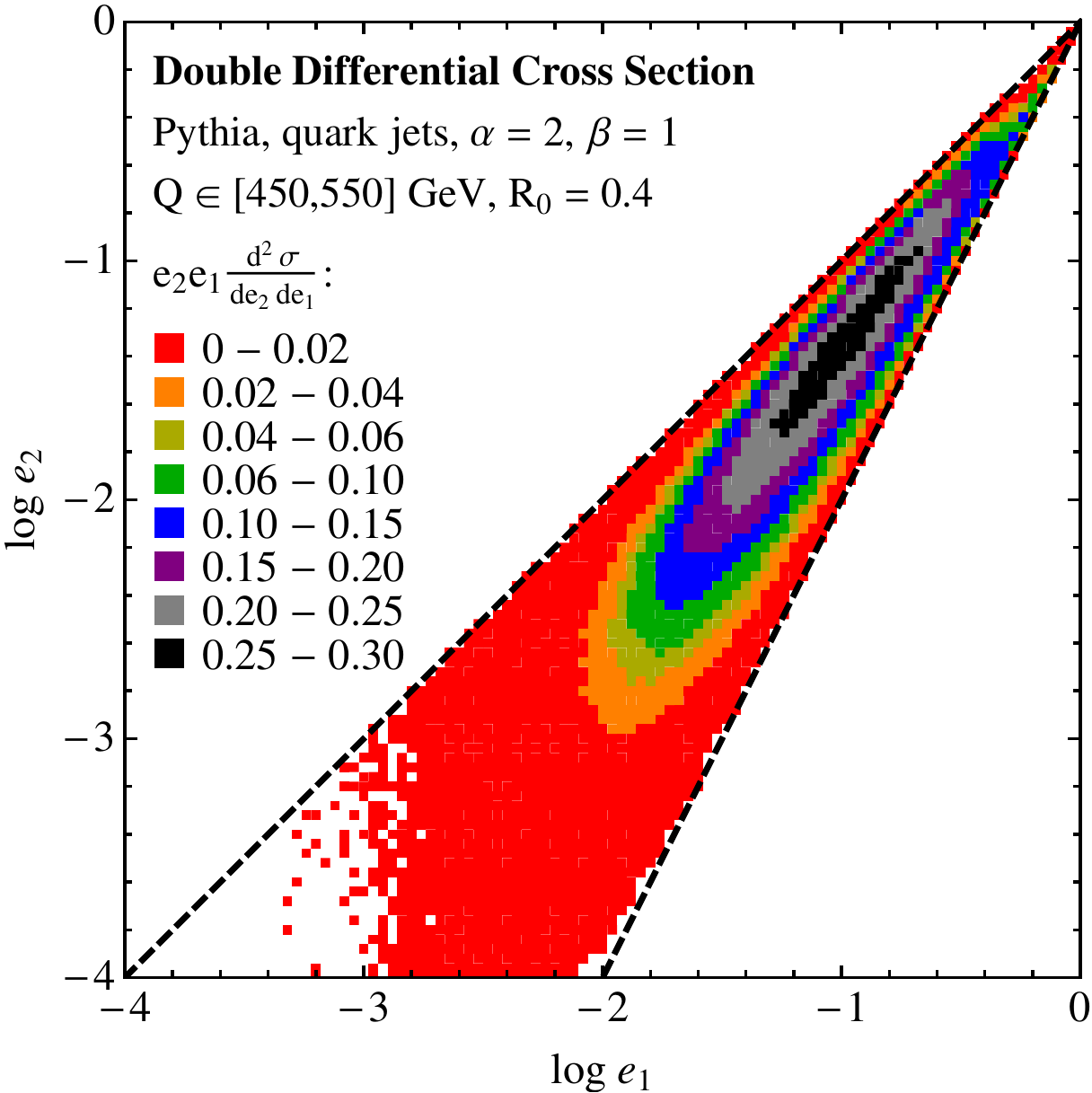}
}
$\quad$
\subfloat[]{\label{fig:anal_2010} 
\includegraphics[width=7.0cm]{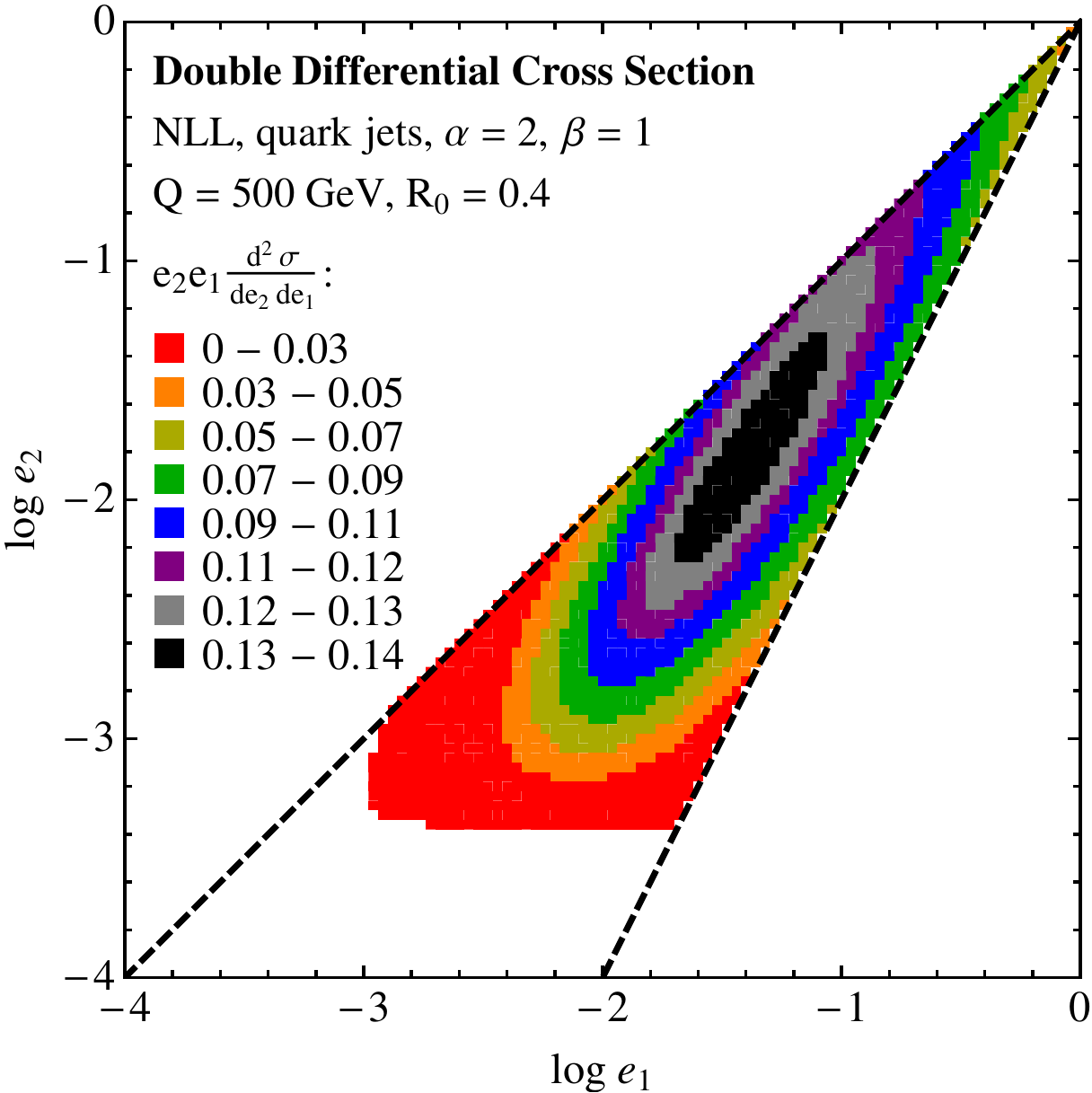}
}
\end{center}
\caption{
Plots of the double differential cross section defined from an analytic NLL interpolation measured on quark jets with one angularity fixed to be thrust ($\alpha=2$) and scanning over the other angularity: $\beta = 1.5,1,0.5,0.2$.  The energy of the jets is $Q=500$ GeV and the jet radius is $R_0=0.4$.  The dashed lines on the plot correspond to the expected phase space boundary (figures from \cite{Larkoski:2014tva}).
}
\label{fig:anal_BAng2_DD_plot}
\end{figure}

In the paper "Toward Multi-Differential Cross Sections: Measuring Two Angularities on a Single Jet" \cite{Larkoski:2014tva}, we initiated the study of the formal factorization and resummation properties of multi-differential cross-sections in QCD, by calculating the double differential cross section of two angularities $\ecf{2}{\alpha}$ and $\ecf{2}{\beta}$ measured on a single jet. A comparison of our calculation with Monte Carlo predictions is shown in \Fig{fig:anal_BAng2_DD_plot}. While these measurements do not resolve a two-prong substructure within the jet, as is required for the calculation of an observable such as $D_2$, they provide a simple example to understand how multi-differential cross sections can be studied in SCET. For example, in this paper, we noted that a new mode would need to be introduced into the effective theory, which is simultaneously soft and collinear. There has since been a number of works using soft-collinear modes in SCET, and studying multi-differential cross sections \cite{Procura:2014cba,Chien:2015cka,Pietrulewicz:2016nwo}. This work has also been used to study quark vs gluon tagging at the LHC \cite{Larkoski:2014pca}.

\begin{figure}
\begin{center}
\subfloat[]{\label{fig:twosoft_intro}
\includegraphics[scale = .2225]{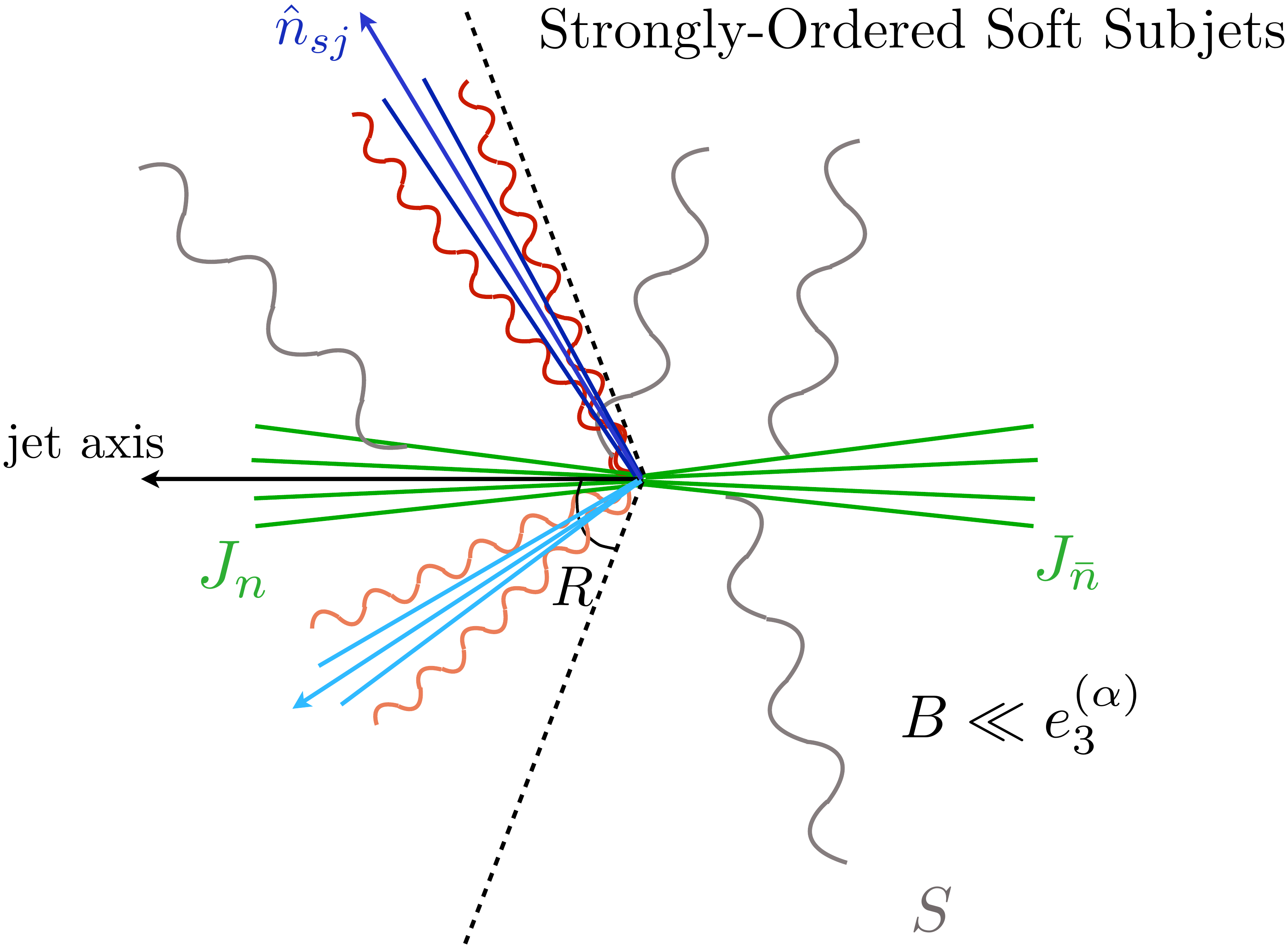}
}
$\qquad$
\subfloat[]{\label{fig:twosoft_scales_intro}
\includegraphics[scale = 0.225]{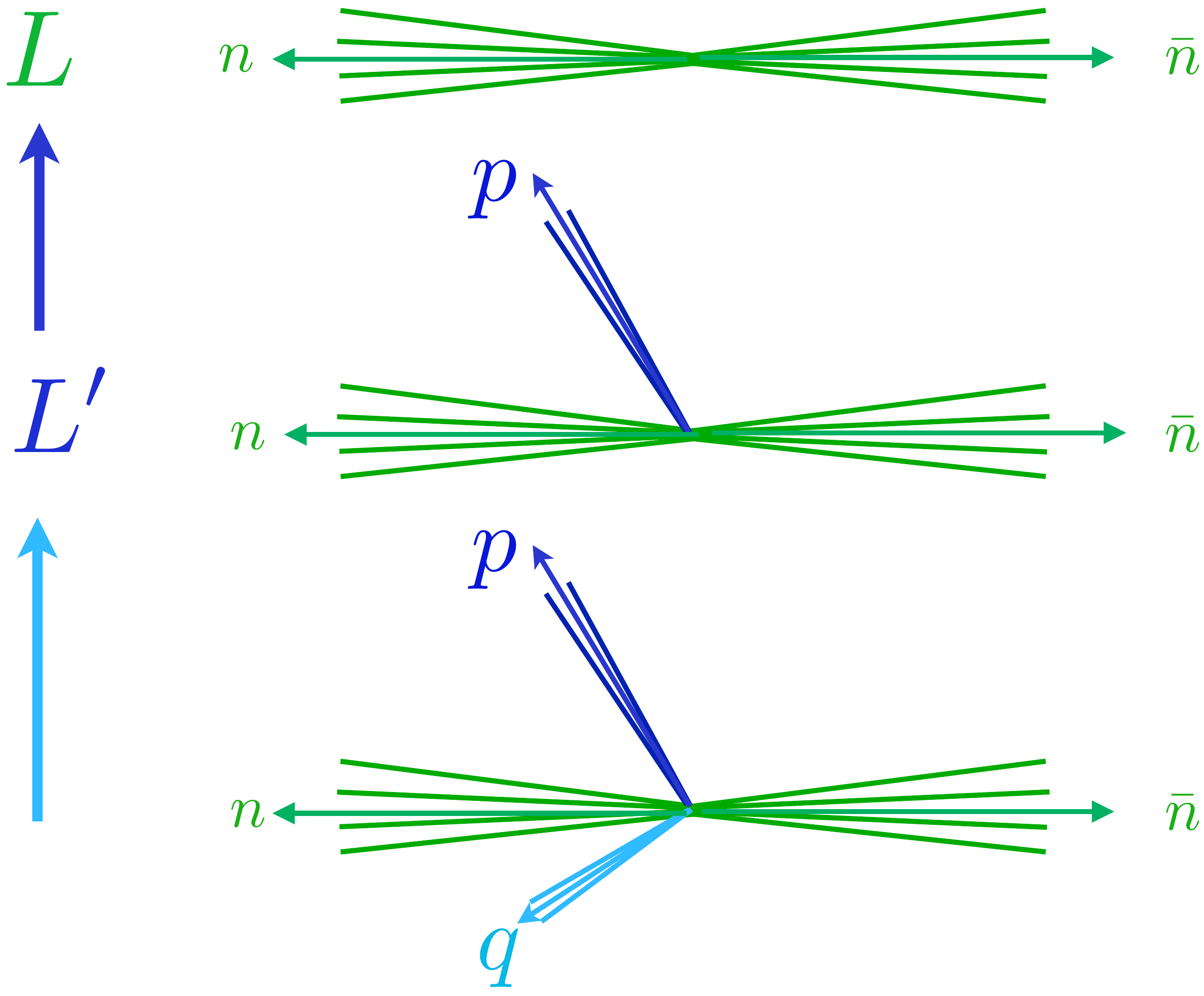}
}
\end{center}
\caption{ (a) Schematic depiction of the region of phase space defined by two strongly-ordered soft subjets. (b) Illustration of the resolved subjets as a function of the resolution scale, as implemented by the matching procedure in this region of phase space
(figures from \cite{Larkoski:2015zka}).}
\label{fig:doublesoft_intro}
\end{figure}


In the paper, "Analytic Boosted Boson Discrimination" \cite{Larkoski:2015kga}, discussed in \Chap{chap:D2_anal} we performed the first rigorous factorization based analysis of a two prong jet substructure observable, and performed an explicit calculation for the case of the $D_2$ observable in $e^+e^-\to$ dijets, for both signal (boosted $Z$) and background (QCD) jets. This included a thorough analysis of the effective field theory descriptions of all phase space regions in the multi-differential phase space, a description of the non-perturbative effects on the $D_2$ distribution and an extensive comparison with a variety of Monte Carlo generators.

The structure of the $D_2$ observable derived from a power counting analysis played a central role in the calculability of the observable. Indeed, the form of the observable was chosen so that all phase space regions are parametrically separated, and so that contours of constant value of the $D_2$ observable are entirely contained in a single region of phase space. This allows the value of the differential cross section at each value of $D_2$ to be computed within a single effective field theory, greatly simplifying the calculation.

\begin{figure}
\begin{center}
\subfloat[]{\label{fig:D2_anal_summary_a}
\includegraphics[width = 7.25cm]{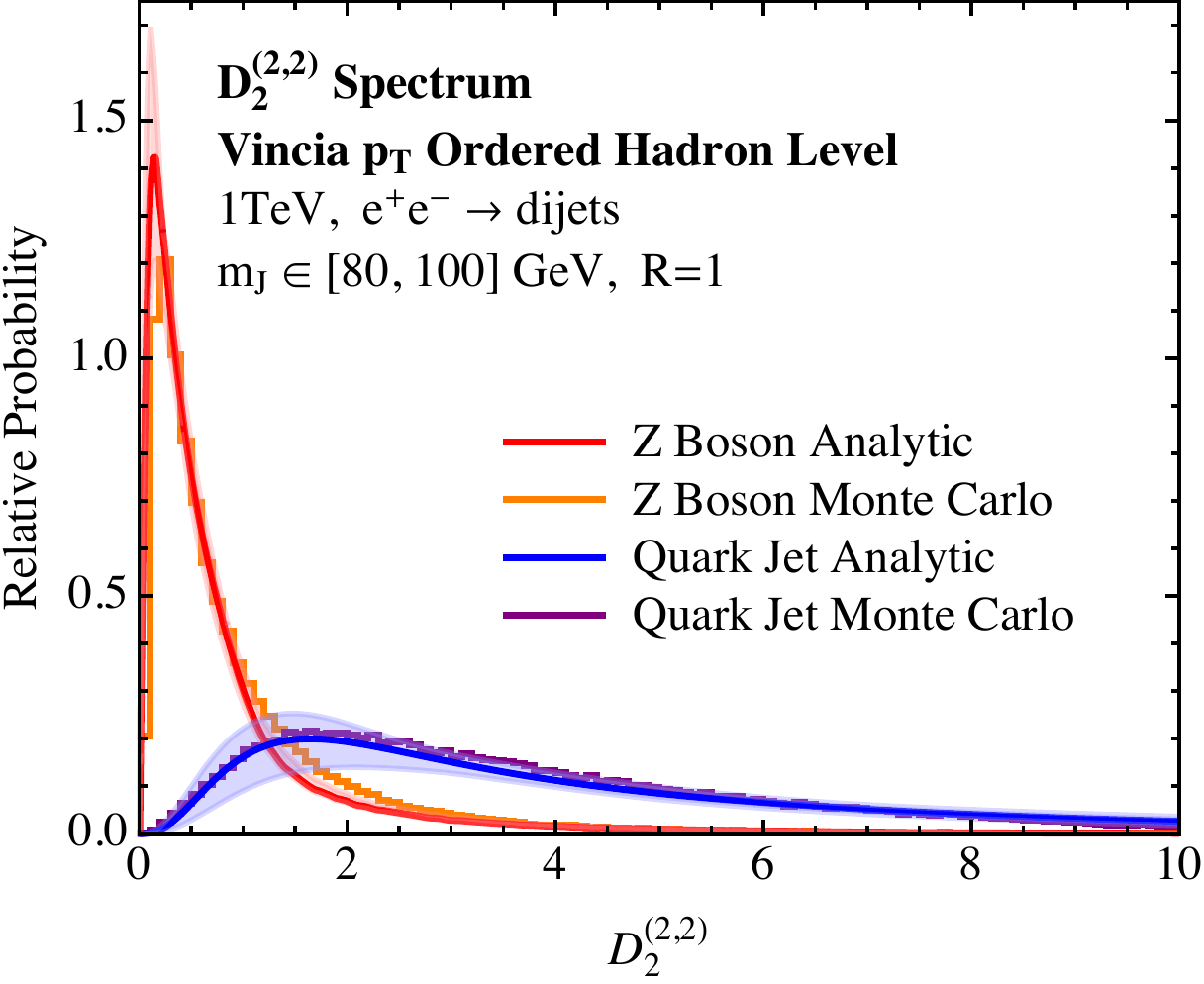}
}
\subfloat[]{\label{fig:D2_anal_summary_b}
\includegraphics[width = 7.25cm]{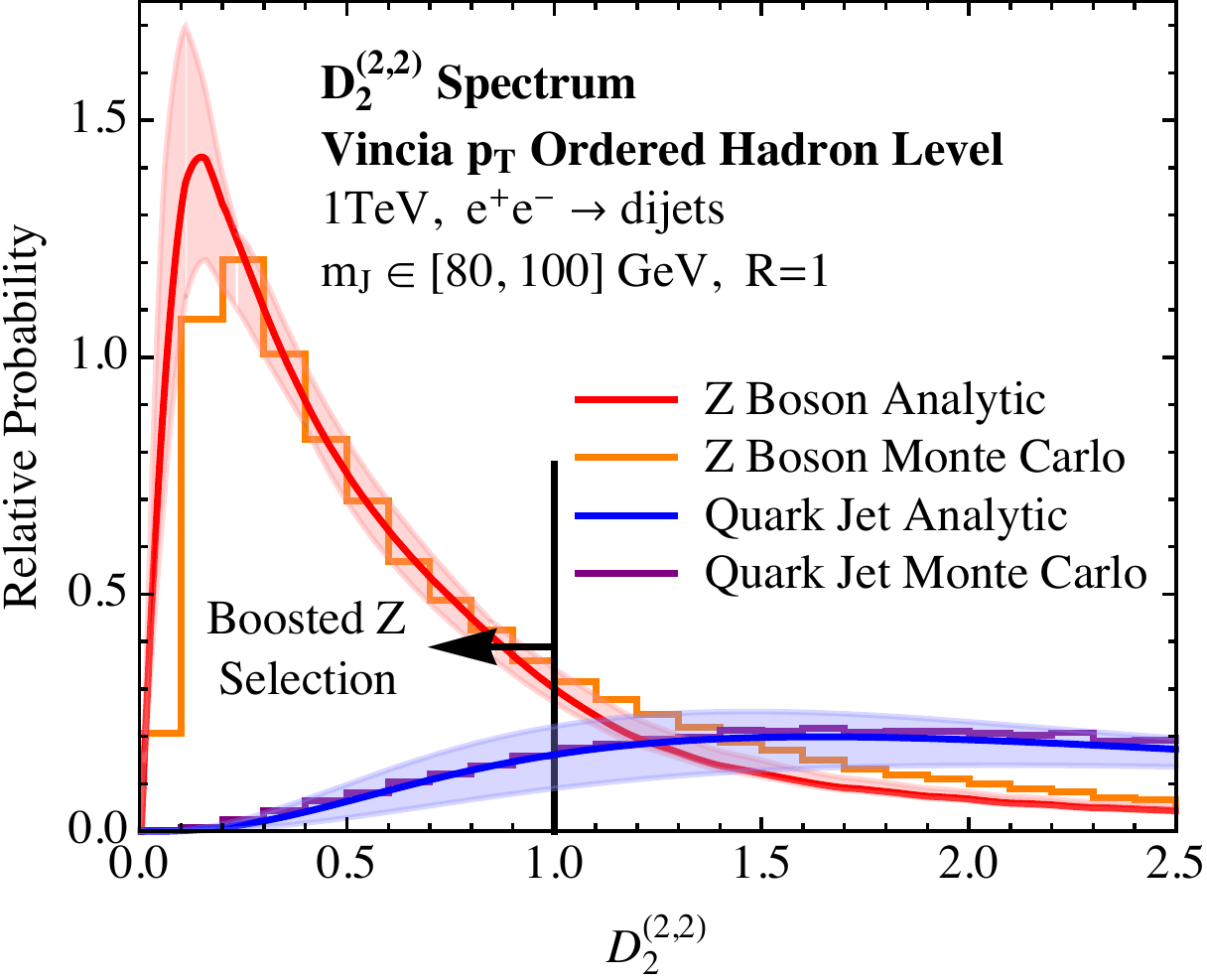}
}
\end{center}
\vspace{-0.2cm}
\caption{
A comparison of signal and background $\Dobs{2}{2,2}$ distributions for the four different Monte Carlo generators and our analytic calculation, including hadronization. Here we show the complete distributions, including the long tail for the background distribution. Although we extend the factorization theorem beyond its naive region of applicability into the tail, excellent agreement with Monte Carlo is found (figures from \cite{Larkoski:2015kga}).
}
\label{fig:D2_anal_summary}
\end{figure}

Our analysis indicated some discrepancies between the description of wide angle soft radiation in different Monte Carlo generators. The effective field theory description of this region of phase space was introduced in our paper "Non-Global Logarithms, Factorization, and the Soft Substructure of Jets" \cite{Larkoski:2015zka}.  It allows for a description of the region of phase space where the jet consists of subjets with small energy, and wide angle with respect to the hard core of the jet, and allows a systematic separation of the different scales involved in this region of phase space. This is shown schematically in \Fig{fig:doublesoft_intro}. This region of phase space also plays an important role in understanding Non-Global Logarithms \cite{Dasgupta:2001sh}. We also proposed an event shape which specifically probes these non-global correlations, and performed an analytic calculation in $e^+e^-$ \cite{Larkoski:2015npa}.

\begin{figure}
\begin{center}
\subfloat[]{\label{fig:D2sumROC_a}
\includegraphics[width= 7.2cm]{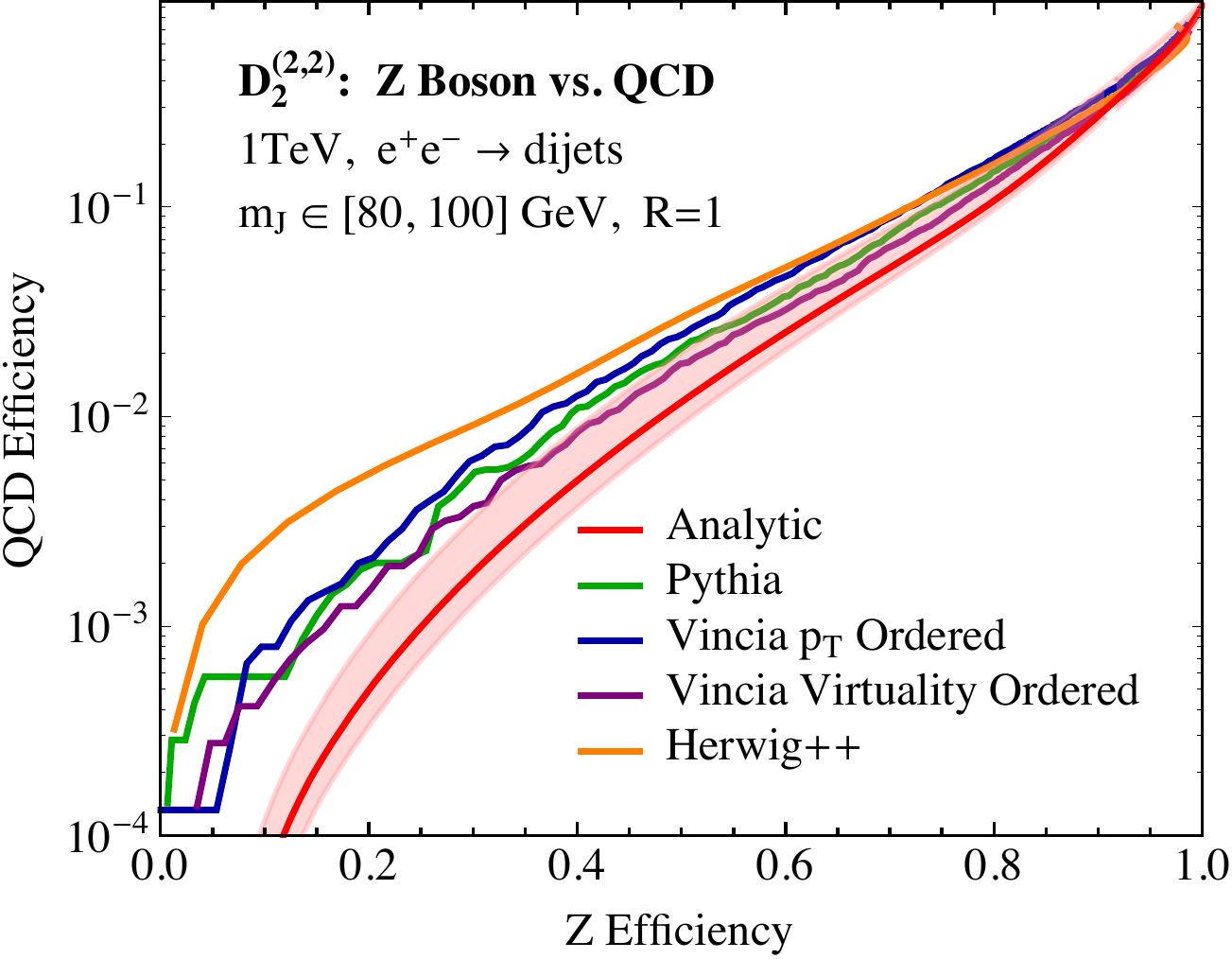}
}
\subfloat[]{\label{fig:D2sumROC_b}
\includegraphics[width = 7.0cm]{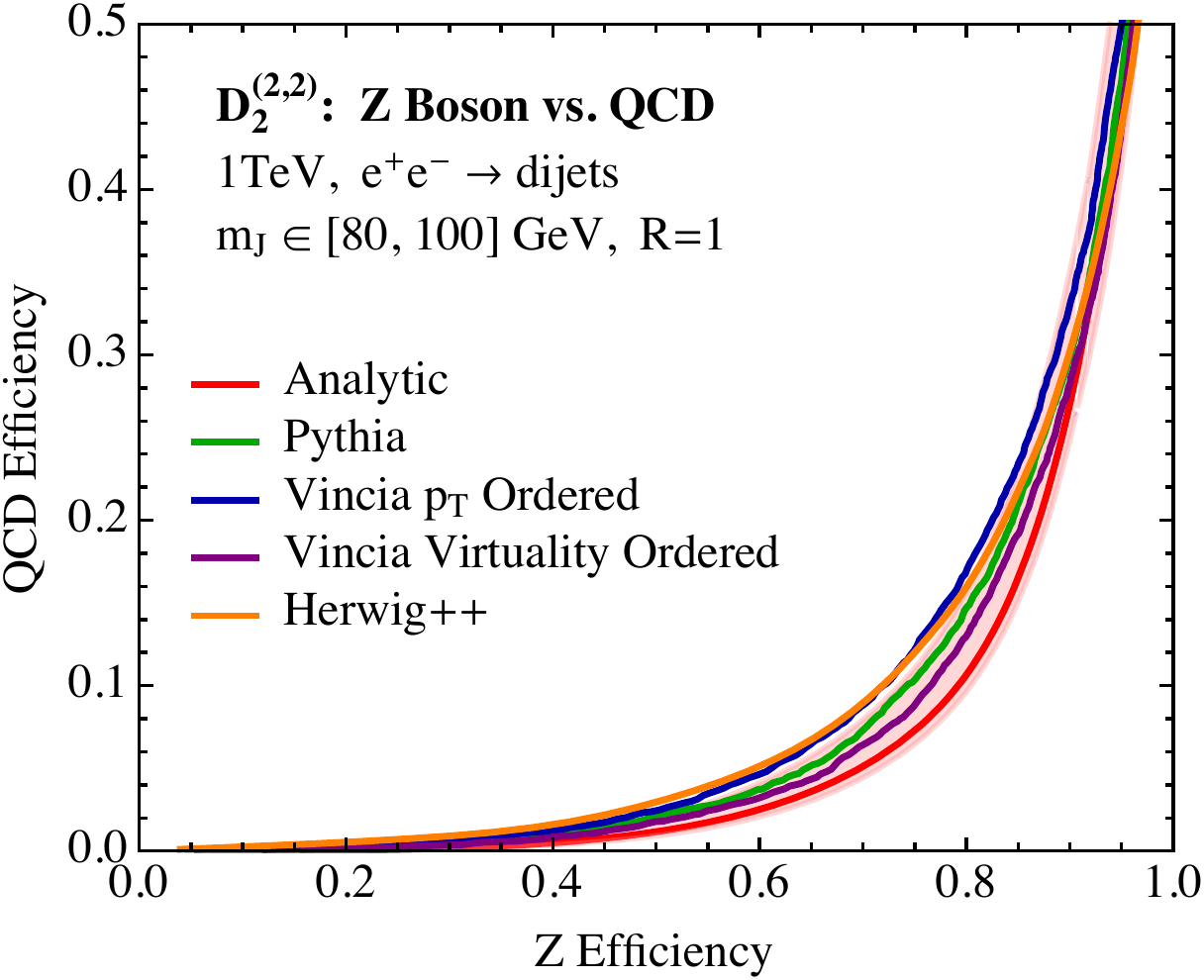}
}\end{center}
\caption{ Signal vs.~background efficiency curves for $\Dobs{2}{2,2}$ for the Monte Carlo samples as compared to our analytic prediction on a a) logarithmic scale plot and b) linear scale plot.  The band of the analytic prediction is representative of the perturbative scale uncertainty. Good agreement between the analytic calculation and Monte Carlo is observed (figures from \cite{Larkoski:2015kga}).
}
\label{fig:D2sumROC}
\end{figure}

In \Fig{fig:D2_anal_summary} a comparison of our analytic calculation including the effect of hadronization for the $D_2$ spectrum for boosted $Z$ jets and QCD jets is shown and compared with Monte Carlo. Excellent agreement is observed. The uncertainties on the analytic prediction come from scale variations of the different functions appearing in the factorized formula for the $D_2$ cross sections. Details are given in \Chap{chap:D2_anal}.  Having an explicit analytic prediction for both background and signal jets allowed us to make the first analytic ROC curves for a two-prong jet substructure observable, which are shown in \Fig{fig:D2sumROC}. Again, overall good agreement is observed, with the disagreement being driven by the signal distribution. It would be interesting to understand if higher order calculations would increase or reduce this discrepancy.

It would be of considerable practical and theoretical interest to extend this calculation to the LHC, where the $D_2$ observable is used extensively for boosted $W/Z/H$ tagging. There are two main obstructions to this calculation, one of which is experimental, and one of which is theoretical. On the experimental side, the $D_2$ observable is used in conjuction with a grooming procedure to remove the effects of pile-up. The effects of this grooming procedure would need to be incorporated in the theory calculation. On the theory side, the calculation in $pp$ is considerably more complicated than in $e^+e^-$ due to the color structure. For $e^+e^- \to$ dijets, even with an additional resolved jet, there are at most three jets, and a unique color structure, implying a color diagonal renormalization group evolution. For the LHC, even for the simplest process, such as $pp\to Z/W/H+$ jet, with an extra resolved subjet, there are four jets. This implies many partonic channels, each with non-diagonal color renormalization group evolution. It is interesting that grooming procedures, which are used to remove pile-up, also have the effect of removing wide angle color correlations. From the theory perspective, they therefore have the potential to simplify calculations. Because of these two considerations, it seems likely that the most interesting, and practical goal would be a calculation of a groomed $D_2$ at the LHC. This should be possible in the future using the groomer Soft Drop \cite{Larkoski:2014wba}, which exhibits several nice theoretical properties \cite{Frye:2016okc,Frye:2016aiz}. This would allow for first principles precision predictions for jet substructure variables of phenomenological importance at the LHC.

\section{Precision Predictions for LHC Processes Involving Jets}

\begin{figure}
\begin{center}
\subfloat[]{
\includegraphics[width=6.0cm]{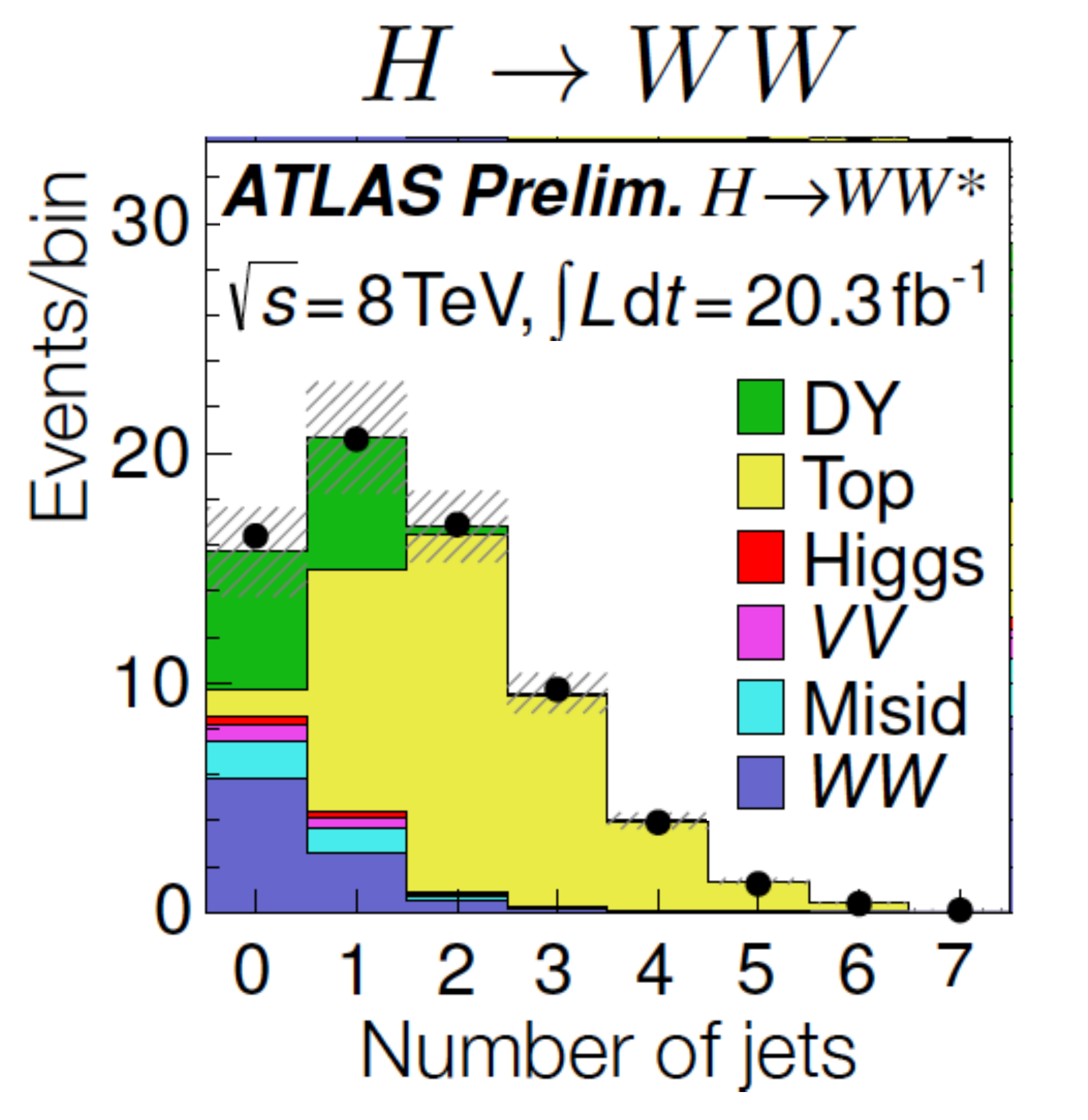}
}
$\qquad$
\subfloat[]{
\includegraphics[width=8.0cm]{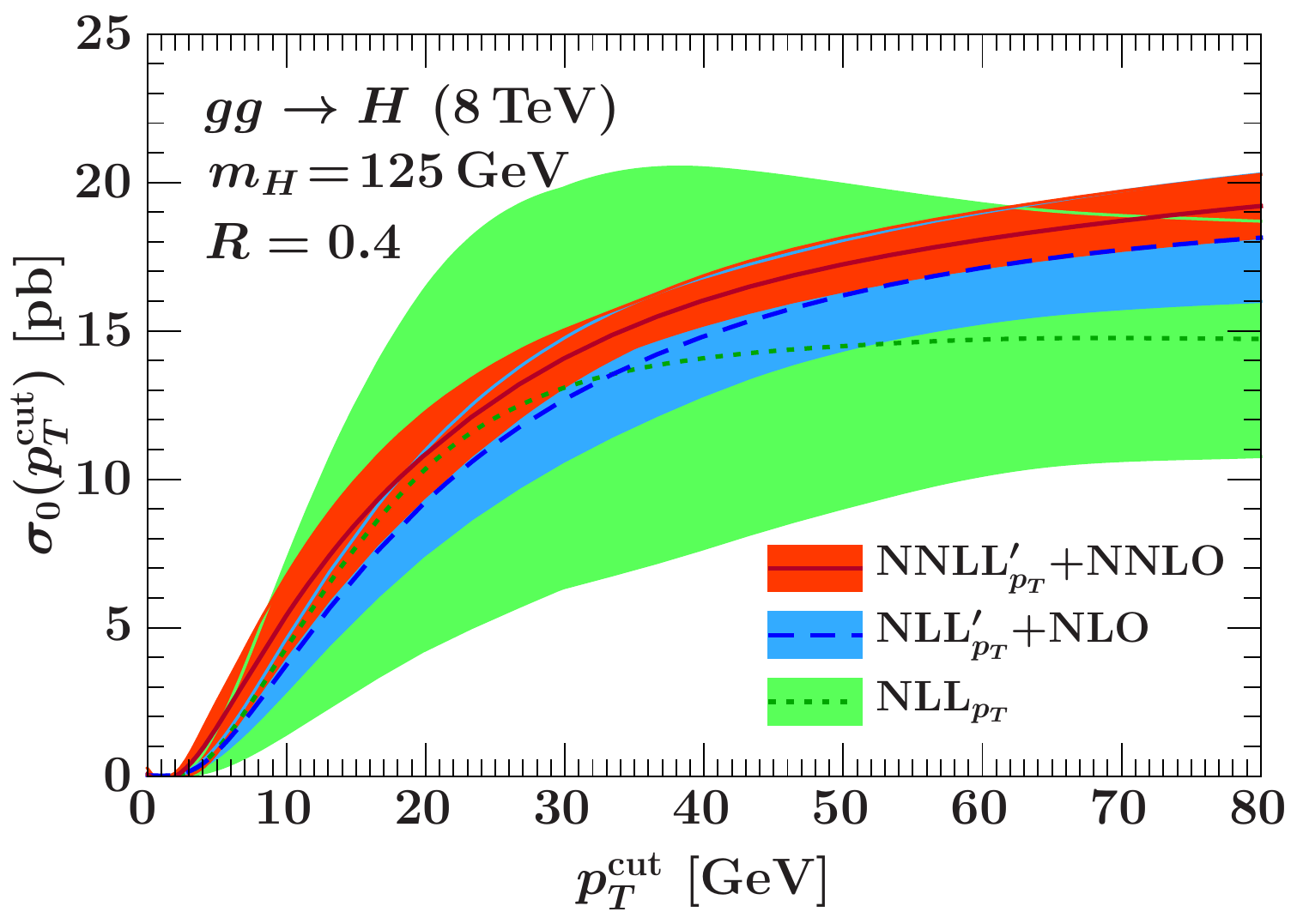}
}
\end{center}
\caption{The distribution of signal and background processes in exclusive jet bins for the process $H\to WW$ (left). The exclusive zero jet spectrum for $gg\to H$ computed in resummed perturbation theory as a function of the $\ptveto$ used to define a jet (right)(figure from \cite{Stewart:2013faa}).}
 \label{fig:jet_vetos}
\end{figure}

One of the main goals of the LHC is the study of the newly discovered Higgs boson, as well as the $W/Z$ bosons, and the top quark. Since the LHC is a proton-proton collider, these states are often produced in association with jets. The proper treatment of final states involving jets is therefore essential for achieving precision predictions for the LHC. As an example, consider the process $H\to WW$ at the LHC. In \Fig{fig:jet_vetos}, we show the distribution of signal and background process in exclusive jet bins. To separate the $H\to WW$ signal from the large background, particularly from top production, one can restrict to the exclusive zero jet bin where the signal to background ratio is highest. From a theoretical perspective, this restriction to the exclusive zero jet bin places significant constraints on the radiation in the final state, and introduces a new scale into the problem, namely $\ptveto$, the scale above which a jet is defined as being a jet. The presence of this additional scale in the problem introduces large logarithms into the perturbative expansion, which need to be resummed to all orders to have precision predictions, as are required to measure properties of the Higgs boson. Such calculations have been performed in SCET for on-shell Higgs production, and the exclusive zero jet cross section is shown in \Fig{fig:jet_vetos} as a function of $\ptveto$ \cite{Stewart:2013faa}.

In performing such a calculation for a process involving jets at the LHC, many of the ingredients describing the dynamics of the jets are universal, and are described within the effective theory. The process dependence is carried by a matching coefficient describing the hard virtual fluctuations at the scale of the underlying hard process.  The full QCD process is matched onto hard scattering operators in the effective field theory. To perform such a matching procedure, two ingredients are necessary. First, a complete basis of operators is required in the effective field theory, and second, the hard scattering amplitudes computed in fixed order QCD are required. For complicated processes, particularly those involving multiple jets, the construction of a complete operator basis in the effective field theory can be quite difficult, and furthermore, the fixed order amplitudes can be complicated, and often not in a form which can be easily interfaced with the SCET operators.

Many modern fixed order QCD calculations are performed using spinor helicity techniques for on-shell states of definite helicities. This has been shown to greatly simplify results, removing large gauge redundancies, and giving very compact expressions. The notion of helicity is also very natural in SCET, where external reference vectors identifying the directions of the jets are already present in the effective field theory. In the paper "Employing Helicity Amplitudes for Resummation" \cite{Moult:2015aoa}, discussed in \Chap{chap:helops}, we introduced a convenient operator basis in SCET, which uses operators of definite helicity. These helicity operators make enumerating operator bases trivial, and the Wilson coefficients of the hard scattering operators are directly given by the (IR finite) pieces of color stripped helicity amplitudes. They allow state of the art predictions for the hard scattering process to be easily combined with an effective theory description of the dynamics of QCD jets. Hopefully this will enable more complex final states to be studied in SCET. Recently, we have also extended this formalism to subleading power in the SCET expansion \cite{Kolodrubetz:2016uim}.

\begin{figure}
\begin{center}
\subfloat[]{\label{fig:HiggsWW_sumb}
\includegraphics[width=6cm]{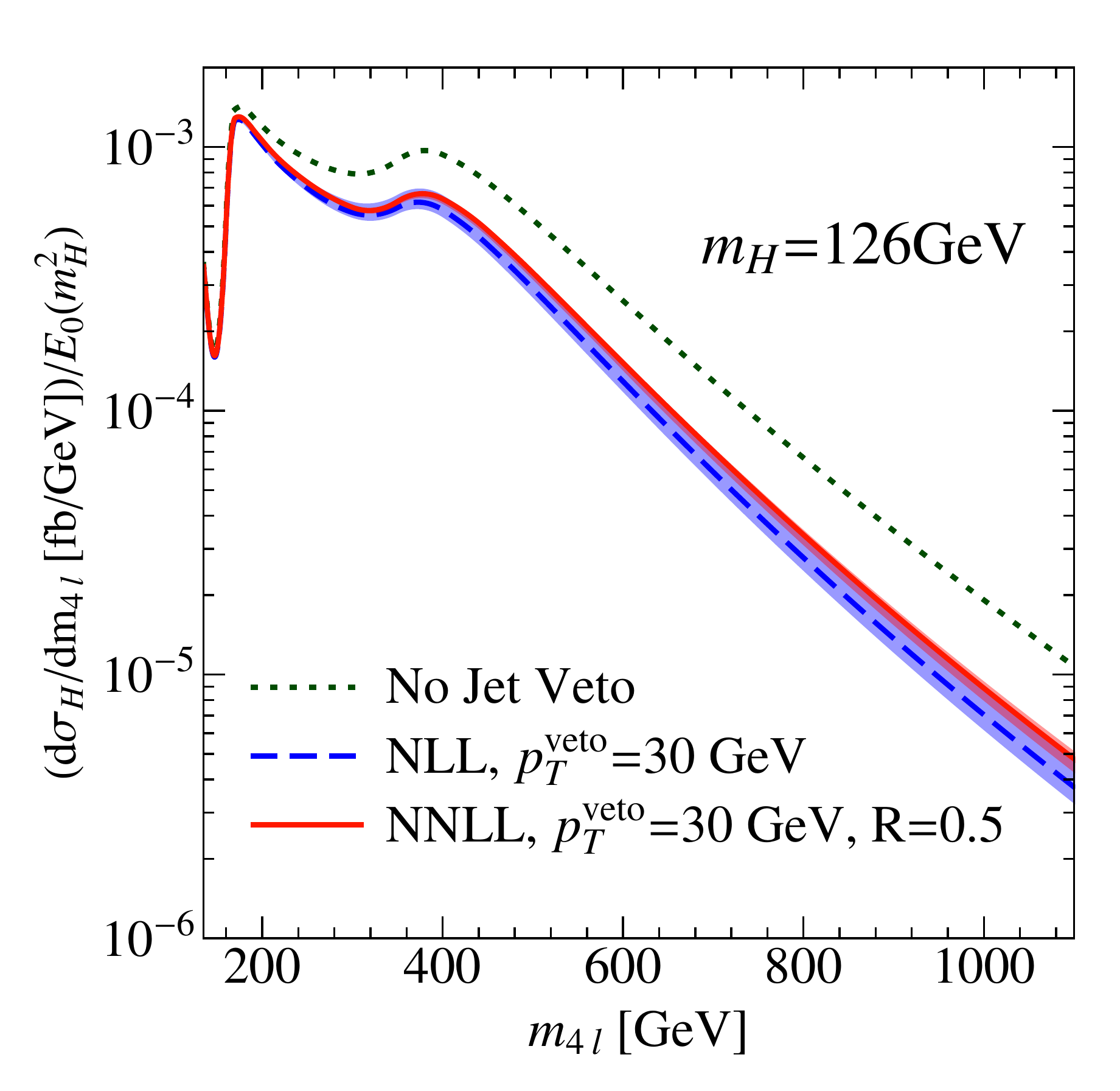}
}
$\qquad$
\subfloat[]{\label{fig:HiggsWW_suma}
\includegraphics[width=6cm]{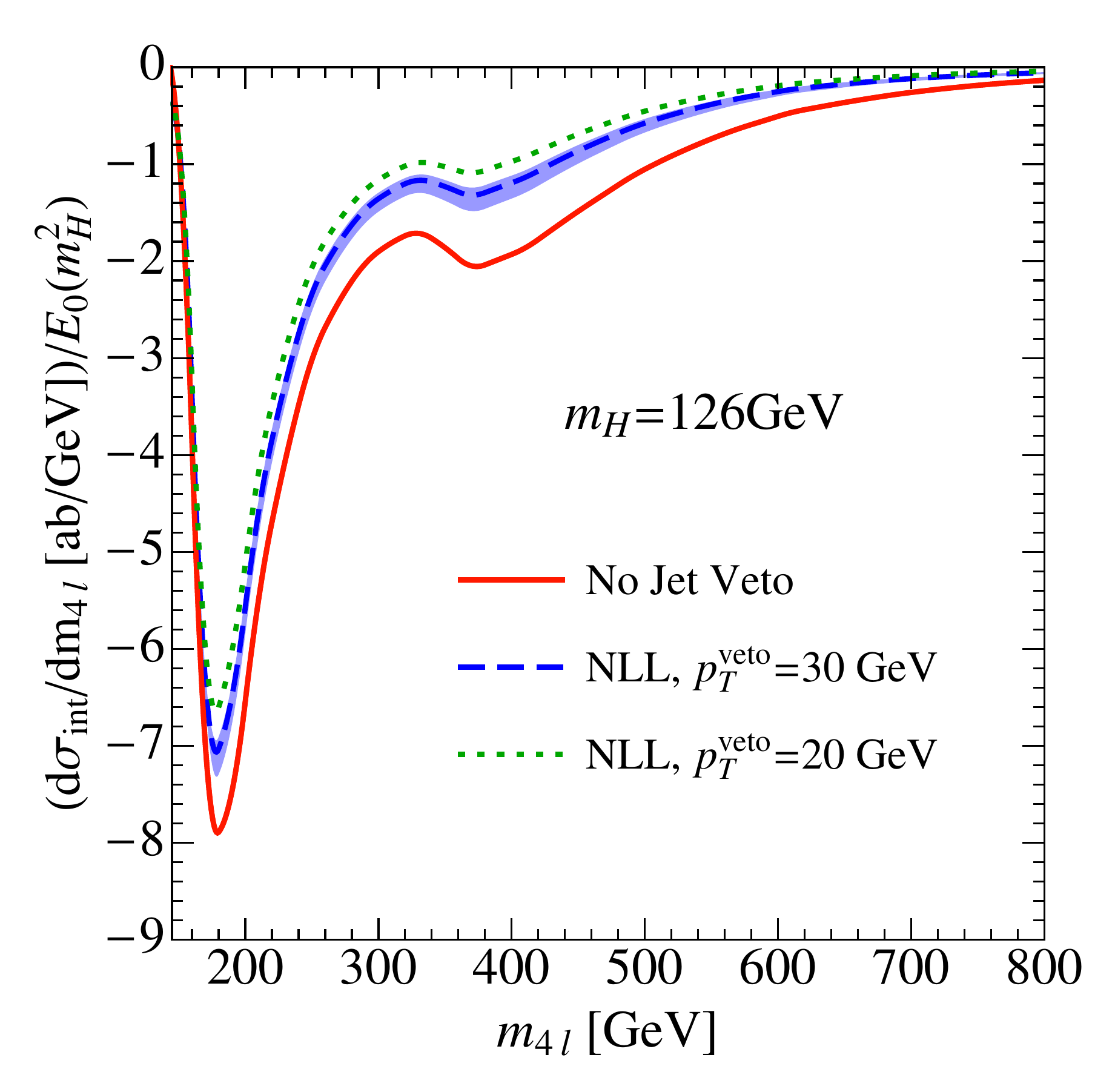}
}
\end{center}
\caption{The off-shell Higgs cross section in the exclusive zero jet bin for the Higgs mediated process in (a) and the interference in (b) . Results are normalized by the jet veto suppression at the Higgs mass, such that the on-shell cross section is the same in all cases, allowing one to focus on the modification to the shape of the distribution. NLL and NNLL results are similar, with a small modification due to the finite jet radius, which is not present in the NLL calculation (figures from \cite{Moult:2014pja}). }
\label{fig:HiggsWW_sum}
\end{figure}

To demonstrate the applicability of the helicity operators in SCET, in the paper "Jet Vetoes Interfering with $H\to WW$" \cite{Moult:2014pja}, discussed in \Chap{chap:WW}, I considered the resonance and continuum production of $pp \to l\nu \bar l \bar \nu$, which can proceed with through the Higgs boson, or through $WW$ production. This process represents an interesting application of the helicity operators, as the amplitudes for $pp \to l\nu \bar l \bar \nu$ are complicated, and only available in spinor helicity form, and the matching to SCET can be greatly simplified. Furthermore, the process is interesting as a probe of the properties of the Higgs. With the discovery of a Higgs like boson by both the ATLAS and CMS experiments at the LHC, an important goal of the LHC program will be the characterization of the resonance. Of particular importance is the width of the Higgs like boson. The width of the Standard Model boson is extremely narrow due to the fact that it is light, and couples proportional to mass. Other decay channels of the Higgs, could significantly modify the Higgs width. The experimental resolution on the direct lineshape is of the order of $1$ GeV, while the predicted Standard Model value is $\sim 4$ MeV. Therefore, it is an important problem to come up with ways of measuring the Higgs width at the LHC. The width is observable at an $e^+e^-$ machine through Higgstralung.  It was noted that a (somewhat model dependent) way of bounding the Higgs width is to study its off shell production in the $WW$ and $ZZ$ channels and interference with continuum $WW$ or $ZZ$ production, due to the difference in scaling of the cross section with the width in the on-shell and off-shell region \cite{Caola:2013yja}. For the $WW$ channel, the reduction of the background makes important use of jet vetoes. It is therefore an interesting question to understand how these jet vetoes effect the off-shell cross section. 

\begin{figure}
\begin{center}
\subfloat[]{\label{fig:CMS_width}
\includegraphics[width=7cm]{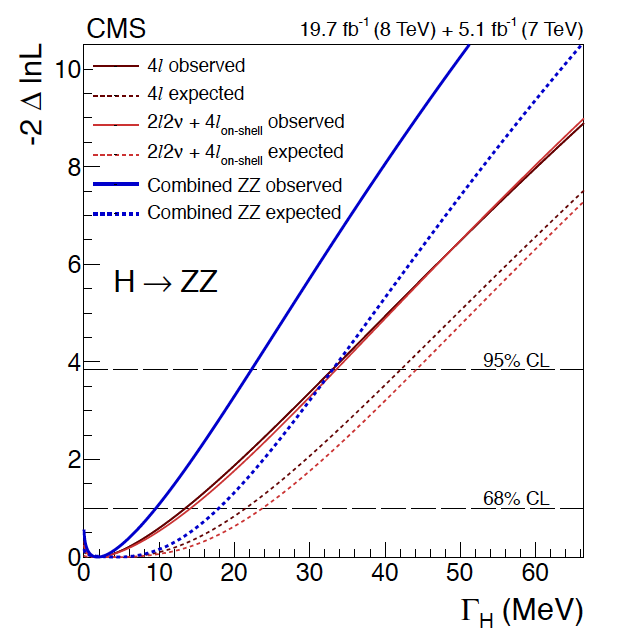}
}
\end{center}
\caption{The constraints on the Higgs width as measured by CMS using the off-shell cross section for $H\to ZZ$ (figure from \cite{Khachatryan:2014iha}).
}
\label{fig:CMS_width}
\end{figure}

In \Chap{chap:WW} we present a calculation of the exclusive zero jet cross section for both Higgs and continuum production, as well as their interference, in the far off-shell region. An interesting feature of this calculation, is that the size of the logarithm is a function of the invariant mass of the $WW$ pair, and therefore changes throughout the invariant mass spectrum. This implies that the effect of the jet veto is not merely a rescaling of the cross section, but a reshaping of the distribution. This reshaping of the distribution is shown in \Fig{fig:HiggsWW_sum}. A precise calculation of this reshaping of the distribution is important in interpreting any possible new physics contributions to the spectrum.  This measurement of the off-shell cross section has been performed at the LHC for the $ZZ$ channel, and used to put a bound on the Higgs cross section. The result is shown in \Fig{fig:CMS_width} \cite{Khachatryan:2014iha}.

%% file: chap2.tex


%






\newcommand{\cA}{{\cal I}}

\chapter{Power Counting to Better Jet Observables}\label{chap:PC}

Optimized jet substructure observables for identifying boosted topologies will play an essential role in maximizing the physics reach of the Large Hadron Collider. Ideally, the design of discriminating variables would be informed by analytic calculations in perturbative QCD. Unfortunately, explicit calculations are often not feasible due to the complexity of the observables used for discrimination, and so many validation studies rely heavily, and solely, on Monte Carlo. In this chapter we show how methods based on the parametric power counting of the dynamics of QCD, familiar from effective theory analyses, can be used to design, understand, and make robust predictions for the behavior of jet substructure variables.  As a concrete example, we apply power counting for discriminating boosted $Z$ bosons from massive QCD jets using observables formed from the $n$-point energy correlation functions. We show that power counting alone gives a definite prediction for the observable that optimally separates the background-rich from the signal-rich regions of phase space.  Power counting can also be used to understand effects of phase space cuts and the effect of contamination from pile-up, which we discuss.  As these arguments rely only on the parametric scaling of QCD, the predictions from power counting must be reproduced by any Monte Carlo, which we verify using \pythia{8} and \herwigpp.  We also use the example of quark versus gluon discrimination to demonstrate the limits of the power counting technique.


\section{Introduction}
\label{sec:intro}

Over the past several years there has been an explosion in the number of jet observables and techniques developed for discrimination and grooming \cite{Abdesselam:2010pt,Altheimer:2012mn,Altheimer:2013yza}.  Several of these are used by the ATLAS and CMS experiments, and their performance has been validated directly on data \cite{CMS:2011xsa,Miller:2011qg,ATLAS-CONF-2012-066,Chatrchyan:2012mec,ATLAS:2012jla,Aad:2012meb,ATLAS:2012kla,ATLAS:2012am,Aad:2013gja,Aad:2013fba,TheATLAScollaboration:2013tia,TheATLAScollaboration:2013sia,TheATLAScollaboration:2013ria,TheATLAScollaboration:2013pia,CMS:2013uea,CMS:2013kfa,CMS:2013wea,CMS-PAS-JME-10-013,CMS-PAS-QCD-10-041,Aad:2014gea,LOCH:2014lla,CMS:2014fya} and employed in new physics searches in highly boosted regimes \cite{CMS:2011bqa,Fleischmann:2013woa,Pilot:2013bla,TheATLAScollaboration:2013qia,Chatrchyan:2012ku,Chatrchyan:2012sn,CMS:2013cda,CMS:2014afa,CMS:2014aka}.  Analyses using jets will become increasingly important at the higher energies and luminosities of Run 2 of the LHC.

While the proliferation of jet observables is exciting for the field, the vast majority of proposed observables and procedures have been analyzed exclusively in Monte Carlo simulation.  Monte Carlos are vital for making predictions at the LHC, but should not be a substitute for an analytical understanding, where possible.  Because Monte Carlos rely on tuning the description of non-perturbative physics to data, this can obscure what the robust perturbative QCD predictions are and hide direct insight into the dependence of the distributions on the parameters of the observable.  This is especially confusing when different Monte Carlo programs produce different results.

Perturbative predictions of distributions have traditionally been constrained to only the simplest observables, such as the jet mass \cite{Catani:1991bd,Chien:2010kc,Chien:2012ur,Dasgupta:2012hg,Jouttenus:2013hs}, but to high accuracy. Such high-order calculations are important for reducing the systematic theoretical uncertainties. More recently, resummation has been applied to some simple jet substructure variables \cite{Feige:2012vc,Dasgupta:2013ihk,Dasgupta:2013via,Larkoski:2014pca}, and an understanding of some of the subtleties of resummation for ratio observables, as often used in jet substructure, has been developed \cite{Larkoski:2013paa,Larkoski:2014tva,Larkoski:2014bia}. Even the simplest calculations have suggested new, improved techniques, like the modified Mass Drop Tagger \cite{Dasgupta:2013ihk,Dasgupta:2013via}, or uncovered unexpected structures in perturbative QCD, like Sudakov Safety \cite{Larkoski:2013paa,Larkoski:2014wba,Larkoski:2014bia}.  For more complex observables, however, an analytic calculation may be essentially impossible, and we must rely on Monte Carlo simulations. Because of the wide variety of jet observables, some of which can be calculated analytically and some that cannot, it is necessary to find an organizing principle that can be used to identify the robust predictions of QCD, without requiring a complete calculation to a given perturbative accuracy.

In this chapter we show how power counting methods can be used to design and understand the behavior of jet substructure variables. With minimal computational effort, power counting accurately captures the parametric predictions of perturbative QCD. The dynamics of a QCD jet are dominated by soft and collinear emissions and so by identifying the parametric scaling of soft and collinear contributions to a jet observable, we are able to make concrete and justified statements about the performance of jet substructure variables.  Formal parametric scaling, or power counting, is widely used in the formalism of soft-collinear effective theory (SCET) \cite{Bauer:2000ew,Bauer:2000yr,Bauer:2001ct,Bauer:2001yt}, an effective field theory of QCD in the soft and collinear limits. However, in this chapter, we will not rely on any results from SCET so as to make the discussion widely accessible.  Similar techniques were employed in \Ref{Walsh:2011fz}, but with the goal of determining which jet observables are calculable.

As a concrete application of the soft and collinear power counting method, we will focus on observables formed from the generalized $n$-point energy correlation functions $\ecf{n}{\beta}$ \cite{Larkoski:2013eya}, relevant for discriminating massive QCD jets from boosted, heavy objects.  Measuring multiple energy correlation functions on a jet defines a multi-dimensional phase space populated by signal and background jets.  By appropriately power counting the dominant regions of phase space, we are able to identify the signal- and background-rich regions and determine powerful observables for discrimination.  In addition, from power counting arguments alone, we are able to predict the effect of pile-up contamination on the different regions of phase space.  We apply power counting to the following:
\begin{itemize}

\item {\bf Boosted Z Bosons vs.~QCD} The two- and three-point energy correlation functions, $\ecf{2}{\beta}$ and $\ecf{3}{\beta}$, have been shown to be among the most powerful observables for identifying the hadronic decays of boosted $Z$ bosons \cite{Larkoski:2013eya}.  We  discuss the phase space defined by $\ecf{2}{\beta}$ and $\ecf{3}{\beta}$, and determine which regions are populated by signal and background jets. Using this understanding of the phase space, we propose a powerful discriminating variable to identify boosted two prong jets, given by
\begin{equation}
\Dobs{2}{\beta}= \frac{\ecf{3}{\beta}}{(\ecf{2}{\beta})^3}\,.
\end{equation}
This should be contrasted with the variable $\Cobs{2}{\beta}=\ecf{3}{\beta}/(\ecf{2}{\beta})^2$ originally proposed in \Ref{Larkoski:2013eya}. We also show that power counting can be used to understand the impact of pile-up radiation on the different regions of phase space, and in turn to understand the susceptibility of signal and background distributions to pile-up.

\item {\bf Quarks vs.~Gluons} Quark versus gluon jet discrimination is somewhat of a non-example for the application of power counting because there is nothing parametrically distinct between quark and gluon jets.  However, this will illustrate why quark versus gluon discrimination is such a hard problem, and why different Monte Carlos can have wildly different predictions \cite{Larkoski:2014pca}.

\end{itemize}

The outline of this chapter is as follows.  In \Sec{sec:modes} we will precisely define what we mean by ``collinear'' and ``soft'' modes of QCD and introduce the observables used throughout this chapter.  While we will mostly focus on the energy correlation functions, we will also discuss the $N$-subjettiness observables \cite{Thaler:2010tr,Thaler:2011gf} as a point of reference.  In \Sec{sec:boostz}, we apply power counting to the study of $Z$ versus QCD discrimination using the two- and three-point energy correlation functions $\ecf{2}{\beta}$ and $\ecf{3}{\beta}$.  We argue that the single most powerful observable for discrimination is $\ecf{3}{\beta}/(\ecf{2}{\beta})^3$. Power counting is used to understand how the addition of pile-up radiation effects the distributions of this variable, and show that they are more robust to pile-up than for previously proposed variables formed from the energy correlation functions.\footnote{The CMS study of \Ref{CMS:2013uea} found that the observable $C_2^{(\beta)}\equiv\ecf{3}{\beta}/(\ecf{2}{\beta})^2$ suggested in \Ref{Larkoski:2013eya} for boosted $Z$ identification is very sensitive to pile-up contamination.} We verify that these predictions are borne out in Monte Carlo.  In \Sec{sec:qvg}, we attempt to apply power counting to quark versus gluon jet discrimination.  Na\"ively, this should be the simplest case, however, power counting arguments are not applicable because all qualities of quarks and gluons only differ by order-1 numbers.  Finally, we conclude in \Sec{sec:conc} by re-emphasizing that power counting is a useful predictive tool for jet observables that are too complicated for direct analytic calculations, and suggest some problems to which it may prove fruitful.

\section{Observable Basis and Dominant Physics of QCD}
\label{sec:modes}

\subsection{Observables}

Throughout this chapter, our analyses will be focused around the (normalized) $n$-point energy correlation functions $\ecf{n}{\beta}. $\footnote{The notation $\ecf{n}{\beta}$ differs from the original notation $\text{ECF}(n,\beta)$ presented in \Ref{Larkoski:2013eya} where the energy correlation functions were defined, but we hope that this notation used here is more compact.  Specifically, the relationship is
\begin{equation}
\ecf{n}{\beta} = \frac{\text{ECF}(n,\beta)}{\left(
\text{ECF}(1,\beta)
\right)^n} \ .
\end{equation}
} The two-, and three-point energy correlation functions are defined as
\begin{align}
\ecf{2}{\beta} &= \frac{1}{p_{TJ}^2}\sum_{1\leq i<j\leq n_J} p_{Ti}p_{Tj} R_{ij}^\beta \ ,\nonumber \\
\ecf{3}{\beta} &= \frac{1}{p_{TJ}^3}\sum_{1\leq i<j<k\leq n_J} p_{Ti}p_{Tj}p_{Tk} R_{ij}^\beta R_{ik}^\beta R_{jk}^\beta \ , \nonumber \\
\end{align}
where $p_{TJ}$ is the transverse momentum of the jet with respect to the beam, $p_{Ti}$ is the transverse momentum of particle $i$, and $n_J$ is the number of particles in the jet.  The boost-invariant angle $R_{ij}^2 = (\phi_i-\phi_j)^2+(y_i-y_j)^2$ is the Euclidean distance in the azimuth-rapidity plane and for infrared and collinear (IRC) safety, the angular exponent $\beta>0$.  In this chapter we will only study up through $\ecf{3}{\beta}$, but higher-point energy correlation functions are defined as the natural generalization. We will often omit the explicit dependence on $\beta$, denoting the $n$-point energy correlation function simply as $\ecfnobeta{n}$.

The energy correlation functions have many nice properties that make them ideal candidates for defining a basis of jet observables.  First, the energy correlation functions are defined such that $\ecf{n}{\beta}\to0$ in any of the soft or collinear limits of a configuration of $n$ particles.  Second, because all angles in the energy correlation functions are measured between pairs of particles, $\ecf{n}{\beta}$ is insensitive to recoil or referred to as  ``recoil-free'' \cite{Catani:1992jc,Dokshitzer:1998kz,Banfi:2004yd,Larkoski:2013eya,Larkoski:2014uqa}. This means that it is not sensitive to the angular displacement of the hardest particle (or jet core) from the jet momentum axis due to soft, wide angle radiation in the jet.  The effects of recoil decrease the sensitivity of an observable to the structure of radiation about the hard core of the jet, making it less efficient for discrimination purposes.

Depending on the application, different energy correlation functions are useful as discriminating observables.  As discussed in \Ref{Larkoski:2013eya}, the two-point energy correlation function is sensitive to radiation about a single hard core, and so is useful for quark versus gluon discrimination. Similarly, the three- and four-point energy correlation functions are useful for 2- or 3-prong jet identification, respectively, corresponding to boosted electroweak bosons ($W/Z/H$) or hadronically decaying top quarks.  By measuring appropriate energy correlation functions we define a phase space, populated by signal and background jets.

As a point of reference, we will also study the $N$-subjettiness observables and compare the structure of their phase space with that of the energy correlation functions.  The (normalized) $N$-subjettiness observable $\Nsub{N}{\beta}$ is defined as
\begin{equation}\label{eq:nsubdef}
\Nsub{N}{\beta} = \frac{1}{p_{TJ}}\sum_{1\leq i \leq n_J} p_{Ti}\min\left\{
R_{i1}^\beta,\dotsc,R_{iN}^\beta
\right\} \ .
\end{equation}
The angle $R_{iK}$ is measured between particle $i$ and subjet axis $K$ in the jet.  Thus, $N$-subjettiness partitions a jet into $N$ subjet regions and measures the $p_T$-weighted angular distribution with respect to the subjet axis of each particle.  There are several different choices for how to define the subjet axes; here, we will define the subjet axes by the exclusive $k_T$ jet algorithm \cite{Cacciari:2008gp} with the winner-take-all (WTA) recombination scheme \cite{Bertolini:2013iqa,Larkoski:2014uqa,Salambroadening}.  In contrast to the traditional $E$-scheme recombination \cite{Blazey:2000qt}, which defines the (sub)jet axis to coincide with the net momentum direction, the WTA recombination scheme produces (sub)jet axes that are recoil-free and nearly identical to the $\beta = 1$ minimized axes.\footnote{The $\beta = 1$ minimized axes are also referred to as ``broadening axes'' \cite{Thaler:2011gf,Larkoski:2014uqa} as they correspond to axes that minimize the value of broadening \cite{Rakow:1981qn,Ellis:1986ig,Catani:1992jc}.}  With this definition, the observables $\ecf{2}{\beta}$ and $\Nsub{1}{\beta}$ are identical through NLL accuracy for all $\beta > 0$ \cite{Larkoski:2014uqa}.

Since $N$-subjettiness directly identifies $N$ subjet directions in a jet, it is a powerful variable for $N$-prong jet discrimination.  In particular, the $N$-subjettiness ratios
$$
\Nsub{2,1}{\beta}\equiv \frac{\Nsub{2}{\beta}}{\Nsub{1}{\beta}} \quad\text{and}\quad \Nsub{3,2}{\beta}\equiv \frac{\Nsub{3}{\beta}}{\Nsub{2}{\beta}} \ ,
$$
relevant for boosted $W/Z/H$ and top quark identification, respectively, are widely-used in jet studies at the ATLAS and CMS experiments.  Numerical implementations of the energy correlation functions and $N$-subjettiness are available in the \texttt{EnergyCorrelator} and \texttt{Nsubjettiness} \fastjet{contrib}s \cite{Cacciari:2011ma,fjcontrib}.

\subsection{Soft and Collinear Modes of QCD}
\label{sec:softcollqcd}

At high energies, QCD is approximately a weakly-coupled conformal gauge theory and so jets are dominated by soft and collinear radiation.  Because it is approximately conformal, there is no intrinsic energy or angular scale associated with this radiation.  To introduce a scale, and so to determine the dominant soft and collinear emissions, we must break the conformal invariance by making a measurement on the jet.  The scale of the soft and collinear emissions is set by the measured value of the observable.\footnote{It is important to note that since QCD is not a conformal field theory, we can only use the power counting presented here to study the phase space defined by a set of IRC safe observables.  If we considered IRC unsafe observables, then generically, we would need to power count contributions from non-perturbative physics such as hadronization.
}

This observation can be exploited to make precise statements about the energy and angular structure of a jet, depending on the value of observables measured on that jet. This reasoning is often implicitly understood in the jet community and literature, and is formalized in SCET.  Nevertheless, these precise power-counting arguments are not widely used outside of SCET, and so we hope that the applications in this chapter illustrate their effectiveness and relative simplicity.

We begin by defining a soft emission, $s$, as one for which
\begin{equation}
z_s \equiv \frac{p_{Ts}}{p_{TJ}}\ll1 \ , \qquad R_{sj}\sim 1 \ ,
\end{equation}
where $j$ is any other particle in the jet and $R_{sj}\sim 1$ means that $R_{sj}$ is not associated with any parametric scaling.  Similarly, a collinear emission, $c$, is defined as having a $p_T$ fraction
\begin{equation}
\frac{p_{Tc}}{p_{TJ}}\sim1 \ ,
\end{equation}
but with an angle to other particles which depends on whether they are also collinear or soft:
\begin{equation}
R_{cc}\ll 1\ , \qquad R_{cs}\sim 1 \ .
\end{equation}
Here, $R_{cc}$ is the angle between two collinear particles, while $R_{cs}$  is the angle between a soft particle and a collinear particle.  The precise scalings of $R_{cc}$ and $z_s$ will depend on the observable in question, as will be explained shortly.  Soft emissions also implicitly include radiation that is simultaneously both soft and collinear.

To introduce these ideas concretely, we use the example of the two-point energy correlation function:
\begin{equation}\label{eq:2pt_ex}
\ecf{2}{\beta} =  \frac{1}{p_{TJ}^2}\sum_{1\leq i<j\leq n_J} p_{Ti}p_{Tj} R_{ij}^\beta \ .
\end{equation}
Consider performing a measurement of $\ecf{2}{\beta}$ on a jet and further requiring $\ecf{2}{\beta}\ll1$. Because the energy flow in jets is in general collimated, this defines a non-trivial region of phase space, with a large fraction of jets satisfying this requirement.  A large value of $\ecf{2}{\beta}$ would mean that there is a hard, perturbative splitting in the jet which is suppressed by the small value of $\alpha_s$. From the definition of $\ecf{2}{\beta}$ in \Eq{eq:2pt_ex}, we see that a measurement of $\ecf{2}{\beta}\ll1$ forces all particles in the jet to either have small $p_{Ti}$ or small $R_{ij}$. In other words, the observable is dominated by soft and collinear emissions. The precise scaling of $p_{Ti}$ and $R_{ij}$ is then determined  by the measured value of $\ecf{2}{\beta}$.\footnote{In SCET, $R_{cc}$ and $z_s$ are often immediately assigned a related scaling. While this is true for this example, it is not in general true in the case of multiple measurements, and we wish to emphasize in this section how the measurement sets both scalings.}

There are three possible configurations that contribute to $\ecf{2}{\beta} \ll 1$: soft-soft correlations, soft-collinear correlations, and collinear-collinear correlations.  Therefore, $\ecf{2}{\beta}$ can be expressed as
\begin{equation}
\ecf{2}{\beta} \sim \frac{1}{p_{TJ}^2}\sum_{s} p_{Ts}p_{Ts} R_{ss}^\beta + \frac{1}{p_{TJ}^2}\sum_{s,c} p_{Ts}p_{Tc} R_{cs}^\beta + \frac{1}{p_{TJ}^2}\sum_{c} p_{Tc}p_{Tc} R_{cc}^\beta  \ ,
\end{equation}
where we have separated the contributions to $\ecf{2}{\beta}$ into the three different correlations.  To determine the dominant contributions to $\ecf{2}{\beta}$, we will throw away those contributions that are parametrically smaller, according to our definitions of soft and collinear above.  First, $p_{Ts} \ll p_{Tc}$, and so we can ignore the first term to leading power.  Because $R_{cs}\sim 1$, we set $R_{cs}= 1$ in the second term, for the purpose of scaling.  Also, note that $p_{Tc}\sim p_{TJ}$ and so we can replace the instances of $p_{Tc}$ with $p_{TJ}$ in the second and third terms.  Making these replacements, we find
\begin{equation}\label{eq:fact_e2}
\ecf{2}{\beta} \sim \sum_{s}  z_s +\sum_{c}  R_{cc}^\beta  \ ,
\end{equation}
where we have ignored any corrections arising at higher power in the soft and collinear emissions' energies and angles. We wish to emphasize with the explicit summation symbols that we have not restricted to a single soft or collinear emission, but consider an arbitrary number of emissions. Furthermore, we do not assume a strongly ordered limit, but instead explore the complete phase space arising from soft and collinear emissions, including regions where such ordering is explicitly broken.

\Eq{eq:fact_e2} demonstrates the dominant structure of a jet on which we have measured $\ecf{2}{\beta}\ll 1$.  The contribution to $\ecf{2}{\beta}$ from soft and collinear emissions do not mix to this accuracy; that is, they factorize from one another.  Also, because there is no measurement to distinguish the soft and collinear contributions to $\ecf{2}{\beta}$, we then have that 
\begin{equation}
\ecf{2}{\beta} \sim  z_s \sim R_{cc}^\beta  \,.
\end{equation}
That is, the measured value of $\ecf{2}{\beta}$ sets the $p_T$ of the soft particles relative to the jet and the splitting angle of the collinear particles, and therefore defines the structure of the jet. 

In \Eq{eq:fact_e2}, we have explicitly written a summation over the particles with soft and collinear scalings. To determine the scalings of the different contributions, it is clearly sufficient to consider the scaling of an individual term in each sum. In the remainder of this chapter we will drop the explicit summation for notational simplicity.

Scaling arguments similar to the power counting approach discussed here are often used in other approaches to QCD resummation to identify the relevant soft and collinear scales, and could also be used to analyze observables. For example, in the method of regions \cite{Beneke:1997zp,Smirnov:2001in}, the regions of integration over QCD matrix elements which contribute dominantly to a given observable are determined, and an expansion about each of these regions is performed. These regions of integration, and the scaling of the momenta in these regions, basically correspond to the modes of the effective theory determined through the power counting approach.\footnote{More precisely, only on-shell modes appear as degrees of freedom in the effective theory.}

Similarly, in the CAESAR approach to resummation \cite{Banfi:2004yd}, implemented in an automated computer program, the first step of the program is the identification of the relevant soft and collinear scales. This is performed by expanding a given observable in the soft and collinear limits, and considering the region of integration for a single emission. This procedure is similar to that used in the case of $\ecf{2}{\beta}$ just discussed, and would identify the same dominant contributions and scalings. Using the knowledge of the behavior of the QCD splitting functions, CAESAR then performs a resummed calculation of the observable. However, the CAESAR computer program is currently restricted to observables for which the relevant scales, and hence the logarithmic structure, is determined by a single emission, and further, to single differential distributions. 

When considering observables relevant for jet substructure, one is interested in variables such as $\ecf{n}{\beta}$, $n>2$, whose behavior is not determined by the single emission phase space. For such observables, the 
 single-emission analysis is not sufficient
 and the explicit analysis of QCD matrix elements to determine the dominant regions of integration which contribute becomes quite complicated. For these cases, we find the power counting approach of the effective field theory paradigm to be a particularly convenient organizing principle. Using the knowledge that on-shell soft and collinear modes dominate, a consistent power counting can be used to determine the relevant scalings of these modes in terms of the measured observables, which is reduced to a simple algebraic exercise. Although the evaluation of QCD matrix elements in these scaling limits is of course required for a complete calculation, it is not required to determine the power counting, and we will see that power counting alone will often be sufficient for constructing discriminating observables for jet substructure studies.

Throughout the rest of this chapter, we will employ these power-counting arguments to determine the dominant structure of jets on which multiple measurements have been made, for example $\ecf{2}{\beta}$ and $\ecf{3}{\beta}$.  In this case, the phase space that results is much more complicated than the example of $\ecf{2}{\beta}$ discussed above, but importantly, appropriate power counting of the contributions from soft and collinear emissions will organize the phase space into well-defined regions automatically.

\section{Power Counting Boosted Z Boson vs. QCD Discrimination}
\label{sec:boostz}

As a detailed example of the usefulness of power counting, we consider the problem of discriminating hadronically-decaying, boosted $Z$ bosons from massive QCD jets.  Because $Z$ boson decays have a 2-prong structure, we will measure the two- and three-point energy correlation functions, $\ecfnobeta{2}$ and $\ecfnobeta{3}$, on the jets, defining a two-dimensional phase space.  We will find that there are two distinct regions of this phase space corresponding to jets with one or two hard prongs.  QCD jets exist dominantly in the former region while boosted $Z$ bosons exist dominantly in the latter.  Power counting these phase space regions will allow us to determine the boundaries of the regions and to define observables that separate the signal and background regions most efficiently.

Both because it is a non-trivial application, as well as still being tractable, we will present a detailed analysis of the phase space regions for boosted $Z$ identification.  This will require several pieces.  First, we will study the full phase space of perturbative jets defined by $\ecfnobeta{2}$ and $\ecfnobeta{3}$ and identify signal and background regions via power counting. This will lead us to define a discriminating variable, $\Dobs{2}{\beta}$. Second, any realistic application of a boosted $Z$ tagger includes a cut on the jet mass in the window around $m_Z$, and the effect of the mass cut on the discrimination power can also be understood by a power counting analysis of the phase space.  Third, at the high luminosities of the LHC, contamination from pile-up is important and can substantially modify distributions for jet substructure variables.  By appropriate power counting of the pile-up radiation, we can understand the effect of pile-up on the perturbative phase space and determine how susceptible the distributions of different discrimination variables are to pile-up contamination.  As a reference, throughout this section we will contrast the energy correlation functions to the $N$-subjettiness observables $\Nsub{1}{\beta}$ and ${\Nsub{2}{\beta}}$ \cite{Thaler:2010tr,Thaler:2011gf}.  A full effective theory analysis and analytic calculation of $\Dobs{2}{\beta}$ will be presented in  \Chap{chap:D2_anal}.

\subsection{Perturbative Radiation Phase Space}\label{sec:pert}

\begin{figure}
\begin{center}
\subfloat[]{\label{fig:unresolved_chap2}
\includegraphics[width=6cm]{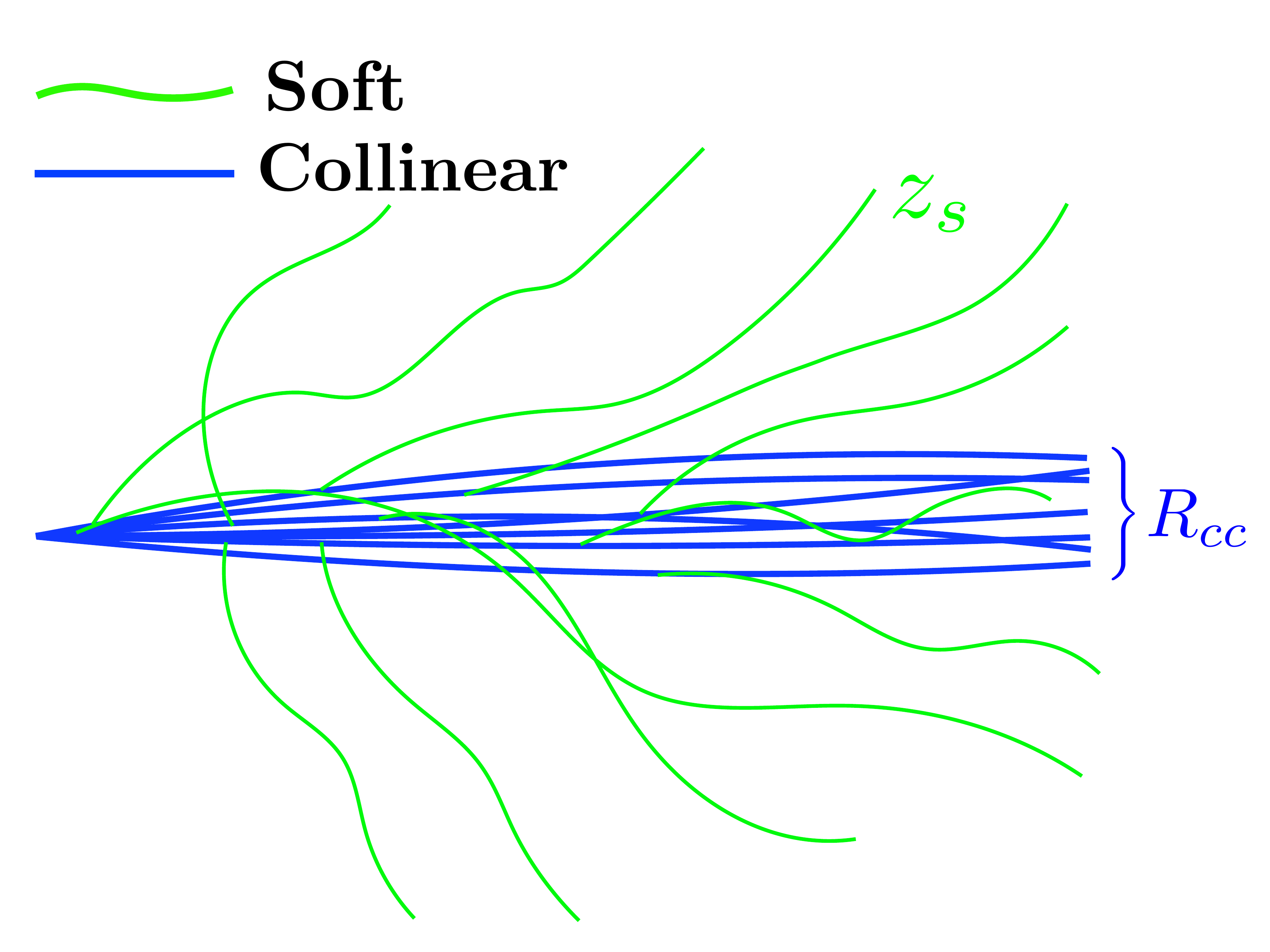}    
}\qquad
\subfloat[]{\label{fig:NINJA_chap2}
\includegraphics[width=6cm]{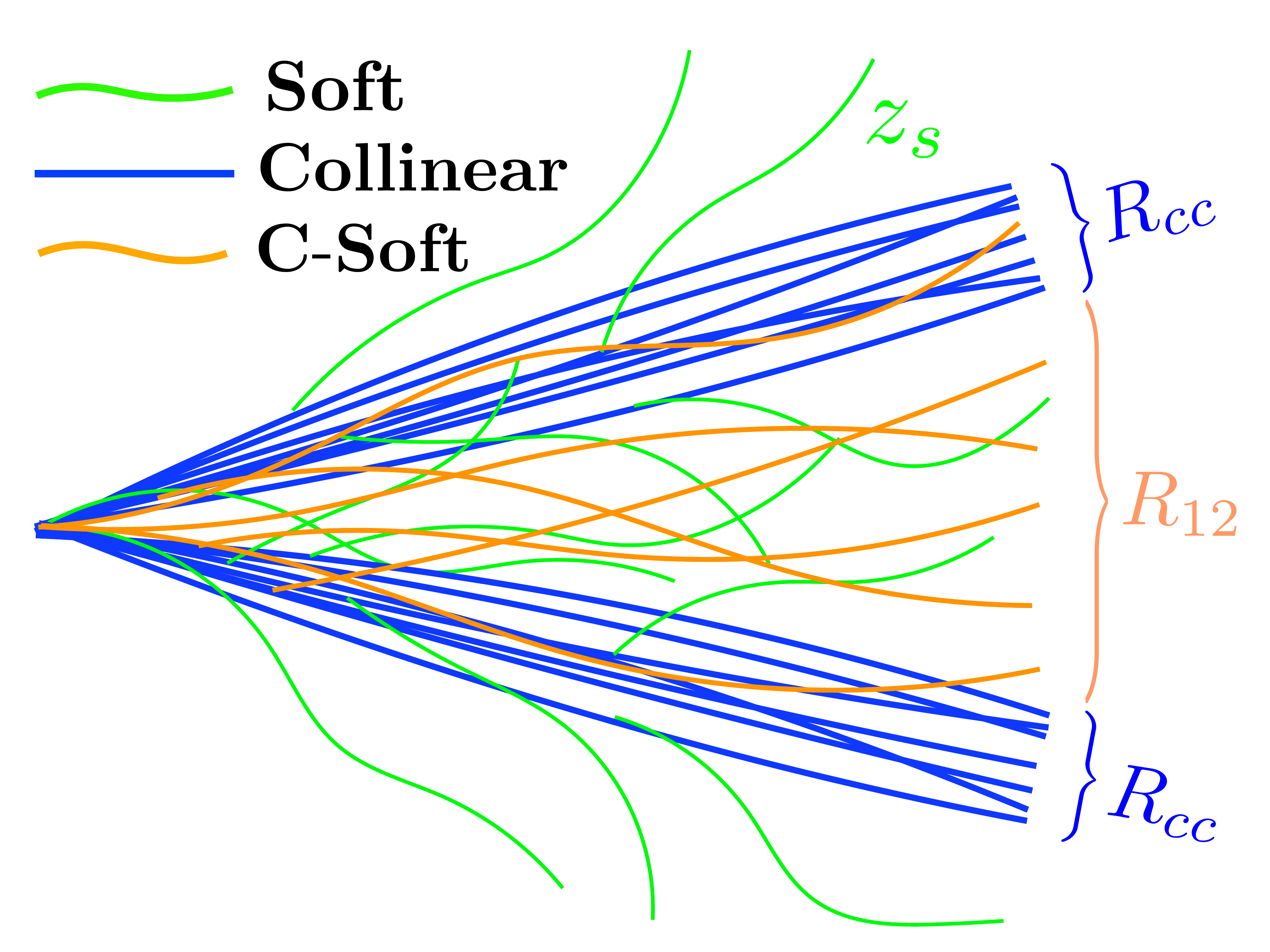}
}
\end{center}
\caption{a) 1-prong jet, dominated by collinear (blue) and soft (green) radiation. The angular size of the collinear radiation is $R_{cc}$ and the $p_T$ fraction of the soft radiation is $z_s$.  b) 2-prong jet resolved into two subjets, dominated by collinear (blue), soft (green), and collinear-soft (orange) radiation emitted from the dipole formed by the two subjets.  The subjets are separated by an angle $R_{12}$  and the $p_T$ fraction of the collinear-soft radiation is $z_{cs}$.
}
\label{fig:pics_jets_chap2}
\end{figure}

We begin by studying the $(\ecfnobeta{2}, \ecfnobeta{3})$ phase space arising from perturbative radiation from the jet. The measurement of $\ecfnobeta{2}$ and $\ecfnobeta{3}$ on a jet can resolve at most two hard subjets. The phase space for the variables $\ecfnobeta{2}$ and $\ecfnobeta{3}$ is therefore composed of jets which are unresolved by the measurement, dominantly from the QCD background, and shown schematically in \Fig{fig:unresolved_chap2}, and jets with a resolved 2-prong structure, as from boosted $Z$ decays, shown schematically in \Fig{fig:NINJA_chap2}. We will find that the resolved and unresolved jets live in parametrically different regions of the phase space, and the boundary between the two regions can be understood from a power counting analysis.

First, consider the case of the measurement of $\ecf{2}{\beta}$ and $\ecf{3}{\beta}$ on a jet with a single hard core of radiation, as in \Fig{fig:unresolved_chap2}, which is dominated by soft radiation with characteristic $p_T$ fraction $z_s \ll 1$, and collinear radiation with a characteristic angular size $R_{cc}\ll1$.  All other scales are order-1 numbers that we will assume are equal to 1 without further discussion. With these assumptions, we are able to determine the scaling of the contributions to $\ecf{2}{\beta}$ and $\ecf{3}{\beta}$ from collections of soft and collinear particles.  The scalings are given in \Tab{tab:pc_chap2} for contributions from three collinear particles ($CCC$), two collinear and one soft particle ($CCS$), one collinear and two soft particles ($CSS$), and three soft particles ($SSS$).\footnote{The contributions in \Tab{tab:pc_chap2} are from \emph{any} subset of three particles in the jet. We do not single out an initial parton from which the others arise  as in a showering picture. }

\begin{table}[t]
\begin{center}
\begin{tabular}{c|c|c}
modes&$\ecf{2}{\beta}$ &$\ecf{3}{\beta}$ \\ 
\hline
$CCC$ & $R_{cc}^{\beta}$ & $R_{cc}^{3\beta}$ \\
$CCS$ & $R_{cc}^{\beta}+z_s$ &$z_s R_{cc}^{\beta}$ \\
$CSS$ & $z_s+z_s^2$ & $z_s^2$ \\
$SSS$ & $z_s^2$  & $z_s^3$
\end{tabular}
\end{center}
\caption{Scaling of the contributions of 1-prong jets to $\ecf{2}{\beta}$ and $\ecf{3}{\beta}$ from the different possible configurations of soft ($S$) and collinear ($C$) radiation.
}
\label{tab:pc_chap2}
\end{table}

Dropping those contributions that are manifestly power-suppressed, the two- and three-point energy correlation functions measured on 1-prong jets therefore scale like
\begin{align}
\ecf{2}{\beta} & \sim  R_{cc}^\beta + z_s \,, \\
\ecf{3}{\beta} & \sim R_{cc}^{3\beta}+z_s^2 + R_{cc}^\beta z_s\,.
\end{align}
To go further, we must determine the relative size of $z_s$ and $R_{cc}^\beta$.  There are two possibilities, depending on the region of phase space identified by the measurement: either $z_s$ makes a dominant contribution to $\ecfnobeta{2}$, or its contribution is power suppressed with respect to $R_{cc}^\beta$.  In the case that $z_s$ contributes to $\ecfnobeta{2}$, this immediately implies that $\ecfnobeta{3}\sim (\ecfnobeta{2})^2$, regardless of the precise scaling of $R_{cc}^\beta$.\footnote{Note that on the true upper boundary of the phase space, the assumption of strong ordering of emissions is broken.}  If instead $z_s$ gives a  subleading contribution compared to $R_{cc}^\beta$ in $\ecfnobeta{2}$, then $\ecfnobeta{3}\sim (\ecfnobeta{2})^3$.\footnote{The existence of a consistent power counting does not guarantee a factorization theorem. Indeed, while a factorization theorem exists on the quadratic boundary, it does not exist at leading power on the cubic boundary.} Therefore, from this simple analysis, we have shown that 1-prong jets populate the region of phase space defined by $(\ecfnobeta{2})^3 \lesssim \ecfnobeta{3} \lesssim (\ecfnobeta{2})^2$. Fascinatingly, this also implies that the relative values of $\ecfnobeta{2}$ and $\ecfnobeta{3}$ provide a direct probe of the ordering of emissions inside the jet, so that assumptions about the measured values of $\ecfnobeta{2}$ and $\ecfnobeta{3}$ are observable proxies for the ordering of emissions. The scaling of $R_{cc}$ and $z_s$ on each boundary of the phase space can then easily be determined, but will not be important for our discussion.

\begin{figure}
\begin{center}
\includegraphics[width=6.5cm]{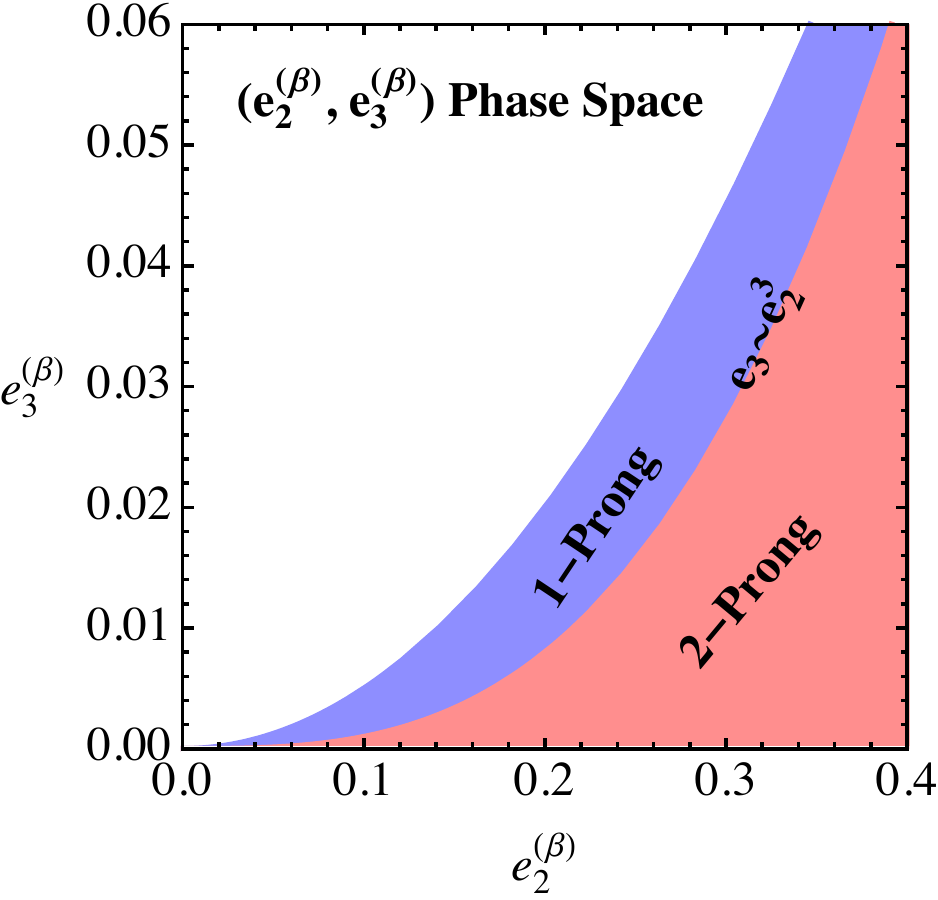}
\end{center}
\caption{ 
Phase space defined by the measurement of the energy correlation functions $ \ecfnobeta{2}$ and $\ecfnobeta{3}$. The phase space is divided into 1- and 2-prong regions with a boundary corresponding to the curve $ \ecfnobeta{3}\sim (\ecfnobeta{2})^3$.
}
\label{fig:ps_nolines}
\end{figure}

This analysis shows that 1-prong jets fill out a non-trivial region in the  $(\ecfnobeta{2}, \ecfnobeta{3})$ phase space, and of particular interest for the design of discriminating observables is the fact that this region of phase space has a lower boundary. This region is shown in blue in \Fig{fig:ps_nolines}. To understand the region of phase space for $\ecfnobeta{3} \ll  (\ecfnobeta{2}) ^3$ we must consider the case in which the measurement of $\ecfnobeta{2}$ and $\ecfnobeta{3}$ resolves two subjets within the jet.

The setup for the power counting of 2-prong jets is illustrated in \Fig{fig:NINJA_chap2}.  We consider a jet with two subjets, each of which carry ${\cal O}(1)$ of the jet $p_T$ and are separated by an angle $R_{12}\ll1$.  Each of the subjets has collinear emissions at a characteristic angle $R_{cc}\ll R_{12}$.  Because $R_{12}\ll 1$, there is in general global soft radiation at large angles with respect to the subjets with characteristic $p_T$ fraction $z_s\ll 1$.  For color-singlet jets, like boosted $Z$ bosons, this global soft radiation contribution comes purely from initial state radiation (ISR).\footnote{While this background is to a certain extent irreducible, given the important feature of ISR is its color-uncorrelated and soft nature, many of our observations about the effects of pile-up in \Sec{sec:pu} will be applicable to ISR.}  Finally, there is radiation from the dipole formed from the two subjets (called ``collinear-soft'' radiation), with characteristic  angle $R_{12}$ from the subjets, and with $p_T$ fraction $z_{cs}$.  The effective theory of this phase space region for the observable $N$-jettiness \cite{Stewart:2010tn} was studied in \Ref{Bauer:2011uc}.

We now consider the power counting of $\ecf{2}{\beta}$ and $\ecf{3}{\beta}$ for 2-prong jets. By the definition of this region of phase space, the hard splitting sets the value of $\ecf{2}{\beta}$. That is, we have $\ecf{2}{\beta}\sim R_{12}^\beta$, with all other contributions suppressed. For $\ecf{3}{\beta}$, it is clear that the leading contributions must arise from correlations between the two hard subjets with either the global soft, collinear or collinear-soft modes.  The scaling of these different contributions to $\ecf{3}{\beta}$ is given in \Tab{tab:pc_ninja}, from which we find that the scaling of the two- and three-point energy correlation functions for 2-pronged jets is
\begin{align}
\ecf{2}{\beta} & \sim  R_{12}^\beta \,, \\
\ecf{3}{\beta} & \sim R_{12}^{\beta}z_s+R_{12}^{2\beta} R_{cc}^{\beta}+R_{12}^{3\beta}z_{cs}\,.
\end{align}
There is no measurement performed to distinguish the three contributions to $\ecf{3}{\beta}$ and so we must assume that they all scale equally.

\begin{table}[t]
\begin{center}
\begin{tabular}{c|c}
modes &$\ecf{3}{\beta}$ \\ 
\hline
$C_1C_2\,S$ & $R_{12}^{\beta}z_s$  \\
$C_1C_2\, C$ & $R_{12}^{2\beta} R_{cc}^{\beta}$  \\
$C_1C_2\,C_s$ & $R_{12}^{3\beta}z_{cs}$  
\end{tabular}
\end{center}
\caption{Scaling of the contributions from global soft ($S$), collinear ($C$), and collinear-soft ($C_s$) radiation correlated with the two hard subjets (denoted by $C_1$ and $C_2$) in 2-prong jets to  $\ecf{3}{\beta}$ from the different possible configurations.
}
\label{tab:pc_ninja}
\end{table}

This result is sufficient to set the relative scaling of $\ecf{2}{\beta}$ and $\ecf{3}{\beta}$.  As we assume that the jet only has two hard subjets, we have that $z_{cs}\ll1$ and so
\begin{equation}
(\ecfnobeta{2})^3 \sim R_{12}^{3\beta} \gg R_{12}^{3\beta}z_{cs} \sim \ecfnobeta{3} \ ,
\end{equation}
which defines the 2-prong jet region of phase space as that for which $\ecfnobeta{3} \ll (\ecfnobeta{2})^3$.  With this identification, note the scaling of the various modes:
\begin{align}
R_{12}^\beta \sim \ecfnobeta{2}\, , \quad z_s \sim \frac{\ecfnobeta{3}}{\ecfnobeta{2}} \, , \quad R_{cc}^\beta \sim \frac{\ecfnobeta{3}}{(\ecfnobeta{2})^2} \, , \quad z_{cs} \sim \frac{\ecfnobeta{3}}{(\ecfnobeta{2})^3} \, .
\end{align}
While not important for our goals here, the fact that the energy correlation functions parametrically separate the scaling of the modes that contribute to the observables is vital for an effective theory analysis and calculability. This will be discussed in detail in \Chap{chap:D2_anal}, where an explicit calculation of the $D_2$ observable will be presented.  Note that because $\ecfnobeta{2}$ is first non-zero at a lower order in perturbation theory than $\ecfnobeta{3}$, $\ecfnobeta{3}$ can be zero while $\ecfnobeta{2}$ is non-zero.  Therefore, this 2-prong region of phase space extends down to the kinematic limit of $\ecfnobeta{3}  = 0$, as shown in red in \Fig{fig:ps_nolines}.

This power counting analysis, although very simple in nature, provides a powerful picture of the phase space defined by the measurement of $\ecfnobeta{2}$ and $\ecfnobeta{3}$, which is shown in \Fig{fig:ps_nolines}.  The 1- and 2-prong jets are defined to populate the phase space regions where
\begin{align*}
\text{1-prong jet: }& (\ecfnobeta{2})^3\lesssim \ecfnobeta{3} \lesssim (\ecfnobeta{2})^2\, ,\\
\text{2-prong jet: }& 0< \ecfnobeta{3} \ll (\ecfnobeta{2})^3\, .
\end{align*}
Background QCD jets dominantly populate the 1-prong region of phase space, while signal boosted $Z$ decays dominantly populate the 2-prong region of phase space.  This has important consequences for the optimal discrimination observable.  

An interesting observation about the boundary between 1- and 2-prong jets, defined by $ \ecfnobeta{3}\sim (\ecfnobeta{2})^3$, is that it is approximately invariant to boosts along the jet direction.  For a narrow jet, a boost along the jet direction by an amount $\gamma$ scales $p_T$s and angles as
\begin{equation}
p_T\to \gamma p_T \ ,\qquad R \to \gamma^{-1} R \ .
\end{equation}
Therefore, under a boost, $\ecf{2}{\beta}$ and $\ecf{3}{\beta}$ scale as
\begin{equation}
\ecf{2}{\beta} \to \gamma^{-\beta}\ecf{2}{\beta} \ , \qquad \ecf{3}{\beta} \to \gamma^{-3\beta} \ecf{3}{\beta} \ .
\end{equation}
Thus, the boundary between 1- and 2-prong jets, where $ \ecfnobeta{3}\sim (\ecfnobeta{2})^3$, is invariant to  boosts along the jet direction.  That is, under boosts, a jet will move along a contour of constant $ \ecfnobeta{3}/ (\ecfnobeta{2})^3$ in the ($\ecfnobeta{2},\ecfnobeta{3}$) plane.  

The analysis presented in this section is also the initial step in establishing rigorous factorization theorems in the different regions of phase space, allowing for analytic resummation of the double differential cross section of $\ecfnobeta{2}$ and $\ecfnobeta{3}$. Such a factorization and resummation will be presented in \Chap{chap:D2_anal}, building on the analysis of this chapter.

\subsubsection{Optimal Discrimination Observables}
\label{sec:opte2e3}

\begin{figure}
\begin{center}
\subfloat[]{\label{fig:e2e3_ps}
\includegraphics[width=6.5cm]{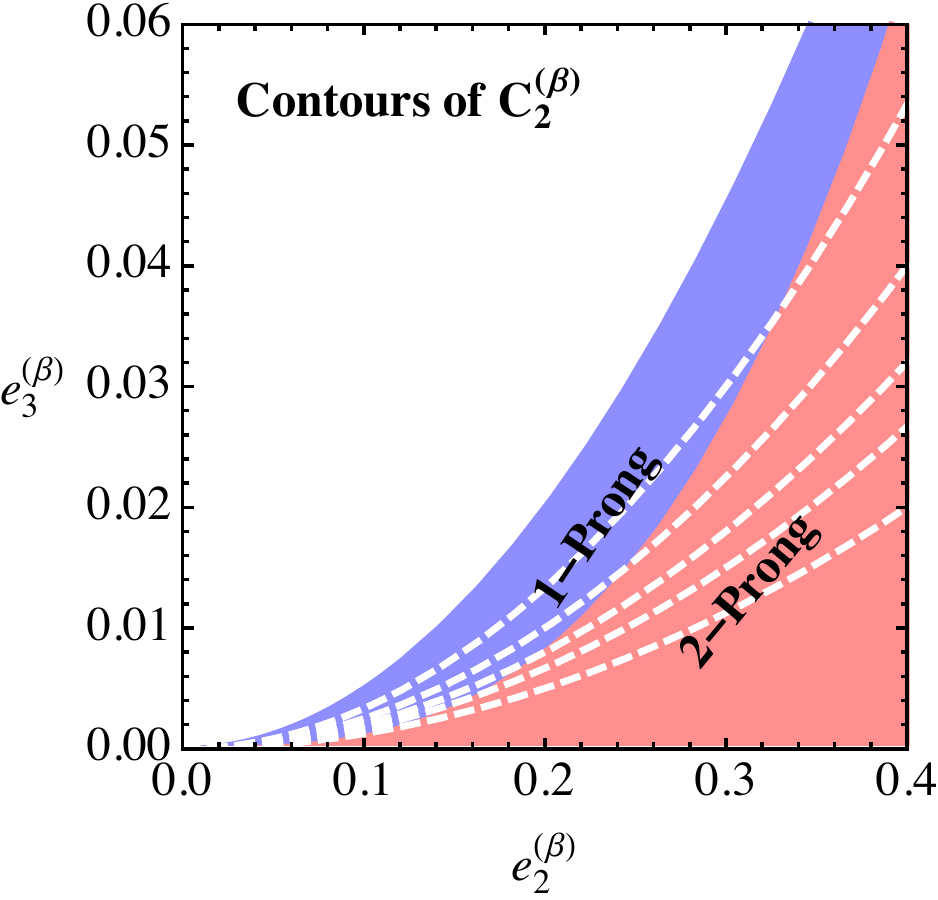}
}\qquad
\subfloat[]{\label{fig:contours_ps}
\includegraphics[width=6.5cm]{figures/D2_ps.pdf}
}
\end{center}
\caption{ 
Contours of constant $\Cobs{2}{\beta}$ (left) and $\Dobs{2}{\beta}$ (right) in the phase space defined by $\ecf{2}{\beta},\ecf{3}{\beta}$.  The 1- and 2-prong regions of phase space are labeled, with their boundary corresponding to the curve $ \ecfnobeta{3}\sim (\ecfnobeta{2})^3$.
}
\label{fig:ps}
\end{figure}

The fact that the signal and background regions of phase space are parametrically separated implies that from power counting alone, we can determine the optimal observable for separating signal from background.  Because the boundary between the background-rich and signal-rich regions is $ \ecfnobeta{3}\sim (\ecfnobeta{2})^3$, this suggests that the optimal observable for discriminating boosted $Z$ bosons from QCD jets is\footnote{We thank Jesse Thaler for suggesting the notation ``$D$'' for these observables. Unlike $\Cobs{2}{\beta}$, whose name was motivated by its relation to the classic $e^+e^-$ event shape parameter C, $\Dobs{2}{\beta}$ is not related to the D parameter. }
\begin{equation}
\Dobs{2}{\beta} \equiv \frac{\ecf{3}{\beta}}{\left(
\ecf{2}{\beta}
\right)^3} \, .
\end{equation}
Signal jets will be characterized by a small value of $\Dobs{2}{\beta}$, while background jets will predominantly have large $\Dobs{2}{\beta}$.  With this observable, parametrically there is no mixing of the signal-rich and background-rich regions. Contours of constant $D_2^{(\beta)}$ lie entirely in the signal or background region, as is shown schematically in \Fig{fig:ps}.  Determining the precise discrimination power of $\Dobs{2}{\beta}$ requires an understanding of the ${\cal O}(1)$ details of the distributions of signal and background, beyond any purely power counting analysis.

The observation that the scaling relation $ \ecfnobeta{3}\sim (\ecfnobeta{2})^3$ is boost invariant provides further motivation for the variable $\Dobs{2}{\beta}$. Under boosts along the jet axis, jets can move along curves of constant $\Dobs{2}{\beta}$, but cannot cross the boundary between the 2-prong and 1-prong regions of phase space. This can be used to give a boost invariant definition of a 2-prong jet, as a jet with a small value of $\Dobs{2}{\beta}$, and a 1-prong jet, as a jet with large $\Dobs{2}{\beta}$.

\Ref{Larkoski:2013eya} used the two- and three-point energy correlation functions in the combination
\begin{equation}
C_2^{(\beta)} \equiv  \frac{\ecf{3}{\beta}}{\left({\ecf{2}{\beta}}\right)^2}
\end{equation}
for boosted $Z$ boson discrimination.  From the power counting analysis in this section, this variable is not a natural choice.  In particular, contours of constant $C_2^{(\beta)}$ pass through both the 1-prong and 2-prong regions of phase space, mixing the signal and background for any value of $C_2^{(\beta)}$, as shown in \Fig{fig:ps}. Therefore, from the power counting perspective, we would expect that $C_2^{(\beta)}$ is a poor boosted $Z$ boson discriminating observable.  Nevertheless, \Ref{Larkoski:2013eya} found that with a tight jet mass cut, and in the absence of pile-up, $C_2^{(\beta)}$ is a powerful boosted $Z$ discriminant.  A mass cut constrains the phase space significantly, which we will discuss in detail in \Sec{sec:masscute2e3}, allowing us to understand the result of \Ref{Larkoski:2013eya}. Pile-up will be addressed in \Sec{sec:pu}. 

It is important to recall that while $\ecf{2}{\beta}$ and $\ecf{3}{\beta}$ are IRC safe observables, so that their phase space can be analyzed with power counting techniques, ratios of IRC safe observables are not in general IRC safe \cite{Soyez:2012hv,Larkoski:2013paa,Larkoski:2014bia}. The observables $\Cobs{2}{\beta}$ and $\Dobs{2}{\beta}$ are however Sudakov safe \cite{Larkoski:2013paa,Larkoski:2014bia}, and therefore can be reliably studied with Monte Carlo simulation without applying any form of additional cut, such as a jet mass cut, on the phase space.

\subsubsection{Contrasting with $N$-subjettiness}
\label{sec:nsubcomp}

At this point, it is interesting to apply the power counting analysis to other observables for boosted $Z$ discrimination and see what conclusions can be made.  For concreteness, we will contrast the energy correlation functions with the $N$-subjettiness observables $\Nsub{1}{\beta}$ and $\Nsub{2}{\beta}$, defined as
\begin{equation}
\Nsub{N}{\beta} = \frac{1}{p_{TJ}}\sum_{1\leq i \leq n_J} p_{Ti}\min\left\{
R_{i1}^\beta,\dotsc,R_{iN}^\beta
\right\} \ .
\end{equation}
As with the energy correlation functions, we will consider $\Nsub{1}{\beta}$ and $\Nsub{2}{\beta}$ as measured on 1-prong and 2-prong jets and determine the regions of phase space where background and signal jets populate.  This can then be used to determine the optimal observable for boosted $Z$ discrimination from the $N$-subjettiness observables. We use the same notation for the scalings of the modes as in \Sec{sec:pert}.

\begin{table}[t]
\begin{center}
\begin{tabular}{c|c|c}
modes&$\Nsub{1}{\beta}$ &$\Nsub{2}{\beta}$ \\ 
\hline
$CCC$ & $R_{cc}^{\beta}$ & $R_{cc}^{\beta}$ \\
$CCS$ & $R_{cc}^{\beta}+z_s$ &$R_{cc}^\beta+z_s$ \\
$CSS$ & $z_s$ & $z_s$ \\
$SSS$ & $z_s$  & $z_s$
\end{tabular}
\end{center}
\caption{Scaling of the contributions of 1-prong jets to $\Nsub{1}{\beta}$ and $\Nsub{2}{\beta}$ from the different possible configurations of soft ($S$) and collinear ($C$) radiation.
}
\label{tab:pc_nsub}
\end{table}

Starting with 1-prong jets, and repeating the analysis of \Sec{sec:pert}, we find the dominant contributions to $\Nsub{1}{\beta}$ and $\Nsub{2}{\beta}$ as given in \Tab{tab:pc_nsub}.  For the configuration of two collinear particles and a soft particle ($CCS$), $\Nsub{2}{\beta}$ is either dominated by $z_s$ or by $R_{cc}^\beta$.  In this configuration, the two subjet axes can either lie on the two collinear particles or one axis can be on a collinear particle and the other on a soft particle.  Importantly, the measurement of $\Nsub{1}{\beta}$ and $\Nsub{2}{\beta}$ cannot distinguish these two possibilities and therefore cannot determine if the second axis in the 1-prong jet is at a small or large angle with respect to the first.\footnote{For this reason, soft and collinear contributions to $\Nsub{2}{\beta}$ on 1-prong jets do not factorize and therefore cannot be computed in SCET.}  With either configuration, $\Nsub{1}{\beta}$ and $\Nsub{2}{\beta}$ scale as
\begin{align}
\Nsub{1}{\beta} & \sim  R_{cc}^\beta + z_s \,, \\
\Nsub{2}{\beta} & \sim R_{cc}^\beta + z_s\,.
\end{align}
That is, for 1-prong jets, $\Nsub{1}{\beta}\sim\Nsub{2}{\beta}$.

\begin{table}[t]
\begin{center}
\begin{tabular}{c|c}
modes &$\Nsub{2}{\beta}$ \\ 
\hline
$C_1C_2\,S$ & $z_s$  \\
$C_1C_2\, C$ & $R_{cc}^{\beta}$  \\
$C_1C_2\,C_s$ & $R_{12}^{\beta}z_{cs}$  
\end{tabular}
\end{center}
\caption{Scaling of the contributions from global soft ($S$), collinear ($C$), and collinear-soft ($C_s$) radiation correlated with the two hard subjets (denoted by $C_1$ and $C_2$) in 2-prong jets to  $\Nsub{2}{\beta}$ from the different possible configurations.
}
\label{tab:pc_ninja_nsub}
\end{table}

For 2-prong jets, $\Nsub{1}{\beta}$ is dominated by the hard splitting, as was the case with the two-point energy correlation function, hence $\Nsub{1}{\beta}\sim R_{12}^\beta$.  For $\Nsub{2}{\beta}$, the two axes lie along the two hard prongs, so, just like with the three-point energy correlation function, $\Nsub{2}{\beta}$ is set by the radiation about those two hard prongs: global soft, collinear, or collinear soft.  \Tab{tab:pc_ninja_nsub} lists the contributions to $\Nsub{2}{\beta}$ from each of these modes, leading to the scaling
\begin{align}
\Nsub{1}{\beta} &\sim R_{12}^\beta \,, \\
\Nsub{2}{\beta} &\sim z_s +R_{cc}^\beta + R_{12}^\beta z_{cs}\, .
\end{align}
Demanding that the jet only has two hard prongs implies that $\Nsub{2}{\beta}\sim z_s\sim R_{cc}^\beta\sim R_{12}^\beta z_{cs} \ll R_{12}^\beta\sim \Nsub{1}{\beta}$, but no other conclusions can be made from power counting alone.  Unlike the well-defined division of phase space by the energy correlation functions, $N$-subjettiness has a much weaker division of 
\begin{align*}
\text{1-prong jet: }& \Nsub{2}{\beta}\sim \Nsub{1}{\beta}\, ,\\
\text{2-prong jet: }& \Nsub{2}{\beta}\ll \Nsub{1}{\beta}\, .
\end{align*}
This does suggest, however, that the optimal discrimination variable using $N$-subjettiness is $\Nsub{2,1}{\beta}\equiv\Nsub{2}{\beta}/ \Nsub{1}{\beta}$, which is what is widely used experimentally.  Nevertheless, the weaker phase space separation of $N$-subjettiness compared with that for the energy correlation functions would na\"ively imply that $\ecfnobeta{2}$ and $\ecfnobeta{3}$ provides better discrimination than $\Nsub{1}{\beta}$ and $\Nsub{2}{\beta}$; however, this statement requires an understanding of $\mathcal{O}(1)$ numbers, which is beyond the scope of a power counting analysis.

\subsubsection{Effect of a Mass Cut}
\label{sec:masscute2e3}

\begin{figure}
\begin{center}
\subfloat[]{\label{fig:beq2_C2}
\includegraphics[width=6.5cm]{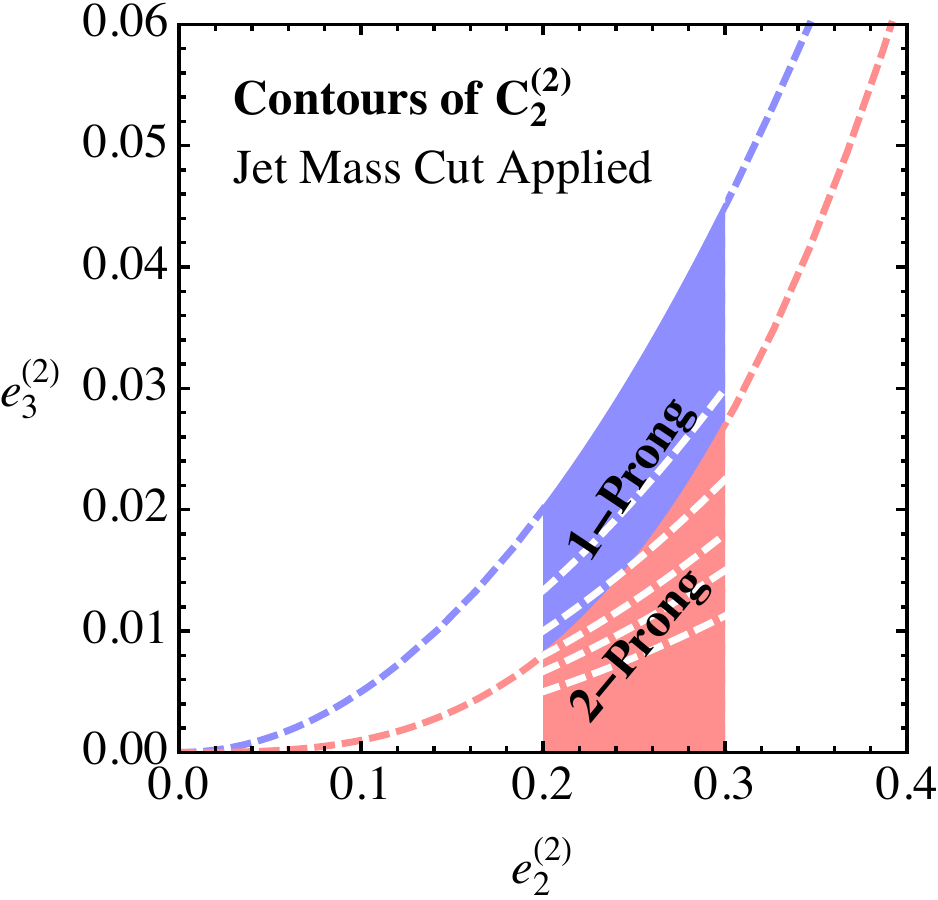}
}\qquad
\subfloat[]{\label{fig:beq2_D2}
\includegraphics[width=6.5cm]{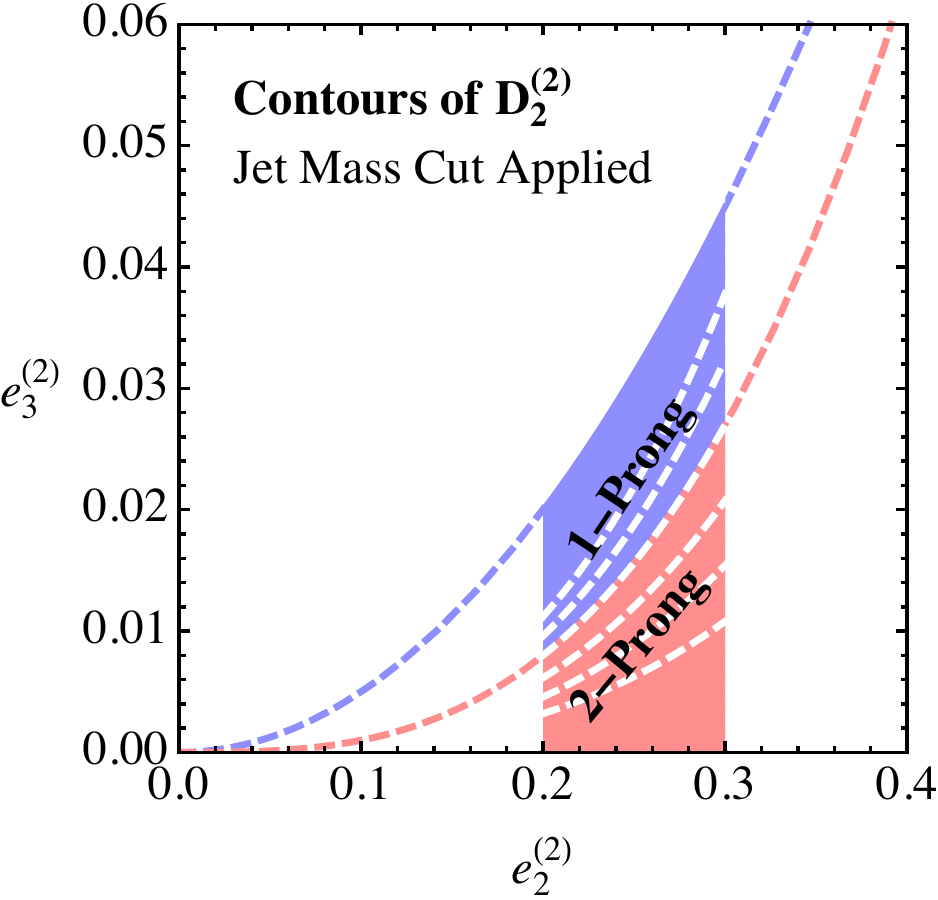}
}
\end{center}
\caption{Phase space defined by the energy correlation functions $\ecf{2}{2},\ecf{3}{2}$ in the presence of a mass cut. Contours of constant $\Cobs{2}{2}$ (left) and $\Dobs{2}{2}$ (right) are shown for reference. 
}
\label{fig:mcut}
\end{figure}

In an experimental application of $\Dobs{2}{\beta}$ to boosted Z discrimination, a mass cut is performed on the jet around the mass of the $Z$ boson.  In addition to removing a large fraction of the background, this cut also guarantees that the identified jets are actually generated from boosted $Z$ decays.  To fully understand the effect of the mass cut on the phase space requires analyzing the three-dimensional phase space of the mass, $\ecfnobeta{2}$, and $\ecfnobeta{3}$.  While complete, this full analysis would be distracting to the physics points that we wish to make in this section, and the impact of the mass cut can be understood without performing this analysis.  For $\beta = 2$, the two-point energy correlation function is simply related to the jet mass $m$ at fixed jet $p_T$:
\begin{equation}
\ecf{2}{2} \simeq \frac{m^2}{p_T^2} \ ,
\end{equation}
for central jets assuming that $m \ll p_T$ and up to overall factors of order 1.  Therefore, a cut on the jet mass is a cut on $\ecf{2}{2}$.  In this section, we will begin by discussing the simpler case of $\beta=2$, and then proceed to comment on the effect of a mass cut for general $\beta$.

The phase space in the $\ecf{2}{2},\ecf{3}{2}$ plane with the jet mass constrained to a window, and for some finite range of jet $p_T$ is shown schematically in \Fig{fig:mcut}.  Jets of a given mass can have that mass generated either by substantial soft radiation (for 1-prong jets) or by a hard splitting in the jet (a 2-prong jet), and so we want a discrimination observable that separates these two regions cleanly.  The boundary between the 1-prong and 2-prong jet regions is still defined by $\ecf{3}{2}\sim (\ecf{2}{2})^3$, and so we expect $\Dobs{2}{2}$ to be the most powerful discriminant.  However, by making a mass cut, the region of phase space at small masses, dominated by 1-prong jets, is removed.  Therefore, the fact that contours of the observable $\Cobs{2}{2}$ mix both 1- and 2-prong jets is much less of an issue.  Except at very high signal efficiencies, when one is sensitive to the functional form of the boundary between the signal and background regions, the discrimination performance of $\Cobs{2}{2}$ should be similar to that of $\Dobs{2}{2}$ when a tight mass cut is imposed. Indeed, in a sufficiently narrow window, any variable of the form $\ecf{3}{2}/(\ecf{2}{2})^n$, would provide reasonable discrimination, with all the discrimination power coming from $\ecf{3}{2}$ alone.  However, $\Dobs{2}{2}$ has the advantage that its discrimination power does not suffer from significant dependence on the value of the lower mass cut.

\begin{figure}
\begin{center}
\subfloat[]{\label{fig:bneq2_C2}
\includegraphics[width=6.5cm]{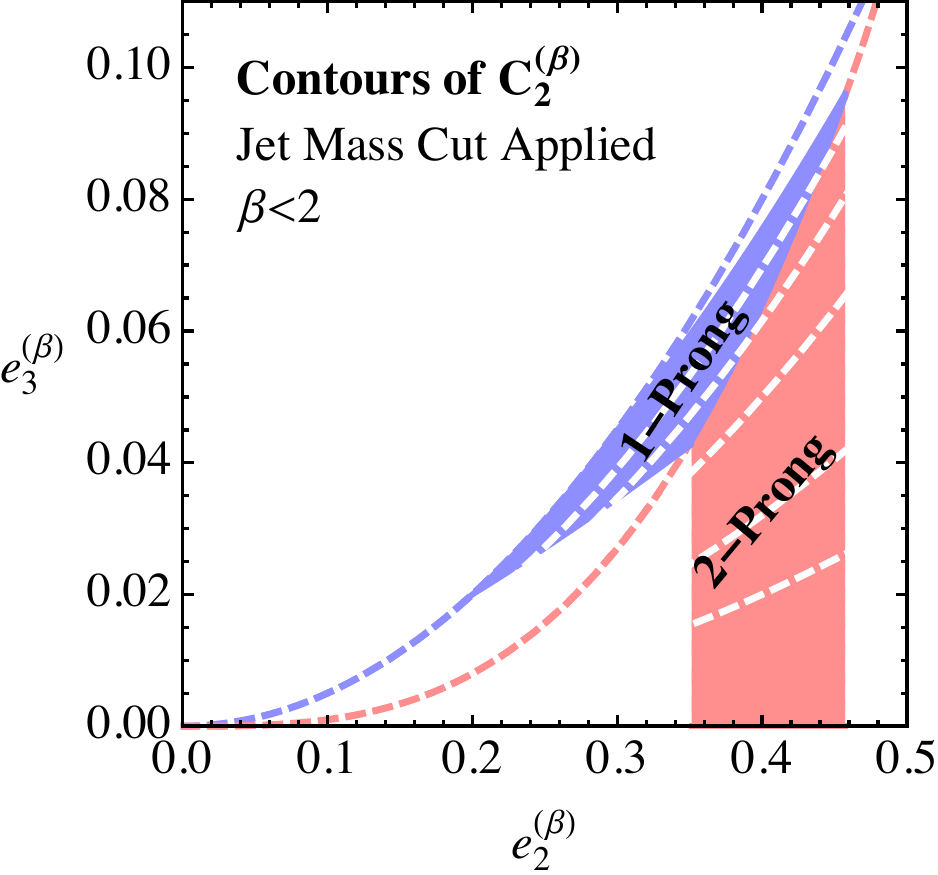}
}\qquad
\subfloat[]{\label{fig:bneq2_D2}
\includegraphics[width=6.5cm]{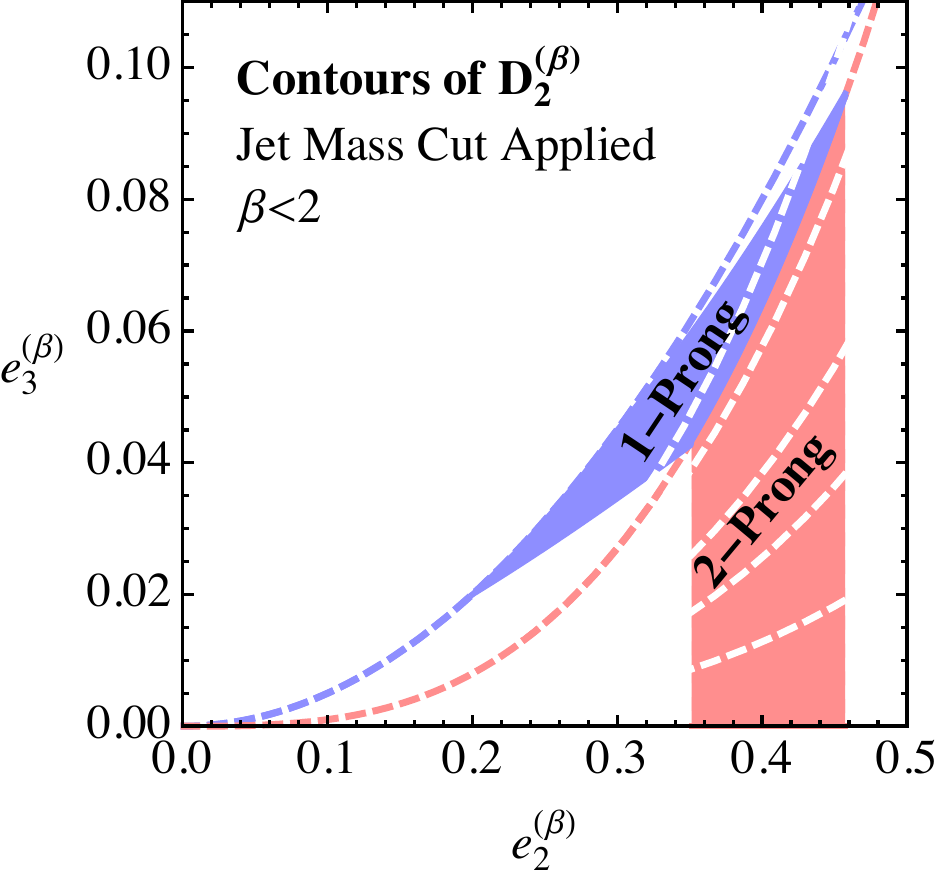}
}
\end{center}
\caption{Phase space defined by the energy correlation functions $\ecf{2}{\beta},\ecf{3}{\beta}$, for $\beta<2$, in the presence of a mass cut. Contours of constant $\Cobs{2}{2}$ (left) and $\Dobs{2}{2}$ (right) are shown for reference.
}
\label{fig:mcut_bneq2}
\end{figure}

While a lower mass cut is important for removing 1-prong background jets, an upper mass cut is also necessary for powerful discrimination.  The mass distribution of QCD jets has a long tail extending to masses of order the $p_T$ of the jet.  For these jets, the mass is generated by an honest hard splitting, and so these background jets look exactly like the signal from their substructure.  While the cross section for these high mass QCD jets is suppressed by $\alpha_s$, they can still be a significant background and therefore should be removed.  

Let's now consider the general $\beta$ case.  We will first consider the effect of a mass cut in the 2-prong region of the $\ecf{2}{\beta},\ecf{3}{\beta}$ plane.  Recall that in this region of phase space
\begin{equation}
\ecf{2}{\beta} \sim R_{12}^\beta \ ,
\end{equation}
where $R_{12}$ is the angle between the hard subjets.  Therefore, in this region of phase space $\ecf{2}{\beta}$ is simply related to the mass:
\begin{equation}
\ecf{2}{\beta} \sim \left(
\frac{m^2}{p_T^2}
\right)^{\beta/2} \ .
\end{equation}
A cut on the jet mass is therefore equivalent to an appropriate cut on $\ecf{2}{\beta}$ for 2-prong jets.

A mass cut in the 1-prong region of phase space is more subtle, as the dominant contributing mode to $\ecf{2}{\beta}$ changes throughout the phase space.  Recall that in this region, $\ecf{2}{\beta}$ has the scalings
\begin{align}
\ecf{2}{\beta}&\sim R_{cc}^\beta +z_s\,.
\end{align}
while
\begin{align}
 \ecf{2}{2}&\sim\frac{m^2}{p_T^2}\sim R_{cc}^2 +z_s\,.
\end{align}
While the soft contributions have the same scaling for both variables, the collinear contributions do not. There are two possibilities as for the relative scalings of $\ecf{2}{\beta}$ and the mass: if soft emissions do not contribute, then
\begin{equation}\label{eq:nosoft1prong}
\ecf{2}{\beta} \sim \left(
\frac{m^2}{p_T^2}
\right)^{\beta/2} \ ,
\end{equation}
which matches onto the relative scaling in the 2-prong region of phase space.  If instead soft emissions do contribute, then
\begin{equation}\label{eq:yessoft1prong}
\ecf{2}{\beta} \sim 
\frac{m^2}{p_T^2}\ ,
\end{equation}
which defines the upper boundary of the 1-prong phase space.  These phase space boundaries for jets on which two two-point energy correlation functions with different angular exponents (or recoil-free angularities \cite{Larkoski:2014uqa}) are measured is discussed in detail in \Ref{Larkoski:2014tva}.

Depending on whether $\beta$ is less than or greater than 2, the mass cut manifests itself differently.  For $\beta < 2$, note that from \Eqs{eq:nosoft1prong}{eq:yessoft1prong}, $\ecf{2}{\beta} > \ecf{2}{2}$ in the two-prong region, and so smaller values of $\ecf{2}{\beta}$ can correspond to the same mass.  Conversely, for $\beta > 2$, $\ecf{2}{\beta} < \ecf{2}{2}$ and so larger values of $\ecf{2}{\beta}$ can correspond to the same mass.  The effect of a mass cut on the allowed phase space for $\beta < 2$ is illustrated schematically in \Fig{fig:mcut}.  Because in this case small values of $\ecf{2}{\beta}$ can satisfy the mass cut, contours of $\Cobs{2}{\beta}$ can pass through the background region of phase space and significantly reduce the discrimination power.  Again, because it respects the parametric scaling of the phase space boundaries, we expect the discrimination power of $\Dobs{2}{\beta}$ to be more robust as $\beta$ decreases from 2.  However, the precise discrimination power depends on understanding the ${\cal O}(1)$ region around the 1-prong and 2-prong jet boundary as $\beta$ moves away from 2.  This observation also explains why \Ref{Larkoski:2013eya} found that the optimal choice for boosted $Z$ boson discrimination using $\Cobs{2}{\beta}$ with a tight mass cut was $\beta \simeq 2$.

\subsubsection{Summary of Power Counting Predictions}
\label{sec:sumpredict}

Here, we summarize the main predictions from our power counting analysis of boosted $Z$ discrimination, before a Monte Carlo study in \Sec{sec:mc_e2e3}.  We have:
\begin{itemize}

\item The parametric scaling of the boundary between 1-prong and 2-prong jets in the $(\ecfnobeta{2},\ecfnobeta{3})$ phase space is $\ecfnobeta{3}\sim (\ecfnobeta{2})^3$.  Therefore, $\Dobs{2}{\beta}$ should be a more powerful discrimination observable than $\Cobs{2}{\beta}$ because contours of constant $\Dobs{2}{\beta}$ do not mix signal and background regions, while contours of $\Cobs{2}{\beta}$ do mix signal and background regions.

\item When a mass cut is imposed on the jet, $\Cobs{2}{\beta}$ should have similar discrimination power to $\Dobs{2}{\beta}$, for $\beta \simeq 2$, except at high signal efficiency when the observable is sensitive to the boundary between the signal and background regions.  At high signal efficiency, $\Dobs{2}{\beta}$ should be a slightly better discriminant than $\Cobs{2}{\beta}$ for $\beta \simeq 2$.

\item The discrimination power of $\Cobs{2}{\beta}$ should decrease substantially as $\beta$ decreases from 2 when there is a mass cut on the jets.  By contrast, the discrimination power of $\Dobs{2}{\beta}$ should be more robust as $\beta$ decreases from 2.

\item The power counting predictions stated above should be robust to Monte Carlo tuning and reproduced by any Monte Carlo simulation, e.g. \herwigpp\ or \pythia{8}, since they are determined by parametric scaling of QCD dynamics.
\end{itemize}

\subsubsection{Monte Carlo Analysis}
\label{sec:mc_e2e3}

\begin{figure}
\begin{center}
\subfloat[]{
\includegraphics[width=6.5cm]{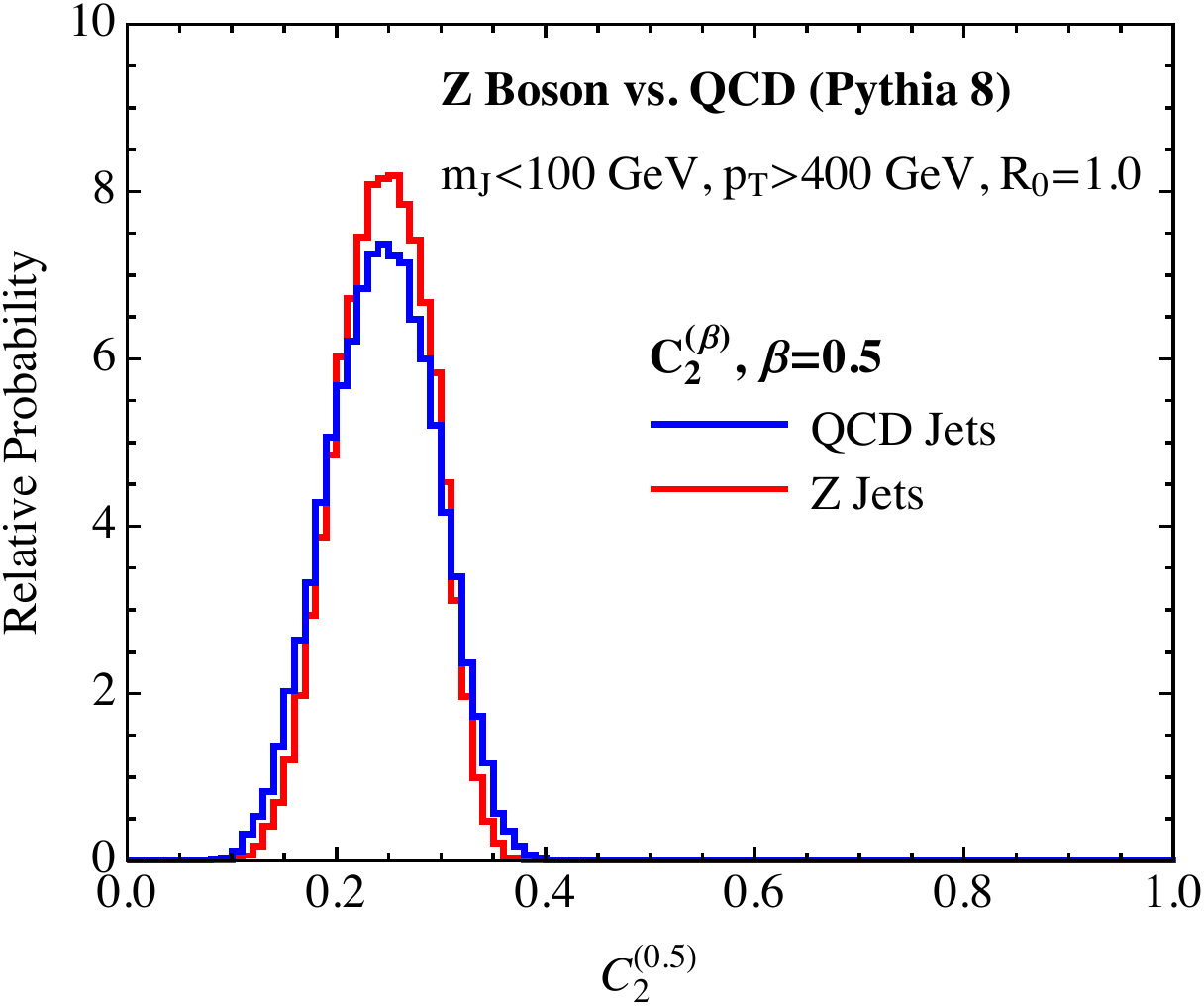}
}\qquad
\subfloat[]{
\includegraphics[width=6.5cm]{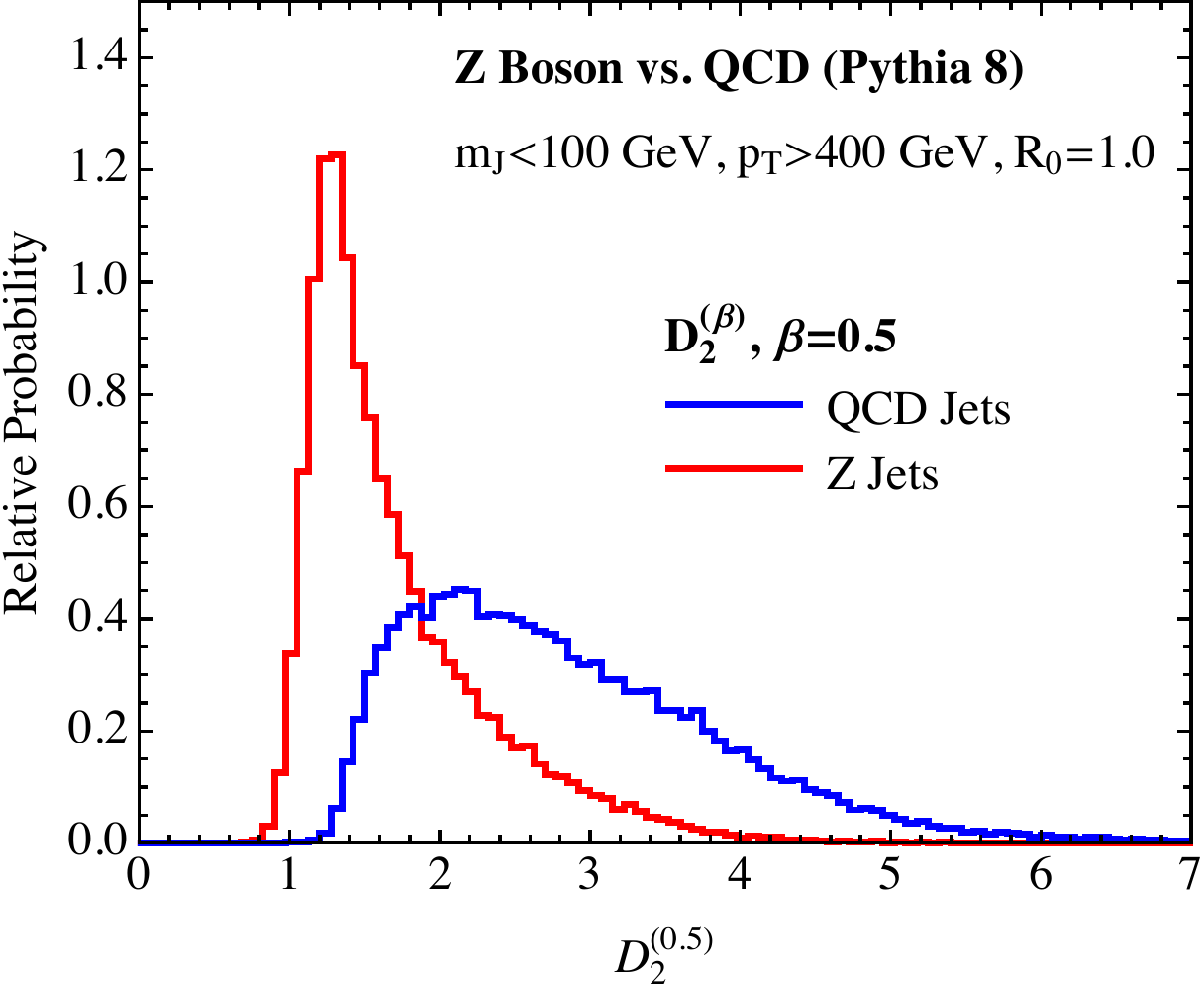}
}\qquad
\subfloat[]{
\includegraphics[width=6.5cm]{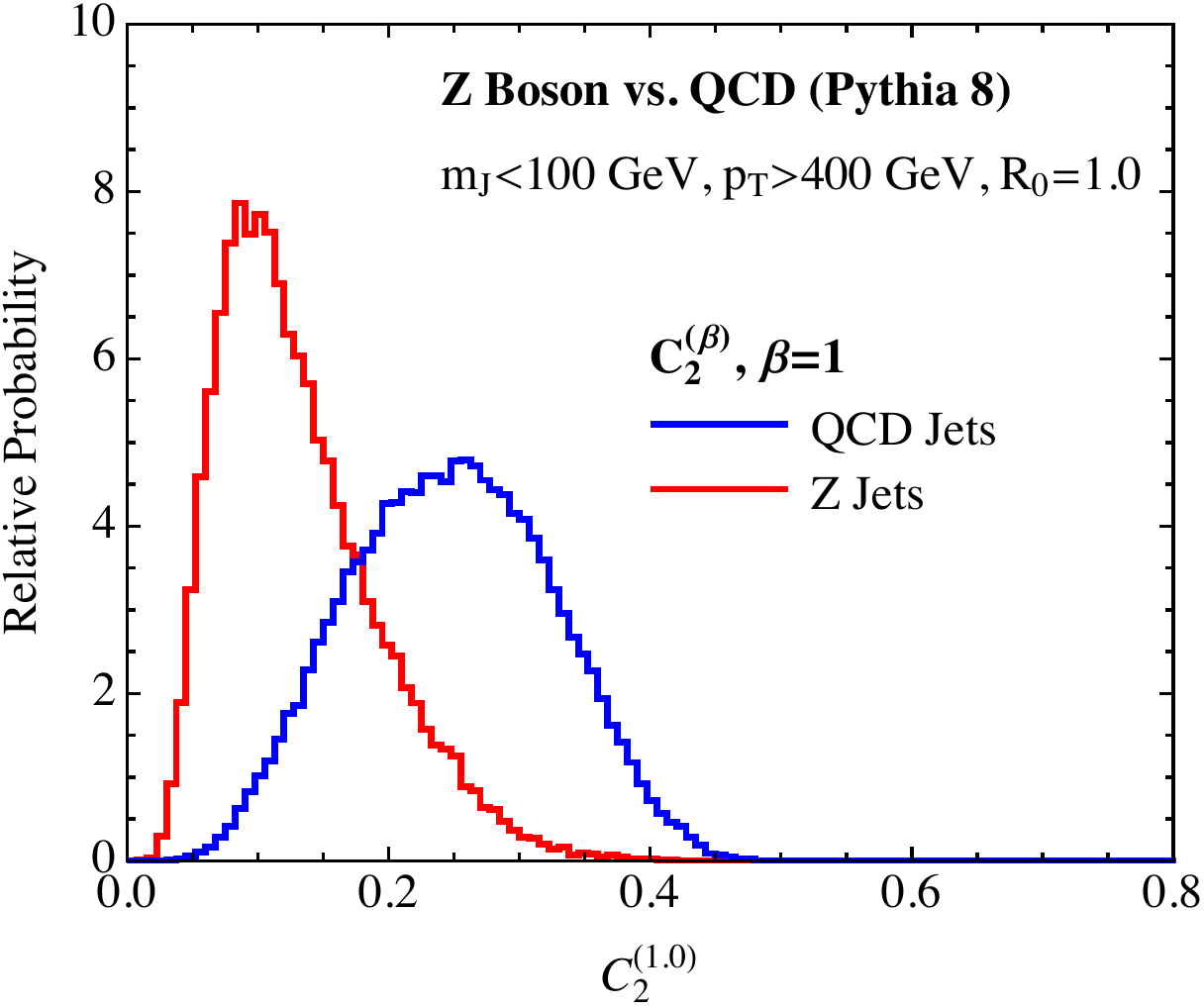}
}\qquad
\subfloat[]{
\includegraphics[width=6.5cm]{figures/newC2_nmcut_b1.pdf}
}\qquad
\subfloat[]{
\includegraphics[width=6.5cm]{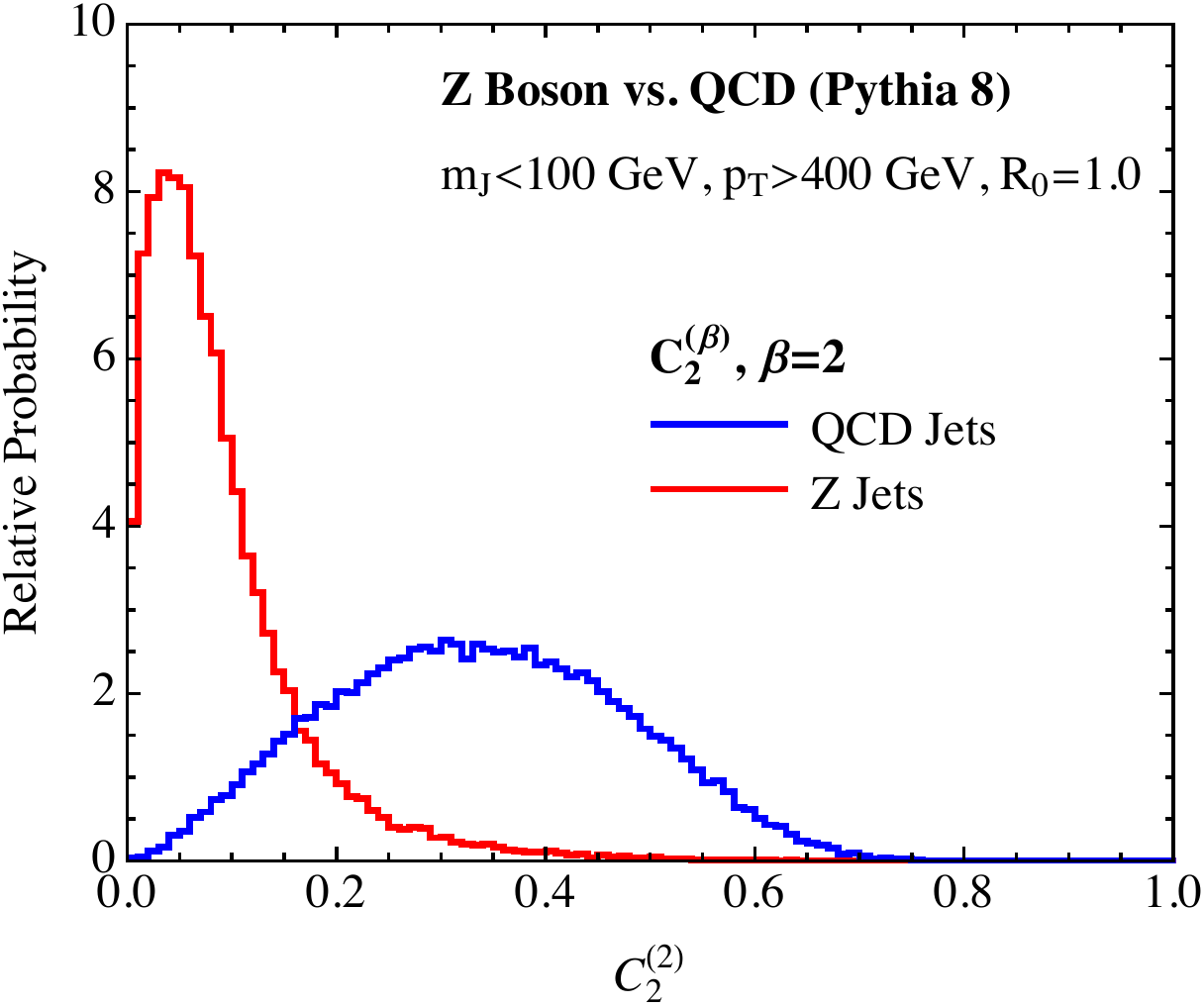}
}\qquad
\subfloat[]{
\includegraphics[width=6.5cm]{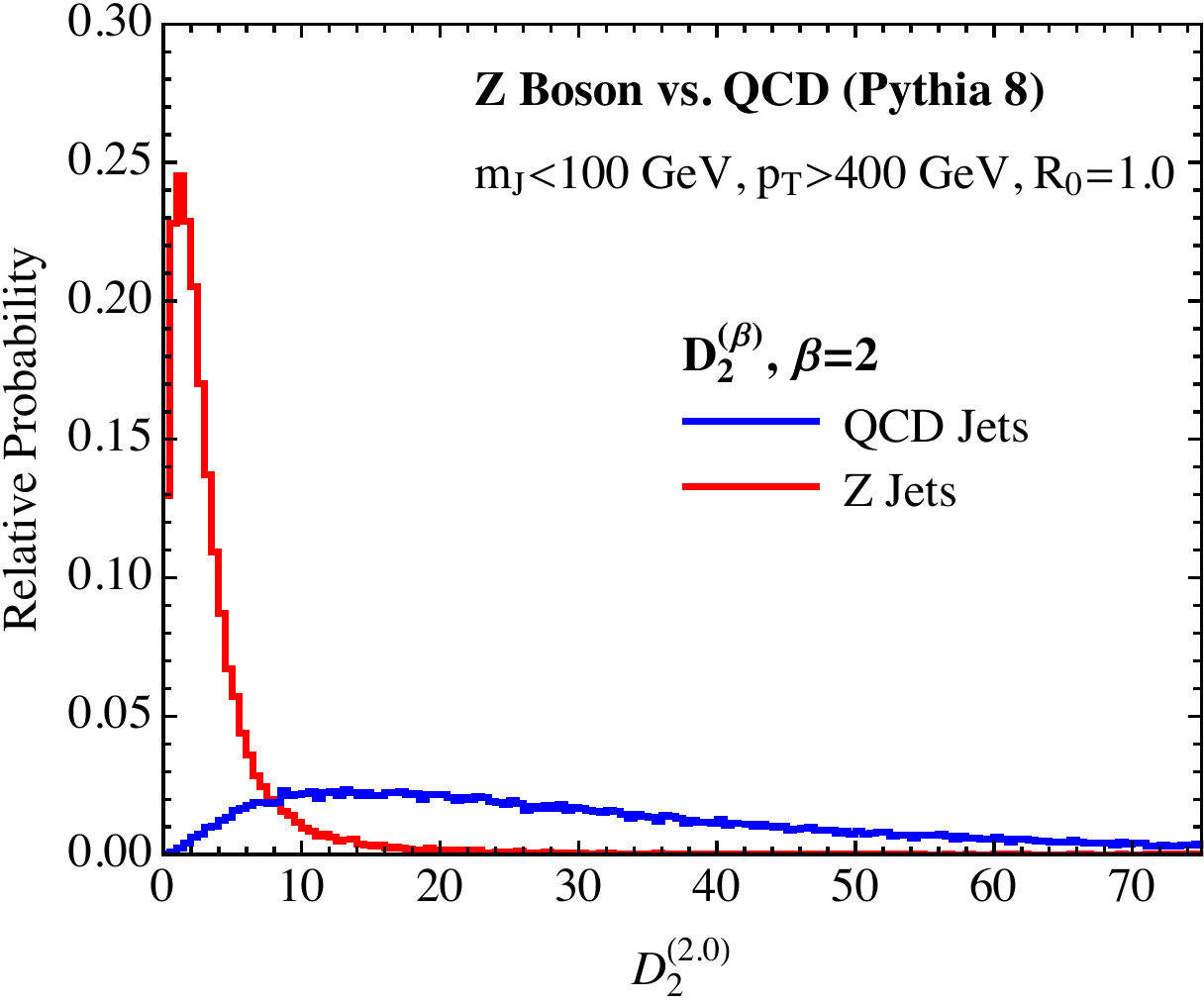}
}
\end{center}
\caption{ Signal and background distributions for the ratio observables $\Cobs{2}{\beta}$ (left) and $\Dobs{2}{\beta}$ (right) for $\beta=0.5,1,2$ from the \madgraph~and \pythia{8} samples. No lower mass cut on the jets is applied but we take $m_J<100$ GeV. 
}
\label{fig:nomasscut}
\end{figure}

\begin{figure}
\begin{center}
\subfloat[]{
\includegraphics[width=6.5cm]{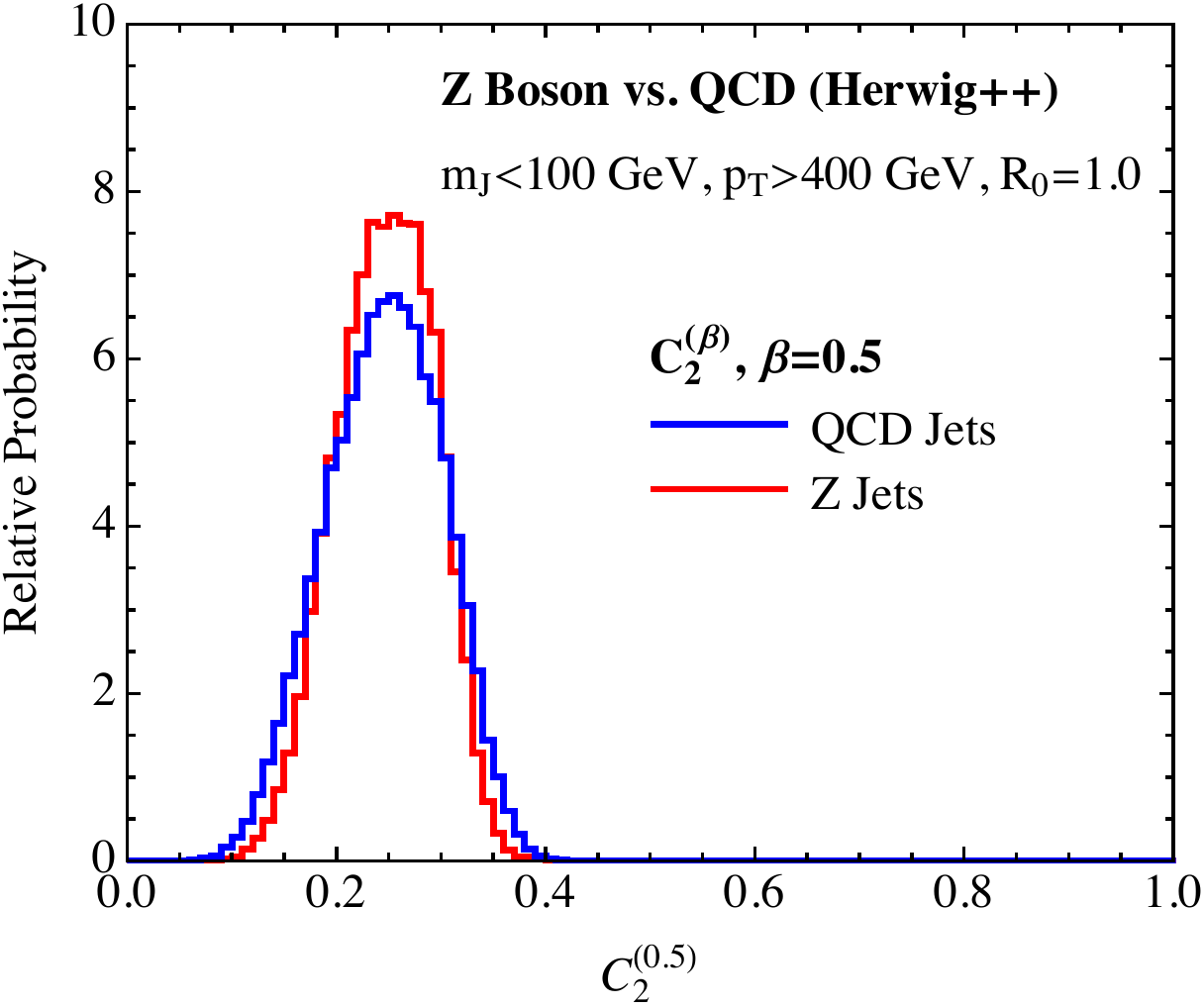}
}\qquad
\subfloat[]{
\includegraphics[width=6.5cm]{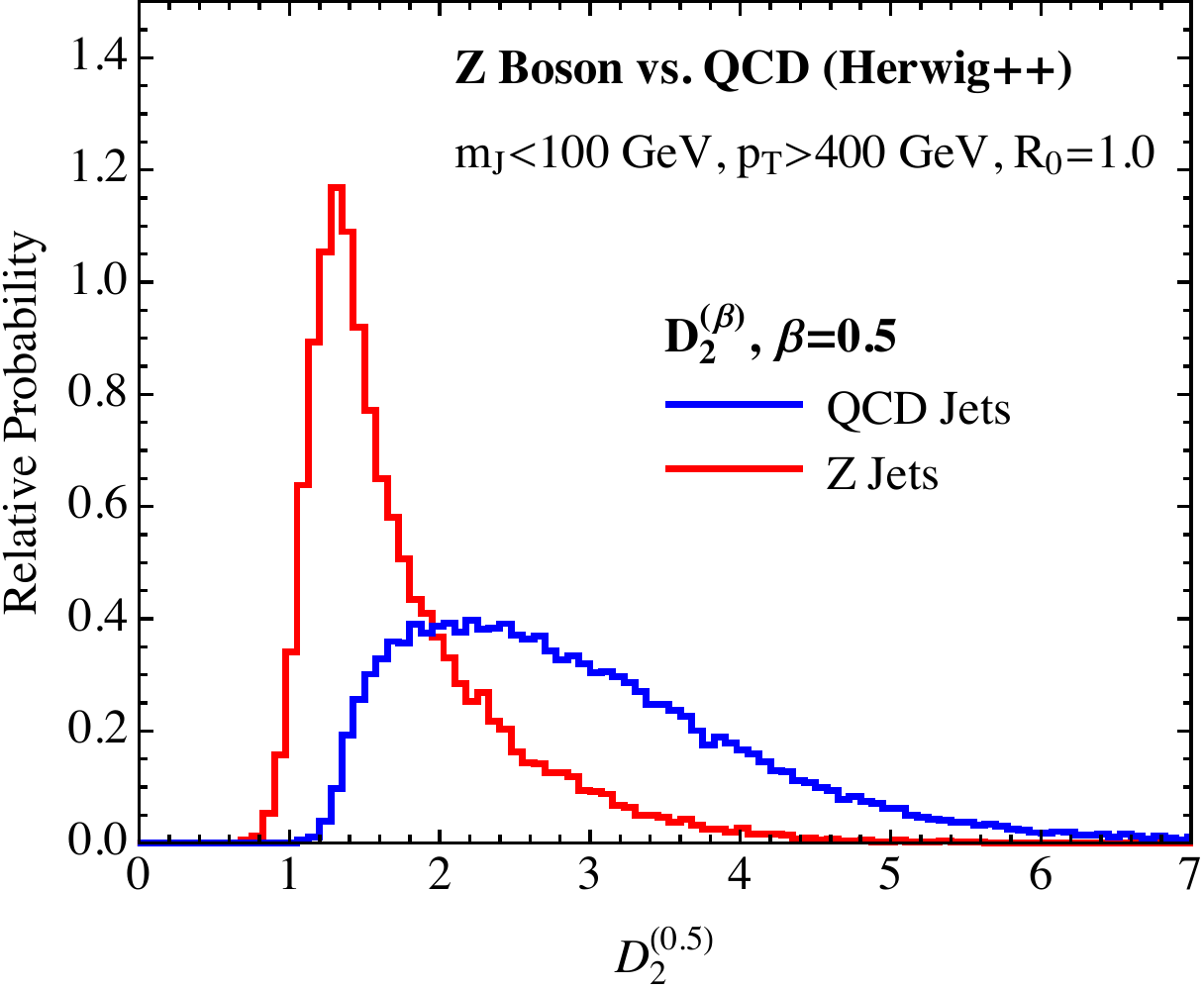}
}\qquad
\subfloat[]{
\includegraphics[width=6.5cm]{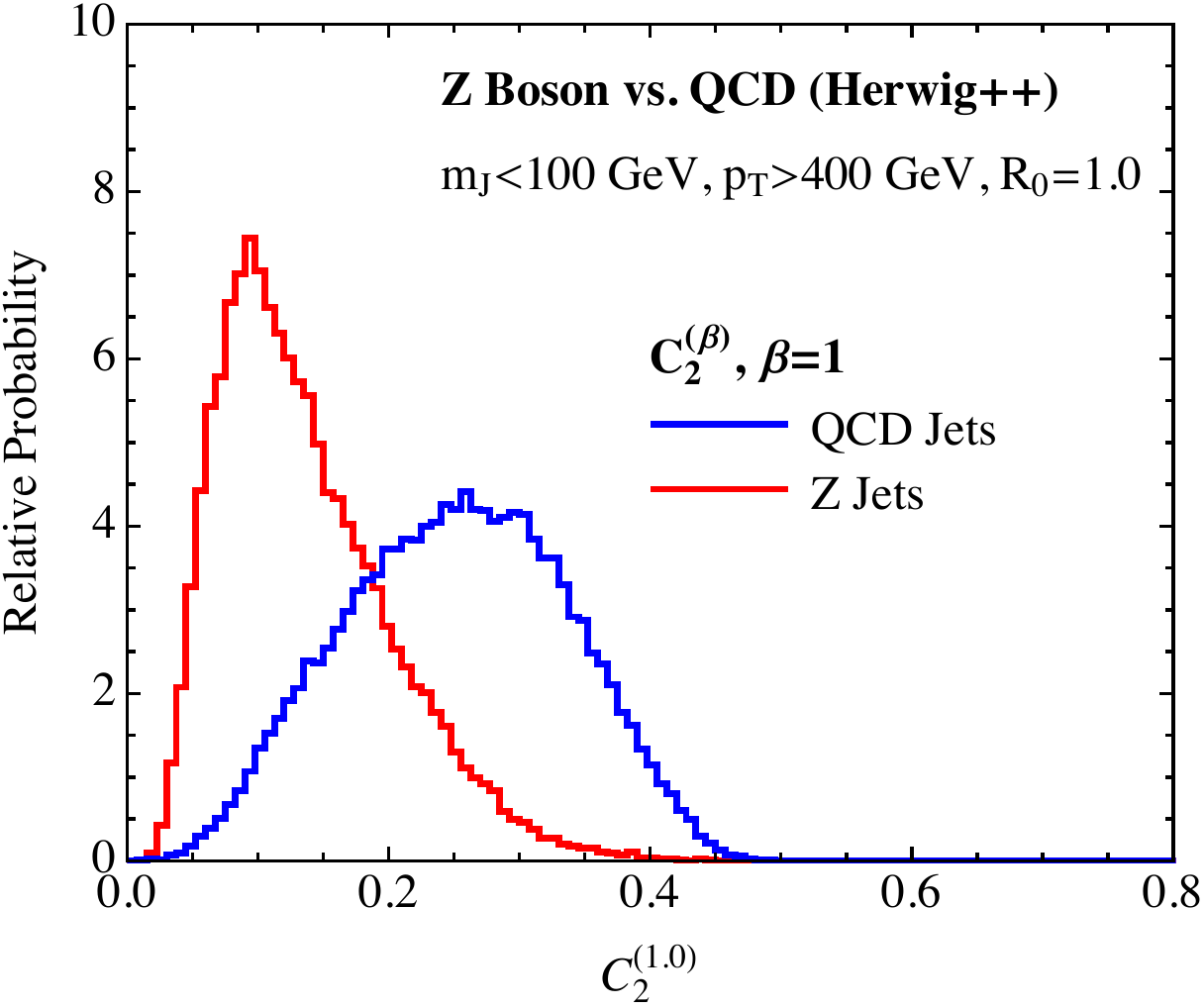}
}\qquad
\subfloat[]{
\includegraphics[width=6.5cm]{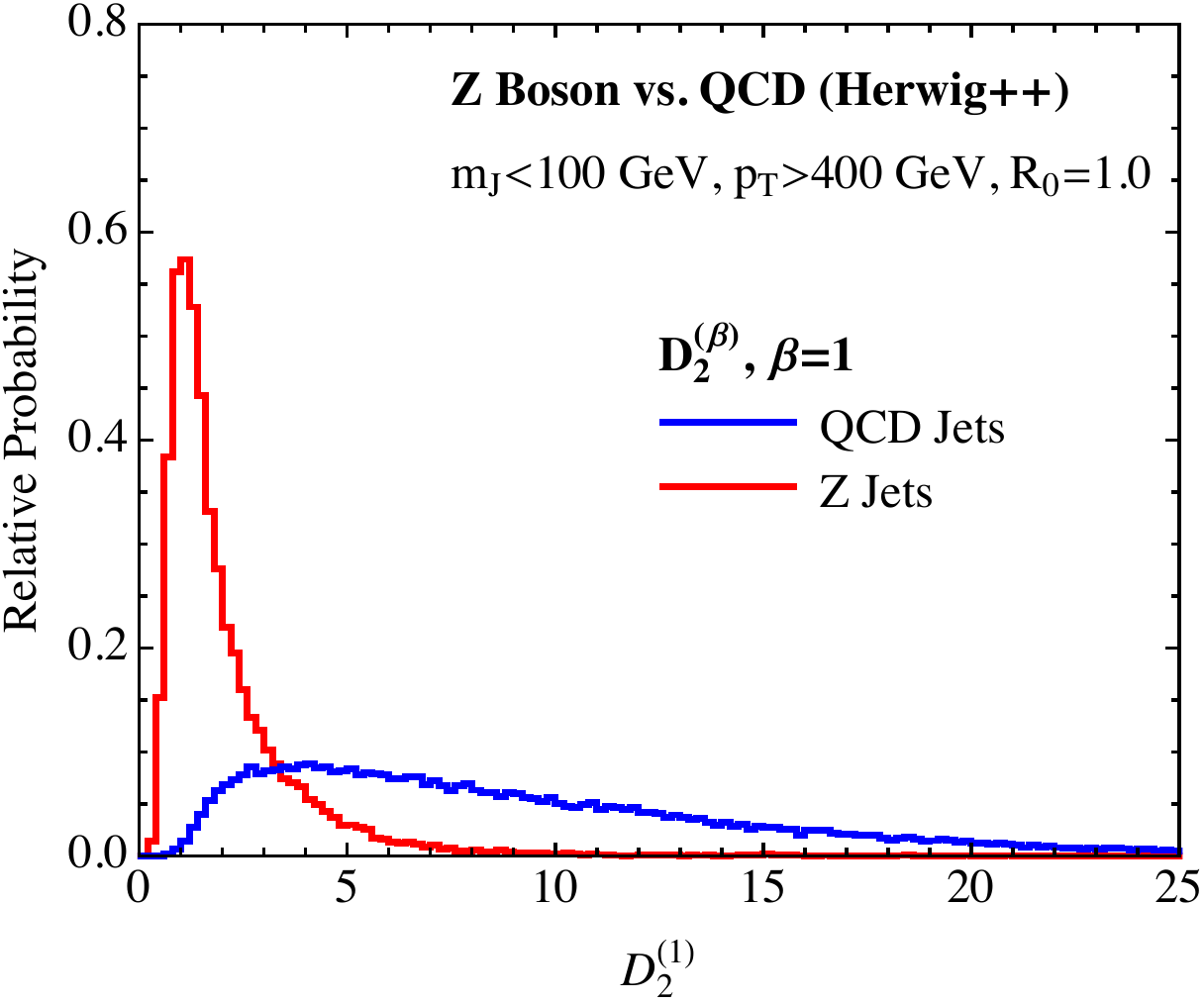}
}\qquad
\subfloat[]{
\includegraphics[width=6.5cm]{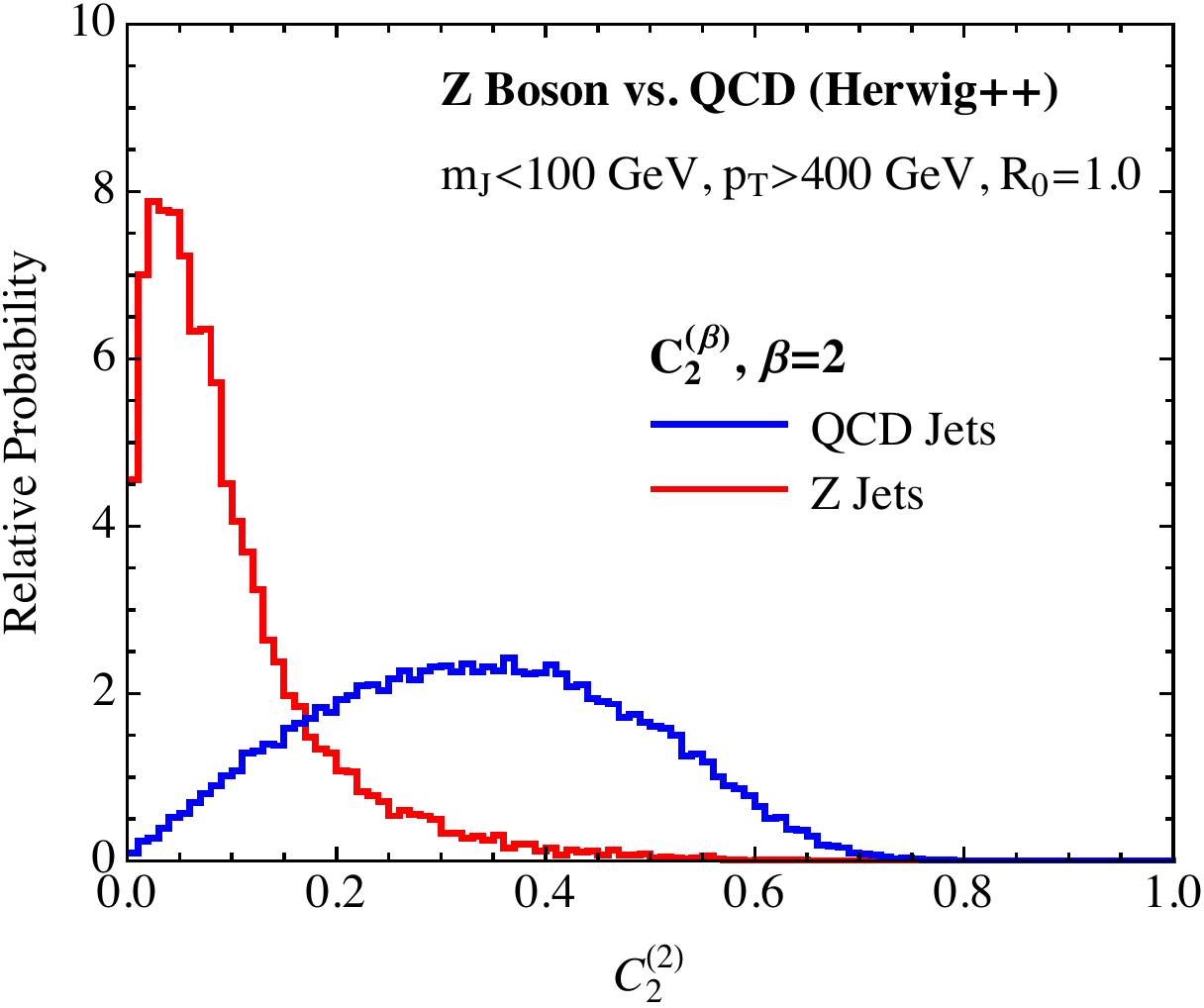}
}\qquad
\subfloat[]{
\includegraphics[width=6.5cm]{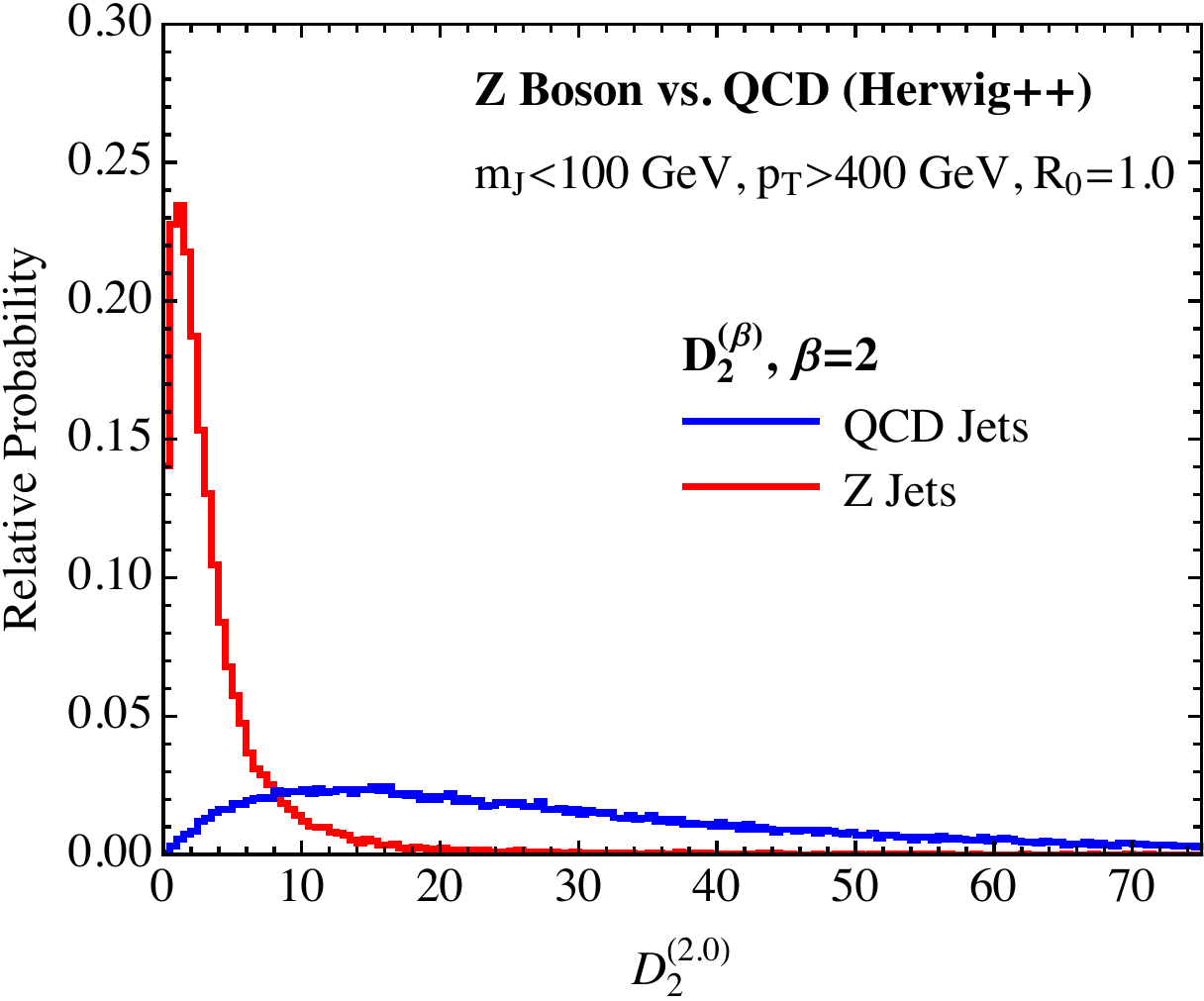}
}
\end{center}
\caption{ 
Same plots as in \Fig{fig:nomasscut}, from the \herwigpp~samples.
}
\label{fig:nomasscut_Herwig}
\end{figure}

\begin{figure}
\begin{center}
\subfloat[]{
\includegraphics[width=6.5cm]{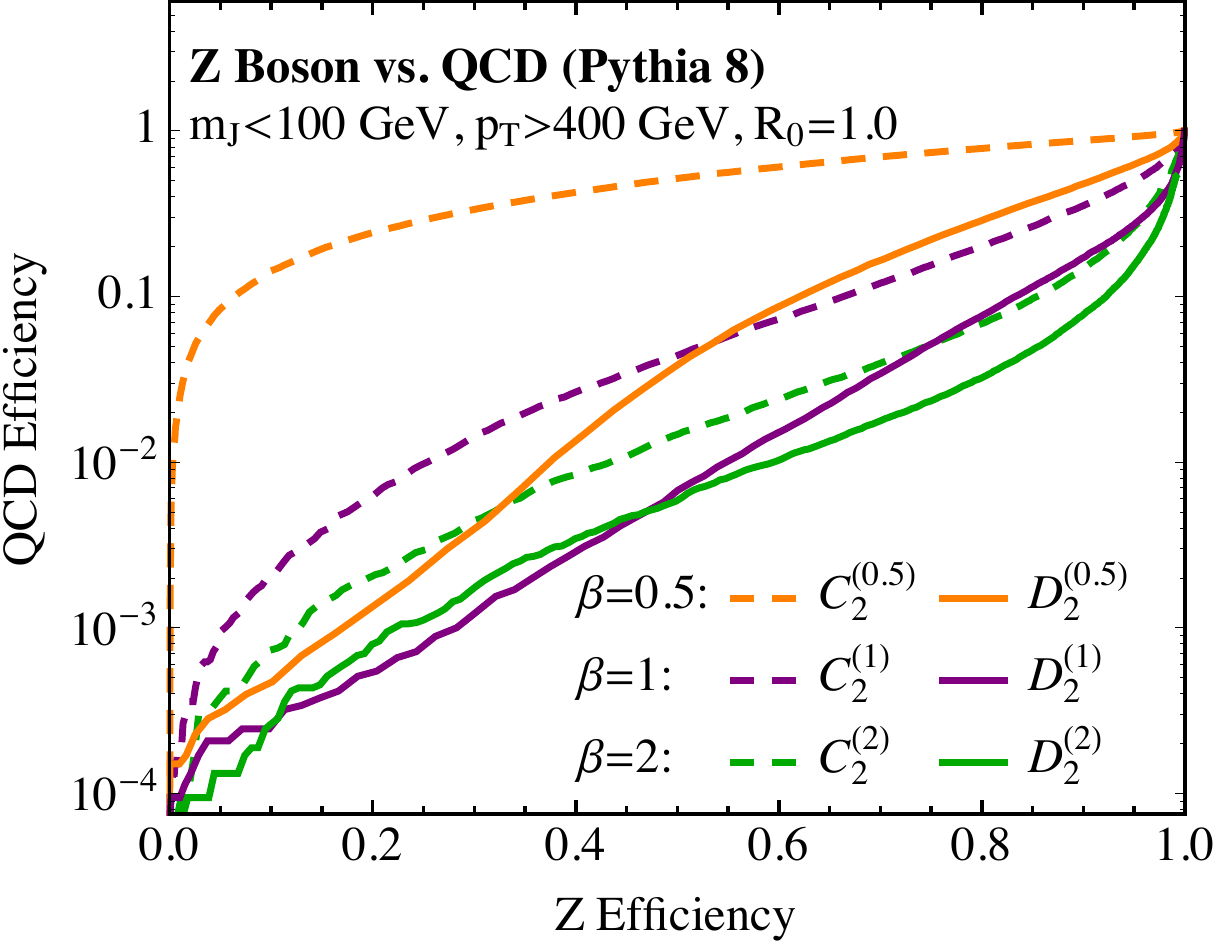}
}\qquad
\subfloat[]{
\includegraphics[width=6.5cm]{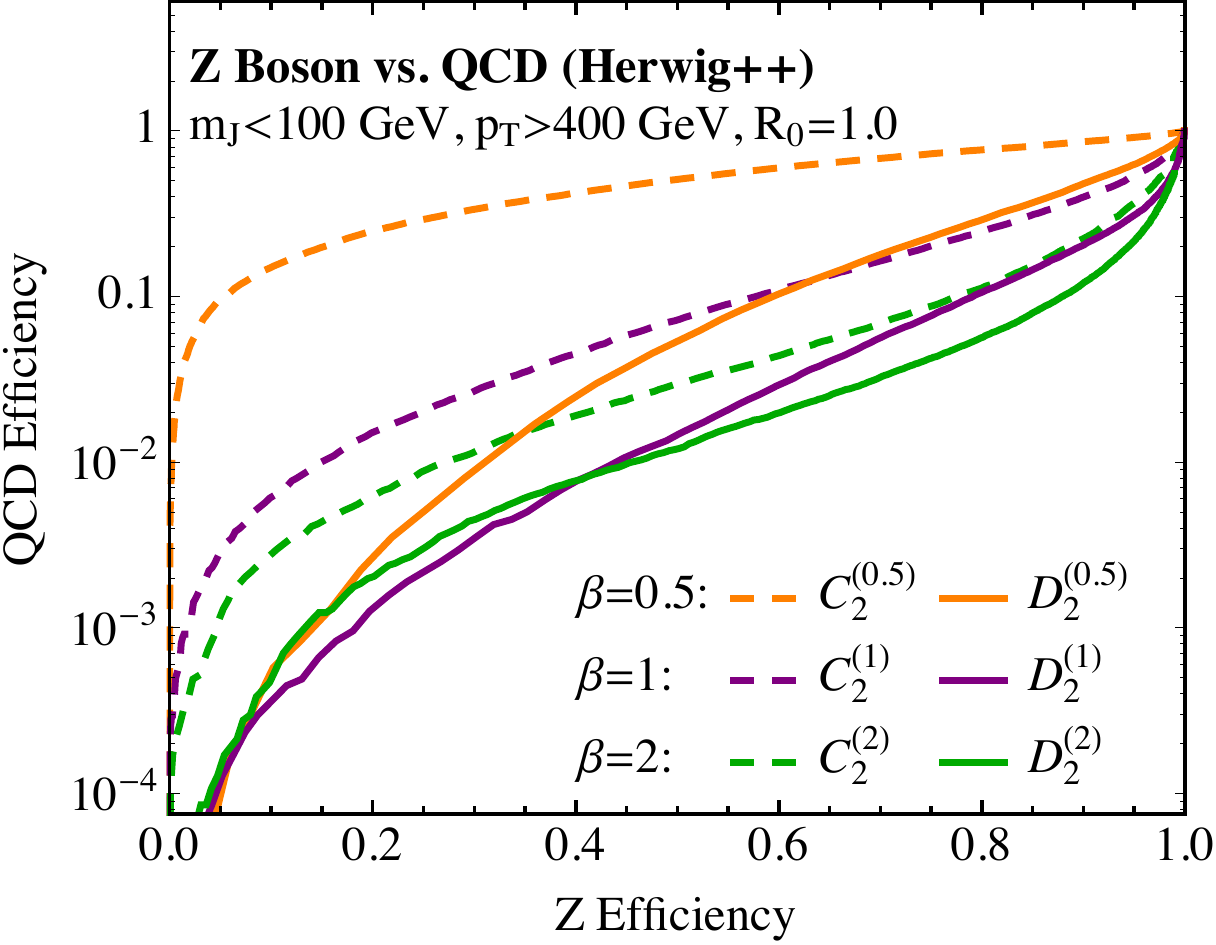}
}
\end{center}
\caption{Signal vs.~background efficiency curves (ROC curves) for $\Cobs{2}{\beta}$ and $\Dobs{2}{\beta}$ for $\beta=0.5,1,2$ for jets with $m_J<100$ GeV, showered with \pythia{8} (left) and \herwigpp~(right). Power counting predictions for the behavior of the ROC curves are robustly reproduced by both Monte Carlo generators. 
}
\label{fig:ROC_C2_nmcut}
\end{figure}

To test these predictions, we will study the different ratio observables formed from $\ecf{2}{\beta}$ and  ${\ecf{3}{\beta}}$ in Monte Carlo simulation. We generated background QCD jets from $pp\to Zj$ events, with the $Z$ decaying leptonically, and boosted $Z$ decays from $pp\to ZZ$ events, with one $Z$ decaying leptonically, and the other to quarks. Events were generated with \madgraph{2.1.2} \cite{Alwall:2014hca} at the $8$ TeV LHC, and showered with either \pythia{8.183} \cite{Sjostrand:2006za,Sjostrand:2007gs} or \herwigpp{2.6.3} \cite{Marchesini:1991ch,Corcella:2000bw,Corcella:2002jc,Bahr:2008pv}, to test the robustness of our predictions to the details of the Monte Carlo generator. Anti-$k_T$ \cite{Cacciari:2008gp} jets with radius $R=1.0$ and $p_T>400$ GeV were clustered in \fastjet{3.0.3} \cite{Cacciari:2011ma} using the Winner Take All (WTA) recombination scheme \cite{Larkoski:2014uqa,Larkoski:2014bia}. The energy correlation functions and $N$-subjettiness ratio observables were calculated using the \texttt{EnergyCorrelator} and \texttt{Nsubjettiness} \fastjet{contrib}s \cite{Cacciari:2011ma,fjcontrib}.

We first compare the discrimination power of $\Cobs{2}{\beta}$ to $\Dobs{2}{\beta}$ with no lower mass cut on the jets for several values of the angular exponent.  We require that $m_J < 100$ GeV which removes a significant fraction of QCD jets that have honest 2-prong structure.  Therefore, we are testing the power of $\Cobs{2}{\beta}$ and $\Dobs{2}{\beta}$ to discriminate between 1-prong and 2-prong jets.  In \Fig{fig:nomasscut}, we show the raw distributions of $\Cobs{2}{\beta}$ and $\Dobs{2}{\beta}$ measured on signal and background for $\beta = 0.5,1,2$.  Especially at small $\beta$, $\Dobs{2}{\beta}$ is much more efficient at separating boosted $Z$s from QCD jets than is $\Cobs{2}{\beta}$.  This is exactly as predicted by the power counting, because $\Cobs{2}{\beta}$ mixes the signal and background regions of phase space, an effect that is magnified at smaller $\beta$.  The discrimination power is quantified in \Fig{fig:ROC_C2_nmcut} where we show the signal vs.~background efficiency curves (ROC curves) for the three choices of $\beta$ for $\Cobs{2}{\beta}$ and $\Dobs{2}{\beta}$.  At low signal efficiency, every $\Dobs{2}{\beta}$ is a better discriminant than any $\Cobs{2}{\beta}$, and the performance of $\Dobs{2}{\beta}$ is much more stable as a function of $\beta$ than $\Cobs{2}{\beta}$.

\begin{figure}
\begin{center}
\subfloat[]{\label{fig:90p_mcut_pythia}
\includegraphics[width=6cm]{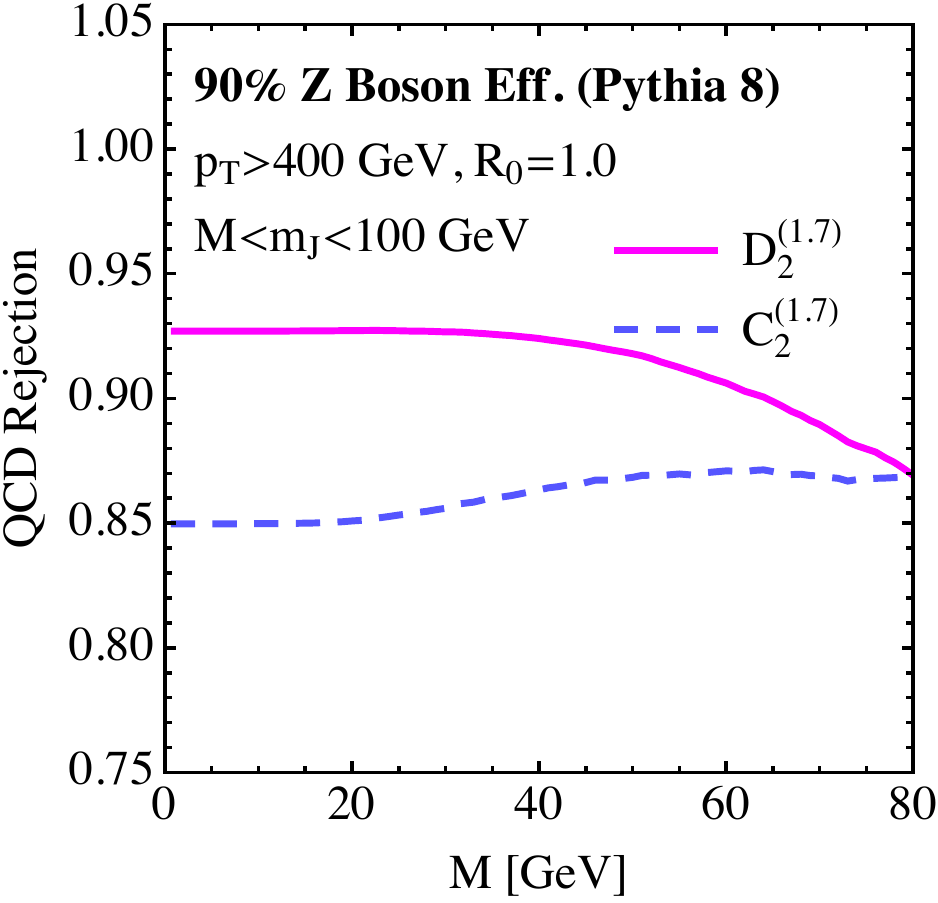}
}\qquad
\subfloat[]{\label{fig:90p_mcut_Herwig}
\includegraphics[width=6cm]{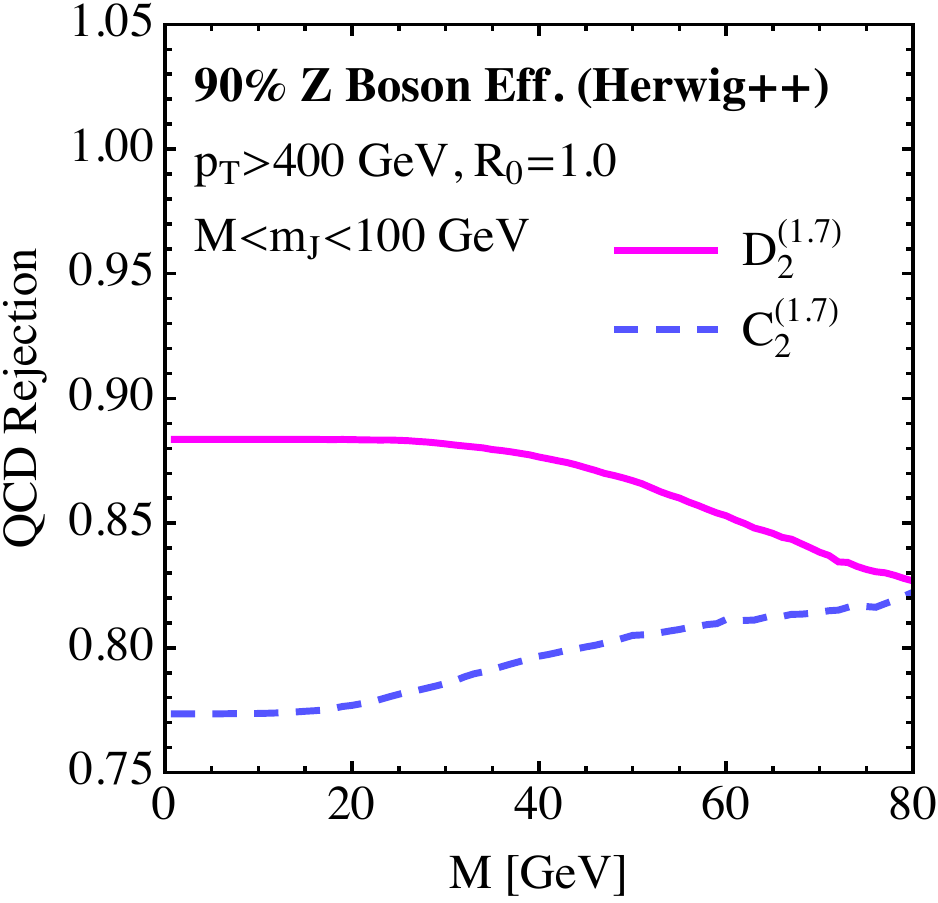}
}
\end{center}
\caption{ QCD rejection efficiency at $90\%$ signal efficiency as a function of the lower mass cut, as predicted by \pythia{8} (left), and \herwigpp (right).  The plots compare the efficiencies of $\Cobs{2}{1.7}$ and $\Dobs{2}{1.7}$.
}
\label{fig:m_dep}
\end{figure}

In the presence of a narrow mass cut window, the power counting analysis of \Sec{sec:masscute2e3} predicted that for $\beta$ near 2, the discrimination power of $\Cobs{2}{\beta}$ and $\Dobs{2}{\beta}$ should be comparable except at high signal efficiency when $\Dobs{2}{\beta}$ should be more discriminating.  To show that this is borne out in Monte Carlo, in \Fig{fig:m_dep} we first plot the rejection efficiency of $\Cobs{2}{1.7}$ and $\Dobs{2}{1.7}$ at 90\% signal efficiency, as a function of the lower mass cut on the jets.\footnote{We use $\beta = 1.7$ as this value was shown in \Ref{Larkoski:2013eya} to be the optimal choice for boosted $Z$ identification.}  When the lower mass cut is near zero, $\Dobs{2}{1.7}$ is significantly more efficient at rejecting QCD background than is $\Cobs{2}{1.7}$, as observed earlier.  As the lower mass cut increases, however, the difference in discrimination power between the two observables decreases in both \pythia{8} and \herwigpp~Monte Carlos. This dependence on the lower mass cut shows that $\Dobs{2}{\beta}$ captures the correct underlying physics of the $(\ecf{2,}{\beta},{\ecf{3}{\beta}})$ phase space, while $\Cobs{2}{\beta}$ does not. The light QCD jets that are added as the mass cut is lowered should be rejected by a variable that partitions the phase space into regions of 1-prong and 2-prong jets, increasing the observed rejection efficiency. This is true for $\Dobs{2}{\beta}$; however, exactly the opposite is true for $\Cobs{2}{\beta}$. 

\begin{figure}
\begin{center}
\subfloat[]{\label{fig:ROC_mcut_pythia_Log}
\includegraphics[width=6cm]{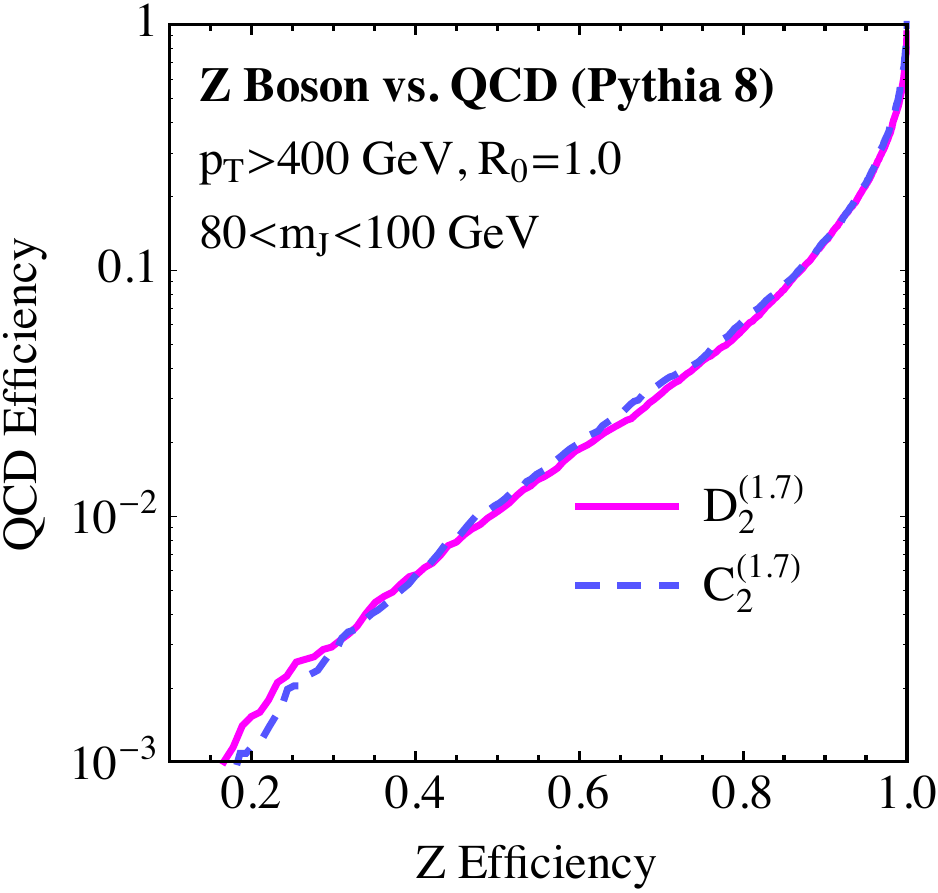}
}\qquad
\subfloat[]{\label{fig:ROC_mcut_Herwig_Log}
\includegraphics[width=6cm]{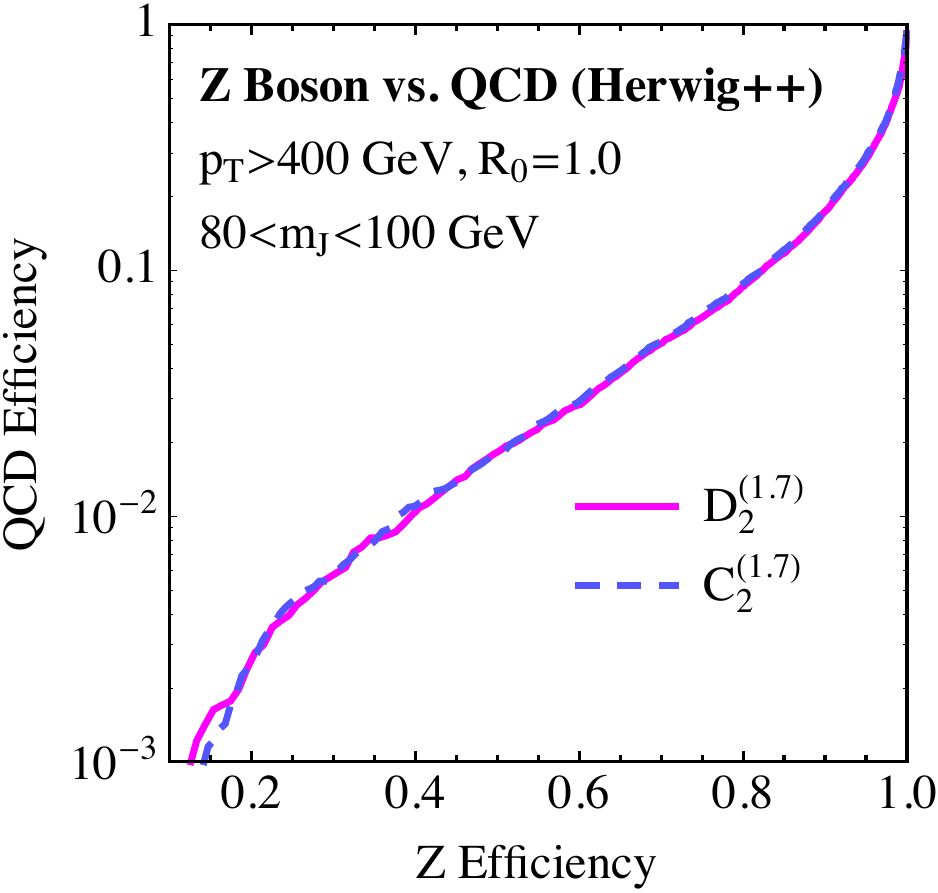}
}\qquad
\subfloat[]{\label{fig:ROC_mcut_pythia}
\includegraphics[width=6cm]{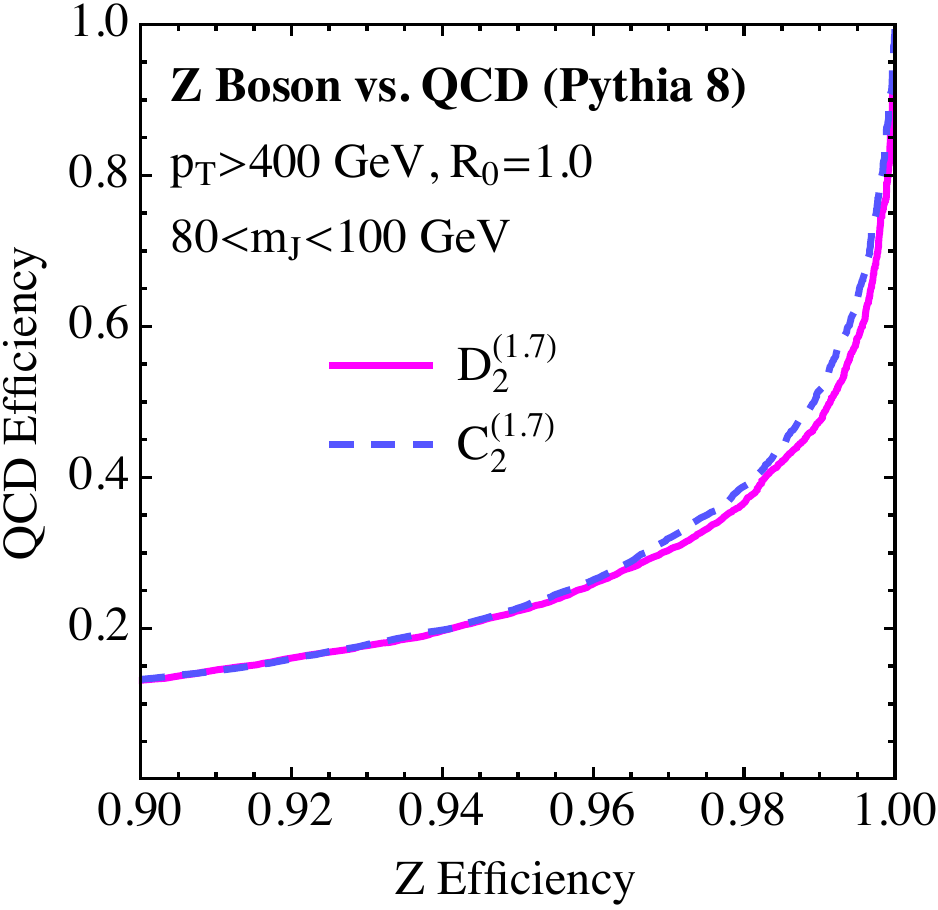}
}\qquad
\subfloat[]{\label{fig:ROC_mcut_Herwig}
\includegraphics[width=6cm]{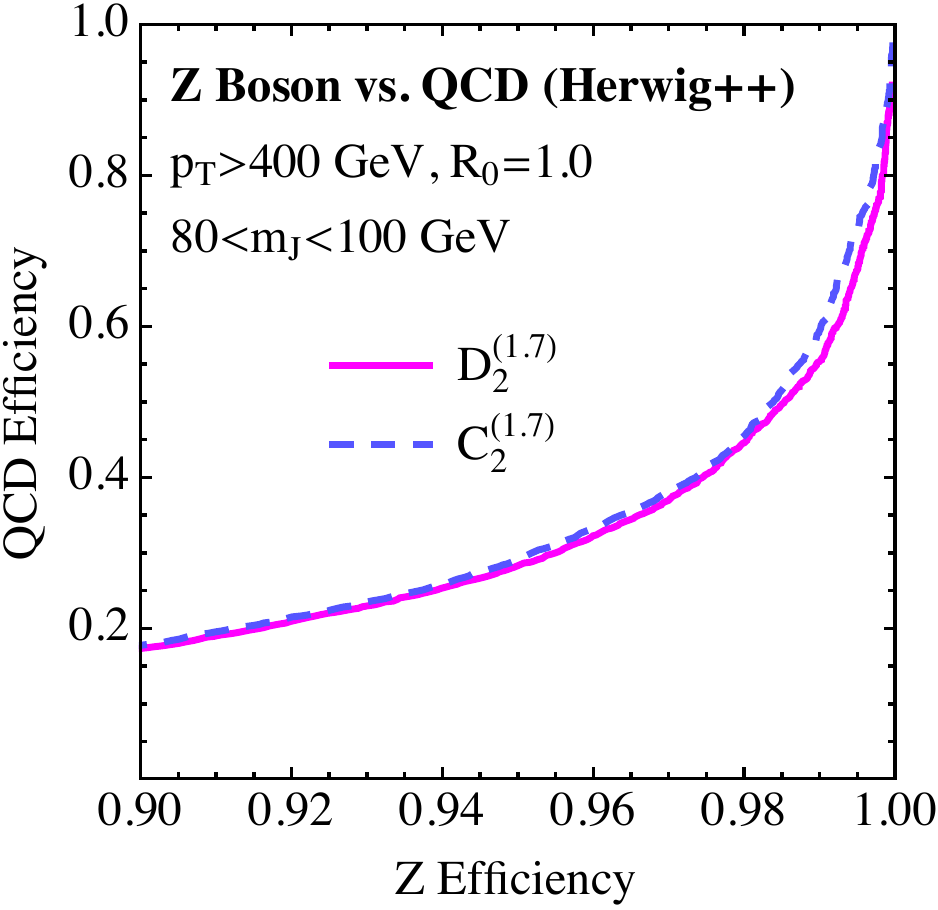}
}\qquad
\subfloat[]{
\includegraphics[width=6cm]{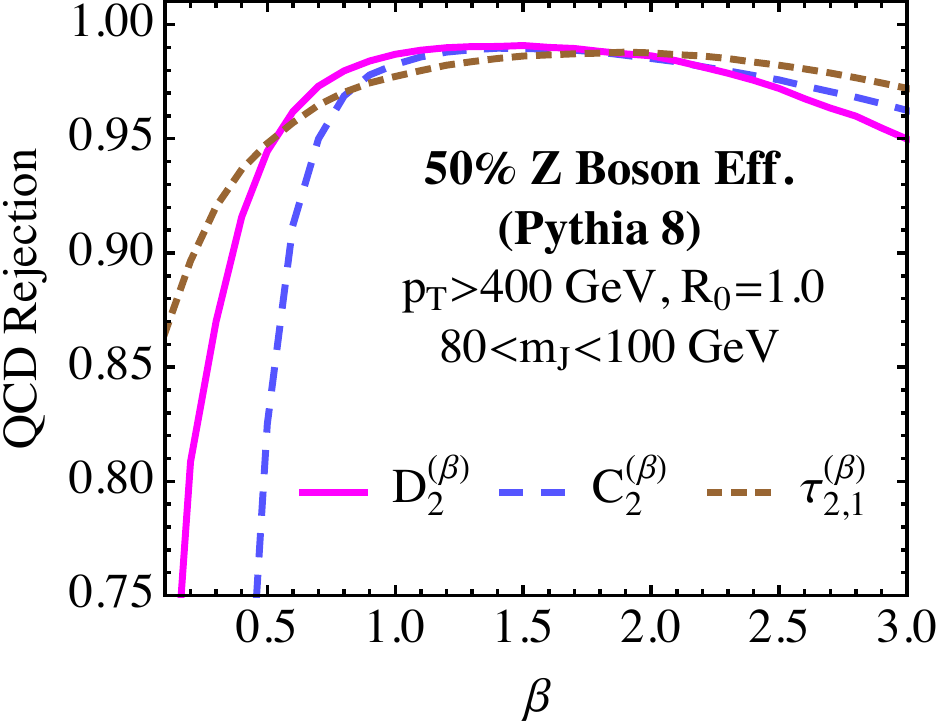}
}\qquad
\subfloat[]{
\includegraphics[width=6cm]{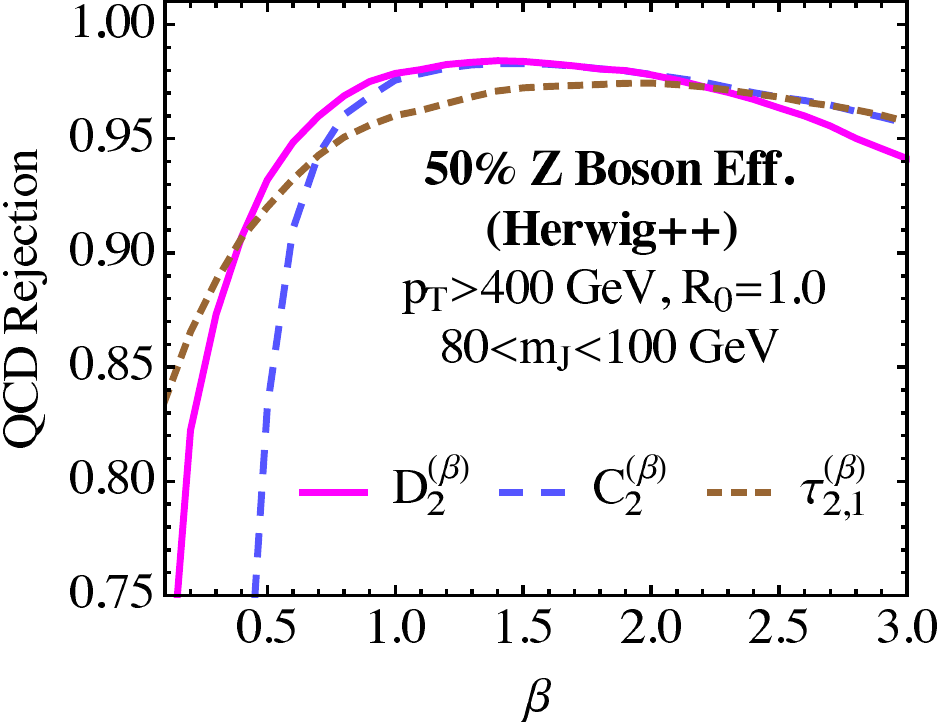}
}
\end{center}
\caption{ Comparison of $\Cobs{2}{\beta}$ and $\Dobs{2}{\beta}$ in the presence of a tight mass cut, $80<m_J<100$ GeV, with the \pythia{8} (left) and \herwigpp (right) samples. ROC curves for $\Cobs{2}{1.7}$ and $\Dobs{2}{1.7}$ demonstrate that with a tight mass cut, both observables perform comparably over a large range of signal efficiencies (top), with $\Dobs{2}{1.7}$ performing slightly better at high signal efficiencies (middle), behavior which is reproduced by both Monte Carlo generators. The QCD rejection rate at $50\%$ signal efficiency as a function of $\beta$ is shown at bottom for $\Cobs{2}{\beta}$, $\Dobs{2}{\beta}$ and $\Nsub{2,1}{\beta}$. 
}
\label{fig:ROC_mcut}
\end{figure}

We now study in more detail the case in which we have constrained the jet mass to lie in the tight mass cut window of $80 < m_J < 100$ GeV.  Over the whole signal efficiency range, $\Cobs{2}{1.7}$ and $\Dobs{2}{1.7}$ have nearly identical ROC curves in both \pythia{8} and \herwigpp, as exhibited in \Fig{fig:ROC_mcut}.  However, focusing in on the high signal efficiency region, we see that indeed $\Dobs{2}{1.7}$ has a slightly better rejection rate than $\Cobs{2}{1.7}$.  This behavior is manifest in both \pythia{8} and \herwigpp, showing that this prediction from the power counting analysis of \Sec{sec:masscute2e3} is robust to the precise details of the parton shower in the Monte Carlo generator. This should be contrasted with the actual numerical value of the QCD rejection, which depends on the generator.  For $\beta\simeq 2$ with a tight mass cut window of $80<m_J<100$ GeV, any discriminating variable of the form $\ecf{3}{\beta}/{\ecf{2}{\beta}}^n$, for $n>0$, provides reasonable discrimination power. The jet mass cut fixes $\ecf{2}{2}$ to a narrow window, and all discrimination power comes from $\ecf{3}{2}$ alone. This demonstrates why  \Ref{Larkoski:2013eya} observed near-optimal discrimination power using $\Cobs{2}{2}$, with $80<m_J<100$ GeV.

The final power counting prediction for the behavior of the observables was that the discrimination power of $\Dobs{2}{\beta}$ should be much more robust than $\Cobs{2}{\beta}$ as $\beta$ decreases from 2, with a mass cut on the jets.  This behavior is reproduced in Monte Carlo, as shown in \Fig{fig:ROC_mcut}, where we have plotted the QCD rejection efficiency at 50\% signal efficiency as a function of $\beta$.  We have also included in these plots the $N$-subjettiness ratio $\Nsub{2,1}{\beta}$ for comparison.  As $\beta\to 0$, the discrimination power of both $\Cobs{2}{\beta}$ and $\Dobs{2}{\beta}$ decrease; however, $\Dobs{2}{\beta}$ maintains high discrimination power to much smaller values of $\beta$ than $\Cobs{2}{\beta}$.  Nevertheless, note that as $\beta\to 0$, $\Nsub{2,1}{\beta}$, while not the optimal discrimination observable, has even more robust discrimination power than $\Dobs{2}{\beta}$.  An understanding of this behavior requires an analyses of ${\cal O}(1)$ numbers, which is beyond what power counting alone can predict.

Thus, we see that all the power counting predictions are realized in both Monte Carlo simulations, demonstrating that parametric scalings are indeed determining the behavior of the substructure observables. We emphasize that the level of agreement between the Monte Carlo generators for the power counting predictions is quite remarkable, given that numerical values for rejection or acceptance efficiencies, for example, do not agree particularly well between the generators. Power counting has allowed us to identify the robust predictions of perturbative QCD.

\subsection{Including Pile-Up}\label{sec:pu}

The power counting analysis of the previous section included only perturbative radiation.  At a high luminosity hadron collider such as the LHC, also important is the effect of multiple proton collisions per bunch crossing, referred to as pile-up.  Pile-up radiation is uncorrelated with the hard scattering event, and as such, has an energy scale that is independent of the hard parton collision energy.  Thus, pile-up can produce a significant amount of contaminating radiation in the event and substantially change jet $p_T$s, masses, or observables from their perturbative values.  An important problem in jet substructure is both to define observables that are less sensitive to the effects of pile-up, as well as to remove or ``groom'', to the greatest extent possible, radiation in a jet or event that most likely is from pile-up.  Several methods for jet grooming and pile-up subtraction have been presented  \cite{Butterworth:2008iy,Ellis:2009su,Ellis:2009me,Krohn:2009th,Soyez:2012hv,Krohn:2013lba,Larkoski:2014wba,Berta:2014eza,Cacciari:2014gra,Bertolini:2014bba}, and are used by the experiments \cite{CMS:2013uea,CMS-PAS-JME-10-013,TheATLAScollaboration:2013ria,TheATLAScollaboration:2013pia,ATLAS-CONF-2012-066}, but we will not consider them here.\footnote{The effects of jet grooming techniques can be understood using power counting techniques, and have been considered in \cite{Walsh:2011fz}.}  Instead, we will demonstrate that power counting can be used to understand the effect of pile-up radiation on the $(\ecf{2}{\beta} ,\ecf{3}{\beta})$ phase space, and therefore on signal and background distributions for observables formed from the energy correlation functions. We envision that similar techniques could be used to develop jet substructure variables with improved resilience to pile-up, but in this chapter we will restrict ourselves to an understanding of the behavior of $\Cobs{2}{\beta}$ and $\Dobs{2}{\beta}$.\footnote{While the following analysis is quite general, it is restricted to recoil-free observables defined with a recoil-free jet algorithm, as used in this chapter. In the case of a recoil sensitive observable, there is a non-linear response to pile-up due to the displacement of soft and collinear modes with respect to the jet axis. In this case the power counting analysis described here does not apply directly, and a more thorough analysis is required.}

To incorporate pile-up radiation into the power counting analysis, we must make some simplifying assumptions. Because pile-up is independent of the hard scattering event, we will assume that pile-up radiation is uniformly distributed over the jet area.\footnote{This model of pile-up would be removed by area subtraction \cite{Soyez:2012hv}. However, this would also remove perturbative soft radiation depending on the region of phase space. This could be studied in detail using power counting.}  This assumption essentially defines pile-up as another soft mode in the jet, with all angles associated with pile-up scaling as ${\cal O}(1)$. We will denote the $p_T$ fraction of pile-up radiation in the jet as
\begin{equation}
z_{pu} \equiv \frac{p_{T\, pu}}{p_{TJ}} \ .
\end{equation}
No assumption of the relative size of the perturbative soft radiation energy fraction $z_s$ with respect to $z_{pu}$ is made at this point, and indeed the impact of pile-up on the phase space will depend on this relation.

Assuming only that the pile-up $p_T$ fraction $z_{pu} \ll 1$, the two- and three-point correlation functions for 1-prong jets have the scaling
\begin{align}
\ecf{2}{\beta} & \sim  R_{cc}^\beta + z_s + z_{pu} \,, \\
\ecf{3}{\beta} & \sim R_{cc}^{3\beta}+z_s^2 + R_{cc}^\beta z_s +z_{pu}^2 + R_{cc}^\beta z_{pu} \,.
\end{align}
For 2-prong jets, the correlation functions have the scaling 
\begin{align}
\ecf{2}{\beta} & \sim  R_{12}^\beta + z_{pu} \,, \\
\ecf{3}{\beta} & \sim R_{12}^{\beta}z_s+R_{12}^{2\beta} R_{cc}^{\beta}+R_{12}^{3\beta}z_{cs} + R_{12}^\beta z_{pu} + z_{pu}^2\,.
\end{align}
From these scalings, we will be able to understand how pile-up radiation impacts jets in different regions of phase space, and hence the distributions in $\Cobs{2}{\beta}$ and $\Dobs{2}{\beta}$. Note that $z_{pu}$ is a fixed quantity measuring the fraction of pile-up radiation in the jet, and unlike the scalings for the soft, collinear and collinear-soft modes, its scaling is constant throughout the phase space. To understand the impact of the pile-up radiation on different regions of phase space, we will therefore need to understand how the values of $\ecf{2}{\beta}$ and $\ecf{3}{\beta}$ are modified by $z_{pu}$, depending on the different scalings of the contributing modes.

We begin the study of the phase space at small $\ecf{2}{\beta}$.  In the limit when $z_{pu} \gg  z_s$, pile-up dominates the structure of the jet.  In this limit, both 1-prong and 2-prong jets are forced into the region of phase space where $\ecf{3}{\beta} \sim (\ecf{2}{\beta})^2$. Note however that the scaling of the upper boundary of the phase space is robust. We must assume, as we will in what follows, that the value of $z_{pu}$ is such that this region does not extend far into the phase space, or else the energy correlation functions cannot be used to discriminate 1- and 2-prong jets, as their structure is completely dominated by pile-up radiation.

\begin{figure}
\begin{center}
\subfloat[]{
\includegraphics[width=6.5cm]{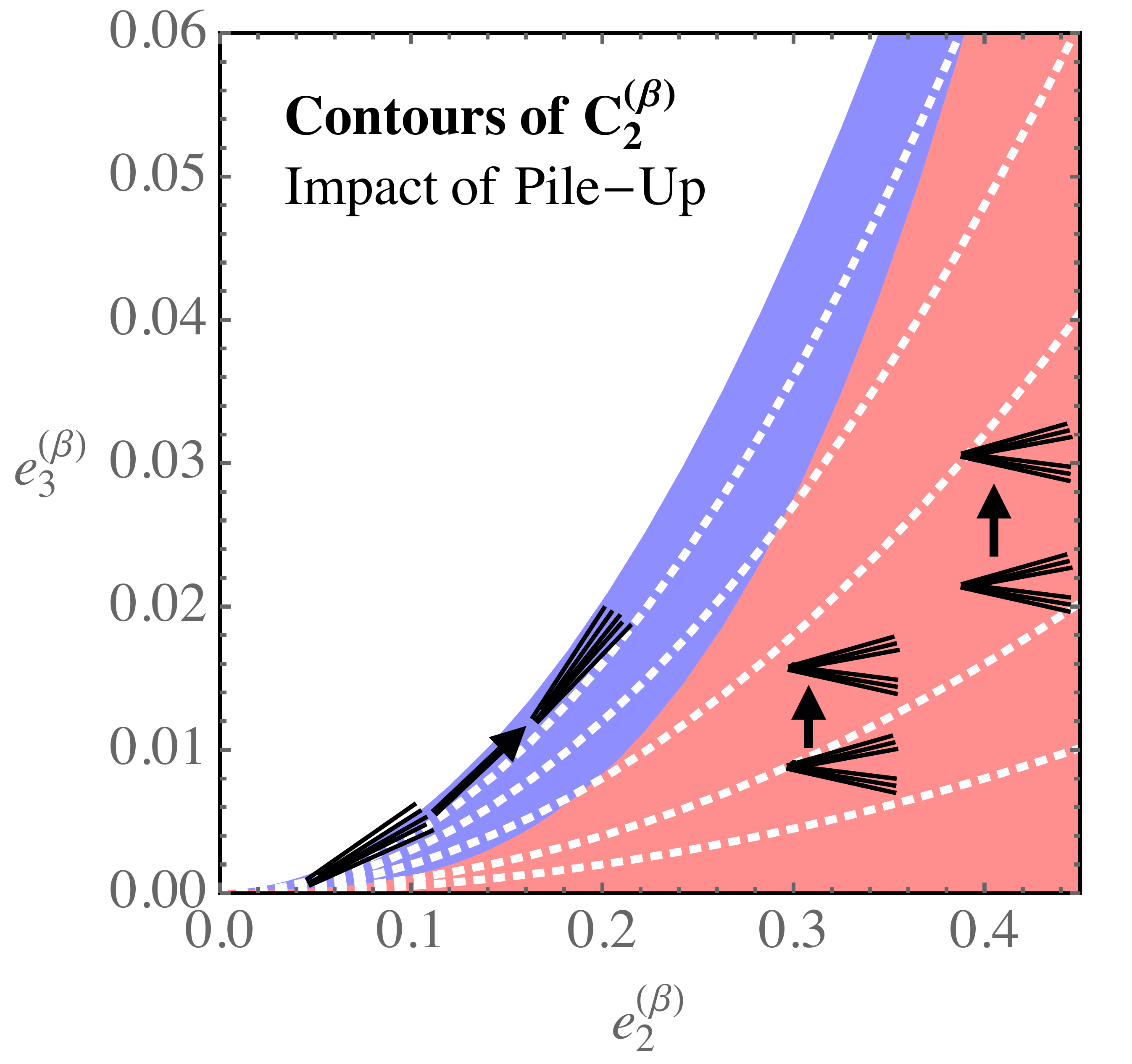}
}\qquad
\subfloat[]{
\includegraphics[width=6.5cm]{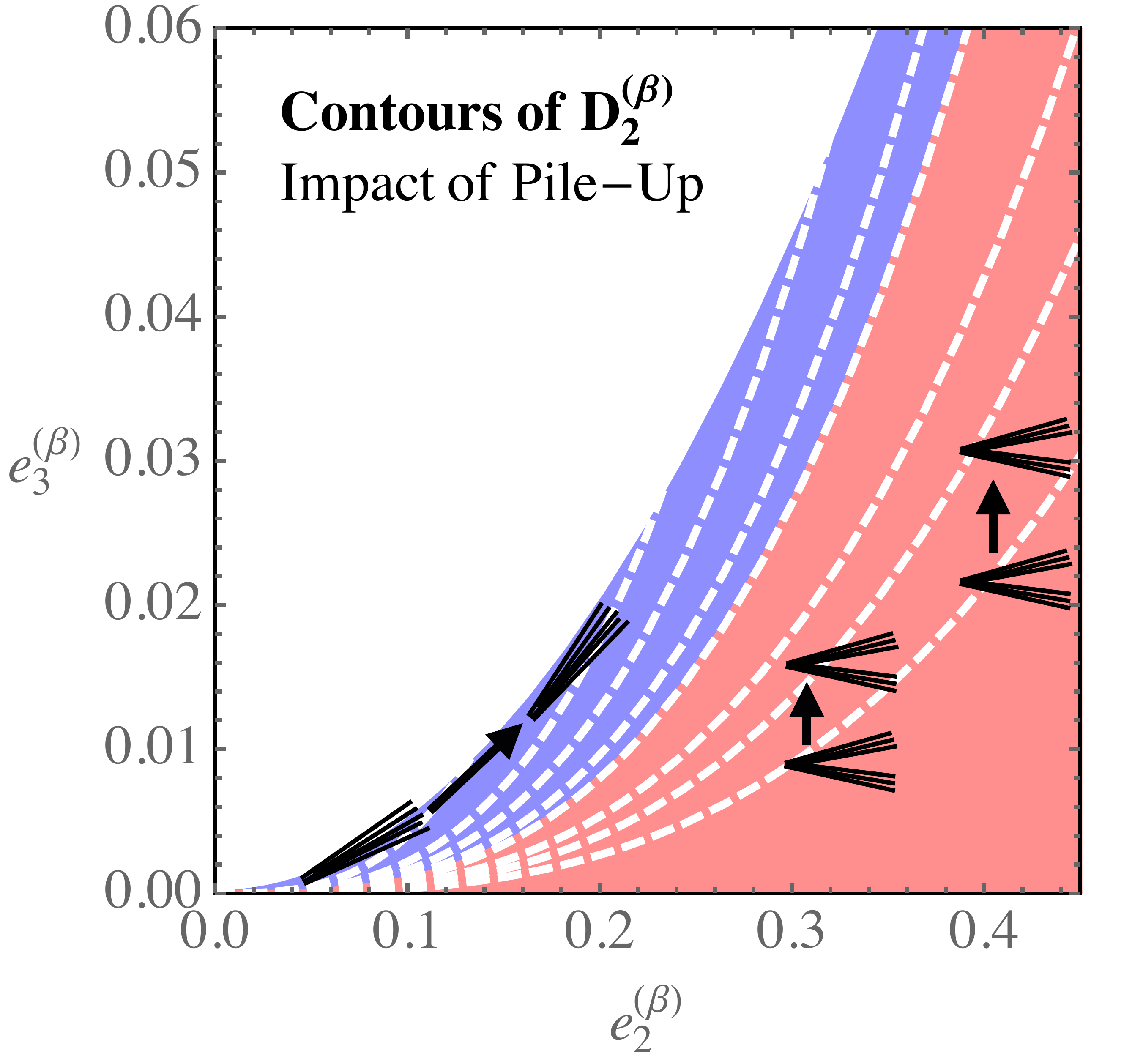}
}
\end{center}
\caption{Illustration of the effect of pile-up on the $\ecfnobeta{2} ,\ecfnobeta{3}$ phase space.   The 1- and 2-prong regions of phase space are denoted by blue or red, respectively, and the arrows show the direction that the jets move in the phase space with the addition of pile-up.  Contours of constant $\Cobsnobeta{2}$ (left) and $\Dobsnobeta{2}$ (right) are shown for reference.
}
\label{fig:ps_pu}
\end{figure}

Moving to slightly larger values of $\ecf{2}{\beta}$, we encounter a region dominated by 1-prong background jets, where $z_{pu} \sim z_s$. Under the addition of pile-up radiation, the two- and three-point correlation functions for 1-prong jets are modified as
\begin{align}
\ecf{2}{\beta} & \to \ecf{2}{\beta} + z_{pu} \,, \\
\ecf{3}{\beta} & \to \ecf{3}{\beta} + \ecf{2}{\beta} z_{pu} +z_{pu}^2 \,.
\end{align}
For $z_{pu}\sim z_s$, the addition of pile-up radiation therefore pushes all 1-prong jets towards the boundary $\ecf{3}{\beta} \sim (\ecf{2}{\beta})^2$. Jets that already satisfy this scaling, maintain it under the addition of pile-up, but move to larger values of $\ecf{2}{\beta}$. This behavior is illustrated in \Fig{fig:ps_pu}, and will imply a very different behavior for the distributions of $\Cobs{2}{\beta}$ and $\Dobs{2}{\beta}$  for background.

At larger values of $\ecf{2}{\beta}$, populated primarily by jets with two hard prongs, pile-up is a power-suppressed contribution to $\ecfnobeta{2}$, but still contributes to $\ecfnobeta{3}$.  That is, pile-up affects the two- and three-point correlation functions measured on 2-prong jets as
\begin{align}
\ecf{2}{\beta} & \to \ecf{2}{\beta} \,, \\
\ecf{3}{\beta} & \to \ecf{3}{\beta} + \ecf{2}{\beta} z_{pu}  \,.
\end{align}
Therefore, pile-up shifts 2-prong jets vertically in the $(\ecfnobeta{2} ,\ecfnobeta{3})$ phase space plane by an amount proportional to the perturbative value of $\ecf{2}{\beta}$.  This behavior is illustrated schematically in \Fig{fig:ps_pu}. At even larger values of $\ecf{2}{\beta}$, we enter a regime where $z_{pu} \ll z_s$. Here the scale of the pile-up radiation is parametrically smaller than the soft perturbative radiation, and so to leading power pile-up can be ignored.

We will now use this understanding of the effect of pile-up radiation on different regions of the $(\ecfnobeta{2} ,\ecfnobeta{3})$ phase space to understand its impact on the distributions for the observables $\Cobs{2}{\beta}$ and $\Dobs{2}{\beta}$ for both the signal and background. We begin by discussing the impact of pile-up radiation on the background distribution of $\Dobs{2}{\beta}$. Recall that at small $\ecf{2}{\beta}$, both $\ecf{2}{\beta}$ and $\ecf{3}{\beta}$ are shifted by the addition of the pile-up radiation, but maintain the parametric scaling $\ecf{3}{\beta} \sim (\ecf{2}{\beta})^2$. This has an interesting effect on the $\Dobs{2}{\beta}$ distribution due to the fact that the functional form of the contours, which are cubic, does not match the quadratic scaling of the upper boundary of the phase space.  The addition of pile-up pushes jets out of the small $\ecf{2}{\beta}$ region of phase space, where $\Dobs{2}{\beta}$ takes large values.  Therefore, an effect of pile-up is to reduce the value the $\Dobs{2}{\beta}$ measured on a jet, compressing the long tail of the perturbative $\Dobs{2}{\beta}$ distribution (exhibited in \Fig{fig:nomasscut}, for example) toward a central value.

We can predict the value of $\Dobs{2}{\beta}$ for background jets in the limit of infinite pile-up.  In this limit, the jet has a single hard core of radiation surrounded by perfectly uniform pile-up radiation. If the energy fraction of each of the $n$ pile-up particles is $z_{pu}$, then the two- and three-point correlation functions take the values
\begin{align}
\ecf{2}{\beta}=n z_{pu} \, , \qquad \text{and} \qquad \ecf{3}{\beta}={n\choose 2}z_{pu}^2\,,
\end{align}
so that, as $n\to \infty$, we have the relation
\begin{align}\label{eq:pure_pu_rel}
\ecf{3}{\beta}=\frac{1}{2}(\ecf{2}{\beta})^2\,. 
\end{align}
Using the definition of $\Dobs{2}{\beta}$, we find that in this limit,
\begin{align}
\Dobs{2}{\beta}=\frac{1}{2\ecf{2}{\beta}}\,.
\end{align}
As the amount of pile-up increases, we expect that the distribution of $\Dobs{2}{\beta}$ accumulates about this value, with a minimal change in the mean, but significant decrease in the width of the distribution. This  behavior is relatively distinct from that of most event shapes under pile-up, which tend to have a shift of the mean as pile-up is increased. The reason that this behavior is pronounced with $\Dobs{2}{\beta}$ is because it is both infrared and collinear unsafe and because the scalings of $\ecf{2}{\beta}$ and $\ecf{3}{\beta}$ in the observable and the upper boundary of the phase space are different. 

For the background distribution of $\Cobs{2}{\beta}$, on the other hand, the parametric scaling of the observable is unaffected by the addition of pile-up, but we expect an ${\cal O}(1)$ shift of the mean of the distribution to larger values. As pile-up increases, from \Eq{eq:pure_pu_rel} we expect the distribution should accumulate about the infinite pile-up limit of $\Cobs{2}{\beta}=1/2$.  We therefore predict that as the pile-up increases, the distribution of $\Cobs{2}{\beta}$ on background jets becomes independent of $\beta$.

We can also understand the behavior of the signal distribution under the addition of soft pile-up radiation. As was discussed, and is shown schematically in \Fig{fig:ps_pu}, signal jets at larger $\ecf{2}{\beta}$ are shifted vertically in the $(\ecf{2}{\beta}$, $\ecf{3}{\beta})$ phase space by pile-up radiation. This predicts that for both $\Cobs{2}{\beta}$ and $\Dobs{2}{\beta}$, the primary effect of the addition of pile-up radiation will be to shift the mean of the distribution to larger values, with a limited modification to its shape. Furthermore, due to the cubic contours for $\Dobs{2}{\beta}$, the shift of the mean of the distribution will be smaller for $\Dobs{2}{\beta}$ than for $\Cobs{2}{\beta}$, implying reduced sensitivity of $\Dobs{2}{\beta}$ to pile-up radiation.

Note that a similar analysis can be straightforwardly applied to the $N$-subjettiness observables $\Nsub{1}{\beta}$ and $\Nsub{2}{\beta}$. Because the analysis proceeds identically, we simply state the result. Under the addition of pile-up radiation, single prong jets at small $\Nsub{1}{\beta}$ experience a shift of both observables, but their parametric scaling remains the same: 
\begin{align}
\Nsub{1}{\beta} & \to  \Nsub{1}{\beta}+z_{pu} \,, \\
\Nsub{2}{\beta} & \to \Nsub{2}{\beta}+z_{pu}\,.
\end{align}
That is, under the addition of pile-up, background jets move along the upper boundary of the phase space, where $\Nsub{2}{\beta}\sim\Nsub{1}{\beta}$.

For jets with two hard subjets, the value of $\Nsub{1}{\beta}$ is not affected, while $\Nsub{2}{\beta}$ shifts as
\begin{align}
\Nsub{1}{\beta} &\to \Nsub{1}{\beta} \,, \\
\Nsub{2}{\beta} &\to \Nsub{2}{\beta}+z_{pu}\, .
\end{align}
This corresponds to a vertical movement in the $\Nsub{1}{\beta},\Nsub{2}{\beta}$ phase space under the addition of pile-up, as was the case for $\ecf{2}{\beta} ,\ecf{3}{\beta}$. We therefore expect a similar behavior for $\Nsub{2,1}{\beta}$ with the addition of pile-up, with a shift of the mean value for the signal distributions and accumulation near $1$ for the background distributions. Unfortunately, power counting alone does not allow us to compare the expected shifts in $\Nsub{2,1}{\beta}$ as compared with those in $\Cobs{2}{\beta}$ and $\Dobs{2}{\beta}$.

\subsubsection{Summary of Power Counting Predictions}
\label{sec:sumpredict_pu}

Here, we summarize the main predictions from our power counting analysis of the impact of pile-up radiation on the $\Cobs{2}{\beta}$ and $\Dobs{2}{\beta}$ distributions. We have:
\begin{itemize}

\item For background $\Dobs{2}{\beta}$ distributions the primary effect of the addition of pile-up radiation is to narrow the distribution; in particular, the long tail of the $\Dobs{2}{\beta}$ distribution is truncated because pile-up moves those jets in the 1-prong region of phase space out of the region of small $e_2$. The peak of the $\Dobs{2}{\beta}$ background distribution should be relatively insensitive to the addition of pile-up, with $\Dobs{2}{\beta}$ accumulating around $\Dobs{2}{\beta} = 1/(2 \ecf{2}{\beta})$ in the limit that uniform pile-up dominates. 

\item For background $\Cobs{2}{\beta}$ distributions the primary effect of the addition of pile-up radiation is a shift of the peak to larger values by an $\mathcal{O}(1)$ amount proportional to the pile-up. The distribution also becomes compressed, and accumulates around $\Cobs{2}{\beta}= 1/2$ in the limit that uniform pile-up dominates.

\item For signal, the primary effect of the addition of pile-up radiation is to translate the mean of the distribution. The displacement of the mean is expected to be smaller for $\Dobs{2}{\beta}$ than for $\Cobs{2}{\beta}$ because of the different scalings for contours of constant $\Cobs{2}{\beta}$ and $\Dobs{2}{\beta}$.

\end{itemize}

\subsubsection{Monte Carlo Analysis}
\label{sec:mc_e2e3_pu}

We now study these predictions in Monte Carlo using the \pythia{8} event samples described in \Sec{sec:mc_e2e3}.\footnote{In this section, we restrict to the \pythia{8} generator, having satisfied ourselves in \Sec{sec:mc_e2e3} that the parametrics of the perturbative phase space are well described by both generators.}  Pile-up was simulated by adding $N_{PV}$ minimum bias events at the 8 TeV LHC, generated with \pythia{8}, to the $pp\to Zj$ and $pp\to ZZ$ samples. To demonstrate the resilience of the distributions to pile-up, we wish to add pile-up radiation to a set of jets with well-defined perturbative properties.  To do this, we cluster jets with the WTA recombination scheme \cite{Larkoski:2014uqa,Larkoski:2014bia} and require that the mass of the jets in the absence of pile-up is $m_J<100$ GeV.  It was shown in \Ref{Larkoski:2014bia} that the jet axis found by the WTA recombination scheme is robust to pile-up and so, when pile-up is included, the perturbative content of the jets will be unaffected. This procedure, although clearly not related to an experimental analysis, provides a measure of the sensitivity of the distributions to soft pile-up radiation. This procedure is similar to that used in \Ref{Soyez:2012hv} to assess the impact of pile-up and pile-up subtraction techniques on a variety of different jet shapes. 

\begin{figure}
\begin{center}
\subfloat[]{
\includegraphics[width=6.5cm]{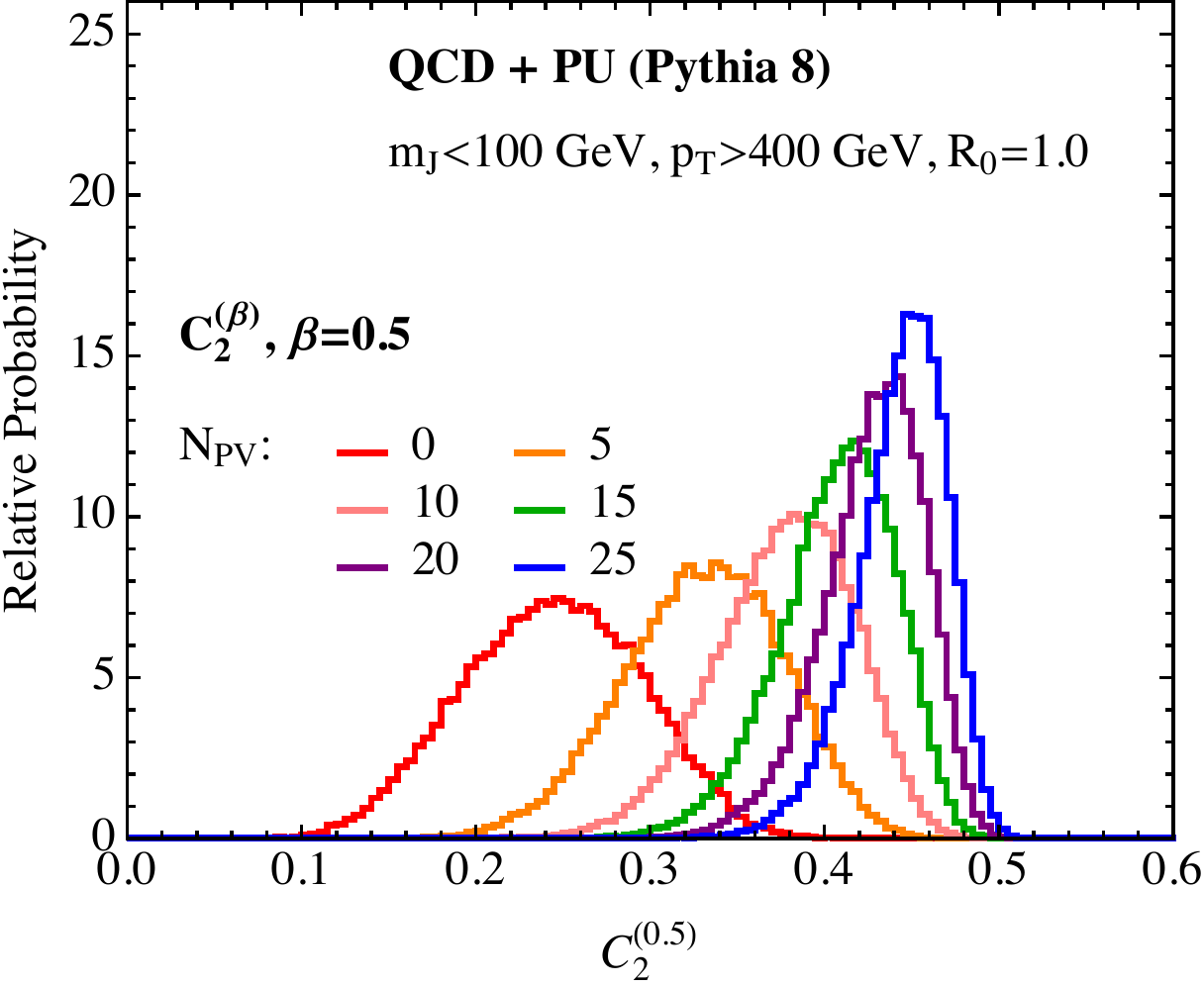}
}\qquad
\subfloat[]{
\includegraphics[width=6.5cm]{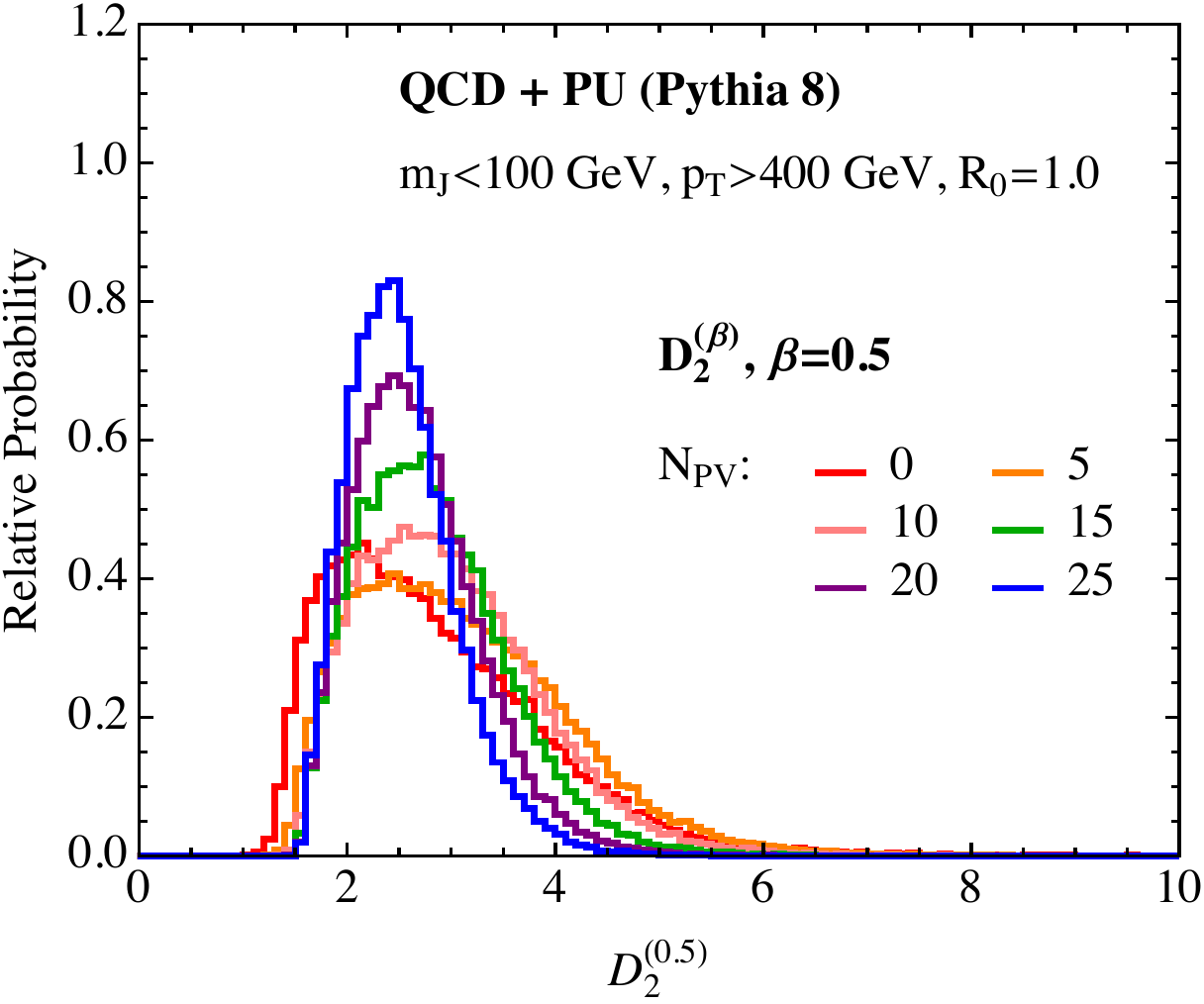}
}\qquad
\subfloat[]{
\includegraphics[width=6.5cm]{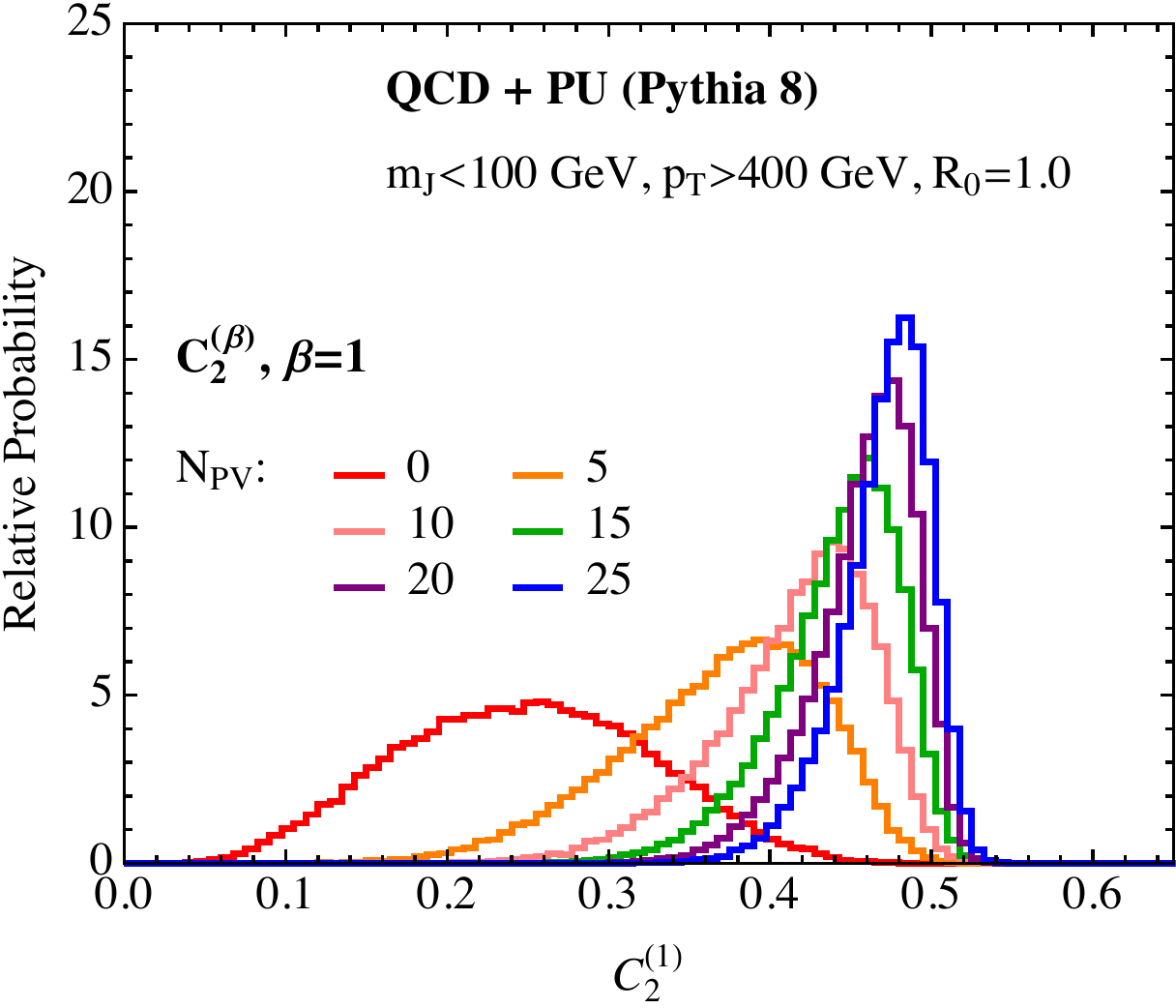}
}\qquad
\subfloat[]{
\includegraphics[width=6.5cm]{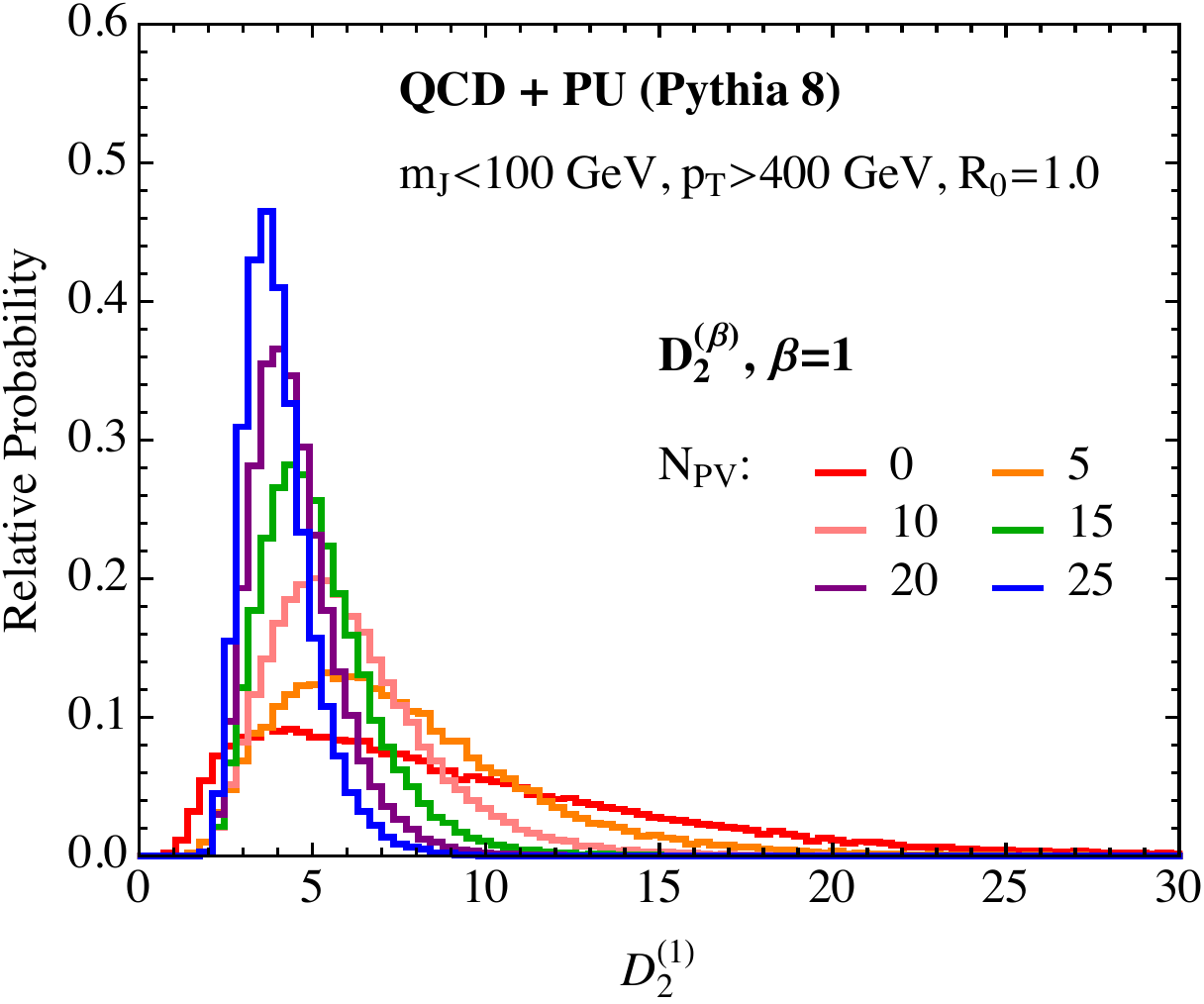}
}\qquad
\subfloat[]{
\includegraphics[width=6.5cm]{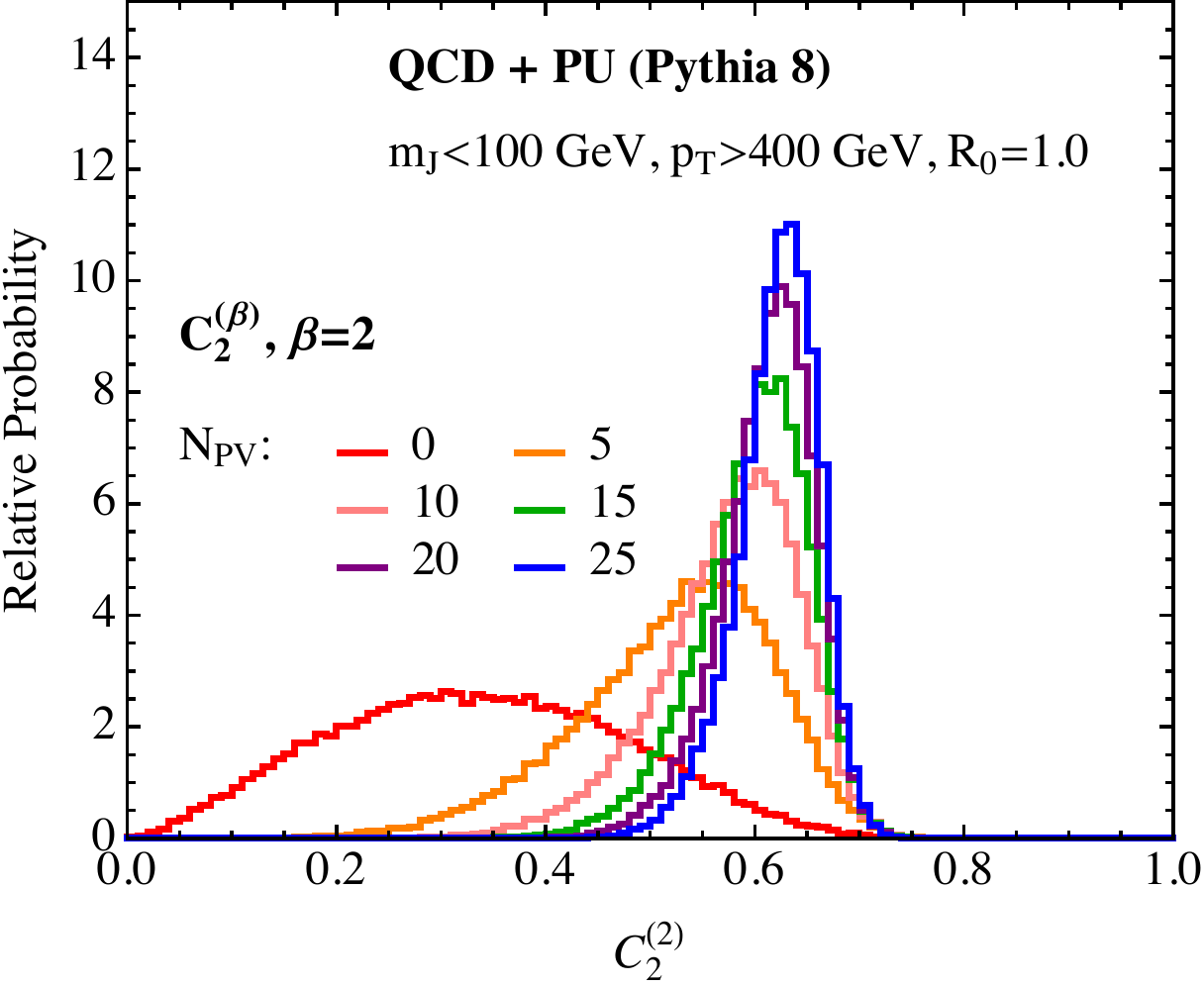}
}\qquad
\subfloat[]{
\includegraphics[width=6.5cm]{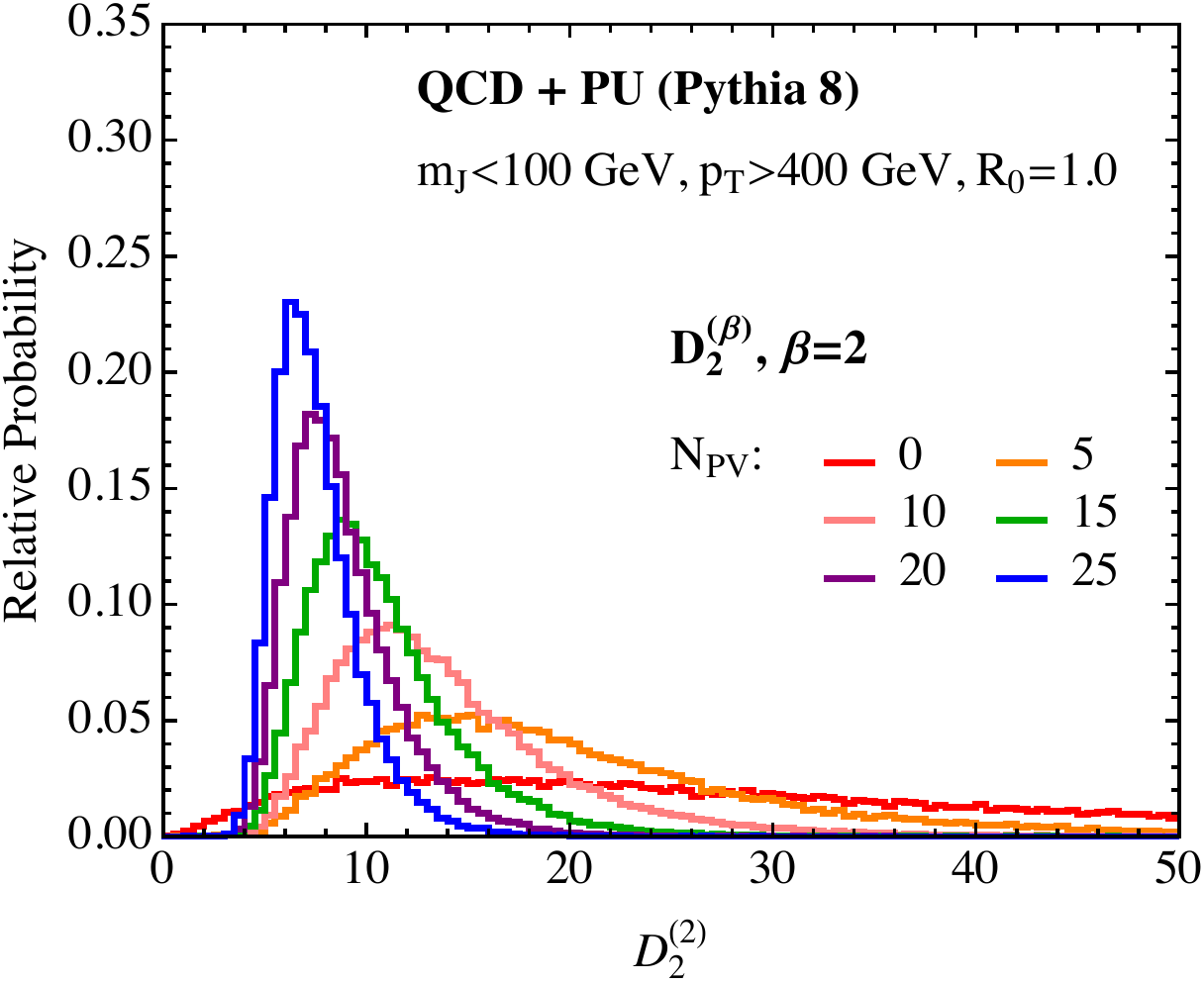}
}
\end{center}
\caption{Effect of pile-up on the distributions of $\Cobs{2}{\beta}$ (left) and $\Dobs{2}{\beta}$ (right) for QCD jets for $\beta = 0.5,1,2$ as measured on the \pythia{8} samples.  The number of pile-up vertices ranges from $N_{PV} = 0 $ (no pile-up) to $ N_{PV}=25$.
}
\label{fig:Bkg_pu}
\end{figure}

\begin{figure}
\begin{center}
\subfloat[]{
\includegraphics[width=6.5cm]{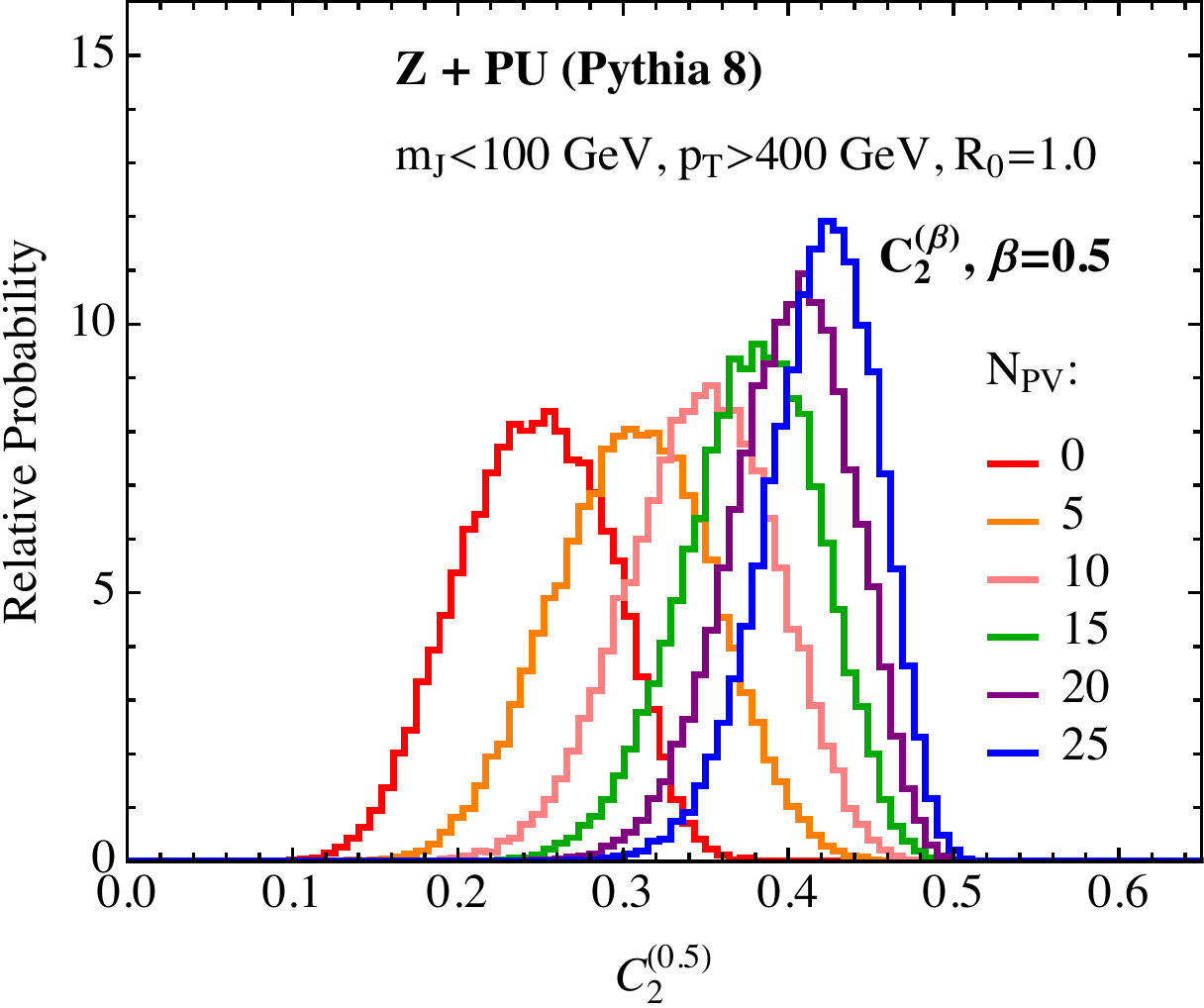}
}\qquad
\subfloat[]{
\includegraphics[width=6.5cm]{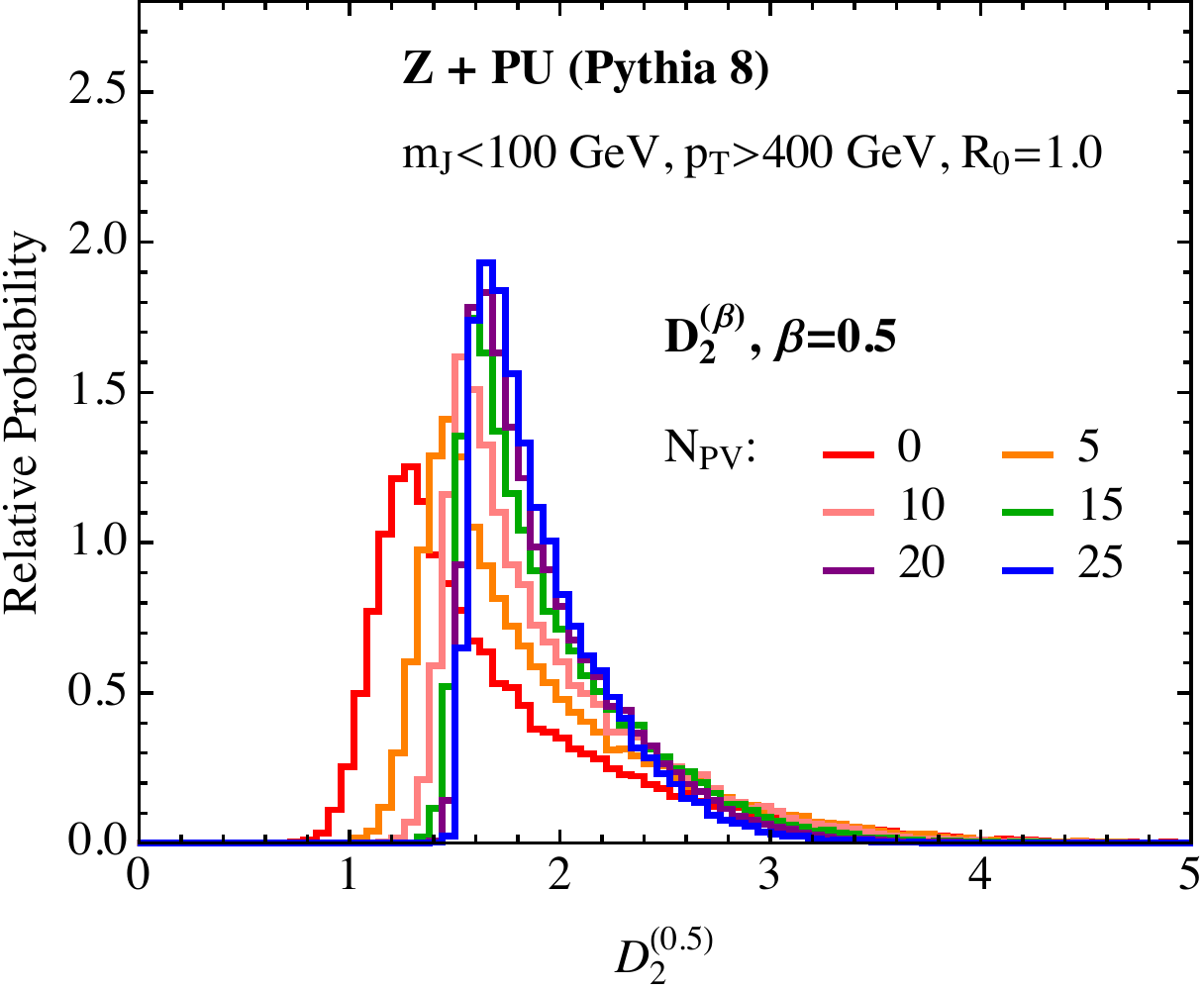}
}\qquad
\subfloat[]{
\includegraphics[width=6.5cm]{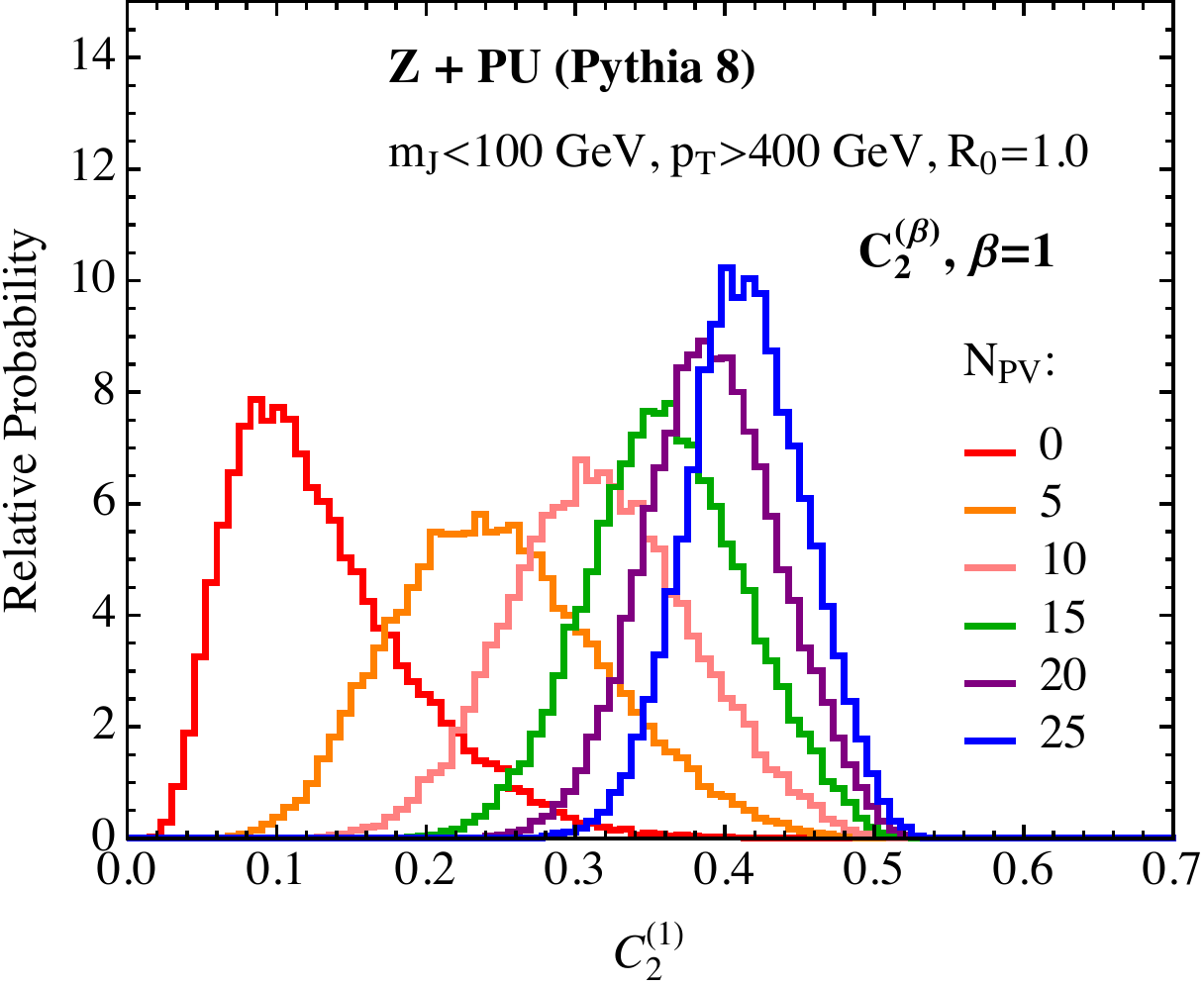}
}\qquad
\subfloat[]{
\includegraphics[width=6.5cm]{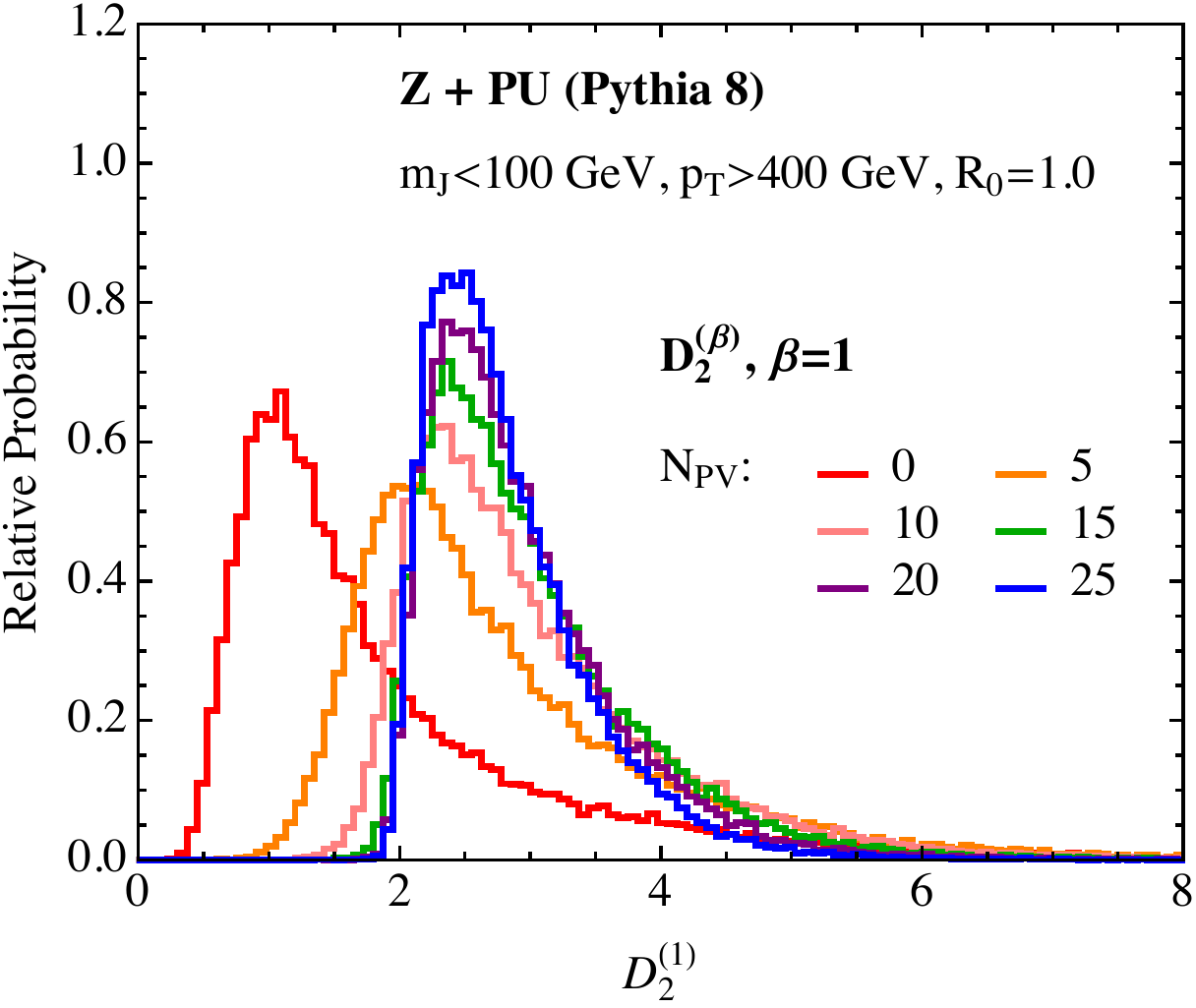}
}\qquad
\subfloat[]{
\includegraphics[width=6.5cm]{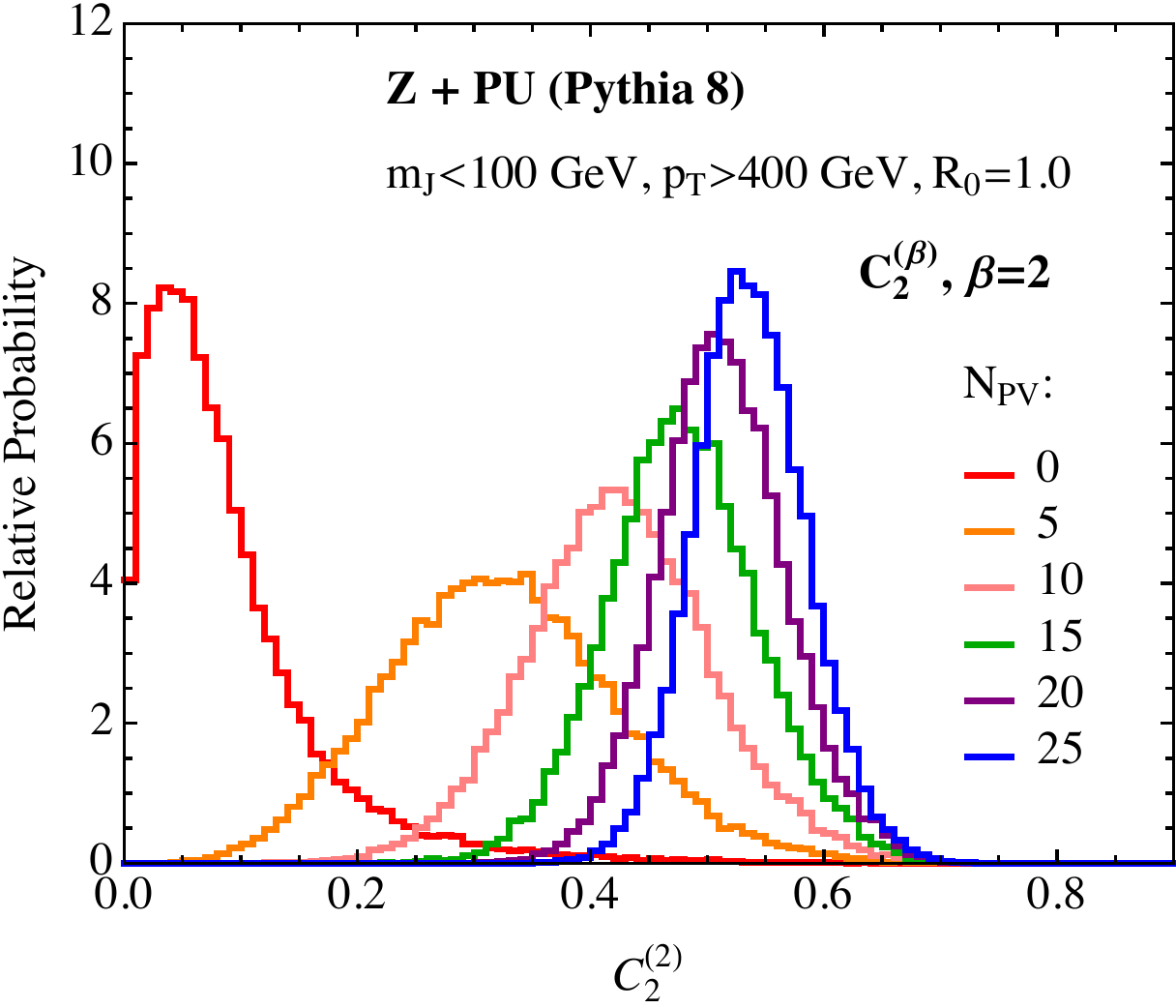}
}\qquad
\subfloat[]{
\includegraphics[width=6.5cm]{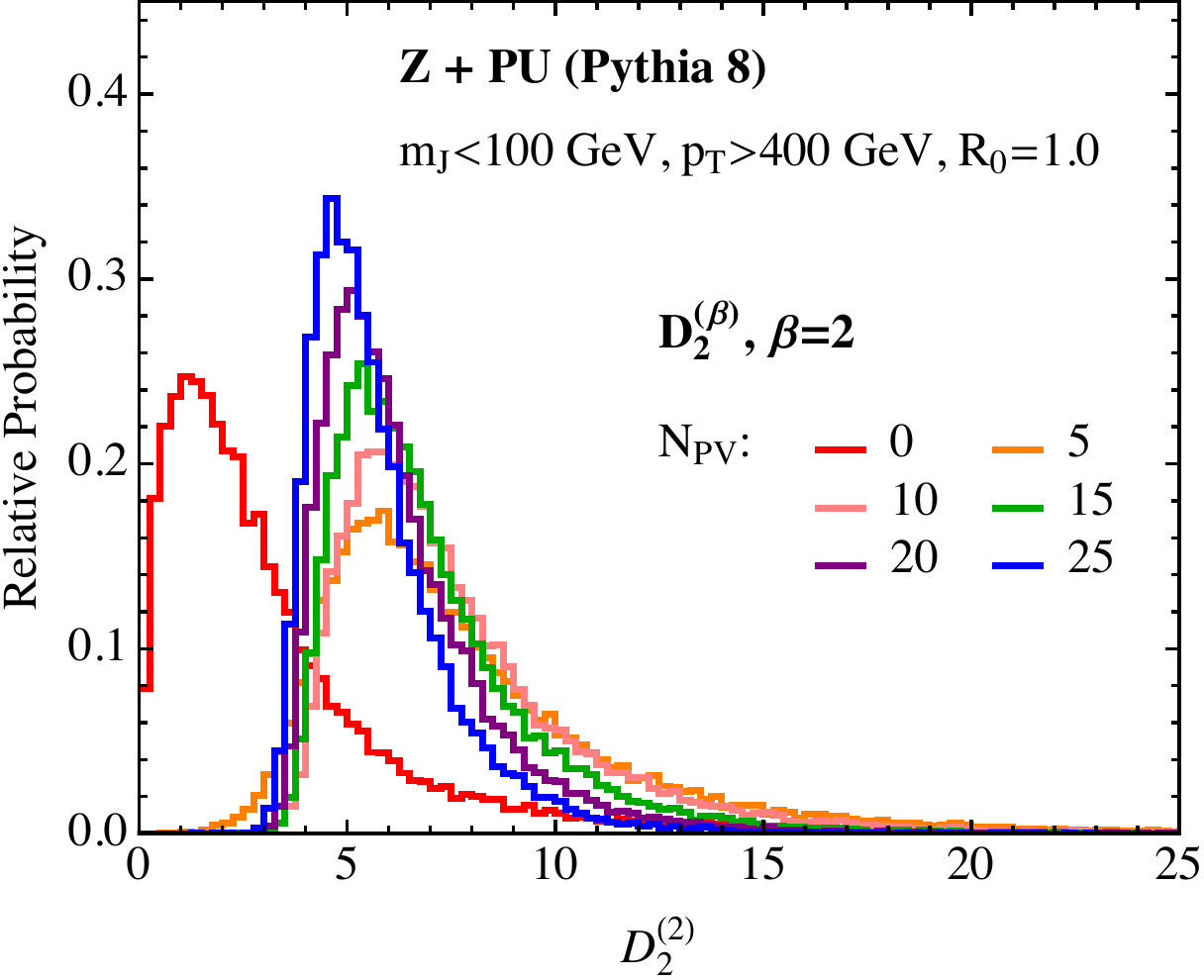}
}
\end{center}
\caption{
The same as \Fig{fig:Bkg_pu}, but measured on boosted $Z$ jets.
}
\label{fig:Sig_pu}
\end{figure}

\begin{figure}
\begin{center}
\subfloat[]{
\includegraphics[width=6.5cm]{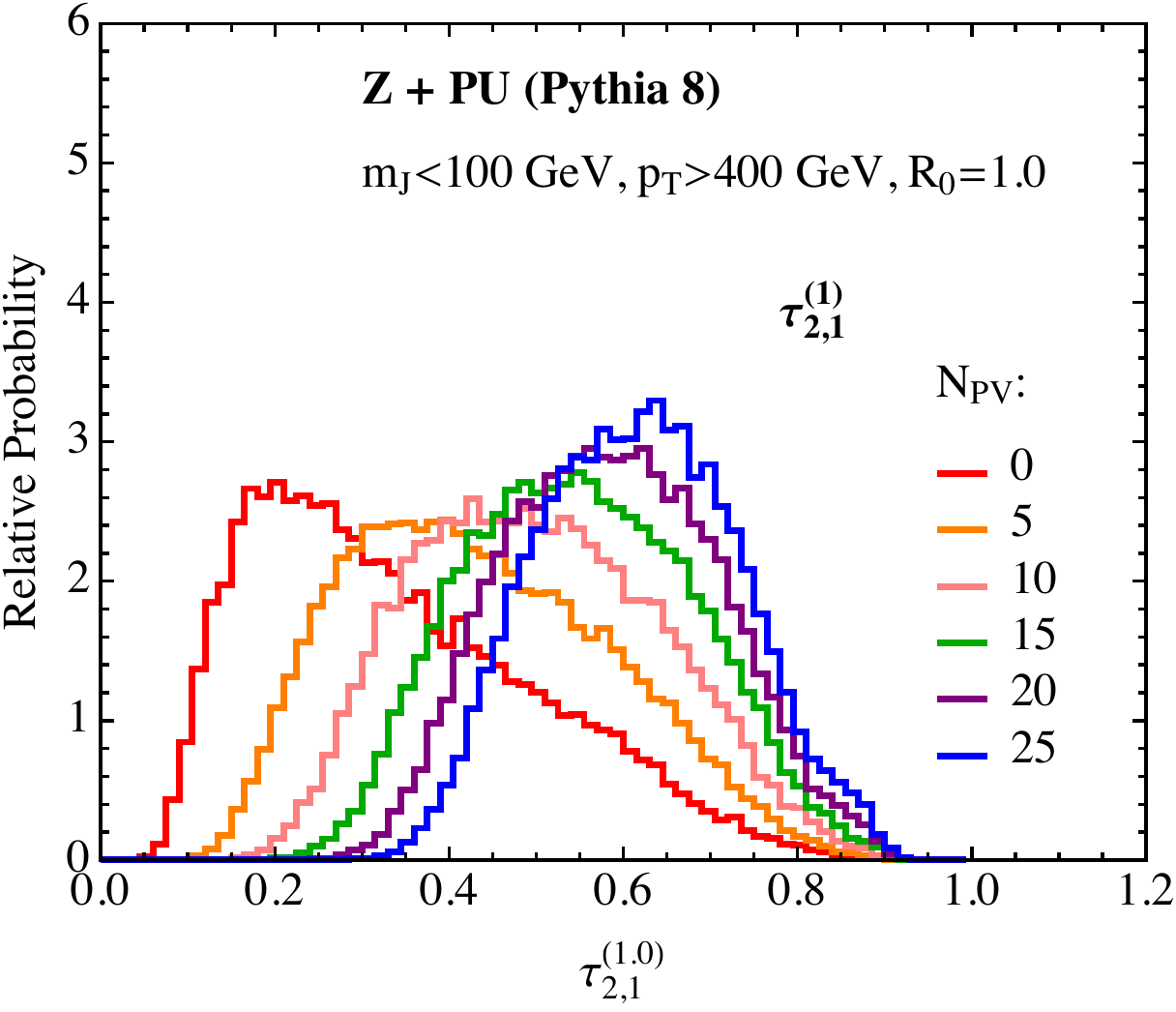}
}\qquad
\subfloat[]{
\includegraphics[width=6.5cm]{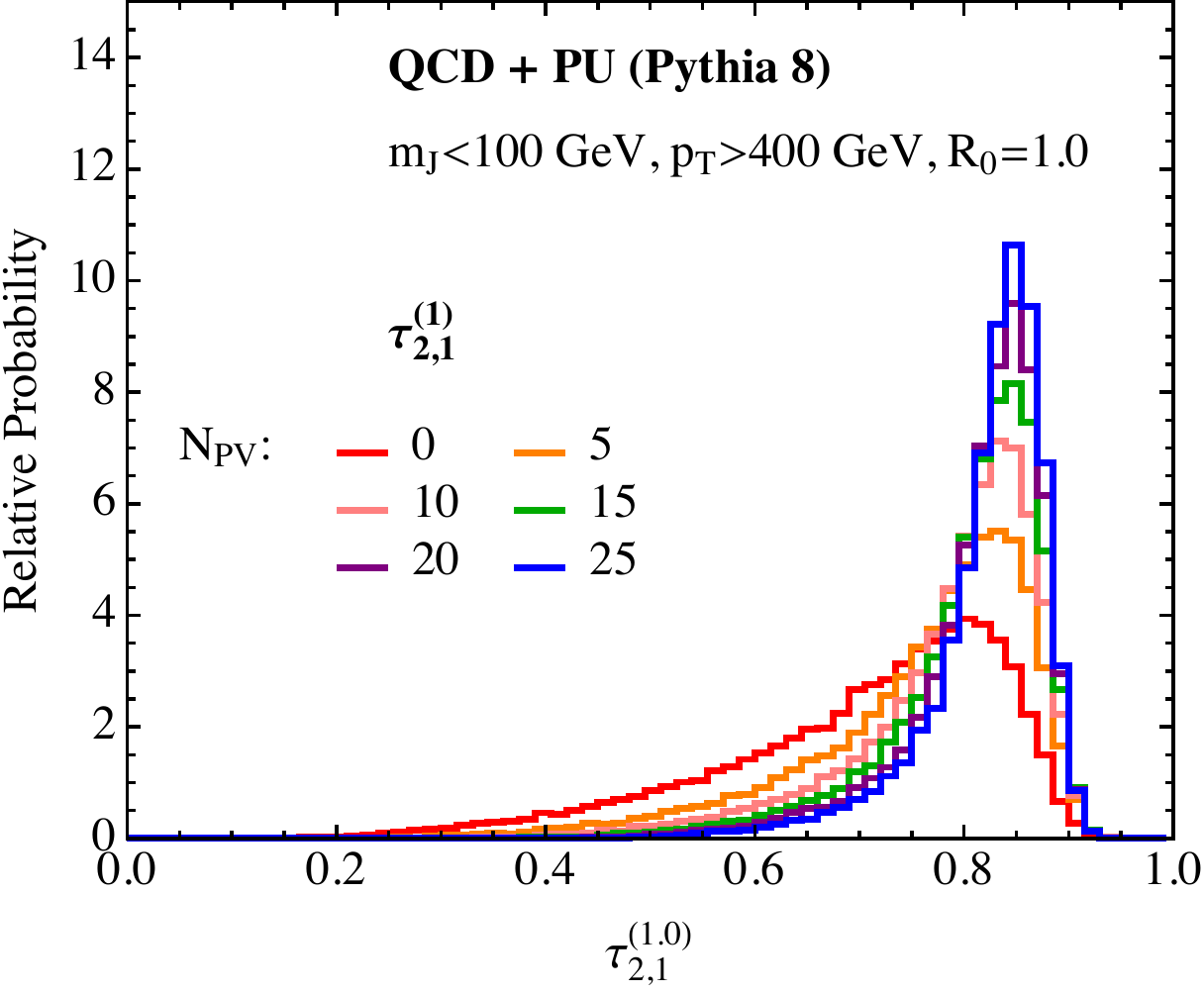}
}
\end{center}
\caption{Effect of pile-up contamination on the measured value of $\Nsub{2,1}{1}$ for signal (left) and background (right) jets from the \pythia{8} samples.  The number of pile-up vertices ranges from $N_{PV}=0$ (no pile-up) to $N_{PV} = 25$. 
}
\label{fig:nsub_pu}
\end{figure}

Using this sample, we can assess the degree to which the power counting predictions of \Sec{sec:pu} are realized in the Monte Carlo simulation. We begin by considering the effect of pile-up on the background distributions. In \Fig{fig:Bkg_pu}, we plot background distributions for $\Cobs{2}{\beta}$ and $\Dobs{2}{\beta}$ with the addition of up to $N_{PV}=25$ pile-up vertices for a few values of $\beta$. For the variable $\Dobs{2}{\beta}$, the power counting analysis of \Sec{sec:pu} predicted that the dominant effect of the addition of pile-up would be a compression of the long tail of the distribution into a peak around $1/2\ecf{2}{\beta}$, with relatively little shift in the mean. This behavior is manifest for all three values of $\beta$ shown. The peak value of the distribution is remarkably stable under the addition of pile-up. On the other hand, for the observable $\Cobs{2}{\beta}$, the mean of the distribution is highly unstable to the addition of pile-up. The dominant effects of pile-up on the distribution of $\Cobs{2}{\beta}$ is a displacement of the mean and the accumulation near the value $\Cobs{2}{\beta}=1/2$. 

In \Fig{fig:Sig_pu} we consider the same set of distributions as for \Fig{fig:Bkg_pu}, but for the signal boosted $Z$ boson sample. In this case, the analysis of the phase space predicted that the dominant effect of the pile-up on the distributions is a shift for both $\Cobs{2}{\beta}$ and $\Dobs{2}{\beta}$, with the shift being smaller for $\Dobs{2}{\beta}$. This behavior is manifest in \Fig{fig:Sig_pu}. The difference in the stability of the mean between $\Cobs{2}{\beta}$ and $\Dobs{2}{\beta}$ is particularly pronounced at small $\beta$. At larger $\beta$, the $\Dobs{2}{\beta}$ distribution exhibits a jump at small amounts of pile-up, and then remains stable as pile-up increases. Unfortunately, we have not been able to understand this behavior completely from power counting. Nevertheless, the improved stability of the distributions of $\Dobs{2}{\beta}$ as compared with $\Cobs{2}{\beta}$ is promising.

For comparison, in \Fig{fig:nsub_pu}, we consider the impact of pile-up on signal and background distributions for $\Nsub{2,1}{\beta}$, for the representative value $\beta=1$. As for $\Cobs{2}{\beta}$ and $\Dobs{2}{\beta}$, we expect that the dominant effect of pile-up on the signal distributions is a shift of the peak value, while for the background distributions, we expect a small shift of the mean and an accumulation of the distribution near $\Nsub{2,1}{\beta}=1$. This is exhibited in the Monte Carlo.

\begin{figure}
\begin{center}
\subfloat[]{
\includegraphics[width=6.5cm]{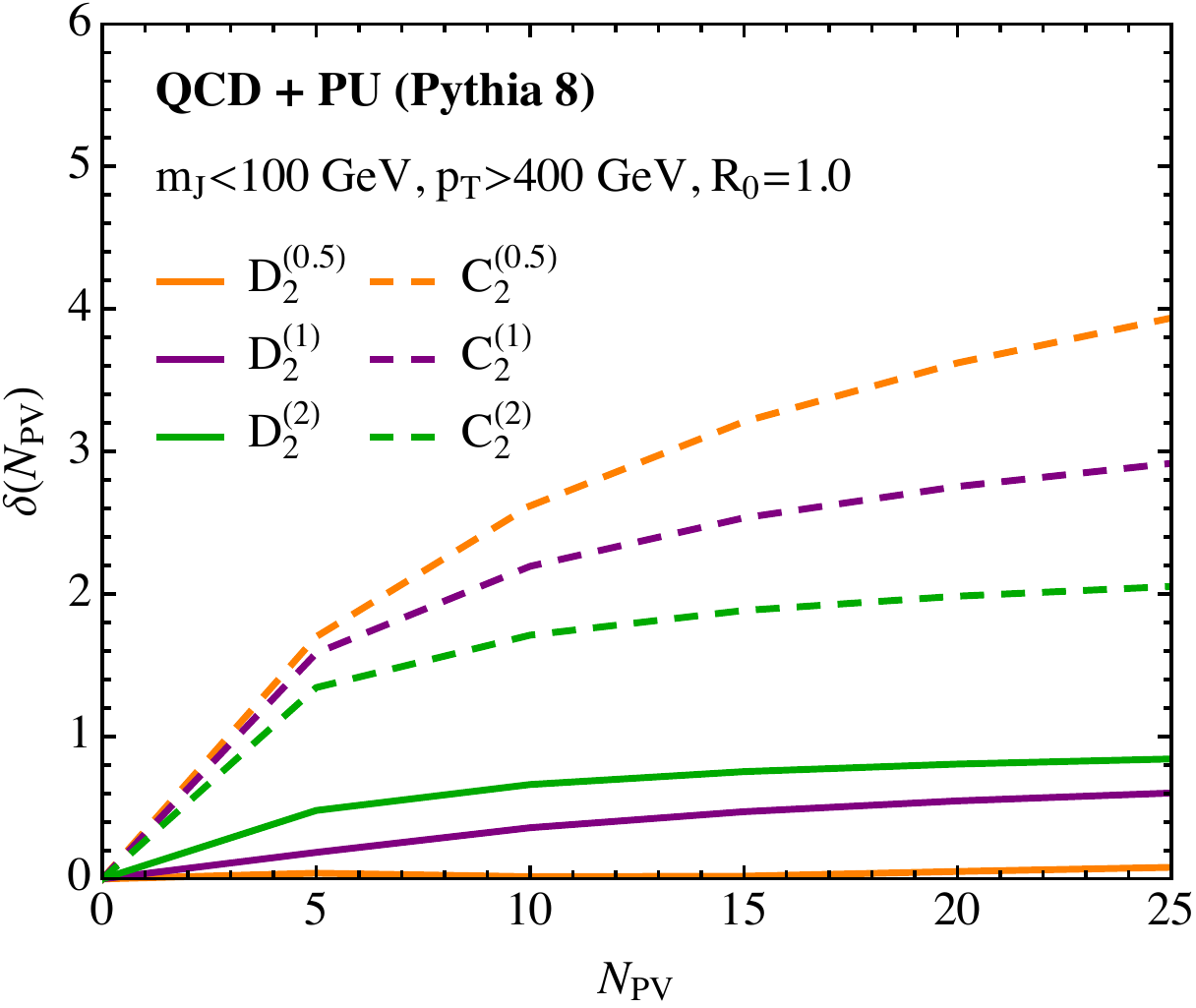}
}\qquad
\subfloat[]{
\includegraphics[width=6.5cm]{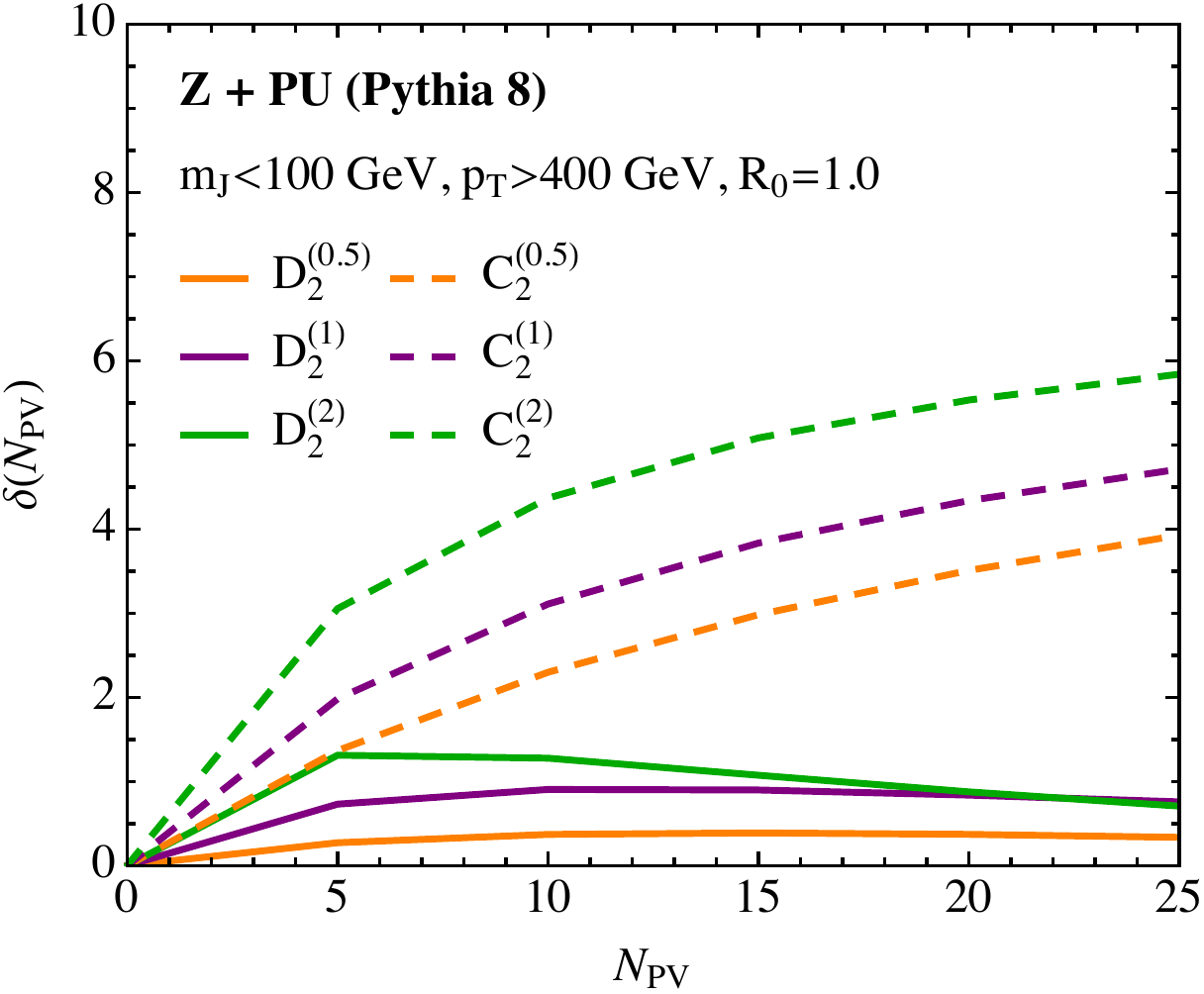}
}
\end{center}
\caption{A comparison of the susceptibility as a function of the number of pile-up vertices $N_{PV}$ of background (left) and signal (right) distributions for $\Cobs{2}{\beta}$ and $\Dobs{2}{\beta}$ to pile-up using the measure $\delta (\text{$N_{PV}$})$ for $\beta = 0.5,1,2$. 
}
\label{fig:C2_pu_mean}
\end{figure}

For a more quantitative study of the stability of the distributions to pile-up, we define
\begin{equation}\label{eq:delta_$N_{PV}$}
\delta_{\Xobs{2}{\beta}}(\text{$N_{PV}$})=\frac{\langle \Xobs{2}{\beta} (\text{$N_{PV}$}) \rangle -\langle \Xobs{2}{\beta} (\text{$N_{PV}$}=0) \rangle}{\sigma_{\Xobs{2}{\beta}}(\text{$N_{PV}$}=0)}\,,
\end{equation}
where $\Xobs{2}{\beta}$ stands for either $\Cobs{2}{\beta}$ or $\Dobs{2}{\beta}$ and $\sigma$ denotes the standard deviation.  This quantity is a measure of how much the mean of the distribution is affected by pile-up, normalized by the width of the distribution, which is important since the observables $\Cobs{2}{\beta}$ and $\Dobs{2}{\beta}$ have support over very different ranges.  While it is clear from \Fig{fig:Bkg_pu} that the dominant effect of pile-up on the background distributions for $\Dobs{2}{\beta}$ is not a shift of the mean, and so the change of the distribution is not accurately captured by the measure of \Eq{eq:delta_$N_{PV}$}, the deviation of the mean is a commonly studied measure of an observable's susceptibility to pile-up. In \Fig{fig:C2_pu_mean} we plot $\delta(\text{$N_{PV}$})$ for the variables $\Cobs{2}{\beta}$ and $\Dobs{2}{\beta}$. As was demonstrated in \Figs{fig:Bkg_pu}{fig:Sig_pu}, the mean of the distributions of $\Dobs{2}{\beta}$ is considerably more stable for both the signal and background distributions.

\section{Power Counting Quark vs. Gluon Discrimination}
\label{sec:qvg}

Unlike the case of boosted $Z$ bosons vs.~massive QCD jets, applying a power counting analysis to quark vs.~gluon jet discrimination demonstrates the limitations of the technique.  Both quark and gluon jets dominantly have only a single hard core, and so the natural discrimination observables are the two-point energy correlation functions, $\ecf{2}{\beta}$.\footnote{To next-to-leading logarithmic accuracy, $\ecf{2}{\beta}$ are identical to the recoil-free angularities \cite{Larkoski:2014uqa}.}  As shown in \Sec{sec:softcollqcd}, power counting the two-point energy correlation functions constrains the soft and collinear radiation as:
\begin{equation}
\ecf{2}{\beta}\sim z_s \sim R_{cc}^\beta \ .
\end{equation}
With power counting alone, this is as far as our analysis can go.  $\ecf{2}{\beta}$ does not parametrically separate quark and gluon jets from one another.

This result is not surprising, however, because there are no qualities of quark and gluon jets that are parametrically different.  Indeed,
\begin{align*}
C_A&\sim C_F \ ,\\
N_C&\sim n_f \ ,\\
\text{spin } 1 &\sim \text{spin }1/2 \ ,
\end{align*}
where $C_A$ and $C_F$ are the color factors for gluons and quarks, $N_C$ is the number of colors, and $n_f$ the number of active fermions.  Predictions of what the best observable for quark vs.~gluon discrimination is requires a detailed analysis of the effects of these order-1 parameters, which has been studied in several chapters \cite{Gallicchio:2011xq,Gallicchio:2012ez,Larkoski:2013eya,Larkoski:2014pca}.  However, with the additional input of the form of the splitting functions for quarks and gluons, we can predict that the discrimination power of $\ecf{2}{\beta}$ improves as $\beta$ decreases because smaller $\beta$ emphasizes the collinear region of phase space over soft emissions.  Collinear emissions are sensitive to the spin of the parton in addition to the total color of the jet, and thus are more distinct between quark and gluon jets.  This prediction is borne out by explicit calculation to next-to-leading logarithmic accuracy \cite{Larkoski:2013eya}. An analytic calculation of the improved discrimination power from simultaneous measurement of the recoil-free angularities for two different powers of the angular exponent was calculated in \cite{Larkoski:2014pca}.

Nevertheless, this suggests that power counting does make a definite prediction of quark vs.~gluon discrimination performance.  Because all the physics of quark vs.~gluon jet discrimination is controlled by order-1 numbers, the predicted discrimination should be sensitive to the tuning of order-1 numbers in a Monte Carlo.  It has been observed that \pythia{8} and \herwigpp\  give wildly different predictions for quark vs.~gluon discrimination power \cite{Larkoski:2013eya,Larkoski:2014pca,CMS:2013kfa,Aad:2014gea}, and presumably the difference is dominated by the tuning of the Monte Carlos.  However, isolating pure samples of quark and gluon jets is challenging experimentally 
\cite{CMS:2013kfa,CMS:2013wea,Aad:2014gea,Gallicchio:2011xc} 
and most of the subtle differences between quarks and gluons only appear at an order formally beyond the accuracy of a Monte Carlo.  Therefore, to solve this issue will require significant effort from experimentalists, Monte Carlo authors, and theorists to properly define quark and gluon jets, to identify the dominant physics, and to isolate pure samples for tuning.

\section{Conclusions}
\label{sec:conc}

In this chapter we have demonstrated that power counting techniques can be a powerful guiding principle when constructing observables for jet substructure and for understanding their behavior. Since power counting captures the parametric physics of the underlying theory, its predictions should be robust to Monte Carlo tunings. Using the simple example of discriminating boosted $Z$ bosons from QCD jets with the energy correlation functions, we showed that a power counting analysis identified $\Dobs{2}{\beta}$ as the natural discrimination observable. The scaling of this observable parametrically separates regions of the $(\ecf{2}{\beta},\ecf{3}{\beta})$ phase space dominated by 1- and 2-prong jets.  The distinction between 1- and 2-prong jets is invariant to boosts along the jet direction.

To verify the power counting predictions, we performed a Monte Carlo analysis comparing $\Dobs{2}{\beta}$ with a previously proposed observable, $\Cobs{2}{\beta}$, also formed from the energy correlation functions. We showed that $\Dobs{2}{\beta}$ is a superior observable for discrimination because $\Cobs{2}{\beta}$ inextricably mixes signal-rich and background-rich regions of phase space. All power counting predictions were confirmed by both \herwigpp\ and \pythia{8}, showing that the dominant behavior of the observables is governed by parametric scalings and not by $\mathcal{O}(1)$ numbers. This was contrasted with the case of quark vs.~gluon discrimination for which no parametric differences exist, leading to large discrepancies when simulating quark vs.~gluon discrimination with different Monte Carlo generators.

We also demonstrated that power counting can be used to understand the impact of pile-up on different regions of the phase space, and hence on the distributions of discriminating variables. The distributions for $\Dobs{2}{\beta}$ exhibited improved stability compared with those of $\Cobs{2}{\beta}$, while the background distributions have the interesting feature of being compressed to a central value by the addition of pile-up radiation.

We anticipate many directions to which the power counting approach could be applied. We have restricted ourselves in this chapter to a study of observables formed from ratios of energy correlation functions with the same angular exponent. A natural generalization is to ratios of energy correlation functions with different angular exponents, where the optimal observable is given by $\Dobs{2}{\alpha,\beta}=\ecf{3}{\beta}/(\ecf{2}{\alpha})^{3\beta/\alpha}$. Such variables could be useful when considering pile-up in the presence of mass cuts, which are required experimentally. In the presence of a mass cut, an angular exponent of $\ecfnobeta{2}$ near $2$ provides a simple restriction on the phase space, while lowering the angular exponent of $\ecfnobeta{3}$ reduces the effect of soft wide angle radiation. Along these lines, the impact of grooming techniques on the phase space is also simple to understand by power counting \cite{Walsh:2011fz}, and could be used to motivate the design of variables with desirable behavior under grooming.

As another example of considerable interest, the power counting analysis can be extended to the study of top quark discrimination variables by considering the phase space for 1-, 2- and 3-prong jets defined by the two-, three- and four-point energy correlation functions. While a complete analytic calculation for this case is not feasible, a power counting analysis is, and can be used to predict discriminating observables with considerably improved performance compared to those originally proposed in \cite{Larkoski:2013eya}. In the case of a three dimensional phase space, a cut on the jet mass only reduces the phase space to a two dimensional subspace, so that the functional form of the observable remains important. This case was studied in detail in \cite{Larkoski:2014zma}, where an observable $D_3$ was proposed, and studied in Monte Carlo, where it showed improved performance over several other top tagging observables.

Our observation that boost-invariant combinations of the energy correlation functions are the most powerful discriminants can also be exploited for discrimination: we can use boost invariance as a guide for defining the best observables. Together with power counting, this gives a simple but powerful analytic handle to understand and design jet substructure observables.

%% file: chap8.tex

%
%
%
%
%

\renewcommand{\textfraction}{0.10}
\renewcommand{\topfraction}{0.90}
\renewcommand{\bottomfraction}{0.90}
\renewcommand{\floatpagefraction}{0.65}

\newcommand{\oXn}{\varmathbb{X}_{\hat n}}
\newcommand{\Xn}{X_{\hat n}}
\newcommand{\half}{\frac{1}{2}}
\newcommand{\ecflp}[2]{\tilde e_{#1}^{(#2)}}
\newcommand{\ecfop}[2]{\mathbf{E_{#1}}^{(#2)}}
\newcommand{\anglediff}{\text{tan}^2\left(\frac{R}{2}\right)-\text{tan}^2\left(\frac{\theta_{SJ}}{2}\right)}
\newcommand{\sja}{n_{sj}}
\newcommand{\sjabar}{\bar{n}_{sj}}
\newcommand{\outj}{B}

\def\ea{e_\alpha}
\def\eb{e_\beta}
\def\eaba{e_\alpha^{\beta/\alpha}}
\def\ebab{e_\beta^{\alpha/\beta}}
\def\log{\text{log}}
\def\muss{\mu_{S\rightarrow S}}
\def\musj{\mu_{S\rightarrow J}}

\def \thetac {\theta_c}
\def \thetacs {\theta_{cs}}
\def \thetasj {\theta_{sj}}
\def \zs {z_{s}}
\def \zcs {z_{cs}}
\def \zsj {z_{sj}}

\def\tauo{{\tau_1}}
\def\taut{{\tau_2}}

\def\be{\begin{equation}}
\def\ee{\end{equation}}

\newcommand{\img}{\mathrm{i}}

\def\nslash{n\hspace{-2mm}\slash}
\def\nbarslash{\bar n\hspace{-2mm}\slash}
\def\nslashinline{n\!\!\!\slash}
\def\nbarslashinline{\bar n\!\!\!\slash}
\def\nbar{\bar n}

\newcommand{\eeclp}[2]{\tilde e_{#1}^{(#2)}}
\newcommand{\ecfLa}{e_{2}^{(\alpha)}}
\newcommand{\ecfLb}{e_{2}^{(\beta)}}
\newcommand{\ecfres}{e_{3}^{(\alpha)}}
\newcommand{\ecfresgen}{e_\text{res}}
\newcommand{\ecfresgenLP}{\tilde e_\text{res}}
\newcommand{\ecfreslp}{\tilde e_{3}^{(\alpha)}}
\newcommand{\ecfR}{e_{R\alpha}^{(2)}}

\newcommand{\sje}{z_{sj}}
\newcommand{\sjtheta}{\theta_{sj}}
\newcommand{\sjOmega}{\Omega_{sj}}

\DeclareRobustCommand{\Sec}[1]{Sec.~\ref{#1}}
\DeclareRobustCommand{\Secs}[2]{Secs.~\ref{#1} and \ref{#2}}
\DeclareRobustCommand{\App}[1]{App.~\ref{#1}}
\DeclareRobustCommand{\Tab}[1]{Table~\ref{#1}}
\DeclareRobustCommand{\Tabs}[2]{Tables~\ref{#1} and \ref{#2}}
\DeclareRobustCommand{\Fig}[1]{Fig.~\ref{#1}}
\DeclareRobustCommand{\Figs}[2]{Figs.~\ref{#1} and \ref{#2}}
\DeclareRobustCommand{\Eq}[1]{Eq.~(\ref{#1})}
\DeclareRobustCommand{\Eqs}[2]{Eqs.~(\ref{#1}) and (\ref{#2})}
\DeclareRobustCommand{\Ref}[1]{Ref.~\cite{#1}}
\DeclareRobustCommand{\Refs}[1]{Refs.~\cite{#1}}

\DeclareRobustCommand{\order}[1]{{\cal O}(#1)}

\newcommand{\nlojet}{\textsc{NLOJet++}}
\newcommand{\sherpa}{\textsc{Sherpa}}
\newcommand{\ariadne}{\textsc{Ariadne}}
\newcommand{\geneva}{\textsc{Geneva}}
\newcommand{\dire}{\textsc{Dire}}

\DeclareRobustCommand{\order}[1]{{\cal O}(#1)}

%
%

%
%
%
%
%

\chapter{Analytic Boosted Boson Discrimination}\label{chap:D2_anal}

Observables which discriminate boosted topologies from massive QCD jets are of great importance for the success of the jet substructure program at the Large Hadron Collider. Such observables, while both widely and successfully used, have been studied almost exclusively with Monte Carlo simulations. In this chapter we present the first all-orders factorization theorem for a two-prong discriminant based on a jet shape variable, $D_2$, valid for both signal and background jets. Our factorization theorem simultaneously describes the production of both collinear and soft subjets, and we introduce a novel zero-bin procedure to correctly describe the transition region between these limits. By proving an all orders factorization theorem, we enable a systematically improvable description, and allow for precision comparisons between data, Monte Carlo, and first principles QCD calculations for jet substructure observables. Using our factorization theorem, we present numerical results for the discrimination of a boosted $Z$ boson from massive QCD background jets. We compare our results with Monte Carlo predictions which allows for a detailed understanding of the extent to which these generators accurately describe the formation of two-prong QCD jets, and informs their usage in substructure analyses. Our calculation also provides considerable insight into the discrimination power and calculability of jet substructure observables in general.


\section{Introduction}\label{sec:intro}

The last several years has seen a surge of interest in the field of jet substructure \cite{Abdesselam:2010pt,Altheimer:2012mn,Altheimer:2013yza,Adams:2015hiv}, both as an essential tool for extending new physics searches at the Large Hadron Collider (LHC) into the TeV energy regime, and as an important playground for improving our understanding of high energy QCD, both perturbative and non-perturbative. Of particular phenomenological interest are substructure observables that are sensitive to hard subjets within a jet.  In the highly boosted regime, the hadronic decay products of electroweak-scale particles can become collimated and each appear as a jet in the detector.  Unlike typical massive QCD jets, however, these boosted electroweak jets exhibit a multi-prong substructure that can be identified by the measurement of appropriate observables.  Many such observables have been proposed and studied on LHC simulation or data \cite{CMS:2011xsa,Miller:2011qg,Chatrchyan:2012mec,ATLAS:2012jla,Aad:2012meb,ATLAS:2012kla,ATLAS:2012am,Aad:2013gja,Aad:2013fba,TheATLAScollaboration:2013tia,TheATLAScollaboration:2013sia,TheATLAScollaboration:2013ria,TheATLAScollaboration:2013pia,CMS:2013uea,CMS:2013kfa,CMS:2013wea,CMS-PAS-JME-10-013,CMS-PAS-QCD-10-041,Aad:2014gea,LOCH:2014lla,CMS:2014fya,CMS:2014joa,Aad:2014haa} or used in new physics searches  \cite{CMS:2011bqa,Fleischmann:2013woa,Pilot:2013bla,TheATLAScollaboration:2013qia,Chatrchyan:2012ku,Chatrchyan:2012sn,CMS:2013cda,CMS:2014afa,CMS:2014aka,Khachatryan:2015axa,CMS:1900uua,Khachatryan:2015bma,Aad:2015owa}.

The vast majority of proposed jet substructure observables, however, have been analyzed exclusively within Monte Carlo simulation.  While Monte Carlos play an essential role in the simulation of realistic hadron collision events, they can often obscure the underlying physics that governs the behavior of a particular observable.  Additionally, it is challenging to disentangle perturbative physics from the tuning of non-perturbative physics so as to understand how to systematically improve the accuracy of the Monte Carlo.  Recently, there has been an increasing number of analytical studies of jet substructure observables, including the calculation of the signal distribution for $N$-subjettiness to next-to-next-to-next-to-leading-log order \cite{Feige:2012vc}, a fixed-order prediction for planar flow \cite{Field:2012rw}, calculations of groomed jet masses \cite{Dasgupta:2013ihk,Dasgupta:2013via,Larkoski:2014pca,Dasgupta:2015yua} and the jet profile/ shape \cite{Seymour:1997kj,Li:2011hy,Larkoski:2012eh,Jankowiak:2012na,Chien:2014nsa,Chien:2014zna,Isaacson:2015fra} for both signal and background jets, an analytic understanding of jet charge \cite{Krohn:2012fg,Waalewijn:2012sv}, predictions for fractional jet multiplicity \cite{Bertolini:2015pka}, and calculations of the associated subjet rate \cite{Bhattacherjee:2015psa}.  Especially in the case of the groomed jet observables, analytic predictions informed the construction of more performant and easier to calculate observables.  With the recent start of Run 2 of the LHC, where the phase space for high energy jets only grows, it will be increasingly important to have analytical calculations to guide experimental understanding of jet dynamics.

It is well known that the measurement of observables on a jet can introduce ratios of hierarchical scales appearing in logarithms at every order in the perturbative expansion. Accurate predictions over all of phase space require resummation of these large logarithms to all orders in perturbation theory. While this resummation is well understood for simple observables such as the jet mass \cite{Catani:1991bd,Chien:2010kc,Chien:2012ur,Dasgupta:2012hg,Jouttenus:2013hs}, where it has been performed to high accuracy, a similar level of analytic understanding has not yet been achieved for more complicated jet substructure observables. Jet substructure observables are typically sensitive to a multitude of scales, corresponding to characteristic features of the jet, resulting in a much more subtle procedure for resummation.  

A ubiquitous feature of some of the most powerful observables used for identification of jet substructure is that they are formed from the ratio of infrared and collinear (IRC) safe observables.  Examples of such observables include ratios of $N$-subjettiness variables \cite{Thaler:2010tr,Thaler:2011gf}, ratios of energy correlation functions \cite{Larkoski:2013eya,Larkoski:2014gra,Larkoski:2014zma}, or planar flow \cite{Almeida:2008yp}.  In general, ratios of IRC safe observables are not themselves IRC safe \cite{Soyez:2012hv} and cannot be calculated to any fixed order in perturbative QCD.  Nevertheless, it has been shown that these ratio observables are calculable in resummed perturbation theory and are therefore referred to as Sudakov safe \cite{Larkoski:2013paa,Larkoski:2014wba,Larkoski:2014bia,Larkoski:2015lea}.  Distributions of Sudakov safe observables can be calculated by appropriately marginalizing resummed multi-differential cross sections of IRC safe observables.  An understanding of the factorization properties of multi-differential jet cross sections has been presented in \Refs{Larkoski:2014tva,Procura:2014cba,Larkoski:2015zka} by identifying distinct factorization theorems in parametrically separated phase space regions defined by the measurements performed on the jet.  Combining this understanding of multi-differential factorization with the required effective field theories, all ingredients are now available for analytic resummation and systematically improvable predictions.

As an explicit example, observables that resolve two-prong substructure are sensitive to both the scales characterizing the subjets as well as to the scales characterizing the full jet.  A study of the resummation necessary for describing jets with a two-prong substructure was initiated in \Ref{Bauer:2011uc} which considered the region of phase space with two collinear subjets of comparable energy, and introduced an effective field theory description capturing all relevant scales of the problem.   Recently, an effective field theory description for the region of two-prong jet phase space with a hard core and a soft, wide angle subjet was developed in \Ref{Larkoski:2015zka}, where it was applied to the resummation of non-global logarithms \cite{Dasgupta:2001sh}. Combined, the collinear subjet and soft subjet factorization theorems allow for a complete description of the dominant dynamics of jets with two-prong substructure.

In this chapter we will study the factorization and resummation of the jet substructure observable $\Dobsnobeta{2}$ \cite{Larkoski:2014gra}, a ratio-type observable formed from the energy correlation functions.  We will give a detailed effective theory analysis using the language of soft-collinear effective theory (SCET) \cite{Bauer:2000yr,Bauer:2001ct,Bauer:2001yt,Bauer:2002nz} in all regions of phase space required for the description of a one or two-prong jet, and will prove all-orders leading-power factorization theorems in each region. We will then use these factorization theorems to calculate the $\Dobsnobeta{2}$ distribution for jets initiated by boosted hadronic decays of electroweak bosons or from light QCD partons and compare to Monte Carlo simulation.  These calculations will also allow us to make first-principles predictions for the efficiency of the observable $\Dobsnobeta{2}$ to discriminate boosted electroweak signal jets from QCD background jets.

\begin{figure}
\begin{center}
\subfloat[]{\label{fig:intro_plota}
\includegraphics[width= 7.25cm]{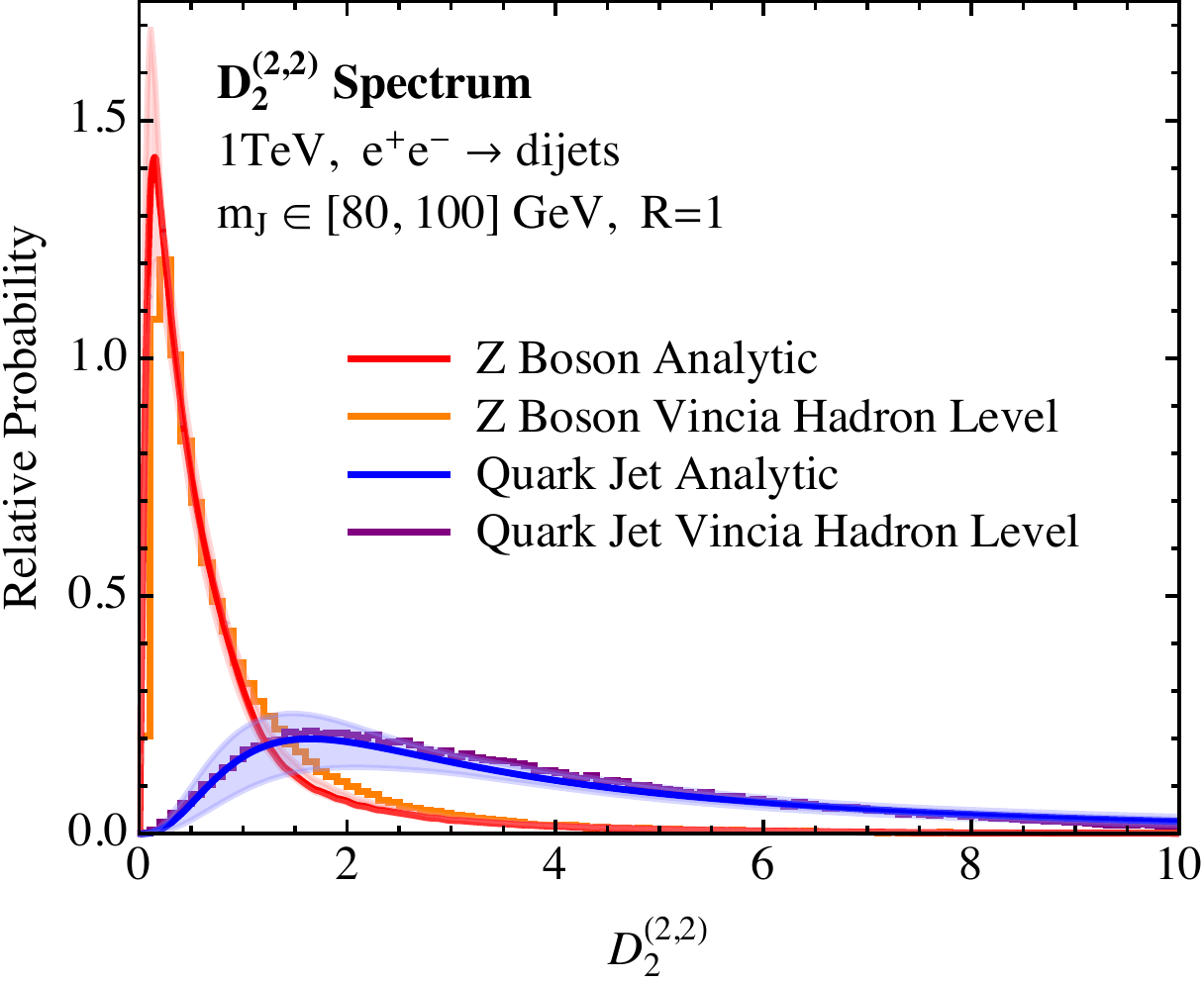}
}
\subfloat[]{\label{fig:intro_plotb}
\includegraphics[width = 7.25cm]{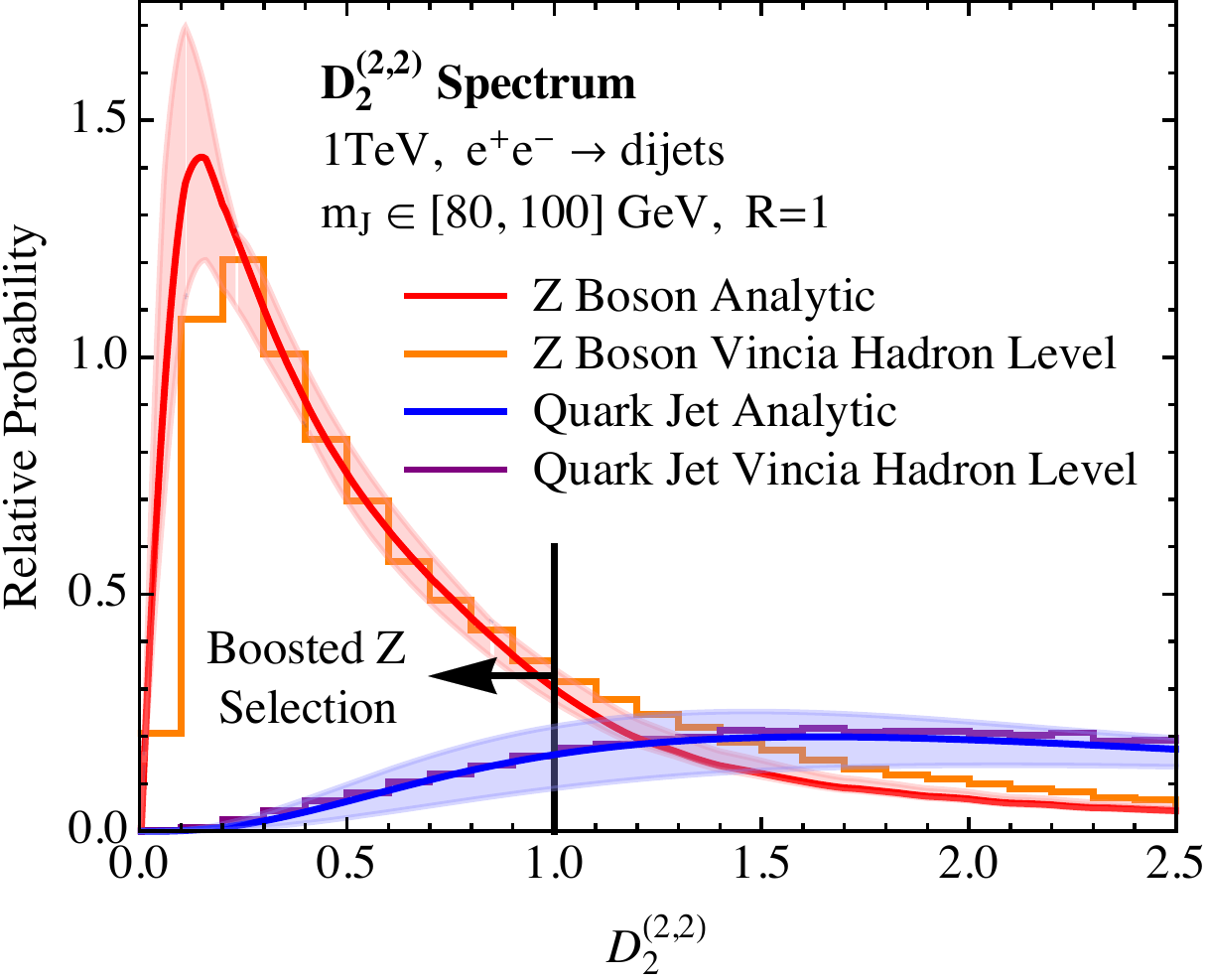}
}
\end{center}
\vspace{-0.2cm}
\caption{Comparison of our analytic calculation with \vincia Monte Carlo predictions for the two prong discriminant, $D_2$. Predictions for both boosted $Z$ bosons and massive QCD jets at a 1 TeV $e^+e^-$ collider are shown. The Monte Carlo is fully hadronized, and non-perturbative effects have been included in the analytic calculation through a shape function. In a) we show the complete distribution, and in b) we zoom in to focus on the region relevant for boosted $Z$ discrimination.
}
\label{fig:intro_plot}
\end{figure}

Our factorized description is valid to all orders in $\alpha_s$, expressing the cross section as a product of field theoretic matrix elements, each of which is calculable order by order in perturbation theory, allowing for a systematically improvable description of the $D_2$ observable. Furthermore, the factorization theorem enables a clean separation of perturbative and non-perturbative physics, allowing for non-perturbative contributions to the observable to be included in the analytic calculation through the use of shape functions \cite{Korchemsky:1999kt,Korchemsky:2000kp}.  In this chapter we work to next-to-leading logarithmic (NLL) accuracy to demonstrate all aspects of the required factorization theorems necessary for precision jet substructure predictions. We will see that even at this first non-trivial order, we gain insight into qualitative and quantitative features of the $D_2$ distribution.  While we will give an extensive discussion of our numerical results and comparisons with a variety of Monte Carlo programs in this chapter, in \Fig{fig:intro_plot} we compare our analytic predictions for the $D_2$ observable, including non-perturbative effects, for hadronically-decaying boosted $Z$ bosons and QCD jets in $e^+e^-$ collisions with the distributions predicted by the \vincia{} \cite{Giele:2007di,Giele:2011cb,GehrmannDeRidder:2011dm,Ritzmann:2012ca,Hartgring:2013jma,Larkoski:2013yi} Monte Carlo program at hadron level.  Excellent agreement between analytic and Monte Carlo predictions is observed, demonstrating a quantitative understanding of boosted jet observables from first principles.

\subsection{Overview of the chapter}\label{sec:approach}

While there exists a large number of two-prong discriminants in the jet substructure literature, any of which would be interesting to understand analytically, we will use calculability and factorizability as guides for constructing the observable to study in this chapter.  This procedure will ultimately lead us to the observable $\Dobsnobeta{2}$ and will demonstrate that $\Dobsnobeta{2}$ has particularly nice factorization and calculability properties.  This approach will proceed in the following steps:
\begin{enumerate}

\item Identify the relevant subjet configurations for the description of a two-prong discriminant.

\item Isolate each of these relevant regions by the measurement of a collection of IRC safe observables.

\item Study the phase space defined by this collection of IRC safe observables, and prove all-orders factorization theorems in each parametrically-defined region of phase space.

\item Identify a two-prong discriminant formed from the collection of IRC safe observables which respects the parametric factorization theorems of the phase space.

\end{enumerate}
A detailed analysis of each of these steps will be the subject of this chapter. Here, we provide a brief summary so that the logic of the approach is clear, and so that the reader can skip technical details in the different sections without missing the general idea of the approach.

\begin{figure}
\begin{center}
\subfloat[]{\label{fig:soft_haze}
\includegraphics[width=3.95cm, trim =0 -0.5cm 0 0]{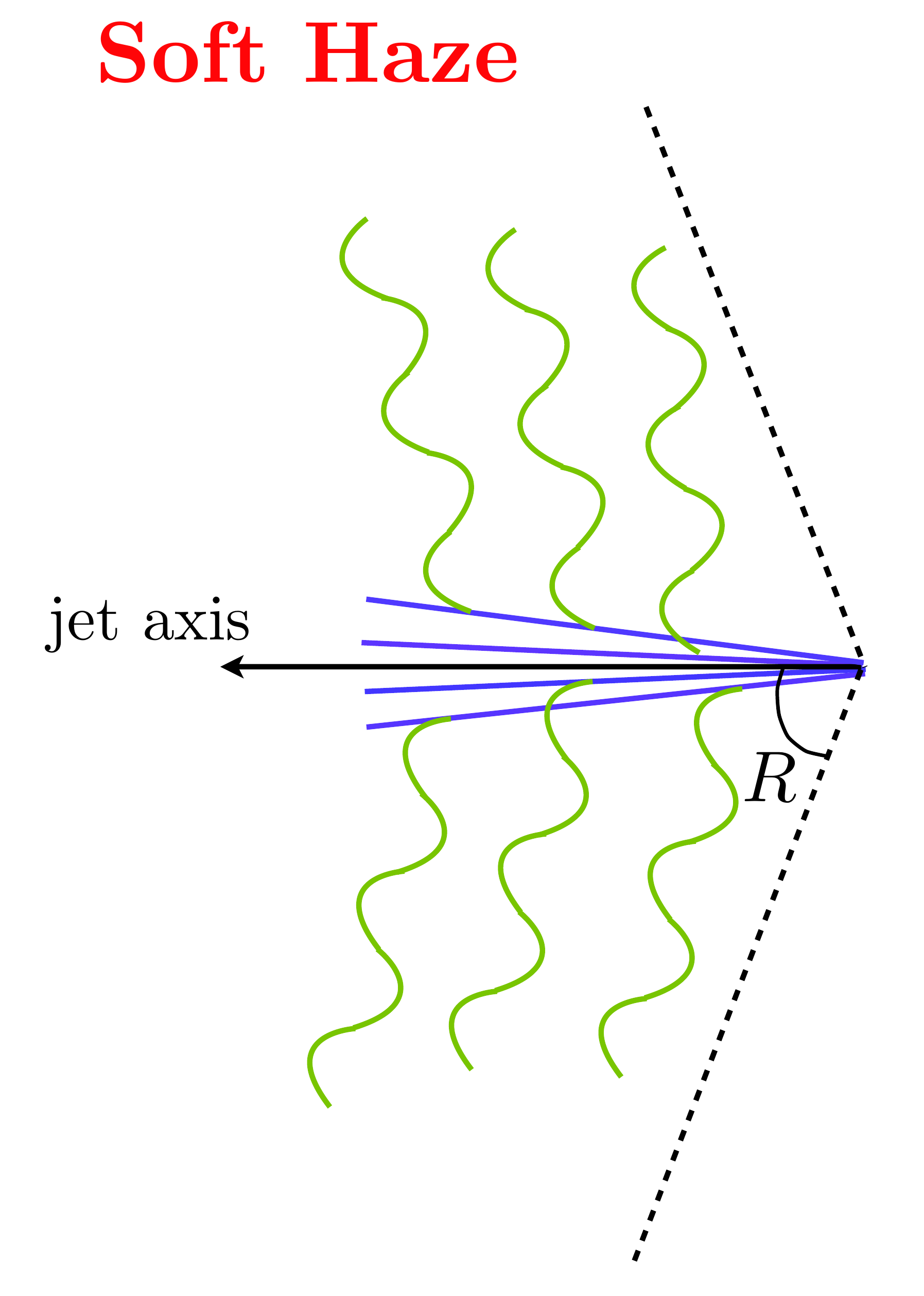}
}\qquad
\subfloat[]{\label{fig:ninja}
\includegraphics[width=4.1cm, trim =0 0 0 0]{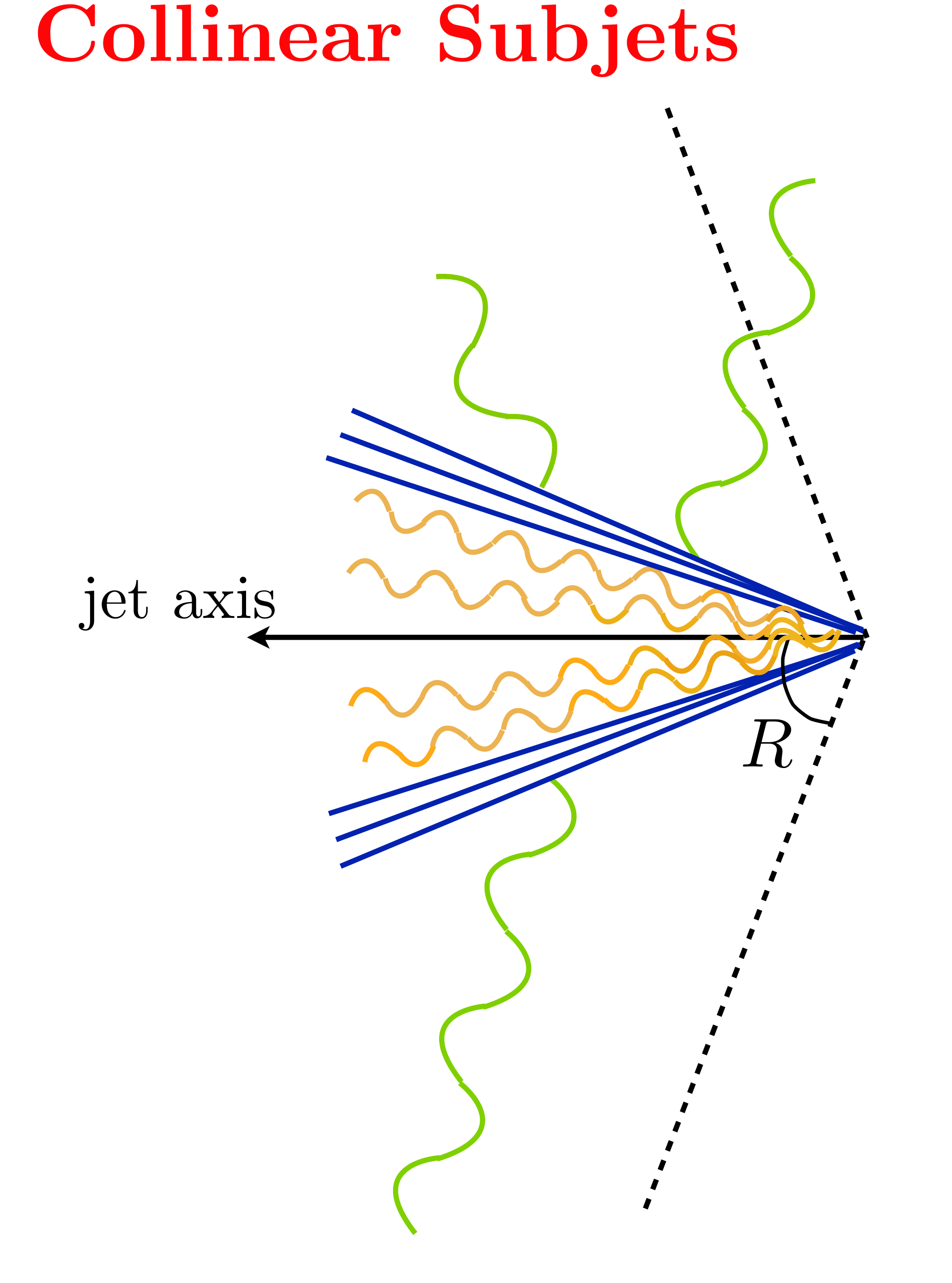}
}\qquad
\subfloat[]{\label{fig:soft_jet}
\includegraphics[width=3.9cm, trim =0 -0.75cm 0 0]{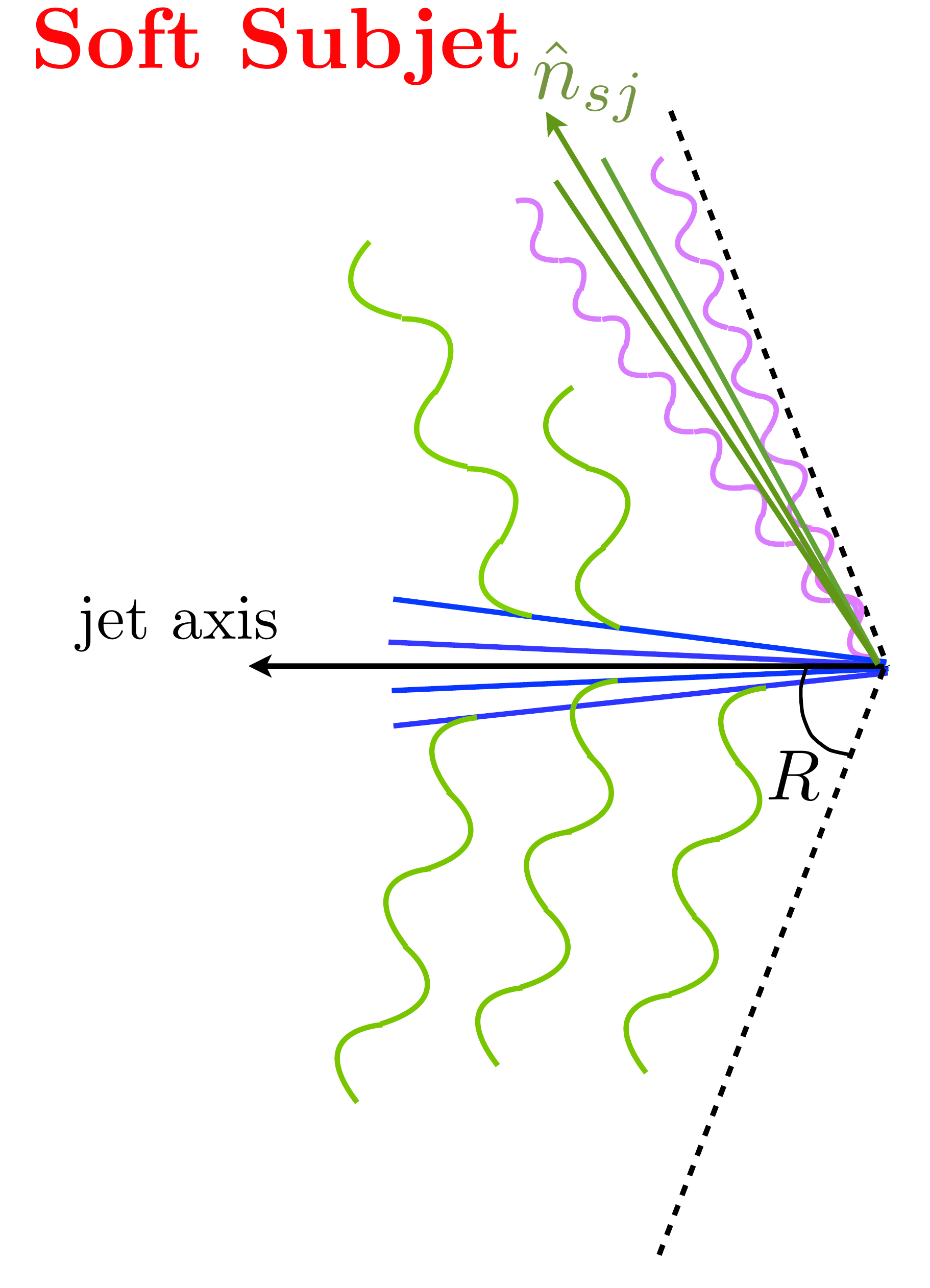}
}
\end{center}
\caption{ Regions of interest for studying the two-prong substructure of a jet. a) Soft haze region in which no subjets are resolved. b) Collinear subjets with comparable energy and a small opening angle. c) Soft subjet carrying a small fraction of the total energy, and at a wide angle from the hard subjet.}
\label{fig:diff_jets}
\end{figure}

The complete description of an observable capable of discriminating one- from two-prong substructure requires the factorized description of the following three relevant subjet configurations, shown schematically in \Fig{fig:diff_jets}:

\begin{itemize}

\item {\bf{Soft Haze}}: \Fig{fig:soft_haze} shows a jet in what we will refer to as the soft haze region of phase space. In the soft haze region there is no resolved subjet, only a single hard core with soft wide angle emissions. This region of phase space typically contains emissions beyond the strongly ordered limit, but is the dominant background region for QCD jets, for which a hard splitting is $\alpha_s$ suppressed.

\item {\bf{Collinear Subjets}}: \Fig{fig:ninja} shows a jet with two hard, collinear subjets. Both subjets carry approximately half of the total energy of the jet, and have a small opening angle. This region of phase space, and its corresponding effective field theory description, has been studied in \Ref{Bauer:2011uc}.

\item {\bf{Soft Subjet}}: \Fig{fig:soft_jet}  shows the soft subjet region of phase space which consists of jets with two subjets with hierarchical energies separated by an angle comparable to the jet radius $R$.  The soft subjet probes the boundary of the jet and we take $R\sim 1$. An effective field theory description for this region of phase space was presented in \Ref{Larkoski:2015zka}.

\end{itemize}

As a basis of IRC safe observables for isolating these three subjet configurations, we use the energy correlation functions \cite{Larkoski:2013eya}, which we define in \Sec{sec:obs_def}. In particular, we will show that the measurement of three energy correlation functions, two 2-point energy correlation functions $\ecf{2}{\alpha}, \ecf{2}{\beta}$, and a 3-point energy correlation function $\ecf{3}{\alpha}$, allows for parametric separation of the different subjet configurations.  While we will focus on the particular case of observables formed from the energy correlation functions, we believe that this approach is more general and could be applied to other IRC safe observable bases. 

With the energy correlation functions as our basis, we study the multi-differential phase space defined by the simultaneous measurement of these observables on a jet in \Sec{sec:phase_space}. Using the power counting technique of \Refs{Larkoski:2014gra,Larkoski:2014zma}, we show that the angular exponents of the energy correlation functions, $\alpha$ and $\beta$, can be chosen such that the different subjet configurations occupy parametrically separated regions of this phase space, and extend to all boundaries of the phase space. This parametric separation allows for each region to be separately described by its own effective field theory. The required effective field theories are described in \Sec{sec:Fact}, and are formulated in the language of SCET.  The formulation in SCET allows us to prove all-orders factorization theorems valid at leading-power in each of the phase space regions, and to resum logarithms to arbitrary accuracy using renormalization group techniques. 

Having understood in detail both the structure of the phase space defined by the IRC safe measurements as well as the factorization theorems defined in each region, we will show in \Sec{sec:def_D2} that this leads unambiguously to the definition of a two-prong discriminant observable which is amenable to factorization. This observable will be a generalized form of $\Dobsnobeta{2}$ \cite{Larkoski:2014gra} which will depend on both angular exponents $\alpha$ and $\beta$.  Calculating the distribution of $\Dobsnobeta{2}$ is accomplished by appropriate marginalization of the multi-differential cross section.  Depending on the phase space cuts that have been made, $\Dobsnobeta{2}$ may or may not be IRC safe itself, and so the marginalization will in general only be defined within resummed perturbation theory.

The outline of this chapter is as follows. In \Sec{sec:phase_space} we define the energy correlation functions used in this chapter and describe how the different subjet configurations shown schematically in \Fig{fig:diff_jets} can be isolated by demanding parametric relations between the measured values of these observables.  In \Sec{sec:Fact} we discuss the effective field theory descriptions in the different phase space regions, and present the factorization theorems that describe their dynamics. Although some of the relevant effective field theories have been presented elsewhere, we attempt to keep the discussion self-contained by providing a brief review of their most salient features.  All field theoretic definitions of the functions appearing in the factorization theorems, as well as their calculations to one-loop accuracy, are provided in appendices. 
 
 In \Sec{sec:friendly}  we show how the detailed understanding of the multi-differential phase space leads to the definition of $\Dobsnobeta{2}$ as a powerful one- versus two-prong jet discriminant.  In \Sec{sec:sudsafe} we emphasize that without a mass cut, $\Dobsnobeta{2}$ is not IRC safe but is Sudakov safe and whose all-orders distribution exhibits paradoxical dependence on $\alpha_s$.  In \Sec{sec:fixed_order} we study the fixed-order distribution of $\Dobsnobeta{2}$ in the presence of a mass cut to understand its behavior in singular limits. In \Sec{sec:merging} we discuss how the different effective field theories can be consistently merged to give a factorized description of the $\Dobsnobeta{2}$ observable, and introduce a novel zero-bin procedure to implement this merging. 
 
 In \Sec{sec:results} we present numerical results for both signal and background distributions for $\Dobsnobeta{2}$ as measured in $e^+e^-$ collisions and compare our analytic calculation with several Monte Carlo generators. We emphasize many features of the calculation which provide considerable insight into two-prong discrimination, and the ability of current Monte Carlo generators to accurately describe substructure observables. In \Sec{sec:LEP} we discuss numerical results for the $D_2$ observable at $e^+e^-$ collisions at the $Z$ pole at LEP, and demonstrate that being sensitive to correlations between three emissions, the $D_2$ observable can be used as a more differential probe of the perturbative shower for tuning Monte Carlo generators. In \Sec{sec:LHC} we discuss how to extend our calculations to $pp$ collisions at the LHC.  We conclude in \Sec{sec:conc}, and discuss future directions for further improving the analytic understanding of jet substructure.

\section{Characterizing a Two-Prong Jet}
\label{sec:phase_space}

In this section, we develop the framework necessary to construct the all-orders factorization theorems for analytic two-prong discrimination predictions.  We begin in \Sec{sec:obs_def} by defining the energy correlation functions, which we will use to isolate the three subjet configurations discussed in the introduction.  Using the power counting analysis of \Refs{Larkoski:2014gra,Larkoski:2014zma}, we study the phase space defined by measuring the energy correlation functions in \Sec{sec:power_counting}.  Throughout this chapter, our proxy for a two-prong jet will be a boosted, hadronically decaying $Z$ boson, but our analysis holds for $W$ or $H$ bosons, as well.

\subsection{Observable Definitions}\label{sec:obs_def}

To distinguish the three different subjet configurations of \Fig{fig:diff_jets} with IRC safe measurements, observables which are sensitive to both one- and two-prong structure are required. Although many possible observable bases exist, in this chapter we will use the energy correlation functions \cite{Larkoski:2013eya,Larkoski:2014gra}, as we will find that they provide a convenient basis.

The $n$-point energy correlation function is an IRC safe observable that is sensitive to $n$-prong structure in a jet.  For studying the two-prong structure of a jet, we will need the 2- and 3-point energy correlation functions, which we define for $e^+e^-$ collisions as \cite{Larkoski:2013eya}\footnote{For massive hadrons, there exist several possible definitions of the energy correlation functions depending on the particular mass scheme \cite{Salam:2001bd,Mateu:2012nk}.  The definition in \Eq{eq:ecf_def} is an $E$-scheme definition.  A $p$-scheme definition will be presented in \Sec{sec:LEP} when we discuss the connection to LEP. Since the different definitions are equivalent for massless partons, their perturbative calculations are identical. The different definitions differ only in their non-perturbative corrections.}
\begin{align}\label{eq:ecf_def}
\ecf{2}{\alpha}&= \frac{1}{E_J^2} \sum_{i<j\in J} E_i E_j \left(
\frac{2p_i \cdot p_j}{E_i E_j}
\right)^{\alpha/2} \,, \\
\ecf{3}{\alpha}&= \frac{1}{E_J^3} \sum_{i<j<k\in J} E_i E_j E_k \left(
\frac{2p_i \cdot p_j}{E_i E_j}
\frac{2p_i \cdot p_k}{E_i E_k}
\frac{2p_j \cdot p_k}{E_j E_k}
\right)^{\alpha/2} \,. \nonumber
\end{align}
Here $J$ denotes the jet, $E_i$ and $p_i$ are the energy and four momentum of particle $i$ in the jet and $\alpha$ is an angular exponent that is required to be greater than 0 for IRC safety.  The 4-point and higher energy correlation functions are defined as the natural generalizations of \Eq{eq:ecf_def}, although we will not use them in this chapter.

While we will mostly focus on the case of an $e^+e^-$ collider, the energy correlation functions have natural generalizations to hadron colliders, by replacing $E$ by $p_T$ and using hadron collider coordinates, $\eta$ and $\phi$. This definition is given explicitly in \Eq{eq:pp_def_ecf}. At central rapidity, this modification does not change the behavior of the observables, or any of the conclusions presented in the next sections.  Of course, the hadron collider environment has other effects not present in an $e^+e^-$ collider, like initial state radiation or underlying event, that will affect the energy correlation functions.  A brief discussion of the behavior of the energy correlation functions in $pp$ colliders will be given in \Sec{sec:LHC}. Numerical implementations of the energy correlation functions for both $e^+e^-$ and hadron colliders are available in the \texttt{EnergyCorrelator} \fastjet{contrib} \cite{Cacciari:2011ma,fjcontrib}.

\subsection{Power Counting the $\ecf{2}{\alpha}, \ecf{2}{\beta}, \ecf{3}{\alpha}$ Phase Space}\label{sec:power_counting}

With a basis of IRC safe observables identified, we now demonstrate that the measurement of multiple energy correlation functions parametrically separates the three different subjet configurations identified in \Fig{fig:diff_jets}. In particular, the simultaneous measurement of $\ecf{2}{\alpha}$, $\ecf{2}{\beta}$, and $\ecf{3}{\alpha}$ is sufficient for this purpose, and we will study in detail the phase space defined by their measurement.  From this analysis, we will be able to determine for which values of the angular exponents $\alpha$ and $\beta$ the three subjet configurations are parametrically separated within this phase space. The power counting parameters that define ``parametric'' will be set by the observables themselves, as is typical in effective field theory.

We begin by considering how the energy correlation functions can be used to separate one- and two-prong jets. This has been previously discussed in \Ref{Larkoski:2014gra} by measuring $\ecf{2}{\alpha}$ and $\ecf{3}{\alpha}$, but here we consider the phase space defined by $\ecf{2}{\beta}$ and $\ecf{3}{\alpha}$ with $\alpha$ and $\beta$ in general different.  A minimal constraint on the angular exponents, both for calculability and discrimination power, is that the soft haze and collinear subjets configurations are parametrically separated by the measurements.  A power counting analysis of the soft subjet region yields no new constraints beyond those from the soft haze or collinear subjets.

\begin{figure}
\begin{center}
\subfloat[]{\label{fig:unresolved}
\includegraphics[width=6cm]{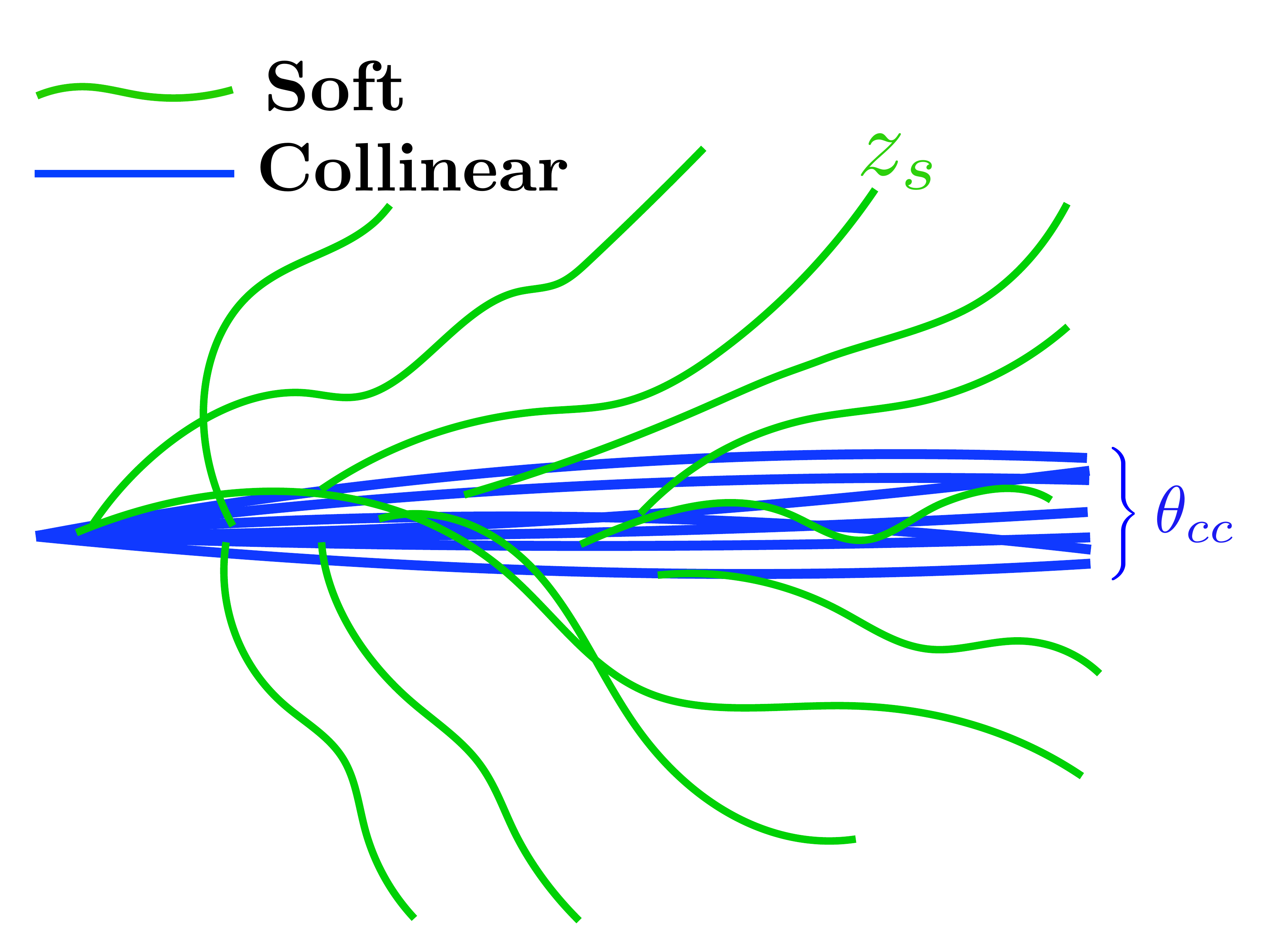}    
}\qquad
\subfloat[]{\label{fig:NINJA}
\includegraphics[width=6cm, trim =0 -0.5cm 0 0]{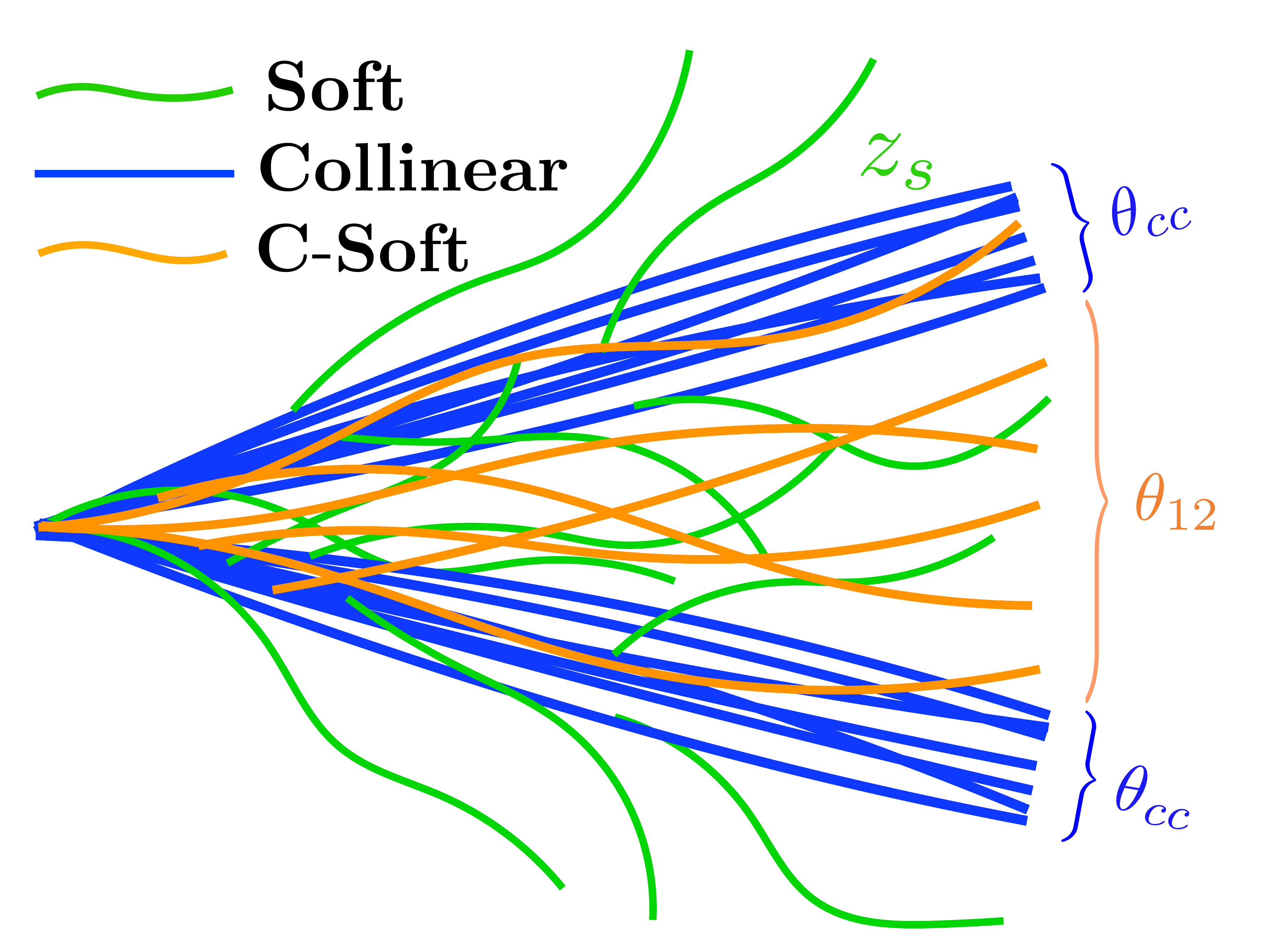}
}
\end{center}
\caption{a) Schematic of a one-prong soft haze jet, dominated by collinear (blue) and soft (green) radiation. The angular size of the collinear radiation is $\theta_{cc}$ and the energy fraction of the soft radiation is $z_s$.  b) Schematic of a jet resolved into two collinear subjets, dominated by collinear (blue), soft (green), and collinear-soft (orange) radiation emitted from the dipole formed by the two subjets.  The subjets are separated by an angle $\theta_{12}$  and the energy fraction of the collinear-soft radiation is $z_{cs}$. 
}
\label{fig:pics_jets}
\end{figure}

The setup for the power counting analysis of the soft haze and collinear subjets configurations is shown in \Fig{fig:pics_jets}, where all relevant modes are indicated. The one-prong jet illustrated in \Fig{fig:unresolved} is described by soft modes with energy fraction $z_s$ emitted at ${\cal O}(1)$ angles, and collinear modes with characteristic angular size $\theta_{cc}$ with ${\cal O}(1)$ energy fraction. The collinear subjets configuration illustrated in \Fig{fig:NINJA} consists of two subjets, each of which carry an ${\cal O}(1)$ fraction of the jet's energy and are separated by an angle $ \theta_{12}\ll1$.  Each of the subjets has collinear emissions at a characteristic angle $ \theta_{cc}\ll  \theta_{12}$, and global soft radiation at large angles with respect to the subjets, with characteristic energy fraction $z_s\ll 1$. In the case of two collinear subjets arising from the decay of a color singlet particle, the long wavelength global soft radiation is not present due to color coherence, but the power counting arguments of this section remain otherwise unchanged. Finally, there is radiation from the dipole formed from the two subjets (called ``collinear-soft'' radiation), with characteristic  angle $ \theta_{12}$ from the subjets, and with energy fraction $z_{cs}$.  The effective theory of this phase space region for the observable $N$-jettiness \cite{Stewart:2010tn} was studied in \Ref{Bauer:2011uc}.\footnote{It is of historical interest to note that the generalization of two-prong event shapes, such as thrust, to event shapes for characterizing three jet structure was considered early on, for example with the introduction of the triplicity event shape \cite{Brandt:1978zm}. However, it was not until more recently, with the growth of the jet substructure field at the LHC, that significant theoretical study was given to such observables.}

We are now able to determine the parametric form of the dominant contributions to the observables $\ecf{2}{\beta}$ and $\ecf{3}{\alpha}$.  In the soft haze region, the dominant contributions to the energy correlation functions are\footnote{It is important to understand that this relationship is valid to an arbitrary number of emissions. When performing the power counting, a summation over all the particles with soft and collinear  scalings in the jet must be considered. However, to  determine  the  scalings  of  the observable,  it  is  sufficient  to consider the scaling of the different types of individual terms in the sum. For example, the three terms contributing to the expression for $\ecf{3}{\alpha}$ arise from correlations between subsets of three collinear particles, one collinear particle and two soft particles, and two collinear particles and a soft particle, respectively. Contributions from other combinations of particles are power suppressed. Because of this simplification, in this chapter we will never write explicit summations when discussing the scaling of observables.}
\begin{align}
\ecf{2}{\beta}&\sim \zs+\thetac^\beta\,, \nonumber \\
\ecf{3}{\alpha}&\sim \thetac^{3\alpha}+\zs^2 +\thetac^\alpha \zs\,.
\end{align}
From these parametrics, it is straightforward to show that one-prong jets live in a region of the $\ecf{2}{\beta}, \ecf{3}{\alpha}$ phase space bounded from above and below, whose precise scaling depends on the relative size of the angular exponents $\alpha$ and $\beta$.  The scaling of upper and lower boundaries of the one-prong region of phase space for all $\alpha$ and $\beta$ are listed in \Tab{tab:pc}.  For $\alpha = \beta$, as studied in \Ref{Larkoski:2014gra}, one-prong jets live in the region defined by $\left( \ecf{2}{\beta}\right)^{3} \lesssim \ecf{3}{\beta} \lesssim \left( \ecf{2}{\beta}\right)^{2}$.

\begin{table}[t]
\begin{center}
\begin{tabular}{c|c|c|c|c}
&$\alpha\leq \beta/2$ & $\beta/2 <\alpha < 2\beta/3$ & $ 2\beta/3 \leq \alpha < \beta$ & $\alpha \geq \beta$ \\ 
\hline
upper& $\ecf{3}{\alpha}\sim \left( \ecf{2}{\beta}\right)^{3\alpha/\beta}$ & $\ecf{3}{\alpha} \sim \left (\ecf{2}{\beta} \right )^{\alpha/\beta +1}$&$\ecf{3}{\alpha} \sim \left (\ecf{2}{\beta} \right )^{\alpha/\beta +1}$&$\ecf{3}{\alpha} \sim \left (\ecf{2}{\beta} \right )^{2}$\\
lower  & $\ecf{3}{\alpha}\sim \left( \ecf{2}{\beta}\right)^2$ &$\ecf{3}{\alpha}\sim \left( \ecf{2}{\beta}\right)^{2}$&$\ecf{3}{\alpha}\sim \left( \ecf{2}{\beta}\right)^{3\alpha/\beta}$&$\ecf{3}{\alpha}\sim \left( \ecf{2}{\beta}\right)^{3\alpha/\beta}$
\end{tabular}
\end{center}
\caption{Parametric scaling of the upper and lower boundaries of the one-prong region of the $\ecf{2}{\beta}, \ecf{3}{\alpha}$ phase space as a function of the angular exponents $\alpha$ and $\beta$.
}
\label{tab:pc}
\end{table}

For the collinear subjets configuration, the dominant contributions to the observables $\ecf{2}{\beta}$ and $\ecf{3}{\alpha}$ are
\begin{align}
\ecf{2}{\beta}&\sim \theta_{12}^\beta\,, \nonumber \\
\ecf{3}{\alpha}&\sim \theta_{cc}^{\alpha} \theta_{12}^{2\alpha}+\theta_{12}^\alpha z_s +\theta_{12}^{3\alpha}z_{cs}+z_s^2\,.
\end{align}
The 2-point energy correlation function $\ecf{2}{\beta}$ is set by the angle of the hard splitting, $ \theta_{12}$, and the scaling of all other modes (soft, collinear, or collinear-soft) are set by the $\ecf{3}{\alpha}$ measurement. The requirement
\begin{align}\label{eq:tpb}
z_{cs} \sim \frac{\ecf{3}{\alpha}}{  \left(  \ecf{2}{\beta}  \right)^{3\alpha/\beta} } \ll 1\,,
\end{align}
then implies that the two-prong jets occupy the region of phase space defined by  $\ecf{3}{\alpha} \ll  \left(  \ecf{2}{\beta}  \right)^{3\alpha/\beta}$.  

For optimal discrimination, the one- and two-prong regions of this phase space should not overlap.  Since they are physically distinct, a proper division of the phase space will allow distinct factorizations, simplifying calculations.  Comparing the boundaries of the one-prong region listed in \Tab{tab:pc} with the  upper boundary of the two-prong region from \Eq{eq:tpb}, we find that the one- and two-prong jets do not overlap with the following restriction on the angular exponents $\alpha$ and $\beta$:
\be\label{eq:restriction_1}
 3\alpha \geq 2\beta\,.
\ee
Note that when $\alpha=\beta$ this is satisfied, consistent with the analysis of  \Ref{Larkoski:2014gra}.

\begin{figure}
\begin{center}
\subfloat[]{\label{fig:D2_2D_a}
\includegraphics[width= 6.5cm]{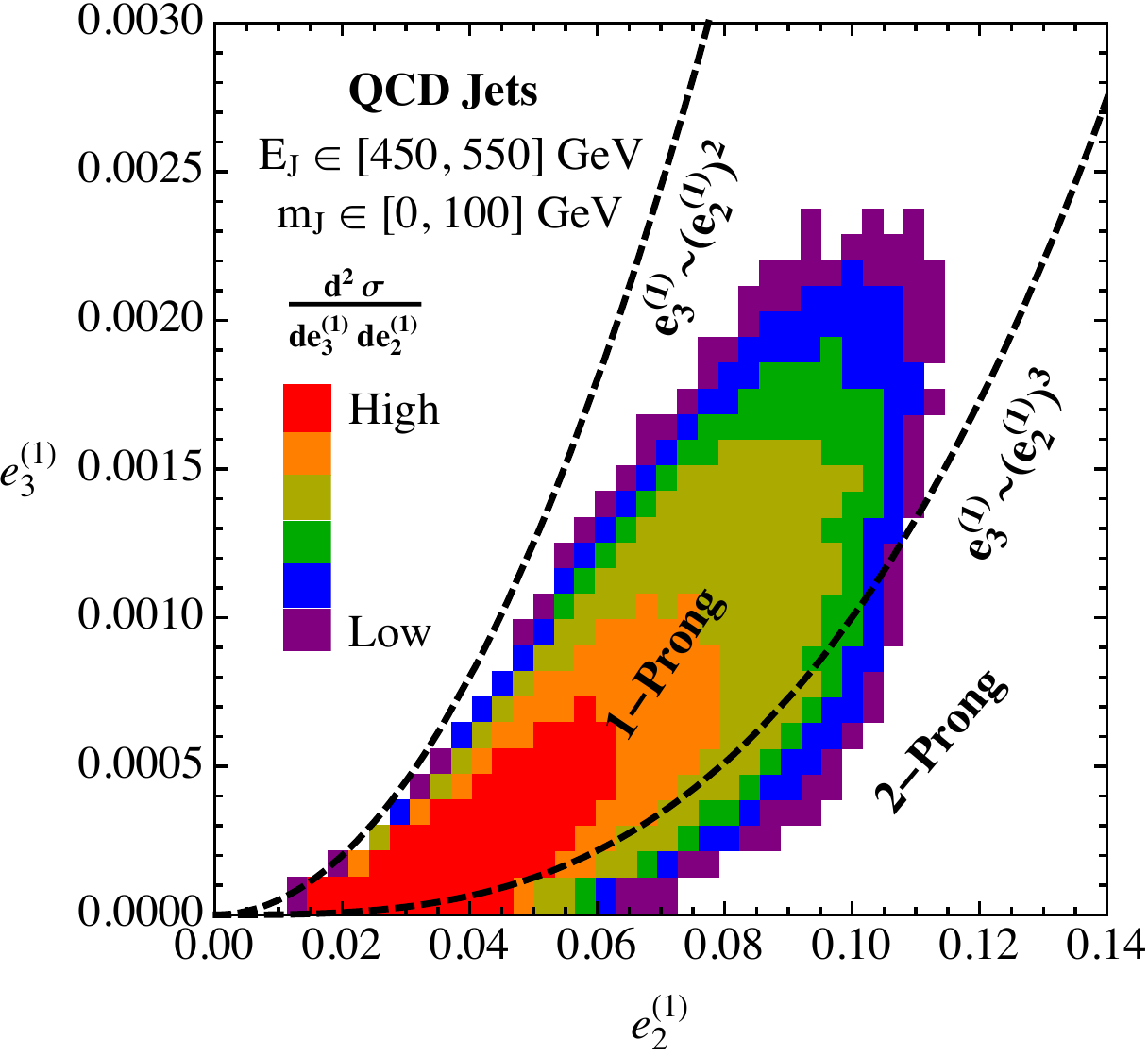}
}
\qquad
\subfloat[]{\label{fig:D2_2D_b}
\includegraphics[width = 6.5cm]{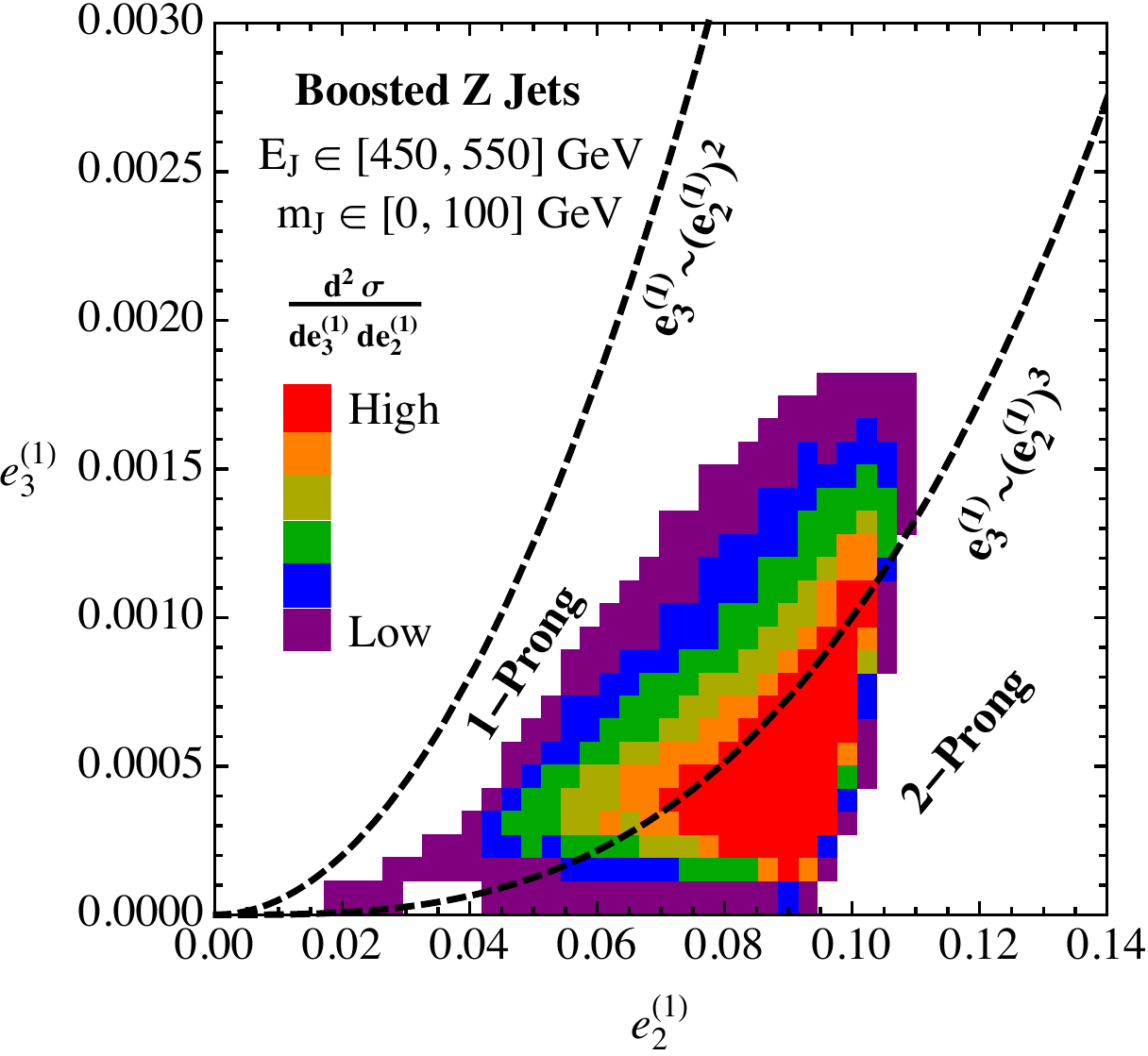}
}
\end{center}
\vspace{-0.2cm}
\caption{Monte Carlo distributions in the $\ecf{2}{1},\ecf{3}{1}$ plane, for QCD quark jets (left) and boosted $Z\to q\bar q$ jets (right). The parametric scalings predicted by the power counting analysis are shown as dashed lines, and the one- and two-prong regions of phase space are labelled, and extend between the parametric boundaries. Note the upper boundary is constrained to have a maximal value of $\frac{1}{2}(\ecf{2}{\alpha})^2=\ecf{3}{\alpha}$.
}
\label{fig:D2_2D}
\end{figure}

Because these power counting arguments rely exclusively on the parametric behavior of QCD in the soft and collinear limits, they must be reproduced by any Monte Carlo simulation, regardless of its shower and hadronization models.  To illustrate the robust boundary between the one- and two-prong regions of phase space predicted in \Eq{eq:restriction_1}, in \Fig{fig:D2_2D}, we plot the distribution in the $\ecf{2}{\beta},\ecf{3}{\alpha}$ plane of jets initiated by light QCD partons and those from boosted hadronic decays of $Z$ bosons as generated in $e^+e^-$ collisions in \pythia{} \cite{Sjostrand:2006za,Sjostrand:2007gs}.  Details of the Monte Carlo generation are presented in \Sec{sec:results}.  QCD jets are dominantly one-pronged, while jets from $Z$ decays are dominantly two-pronged.  We have chosen to use angular exponents $\alpha=\beta=1$ for this plot, as the small value of the angular exponent allows the structure of the phase space to be seen in a non-logarithmic binning.  The predicted behavior persists for all values of $\alpha$ and $\beta$ consistent with \Eq{eq:restriction_1}, while the choice made here is simply for illustrative aesthetics.  On these plots, we have added dashed lines corresponding to the predicted one- and two-prong phase space boundaries to guide the eye.  The one-prong QCD jets and the two-prong boosted $Z$ jets indeed dominantly live in their respective phase space regions as predicted by power counting.

\begin{figure}
\begin{center}
\subfloat[]{\label{fig:tab_pc}
\begin{tabular}{|c|c|}
\hline
Subjet Configuration & Defining Relation\\
\hline
\hline
Soft Haze&   $\left (\ecf{2}{\beta} \right )^{3\alpha/\beta} \lesssim \ecf{3}{\alpha} \lesssim \left (\ecf{2}{\beta} \right )^{2}$   \\
\hline
Collinear Subjets&   $\ecf{2}{\alpha} \sim  \left (\ecf{2}{\beta}   \right) ^{\alpha/\beta}$ and $\ecf{3}{\alpha}\ll \left( \ecf{2}{\beta}\right)^{3\alpha/\beta}$ \\
\hline
Soft Subjet&  $\ecf{2}{\alpha} \sim  \ecf{2}{\beta}  $ and $\ecf{3}{\alpha}\ll \left( \ecf{2}{\beta}\right)^{3\alpha/\beta}$   \\
\hline
\end{tabular}
}\\
\subfloat[]{\label{fig:2pt3ptps}
\includegraphics[width= 5.5cm]{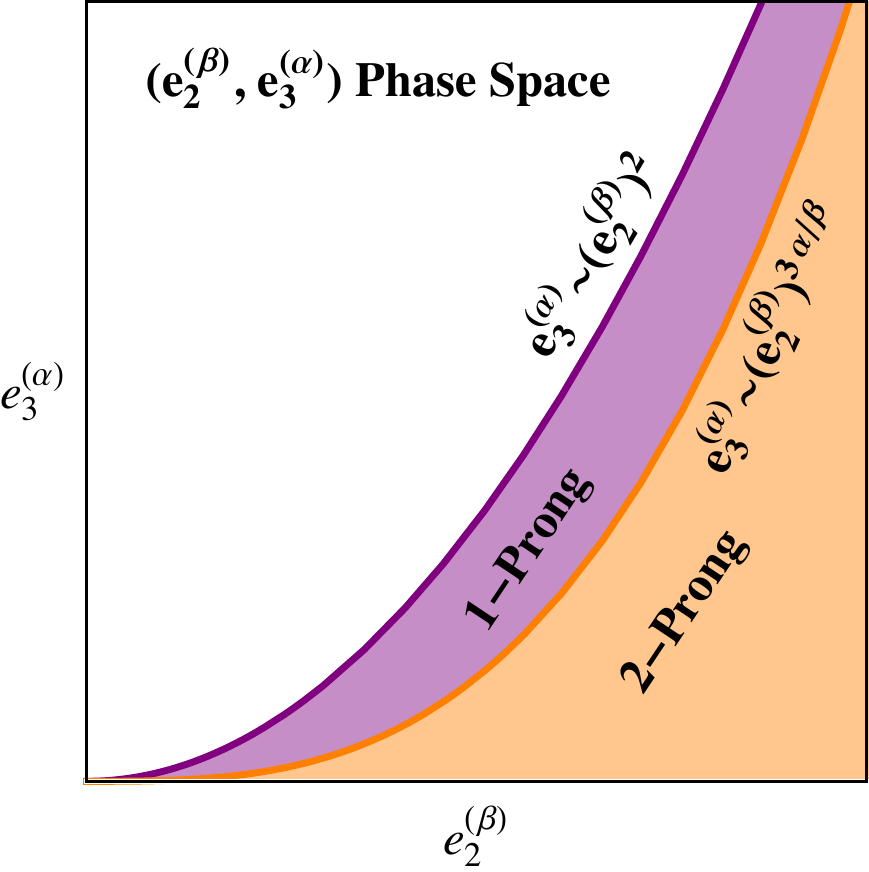}
}
$\qquad$
\subfloat[]{\label{fig:2ptps}
\includegraphics[width= 5.5cm]{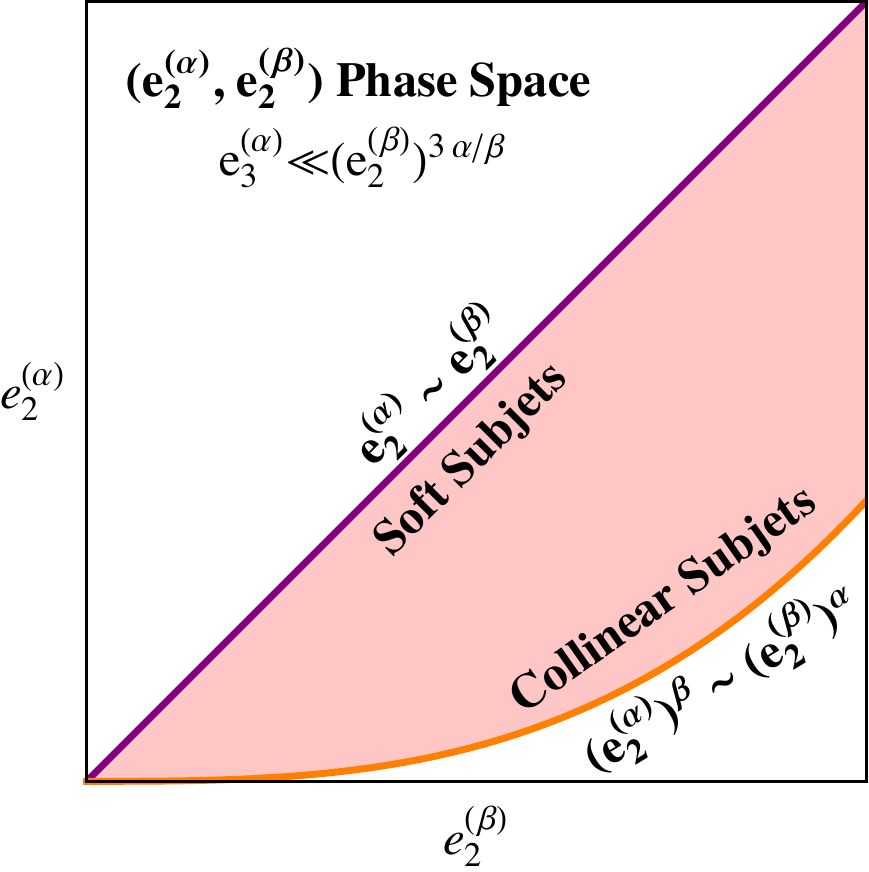}
}
\end{center}
\vspace{-0.2cm}
\caption{ a) Table summarizing the defining relations for the different subjet configurations in terms of the energy correlation functions $\ecf{2}{\alpha}, \ecf{2}{\beta},\ecf{3}{\alpha}$.  b) The one- and two-prong jets regions in the  $\ecf{2}{\beta}, \ecf{3}{\alpha}$ phase space. Jets with a two-prong structure lie in the lower (orange) region of phase space, while jets with a one-prong structure lie in the upper (purple) region of phase space. c) The projection onto the $\ecf{2}{\alpha}, \ecf{2}{\beta}$ phase space in which the soft subjet and collinear subjets are separated. 
}
\label{fig:phasespace}
\end{figure}

The measurement of $\ecf{2}{\beta}$ and $\ecf{3}{\alpha}$ alone is sufficient to separate one- and two prong jets.  However, the two-prong jets can exhibit either collinear subjets or a soft, wide angle subjet. To separate the collinear and soft subjet two-prong jets, we make an additional IRC safe measurement on the full jet.  Following \Ref{Larkoski:2015zka}, in addition to $\ecf{2}{\beta}$ and $\ecf{3}{\alpha}$, we measure $\ecf{2}{\alpha}$, with $\alpha \neq \beta$. In particular, the soft subjet and collinear subjet regions of phase space are defined by the simple conditions
\begin{align}
&\text{Collinear Subjet:} \qquad \ecf{2}{\alpha} \sim  \left (\ecf{2}{\beta}   \right) ^{\alpha/\beta}\,, \\
&\text{Soft Subjet:} \qquad \hspace{0.9cm} \ecf{2}{\alpha} \sim  \ecf{2}{\beta}\,.
\end{align}
For $\alpha \neq \beta$ and $\ecf{2}{\beta}\ll 1$, these two regions are parametrically separated.  Equivalently, in the two-prong region of phase space the measurement of both $\ecf{2}{\alpha}$ and $\ecf{2}{\beta}$ can be used to give IRC safe definitions to the subjet energy fraction and splitting angle, allowing the soft subjet and collinear subjets to be distinguished.  In \Fig{fig:phasespace} we summarize and illustrate the measurements that we make on the jet and the parametric relations between the measured values of the energy correlation functions that define the three phase space regions.  The phase space plots of \Figs{fig:2pt3ptps}{fig:2ptps} were also presented in \Ref{Larkoski:2015zka}.

\subsubsection{Jet Mass Cuts}\label{sec:mass_cuts}

In addition to discriminating QCD jets from boosted $Z$ bosons by their number of resolved prongs, we must also impose a mass cut on the jet to ensure that the jet is compatible with a $Z$ decay.  To include a mass cut in our analysis, for general angular exponents $\alpha$ and $\beta$, we would need to measure four observables on the jet: $\ecf{2}{\alpha}$, $\ecf{2}{\beta}$, $\ecf{3}{\alpha}$ and the jet mass.  This would significantly complicate calculations and introduce new parametric phase space regions that would need to be understood.  To avoid this difficulty, we note that, for our definition of $\ecf{2}{\alpha}$ from \Eq{eq:ecf_def}, if all final state particles are massless, then
\begin{equation}\label{eq:ecf_mass_relation}
\ecf{2}{2} = \frac{m_J^2}{E_J^2} \,,
\end{equation}
where $m_J$ is the mass of the jet.  Therefore, choosing $\beta = 2$ we can trivially impose a mass cut within the framework developed here.  Throughout the rest of this chapter, we will set $\beta = 2$ for this reason.  Importantly, from Monte Carlo studies it has been shown that $\beta \sim 2$ provides optimal discrimination power \cite{Larkoski:2013eya,Larkoski:2014gra}, so this restriction does not limit the phenomenological relevance of our results.

Substituting the value $\beta=2$ into the power counting condition of \Eq{eq:restriction_1}, we find that the one- and two-prong regions of phase space are separated if 
\be
\alpha \geq \frac{4}{3}\,.
\ee
To achieve a parametric separation of the one- and two-prong regions of phase space, we will demand that the scalings defining the different regions be separated by at least a single power of $\ecf{2}{\beta}$.  For example, choosing $\alpha=\beta=2$, the scalings of the one-prong and two-prong regions are $\ecf{3}{\alpha}\sim \left( \ecf{2}{\beta}\right)^{3}$ and $\ecf{3}{\alpha}\sim \left( \ecf{2}{\beta}\right)^{2}$, which are parametrically different.  We therefore restrict ourselves to the range of angular exponents
\begin{align}\label{eq:final_constraint_exponents}
\beta=2, \qquad \alpha \gtrsim 2\,.
\end{align}
We expect that for $\alpha<2$ our effective field theory description will begin to break down, while as $\alpha$ is increased above $2$ it should improve.

\section{Factorization and Effective Field Theory Analysis}\label{sec:Fact}

In each region of phase space identified in \Sec{sec:phase_space}, hierarchies of scales associated with the particular kinematic configuration of the jet appear. These include the soft subjet energy fraction $z_{sj}$ in the soft subjet region of phase space, or the splitting angle $\theta_{12}$ of the collinear subjets. Logarithms of these scales appear at each order in perturbation theory, and need to be resummed to all orders to achieve reliable predictions. To perform this resummation, we will prove factorization theorems in each region of phase space by developing an effective field theory description which captures all the scales relevant to that particular region of phase space. These effective field theories are formulated in the language of SCET \cite{Bauer:2000yr,Bauer:2001ct,Bauer:2001yt,Bauer:2002nz}, but include additional modes which are required to describe the dynamics of the scales associated with the jet's particular substructure. Resummation is then achieved by renormalization group evolution within the effective theory.

In this section we discuss each of the effective theories required for a description of the $\ecf{2}{\alpha}, \ecf{2}{\beta}, \ecf{3}{\alpha}$ phase space. For each region of the phase space, we present an analysis of the modes required in the effective field theory description and present the factorization theorem. We also provide a brief discussion of the physics described by each of the functions appearing in the factorization theorem. Field theoretic operator definitions of the functions, as well as their calculation to one-loop accuracy, are presented in appendices.

\subsection{QCD Background}\label{sec:QCD_back}

Three distinct factorization theorems are required to describe the full phase space for massive QCD jets, corresponding to the soft haze, collinear subjets, and soft subjet configurations. Detailed expositions of the factorization theorems for the collinear subjets and soft subjet configurations have been presented in \Refs{Bauer:2011uc,Larkoski:2015zka}, but here we review the important features of the factorization theorems to keep the discussion self-contained.

Throughout this section, all jets are defined using the $e^+e^-$ anti-$k_T$ clustering metric \cite{Cacciari:2008gp,Cacciari:2011ma} with the Winner-Take-All (WTA) recombination scheme \cite{Larkoski:2014uqa,Larkoski:2014bia}. To focus on the aspects of the factorization relevant to the jet substructure, we will present the factorization theorems for the specific case of $e^+e^- \to q \bar q$. The factorization theorem for gluon initiated jets is identical to the quark case, and can be performed using the ingredients in the appendices. The extension to the production of additional jets or $pp$ colliders will be discussed in \Sec{sec:LHC}. 

\subsubsection{Collinear Subjets}\label{sec:ninja}

An effective field theory describing the collinear subjets configuration was first presented in \Ref{Bauer:2011uc} and is referred to as SCET$_+$. We refer the interested reader to \Ref{Bauer:2011uc} for a more detailed discussion, as well as a formal construction of the effective theory. To our knowledge, our calculation is the first, other than that of \Ref{Bauer:2011uc}, to use this effective theory.

\subsubsection*{Mode Structure}\label{sec:ninja_modes}

\begin{figure}
\begin{center}
\subfloat[]{\label{fig:collinear_subjets_a}
\includegraphics[width=5cm]{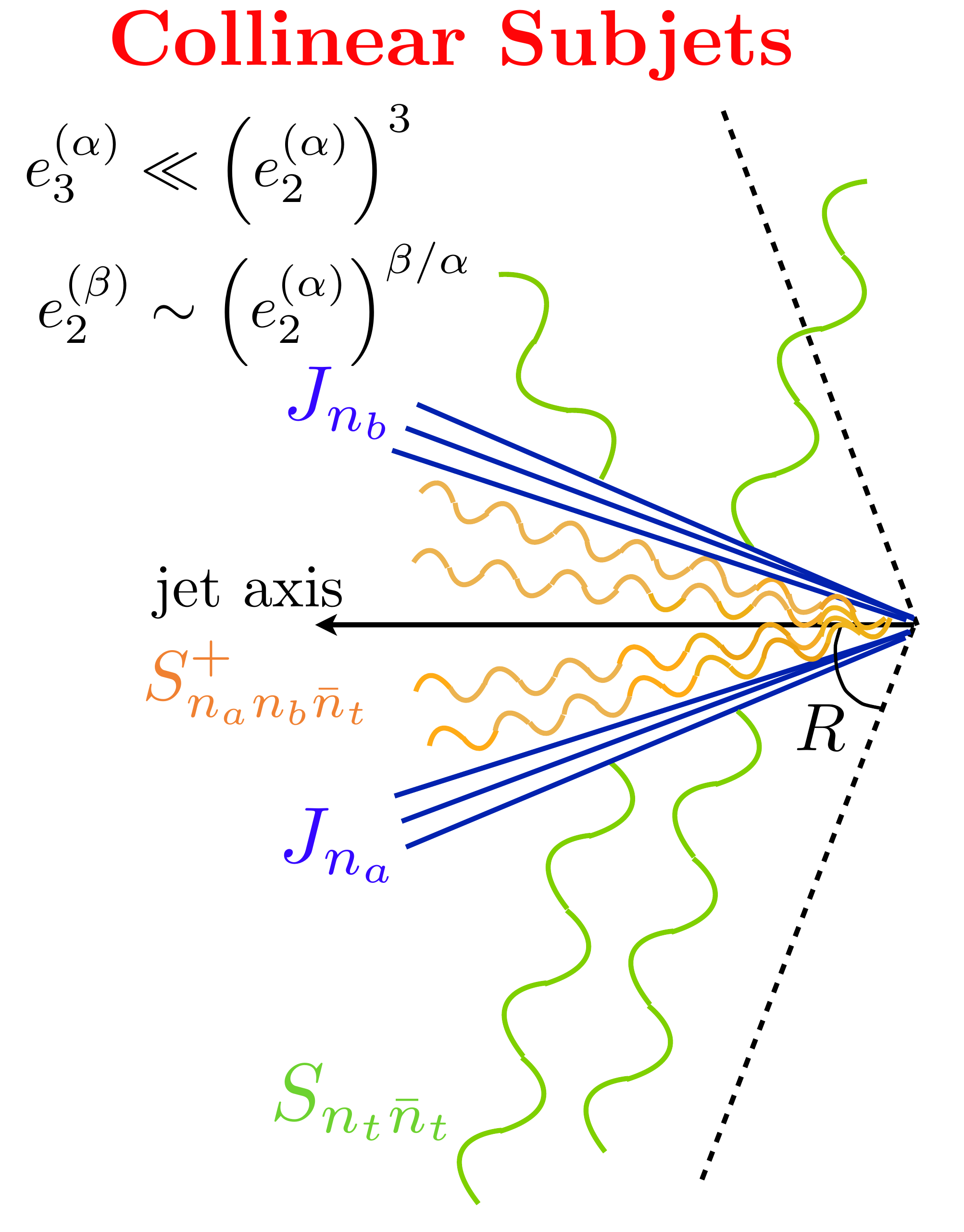}
}
\subfloat[]{\label{fig:collinear_subjets_b}
\includegraphics[width = 10cm]{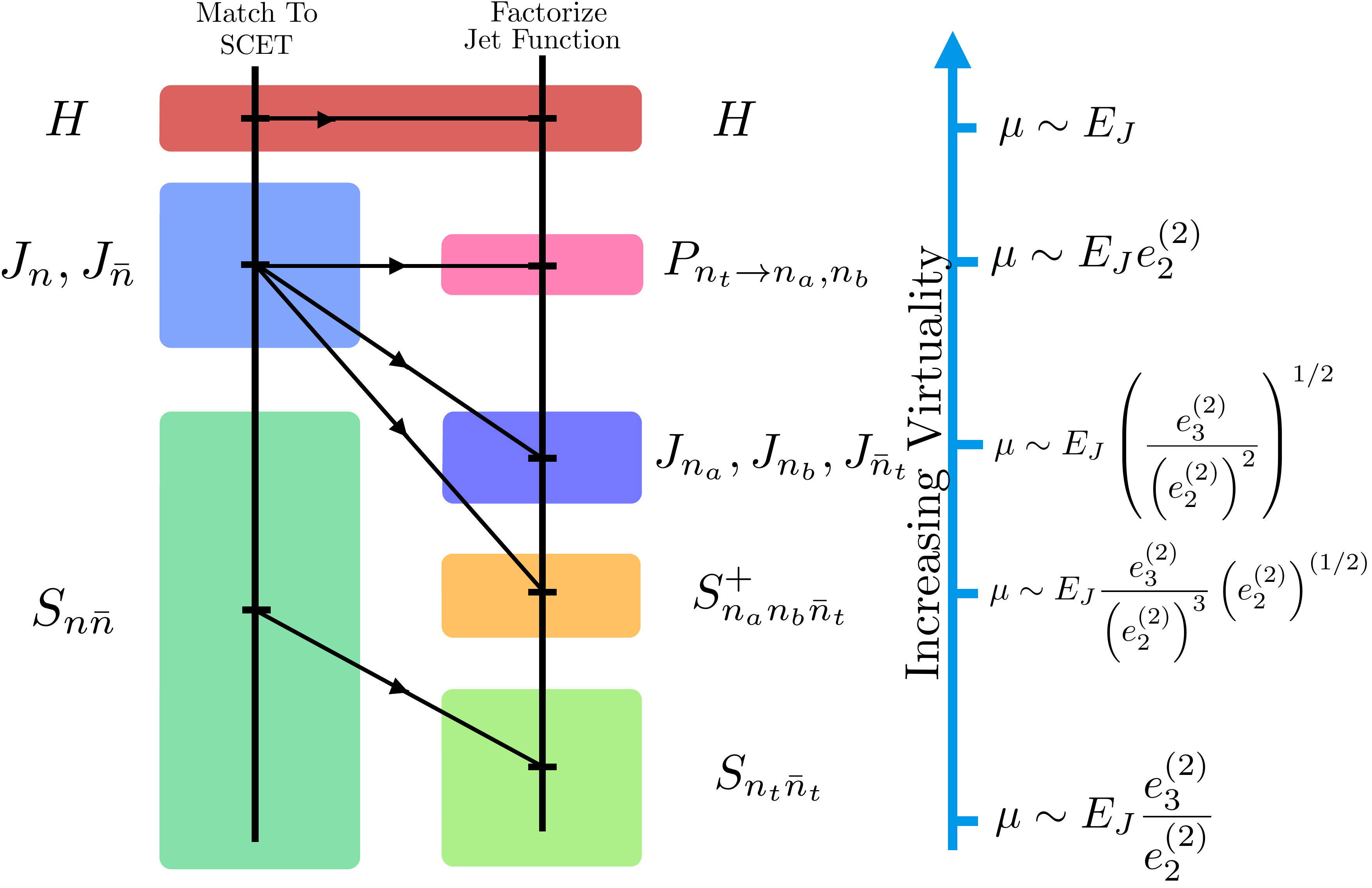}
}
\end{center}
\caption{A schematic depiction of the collinear subjets configuration with dominant QCD radiation and the functions describing its dynamics in the effective field theory is shown in a).   The matching procedure and relevant scales are shown in b), where we have restricted to the case $\alpha=\beta =2$ for simplicity.
}
\label{fig:collinear_subjets}
\end{figure}

The modes of SCET$_+$ are global soft modes, two collinear sectors describing the radiation in each of the collinear subjets, and collinear-soft modes from the dipole of the subjet splitting. These are shown schematically in \Fig{fig:collinear_subjets}. The additional collinear-soft modes, as compared with traditional SCET, are necessary to resum logarithms associated with the subjets' splitting angle. This angle, which is taken to be small, is not resolved by the long wavelength global soft modes.

The parametric scalings of the observables in the collinear subjets region were given in \Sec{sec:power_counting} and are:
\begin{align}\label{eq:collinear_subjets_scaling}
\ecf{2}{\alpha}&\sim \theta_{12}^\alpha \,,  \\
\ecf{2}{\beta}&\sim  \theta_{12}^\beta \,, \\
\ecf{3}{\alpha}&\sim
\theta_{12}^\alpha(\thetac^\alpha \theta_{12}^{\alpha}
+ \zs +\theta_{12}^{2\alpha}\zcs) \,.
\end{align}
Although the measurement of two 2-point energy correlation functions is required to be able to distinguish the soft and collinear subjets, they are redundant in the collinear subjets region from a power counting perspective, due to the relation $\ecf{2}{\beta}\sim \left ( \ecf{2}{\alpha} \right)^{\beta/\alpha}$.  We will therefore always write the scaling of the modes in terms of $\ecf{2}{\alpha}$ and $\ecf{3}{\alpha}$ to simplify expressions.

From \Eq{eq:collinear_subjets_scaling}, we see that $\ecf{2}{\alpha}$ sets the hard splitting scale, while the scalings of all the modes are set by the measurement of $\ecf{3}{\alpha}$. In particular, the scaling of the momenta of the collinear and soft modes are given by
\begin{align}\label{eq:cs_collinear_and_soft}
p_c &\sim E_J\left (\left (\frac{\ecf{3}{\alpha}}{ \left ( \ecf{2}{\alpha} \right )^{2}  }   \right )^{2/\alpha},1,   \left( \frac{\ecf{3}{\alpha}}{ \left ( \ecf{2}{\alpha} \right )^{2}  } \right)^{1/\alpha}   \right )_{n_a\bar n_a, n_b \bar n_b}\,, \\
\label{eq:soft_scale_coll}
p_s &\sim E_J \frac{\ecf{3}{\alpha}}{  \ecf{2}{\alpha}  } \left ( 1,1,1  \right )_{n\bar n} \,,
\end{align}
while the scaling of the collinear-soft mode is given by
\begin{align}\label{eq:cs_cs}
p_{cs} &\sim E_J \frac{\ecf{3}{\alpha}}{ \left ( \ecf{2}{\alpha} \right )^{3}  }    \left (  \left ( \ecf{2}{\alpha} \right)^{2/\alpha},1,    \left ( \ecf{2}{\alpha} \right)^{1/\alpha} \right )_{n\bar n}\,.
\end{align}
Here $E_J$ is the energy of the jet, and the subscripts denote the light-like directions with respect to which the momenta is decomposed.  In the expressions above, the momenta are written in the $(+,-,\perp)$ component basis with respect to the appropriate light-like directions. The subjet directions are labelled by $n_a$ and $n_b$, while the fat jet (containing the two subjets) and the recoiling jet are labelled by $n$ and $\bar n$. The relevant modes and a schematic depiction of the hierarchy of their virtualities is shown in \Fig{fig:collinear_subjets}.  

To have a valid soft and collinear expansion, the scalings of the modes in \Eqs{eq:cs_collinear_and_soft}{eq:cs_cs} imply that
\be
\ecf{2}{\alpha}\sim \left ( \ecf{2}{\beta} \right )^{\alpha/\beta}   \ll 1 \qquad \text{and} \qquad \frac{\ecf{3}{\alpha}}{ \left ( \ecf{2}{\alpha} \right )^{3}  } \sim \frac{\ecf{3}{\alpha}}{ \left ( \ecf{2}{\beta} \right )^{3\alpha/\beta}  } \ll 1 \,.
\ee
This agrees with the boundaries of the phase space found in \Sec{sec:power_counting}.

\subsubsection*{Factorization Theorem}\label{sec:fact_ninja}

In the collinear subjets region of phase space, the values of the 2-point energy correlation functions $\ecf{2}{\alpha}$ and $\ecf{2}{\beta}$ are set by the hard splitting. To leading power, these observables can be used to provide IRC safe definitions of the subjet energy fractions and the angle between the subjets. We therefore write the factorization theorem in terms of $\ecf{2}{\alpha}$, $\ecf{3}{\alpha}$, and the energy fraction of one of the subjets, which we denote by $z$. We further assume that an IRC safe observable, $B$, is measured in the out-of-jet region. Dependence on $B$ enters only into the out-of-jet jet function, and the out-of-jet contribution to the soft function.

The factorization theorem formulated in SCET$_+$ for the collinear subjets region of phase space is given by
\begin{align}\label{eq:NINJA_fact}
\frac{d^3\sigma}{dz\,d\ecf{2}{\alpha}d\ecf{3}{\alpha}}&=\sum_{f,f_a,f_b}H_{n\bar{n}}^{f}J_{\bar{n}}(B) P_{n_t\rightarrow n_a,n_b}^{f\rightarrow f_a f_b}\Big(z;\ecf{2}{\alpha}\Big)\int  de_{3}^{c}de_{3}^{\bar{c}}de_{3}^{s}de_{3}^{cs} \\
& \hspace{-0.5cm} \times\delta\Big(\ecf{3}{\alpha}-e_{3}^c-e_{3}^{\bar{c}}-e_{3}^{s}-e_{3}^{cs}\Big) 
 J_{n_a}^{f_a}\Big(z;e_{3}^{c}\Big)J_{n_b}^{f_b}\Big(1-z;e_{3}^{\bar{c}}\Big)S_{n \bar{n}}\Big(e_3^{s}, B;R\Big)S_{ n_a n_b \bar{n}}^+\Big(e_{3}^{cs}\Big) \,, \nonumber
\end{align}
where we have suppressed the convolution over the out-of-jet measurement, $B$, for simplicity.
Here the $n_a$, $n_b$ denote the collinear directions of the subjets, and we assume that $z\sim 1-z\sim \frac{1}{2}$.  The sum runs over all possible quark flavors that could be produced in an $e^+e^-$ collision. A brief description of the functions entering the factorization theorem of \Eq{eq:NINJA_fact} is as follows:

\begin{itemize}

\item $H_{n\bar{n}}^{f}$ is the hard function describing the underlying short distance process. In this case we consider $e^+e^- \to q \bar q$.

\item $P_{n\rightarrow n_a,n_b}^{f\rightarrow f_a f_b}\Big(z;\ecf{2}{\alpha}\Big)$ is the hard function arising from the matching for the hard splitting into subjets. In this case the partonic channel $f\to f_a f_b$ is restricted to $q\to q g$.

\item  $J_{n_a}^{f_a}\Big(z;e_{3}^{c}\Big)$, $J_{n_b}^{f_b}\Big(1-z;e_{3}^{\bar{c}}\Big)$ are jet functions describing the collinear dynamics of the subjets along the directions $n_a$, $n_b$.

\item  $S_{n \bar{n}}(e_3^{s},B;R)$ is the global soft function. The global soft modes do not resolve the subjet splitting, and are sensitive only to two eikonal lines in the $n$ and $\bar n$ directions. The soft function depends explicitly on the jet radius, $R$.

\item $S_{ n_a n_b \bar{n}}^+\Big(e_{3}^{cs} \Big)$ is the collinear-soft function. The collinear-soft modes resolve the subjet splitting, and hence the function depends on three eikonal lines, namely $n_a, n_b, \bar n$. Although these modes are soft, they are also boosted, and therefore do not resolve the jet boundary, so that the collinear soft function is independent of the jet radius, $R$.

\end{itemize}
This factorization theorem is shown schematically in \Fig{fig:collinear_subjets}, which highlights the radiation described by each of the functions in \Eq{eq:NINJA_fact}, as well as their virtuality scales. The two stage matching procedure onto the SCET$_+$ effective theory, which proceeds through a refactorization of the jet function, is also shown. The fact that the refactorization occurs in the jet function is important in that it implies that it is independent of the global color structure of the event, making it trivial to extend the factorization theorem to events with additional jets. This matching procedure is discussed in detail in \Ref{Bauer:2011uc}.

Operator definitions, and one-loop calculations for the operators appearing in the factorization theorem of \Eq{eq:NINJA_fact}  are given in \App{sec:ninja_app}.


\subsubsection{Soft Subjet}\label{sec:soft_jet}

\begin{figure}
\begin{center}
\subfloat[]{\label{fig:soft_subjets_a}
\includegraphics[width= 5.4cm]{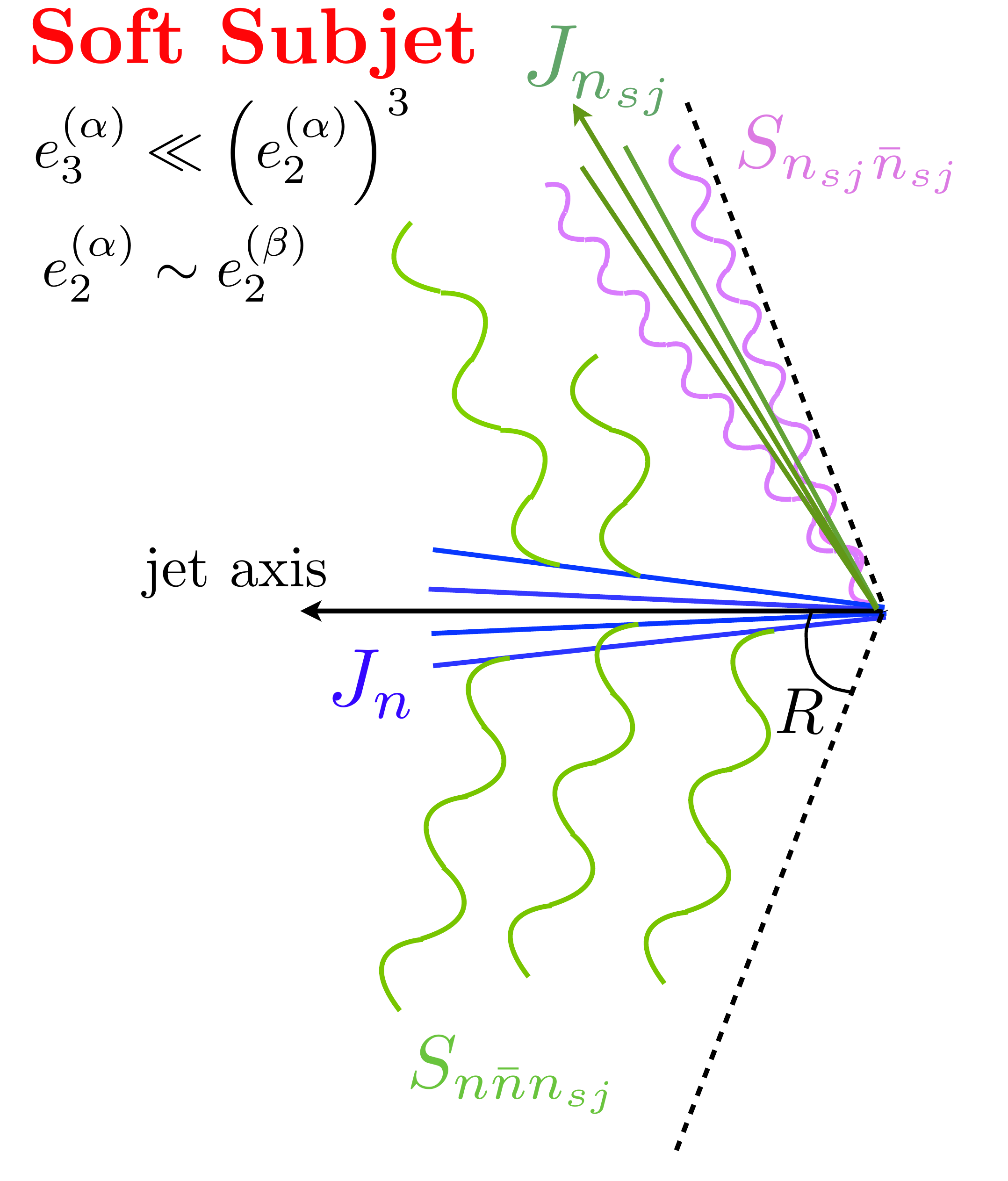}
}
\subfloat[]{\label{fig:soft_subjets_b}
\includegraphics[width =9.5cm]{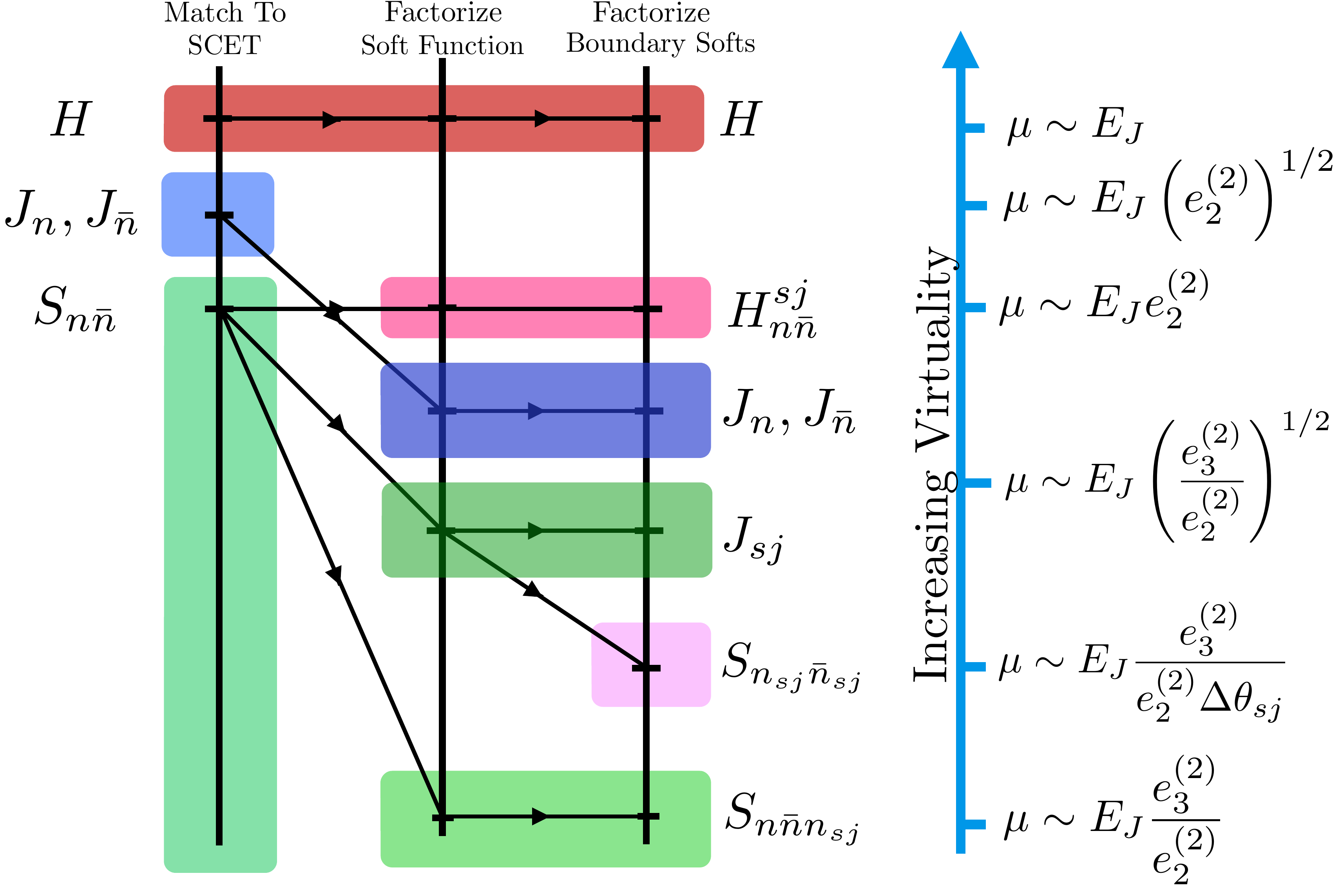}
}
\end{center}
\caption{ A schematic depiction of the soft subjet configuration with dominant QCD radiation and the functions describing its dynamics in the effective field theory is shown in a).   The matching procedure and relevant scales are shown in b), where we have restricted to the case $\alpha=\beta =2$ for simplicity.
}
\label{fig:soft_subjets}
\end{figure}

A factorization theorem describing the soft subjet region of phase space was recently presented in \Ref{Larkoski:2015zka}. In this section we review the basic features of this factorization theorem, but we refer the reader to \Ref{Larkoski:2015zka} for a more detailed discussion.

Unlike for the case of collinear subjets, in the soft subjet configuration, the wide angle soft subjet probes the boundary of the jet. This introduces sensitivity to the details of the jet algorithm used to define the jet, as well as to the measurement made in the region outside the jet. The factorization theorem of \Ref{Larkoski:2015zka} is valid under the assumption that an additive IRC safe observable, $B$, is measured in the out-of-jet region, and that the soft scale associated with this observable, $\Lambda$, satisfies $\Lambda/E_J \ll \ecf{2}{\alpha}$. We will therefore assume that this condition is satisfied throughout this section. However, we will see that the numerical results are fairly insensitive to the details of the choice of scale $\Lambda$. \Ref{Larkoski:2015zka} also used a broadening axis \cite{Larkoski:2014uqa} cone algorithm to define jets, whereas here we use the anti-$k_T$ algorithm, as relevant for phenomenological applications. We will argue that the structure of the factorization theorem is in fact identical in the two cases, to leading power.

\subsubsection*{Mode Structure}\label{sec:soft_jet_modes}

 In the soft subjet region of phase space there are two subjets with an energy hierarchy. We denote the energy of the soft subjet by $z_{sj}$ and the angle from the $n$ axis by $\theta_{sj}$. We also use the notation $\Delta \theta_{sj}=R-\theta_{sj}$ to denote the angle from the soft subjet axis to the jet boundary. The modes of the soft subjet are collinear-soft modes, being both soft and collimated, and we will therefore denote the characteristic angle between them as $\theta_{cs}$. Straightforward power counting can be applied to determine the scaling of the modes for both the energetic jet and the soft subjet. Their contributions to the observable are given by
\begin{align}\label{eq:soft_subjet_scaling}
\ecf{2}{\alpha} &\sim z_{sj}\,, \\ \ecf{2}{\beta} &\sim z_{sj}\,, \\ \ecf{3}{\alpha}&\sim z_{sj} (\theta_c^\alpha +z_{sj} \theta_{cs}^\alpha +z_s)\,.
\end{align}
In the soft subjet region of phase space, we have the relation $\ecf{2}{\alpha} \sim \ecf{2}{\beta}$, and therefore these two observables are redundant from a power counting perspective. We will therefore write the power counting of the modes in terms of $\ecf{2}{\alpha} $ and $\ecf{3}{\alpha} $.

From the contributions to the observables above, we find that the momentum of the collinear and global soft radiation scales like
\begin{align}
p_c &\sim E_J \left ( \left (  \frac{\ecfres}{  \ecfLa }  \right)^{2/\alpha},1,\left (  \frac{\ecfres}{  \ecfLa }  \right)^{1/\alpha}    \right )_{n\bar{n}}\,, \\
p_s &\sim  E_J \,\frac{\ecfres}{  \ecfLa } \left ( 1 ,1,1  \right )_{n\bar{n}}    \,,\nonumber 
\end{align}
where $E_J$ is the energy of the jet and $n$ and $\bar n$ are the light-like directions of the jet of interest and the other jet in the event, respectively.  The soft subjet mode's momentum scales like 
\begin{align}
p_{sj} \sim E_J\, \ecfLa \left ( \left (  \frac{\ecfres}{  \left (\ecfLa\right)^2 }  \right)^{2/\alpha},1,\left (  \frac{\ecfres}{  \left (\ecfLa\right )^2 }  \right)^{1/\alpha}    \right )_{\sja\sjabar}   \,,
\end{align}
in the light-cone coordinates defined by the direction of the soft subjet, $n_{sj}$. These are the complete set of modes defined by the scales set by the measurements of $\ecfLa ,\ecfLb$, and $\ecfres$ alone.

Unlike in the collinear subjet region of phase space there are no collinear-soft modes required in the effective field theory description, since the soft subjet is at a wide angle from the jet axis. However, in this region there is an additional mode, termed a \emph {boundary soft mode} in \Ref{Larkoski:2015zka}, whose appearance is forced by the jet boundary and the energy veto in the region of phase space outside the jet. These modes do not contribute to the $e_2$ observables, but are effectively a collinear-soft mode whose angle with respect to the soft subjet axis is set by the angle to the boundary.  The boundary soft mode's momentum components scale like
\begin{align}\label{eq:boundary_soft_scaling}
p_{bs} &\sim E_J\frac{  \ecfres   }{ \ecfLa   \left(\Delta \sjtheta\right)^\alpha }  \left ( \left(\Delta  \sjtheta\right)^2,1,\Delta  \sjtheta \right )_{\sja\sjabar}   \,, 
\end{align}
written in the light-cone coordinates defined by the soft subjet axis.
The boundary soft modes are required to have a single scale in the soft subjet function. For consistency of the factorization, we must enforce that the soft subjet modes cannot resolve the jet boundary and that the boundary soft modes are localized near the jet boundary.  That is, the angular size of the soft subjet modes, $\theta_{cs}$, must be parametrically smaller than that of the boundary soft modes, namely $\Delta \theta_{sj}$. We therefore find the condition
\begin{equation}\label{eq:applicability_softjet}
\left(\Delta\sjtheta\right)^\alpha \gg  \left(\theta_{cs}\right)^\alpha\sim \frac{\ecfres}{\left (\ecfLa \right )^2}\,, \qquad \text{and}\qquad \Delta\sjtheta \ll 1\,.
\end{equation}
Therefore, the factorization theorem applies in a region of the phase space where the soft subjet is becoming pinched against the boundary of the  jet, but lies far enough away that the collinear modes of the soft subjet do not touch the boundary. A schematic depiction of this region of phase space, along with a summary of all the relevant modes which appear in the factorization theorem is shown in \Fig{fig:soft_subjets}.

In the soft subjet region of phase space, the choice of jet algorithm plays a crucial role, since the soft subjet probes the boundary of the jet. In \Ref{Larkoski:2015zka} the factorization theorem in the soft subjet region of phase space was presented using a broadening axis cone algorithm with radius $R$. We now show that up to power corrections, the factorization theorem in the soft subjet region of phase space is identical with either the anti-$k_T$ or broadening axis cone algorithm. In particular, with the anti-$k_T$ algorithm, the jet boundary is not deformed by the soft subjet, and can be treated as a fixed cone of radius $R$.  This is not true for other jet algorithms, such as such as $k_T$ \cite{Catani:1993hr,Ellis:1993tq} or Cambridge-Aachen \cite{Dokshitzer:1997in,Wobisch:1998wt,Wobisch:2000dk}, where the boundary is deformed by the clustering of soft emissions, a point which has been emphasized elsewhere (see, e.g., \Refs{Appleby:2002ke,Banfi:2005gj,Banfi:2010pa,Kelley:2012kj}).

The validity of the factorization theorem requires the following two conditions, which will put constraints on the power counting in the soft subjet region of phase space. First, the soft subjet must be clustered with the jet axis, rather than with the out-of-jet radiation. This is guaranteed as long as the soft subjet axis satisfies $\theta_{sj} < R$. Second, the radiation clustered with the soft subjet from the out-of-jet region should not distort the boundary of the jet. More precisely, the distortion of the boundary must not modify the value of $\ecf{3}{\alpha}$ at leading power (note that the power counting guarantees that it does not modify $\ecf{2}{\alpha}$). The contribution to $\ecf{3}{\alpha}$ from a soft out-of-jet emission is given by
\be\label{eq:out_jet_estimate}
\ecf{2}{\alpha}   \frac{\Lambda}{E_J}  \ll \ecf{3}{\alpha} \implies  \frac{\Lambda}{E_J}  \ll \frac{  \ecf{3}{\alpha}  }{  \ecf{2}{\alpha}   }\sim\Big(\ecf{2}{\alpha}  \Big)^2\,.
\ee
Since the out-of-jet scale is in principle a free parameter, we can formally enforce this condition in our calculations. Corrections due to a deformation of the jet boundary would enter as power corrections in this region of phase space. The jet boundary therefore acts as a hard boundary of radius $R$, and the factorization theorem is identical to that presented in \Ref{Larkoski:2015zka}.

\subsubsection*{Factorization Theorem}\label{sec:fact_soft_jet}

With an understanding of the precise restrictions on the power counting required for the validity of the soft subjet factorization theorem, we now discuss its structure. Since we have argued that the relevant factorization theorem is identical to that presented in \Ref{Larkoski:2015zka}, we will only state the result.
The factorization theorem in the soft subjet region with the out-of-jet scale satisfying $\Lambda \ll \ecf{2}{\alpha} E_J$, and with jets defined by the anti-$k_T$ jet algorithm, is given by
\begin{align}\label{fact_inclusive_form_1}
&\frac{d\sigma(\outj;R)}{d\ecfLa d\ecfLb d\ecfres }= \\
&\hspace{0.6cm} \int dB_S dB_{J_{\bar n}} \int de_3^{J_n} de_3^{J_{{sj}}} de_3^S  de_3^{S_{sj}}        \delta (B-B_{J_{\bar n}}- B_S)  \delta(\ecfres-e_3^{J_n}-e_3^{J_{{sj}}}-e_3^S-e_3^{S_{sj}}) \nonumber \\
& \hspace{0.6cm}\times H_{n\bar n}(Q^2) H^{sj}_{n\bar{n}}\Big(\ecfLa,\ecfLb\Big)    J_{n}\left(e_3^{J_n}\right)J_{\bar{n}}(B_{J_{\bar n}}) 
 S_{n\bar{n}\sja }\Big(e_3^S;B_{S};R\Big) J_{\sja}\left(e_3^{J_{sj}} \right) S_{\sja\sjabar}\Big(e_3^{S_{sj}};R\Big)\,. \nonumber
\end{align}
In this expression we have explicitly indicated the dependence on the jet boundaries with the jet radius $R$. A brief description of the functions appearing in \Eq{fact_inclusive_form_1} is as follows:
\begin{itemize}
\item $H_{n\bar n}(Q^2)$ is the hard function describing the underlying short distance process. In this case we consider $e^+e^- \to q \bar q$.
\item $H_{n\bar{n}}^{sj}\Big(\ecfLa,\ecfLb\Big) $ is the hard function describing the production of the soft subjet coherently from the initial $q\bar{q}$ dipole, and describes dynamics at the scale set by $\ecfLa,\ecfLb$.
\item $J_{n}\Big(\ecfres\Big)$ is a jet function at the scale $\ecfres$ describing the hard collinear modes of the identified jet along the $n$ direction.
\item  $J_{\bar{n}}(\outj)$ is a jet function describing the collinear modes of the out-of-jet region of the event.
\item $S_{n\bar{n}\sja }\Big(\ecfres;\outj;R\Big)$ is the global soft function involving three Wilson line directions, $n, \bar n, \sja$. The global soft function depends explicitly on both the out-of-jet measurement and the jet radius.
\item $J_{\sja}\Big(\ecfres\Big)$ is a jet function describing the dynamics of the soft subjet modes, which carry the bulk of the energy in the soft subjet.
\item $S_{\sja\sjabar}(\ecfres;R)$ is a soft function describing the dynamics of the boundary soft modes. It depends only on two Wilson line directions $\sja, \sjabar$.
\end{itemize}
These functions, and a schematic depiction of the radiation which they define, are indicated in \Fig{fig:soft_subjets}, along with a schematic depiction of the multistage matching procedure from QCD onto the effective theory, as described in detail in \Ref{Larkoski:2015zka}. Although we will not discuss any details of the matching procedure, it is important to note that it occurs through a refactorization of the soft function, and hence the soft subjet factorization theorem is sensitive to the global color structure of the event, since the soft subjet is emitted coherently from all eikonal lines. This should be contrasted with the case of the collinear subjets factorization theorem, where the matching occurs through a refactorization of the jet function.

In the soft subjet region of phase space, we can relate the variables $\ecfLa\,, \ecfLb$ to the physically more transparent $\sje, \sjtheta$ variables with a simple Jacobian factor, giving the factorization theorem
\begin{align}\label{fact_inclusive_form_2}
&\frac{d\sigma(\outj;R)}{d\sje\, d\sjtheta\, d\ecfres}=\\
&\hspace{0.6cm} \int dB_S dB_{J_{\bar n}} \int de_3^{J_n} de_3^{J_{{sj}}} de_3^S  de_3^{S_{sj}}        \delta (B-B_{J_{\bar n}}- B_S)  \delta(\ecfres-e_3^{J_n}-e_3^{J_{{sj}}}-e_3^S-e_3^{S_{sj}}) \nonumber \\
& \hspace{0.6cm}\times H_{n\bar n}(Q^2) H^{sj}_{n\bar{n}}\left(z_{sj}, \theta_{sj}\right)    J_{n}\left(e_3^{J_n}\right)J_{\bar{n}}(B_{J_{\bar n}}) 
 S_{n\bar{n}\sja }\Big(e_3^S;B_{S};R\Big) J_{\sja}\left(e_3^{J_{sj}} \right) S_{\sja\sjabar}(e_3^{S_{sj}};R)\,. \nonumber
\end{align} 

Operator definitions, and one-loop calculations for the operators appearing in the factorization theorem of \Eqs{fact_inclusive_form_1}{fact_inclusive_form_2}  are given in \App{sec:softjet_app}.

\subsubsection{Soft Haze}\label{sec:soft_haze}

\begin{figure}
\begin{center}
\subfloat[]{\label{fig:soft_hazes_a}
\includegraphics[width= 5cm]{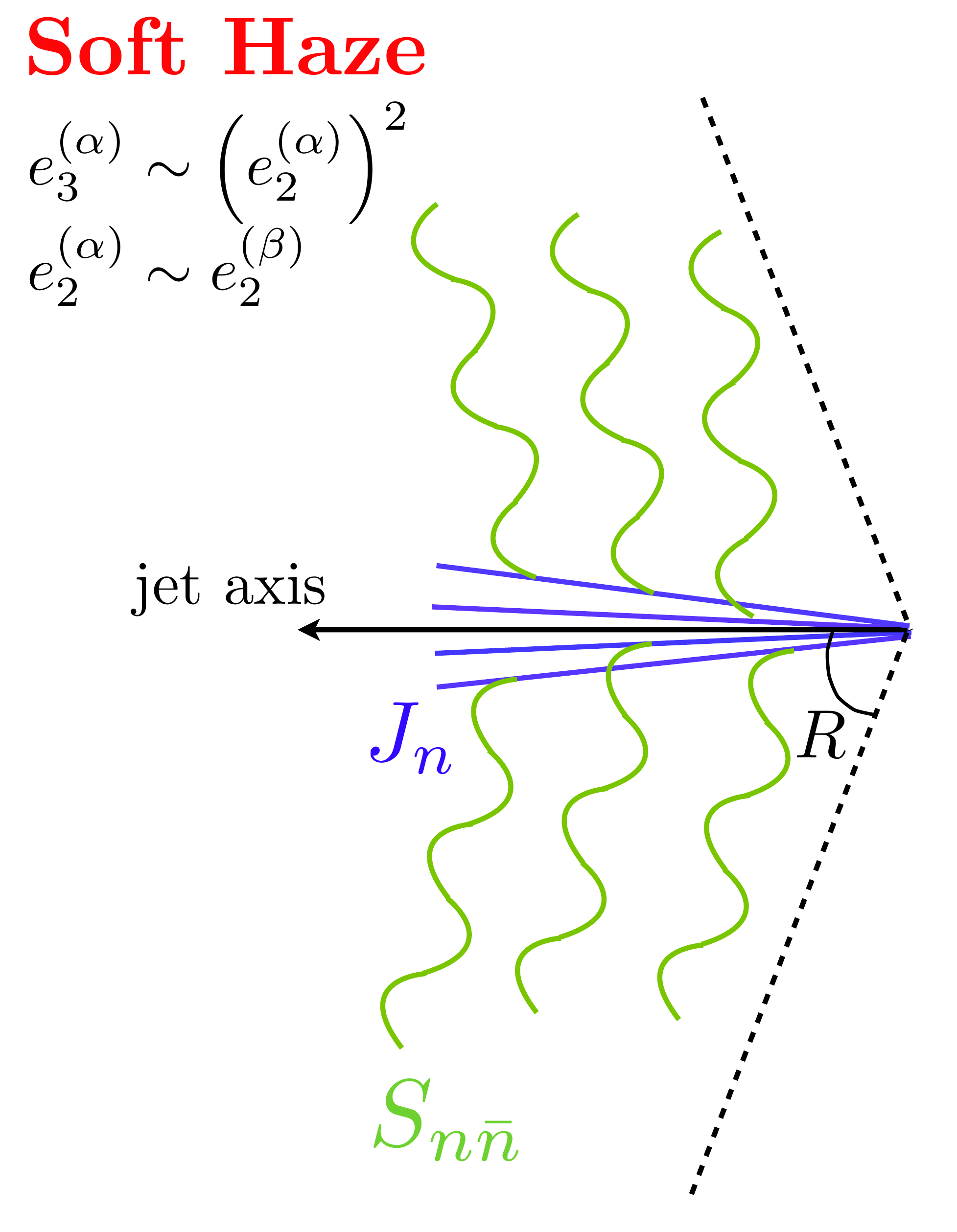}
}
\subfloat[]{\label{fig:soft_hazes_b}
\includegraphics[width = 10cm]{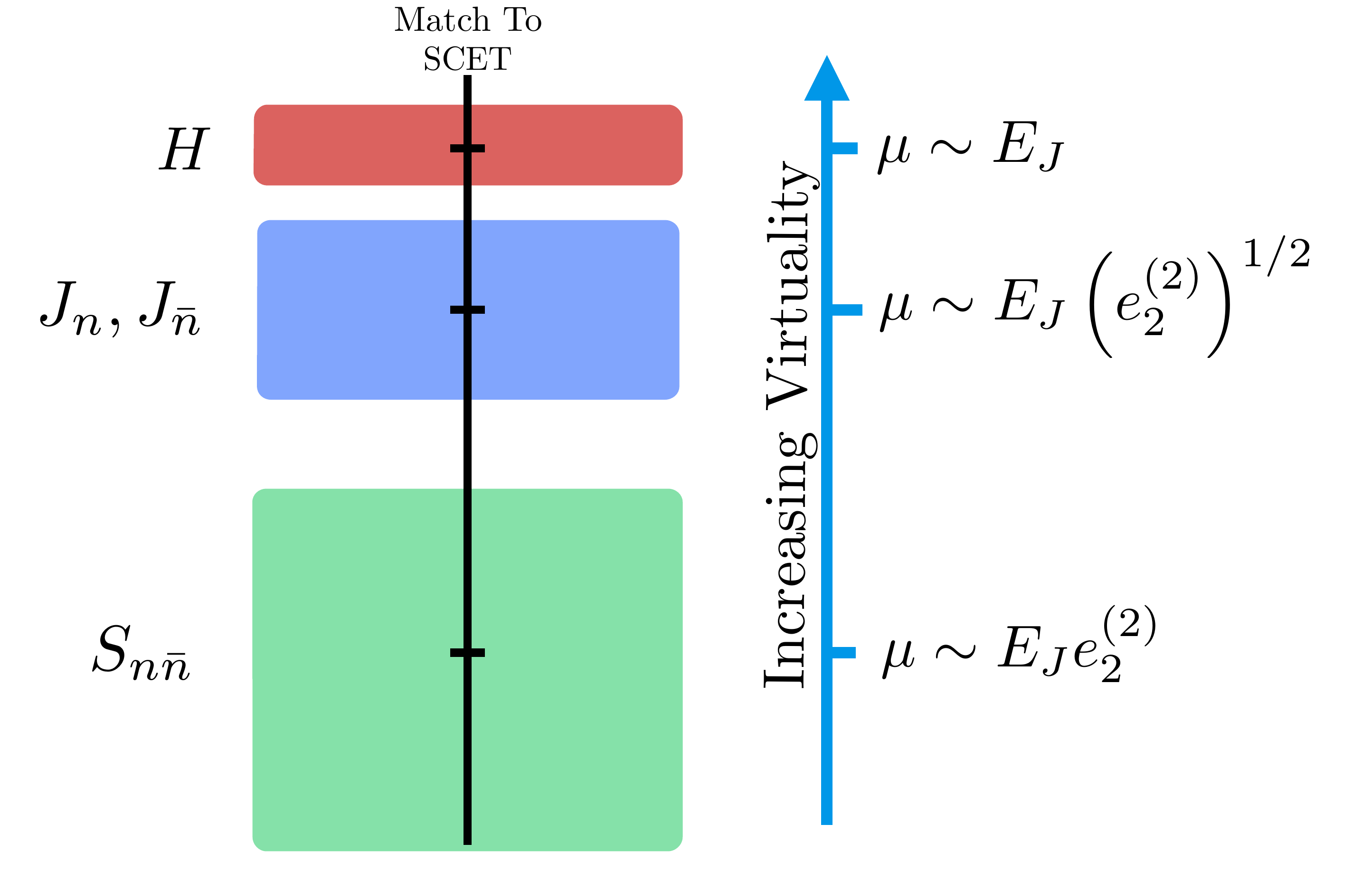}
}
\end{center}
\caption{A schematic depiction of the soft haze configuration where no subjets are resolved, with dominant QCD radiation and the functions describing its dynamics in the effective field theory is shown in a).   The relevant scales in the effective field theory are shown in b), where we have restricted to the case $\alpha=\beta =2$ for simplicity.
}
\label{fig:soft_hazes}
\end{figure}

The soft haze region defines the upper boundary of the $\ecf{2}{\beta},\ecf{3}{\alpha}$ phase space. In this region of phase space  jets consist of a single hard core, with no resolved subjets. A factorization theorem describing this region of phase space has not been presented elsewhere, but can be straightforwardly formulated in standard SCET involving only $n$ and $\bar n$ collinear sectors.

As discussed in \Sec{sec:power_counting}, the power counting in the soft haze region depends sensitively on the relative values of $\alpha$ and $\beta$, and therefore so does the structure of the factorization theorem. Since, from \Eq{eq:final_constraint_exponents}, we restrict ourself to $\alpha \geq \beta$, we will for simplicity only discuss the factorization theorems valid in this case. Factorization theorems for other values of $\alpha$ and $\beta$ can be determined by performing a similar analysis.

\subsubsection*{Mode Structure}\label{sec:soft_modes}

In the soft haze region the observables have the power counting
\begin{align}
\ecf{2}{\alpha}&\sim z_s+ \theta^\alpha_c\,, \\ 
\ecf{2}{\beta}&\sim z_s+ \theta^\beta_c\,, \\ 
\ecf{3}{\alpha}&\sim \zs^2 +\thetac^\alpha \zs+\thetac^{3\alpha}\,,
\end{align}
where we have not yet dropped power suppressed terms. We are interested in the factorization theorem on the upper boundary, with the scaling $\ecf{3}{\alpha} \sim \left (\ecf{2}{\beta} \right )^{2}$.\footnote{There is another parametric choice for the relative scaling of the 2-point energy correlation functions \cite{Larkoski:2014tva}, though it does not extend to the upper boundary of the phase space.  If $(\ecf{2}{\alpha})^\beta\sim (\ecf{2}{\beta})^\alpha$, then the power counting is
\begin{align*}
\ecf{2}{\alpha}&\sim z_s+ \theta^\alpha_c\,, \\ 
\ecf{2}{\beta}&\sim \theta^\beta_c\,, \\ 
\ecf{3}{\alpha}&\sim \zs^2 +\thetac^\alpha \zs\,,
\end{align*}
with both 2-point correlation functions dominated by collinear physics.  For $\alpha>\beta$, this region has the scaling $\ecf{3}{\alpha} \sim \left (\ecf{2}{\beta} \right )^{2\alpha/\beta}$ which does not extend to the upper boundary.} We now assume $\alpha > \beta$. In this case, dropping power suppressed terms, the appropriate power counting is
\begin{align}
\ecf{2}{\alpha}&\sim z_s\,, \\ 
\ecf{2}{\beta}&\sim z_s+ \theta^\beta_c\,, \\ 
\ecf{3}{\alpha}&\sim \zs^2\,.
\end{align}
It is also interesting to consider the case $\alpha=\beta$ because in the soft haze region it is not necessary to measure two different 2-point energy correlation functions, unlike in the two-prong region of phase space.
In the case that $\alpha=\beta$, we have instead,
\begin{align}\label{eq:power_count_softhaze_alpha}
\ecf{2}{\alpha}&\sim z_s+ \theta^\alpha_c\,, \\ 
\ecf{3}{\alpha}&\sim \zs^2 +\thetac^\alpha \zs\,,
\end{align}
where the second term in the expression for $\ecf{3}{\alpha}$ is no longer power suppressed. This will modify the factorization theorem between the two cases.

In both cases, the scaling of the modes is then given by
\begin{align}
p_c &\sim E_J \left (\left (\ecf{2}{\beta}\right )^{2/\beta}, 1, \left (\ecf{2}{\beta}\right )^{1/\beta}\right )_{n \bar n}, \\ p_s &\sim \ecf{2}{\beta}E_J \left (1,1,1\right)_{n \bar n},
\end{align}
with $\beta=\alpha$ in the second case.
Here $E_J$ is the energy of the jet and the subscripts denote the light-like directions with respect to which the momenta is decomposed.
This scaling should be recognized as the usual power counting of the collinear and soft modes for the angularities with angular exponent $\beta$ \cite{Ellis:2010rwa,Larkoski:2014tva}.

\subsubsection*{Factorization Theorem}\label{sec:fact_soft_haze}

The factorization theorem in the soft haze region of phase space can now be straightforwardly read off from the power counting expressions of the previous sections.  We state it both for the case $\alpha=\beta$ and $\alpha > \beta$. For $\alpha>\beta$, we have
\begin{align} \label{eq:fact_soft_haze}
\hspace{-0.5cm}\frac{d\sigma}{d\ecf{2}{\alpha}d\ecf{2}{\beta}d\ecf{3}{\alpha}}&=H_{n\bar{n}}(Q^2)    J_{\bar{n}}(B)\int de_2^{c}de_2^{s}    \delta\left(\ecf{2}{\beta}-  e_2^c  -e_2^{s}\right)  J_{n}\left(e_{2}^{c}\right)S_{n \bar{n}}\left(e_2^{s},\ecf{2}{\alpha},\ecf{3}{\alpha},R,\outj\right)\,,
\end{align}
where we have suppressed the convolution over the out-of-jet measurement $B$, to focus on the structure of the in-jet measurements.
For $\alpha=\beta$, the factorization theorem takes an interesting form\footnote{When calculating the tail of the $D_2$ distribution, one might be tempted to marginalize over $\ecf{2}{\beta}$ in \Eq{eq:fact_soft_haze}. This na\"ive marginalization does not yield the correct result. Rather, if one started the derivation of the factorization theorem with only the measurements of $\ecf{2}{\alpha}$ and  $\ecf{3}{\alpha}$ imposed, so that all possible $\ecf{2}{\beta}$ configurations are integrated over, then \Eq{eq:fact_soft_haze2} would be obtained. Thus \Eq{eq:fact_soft_haze2} is the correct marginalization over $\ecf{2}{\beta}$ in \Eq{eq:fact_soft_haze}.}
\begin{align} \label{eq:fact_soft_haze2}
\frac{d\sigma}{d\ecf{2}{\alpha}d\ecf{3}{\alpha}}&=H_{n\bar{n}}(Q^2)    J_{\bar{n}} (B)\int de_2^{c}de_2^{s} de_3^{s}   \delta\left(\ecf{2}{\alpha}-e_2^{c}-e_2^{s}\right) \delta \left ( \ecf{3}{\alpha}- e_2^c \,e_2^{s}-e_3^{s}  \right)    \\  
&\hspace{8cm} \times J_{n}\left(e_{2}^{c}\right)S_{n \bar{n}}\left(e_2^{s},e_3^{s},R,\outj\right)\,, \nonumber
\end{align}
where again the convolution over $B$ has been suppressed.
A brief description of the functions appearing in the factorization theorems is as follows:
\begin{itemize}

\item $H_{n\bar n}\left(Q^2\right)$  is the hard function describing the underlying short distance process. In this case we consider $e^+e^- \to q \bar q$.

\item $J_{\bar{n}} (B)$ is the jet function describing the collinear modes for the recoiling jet.

\item $J_{n}\left(e_{2}^{c} \right)$ is the jet function describing the collinear modes for the jet in the $n$ direction.

\item $S_{n \bar{n}}\left(e_2^{s},e_3^{s},R,\outj\right)$ and $S_{n \bar{n}}\left(e_2^{s},\ecf{2}{\alpha},\ecf{3}{\alpha},R,\outj\right)$ are soft functions describing the global soft radiation from the $n \bar n$ dipole.  These also carry the jet algorithm constraints denoted by $R$, and any out-of-jet measurements $\outj$.

\end{itemize}
These functions, and a schematic depiction of the radiation which they define are indicated in \Fig{fig:soft_hazes}.  In \App{app:softhaze}, we give operator definitions of these functions and the leading-power expression for the $\ecf{3}{\alpha}$ measurement operator in the soft function.

There are several interesting features about the factorization theorems of \Eqs{eq:fact_soft_haze}{eq:fact_soft_haze2}. First, the soft functions are multi-differential, in that they require the simultaneous measurement of multiple quantities. Such multi-differential jet and soft functions have been discussed in detail in \Ref{Larkoski:2014tva,Procura:2014cba}. One other interesting feature of the factorization theorem of \Eq{eq:fact_soft_haze2}, for the case of equal angular exponents, is the appearance of the product structure in the $\delta$-function defining the value of $\ecf{3}{\alpha}$. This product structure is consistent with the power counting of \Eq{eq:power_count_softhaze_alpha} which describes the properties of the 3-point energy correlation function in the soft and collinear limits. It is important to note that this product form does not violate soft-collinear factorization, since only the knowledge of the total $\ecf{2}{\alpha}$ of the soft or collinear sector is required.

The soft contribution to the 3-point energy correlation is first non-vanishing with two real emissions. Therefore at one-loop, the factorization theorem of \Eq{eq:fact_soft_haze} reduces exactly to the factorization theorem for the multi-differential angularities studied in \Refs{Larkoski:2014tva,Procura:2014cba}, whereas the factorization theorem of \Eq{eq:fact_soft_haze2} reduces to the factorization theorem for a single angularity. In this chapter, we will not perform the two-loop calculation necessary to obtain a non-trivial contribution to the three point energy correlation function. Instead, we will obtain an approximation to the cross section in this region by taking a limit of our factorization theorems in the two-prong region of phase space.  This is possible, because as we will show in \Sec{sec:fixed_order} by studying the fixed order distributions for the observable $D_2$, there is no fixed order singularity in the soft haze region of phase space in the presence of a mass cut. This implies that the resummation is not needed to regulate a fixed order singularity. This will be discussed in \Sec{sec:soft_haze_match}. The field theoretic definitions of the functions appearing in the factorization theorem of  \Eq{eq:fact_soft_haze} as well as power expansions of the measurement operators are collected in \App{app:softhaze}. However, because of the fact that we do not explicitly use the results of the soft haze factorization theorem in our calculation, we simply refer the reader to \Refs{Larkoski:2014tva,Procura:2014cba} for the calculations of the one-loop functions relevant to the factorization theorems of \Eqs{eq:fact_soft_haze}{eq:fact_soft_haze2}, and leave for future work the full two-loop calculation.

\subsubsection{Refactorization of the Global Soft Function}\label{sec:refac_soft}

In each of the factorization theorems required for the description of QCD background jets, namely the collinear subjets, soft subjet, and soft haze factorization theorems, there is a global soft function, which is sensitive to both the in-jet measurement of the energy correlation functions,  as well as the out-of-jet measurement $B$. To ensure that all large logarithms are resummed by the renormalization group evolution, we must perform a refactorization of the soft function \cite{Fleming:2007xt,Ellis:2010rwa,Kelley:2011aa,Chien:2012ur,Jouttenus:2013hs}. This ensures that the only logarithms which appear in a given soft function that are sensitive to both in-jet and out-of-jet scales are true non-global logarithms (NGLs) \cite{Dasgupta:2001sh}, which first appear at two-loop order in the calculation of a particular soft function.\footnote{It is important to emphasize that throughout this section we refer to the NGLs which appear in the soft function of a given factorization theorem, and the order in $\alpha_s$ at which they will appear in this particular soft function. Because we combine distinct factorization theorems, some of which include hard splitting functions, or eikonal emission functions, this order is in general distinct from the order at which they will appear in the total cross section, which can be different for each factorization theorem. This combination of the factorization theorems is completely independent from the refactorization of the soft function in a particular factorization theorem.}  Here we focus on the refactorization of the soft subjet and collinear subjets factorization theorems of \Secs{sec:ninja}{sec:soft_jet}, which will be used in our numerical calculation. For both of these factorization theorems, we can write the soft function to all orders in $\alpha_s$ as
\begin{equation}
S\left(   \ecf{3}{\alpha}, B;R,\mu\right)=S^{(\text{out})}  \left( B;R,\mu  \vphantom{ \ecf{3}{\alpha}}  \right)S^{(\text{in})}\left(   \ecf{3}{\alpha};R,\mu\right)S_{\text{NGL}}\left(   \ecf{3}{\alpha}, B;R\right)\,,
\end{equation}
where we have explicitly indicated the renormalization scale $\mu$ dependence \cite{Hornig:2011tg}.  The non-global part of the soft function $S_{\text{NGL}}\left(   \ecf{3}{\alpha}, B;R\right)$ is first non-trivial at two-loop order, beyond the accuracy to which we explicitly calculated the soft functions in this chapter. Furthermore, the anomalous dimension of the soft function factorizes to all orders in perturbation theory as
\begin{align}
\gamma_{S}  \left(   \ecf{3}{\alpha}, B;R;\mu \right)=\gamma^{(\text{out})}_{S}  \left( B;R;\mu  \vphantom{ \ecf{3}{\alpha}} \right)   +\gamma^{(\text{in})}_{S}  \left(   \ecf{3}{\alpha};R;\mu \right)  \,,
\end{align}
and therefore the renormalization group kernels factorize as well. Briefly, this occurs because renormalization group consistency relates the soft anomalous dimension to the sum of all the other anomalous dimensions, each of which can be associated with the in-jet or out-of-jet contributions.\footnote{As discussed in \Ref{Jouttenus:2013hs} there is some ambiguity in how the hard function, for example, is associated with the in-jet or out-of-jet anomalous dimensions, but this does not affect the above argument.}

 While similar refactorizations of the global soft function have been discussed previously, and used in numerical calculations (see especially \Ref{Jouttenus:2013hs} for a detailed discussion), we will discuss it here for completeness.  The refactorization of the global soft function plays a role in our numerical results and is particularly important in appropriately separating scales in the global soft function of the soft subjet factorization theorem of \Sec{sec:soft_jet}.  In \Ref{Larkoski:2015zka} the structure of the one-loop calculation of the soft subjet factorization theorem was discussed in detail, with a particular focus on the dependence on the angle $\Delta \theta_{sj}$ between the soft subjet and the boundary. There it was found that the while the out-of-jet soft function contained dependence on the angle between the soft subjet and the boundary, $\Delta \theta_{sj}$, this dependence vanishes in the in-jet contribution to the soft function due to a zero bin subtraction. Renormalization group consistency is achieved since the $\Delta \theta_{sj}$ dependence associated with the in-jet region is carried by  the boundary soft function. Therefore, the refactorization of the global soft function for the soft subjet factorization theorem allows the soft function to be separated into a piece with $\Delta \theta_{sj}$ dependence, and a piece with no $\Delta \theta_{sj}$ dependence, and is crucial for resumming all large logarithms associated with this scale.  The one-loop anomalous dimensions, split into out-of-jet and in-jet contributions, as well as canonical scales for both the in-jet and out-of-jet soft functions are given in \App{sec:ninja_app}, \App{sec:softjet_app}, and \App{sec:soft_subjet_cbin}. Further details of this refactorization, and in particular a discussion on the dependence on $\Delta \theta_{sj}$ is also given.

For completeness, we also give the final refactorized expressions for the factorization theorems for the collinear subjets and soft subjet factorization theorems that will be used when presenting numerical results. For the collinear subjets factorization theorem, we have 
 \begin{align}\label{eq:NINJA_fact_refac}
\frac{d^3\sigma}{dz\,d\ecf{2}{\alpha}d\ecf{3}{\alpha}}&=\sum_{f,f_a,f_b}H_{n\bar{n}}^{f}(Q^2)     P_{n_t\rightarrow n_a,n_b}^{f\rightarrow f_a f_b}\Big(z;\ecf{2}{\alpha}\Big)    \int dB_S dB_{J_{\bar n}}        \int  de_{3}^{c}de_{3}^{\bar{c}}de_{3}^{s}de_{3}^{cs} \\
& \hspace{-0.5cm} \times     \delta (B-B_{J_{\bar n}}- B_S)    \delta\Big(\ecf{3}{\alpha}-e_{3}^c-e_{3}^{\bar{c}}-e_{3}^{s}-e_{3}^{cs}\Big)     \nonumber\\
& \hspace{-0.5cm} \times J_{\bar{n}}(B_{J_{\bar n}} )  J_{n_a}^{f_a}\Big(z;e_{3}^{c}\Big)J_{n_b}^{f_b}\Big(1-z;e_{3}^{\bar{c}}\Big)    S^{(\text{out})}_{n \bar{n}}\Big(B_S;R\Big)   S^{(\text{in})}_{n \bar{n}}\Big(e_3^{s};R\Big)    S_{ n_a n_b \bar{n}}^+\Big(e_{3}^{cs}\Big) \,, \nonumber
\end{align}
 while for the soft subjet factorization theorem, we have
 \begin{align}\label{fact_inclusive_form_2_refac}
\frac{d\sigma(\outj;R)}{d\sje\, d\sjtheta\, d\ecfres}&= H_{n\bar n}(Q^2) H^{sj}_{n\bar{n}}\left(z_{sj}, \theta_{sj}\right)   \int dB_S dB_{J_{\bar n}} \int de_3^{J_n} de_3^{J_{{sj}}} de_3^S  de_3^{S_{sj}}     \\
&\hspace{-0.5cm}  \times      \delta (B-B_{J_{\bar n}}- B_S)  \delta(\ecfres-e_3^{J_n}-e_3^{J_{{sj}}}-e_3^S-e_3^{S_{sj}}) \nonumber \\
& \hspace{-0.5cm}\times     J_{n}\left(e_3^{J_n}\right)J_{\bar{n}}(B_{J_{\bar n}}) 
 S^{(\text{out})}_{n\bar{n}\sja }\Big(B_{S};R\Big)     S^{(\text{in})}_{n\bar{n}\sja }\Big(e_3^S;R\Big)     J_{\sja}\left(e_3^{J_{sj}} \right) S_{\sja\sjabar}(e_3^{S_{sj}};R)\,. \nonumber
\end{align} 
In this form, each function in \Eqs{eq:NINJA_fact_refac}{fact_inclusive_form_2_refac} contains logarithms of a single scale, which can be resummed through renormalization group evolution.

\subsection{Boosted Boson Signal}\label{sec:signal_fact}

\begin{figure}
\begin{center}
\subfloat[]{\label{fig:boosted_boson_a}
\includegraphics[width=6.2cm]{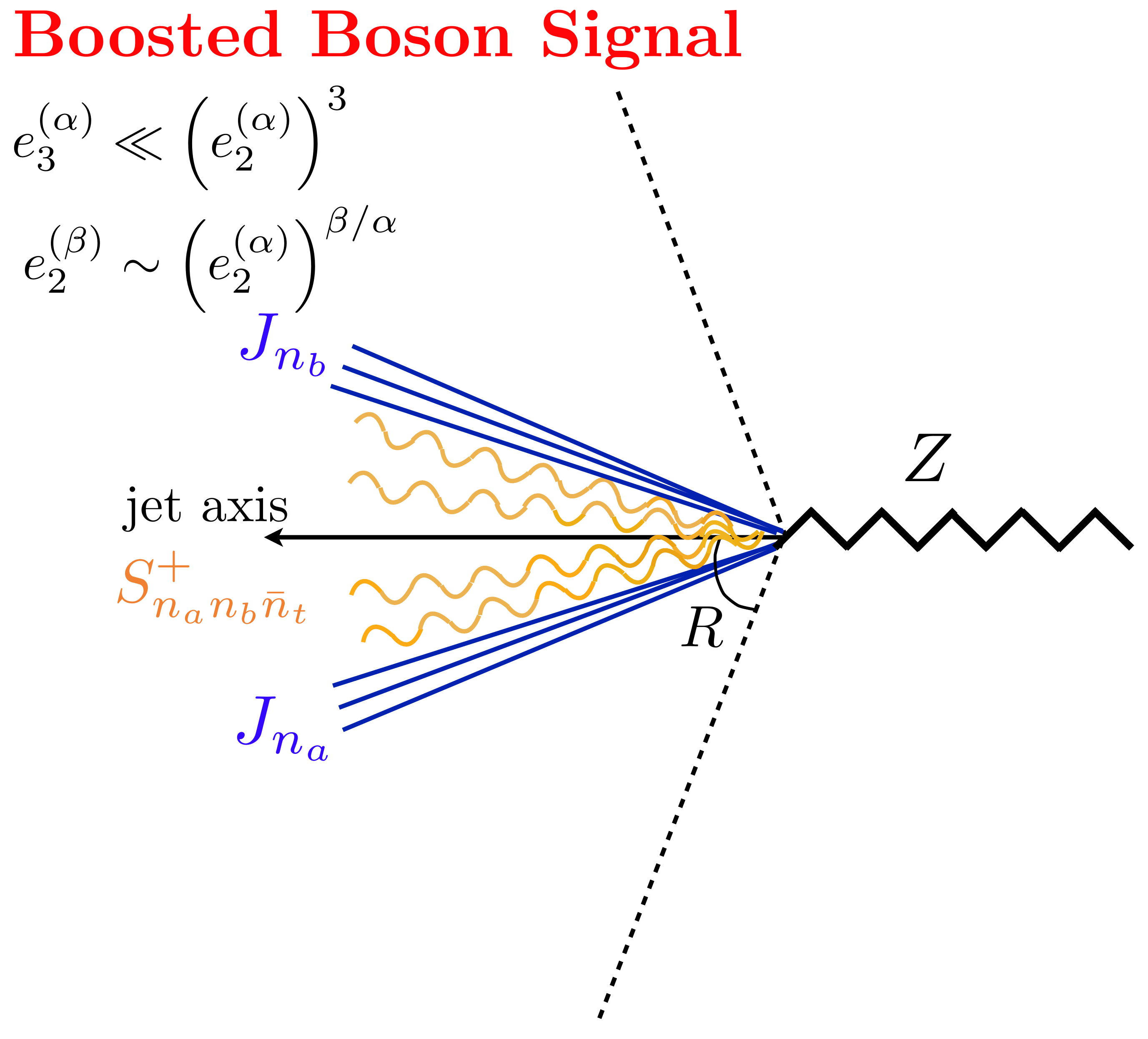}
}
\subfloat[]{\label{fig:boosted_boson_b}
\includegraphics[width = 8.5cm]{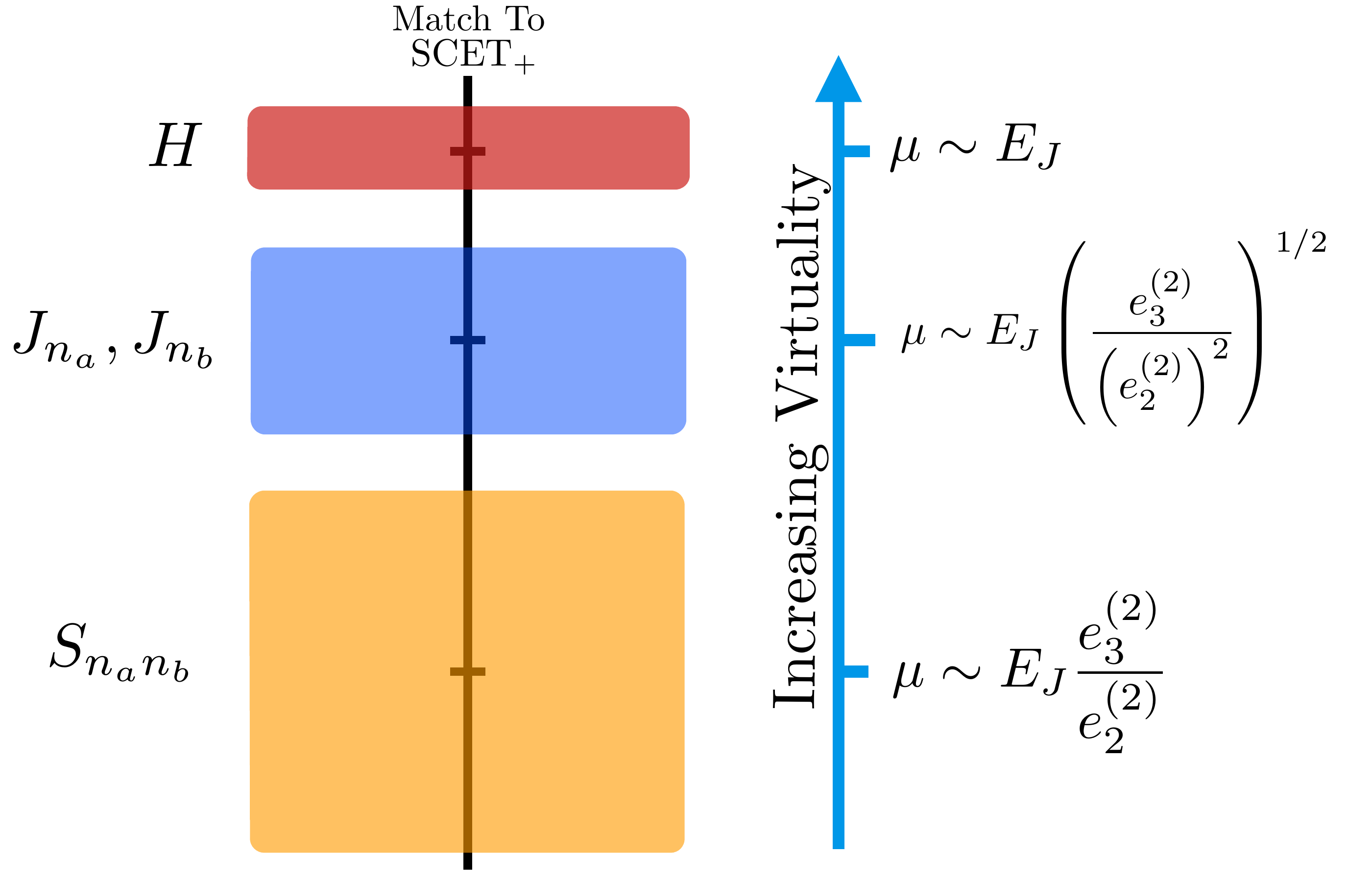}
}
\end{center}
\caption{A schematic depiction of the boosted $Z$ boson configuration with dominant QCD radiation and the functions describing its dynamics in the effective field theory is shown in a).  The relevant scales, ordered in virtuality, are summarized in b), where we have restricted to the case $\alpha=\beta =2$ for simplicity.
}
\label{fig:boosted_boson}
\end{figure}

In this section we discuss the effective field theory and factorization theorem relevant for the hadronically-decaying boosted boson signal. For concreteness, we will consider the case of a boosted $Z$ boson decaying to a massless $q\bar q$ pair; however, the extension to other color-neutral boosted particles is trivial. We will work in the narrow width approximation, setting the width of the $Z$ boson $\Gamma_Z=0$. Corrections to this approximation are trivial to implement, as they do not modify the structure of the factorization, and are expected to have a minimal effect.

A factorization theorem for the $N$-subjettiness observable $\Nsub{1,2}{\beta}$ \cite{Stewart:2010tn,Thaler:2010tr,Thaler:2011gf} measured on boosted $Z$ jets was presented in \Ref{Feige:2012vc}. This factorization theorem was obtained by boosting an appropriately chosen $e^+e^-$ event shape. A factorization theorem can also be formulated using the SCET$_+$ effective theory,\footnote{Here we have slightly extended the usage of the SCET$_+$ nomenclature beyond that which it was originally used in \Ref{Bauer:2011uc}. In particular, in the case of the signal distribution, there are no global soft modes, and the matching to the effective theory proceeds in quite a different way than for the case of a two prong QCD jet as originally considered in \Ref{Bauer:2011uc}. Nevertheless, because the effective theory contains a collinear-soft mode, we will refer to it as SCET$_+$. } where the collinear-soft mode, which was described in \Sec{sec:ninja}, corresponds to the boosted soft mode of the $e^+e^-$ event shape. We will take this second approach, as it is in line with the general spirit of this chapter, of developing effective field theory descriptions of jet substructure configurations. However, the approach of relating to boosted $e^+e^-$ event shape variables is useful for relating results to higher order calculations known in the literature. Despite the fact that the factorization for the energy correlation functions in the signal region follows straightforwardly from that of \Ref{Feige:2012vc}, or from the SCET$_+$ factorization theorem of \Sec{sec:ninja},  we will discuss it here for completeness.

We assume the process $e^+e^-\to ZZ \to q \bar q l\bar l$, where $l$ is a lepton to avoid having to describe additional jets, although the extension to two hadronically-decaying $Z$ bosons is trivial. The factorization theorem is then similar to that presented in \Sec{sec:ninja}, however, there are no global soft modes since the $Z$ is a color singlet. The scaling of the collinear and collinear-soft modes are identical to those given in \Sec{sec:ninja}, so we do not repeat them here.  The factorization theorem is given by
\begin{align}\label{eq:signal_fact}
\frac{d\sigma}{dz\,d\ecf{2}{\alpha}d\ecf{3}{\alpha}}&=H(Q^2) P_{n\rightarrow n_a,n_b}^{Z\rightarrow q \bar q}\Big(z;\ecf{2}{\alpha}\Big)\int  de_{3}^{c}de_{3}^{\bar{c}}de_{3}^{s}de_{3}^{cs} \\
& \hspace{-0.5cm} \times \delta\Big(\ecf{3}{\alpha}-e_{3}^c-e_{3}^{\bar{c}}-e_{3}^{cs}\Big) 
 J_{n_a}^{q}\Big(z;e_{3}^{c}\Big)J_{n_b}^{\bar q}\Big(1-z;e_{3}^{\bar{c}}\Big)S_{ n_a n_b }^+\Big(e_{3}^{cs}\Big) \,. \nonumber
\end{align}
As with the factorization theorem in \Sec{sec:ninja}, we have chosen to write the factorization theorem in terms of $\ecf{2}{\alpha}$, $\ecf{3}{\alpha}$, and the energy fraction of one of the subjets, $z$. A brief description of the functions appearing in \Eq{eq:signal_fact} is as follows:
 \begin{itemize}
 \item $H(Q^2)$ is the hard function describing the production of the on-shell $Z$ bosons in an $e^+e^-$ collision. It also includes the leptonic decay of the $Z$ boson. Following \Ref{Feige:2012vc} we assume that the $Z$ boson is unpolarized and so its decay matrix element is flat in the cosine of the boost angle. Non-flat distributions corresponding to some particular decay or production mechanism are straighforward to include.
 \item $P_{n\rightarrow n_a,n_b}^{Z\rightarrow q\bar q}\Big(z;\ecf{2}{\alpha}\Big)$ describes the decay of the on-shell $Z$ boson into a $q\bar q$ pair with momenta along the $n_a$ and $n_b$ axes.  
 \item  $J_{n_a}^{q}\Big(z;e_{3}^{c}\Big)$, $J_{n_b}^{\bar q}\Big(1-z;e_{3}^{\bar{c}}\Big)$ are the jet functions describing the collinear radiation associated with the two collinear subjets.
 \item  $S_{ n_a n_b }^+\Big(e_{3}^{cs}\Big)$ is the collinear-soft function describing the radiation from the $q\bar q $ dipole formed by the two collinear subjets.
 \end{itemize}
The basic structure of the factorization theorem, and the radiation described by the different functions, as well as their scalings, are shown schematically in \Fig{fig:boosted_boson}. Operator definitions, and one-loop calculations for the operators appearing in the factorization theorem of \Eq{eq:signal_fact}  are given in \App{sec:signal_app}. Because the collinear soft modes are boosted, the collinear soft function does not require a refactorization, as was necessary for the global soft functions, in \Sec{sec:refac_soft}.

It is important to emphasize the distinction between our treatment of a boosted $Z$ jet, where we presented a single factorization theorem, and a massive QCD jet, where three distinct factorization theorems were required. While it is obvious that the soft haze region does not exist for a boosted $Z$ jet, the soft subjet region does. However, unlike the case of a massive QCD jet, where the soft subjet region is enhanced by a factor of $1/z_{sj}$ from the eikonal emission factor, no such enhancement exists for the $Z$ decay. Indeed, it was shown in \Ref{Feige:2012vc} that the effect of the jet boundary, which would arise from the soft subjet configuration, is power suppressed by $1/Q$. While it would be potentially interesting to analytically study the jet radius dependence for the signal distribution using the soft subjet factorization theorem, this is beyond the scope of this chapter. We will therefore neglect jet radius effects and write the factorization theorem in \Eq{eq:signal_fact} with no $R$ dependence.  

The factorization theorem of \Eq{eq:signal_fact} provides an accurate description of the boosted boson signal in the two-prong region of phase space, where $\ecf{3}{\alpha} \ll \left ( \ecf{2}{\alpha} \right )^{3}$. However, to be able to compare the signal and background distributions, a valid description of the region $\ecf{3}{\alpha} \gtrsim \left ( \ecf{2}{\alpha} \right )^{3}$ is also required. Unlike for the case of a massive QCD jet, where this region is described by the soft haze factorization theorem, for a boosted $Z$ boson, an accurate description of this region requires matching to the fixed order $Z\to q\bar q g$ matrix element. Since the boost of the $Z$ boson is fixed, this corresponds to a hard gluon emission from the $q\bar q$ dipole. In the numerical results shown throughout the chapter, we have performed this matching to fixed order, directly within the SCET$_+$ effective theory. The fixed order cross section for $\Dobs{2}{\alpha,\beta}$ onto which the result of the factorization theorem was matched, was calculated numerically by boosting the leading order $e^+e^-\to q\bar q g$ matrix element and performing a Monte Carlo integration. This allows for the consideration of general angular exponents $\alpha$ and $\beta$ in which case the required integrals are difficult, if not impossible, to evaluate analytically.

\section{A Factorization Friendly Two-Prong Discriminant}\label{sec:friendly}

The approach to two-prong discrimination taken in this chapter is to use calculability and factorizability constraints to guide the construction of an observable. Having understood in detail the structure of the $\ecf{2}{\alpha}, \ecf{2}{\beta}, \ecf{3}{\alpha}$ phase space, along with the effective field theories describing each parametric region, we now show how a powerful two-prong discriminant, $D_2$, emerges from this analysis naturally.  After defining the $D_2$ observable, we discuss some of its interesting properties, and show that the factorization theorems of \Sec{sec:Fact} can be combined to give a factorized description of the observable over the entire phase space.

\subsection{Defining $D_2$}\label{sec:def_D2}

The goal of boosted boson discrimination is to define observables which distinguish between one- and two-prong jets. As a simplification, we will take the view that both collinear and soft subjets should be treated as two-pronged by the discriminant, while soft haze jets should be treated as one-pronged.  Treating both the collinear and soft subjets as two-pronged immediately implies that a marginalization over the soft subjet and collinear subjet factorization theorems will need to be performed to obtain a prediction for the two-prong discriminant. This will be discussed in \Sec{sec:merging}. 
A more sophisticated observable could take advantage of the different fraction of signal and QCD jets in the soft subjet and collinear subjets regions of phase space, and we will give a simple example of such an observable in \Sec{sec:2insight}. 

\begin{figure*}[t]
\begin{center}
\subfloat[]{\label{fig:D2_contours}
\includegraphics[width=6.5cm]{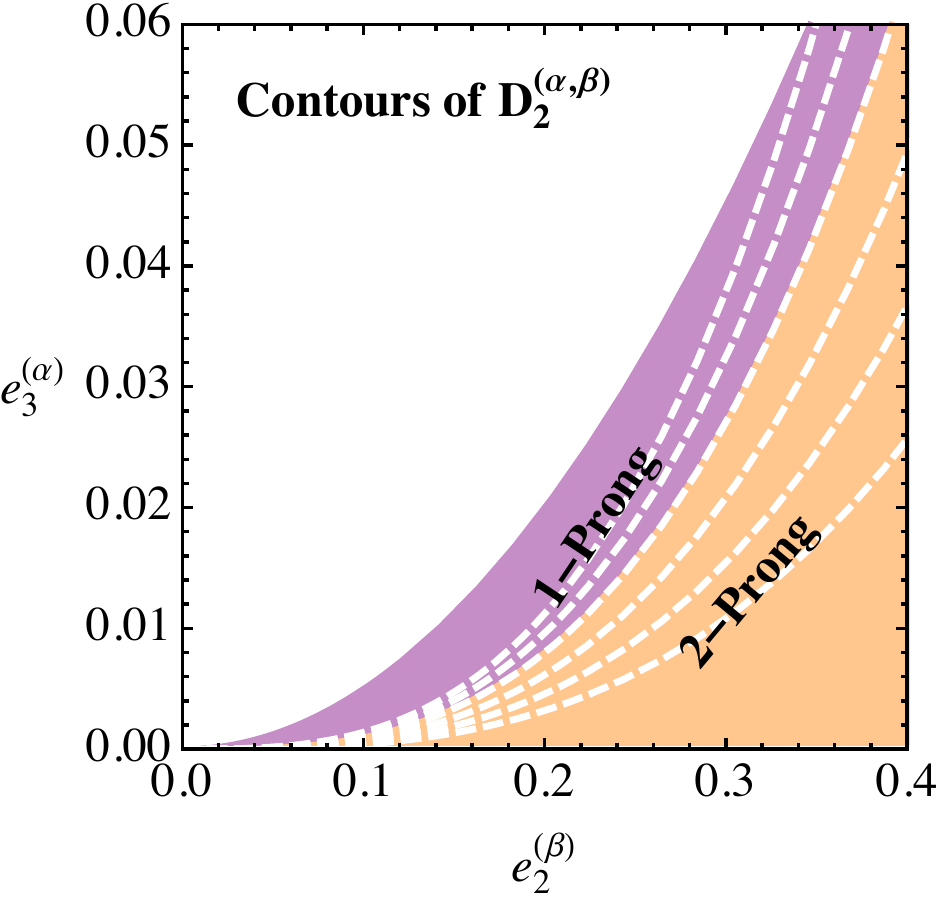}
}\ 
\subfloat[]{\label{fig:D2_plot}
\includegraphics[width=7.5cm,trim =0 0cm 0 0]{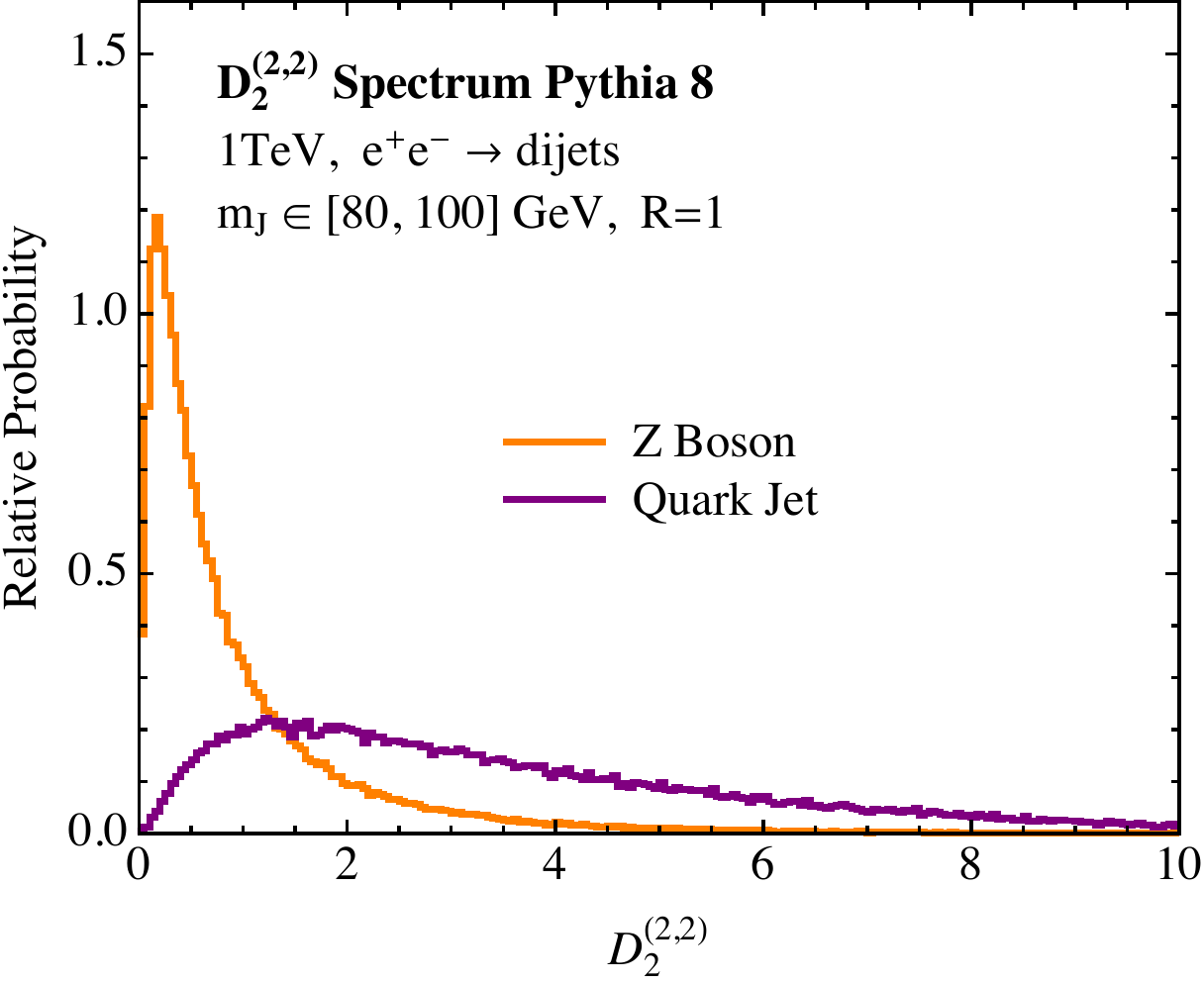}
}
\end{center}
\caption{ a) Contours of the observable $D_2$ in the $\ecf{2}{\beta},\ecf{3}{\alpha}$ plane. b) Sample $D_2$ spectra for boosted $Z$ bosons and QCD jets, generated in Monte Carlo. Angular exponents $\alpha=\beta =2$ have been used. 
}
\label{fig:C2vD2}
\end{figure*}

We will consider discriminants, which we denote $\Dobs{2}{\alpha, \beta}$, which parametrize a family of contours in the $\ecf{2}{\beta},\ecf{3}{\alpha}$ plane, as shown schematically in \Fig{fig:C2vD2}. Such observables can be calculated by marginalizing the double differential cross section \cite{Larkoski:2013paa}
\be \label{eq:marginalization}
\frac{d\sigma}{d\Dobs{2}{\alpha, \beta}}=\int d\ecf{2}{\beta} d\ecf{3}{\alpha} \delta \left(\Dobs{2}{\alpha, \beta}-\Dobs{2}{\alpha, \beta}(\ecf{2}{\beta},\ecf{3}{\alpha}) \right)   \frac{d^2 \sigma}{d\ecf{2}{\beta} d\ecf{3}{\alpha}}\,.
\ee
For the observable $\Dobs{2}{\alpha, \beta}$ to be calculable using the factorization theorems of \Sec{sec:Fact}, the curves over which the marginalization is performed in \Eq{eq:marginalization} must lie entirely in a region of phase space in which there is a description in terms of a single effective field theory (up to the marginalization over the collinear and soft subjets). Stated another way, the contours of $\Dobs{2}{\alpha, \beta}$ must lie either entirely in the one-prong region of phase space, or entirely in the two-prong region of phase space. This condition is also natural from the perspective that $\Dobs{2}{\alpha, \beta}$  provide good discrimination power, a point which has been emphasized in \Refs{Larkoski:2014gra,Larkoski:2014zma}. If the contours do not respect the parametric scalings of the phase space, the marginalization cannot be performed within a single effective field theory. A more sophisticated interpolation between the different effective field theories, along the lines of \Refs{Larkoski:2014tva,Procura:2014cba} is then required.

In \Sec{sec:phase_space}, a power counting analysis was used to show that for $3\alpha/\beta > 2$, the one- and two-prong regions of phase space are parametrically separated, with the contour separating them scaling as $\ecf{3}{\alpha} \sim \left(\ecf{2}{\beta}\right)^{3\alpha/\beta}$. This implies that, parametrically, the optimal two-prong discriminant formed from $\ecf{2}{\beta}$ and $\ecf{3}{\alpha}$ is
\begin{equation}\label{eq:D2_gen}
\Dobs{2}{\alpha, \beta}= \frac{\ecf{3}{\alpha}}{(\ecf{2}{\beta})^{3\alpha/\beta}}\,.
\end{equation}
This extends the definition of \Ref{Larkoski:2014gra}, which considered the observable $\Dobs{2}{\alpha, \alpha}$, with equal angular exponents.
To simplify our notation, we will often not explicitly write the angular exponents $\alpha$ and $\beta$, referring to the observable simply as $D_2$.

The $D_2$ observable takes small values for a two-prong jet and large values for a one-prong jet.  Its contours in the $\ecf{2}{\beta},\ecf{3}{\alpha}$ phase space are shown schematically in \Fig{fig:C2vD2}, along with illustrative Monte Carlo generated spectra for both boosted $Z$ jets and massive QCD jets in $e^+e^-$ collisions. A more detailed discussion of the discrimination power of $D_2$, as well as the details of the Monte Carlo generation, will be given in \Sec{sec:results}.

\subsection{Sudakov Safety of $D_2$}\label{sec:sudsafe}

One interesting feature of the $D_2$ observable is that it is not IRC safe without an explicit cut on $\ecf{2}{\beta}$. For every value of $D_2$, the contour over which the double differential cross section is marginalized passes through the origin of the phase space, where the soft and collinear singularities are located. This feature is shown in \Fig{fig:D2_contours}. At every fixed order in perturbation theory, this gives rise to an ill-defined (divergent) cross section. However, a resummed calculation of the double differential cross section regularizes the singular region of phase space, and leads to a finite distribution for the $D_2$ observable. This property is referred to as Sudakov safety \cite{Larkoski:2013paa,Larkoski:2015lea}.  Because Sudakov safe observables are not calculable in fixed order perturbation theory, they do not generically have an $\alpha_s$ expansion, and we will show that the $D_2$ spectrum exhibits a particularly interesting dependence on $\alpha_s$.

The regularization of the fixed order singularity in the double differential cross section is achieved by the all orders resummation of logarithmically enhanced terms in the perturbative expansion. In the effective field theory description, this resummation is achieved by renormalization group evolution, and its properties are therefore determined by the form of the SCET anomalous dimensions. To illustrate how the $\alpha_s$ dependence arises from the structure of the renormalization group evolution in SCET, we consider the soft subjet factorization theorem of \Sec{sec:soft_jet} in the leading logarithmic (LL) approximation. The cusp pieces of the anomalous dimensions for the different functions appearing in the factorization are given in Laplace space by (see \App{sec:softjet_app})  
\begin{align}
\mu\frac{d}{d\mu}\log\, H^{sj}_{n\bar{n}}(\sje,\sja,\mu) &=-\frac{\alpha_s C_A}{\pi}\log \left[\frac{\mu^2}{Q^2} z_{sj}^{-2}\right]\,,\\
\mu\frac{d}{d\mu}\log\, J_{\sja }\Big(\eeclp{3}{\alpha}\Big) &=-2\frac{\alpha_s C_A}{\pi(1-\alpha)}\log \left[\eeclp{3}{\alpha}\frac{\mu^\alpha}{Q^\alpha}  z_{sj}^{2-\alpha}     \right] \,,\\
\mu\frac{d}{d\mu}\log\,  S_{\sja \,\sjabar }\Big(\eeclp{3}{\alpha};R\Big)&=\frac{\alpha_s C_A}{\pi(1-\alpha)}\log \left[\eeclp{3}{\alpha}\frac{\mu}{Q} z_{sj}\right]\,,\\
\mu\frac{d}{d\mu}\log\,  S_{\sja \,n\,\bar{n}}\Big(\eeclp{3}{\alpha},B;R\Big)&=\frac{\alpha_s C_A}{\pi(1-\alpha)}\log \left[\eeclp{3}{\alpha}\frac{\mu}{Q} z_{sj}\right] \,,
\end{align}
where we have used $\eeclp{3}{\alpha}$ to denote the Laplace conjugate to $\ecf{3}{\alpha}$, and we have kept only IR scales in the logs. Furthermore, we have kept only the terms proportional to $C_A$ so as to resum only the physics associated with the soft subjet.
The hard matching coefficient for the soft subjet production is given by the tree level eikonal emission factor
\begin{align}
H^{sj(\text{tree})}_{n\bar{n}}(\sje,\sja)&=\frac{\alpha_s C_F}{\pi\sje}\frac{n\cdot \bar{n}}{n\cdot\sja\,\sja\cdot\bar{n}}\,.
\end{align}
Solving the renormalization group equations, and running all functions to the hard scale $Q$, we then find that in the soft subjet region of phase space the multi-differential cross section can be written to LL accuracy as
\begin{align}\label{eq:LL_sudakov}
&\frac{d\sigma}{d\ecf{3}{\alpha}    dz_{sj} d\theta_{sj}  }= - \frac{\alpha_s^2 C_F C_A}{\alpha\pi^2}\frac{4}{n\cdot\sja\,\sja\cdot\bar{n}}\frac{\log \left[\ecf{3}{\alpha}z_{sj}^{-2}\right]}{\sje\ecf{3}{\alpha}} e^{-\frac{\alpha_s}{\pi}\frac{C_A}{\alpha}\log^2\left[
\ecf{3}{\alpha}z_{sj}^{-2}
\right]}\,,
\end{align}
exhibiting a familiar Sudakov form.

A complete calculation of the $D_2$ spectrum requires marginalizing over both the soft subjet and collinear subjet configurations, which we discuss in \Sec{sec:merging}. However, to demonstrate the $\alpha_s$ behavior in the simplest manner, we will consider just the soft subjet effective theory. In particular, we will fix the angle of the soft subjet, but allow it to be arbitrarily soft, so as to probe the singular region of phase space. The result is then representative of the contribution from the soft subjet region of phase space. An exactly analogous behavior occurs for the contribution from the collinear subjets region of phase space.

Fixing $\theta_{sj}$ to satisfy $n \cdot n_{sj}=1/2$ (and therefore $\bar n\cdot n_{sj}=3/2$), and restricting to $\alpha=\beta$ for simplicity, the 2-point energy correlation function in the soft subjet region of phase space is simply
\begin{equation}
\ecf{2}{\alpha} = z_{sj} \,.
\end{equation}
The corresponding $D_2$ distribution is then obtained by marginalizing the multi-differential cross section of \Eq{eq:LL_sudakov}
\begin{align}
\frac{d\sigma}{dD_2}&=\int dz_{sj}\, d\theta_{sj}\, d\ecf{3}{\alpha}\, \delta \left ( D_2 -\frac{ \ecf{3}{\alpha}  }{(\ecf{2}{\alpha})^3   }   \right )    \frac{d\sigma}{d\ecf{3}{\alpha}    dz_{sj} d\theta_{sj}  }\\
&
\hspace{1cm}
\to\int dz_{sj} \,d\ecf{3}{\alpha}\, \delta \left ( D_2 -\frac{ \ecf{3}{\alpha}  }{z_{sj}^3   }   \right )    \frac{d\sigma}{d\ecf{3}{\alpha}    dz_{sj} d\theta_{sj}  }  \nonumber \,,
\end{align}
where, in the second line, we have fixed $\theta_{sj}$ and so we do not integrate over it.  Inserting the multi-differential cross section and fixing $\theta_{sj}$, we then have
\begin{align}\label{eq:sudsaferes}
\frac{d\sigma^{sj}}{dD_2}&=-\frac{16}{3}\frac{\alpha_s^2 C_F C_A}{\alpha\pi^2}\int_0^{1} dz_{sj}   \frac{\log \left[D_2z_{sj}\right]}{D_2\sje } e^{-\frac{\alpha_s}{\pi}\frac{C_A}{\alpha}\log^2\left[
D_2 z_{sj}
\right]}\\
&=\frac{8}{3}\frac{\alpha_s C_F}{\pi}\frac{e^{-\frac{\alpha_s}{\pi}\frac{C_A}{\alpha} \log^2 D_2}}{D_2} \,, \nonumber
\end{align}
where the $sj$ superscript denotes that this is representative of a contribution from the soft subjet region of phase space.  Importantly, because the soft subjet is defined by requirements on IRC safe measurements, the cross section in \Eq{eq:sudsaferes} is a well-defined and in principle measurable quantity.

The $\alpha_s$ dependence in this distribution of $D_2$ is very surprising.  Because $D_2$ is defined with respect to the 3-point energy correlation function, one would na\"ively expect that $D_2$ only makes sense for a jet with at least three partons.  Indeed, if we make an explicit cut on $z_{sj}$, for example, then $D_2$ is IRC safe, and first non-zero for a jet with three partons at ${\cal O}(\alpha_s^2)$.  However, because $D_2$ without a cut on $z_{sj}$ is not IRC safe, this intuition fails, and in a fascinating way.  By resumming the large logarithms of $z_{sj}$ to all orders and then marginalizing, the $D_2$ distribution calculated in \Eq{eq:sudsaferes} actually starts at ${\cal O}(\alpha_s)$!  Including emissions to all orders has effectively generated a non-trivial distribution for $D_2$ at one order {\it lower} in $\alpha_s$ than when it is first, na\"ively, non-zero.  Other examples of Sudakov safe observables in the literature have expansions in $\sqrt{\alpha_s}$ \cite{Larkoski:2013paa,Larkoski:2015lea} or are even independent of $\alpha_s$ \cite{Larkoski:2014wba,Larkoski:2014bia,Larkoski:2015lea}.  To our knowledge, $D_2$ is the first example of a Sudakov safe observable for which all-orders resummation reduces the order in $\alpha_s$ when the observable's distribution is first non-zero.\footnote{For observables that do not have universal behavior in the ultraviolet \cite{Larkoski:2015lea}.}  We re-emphasize that though the distribution of $D_2$ in \Eq{eq:sudsaferes} is a Taylor series in $\alpha_s$, it is impossible in purely fixed-order perturbation theory to systematically calculate it.

\subsection{Fixed-Order $D_2$ Distributions with a Mass Cut}\label{sec:fixed_order}

\begin{figure}
\begin{center}
\subfloat[]{\label{fig:fixed_order_a}
\includegraphics[width=6.8cm]{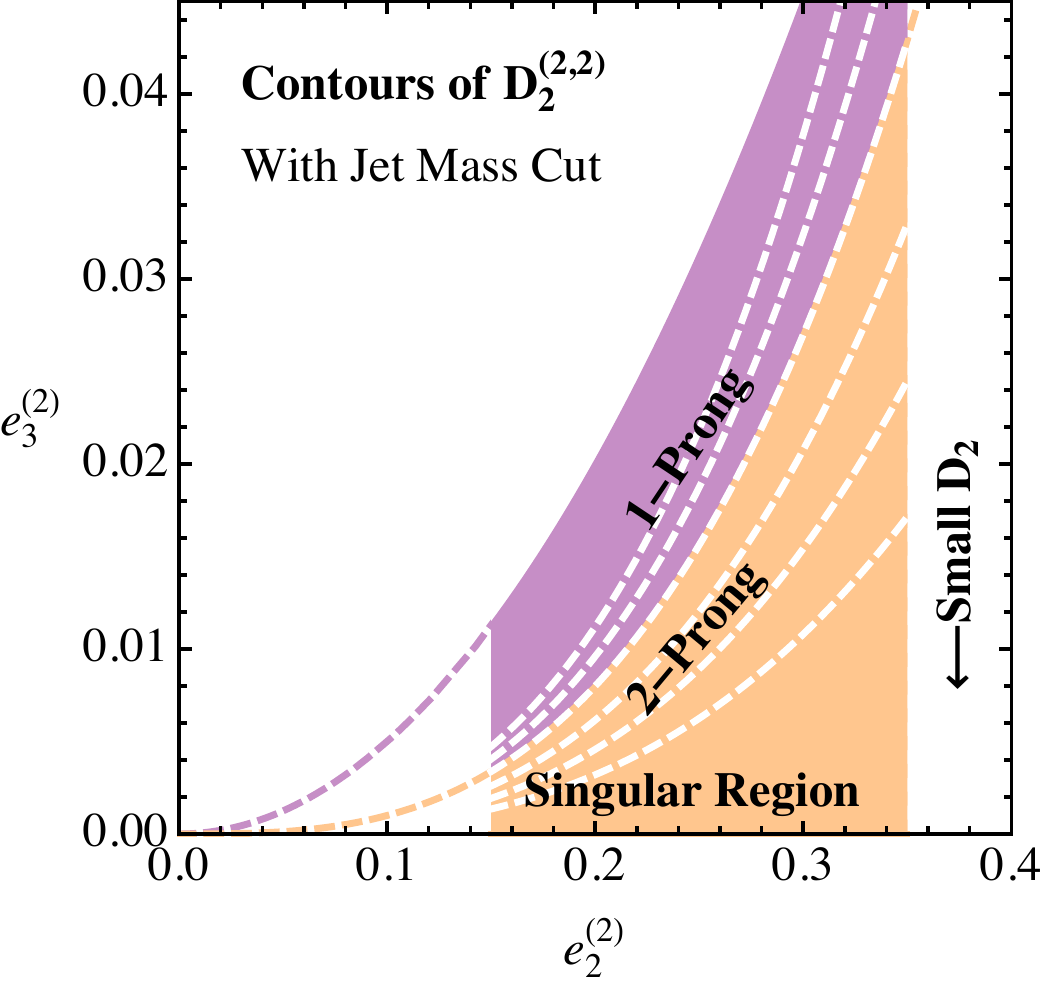}
}\ 
\subfloat[]{\label{fig:fixed_order_b}
\includegraphics[width=7.15cm, trim =0 0cm 0 0]{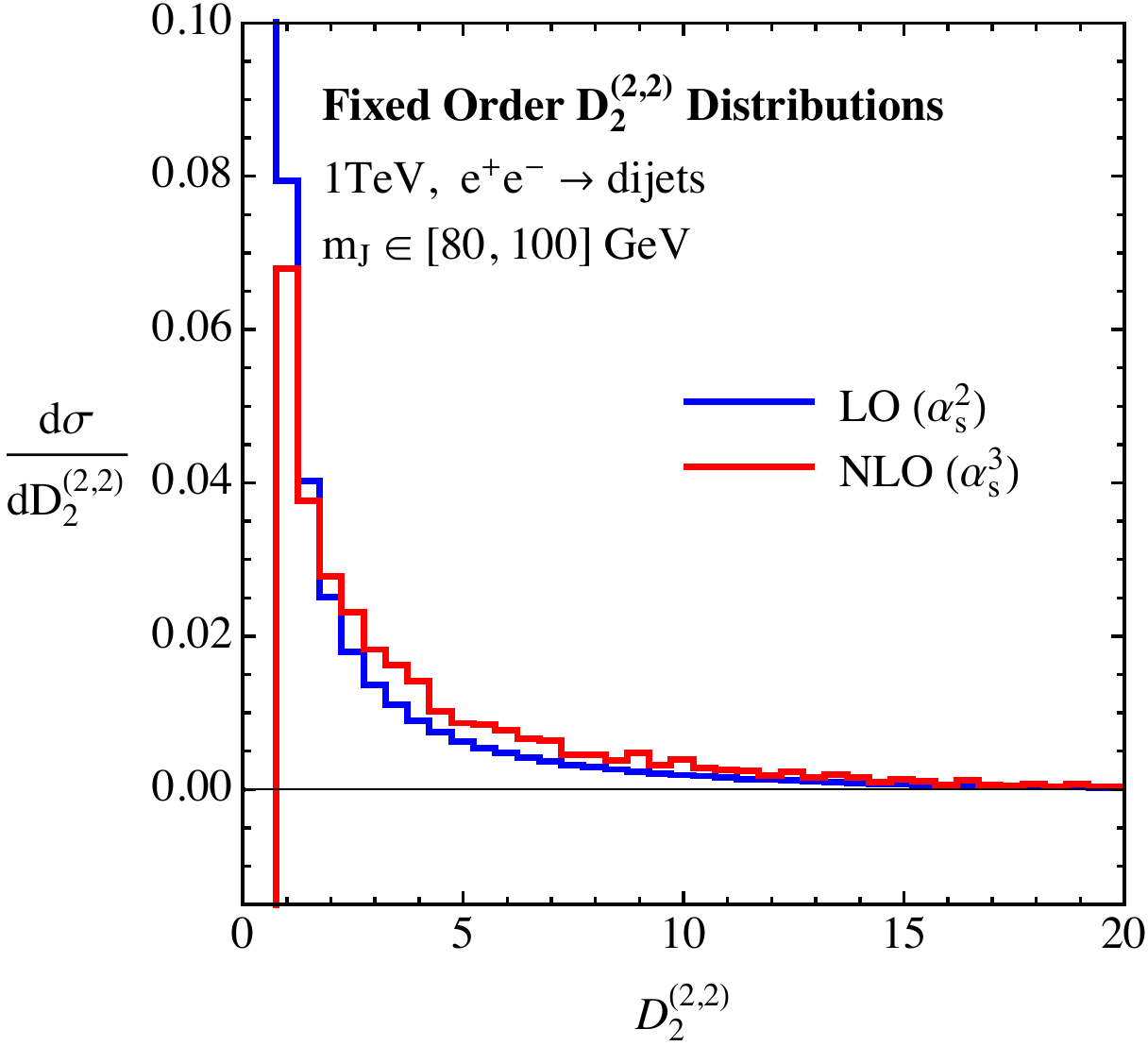}
}
\end{center}
\caption{ a) A schematic depiction of the $\ecf{2}{2}, \ecf{3}{2}$ phase space in the presence of a mass cut, along with contours of the $D_2$ observable.  b) Leading order (through $\alpha_s^2$) and next-to-leading order (through $\alpha_s^3$) distributions for the $D_2$ observable in the presence of a mass cut as measured on hemisphere jets in $e^+e^-$ collisions. 
}
\label{fig:fixed_order}
\end{figure}

Although $D_2$ is not IRC safe without a cut on $\ecf{2}{\beta}$, leading to its interesting Sudakov safe behavior, in experimental analyses a jet mass cut will be always be applied. We will therefore be most interested in this case. In \Fig{fig:fixed_order_a} we show a schematic depiction of the $\ecf{2}{\alpha}, \ecf{3}{\beta}$ phase space in the presence of a mass cut for $\alpha=\beta =2$, along with contours of the $D_2$ observable. As is indicated in the figure, the mass cut removes the origin of the phase space, making $D_2$ IRC safe and calculable in fixed-order perturbation theory. It is therefore interesting to study the singularity structure of the fixed-order perturbative expansion of $D_2$ in the presence of a mass cut.

In \Fig{fig:fixed_order_b} we show both the leading order $(\alpha_s^2)$ (LO) and the next-to-leading order $(\alpha_s^3)$ (NLO)  fixed-order distributions of the $D_2^{(2,2)}$ observable as measured on the most energetic hemisphere jet in $e^+e^- \to $ dijets events at $1$ TeV center of mass energy, and with a jet mass cut of $m_J\in [80,100]$ GeV, in anticipation of our application to boosted $Z$ boson discrimination. However, the detailed range of the mass cut window is irrelevant to the arguments of this section. \nlojet~\cite{Nagy:1997yn,Nagy:1998bb,Nagy:2001fj,Nagy:2001xb,Nagy:2003tz} was used to generate the distributions. The fixed-order $D_2$ distribution diverges at small values, and its sign in this region flips order-by-order, characteristic of the Sudakov region. This behavior makes clear the necessity of resummation in the small $D_2$ region. However, importantly, there is no divergence or other structure at large values of $D_2$. Instead, the distribution exhibits a tail extending to large values both at LO and NLO, and this behavior is expected to persist to higher orders.  This long tail arises from the fact that the upper boundary of the phase space is parametrically far, of distance $\sim1/\ecf{2}{\alpha}$, from the two-prong region of phase space. A schematic depiction of the singularity structure in the $\ecf{2}{2}, \ecf{3}{2}$ phase space is shown in \Fig{fig:fixed_order_a}.  The observation that a fixed-order singularity exists only at small values of $D_2$ is important for the resummation of the observable in the presence of a mass cut. In particular, while resummation in the soft subjet and collinear subjet factorization theorems are necessary to regulate a fixed-order singularity, the soft haze factorization theorem presented in \Sec{sec:soft_haze} is not.

The fixed-order behavior of the $D_2$ observable is in some ways much more similar to that of a traditional jet or event shape than might na\"ively be expected. However, there are some important differences. In particular, a mass cut of $80 < m_J <100 $ GeV has been applied, which is comparable to the location of the Sudakov peak in the mass for a jet of energy $500$ GeV. Therefore, unlike in the case of a traditional jet shape, where there is a transition from a region where resummation is important to a far tail region where a fixed order calculation provides an accurate description, in this case, for all values of $D_2$, there is an overall Sudakov suppression due to the mass cut, in addition to the divergence at small values of $D_2$. This is however, a small effect in the fixed order distribution compared to the divergence at smaller values, and most importantly, does not require regularization, as it is regulated by the mass cut.

\subsection{Merging Factorization Theorems}\label{sec:merging}

A complete description of the $D_2$ observable for background jets requires combining the three factorization theorems presented in \Sec{sec:Fact}. This involves both the merging of the soft subjet and collinear subjets factorization theorems, which must be performed before the marginalization over the $D_2$ contours, as well as the matching between the small $D_2$ description of the resolved two-prong region and large $D_2$ description of the unresolved region.  We will discuss how the matching is accomplished for these two cases in turn.

\subsubsection{Merging Soft and Collinear Subjets}\label{sec:soft_NINJA_match}

The region of phase space in which two subjets are resolved by the measurement is described by two distinct factorization theorems.  These two regions of phase space are separated by the measurement of the two 2-point energy correlation functions, $\ecf{2}{\alpha},\ecf{2}{\beta}$. However, in the calculation of $D_2$, both regions are treated as two-pronged, and the additional 2-point energy correlation function must be marginalized over. Since each effective theory can only be used within its regime of validity, a merged description, valid in both the soft subjets and collinear subjets region of phase space, is required. To accomplish this, we introduce a novel procedure for merging the two factorization theorems.

At a fixed $\ecf{3}{\alpha}$, the soft subjet and collinear subjets fill out the $\ecf{2}{\alpha}, \ecf{2}{\beta}$ phase space, which was shown in \Fig{fig:2ptps}. This phase space has also been studied in the context of two angularities measured on a single jet in \Refs{Larkoski:2014tva,Procura:2014cba}. In this case factorization theorems involving only collinear and soft modes exist on the boundaries of phase space, and an additional collinear-soft mode is required in the bulk of phase space. New logarithms exist in the bulk of the phase space, so called $k_T$ logarithms \cite{Larkoski:2014tva}, which can either be captured by the additional collinear-soft mode proposed in \Ref{Procura:2014cba}, or by the interpolation procedure of \Ref{Larkoski:2014tva}. In this case, the factorization theorems involving only the collinear and soft modes do not extend beyond the boundaries of the phase space, and they cannot be directly matched onto one another, as this would neglect the resummation of the $k_T$ logarithms, which are not present in either factorization theorem. We will now argue that the case of interest in this chapter, namely of two resolved subjets, is different. In particular, the soft subjet and collinear subjets factorization theorems extend from the boundaries of phase space, and already contain all the modes required for a description in the bulk of the phase space. In particular no additional modes exist in the bulk region of the phase space. This implies in particular that a description of the entire phase space region can be obtained by a proper merging of the collinear subjets and soft subjet factorization theorems, which is the approach that we will take.

\begin{figure}
\begin{center}
\subfloat[]{\label{fig:soft_collinear_match_a}
\includegraphics[width= 6.25cm]{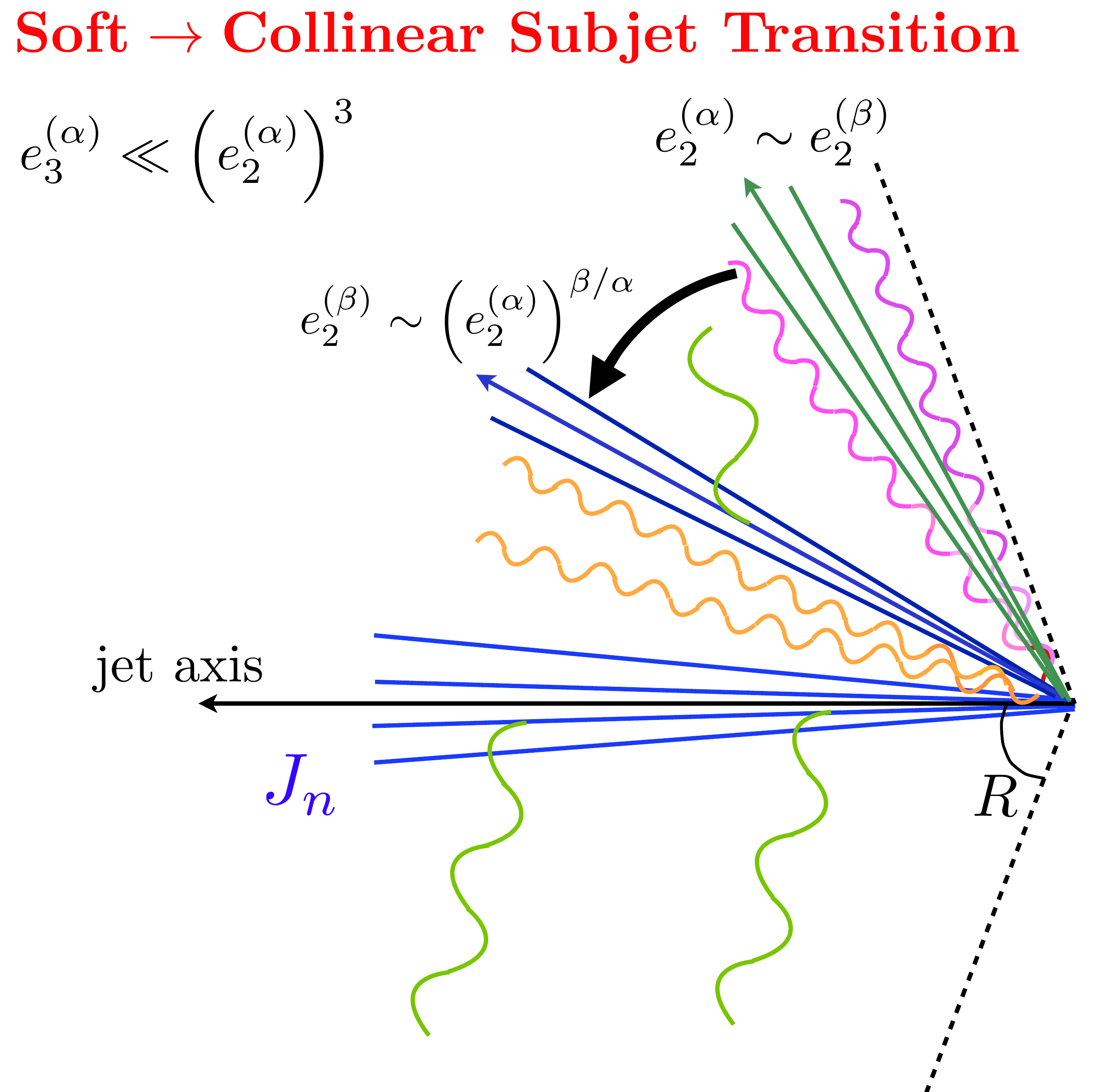}
}
\ 
\subfloat[]{\label{fig:soft_collinear_match_b}
\includegraphics[width = 7.3cm]{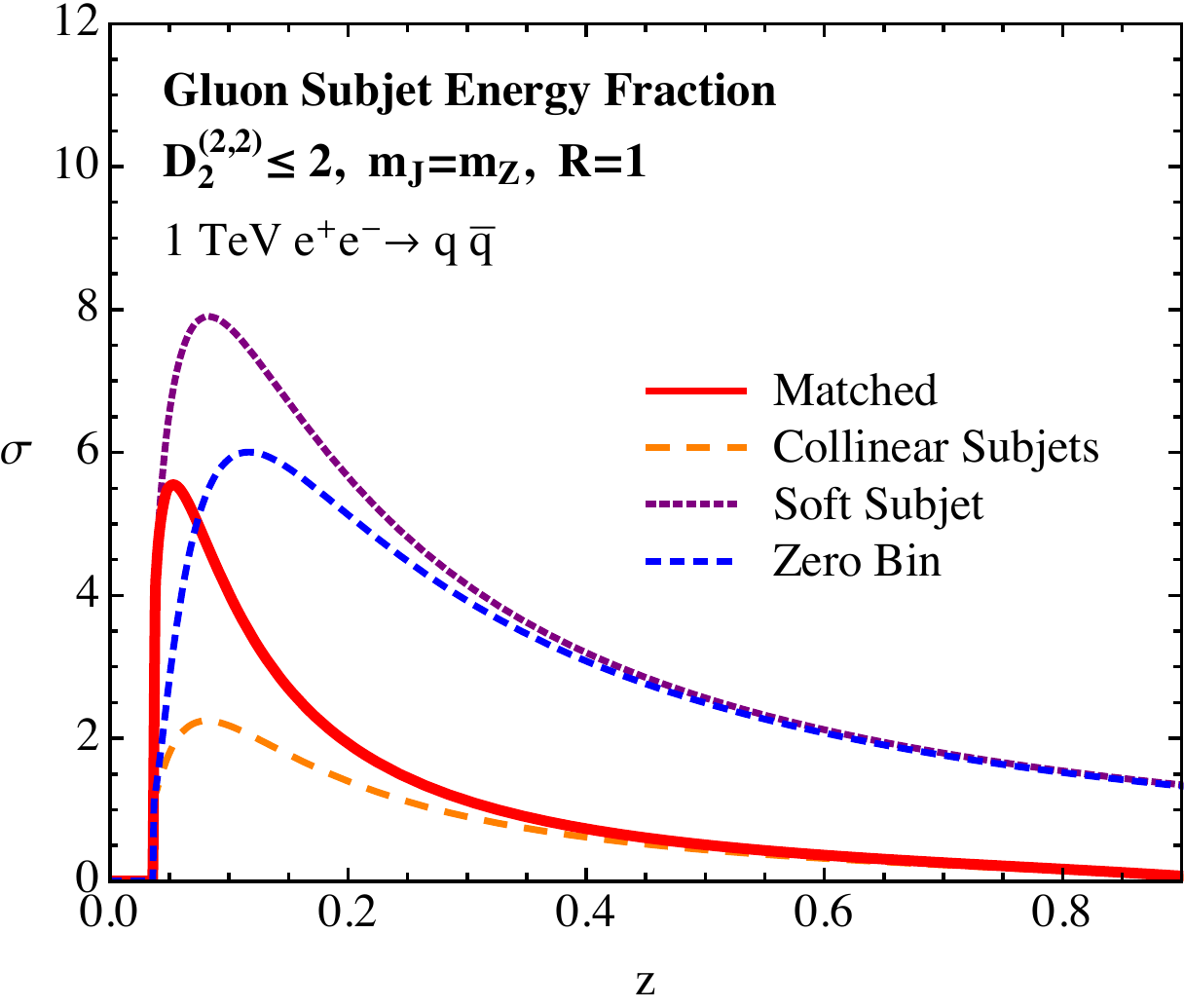}
}
\end{center}
\vspace{-0.2cm}
\caption{ a) A schematic depiction of the transition between the soft subjet and collinear subjets regions of phase space.   b) Distribution of the energy fraction of the gluon subjet as predicted by the collinear subjets effective theory, the soft subjet effective theory, and the merged description. The collinear zero bin of the soft subjet is also shown.
}
\label{fig:soft_collinear_match}
\end{figure}

To see that no additional modes are present in the bulk of the phase space, it is sufficient to look for modes which transition between the modes present in the effective theory descriptions in the soft subjet and collinear subjets regions of phase space, and which contribute at leading power. When transitioning from the collinear subjets region of phase space to the soft subjet region of  phase space, as is shown schematically in \Fig{fig:soft_collinear_match_a}, the collinear modes of one of the jets become the soft subjet and boundary soft modes of the soft subjet factorization theorem. On the other hand, the collinear-soft modes transition to the global soft modes. However, one could possibly be concerned that there exist additional modes which appear as collinear-soft modes on the boundary of phase space where the collinear subjets exist, but which transition to soft subjet modes instead of global soft modes. However, one can immediately see that such modes cannot exist, since the energy fraction of the soft subjet modes is set by the $\ecfnobeta{2}$ measurement, while the energy fraction of the collinear-soft modes is set by the $\ecfnobeta{3}$ measurement. Since $e_3$ is fixed, and the transition is occurring only in the $\ecf{2}{\alpha}, \ecf{2}{\beta}$ phase space, such modes cannot exist. This implies that all contributing modes already exist in either the soft subjet, or collinear subjets factorization theorems. This is a crucial difference from the case of the double differential angularities, which in some sense simplifies the analysis. Since no additional modes exist in the bulk of the phase space, the factorization theorems can be extended from the boundaries, and can be matched onto each other. This will allow for the resummation of all large logarithms. We will now discuss in more detail our implementation of this matching, after which we will see that our argument, presented here based on power counting, for the absence of additional modes, is explicitly realized through our merging procedure.

This suggests then the procedure we will use for interpolating between the collinear subjets and soft subjet factorization theorem, as sketched in \Ref{Larkoski:2015zka}, where the soft subjet factorization theorem was originally introduced. It proceeds by implementing a zero bin subtraction \cite{Manohar:2006nz} in factorization theorem space (the meaning of this will become clear shortly) to remove double counting in the overlapping region between the effective theories. This is a non-trivial and novel example of the zero bin procedure, and demonstrates the general utility of its approach. 

Recall that in a standard SCET factorization, the cross section is written as a convolution of a jet function, which describes the collinear physics, and a soft function, which describes the soft physics. To achieve this mode separation without introducing a double counting, the soft limit of the jet function must be subtracted, which is referred to in the literature as a zero bin subtraction. Here we extend this approach to the case of two distinct factorization theorems  which describe different regions of a multi-differential phase space, the soft subjet and collinear subjets effective field theories, but which overlap in the bulk of the two-prong phase space. It is important that here we only focus on the two-prong region of phase space; the matching to the one-prong region of phase space will be discussed in \Sec{sec:soft_haze_match}.  To perform the matching in the two-prong region of phase space, inspired by the zero-bin procedure,  we will write the cross section as a sum of the contributions from the soft subjet factorization theorem and the collinear subjets factorization theorem, with a zero bin contribution to remove the overlap between the effective theories. Explicitly, we write
\begin{align}\label{eq:matched}
\sigma=\left( \sigma_{sj}-\sigma_{sj}|_{cs} \right)+\sigma_{cs}\,,
\end{align}
where we have suppressed that at this stage the cross section is still differential in the kinematics of the subjets, so that our notation is not overly cumbersome.  The cross section in the soft subjet or collinear subjets regions of phase space are denoted by $sj$ and $cs$ subscripts, respectively.
Here the zero bin contribution, which removes the double counting, is given by $\sigma_{sj}|_{cs}$. Explicitly, $\sigma_{sj}|_{cs}$ is obtained by taking the limit of the soft subjet factorization theorem in the power counting of the collinear subjets factorization theorem. The anomalous dimensions and one-loop matrix elements for the collinear zero bin of the soft subjet factorization theorem are given in \App{sec:soft_subjet_cbin}. Each of the three contributions to the cross section given in \Eq{eq:matched} are associated with their own factorization theorem. However, the contributions to the cross section with the clearest physical interpretation are $\sigma_{cs}$ and the combined term $\left( \sigma_{sj}-\sigma_{sj}|_{cs} \right)$, which we will refer to the as the zero bin subtracted soft subjet contribution. It is the contribution which can be interpreted over the entire phase space as the contribution from a soft subjet, and all logarithms contained in this expression are of soft scales.

We specifically subtract the collinear-bin of the soft subjet factorization, and not the soft-bin of the collinear factorization. This is due to the need to cancel the contributions from the boundary soft modes of the soft subjet factorization in the collinear region. Since no analogous mode to the boundary softs is found in the collinear resummation, any soft expansion would miss this contribution, resulting in a logarithm being resummed in an inappropriate \emph{collinear} region of phase space. This is in contrast to what happens when comparing the two subtractions \emph{in the soft region}. So long as one uses the relative transverse momentum of the subjets as the splitting scale of the collinear factorization, the collinear-bin of the soft subjet does match the soft-bin of the collinear factorization in the soft region. This is the result of the merging of various soft scales. In the soft jet collinear-bin, the expanded boundary softs and global soft scales naturally merge, and in the soft-bin of the collinear jets, the global softs and collinear-softs also naturally merge in the soft region. This can be explicitly verified with the canonical scales given in \App{sec:canonical_merging_scales}. Thus the collinear-bin of the soft subjet is the appropriate subtraction throughout phase space, to remove double counting at all points.

\begin{figure}
\begin{center}
\subfloat[]{\label{fig:soft_collinear_match_a_pt6}
\includegraphics[width= 7.0cm]{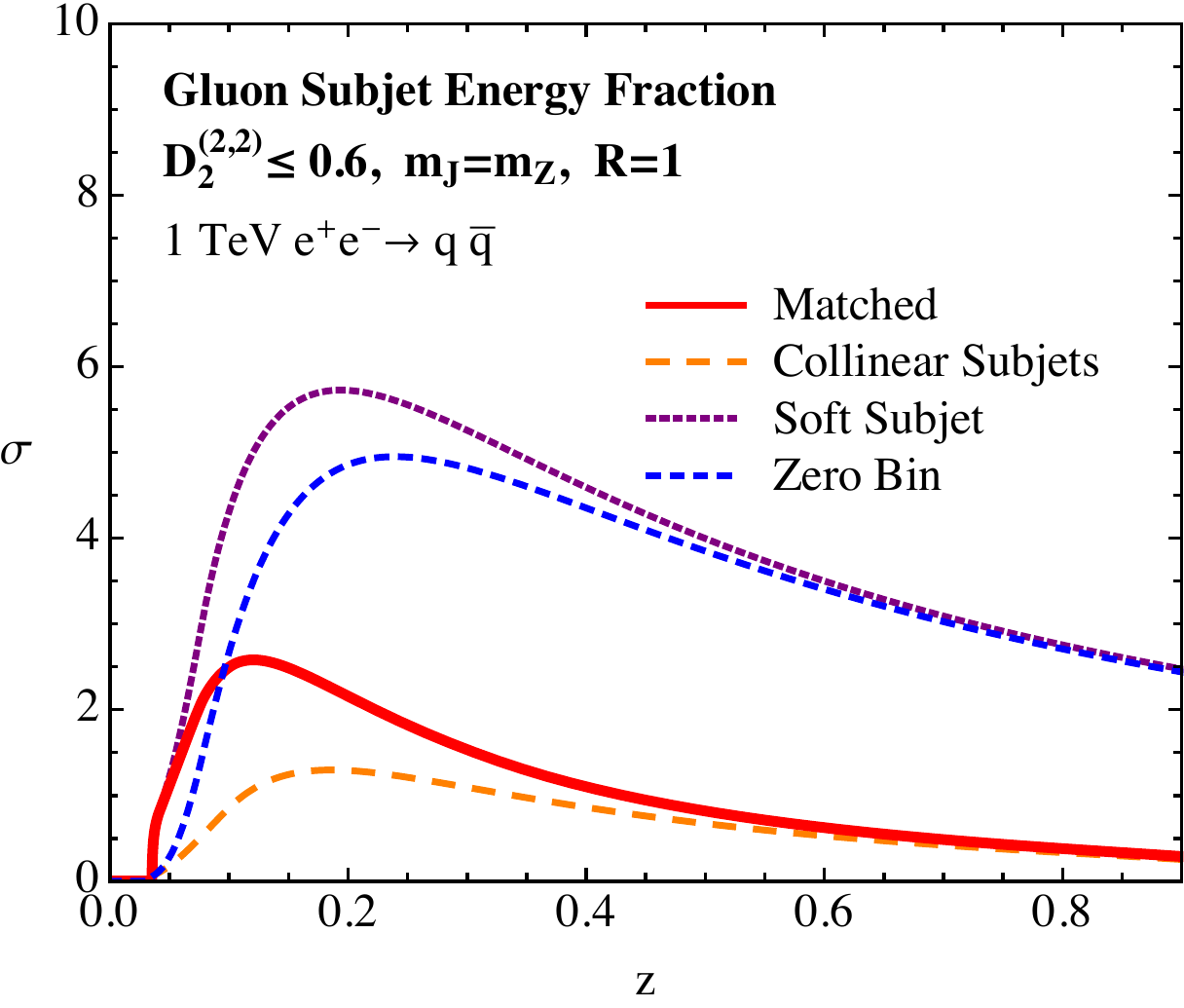}
}
\ 
\subfloat[]{\label{fig:soft_collinear_match_b_pt6}
\includegraphics[width = 6.9cm]{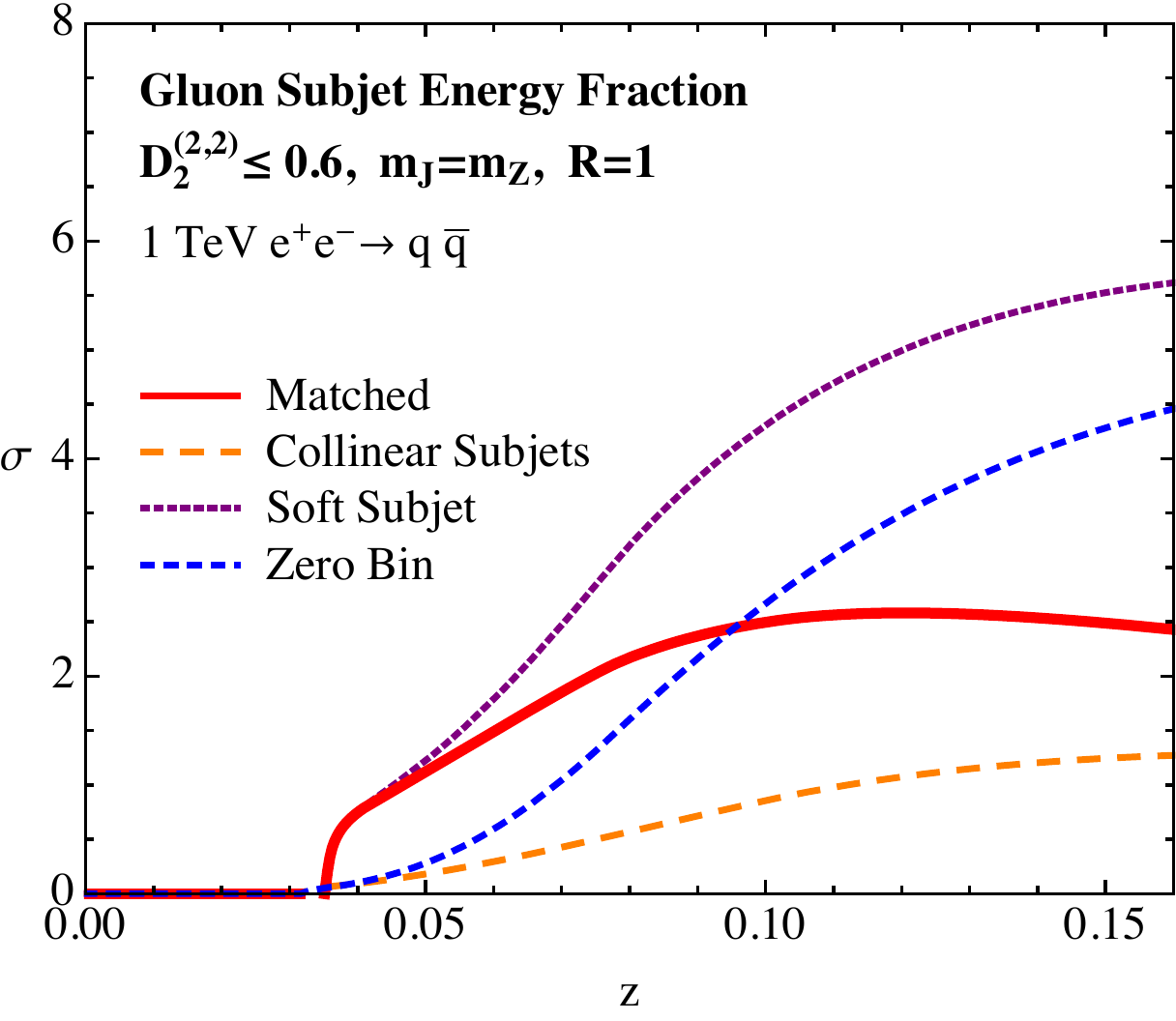}
}
\end{center}
\vspace{-0.2cm}
\caption{ a) Distribution of the energy fraction of the gluon subjet as predicted by the collinear subjets effective theory, the soft subjet effective theory, the collinear zero bin, and the matched description. A zoomed version at small $z$ is shown in b).
}
\label{fig:soft_collinear_match_pt6}
\end{figure}

Having defined our merging procedure, implemented through the zero bin, we can now revisit our argument for the absence of additional modes, previously given by power counting,  which can be verified from an explicit calculation. Taking the collinear-bin of the soft subjet factorization, and the soft-bin of the collinear subjet factorization, one finds identical fixed order expressions, as well as a one-to-one mapping of the anomolous dimensions between these two re-expanded factorizations. With the merging of the soft scales in the ``bins''  of the primary factorizations as one enters the soft region then implies they are numerically equivalent. No new logarithms appear in the bulk of phase space, unlike the case of two angularities \cite{Larkoski:2014tva}. This emphasizes that the collinear-soft region is a genuine overlap between the factorizations, with no new structures not already found in the factorizations.

To see visually the effect that this matching has, it is interesting to look at the distribution of the energy fraction of the one of the subjets. In \Fig{fig:soft_collinear_match_b}, we plot the distribution of the gluon subjet's energy fraction as computed in the collinear subjets and soft subjet factorization theorems, as well as the energy spectrum for the matched cross section of \Eq{eq:matched} and zero bin contribution. The energy spectrum is cumulative $D_2 \leq 2$, which is the majority of the two-prong region, and for simplicity we have fixed the jet mass $m_J=m_Z$. The matched contribution smoothly interpolates between the spectrum for the collinear subjets at large values of $z$, where the collinear subjets factorization theorem is valid, and captures all logarithms of the splitting angle, and that for the soft subjet factorization theorem at small values of $z$, accurately resumming large logarithms of $z$. It is also important to note that for large $z$, the zero bin contribution matches exactly onto the soft subjet contribution, removing its contribution in this region.  One can also see that the collinear-bin of the soft subjets cancels the collinear contribution to the soft region, up to power corrections, as argued above. We find that the collinear subjets provides a good description over a large range of values, with the soft subjet factorization theorem only required at small values of $z$. 

In \Fig{fig:soft_collinear_match_a_pt6}, we show the energy spectra at cumulative $D_2 \leq 0.6$, along with a zoomed version at small values of $z$, in \Fig{fig:soft_collinear_match_b_pt6}. This figures makes clear that our matched prediction, computed using our zero-bin approach, reproduces correctly the behavior of the collinear subjets at large values of $z$, and the soft subjet factorization theorem at small values of $z$. In particular, in \Fig{fig:soft_collinear_match_b_pt6}, we see that below $z\sim 0.05$, the soft subjet and matched predictions are indistinguishable.

Although we will not study this case explicitly in this chapter, we have also performed the matching for gluon jets, where the dominant contribution comes from $g \to gg$ splitting. This case is somewhat interesting due to the fact that the Bose symmetry of the final gluons guarantees that the $z$ distribution is symmetric about $z=0.5$, leading to peaks in the $z$ distribution due to soft singularities at both $z=0$ and $z=1$. Nevertheless, the same matching procedure works identically in this case, and this procedure could therefore also be straightforwardly applied for studying substructure in gluon jets, as would be required for a complete calculation at the LHC.

We have shown here the matched subjet energy spectra for the particular choice of jet radius $R=1$ at a center of mass energy of $1$ TeV for quark jets, as this is the particular case that we will focus on throughout the rest of the chapter. However, we have investigated the properties of the matching away from these parameters. It is important to note that our procedure for merging factorization theorem must be carefully treated at small $R$. This manifests itself as a breakdown in the zero bin procedure. In particular, for a fixed value of $\ecf{2}{\alpha}$, if $R$ is small, then the power counting $\ecf{2}{\alpha}\sim z_{sj}$ is invalidated. In other words, for small $R$ there does not exist a region of phase space which contributes to $\ecf{2}{\alpha}$ for which $z_{sj}$ is sufficiently small that the soft subjet expansion is valid. 

We can bound the specific $R$ that eliminates the soft subjet region by considering the minimum energy fraction accessible to a subjet at a fixed $\ecf{2}{\alpha}$:
\begin{align}\label{eq:min_z}
z_{\min}\approx \frac{\ecf{2}{\alpha}}{\left(2\, \text{sin} \frac{R}{2}\right)^{\alpha}}\,.
\end{align}
As a necessary condition for a soft subjet, one must fulfill the condition:
\begin{align}\label{eq:min_z_sj}
z_{\min}\sim \ecf{2}{\alpha} \rightarrow 1\sim \left(2\, \text{sin} \frac{R}{2}\right)^{\alpha}\,,
\end{align}
and so $R\sim 1$ for the soft subjet to contribute.  To eliminate the soft subjet then requires $R \ll 1$ and to still have valid collinear subjet regions requires that $R$ and $\ecf{2}{\alpha}$ are related as:
\begin{align}\label{eq:no_sj_R}
1\gg R^\alpha\gg \ecf{2}{\alpha}.
\end{align}
Finally, one should distinguish a fixed mass jet from a fixed $\ecf{2}{\alpha}$. In the case $\alpha=2$, since $\ecf{2}{2}= \frac{m_J^2}{E_J^2}$, by varying $E_J$ or $R$, we can open or close the soft subjet region. 

This appears in the zero bin by the fact that the zero bin subtraction is greater in all regions than the soft subjet, leading to a negative total cross section. We find numerically that this occurs for $R< 0.5$ for the case of $m_J=90$ GeV, and $Q=1$ TeV. This value depends fairly sensitively on $m_J$ and $Q$, or equivalently $\ecf{2}{\alpha}$. In this case, only the collinear subjets factorization theorem should be used, and it is valid throughout the entire available phase space. In this chapter we focus primarily on the case of fat jets, defined with $R=1$, and therefore it is necessary to perform the matching between the soft subjet region and the collinear subjets region for jets of energy $500$ GeV. However, in \Sec{sec:R_dependence}, we perform a brief survey of different $R$ values, comparing our analytic predictions with distributions from Monte Carlo generators. A more phenomenological study of the importance of the matching for different physics processes of interest for an $e^+e^-$ collider, the LHC, or even a possible $100$ TeV collider, where even higher boosts can be achieved, would be interesting, but is well beyond the scope of our initial investigation and can be straightforwardly treated using our techniques.

While we have used a zero bin procedure to perform the matching between the collinear subjets and soft subjet factorization theorems, it is also possible to develop a dedicated effective field theory valid when the soft subjet becomes collinear. This effective field theory is related to our zero bin contribution, and has been developed in \cite{Pietrulewicz:2016nwo}. While we believe that this approach is nice in principle, for the observable $D_2$, we find that such an effective field theory has a vanishing region of validity, as can be seen from the zero bin contribution in \Fig{fig:soft_collinear_match_b}, and \Figs{fig:soft_collinear_match_a_pt6}{fig:soft_collinear_match_b_pt6}. We therefore believe that our use of the zero bin, as generalized to distinct factorization theorems, represents a natural approach to the merging of the distinct factorization theorems. However, we acknowledge that this is an observable dependent statement, and there may be cases where there is a sufficiently large region of overlap between the soft subjet and collinear subjets effective theories, and in this case it might prove useful to have a separate effective field theory description which is valid in the case that the soft subjet becomes collinear.

\subsubsection{Matching Resolved to Unresolved Subjets}\label{sec:soft_haze_match}

An important feature of the $D_2$ observable is that its contours respect the parametric scaling of the phase space, as emphasized in \Fig{fig:C2vD2}. This implies that the marginalization over the contours defining the observable can be performed at small $D_2$ entirely within the merged effective theory of \Sec{sec:soft_NINJA_match}, and at large $D_2$ within the soft haze effective field theory. Hence the matching between these two different descriptions can be performed at the level of the $D_2$ distribution instead of at the level of the double differential cross section, which is a great simplification, and primary feature of the $D_2$ observable.

The soft haze factorization theorem presented in \Sec{sec:soft_haze} first contributes to the shape of the $D_2$ distribution at two emissions, the first order at which $\ecf{3}{\alpha}$ can be non-zero (technically at next-to-next-to-leading logarithmic prime order, NNLL$'$, in the logarithmic counting). Since our focus is on an initial investigation of the factorization properties of two-prong discriminants, the necessary two-loop calculation is beyond the scope of this chapter.  Na\"ively, this implies that since the merged effective field theory describing the two-prong region of phase space is only valid for $D_2 \lesssim 1$, our predictions should not be extended beyond $D_2 \lesssim 1$. However, we will argue that because of the structure of fixed order singularities for the $D_2$ observable, extending our two-prong factorization theorems to large $D_2$ will provide an accurate description of the $D_2$ distribution for a wide range of $E_J$ and $R$.

As shown in \Sec{sec:fixed_order}, there does not exist a fixed order singularity at large $D_2$. In particular, this implies that if extended into this region, the factorization theorems valid at small $D_2$ will not diverge. Furthermore, one in fact expects that they provide a reasonable description of the shape. They contain both an overall Sudakov factor for the $\ecf{2}{\beta}$ scale of the jet, and also provide a description of the internal structure of the jet in terms of splitting functions (in the case of the collinear subjets factorization). While the splitting function does not exactly reproduce the matrix elements in the soft haze factorization theorem, it provides a good description of them. We believe that this is a consistent approach which suffices for this initial investigation. 

Perhaps the most important fixed order correction not captured in the subjet factorition for $D_2$ is simply the endpoint of the distribution, which arises from the kinematic boundaries of the phase space. Since we will normalize our distributions to $1$, in order to compare to the Monte Carlo generators, the height of the peak is correlated with the endpoint. Matching to the soft haze region would give the resummed distribution the correct endpoint in the tail, and thus can shift the peak up in general. This endpoint is sensitive to the specific $R$ and $E_J$ of the jet, as well as to the values of the angular exponents $\alpha$ and $\beta$. Recall that since the Monte Carlo generators respect momentum conservation, they always terminate their distributions before the physical endpoint of the spectrum. We will also see how this disagreement in the tail region changes as a function of $R$ and $E_J$ in \Secs{sec:R_dependence}{sec:jet_energy} respectively.  However, for the case of dijets produced at a center of mass energy of $1$ TeV, with a jet mass cut of $80< m_J <100$ GeV, as is relevant for boosted boson discrimination, and on which we primarily focus throughout this chapter, we will see that this discrepancy in the tail region is minimal, and we will find good agreement between our analytic calculations and the Monte Carlo predictions. It would of course be interesting to perform the complete two-loop calculation in the soft haze region of phase space; however, we believe that this would have a minor effect for a substantial range of parameter space. Nevertheless, the proper inclusion of this region of phase space would also be interesting from a resummation perspective, as it would require matching between two distinct factorization theorems involving a different number of resolved jets, instead of the more familiar case of matching a resummed distribution to a fixed order calculation. We leave further investigations of this to future work.

\section{Numerical Results and Comparison with Monte Carlo}\label{sec:results}

We now present numerical results for signal and background distributions for the $D_2$ observable in $e^+e^-$ collisions. We give a detailed comparison with Monte Carlo, at parton level in Secs.~\ref{sec:MC} through \ref{sec:jet_energy} and including hadronization in \Sec{sec:Hadronization}. We then study the discrimination power of $D_2$ analytically in \Sec{sec:ROC}, and comment on the optimal choice of angular exponents. In \Sec{sec:2insight} possible observables which go beyond $D_2$, and separately resolve the soft subjet, and collinear subjets region of phase space, and how these could be used for possible improvements to boosted boson discrimination.

Throughout this section we use \fastjet{3.1.2} \cite{Cacciari:2011ma} and the \texttt{EnergyCorrelator} \fastjet{contrib} \cite{Cacciari:2011ma,fjcontrib} for jet clustering and analysis. All jets are clustered using the $e^+e^-$ anti-$k_T$ metric \cite{Cacciari:2008gp,Cacciari:2011ma} using the WTA recombination scheme \cite{Larkoski:2014uqa,Larkoski:2014bia}, with an energy metric.\footnote{We thank Jesse Thaler for use of a preliminary version of his code for WTA in $e^+e^-$ collisions. This code is now available in the \fastjet{contrib}.}

\subsection{Comparison with Parton-Level Monte Carlo}\label{sec:MC}

Previous studies of boosted boson discrimination with ratios of IRC safe jet observables have relied entirely on Monte Carlo simulations. While the implementation of both the perturbative shower and hadronization are well-tuned to describe simple event-wide observables, jet substructure observables probe significantly more detailed correlations.  For the particular case of observables sensitive to two-prong structure, their discrimination power is sensitive to the description of massive QCD jets in the phase space region where the jets are dominated by a resolved splitting.  One might na\"ively expect that this region of phase space is sensitive to the implementation of the parton shower model, and we will see that this is indeed the case.

 While a comparison to recent LHC data on jet substructure observables (for example: \cite{ATLAS:2012am,Aad:2013gja,TheATLAScollaboration:2013tia,TheATLAScollaboration:2013qia,CMS:2014fya,CMS:2014joa}) is possible, the lack of analytic calculations means that it is difficult to disentangle perturbative from non-perturbative effects. In this section we compare the results of our analytic calculation for $D_2$ with a number of Monte Carlo generators at parton level, focusing in particular on the small $D_2$ region.\footnote{One should always be wary of comparisons of Monte Carlo generators at parton level which employ different hadronization models.  Our comparisons at parton level presented in this section are to set the stage for fully hadronized comparisons in the following section. However, we take the view that a parton shower should achieve, to the greatest extent possible, a clean separation between perturbative and non-perturbative physics, and therefore should provide an accurate description of observables both at parton and hadron level.} This allows for a detailed probe of the simulation of two-prong jets in QCD by the perturbative shower (for a discussion of some other variables, see \Ref{Fischer:2014bja,Fischer:2015pqa}).  A large number of implementations of the perturbative shower exist, and are implemented in popular Monte Carlo generators (for reviews, see e.g. \cite{Buckley:2011ms,Skands:2012ts,Seymour:2013ega,Gieseke:2013eva,Hoche:2014rga}). Some examples include \pythia{} \cite{Sjostrand:2006za,Sjostrand:2007gs}, a $p_T$-ordered dipole shower; \vincia{} \cite{Giele:2007di,Giele:2011cb,GehrmannDeRidder:2011dm,Ritzmann:2012ca,Hartgring:2013jma,Larkoski:2013yi}, \sherpa{} \cite{Gleisberg:2003xi,Gleisberg:2008ta}, \ariadne{} \cite{Lonnblad:1992tz}, and \dire{} \cite{Hoche:2015sya}, dipole-antenna showers; and \herwigpp{} \cite{Marchesini:1991ch,Corcella:2000bw,Corcella:2002jc,Bahr:2008pv}, an angular-ordered dipole shower.\footnote{\herwigpp{} also has the option for a dipole-antenna shower implementation \cite{Platzer:2011bc} though we will not use it here.}

As representative of these different Monte Carlo shower implementations, we will use the following Monte Carlo generators throughout this section:
\begin{itemize}
\item \pythia{8.205}
\item \vincia{1.2.01} with a $p_T$-ordered shower
\item \vincia{1.2.01} with a virtuality-ordered shower
\item \herwigpp{2.7.1}
\end{itemize}
All Monte Carlos were showered with default settings except for the caveats listed below and requiring two-loop running of $\alpha_s$ with $\alpha_s(m_Z)=0.118$.  The different shower evolution variables within the \vincia{} Monte Carlo enables a study of their effects.  For background distributions, we generate $e^+e^-\to $ dijets at 1 TeV center of mass energy and study the highest energy $R=1.0$ anti-$k_T$ jet in the event.  For signal distributions in \pythia{} and \vincia{}, we generate $e^+e^-\to ZZ$ events with both $Z$s decaying hadronically.  For \herwig{}, the fixed-order signal distributions are generated in \madgraph{2.1.2} \cite{Alwall:2014hca} and showered in \herwig{}.  All jets are required to have a mass in the window $m_J\in[80,100]$ GeV.  In all plots shown in this section, hadronization has been turned off in all Monte Carlos.  Fixed-order matching was also turned off in \vincia{}.

\begin{figure}
\begin{center}
\subfloat[]{\label{fig:D2_ee_bkg}
\includegraphics[width= 7.15cm]{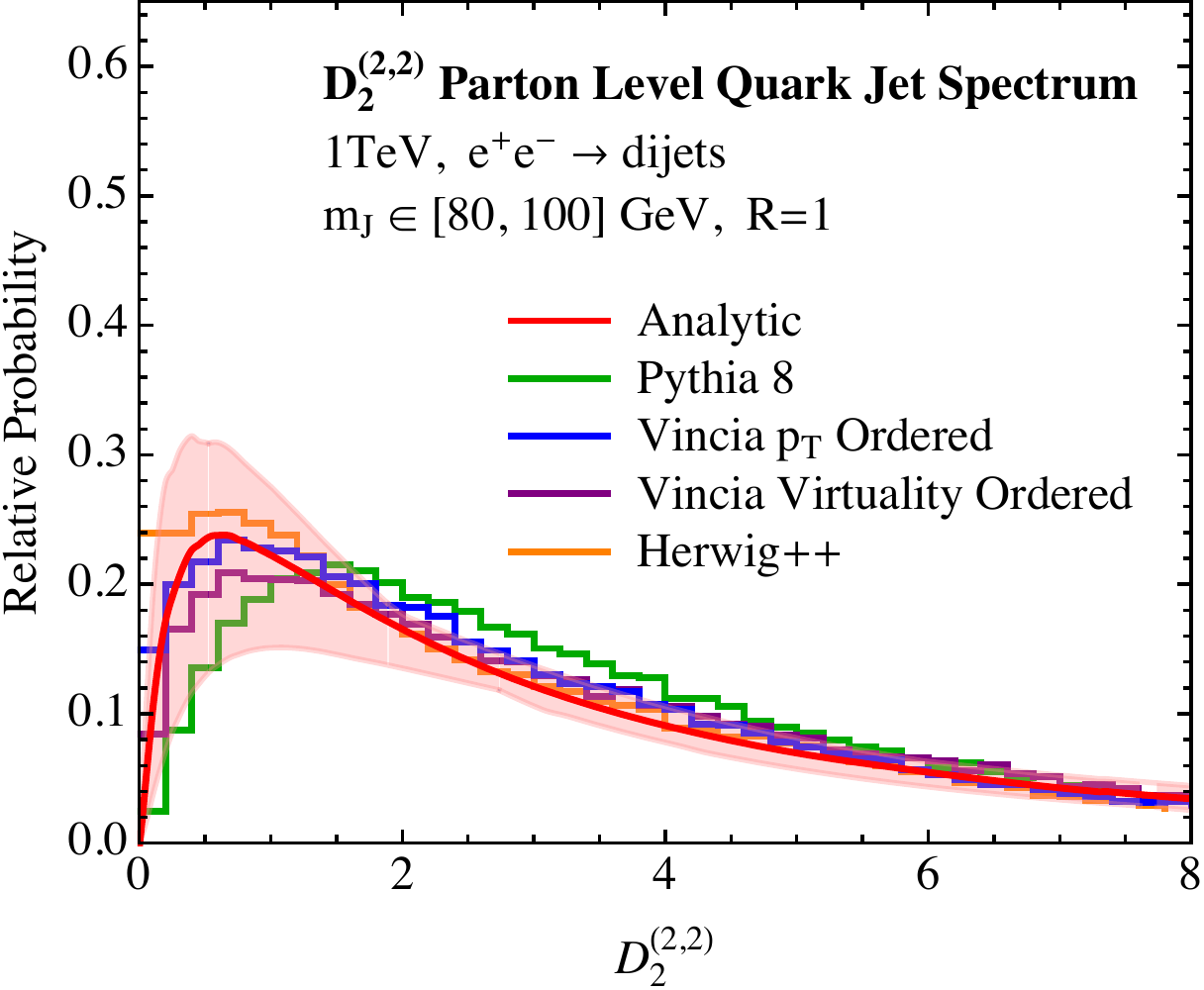}
}
\ 
\subfloat[]{\label{fig:D2_ee_sig}
\includegraphics[width = 7.15cm]{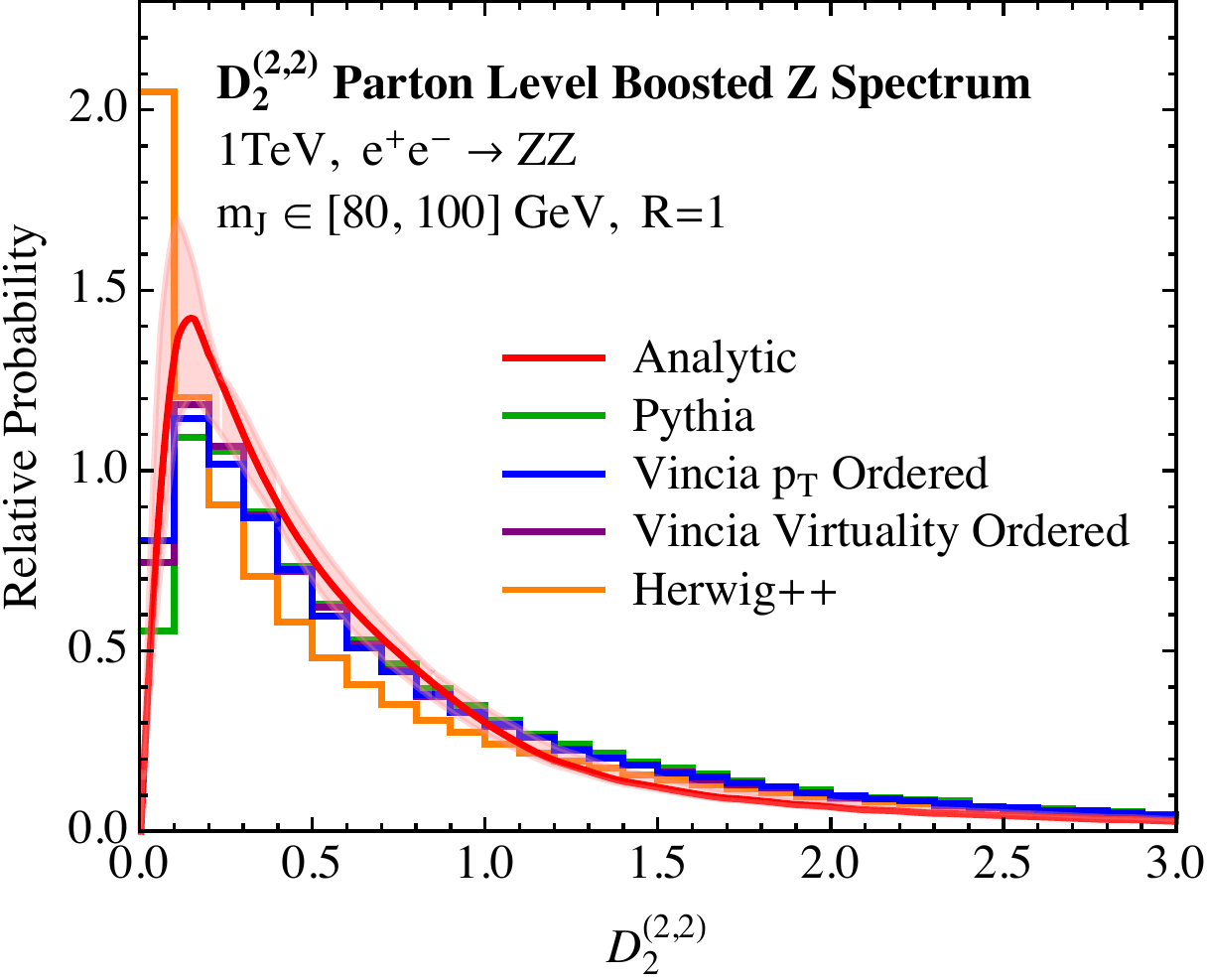}
}
\end{center}
\vspace{-0.2cm}
\caption{A comparison of our analytic prediction for $\Dobs{2}{2,2}$ compared with the parton-level predictions of the \pythia{}, \vincia{} and \herwig{} Monte Carlos. a) The $D_2$ distributions as measured on QCD background jets.  b) The $D_2$ distributions as measured on boosted $Z$ boson signal jets. The solid line is the central value of our analytic calculation and the shaded bands are representative of perturbative scale variations.
}
\label{fig:MC_compare}
\end{figure}

\Fig{fig:MC_compare} compares our analytic prediction for the $\Dobs{2}{2,2}$ spectrum to the parton-level Monte Carlo simulations in both background (\Fig{fig:D2_ee_bkg}) and signal (\Fig{fig:D2_ee_sig}) samples. The details of the scale variations used to make the uncertainty bands will be explained in \Sec{sec:scales}, but the pinch in the uncertainties should not be taken as physical. All Monte Carlos have similar distributions as measured on signal jets, though \herwig{} is more peaked at small values than the other generators.  Our analytic prediction, shown with perturbative scale variation, agrees well with the Monte Carlo generators. On background jets, however, the distributions are distinct, especially at small values of $D_2$.  Small $D_2$ is the region where the jet has a two-prong structure, but unlike for signal jets, for background jets that structure is not generated by a hard matrix element. In the case of collinear subjets, it is generated by a hard splitting function, while for a soft subjet, it is generated by an eikonal emission. Because it is an antenna shower, and therefore accurately describes the emission of wide-angle, soft gluons from leading-color dipoles, the \vincia{} distribution agrees the best with our calculation.  Also, $p_T$-ordering agrees better than virtuality-ordering, as expected because the scales at which the functions in the factorization theorem for $D_2$ are evaluated are essentially the relative $p_T$ of the dominant emissions in the jet. This agrees with the conclusions of \Ref{Platzer:2009jq,Hartgring:2013jma}, where it was shown that for antenna showers, the choice of  $p_T$ as a shower ordering variable absorbed all logarithms to $\mathcal{O}(\alpha_s^2)$. The \pythia{} distribution is shifted right as compared to \vincia{} and our calculation, which appears to be due to the way that \pythia{} populates the soft, wide angle region of phase space.\footnote{We have checked that this conclusion remains true with and without including matrix element corrections in the parton shower. We find little difference between the two cases.} \herwig{}, though a dipole shower like \pythia{}, 
agrees well with our calculation as shown in \Fig{fig:D2_ee_bkg}.

For reference, in \App{app:twoemissionMC} we show a collection of $\ecf{2}{2}$ distributions at both parton and hadron level for each of the different Monte Carlo generators. Since $\ecf{2}{2}$, which is related to the jet mass by \Eq{eq:ecf_mass_relation}, is set by a single emission, the agreement between the different generators, particularly at parton level, is significantly better than for the $D_2$ observable. This further emphasizes the fact that the $D_2$ observable offers a more differential probe of the perturbative shower, going beyond the one emission observables on which Monte Carlo generators have primarily been tuned.

In the following sections we will study the partonic $D_2$ distributions in more detail.  We will restrict ourselves to comparing and contrasting $p_T$-ordered \vincia{} and \pythia{} for a few reasons.  First, as exhibited in \Fig{fig:D2_ee_bkg}, these Monte Carlos represent the largest spread in their predicted $D_2$ spectra.  \herwig{}, while it performs very similarly to \vincia{}, has a different hadronization model than \pythia{} and \vincia{}.  So, directly comparing \pythia{} and \vincia{} minimizes any implicit hadronization effects 
when comparing the Monte Carlos at parton level. There are still differences due to the cutoff of the perturbative shower, which will be discussed in \Sec{sec:scales}.

\subsection{Monte Carlos and Perturbative Scale Variation}\label{sec:scales}

The fact that, in particular, the $p_T$-ordered \vincia{} distribution for $D_2$ as measured on background agreed with our calculation while the \pythia{} distribution disagreed in the small $D_2$ region can be understood and quantified further.  The bulk of the disagreement between our analytic calculation and \pythia{}, illustrated in \Fig{fig:D2_ee_bkg}, occurs near the peak of the $D_2$ distribution.  It is well-known that for many observables perturbative uncertainties tend to be significant in the peak region of the distribution.  Therefore, it is possible that the difference between the $p_T$-ordered \vincia{} and \pythia{} $D_2$ distributions can fully be explained by large perturbative uncertainties.  In this section, we will provide evidence showing that this is not the case and that this discrepancy arises from the different way that the two Monte Carlos treat the soft subjet region of phase space.  

To estimate perturbative uncertainties in our resummed analytic calculation, the standard procedure is to vary the scales that appear in the calculation by factors of 2.  This is at the very least a proxy for the sensitivity of the cross section on these scales.  Because our factorization theorems contain many functions, as well as merging of distinct factorization theorems, in principle there are numerous scales that could be varied, a complete analysis of which is beyond the scope of this chapter.  A complete list of the variations considered as well as the resummation procedure can be found in \App{sec:canonical_merging_scales}, while here we only summarize. In all factorizations theorems, we vary the subjet splitting scales, the in-jet soft radiation scales, the out-of-jet soft radiation scales, as well as where the freeze-out for the Landau pole occurs in the running of $\alpha_s$. We do not separately vary the scale in the soft subjet factorization theorem and the collinear zero bin to ensure that the zero bin subtraction is implemented correctly. The scale variation band for the total cross section is then taken as the combined band for all possible combinations of these scale variations. The soft subjet cross section displays a particular sensitivity to the out-of-jet scale setting, since the running between the boundary soft modes and the out-of-jet modes forces the soft subjet energy spectrum to vanish at the jet boundary,\footnote{As explained in \Ref{Larkoski:2015zka}, this is connected with the buffer region of \Ref{Dasgupta:2002bw}.} though the fixed order cross section probes the soft divergence in this region. Thus we also consider several different schemes for handling the out-of-jet scale setting. We believe that our scale variation bands are representative, and this is supported by the agreement with the Monte Carlo.

\begin{figure}
\begin{center}
\includegraphics[width= 7.2cm]{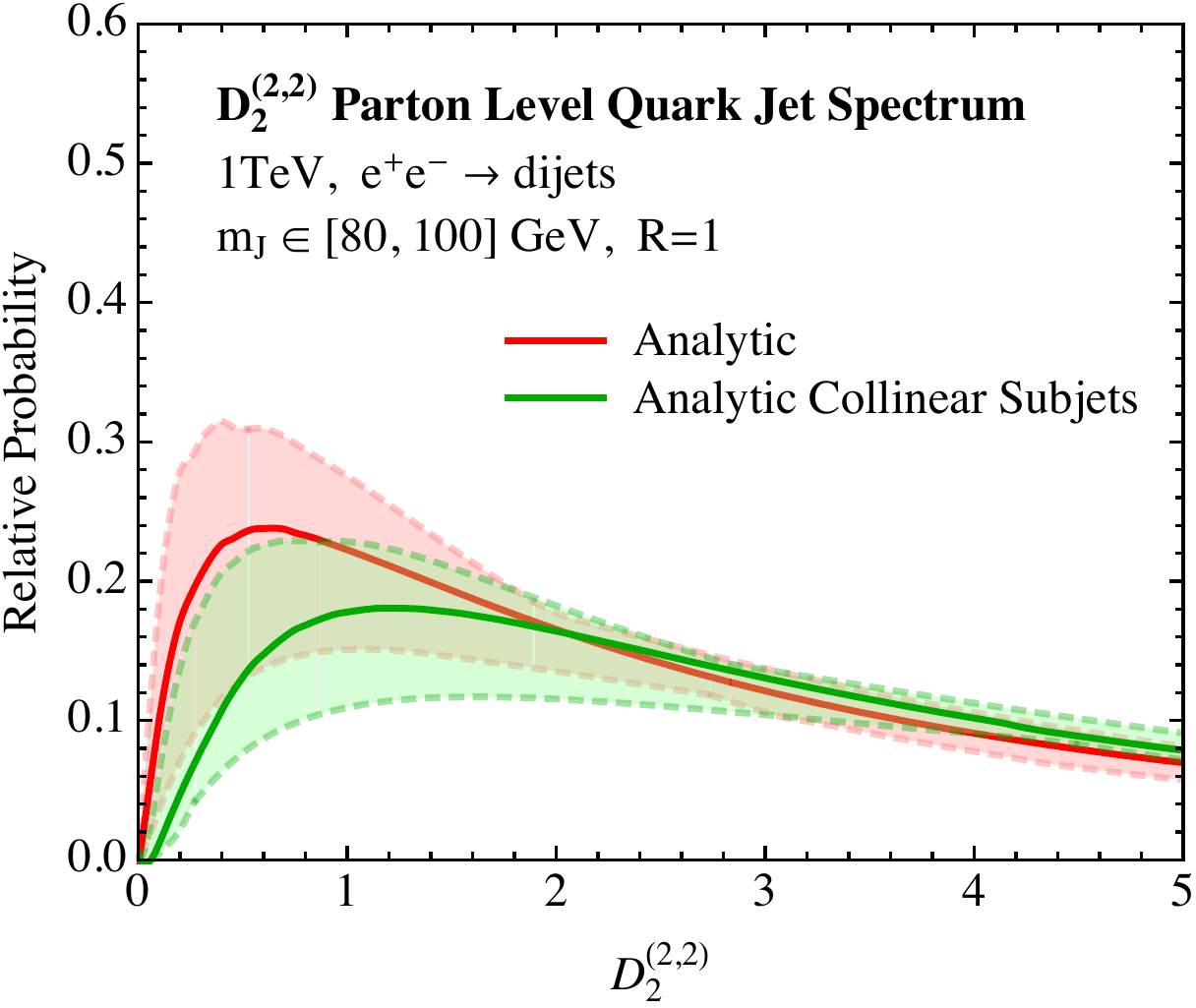}
\end{center}
\caption{
Analytic prediction for the $D_2$ distribution for background QCD jets including the envelope of the perturbative scale variation, as compared with an analytic calculation including just the collinear subjets region of phase space. The effect of the soft subjet region of phase space is clearly visible at small values of $D_2$.
}
\label{fig:D2_scale}
\end{figure}

Having understood the perturbative uncertainty bands, we now discuss in more detail the discrepancy between the different Monte Carlo generators arising at small values of $D_2$, as exemplified by the difference between the $p_T$-ordered \vincia{} and \pythia{} distributions. To understand the origin of this discrepancy, we begin by understanding the effect of the soft subjet region of phase space in our analytic calculation. This is possible due to our complete separation of the phase space using the energy correlation functions. In particular, because we have formulations of distinct factorization theorems in the soft subjet and collinear subjets regions of phase space, we can make an analytic prediction for the contribution arising just from the collinear subjets region of phase space. In \Fig{fig:D2_scale} we show a comparison of the $D_2$ distribution for background QCD jets as computed using our complete factorization theorem, incorporating both the soft subjet and collinear subjets region of phase space, as compared with the calculation incorporating only the collinear subjets region of phase space. Comparing the two curves, we are able to understand the effect of the soft subjet region of phase space. In particular, we see that the soft subjet has a considerable effect on the distribution at small values of $D_2$, giving rise to a more peaked distribution, with the peak at smaller values of $D_2$, as compared to the result computed using only the collinear subjets region of phase space. Although the perturbative error bands are large, the systematic effect of the soft subjet region of phase space is clear. 

One further feature of the $D_2$ distributions, which is made clear by \Fig{fig:D2_scale}, is that the full $D_2$ distribution is not the result of a single Sudakov peak, and therefore our intuition about the behavior of different orders in the perturbative expansion, and the behavior of scale variations from traditional event shapes fails. In particular, while it is generically the case for traditional event shape distributions that lower order resummed results overshoot in the peak region, it is not at all clear that this behavior should be true for $D_2$, and indeed it is not observed. Instead, the contribution from the collinear subjets alone is expected to undershoot the peak of the $D_2$ distribution, since it does not incorporate the soft subjet region of phase space. The final contribution is then obtained as a superposition of two distinct Sudakov peaks, and can therefore behave quite differently from traditional event shapes.

\begin{figure}
\begin{center}
\subfloat[]{\label{fig:cutoff_Rpt5}
\includegraphics[width= 7.2cm]{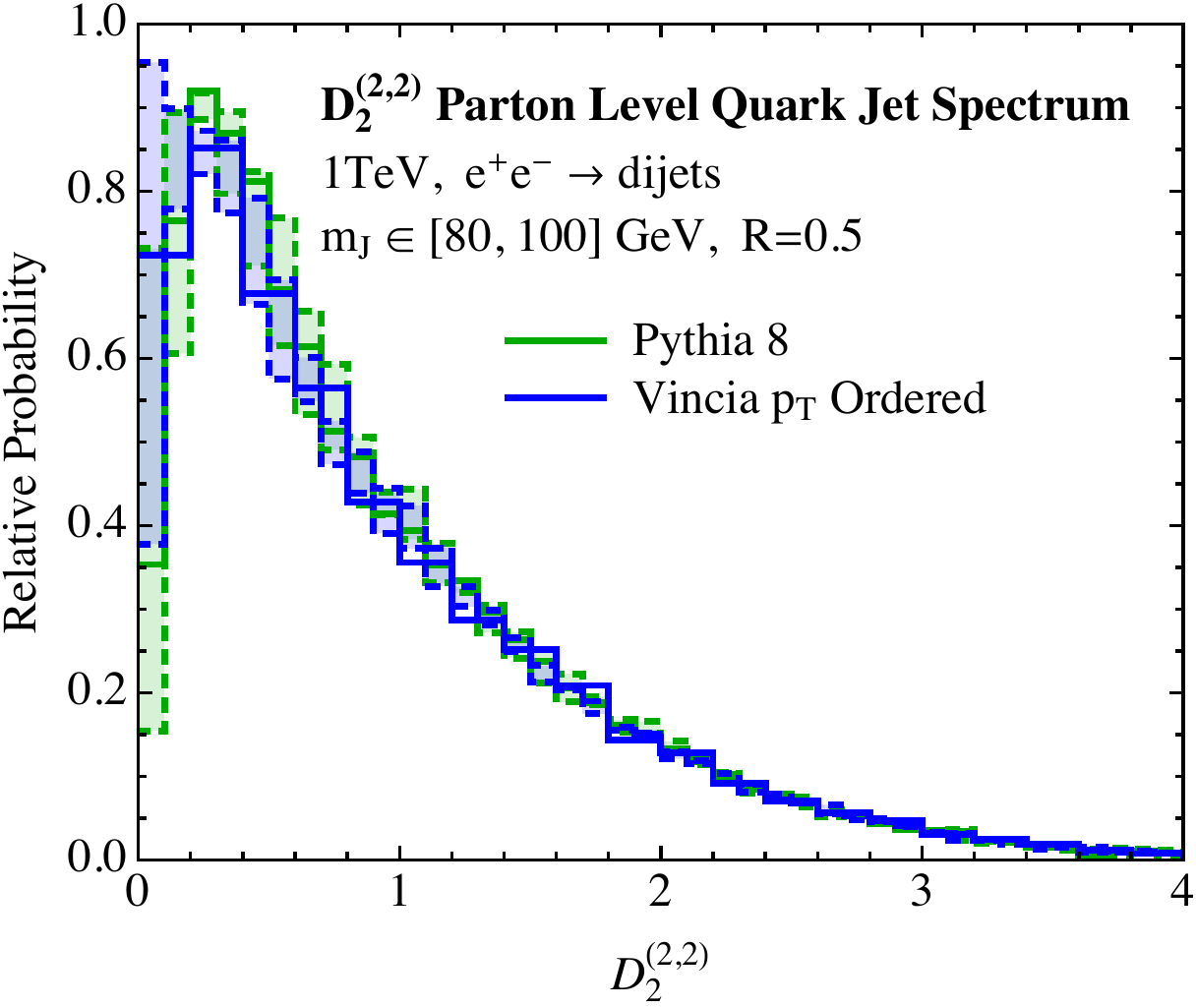}
}
\ 
\subfloat[]{\label{fig:cutoff_Rpt7}
\includegraphics[width = 7.2cm]{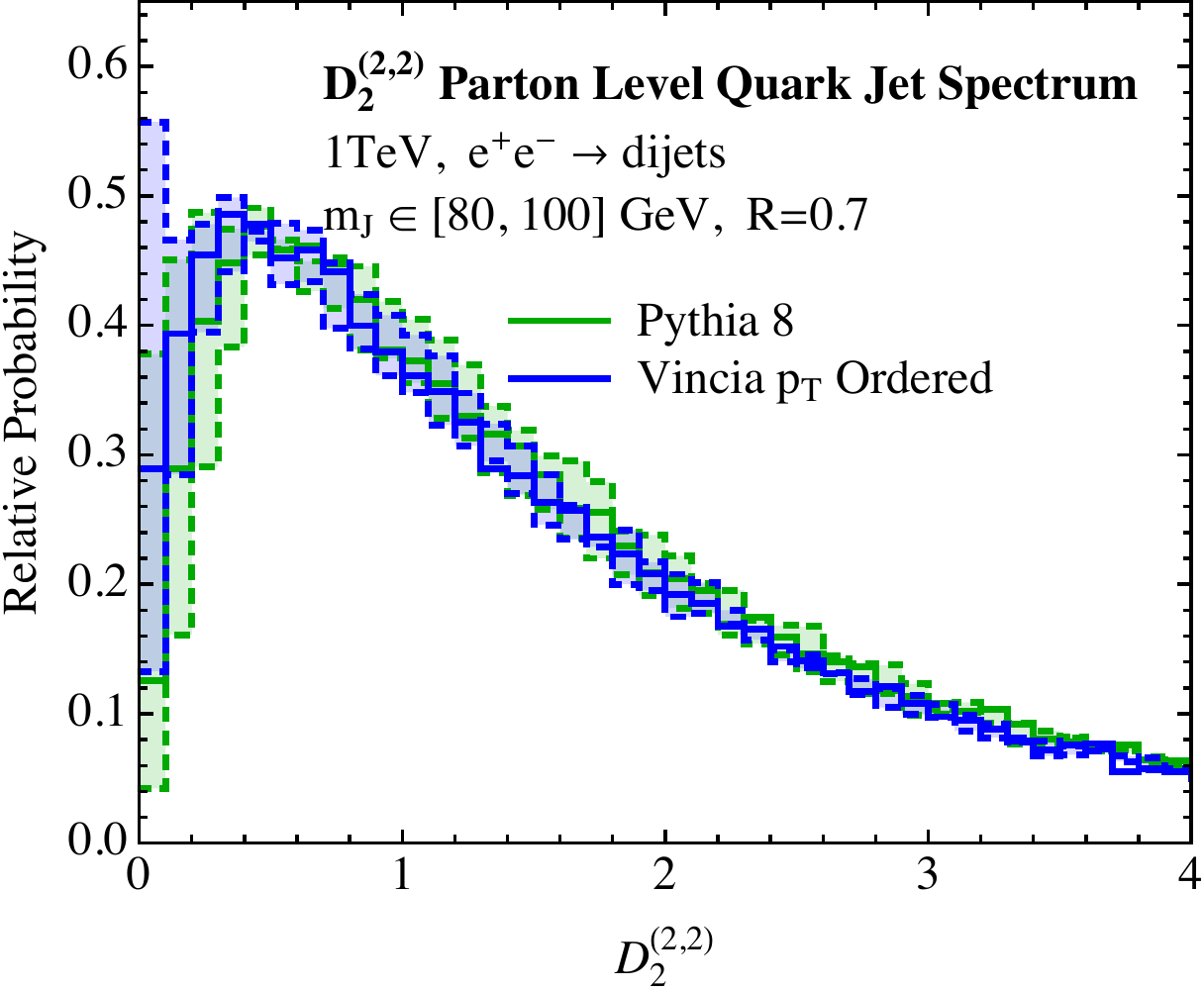}
}\\
\subfloat[]{\label{fig:cutoff_R1}
\includegraphics[width= 7.2cm]{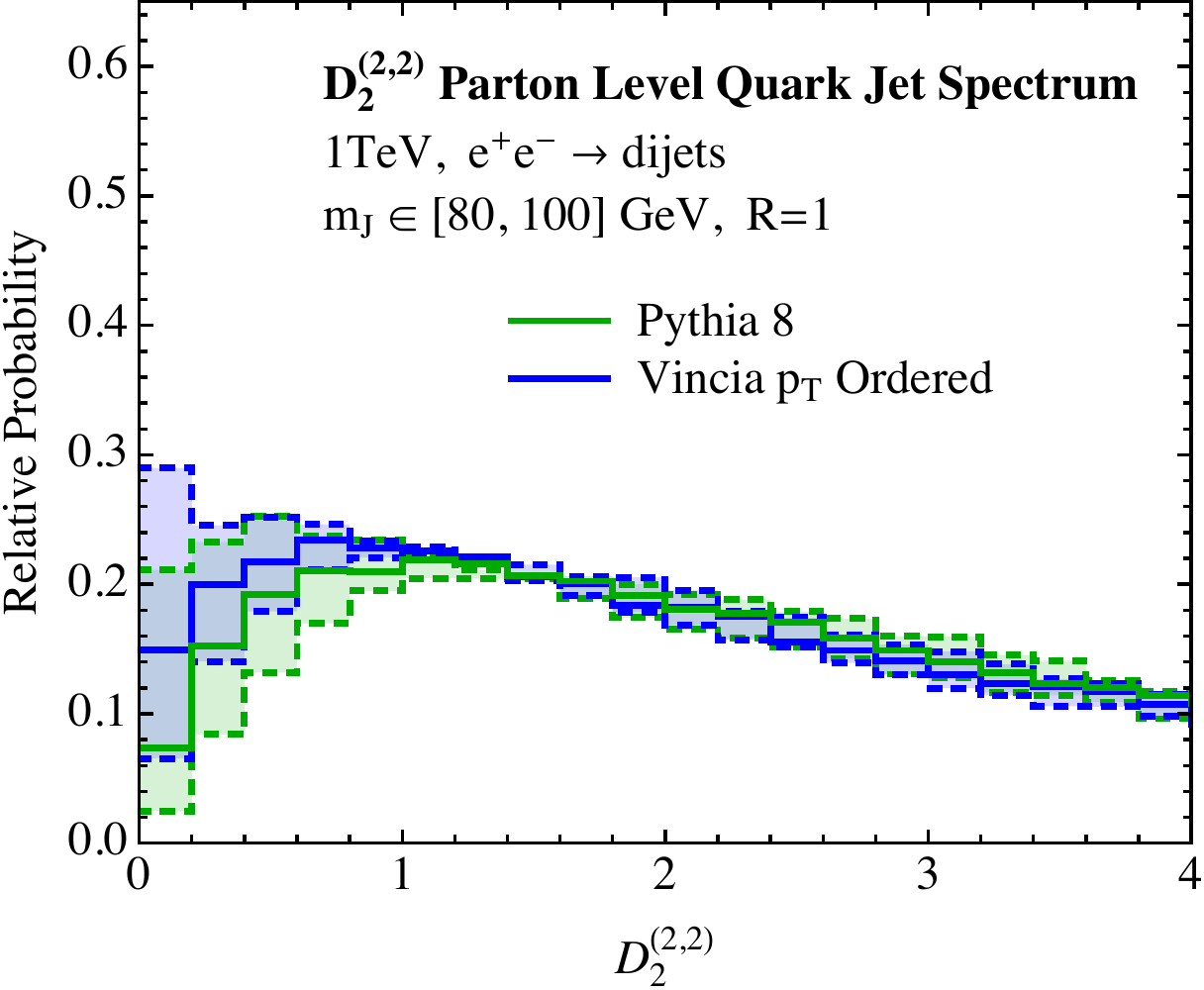}
}
\ 
\subfloat[]{\label{fig:cutoff_R1pt2}
\includegraphics[width = 7.2cm]{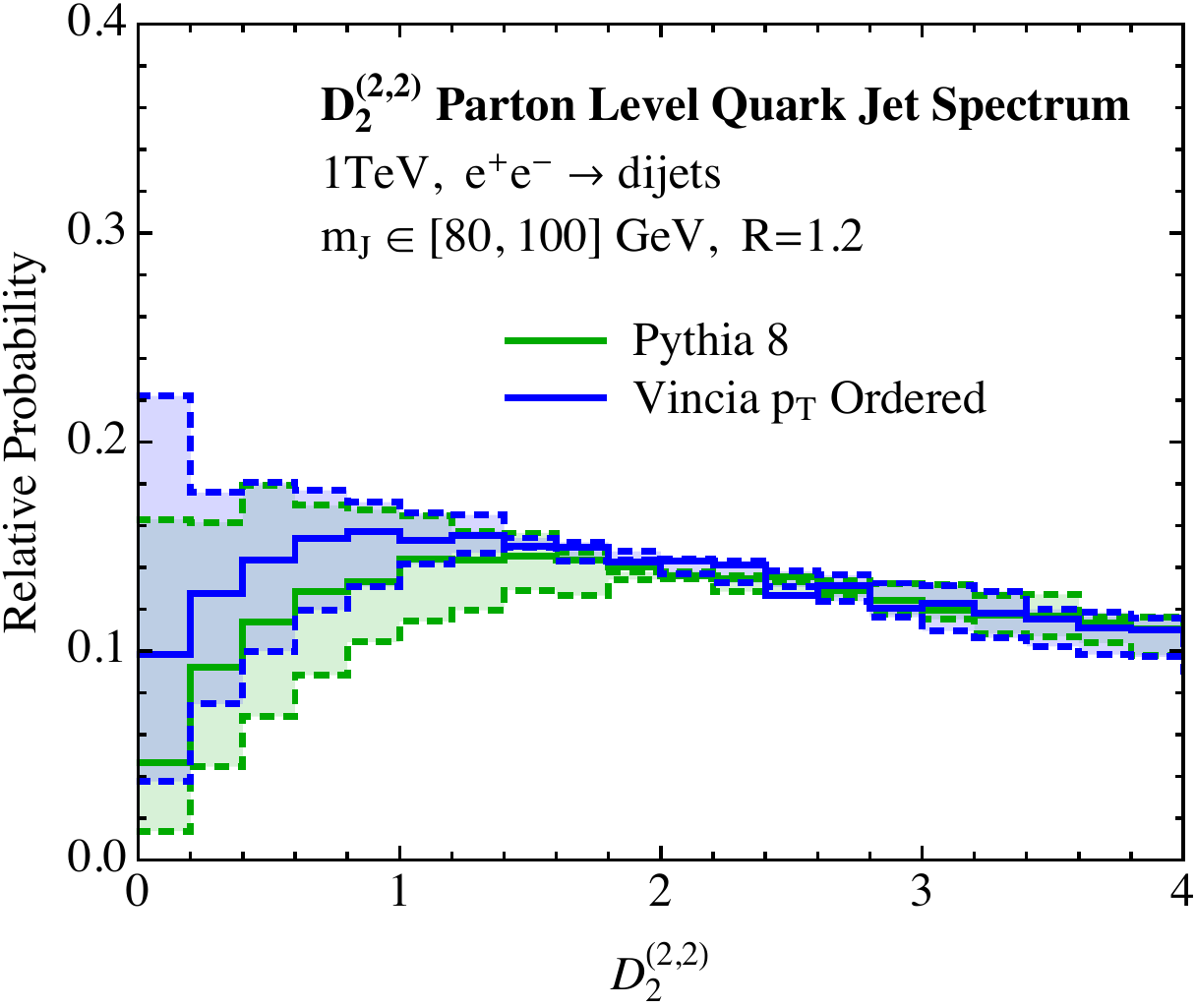}
}
\end{center}
\caption{
Comparison of the $D_2$ distribution for background QCD jets in \pythia{} and $p_T$-ordered \vincia{} for different jet radii. In each plot, the central value is obtained using a shower cutoff of $0.8$ GeV, and the uncertainty bands are generated by varying this cutoff between $0.4$ GeV, and $1.2$ GeV. 
}
\label{fig:cutoff_test}
\end{figure}

Monte Carlo descriptions of the perturbative shower should provide a similar description of collinear physics, but can differ in their description of soft wide angle radiation. Some of these differences were discussed in \Sec{sec:MC}. As discussed earlier, because \vincia{} is a dipole-antenna shower, it should accurately describe both the hard collinear and soft wide-angle regions of phase space.  
Because small values of $D_2$ are sensitive to both collinear and soft physics, the fact that the \pythia{} distribution at small $D_2$ is distinct suggests that its description of soft wide-angle physics is the reason.\footnote{Part of the reason for why \pythia{} seems to not correctly describe the soft, wide-angle region of phase space may be due to the fact that while it uses kinematics of dipoles in its shower, it still uses the Altarelli-Parisi splitting functions as an approximation of the squared matrix element.  The dipole and its emission is then boosted to the appropriate frame, which may over-populate the soft wide-angle region of phase space as compared to the eikonal matrix element. We thank Torbj\"orn Sj\"ostrand and Peter Skands for detailed discussions of this point.}  The difference observed in our analytic calculation arising from the soft subjet region of phase space is similar to that observed between the $p_T$-ordered \vincia{} and \pythia{} Monte Carlo distributions. It is therefore interesting to investigate whether the difference in Monte Carlo distributions can arise from different descriptions of wide angle soft radiation. We will now provide evidence that this is indeed the case. It is important to emphasize, however, that since we perform this comparison at parton level, there is some ambiguity in effects due to the perturbative cutoff of the shower, and those arising from different descriptions of wide angle soft radiation.  In particular, the Monte Carlos will in general have different low-scale $p_T$ cutoffs at which the perturbative parton shower is terminated.  Varying this scale can potentially greatly increase or decrease the number of soft emissions because the value of $\alpha_s$ in this region is large.  In particular, for the versions of \pythia{} and \vincia{} that were use to generate events in \Fig{fig:MC_compare}, the cutoff in \pythia{} is $0.4$ GeV, while the cutoff in \vincia{} is $0.8$ GeV. Indeed, these are the default values for these showers.  Therefore, we expect that the \pythia{} parton shower produces more soft emissions than \vincia{}, which would increase the value of $D_2$, and potentially also contribute to the observed difference.

To attempt to disentangle the effects of the shower cutoff from differences in the modeling of soft radiation, in \Fig{fig:cutoff_test}, we consider Monte Carlo predictions of the $D_2$ distribution as measured on QCD jets, with different jet radii, namely $R=0.5,0.7,1.0,1.2$. By using different jet radii, we can control the importance of the soft subjet region of phase space. With small jet radii, the soft subjet region of phase space does not exist, while it becomes increasingly important as the jet radius is increased.  Analytic predictions for the $D_2$ distribution for different values of the jet radius, $R$, will be given in \Sec{sec:R_dependence}. Here we compare the $p_T$-ordered \vincia{} and \pythia{} Monte Carlo. To generate the central values of the curves, we have used a cutoff of $0.8$ GeV, and the uncertainty bands are generated by varying this cutoff from $0.4$ GeV to $1.2$ GeV, to understand its effect. From \Fig{fig:cutoff_test} we see that while there is a relatively large uncertainty band from varying the perturbative cutoff of the shower, the uncertainty bands for the two generators become systematically more offset as the jet radius is increased. At $R=0.5$ or $0.7$, there is very small distinction, while for $R=1.0$ or $1.2$, there is a relatively large offset. This suggests that while some of the distinction seen between the Monte Carlo distributions in \Fig{fig:MC_compare} is due to the different treatment of the cutoff of the perturbative shower, there is a difference arising from the different description of perturbative wide angle soft radiation. The dipole-antenna based shower, \vincia{}, which is expected to provide an accurate description of wide angle soft radiation seems to agree best with our analytic calculation, suggesting that \pythia{} is not providing a complete description of wide angle soft radiation, which can play an important role in jet substructure studies.

This analysis also shows some of the difficulties in disentangling perturbative from non-peturbative effects, and the importance of having analytic calculations and precise theoretical control of different phase space regions to do so. However, by measuring sufficiently many observables on a jet, we are able to isolate distinct phase space regions and study in detail the extent to which Monte Carlo parton showers reproduce the physics in the different regions.  $D_2$, or similar jet substructure observables, could therefore be powerful tools for tuning Monte Carlos, both to formally-accurate perturbative calculations, as well as data. In the remaining sections of the chapter, we will use the default shower cutoffs in the Monte Carlo generators, as was done in \Fig{fig:MC_compare}, and will not show uncertainty bands on our Monte Carlo distributions from varying this parameter.

\subsection{Analytic Jet Radius Dependence}\label{sec:R_dependence}

As demonstrated in the previous section, the region of small $D_2$ is a sensitive probe of the dominant soft or collinear structure in the jet.  It is therefore interesting to study the jet radius dependence of $D_2$ analytically, because the relative size of soft subjet and collinear subjets contributions to $D_2$ will depend on the jet radius.  At large jet radius, as shown earlier, the soft subjet region is an important contribution at small $D_2$, but as the jet radius decreases, the collinear subjets should dominate.  In this section, we will study the jet radius dependence of $D_2$ and compare our analytic calculation to Monte Carlo. This will also demonstrate that our analytic calculation accurately describes the $R$ dependence of the $D_2$ distribution.  As in the previous section, we will restrict this study to $p_T$-ordered \vincia{} and \pythia{} showers, and will take the jet radius to be $R=0.5,0.7,1.0,1.2$, which are representative of a wide range of values of experimental interest. Larger values of $R$ can be straightforwardly studied with our approach, but are of less phenomenological interest.  It is expected that for smaller values of $R$ logarithms of $R$ may become numerically important \cite{Seymour:1997kj,Tackmann:2012bt,Dasgupta:2014yra}, so we do not consider them here.

\begin{figure}
\begin{center}
\subfloat[]{\label{fig:Rdep_a}
\includegraphics[width= 7.0cm]{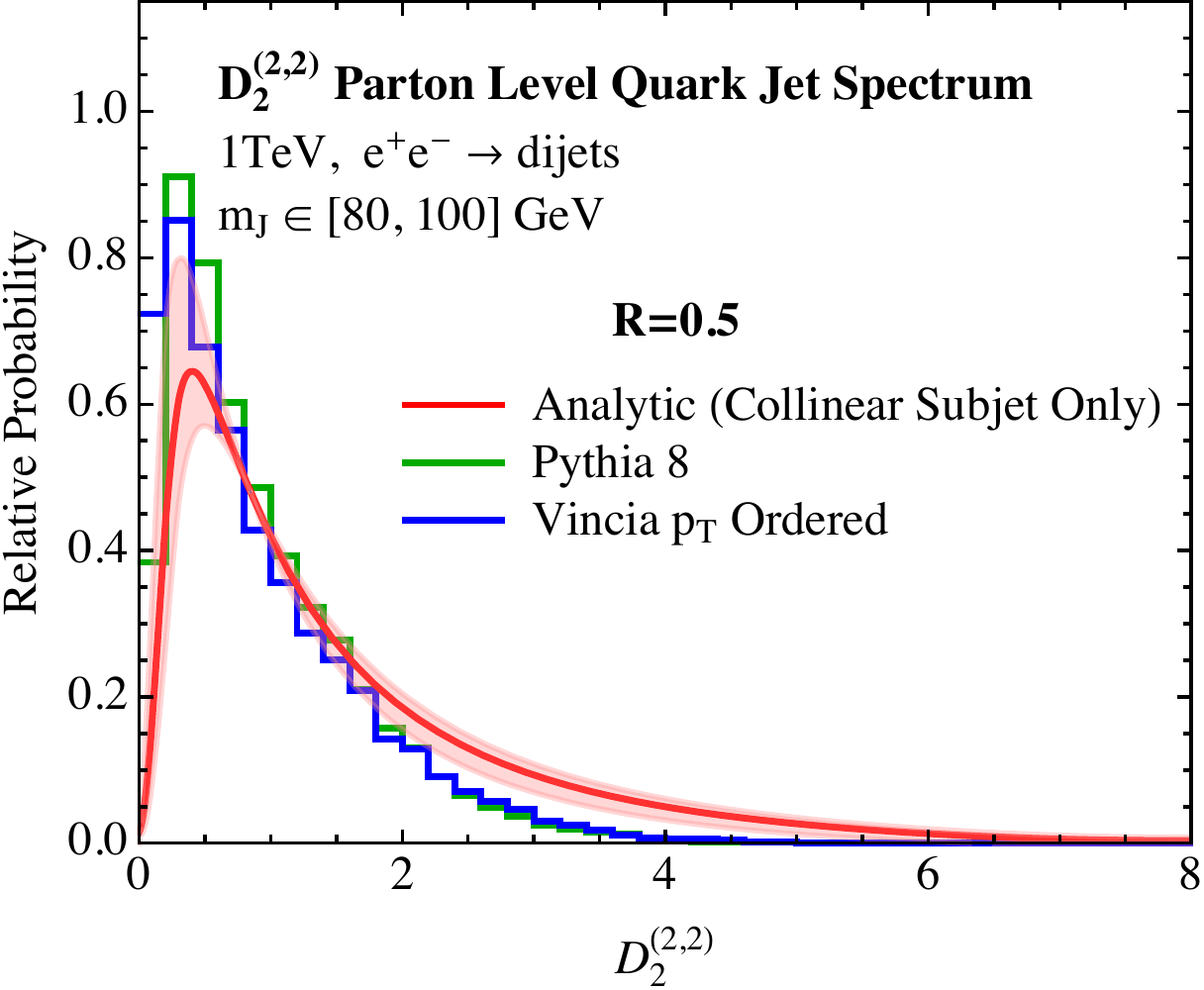}
}
\subfloat[]{\label{fig:Rdep_b}
\includegraphics[width = 7.0cm]{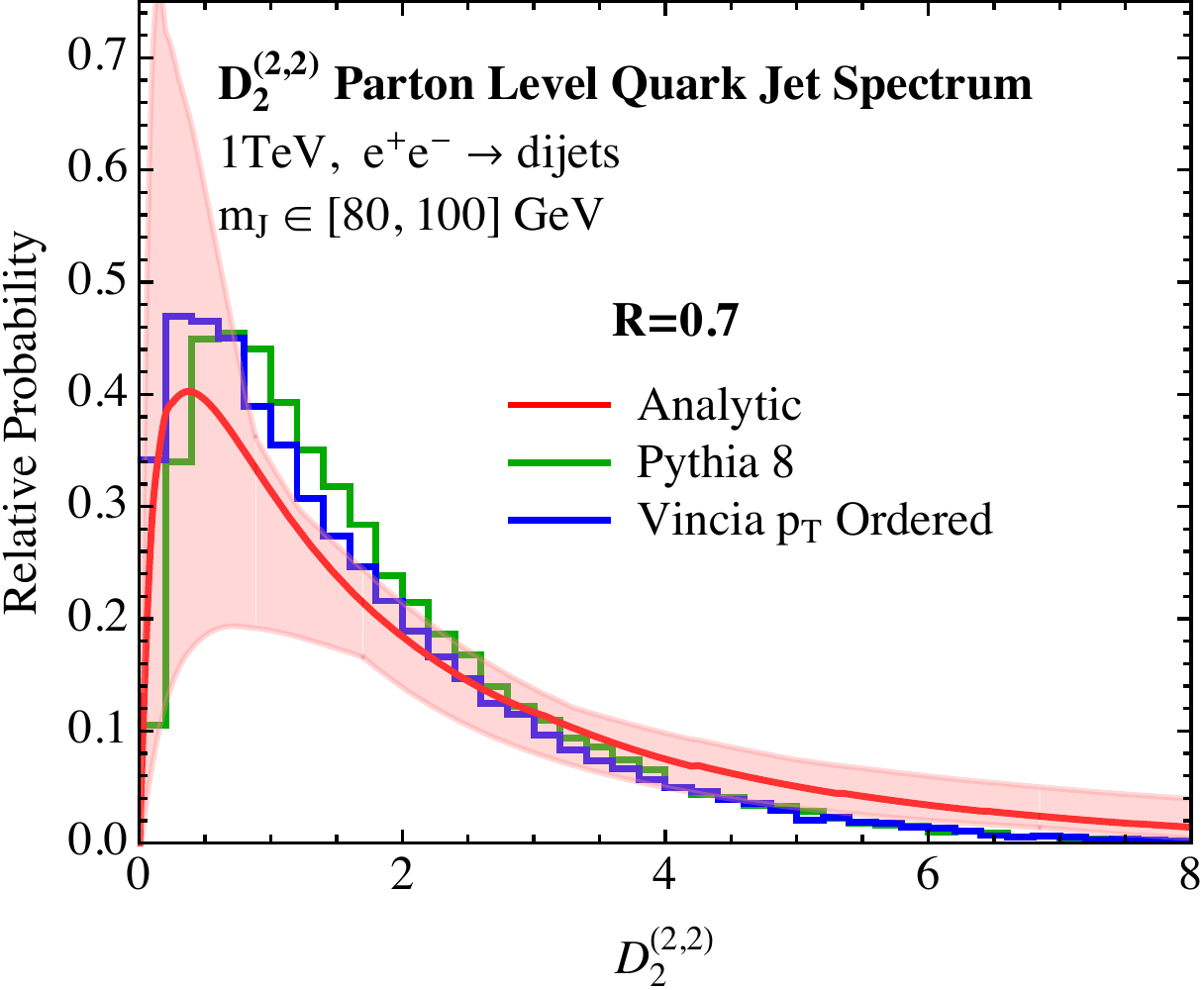}
}\\
\subfloat[]{\label{fig:Rdep_c}
\includegraphics[width= 7.0cm]{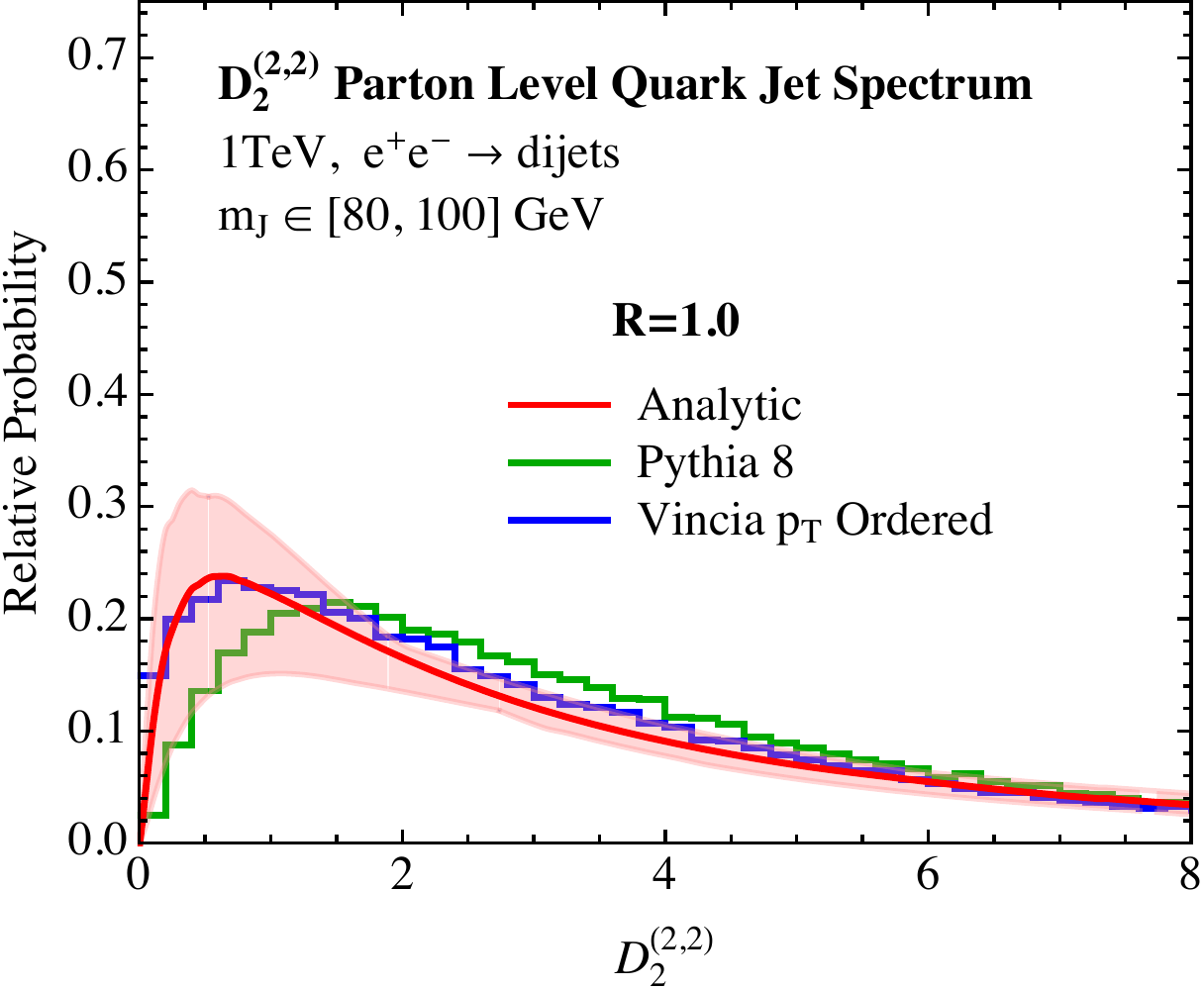}
}
\subfloat[]{\label{fig:Rdep_d}
\includegraphics[width = 7.0cm]{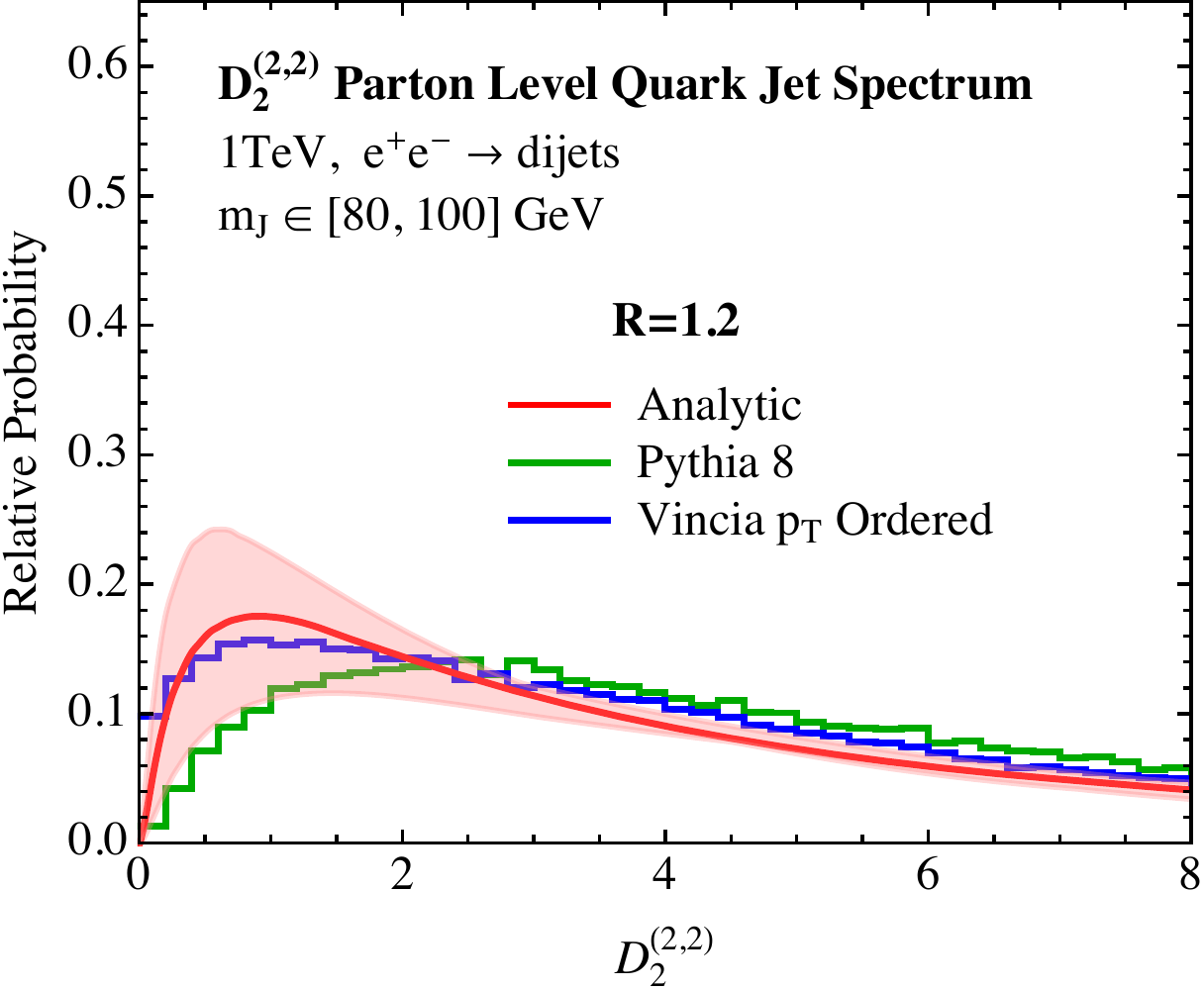}
}
\end{center}
\caption{ Comparison of QCD background $D_2$ distributions from $p_T$-ordered \vincia{} and \pythia{} to our analytic prediction as a function of the jet radius, $R$. The values $R=0.5,0.7,1.0,1.2$ are shown in Figures a)-d), respectively. In the analytic prediction for $R=0.5$, only the collinear subjets factorization theorem is used, while for all other values of the jet radius the analytic calculation includes contributions from both the collinear subjets and soft subjet factorization theorems. 
}
\label{fig:R_dependence}
\end{figure}

Comparisons of parton level Monte Carlo results from both $p_T$-ordered \vincia{} and \pythia{} to our analytic calculation are shown in \Fig{fig:R_dependence}. Since we scan over a range of jet radii, perturbative uncertainties for each $R$ value are not as extensively explored as earlier with $R=1$, and are only meant as a rough estimate of the perturbative uncertainty.  Our focus here is simply to show that the scaling behavior with $R$ between our analytic calculation and the Monte Carlos agree. 
There is excellent agreement between the Monte Carlo results and our analytic calculations over the entire range of $R$ values, with $p_T$-ordered \vincia{} providing a more accurate description of the analytic calculation than \pythia{}. The size of the region over which there is a large disagreement between the Monte Carlos increases monotonically with $R$, and for  $R\gtrsim 1$, there is even considerable disagreement in the position of the peak of the distribution between the generators. However, for smaller values of $R$, the discrepancy between \vincia{} and \pythia{} at small $D_2$ is reduced.  Indeed the position of the peak of the distribution coincides between the different generators for $R=0.5$, with only a small discrepancy in the bin at lowest $\Dobsnobeta{2}$. However, here hadronization will play an important role, smearing out this effect. The effect of hadronization, and its implementation in our analytic calculation, will be discussed in \Sec{sec:Hadronization}.

For jet radii of $R=0.7,1.0,1.2$ our analytic calculation consists of both collinear subjets and soft subjet contributions.  For $R=0.5$, however, we only include the contribution from collinear subjets, which is guided by our matching procedure between the collinear subjets and soft subjet factorization theorems, as discussed in \Sec{sec:soft_NINJA_match}.  For a fixed jet mass, as the value of $R$ is decreased, the region of validity of the soft subjet factorization theorem vanishes rapidly.  For jet masses in the range $80 < m_J < 100$ GeV, and $Q=1$ TeV, we find that between $R=0.7$ and $R=0.5$  the region of validity of the soft subjet rapidly shrinks to zero, and there should not be a transition between the collinear subjets factorization theorem and the soft subjet factorization theorem. Because of this, for the value of $R=0.7$, our perturbative error bands are more extensive, and are taken as the envelope of curves both that include the matched soft jet, and curves that do not. While this is certainly over conservative in the error estimate, we have included this to emphasize this point. This feature is also seen explicitly in the plots of \Fig{fig:R_dependence}, where the region of disagreement between the different Monte Carlo generators is squeezed towards zero. A similar effect occurs as the energy (or $p_T$) of the jet is increased with a fixed jet mass, which will be discussed in \Sec{sec:jet_energy}.

As discussed in \Sec{sec:scales}, this observation of the $R$ dependence provides support that the discrepancy between the Monte Carlo generators, as represented by \pythia{} and \vincia{}, is dominantly due to \pythia{}'s description of wide angle soft radiation.
The disagreements between Monte Carlo generators is thus highly sensitive to the degree to which soft subjets can impact the distributions, a feature which should be taken into account when performing jet substructure studies.  Throughout the remainder of this chapter, we will study the case $R=1$ exclusively, because both collinear subjets and soft subjet regions of phase space must be included and that radius is relevant to a large number of jet substructure studies using fat jets.

\subsection{Analytic Jet Energy Dependence}\label{sec:jet_energy}

In addition to studying the dependence on the jet radius as a probe of the importance of the soft subjet and of the Monte Carlo description of the shower, it is also interesting to study the dependence of the $D_2$ distribution on the energy of the jet, with a fixed mass cut. For highly energetic jets, one expects that the soft subjet will play a negligible role, as the region of validity of the soft subjet factorization theorem shrinks as the energy of the jet is increased, as long as the mass of the jet is kept fixed. On the other hand, since we keep the jet radius used in the clustering fixed, the angular separation of the collinear particles decreases with energy, but the phase space for wide angle global soft radiation increases considerably.  This radiation is present both in the collinear subjets and soft haze factorization theorems. It is also of course present in the soft subjet factorization theorem, although we have argued that we expect this to give a small contribution. Studying the jet energy dependence therefore probes the behavior of the generators in a fashion complementary to the $R$ dependence.

In this section, we study the perturbative $D_2$ distribution for center of mass energies ranging from $500$ GeV to $2$ TeV, for a fixed jet radius of $R=1$, and with a fixed mass cut of $80<m_J<100$ GeV. This region of energies covers the majority of the phenomenologically interesting phase space available at the LHC. We will also perform a more detailed study at LEP energies in \Sec{sec:LEP}. For our resummation, we require (amongst other things), that $\ecf{2}{\alpha} \ll 1$. For the case of $\alpha=2$ for which we will be most interested, this corresponds to the assumption $\ecf{2}{2} = m_J^2/E_J^2 \ll 1$. For a mass cut around the $Z$ pole mass, this expansion is valid throughout the range of energies we consider.   The case when $\ecf{2}{2}\lesssim 1$, but not parametrically so, is outside the scope of this chapter.

\begin{figure}
\begin{center}
\subfloat[]{\label{fig:Edep_a}
\includegraphics[width= 7.0cm]{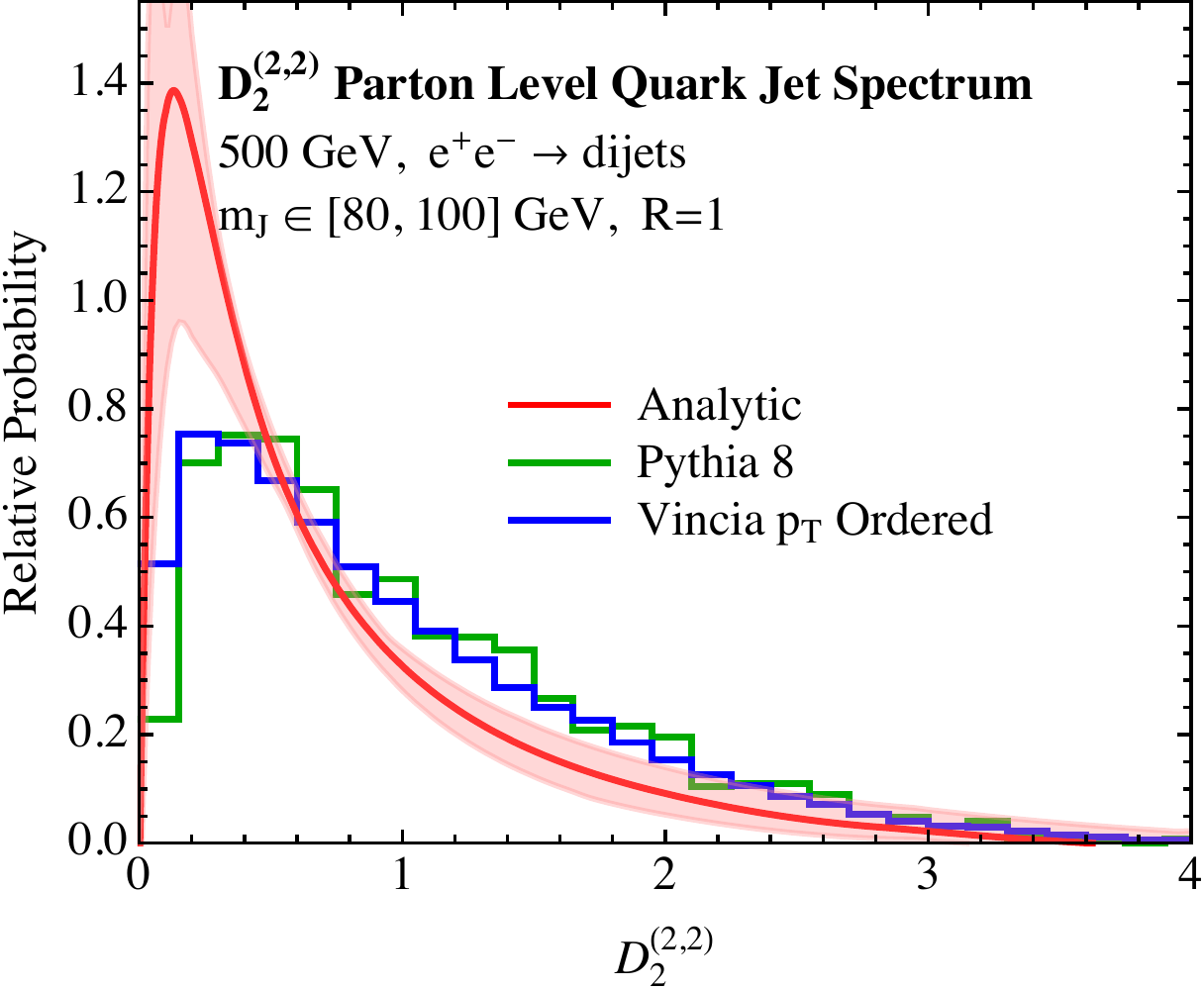}
}
\subfloat[]{\label{fig:Edep_b}
\includegraphics[width = 7.0cm]{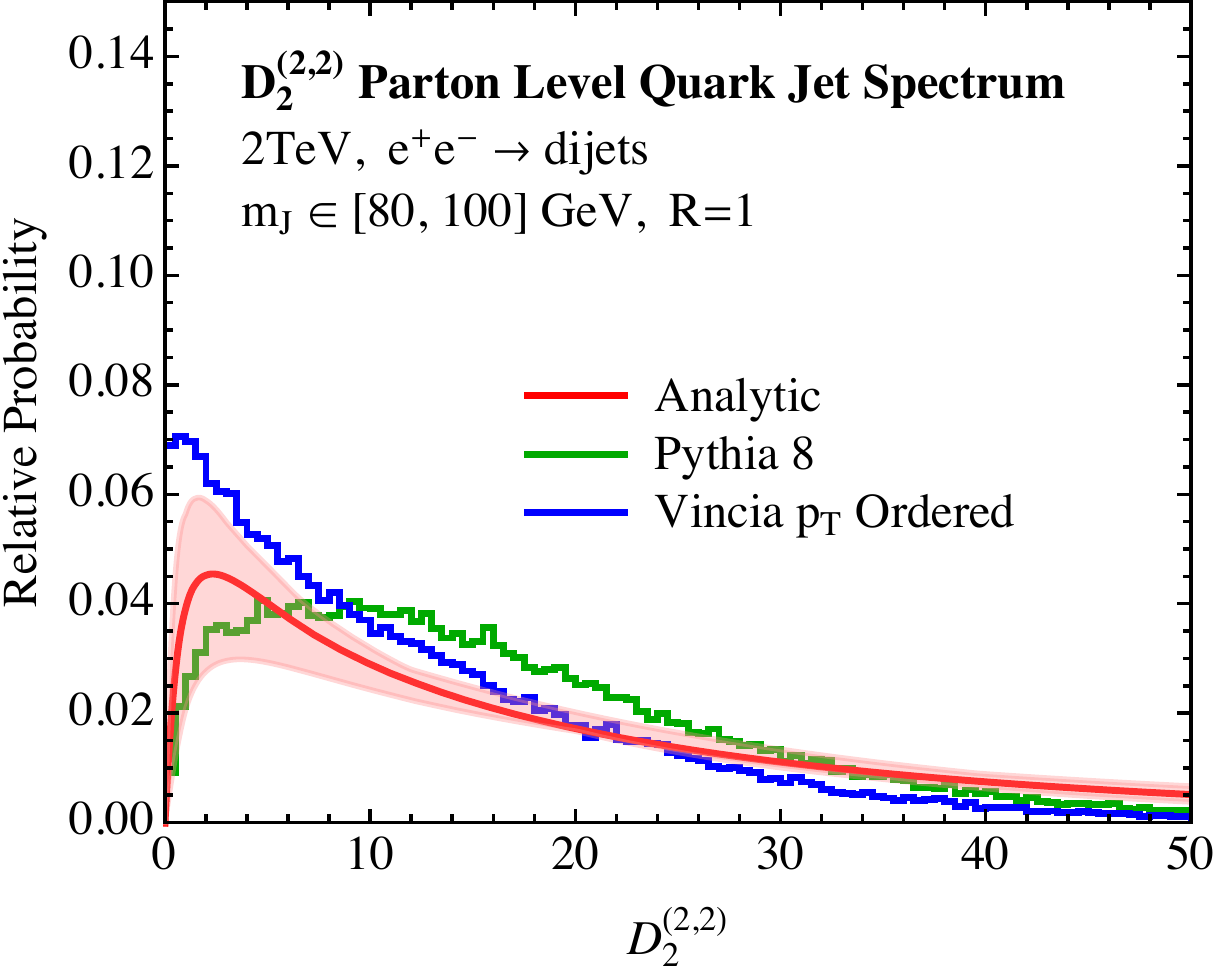}
}
\end{center}
\caption{ Comparison of QCD background $D_2$ distributions from $p_T$-ordered \vincia{} and \pythia{} to our analytic prediction as a function of the jet energy, $E_J$. The values $E_J=500$ GeV, and $2$ TeV are shown in Figures a) and b), respectively. A jet radius of $R=1$ is used for all values of the jet energy.
}
\label{fig:E_dependence}
\end{figure}

In \Fig{fig:E_dependence} we show distributions for the $D_2$ observable as obtained from Monte Carlo simulation, and compared with our analytic calculation. As in \Sec{sec:R_dependence}, we restrict to $p_T$-ordered \vincia{} and \pythia{} at parton level. The perturbative scale variations for each energy value are less extensively explored and are only meant to provide a rough estimate of the perturbative uncertainty.  The evolution of the difference between the \vincia{} and \pythia{} generators is again quite fascinating, with the discrepancy between the generators increasing significantly with energy, to the point that at $2$ TeV the qualitative shape of the distributions doesn't agree. In particular, the behavior at small $D_2$ is completely different between the two generators, with \vincia{} having a large peak, which is not present in \pythia{}.

As discussed in \Sec{sec:scales}, this discrepancy between the generators is evidence that wide angle soft radiation is not accurately modeled by the \pythia{} generator, as compared with \vincia{}.   The phase space for wide angle soft radiation drastically increases as the energy of the jet is increased, with a fixed jet mass and radius. As evidence that this is indeed the cause of the discrepancy, we have checked that the conclusions of \Sec{sec:R_dependence} remains true at higher energy, as long as the jet radius is taken to scale as $R \sim 2m_J/p_T$, so that it constrains the wide angle soft radiation. For example, for $R=0.2$ at $2$ TeV, we find excellent agreement between the $D_2$ distributions as generated by \pythia{} and \vincia{}.\footnote{We include this plot in \App{app:twoemissionMC}, \Fig{fig:D2_app_R021}, for reference.} However, as was emphasized in \Sec{sec:scales}, since this study is performed at parton level, differences in the cutoff of the perturbative shower between the different generators also contribute to this difference. Because the fact that the disagreement is so large between the generators, and is arising from the modeling of soft radiation, this may be an excellent observable to study soft radiation and color coherence in parton showers.

As a reference, in \App{app:twoemissionMC} we show distributions of the $\ecf{2}{2}$ observable, measured at both $500$ GeV, and $2$ TeV for both the \vincia{} and \pythia{} Monte Carlos, and at both parton and hadron level. Unlike for the $D_2$ observable, since $\ecf{2}{2}$ is set by a single emission, excellent agreement is observed for the $\ecf{2}{2}$ observables between \pythia{} and \vincia{} both at parton level, emphasizing that $D_2$ offers a more differential probe of the perturbative shower than single emission observables.

Our analytic predictions at $2$ TeV, as shown in \Fig{fig:E_dependence}, are intermediate between the \pythia{} and \vincia{} results. They exhibit a peaked structure at small values of $D_2$, but not to the extent seen in the \vincia{} distribution. We believe that this is largely due to the normalization of the distributions, and the fact that we do not match to fixed order in the tail of the distribution. Since this tail becomes longer at higher energies, a larger disagreement in the peak region is also seen. However, the shape agrees qualitatively with the \vincia{} result. On the other hand, at $500$ GeV, our analytic prediction has a large peak. This is evidence that because the $D_2$ spectrum is much more sharply peaked at $500$ GeV, higher order resummation may be more important in the peak region. However, the relatively good agreement between analytics and Monte Carlo shows that our factorization theorem is able to accurately capture the energy dependence over a large range of energies.

The results for both the jet radius and jet energy dependence of the $D_2$ distributions demonstrate that the extent to which the perturbative parton shower is able to describe QCD jets in the two-prong regime depends strongly on the parameters of the jet.  In some regions of the phase space, there is not even a qualitative agreement between different generators. This is important to keep in mind for jet substructure studies based solely on a single Monte Carlo generator, and also emphasizes the importance of analytic calculations for jet substructure, in particular in regions of phase space where there is considerable disagreement between the Monte Carlo generators. It is important to note that hadronization will remove some of the discrepancies in the $D_2$ distributions between the \vincia{} and \pythia{} generators, especially at high energies, where it will smear out the peak at low values of $D_2$. While this improves qualitatively the behavior of the distributions, discrepancies in the shape still remain. This will be discussed in detail in \Sec{sec:Hadronization}, along with its incorporation into our analytic calculation.  For comparison to precision analytic calculations and interpreting data, it is vital that Monte Carlo generators provide accurate descriptions of both the perturbative and non-perturbative aspects of QCD jets, and not compensating for perturbative discrepancies by the tuning of non-perturbative parameters. This is especially important for disentangling non-perturbative effects from perturbative effects, the latter of which should in principle be under much better control, and for extracting reliable information about non-perturbative QCD from jet physics.

Throughout the rest of the chapter we will focus on jets with radius $R=1$ at a center of mass energy of $1$ TeV.

\subsection{Impact of Hadronization}\label{sec:Hadronization}

Hadronization plays an important role in a complete description of any jet observable, and a description of non-perturbative effects, preferably from a field-theoretic approach, is required to compare with experimental data. An advantage of the factorization approach taken in this chapter is that it allows for a clean separation of perturbative and non-perturbative physics. Non-perturbative effects enter the factorization theorems presented in \Sec{sec:Fact} through the soft function, which describes the dynamics of soft radiation, both perturbative and non-perturbative, between the jets. For a large class of additive observables, the treatment of non-perturbative physics in the soft function is well-understood, and can be incorporated using shape functions \cite{Korchemsky:1999kt,Korchemsky:2000kp,Bosch:2004th,Hoang:2007vb,Ligeti:2008ac}. Shape functions have support over a region of size $\Lambda_{\text{QCD}}$, and are convolved with the perturbative soft function. In the tail region of the distribution, where the observable is dominated by perturbative emissions, they reduce to a shift.  For a large class of observables, this shift is determined by a universal \cite{Akhoury:1995sp,Dokshitzer:1995zt} non-perturbative parameter multiplied by a calculable, observable dependent number \cite{Dokshitzer:1995zt,Lee:2006fn,Lee:2007jr}. Similar shape functions have also been used to incorporate the effects of pile-up and the underlying event at hadron colliders \cite{Stewart:2014nna}.

The effect of non-perturbative physics on multi-differential cross sections has not been well-studied. For the double differential cross section of two angularities,  \Ref{Larkoski:2013paa} considered using uncorrelated shape functions for each angularity individually, but it is expected that a complete description would require a shape function incorporating non-perturbative correlations between observables. For the particular case of the $D_2$ observable, we will argue that a single parameter shape function can be used to accurately describe the dominant non-perturbative effects, and in particular, that a study of multi-differential shape functions with non-perturbative correlations, is not required.

In \Sec{sec:fixed_order} we performed a study of the fixed order singular structure of the $D_2$ observable in the presence of a jet mass cut. Importantly, we showed that $D_2$ only has a singularity at $D_2=0$, with its behavior at all other values regulated by the mass cut. Non-perturbative corrections to the $D_2$ observable will play an important role only when the soft scale becomes non-perturbative, which as just argued, for a perturbative mass cut of the form studied in this chapter, only occurs as $D_2 \to 0$. Recall that the $D_2$ observable is defined as 
\begin{equation}
\Dobs{2}{\alpha, \beta}= \frac{\ecf{3}{\alpha}}{(\ecf{2}{\beta})^{3\alpha/\beta}}\,,
\end{equation}
which is not additive. However, in the two-prong region of phase space, namely $D_2\to 0$, the value of $\ecf{2}{\beta}$ is set to leading power by the hard splitting, and so $D_2$ effectively reduces to an additive observable. In this region of phase space the description of non-perturbative effects in terms of a shape function can therefore be rigorously justified from our factorization theorem, and it can be applied directly to the $D_2$ distribution.  For large values of $D_2$, it is not additive, and the use of a shape function cannot be formally justified. However, in this region, a shape function is not required, as any singular behavior is regulated by a mass cut. We therefore will use a shape function that falls off exponentially at large values of $D_2$.  We believe that this is a self-consistent approach until non-perturbative corrections to multi-differential cross sections are better understood.

In the two-prong region of phase space, we have shown that two distinct factorization theorems, namely the soft subjet and collinear subjets, are required, and in \Sec{sec:soft_NINJA_match} we showed how these two descriptions can be merged to provide a complete description of the two-prong region of phase space. Importantly, the two factorization theorems describing the two-prong region of phase space have soft functions with different numbers of Wilson lines. The collinear subjets soft function is a two-eikonal line soft function, while the soft subjet soft function has three eikonal lines. Since the shape function describes the non-perturbative contribution to the soft function, in general we should allow for two distinct shape functions, with independent parameterizations. The zero-bin merging procedure in \Sec{sec:soft_NINJA_match} would then be performed on the non-perturbative cross sections, after convolution with the appropriate shape function. However, at the level of perturbative accuracy which we work, and because we will simply be extracting our shape function parameters by comparing to Monte Carlo, the use of distinct parameterizations of different shape functions for both the soft subjet and collinear subjets soft functions would introduce many redundant parameters. To simplify the situation in this initial investigation, we will choose to use the same parametrization of the shape function, and the same non-perturbative parameters for both soft functions. This allows for the non-perturbative corrections to be described by a single parameter, and as we will see provides an excellent description of the Monte Carlo data. Because we use the same shape function for both the soft subjet, and collinear subjets soft functions, it also implies that the shape function can be applied after the zero bin merging procedure, namely, directly at the level of the $D_2$ distribution.

As a simple parametrization of a shape function for $D_2$, we follow \Ref{Stewart:2014nna} and consider
\begin{align}\label{eq:shape_func}
F(\epsilon)=\frac{4\epsilon}{\Omega_D^2}e^{-2\epsilon/\Omega_D}\,,
\end{align}
where $\epsilon$ is the energy and $\Omega_D\sim \Lambda_{\text{QCD}}$ is a non-perturbative scale. Note that while we will use the same value of $\Omega_D$ for the signal and background distributions, it will have very different effects on the two distributions, which will arise naturally from the power counting in the different factorization theorems, as will be shown in this section. The function of \Eq{eq:shape_func} satisfies the required properties that it is normalized to $1$, has a finite first moment $\Omega_D$, vanishes at $\epsilon=0$, and falls off exponentially at high energies \cite{Hoang:2007vb}. More general bases of shape functions are discussed in \Ref{Ligeti:2008ac}, although we find that the single parameter shape function of \Eq{eq:shape_func} is sufficient to describe the dominant effects of hadronization.  

As discussed above, we will use the shape function of \Eq{eq:shape_func} for both the collinear subjets and soft subjets factorization theorems, with the same value of $\Omega_D$ in both cases. Because we have enforced this simplification to reduce the number of parameters, it is then most interesting to focus on $\Omega_D$ for the collinear subjets factorization theorem, which has two eikonal lines. In this case, we can show that we can relate the $\Omega_D$ parameter to universal non-perturbative parameters appearing in $e^+e^-\to$ dijet factorization theorems, which have been measured in experiment. Therefore, throughout the rest of this section, we will focus on deriving scaling relations for $\Omega_D$, assuming we are working in the collinear subjets factorization theorem. Again, we wish to emphasize that this is merely a simplification we have made to reduce the number of parameters, and a more general treatment could be performed, but we will see that with only the single $\Omega_D$, with properties derived assuming the collinear subjets factorization theorem, excellent agreement with Monte Carlo is observed.

The effect of non-perturbative physics as modeled by the shape function is very different for background or signal distributions.  For background, when $D_2$ is small, the contribution to $\ecf{3}{\alpha}$ from a non-perturbative soft emission is
\begin{equation}
\left.\ecf{3}{\alpha}\right|_\text{np}\sim \frac{\epsilon}{E_J} \ecf{2}{\alpha}\,,
\end{equation}
where $\epsilon$ is the energy of the non-perturbative emission and $E_J$ is the energy of the jet, as shown in \Eq{eq:soft_scale_coll}.  The non-perturbative contribution to $D_2$ is therefore
\begin{equation}
\left.\Dobs{2}{\alpha,\beta}\right|_\text{np} = \frac{\left.\ecf{3}{\alpha}\right|_\text{np}}{(\ecf{2}{\beta})^{3\alpha/\beta}} \sim \frac{\epsilon}{E_J}\frac{ \ecf{2}{\alpha}}{(\ecf{2}{\beta})^{3\alpha/\beta}} \,.
\end{equation}
In terms of the shape function, the non-perturbative distribution of $D_2$ for background jets can then be written as a convolution:\footnote{In this initial investigation we do not include a gap in our shape function, which would implement a minimum hadronic energy deposit, as expected physically \cite{Hoang:2007vb}. Such gapped shape functions, and their associated renormalon \cite{Beneke:1998ui} ambiguity \cite{Gardi:2000yh} have been studied for arbitrary angular exponents \cite{Hornig:2009vb}, and could be straightforwardly incorporated in our analysis. However, we observe excellent agreement with our single parameter shape function, which we therefore find to be sufficient for our purposes.   }
\begin{equation}\label{eq:shape_bkgd}
\frac{d\sigma_{\text{np}}  }{d\Dobs{2}{\alpha, \beta}}=\int_0^\infty d\epsilon \,   F(\epsilon)\,\frac{d\sigma_{\text{p}} \left ( \Dobs{2}{\alpha, \beta}-\frac{\epsilon}{E_J}\frac{\ecf{2}{\alpha}}{(\ecf{2}{\beta})^{3\alpha/\beta}} \right)  }{d\Dobs{2}{\alpha, \beta}}  \,,
\end{equation}
where $\sigma_{\text{np}}$ and $\sigma_{\text{p}}$ denote the non-perturbative and perturbative cross sections, respectively.

We can estimate the scale at which the global softs of the collinear subjets factorization theorem become non-perturbative from the scaling of the modes given in \Eq{eq:cs_collinear_and_soft}. Rewriting this scaling in terms of the center of mass energy of the $e^+e^-$ collision, $Q$, and $D_2$, we find that the global soft scale of the collinear subjets factorization theorem has virtuality
\begin{align}
\mu_S=2^3\, D_2\, m_Z \left(   \frac{m_Z}{Q} \right)^3\,,
\end{align}
where we have assumed a jet mass, $m_J=m_Z$, as relevant for boosted $Z$ discrimination. Taking $\Lambda_{\text{QCD}}=500$ MeV, we find that the global soft scale enters the non-perturbative regime at $D_2 \simeq 1$.   

Restricting to $\beta = 2$, in the collinear subjets region of the background jet phase space, the non-perturbative distribution of $\Dobs{2}{\alpha, 2}$ is then
\begin{equation}\label{eq:shape_function_eq1}
\frac{d\sigma_{\text{np}}  }{d\Dobs{2}{\alpha, 2}}=\int_0^\infty d\epsilon \,   F(\epsilon)\,\frac{d\sigma_{\text{p}} \left ( \Dobs{2}{\alpha, 2}-2^{\alpha -2}\frac{\epsilon}{E_J}\frac{E_J^{2\alpha}}{  m_J^{2\alpha}  } \right)  }{d\Dobs{2}{\alpha, 2}}  \,,
\end{equation}
where we have used 
\begin{equation}
\ecf{2}{2} = \frac{m_J^2}{E_J^2} \,,
\end{equation}
and that, in the collinear subjets region of phase space,
\begin{equation}
\ecf{2}{\alpha} \simeq 2^{\alpha-2}\left(
\ecf{2}{2}
\right)^{\alpha/2}\,.
\end{equation}
Because we consider fixed-energy jets with masses in a narrow window, $\ecf{2}{2}$ is just a number and can be removed by appropriate change of variables.  Making this change, we then have
\begin{equation}\label{eq:np_background}
\frac{d\sigma_{\text{np}}  }{d\Dobs{2}{\alpha, 2}}=\int_0^\infty d\epsilon \,   F(\epsilon)\,\frac{d\sigma_{\text{p}} \left ( \Dobs{2}{\alpha, 2}-\frac{\epsilon}{E_J} \right)  }{d\Dobs{2}{\alpha, 2}}  \,,
\end{equation}
where the non-perturbative parameter in the shape function is effectively modified to
\begin{equation}
\tilde \Omega_D =  2^{\alpha - 2}\frac{    \Omega_D }{ \frac{m_J^{2\alpha}}{E_J^{2\alpha}}  }\,.
\end{equation}

The non-perturbative parameter $\Omega_D$ still has implicit dependence on the angular exponent $\alpha$.  Because the global soft modes have the lowest virtuality and can only resolve the back-to-back soft Wilson lines in the $n$ and $\bar n$ directions, we can use the results of \Refs{Lee:2006fn,Lee:2007jr} to extract the $\alpha$ dependence.  By the boost invariance of the soft function\footnote{This boost invariance holds strictly only for a soft function with no jet algorithm restrictions. However, since we are considering fat jets close to hemispherical, we expect corrections to the boost invariance of the soft function to be small.} along the $n-\bar n$ directions and the form of the observable $\ecf{3}{\alpha}$ as measured on soft particles, it follows that $\Omega_D$ takes the form
\begin{equation}\label{eq:alpha_np}
\Omega_D = \frac{3}{2\alpha-1}\Sigma \,,
\end{equation}
where $\Sigma$ is a universal non-perturbative matrix element of two soft Wilson lines and all dependence on $\alpha$ has been extracted.\footnote{In this section we ignore the effects of hadron masses, and their associated power corrections of $\mathcal{O}(m_H/Q)$, where $m_H$ is the mass of a stable hadron in the jet. While these power corrections can also be incorporated through the shape function, in general, they break the universality of the non-perturbative matrix element, $\Sigma$ \cite{Salam:2001bd,Mateu:2012nk}. In particular, \Eq{eq:alpha_np} is no longer in general true, for a $\Sigma$ that is independent of the angular exponent $\alpha$ \cite{Salam:2001bd,Mateu:2012nk}. This depends on the precise definition of the energy correlation functions for massive particles. However, the value of $\Sigma$ can still be extracted from dijet event shapes in the same universality class as a particular angularity \cite{Mateu:2012nk}. Furthermore, $\Omega_D$ has a scale dependence from renormalization group evolution, $\Omega_D=\Omega_D(\mu)$, although this dependence is logarithmic, and is therefore small compared to our uncertainties. We will discuss briefly the impact of hadron masses and the renormalization group evolution of $\Omega_D$ in \Sec{sec:LEP}, and in \App{app:shape_RGE}.}  We have normalized the matrix element such that the coefficient is unity for $\alpha = 2$. We will shortly discuss the extent to which the values of $\Omega_D$ we obtain from comparison with the parton shower agree with the known values of this universal non-perturbative matrix element.

For signal jets, the lowest virtuality mode in the jet are the collinear-soft modes. Unlike the global soft modes of the collinear subjets factorization theorem, which did not resolve the substructure of the jets, allowing us to relate the non-perturbative parameter appearing in the shape function to that appearing in dijet event shapes, the collinear soft modes in the signal factorization theorem resolve the jet substructure. However, since the decaying boson is a color singlet, there are still only two eikonal lines present in the factorization theorem. Boost invariance of the soft function will therefore again allow us to relate the non-perturbative parameter for the signal distribution to that appearing in dijet event shapes. This is similar to the argument used in \Ref{Feige:2012vc} to calculate the signal distribution for $2$-subjettiness.

A non-perturbative collinear-soft emission contributes to $\ecf{3}{\alpha}$ as
\begin{equation}
\left.\ecf{3}{\alpha}\right|_\text{np}\sim \frac{\epsilon}{E_J} (\ecf{2}{\alpha})^3\,,
\end{equation}
where now $\epsilon$ is the energy of the non-perturbative collinear-soft emission, as shown in \Eq{eq:cs_cs}.  The non-perturbative contribution to $D_2$ for signal jets is therefore
\begin{align}
\left.\Dobs{2}{\alpha,\beta}\right|_\text{np} &= \frac{\left.\ecf{3}{\alpha}\right|_\text{np}}{(\ecf{2}{\beta})^{3\alpha/\beta}} \sim \frac{\epsilon}{E_J}\frac{ (\ecf{2}{\alpha})^3}{(\ecf{2}{\beta})^{3\alpha/\beta}} \\
&\simeq 2^{3(\alpha-\beta)}\frac{\epsilon}{E_J}\nonumber \,,
\end{align}
where in the second line we have used the parametric relationship between $\ecf{2}{\alpha}$ and $\ecf{2}{\beta}$ in the collinear subjets region.
Convolving with the shape function, the non-perturbative distribution for signal jets is then
\begin{equation}\label{eq:np_signal}
\frac{d\sigma_{\text{np}}  }{d\Dobs{2}{\alpha, \beta}}=\int_0^\infty d\epsilon \,   F(\epsilon)\,\frac{d\sigma_{\text{p}} \left ( \Dobs{2}{\alpha, \beta}-2^{3(\alpha-\beta)}\frac{\epsilon}{E_J} \right)  }{d\Dobs{2}{\alpha, \beta}}  \,.
\end{equation}

It is important to note how the different scales for the soft radiation in the case of the signal and background jets leads to different behavior of the $D_2$ distributions after hadronization. In particular, from \Eqs{eq:np_background}{eq:np_signal} one can determine the shift in the first moment of the $D_2$ distribution caused by hadronization, which we will denote by $\Delta_D$. Restricting to the case $\alpha=\beta=2$ for simplicity, we find that for the background distribution, 
\begin{equation}\label{eq:bkg_shift_np}
\Delta_D=\frac{\Omega_D}{ E_J \left(  \frac{m_J}{E_J} \right)^4}\,,
\end{equation}
whereas for the signal jets, we have
\begin{equation} \label{eq:signal_nonpert}
\Delta_D=\frac{\Omega_D}{ E_J }\,.
\end{equation}
Since $\Omega_D$ should be of the scale $1$ GeV, we see that for signal jets, the shift in the first moment due to hadronization is highly suppressed, and behaves differently than a traditional event shape due to the boost factor, while for background jets, since $m_J \ll E_J$, the effect of hadronization is significant. We will see that both of these features, which are consequences of the power counting of the dominant modes, are well reproduced in the Monte Carlo simulations.

\begin{figure}
\begin{center}
\subfloat[]{\label{fig:D2_ee_hadbkg_a}
\includegraphics[width= 7.25cm]{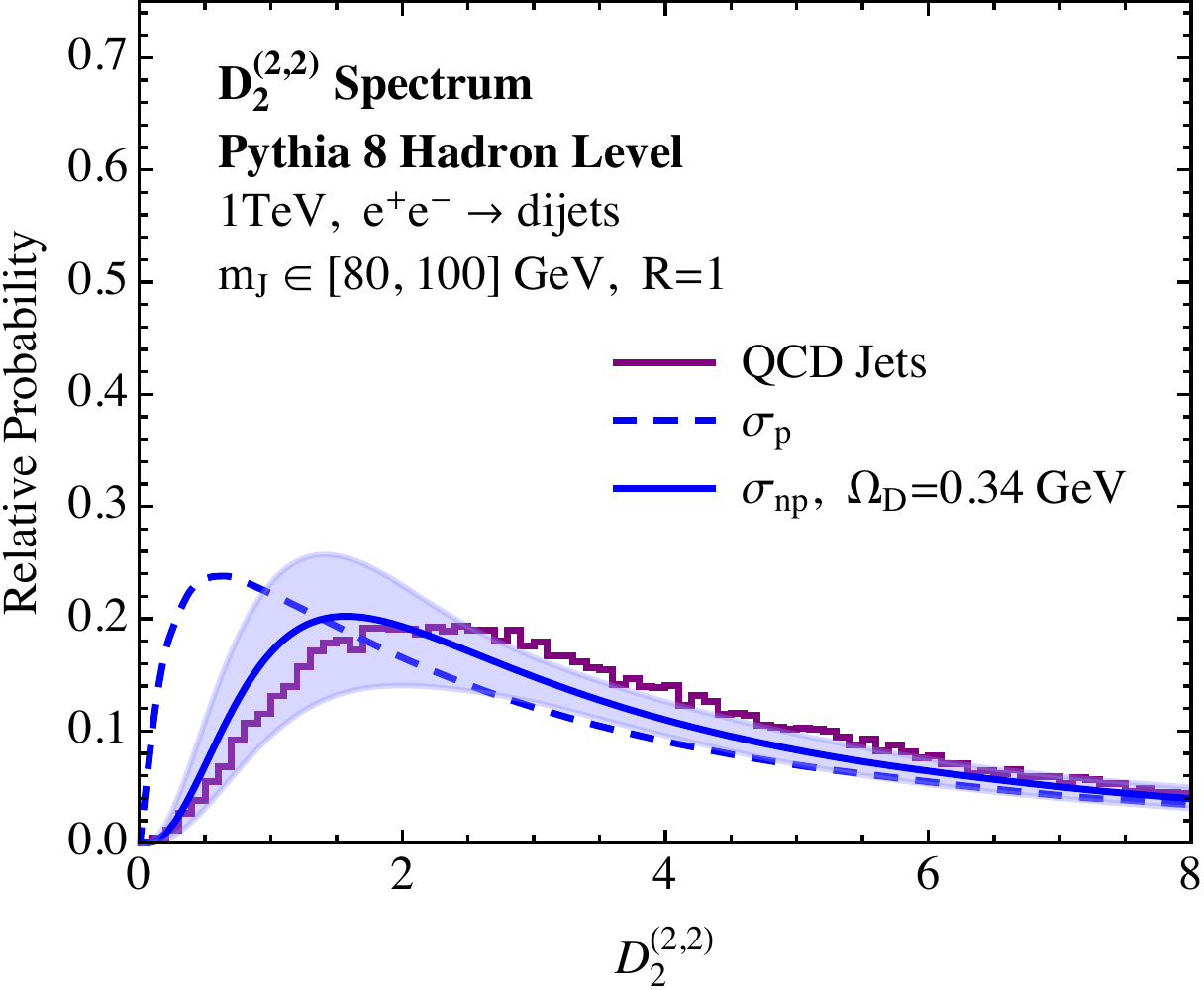}
}
\subfloat[]{\label{fig:D2_ee_hadbkg_b}
\includegraphics[width= 7.25cm]{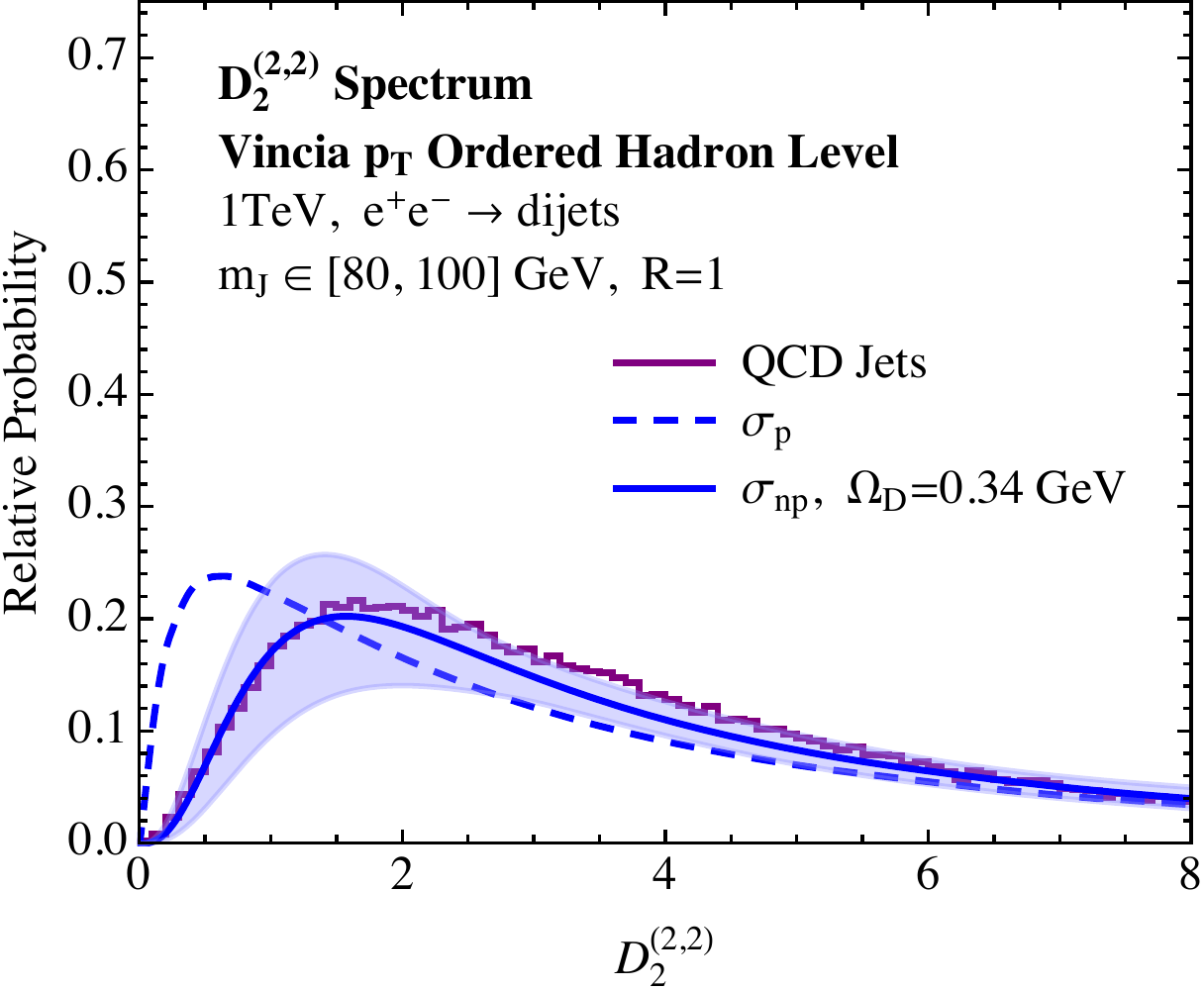}
}\\
\subfloat[]{\label{fig:D2_ee_hadbkg_c}
\includegraphics[width= 7.25cm]{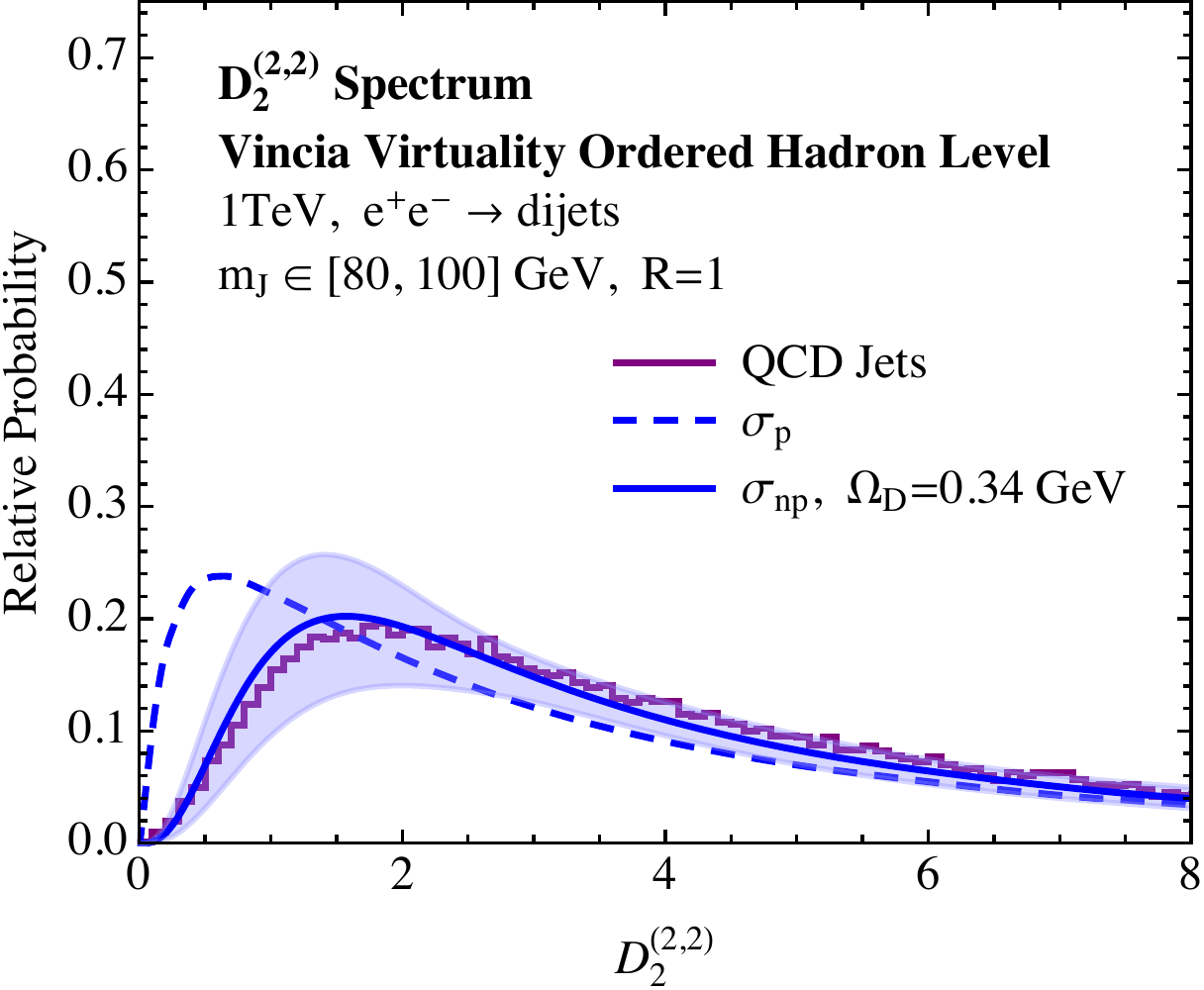}
}
\subfloat[]{\label{fig:D2_ee_hadbkg_d}
\includegraphics[width= 7.25cm]{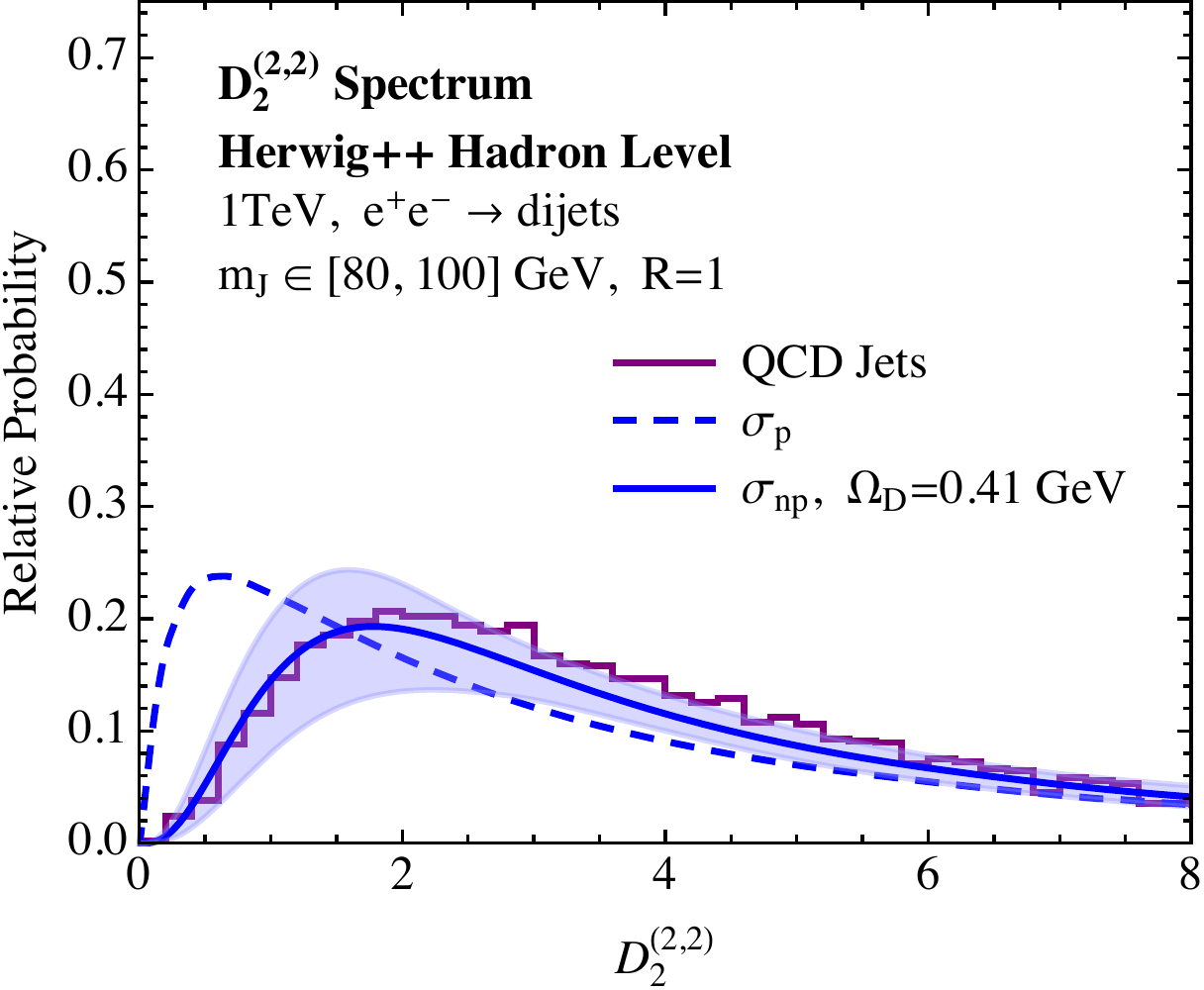}
}
\end{center}
\vspace{-0.2cm}
\caption{A comparison of the $\Dobs{2}{2,2}$ distributions for background QCD jets from our analytic prediction and the various hadron-level Monte Carlos.  $\sigma_p$ denotes the parton level perturbative prediction for the distribution and $\sigma_{np}=\sigma_p\otimes F_D$ is the perturbative prediction convolved with the non-perturbative shape function.  The values of the non-perturbative parameter $\Omega_D$ used are also shown.
}
\label{fig:D2_hadr}
\end{figure}

\begin{figure}
\begin{center}
\subfloat[]{\label{fig:D2_ee_hadsig_a}
\includegraphics[width = 7.25cm]{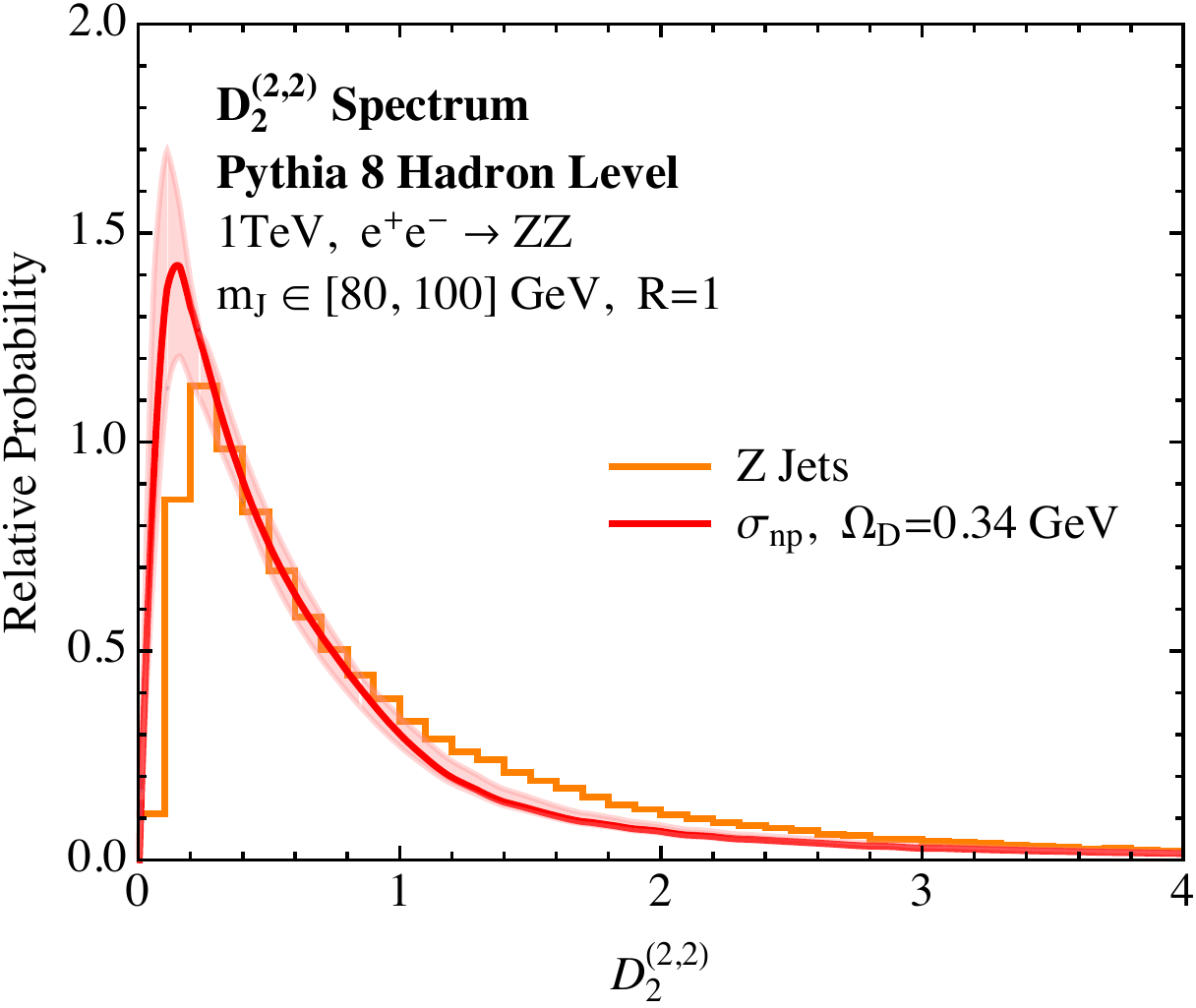}
}
\subfloat[]{\label{fig:D2_ee_hadsig_b}
\includegraphics[width = 7.25cm]{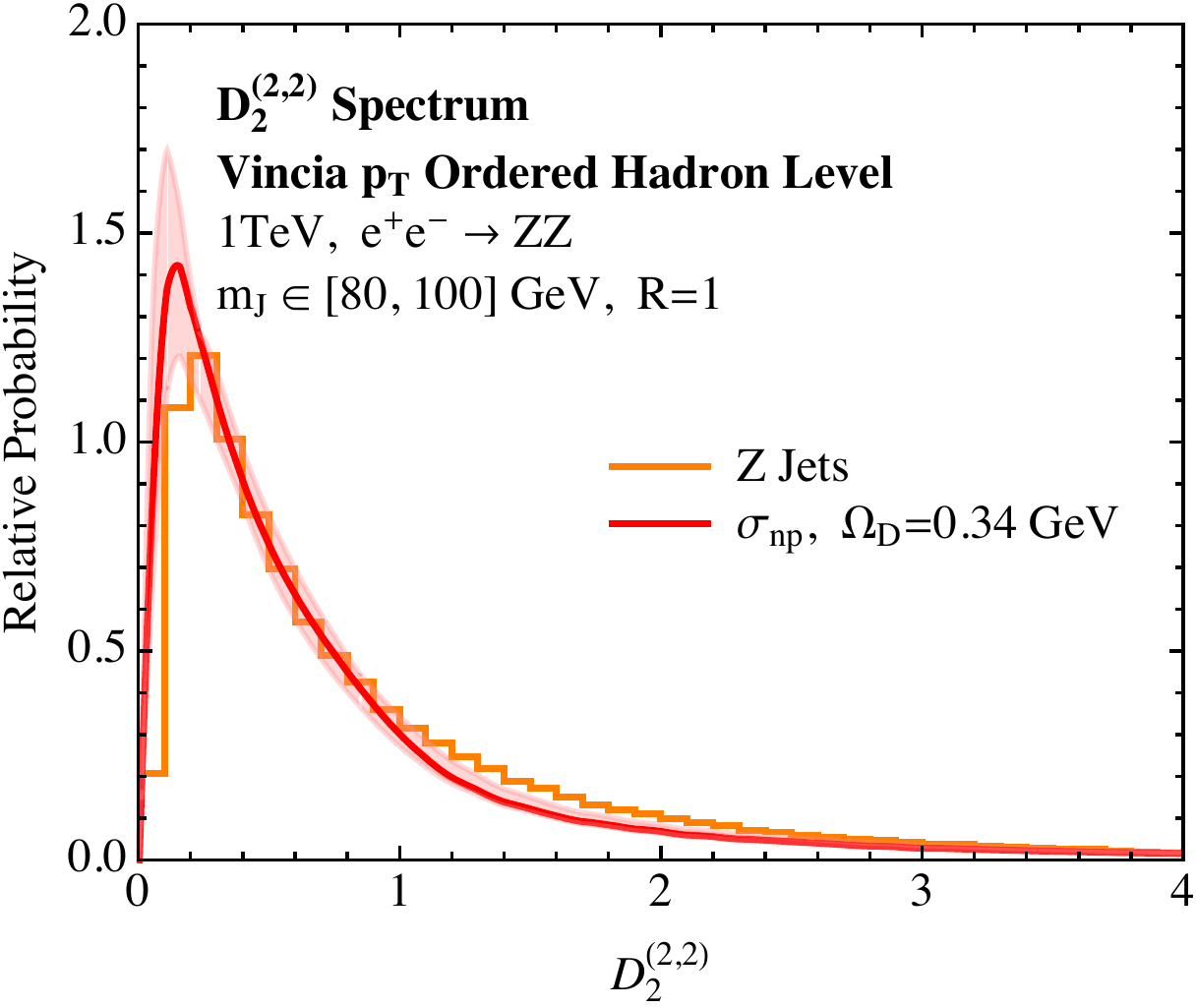}
}\\
\subfloat[]{\label{fig:D2_ee_hadsig_c}
\includegraphics[width = 7.25cm]{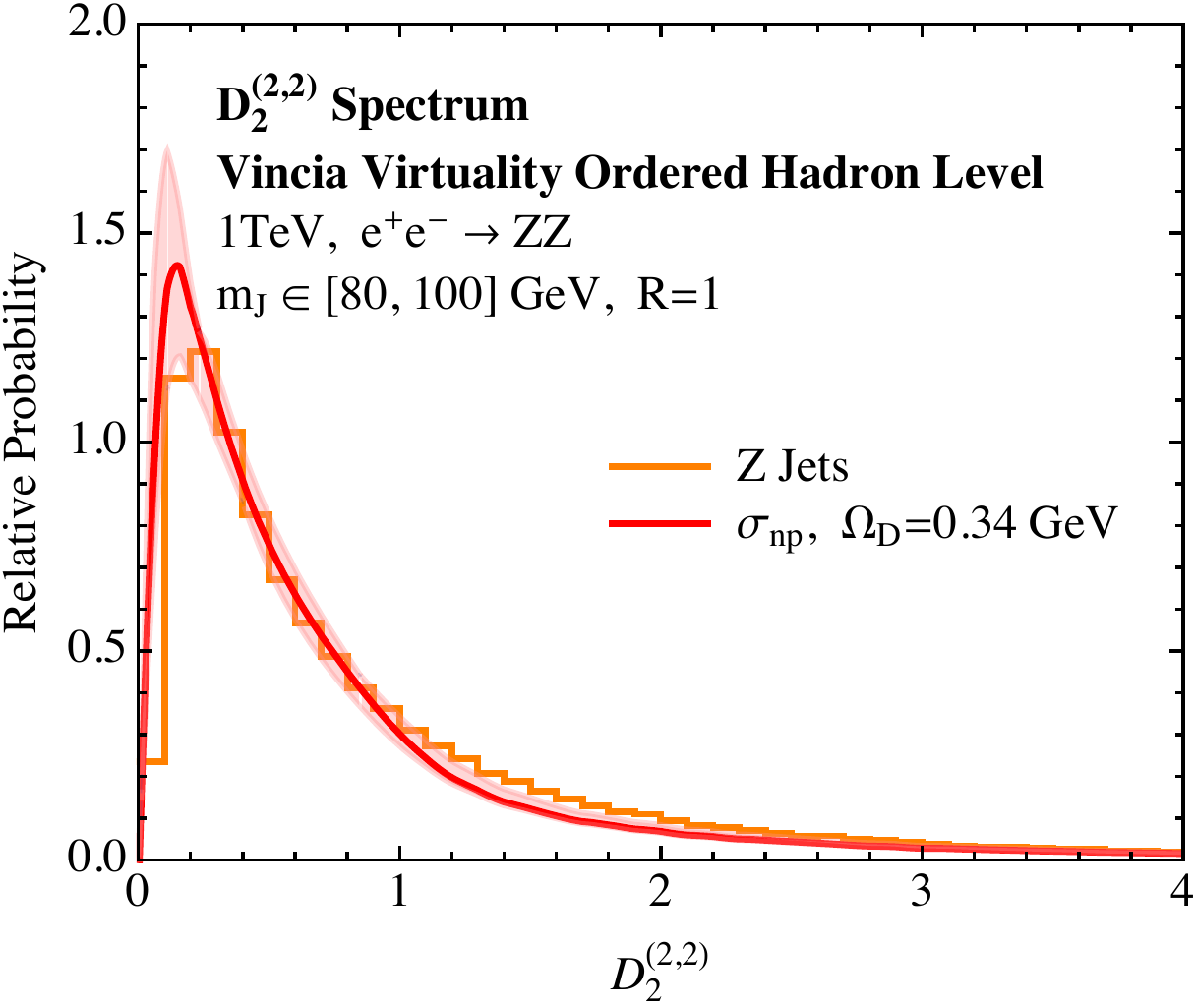}
}
\subfloat[]{\label{fig:D2_ee_hadsig_d}
\includegraphics[width = 7.25cm]{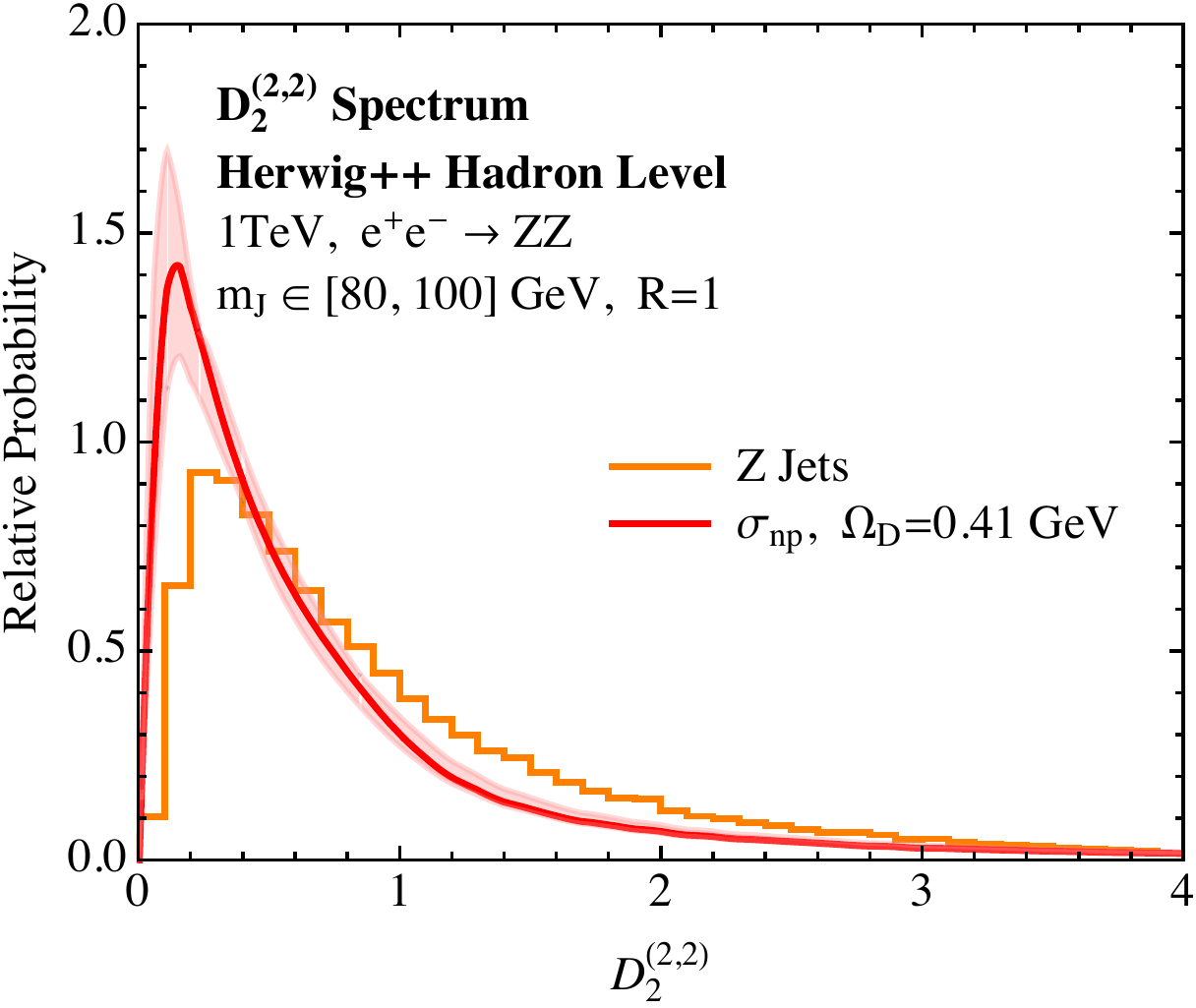}
}
\end{center}
\vspace{-0.2cm}
\caption{
A comparison of the $\Dobs{2}{2,2}$ distributions for signal boosted $Z$ jets from our analytic prediction and the various hadron-level Monte Carlos.  $\sigma_p$ denotes the parton level perturbative prediction for the distribution and $\sigma_{np}=\sigma_p\otimes F_D$ is the perturbative prediction convolved with the non-perturbative shape function, although for the signal this has a negligible effect.  The values of the non-perturbative parameter $\Omega_D$ used are also shown.
}
\label{fig:D2_hadr_signal}
\end{figure}

Comparisons between the hadron-level distributions of $\Dobs{2}{2,2}$ from our analytic calculations and the Monte Carlos are presented in \Fig{fig:D2_hadr} for background and \Fig{fig:D2_hadr_signal} for signal jets.  For background distributions, we compare our perturbative calculation convolved with the shape function, as defined in \Eq{eq:shape_bkgd}. Both \vincia{} and \pythia{} use the same hadronization model, but \herwigpp{} uses a distinct hadronization model, and therefore we allow for a different shape parameter, $\Omega_D$, for the two cases. For the case of \pythia{} and \vincia{}, because we find the best agreement in the shape of the perturbative spectrum, with parton level \vincia{} with $p_T$ ordering, we choose to extract the value of $\Omega_D$ by fitting to the hadronized distribution for $p_T$ ordered \vincia{}. However, we will shortly discuss the level of ambiguity in $\Omega_D$ arising from this extraction.  For jets with an energy of 500 GeV and mass of 90 GeV, we find that the choice $\Omega_D = 0.34\pm 0.03$ GeV provides the best agreement of our perturbative calculation with $p_T$ ordered \vincia{}, while $\Omega_D=0.41 \pm 0.03$ GeV provides the best agreement with \herwigpp{}. The errors assigned here come only from the fitting itself, and are due to the statistical uncertainties of the Monte Carlo distributions due to the finite width of the histogram bins.  These errors do not take into account any other uncertainties; for example, whether one should perform the fit to hadron level \vincia{} or \pythia{}. This level of agreement between the non-perturbative parameters extracted from \pythia{} and \herwigpp{} is comparable to more detailed studies, such as \Ref{Mateu:2012nk}.  A comparison of the distributions of \Fig{fig:D2_hadr} before and after hadronization shows that hadronization has a considerable effect on the background distributions, particularly at small values of $D_2$, as expected from \Eq{eq:bkg_shift_np}. This effect, which in the Monte Carlos is realized through tuned hadronization models,  is well described by the single parameter shape function. Importantly, as discussed above, if different shape parameters were used for the collinear subjets and soft subjets factorization theorems, they would be nearly degenerate in the fit at the level of perturbative accuracy that we work, which is why we have made the simplification of working with a single non-perturbative shape parameter.

We have argued that the non-perturbative parameter $\Omega_D$ in the collinear subjets factorization theorem can be related to a universal non-perturbative matrix element of two soft Wilson lines.  Such non-perturbative matrix elements appear in the factorization theorems of a large class of $e^+e^-$ event observables, and has therefore been measured from data at LEP.\footnote{An extremely large literature exists on such measurements, and their theoretical interpretation, to which we cannot do justice in this brief section. We refer the reader to, for example, \Refs{Korchemsky:1999kt,Korchemsky:2000kp,Achard:2004sv,Gehrmann:2009eh,Abbate:2010xh,Abbate:2012jh,Hoang:2014wka,Hoang:2015hka} and references therein.} While the value of $\Omega_D$ that we have determined for the two parton showers is by no means precise, it is interesting to compare our value with those extracted from precision studies of $e^+e^-$ collider observables which have been performed in the literature. Using the particular case of $\alpha=\beta=2$, and converting to our normalization, a recent extraction of the non-perturbative parameter from an N$^3$LL$'$ analysis of the $C$-parameter event shape using LEP data, and including power corrections and hadron mass effects \cite{Salam:2001bd,Mateu:2012nk}, gives a value of $\Omega_D = 0.28$ GeV \cite{Hoang:2014wka,Hoang:2015hka}. This agrees well with our values extracted through comparison with Monte Carlo.
Going forward, with the goal of increasing both the precision and understanding of jet substructure, the ability to relate the dominant non-perturbative corrections to the $D_2$ observable to known non-perturbative parameters measured in $e^+e^-$ is a valuable feature, and that further study on the non-perturbative corrections to multi-differential cross sections is of great importance.

\begin{figure}
\begin{center}
\subfloat[]{\label{fig:D2_ee_hadbkg_a2}
\includegraphics[width= 7.25cm]{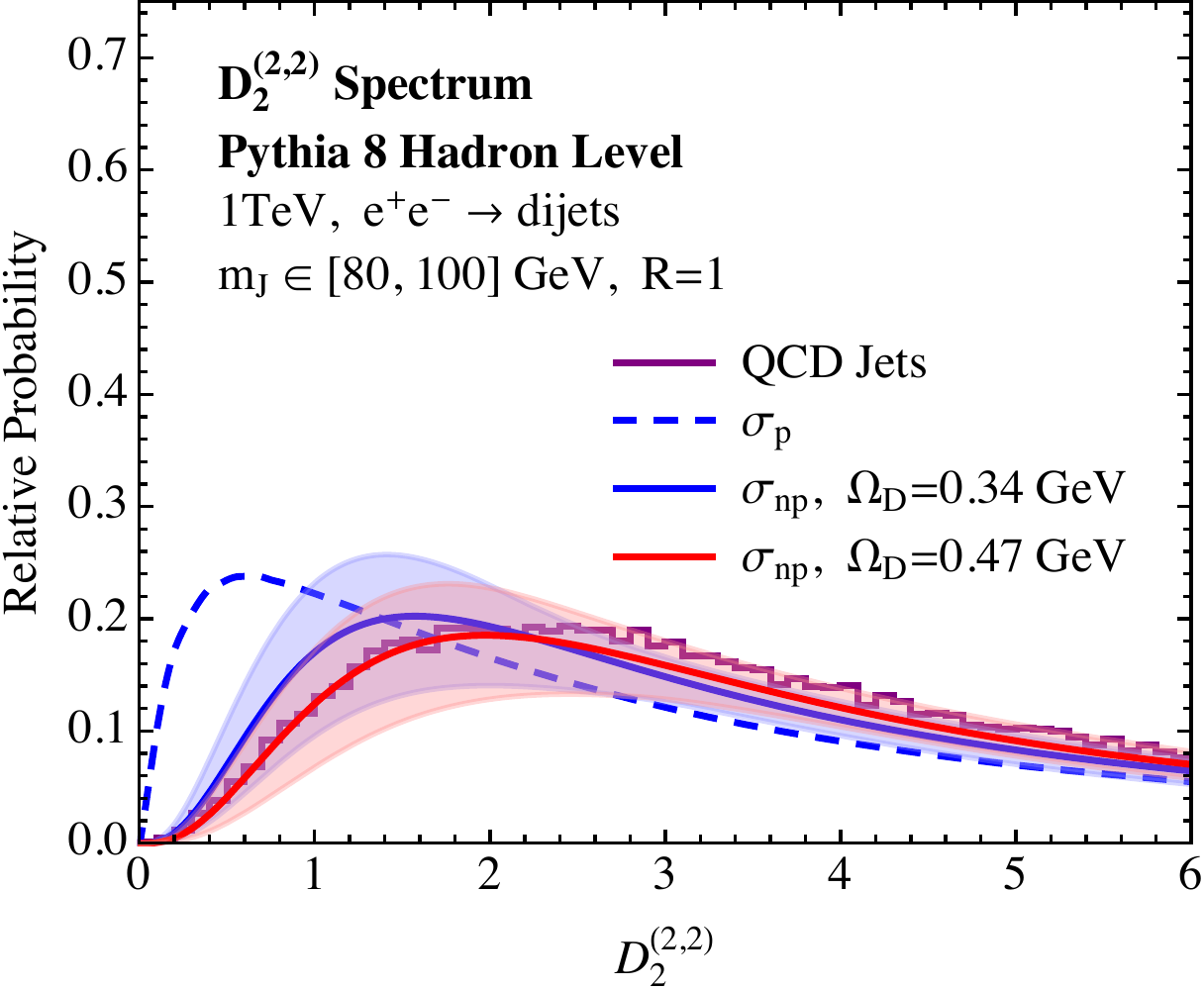}
}
\subfloat[]{\label{fig:D2_ee_hadbkg_b2}
\includegraphics[width= 7.25cm]{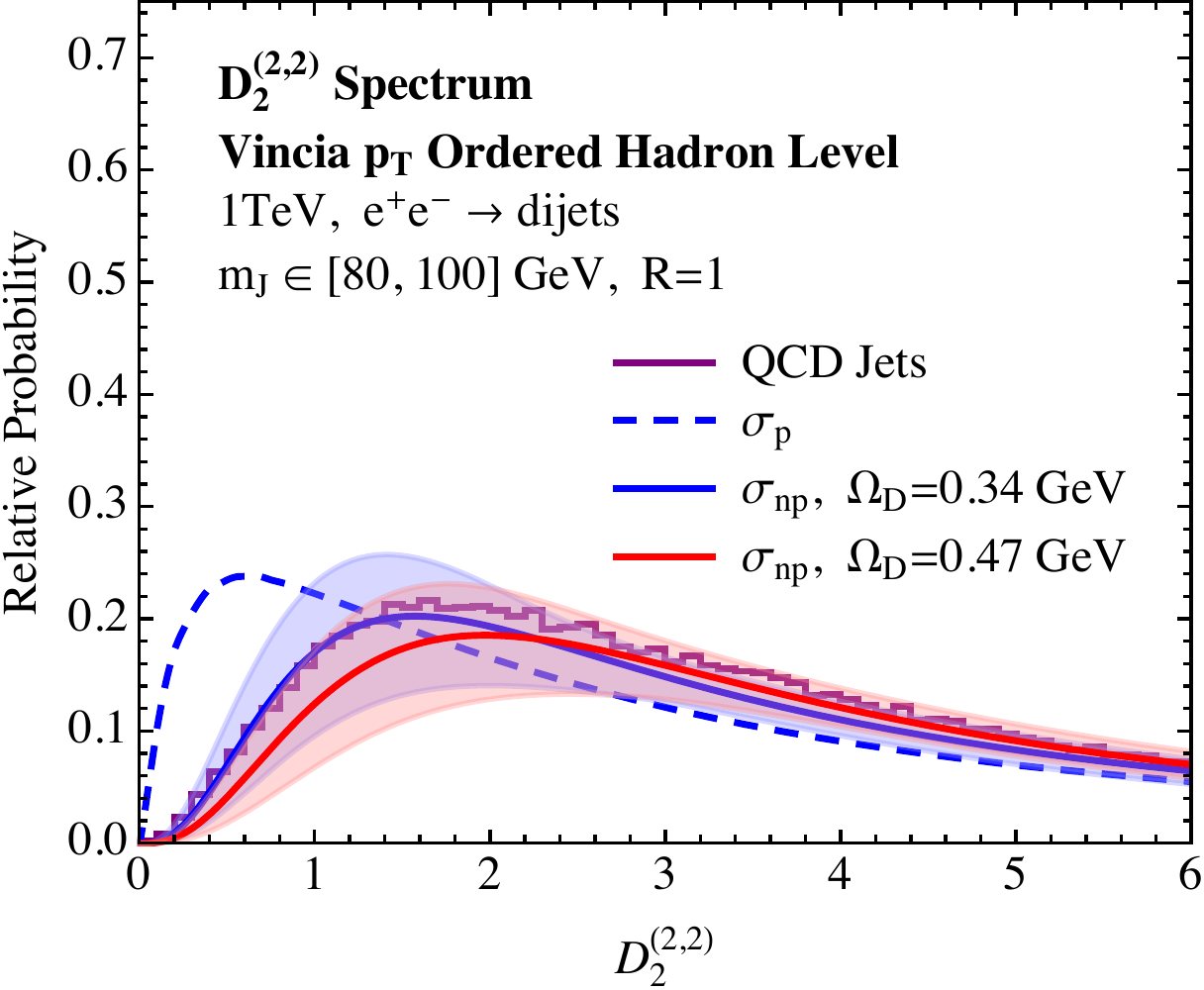}
}
\end{center}
\vspace{-0.2cm}
\caption{A comparison of the $\Dobs{2}{2,2}$ distributions for background QCD jets from our analytic prediction and \pythia{} and $p_T$ ordered \vincia{} Monte Carlos in a) and b).  Analytic predictions for different values of the non-perturbative shape parameter $\Omega_D$ are shown. 
}
\label{fig:D2_hadr_doubleshape}
\end{figure}

Many of the features of the background distributions which were present before hadronization in \Fig{fig:D2_ee_bkg} persist after convolution with the shape function.  However, they are greatly reduced, and they become difficult to disentangle from modifications to the non-perturbative shape parameter at the order we work. In particular, from \Fig{fig:D2_hadr}, we see that for the choices of $\Omega_D$ that we have used, both \vincia{} showers agree well with our analytic calculation.  On the other hand, there is significant disagreement between our calculation with the chosen values of $\Omega_D$ and \pythia{}. The $D_2$ distribution in \pythia{} is systematically pushed to higher values as compared with our calculation.

To try and asses the extent to which this can be accommodated for by adjusting the value of $\Omega_D$, in \Fig{fig:D2_hadr_doubleshape} we show plots of both \pythia{} and \vincia{} with $p_T$ ordering compared with our analytic results for two different values of the shape parameter. The values $\Omega_D=0.34$ GeV and $\Omega_D=0.47$ GeV were chosen to give best agreement with the \vincia{} and \pythia{} distributions, respectively. This figure makes clear that the disagreement between the $D_2$ distributions as generated by the two Monte Carlo generators can largely be remedied by using different values of the non-perturbative parameter. We note also that the effect of changing the non-perturbative parameter is of course similar to that of changing the perturbative cutoff of the shower, as was discussed in \Sec{sec:scales}, making it difficult to disentangle these two effects. However, we have argued that there are also legitimate differences in the modeling of soft radiation between the generators. While these can be partially compensated for by non-perturbative physics, with higher perturbative accuracy, one could begin to disentangle these effects.

This plot also gives a feel for the extent to which $\Omega_D$ can be varied before significant disagreement is seen between the analytic calculation and a given Monte Carlo distribution. Performing the perturbative calculation to higher accuracy would help to resolve some of these ambiguities in the value of the shape parameter, by reducing the perturbative uncertainty on the shape of the distribution, as well as its normalization. Throughout the rest of this chapter, when comparing our analytic predictions with \vincia{} or \pythia{}, we will use the value $\Omega_D=0.34$ GeV as obtained from our fit to hadron level $p_T$-ordered \vincia{}.  From comparison to our analytic calculations, we believe that \vincia{} best describes the perturbative distribution and implements a clean separation between perturbative and non-perturbative physics, as in our analytic calculation. However, one should keep in mind the level of sensitivity to this parameter. In particular, for the application of boosted $Z$ discrimination, we will see that the discrimination power of the observable will depend sensitively on the shape of the $D_2$ distribution below the peak, and will therefore exhibit great sensitivity to the value of the shape parameter.

For the signal distributions, shown in \Fig{fig:D2_hadr_signal}, we use the same choice of non-perturbative parameters as for the background distributions. From \Eq{eq:signal_nonpert}, we have seen that for the jets with $E_J=500$ GeV, the non-perturbative shift is expected to be of the order $1/500$, and is therefore completely negligible to the level of accuracy that we work, and the equality of the non-perturbative parameters between the signal and background distributions is not tested. For the signal distributions, we see excellent agreement between the theory prediction and all the Monte Carlo generators. Due to the sharp peak in the distribution, we expect higher order resummation is necessary to provide a more accurate description right in the peak region, where the perturbative uncertainty in our calculation becomes large. Due to the fact that the distributions are normalized, this uncertainty also manifests itself in the tail of the distribution. It is known how to calculate the signal distribution to higher accuracy \cite{Feige:2012vc}, and so we do not consider this issue further here.  The effect of the shape function on our analytic results are consistent with all of the Monte Carlos, whose signal $D_2$ distribution is changed only slightly (i.e., only in the lowest bins) after hadronization.

We conclude this section by emphasizing how the choice of variable can greatly facilitate comparisons with Monte Carlos. An important feature of the $D_2$ observable is that it cleanly separates phase space regions dominated by different physics. In particular, it separates the region of phase space where a subjet is formed from that where no subjet is formed, as well as separating the regions of phase space where hadronization is important from those where it plays a minor role. This enables these effects to be cleanly disentangled, and provides a sensitive probe of their modeling. We therefore believe that the observable $D_2$ could play an important role in the tuning of Monte Carlo generators for jet substructure studies, and could be used to complement some of the observables proposed and studied in \Refs{Fischer:2014bja,Fischer:2015pqa}.\footnote{Note that \Refs{Fischer:2014bja,Fischer:2015pqa} used the observable $C_2$, also formed from the energy correlation functions, which was proposed in \Ref{Larkoski:2013eya}. Unlike $D_2$, $C_2$ does not cleanly separate the two-prong region of phase space from the one-prong region of phase space. A detailed discussion of this point can be found in \Ref{Larkoski:2014gra}. The clean separation of the one- and two-prong regions of phase space is the essential feature of the $D_2$ observable, which allows for its precise theoretical calculation and its sensitivity to the shower implementation. } Furthermore, the observable $D_3$ \cite{Larkoski:2014zma}, which is sensitive to three-prong substructure within a jet also provides a clean separation of two- and three-prong regions, and could be used to provide an even more detailed understanding of jet substructure and the perturbative shower evolution.

\subsection{Analytic Boosted $Z$ Discrimination with $D_2$}\label{sec:ROC}

In this section, we use our analytic calculation, combined with the non-perturbative shape functions of \Sec{sec:Hadronization}, to make complete predictions for the discrimination power of $\Dobs{2}{2,2}$ for hadronically-decaying boosted $Z$ bosons versus QCD quark jets at an $e^+e^-$ collider.  We present comparisons of our calculation to the results of fully hadronized \pythia{}, \vincia{}, and \herwig{} Monte Carlos.  Here, we also present Monte Carlo results from scanning over a range of values for the angular exponent $\alpha$ that is consistent with our factorization theorem. Analytic results for boosted boson discrimination were also presented recently in \Ref{Dasgupta:2015yua} for groomed mass taggers, as well as an analytic study of the optimal parameters.

\begin{figure}
\begin{center}
\subfloat[]{\label{fig:D2_disc_py}
\includegraphics[width= 7.25cm]{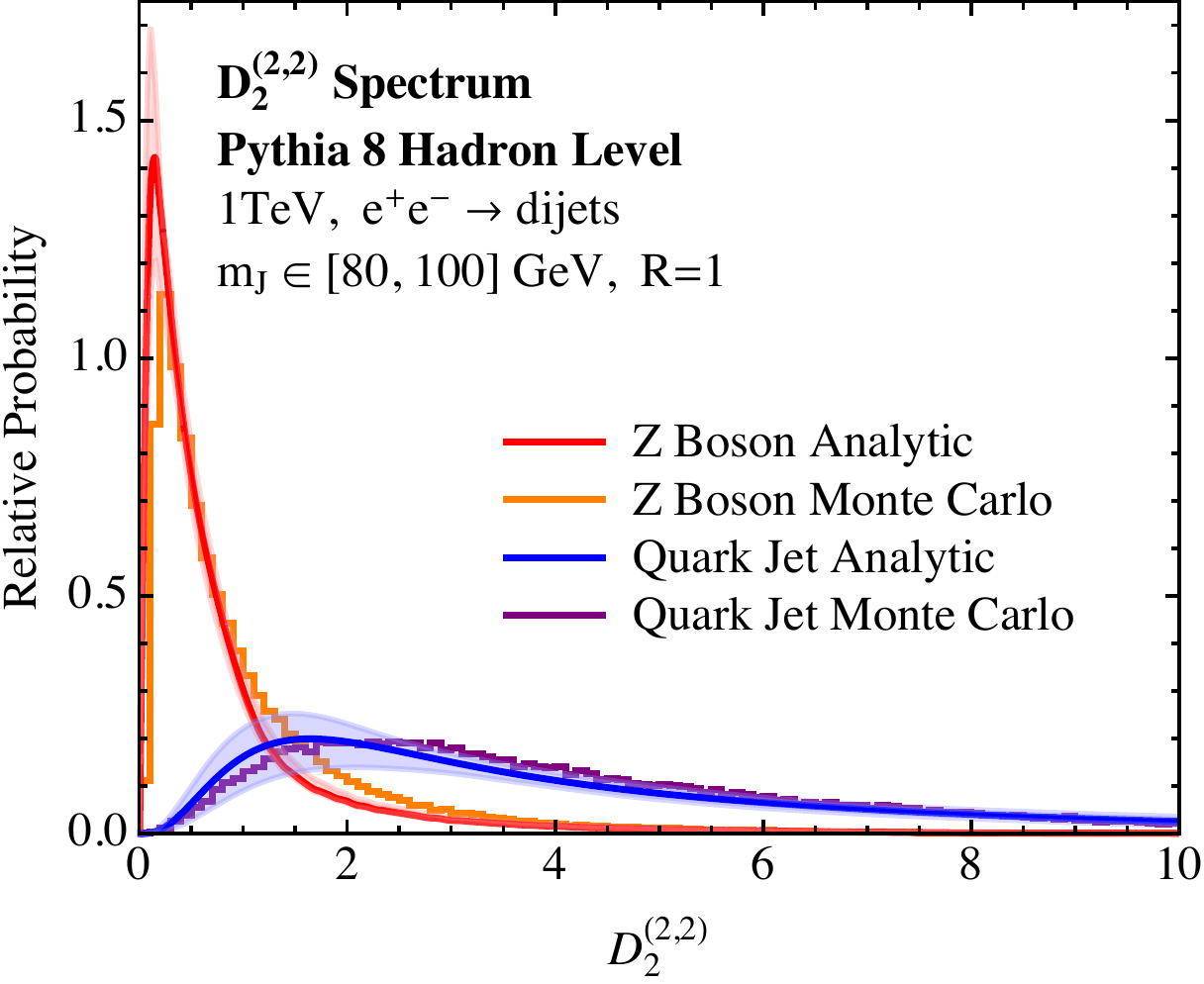}
}
\subfloat[]{\label{fig:D2_disc_vinp}
\includegraphics[width = 7.25cm]{figures/fig16_vinciapt.pdf}
}\\
\subfloat[]{\label{fig:D2_disc_vinv}
\includegraphics[width= 7.25cm]{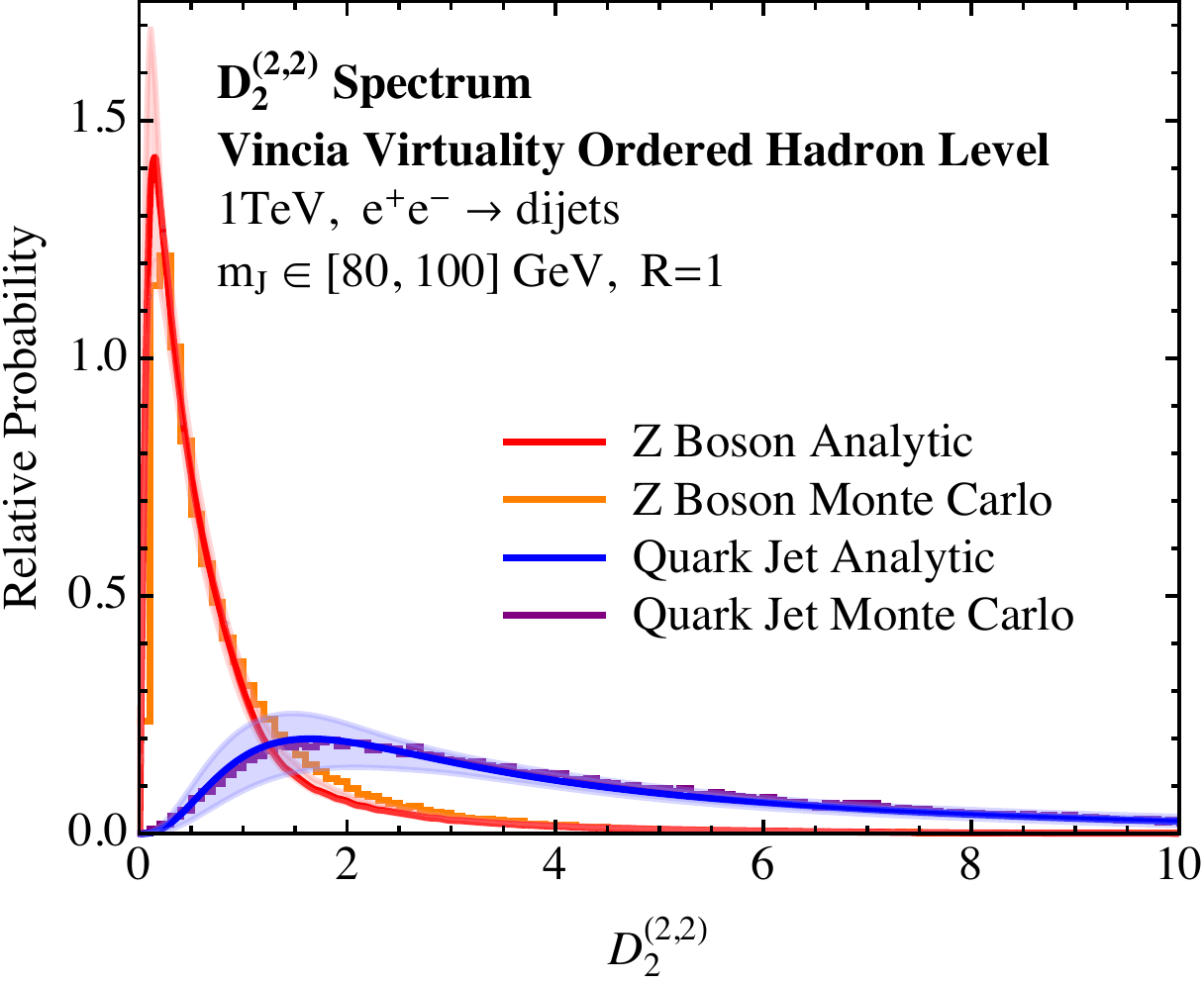}
}
\subfloat[]{\label{fig:D2_disc_her}
\includegraphics[width = 7.25cm]{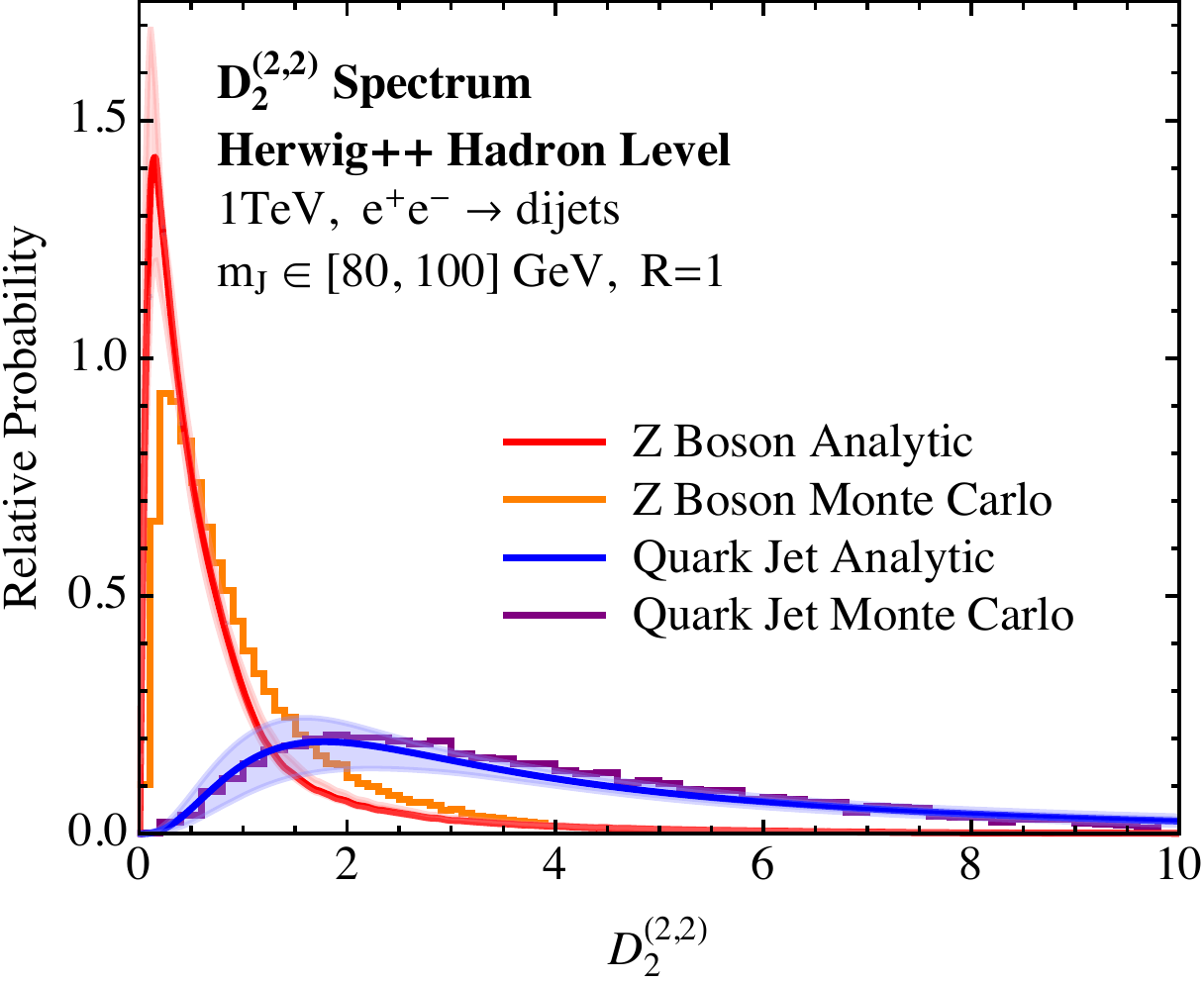}
}
\end{center}
\vspace{-0.2cm}
\caption{
A comparison of signal and background $\Dobs{2}{2,2}$ distributions for the four different Monte Carlo generators and our analytic calculation, including hadronization. Here we show the complete distributions, including the long tail for the background distribution. Although we extend the factorization theorem beyond its naive region of applicability into the tail, excellent agreement with Monte Carlo is found.
}
\label{fig:D2_disc}
\end{figure}

\begin{figure}
\begin{center}
\subfloat[]{\label{fig:D2_disc_py_zoom}
\includegraphics[width= 7.25cm]{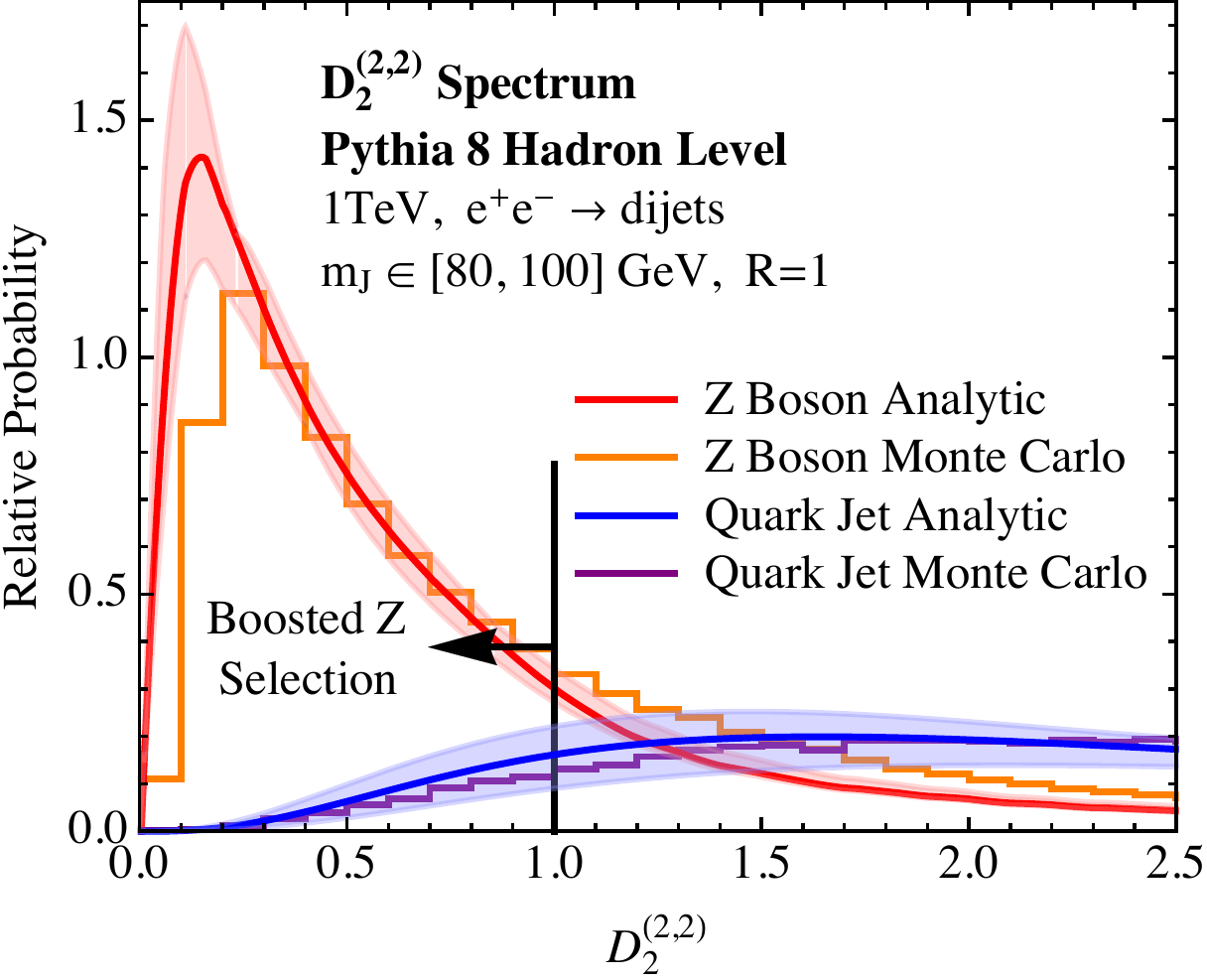}
}
\subfloat[]{\label{fig:D2_disc_vinp_zoom}
\includegraphics[width = 7.25cm]{figures/fig16_vinciapt_zoom.pdf}
}\\
\subfloat[]{\label{fig:D2_disc_vinv_zoom}
\includegraphics[width= 7.25cm]{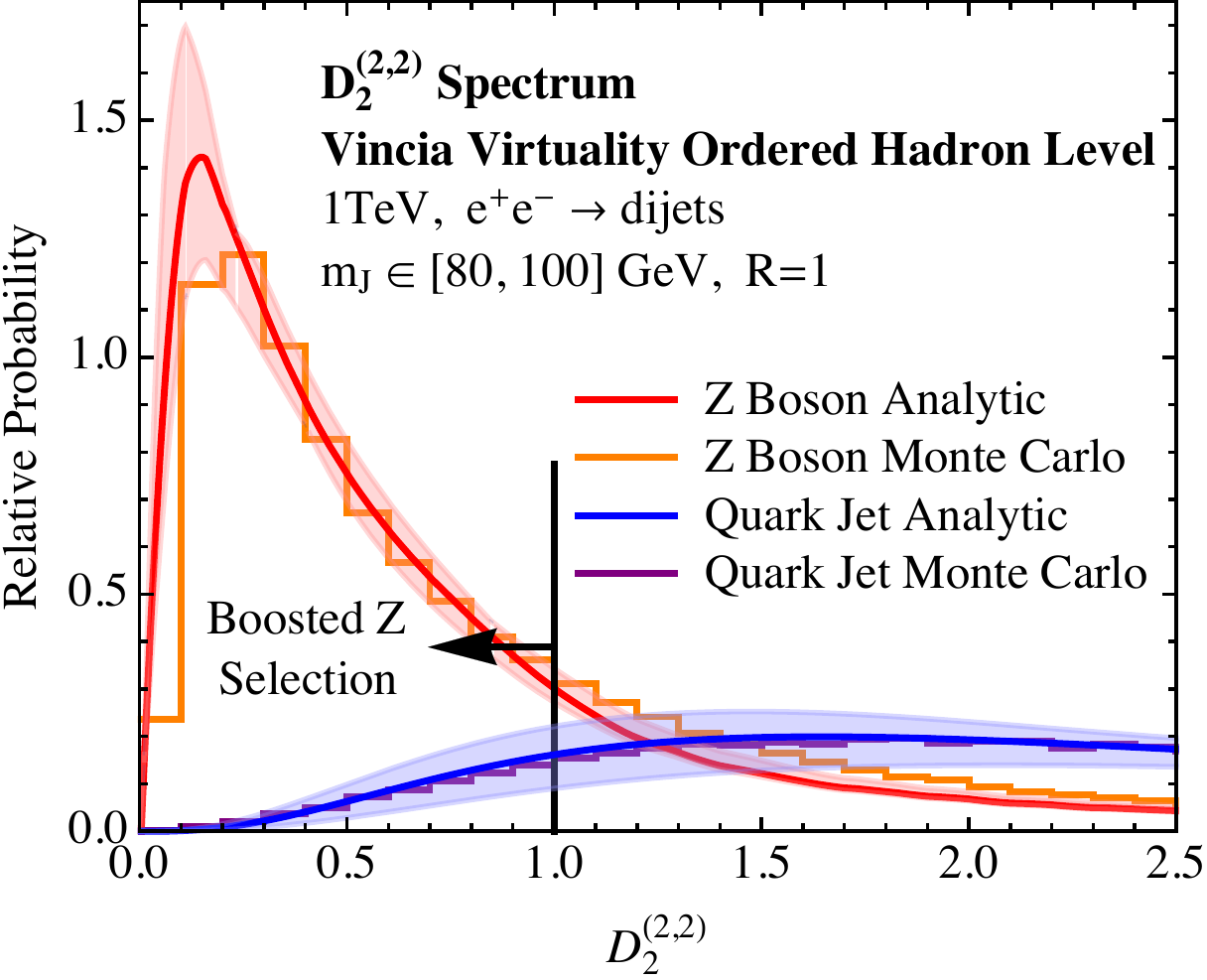}
}
\subfloat[]{\label{fig:D2_disc_her_zoom}
\includegraphics[width = 7.25cm]{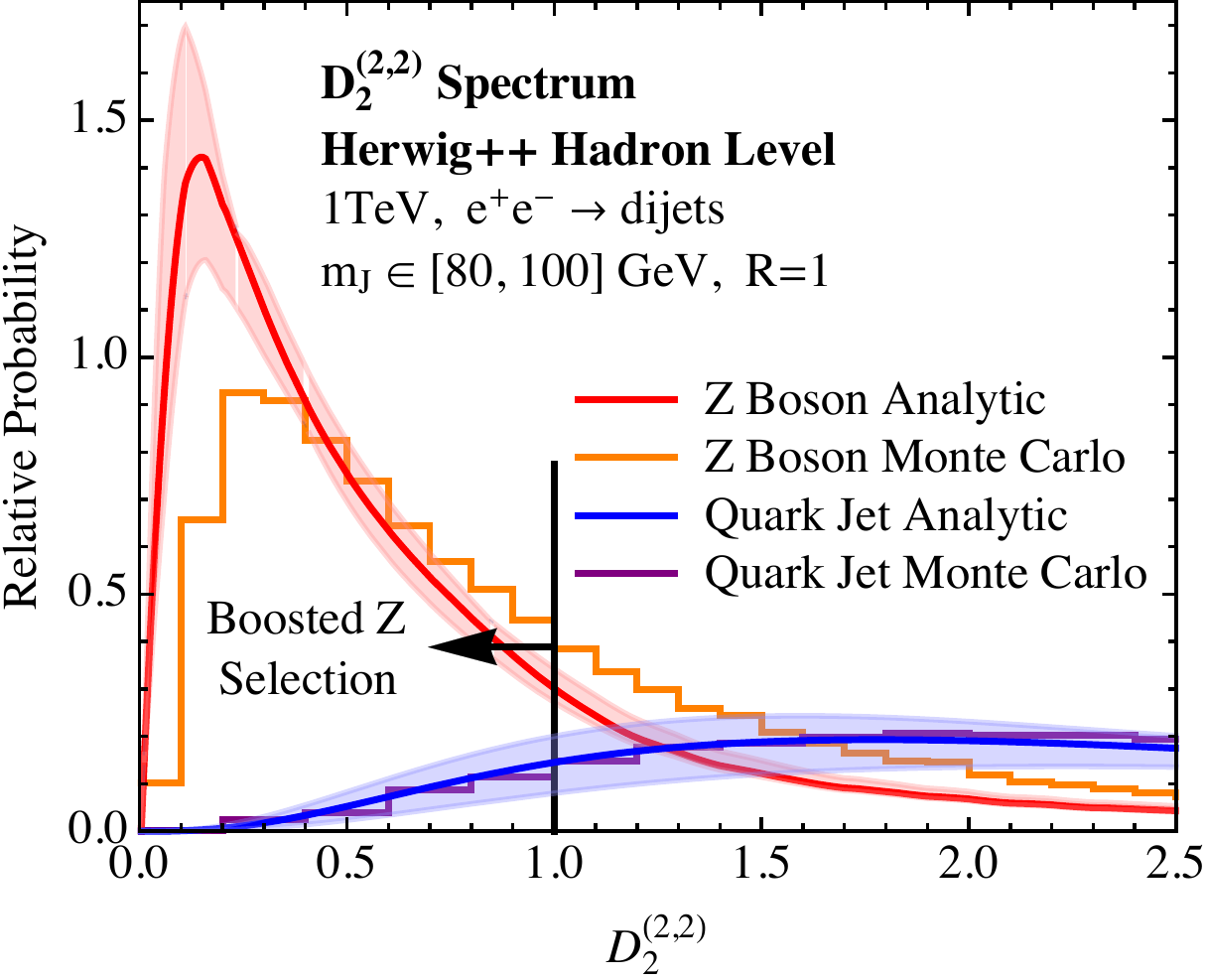}
}
\end{center}
\vspace{-0.2cm}
\caption{
A comparison of signal and background $\Dobs{2}{2,2}$ distributions for the four different Monte Carlo generators and our analytic calculation, including hadronization. Here we show a zoomed in view of the distributions at small $D_2$, along with a representative cut that could be used to select a relatively pure sample of boosted $Z$ bosons. Relevant cuts for boosted $Z$ discrimination are to the left of the perturbative peak for the background distributions.
}
\label{fig:D2_disc_zoom}
\end{figure}

In \Figs{fig:D2_disc}{fig:D2_disc_zoom} we overlay the distributions for $\Dobs{2}{2,2}$ as measured on signal and background for each Monte Carlo sample, and compare with our analytical calculations including the non-perturbative shape function contributions. \Fig{fig:D2_disc} shows the complete $D_2$ distributions, including the long tail of the background distribution, while \Fig{fig:D2_disc_zoom} shows a zoomed in version, focusing on small values of $D_2$, as is most relevant for signal versus background discrimination. A representative cut on the $D_2$ distribution, as could be used to select a relatively pure sample of boosted $Z$ bosons, is also indicated. In general, the agreement between the Monte Carlos, for both signal and background distributions, and our calculation is impressive. This holds true both for the overall shape of the distributions, including the long tail of the background distribution, and for the detailed shape at small values of $D_2$. It is also important to note that the perturbative uncertainties remain under control, even in the small $D_2$ region, as seen in \Fig{fig:D2_disc_zoom}. The uncertainty bands do not incorporate variations in the non-perturbative parameter $\Omega_D$. There are however, some small deviations between the analytic predictions and the Monte Carlo distributions.  The background distribution in \pythia{} is pushed to slightly higher values than our calculation.  This implies that the signal versus background discrimination power as predicted with \pythia{} will be overestimated.  The most conservative prediction for the signal versus background discrimination power is from \herwig{}, whose background distribution is nearly identical to our calculation.  That \pythia{} tends to be optimisitic and \herwig{} tends to be pessimistic with respect to discrimination power has been observed in several other jet substructure analyses \cite{Larkoski:2013eya,Larkoski:2014gra,Larkoski:2014zma,Aad:2014gea}.

An important feature of the $D_2$ distributions, made clear by \Fig{fig:D2_disc_zoom}, is that in the region of interest relevant for boosted $Z$ discrimination, the background distribution is deep in the non-perturbative regime. Therefore, although the perturbative uncertainties are small, the effect of the shape function, and variations of the non-perturbative parameter $\Omega_D$, is large.  Estimates of the uncertainties due to the form of the shape function, or the use of more complicated functional forms, along the lines of \Ref{Ligeti:2008ac} are well beyond the scope of this chapter.  An advantage of our factorization approach is that we are able to achieve a clean separation of perturbative and non-perturbative effects, and demonstrate relations between the non-perturbative matrix elements appearing in our factorization theorems and non-perturbative matrix elements which have been measured with other event shapes, by using their field theoretic definitions. This separation is essential for understanding discrimination performance in the non-perturbative region, which we see is required for jet substructure studies related to boosted boson discrimination.  Importantly, though, $D_2$ seems to take advantage of the different hadronization corrections to signal and background jets, and the overlap of the signal and background regions of $D_2$ decreases significantly in going from parton-level to fully hadronized jets.

\begin{figure}
\begin{center}
\subfloat[]{\label{fig:D2_roca}
\includegraphics[width= 7.2cm]{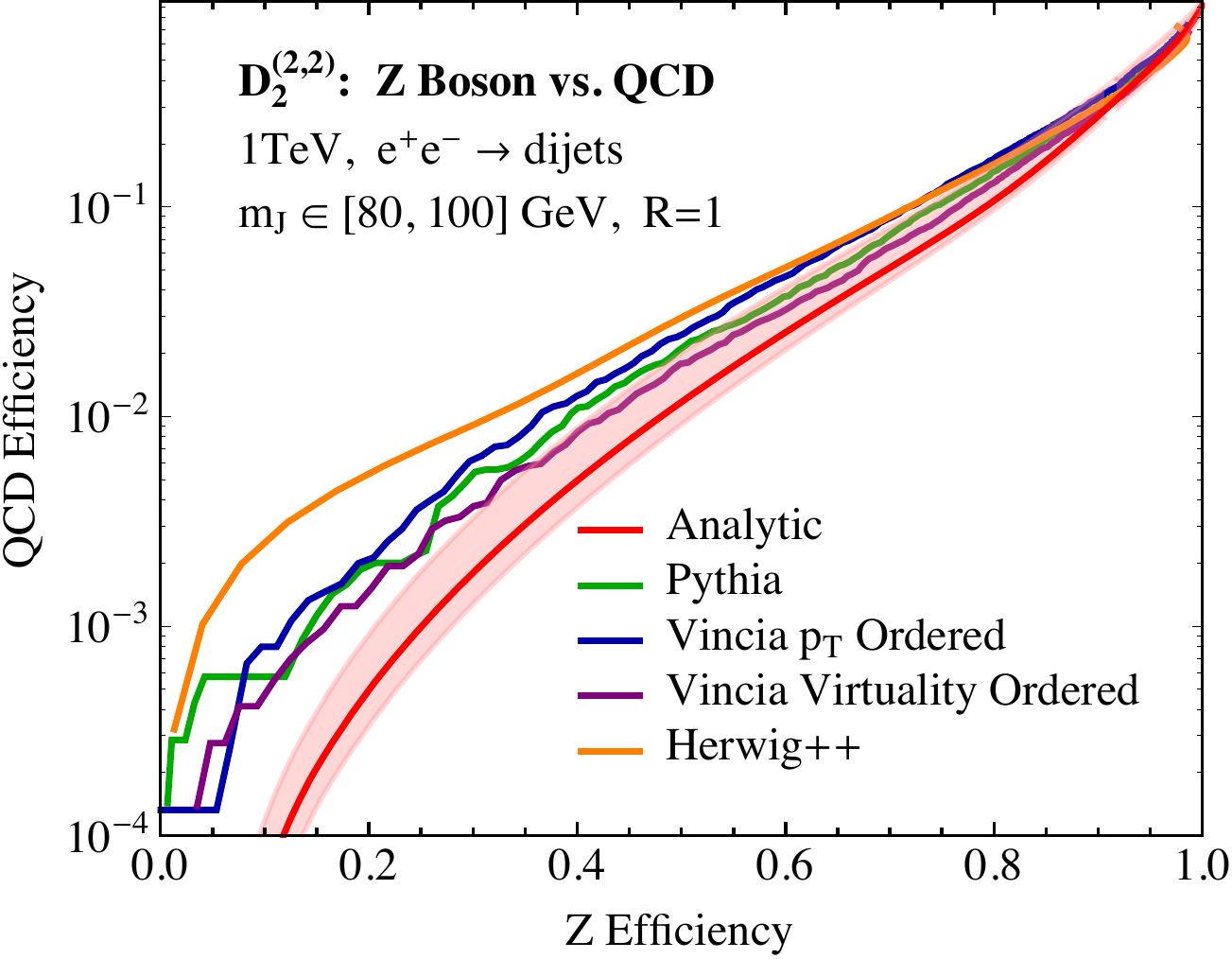}
}
\ 
\subfloat[]{\label{fig:D2_rocb}
\includegraphics[width = 7.0cm]{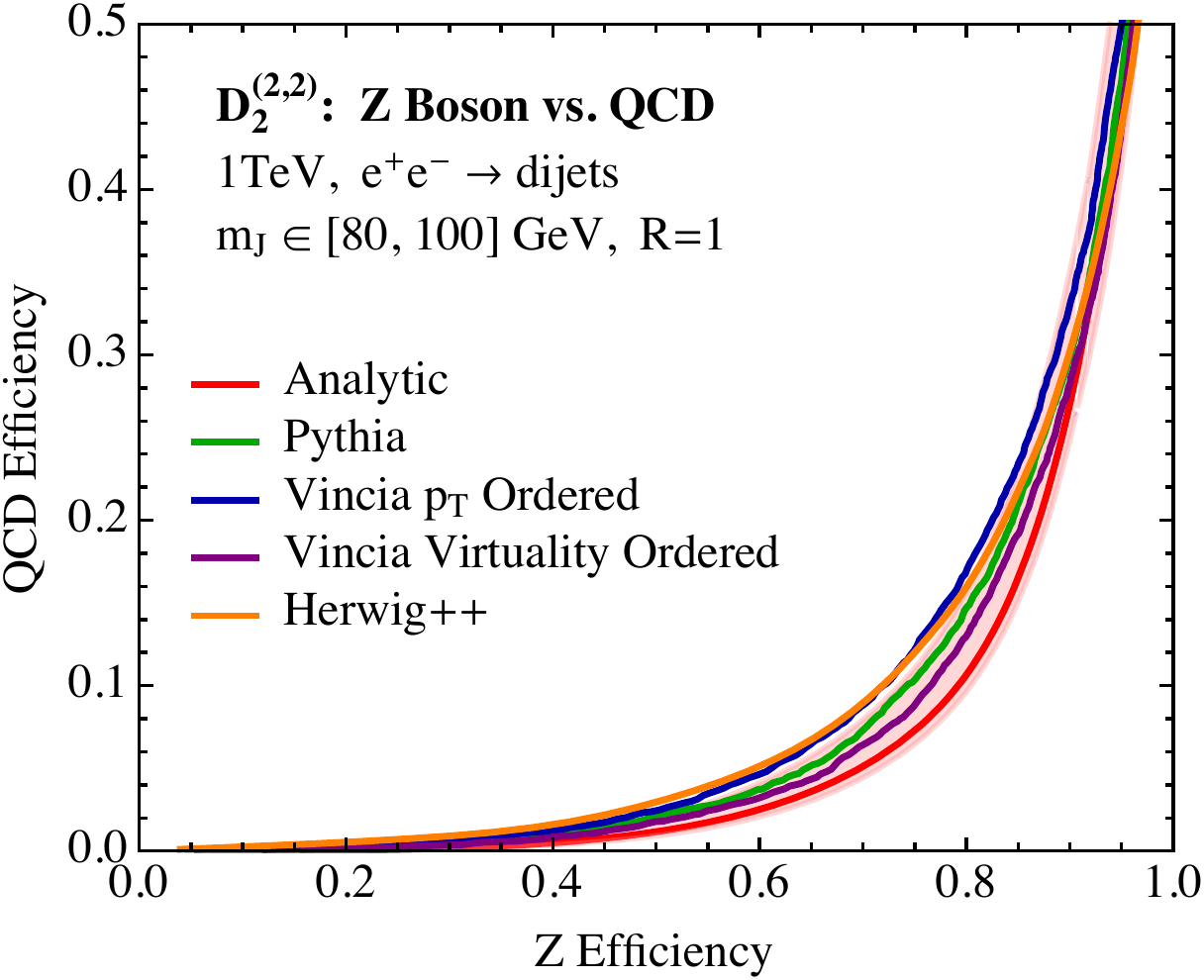}
}\end{center}
\caption{ Signal vs.~background efficiency curves for $\Dobs{2}{2,2}$ for the Monte Carlo samples as compared to our analytic prediction on a a) logarithmic scale plot and b) linear scale plot.  The band of the analytic prediction is representative of the perturbative scale uncertainty.
}
\label{fig:ROC}
\end{figure}

In \Fig{fig:ROC}, we have used these raw distributions to produce signal versus background efficiency curves (ROC curves) by making a sliding cut in $D_2$.  The ROC curve from each Monte Carlo sample as well as our analytic prediction from our calculated signal and background distributions are shown in both logarithmic plot and linear plot in \Figs{fig:D2_roca}{fig:D2_rocb}, respectively. The band around our analytic prediction should be taken as representative of the signal versus background efficiency range from varying the perturbative scales. \footnote{Note that ROC curves only make sense for normalized distributions, and therefore the envelopes from scale variation cannot be used. Instead, ROC curves are generated from normalized signal and background distributions made with a variety of scale choices, with scales varied separately in the signal and background distributions. We then take the envelope of these ROC curves to generate the uncertainty bands for the ROC curves. } For the analytic predictions, we use $\Omega_D=0.34$, as obtained from our fit to the $p_T$ ordered \vincia{} shower.  Consistent with the distributions in \Fig{fig:D2_disc}, the Monte Carlos are in qualitative agreement with our analytic prediction for the ROC curve.  In general, our analytic prediction seems to give an optimistic prediction for the discrimination power, however, this is driven by the fact that our resummed prediction for the signal distribution is more peaked. It would be interesting to perform the NNLL resummation for the signal, which should significantly reduce the uncertainty in the signal calculation, particularly in the peak region, where the perturbative uncertainties in our present calculation are quite large. Because of the fact that the distributions are normalized, an improved behavior in the peak of the distribution could also improve the agreement in the tail of the signal distribution, which is currently systematically low, due to the fact that the peak is systematically high. This could enable a conclusive understanding as to the discrepancy between the different Monte Carlo generators for both signal and background distributions. In particular, our analytic calculations suggest that the \herwigpp{} generator provides pessimistic predictions for the discrimination power of the $D_2$ observable due to the underestimation of the peak height for the signal distribution, and it would be interesting to understand this further. Due to the importance of analytically understanding the discrimination power of jet substructure observables, such a calculation is well motivated. For the case of $\alpha=\beta=2$, the required perturbative components could be obtained following relations to $e^+e^-$ event shapes as were used in \Ref{Feige:2012vc}.

\begin{figure}
\begin{center}
\subfloat[]{\label{fig:D2_roca2}
\includegraphics[width= 7.2cm]{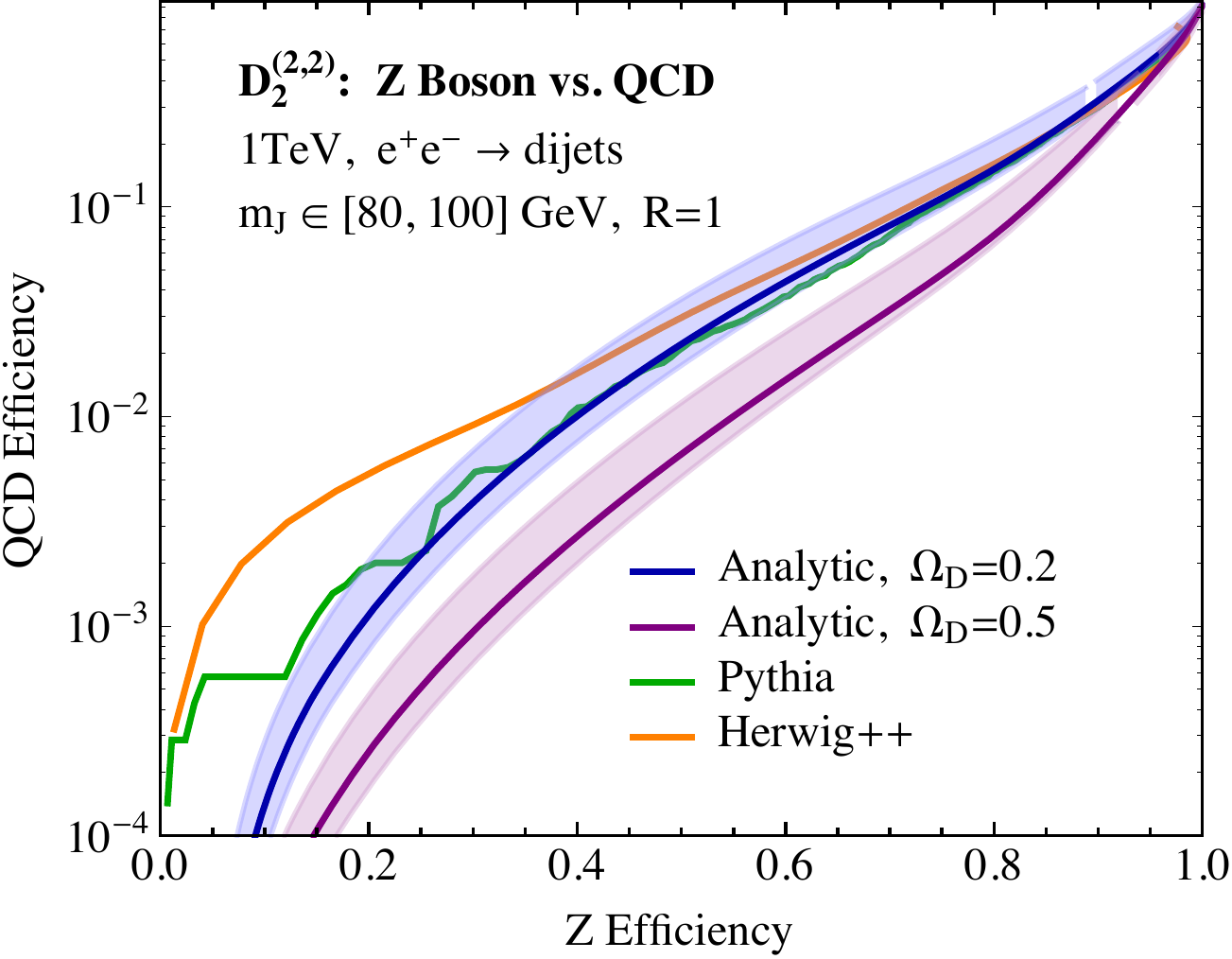}
}
\ 
\subfloat[]{\label{fig:D2_rocb2}
\includegraphics[width = 7.0cm]{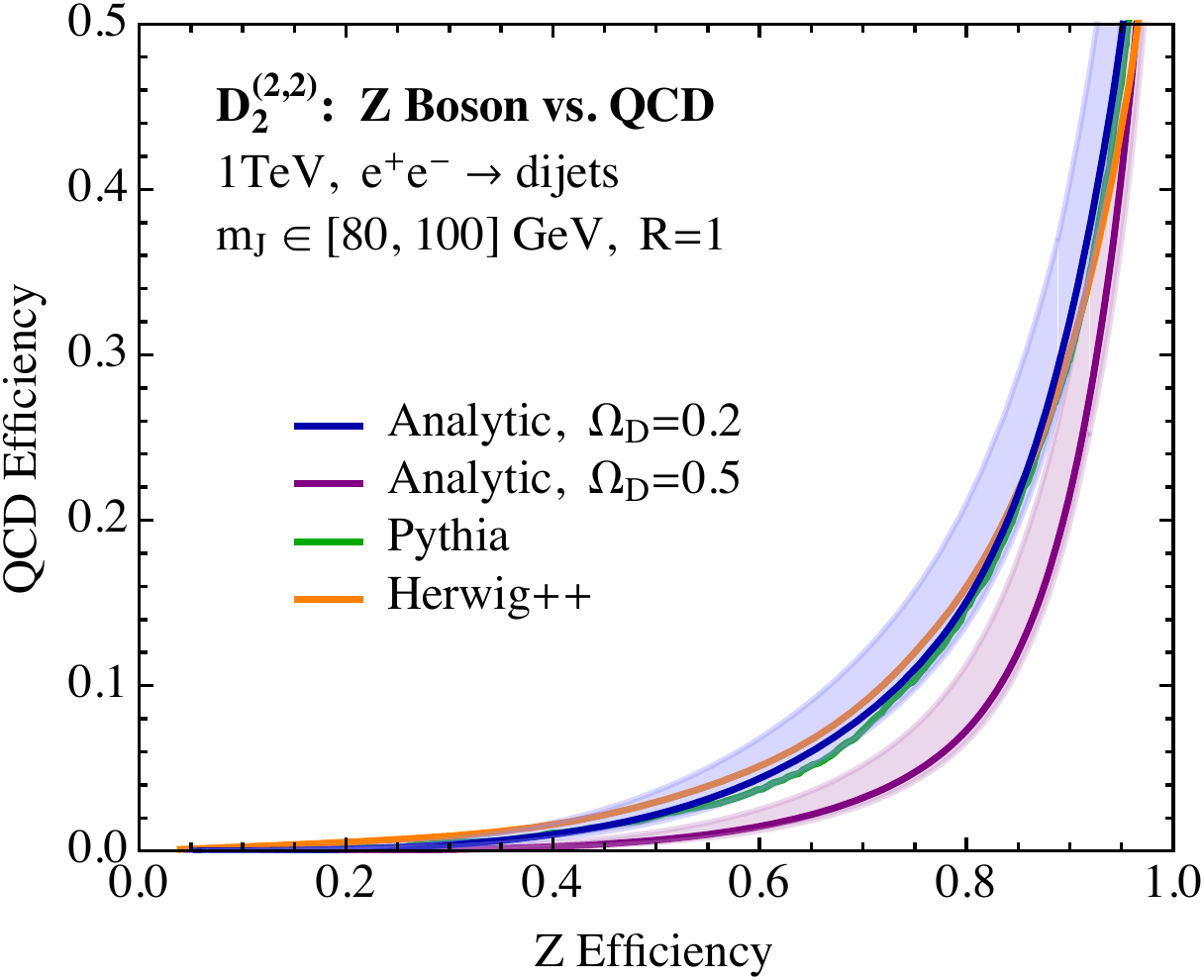}
}\end{center}
\caption{ Signal vs.~background efficiency curves for $\Dobs{2}{2,2}$ for the Monte Carlo samples as compared to our analytic prediction for two different values of the non-perturbative shape parameter, chosen by varying our central value by $\pm0.15$ GeV. Results are shown on a  logarithmic scale in a) and a linear scale in b).  Perturbative scale uncertainties are also shown.
}
\label{fig:ROC_doubleshape}
\end{figure}

One feature made clear by the linear ROC curve in \Fig{fig:D2_rocb} is the increase in perturbative uncertainty with increasing $Z$ efficiency. As emphasized earlier, this is due to the fact that for the region of interest for $Z$ discrimination, one is probing values of $D_2$ which are below the peak of the background distribution, and therefore in the non-perturbative regime. As the $Z$ efficiency is increased, one enters the peak region of the background distribution, where the perturbative uncertainty is largest, causing a corresponding increase in the uncertainty band for the ROC curve. However, we do not include uncertainties due to the non-perturbative parameter $\Omega_D$ or from the shape function, in \Fig{fig:D2_rocb}, which are the dominant sources of uncertainty in this region.

To demonstrate that is indeed the case, in \Fig{fig:ROC_doubleshape} we show ROC curves in both linear and log scales for two different values of the non-perturbative shape parameter. The values of $\Omega_D$ where chosen by varying our central value of $\Omega_D=0.34$ GeV by $\pm 0.15$ GeV (and rounding to nice numbers). We have also shown the distributions from the \herwigpp{} and \pythia{} generators as representative of the ROC curves generated by the Monte Carlo generators. This figure makes clear that in the region of efficiencies of interest for boosted $Z$ tagging, one is extremely sensitive to the $D_2$ distribution in the deeply non-perturbative region, and this uncertainty swamps the perturbative uncertainty. To be able to improve the accuracy in this region will require detailed comparisons with Monte Carlo, data, and analytic calculations, to allow for a clean separation of the non-perturbative parameter from perturbative modifications to the shape of the distribution.

\begin{figure}
\begin{center}
\subfloat[]{\label{fig:D2_50sig}
\includegraphics[width= 7.2cm]{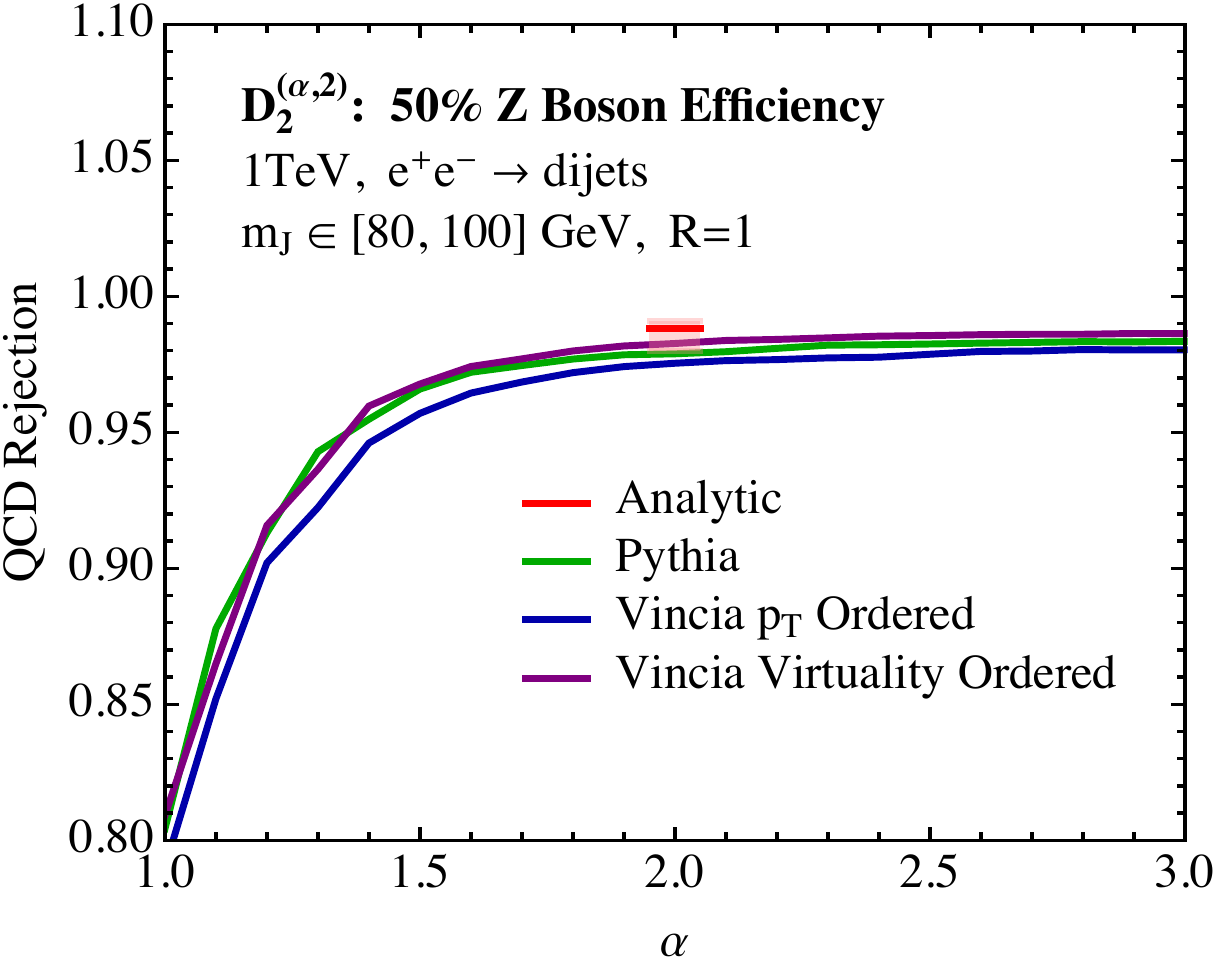}
}
\ 
\subfloat[]{\label{fig:D2_75sig}
\includegraphics[width = 7.0cm]{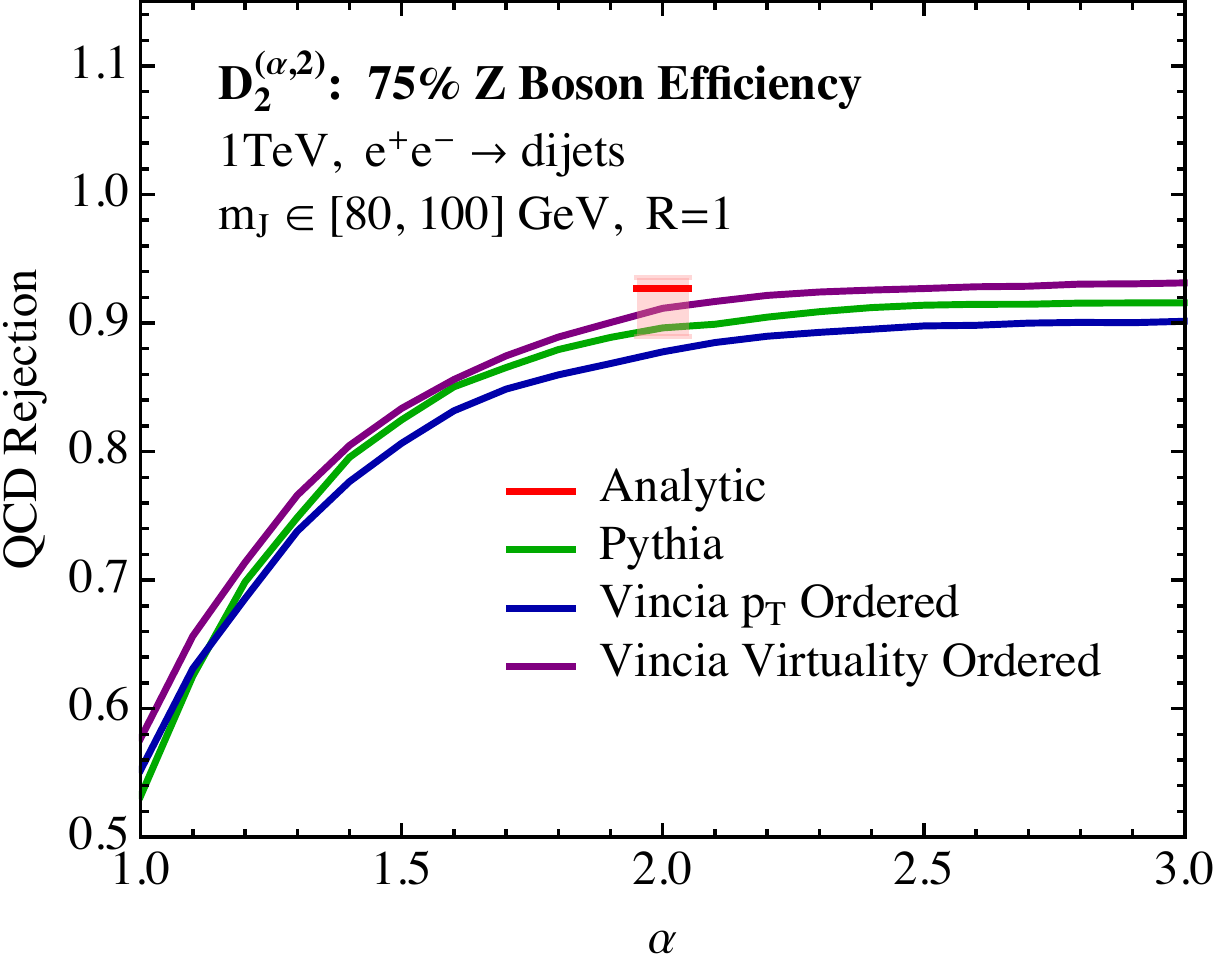}
}\end{center}
\caption{Background rejection rate at fixed a) 50\% and b) 75\% signal efficiency as a function of the angular exponent of the 3-point energy correlation function in $D_2$, and a comparison to our analytic prediction for $\alpha=2$.
}
\label{fig:effs}
\end{figure}

To further understand the discrimination power of the $D_2$ observable, in \Figs{fig:D2_50sig}{fig:D2_75sig} we show the background rejection rate at 50\% and 75\% signal efficiency as a function of $\alpha$, the angular exponent of the 3-point energy correlation function in $D_2$.  Below about $\alpha=4/3$, all rejection rates dramatically decrease as $\alpha$ decreases, while above about $\alpha = 4/3$, the QCD rejection rate in all Monte Carlo samples is impressively flat.  This is consistent with our power counting analysis of the $\ecf{2}{2},\ecf{3}{\alpha}$ phase space plane in \Sec{sec:mass_cuts} and is a powerful verification that the Monte Carlos respect the parametric dynamics of QCD.

Although our factorization theorem is valid in the region $\alpha\gtrsim 2$, for $\beta=2$, in \Figs{fig:D2_50sig}{fig:D2_75sig}  we have only shown the analytic prediction for the value $\alpha=2$, where we find that it agrees well with the Monte Carlo results, as expected from the agreement of the distributions and ROC curves. For $\alpha>2$, while our prediction for the background distribution remains accurate (indeed our power counting becomes more valid in this region), the signal distribution becomes extremely sharply peaked, which is difficult to describe, and sensitive to normalization. Due to the fact that this region is also of less phenomenological interest, both because the large angular exponent makes the observable sensitive to pile up contamination, and because both power counting and Monte Carlo analyses indicate that optimal performance is achieved for $\alpha=2$, we have decided not to focus on this region. It would be potentially interesting to see if higher order resummation would be sufficient to describe the sharply peaked signal distribution in this region, as well as to test the universality of the non-perturbative power corrections.

One further interesting feature of \Figs{fig:D2_50sig}{fig:D2_75sig} is the correspondence between the perturbative scale variations, and the spread in the curves from the different Monte Carlo generators, which agree well at both 50\% and 75\%. For the case of $p_T$-ordered \vincia{} as compared with virtuality ordered \vincia{}, this correspondence is precise, as the difference between the Monte Carlos can be viewed as a scale variation, and identical hadronization models are used.

\subsection{Discrimination in the Two-Prong Regime}\label{sec:2insight}

\begin{figure}
\begin{center}
\subfloat[]{\label{fig:broadening_a}
\includegraphics[width= 7.25cm]{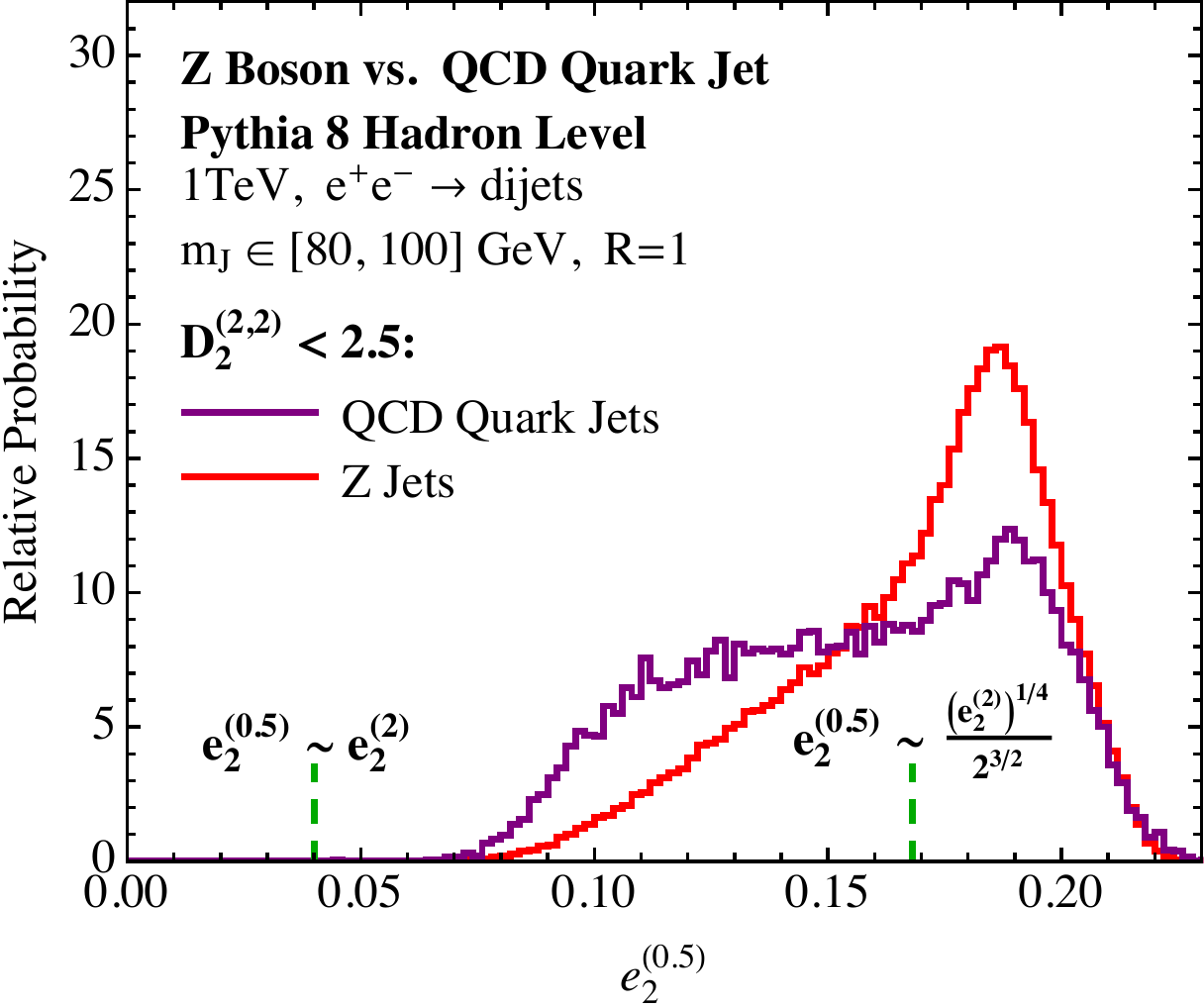}
}
\subfloat[]{\label{fig:broadening_b}
\includegraphics[width= 7.25cm]{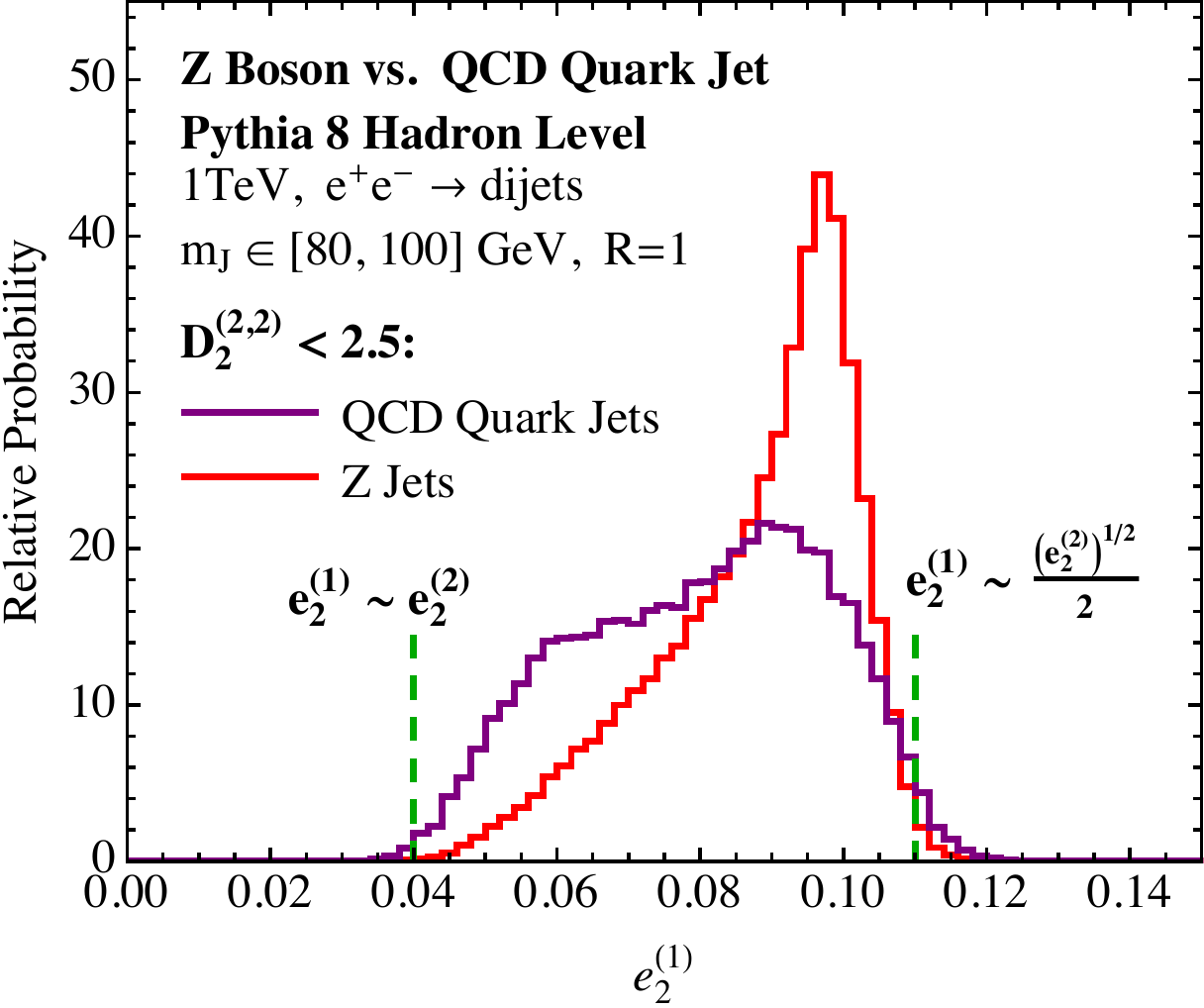}
}
\end{center}
\caption{Distributions for $\ecf{2}{0.5}$ (left) and $\ecf{2}{1.0}$ (right) from the signal and background \pythia{} Monte Carlo samples.  In addition to the mass cut $m_J\in [80,100]$ GeV, these jets are also required to have $\Dobs{2}{2,2}<2.5$ to guarantee that these jets are dominated by two-prong structure.  The parametric boundaries of $\ecf{2}{\beta}$ from \Eq{eq:e2bounds} are shown with the green dashed lines.
}
\label{fig:D2_ee_further}
\end{figure}

Throughout this chapter, we have emphasized that the discrimination of boosted hadronically decaying $Z$ bosons (or $W$ or $H$ bosons) from massive QCD jets is effectively a problem of discriminating one- from two-prong jets.  We have demonstrated that the observable $D_2$ is powerful for this goal.  However, in the formulation of our factorization theorem for calculating the distribution of $D_2$, we needed to perform additional 2-point energy correlation function measurements on the jet to separate contributions from soft subjet and collinear subjets contributions to background.  While indeed the signal jets are dominantly two-pronged, we further know that those prongs are dominantly collinear, and do not have parametrically different energies.  Therefore, we are able to further discriminate signal from background jets in the two-prong region of phase space by exploiting additional measurements that can isolate the soft subjet and collinear subjet configurations.  A detailed analysis of this is beyond the scope of this chapter, but here, we will demonstrate in Monte Carlo that such a procedure is viable.

To investigate this, we measure the observable $\Dobs{2}{2,2}$ on jets on which a tight mass window cut has been applied.  Other angular exponents for $D_2$ can be used also, but here we only measure $D_2$ to define two-prong jets.  We restrict to the two-prong region of phase space by requiring that $\Dobs{2}{2,2} < 2.5$.  Then, on the jets that pass these cuts, we measure two, 2-point energy correlation functions, $\ecf{2}{2}$ and $\ecf{2}{\beta}$, where $\beta < 2$.  As discussed in \Sec{sec:Fact}, the measurement of the two 2-point energy correlation functions provides an IRC safe definition of the subjets' energy fractions and splitting angle.  Because we make a tight mass cut on the jets, $\ecf{2}{2}$ is essentially fixed, and only $\ecf{2}{\beta}$ is undefined.  We will study the distribution of $\ecf{2}{\beta}$ for both signal and background jets in this region of phase space.

For a fixed value of $\ecf{2}{2}$ and $\beta < 2$, $\ecf{2}{\beta}$ is parametrically bounded as
\begin{equation}\label{eq:e2bounds}
\ecf{2}{2} \lesssim \ecf{2}{\beta} \lesssim 2^{\beta-2}(\ecf{2}{2})^{\beta / 2}\,.
\end{equation}
In the two-prong region, the lower bound is set by the soft subjet while the upper bound is set by collinear subjets.  Therefore, $\ecf{2}{\beta}$ for signal jets will peak near $2^{\beta-2}(\ecf{2}{2})^{\beta / 2}$, while background QCD jets will fill out the full range.  We illustrate this in \Fig{fig:D2_ee_further} on the hadronized \pythia{} sample with the appropriate cuts applied.  We show plots of the distributions of $\ecf{2}{0.5}$ and $\ecf{2}{1}$ on both signal and background jets and have added dotted lines to denote the parametric upper and lower boundaries.  As expected, signal peaks near the upper boundary and background fills out the entire allowed region and so this additional information could be used for discrimination.  For the very small values of $\beta=0.5$, an $\mathcal{O}(1)$ drift is observed with respect to the parametric boundaries, while for $\beta=1$, the parametric boundaries are extremely well respected. 

This demonstrates a simple example of an observable which goes beyond the simple one vs. two prong picture of jet substructure, asking more differential questions about the subjets themselves. In particular, it could be used both to further improve the discrimination power of boosted boson discriminants, and to study in detail the QCD properties of subjets.

\section{Looking Back at LEP}\label{sec:LEP}

In this section, we consider the $D_2$ distribution for QCD jets in $e^+e^-$ collisions at the $Z$ pole at LEP, for which a large amount of data exists.   While the use of $D_2$ for boosted boson discrimination is not possible, nor relevant, at LEP, this will emphasize the sensitivity of $D_2$ as a probe of two-prong structure in jets. We will suggest the importance of using variables sensitive to two emissions off of a primary quark in tuning Monte Carlo generators to LEP data.

Our definition of the energy correlation functions in \Eq{eq:ecf_def} makes implicit assumptions about the treatment of hadron masses, which we have ignored to this point.  The definition given there is an $E$-scheme treatment of hadron masses \cite{Salam:2001bd,Mateu:2012nk}, but we could equally well define $p$-scheme energy correlation functions as:
\begin{align}\label{eq:ecf_pscheme}
\ecf{2}{\beta}&=\frac{1}{E_J^2}\sum_{ i<j\in J} |\vec p_{i}|\,   |\vec p_{j}| \left[2(1-\cos\theta_{ij})\right]^{\beta/2}\,, \\
\ecf{3}{\beta}&=\frac{1}{E_J^3}\sum_{ i<j<k \in J} |\vec p_{i}|\,  |\vec p_{j}|  \,|\vec p_{k}| \left[2(1-\cos\theta_{ij})2(1-\cos\theta_{jk})2(1-\cos\theta_{ik})\right]^{\beta/2}\,, \nonumber
\end{align}
where $\vec p_i$ denotes the three-momenta of particle $i$. For massless particles, this definition is identical to that of \Eq{eq:ecf_def}, and so our perturbative analytics would be unchanged by using this definition or the definition of \Eq{eq:ecf_def}.\footnote{As will be discussed shortly, the differences in our analytic calculation due to hadron masses will arise through non-perturbative effects, namely the shape function.}  The definitions of \Eq{eq:ecf_def} and \Eq{eq:ecf_pscheme} differ for massive particles. In particular, the energy correlation functions as defined in \Eq{eq:ecf_pscheme} have the advantage that they vanish for low momentum or collimated particles regardless of whether these particles are massless or massive, which is not true of the definition in \Eq{eq:ecf_def}. Because of this, we expect that the energy correlation functions as defined in \Eq{eq:ecf_pscheme} are less sensitive to hadron mass effects and that kinematic restrictions on the energy correlation functions remain the same before and after hadronization, so that the phase space studied in \Sec{sec:power_counting} assuming massless particles is not significantly modified.

At LEP energies, hadronization will also have a larger effect on the $D_2$ spectrum than at $1$ TeV. However, a particularly important aspect of our all orders factorization theorem is that it isolates perturbative and non-perturbative physics contributions. In this section we will again implement non-perturbative effects into our analytic calculation using the shape function defined in \Eq{eq:shape_func}. There are two effects which determine how the shape function depends on the jet mass, $m_J$, and the center of mass energy, $Q$. First, for a fixed valued of $\Omega_D$, the 
shift in the first moment of the $D_2$ distribution was given in \Eq{eq:bkg_shift_np}, which we recall here for convenience, by
\begin{equation}\label{eq:LEP_shape}
\Delta_D=\frac{\Omega_D}{ E_J \left(  \frac{m_J}{E_J} \right)^4}\,.
\end{equation}
This has dependence on both $m_J$ and $Q$ (through $E_J$), and for the jets we consider at LEP, this is a considerably larger shift than for the $1$ TeV jets studied in \Sec{sec:Hadronization}. This scaling is a non-trivial prediction of our factorization framework, and we will see that it is well respected when we perform a comparison of our analytic results with Monte Carlo. Furthermore, the parameter $\Omega_D$ has a logarithmic dependence on a renormalization scale, $\Omega_D=\Omega_D(\mu)$, through renormalization group evolution \cite{Mateu:2012nk}, which is briefly reviewed in \App{app:shape_RGE}. However, this effect is small compared with the linear change in the first moment with $E_J$ for a fixed $m_J/ E_J$. A numerical estimate for the effect of the running of $\Omega_D(\mu)$ is given in \App{app:shape_RGE}. At the level of accuracy to which we work in this chapter, we cannot probe this logarithmic running, although we will see that our results are consistent with it.

The definition of the energy correlation functions given in \Eq{eq:ecf_pscheme} also has an effect on the universality of the non-perturbative parameter $\Omega_D$, when hadron mass effects are included. Power corrections due to hadron mass effects are of order $\mathcal{O}(m_H/Q)$, where $m_H$ is a light hadron mass, and are therefore of the same order as the leading $\mathcal{O}(\Lambda_{\text{QCD}}/Q)$ power corrections. In the $p$-scheme definition of the energy correlation functions which we have chosen in \Eq{eq:ecf_pscheme}, it is no longer possible to extract the dependence on the angular exponent alpha from $\Omega_D$, as was done in \Eq{eq:alpha_np}. However, to the accuracy to which we work, we expect this to be a negligible effect, and furthermore, the case $\alpha=2$ is of most phenomenological interest, and is the case we have focused on exclusively in this chapter. Furthermore, even in the presence of hadron mass effects, it is still possible to extract the parameter $\Omega_D$ from dijet event shapes in the same universality class \cite{Mateu:2012nk}. This exhibits the benefits of the factorization approach both for separating perturbative and non-perturbative effects, and for relating non-perturbative parameters to maintain predictivity.

One further distinction between the case of boosted $Z$ discrimination and the measurement of QCD jet shapes at the $Z$ pole is that while a tight mass cut is natural for boosted $Z$ discrimination, it is not natural in jet shape analyses. However, our shape function analysis, as derived in \Sec{sec:Hadronization}, is valid at a fixed jet mass (or correspondingly fixed value of $\ecf{2}{\beta}$). This is clear from both \Eq{eq:shape_function_eq1} and from the equation for the shift in the first moment in \Eq{eq:bkg_shift_np}. However, we emphasize that the non-perturbative parameter $\Omega_D$ is unique, and the scaling of the non-perturbative shift with the jet mass is fully determined. To achieve an analytic prediction for the non-perturbative $D_2$ spectrum inclusive over the jet mass mass, one must calculate the perturbative $D_2$ spectra differentially in the jet mass, convolve with a shape function for each value of the jet mass, and then integrate over the jet mass. While this is in principle straightforward, it is computationally intensive, and is beyond the scope of this chapter. Instead, we will enforce a jet mass cut of $8< m_J <16$ GeV. This mass cut was chosen because it is near to the Sudakov peak of the jet mass distribution for this jet energy and the $m_J$ in this range are set by low scale, but still perturbative, emissions.

\begin{figure}
\begin{center}
\subfloat[]{\label{fig:D2_LEP1_a}
\includegraphics[width= 7.2cm]{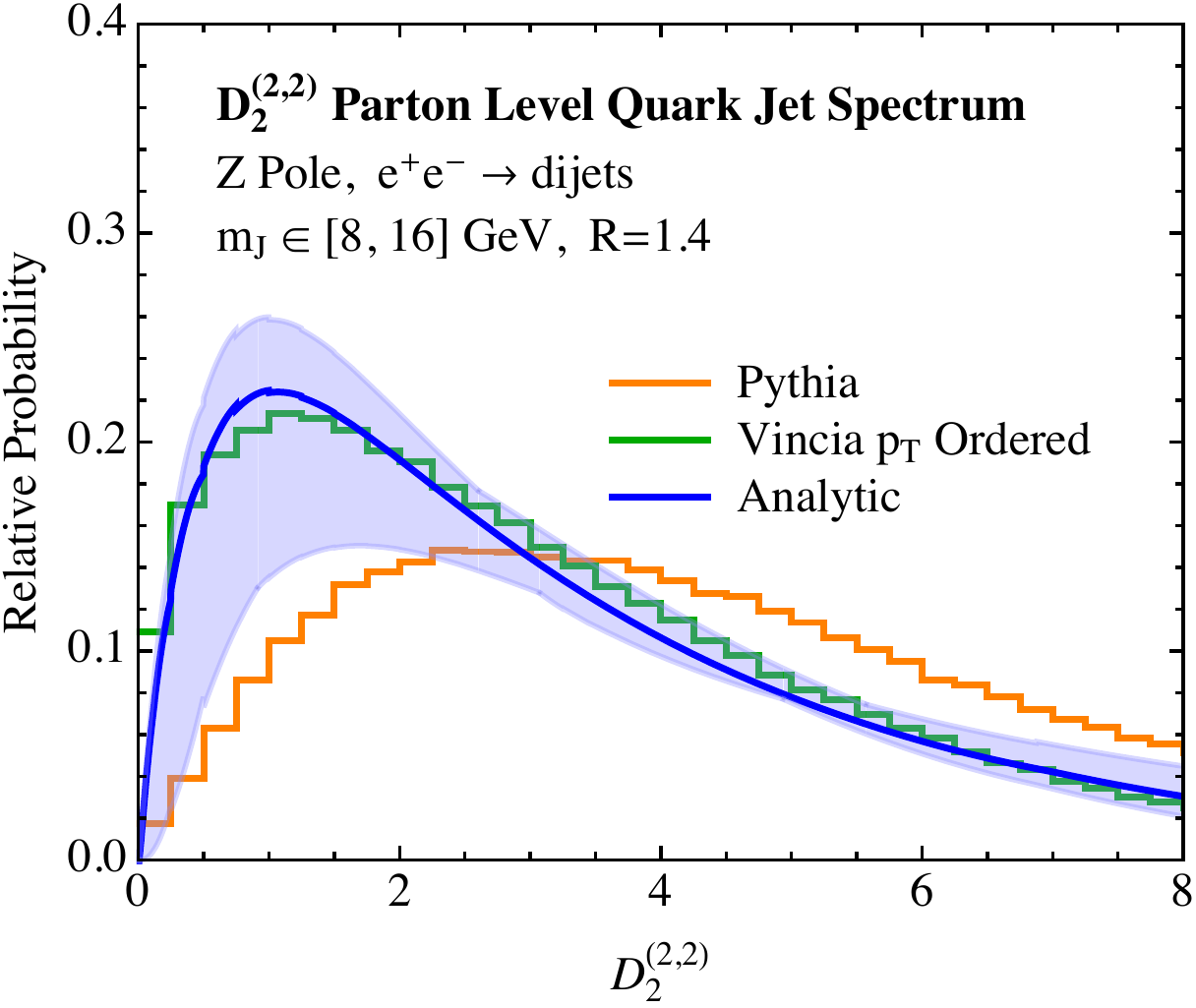}
}
\ 
\subfloat[]{\label{fig:D2_LEP1_b}
\includegraphics[width = 7.2cm]{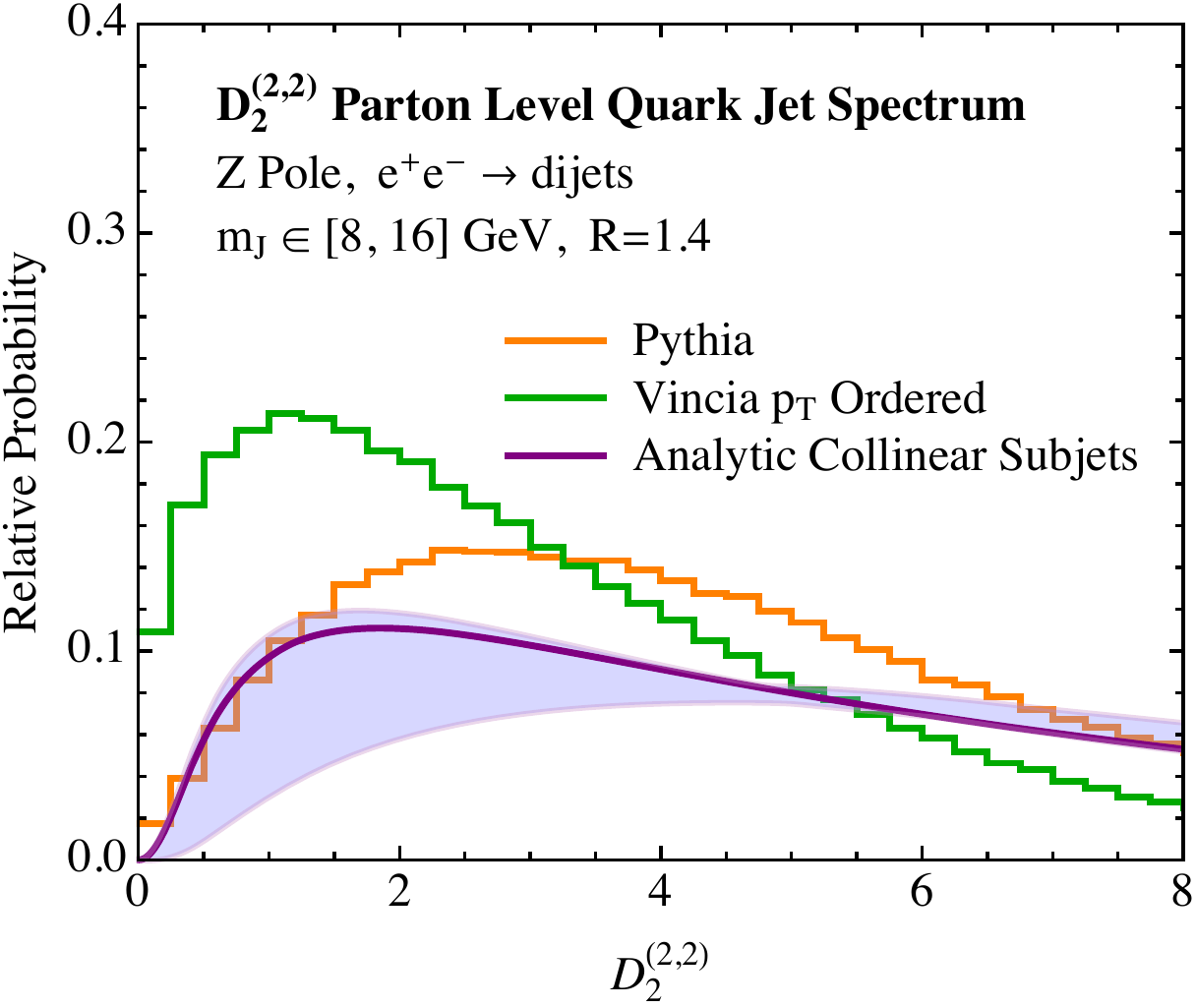}
}
\end{center}
\vspace{-0.2cm}
\caption{A comparison of the $D_2$ spectrum as measured on quark initiated jets at the $Z$ pole from the \pythia{} and $p_T$-ordered \vincia{} Monte Carlo generators to our analytic predictions. a) Comparison of our complete analytic calculation including both the soft subjet and collinear subjets region of phase space with the predictions of the Monte Carlo generators. b) Comparison of our analytic calculation including only the collinear subjets region of phase space compared with the predictions of the Monte Carlo generators.  
}
\label{fig:D2_LEP1}
\end{figure}

Similar to what we did in our numerical analysis at $1$ TeV, we begin in \Fig{fig:D2_LEP1} by comparing our analytic prediction for the $D_2$ spectrum with the distributions from parton level Monte Carlo. In \Fig{fig:D2_LEP1_a}, we show a comparison of our complete analytic calculation, including perturbative scale variations, along with Monte Carlo predictions from both  \pythia{} and $p_T$-ordered \vincia{}, which we take as representative of the different Monte Carlo generators. We use a jet radius of $R=1.4$ to approximate hemisphere jets. We find good agreement with the prediction with the \vincia{} Monte Carlo, and much worse agreement in the shape of the distribution with the \pythia{} generator. We believe that this is again due to the effects of wide angle soft radiation, which are not well modeled by the implementation of the perturbative shower in \pythia{}, but which are well modeled by the antenna dipole shower of the \vincia{} Monte Carlo generator, and are explicitly included in our analytic calculation through the soft subjet factorization theorem.\footnote{Again it is important to emphasize, that, as was discussed in \Sec{sec:scales}, there are also large uncertainties due to the treatment of the shower cutoff.} To demonstrate that this is indeed the case, in \Fig{fig:D2_LEP1_b}, we show a comparison of our analytic prediction, including only the collinear subjets region of phase space, with both Monte Carlo generators. Excellent agreement is seen with the \pythia{} generator, particularly at small $D_2$, while the agreement between the prediction including only the collinear subjets region of phase space and the \vincia{} prediction are completely different. This emphasizes the large effect played by the soft subjet at LEP energies.

\begin{figure}
\begin{center}
\subfloat[]{\label{fig:D2_LEP_a}
\includegraphics[width= 7.2cm]{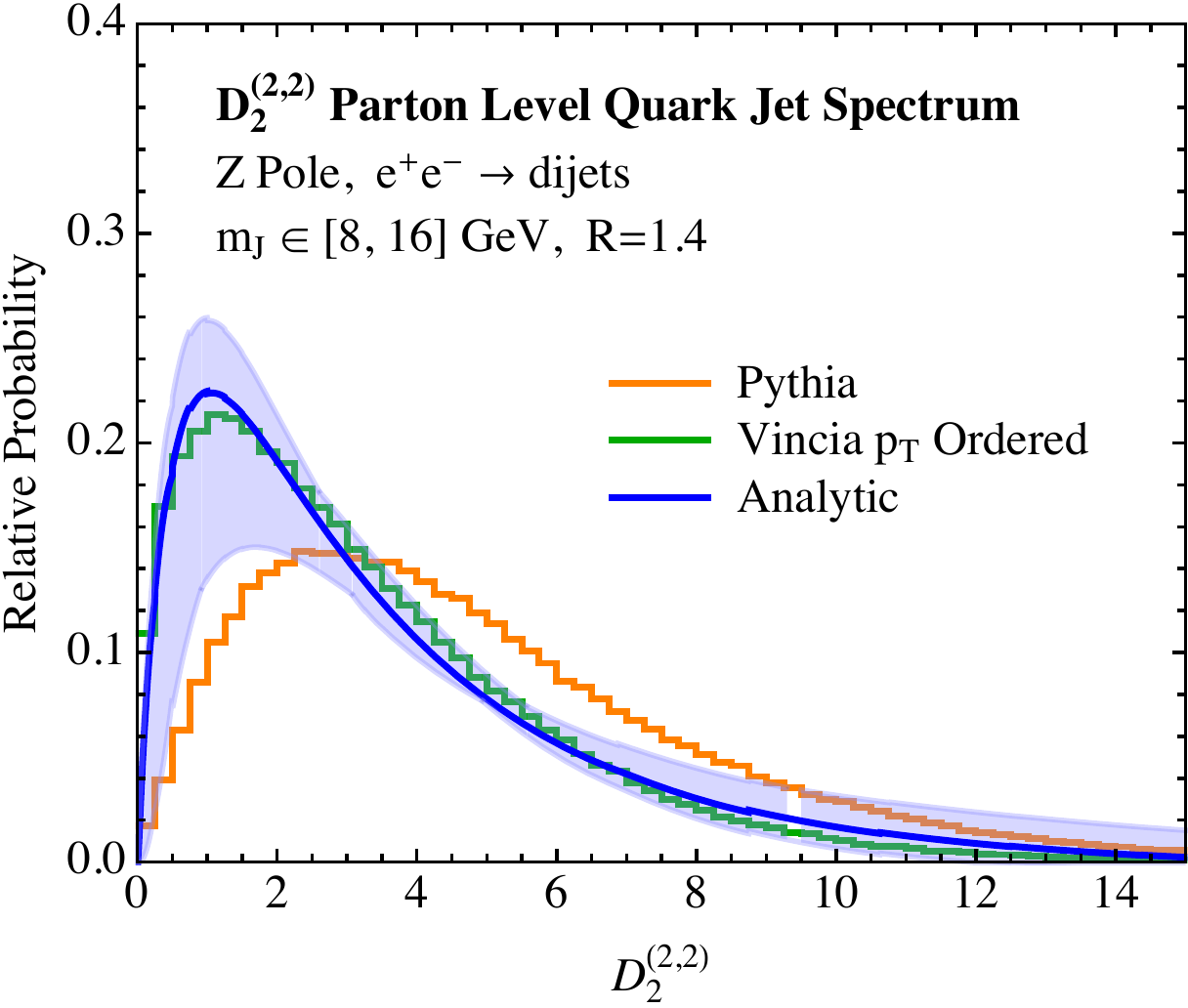}
}
\ 
\subfloat[]{\label{fig:D2_LEP_b}
\includegraphics[width = 7.2cm]{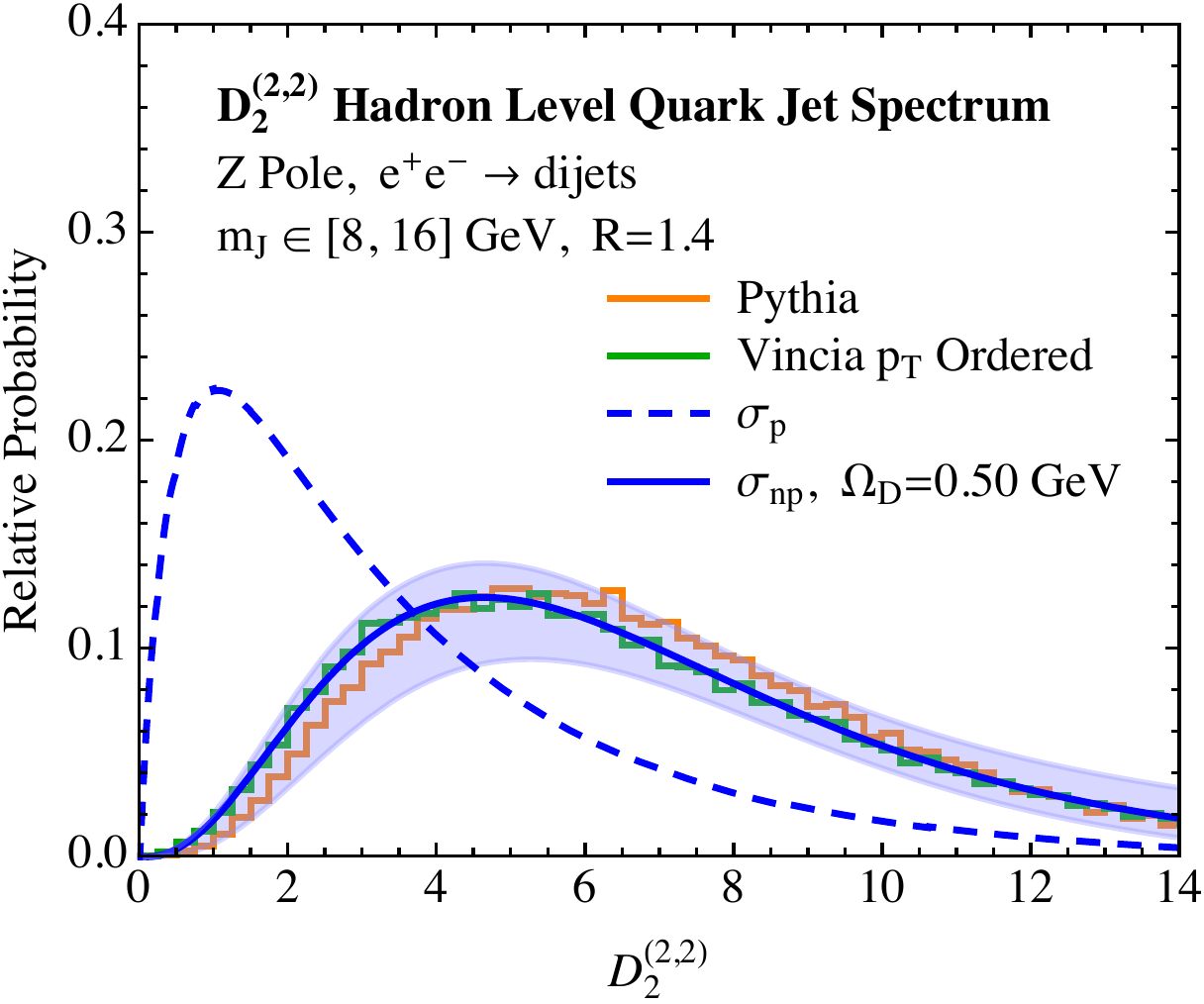}
}
\end{center}
\vspace{-0.2cm}
\caption{A comparison of the $D_2$ spectrum as measured on quark initiated jets at the $Z$ pole from the \pythia{} and $p_T$-ordered \vincia{} Monte Carlo generators to our analytic predictions. Results are shown both for parton level Monte Carlo compared with perturbative analytics in a), and for hadron level Monte Carlo compared with non-perturbative analytics in b). 
}
\label{fig:D2_LEP}
\end{figure}

In \Fig{fig:D2_LEP_b} we show our analytic prediction for the non-perturbative spectra using the shape function. An alternate view of the perturbative spectrum is shown in  \Fig{fig:D2_LEP_a} for reference, and to show the overall shape of the perturbative distribution.  We have used a valued of $\Omega_D=0.50$ GeV, which was obtained by fitting to the \vincia{} Monte Carlo, since we obtained the best agreement with the shape of the \vincia{} Monte Carlo at parton level with our perturbative spectrum. There is considerable uncertainty on this value, probably of the order $\pm0.3$ GeV due to the wide mass window, which is probably slightly large for the na\"ive application of our shape function. Furthermore, as demonstrated in \Sec{sec:Hadronization}, there is some ambiguity in the value of $\Omega_D$, depending on whether it is extracted from hadron level \pythia{} or \vincia{}, which is of this same order. However, this value is consistent with $\Omega_D=0.34$ GeV as extracted from our analysis at $1$ TeV. Although it is expected that the logarithmic running of the $\Omega_D$ parameter will decrease its value slightly, this effect is expected to be small.  The amount by which it is expected to decrease depends on another non-perturbative parameter, but is estimated in \App{app:shape_RGE} that $\Omega_D$ should decrease by approximately $0.1$ GeV between our predictions at $1$ TeV and those at LEP energies.  This is an important consistency check on our results, but due to the large uncertainty, we cannot claim to probe this running over the scales that we have considered. The analytic perturbative spectrum is also shown for reference. Good overall agreement with both Monte Carlo generators is observed, and the discrepancy between the \pythia{} and \vincia{} generators which was present at parton level is reduced, although still non-negligible. As was discussed in \Sec{sec:Hadronization}, it could also be compensated for by a modification of the non-perturbative shape parameter. In particular, the effect of hadronization is well captured by non-perturbative shape function. Hadronization has a significantly larger effect on the $D_2$ observable at $Z$ pole energies than at $1$ TeV. This demonstrates the consistency of our implementation of the non-perturbative corrections through the shape function, which predicts the scaling of the shift in the first moment through \Eq{eq:LEP_shape}. 

Unlike for the $D_2$ distributions at $1$ TeV, where the effect of hadronization was well described only by a shift in the first moment, at LEP energies the hadronization also has a non-trivial effect on the shape of the distribution. This can clearly be seen by comparing the dashed perturbative spectrum and the non-perturbative results in \Fig{fig:D2_LEP_b}. While our factorization of non-perturbative effects in terms of a shape function is completely generic, it is only the first moment of the shape function which is universal, with the full non-perturbative shape function being in general observable dependent. However, the modification in the shape of the $D_2$ spectrum due to hadronization effects seems to be quite well captured by the shape function of \Eq{eq:shape_func}. In our plots we do not include any uncertainties due to the form of the non-perturbative shape function, despite the fact that they are the dominant effect throughout most of the hadronized distribution. More general shape functions, and a study of their associated uncertainties could be studied along the lines of \Ref{Ligeti:2008ac}, although this is beyond the scope of this chapter, and could only be justified if the perturbative components of our calculation were computed to a higher level of accuracy.

Since the $D_2$ spectrum is sensitive to the emissions from the gluon subjet, it is sensitive to the radiation pattern generated by a gluon, and could potentially be used to improve the Monte Carlo description of gluons and the modeling of color coherence effects. In contrast to most observables which have been used for tuning Monte Carlos to LEP data, such as the jet mass which is set by a single emission, $D_2$ requires two emissions off of the initiating quark to be non-zero, and therefore can be used as a more detailed probe of the perturbative shower. Although non-perturbative effects play a large role for jets in this energy range, we have shown that our factorization theorem allows us to cleanly separate perturbative from non-perturbative effects, which could be useful when tuning Monte Carlo generators, allowing one to disentangle genuine perturbative effects which should be well described by the Monte Carlo shower, from effects which should be captured by the hadronization model. We believe that higher order calculations of QCD jet shapes sensitive to three particle correlations, such as $D_2$, and their use in Monte Carlo tunings is therefore well motivated.

For reference, in \App{app:twoemissionMC} we show a collection of $\ecf{2}{2}$ distributions measured at the $Z$ pole, at both parton and hadron level for both the \vincia{} and \pythia{} event generators. Unlike for the $D_2$ observable, the \vincia{} and \pythia{} generators agree both at parton and hadron level to an excellent degree. This is of course expected due to the fact that these Monte Carlos have been tuned to LEP event shapes, but further emphasizes the fact that $D_2$, and other observables sensitive to additional emissions, provide a more detailed probe of the perturbative shower.

\section{Looking Towards the LHC}\label{sec:LHC}

Throughout this chapter, we have restricted our analysis to $e^+e^-$ colliders so that we could ignore subtleties with initial state radiation, pile-up and other features important at hadron colliders. However, it is precisely for including these effects that a rigorous factorization based approach to jet substructure, such as that presented in this chapter, will prove most essential.  In this section, we discuss the extension to the LHC and in particular to what extent conclusions for $e^+e^-$ colliders holds for the LHC.

The energy correlation functions have a natural longitudinally-invariant generalization relevant for $pp$ colliders, which is given by \cite{Larkoski:2013eya,Larkoski:2014gra}
\begin{align}\label{eq:pp_def_ecf}
\ecf{2}{\beta} &= \frac{1}{p_{TJ}^2}\sum_{1\leq i<j\leq n_J} p_{Ti}p_{Tj} R_{ij}^\beta \ ,\nonumber \\
\ecf{3}{\beta} &= \frac{1}{p_{TJ}^3}\sum_{1\leq i<j<k\leq n_J} p_{Ti}p_{Tj}p_{Tk} R_{ij}^\beta R_{ik}^\beta R_{jk}^\beta \ .
\end{align}
Here $p_{TJ}$ is the transverse momentum of the jet with respect to the beam, $p_{Ti}$ is the transverse momentum of particle $i$, and $n_J$ is the number of particles contained in the jet.  The boost-invariant angle $R_{ij}^2 = (\phi_i-\phi_j)^2+(y_i-y_j)^2$ is defined as the Euclidean distance in the azimuth-rapidity plane. For central rapidity jets, which we will restrict ourselves to in this section, the power counting discussion of \Sec{sec:phase_space} is unmodified. Therefore, the same conclusions for the form of the optimal observable, $D_2$, as well as the range of angular exponents, apply. A simplified version of the $\Dobs{2}{\alpha, \beta}$ variable, restricted to have equal angular exponents $\alpha=\beta$, was used in \Ref{Larkoski:2014gra}, for jet substructure studies at the LHC.

It is in principle straightforward to extend the factorization theorems for $D_2$ to hadronic colliders, where $D_2$ is measured on a single jet in an exclusive $N$-jet event. Factorization theorems for exclusive $N$-jet production defined using $N$-jettiness \cite{Stewart:2009yx,Stewart:2010tn} or with a $p_T$-veto \cite{Liu:2012sz,Liu:2013hba} on additional radiation exist and could be combined with the factorization theorems of \Sec{sec:Fact} to describe the jet substructure. We now briefly discuss how each of these factorization theorems can be interfaced with the presence of additional eikonal lines, representing either additional jets or beam directions in $pp$ collisions.

Recall from \Sec{sec:ninja}, that the collinear subjets factorization theorem is formulated as a refactorization of the jet function for a particular jet in the $n$ direction,  and it is therefore insensitive to the global color structure of the event, seeing only the total color. Intuitively, the collinear-soft modes are boosted, and therefore all additional Wilson lines in the event are grouped in the $\bar n$ direction. Furthermore, the global soft modes, which resolve the global color structure of the event do not resolve the jet substructure. This property of the collinear subjets factorization theorem has the feature that it can be trivially combined with a factorization theorem with an arbitrary number of eikonal lines, without complicating the color structure. All that is then required, apart from the substructure components, is the addition of an additional measurement function in to the global soft function. Indeed, this extension has been discussed in detail in \Ref{Bauer:2011uc}. This same property is of course also true for the soft haze factorization theorem, as no additional Wilson lines are required to describe the jet substructure in the first place. 

However, for the soft subjet factorization theorem, the presence of additional Wilson lines does significantly complicate the factorization from a calculational perspective. In particular, since the subjet is soft, arising from a refactorization of the soft function, it is emitted coherently from the $N$-eikonal line structure as a whole, requiring a proper treatment of all color correlations, which becomes complicated with even a few additional Wilson lines. A conjectural proposal for the all orders soft subjet factorization theorem with $N$-eikonal lines was given in \Ref{Larkoski:2015zka}, where the soft subjet factorization theorem was first proposed and studied in the large $N_c$ limit. However, more work is required to understand its structure, and an efficient organization of the color correlations at finite $N_c$. Furthermore for the soft subjet factorization theorem, the final soft function has an additional eikonal line, since the jet substructure is resolved by the long wavelength global soft modes, further complicating the calculation (although there has recently been some progress in the computation of soft functions \cite{guido_talk,Boughezal:2015eha}). We emphasize however, that these are purely technical complications, and believe that the extension to a calculation of jet substructure in $pp$ would be well worthwhile for improving our understanding of analytic jet substructure. Furthermore, depending on the relevant boosts and jet radii, the techniques of this chapter could be used to identify whether the soft subjet factorization theorem plays an important role, or could be formally neglected, simplifying the calculation in more complicated cases.

\begin{figure}
\begin{center}
\subfloat[]{\label{fig:D2_LHC_a}
\includegraphics[width= 7cm]{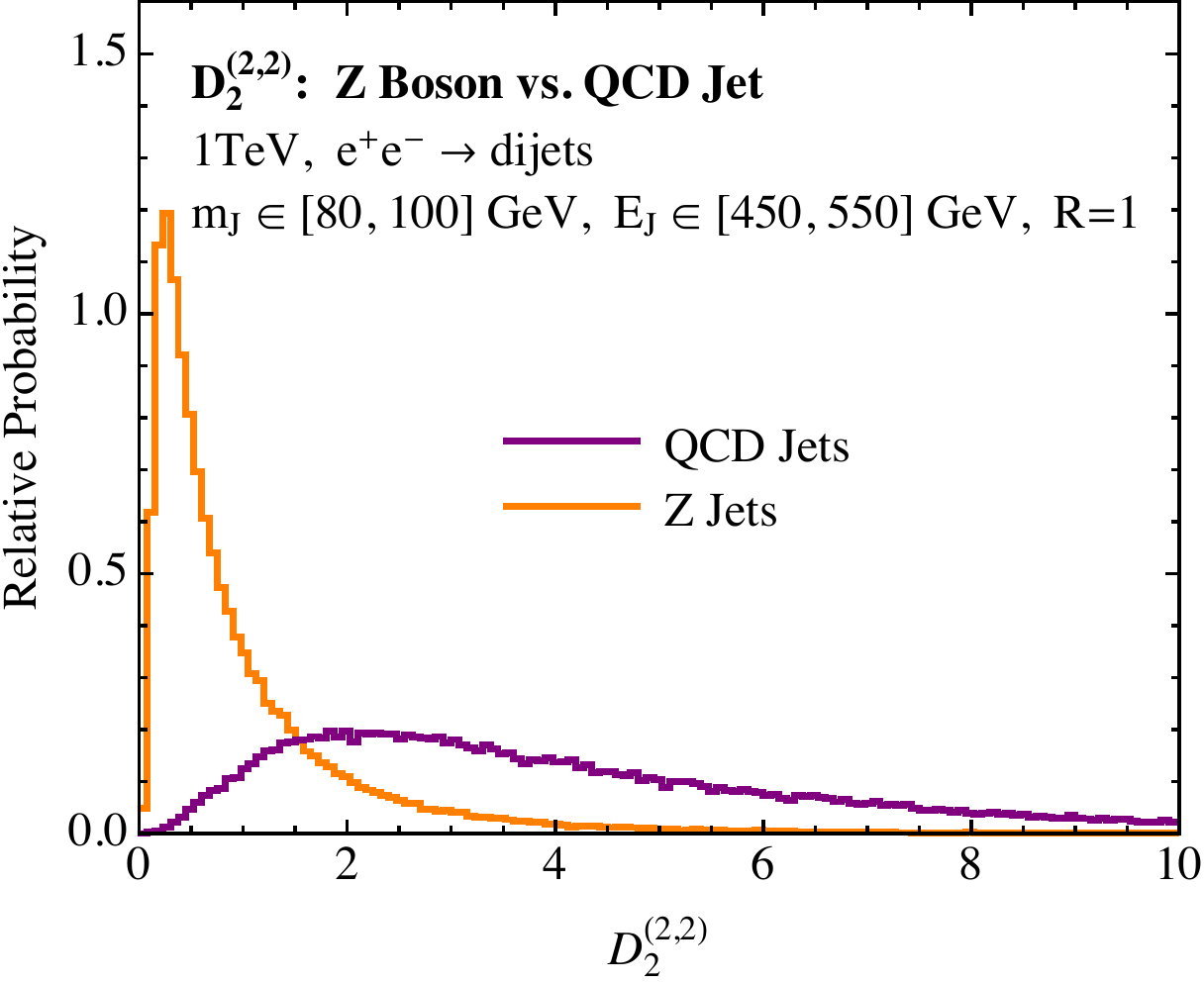}
}
\ 
\subfloat[]{\label{fig:D2_LHC_b}
\includegraphics[width = 7cm]{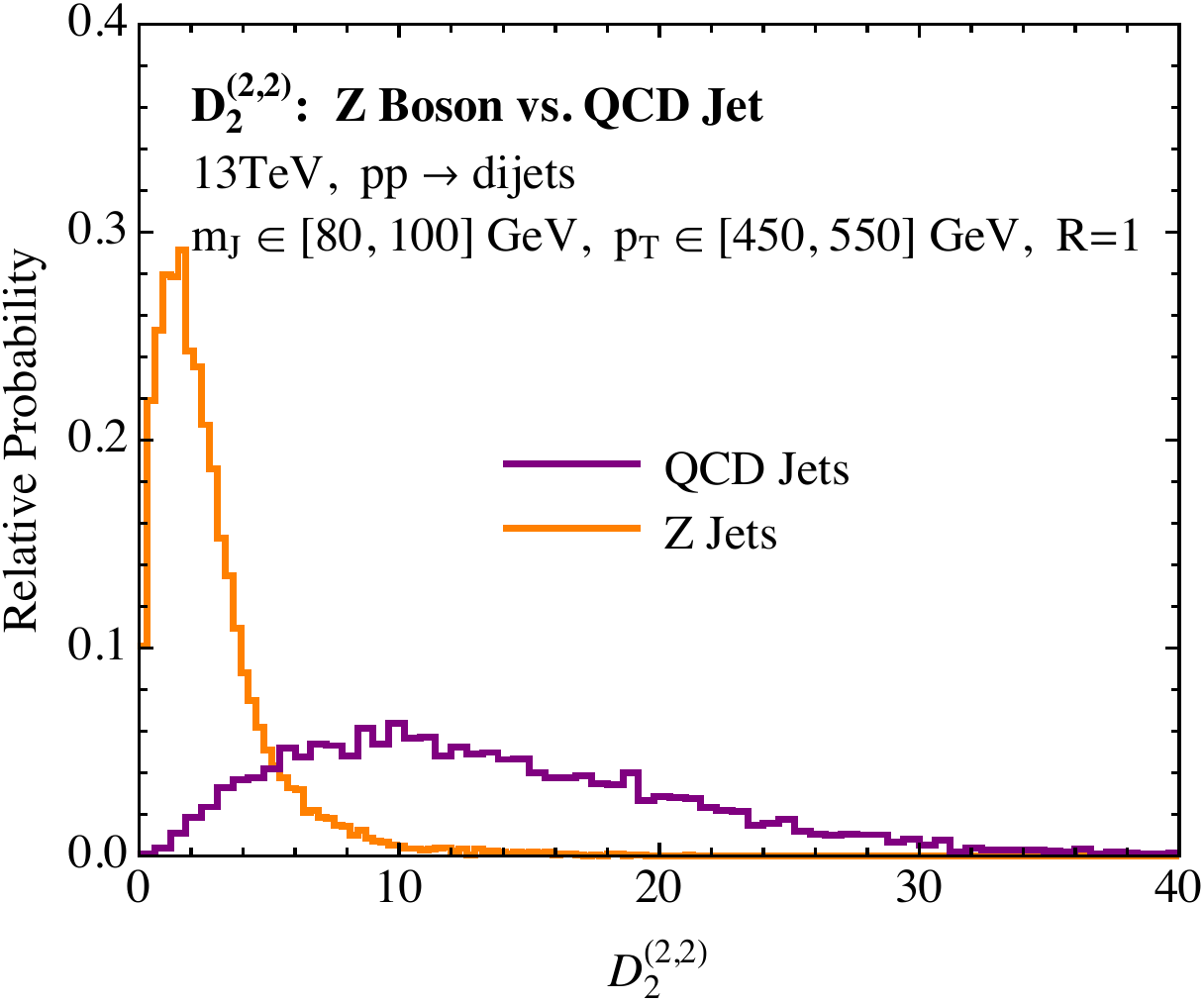}
}
\end{center}
\vspace{-0.2cm}
\caption{A comparison of the $\Dobs{2}{2,2}$ distributions for signal and background jets.  a) Distributions for $R=1$ jets at a 1 TeV $e^+e^-$ collider.  b) Distributions for $R=1$ jets at the 13 TeV LHC, for jets with transverse momenta in the range $p_T\in[450,550]$ GeV.
}
\label{fig:D2_LHC}
\end{figure}

For these reasons, a full calculation in $pp$ is well beyond the current scope of this initial investigation. We will instead restrict ourselves to a brief Monte Carlo study comparing the properties of $D_2$ in $e^+e^-$ and $pp$ to show that the distributions exhibit similar features. In \Fig{fig:D2_LHC} we compare the Monte Carlo predictions for $\Dobs{2}{2,2}$ as measured in $e^+e^-$ collisions, shown in \Fig{fig:D2_LHC_a}, and $pp$ collisions, shown in \Fig{fig:D2_LHC_b}.  For $e^+e^-$ collisions, the event selection is identical to earlier.  For $pp$ collisions, we generate background events from the parton-level process $pp \to q \bar q$ and signal events from $pp \to ZZ \to q\bar q q\bar q$ events, where $q$ denotes a massless quark, with \pythia{8.205} at the $13$ TeV LHC.\footnote{Since we only briefly mention the case of $pp$ colliders, we do not perform a systematic study of the variation of the $D_2$ distribution in $pp$ with different Monte Carlo generators, as we did for the case of $e^+e^-$. However, we believe that this is essential in any jet substructure study at $pp$, as we expect large variations will be present, as in the $e^+e^-$ case. It would be particularly interesting to compare a $p_T$-ordered dipole-antenna shower, such as was recently implemented for $pp$ in \dire{} \cite{Hoche:2015sya}, with the \pythia{} and \herwigpp{} generators which are more commonly used in jet substructure studies at the LHC. As was found for the case of $e^+e^-$, we expect a $p_T$-ordered dipole-antenna shower to provide the most accurate description of substructure observables at a $pp$ collider, in particular, those sensitive to wide angle soft radiation.} Jets are clustered with the anti-$k_T$ algorithm with radius $R=1.0$, and using the WTA recombination scheme, with a $p_T$ metric. We cut on the transverse momentum of the hardest jet, requiring $p_T \in[450,550]$ GeV, and on the jet mass requiring $m_J \in [80,100]$ GeV. These are chosen to be similar to the cuts on the jets for the case of $e^+e^-$, although they are of course not identical, and strict comparisons should not be made between the two cases. The shapes and general features of the $D_2$ distributions at the two colliders are very similar.  There is a relative scaling between the $D_2$ distributions in $e^+e^-$ and $pp$ due to the different observable definitions. The $e^+e^-$ definition uses the $1-\cos(\theta_{ij})$ measure of \Eq{eq:ecf_def}, while the $pp$ definition uses the boost invariant definition in terms of $R_{ij}$, as in \Eq{eq:pp_def_ecf}. Since the $\ecf{3}{\alpha}$ observable correlates particles of separation up to $2R$, where $R$ is the jet radius, for $\alpha=\beta=2$, this gives an expected factor of $4$ difference between the two cases, as is approximately observed in \Fig{fig:D2_LHC}.

The similar behavior of the $e^+e^-$ and $pp$ distributions suggests that a complete a calculation using our techniques would provide an excellent description of the $D_2$ distribution at a hadron collider, as we have found for $e^+e^-$. Such a calculation would also be interesting to better understand the effects of initial state radiation on the $D_2$ distribution. A simple setting where this calculation would be feasible, for example, would be to consider measuring the $D_2$ distribution on a jet recoiling against a color-singlet such as a $W$, $Z$ or $H$ boson, as was used in \Ref{Jouttenus:2013hs} to perform a NNLL calculation of the jet mass. Although the effects of non-global logarithms would need to be understood, and could play an important role, recent progress in this area suggests that this issue could be addressed, either by direct resummation of the NGLs \cite{Banfi:2002hw,Weigert:2003mm,Hatta:2013iba,Caron-Huot:2015bja,Larkoski:2015zka}, or through the use of jet grooming algorithms which remove NGLs \cite{Dasgupta:2013ihk,Dasgupta:2013via,Larkoski:2014wba}.  While it is truly uncorrelated with the jet, the effect of radiation from pile-up on $D_2$ could also be mitigated using similar jet grooming algorithms.

\section{Conclusions}\label{sec:conc}

In this chapter we have presented a novel approach to the factorization of jet substructure observables, and applied it to the identification of two-prong substructure.  Instead of starting  with a given two-prong discriminant, we used the energy correlation functions as a basis of IRC safe observables to isolate the possible subjet configurations. We then studied the phase space defined by these IRC safe observables and proved all orders factorization theorems in each region of phase space.  This procedure naturally identified an observable, $D_2$, which we argued provided optimal discrimination power, and which preserved the factorization properties of the individual factorization theorems describing different regions of the phase space defined by our basis of observables. We showed that a factorized description of this observable could be obtained by merging the different factorization theorems, and introduced a novel zero bin procedure in factorization theorem space to implement this merging. An important benefit of this approach is that our factorization theorems are valid to all orders in $\alpha_s$ at leading power and therefore provide a systematically improvable description of $D_2$.

Using our factorized description of the $D_2$ observable, we presented a numerical study of our results at an $e^+e^-$ collider, for both the signal and background distributions, resulting in analytic boosted $Z$ boson versus massive QCD jet discrimination predictions. We compared with a variety of Monte Carlo generators, and demonstrated that the low $D_2$ region, where a hard two-prong substructure is resolved, is a sensitive probe of the Monte Carlo parton shower description.  We also studied the effect of non-perturbative corrections, showing that they can be well-described using a simple shape function, and related the single parameter of this shape function to a universal non-perturbative matrix element measured at LEP. This is vital for comparing our calculation with data. 

Because our calculation presents the first factorized description of a two-prong discriminant jet observable in both signal and background regions, there are a large number of directions for future study which are of great interest. First, our calculation was presented in the context of jets produced in $e^+e^-$ collisions. For applications at the LHC, where jet substructure plays a vital role, it is important to extend the calculation to jets produced at a $pp$ collider. The factorization theorem we presented straightforwardly generalizes to $pp$ colliders with only complications due to soft radiation from the beams and the more complicated color structure of the hard interaction. The treatment of both these effects are well-understood and their inclusion in a jet substructure calculation would allow the first precision comparisons of calculations with data.

An interesting potential application of our factorization theorems, and merging procedures, which describe in a more differential way the substructure of jets, is to improve jet shape based subtraction schemes for QCD calculations at NNLO and beyond. Quite recently, subtractions based on the $N$-jettiness observable \cite{Stewart:2010tn} have been used to perform NNLO calculations in QCD \cite{Boughezal:2015aha,Boughezal:2015dva,Gaunt:2015pea}. This allowed, in particular, the calculation of $W$, $H+1$ jet at NNLO  \cite{Boughezal:2015aha,Boughezal:2015dva} ($H+1$ jet at NNLO was also calculated using more traditional subtraction techniques in \cite{Boughezal:2015dra}). The use of more differential subtractions based on more differential factorization theorems would allow for more local, and potentially numerically more efficient subtractions.

It would also be interesting to apply our calculation approach to other observables. For example, the $N$-subjettiness observables \cite{Thaler:2010tr,Thaler:2011gf} are used extensively in jet substructure studies at the LHC, and it would be of significant phenomenological relevance to obtain a factorized description of these observables. The approach presented in this chapter could also be extended to study more differential observables, such as those used for boosted top discrimination, which can resolve three subjets. A generalization of the $\Dobs{2}{\alpha, \beta}$ observable, $\Dobs{3}{\alpha, \beta, \gamma}$, which resolves three prong structure was introduced in \Ref{Larkoski:2014zma} (see also \Ref{Larkoski:2015yqa} where it was used for boosted top discrimination at a 100 TeV collider). The $\Dobs{3}{\alpha, \beta, \gamma}$ observable should exhibit similar factorization properties to that of $\Dobs{2}{\alpha, \beta}$, and hence should be calculable with similar techniques. A rigorous factorization will also prove essential in this case, allowing for the separation of perturbative and non-perturbative physics, as well as effects associated with the finite top width \cite{Fleming:2007qr,Fleming:2007xt}. More generally, we anticipate that the approach to the factorization of jet substructure observables presented in this chapter will allow for the construction of more powerful jet substructure discriminants and will enable a more detailed analytic understanding of the substructure of high energy QCD jets.

%% file: chap4.tex
\newcommand{\eq}[1]{Eq.~\eqref{eq:#1}}
\newcommand{\eqs}[2]{Eqs.~\eqref{eq:#1} and \eqref{eq:#2}}
\newcommand{\eqss}[3]{Eqs.~\eqref{eq:#1}, \eqref{eq:#2}, and \eqref{eq:#3}}
\renewcommand{\sec}[1]{Sec.~\ref{sec:#1}}
\newcommand{\secs}[2]{Secs.~\ref{sec:#1} and \ref{sec:#2}}
\newcommand{\secss}[3]{Secs.~\ref{sec:#1}, \ref{sec:#2}, and \ref{sec:#3}}
\newcommand{\subsec}[1]{Sec.~\ref{subsec:#1}}
\newcommand{\subsecs}[2]{Secs.~\ref{subsec:#1} and \ref{subsec:#2}}
\newcommand{\fig}[1]{Fig.~\ref{fig:#1}}
\newcommand{\figs}[2]{Figs.~\ref{fig:#1} and \ref{fig:#2}}
\newcommand{\app}[1]{Appendix~\ref{app:#1}}
\newcommand{\apps}[2]{Apps.~\ref{app:#1} and \ref{app:#2}}

\newcommand{\abs}[1]{\lvert#1\rvert}
\newcommand{\vev}[1]{\langle #1 \rangle}
\newcommand{\Vev}[1]{\bigl\langle #1 \bigr\rangle}
\newcommand{\mae}[3]{\langle#1\rvert#2\rvert#3\rangle}
\newcommand{\Mae}[3]{\bigl\langle#1\bigr\rvert#2\bigr\rvert#3\bigr\rangle}
\newcommand{\MAe}[3]{\Bigl\langle#1\Bigr\rvert#2\Bigr\rvert#3\Bigr\rangle}
\newcommand{\ang}[1]{\langle #1 \rangle}

\newcommand{\inte}[1]{\int\! \df #1 \,}
\newcommand{\intlim}[3]{\int_{#1}^{#2}\! \df #3 \,}

\newcommand{\Li}{\mathrm{Li}}

\newcommand{\fr}{\frac}
\newcommand{\ra}{\rightarrow}
\newcommand{\lra}{\leftrightarrow}

\newcommand{\al}{\alpha}
\newcommand{\bt}{\beta}
\newcommand{\ga}{\gamma}
\newcommand{\Ga}{\Gamma}
\newcommand{\de}{\delta}
\newcommand{\e}{\epsilon}
\newcommand{\eps}{\epsilon}
\newcommand{\ve}{\varepsilon}
\newcommand{\La}{\Lambda}
\newcommand{\si}{\sigma}
\newcommand{\om}{\omega}
\newcommand{\balpha}{{\bar \alpha}}
\newcommand{\bbeta}{{\bar \beta}}
\newcommand{\bgamma}{{\bar \gamma}}
\newcommand{\bdelta}{{\bar \delta}}

\newcommand{\cBslash}{\cB\!\!\!\!\slash}
\newcommand{\cH}{{\mathcal H}}
\newcommand{\cI}{{\mathcal I}}
\newcommand{\cM}{{\mathcal M}}
\newcommand{\cV}{{\mathcal V}}
\newcommand{\bq}{{\bar{q}}}
\newcommand{\mcdot}{{\!\cdot\!}}

\newcommand{\vC}{\vec{C}}

\newcommand{\vT}{\bar{T}}

\newcommand{\hD}{\widehat{D}}
\newcommand{\hatt}{\hat{t}}
\newcommand{\hT}{\widehat{T}}
\newcommand{\hU}{\widehat{U}}
\newcommand{\hX}{\widehat{X}}
\newcommand{\hZ}{\widehat{Z}}
\newcommand{\hga}{\widehat{\gamma}}
\newcommand{\hw}{\widehat{\omega}}
\newcommand{\hDe}{\widehat{\Delta}}

\newcommand{\Bslash}{B\!\!\!\!\slash}
\newcommand{\Dslash}{D\!\!\!\!\slash}
\newcommand{\kslash}{k\!\!\!\slash}
\newcommand{\bnslash}{\bar{n}\!\!\!\slash}
\newcommand{\pslash}{p\!\!\!\slash}
\newcommand{\lpslash}{\lp\!\!\!\slash}
\newcommand{\qslash}{q\!\!\!\slash}
\newcommand{\vslash}{v\!\!\!\slash}

\newcommand{\GeV}{\:\mathrm{GeV}}


\newcommand{\lqcd}{\Lambda_\mathrm{QCD}}

\newcommand{\ldel}{\tilde \delta} 

\newcommand{\Ecm}{E_\mathrm{cm}}

\newcommand{\soft}{{s}}

\renewcommand{\P}{\mathrm{P}}       
\newcommand{\C}{\mathrm{C}}       

\newcommand{\IR}{\mathrm{IR}}
\newcommand{\UV}{\mathrm{UV}}
\newcommand{\eff}{\mathrm{eff}}
\newcommand{\fin}{\mathrm{fin}}
\renewcommand{\div}{\mathrm{div}}
\newcommand{\cusp}{\mathrm{cusp}}

\newcommand{\id}{\mathbf{1}}

\newcommand{\Tau}{\mathrm{T}}

\newcommand{\Op}{\vec{O}}

\newcommand{\zero}{{(0)}}
\newcommand{\one}{{(1)}}
\newcommand{\two}{{(2)}}

\setcounter{MaxMatrixCols}{22}

\renewcommand{\arraystretch}{1.0}
\tabcolsep 5pt

\allowdisplaybreaks[4]

\chapter{Employing Helicity Amplitudes for Resummation}\label{chap:helops}

\section{Introduction}
\label{sec:intro}

The production of hadronic jets is one of the most basic processes at particle colliders. Processes including a vector boson ($W$, $Z$, $\gamma$) or Higgs boson together with jets provide probes of the Standard Model (SM), and are also dominant backgrounds for many new-physics searches. Optimizing the precision and discovery potential of these channels requires accurate predictions of the SM backgrounds. Furthermore, the growth of the jet substructure field has sparked a renewed interest in the study of jets themselves, both for an improved understanding of QCD, and for applications to identify boosted heavy objects in and beyond the SM.

Precise predictions for jet production require perturbative calculations including both fixed-order corrections as well as logarithmic resummation. QCD corrections to processes with jets are typically enhanced due to phase space restrictions. Such restrictions often introduce sensitivity to low momentum scales, $p$, of order a few tens of GeV, in addition to the hard scale, $Q$, which is of order the partonic center-of-mass energy. In this case, the perturbative series contains large double logarithms $\alpha_s^n\ln^m(p/Q)$ with $m \leq 2n$. To obtain the best possible perturbative predictions, these logarithms should be resummed to all orders in $\alpha_s$.

There has been tremendous progress in the calculation of fixed-order perturbative amplitudes in QCD using the spinor helicity formalism \cite{DeCausmaecker:1981bg,Berends:1981uq,Gunion:1985vca,Xu:1986xb}, color ordering techniques \cite{Berends:1987me,Mangano:1987xk,Mangano:1988kk,Bern:1990ux} and unitarity based methods \cite{Bern:1994zx,Bern:1994cg}. NLO predictions are now available for a large number of high multiplicity final states, including $pp\rightarrow V+$ up to 5 jets \cite{Giele:1991vf, Arnold:1988dp, Giele:1993dj, Bern:1997sc, Ellis:2008qc, Berger:2009zg, Berger:2009ep, Berger:2010vm, Berger:2010zx, Ita:2011wn, Bern:2013gka}, $pp\rightarrow$ up to 5 jets \cite{Bern:1993mq, Kunszt:1993sd, Kunszt:1994tq, Bern:1994fz, Nagy:2001fj, Bern:2011ep, Badger:2012pf, Badger:2013yda}, and $pp\rightarrow  H+$ up to 3 jets \cite{Campbell:2006xx, Kauffman:1996ix, Schmidt:1997wr, Dixon:2009uk, Badger:2009hw, Campbell:2010cz, vanDeurzen:2013rv, Cullen:2013saa, Campanario:2013fsa}, and there are many efforts~\cite{Binoth:2010ra,AlcarazMaestre:2012vp, Ossola:2007ax, Berger:2008sj,Binoth:2008uq, Mastrolia:2010nb, Badger:2010nx, Fleischer:2010sq, Cullen:2011kv, Hirschi:2011pa, Bevilacqua:2011xh, Cullen:2011ac, Cascioli:2011va, Badger:2012pg, Actis:2012qn, Peraro:2014cba, Alwall:2014hca, Cullen:2014yla} to fully automatize the computation of one-loop corrections to generic helicity amplitudes.

For high-multiplicity jet events, the resummation of large logarithms is typically achieved with parton shower Monte Carlo programs. Here, the hard process enters through tree-level (and also one-loop) matrix elements and the QCD corrections due to final-state and initial-state radiation are described by the parton shower.
The parton shower resums logarithms at the leading logarithmic (LL) accuracy, with some subleading improvements, but it is difficult to reliably assess and systematically improve its logarithmic accuracy. 

The approach we will take in this chapter is to match onto soft-collinear effective theory (SCET)~\cite{Bauer:2000ew, Bauer:2000yr, Bauer:2001ct, Bauer:2001yt}, the effective theory describing the soft and collinear limits of QCD. In SCET, the QCD corrections at the hard scale are captured by process-dependent Wilson coefficients. The low-energy QCD dynamics does not depend
on the details of the hard scattering (other than the underlying Born kinematics), similar to the parton shower picture. Resummation in SCET is achieved analytically through renormalization group evolution (RGE) in the effective theory, allowing one to systematically improve the logarithmic accuracy and assess the associated perturbative uncertainties.
For example, for dijet event shape variables in $e^+e^-$ collisions, SCET has enabled resummation to N$^3$LL accuracy and global fits for $\alpha_s(m_Z)$~\cite{Becher:2008cf, Chien:2010kc, Abbate:2010xh, Abbate:2012jh, Hoang:2014wka, Hoang:2015hka}. The analytic higher-order resummation can also be used to improve the Monte Carlo parton-shower description~\cite{Alioli:2012fc, Alioli:2013hqa, Alioli:2015toa}.
Furthermore, SCET allows for the direct calculation of exclusive jet cross sections, eliminating the need for numerical subtraction schemes for real emissions up to power corrections.

An important prerequisite for employing SCET is to obtain the hard matching coefficients, which are extracted from the fixed-order QCD amplitudes. The matching for $V+2$ parton and $H+2$ parton processes is well known from the QCD quark and gluon form factors, and is known to three loops~\cite{Baikov:2009bg, Gehrmann:2010ue, Abbate:2010xh}. The matching for $V+3$ partons~\cite{Becher:2009th, Becher:2011fc, Becher:2012xr, Becher:2013vva}, and $H+3$ partons~\cite{Liu:2012sz, Jouttenus:2013hs, Huang:2014mca, Becher:2014tsa},  has been performed at both NLO and NNLO. Partonic processes with four external quarks have been studied in SCET in Refs.~\cite{Chiu:2008vv, Zhu:2010mr, Kelley:2010qs, Wang:2010ue, Ahrens:2010zv, Ahrens:2011mw, Li:2014ula, Liu:2014oog}, and the matching for all massless $2\to2$ processes has been obtained at NLO in Ref.~\cite{Kelley:2010fn} and recently at NNLO in Ref.~\cite{Broggio:2014hoa}.

For high-multiplicity processes, the usual approach to constructing an operator basis with explicit Lorentz indices and gamma matrices is laborious. In this chapter, we introduce a convenient formalism, based on helicity operators, which allows for a seamless matching for higher multiplicity processes onto SCET. A first look at the formalism discussed here was already given in Ref.~\cite{Stewart:2012yh}.  Indeed, results for helicity amplitudes are already employed in the SCET matching calculations mentioned above, though without the construction of corresponding SCET operators.

In the spinor helicity formalism, the individual helicity amplitudes (i.e. the amplitudes for given fixed external helicities) are calculated, as opposed to calculating the amplitude for arbitrary external spins in one step and then summing over all spins at the end. One advantage is that the individual helicity amplitudes typically yield more compact expressions. And since they correspond to distinct external states, they can be squared and summed at the end. Helicity amplitudes remove the large redundancies in the usual description of (external) gauge fields, allowing for much simplified calculations particularly for amplitudes with many external gluons.

As we will see, this helicity-based approach is also advantageous in SCET.
In SCET, as we will review in \subsec{scet}, collinear fields carry label directions corresponding to the directions of jets in the process, which provide natural lightlike vectors with which to define fields of definite helicity. As we will demonstrate, the construction of an appropriate operator basis becomes simple when using operators built out of fields with definite helicity. Furthermore, using such a helicity operator basis greatly facilitates the matching of QCD onto SCET, because one can directly utilize the known QCD helicity amplitudes for the matching. Together, this substantially simplifies the study of high-multiplicity jet processes with SCET.

\subsection{Overview}

Consider a process with $N$ final-state jets and $L$ leptons, photons, or other nonstrongly interacting particles, with the  underlying hard Born process
\begin{equation} \label{eq:interaction}
\kappa_a (q_a)\, \kappa_b(q_b) \to \kappa_1(q_1) \dotsb \kappa_{N+L}(q_{N+L})
\,,\end{equation}
where $\kappa_{a,b}$ denote the colliding partons, and $\kappa_i$ denote the outgoing quarks, gluons, leptons, and other particles with momenta $q_i$. The incoming partons are along the beam directions, $q_{a,b}^\mu = x_{a,b} P_{a,b}^\mu$, where $x_{a,b}$ are the momentum fractions and $P^\mu_{a,b}$ the (anti)proton momenta.  For definiteness, we consider two colliding partons, but our discussion of the matching will be completely crossing symmetric, so it applies equally well to $ep$ and $ee$ collisions.

In SCET, the active-parton exclusive jet cross section corresponding to \eq{interaction} can be proven to factorize for a variety of jet resolution variables.\footnote{Here active parton refers to initial-state quarks or gluons. Proofs of factorization with initial-state hadrons must also account for effects due to Glaubers~\cite{Collins:1988ig}, which may or may not cancel, and whose relevance depends on the observable in question~\cite{Gaunt:2014ska,Zeng:2015iba}.} The factorized expression for the exclusive jet cross section can be written schematically in the form
\begin{align} \label{eq:sigma}
\df\sigma &=
\int\!\df x_a\, \df x_b\, \df \Phi_{N+L}(q_a \!+ q_b; q_1, \ldots)\, M(\{q_i\})
\\\nn &\quad \times
\sum_{\kappa} \tr\,\bigl[ \hH_{\kappa}(\{q_i\}) \hS_\kappa \bigr] \otimes
\Bigl[ B_{\kappa_a} B_{\kappa_b} \prod_J J_{\kappa_J} \Bigr]
+ \dotsb
\,.\end{align}
Here, $\df \Phi_{N+L}(q_a\!+ q_b; q_1, \ldots)$ denotes the Lorentz-invariant phase space for the Born process in \eq{interaction}, and $M(\{q_i\})$ denotes the measurement made on the hard momenta of the jets (which in the factorization are approximated by the Born momenta $q_i$). The dependence on the underlying hard interaction is encoded in the hard function $\hH_{\kappa}(\{q_i\})$, where $\{q_i\} \equiv \{q_1, \ldots, q_{N+L}\}$, the sum over $\kappa \equiv \{\kappa_a, \kappa_b, \ldots \kappa_{N+L}\}$ is over all relevant partonic processes, and the trace is over color. Any dependence probing softer momenta, such as measuring jet masses or low $p_T$s, as well as the choice of jet algorithm, will affect the precise form of the factorization, but not the hard function $\hH_\kappa$. This dependence enters through the definition of the soft function $\hS_\kappa$ (describing soft radiation), jet functions $J_{\kappa_J}$ (describing energetic final-state radiation in the jets) and the beam functions $B_i$ (describing energetic initial-state radiation along the beam direction). More precisely, the beam function is given by $B_i = \sum_{i'} {\cal I}_{i i'} \otimes f_{i'}$ with $f_i$ the parton distributions of the incoming protons, and ${\cal I}_{i i'}$ a perturbatively calculable matching coefficient depending on the measurement definition~\cite{Stewart:2009yx}. The ellipses at the end of \eq{sigma} denote power-suppressed corrections. All functions in the factorized cross section depend only on the physics at a single scale. This allows one to evaluate all functions at their own natural scale, and then evolve them to a common scale using their RGE. This procedure resums the large logarithms of scale ratios appearing in the cross section to all orders in perturbation theory.

The explicit form of the factorization theorem in \eq{sigma}, including field-theoretic definitions for the jet, beam, and soft functions is known for a number of exclusive jet cross sections and measurements of interest. For example, factorization theorems exist for the $N$-jet cross section defined using $N$-jettiness~\cite{Stewart:2009yx, Stewart:2010tn, Stewart:2010pd, Berger:2010xi, Jouttenus:2011wh, Jouttenus:2013hs, Kang:2013nha, Kang:2013lga, Stewart:2015waa}. These have also been utilized to include higher-order resummation in Monte Carlo programs~\cite{Alioli:2012fc, Alioli:2013hqa, Alioli:2015toa}, and are the basis of the $N$-jettiness subtraction method for fixed-order calculations~\cite{Boughezal:2015dva, Gaunt:2015pea}. In addition, there has been a focus on color-singlet production at small $q_T$~\cite{Catani:2000vq, Becher:2010tm, GarciaEchevarria:2011rb, Chiu:2012ir, Catani:2013tia}, as well as the factorization of processes defined with jet algorithms~\cite{Delenda:2006nf, Bauer:2008jx, Ellis:2010rwa, Walsh:2011fz, Kelley:2012zs, Banfi:2012yh, Becher:2012qa, Tackmann:2012bt, Banfi:2012jm, Liu:2012sz, Becher:2013xia, Stewart:2013faa, Shao:2013uba, Li:2014ria, Jaiswal:2014yba, Gangal:2014qda, Becher:2014aya}, jet shape variables~\cite{Hornig:2009vb, Ellis:2009wj, Bauer:2011uc, Feige:2012vc, Chien:2012ur, Larkoski:2014tva, Larkoski:2014pca, Chien:2014nsa, Procura:2014cba, Becher:2015gsa, Larkoski:2015zka, Larkoski:2015kga}, or fragmentation properties~\cite{Procura:2009vm, Liu:2010ng, Jain:2011xz, Procura:2011aq, Krohn:2012fg, Waalewijn:2012sv, Chang:2013rca, Bauer:2013bza, Ritzmann:2014mka} for identified jets. The same hard functions also appear in threshold resummation factorization formulas, which are often used to obtain an approximate higher order result for inclusive cross sections.

The focus of our chapter is the hard function $\hH_{\kappa}(\{q_i\})$ in \eq{sigma}, which contains the process-dependent underlying hard interaction of \eq{interaction}, but is independent of the particular measurement. In SCET, the dependence on the hard interaction is encoded in the Wilson coefficients, $\vC$, of a basis of operators built out of SCET fields. The Wilson coefficients can be calculated through a matching calculation from QCD onto the effective theory. The hard function appearing in the factorization theorem is then given by
\begin{align}
\hH_\kappa(\{q_i\})
&= \sum_{\{\la_i\}}
  \vC_{\la_1\cdot\cdot(\cdot\cdot\la_n)}(\{q_i\})\,
  \vC_{\la_1 \cdot\cdot(\cdot\cdot\la_n)}^\dagger (\{q_i\})
\,.\end{align}
Here, the $\{ \lambda_i \}$ denote helicity labels and the sum runs over all relevant helicity configurations. The $\vC$ are vectors in color space, and the hard function is therefore a matrix in color space.

For processes of higher multiplicities, the construction of a complete basis of SCET operators, and the subsequent matching calculation, becomes laborious due to the proliferation of Lorentz and color structures, similar to the case of high-multiplicity fixed-order calculations using standard Feynman diagrams. The use of SCET helicity fields introduced in this chapter, combined with analogous color management techniques as used in the calculation of amplitudes, makes the construction of an operator basis extremely simple, even in the case of high-multiplicity processes. Furthermore, with this basis choice, the SCET Wilson coefficients are precisely given by the IR-finite parts of the color-ordered QCD helicity amplitudes, rendering the matching procedure almost trivial. Combining the results for the hard function with known results for the soft, jet, and beam functions, then allows for the resummation of jet observables in higher multiplicity processes, which are ubiquitous at the LHC.

The remainder of this chapter is organized as follows.
In \subsec{helicity}, we review the notation for the spinor-helicity formalism. Additional useful helicity and color identities can be found in \app{useful}. We provide a brief summary of SCET in \subsec{scet}. In \sec{basis}, we introduce SCET helicity fields and operators, and describe the construction of the helicity and color basis, as well as its symmetry properties. In \sec{matching}, we discuss the matching from QCD onto the SCET helicity operators, including a discussion of the dependence on the regularization and renormalization scheme. We then demonstrate the matching explicitly for $H+0,1,2$ jets in \sec{higgs}, $V+0,1,2$ jets in \sec{vec}, and $pp\to2,3$ jets in \sec{pp}. Explicit results for the required helicity amplitudes are collected in the appendices. In \sec{running}, we discuss the general renormalization group evolution of the hard coefficients, which involves mixing between different color structures, to all orders. We give explicit results for the anomalous dimensions for up to $4$ colored particles plus an arbitrary number of uncolored particles. We conclude in \sec{conclusions}.

\section{Notation}
\label{sec:helicity}

\subsection{Helicity Formalism}
\label{subsec:helicity}

We will use the standard notation for the spinor algebra (for a review see for example Refs.~\cite{Dixon:1996wi, Dixon:2013uaa}).
Consider the four-component spinor $u(p)$ of a massless Dirac particle with momentum $p$, satisfying the massless Dirac equation,
\begin{equation} \label{eq:Dirac}
\pslash\, u(p)=0
\,, \qquad
p^2 = 0
\,.\end{equation}
The charge conjugate (antiparticle) spinor $v(p)$ also satisfies \eq{Dirac}, and we can choose a representation such that $v(p) = u(p)$. The spinors and conjugate spinors for the two helicity states are denoted by
\begin{align} \label{eq:braket_def}
\ket{p\pm} &= \frac{1 \pm \ga_5}{2}\, u(p)
\,,\nn\\
\bra{p\pm} &= \mathrm{sgn}(p^0)\, \bar{u}(p)\,\frac{1 \mp \ga_5}{2}
\,.\end{align}
For massless particles chirality and helicity agree while for antiparticles they are opposite, so $\ket{p+} = u_+(p) = v_-(p)$ corresponds to positive (negative) helicity for particles (antiparticles). The spinors $\ket{p\pm}$ are defined by \eqs{Dirac}{braket_def} for both physical ($p^0 >0$) and unphysical ($p^0<0$) momenta. Their explicit expression, including our overall phase convention, is given in \app{helicity}.

The spinor products are denoted by
\begin{equation}
\langle p q \rangle = \braket{p-}{q+}
\,,\qquad
[p q] = \braket{p+}{q-}
\,.\end{equation}
They satisfy
\begin{align}
\langle pq \rangle = - \langle qp \rangle
\,,\quad
[pq] = - [qp]
\,,\quad
\langle pq \rangle [qp] = 2p\cdot q
\,.\end{align}
Additional relations are collected in \app{helicity}. The minus sign for $p^0 < 0$ in \eq{braket_def} is included so the spinor relations are invariant under inverting the signs of momenta, $p\to -p$, when crossing particles between the initial and final state, e.g. $\langle (-p)q \rangle [q(-p)] = 2(-p) \cdot q$.

If there are several momenta $p_i$, it is common to abbreviate
\begin{equation}
\ket{p_i\pm} = \ket{i\pm}
\,,\qquad
\langle p_i p_j \rangle = \langle ij \rangle
\,,\qquad
[p_i p_j] = [ij]
\,.\end{equation}

The polarization vectors of an outgoing gluon with momentum $p$ are given in the helicity formalism by
\begin{equation}
 \ve_+^\mu(p,k) = \frac{\mae{p+}{\ga^\mu}{k+}}{\sqrt{2} \langle kp \rangle}
\,,\quad
 \ve_-^\mu(p,k) = - \frac{\mae{p-}{\ga^\mu}{k-}}{\sqrt{2} [kp]}
\,,\end{equation}
where $k$ is an arbitrary reference vector with $k^2=0$, which fixes the gauge of the external gluons.
Using the relations in \app{helicity}, it is easy to check that
\begin{align}
p\cdot \ve_\pm(p,k) &= k\cdot \ve_\pm(p,k) = 0
\,,\nn\\
\ve_\pm(p,k) \cdot \ve_\pm(p,k) &= 0
\,,\nn\\
\ve_\pm(p,k) \cdot \ve_\mp(p,k) &= -1
\,,\nn\\
\ve_\pm^*(p,k) &= \ve_\mp(p,k)
\,,\end{align}
as is required for physical polarization vectors. With $p^\mu = E (1,0,0,1)$, the choice $k^\mu = E (1,0,0,-1)$ yields the conventional 
\begin{equation}
\ve_\pm^\mu(p, k) =
\frac{1}{\sqrt{2}}\, (0,1,\mp\img,0)\,.
\end{equation}
\subsection{SCET}
\label{subsec:scet}

We now briefly review the SCET concepts and notation that will be required for this section. As has been discussed, 
Soft-collinear effective theory is an effective field theory of QCD that
describes the interactions of collinear and soft particles~\cite{Bauer:2000ew,
  Bauer:2000yr, Bauer:2001ct, Bauer:2001yt} in the presence of a hard
interaction.\footnote{Throughout this chapter, we will for simplicity use the notation of SCET$_{\rm I}$. The theory SCET$_{\rm II}$ \cite{Bauer:2002aj} is required for a certain class of observables, for example $p_T$-dependent measurements or vetoes. The helicity operator formalism presented here applies identically to constructing SCET$_{\rm II}$ operators. The collinear operators and matching coefficients are the same for both cases. } Collinear particles are characterized by having large energy and
small invariant mass. To separate the large and small momentum components, it is
convenient to use light-cone coordinates. We define two light-cone vectors
\begin{equation}
n^\mu = (1, \vec{n})
\,,\qquad
\bn^\mu = (1, -\vec{n})
\,,\end{equation}
with $\vec{n}$ a unit three-vector, which satisfy $n^2 = \bn^2 = 0$ and  $n\cdot\bn = 2$.
Any four-momentum $p$ can be decomposed as
\begin{equation} \label{eq:lightcone_dec}
p^\mu = \bn\sdt p\,\frac{n^\mu}{2} + n\sdt p\,\frac{\bn^\mu}{2} + p^\mu_{n\perp}
\,.\end{equation}
An ``$n$-collinear'' particle has momentum $p$ close to the $\vec{n}$ direction,
so that $p$ scales as $(n\!\cdot\! p, \bn \!\cdot\! p, p_{n\perp}) \sim
\bn\!\cdot\! p$ $\,(\la^2,1,\la)$, with $\la \ll 1$ a small parameter. For example, for
a jet of collinear particles in the $\vec{n}$ direction with total momentum 
$p_J$, $\bn \sdt p_J \simeq 2E_J$ corresponds to the large energy of the jet, 
while $n \sdt p_J \simeq m_J^2/E_J \ll E_J$, where $m_J$ is the jet mass, so $\la^2 \simeq m_J^2/E_J^2 \ll 1$.

To construct the fields of the effective theory, the momentum of $n$-collinear particles is written as
\begin{equation} \label{eq:label_dec}
p^\mu = \lp^\mu + k^\mu = \bn\sdt\lp\, \frac{n^\mu}{2} + \lp_{n\perp}^\mu + k^\mu\,,
\,\end{equation}
where $\bn\cdot\lp \sim Q$ and $\lp_{n\perp} \sim \la Q$ are the large momentum
components, while $k\sim \la^2 Q$ is a small residual momentum. Here, $Q$ is the scale of the hard interaction,
and the effective theory expansion is in powers of $\la$.

The SCET fields for $n$-collinear quarks and gluons, $\xi_{n,\lp}(x)$ and
$A_{n,\lp}(x)$, are labeled by the collinear direction $n$ and their large
momentum $\lp$. They are written in position space with respect to the residual
momentum and in momentum space with respect to the large momentum components.
Derivatives acting on the fields pick out the residual momentum dependence,
$\img\partial^\mu \sim k \sim \la^2 Q$. The large label momentum is obtained
from the label momentum operator $\cP_n^\mu$, e.g. $\cP_n^\mu\, \xi_{n,\lp} =
\lp^\mu\, \xi_{n,\lp}$. If there are several fields, $\cP_n$ returns the sum of
the label momenta of all $n$-collinear fields. For convenience, we define
$\bnP_n = \bn\cdot\cP_n$, which picks out the large momentum component.  Frequently,
we will only keep the label $n$ denoting the collinear direction, while the
momentum labels are summed over (subject to momentum conservation) and are
suppressed in our notation.

Collinear operators are constructed out of products of fields and Wilson lines
that are invariant under collinear gauge
transformations~\cite{Bauer:2000yr,Bauer:2001ct}.  The smallest building blocks
are collinearly gauge-invariant quark and gluon fields, defined as
\begin{align} \label{eq:chiB}
\chi_{n,\w}(x) &= \Bigl[\delta(\w - \bnP_n)\, W_n^\dagger(x)\, \xi_n(x) \Bigr]
\,,\nn\\
\cB_{n,\w\perp}^\mu(x)
&= \frac{1}{g}\Bigl[\delta(\w + \bnP_n)\, W_n^\dagger(x)\,\img D_{n\perp}^\mu W_n(x)\Bigr]
\,.\end{align}
With this definition of $\chi_{n,\w}$, we have $\w > 0$ for an incoming quark and $\w < 0$ for an outgoing antiquark. For $\cB_{n,\w\perp}$, $\w > 0$ ($\w < 0$) corresponds to an outgoing (incoming) gluon.
In \eq{chiB}
\begin{equation}
\img D_{n\perp}^\mu = \cP^\mu_{n\perp} + g A^\mu_{n\perp}\,,
\end{equation}
is the collinear covariant derivative and
\begin{equation} \label{eq:Wn}
W_n(x) = \biggl[\sum_\text{perms} \exp\Bigl(-\frac{g}{\bnP_n}\,\bn\sdt A_n(x)\Bigr)\biggr]
\end{equation}
is a Wilson line of $n$-collinear gluons in label momentum space. 
The label operators $\bnP_n$ in \eqs{chiB}{Wn} only act inside the square brackets. 
$W_n(x)$ sums up arbitrary emissions of $n$-collinear gluons from an $ \bar n$-collinear quark
or gluon, which are $\ord{1}$ in the power counting. Since $W_n(x)$ is
localized with respect to the residual position $x$, we can treat
$\chi_{n,\w}(x)$ and $\cB_{n,\w}^\mu(x)$ like local quark and gluon fields. For later use, we give the expansion of the collinear gluon field
\begin{align}\label{eq:gluon_expand}
\cB^\mu_{n,\perp}=A^\mu_{n\perp}-\frac{p^\mu_\perp}{\bar n\cdot p}\bar n \cdot A_{n,p} +\cdots.
\end{align}
Here the ellipses denote terms in the expansion with more than 2 collinear gluon fields, which are not required for our matching calculations.

In our case the effective theory contains several collinear sectors, $n_1, n_2,
\ldots$~\cite{Bauer:2002nz}, where the collinear fields for a given sector
$n_i^\mu=(1,\vec n_i)$ describe a jet in the direction $\vec n_i$, and we also
define $\bar n_i^\mu=(1,-\vec n_i)$.  A fixed-order QCD amplitude with $N$
colored legs is then matched onto operators in SCET with $N$ different collinear
fields.  The different collinear directions have to be well separated, which
means
\begin{equation} \label{eq:nijsep}
  n_i\sdt n_j \gg \la^2 \qquad\text{for}\qquad i\neq j
\,.\end{equation}
The infrared singularities associated with collinear or soft limits of legs in
QCD are entirely described by the Lagrangian and dynamics of SCET itself, so the
QCD amplitudes are only used to describe the hard kinematics away from infrared
singular limits.

Two different $n_i$ and $n_i'$ with $n_i\cdot n_i' \sim \lambda^2$ both
describe the same jet and corresponding collinear physics. Thus, each collinear
sector can be labeled by any member of a set of equivalent vectors, $\{n_i\}$, which are related by reparametrization invariance~\cite{Manohar:2002fd}.
The simplest way to perform the matching is to choose $n_i$ such that the large
label momentum is given by
\begin{equation} \label{eq:pdefault}
\lp_i^\mu = \w_i\,\frac{n_i^\mu}{2} \,,
\end{equation}
with $\lp_{n_i \perp}^\mu =0$.

In general, operators will have sums over distinct equivalence classes, $\{n_i\}$, and matrix elements select a representative vector to describe particles in a particular collinear direction. For many leading power applications there is only a single collinear field in each sector, and we may simply set the large label momentum of that building block field to that of the external parton using the following simple relation,  
\begin{equation}
\int\!\df\lp\, \ldel(\lp - p)\,f(\lp) = f\Bigl(\bn_i\cdot p\,\frac{n_i}{2}\Bigr)
\,,\end{equation}
where $p$ is collinear with the $i$'th jet.
Here the tildes on the integration measure and delta function ensure that the integration over equivalence classes is properly implemented.\footnote{The precise definition of this delta function and measure are
\begin{align} \label{eq:labelsums}
\ldel(\lp_i - p) &\equiv \delta_{\{n_i\},p}\,\delta(\w_i - \bn_i\cdot p)
\,,\nn\\
\int\!\df\lp &\equiv \sum_{\{n_i\}} \int\!\df\w_i
\,,\end{align}
where 
\begin{equation}
\delta_{\{n_i\},p} =
\begin{cases}
   1 &\quad n_i\cdot p = \ord{\lambda^2}
   \,,\\
   0 &\quad \text{otherwise}
\,.\end{cases}
\end{equation}
The Kronecker delta is nonzero if the collinear momentum $p$ is in the $\{ n_i\}$
equivalence class, i.e. $p$ is close enough to be considered as collinear with
the $i$th jet.  The sum in the second line of \eq{labelsums} runs over the different
equivalence classes. }
Because of this, at leading power, the issue of equivalence classes can largely be ignored. 

Particles that exchange large momentum of $\ord{Q}$ between different jets are
off shell by $\ord{n_i\cdot n_j \,Q^2}$. They are integrated out by matching
QCD onto SCET.  Before and after the hard interaction the jets described by the
different collinear sectors evolve independently from each other, with only soft
radiation between the jets.  The corresponding soft degrees of freedom are
described in the effective theory by soft quark and gluon fields, $q_\soft(x)$
and $A_\soft(x)$, which only have residual soft momentum dependence
$\img\partial^\mu \sim \la^2Q$.  They couple to the collinear sectors via the
soft covariant derivative
\begin{equation}
\img D_\soft^\mu = \img \partial^\mu + g A_\soft^\mu\,,
\end{equation}
acting on the collinear fields. At leading power in $\la$, $n$-collinear
particles only couple to the $n\sdt A_\soft$ component of soft gluons, so the
leading-power $n$-collinear Lagrangian only depends on $n\sdt D_\soft$. For example, for
$n$-collinear quarks~\cite{Bauer:2000yr, Bauer:2001ct}
\begin{equation} \label{eq:L_n}
\cL_n = \bar{\xi}_n \Bigl(\img n\sdt D_\soft + g\,n\sdt A_n 
 + \img\Dslash_{n\perp} W_n \frac{1}{\bnP_n}\, W_n^\dagger\,\img\Dslash_{n\perp}
 \Bigr)\frac{\bnslash}{2} \xi_n
\,.\end{equation}
The leading-power $n$-collinear Lagrangian for gluons is given in
Ref.~\cite{Bauer:2001yt}.

\section{SCET Operator Basis}
\label{sec:basis}

In this section, we describe in detail how to construct a basis of helicity and color operators in SCET, which greatly simplifies the construction of a complete operator basis and also facilitates the matching process. Usually, a basis of SCET operators obeying the symmetries of the problem is constructed from the fields $\chi_{n,\omega}$, $\cB^\mu_{n,\omega \perp}$, as well as Lorentz and color structures. This process becomes quite laborious due to the large number of structures which appear for higher multiplicity processes, and the reduction to a minimal basis of operators quickly becomes nontrivial. Instead, we work with a basis of operators with definite helicity structure constructed from scalar SCET building blocks, which, as we will show, has several advantages. First, this simplifies the construction of the operator basis, because each independent helicity configuration gives rise to an independent
helicity operator. In this way, we automatically obtain the minimal number of
independent operators as far as their Lorentz structure is concerned.  Second, operators with distinct helicity structures do not mix under renormalization group evolution, as will be discussed in detail in \sec{running}.  The reason is that
distinct jets can only exchange soft gluons in SCET, which at leading order in
the power counting means they can transfer color but not spin [see \eq{L_n}]. Therefore, the only nontrivial aspect of the operator basis is the color degrees of freedom. The different color structures mix under renormalization group evolution, but their mixing only depends on the color representations and not on the specific helicity configuration.

\subsection{Helicity Fields}
\label{subsec:fields}

We start by defining quark and gluon fields of definite helicity, out of which we
can build operators with a definite helicity structure. To simplify our
discussion we will take all momenta and polarization vectors as outgoing, and
label all fields and operators by their outgoing helicity and momenta. Crossing symmetry, and crossing relations are discussed in \subsec{crossing}.

We define a gluon field of definite helicity%
\footnote{The label $\pm$ on $\cB_\pm$ refers to helicity and should not be confused with light-cone components.}
\begin{equation} \label{eq:cBpm_def}
\cB^a_{i\pm} = -\ve_{\mp\mu}(n_i, \bn_i)\,\cB^{a\mu}_{n_i,\w_i\perp_i}
\,,\end{equation}
where $a$ is an adjoint color index. For $n_i^\mu=(1,0,0,1)$, 
we have
\begin{equation}
\ve_\pm^\mu(n_i, \bn_i) = \frac{1}{\sqrt{2}}\, (0,1,\mp\img,0)
\,,\end{equation}
in which case
\begin{equation}
\cB^a_{i\pm} = \frac{1}{\sqrt{2}} \bigl(\cB^{a,1}_{n_i,\w_i\,\perp_i}
   \pm \img \cB^{a,2}_{n_i,\w_i\,
  \perp_i} \bigr)
\,.\end{equation}

For an external gluon with outgoing polarization vector $\ve(p,k)$ and
outgoing momentum $p$ in the $n_i$-collinear direction, the contraction with
the field $\cB_{i\pm}^a$ contributes 
\begin{equation}\label{eq:pol_A}
 -\ve_{\mp\mu}(n_i, \bn_i)\Bigl[\ve_{\perp_i}^\mu(p,k)
 - \frac{p_{\perp_i}^\mu}{\bn_i\cdot p}\, \bn_i\cdot \ve(p,k)\Bigr]
\,,\end{equation}
where we have used the expansion of the collinear gluon field given in \eq{gluon_expand}.
Since $\ve_{\mp}(n_i,\bn_i)$ is perpendicular to both $n_i$ and $\bn_i$, we
can drop the $\perp_i$ labels in brackets. A convenient choice for the reference
vector is to take $k = \bn_i$, for which the second term in brackets vanishes. Equation~\eqref{eq:pol_A} then becomes
\begin{equation}
-\ve_\mp(n_i, \bn_i)\cdot\ve(p,\bn_i)
\,,\end{equation}
which is equal to 0 or 1 depending on the helicity of $\ve(p,\bn_i)$.
Adopting this choice, the tree-level Feynman rules for
an outgoing gluon with polarization $\pm$ (so $\ve =\ve_\pm$), momentum
$p$ (with $p^0 > 0$), and color $a$ are
\begin{align}
 \Mae{g_\pm^a(p)}{\cB_{i\pm}^b}{0} &= \delta^{a b}\, \ldel(\lp_i - p)
\,,\nn\\
 \Mae{g_\mp^a(p)}{\cB_{i\pm}^b}{0} &= 0
\,.\end{align}
Note that $\cB_{i\pm}^b=\cB_{i\pm}^b(0)$, so we do not get a phase from the
residual momentum.  Similarly, for an incoming gluon with incoming polarization
$\mp$ ($\ve = \ve_\mp$, so $\ve^* = \ve_\pm$), incoming momentum $-p$
(with $p^0 < 0$), and color $a$, we have
\begin{align}
 \Mae{0}{\cB_{i\pm}^b}{g_\mp^a(-p)} &= \delta^{ab}\, \ldel(\lp_i - p)
\,,\nn\\
 \Mae{0}{\cB_{i\pm}^b}{g_\pm^a(-p)} &= 0
\,.\end{align}

We define quark fields with definite helicity\footnote{Technically speaking chirality, although we work in a limit where all external quarks can be treated as massless.} as
\begin{equation} \label{eq:chipm_def}
\chi^\alpha_{i\pm} = \frac{1\pm\gamma_5}{2}\,\chi_{n_i,-\w_i}^\alpha
\,,\qquad
\bar\chi^{\bar \alpha}_{i\pm} = \bar\chi_{n_i,\w_i}^{\bar \alpha}\,\frac{1\mp\gamma_5}{2}
\,,\end{equation}
where $\alpha$ and $\bar \alpha$ are fundamental and antifundamental color indices respectively.

For external quarks with
$n_i$-collinear momentum $p$, the fields contribute factors of the form
\begin{align}
\frac{1\pm\gamma_5}{2}\,\frac{\nslash_i\bnslash_i}{4}\,u(p)
&= \frac{\nslash_i\bnslash_i}{4}\,\ket{p\pm}
= \ket{p\pm}_{n_i}
\,,\end{align}
where in the last equality, we have defined a shorthand notation $\ket{p\pm}_{n_i}$ for the SCET projected spinor. The spinor $\ket{p\pm}_{n_i}$ is proportional to $\ket{n\pm}$; see \eq{ketn}.

The tree-level Feynman rules for incoming ($p^0 < 0$)
and outgoing ($p^0 > 0$) quarks with helicity $+/-$ and color $\alpha$ are then given by
\begin{align} \label{eq:chipm_q}
\Mae{0}{\chi^\beta_{i+}}{q_+^\balpha(-p)}
&= \delta^{\beta\balpha}\,\ldel(\lp_i - p)\, \ket{(-p_i)+}_{n_i}
\,,\nn\\
\Mae{0}{\chi^\beta_{i-}}{q_-^\balpha(-p)}
&= \delta^{\beta\balpha}\,\ldel(\lp_i - p)\, \ket{(-p_i)-}_{n_i}
\,,\nn\\
\Mae{q_+^\alpha(p)}{\bar\chi^\bbeta_{i+}}{0}
&= \delta^{\alpha\bbeta}\,\ldel(\lp_i - p)\, {}_{n_i\!}\bra{p_i+}
\,,\nn\\
\Mae{q_-^\alpha(p)}{\bar\chi^\bbeta_{i-}}{0}
&= \delta^{\alpha\bbeta}\,\ldel(\lp_i - p)\, {}_{n_i\!}\bra{p_i-}
\,,\end{align}
and similarly for antiquarks
\begin{align} \label{eq:chipm_qbar}
\Mae{0}{\bar\chi^\bbeta_{i+}}{\bar{q}_-^\alpha(-p)}
&= \delta^{\alpha\bbeta}\,\ldel(\lp_i - p)\,{}_{n_i\!}\bra{(-p_i)+}
\,, \nn \\
\Mae{0}{\bar\chi^\bbeta_{i-}}{\bar{q}_+^\alpha(-p)}
&= \delta^{\alpha\bbeta}\,\ldel(\lp_i - p)\,{}_{n_i\!}\bra{(-p_i)-}
\,, \nn \\
\Mae{\bar{q}_-^\balpha(p)}{\chi^\beta_{i+}}{0}
&= \delta^{\beta\balpha}\,\ldel(\lp_i - p)\, \ket{p_i+}_{n_i}
\,, \nn \\
\Mae{\bar{q}_+^\balpha(p)}{\chi^\beta_{i-}}{0}
&= \delta^{\beta\balpha}\,\ldel(\lp_i - p)\, \ket{p_i-}_{n_i}
\,.\end{align}
The corresponding Feynman rules with the helicity of the external (anti)quark
flipped vanish.

To avoid the explicit spinors in \eqs{chipm_q}{chipm_qbar}, and exploit the fact
that fermions come in pairs, we also define fermionic vector currents of definite helicity
\begin{align} \label{eq:jpm_def}
J_{ij+}^{\balpha\beta}
&=  \frac{\sqrt{2}\, \ve_-^\mu(n_i, n_j)}{\sqrt{\phantom{2}\!\!\omega_i \,  \omega_j }}\, \frac{\bar{\chi}^\balpha_{i+}\, \gamma_\mu \chi^\beta_{j+}}{\langle n_i n_j\rangle}\,, \nn \\
J_{ij-}^{\balpha\beta}
&= -\, \frac{ \sqrt{2}\, \ve_+^\mu(n_i, n_j)}{\sqrt{\phantom{2}\!\! \omega_i \,  \omega_j }}\, \frac{\bar{\chi}^\balpha_{i-}\, \gamma_\mu \chi^\beta_{j-}}{[n_i n_j]}
\,,\end{align}
where $\omega_i=\bn_i\cdot \tilde p_i$ from \eq{pdefault}, as well as a scalar current
\begin{align} \label{eq:jS_def}
J_{ij0}^{\balpha\beta}&=\frac{2}{\sqrt{\phantom{2}\!\! \omega_i \,  \omega_j}}\frac{\bar \chi^\balpha_{i+}\chi^\beta_{j-}}{ [n_i n_j]   }\,, \nonumber \\
(J^\dagger)_{ij0}^{\balpha\beta}&=  \frac{2}{\sqrt{\phantom{2}\!\! \omega_i \,  \omega_j}}  \frac{\bar \chi^\balpha_{i-}\chi^\beta_{j+}} { \langle n_i  n_j \rangle }
\,.\end{align}
In \eqs{jpm_def}{jS_def} the flavor labels of the quarks have not been made explicit, but in general the two quark fields in a current can have different flavors (for example in $W$ production). Since we are using a basis of physical polarization states it is not necessary to introduce more complicated Dirac structures. For example, pseudovector and pseudoscalar currents, which are usually introduced using $\gamma^5$, are incorporated through the relative coefficients of operators involving $J_+$, $J_-$ or $J_0$, $J_0^\dagger$. As we shall see, this greatly simplifies the construction of the operator basis in the effective theory.

At leading power, there is a single collinear field in each collinear sector, so we can choose $n_i^\mu=p_i^\mu/p_i^0$ to represent the equivalence class $\{n_i\}$, so that $p_i^\mu = \frac12 \bn\cdot p_i\, n_i^\mu$ which gives
\begin{equation}
\ket{p\pm}_{n_i} = \ket{p\pm} = \Bigl\lvert\bn_i\cdot p\, \frac{n_i}{2}\pm\Bigr\rangle
  = \sqrt{\frac{\bn_i\cdot p}{2}}\: \ket{n_i\pm} \,.
\end{equation}
Since we always work at leading power in this chapter, we will always make this choice to simplify the matching.  With this choice, the tree-level Feynman rules for the fermion currents are
\begin{align}\label{eq:feyn_tree}
&\Mae{q_+^{\alpha_1}(p_1)\,\bar{q}_-^{\balpha_2}(p_2)}{J_{12+}^{\bbeta_1\beta_2}}{0}
\,\nn\\ & \quad
= \delta^{\alpha_1\bbeta_1}\,\delta^{\beta_2\balpha_2}\, \ldel(\lp_1 - p_1)\,\ldel(\lp_2 - p_2)
\,, \nn \\
&\Mae{q_-^{\alpha_1}(p_1)\,\bar{q}_+^{\balpha_2}(p_2)}{J_{12-}^{\bbeta_1\beta_2}}{0}
\,\nn\\ & \quad
= \delta^{\alpha_1\bbeta_1}\,\delta^{\beta_2\balpha_2}\, \ldel(\lp_1 - p_1)\,\ldel(\lp_2 - p_2)
\,, \nn \\
&\Mae{q_+^{\alpha_1}(p_1)\,\bar{q}_+^{\balpha_2}(p_2)}{J_{12\,0}^{\bbeta_1\beta_2}}{0}
\,\nn\\ & \quad
= \delta^{\alpha_1\bbeta_1}\,\delta^{\beta_2\balpha_2}\, \ldel(\lp_1 - p_1)\,\ldel(\lp_2 - p_2)
\,, \nn \\
&\Mae{q_-^{\alpha_1}(p_1)\,\bar{q}_-^{\balpha_2}(p_2)}{(J^\dagger)_{12\,0}^{\bbeta_1\beta_2}}{0}
\,\nn\\ & \quad
= \delta^{\alpha_1\bbeta_1}\,\delta^{\beta_2\balpha_2}\, \ldel(\lp_1 - p_1)\,\ldel(\lp_2 - p_2)
\,.\end{align}
The simplicity of these Feynman rules arises due to the unconventional normalization of the operators in \eqs{jpm_def}{jS_def}. This normalization has been chosen to simplify the matching of QCD amplitudes onto SCET operators, as will be seen in \sec{matching}.

We will also make use of leptonic versions of the above currents. These are defined analogously,
\begin{align} \label{eq:jpm_lep_def}
J_{ij+}&=  \frac{\sqrt{2}\, \ve_-^\mu(n_i, n_j)}{\sqrt{\phantom{2}\!\!\omega_i \,  \omega_j }}\,  \frac{\bar{\ell}_{i+}\, \gamma_\mu \ell_{j+}}{\langle n_i n_j \rangle}\,, \nonumber \\
J_{ij-}&=-\, \frac{\sqrt{2}\, \ve_+^\mu(n_i, n_j)}{\sqrt{\phantom{2}\!\! \omega_i \,  \omega_j}}\, \frac{\bar{\ell}_{i\pm}\, \gamma_\mu \ell_{j\pm}}{[n_i n_j]}
\,.\end{align}
Unlike the collinear quark field $\chi$, the leptonic field $\ell$ does not carry color and so does not contain a strong-interaction Wilson line.

All couplings in the SM, except to the Higgs boson, preserve chirality. This limits the need for the scalar current, especially when considering only massless external quarks. In the SM the scalar current can arise through explicit couplings to the Higgs, in which case, even though we still treat the external quarks as massless, the Wilson coefficient for the scalar operator will contain the quark Yukawa coupling. This is relevant for example for $H b\bar b$ processes. The scalar current can also arise through off-diagonal CKM-matrix elements connecting two massless external quarks through a massive quark appearing in a loop. This can occur in multiple vector boson production, or from electroweak loop corrections, neither of which will be discussed in this chapter. When constructing an operator basis in \subsec{helicityops}, we ignore the scalar current, as it is not relevant for the examples that we will treat in this chapter. However, it should be clear that the construction of the basis in \subsec{helicityops} can be trivially generalized to incorporate the scalar current if needed.

\subsection{Helicity Operator Basis}
\label{subsec:helicityops}

Using the definitions for the gluon and quark helicity fields in \eqs{cBpm_def}{jpm_def}, we can construct operators for a given number of external partons with definite helicities and color. (As discussed at the end of the previous section, for the processes we consider in this chapter we do not require the scalar current $J_S$.) In the general case with CKM-matrix elements, we must allow for the two quark flavors within a single current to be different. The situation is simplified in QCD processes, where one can restrict to currents carrying a single flavor label.

For an external state with $n$ particles of definite helicities $\pm$, colors $a_i$, $\alpha_i$, $\bar \alpha_i$, and flavors $f$, $f'$, ..., a complete basis of operators is given by
\begin{align} \label{eq:Opm_gen}
&O_{\pm\pm\dotsb(\pm\dotsb;\dotsb\pm)}^{a_1 a_2 \dotsb \balpha_{i-1} \alpha_i \dotsb \balpha_{n-1} \alpha_n}
(\lp_1, \lp_2, \ldots, \lp_{i-1}, \lp_i, \ldots, \lp_{n-1}, \lp_n)
\nn\\ & \qquad
= S\, \cB_{1\pm}^{a_1}\,\cB_{2\pm}^{a_2}\dotsb J_{f\,i-1,i\pm}^{\balpha_{i-1}\alpha_i}
\dotsb J_{f'\,n-1,n\pm}^{\balpha_{n-1}\alpha_n}
\,.\end{align}
For example, $f=q$ indicates that both quark fields in the current have flavor $q$. When it is necessary to distinguish different flavors with the same current, for example when we consider processes involving W bosons in \sec{vec}, we use a label $f=\bar u d$ such as $J_{\bar u d 12-}$. For simplicity, we will also often suppress the dependence of the operator on the label momenta $\lp_i$. For the operator subscripts, we always put the helicity labels of the gluons first and those of the quark currents in brackets, with the labels for quark currents with different flavor labels $f$ and $f'$ separated by a semicolon, as in \eq{Opm_gen}. The $\pm$ helicity labels of the individual gluon fields and quark currents can all vary independently. Operators with nonzero matching coefficients are restricted to the color-conserving subspace. We will discuss the construction of the color basis in \subsec{color}.

The symmetry factor $S$ in \eq{Opm_gen} is included to simplify the matching. It is given by
\begin{equation} \label{eq:Opm_S}
S = \frac{1}{\prod\limits_i n_{i}^+!\, n_{i}^-!}
\,,\end{equation}
where $n_i^\pm$ denotes the number of fields of type $i=g,u,\bar u, d, \bar d, \dots$ with helicity $\pm$. We also use
\begin{equation}
n = \sum_i (n_i^+ + n_i^-)\,,
\end{equation}
to denote the total number of fields in the operator. Each ${\cal B}_i$ counts as one field, and each $J$ has two fields.

For each set of external particles of definite helicities, colors, and flavors, there is only one independent operator, since the physical external states have been completely specified. All Feynman diagrams contributing to this specific external state will be included in the Wilson coefficient of that specific operator. For the case of pure QCD, quarks always appear in pairs of the same flavor and same chirality, and therefore can be assembled into quark currents labeled by a single flavor. In this case, to keep track of the minimal number of independent operators, we can simply order the helicity labels, and only consider operators of the form
\begin{align} \label{eq:op_struct}
O_{+\cdot\cdot(\cdot\cdot-)}
&= O_{\underbrace{+\dotsb+}\,\underbrace{-\dotsb-}\,(\underbrace{+\dotsb+}\,\underbrace{-\dotsb-})}
\,,
\\\nn &\qquad\quad
n_g^+ \qquad\, n_g^- \,\,\qquad n_q^+ \,\qquad n_q^-
\end{align}
and analogously for any additional quark currents with different quark flavors.\footnote{In the general case with off-diagonal CKM-matrix elements, there is some more freedom in the choice of the operator basis, because quarks of the same flavor do not necessarily appear in pairs. However, it is still true that only a single operator is needed for a specific external state. For example, for external quarks $u_-, \bar d_+, \bar s_+, c_-$, one could either use the operators $J_{us-}J_{cd-}, \text{or the operators }J_{cs-}J_{ud-}$ (where the color structures have been suppressed). Since different helicity combinations are possible, a single flavor assignment does not suffice to construct a complete helicity basis, and one must sum over a basis of flavor assignments. As an example explicitly demonstrating this, we will consider the case of $pp \to W +$ jets in \sec{vec}.}

With the operator basis constructed, for a given $n$-parton process we can match hard scattering processes in QCD onto the leading-power hard-scattering Lagrangian
\begin{equation} \label{eq:Leff}
\cL_\hard \!=\!\! \int\!\! \prod_{i=1}^n\df \lp_i\,
C_{+\cdot\cdot(\cdot\cdot-)}^{a_1\dotsb \alpha_n}(\lp_1, \ldots, \lp_n)\,
O_{+\cdot\cdot(\cdot\cdot-)}^{a_1 \dotsb \alpha_n}(\lp_1, \ldots, \lp_n)
 ,
\end{equation}
where a sum over all color indices is implicit. Lorentz invariance implies that the Wilson coefficients only depend on Lorentz invariant combinations of the momenta. This hard Lagrangian is used in conjunction with the collinear and soft Lagrangians that describe the dynamics of the soft and collinear modes; see for example \eq{L_n}.

We emphasize that \eq{Leff} provides a complete basis in SCET for well-separated jets and additional nonhadronic particles at leading power.  We will discuss in more detail in \sec{matching} the matching and regularization schemes, and demonstrate that no evanescent operators are generated for this case. At subleading power, the SCET operators would involve additional derivative operators, soft fields, or multiple SCET building blocks from the same collinear sector.

\subsection{Example with a $Z$-Boson Exchange}

It is important to note that all kinematic dependence of the hard process, for example, its angular distributions, is encoded in the Wilson coefficients. Since the Wilson coefficients can (in principle) carry an arbitrary kinematic dependence, our choice of helicity basis imposes no restriction on the possible structure or mediating particles of the hard interaction. For example, the spin of an intermediate particle may modify the angular distribution of the decay products, and hence the Wilson coefficients, but this can always be described by the same basis of helicity operators.

As a simple example to demonstrate this point we consider $e^+e^-\to e^+e^- $ at tree level. This process can proceed through either an off-shell $\gamma$ or $Z$ boson. Because the SM couplings to both of these particles preserve chirality, a basis of operators for this process is given by
\begin{align}
O_{(++)}
&= \frac{1}{4}\, J_{e\,12+}\, J_{e\,34+}
\,, \nn \\
O_{(+-)}
&= J_{e\,12+}\, J_{e\,34-}
\,, \nn \\
O_{(--)}
&= \frac{1}{4}\, J_{e\,12-}\, J_{e\,34-}
\,,\end{align}
where the leptonic current is defined in \eq{jpm_lep_def}. The fact that this is a complete basis relies only on the fact that the couplings preserve chirality, and is independent of e.g.~the possible number of polarizations of the mediating $Z$ or $\gamma$.

We now consider the calculation of the Wilson coefficients for the matching to these operators (the matching procedure is discussed in detail in \sec{matching}). At tree level, the Wilson coefficients are easily calculated, giving
\begin{align} \label{eq:Cexample}
C_{(++)}&=- e^2  \bigl[1 + v_{R}^e v_{R}^e P_Z(s_{12})\bigr]
 \frac{2[13]\ang{24}}{s_{12}} + (1 \lra 3) \,,
 \nn \\
C_{(+-)}&=-e^2  \bigl[1+ v_{R}^e v_{L}^e P_Z(s_{12})\bigr]
\frac{2[14]\ang{23}}{s_{12}}  \,,
\nn \\
C_{(--)}&=-e^2  \bigl[1 + v_{L}^ev_{L}^e P_Z(s_{12})\bigr]
\frac{2[24]\ang{13}}{s_{12}} + (1 \lra 3) \,.
\end{align}
Here $s_{12}=(p_1+p_2)^2$, $P_Z$ is the ratio of the $Z$ and photon propagators,
\begin{equation}
P_{Z}(s) = \frac{s}{s-m_Z^2 + \img \Ga_{Z} m_{Z}}
\,,\end{equation}
and the couplings $v_{L,R}$ to the $Z$ boson are
\begin{align}
 v_L^e = \frac{1 - 2 \sin^2 \theta_W}{\sin(2\theta_W)}
 \,, \quad
 v_R^e= - \frac{2 \sin^2 \theta_W}{\sin(2\theta_W)}
\,.\end{align}
Note that the presence of the spinor factors in \eq{Cexample} occur due to our normalization conventions for the currents. 

Now, consider calculating the scattering amplitude in the effective theory, for example for the case when both electrons have positive helicity. The matrix element in the effective theory gives
\begin{align}
&\Mae{e^-_+(p_1)\,e^+_-(p_2)\, e^-_+(p_3)\,e^+_-(p_4)}{\img \cL_\hard}{0}
\nn\\ 
& 
= \img \MAe{e^-_+(p_1)\,e^+_-(p_2)\, e^-_+(p_3)\,e^+_-(p_4)}{\int\!\prod_{i=1}^n\df \lp_i\,C_{++} O_{++}}{0}
\nn\\ 
& 
= - \img e^2  \big[1 + v_{R}^e v_{R}^e P_Z(s_{12})\big] 
 \frac{2[13]\ang{24}}{s_{12}} + (1 \lra 3) \,,
\end{align}
using the Feynman rules of \eq{feyn_tree}. The effective theory therefore reproduces the full theory scattering amplitude. The same is true of the other helicity configurations, so the familiar angular distributions for $e^+e^-\to e^+e^- $, as well as the different couplings of the $Z$ to left- and right-handed particles, are entirely encoded in the Wilson coefficients.

\subsection{Color Basis} \label{subsec:color}

In addition to working with a basis of operators with definite helicity, we can also choose a color basis that facilitates the matching. When constructing a basis of operators in SCET, we are free to choose an arbitrary color basis. With respect to color, we can think of \eq{Leff} as having a separate Wilson coefficient for each color configuration. For specific processes the color structure of the Wilson coefficients can be further decomposed as
\begin{equation} \label{eq:Cpm_color}
C_{+\cdot\cdot(\cdot\cdot-)}^{a_1\dotsb\alpha_n}
= \sum_k C_{+\cdot\cdot(\cdot\cdot-)}^k T_k^{a_1\dotsb\alpha_n}
\equiv \vT^{ a_1\dotsb\alpha_n} \vC_{+\cdot\cdot(\cdot\cdot-)}
\,.\end{equation}
Here, $\vT^{ a_1\dotsb\alpha_n}$ is a row vector whose entries  $T_k^{a_1\dotsb\alpha_n}$ are suitable color structures that together
provide a complete basis for all allowed color structures, but which do not necessarily all have to be independent. In other words, the elements of $\vT^{ a_1\dotsb\alpha_n}$ span the color-conserving subspace of the full color space spanned by $\{a_1 \dotsb\alpha_n\}$, and $\vC$ is a vector in this subspace. Throughout this chapter we will refer to the elements of $\vT^{ a_1\dotsb\alpha_n}$ as a color basis, although they will generically be overcomplete, since this allows for simpler choices of color structures. As discussed below, due to the overcompleteness of the bases, some care will be required for their consistent usage.

Using \eq{Cpm_color}, we can rewrite \eq{Leff} as
\begin{equation} \label{eq:Leff_alt}
\cL_\hard = \!\int\!\prod_{i=1}^n\!\df \lp_i\,
\Op^\dagger_{+\cdot\cdot(\cdot\cdot-)}(\lp_1, \ldots, \lp_n)
\vC_{+\cdot\cdot(\cdot\cdot-)}(\lp_1, \ldots, \lp_n)
,\end{equation}
where $\Op^\dagger$ is a conjugate vector defined by
\begin{equation} \label{eq:Opm_color}
\Op^\dagger_{+\cdot\cdot(\cdot\cdot-)}
= O_{+\cdot\cdot(\cdot\cdot-)}^{a_1\dotsb\alpha_n}\, \vT^{ a_1\dotsb\alpha_n}
\,.\end{equation}
While the form $C_{+\cdot\cdot(\cdot\cdot-)}^{a_1\dotsb \alpha_n}\,
O_{+\cdot\cdot(\cdot\cdot-)}^{a_1 \dotsb \alpha_n}$ in \eq{Leff} is more convenient to discuss the matching and the symmetry properties of operators and Wilson coefficients, the alternative form in \eq{Leff_alt} is more convenient to discuss the mixing of the color structures under renormalization.

For low multiplicities of colored particles it can be convenient to use orthogonal color bases, e.g., the singlet-octet basis for $q\bar q q' \bar q'$ is orthogonal. However, using orthogonal bases becomes increasingly difficult for higher multiplicity processes, and the color bases used for many fixed-order calculations are not orthogonal. (See e.g.\ Refs.~\cite{Keppeler:2012ih, Sjodahl:2015qoa} for a discussion of the use of orthogonal bases for $SU(N)$.) The use of a nonorthogonal color basis implies that when written in component form in a particular basis, the conjugate $\vC^\dagger$ of the vector $\vC$ is not just given by the naive complex conjugate transpose of the components of the vector. Instead, we have
\begin{align}\label{eq:def_dagger}
 \vec C^\dagger = \left[ C^{ a_1\dotsb\alpha_n} \right]^* \vT^{a_1\dotsb\alpha_n}
= \vC^{*T}\, \hT
\,,\end{align}
where
\begin{equation}\label{eq:hatT_def}
\hT = \sum_{a_1,\ldots,\alpha_n} (\vT^{a_1\dotsb\alpha_n})^\dagger \vT^{a_1\dotsb\alpha_n}
\end{equation}
is the matrix of color sums for the chosen basis. If the basis is orthogonal (orthonormal), then $\hT$ is a diagonal matrix (identity matrix). Note that \eq{hatT_def} implies that by definition $\hT^{* T} = \hT$.

Similar to \eq{def_dagger}, for an abstract matrix $\widehat X$ in color space, the components of its Hermitian conjugate $\widehat X^\dagger$ when written in a particular basis are given in terms of the components of $\widehat X$ as
\begin{equation} \label{eq:def_daggermatrix}
\widehat X^\dagger = \hT^{-1}\, \widehat X^{*T}\, \hT
\,.\end{equation}

A proper treatment of the nonorthogonality of the color basis is also important in the factorization theorem of \eq{sigma}. Here, the color indices of the Wilson coefficients are contracted with the soft function as
\begin{align} \label{eq:trHS}
\left[C^{ a_1 \dotsb\alpha_n} \right]^*  S_\kappa^{a_1 \dotsb\alpha_n  b_1\dotsb\beta_n }    C^{b_1\dotsb\beta_n}
&= \vC^\dagger\, \hS_\kappa \vC \nonumber \\
& = \vC^{*T} \hT\, \hS_\kappa \vC
\,.\end{align}
At tree level, the soft function is simply the color-space identity
\begin{align}\label{eq:soft_id}
\hS_\kappa = \id \,,
\end{align}
which follows from its color basis independent definition in terms of Wilson lines [see e.g.\ Ref.~\cite{Jouttenus:2011wh} or \eq{softfunction_def}]. Here we have suppressed the dependence of $\hS$ on soft momenta. The action of the identity on an element of the color space is defined by
\begin{align}
(\id \vT)^{\cdots a_i \cdots \alpha_j \cdots} = \vT^{\cdots a_i \cdots \alpha_j \cdots}\,,
\end{align}
and its matrix representation in any color basis is given by $\id = \text{diag}(1,1,\cdots , 1)$.
In the literature, see e.g.\ Refs.~\cite{Kidonakis:1998nf,Kelley:2010fn, Kelley:2010qs,Broggio:2014hoa,Becher:2015gsa}, often a different convention is used, where the $\hT$ matrix is absorbed into the definition of the soft function. In this convention, the soft function becomes explicitly basis dependent and is not the same as the basis-independent color-space identity.  One should be careful to not identify the two.

As an example to demonstrate our notation for the color basis, consider the process $ggq\bq$. A convenient choice for a complete basis of color structures is
\begin{align} \label{eq:ggqqcol}
\vT^{ ab \alpha\bbeta}
&= \Bigl(
   (T^a T^b)_{\alpha\bbeta}\,,\, (T^b T^a)_{\alpha\bbeta} \,,\, \tr[T^a T^b]\, \delta_{\alpha\bbeta}
   \Bigr)
\nn\\
&\equiv \begin{pmatrix}
   (T^a T^b)_{\alpha\bbeta} \\ (T^b T^a)_{\alpha\bbeta} \\ 
   \tr[T^a T^b]\, \delta_{\alpha\bbeta}
\end{pmatrix}^{\!\!\!T}
.\end{align}
For cases with many color structures we will write $\vT$ as the transpose of a column vector as above. The transpose in this case only refers to the vector itself, not to the individual color structures. The color-sum matrix for this particular basis is
\begin{align}
\widehat T_{ggq\bq}
&= (\vT^{ ab \alpha\bbeta})^\dagger \vT^{ ab \alpha\bbeta}
\nn \\
&= \frac{C_F N}{2}
\begin{pmatrix}
   2C_F & 2C_F - C_A & 2T_F \\
   2C_F - C_A & 2C_F & 2T_F \\
   2T_F & 2T_F & 2T_F N
\end{pmatrix}.
\end{align}
Our conventions for color factors are given in \app{color}.
Explicit expressions for $\widehat T$ for the bases used in this chapter are given in \app{treesoft} for up to five partons.

Depending on the application, different choices of color basis can be used.
For example, in fixed-order QCD calculations, color ordering  \cite{Berends:1987me,Mangano:1987xk,Mangano:1988kk,Bern:1990ux} is often used to organize color information and simplify the singularity structure of amplitudes, while the color flow basis \cite{Maltoni:2002mq} is often used to interface with Monte Carlo generators. For a brief review of the color decomposition of QCD amplitudes, see \app{color_decomp}. Choosing a corresponding color basis in SCET has the advantage that the Wilson coefficients are given directly by the finite parts of the color-stripped helicity amplitudes, as defined in \eq{matching_general}, which can be efficiently calculated using unitarity methods. In this chapter we will use color bases corresponding to the color decompositions of the QCD amplitudes when giving explicit results for the matching coefficients, although we emphasize that an arbitrary basis can be chosen depending on the application.

Finally, note that the color structures appearing in the decomposition of a QCD amplitude up to a given loop order may not form a complete basis. The color basis in SCET must be complete even if the matching coefficients of some color structures are zero to a given loop order, since all structures can in principle mix under renormalization group evolution, as will be discussed in \sec{running}. In this case, we always choose a complete basis in SCET such that the color structures appearing in the amplitudes to some fixed order are contained as a subset.

\subsection{Parity and Charge Conjugation}
\label{subsec:discrete}

Under charge conjugation, the fields transform as
\begin{align} \label{eq:Cfield}
\C\, \cB^a_{i\pm}\, T^a_{\alpha\bbeta}\,\C &= - \cB^a_{i \pm} T^a_{\beta\balpha}
\,,\nn\\
\C\, J^{\balpha\beta}_{ij\pm}\,\C &= -J^{\bbeta\alpha}_{ji\mp}
\,.\end{align}
The minus sign on the right-hand side of the second equation comes from
anticommutation of the fermion fields.

Under parity, the fields transform as
\begin{align} \label{eq:Pfield}
\P\, \cB^a_{i\pm}(\lp_i, x)\, \P
&= e^{\pm2\img \phi_{n_i}}\cB^a_{i \mp}(\lp_i^\P, x^\P)
\,,\nn\\
\P\, J^{\balpha\beta}_{ij\pm}(\lp_i, \lp_j, x)\, \P
&= e^{\pm \img (\phi_{n_i}-\phi_{n_j})} J^{\balpha\beta}_{ij\mp}(\lp_i^\P, \lp_j^\P, x^\P)
\,,\end{align}
where we have made the dependence on $\lp_i$ and $x$ explicit, and the
parity-transformed vectors are $\lp_i^\P = \w_i\,\bn_i/2$, $x^\P_\mu = x^\mu$. The $\phi_{n_i}$ are real phases, whose exact definition is given in \app{helicity}. The phases appearing in the parity transformation of the helicity operators exactly cancel the phases appearing in the corresponding helicity amplitude under a parity transformation. This overall phase is determined by the little group scaling (see \app{helicity} for a brief review).

Using the transformations of the helicity fields under parity and charge
conjugation in \eqs{Cfield}{Pfield}, it is straightforward to determine how
these discrete symmetries act on the helicity operators. Parity and charge
conjugation invariance of QCD implies that the effective Lagrangian in \eq{Leff}
must also be invariant. (For amplitudes involving electroweak interactions, parity and charge conjugation invariance are explicitly violated. This is treated by extracting parity and charge violating couplings from the operators and amplitudes. See \sec{vec} for a discussion.) This then allows one to derive corresponding relations for the Wilson coefficients.

To illustrate this with a nontrivial example we consider the $ggq\bq$ process. The operators transform under charge conjugation as
\begin{align}
&\C\, O_{\la_1\la_2(\pm)}^{ab\,\balpha\beta}(\lp_1, \lp_2; \lp_3, \lp_4)\, \vT^{ ab\al\bbeta}\,\C
\nn\\ & \qquad
= \C\, S\, \cB_{1\la_1}^a \cB_{2\la_2}^b J_{34\pm}^{\balpha\beta} \, \vT^{ ab\al\bbeta}\,\C
\nn\\ & \qquad
= - O_{\la_1\la_2(\mp)}^{ba\,\balpha\beta}(\lp_1, \lp_2; \lp_4, \lp_3)\, \vT^{ ab\al\bbeta}
\,,\end{align}
where $\lambda_{1,2}$ denote the gluon helicities, and $\vT^{ ab\al\bt}$ is as given in \eq{ggqqcol}. From the invariance of \eq{Leff} we can infer that the Wilson coefficients must satisfy
\begin{equation} 
C_{\la_1\la_2(\pm)}^{ab\,\alpha\bbeta}(\lp_1, \lp_2; \lp_3, \lp_4)
= -C_{\la_1\la_2(\mp)}^{ba\,\alpha\bbeta}(\lp_1, \lp_2; \lp_4, \lp_3)
\,.\end{equation}
In the color basis of \eq{ggqqcol}, we can write this as
\begin{align}  \label{eq:ggqq_charge_basis}
\vC_{\la_1\la_2(\pm)}(\lp_1, \lp_2; \lp_3, \lp_4)
&= \hV \vC_{\la_1\la_2(\mp)}(\lp_1, \lp_2; \lp_4, \lp_3)
\,,\nn \\
\text{with}\qquad
\hV & =
\begin{pmatrix}
  0 & -1 & 0 \\
  -1 & 0 & 0 \\
  0 & 0 & -1
\end{pmatrix}
\,.\end{align}

Now consider the behavior under parity. For concreteness we consider the case of positive helicity gluons. The operators transform as
\begin{align}
&\P\, O_{++(\pm)}^{ab\,\balpha\beta}(\lp_1, \lp_2; \lp_3, \lp_4)\, \P
= \P\, \frac{1}{2} \cB_{1+}^a \cB_{2+}^b J_{34\pm}^{\balpha\beta} \, \P
\\ &
= e^{ \img (2\phi_{n_1}+2\phi_{n_2}\pm(\phi_{n_3}-\phi_{n_4}))} O_{--(\mp)}^{ab\,\balpha\beta}(\lp_1^P, \lp_2^P; \lp_3^P, \lp_4^P)\, 
\,. \nn\end{align}
The invariance of \eq{Leff}
under parity then implies that the Wilson coefficients satisfy
\begin{align} \label{eq:ggqq_parity}
&\vC_{++(\pm)}(\lp_1, \lp_2; \lp_3, \lp_4) \nn\\
&= \vC_{--(\mp)}(\lp_1^P, \lp_2^P; \lp_3^P, \lp_4^P) e^{ -\img (2\phi_{n_1}+2\phi_{n_2}\pm(\phi_{n_3}-\phi_{n_4}))} \nn \\
&= \vC_{--(\mp)}(\lp_1, \lp_2; \lp_3, \lp_4) \Big|_{\langle..\rangle \leftrightarrow [..]}
\,.\end{align}
Here we have introduced the notation $ \langle .. \rangle \leftrightarrow [..]$ to indicate that all angle and square spinors have been switched in the Wilson coefficient. The fact that the phase appearing in the parity transformation of the operator exactly matches the phase arising from evaluating the Wilson coefficient with parity related momenta is guaranteed by little group scaling, and will therefore occur generically. See \eqs{app_parity1}{app_parity2} and the surrounding discussion for a review.

Below we will use charge conjugation to reduce the number of Wilson
coefficients for which we have to carry out the matching explicitly. We
will use parity only when it helps to avoid substantial repetitions in the
matching.

\subsection{Crossing Symmetry}
\label{subsec:crossing}

Our basis is automatically crossing symmetric, since the gluon fields $\cB_{i\pm}$ can absorb or emit a gluon and the quark current $J_{ij\pm}$ can destroy or produce a quark-antiquark pair, or destroy and create a quark or antiquark. We will first illustrate how to use crossing symmetry in an example and then describe how to technically have crossing symmetric Wilson coefficients.

We will again consider the process $ggq\bq$ as an example. Due to our outgoing conventions, the default Wilson coefficient is for the unphysical processes with all outgoing particles:
\begin{equation}
  0 \to g_+^a(p_1) g_-^b(p_2) q_+^\alpha(p_3) \bq_-^\bbeta(p_4) : C_{+-(+)}^{ab\al\bbeta}(\lp_1,\lp_2;\lp_3,\lp_4)
\,,\end{equation}
where we picked one specific helicity configuration for definiteness.
Crossing a particle from the final state to the initial state flips its helicity, changes the sign of its momentum, and changes it to its antiparticle.
In addition we get a minus sign for each crossed fermion, though in practice these can be ignored as they do not modify the cross section.
This allows one to obtain the Wilson coefficient for any crossing. For example, for the following possible crossings, the Wilson coefficients are given by
\begin{align} \label{eq:crossing}
  g_+^a(p_1) g_-^b(p_2) \to q_+^\al(p_3) \bq_-^\bbeta(p_4) & : C_{+-(+)}^{ba\al\bbeta}(-\lp_2,-\lp_1;\lp_3,\lp_4)\,,
  \nn \\
  g_+^a(p_1) q_+^\balpha(p_2) \to g_+^b(p_3) q_+^\bt(p_4) & : -C_{+-(+)}^{ba\bt\balpha}(\lp_3,-\lp_1;\lp_4,-\lp_2)\,,
  \nn \\
  g_+^a(p_1) \bq_-^\al(p_2) \to g_+^b(p_3) \bq_-^\bbeta(p_4) & : -C_{+-(+)}^{ba\al\bbeta}(\lp_3,-\lp_1;-\lp_2,\lp_4)\,,
  \nn \\
  q_+^\balpha(p_1) \bq_-^\beta(p_2) \to g_+^a(p_3) g_-^b(p_4) & : C_{+-(+)}^{ab\bt\balpha}(\lp_3,\lp_4;-\lp_2,-\lp_1)\,.
\end{align}

Since the signs of momenta change when crossing particles between the final and initial state, care is required in taking the proper branch cuts to maintain crossing symmetry for the Wilson coefficients. In terms of the Lorentz invariants 
\begin{align}
  s_{ij}=(p_i+p_j)^2
\end{align}
this amounts to the choice of branch cut defined by $s_{ij} \to s_{ij} + \img 0$. In particular, we write all logarithms as
\begin{equation} \label{eq:Lij_def}
L_{ij} \equiv \ln\Bigl(-\frac{s_{ij}}{\mu^2} -\img0 \Bigr) = \ln\Bigl(\frac{s_{ij}}{\mu^2}\Bigr) - \img\pi \theta(s_{ij})
\,.\end{equation}
For spinors, crossing symmetry is obtained by defining the conjugate spinors $\bra{p\pm}$ as was done in \eq{braket_def}, resulting in the following relation
\begin{equation}
\bra{p\pm} = \mathrm{sgn}(p^0)\, \overline{\ket{p\pm}}
\,.\end{equation}
The additional minus sign for negative $p^0$ is included to use the same branch (of the square root inside the spinors) for both spinors and conjugate spinors, i.e., for $p^0 > 0$ we have
\begin{align}
\ket{(-p)\pm} &= \img\ket{p\pm}\,,
\nn \\
\bra{(-p)\pm} &=
-(-\img) \bra{p\pm} = \img\bra{p\pm}
\,.\end{align}
In this way all spinor identities are automatically valid for both positive and negative momenta, which makes it easy to use crossing symmetry.

\subsection{Hard Function}
\label{subsec:hard}

In the factorized expression for the cross section given in \eq{sigma},
the dependence on the underlying hard Born process appears through the hard function $\hH_\kappa$. In terms of the Wilson coefficients of the operator basis in the effective theory, the hard function for a particular partonic channel $\kappa$ is given by
\begin{align}\label{eq:hard_C}
  \hH_\kappa(\{\lp_i\})
  = \sum_{\{\la_i\}}
  \vC_{\la_1\cdot\cdot(\cdot\cdot\la_n)}(\{\lp_i\})\,
  \vC_{\la_1 \cdot\cdot(\cdot\cdot\la_n)}^\dagger (\{\lp_i\})
 \,, 
\end{align}
where $\{\lp_i\} \equiv \{\lp_1, \lp_2, \ldots\}$. For unpolarized experiments we simply sum over all helicity operators, so $\hH_\kappa(\{\lp_i\})$ with its sum over helicities in \eq{hard_C} appears as a multiplicative factor. It is important to note that the color indices of the Wilson coefficients are not contracted with each other, rather they are contracted with the color indices of the soft function through the trace seen in \eq{sigma}.

As an explicit example to demonstrate the treatment of both color and helicity indices, we consider the contribution of the $ggq\bq$ partonic channel to the $pp \to 2$ jets process. In this case, the Wilson coefficients are given by $\vC_{\la_1 \la_2 (\la_3)}$, where $\lambda_1,\lambda_2$ denote the helicities of the gluons, $\lambda_3$ denotes the helicity of the quark current, and recall that the vector denotes the possible color structures, which were given explicitly for this case in \eq{ggqqcol}. The hard function for this partonic channel is then given by
\begin{align} \label{eq:ggqqhard}
  \hH_{ggq\bq}(\{\lp_i\})
  & = \!\sum_{\{\la_i\}}\!
  \vC_{\la_1 \la_2 (\la_3)}(\{\lp_i\})\,
  \vC_{\la_1 \la_2 (\la_3)}^\dagger (\{\lp_i\})
  \nn \\
  & = 
  \vC_{++ (+)}
  \vC_{++ (+)}^\dagger +
  \vC_{+- (+)}
  \vC_{+- (+)}^\dagger +
  \nn \\ & \quad
  \vC_{-+ (+)}
  \vC_{-+ (+)}^\dagger +
  \vC_{-- (+)}
  \vC_{-- (+)}^\dagger +
  \nn \\ & \quad
  \vC_{++ (-)}
  \vC_{++ (-)}^\dagger +
  \vC_{+- (-)}
  \vC_{+- (-)}^\dagger +
  \nn \\ & \quad
  \vC_{-+ (-)}
  \vC_{-+ (-)}^\dagger +
  \vC_{-- (-)}
  \vC_{-- (-)}^\dagger
 \,.
\end{align}
Here, explicit expressions are only needed for $\vC_{++(+)}, \vC_{+-(+)}$ and $\vC_{--(+)}$. One can obtain $\vC_{-+(+)}$ using Bose symmetry simply by interchanging the gluons,
\begin{align}
\vC_{-+(+)}^{ab\alpha\bbeta}(\lp_1,\lp_2;\lp_3,\lp_4) &= \vC_{+-(+)}^{ba\alpha\bbeta}(\lp_2,\lp_1;\lp_3,\lp_4)
\,,\end{align}
or equivalently,
\begin{align}
\vC_{-+(+)}(\lp_1,\lp_2;\lp_3,\lp_4) &= \hV \vC_{+-(+)}(\lp_2,\lp_1;\lp_3,\lp_4)
\,,\nn \\
\text{with}\qquad
\hV & =
\begin{pmatrix}
  0 & 1 & 0 \\
  1 & 0 & 0 \\
  0 & 0 & 1
\end{pmatrix}
\,.\end{align}
As explained in \subsec{ggqqbarbasis}, the remaining $\vC_{\la_1\la_2(-)}$ can be obtained from the expressions for the other Wilson coefficients by charge conjugation.

In \eq{ggqqhard}, the Wilson coefficients are vectors in the color basis of \eq{ggqqcol} and thus the hard function is a matrix in this basis. As discussed in \subsec{color}, the tree-level soft function is the color-space identity, i.e.,
\begin{equation}
  S_{ggq\bq}^{\zero\,b_1 b_2 \beta_1 \bbeta_2\, a_1 a_2 \alpha_1 \balpha_2} = \de^{b_1 a_1} \de^{b_2 a_2} \de_{\bt_1\alpha_1} \de_{\bbeta_2\balpha_2}
  \equiv \id
\,.\end{equation}
With the color trace in \eq{sigma} this amounts to contracting the color indices of the Wilson coefficients. In the color basis of \eq{ggqqcol}, this simply becomes
\begin{equation}
  \hS_{ggq\bq}^\zero = \id =
\begin{pmatrix}
  1 & 0 & 0 \\
  0 & 1 & 0 \\
  0 & 0 & 1
\end{pmatrix}
.\end{equation}
The tree-level soft function also has dependence on momenta depending on the measurement being made, which are not shown here.

To demonstrate a complete calculation of the cross section using the factorization theorem of \eq{sigma} together with the hard functions computed using the helicity operator formalism, it is instructive to see how the leading-order cross section is reproduced from \eq{sigma}. We consider the simple case of $H+0$ jets in the $m_t \rightarrow \infty$ limit. For this channel, there is a unique color structure $\delta_{a_1a_2}$, and using the results of \subsec{H0jet} and \app{H0amplitudes}, the lowest order Wilson coefficients are given by
\begin{align}
\vC_{++}(\lp_1,\lp_2;\lp_3)&=\delta_{a_1 a_2} \frac{\alpha_s}{3\pi v} \frac{s_{12}}{2}\frac{[12]}{\langle 1 2\rangle},\\
\vC_{--}(\lp_1,\lp_2;\lp_3)&=\delta_{a_1 a_2} \frac{\alpha_s}{3\pi v} \frac{s_{12}}{2}\frac{\langle12\rangle}{[ 1 2]},\\
\vC_{+-}(\lp_1,\lp_2;\lp_3)&=\vC_{-+}(\lp_1,\lp_2;\lp_3)=0\,,
\end{align}
where $v=\left (  \sqrt{2} G_F  \right )^{-1/2}=246$GeV. Note that these are simply the helicity amplitudes for the process, as will be shown more generally in \sec{matching}.
Analytically continuing to physical momenta, squaring, and summing over helicities, the tree-level hard function is given by
\begin{align}
  H_{ggH}^{\zero\, a_1 a_2\, b_1 b_2}(\lp_1,\lp_2;\lp_3)
  &=  \Bigl| \frac{\al_s}{3\pi v} \frac{s_{12}}{2} \Bigr|^2\, 2\, \de_{a_1 a_2} \de_{b_1 b_2} 
  \nn \\ &=
  \frac{\al_s^2 s_{12}^2}{18\pi^2 v^2} \, \de_{a_1 a_2} \de_{b_1 b_2} 
\,.\end{align}
Note that only 2 of the 4 helicity configurations contribute, hence the factor of 2.

The tree-level gluon beam functions are given by the gluon PDFs. Since there are no jets in the final state, there are no jet functions. The tree-level soft function is the identity in color space%
\footnote{Since there is only one color structure, the tree-level soft function is normally defined as
\begin{equation}
   S^\zero_{gg} = \frac{1}{N^2-1}\,\de_{a_1 a_2} \de_{b_1 b_2}\, \de^{b_1 a_1} \de^{b_2 a_2} = 1
\,.\end{equation}
Here we do not absorb numerical prefactors into our soft functions, because this is not useful for processes with more final-state partons.}
\begin{equation}
  S_{gg}^{\zero\, b_1 b_2\, a_1 a_2} = \de^{b_1 a_1} \de^{b_2 a_2}
\,.\end{equation}
The leading-order cross section is then given by
\begin{align} \label{eq:hard_comp_H}
\si &= \frac{1}{2\Ecm^2} \frac{1}{[2(N^2-1)]^2}
 \int\! \frac{\df x_1}{x_1}\, \frac{\df x_2}{x_2}\, f_g(x_1) f_g(x_2)
\nn \\ & \quad\times
  \int\! \frac{\df^4p_3}{(2\pi)^3}\, \theta(p_3^0)\, \de(p_3^2 \!-\! m_H^2)
\nn \\ & \quad\times
  (2\pi)^4 \delta^4\Bigl(x_1 \Ecm \frac{n_1}{2} \!+\! x_2 \Ecm \frac{n_2}{2} \!-\! p_3 \Bigr)
 \nn \\ & \quad  
\times H_{ggH}^{\zero\, a_1 a_2\, b_1 b_2}(\lp_1,\lp_2;\lp_3)\,  S_{gg}^{\zero\, b_1 b_2\, a_1 a_2}
\nn \\
&= \frac{\al_s^2 m_H^2}{576 \pi v^2 \Ecm^2} \int\! \df Y
 f_g\Bigl(\frac{m_H}{\Ecm} e^Y\Bigr) f_g\Bigl(\frac{m_H}{\Ecm} e^{-Y}\Bigr)
\,.\end{align}
The $1/(2\Ecm^2)$ factor is the flux factor and for each of the incoming gluons we get a $1/[2(N^2-1)]$ from averaging over its spin and color. This is followed by integrals over the gluon PDFs, $f_g$, and the Higgs phase space, where we have restricted to the production of an on-shell Higgs. The final expression in \eq{hard_comp_H} agrees with the standard result, where the first factor is the Born cross section.

We now briefly discuss our choice of normalization. The currents in \eq{jpm_def} were normalized such that
the Wilson coefficients are simply given by the finite part of the QCD helicity amplitudes (see \eq{matching_general} and \sec{matching}). 
This is distinct from the normalization typically used for SCET operators, e.g.~$\bar{\chi_i} \ga^\mu \chi_j$, which is chosen to facilitate the matching to QCD operators. We now show that the extra factors in \eq{jpm_def} arrange themselves to produce the standard normalization for the jet function (or beam function). Starting from the current and its conjugate,
\begin{align}\label{eq:fierzeg}
&J_{ij\pm}^{\balpha\beta}
(J_{ij\pm}^{\bgamma\delta})^\dagger
  \\
&=\frac{\sqrt{2}\ve_\mp^\mu(n_i, n_j)}
  {   \sqrt{ \vphantom{2}    \omega_i\, \omega_j }} \frac{\bar{\chi}^\balpha_{i\pm} \gamma_\mu \chi^\beta_{j\pm}}
   {\langle n_i \!\mp | n_j \pm\rangle}
 \frac{\sqrt{2}\ve_\pm^\nu(n_i, n_j)}
   {\left(\sqrt{\vphantom{2} \omega_i \,  \omega_j }\right)^*}
 \frac{\bar{\chi}^\bdelta_{j\pm}\, \gamma_\nu \chi^\gamma_{i\pm}}
 {\langle n_j\!\pm | n_i \mp\rangle}
\nn \\
&= 4\frac{\de^{\ga\balpha}}{N} \frac{\de^{\beta \bdelta}}{N}\,
\frac{\ve_\pm^\nu(n_i, n_j)\,\ve_\mp^\mu(n_i, n_j)}{2 n_i \sdt n_j\, |\omega_i\, \omega_j|}\,
\tr\Big[\ga_\nu \frac{\nslash_i}{4} \ga_\mu \frac{\nslash_j}{4} \Big] 
\nn \\ & \quad \times
\Big(\bar{\chi}_{i\pm} \frac{\bnslash_i}{2} \chi_{i\pm}\Big)
\Big(\bar{\chi}_{j\pm} \frac{\bnslash_j}{2} \chi_{j\pm}\Big)
 + \ldots
\nn \\
&=2\, \de^{\ga\balpha} \de^{\beta \bdelta}\,
\Big(\frac{1}{2N} \frac{1}{|\w_i|} \bar{\chi}_{i\pm} \frac{\bnslash_i}{2} \chi_{i\pm}\Big)
\Big(\frac{1}{2N} \frac{1}{|\w_j|} \bar{\chi}_{j\pm} \frac{\bnslash_j}{2} \chi_{j\pm}\Big)
\,,\nonumber
\end{align}
where we have rearranged the expression in a factorized form using the SCET Fierz formula in spin
\begin{align}
1 \otimes 1=\frac{1}{2} \biggl[ \frac{\bnslash_i}{2} \otimes \frac{\nslash_i}{2}-\frac{\bnslash_i \gamma_5}{2} \otimes \frac{\nslash_i \gamma_5}{2}-\frac{\bnslash_i \gamma_\perp^\mu}{2} \otimes \frac{\nslash_i \gamma_{\perp\mu}}{2} \biggr]
,\end{align}
which applies for the SCET projected spinors. In the last line of \eq{fierzeg} we have dropped the color nonsinglet terms and terms which vanish when averaging over helicities, which are indicated by ellipses. The delta functions in color space highlight that the jet function does not modify the color structure. The factor $1/\w_{i,j}$, which arises from the normalization of the helicity currents, is part of the standard definition of the jet function and ensures that this operator has the correct mass dimension.

\section{Matching and Scheme Dependence}
\label{sec:matching}

In this section, we discuss the matching of QCD onto the SCET helicity operator basis introduced in the previous section. We start with a discussion of the matching for generic helicity operators in \subsec{genmatch}. In \subsec{renscheme} we discuss in detail the subject of renormalization schemes, and the issue of converting between regularization/renormalization schemes commonly used in spinor-helicity calculations, and those used in SCET. We also demonstrate that evanescent operators are not generated in our basis.

\subsection{Generic Matching} \label{subsec:genmatch}

In this chapter, we work to leading order in the power counting, which means we only require operators that contain exactly one field per collinear sector. That is, different $n_i$ in \eq{Opm_gen} are implicitly restricted to belong to different equivalence classes,  $\{n_i\} \neq \{n_j\}$ for $i\neq j$. Operators with more than one field per collinear direction are power-suppressed compared to the respective leading-order operators that have the same set of collinear directions and the minimal number of fields.

At leading order, the Wilson coefficients can thus be determined by computing matrix elements of \eq{Leff}, with all external particles assigned well-separated momenta, so that they belong to separate collinear sectors. The only helicity operator that contributes in this case is the one that matches the set of external helicities, picking out the corresponding Wilson coefficient. Since we only have one external particle per collinear sector, we can simply choose $n_i = p_i/p_i^0$ in the matching calculation to represent the equivalence class $\{n_i\}$.

To compute the matrix element of $\cL_\hard$, we first note that the helicity operators are symmetric (modulo minus signs from fermion anticommutation) under simultaneously interchanging the label momenta and indices of identical fields, and the same is thus also true for their Wilson coefficients. For example, at tree level
\begin{align}
&\Mae{g_+^{a_1}(p_1)\, g_+^{a_2}(p_2)}{O_{++}^{b_1 b_2}}{0}^{\tree}
\nn\\ & \quad
= \frac{1}{2}\bigl[ \delta^{a_1 b_1}\, \delta^{a_2 b_2} \ldel(\lp_1 - p_1)\,\ldel(\lp_2 - p_2)
\nn\\ &\qquad
+ \delta^{a_1 b_2}\, \delta^{a_2 b_1} \ldel(\lp_1 - p_2)\,\ldel(\lp_2 - p_1) \bigr]
\end{align}
so the tree-level matrix element of $\cL_\hard$ gives
\begin{align}
&\Mae{g_+^{a_1}(p_1)\, g_+^{a_2}(p_2)}{\cL_\hard}{0}^{\tree}
\\\nn &\quad
= \frac{1}{2} \bigl[C_{++}^{a_1 a_2}(\lp_1, \lp_2) + C_{++}^{a_2 a_1}(\lp_2, \lp_1)\bigr]
= C_{++}^{a_1 a_2}(\lp_1, \lp_2)
\,.\end{align}
By choosing $n_i = p_i/p_i^0$, the label momenta $\lp_i$ on the right-hand side simply become $\lp_i \equiv \bn\cdot p_i\, n_i/2 = p_i$.

Taking into account the symmetry factor in \eq{Opm_S}, one can easily see that this result generalizes to more than two gluons or quark currents with the same helicity. In the case of identical fermions, the various terms in the operator matrix element have relative minus signs due to fermion anticommutation which precisely match the (anti)symmetry properties of the Wilson coefficients. Hence, the tree-level matrix element of $\cL_\hard$ is equal to the Wilson coefficient that corresponds to the configuration of external particles,
\begin{align} \label{eq:Leff_me}
&\Mae{g_1g_2\dotsb q_{n-1}\bar{q}_n}{\cL_\hard}{0}^{\tree}
\nn\\ & \qquad
= C_{+\cdot\cdot(\cdot\cdot-)}^{a_1 a_2\dotsb\alpha_{n-1}\balpha_n}(\lp_1,\lp_2,\ldots,\lp_{n-1},\lp_n)
\,.\end{align}
Here and below, $g_i \equiv g_\pm^{a_i}(p_i)$ stands for a gluon with helicity $\pm$, momentum $p_i$, color $a_i$, and analogously for (anti)quarks. From \eq{Leff_me} we obtain the generic tree-level matching equation
\begin{equation} \label{eq:matching_LO}
C_{+\cdot\cdot(\cdot\cdot-)}^{a_1\dotsb\balpha_n}(\lp_1,\ldots,\lp_n)
= -\img \cA^\tree(g_1 \dotsb \bar{q}_n)
\,,\end{equation}
where $\cA^\tree$ denotes the tree-level QCD helicity amplitude. Intuitively, since all external particles are energetic and well separated, we are away from any soft or collinear limits and so all propagators in the QCD tree-level diagram are far off shell and can be shrunk to a point. Hence, the tree-level diagram simply becomes the Wilson coefficient in SCET.

The above discussion can be extended to higher orders in perturbation theory. In pure dimensional regularization (where $\epsilon$ is used to simultaneously regulate UV and IR divergences) all bare loop graphs in SCET are scaleless and vanish. Here the UV and IR divergences precisely cancel each other, and the bare matrix elements are given by their tree-level expressions, \eq{Leff_me}. Including the counterterm $\delta_O(\eps_\mathrm{UV})$ due to operator renormalization removes the UV divergences and leaves the IR divergences. Schematically, the renormalized loop amplitude computed in SCET using $\cL_\hard$ is
\begin{equation}
\cA_\mathrm{SCET}
= \int\! (\vev{\vec O^\dagger}^\tree 
   \!+\! \vev{\vec O^\dagger}^\mathrm{loop})\, \img \vec C
= \bigl[1 + \delta_O(\eps_\IR) \bigr] \img \vec C 
,
\end{equation}
where we used that the loop contribution is a pure counterterm and thus proportional to the tree-level expression. In general, the counterterm $\delta_O$ is a matrix in color space, as we will see explicitly in \sec{running} and \app{IRdiv}. By construction, the $1/\epsilon$ IR divergences in the effective theory, $C\,\delta_O(\eps_\IR)$, have to exactly match those of the full theory. Therefore, beyond tree level the matching coefficients in $\overline{\mathrm{MS}}$ are given by the infrared-finite part of the renormalized full-theory amplitude, ${\cal A}_{\rm ren}$, computed in pure dimensional regularization. The IR-finite part is obtained by multiplying ${\cal A}_{\rm ren}$ by SCET $\overline{\rm MS}$ renormalization factors, which cancel the full theory $1/\epsilon_\IR$ poles.  Decomposing the renormalized QCD amplitude in a color basis so that ${\cal A}_{\rm ren}^{a_1 \dotsb\alpha_n}= \vT^{ a_1\dotsb\bar \alpha_n} \vec {\cal A}_{\rm ren}(g_1\dotsb \bar q_n)$, the all-orders form of \eq{matching_LO} becomes
\begin{align}  \label{eq:matching_general}
C_{+\cdot\cdot(\cdot\cdot-)}^{a_1\dotsb\balpha_n}(\lp_1,\ldots,\lp_n)
&= -\img \cA_\fin(g_1\dotsb \bar{q}_n) 
 \\
&\equiv \frac{-\img \,
  \vT^{ a_1\dotsb\bar\alpha_n}   \widehat Z_C^{-1} \vec {\cal A}_{\rm ren}(g_1\dotsb \bar{q}_n)}{ Z_\xi^{n_q/2} Z_A^{n_g/2} }
  \nn
\,.\end{align}
The SCET renormalization factors $\widehat Z_C$, $Z_\xi$, and $Z_A$ are discussed in \subsec{running_general}. At one-loop order this corresponds to taking $(-\img {\cal A}_{\rm ren}^{a_1 \dotsb\bar\alpha_n})$ and simply dropping the $1/\epsilon_\IR$ terms. In \subsec{renscheme} we will discuss in more detail the use of different renormalization schemes to compute $\vec {\cal A}_{\rm ren}(g_1\dotsb \bar{q}_n)$.

If the same color decomposition is used for the QCD amplitude as for the Wilson coefficients in \eq{Cpm_color}, we can immediately read off the coefficients $\vC$ in this color basis from \eq{matching_general}. As an example, consider for simplicity the leading color $n$ gluon amplitude, which has the color decomposition (see \app{color_decomp})
\begin{align}
\cA_n(g_{1}\dotsb g_{n})
&= \img g_s^{n-2} \!\!\sum_{\si \in S_n/Z_n}\!\! \tr[T^{a_{\si(1)}} \dotsb T^{a_{\si(n)}}]
\nn\\ & \quad\times
\sum_i g_s^i\, A_n^{(i)}(\si(1),\ldots,\si(n))
\,,\end{align}
where the first sum runs over all permutations $\sigma$ of $n$ objects ($S_n$) excluding cyclic permutations ($Z_n$). The $A_n^{(i)}$ are the color-ordered or partial amplitudes at $i$ loops. Each is separately gauge invariant and only depends on the external momenta and helicities $(p_i\pm) \equiv (i^\pm)$. If we choose
\begin{equation}
T_k^{a_1\dotsb a_n} = \tr[T^{a_{\si_k(1)}} \dotsb T^{a_{\si_k(n)}}]\,,
\end{equation}
as the color basis in \eq{Cpm_color}, where $\sigma_k$ is the $k$th permutation in $S_n/Z_n$, then the Wilson coefficients in this color basis are given directly by
\begin{align}
C^k_{\lambda_1\cdots\lambda_2}&(\lp_1,\ldots,\lp_n) =\nonumber \\
& g_s^{n-2} \sum_i g_s^i\,  A_{n,\fin}^{(i)}(\si_k(1^{\lambda_1}),\ldots,\si_k(n^{\lambda_2}))
\,,\end{align}
where the subscript ``$\fin$'' denotes the IR-finite part of the helicity amplitude, as defined in \eq{matching_general}. This is easily extended beyond leading color, given a valid choice of subleading color basis. Our basis therefore achieves seamless matching from QCD helicity amplitudes onto SCET operators.

\subsection{Renormalization Schemes}
\label{subsec:renscheme}

In this section we discuss in more detail the issue of renormalization/regularization schemes in QCD and in SCET. In particular, the construction of a basis of helicity operators discussed in \sec{basis} relied heavily on massless quarks and gluons having two helicity states, which is a feature specific to 4 dimensions. We clarify this issue here and discuss the conversion between various schemes.

In dimensional regularization, divergences are regularized by analytically continuing the particle momenta to $d$ dimensions. In a general scheme, the helicities of quarks and gluons live in $d^g_s$, $d^q_s$ dimensional spaces respectively. We shall here restrict ourselves to schemes where quarks have two helicities, but $d^g_s$ is analytically continued. This is true of most commonly used regularization schemes, but is not necessary \cite{Catani:1996pk}. Different schemes within dimensional regularization differ in their treatment of $d^g_s$ for internal (unobserved) and external (observed) particles. In the conventional dimensional regularization (CDR), 't Hooft-Veltman (HV)~\cite{tHooft:1972fi}, and four-dimensional helicity (FDH)~\cite{Bern:1991aq, Bern:2002zk} schemes the internal/external polarizations are treated in $d/d$ (CDR), $d/4$ (HV), $4/4$ (FDH) dimensions. 

For helicity-based computations, the FDH scheme has the advantage of having all helicities defined in 4 dimensions, where the spinor-helicity formalism applies, as well as preserving supersymmetry. Indeed, most of the recent one-loop computations of helicity amplitudes utilize on-shell methods and therefore employ the FDH scheme. However, most existing calculations of SCET matrix elements (jet, beam, and soft functions) use $d$-dimensional internal gluons, corresponding to the CDR/HV schemes.\footnote{Recently while this chapter was being finalized, a calculation of the inclusive jet and soft functions in both FDH and dimensional reduction (DRED)~\cite{Siegel:1979wq} appeared in Ref.~\cite{Broggio:2015dga}. The conclusions of this section agree with their study of the regularization scheme dependence of QCD amplitudes.} As we will discuss below, CDR and HV are identical for matching onto SCET.

Although the FDH scheme is convenient for helicity amplitude computations, it leads to subtleties beyond NLO~\cite{Kilgore:2011ta,Boughezal:2011br}. As explained in Ref.~\cite{Boughezal:2011br}, this discrepancy arises due to the different number of dimensions for the momenta in the loop integral and the spin space, leading to components of the gluon field whose couplings to quarks are not protected by gauge invariance and require separate renormalization. Nevertheless, it has been shown that FDH is a consistent regularization scheme to NNLO \cite{Broggio:2015dga}. The presence of these extra degrees of freedom in the FDH scheme is quite inconvenient in the formal construction of SCET, especially when working to subleading power. Because of this fact, and because most SCET calculations are performed in CDR/HV, our discussion of SCET schemes will focus on regularization schemes where the dimension of the gluon field and the momentum space are analytically continued in the same manner. We will also discuss how full-theory helicity amplitudes in the FDH scheme are converted to CDR/HV for the purposes of matching to SCET.

We will now describe how helicity amplitudes in the FDH scheme can be converted to CDR/HV.
To get a finite correction from the $\ord{\eps}$ part of the gluon polarization requires a factor from either ultraviolet (UV) or infrared (IR) $1/\eps$ divergences. Although the regularization of UV and IR divergences is coupled in pure dimensional regularization schemes by use of a common $\epsilon$, they can in principle be separately regulated, and we discuss their role in the scheme conversion separately below.

When matching to SCET, the UV regulators in the full and effective theory need not be equal. Indeed, the effective theory does not reproduce the UV of the full theory. In massless QCD, scheme dependence due to the UV divergences only affects the coupling constant through virtual (internal) gluons. Therefore, the CDR and HV schemes have the same standard $\overline{\text{MS}}$ coupling, $\alpha_s(\mu)$, while FDH has a different coupling, $\alpha_s^\mathrm{FDH}(\mu)$. The conversion between these couplings is achieved by a perturbatively calculable shift, known to two loops~\cite{Altarelli:1980fi, Kunszt:1993sd, Bern:2002zk}
\begin{align} \label{eq:DR_HV_UV}
 \al_s^\mathrm{FDH}(\mu) &= \al_s(\mu) \biggl[1 + \frac{C_A}{3} \frac{\al_s(\mu)}{4\pi}
 \nn \\ & \quad
 + \Bigl(\frac{22}{9}\,C_A^2 - 2 C_F T_F n_f\Bigr) \bigg(\frac{\al_s(\mu)}{4\pi}\bigg)^2 \biggr]
\,.\end{align}
This replacement rule for the coupling captures the effect of the scheme choice from UV divergences. One can therefore perform a matching calculation, treating $\alpha_s$ in the full and effective theories as independent parameters that can be defined in different schemes. A conversion between schemes can then be used to ensure that the matching coefficients are written entirely in terms of $\alpha_s$ defined in one scheme, for example using \eq{DR_HV_UV}. The issue of UV regularization is therefore simple to handle in the matching.

The structure of $1/\eps^2$ and $1/\eps$ IR divergences in one-loop QCD amplitudes is well known, and allows one to determine their effect on converting amplitudes from FDH to CDR/HV. For a QCD amplitude involving $n_q$ (anti)quarks and $n_g$ gluons the FDH and HV one-loop amplitudes $\cA^\one$ are related by~\cite{Kunszt:1993sd, Catani:1996pk}
\begin{equation} \label{eq:DR_HV_IR}
 \cA_\mathrm{HV}^\one = \cA_\mathrm{FDH}^\one
 - \frac{\alpha_s}{4\pi} \Bigl(\frac{n_q}{2} C_F + \frac{n_g}{6} C_A\Bigr)  \cA^\zero
\,,\end{equation}
where $\cA^\zero$ denotes the tree-level amplitude, and the precise scheme of the $\alpha_s$ entering here is a two-loop effect. At one loop, the FDH scheme can therefore be consistently used when calculating full-theory helicity amplitudes and results can easily be converted to HV with \eqs{DR_HV_UV}{DR_HV_IR} for use in SCET Wilson coefficients.

We will now compare CDR and HV schemes for SCET calculations and the construction of the operator basis. In the HV scheme, all external polarizations are 4 dimensional, so that one can use a basis of helicity operators, as was constructed in \sec{basis}. However, in CDR external polarizations are $d$ dimensional, with the limit $d\to 4$ taken. In particular, this implies that one must work with $d-2$ gluon polarizations at intermediate steps, potentially allowing for the presence of evanescent operators corresponding to operators involving the additional components of the gluon field, so-called $\epsilon$-helicities. However, we will now argue that there is no real distinction between the two schemes, and that one does not need to consider evanescent operators in SCET at leading power.

First consider the Wilson coefficients and matching. In the case of CDR, the operator basis must be extended to include operators involving the $\epsilon$-helicities. However, their presence does not affect the matching coefficients for operators with physical helicities, since they do not contribute at tree level and all loop corrections are scaleless and vanish.
Additionally, in \sec{running}, we will discuss the fact that the SCET renormalization of the operators is spin independent at leading power, and therefore there is no mixing under renormalization group evolution between the physical and evanescent operators. For the beam and jet functions, azimuthal symmetry implies that the difference between a field with 2 or $2-2\eps$ polarizations is simply an overall factor of $1-\eps$ and thus can be easily taken into account. The independence of the soft function to the differences in the CDR/HV regularization schemes follows from the insensitivity of the soft emission to the polarization of the radiating parton, which is made manifest by the SCET Lagrangian and the fact that the soft function can be written as a matrix element of Wilson lines. Thus there is no difference between CDR and HV and the helicity operator basis suffices.

\section{Higgs + Jets}
\label{sec:higgs}

In this section, we consider the production of an on-shell Higgs + jets. We give the helicity operator basis and matching relations for $H + 0,1,2$ jets, and the corresponding helicity amplitudes are collected in \app{Hamplitudes}.

\subsection{\boldmath $H + 0$ Jets}
\label{subsec:H0jet}

The $ggH$ and $q\bq H$ processes contribute to the $H+0$ jets process. For $q\bq H$, the scalar current in \eq{jS_def} is required, and the helicity operator basis is given by
\begin{align} \label{eq:qqH_basis}
O_1^{\balpha\beta}
&= J_{12\,0}^{\balpha\beta}\, H_3\,,
\nn \\
O_2^{\balpha\beta}
&= (J^\dagger)_{12\,0}^{\balpha\beta}\, H_3
\,,\end{align}
with the unique color structure
\begin{equation} \label{eq:H0qq_color}
\vT^{ \alpha\bbeta} = \begin{pmatrix} \de_{\alpha\bbeta} \end{pmatrix}
\,.\end{equation}
These operators are relevant when considering Higgs decays to massive quarks, for example $H\to \bar b b$. However, we will not consider this case further since for Higgs production the $b\bar bH$ and $t\bar tH$ contributions are much smaller than the dominant gluon-fusion hard scattering process.

For $ggH$, the basis of helicity operators is given by
\begin{align} \label{eq:ggH_basis}
O_{++}^{ab}
&= \frac{1}{2}\, \cB_{1+}^a\, \cB_{2+}^b\,  H_3
\,,\nn\\
O_{--}^{ab}
&= \frac{1}{2}\, \cB_{1-}^a\, \cB_{2-}^b\, H_3
\,.\end{align}
The operator $O_{+-}$ is not allowed by angular momentum conservation. Similar helicity operators, extended to include the decay of the Higgs, were used in Ref.~\cite{Moult:2014pja}.
There is again a unique color structure for this process,
\begin{equation} \label{eq:H0_color}
\vT^{ ab} = \begin{pmatrix} \de^{ab} \end{pmatrix}
\,.\end{equation}
Writing the QCD helicity amplitudes as
\begin{align}
\cA(g_1 g_2 H_3) &= \img \delta^{a_1 a_2}\, A(1, 2; 3_H)
\,,\end{align}
the Wilson coefficients for $ggH$ are given by
\begin{align} \label{eq:ggH_coeffs}
\vC_{++}(\lp_1, \lp_2; \lp_3) &= A_\fin(1^+, 2^+; 3_H)
\,, \nn \\
\vC_{--}(\lp_1, \lp_2; \lp_3) &= A_\fin(1^-, 2^-; 3_H)
\,.\end{align}
The subscript ``$\fin$'' in \eq{ggH_coeffs} denotes the IR-finite part of the helicity amplitudes, as discussed in \sec{matching}.
Note that the two amplitudes appearing in \eq{ggH_coeffs} are related by parity.
The results for the gluon amplitudes up to NNLO are given in \app{H0amplitudes}. They correspond to the usual gluon-fusion process, where the Higgs couples to a (top) quark loop at leading order. The LO amplitude including the dependence on the mass of the quark running in the loop is well known. The NLO amplitudes are also known including the full quark-mass dependence~\cite{Dawson:1990zj, Djouadi:1991tka, Spira:1995rr, Harlander:2005rq, Anastasiou:2006hc}, while the NNLO~\cite{Harlander:2000mg, Harlander:2009bw, Pak:2009bx} and N$^3$LO~\cite{Baikov:2009bg,Gehrmann:2010ue} amplitudes are known in an expansion in $m_H/m_t$. 

\subsection{\boldmath $H + 1$ Jet}

The $gq\bar qH$ and $gggH$ processes contribute to the $H+1$ jet process. For $gq\bar q$, the basis of helicity operators is given by
\begin{align}
O_{+(+)}^{a\, \balpha\beta}
&= \cB_{1+}^a\, J_{23+}^{\balpha\beta}\, H_4
\,,\nn\\
O_{-(+)}^{a\, \balpha\beta}
&= \cB_{1-}^a\, J_{23+}^{\balpha\beta}\, H_4
\,,\nn\\
O_{+(-)}^{a\, \balpha\beta}
&= \cB_{1+}^a\, J_{23-}^{\balpha\beta}\, H_4
\,,\nn\\
O_{-(-)}^{a\, \balpha\beta}
&= \cB_{1-}^a\, J_{23-}^{\balpha\beta}\, H_4
\,.\end{align}
Note that we consider only QCD corrections to the $ggH$ process, so the $q\bar q$ pair is described by $J_{ij\pm}$.
For $ggg$, the helicity operator basis is
\begin{align} \label{eq:H1_basis}
O_{+++}^{abc}
&= \frac{1}{3!}\, \cB_{1+}^a\, \cB_{2+}^b\, \cB_{3+}^c\, H_4
\,,\nn\\
O_{++-}^{abc}
&= \frac{1}{2}\, \cB_{1+}^a\, \cB_{2+}^b\, \cB_{3-}^c\, H_4
\,,\nn\\
O_{--+}^{abc}
&= \frac{1}{2}\, \cB_{1-}^a\, \cB_{2-}^b\, \cB_{3+}^c\, H_4
\,,\nn\\
O_{---}^{abc}
&= \frac{1}{3!}\, \cB_{1-}^a\, \cB_{2-}^b\, \cB_{3-}^c\, H_4
\,.\end{align}
For both cases the color space is one dimensional and we use the respective color structures as basis elements
\begin{equation} \label{eq:H1_color}
\vT^{ a\alpha\bbeta} = \begin{pmatrix} T^a_{\alpha\bbeta} \end{pmatrix}
\,, \qquad
\vT^{ abc} = \begin{pmatrix} \img f^{abc} \end{pmatrix}
\,.\end{equation}

In principle, there could be another independent color structure, $d^{abc}$, for $gggH$.
The $gggH$ operators transform under charge conjugation as
\begin{align}
&\C\, O_{\la_1\la_2\la_3}^{abc}(\lp_1, \lp_2, \lp_3; \lp_4)\, \vT^{ abc} \,\C
\nn\\ & \qquad
= -O_{\la_1\la_2\la_3}^{cba}(\lp_1, \lp_2, \lp_3; \lp_4)\, \vT^{ abc}
\,.\end{align}
Charge conjugation invariance of QCD thus leads to
\begin{equation} \label{eq:H1_charge}
C_{\la_1\la_2\la_3}^{abc}(\lp_1, \lp_2, \lp_3; \lp_4)
= - C_{\la_1\la_2\la_3}^{cba}(\lp_1, \lp_2, \lp_3; \lp_4)
\,,\end{equation}
which implies that the $d^{abc}$ color structure cannot arise to all orders in perturbation theory, so it suffices to consider $\img f^{abc}$ as in \eq{H1_color}. This also means that the $d^{abc}$ color structure cannot be generated by mixing under renormalization group evolution, which will be seen explicitly in \eq{mix23gluons}.

Using \eq{H1_color}, we write the QCD helicity amplitudes as
\begin{align}
\cA(g_1 g_2 g_3 H_4) &= \img\, (\img f^{a_1 a_2 a_3})\, A(1, 2, 3; 4_H)
\,,\nn\\
\cA(g_1 q_{2} \bq_{3} H_4) &= \img\, T^{a_1}_{\al_2 \balpha_3}\, A(1; 2_q, 3_\bq; 4_H)
\,.\end{align}
The Wilson coefficients for $gq\bq H$ are then given by
\begin{align} \label{eq:gqqH_coeffs}
\vC_{+(+)}(\lp_1;\lp_2,\lp_3;\lp_4) &= A_\fin(1^+; 2_q^+, 3_\bq^-; 4_H)
\,, \nn \\
\vC_{-(+)}(\lp_1;\lp_2,\lp_3;\lp_4) &= A_\fin(1^-; 2_q^+, 3_\bq^-; 4_H)
\,, \nn \\[0.5ex]
\vC_{+(-)}(\lp_1; \lp_2, \lp_3; \lp_4) &= \vC_{+(+)}(\lp_1; \lp_3, \lp_2; \lp_4)
\,, \nn \\
\vC_{-(-)}(\lp_1; \lp_2, \lp_3; \lp_4) &= \vC_{-(+)}(\lp_1; \lp_3, \lp_2; \lp_4)
\,,\end{align}
where the last two coefficients follow from charge conjugation invariance.
The Wilson coefficients for $gggH$ are given by
\begin{align} \label{eq:gggH_coeffs}
\vC_{+++}(\lp_1,\lp_2,\lp_3;\lp_4) &= A_\fin(1^+, 2^+, 3^+; 4_H)
\,,\nn\\
\vC_{++-}(\lp_1,\lp_2,\lp_3;\lp_4) &= A_\fin(1^+, 2^+, 3^-; 4_H)
\,,\nn\\[0.5ex]
\vC_{--+}(\lp_1, \lp_2, \lp_3; \lp_4)
&= \vC_{++-}(\lp_1, \lp_2, \lp_3; \lp_4) \Big|_{\langle..\rangle \leftrightarrow [..]}
\,,\nn\\
\vC_{---}(\lp_1, \lp_2, \lp_3; \lp_4)
&= \vC_{+++}(\lp_1, \lp_2, \lp_3; \lp_4) \Big|_{\langle..\rangle \leftrightarrow [..]}
\,,\end{align}
where the last two relations follow from parity invariance. As before, the subscript ``$\fin$'' in \eqs{gqqH_coeffs}{gggH_coeffs} denotes the finite part of the IR divergent amplitudes. 
The NLO helicity amplitudes were calculated in Ref.~\cite{Schmidt:1997wr}, and are given in \app{H1amplitudes}, and the NNLO helicity amplitudes were calculated in Ref.~\cite{Gehrmann:2011aa}. Both calculations were performed in the $m_t\to\infty$ limit.
At NLO, the first corrections in $m_H^2/m_t^2$ were obtained in Ref.~\cite{Neill:2009mz}.

\subsection{\boldmath $H + 2$ Jets}

For $H+2$ jets, the $q\bq\, q'\bq'H$, $q\bq\, q\bq H$, $ggq\bq H$, and $ggggH$ processes contribute, each of which we discuss in turn. Again, we consider only QCD corrections to the $ggH$ process, so $q\bar q$ pairs are described by the helicity currents $J_{ij\pm}$. The LO helicity amplitudes for $H + 2$ jets in the $m_t\to\infty$ limit were calculated in Refs.~\cite{Dawson:1991au, Kauffman:1996ix} and are collected in \app{H2amplitudes} for each channel. The LO amplitudes including the $m_t$ dependence were calculated in \cite{DelDuca:2001fn} (but explicit expressions for $ggggH$ were not given due to their length). The NLO helicity amplitudes were computed in Refs.~\cite{Berger:2006sh, Badger:2007si, Glover:2008ffa, Dixon:2009uk, Badger:2009hw, Badger:2009vh}.

\subsubsection{$q\bq\, q'\bq' H$ and $q\bq\, q\bq H$}

For the case of distinct quark flavors, $q\bq\, q'\bq'H$, the helicity basis consists of four independent operators,
\begin{align} \label{eq:qqQQH_basis}
O_{(+;+)}^{\balpha\bt\bgamma\de}
&= J_{q\, 12+}^{\balpha\bt}\, J_{q'\, 34+}^{\bgamma\de}\, H_5
\,,\nn\\
O_{(+;-)}^{\balpha\bt\bgamma\de}
&= J_{q\, 12+}^{\balpha\bt}\, J_{q'\, 34-}^{\bgamma\de}\, H_5
\,,\nn\\
O_{(-;+)}^{\balpha\bt\bgamma\de}
&= J_{q\, 12-}^{\balpha\bt}\, J_{q'\, 34+}^{\bgamma\de}\, H_5
\,,\nn\\
O_{(-;-)}^{\balpha\bt\bgamma\de}
&= J_{q\, 12-}^{\balpha\bt}\, J_{q'\, 34-}^{\bgamma\de}\, H_5
\,,\end{align}
where the additional labels on the quark currents indicate the quark flavors. For the case of identical quark flavors, $q\bq\,q\bq H$, the basis only has three independent helicity operators,
\begin{align} \label{eq:qqqqH_basis}
O_{(++)}^{\balpha\bt\bgamma\de}
&= \frac{1}{4}\, J_{12+}^{\balpha\bt}\, J_{34+}^{\bgamma\de}\, H_5
\,,\nn\\
O_{(+-)}^{\balpha\bt\bgamma\de}
&= J_{12+}^{\balpha\bt}\, J_{34-}^{\bgamma\de}\, H_5
\,,\nn\\
O_{(--)}^{\balpha\bt\bgamma\de}
&= \frac{1}{4}\, J_{12-}^{\balpha\bt}\, J_{34-}^{\bgamma\de}\, H_5
\,,\end{align}
since both quark currents have the same flavor. In both cases we use the color basis
\begin{equation} \label{eq:qqqqH_color}
\vT^{ \al\bbeta\ga\bdelta} =
2T_F\Bigl(
  \de_{\al\bdelta}\, \de_{\ga\bbeta}\,,\, \delta_{\al\bbeta}\, \de_{\ga\bdelta}
\Bigr)
\,.\end{equation}

The QCD helicity amplitudes for $q\bq\,q'\bq'H$ can be color decomposed in the basis of \eq{qqqqH_color} as
\begin{align} \label{eq:qqQQH_QCD}
 \cA(q_{1} \bq_{2} q_3' \bq_4' H_5)
&= 2 \img T_F  \Bigl[
\de_{\al_1\balpha_4} \de_{\al_3\balpha_2}  A(1_q,2_\bq;3_{q'},4_{\bq'};5_H)
\nn\\ & \quad
+ \frac{1}{N}\,\de_{\al_1 \balpha_2} \de_{\al_3 \balpha_4} B(1_q,2_\bq;3_{q'},4_{\bq'};5_H) \Bigr]
\,,\end{align}
where we have included a factor of $1/N$ for convenience. The amplitude vanishes when the quark and antiquark of the same flavor have the same helicity, in accordance with the fact that the operators of \eq{qqQQH_basis} provide a complete basis of helicity operators.
For identical quark flavors, $q\bq\,q\bq H$, the amplitudes can be obtained from the $q\bq\,q'\bq'H$ amplitudes using the relation
\begin{align}
\cA(q_{1} \bq_{2} q_3 \bq_4 H_5)= \cA(q_{1} \bq_{2} q_3' \bq_4' H_5)-\cA(q_{1} \bq_{4} q_3' \bq_2' H_5)
\,.\end{align}
The Wilson coefficients for $q\bq\,q'\bq'H$ are then given by
\begin{align} \label{eq:qqQQH_coeffs}
\vC_{(+;+)}(\lp_1,\lp_2;\lp_3,\lp_4;\lp_5)
&= \begin{pmatrix}
  A_\fin(1_q^+,2_\bq^-; 3_{q'}^+, 4_{\bq'}^-; 5_H) \\
  \tfrac{1}{N} B_\fin(1_q^+,2_\bq^-; 3_{q'}^+, 4_{\bq'}^-; 5_H)
\end{pmatrix}
,\nn\\
\vC_{(+;-)}(\lp_1,\lp_2;\lp_3,\lp_4;\lp_5)
&= \begin{pmatrix}
  A_\fin(1_q^+,2_\bq^-; 3_{q'}^-, 4_{\bq'}^+; 5_H) \\
 \tfrac{1}{N} B_\fin(1_q^+,2_\bq^-; 3_{q'}^-, 4_{\bq'}^+; 5_H)
\end{pmatrix}
,\nn\\
\vC_{(-;+)}(\lp_1,\lp_2;\lp_3,\lp_4;\lp_5) &= \vC_{(+;-)}(\lp_2,\lp_1;\lp_4,\lp_3;\lp_5)
\,,\nn\\
\vC_{(-;-)}(\lp_1,\lp_2;\lp_3,\lp_4;\lp_5) &= \vC_{(+;+)}(\lp_2,\lp_1;\lp_4,\lp_3;\lp_5)
\,,\end{align}
and for $q\bq\,q\bq H$ they are given in terms of the amplitudes $A_{\rm fin}$ and $B_{\rm fin}$ for $q\bq\,q'\bq'H$ by
\begin{align} \label{eq:qqqqH_coeffs}
\vC_{(++)}(\lp_1,\lp_2;\lp_3,\lp_4;\lp_5)
&= \begin{pmatrix}
  A_\fin(1_q^+,2_\bq^-; 3_{q}^+, 4_{\bq}^-; 5_H)
  -\frac{1}{N} B_\fin(1_q^+, 4_{\bq}^-; 3_{q}^+, 2_{\bq}^-; 5_H) \\
  \frac{1}{N} B_\fin(1_q^+,2_\bq^-; 3_{q}^+, 4_{\bq}^-; 5_H) -A_\fin(1_q^+,4_\bq^-; 3_{q}^+, 2_{\bq}^-; 5_H)
\end{pmatrix}
, \nn \\[1ex]
\vC_{(+-)}(\lp_1,\lp_2;\lp_3,\lp_4;\lp_5)
&= \begin{pmatrix}
  A_\fin(1_q^+,2_\bq^-; 3_{q}^-, 4_{\bq}^+; 5_H) \\
  \tfrac{1}{N} B_\fin(1_q^+,2_\bq^-; 3_{q}^-, 4_{\bq}^+; 5_H)
\end{pmatrix}
,\nn \\
\vC_{(--)}(\lp_1,\lp_2;\lp_3,\lp_4;\lp_5) &= \vC_{(++)}(\lp_2,\lp_1;\lp_4,\lp_3;\lp_5)
\,.\end{align}
The relations for $\vC_{(-;\pm)}$ and $\vC_{(--)}$ follow from charge conjugation invariance.
Note that there is no exchange term for $\vC_{(+-)}$, since the amplitude vanishes when the quark and antiquark of the same flavor have the same helicity (both $+$ or both $-$). Also, recall that the symmetry factors of $1/4$ in \eq{qqqqH_basis} already take care of the interchange of identical (anti)quarks, so there are no additional symmetry factors needed for $\vC_{(++)}$. Explicit expressions for the required amplitudes at tree level are given in \app{qqqqH}.

\subsubsection{$gg q\bar q H$}

For $gg q\bar q H$, the helicity basis consists of a total of six independent operators,
\begin{align} \label{eq:ggqqH_basis}
O_{++(+)}^{ab\, \balpha\beta}
&= \frac{1}{2}\, \cB_{1+}^a\, \cB_{2+}^b\, J_{34+}^{\balpha\beta}\, H_5
\,,\nn\\
O_{+-(+)}^{ab\, \balpha\beta}
&= \cB_{1+}^a\, \cB_{2-}^b\, J_{34+}^{\balpha\beta}\, H_5
\,,\nn\\
O_{--(+)}^{ab\, \balpha\beta}
&= \frac{1}{2} \cB_{1-}^a\, \cB_{2-}^b\, J_{34+}^{\balpha\beta}\, H_5
\,,\nn\\
O_{++(-)}^{ab\, \balpha\beta}
&= \frac{1}{2}\, \cB_{1+}^a\, \cB_{2+}^b\, J_{34-}^{\balpha\beta}\, H_5
\,,\nn\\
O_{+-(-)}^{ab\, \balpha\beta}
&= \cB_{1+}^a\, \cB_{2-}^b\, J_{34-}^{\balpha\beta}\, H_5
\,,\nn\\
O_{--(-)}^{ab\, \balpha\beta}
&= \frac{1}{2} \cB_{1-}^a\, \cB_{2-}^b\, J_{34-}^{\balpha\beta}\, H_5
\,.\end{align}
We use the color basis already given in \eq{ggqqcol},
\begin{equation} \label{eq:ggqqH_color}
\vT^{ ab \alpha\bbeta}
= \Bigl(
   (T^a T^b)_{\alpha\bbeta}\,,\, (T^b T^a)_{\alpha\bbeta} \,,\, \tr[T^a T^b]\, \delta_{\alpha\bbeta}
   \Bigr)
\,.\end{equation}

Using \eq{ggqqH_color}, the color decomposition of the QCD helicity amplitudes into partial amplitudes is
\begin{align} \label{eq:ggqqH_QCD}
&\cA\bigl(g_1 g_2\, q_{3} \bq_{4} H_5 \bigr)
\nn\\ & \quad
= \img \sum_{\sigma\in S_2} \bigl[T^{a_{\sigma(1)}} T^{a_{\sigma(2)}}\bigr]_{\alpha_3\balpha_4}
\,A(\sigma(1),\sigma(2); 3_q, 4_\bq; 5_H)
\nn\\ & \qquad
+ \img\, \tr[T^{a_1} T^{a_2}]\,\delta_{\alpha_3\balpha_4}\, B(1,2; 3_q, 4_\bq; 5_H)
\,.\end{align}
The $B$ amplitudes vanish at tree level. From \eq{ggqqH_QCD} we can read off the Wilson coefficients,
\begin{align} \label{eq:ggqqH_coeffs}
\vC_{+-(+)}(\lp_1,\lp_2;\lp_3,\lp_4;\lp_5) &=
\begin{pmatrix}
   A_\fin(1^+,2^-;3_q^+,4_\bq^-; 5_H) \\
   A_\fin(2^-,1^+;3_q^+,4_\bq^-; 5_H) \\
   B_\fin(1^+,2^-;3_q^+,4_\bq^-; 5_H) \\
\end{pmatrix}
,\nn\\
\vC_{++(+)}(\lp_1,\lp_2;\lp_3,\lp_4;\lp_5)
&= \begin{pmatrix}
   A_\fin(1^+,2^+;3_q^+,4_\bq^-; 5_H) \\
   A_\fin(2^+,1^+;3_q^+,4_\bq^-; 5_H) \\
   B_\fin(1^+,2^+;3_q^+,4_\bq^-; 5_H) \\
\end{pmatrix}
,\nn\\
\vC_{--(+)}(\lp_1,\lp_2;\lp_3,\lp_4;\lp_5)
&= \begin{pmatrix}
   A_\fin(1^-,2^-;3_q^+,4_\bq^-; 5_H) \\
   A_\fin(2^-,1^-;3_q^+,4_\bq^-; 5_H) \\
   B_\fin(1^-,2^-;3_q^+,4_\bq^-; 5_H) \\
\end{pmatrix}
.\end{align}
The Wilson coefficients of the last three operators in \eq{ggqqH_basis} are obtained by charge conjugation as discussed in \subsec{discrete}. Under charge conjugation, the operators transform as
\begin{align}
&\C\, O_{\la_1\la_2(\pm)}^{ab\,\balpha\beta}(\lp_1, \lp_2; \lp_3, \lp_4;\lp_5)\, \vT^{ab\,\al\bbeta}\,\C
\nn\\ & \qquad
= - O_{\la_1\la_2(\mp)}^{ba\,\balpha\beta}(\lp_1, \lp_2; \lp_4, \lp_3;\lp_5)\, \vT^{ab\,\al\bbeta}
\,,\end{align}
so charge conjugation invariance of QCD implies
\begin{align}
\vC_{\la_1\la_2(-)}(\lp_1,\lp_2;\lp_3,\lp_4;\lp_5)
&= \hV \vC_{\la_1\la_2(+)}(\lp_1,\lp_2;\lp_4,\lp_3;\lp_5)
\nn\\
\text{with}\qquad
\hV & =
\begin{pmatrix}
  0 & -1 & 0 \\
  -1 & 0 & 0 \\
  0 & 0 & -1
\end{pmatrix}
.\end{align}
Explicit expressions for the required amplitudes at tree level are given in \app{ggqqH}.

\subsubsection{$ggggH$}

For $ggggH$, the helicity basis consists of five independent operators,
\begin{align} \label{eq:ggggH_basis}
O_{++++}^{abcd} &= \frac{1}{4!}\, \cB_{1+}^a \cB_{2+}^b \cB_{3+}^c \cB_{4+}^d\, H_5
\,,\nn\\
O_{+++-}^{abcd} &= \frac{1}{3!}\, \cB_{1+}^a \cB_{2+}^b \cB_{3+}^c \cB_{4-}^d\, H_5
\,,\nn\\
O_{++--}^{abcd} &= \frac{1}{4}\, \cB_{1+}^a \cB_{2+}^b \cB_{3-}^c \cB_{4-}^d\, H_5
\,,\nn\\
O_{---+}^{abcd} &= \frac{1}{3!}\, \cB_{1-}^a \cB_{2-}^b \cB_{3-}^c \cB_{4+}^d\, H_5
\,,\nn\\
O_{----}^{abcd} &= \frac{1}{4!}\, \cB_{1-}^a \cB_{2-}^b \cB_{3-}^c \cB_{4-}^d\, H_5
\,.\end{align}
We use the basis of color structures
\begin{equation} \label{eq:ggggH_color}
\vT^{ abcd} =
\frac{1}{2\cdot 2T_F}\begin{pmatrix}
\tr[abcd] + \tr[dcba] \\ \tr[acdb] + \tr[bdca] \\ \tr[adbc] + \tr[cbda] \\
2\tr[ab]\, \tr[cd] \\ 2\tr[ac]\, \tr[db] \\ 2\tr[ad]\, \tr[bc]
\end{pmatrix}^{\!\!\!T}
,\end{equation}
where we have used the shorthand notation
\begin{equation}
\tr[ab] = \tr[T^a T^b]
\,,\qquad
\tr[abcd] = \tr[T^a T^b T^c T^d]
\,.\end{equation}
Note that the three independent color structures with a minus sign instead of the plus sign in the first three lines in \eq{ggggH_color} can be eliminated using charge conjugation invariance, see \subsec{ggggbasis}.

The color decomposition of the QCD helicity amplitudes into partial amplitudes using the color basis in \eq{ggggH_color} is
\begin{align} \label{eq:ggggH_QCD}
\cA(g_1 g_2 g_3 g_4 H_5)
&=\frac{ \img}{2T_F} \biggl[\sum_{\si \in S_4/Z_4}\! \tr[a_{\si(1)} a_{\si(2)} a_{\si(3)} a_{\si(4)}]
\nn \\ & \qquad \times
A\bigl(\si(1),\si(2),\si(3),\si(4); 5_H\bigr)
\nn\\ &\quad
+ \sum_{\si \in S_4/Z_2^3}\! \tr[a_{\si(1)} a_{\si(2)}] \tr[a_{\si(3)} a_{\si(4)}]
\nn\\ & \qquad \times
B\bigl(\si(1),\si(2),\si(3),\si(4); 5_H\bigr) \biggr]
\,,\end{align}
where the $B$ amplitudes vanish at tree level. From \eq{ggggH_QCD} we obtain the Wilson coefficients,
\begin{align} \label{eq:ggggH_coeffs}
\vC_{++--}(\lp_1,\lp_2,\lp_3,\lp_4;\lp_5)
&= \begin{pmatrix}
  2A_\fin(1^+,2^+,3^-,4^-; 5_H) \\
 2A_\fin(1^+,3^-,4^-,2^+; 5_H) \\
  2A_\fin(1^+,4^-,2^+,3^-; 5_H) \\
  B_\fin(1^+,2^+,3^-,4^-; 5_H) \\
  B_\fin(1^+,3^-,4^-,2^+; 5_H) \\
  B_\fin(1^+,4^-,2^+,3^-; 5_H)
\end{pmatrix}
\!,\nn\\
\vC_{+++-}(\lp_1,\lp_2,\lp_3,\lp_4;\lp_5)
&= \begin{pmatrix}
  2A_\fin(1^+,2^+,3^+,4^-; 5_H) \\
  2A_\fin(1^+,3^+,4^-,2^+; 5_H) \\
  2A_\fin(1^+,4^-,2^+,3^+; 5_H) \\
  B_\fin(1^+,2^+,3^+,4^-; 5_H) \\
  B_\fin(1^+,3^+,4^-,2^+; 5_H) \\
  B_\fin(1^+,4^-,2^+,3^+; 5_H) \\
\end{pmatrix}
\!,\nn\\
\vC_{++++}(\lp_1,\lp_2,\lp_3,\lp_4;\lp_5)
&= \begin{pmatrix}
2A_\fin(1^+,2^+,3^+,4^+; 5_H) \\
  2A_\fin(1^+,3^+,4^+,2^+; 5_H) \\
  2A_\fin(1^+,4^+,2^+,3^+; 5_H) \\
  B_\fin(1^+,2^+,3^+,4^+; 5_H) \\
  B_\fin(1^+,3^+,4^+,2^+; 5_H) \\
  B_\fin(1^+,4^+,2^+,3^+; 5_H) \\
\end{pmatrix}
\!,\nn\\
\vC_{---+}(\lp_1, \ldots; \lp_5)
&= \vC_{+++-}(\lp_1, \ldots; \lp_5) \Big|_{\langle..\rangle \leftrightarrow [..]}
\,,\nn\\
\vC_{----}(\lp_1, \ldots; \lp_5)
&= \vC_{++++}(\lp_1, \ldots; \lp_5) \Big|_{\langle..\rangle \leftrightarrow [..]}
\,.\end{align}
The last two coefficients follow from parity invariance. The factors of two in the first three entries of the coefficients come from combining the two traces in the first three entries in \eq{ggggH_color} using charge conjugation invariance. Because of the cyclic symmetry of the traces, the partial amplitudes are invariant under the corresponding cyclic permutations of their first four arguments, which means that most of the amplitudes in \eq{ggggH_coeffs} are not independent. Explicit expressions for the necessary amplitudes at tree level are given in \app{ggggH}.

\section{Vector Boson + Jets}
\label{sec:vec}

In this section, we give the helicity operator basis and the corresponding matching for the production of a $\gamma$, $Z$, or $W$ vector boson in association with up to two jets. The corresponding helicity amplitudes are collected in \app{Zamplitudes}.

We work at tree level in the electroweak coupling and consider only QCD corrections, so any external $q\bar q$ pairs are described by the helicity vector currents $J_{ij\pm}$ in \eq{jpm_def}. We always include the subsequent leptonic decays $\ga/Z \to \ell \bar \ell$, $W^\pm \to \nu \bar \ell/\ell \bar \nu$. In the following, for $\ga/Z$ processes, $\ell$ stands for any charged lepton or neutrino flavor, and $q$ stands for any quark flavor. For $W$ processes, we use $\ell$ to denote any charged lepton flavor and $\nu$ the corresponding neutrino flavor. Similarly, we use $u$ and $d$ to denote any up-type or down-type quark flavor (i.e. not necessarily first generation quarks only).

The operators in the helicity bases satisfy the transformation properties under C and P as discussed in \subsec{discrete}. However, the weak couplings in the amplitudes explicitly violate C and P. Therefore, to utilize the C and P transformations of the operators and minimize the number of required amplitudes and Wilson coefficients, it is useful to separate the weak couplings from the amplitudes.

We define $P_Z$ and $P_W$ as the ratios of the $Z$ and $W$ propagators to the photon propagator,
\begin{equation}
P_{Z,W}(s) = \frac{s}{s-m_{Z,W}^2 + \img \Ga_{Z,W} m_{Z,W}}
\,.\end{equation}
The left- and right-handed couplings $v_{L,R}$ of a particle to the $Z$ boson are, as usual,
\begin{align}
 v_L^i = \frac{2 T_3^i - 2Q^i \sin^2 \theta_W}{\sin(2\theta_W)}
 \,, \quad
 v_R^i = - \frac{2Q^i \sin^2 \theta_W}{\sin(2\theta_W)}
\,,\end{align}
where $T_3^i$ is the third component of weak isospin, $Q^i$ is the electromagnetic charge in units of $\abs{e}$, and $\theta_W$ is the weak mixing angle.

The $\gamma/Z$ amplitudes can then be decomposed as
\begin{align} \label{eq:Z_expand}
&\cA(\dotsb \ell \bar \ell)
\nn \\ & \quad
= e^2
\biggl\{ \bigl[Q^\ell Q^q + v_{L,R}^\ell v_{L,R}^q P_Z(s_{\ell\bar\ell})\bigr] \cA_q(\dotsb \ell \bar \ell)
\nn \\ & \qquad
+\sum_{i=1}^{n_f} \Bigl[Q^\ell Q^i + v_{L,R}^\ell\frac{v_L^i + v_R^i}{2} P_Z(s_{\ell\bar\ell}) \Bigr] \cA_v(\dotsb \ell \bar\ell)
\nn \\ & \qquad
+ \frac{v_{L,R}^\ell}{\sin(2\theta_W)} P_Z(s_{\ell\bar\ell})\, \cA_{a}(\dotsb \ell \bar\ell) \biggr\}
\,.\end{align}
Here, $\cA_q$ corresponds to the usual contribution where the vector boson couples directly to the external quark line with flavor $q$. (There is one such contribution for each external $q\bar q$ pair, and this contribution is absent for pure gluonic amplitudes like $gggZ$.) For $\cA_v$, the $\ga/Z$ couples to an internal quark loop through a vector current and the sum runs over all considered internal quark flavors. For $\cA_a$, the $Z$ boson couples to an internal quark loop through the axial-vector current. This means that when using parity and charge conjugation we have to include an additional relative minus sign for this contribution. We have also made the assumption in \eq{Z_expand} that all quarks, except for the top, are massless. Since $\cA_a$ vanishes when summed over a massless isodoublet, this has the consequence that only the $b,t$ isodoublet contributes to $\cA_a$, hence the lack of sum over flavors. We have made this simplification following the one-loop calculation of Ref.~\cite{Bern:1997sc}, which calculated the amplitude in an expansion in $1/m_t^2$, assuming all other kinematic invariants to be smaller than the top mass. From the point of view of constructing a basis these assumptions are trivial to relax.

The $W^\mp$ amplitudes can be written as
\begin{align}\label{eq:W_expand}
\cA(\dotsb \ell^- \bar \nu^+)
&= \frac{e^2 V_{ud}}{2\sin^2 \theta_W}\, P_W(s_{\ell\bar\nu})\, \cA_q(\dotsb \ell^- \bar \nu^+)
\,, \nn \\
\cA(\dotsb \nu^- \bar \ell^+)
&= \frac{e^2 V_{ud}^\dagger}{2\sin^2 \theta_W}\, P_W(s_{\nu\bar\ell})\, \cA_q(\dotsb \nu^- \bar \ell^+)
\,,\end{align}
where $V_{ud}$ is the appropriate CKM-matrix element. The $\cA_q$ amplitudes are the same in \eqs{Z_expand}{W_expand}, since all electroweak couplings have been extracted, but we have explicitly included the helicity labels (not to be mistaken as charge labels) to emphasize that these are the only possible helicities. The analogs of $\cA_v$ and $\cA_a$ do not exist for $W$ production.

We note again that \eqs{Z_expand}{W_expand} hold at tree level in the electroweak coupling, which is what we consider in this chapter. At this level, the leptons always couple to the vector boson through the currents [see \eq{Fierzetc}]
\begin{equation} \label{eq:lep_reduce}
\mae{p_\ell\pm}{\ga^\mu}{p_{\bar\ell}\pm} = \mae{p_{\bar\ell}\mp}{\ga^\mu}{p_\ell\mp}
\,.\end{equation}
This allows us to obtain the Wilson coefficients for opposite lepton helicities simply by interchanging the lepton momenta.

\subsection{\boldmath $V + 0$ Jets}
\label{subsec:V0Jets}

For $\ga/Z+0$ jets, the partonic process is $q\bar q \ell \bar \ell$, and the basis of helicity operators is
\begin{align} \label{eq:Z0_basis}
O_{(+;\pm)}^{\balpha\bt}
&= J_{q\, 12+}^{\balpha\bt}\, J_{\ell \, 34\pm}
\,,\nn\\
O_{(-;\pm)}^{\balpha\bt}
&= J_{q\, 12-}^{\balpha\bt}\, J_{\ell \, 34\pm}
\,.\end{align}
In principle, the process $gg \ell \bar \ell$ is allowed through the axial anomaly, but its contribution vanishes because in the matching calculation the gluons are taken to be on shell, and we neglect lepton masses.

For $W^\mp+0$ jets, the partonic processes are $u \bar d \ell \bar \nu$ and $d \bar u \nu \bar \ell$, respectively. Since the $W$ only couples to left-handed fields, the helicity basis simplifies to
\begin{align} \label{eq:W0_basis}
O_{(W^-)}^{\balpha\bt}
&= J_{\bar u d\, 12-}^{\balpha\bt}\, J_{\bar\ell \nu\, 34-}
\,,\nn\\
O_{(W^+)}^{\balpha\bt}
&= J_{\bar d u\, 12-}^{\balpha\bt}\, J_{\bar\nu \ell \, 34-}
\,.\end{align}
Here, we have explicitly written out the flavor structure of the currents. However, we use the shorthand subscript $(W^\mp)$ on the operators and Wilson coefficients, since we will not focus any further on the flavor structure. In an explicit calculation, one must of course sum over all relevant flavor combinations.

The unique color structure for $V+0$ jets is
\begin{equation} \label{eq:Z0_color}
\vT^{ \al\bbeta} =  \begin{pmatrix} \de_{\al\bbeta} \end{pmatrix}
,\end{equation}
and extracting it from the amplitudes, we have
\begin{align}
\cA_{q,v,a}(q_{1}\bar q_{2} \ell_{3} \bar \ell_{4})
= \img\, \delta_{\alpha_1 \balpha_2}  \, A_{q,v,a}(1_q,2_{\bar q}; 3_\ell, 4_{\bar \ell})
\,. \end{align}
Here, $A_v$ and $A_a$ first appear at two loops. In addition, $A_a$ is proportional to the top and bottom mass splitting due to isodoublet cancellations. It drops out when both top and bottom are treated as massless (e.g., when the matching scale is much larger than the top mass).

We use the same electroweak decomposition as in \eqs{Z_expand}{W_expand} to write the Wilson coefficients. For $\gamma/Z+0$ jets, we have
\begin{align} \label{eq:Z0_expand_Wilson}
&\vec C_{(\lambda_q;\lambda_\ell)}(\lp_1, \lp_2; \lp_3, \lp_4)
\nn \\ & \quad
= e^2  \,
\biggl\{ \bigl[Q^\ell Q^q + v_{\lambda_\ell}^\ell v_{\lambda_q}^q P_Z(s_{34})\bigr]
\vec C_{q(\lambda_q;\lambda_\ell)}(\dots)
\nn \\ & \qquad
+ \sum_{i=1}^{n_f} \Bigl[Q^\ell Q^i + v_{\lambda_\ell}^\ell\frac{v_L^i + v_R^i}{2} P_Z(s_{34}) \Bigr]
\vec C_{v(\lambda_q;\lambda_\ell)}(\dots)
\nn \\ & \qquad
+ \frac{v_{\lambda_\ell}^\ell}{\sin(2\theta_W)} P_Z(s_{34})\, \vec C_{a(\lambda_q;\lambda_\ell)}(\dots)  \biggr\}
\,,\end{align}
where the weak couplings are determined by the helicity labels of the quark and lepton currents,
\begin{align}
v_{+}^\ell = v_R^\ell
\,,\quad
v_{-}^\ell = v_L^\ell
\,,\qquad
v_{+}^q = v_R^q
\,,\quad
v_{-}^q = v_L^q
\,.\end{align}
For $W + 0$ jets, we simply have
\begin{align}
\vC_{(W^-)}(\lp_1, \lp_2; \lp_3, \lp_4)
&= \frac{e^2 V_{ud}}{2\sin^2 \theta_W}\, P_W(s_{34})\, \vC_{q (-;-)}(\ldots)
\,, \nn \\
\vC_{(W^+)}(\lp_1, \lp_2; \lp_3, \lp_4)
&= \frac{e^2 V_{ud}^{\dagger}}{2\sin^2 \theta_W}\, P_W(s_{34})\, \vC_{q (-;-)}(\ldots)
\,.\end{align}
In all cases, the momentum arguments on the right-hand side are the same as on the left-hand side.
Note that the $\vC_{q(-;-)}$ coefficient is the same in all cases. The Wilson coefficients are given by
\begin{align} \label{eq:qqV_coeffs}
\vC_{x(+;+)}(\lp_1, \lp_2; \lp_3, \lp_4) &= A_{x,\fin}(1_q^+,2_{\bar q}^-;3_\ell^+,4_{\bar \ell}^-)
\,, \nn \\
\vC_{x(+;-)}(\lp_1, \lp_2; \lp_3, \lp_4) &= \vC_{x(+;+)}(\lp_1, \lp_2; \lp_4, \lp_3)
\,, \nn \\
\vC_{q,v(-;\pm)}(\lp_1, \lp_2; \lp_3, \lp_4) &= \vC_{q,v(+;\pm)}(\lp_2, \lp_1; \lp_3, \lp_4)
\,, \nn \\
\vC_{a(-;\pm)}(\lp_1, \lp_2; \lp_3, \lp_4) &= -\vC_{a(+;\pm)}(\lp_2, \lp_1; \lp_3, \lp_4)
\,, \end{align}
where $x = q,v,a$ and as discussed in \sec{matching} the subscript ``$\fin$'' denotes the IR-finite part of the helicity amplitudes.
The second relation follows from \eq{lep_reduce}. The last two relations follow from charge conjugation invariance.
At tree level and one loop only $\vC_q$ receives a nonvanishing contribution. The $A_q$ amplitude is given in \app{V0amplitudes}.

\subsection{\boldmath $V + 1$ Jet}
\label{subsec:V1J}

\subsubsection{$gq\bar qV$}

For $\ga/Z+1$ jet, the partonic process is $g q\bar q \ell \bar\ell$, and the basis of helicity operators is
\begin{align} \label{eq:Z1_basis}
O_{+(+;\pm)}^{a\,\balpha\bt}
&= \cB_{1+}^a\, J_{q\, 23+}^{\balpha\bt}\, J_{\ell \, 45\pm}
\,,\nn\\
O_{+(-;\pm)}^{a\,\balpha\bt}
&= \cB_{1+}^a\, J_{q\, 23-}^{\balpha\bt}\, J_{\ell \, 45\pm}
\,,\nn\\
O_{-(+;\pm)}^{a\,\balpha\bt}
&= \cB_{1-}^a\, J_{q\, 23+}^{\balpha\bt}\, J_{\ell \, 45\pm}
\,,\nn\\
O_{-(-;\pm)}^{a\,\balpha\bt}
&= \cB_{1-}^a\, J_{q\, 23-}^{\balpha\bt}\, J_{\ell \, 45\pm}
\,.\end{align}

For $W^\mp+1$ jet, the partonic processes are $gu \bar d \ell \bar \nu$ and $gd \bar u \nu \bar \ell$, respectively, and the helicity operator basis is
\begin{align} \label{eq:W1_basis}
O_{\pm\, (W^-)}^{a\,\balpha\bt}
&= \cB_{1\pm}^a\, J_{\bar u d\, 23-}^{\balpha\bt}\, J_{\bar\ell \nu\, 45-}
\,,\nn\\
O_{\pm\, (W^+)}^{a\,\balpha\bt}
&= \cB_{1\pm}^a\, J_{\bar d u\, 23-}^{\balpha\bt}\, J_{\bar\nu \ell \, 45-}
\,.\end{align}
The unique color structure for $gq\bar qV$ is
\begin{equation} \label{eq:Z1q_color}
\vT^{ a\, \al\bbeta} = \begin{pmatrix} T^a_{\al\bbeta} \end{pmatrix}
,\end{equation}
and extracting it from each of the amplitudes, we have
\begin{align}
\cA_{x}(g_1 q_{2}\bar q_{3} \ell_{4} \bar \ell_{5})
&=\img\, T^{a_1}_{\alpha_2 \balpha_3} \, A_{x}(1;2_q,3_{\bar q};4_\ell,5_{\bar \ell})
\,,\end{align}
where the subscript $x$ stands for one of $q,v,a$.

As for $V+0$ jets, we write the Wilson coefficients using the electroweak decomposition in \eqs{Z_expand}{W_expand}. For $\gamma/Z+1$ jet, we have
\begin{align}
&\vec C_{\lambda(\lambda_q;\lambda_\ell)}(\lp_1; \lp_2, \lp_3; \lp_4, \lp_5)
\nn \\* & \quad
= e^2  \,
\biggl\{ \bigl[Q^\ell Q^q + v_{\lambda_\ell}^\ell v_{\lambda_q}^q P_Z(s_{45})\bigr]
\vec C_{q\lambda(\lambda_q;\lambda_\ell)}(\ldots)
\nn \\ & \qquad
+ \sum_{i=1}^{n_f} \Bigl[Q^\ell Q^i + v_{\lambda_\ell}^\ell\frac{v_L^i + v_R^i}{2} P_Z(s_{45}) \Bigr]
\vec C_{v\lambda(\lambda_q;\lambda_\ell)}(\ldots)
\nn \\ & \qquad
+ \frac{v_{\lambda_\ell}^\ell}{\sin(2\theta_W)} P_Z(s_{45})\, \vec C_{a\lambda(\lambda_q;\lambda_\ell)}(\ldots) \biggr\}
\,,\end{align}
where the weak couplings are determined by the helicity labels of the quark and lepton currents,
\begin{align}
v_{+}^\ell = v_R^\ell
\,,\quad
v_{-}^\ell = v_L^\ell
\,,\qquad
v_{+}^q = v_R^q
\,,\quad
v_{-}^q = v_L^q
\,.\end{align}
For $W + 1$ jet, we have
\begin{align}
\vC_{\lambda(W^\mp)}(\ldots)
&= \frac{e^2 V_{ud}^{(\dagger)}}{2\sin^2 \theta_W}\, P_W(s_{45})\, \vC_{q \lambda(-;-)}(\ldots)
\,.\end{align}
The Wilson coefficients are given by 
\begin{align} \label{eq:gqqV_coeffs}
\vC_{x+(+;+)}(\lp_1; \lp_2, \lp_3; \lp_4, \lp_5)
&= A_{x,\fin} (1^+;2_q^+,3_{\bar q}^-; 4_\ell^+,5_{\bar \ell}^-)
\,, \nn \\
\vC_{x\lambda(+;-)}(\lp_1; \lp_2, \lp_3; \lp_4, \lp_5)
&= \vC_{x\lambda(+;+)}(\lp_1; \lp_2, \lp_3; \lp_5, \lp_4)
\,, \nn \\
\vC_{q,v\lambda(-;\pm)}(\lp_1; \lp_2, \lp_3; \lp_4, \lp_5)
&= -\vC_{q,v\lambda(+;\pm)}(\lp_1; \lp_3, \lp_2; \lp_4, \lp_5)
\,, \nn \\
\vC_{a\lambda(-;\pm)}(\lp_1; \lp_2, \lp_3; \lp_4, \lp_5)
&= \vC_{a\lambda(+;\pm)}(\lp_1; \lp_3, \lp_2; \lp_4, \lp_5)
\,.\end{align}
The second relation follows from \eq{lep_reduce}, and the last two relations follow from charge conjugation invariance. The Wilson coefficients with a negative helicity gluon follow from parity invariance,
\begin{align}
&\vC_{q,v-(+;\pm)}(\lp_1; \lp_2, \lp_3; \lp_4, \lp_5)
\nn \\ & \qquad
= \vC_{q,v+(-;\mp)}(\lp_1; \lp_2, \lp_3; \lp_4, \lp_5)\Big|_{\langle..\rangle \leftrightarrow [..]}
\,, \nn \\
&\vC_{a-(+;\pm)}(\lp_1; \lp_2, \lp_3; \lp_4, \lp_5)
\nn \\ & \qquad
= - \vC_{a+(-;\mp)}(\lp_1; \lp_2, \lp_3; \lp_4, \lp_5)\Big|_{\langle..\rangle \leftrightarrow [..]}
\,.\end{align}
The helicity amplitudes for $g q \bar q \ell \bar \ell$ were calculated in Ref.~\cite{Giele:1991vf, Arnold:1988dp, Korner:1990sj}. We provide the tree-level and one-loop results in \app{V1amplitudes}. The two-loop amplitudes were computed in Refs.~\cite{Garland:2001tf, Garland:2002ak}.

\subsubsection{$gggV$}

The partonic process $ggg\ell\bar\ell$ first appears at one loop, and thus contributes only at relative $\ord{\alpha_s^2}$ to $\ga/Z+1$ jet. Nevertheless, for the sake of completeness (and curiosity) we briefly discuss it here. The helicity operator basis is
\begin{align}
O_{+++(\pm)}^{abc}
&= \frac{1}{3!}\, \cB_{1+}^a\, \cB_{2+}^b\, \cB_{3+}^c\, J_{\ell \, 45\pm}
\,,\nn\\
O_{++-(\pm)}^{abc}
&= \frac{1}{2}\, \cB_{1+}^a\, \cB_{2+}^b\, \cB_{3-}^c\, J_{\ell \, 45\pm}
\,,\nn\\
O_{+--(\pm)}^{abc}
&= \frac{1}{2}\, \cB_{1+}^a\, \cB_{2-}^b\, \cB_{3-}^c\, J_{\ell \, 45\pm}
\,,\nn\\
O_{---(\pm)}^{abc}
&= \frac{1}{3!}\, \cB_{1-}^a\, \cB_{2-}^b\, \cB_{3-}^c\, J_{\ell \, 45\pm}
\,.\end{align}

The color space is two dimensional. We use the basis
\begin{equation} \label{eq:Z1g_color}
\vT^{abc} =
\Bigl(  \img f^{abc}\,,\, d^{abc} \Bigr)
\,,\end{equation}
in terms of which we can write the $ggg \ell \bar \ell$ amplitudes as
\begin{align}\label{eq:Zggg_amp_color}
\cA_v(g_1 g_2  g_3  \ell_4 \bar \ell_5)
&= \img\, d^{a_1 a_2 a_3} A_v(1,2,3;4_\ell,5_{\bar \ell})
\,, \nn \\
\cA_a(g_1 g_2  g_3  \ell_4 \bar \ell_5)
&= \img\, (\img f^{a_1 a_2 a_3}) A_{a}(1,2,3;4_\ell,5_{\bar \ell})
\,.\end{align}
We will justify shortly that to all orders, only a single color structure appears for each of $\cA_v$, $\cA_a$.
This process can only occur via a closed quark loop, so there is no $\cA_q$ contribution.
The $gggV$ operators transform under charge conjugation as
\begin{align}
&\C\, O_{\la_1\la_2\la_3(\pm)}^{abc}(\lp_1, \lp_2, \lp_3; \lp_4, \lp_5)\, \vT^{ abc} \,\C
\nn\\ & \qquad
= O_{\la_1\la_2\la_3(\mp)}^{cba}(\lp_1, \lp_2, \lp_3; \lp_5, \lp_4)\, \vT^{ abc}
\,.\end{align}
Charge conjugation invariance of QCD thus leads to
\begin{align}
&C_{v\la_1\la_2\la_3(\pm)}^{abc}(\lp_1, \lp_2, \lp_3; \lp_4, \lp_5)
\nn \\ & \quad
= C_{v\la_1\la_2\la_3(\mp)}^{cba}(\lp_1, \lp_2, \lp_3; \lp_5, \lp_4)
\nn \\ & \quad
= C_{v\la_1\la_2\la_3(\pm)}^{cba}(\lp_1, \lp_2, \lp_3; \lp_4, \lp_5)
\,,\end{align}
where we used \eq{lep_reduce} in the last line. This implies that to all orders in the strong coupling, only the fully symmetric color structure $d^{abc}$ can contribute to $\cA_v$ and $\vC_v$. For $\vC_a$ the same relation holds but with an additional minus sign on the right-hand side due to the weak axial-vector coupling in $\cA_a$. This implies that for $\cA_a$ and $\vC_a$ only the fully antisymmetric color structure $\img f^{abc}$ contributes, as given in \eq{Zggg_amp_color}.

We decompose the $ggg\ell \bar \ell$ Wilson coefficients as
\begin{align}
&\vC_{\lambda_1\lambda_2\lambda_3(\lambda_\ell)}
\nn \\ & \quad
= e^2  \,
\biggl\{
\sum_{i=1}^{n_f} \Bigl[Q^\ell Q^i + v_{\lambda_\ell}^\ell\frac{v_L^i + v_R^i}{2} P_Z(s_{45}) \Bigr]
\vC_{v\lambda_1\lambda_2\lambda_3(\lambda_\ell)}
\nn \\ & \qquad
+ \frac{v_{\lambda_\ell}^\ell}{\sin(2\theta_W)} P_Z(s_{45})\, \vC_{a\lambda_1\lambda_2\lambda_3(\lambda_\ell)}
\biggr\}
\,,\end{align}
where
\begin{align}
v_{+}^\ell = v_R^\ell
\,,\qquad
v_{-}^\ell = v_L^\ell
\,,\end{align}
and we have 
\begin{align}
&\vC_{v\lambda_1\lambda_2\lambda_3(+)}(\lp_1, \lp_2, \lp_3; \lp_4, \lp_5)
\nn \\ & \qquad
= \begin{pmatrix} 0 \\ A_{v, \fin} (1^{\lambda_1},2^{\lambda_2}, 3^{\lambda_3}; 4_\ell^+,5_{\bar \ell}^-)    \end{pmatrix}
, \nn \\
&\vC_{a\lambda_1\lambda_2\lambda_3(+)}(\lp_1, \lp_2, \lp_3; \lp_4, \lp_5)
\nn \\ & \qquad
= \begin{pmatrix} A_{a, \fin}(1^{\lambda_1},2^{\lambda_2}, 3^{\lambda_3}; 4_\ell^+,5_{\bar \ell}^-) \\ 0   \end{pmatrix}
\,, \nn \\
&\vC_{v,a\lambda_1\lambda_2\lambda_3(-)}(\lp_1, \lp_2, \lp_3; \lp_4, \lp_5)
\nn \\ & \qquad
= \vC_{v,a\lambda_1\lambda_2\lambda_3(+)}(\lp_1, \lp_2, \lp_3; \lp_5, \lp_4)
\,.\end{align}
For brevity, we have not written out the various gluon helicity combinations.
The one-loop amplitudes for $gggZ$ were calculated in Ref.~\cite{vanderBij:1988ac}, and the two-loop amplitudes were computed in Ref.~\cite{Gehrmann:2013vga}. Since their contribution is very small we do not repeat them here.

\subsection{\boldmath $V + 2$ Jets}
\label{subsec:V2J}

Here we consider the processes $q'\bq' q\bq\, V$, $q\bq\,q\bq\, V$, and $gg\,q\bq\, V$. The $ggggV$ process is allowed as well, but only arises at one loop, so we do not explicitly consider here. It can be treated similarly to $gggV$, but using the $gggg$ color basis analogous to that for $ggggH$ given in \eq{ggggH_color}.

The NLO helicity amplitudes for $V+2$ jets were calculated in Refs.~\cite{Bern:1996ka, Bern:1997sc} assuming that all kinematic scales are smaller than the top mass $m_t$ and including the $1/m_t^2$ corrections. We give the full expressions for the LO results in \app{V2amplitudes}. Since the NLO results are rather long, we do not repeat them, but we show how to convert the results of Refs.~\cite{Bern:1996ka, Bern:1997sc} to our notation.

\subsubsection{$q'\bq' q\bq\, V$ and $q\bq\, q\bq\, V$}

For $q' \bq' q\bq\, \ell \bar \ell$, the helicity operator basis is
\begin{align} \label{eq:Z2_basis_qQ}
O_{(+;+;\pm)}^{\balpha\bt\bgamma\delta}
&= J_{q'\, 12+}^{\balpha\bt}\, J_{q\, 34+}^{\bgamma\delta}\, J_{\ell \, 56\pm}
\,,\nn\\
O_{(+;-;\pm)}^{\balpha\bt\bgamma\delta}
&= J_{q'\, 12+}^{\balpha\bt}\, J_{q\, 34-}^{\bgamma\delta}\, J_{\ell \, 56\pm}
\,,\nn\\
O_{(-;+;\pm)}^{\balpha\bt\bgamma\delta}
&= J_{q'\, 12-}^{\balpha\bt}\, J_{q\, 34+}^{\bgamma\delta}\, J_{\ell \, 56\pm}
\,,\nn\\
O_{(-;-;\pm)}^{\balpha\bt\bgamma\delta}
&= J_{q'\, 12-}^{\balpha\bt}\, J_{q\, 34-}^{\bgamma\delta}\, J_{\ell \, 56\pm}
\,.\end{align}
For identical quark flavors, $q\bq\, q\bq\,\ell \bar \ell$, the basis reduces to
\begin{align}\label{eq:Z2_basis_qq}
O_{(++;\pm)}^{\balpha\bt\bgamma\delta}
&= \frac{1}{4}J_{q\, 12+}^{\balpha\bt}\, J_{q\, 34+}^{\bgamma\delta}\, J_{\ell \, 56\pm}
\,,\nn\\
O_{(+-;\pm)}^{\balpha\bt\bgamma\delta}
&= J_{q\, 12+}^{\balpha\bt}\, J_{q\, 34-}^{\bgamma\delta}\, J_{\ell \, 56\pm}
\,,\nn\\
O_{(--;\pm)}^{\balpha\bt\bgamma\delta}
&= \frac{1}{4}J_{q\, 12-}^{\balpha\bt}\, J_{q\, 34-}^{\bgamma\delta}\, J_{\ell \, 56\pm}
\,.\end{align}
For $W+2$ jets, the corresponding partonic processes are $q \bq\, u \bar d\, \ell \bar \nu$ and $q \bq\, d \bar u\, \nu \bar \ell$,
and the helicity operator basis is
\begin{align} \label{eq:W2_basis_qq}
O_{(\pm; W^-)}^{\balpha\bt\bgamma\delta}
&= J_{q\, 12\pm}^{\balpha\bt}\, J_{\bar u d\, 34-}^{\bgamma\delta}\, J_{\bar\ell \nu\, 56-}
\,,\nn\\
O_{(\pm; W^+)}^{\balpha\bt\bgamma\delta}
&= J_{q\, 12\pm}^{\balpha\bt}\, J_{\bar d u\, 34-}^{\bgamma\delta}\, J_{\bar\nu \ell \, 56-}
\,.\end{align}
We use the color basis
\begin{equation} \label{eq:qqqqV_color}
\vT^{ \al\bbeta\ga\bdelta} =
 2T_F \Bigl(
  \de_{\al\bdelta}\, \de_{\ga\bbeta}\,,\, \delta_{\al\bbeta}\, \de_{\ga\bdelta}
\Bigr)
\,.\end{equation}
For distinct quark flavors, the color decomposition of the amplitudes in this basis is 
\begin{align}\label{eq:qqQQV2Jets}
&\cA_x(q_1' \bq_2' q_3 \bq_4 \ell_5 \bar \ell_6)
\\* & \quad
=2T_F \img\, \de_{\al_1\balpha_4} \de_{\al_3\balpha_2}\, A_x(1_{q'},2_{\bq'};3_q,4_\bq;5_\ell,6_{\bar \ell})
\nn \\ & \qquad
+2T_F \img\,\de_{\al_1 \balpha_2} \de_{\al_3 \balpha_4}\, \frac{1}{N}\, B_x(1_{q'},2_{\bq'};3_q,4_{\bar q};5_\ell,6_{\bar \ell})
\nn\,.\end{align}
For identical quark flavors the amplitudes can be obtained from the distinct flavor amplitudes using
\begin{align}
\cA_x(q_1 \bq_2 q_3 \bq_4 \ell_5 \bar \ell_6)
&= \cA_x(q_1' \bq_2' q_3 \bq_4 \ell_5 \bar \ell_6)
\nn \\ & \quad
- \cA_x(q_1' \bq_4' q_3 \bq_2 \ell_5 \bar \ell_6)
\,,\end{align}
where it is to be understood that the electroweak couplings of $q'$ must also be replaced by those of $q$.

Writing the Wilson coefficients in the decomposition in \eqs{Z_expand}{W_expand}, we have for the $q' \bq' q\bq\, \ell \bar \ell$ channel
\begin{align}
&\vC_{(\lambda_{q'};\lambda_q;\lambda_\ell)}(\lp_1, \lp_2; \lp_3, \lp_4; \lp_5, \lp_6)
\nn \\ & \quad
= e^2  \,
\biggl\{ \bigl[Q^\ell Q^q + v_{\lambda_\ell}^\ell v_{\lambda_q}^q P_Z(s_{56})\bigr]
\nn \\ & \qquad\qquad\times
\vC_{q(\lambda_{q'};\lambda_q;\lambda_\ell)}(\lp_1, \lp_2; \lp_3, \lp_4; \lp_5, \lp_6)
\nn \\ & \qquad
+ \bigl[Q^\ell Q^{q'} + v_{\lambda_\ell}^\ell v_{\lambda_{q'}}^{q'} P_Z(s_{56})\bigr]
\nn \\ & \qquad\quad\times
\vC_{q(\lambda_q;\lambda_{q'};\lambda_\ell)}(\lp_3, \lp_4; \lp_1, \lp_2; \lp_5, \lp_6)
\nn \\ & \qquad
+ \sum_{i=1}^{n_f} \Bigl[Q^\ell Q^i + v_{\lambda_\ell}^\ell\frac{v_L^i + v_R^i}{2} P_Z(s_{56}) \Bigr]
\vC_{v(\lambda_{q'};\lambda_q;\lambda_\ell)}(\ldots)
\nn \\ & \qquad
+ \frac{v_{\lambda_\ell}^\ell}{\sin(2\theta_W)} P_Z(s_{56})\, \vC_{a(\lambda_{q'};\lambda_q;\lambda_\ell)}(\ldots)
\biggr\}
\,,\end{align}
with the weak couplings
\begin{align}
v_{+}^\ell = v_R^\ell
\,,\quad
v_{-}^\ell = v_L^\ell
\,,\qquad
v_{+}^q = v_R^q
\,,\quad
v_{-}^q = v_L^q
\,.\end{align}
The same decomposition is used for the case of identical flavors, $q\bq\,q\bq\,\ell\bar\ell$. 
For the $W^\mp$ channels, $q \bq\, u \bar d\, \ell \bar \nu$ and $q \bq\, d \bar u\, \nu \bar \ell$, we have
\begin{align} \label{eq:W2_expand}
\vC_{(\lambda_q;W^\mp)}(\dots)
&= \frac{e^2 V_{ud}^{(\dagger)}}{2\sin^2 \theta_W}\, P_W(s_{56})\, \vC_{q (\lambda_q;-;-)}(\ldots)
\,.\end{align}

The coefficients for $q'\bq' q\bq\,V$ are given by
\begin{align} \label{eq:qqQQV_coeffs}
\vC_{x(+;+;+)}(\lp_1,\lp_2;\lp_3,\lp_4;\lp_5, \lp_6)
&= \begin{pmatrix}
  A_{x,\fin}(1_{q'}^+, 2_{\bq'}^-; 3_q^+, 4_\bq^-; 5_\ell^+, 6_{\bar \ell}^-) \\
  \tfrac{1}{N} B_{x,\fin}(1_{q'}^+, 2_{\bq'}^-; 3_q^+, 4_\bq^-; 5_\ell^+, 6_{\bar \ell}^-)
\end{pmatrix}
,\nn\\[1ex]
\vC_{x(+;-;+)}(\lp_1,\lp_2;\lp_3,\lp_4;\lp_5, \lp_6)
&= \begin{pmatrix}
  A_{x,\fin}(1_{q'}^+, 2_{\bq'}^-; 3_q^-, 4_\bq^+; 5_\ell^+, 6_{\bar \ell}^-) \\
   \tfrac{1}{N} B_{x,\fin}(1_{q'}^+, 2_{\bq'}^-; 3_q^-, 4_\bq^+; 5_\ell^+, 6_{\bar \ell}^-)
\end{pmatrix}
,\nn\\
\vC_{x(+;\pm;-)}(\lp_1,\lp_2;\lp_3,\lp_4;\lp_5,\lp_6) &= \vC_{x(+;\pm;+)}(\lp_1,\lp_2;\lp_3,\lp_4;\lp_6,\lp_5)
\,,\nn\\
\vC_{q,v(-;+;\pm)}(\lp_1,\lp_2;\lp_3,\lp_4;\lp_5,\lp_6) &= -\vC_{q,v(+;-;\pm)}(\lp_2,\lp_1;\lp_4,\lp_3;\lp_5,\lp_6)
\,,\nn\\
\vC_{a(-;+;\pm)}(\lp_1,\lp_2;\lp_3,\lp_4;\lp_5,\lp_6) &= \vC_{a(+;-;\pm)}(\lp_2,\lp_1;\lp_4,\lp_3;\lp_5,\lp_6)
\,,\nn\\
\vC_{q,v(-;-;\pm)}(\lp_1,\lp_2;\lp_3,\lp_4;\lp_5,\lp_6) &= -\vC_{q,v(+;+;\pm)}(\lp_2,\lp_1;\lp_4,\lp_3;\lp_5,\lp_6)
\,,\nn\\
\vC_{a(-;-;\pm)}(\lp_1,\lp_2;\lp_3,\lp_4;\lp_5,\lp_6) &= \vC_{a(+;+;\pm)}(\lp_2,\lp_1;\lp_4,\lp_3;\lp_5,\lp_6)
\,,\end{align}
and for $q\bq\,q\bq\, V$ they are given in terms of the amplitudes $A_{x,{\rm fin}}$ and $B_{x,{\rm fin}}$ for $q'\bq' q\bq\,V$ by
\begin{align} \label{eq:qqqqV_coeffs}
\vC_{x(++;+)}(\lp_1,\lp_2;\lp_3,\lp_4;\lp_5,\lp_6)
&= \begin{pmatrix}
  A_{x,\fin}(1_{q}^+,2_{\bq}^-; 3_q^+, 4_\bq^-; 5_\ell^+, 6_{\bar \ell}^-)
  -\frac{1}{N} B_{x,\fin}(1_{q}^+, 4_{\bq}^-; 3_q^+,2_\bq^-  ; 5_\ell^+, 6_{\bar \ell}^-) \\
  \frac{1}{N} B_{x,\fin}(1_{q}^+, 2_{\bq}^-; 3_q^+, 4_\bq^-; 5_\ell^+, 6_{\bar \ell}^-)
  - A_{x,\fin}(1_{q}^+, 4_{\bq}^-; 3_q^+,2_\bq^-  ; 5_\ell^+, 6_{\bar \ell}^-)
\end{pmatrix}
, \nn \\[1ex]
\vC_{x(+-;+)}(\lp_1,\lp_2;\lp_3,\lp_4;\lp_5,\lp_6)
&= \begin{pmatrix}
  A_{x,\fin}(1_{q}^+,2_{\bq}^-; 3_q^-, 4_\bq^+; 5_\ell^+, 6_{\bar \ell}^-) \\
  \tfrac{1}{N} B_{x,\fin}(1_{q}^+,2_{\bq}^-; 3_q^-, 4_\bq^+; 5_\ell^+, 6_{\bar \ell}^-)
\end{pmatrix}
,\nn \\
\vC_{x(+\pm;-)}(\lp_1,\lp_2;\lp_3,\lp_4;\lp_5,\lp_6) &= \vC_{x(+\pm;+)}(\lp_1,\lp_2;\lp_3,\lp_4;\lp_6,\lp_5)
\,, \nn \\
\vC_{q,v(--;\pm)}(\lp_1,\lp_2;\lp_3,\lp_4;\lp_5,\lp_6) &= -\vC_{q,v(++;\pm)}(\lp_2,\lp_1;\lp_4,\lp_3;\lp_5,\lp_6)
\,, \nn \\
\vC_{a(--;\pm)}(\lp_1,\lp_2;\lp_3,\lp_4;\lp_5,\lp_6) &= \vC_{a(++;\pm)}(\lp_2,\lp_1;\lp_4,\lp_3;\lp_5,\lp_6)
\,.\end{align}
The various relations for the coefficients with flipped helicities follow from \eq{lep_reduce} and charge conjugation invariance. The tree-level helicity amplitudes are given in \app{V2qqqq}.

\subsubsection{$gg\, q\bq\, V$}

For $gg\, q\bq\, \ell \bar \ell$, the helicity operator basis consists of $12$ independent operators,
\begin{align}\label{eq:Z2_basis_g}
O_{++(+;\pm)}^{ab\, \balpha\bt}
&= \frac{1}{2}\, \cB_{1+}^a\, \cB_{2+}^b\, J_{q\, 34+}^{\balpha\bt}\, J_{\ell \, 56\pm}
\,,\nn\\
O_{++(-;\pm)}^{ab\, \balpha\bt}
&= \frac{1}{2}\, \cB_{1+}^a\, \cB_{2+}^b\, J_{q\, 34-}^{\balpha\bt}\, J_{\ell \, 56\pm}
\,,\nn\\
O_{+-(+;\pm)}^{ab\, \balpha\bt}
&= \cB_{1+}^a\, \cB_{2-}^b\, J_{q\, 34+}^{\balpha\bt}\, J_{\ell \, 56\pm}
\,,\nn\\
O_{+-(-;\pm)}^{ab\, \balpha\bt}
&= \cB_{1+}^a\, \cB_{2-}^b\, J_{q\, 34-}^{\balpha\bt}\, J_{\ell \, 56\pm}
\,,\nn\\
O_{--(+;\pm)}^{ab\, \balpha\bt}
&= \frac{1}{2}\, \cB_{1-}^a\, \cB_{2-}^b\, J_{q\, 34+}^{\balpha\bt}\, J_{\ell \, 56\pm}
\,,\nn\\
O_{--(-;\pm)}^{ab\, \balpha\bt}
&= \frac{1}{2}\, \cB_{1-}^a\, \cB_{2-}^b\, J_{q\, 34-}^{\balpha\bt}\, J_{\ell \, 56\pm}
\,.\end{align}
For $W^\mp$, the corresponding partonic processes are $gg\, u \bar d\, \ell \bar \nu$ and $gg\, d \bar u\, \nu \bar \ell$,
and the helicity operator basis reduces to six independent operators,
\begin{align} \label{eq:W2_basis_g}
O_{++\, (W^-)}^{ab\,\balpha\bt}
&= \frac{1}{2}\, \cB_{1+}^a\, \cB_{2+}^b\, J_{\bar u d\, 34-}^{\balpha\bt}\, J_{\bar\ell \nu\, 56-}
\,,\nn\\
O_{+-\, (W^-)}^{ab\,\balpha\bt}
&= \cB_{1+}^a\, \cB_{2-}^b\, J_{\bar u d\, 34-}^{\balpha\bt}\, J_{\bar\ell \nu\, 56-}
\,,\nn\\
O_{--\, (W^-)}^{ab\,\balpha\bt}
&= \frac{1}{2}\, \cB_{1-}^a\, \cB_{2-}^b\, J_{\bar u d\, 34-}^{\balpha\bt}\, J_{\bar\ell \nu\, 56-}
\,,\nn\\
O_{++\, (W^+)}^{ab\,\balpha\bt}
&= \frac{1}{2}\, \cB_{1+}^a\, \cB_{2+}^b\, J_{\bar d u\, 34-}^{\balpha\bt}\, J_{\bar\nu \ell \, 56-}
\,,\nn\\
O_{+-\, (W^+)}^{ab\,\balpha\bt}
&= \cB_{1+}^a\, \cB_{2-}^b\, J_{\bar d u\, 34-}^{\balpha\bt}\, J_{\bar\nu \ell \, 56-}
\,,\nn\\
O_{--\, (W^+)}^{ab\,\balpha\bt}
&= \frac{1}{2}\, \cB_{1-}^a\, \cB_{2-}^b\, J_{\bar d u\, 34-}^{\balpha\bt}\, J_{\bar\nu \ell \, 56-}
\,.\end{align}

We use the color basis
\begin{equation} \label{eq:ggqqV_color}
\vT^{ ab \alpha\bbeta}
= \Bigl(
   (T^a T^b)_{\alpha\bbeta}\,,\, (T^b T^a)_{\alpha\bbeta} \,,\, \tr[T^a T^b]\, \delta_{\alpha\bbeta}
   \Bigr)
\,,\end{equation}
and the amplitudes are color-decomposed as
\begin{align}
&\cA_x(g_1g_2 q_{3}\bar q_{4} \ell_{5} \bar \ell_{6})
\nn \\ & \quad
= \img \sum_{\sigma\in S_2} \bigl[T^{a_{\sigma(1)}} T^{a_{\sigma(2)}}\bigr]_{\alpha_3\balpha_4}
A_x(\sigma(1),\sigma(2);3_q,4_{\bar q};5_\ell,6_{\bar \ell})
\nn \\ & \qquad
+ \img\, \tr[T^{a_1} T^{a_2}]\,\delta_{\alpha_3\balpha_4}\, B_x(1,2;3_q,4_{\bar q};5_\ell,6_{\bar \ell})
\,.\end{align}

Writing the Wilson coefficients in the decomposition in \eqs{Z_expand}{W_expand}, we have for the $gg\, q\bq\, \ell \bar \ell$ channel
\begin{align}
&\vC_{\lambda_1\lambda_2(\lambda_q;\lambda_\ell)}(\lp_1, \lp_2; \lp_3, \lp_4; \lp_5, \lp_6)
\nn \\ & \quad
= e^2  \,
\biggl\{ \bigl[Q^\ell Q^q + v_{\lambda_\ell}^\ell v_{\lambda_q}^q P_Z(s_{56})\bigr]
\vC_{q \lambda_1\lambda_2(\lambda_q;\lambda_\ell)}(\ldots)
\nn \\ & \qquad
+ \sum_{i=1}^{n_f} \Bigl[Q^\ell Q^i + v_{\lambda_\ell}^\ell\frac{v_L^i + v_R^i}{2} P_Z(s_{56}) \Bigr]
\vC_{v\lambda_1\lambda_2(\lambda_q;\lambda_\ell)}(\ldots)
\nn \\ & \qquad
+ \frac{v_{\lambda_\ell}^\ell}{\sin(2\theta_W)} P_Z(s_{56})\, \vC_{a\lambda_1\lambda_2(\lambda_q;\lambda_\ell)}(\ldots)
\biggr\}
\,,\end{align}
with the weak couplings
\begin{align}
v_{+}^\ell = v_R^\ell
\,,\quad
v_{-}^\ell = v_L^\ell
\,,\qquad
v_{+}^q = v_R^q
\,,\quad
v_{-}^q = v_L^q
\,.\end{align}
For the $W^\mp$ channels $gg\, u \bar d\, \ell \bar \nu$ and $gg\, d \bar u\, \nu \bar \ell$, we have
\begin{align}
\vC_{\lambda_1\lambda_2(W^\mp)}(\dots)
&= \frac{e^2 V_{ud}^{(\dagger)}}{2\sin^2 \theta_W}\, P_W(s_{56})\, \vC_{q\lambda_1\lambda_2(-;-)}(\ldots)
\,.\end{align}

The coefficients for $gg\,q\bq\,V$ are then given by
\begin{align} \label{eq:ggqqV_coeffs}
&\vC_{x\la_1\la_2(+;+)}(\lp_1,\lp_2;\lp_3,\lp_4;\lp_5,\lp_6)
\nn \\ & \qquad
=
\begin{pmatrix}
   A_{x,\fin}(1^{\la_1},2^{\la_2};3_q^+,4_\bq^-; 5_\ell^+, 6_{\bar\ell}^-) \\
   A_{x,\fin}(2^{\la_2},1^{\la_1};3_q^+,4_\bq^-; 5_\ell^+, 6_{\bar\ell}^-) \\
   B_{x,\fin}(1^{\la_1},2^{\la_2};3_q^+,4_\bq^-; 5_\ell^+, 6_{\bar\ell}^-) \\
\end{pmatrix}
,\nn\\
&\vC_{x\la_1\la_2(+;-)}(\lp_1,\lp_2;\lp_3,\lp_4;\lp_5,\lp_6)
\nn \\ & \qquad
= \vC_{x\la_1\la_2(+;+)}(\lp_1,\lp_2;\lp_3,\lp_4;\lp_6,\lp_5)
\,.\end{align}
The remaining Wilson coefficients are obtained by charge conjugation invariance as follows,
\begin{align}
&\vC_{q,v\la_1\la_2(-;\pm)}(\lp_1,\lp_2;\lp_3,\lp_4;\lp_5,\lp_6)
\nn \\ & \qquad
= \hV \vC_{q,v\la_1\la_2(+;\pm)}(\lp_1,\lp_2;\lp_4,\lp_3;\lp_5,\lp_6)
\,,\nn\\
&\vC_{a\la_1\la_2(-;\pm)}(\lp_1,\lp_2;\lp_3,\lp_4;\lp_5,\lp_6)
\nn \\ & \qquad
= - \hV \vC_{a\la_1\la_2(+;\pm)}(\lp_1,\lp_2;\lp_4,\lp_3;\lp_5,\lp_6)\,,
\nn\\
&\text{with}\qquad
\hV =
\begin{pmatrix}
  0 & 1 & 0 \\
  1 & 0 & 0 \\
  0 & 0 & 1
\end{pmatrix}
.\end{align}
The tree-level helicity amplitudes are given in \app{V2ggqq}.

\section{\boldmath $pp \to$ Jets}
\label{sec:pp}

In this section, we give the operator basis and matching relations for $pp \to$ $2, 3$ jets. We consider only the QCD contributions, so that quarks only appear in same-flavor quark-antiquark pairs with the same chirality, and so are described by the currents $J_{ij\pm}$. The helicity amplitudes for each channel are given in \app{helicityamplitudes}.

\subsection{\boldmath $pp \to 2$ Jets}
\label{subsec:pp2jets}

For $pp\to 2$ jets, the partonic channels $q\bq\, q'\bq'$, $q\bq\, q\bq$, $q\bar q gg$, and $gggg$ contribute. We will discuss each in turn. The one-loop helicity amplitudes for all partonic channels were first calculated in Ref.~\cite{Kunszt:1993sd}. The tree-level and one-loop results are given in \app{pp2jets_app}. The two-loop amplitudes have also been calculated, and can be found in Refs.~\cite{Bern:2003ck, Glover:2003cm} for $q\bar q gg$, Refs.~\cite{Anastasiou:2000kg, Anastasiou:2000ue, Glover:2004si, Freitas:2004tk} for $q\bq\, q'\bq'$, $q\bq\, q\bq$ and in Refs.~\cite{Glover:2001af, Bern:2002tk} for $gggg$.

\subsubsection{$q\bq\, q'\bq'$ and $q\bq\, q\bq$}
\label{subsec:qqbarqqbarbasis}

In the case of distinct quark flavors, $q\bq\, q'\bq'$, the helicity basis consists of four independent operators,
\begin{align} \label{eq:qqQQ_basis}
O_{(+;+)}^{\balpha\bt\bgamma\delta}
&= J_{q\, 12+}^{\balpha\bt}\, J_{q'\, 34+}^{\bgamma\delta}
\,,\nn\\
O_{(+;-)}^{\balpha\bt\bgamma\delta}
&= J_{q\, 12+}^{\balpha\bt}\, J_{q'\, 34-}^{\bgamma\delta}
\,,\nn\\
O_{(-;+)}^{\balpha\bt\bgamma\delta}
&= J_{q\, 12-}^{\balpha\bt}\, J_{q'\, 34+}^{\bgamma\delta}
\,,\nn\\
O_{(-;-)}^{\balpha\bt\bgamma\delta}
&= J_{q\, 12-}^{\balpha\bt}\, J_{q'\, 34-}^{\bgamma\delta}
\,.\end{align}
For identical quark flavors, $q\bq\,q\bq$, the helicity basis only has three independent operators,
\begin{align} \label{eq:qqqq_basis}
O_{(++)}^{\balpha\bt\bgamma\delta}
&= \frac{1}{4}\, J_{12+}^{\balpha\bt}\, J_{34+}^{\bgamma\delta}
\,, \nn \\
O_{(+-)}^{\balpha\bt\bgamma\delta}
&= J_{12+}^{\balpha\bt}\, J_{34-}^{\bgamma\delta}
\,, \nn \\
O_{(--)}^{\balpha\bt\bgamma\delta}
&= \frac{1}{4}\, J_{12-}^{\balpha\bt}\, J_{34-}^{\bgamma\delta}
\,.\end{align}
Here we have not made the flavor label explicit, since both quark currents have the same flavor. In both cases we use the color basis
\begin{equation} \label{eq:qqqq_color}
\vT^{ \al\bbeta\ga\bdelta} =
2T_F \Bigl(
  \de_{\al\bdelta}\, \de_{\ga\bbeta}\,,\, \delta_{\al\bbeta}\, \de_{\ga\bdelta}
\Bigr)
\,.\end{equation}

The QCD helicity amplitudes for $q\bq\,q'\bq'$ can be color-decomposed in the basis of \eq{qqqq_color} as
\begin{align} \label{eq:qqQQ_QCD}
\cA(q_{1} \bq_{2} q_3' \bq_4')
&= 2T_F \img \Bigl[
\de_{\al_1\balpha_4} \de_{\al_3\balpha_2}  A(1_q,2_\bq;3_{q'},4_{\bq'}
\\\nn & \quad
+ \frac{1}{N}\,\de_{\al_1 \balpha_2} \de_{\al_3 \balpha_4} B(1_q,2_\bq;3_{q'},4_{\bq'}) \Bigr]
\,,\end{align}
where we have included a factor of $1/N$ for convenience. The amplitude vanishes in the case that the quark and antiquark of the same flavor have the same helicity. This is equivalent to the fact that the operators of \eq{qqQQ_basis} provide a complete basis of helicity operators. For identical quark flavors, the QCD amplitudes can be written in terms of the amplitudes for the distinct flavor case as
\begin{align} 
\cA(q_{1} \bq_{2} q_3 \bq_4) = \cA(q_{1} \bq_{2} q_3' \bq_4')-\cA(q_{1} \bq_4' q_3' \bq_2)
\,.\end{align}
The Wilson coefficients for $q\bq\,q'\bq'$ are then given by
\begin{align} \label{eq:qqQQ_coeffs}
\vC_{(+;+)}(\lp_1,\lp_2;\lp_3,\lp_4)
&= \begin{pmatrix}
  A_\fin(1_q^+,2_\bq^-; 3_{q'}^+, 4_{\bq'}^-) \\
\tfrac{1}{N} B_\fin(1_q^+,2_\bq^-; 3_{q'}^+, 4_{\bq'}^-)
\end{pmatrix}
,\nn\\[1ex]
\vC_{(+;-)}(\lp_1,\lp_2;\lp_3,\lp_4)
&= \begin{pmatrix}
  A_\fin(1_q^+,2_\bq^-; 3_{q'}^-, 4_{\bq'}^+) \\
  \tfrac{1}{N} B_\fin(1_q^+,2_\bq^-; 3_{q'}^-, 4_{\bq'}^+)
\end{pmatrix}
,\nn\\
\vC_{(-;+)}(\lp_1,\lp_2;\lp_3,\lp_4) &= \vC_{(+;-)}(\lp_2,\lp_1;\lp_4,\lp_3)
\,,\nn\\
\vC_{(-;-)}(\lp_1,\lp_2;\lp_3,\lp_4) &= \vC_{(+;+)}(\lp_2,\lp_1;\lp_4,\lp_3)
\,,\end{align}
and for $q\bq\,q\bq$ they are given in terms of the amplitudes $A_{{\rm fin}}$ and $B_{{\rm fin}}$ for $q\bq\,q'\bq'$ by
\begin{align} \label{eq:qqqq_coeffs}
\vC_{(++)}(\lp_1,\lp_2;\lp_3,\lp_4)
&= \begin{pmatrix}
  A_\fin(1_q^+,2_\bq^-; 3_{q}^+, 4_{\bq}^-)
  -\frac{1}{N} B_\fin(1_q^+, 4_{\bq}^-; 3_{q}^+, 2_\bq^-) \\
  \frac{1}{N} B_\fin(1_q^+,2_\bq^-; 3_{q}^+, 4_{\bq}^-)-A_\fin(1_q^+,4_{\bq}^-; 3_{q}^+, 2_\bq^-)
\end{pmatrix}
, \nn \\[1ex]
\vC_{(+-)}(\lp_1,\lp_2;\lp_3,\lp_4)
&= \begin{pmatrix}
  A_\fin(1_q^+,2_\bq^-; 3_{q}^-, 4_{\bq}^+) \\
  \tfrac{1}{N} B_\fin(1_q^+,2_\bq^-; 3_{q}^-, 4_{\bq}^+)
\end{pmatrix}
,\nn \\
\vC_{(--)}(\lp_1,\lp_2;\lp_3,\lp_4) &= \vC_{(++)}(\lp_2,\lp_1;\lp_4,\lp_3)
\,.\end{align}
The relations for $\vC_{(-;\pm)}$ and $\vC_{(--)}$ follow from charge conjugation invariance.
The Wilson coefficient $\vC_{(+-)}$ is equal to $\vC_{(+;-)}$, since the amplitude vanishes when the quark and antiquark of the same flavor have the same helicity (both $+$ or both $-$), so there is no exchange term. The subscript ``$\fin$'' in \eqs{qqQQ_coeffs}{qqqq_coeffs} denotes the IR-finite part of the helicity amplitudes as discussed in \sec{matching}, see \eq{matching_general}. Recall that the symmetry factors of $1/4$ in \eq{qqqq_basis} already take care of the interchange of identical (anti)quarks, so there are no additional symmetry factors needed for $\vC_{(++)}$. Explicit expressions for all required partial amplitudes at tree level and one loop are given in \app{qqqqamplitudes}.

\subsubsection{$gg q\bar q$}
\label{subsec:ggqqbarbasis}

For $gg q\bar q$, the helicity basis has a total of six independent operators,
\begin{align} \label{eq:ggqq_basis}
O_{++(+)}^{ab\, \balpha\beta}
&= \frac{1}{2}\, \cB_{1+}^a\, \cB_{2+}^b\, J_{34+}^{\balpha\beta}
\,,\nn\\
O_{+-(+)}^{ab\, \balpha\beta}
&= \cB_{1+}^a\, \cB_{2-}^b\, J_{34+}^{\balpha\beta}
\,,\nn\\
O_{--(+)}^{ab\, \balpha\beta}
&= \frac{1}{2} \cB_{1-}^a\, \cB_{2-}^b\, J_{34+}^{\balpha\beta}
\,,\nn\\
O_{++(-)}^{ab\, \balpha\beta}
&= \frac{1}{2}\, \cB_{1+}^a\, \cB_{2+}^b\, J_{34-}^{\balpha\beta}
\,,\nn\\
O_{+-(-)}^{ab\, \balpha\beta}
&= \cB_{1+}^a\, \cB_{2-}^b\, J_{34-}^{\balpha\beta}
\,,\nn\\
O_{--(-)}^{ab\, \balpha\beta}
&= \frac{1}{2} \cB_{1-}^a\, \cB_{2-}^b\, J_{34-}^{\balpha\beta}
\,.\end{align}
Note that the use of a helicity basis has made it easy to count the number of required operators.~\footnote{This should be contrasted with the more complicated basis given in Eq.~(126) of Ref.~\cite{Marcantonini:2008qn} which is built from fields $\chi_{n_i}$ and $\cB_{n_i}^{\perp\mu}$ and standard Dirac structures. It can be reduced to a minimal basis using identities such as $O_2 = - O_1$, $O_8 = O_7 +4 t O_3 - 4 t O_4$ and $O_6=O_5-2O_1+{\cal O}(\epsilon)$ where $t =-\w_1 \w_3 n_1 \sdt n_3/2$, and then can be related to the basis used here.}
For the color structure, we use the basis
\begin{equation} \label{eq:ggqq_color}
\vT^{ ab \alpha\bbeta}
= \Bigl(
   (T^a T^b)_{\alpha\bbeta}\,,\, (T^b T^a)_{\alpha\bbeta} \,,\, \tr[T^a T^b]\, \delta_{\alpha\bbeta}
   \Bigr)
\,.\end{equation}

The color decomposition of the QCD helicity amplitudes into partial amplitudes using the color basis of \eq{ggqq_color} is
\begin{align} \label{eq:ggqq_QCD}
&\cA\bigl(g_1 g_2\, q_{3} \bq_{4} \bigr)
\nn\\ & \quad
= \img \sum_{\sigma\in S_2} \bigl[T^{a_{\sigma(1)}} T^{a_{\sigma(2)}}\bigr]_{\alpha_3\balpha_4}
\,A(\sigma(1),\sigma(2); 3_q, 4_\bq)
\nn\\ & \qquad
+ \img\, \tr[T^{a_1} T^{a_2}]\,\delta_{\alpha_3\balpha_4}\, B(1,2; 3_q, 4_\bq)
\,,\end{align}
from which we can read off the Wilson coefficients,
\begin{align} \label{eq:ggqq_coeffs}
\vC_{+-(+)}(\lp_1,\lp_2;\lp_3,\lp_4) &=
\begin{pmatrix}
   A_\fin(1^+,2^-;3_q^+,4_\bq^-) \\
   A_\fin(2^-,1^+;3_q^+,4_\bq^-) \\
   B_\fin(1^+,2^-;3_q^+,4_\bq^-) \\
\end{pmatrix}
,\nn\\
\vC_{++(+)}(\lp_1,\lp_2;\lp_3,\lp_4)
&= \begin{pmatrix}
   A_\fin(1^+,2^+;3_q^+,4_\bq^-) \\
   A_\fin(2^+,1^+;3_q^+,4_\bq^-) \\
   B_\fin(1^+,2^+;3_q^+,4_\bq^-) \\
\end{pmatrix}
,\nn\\
\vC_{--(+)}(\lp_1,\lp_2;\lp_3,\lp_4)
&= \begin{pmatrix}
   A_\fin(1^-,2^-;3_q^+,4_\bq^-) \\
   A_\fin(2^-,1^-;3_q^+,4_\bq^-) \\
   B_\fin(1^-,2^-;3_q^+,4_\bq^-) \\
\end{pmatrix}
.\end{align}
The remaining coefficients follow from charge conjugation as discussed in \subsec{discrete},
\begin{align}\label{eq:ggqq_charge_matrix}
\vC_{\la_1\la_2(-)}(\lp_1,\lp_2;\lp_3,\lp_4)
&= \hV \vC_{\la_1\la_2(+)}(\lp_1,\lp_2;\lp_4,\lp_3)\,,
\nn\\
\text{with}\qquad
\hV & =
\begin{pmatrix}
  0 & -1 & 0 \\
  -1 & 0 & 0 \\
  0 & 0 & -1
\end{pmatrix}
\,.\end{align}
At tree level, the partial amplitudes are well known, and only the first two entries in $\vC_{+-(\pm)}$ are nonzero. Explicit expressions for all amplitudes at tree level and one loop are given in \app{ggqqamplitudes}.

\subsubsection{$gggg$}
\label{subsec:ggggbasis}

For $gggg$, the helicity basis has five independent operators,
\begin{align} \label{eq:gggg_basis}
O_{++++}^{abcd} &= \frac{1}{4!}\, \cB_{1+}^a \cB_{2+}^b \cB_{3+}^c \cB_{4+}^d
\,,\nn\\
O_{+++-}^{abcd} &= \frac{1}{3!}\, \cB_{1+}^a \cB_{2+}^b \cB_{3+}^c \cB_{4-}^d
\,,\nn\\
O_{++--}^{abcd} &= \frac{1}{4}\, \cB_{1+}^a \cB_{2+}^b \cB_{3-}^c \cB_{4-}^d
\,,\nn\\
O_{---+}^{abcd} &= \frac{1}{3!}\, \cB_{1-}^a \cB_{2-}^b \cB_{3-}^c \cB_{4+}^d
\,,\nn\\
O_{----}^{abcd} &= \frac{1}{4!}\, \cB_{1-}^a \cB_{2-}^b \cB_{3-}^c \cB_{4-}^d
\,.\end{align}
We use the color basis
\begin{equation} \label{eq:gggg_color}
\vT^{abcd} =
\frac{1}{2\cdot 2T_F}\begin{pmatrix}
\tr[abcd] + \tr[dcba] \\ \tr[acdb] + \tr[bdca] \\ \tr[adbc] + \tr[cbda] \\
2\tr[ab] \tr[cd] \\ 2\tr[ac] \tr[db] \\ 2\tr[ad] \tr[bc]
\end{pmatrix}^T
\,,\end{equation}
where we have used the shorthand notation
\begin{equation}
\tr[ab] = \tr[T^a T^b]
\,,\qquad
\tr[abcd] = \tr[T^a T^b T^c T^d]
\,.\end{equation}
Under charge conjugation, the operators transform as
\begin{equation}
\C\, O_{\la_1\la_2\la_3\la_4}^{abcd}\,\vT^{abcd}\, \C = O_{\la_1\la_2\la_3\la_4}^{dcba}\vT^{abcd}
\,.\end{equation}
Thus, charge conjugation invariance of QCD leads to
\begin{equation} \label{eq:gggg_charge}
C_{\la_1\la_2\la_3\la_4}^{abcd} = C_{\la_1\la_2\la_3\la_4}^{dcba}
\,.\end{equation}
In principle, there are three more color structures with a minus sign instead of the plus sign in the first three lines in \eq{gggg_color}. Since charge conjugation is a symmetry of QCD, \eq{gggg_charge} holds to all orders, so these additional color structures cannot contribute. In particular, the color structures in \eq{gggg_color} cannot mix into these additional structures at any order. Hence, it is sufficient to consider the reduced basis in \eq{gggg_color} instead of the 9 different color structures, which were used for example in Ref.~\cite{Kelley:2010fn}. Note that for $N=3$ it is possible to further reduce the color basis by one using the relation
\begin{align}
&\tr[abcd+dcba] + \tr[acdb+bdca] + \tr[adbc+cbda]
\nn\\ & \qquad
= \tr[ab]\tr[cd] + \tr[ac]\tr[db] + \tr[ad]\tr[bc]
\,.\end{align}
We refrain from doing so, since it makes the structure of the anomalous dimension matrix less visible, and because there are no such relations for $N>3$.

The color decomposition of the QCD amplitude into partial amplitudes using the color basis in \eq{gggg_color} is 
\begin{align} \label{eq:gggg_QCD}
\cA(g_1 g_2 g_3 g_4)
&= \frac{\img}{2T_F} \biggl[\sum_{\si \in S_4/Z_4}\! \tr[a_{\si(1)} a_{\si(2)} a_{\si(3)} a_{\si(4)}]
 \nn \\ 
& \qquad \times
A\bigl(\si(1),\si(2),\si(3),\si(4)\bigr)
\nn\\ 
&\quad
+ \sum_{\si \in S_4/Z_2^3}\! \tr[a_{\si(1)} a_{\si(2)}] \tr[a_{\si(3)} a_{\si(4)}]
\nn\\ 
& \qquad \times
B\bigl(\si(1),\si(2),\si(3),\si(4)\bigr) \biggr]
\,,
\end{align}
from which we obtain the Wilson coefficients
\begin{align} \label{eq:gggg_coeffs}
\vC_{++--}(\lp_1,\lp_2,\lp_3,\lp_4)
&= \begin{pmatrix}
  2A_\fin(1^+,2^+,3^-,4^-) \\
  2A_\fin(1^+,3^-,4^-,2^+) \\
  2A_\fin(1^+,4^-,2^+,3^-) \\
  B_\fin(1^+,2^+,3^-,4^-) \\
  B_\fin(1^+,3^-,4^-,2^+) \\
  B_\fin(1^+,4^-,2^+,3^-)
\end{pmatrix}
,\nn\\
\vC_{+++-}(\lp_1,\lp_2,\lp_3,\lp_4)
&= \begin{pmatrix}
  2A_\fin(1^+,2^+,3^+,4^-) \\
  2A_\fin(1^+,3^+,4^-,2^+) \\
  2A_\fin(1^+,4^-,2^+,3^+) \\
  B_\fin(1^+,2^+,3^+,4^-) \\
  B_\fin(1^+,3^+,4^-,2^+) \\
  B_\fin(1^+,4^-,2^+,3^+) \\
\end{pmatrix}
,\nn\\
\vC_{++++}(\lp_1,\lp_2,\lp_3,\lp_4)
&= \begin{pmatrix}
  2A_\fin(1^+,2^+,3^+,4^+) \\
  2A_\fin(1^+,3^+,4^+,2^+) \\
  2A_\fin(1^+,4^+,2^+,3^+) \\
  B_\fin(1^+,2^+,3^+,4^+) \\
  B_\fin(1^+,3^+,4^+,2^+) \\
  B_\fin(1^+,4^+,2^+,3^+) \\
\end{pmatrix}
,\nn\\
\vC_{---+}(\lp_1, \lp_2, \lp_3, \lp_4)
&= \vC_{+++-}(\lp_1, \lp_2, \lp_3, \lp_4)\Big|_{\langle..\rangle \leftrightarrow [..]}
\,,\nn\\
\vC_{----}(\lp_1, \lp_2, \lp_3, \lp_4)
&= \vC_{++++}(\lp_1, \lp_2, \lp_3, \lp_4)\Big|_{\langle..\rangle \leftrightarrow [..]}
\,.\end{align}
The last two coefficients follow from parity invariance. 
The factors of two in the first three entries of the coefficients come from combining the two color structures in the first three entries in \eq{gggg_color} using charge conjugation invariance in~\eq{gggg_charge}. 

The tree-level amplitudes are well known. At tree level, only the $A$ amplitudes with two positive and two negative helicity gluons are nonzero. Because the $A$ amplitudes correspond to a single-trace color structure, which possesses a cyclic symmetry, the corresponding partial amplitudes are invariant under the corresponding cyclic permutations of their arguments. Explicit expressions for the required amplitudes at tree level and one loop are given in \app{ggggamplitudes}.

\subsection{\boldmath $pp \to 3$ Jets}

The four partonic channels $g\,q\bq\, q'\bq'$, $g\, q\bq\, q\bq$, $ggg\, q\bar q$, and $ggggg$ contribute to $pp\to 3$ jets, which we discuss in turn. The one-loop partial amplitudes for the different partonic channels were calculated in Refs.~\cite{Kunszt:1994tq, Bern:1994fz, Bern:1993mq}. Tree-level results for the helicity amplitudes for each partonic process are given in \app{pp3jets_app}.

\subsubsection{$g\, q\bq\, q'\bq'$ and $g\, q\bq\, q\bq$}
\label{subsec:gqqbarqqbarbasis}

For the case of distinct quark flavors, $g\, q\bq\, q'\bq'$, the helicity basis consists of eight independent operators,
\begin{align} \label{eq:gqqQQ_basis}
O_{\pm(+;+)}^{a\,\balpha\bt\bgamma\delta}
&=  \cB_{1\pm}^a\, J_{q\, 23+}^{\balpha\bt}\, J_{q'\, 45+}^{\bgamma\delta}
\,,\nn\\
O_{\pm(+;-)}^{a\,\balpha\bt\bgamma\delta}
&=  \cB_{1\pm}^a\, J_{q\, 23+}^{\balpha\bt}\, J_{q'\, 45-}^{\bgamma\delta}
\,,\nn\\
O_{\pm(-;+)}^{a\,\balpha\bt\bgamma\delta}
&=  \cB_{1\pm}^a\, J_{q\, 23-}^{\balpha\bt}\, J_{q'\, 45+}^{\bgamma\delta}
\,,\nn\\
O_{\pm(-;-)}^{a\,\balpha\bt\bgamma\delta}
&=  \cB_{1\pm}^a\, J_{q\, 23-}^{\balpha\bt}\, J_{q'\, 45-}^{\bgamma\delta}
\,.\end{align}
For identical quark flavors, $g\, q\bq\,q\bq$, the basis reduces to six independent helicity operators,
\begin{align} \label{eq:gqqqq_basis}
O_{\pm(++)}^{a\,\balpha\bt\bgamma\delta}
&= \frac{1}{4}\, \cB_{1\pm}^a\, J_{23+}^{\balpha\bt}\, J_{45+}^{\bgamma\delta}
\,,\nn\\
O_{\pm(+-)}^{a\,\balpha\bt\bgamma\delta}
&= \cB_{1\pm}^a\, J_{23+}^{\balpha\bt}\, J_{45-}^{\bgamma\delta}
\,,\nn\\
O_{\pm(--)}^{a\,\balpha\bt\bgamma\delta}
&= \frac{1}{4}\, \cB_{1\pm}^a\, J_{23-}^{\balpha\bt}\, J_{45-}^{\bgamma\delta}
\,.\end{align}
In both cases we use the color basis
\begin{equation} \label{eq:gqqqq_color}
\vT^{a\,\al\bbeta\ga\bdelta} =
2T_F\Bigl(
  T^a_{\al\bdelta}\, \de_{\ga\bbeta},\, T^a_{\ga\bbeta} \, \de_{\al\bdelta},\,
  T^a_{\al\bbeta}\, \de_{\ga\bdelta},\, T^a_{\ga\bdelta} \, \de_{\al\bbeta}
\Bigr)
.\end{equation}

The QCD helicity amplitudes for $g\,q\bq\,q'\bq'$ can be color decomposed into partial amplitudes in the color basis of \eq{gqqqq_color} as
\begin{align} \label{eq:gqqQQ_QCD}
\cA(g_{1} q_{2} \bq_{3} q_4' \bq_5')
&=  2T_F\img \Bigl[T^{a_1}_{\al_2 \balpha_5} \de_{\al_4\balpha_3} A(1;2_q,3_\bq;4_{q'},5_{\bq'}) +
T^{a_1}_{\al_4 \balpha_3} \de_{\al_2 \balpha_5} A(1;4_{q'},5_{\bq'};2_q,3_\bq)
\nn \\ &\quad
+\frac{1}{N} T^{a_1}_{\al_2 \balpha_3} \de_{\al_4\balpha_5} B(1;2_q,3_\bq;4_{q'},5_{\bq'})
+\frac{1}{N} T^{a_1}_{\al_4 \balpha_5} \de_{\al_2 \balpha_3} B(1;4_{q'},5_{\bq'};2_q,3_\bq) \Bigr]
\,,\end{align}
where we have used the symmetry $q\bar q \leftrightarrow q'\bar q'$, and inserted the factors of $1/N$ for later convenience.  The amplitude vanishes when the quark and antiquark of the same flavor have the same helicity (both $+$ or both $-$), in accordance with the fact that the operators of \eq{gqqQQ_basis} provide a complete basis of helicity operators. For identical quark flavors, the amplitudes can be written in terms of the amplitudes for the distinct flavor case as
\begin{align}
\cA(g_{1} q_{2} \bq_{3} q_4 \bq_5)
&=\cA(g_{1} q_{2} \bq_{3} q_4' \bq_5')-\cA(g_{1} q_{2} \bq_5' q_4' \bq_{3})\,.
\end{align}
The Wilson coefficients for $g\,q\bq\,q'\bq'$ are then given by
\begin{align} \label{eq:gqqQQ_coeffs}
\vC_{+(+;+)}(\lp_1;\lp_2,\lp_3;\lp_4,\lp_5)
&= \begin{pmatrix}
  A_\fin(1^+;2_q^+,3_\bq^-;4_{q'}^+,5_{\bq'}^-) \\
  A_\fin(1^+;4_{q'}^+,5_{\bq'}^-;2_q^+,3_\bq^-) \\
 \frac{1}{N} B_\fin(1^+;2_q^+,3_\bq^-;4_{q'}^+,5_{\bq'}^-)  \\
  \frac{1}{N} B_\fin(1^+;4_{q'}^+,5_{\bq'}^-;2_q^+,3_\bq^-)
\end{pmatrix}
,\nn\\
\vC_{+(+;-)}(\lp_1;\lp_2,\lp_3;\lp_4,\lp_5)
&= \begin{pmatrix}
  A_\fin(1^+;2_q^+,3_\bq^-;4_{q'}^-,5_{\bq'}^+) \\
  A_\fin(1^+;4_{q'}^-,5_{\bq'}^+;2_q^+,3_\bq^-) \\
 \frac{1}{N} B_\fin(1^+;2_q^+,3_\bq^-;4_{q'}^-,5_{\bq'}^+) \\
 \frac{1}{N} B_\fin(1^+;4_{q'}^-,5_{\bq'}^+;2_q^+,3_\bq^-)
\end{pmatrix}
\,,\end{align}
and for $g\,q\bq\,q\bq$ they are given in terms of the amplitudes $A_\fin$ and $B_\fin$ for $g\,q\bq\,q'\bq'$ by
\begin{align} \label{eq:gqqqq_coeffs}
\vC_{+(++)}(\lp_1;\lp_2,\lp_3;\lp_4,\lp_5)
&= \begin{pmatrix}
  A_\fin(1^+;2_q^+,3_\bq^-;4_{q}^+,5_{\bq}^-)  -\frac{1}{N} B_\fin(1^+;2_q^+,5_{\bq}^-;4_{q}^+,3_\bq^-) \\
  A_\fin(1^+;4_{q}^+,5_{\bq}^-;2_q^+,3_\bq^-)  -\frac{1}{N} B_\fin(1^+;4_{q}^+,3_\bq^-;2_{q}^+,5_{\bq}^-)\\
  \frac{1}{N} B_\fin(1^+;2_q^+,3_\bq^-;4_{q}^+,5_{\bq}^-) -A_\fin(1^+;2_q^+,5_{\bq}^-;4_{q}^+,3_\bq^-)  \\
  \frac{1}{N} B_\fin(1^+;4_{q}^+,5_{\bq}^-;2_q^+,3_\bq^-) - A_\fin(1^+;4_{q}^+,3_\bq^-;2_q^+,5_{\bq}^-)
\end{pmatrix}
,\nn\\
\vC_{+(+-)}(\lp_1;\lp_2,\lp_3;\lp_4,\lp_5)
&= \begin{pmatrix}
  A_\fin(1^+;2_q^+,3_\bq^-;4_{q}^-,5_{\bq}^+) \\
  A_\fin(1^+;4_{q}^-,5_{\bq}^+;2_q^+,3_\bq^-) \\
\frac{1}{N}  B_\fin(1^+;2_q^+,3_\bq^-;4_{q}^-,5_{\bq}^+) \\
\frac{1}{N}  B_\fin(1^+;4_{q}^-,5_{\bq}^+;2_q^+,3_\bq^-)
\end{pmatrix}
.\end{align}
Charge conjugation invariance of QCD relates the Wilson coefficients,
\begin{align} \label{eq:gqqqq_charge}
\vC_{\la(-;\pm)}(\lp_1;\lp_2,\lp_3;\lp_4,\lp_5) &= \hV \vC_{\la(+;\mp)}(\lp_1;\lp_3,\lp_2;\lp_5,\lp_4)\,,
\nn\\
\vC_{\la(--)}(\lp_1;\lp_2,\lp_3;\lp_4,\lp_5) &= \hV \vC_{\la(++)}(\lp_1;\lp_3,\lp_2;\lp_5,\lp_4)\,,
\end{align}
with
\begin{align}
\hV &=
\begin{pmatrix}
  0 & -1 & 0 & 0 \\
  -1 & 0 & 0 & 0 \\
  0 & 0 & -1 & 0 \\
  0 & 0 & 0 & -1
\end{pmatrix}
.\end{align}
The remaining Wilson coefficients for a negative helicity gluon follow from parity invariance,
\begin{align}
&\vC_{-(+;\pm)}(\lp_1; \lp_2, \lp_3; \lp_4, \lp_5)
\nn \\ & \qquad
=\vC_{+(-;\mp)}(\lp_1; \lp_2, \lp_3; \lp_4, \lp_5)\Big|_{\langle..\rangle \leftrightarrow [..]}
\,, \nn \\
&\vC_{-(++)}(\lp_1; \lp_2, \lp_3; \lp_4, \lp_5)
\nn \\ & \qquad
=\vC_{+(--)}(\lp_1; \lp_2, \lp_3; \lp_4, \lp_5)\Big|_{\langle..\rangle \leftrightarrow [..]}
\,.\end{align}
Explicit expressions for all required partial amplitudes at tree level are given in \app{gqqqqamplitudes}.

\subsubsection{$ggg\, q\bar q$}
\label{subsec:gggqqbarbasis}

For $ggg\, q\bar q$, we have a basis of eight independent helicity operators,
\begin{align} \label{eq:gggqq_basis}
O_{+++(\pm)}^{abc\, \balpha\beta}
&= \frac{1}{3!}\, \cB_{1+}^a\, \cB_{2+}^b\, \cB_{3+}^c\, J_{45\pm}^{\balpha\beta}
\,,\nn\\
O_{++-(\pm)}^{abc\, \balpha\beta}
&= \frac{1}{2}\, \cB_{1+}^a\, \cB_{2+}^b\, \cB_{3-}^c\, J_{45\pm}^{\balpha\beta}
\,,\nn\\
O_{--+(\pm)}^{abc\, \balpha\beta}
&= \frac{1}{2}\, \cB_{1-}^a\, \cB_{2-}^b\, \cB_{3+}^c\, J_{45\pm}^{\balpha\beta}
\,,\nn\\
O_{---(\pm)}^{abc\, \balpha\beta}
&= \frac{1}{3!} \cB_{1-}^a\, \cB_{2-}^b\, \cB_{3-}^c\, J_{45\pm}^{\balpha\beta}
\,,\end{align}
and we use the color basis
\begin{equation} \label{eq:gggqq_color}
\vT^{abc\,\alpha\bbeta} =
\begin{pmatrix}
  [ T^a T^b T^c]_{\alpha\bbeta} \\  [T^b T^c T^a]_{\alpha\bbeta} \\ [T^c T^a T^b]_{\alpha\bbeta} \\ [T^c T^b T^a]_{\alpha\bbeta} \\ [T^a T^c T^b]_{\alpha\bbeta} \\ [T^b T^a T^c]_{\alpha\bbeta} \\
   \tr[T^c T^a] T^b_{\alpha\bbeta} \\  \tr[T^a T^b] T^c_{\alpha\bbeta} \\ \tr[T^b T^c] T^a_{\alpha\bbeta} \\
   \tr[T^a T^b T^c] \delta_{\alpha\bbeta} \\ \tr[T^c T^b T^a] \delta_{\alpha\bbeta}
\end{pmatrix}^T
\,.\end{equation}

The color decomposition of the QCD helicity amplitudes into partial amplitudes using \eq{gggqq_color} is
\begin{align} \label{eq:gggqq_QCD}
\cA\bigl(g_1 g_2 g_3\, q_{4} \bq_{5} \bigr)
&= \img \!\sum_{\sigma\in S_3}\! \bigl[T^{a_{\sigma(1)}} T^{a_{\sigma(2)}} T^{a_{\sigma(3)}}\bigr]_{\alpha_4\balpha_5}
\nn \\ & \qquad\times
A(\sigma(1),\sigma(2),\sigma(3); 4_q, 5_\bq)
\nn\\ & \quad
+ \img \!\sum_{\sigma\in S_3/Z_2}\!\! \tr\bigl[T^{a_{\sigma(1)}} T^{a_{\sigma(2)}}\bigr] T^{a_{\sigma(3)}}_{\alpha_4\balpha_5}
\nn \\ & \qquad\times
B(\sigma(1),\sigma(2),\sigma(3); 4_q, 5_\bq)
\nn \\& \quad
+ \img \!\sum_{\sigma\in S_3/Z_3}\!\! \tr\bigl[T^{a_{\sigma(1)}} T^{a_{\sigma(2)}} T^{a_{\sigma(3)}}\bigr] \de_{\alpha_4\balpha_5}
\nn \\ & \qquad\times
C(\sigma(1),\sigma(2),\sigma(3); 4_q, 5_\bq)
\,,\end{align}
from which we can read off the Wilson coefficients, 
\begin{align} \label{eq:gggqq_coeffs}
\vC_{++\mp(+)}(\lp_1, \dots; \lp_4, \lp_5)
&= \begin{pmatrix}
   A_\fin(1^+,2^+,3^\mp;4_q^+,5_\bq^-) \\
    A_\fin(2^+,3^\mp,1^+;4_q^+,5_\bq^-) \\
    A_\fin(3^\mp,1^+,2^+;4_q^+,5_\bq^-) \\
    A_\fin(3^\mp,2^+,1^+;4_q^+,5_\bq^-) \\
    A_\fin(1^+,3^\mp,2^+;4_q^+,5_\bq^-) \\
    A_\fin(2^+,1^+,3^\mp;4_q^+,5_\bq^-) \\
    B_\fin(3^\mp,1^+,2^+;4_q^+,5_\bq^-) \\
    B_\fin(1^+,2^+,3^\mp;4_q^+,5_\bq^-) \\
    B_\fin(2^+,3^\mp,1^+;4_q^+,5_\bq^-) \\
    C_\fin(1^+,2^+,3^\mp;4_q^+,5_\bq^-) \\
    C_\fin(3^\mp,2^+,1^+;4_q^+,5_\bq^-)
\end{pmatrix}
.\end{align}
Charge conjugation invariance of QCD relates the coefficients with opposite quark helicities,
\begin{align} \label{eq:gggqq_charge}
&\vC_{\la_1\la_2 \la_3(-)}(\lp_1,\lp_2,\lp_3 ; \lp_4, \lp_5)
= \hV \vC_{\la_1\la_2 \la_3(+)}(\lp_1, \lp_2,\lp_3; \lp_5, \lp_4)\,,
\nn \\
&\text{with}\qquad
\hV =
\begin{pmatrix}
  0_{3\times 3} & 1_{3\times3} &  &  \\
  1_{3\times 3} & 0_{3\times3} &  &  \\
   &  & 1_{3\times3} &  &  \\
   &  &  & 0 & 1 \\
   &  &  & 1 & 0
\end{pmatrix}
,\end{align}
where $1_{n\times n}$ denotes the $n$-dimensional identity matrix and the empty entries are all zero.
The remaining coefficients follow from parity invariance
\begin{align}\label{eq:gggqq_parity}
&\vC_{--+(\pm)}(\lp_1,  \lp_2, \lp_3; \lp_4, \lp_5)
\nn \\& \qquad
=\vC_{++-(\mp)}(\lp_1,  \lp_2,\lp_3; \lp_4, \lp_5)\Big|_{\langle..\rangle \leftrightarrow [..]}
\,,\nn \\
&\vC_{---(\pm)}(\lp_1, \lp_2, \lp_3; \lp_4, \lp_5)
\nn \\ & \qquad
=\vC_{+++(\mp)}(\lp_1, \lp_2,\lp_3; \lp_4, \lp_5)\Big|_{\langle..\rangle \leftrightarrow [..]}
\,.\end{align}
At tree level, the partial amplitudes are well known, and only the $A$ amplitudes are nonzero. Furthermore, the partial amplitudes with all negative or all positive helicity gluons vanish.  Combining the charge and parity relations of \eqs{gggqq_charge}{gggqq_parity}, there are only three independent amplitudes at tree level, which we take to be $A(1^+,2^+,3^-;4_q^+,5_{\bq}^-)$, $A(2^+,3^-,1^+;4_q^+,5_{\bq}^-)$, and $A(3^-,1^+,2^+;4_q^+,5_{\bq}^-)$. These amplitudes are given in \app{gggqqamplitudes}.

\subsubsection{$ggggg$}
\label{subsec:gggggbasis}

For $ggggg$, the basis consists of six independent helicity operators,
\begin{align} \label{eq:ggggg_basis}
O_{+++++}^{abcde} &= \frac{1}{5!}\, \cB_{1+}^a \cB_{2+}^b \cB_{3+}^c \cB_{4+}^d \cB_{5+}^e
\,,\nn\\
O_{++++-}^{abcde} &= \frac{1}{4!}\, \cB_{1+}^a \cB_{2+}^b \cB_{3+}^c \cB_{4+}^d \cB_{5-}^e
\,,\nn\\
O_{+++--}^{abcde} &= \frac{1}{2\cdot 3!}\, \cB_{1+}^a \cB_{2+}^b \cB_{3+}^c \cB_{4-}^d \cB_{5-}^e
\,,\nn\\
O_{---++}^{abcde} &= \frac{1}{2 \cdot 3!}\, \cB_{1-}^a \cB_{2-}^b \cB_{3-}^c \cB_{4+}^d \cB_{5+}^e
\,,\nn\\
O_{----+}^{abcde} &= \frac{1}{4!}\, \cB_{1-}^a \cB_{2-}^b \cB_{3-}^c \cB_{4-}^d \cB_{5+}^e
\,,\nn\\
O_{-----}^{abcde} &= \frac{1}{5!}\, \cB_{1-}^a \cB_{2-}^b \cB_{3-}^c \cB_{4-}^d \cB_{5-}^e
\,.\end{align}
As before, we only need one operator for each number of positive and negative helicities.
We use the color basis
\begin{equation} \label{eq:ggggg_color}
\vT^{ abcde} =
\frac{1}{2\cdot 2T_F}\begin{pmatrix}
\tr[abcde]-\tr[edcba] \\
\tr[acdeb]-\tr[bedca] \\
\tr[acbed]-\tr[debca] \\
\tr[abced]-\tr[decba] \\
\tr[abdec]-\tr[cedba] \\
\tr[acbde]-\tr[edbca] \\
\tr[adceb]-\tr[becda] \\
\tr[adcbe]-\tr[ebcda] \\
\tr[aebdc]-\tr[cdbea] \\
\tr[abdce]-\tr[ecdba] \\
\tr[aecbd]-\tr[dbcea] \\
\tr[acebd]-\tr[dbeca] \\
(\tr[ced]-\tr[dec] )\tr[ab] \\ (\tr[abe]-\tr[eba])\tr[cd] \\ (\tr[acd]-\tr[dca])\tr[be] \\ (\tr[bec]-\tr[ceb])\tr[ad] \\ (\tr[adb]-\tr[bda])\tr[ce] \\ (\tr[ace]-\tr[eca])\tr[bd] \\ (\tr[bdc]-\tr[cdb])\tr[ae] \\ (\tr[aed]-\tr[dea])\tr[bc] \\ (\tr[acb]-\tr[bca])\tr[de] \\ (\tr[bed]-\tr[deb])\tr[ac]
\end{pmatrix}^T
,\end{equation}
where we have used the shorthand notation
\begin{equation}
\tr[ab \cdots cd] = \tr[T^a T^b \cdots T^c T^d]
\,.\end{equation}
A priori, there are twice as many color structures as in \eq{ggggg_color} with a relative plus sign instead of a minus sign between the two traces. Under charge conjugation, the operators transform as
\begin{equation}
\C\, O_{\la_1\la_2\la_3\la_4\la_5}^{abcde}\,\vT^{abcde}\, \C = -O_{\la_1\la_2\la_3\la_4\la_5}^{edcba}(\vT^{abcde})
\,.\end{equation}
Therefore, charge conjugation invariance implies for the Wilson coefficients
\begin{equation}
C_{\la_1\la_2\la_3\la_4\la_5}^{abcde} = - C_{\la_1\la_2\la_3\la_4\la_5}^{edcba}
\,,\end{equation}
and hence these additional color structures cannot appear at any order in perturbation theory, either through matching or renormalization group evolution.

The color decomposition of the QCD amplitude into partial amplitudes using the color basis of \eq{ggggg_color} is
\begin{align} \label{eq:ggggg_QCD}
&\cA(g_1 g_2 g_3 g_4g_5)
\nn \\ & \qquad
= \frac{\img}{2T_F} \biggl[\sum_{\si \in S_5/Z_5}\! \tr[a_{\si(1)} a_{\si(2)} a_{\si(3)} a_{\si(4)} a_{\si(5)}]
\nn \\ & \qquad\qquad \times
A\bigl(\si(1),\si(2),\si(3),\si(4), \si(5)\bigr)
\nn\\ & \qquad\quad
+ \sum_{\si \in S_5/(Z_3 \times Z_2)} \tr[a_{\si(1)} a_{\si(2)} a_{\si(3)}]\, \tr[a_{\si(4)} a_{\si(5)}]
\nn\\ & \qquad\qquad\times
B\bigl(\si(1),\si(2),\si(3),\si(4),\si(5)\bigr) \biggr]
\,,\end{align}
from which we obtain the Wilson coefficients 
\begin{align} \label{eq:ggggg_coeffs}
\vC_{+++--}(\lp_1, \dots ,\lp_5)
&= 2\begin{pmatrix}
 A_\fin(1^+,2^+,3^+,4^-,5^-)  \\  A_\fin(1^+,3^+,4^-,5^-,2^+)  \\  A_\fin(1^+,3^+,2^+,5^-,4^-)  \\
 A_\fin(1^+,2^+,3^+,5^-,4^-)  \\  A_\fin(1^+,2^+,4^-,5^-,3^+)  \\  A_\fin(1^+,3^+,2^+,4^-,5^-)  \\
 A_\fin(1^+,4^-,3^+,5^-,2^+)  \\  A_\fin(1^+,4^-,3^+,2^+,5^-)  \\  A_\fin(1^+,5^-,2^+,4^-,3^+)  \\
 A_\fin(1^+,2^+,4^-,3^+,5^-)  \\  A_\fin(1^+,5^-,3^+,2^+,4^-)  \\  A_\fin(1^+,3^+,5^-,2^+,4^-)  \\
 B_\fin(3^+,5^-,4^-,1^+,2^+)  \\  B_\fin(1^+,2^+,5^-,3^+,4^-)  \\  B_\fin(1^+,3^+,4^-,2^+,5^-)  \\
 B_\fin(2^+,5^-,3^+,1^+,4^-)  \\  B_\fin(1^+,4^+,2^-,3^+,5^-)  \\  B_\fin(1^+,3^+,5^-,2^+,4^-)  \\
 B_\fin(2^+,4^-,3^+,1^+,5^-)  \\  B_\fin(1^+,5^-,4^-,2^+,3^+)  \\  B_\fin(1^+,3^+,2^+,4^-,5^-)  \\
 B_\fin(2^+,5^-,4^-,1^+,3^+)
\end{pmatrix}
, \nn \\
\vC_{----\pm}(\lp_1, \ldots, \lp_5)
&= \vC_{++++\mp}(\lp_1, \ldots, \lp_5)\Big|_{\langle..\rangle \leftrightarrow [..]}
 \,,\nn\\
\vC_{---++}(\lp_1, \dots, \lp_5)
&= \vC_{+++--}(\lp_1, \dots, \lp_5)\Big|_{\langle..\rangle \leftrightarrow [..]}
\,.\end{align}
For brevity, we have not written out the coefficients $\vC_{++++-}$ and $\vC_{+++++}$. They have exactly the same structure as $\vC_{+++--}$ with the replacements $4^- \to 4^+$ and $4^-,5^-\to 4^+, 5^+$, respectively, in the arguments of the helicity amplitudes. The remaining Wilson coefficients are given by parity invariance as shown. The overall factor of two comes from combining the two color structures in \eq{ggggg_color}, which are related by charge conjugation. 

At tree level, all the $B$ amplitudes vanish, as do all the amplitudes in $\vC_{++++\pm}$ and $\vC_{----\mp}$. By the parity relations given in \eq{ggggg_coeffs}, only the $A$ amplitudes in $\vC_{+++--}$ are then required for the tree-level matching. Since these amplitudes correspond to single trace color structures, which posses a cyclic symmetry, the required partial amplitudes are invariant under the corresponding cyclic permutations of their arguments. Therefore, at tree level, there are only two independent amplitudes, which we take to be $A_\fin(1^+,2^+,3^+,4^-,5^-)$ and $A_\fin(1^+,2^+,4^-,3^+,5^-)$. These are given in \app{gggggamplitudes}. Simplifications also occur at one loop, since the $B$ amplitudes can be expressed in terms of sums of permutations of the $A$ amplitudes \cite{Bern:1990ux, Bern:1994zx}.

\section{Renormalization Group Evolution}
\label{sec:running}

In this section, we discuss the renormalization group evolution (RGE) of the Wilson coefficients. We start with a general discussion and give the solution of the RGE to all orders in perturbation theory. For completeness, we also explicitly derive the (known) anomalous dimension at one loop. To discuss the RGE, it is convenient to consider the operators $\Op^\dagger$ in \eq{Opm_color}, which are vectors in color space. Lastly, we give explicit results, in a manifestly crossing symmetric form, for the relevant color mixing matrices for the color bases we have used in the previous sections. Since the operators' renormalization is independent of their helicity structure, we drop all helicity labels throughout this section for notational simplicity.

\subsection{General Discussion}
\label{subsec:running_general}

The renormalization of the hard scattering in SCET can either be carried out as operator renormalization, where the relation between bare and renormalized matrix elements is $\vev{\Op^\dagger}^\mathrm{bare}= Z_\xi^{-n_q/2} Z_A^{-n_g/2} \vev{\Op^\dagger}^\mathrm{ren} \widehat Z_{O}$, or with coefficient renormalization where $\vev{\Op^\dagger}^{\mathrm{bare}} \vC^\mathrm{bare} = Z_\xi^{n_q/2} Z_A^{n_g/2} \vev{\Op^\dagger}^\mathrm{bare} \widehat Z_C  \vec C^{\mathrm{ren}}$. The relationship between the two is $\widehat Z_C = \widehat Z_{O}^{-1}$.
Here $Z_\xi$ and $Z_A$ are the wave-function renormalizations of the SCET collinear quark and gluon fields $\xi_n$ and $A_n$, defined in \subsec{scet}, and
\begin{equation}
n_g = n_g^+ + n_g^-
\,,\qquad
n_q = n_q^+ + n_q^-
\end{equation}
are the total number of quark and gluon helicity fields in the operator (recall that there are two quark fields in each of the fermionic helicity currents).
The UV divergences for $\vev{\Op^\dagger}^\mathrm{bare}$ are given in terms of a local product (as opposed to a convolution over label momenta), since we are working at leading power where the operators contain a single field per collinear sector.

Let us consider more explicitly how the renormalization works at one loop. The counterterm Feynman rule at this order is
\begin{equation} \label{eq:SCETctFeyn}
\vev{\Op^\dagger}^\tree \, \Bigl(Z_\xi^{n_q/2}\, Z_A^{n_g/2}\, \hZ_C - 1 \Bigr)
\,.\end{equation}
At one loop, the UV divergences of $\vev{\Op^\dagger}^\mathrm{bare}$ are proportional to the tree-level matrix element as $\vev{\Op^\dagger}^\tree\,\hD$, where $\hD$ is a matrix in color space, which denotes the $1/\epsilon^2$ and $1/\epsilon$ UV divergences (with $\mu$ defined in the $\overline{\mathrm{MS}}$ scheme) of the bare matrix element.  The counterterm has to cancel these UV divergences so
\begin{equation} \label{eq:Z_O}
\vev{\Op^\dagger}^\tree \Bigl(Z_\xi^{n_q/2}\, Z_A^{n_g/2}\, \hZ_C - 1 \Bigr)  = -\vev{\Op^\dagger}^\tree\,\hD
\,,
\end{equation}
which fixes $\widehat Z_C$ at one loop.

Next consider the renormalization group equations, working to all orders in $\alpha_s$. As usual, the $\mu$ independence of the bare operator implies the renormalization group equation for the Wilson coefficient
\begin{equation} \label{eq:RGE}
\mu \frac{\df \vC(\mu)}{\df\mu} = \hga_C(\mu)\,\vC(\mu)
\,,\end{equation}
where the anomalous dimension matrix is defined as
\begin{equation} \label{eq:ga_O}
\hga_C(\mu)
 =  - \hZ_C^{-1}(\mu) \Big[ \frac{\df}{\df\ln\mu}\, \hZ_C(\mu)\Big] 
\,.\end{equation}
The solution of the RGE in \eq{RGE} can be written as
\begin{equation}
\vC(\mu) = \hU(\mu_0,\mu)\, \vC(\mu_0)
\,,\end{equation}
with the evolution matrix
\begin{equation} \label{eq:U_general}
\hU(\mu_0, \mu) = \cP \exp\biggl[\intlim{\ln \mu_0}{\ln \mu}{\ln\mu'}\, \hga_C(\mu')\biggr]
\,.\end{equation}
Here, $\cP$ denotes path ordering along increasing $\mu$, and $\mu > \mu_0$. The path ordering is necessary since $\hga_C(\mu)$ is a matrix in color space.

The anomalous dimension matrix has the general form%
\begin{equation} \label{eq:gaO_general}
\hga_C(\mu)
= \Gamma_\cusp[\alpha_s(\mu)]\,\hDe(\mu^2) + \hga[\alpha_s(\mu)]
\,,\end{equation}
where $\Gamma_\cusp$ is the cusp anomalous dimension and $\hDe(\mu^2)$ is a process-dependent mixing matrix in color space, which does not depend on $\alpha_s$. Its $\mu$ dependence is given by
\begin{equation} \label{eq:hDe_mudep}
\hDe(\mu^2) = \id (n_g C_A + n_q C_F) \ln\Bigl(\frac{\mu_0}{\mu}\Bigr) + \hDe(\mu_0^2)
\,,\end{equation}
which will be demonstrated explicitly at one loop in \subsec{loops}.
We can then perform the integral in \eq{U_general} by using the running of the coupling, $\df\alpha_s(\mu)/\df\ln\mu = \beta(\alpha_s)$, to switch variables from $\ln\mu$ to $\alpha_s$. We find
\begin{align}
\hU(\mu_0,\mu)
&= e^{-(n_g C_A + n_q C_F) K_\Ga(\mu_0,\mu)}
\\\nn & \quad\times
\bar \cP_{\alpha_s} \exp\Bigl[\eta_\Gamma(\mu_0, \mu)\,\hDe(\mu_0^2) + \widehat K_\gamma (\mu_0,\mu)\Bigr]
\,,\end{align}
where $\bar \cP_{\alpha_s}$ now denotes path ordering along decreasing $\alpha_s$, with $\alpha_s(\mu) < \alpha_s(\mu_0)$, and
\begin{align} \label{eq:Kw_def}
K_\Ga (\mu_0, \mu)
&= \intlim{\al_s(\mu_0)}{\al_s(\mu)}{\al_s} \frac{\Gamma_\cusp(\al_s)}{\beta(\al_s)}
   \intlim{\al_s(\mu_0)}{\al_s}{\al_s'} \frac{1}{\beta(\al_s')}
\,,\nn\\
\eta_\Ga(\mu_0, \mu)
&= \intlim{\al_s(\mu_0)}{\al_s(\mu)}{\al_s} \frac{\Gamma_\cusp(\al_s)}{\beta(\al_s)}
\,,\nn\\
\widehat K_\gamma(\mu_0, \mu)
&= \intlim{\al_s(\mu_0)}{\al_s(\mu)}{\al_s} \frac{\hga(\al_s)}{\bt(\al_s)}
\,.\end{align}
Up to two loops, the noncusp piece $\hga(\alpha_s)$ in \eq{gaO_general} is proportional to the identity operator~\cite{MertAybat:2006mz,Aybat:2006wq}
\begin{align}\label{eq:non_cusp_simplify}
\hga(\alpha_s)= \left( n_q \gamma_C^q + n_g \gamma_C^g\right) \id \,.
\end{align}
In this case, the evolution factor simplifies to
\begin{align}
\hU(\mu_0,\mu)
&= e^{-(n_g C_A + n_q C_F) K_\Ga(\mu_0,\mu) + K_\gamma(\mu_0,\mu)}
\nn \\ & \quad \times
\exp\Bigl[\eta_\Gamma(\mu_0, \mu)\,\hDe(\mu_0^2)\Bigr]
\,.\end{align}
Starting at three loops the noncusp anomalous dimension is not color diagonal,
and starts to depend on a conformal cross ratio built from factors of $p_i\cdot p_j$~\cite{Almelid:2015jia}. (For earlier work beyond two loops see~Refs.~\cite{Gardi:2009zv, Gardi:2009qi, Dixon:2009ur, Becher:2009cu, Becher:2009qa, DelDuca:2011ae, Bret:2011xm, DelDuca:2013ara, Caron-Huot:2013fea}.  The result of Ref.~\cite{Almelid:2015jia} implies that the conjectured all-order dipole color structure in Refs.~\cite{Becher:2009cu, Becher:2009qa} is violated.)

The evolution factors $K_\Ga(\mu_0,\mu)$, and $\eta_\Gamma(\mu_0,\mu)$ are universal. Explicit expressions for the integrals in \eq{Kw_def} to NNLL order, together with the required coefficients for $\Gamma_\cusp$ and the $\beta$ function to three loops, are given for reference in \app{RGE_factors}.

\subsection{One-loop Anomalous Dimension}
\label{subsec:loops}

\begin{figure}[t]
\centering
\subfloat[]{\label{fig:coll}
\includegraphics[scale=0.55]{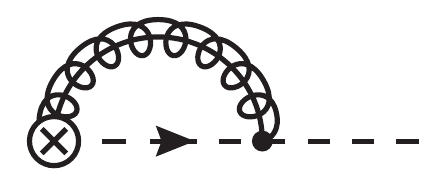}
\hspace{5ex}%
\includegraphics[scale=0.55]{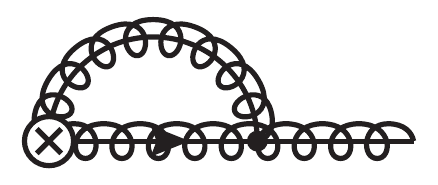}
}
\\
\subfloat[]{\label{fig:soft}%
\includegraphics[scale=0.45]{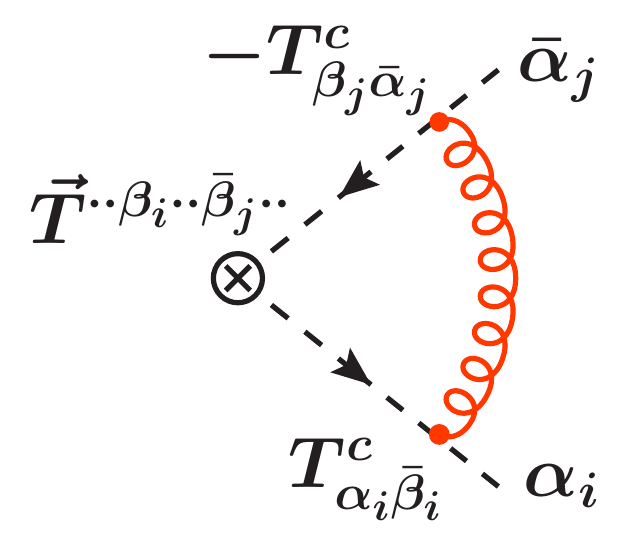}%
\includegraphics[scale=0.45]{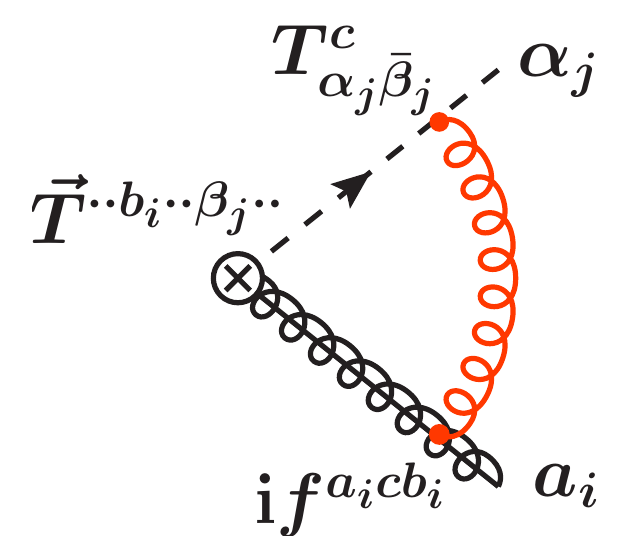}%
\includegraphics[scale=0.45]{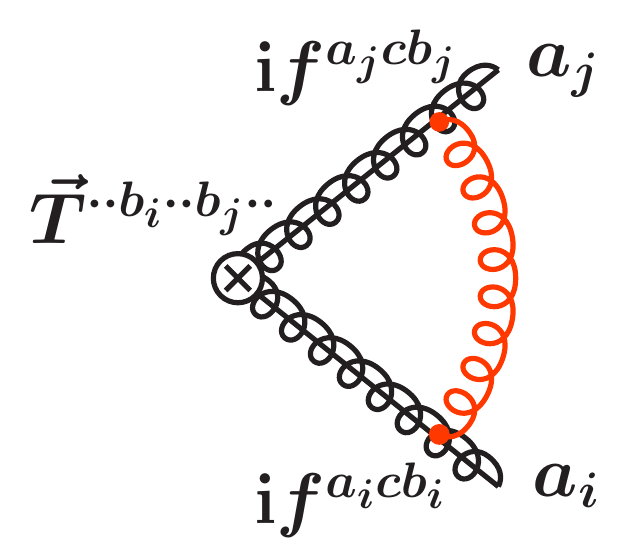}%
}
\caption{(a) Collinear one-loop diagrams. (b) Soft one-loop diagrams connecting two fields $i$ and $j$ in the operator.}
\label{fig:diagrams}
\end{figure}

The anomalous dimension $\hga_C(\mu)$ is process dependent. In this subsection, we derive its general form at one loop.
The anomalous dimension of the operators is determined from the UV divergences in the effective theory. The relevant one-loop diagrams in SCET are shown in \fig{diagrams}. In pure dimensional regularization the UV and IR divergences cancel such that the bare results for the loop diagrams vanish. To extract the UV divergences, we regulate the IR divergences by taking the external particles off shell with $p_i^2 = p_{i\perp}^2 \neq 0$.

Since all fields in the operators correspond to distinct collinear directions, the collinear loop diagrams in \fig{coll} only involve one external line at a time. Different external lines can only interact through the exchange of a soft gluon, shown by the diagrams in \fig{soft}.

When expressing our results, we use the notation [see \eq{Lij_def}]
\begin{align}
L_{{i\perp}} &= \ln \Bigl(-\frac{p_{i\perp}^2}{\mu^2} \Bigr)
\,,\qquad
L_{ij} = \ln \Bigl(-\frac{s_{ij}}{\mu^2} - \img 0\Bigr)
\,,\end{align}
where $s_{ij} = 2p_i\cdot p_j$.

First, we recall the wave function renormalization constants. In Feynman gauge at one loop,
\begin{align} \label{eq:Zi}
Z_\xi &= 1 - \frac{\al_s}{4\pi} \frac{1}{\eps} (C_F + \cdots)
\,,\nn\\
Z_A &= 1 + \frac{\al_s}{4\pi} \frac{1}{\eps} (\beta_0 - 2C_A + \cdots)
\,,\end{align}
where $\beta_0 = 11/3\, C_A - 4/3\, T_F n_f$ is the one-loop beta function coefficient [see \eq{betafunction}],
and $n_f$ is the number of considered quark flavors. Here and below, the ellipses denote possible UV-finite terms, which are irrelevant
for our discussion here. (Using the on-shell scheme for wave function renormalization, the $Z_i$ contain UV-finite pieces,
see \app{IRdiv}.)

The collinear diagrams in \fig{coll} contribute
\begin{align} \label{eq:Ic}
I_c^q = I_c^{\bar{q}}
&= \frac{\al_s C_F}{4\pi} \Bigl(\frac{2}{\eps^2} + \frac{2}{\eps} - \fr{2}{\eps} L_{{i\perp}} + \cdots \Bigr)
\vev{\Op^\dagger}^\tree
\,,\nn\\
I_c^g
&= \frac{\al_s C_A}{4\pi} \Bigl(\frac{2}{\eps^2} + \frac{1}{\eps} - \frac{2}{\eps} L_{{i\perp}}
+ \cdots \Bigr)
\vev{\Op^\dagger}^\tree
\,,\end{align}
where $I_c^i$ denotes the result of the diagram for an external leg of type $i$, either quark or gluon. 

The soft diagrams in \fig{soft} differ from each other only in their color structure. The result of the diagram connecting particles $i$ and $j$ (with $i\neq j$) is given by
\begin{equation} \label{eq:Is}
I_s^{ij}
= \frac{\al_s}{4\pi} \Bigl( \frac{2}{\eps^2} + \frac{2}{\eps} L_{{ij}} - \frac{2}{\eps} L_{{i\perp}}
  - \frac{2}{\eps} L_{{j\perp}}  + \cdots \Bigr)\, \vev{\Op^\dagger}^\tree\, \hatt^c_i\, \hatt^c_j
\,,\end{equation}
where $\hatt^c_i$ and $\hatt^c_j$ are matrices in color space. From \eqs{Ic}{Is} we see explicitly that the operators only mix with respect to the color structure, with no mixing between operators with distinct helicities.

The action of the matrix $\hatt^c_i$ on the color space is to insert a generator acting on the color index of the $i$th particle, i.e.,
\begin{align}
(\vT\, \hatt^c_i)^{\dotsb\alpha_i\dotsb} &= T^c_{\alpha_i\bbeta_i}\,\vT^{\dotsb\beta_i\dotsb}
\,,\nn\\
(\vT\, \hatt^c_i)^{\dotsb\balpha_i\dotsb} &= - \vT^{\dotsb\bbeta_i\dotsb}\,T^c_{\beta_i\balpha_i}
\,,\nn\\
(\vT\, \hatt^c_i)^{\dotsb a_i\dotsb} &= \img f^{a_i c b_i}\, \vT^{\dotsb b_i\dotsb}
\,,\end{align}
for quarks, antiquarks, and gluons, respectively. Our $\hatt^c_i$ is identical to what is usually denoted as $\mathbf{T}_i$ in the notation of Refs.~\cite{Catani:1996jh,Catani:1996vz}.

To give an explicit example, consider $gg\,q\bar{q}$. Then, for quark $i=3$ and antiquark $j=4$ we have
\begin{align}
\Op^\dagger\, \hatt^c_3\, \hatt^c_4
&= O^{a_1 a_2\balpha_3\alpha_4}\,(\vT\, \hatt^c_3\, \hatt^c_4)^{a_1 a_2\alpha_3\balpha_4}
\nn\\
&= O^{a_1 a_2\balpha_3\alpha_4}\, T^c_{\alpha_3 \bbeta_3}\,(-T^c_{\beta_4\balpha_4})
\vT^{a_1 a_2 \beta_3\bbeta_4}
\,,\end{align}
while for gluon $i=1$ and quark $j=3$,
\begin{equation}
\Op^\dagger\, \hatt^c_1\, \hatt^c_3
= O^{a_1 a_2\balpha_3\alpha_4}\, \img f^{a_1 c b_1} T^c_{\alpha_3 \bbeta_3} \vT^{ b_1 a_2 \beta_3\balpha_4}
\,.\end{equation}
Plugging in the explicit basis in \eq{ggqq_color} and using the relations in \app{color}, we can rewrite the resulting color structures above in terms of the basis in \eq{ggqq_color}, which yields 
\begin{align}
\hatt^c_3\, \hatt^c_4
&= - \begin{pmatrix}
   C_F - \tfrac{1}{2}C_A & 0 & 0 \\
   0 & C_F - \tfrac{1}{2}C_A & 0  \\
   T_F & T_F & C_F
\end{pmatrix}
,\nn\\
\hatt^c_1\, \hatt^c_3
&= - \begin{pmatrix}
   \tfrac{1}{2}C_A & 0 & T_F\\
   0 & 0 & -T_F \\
 0 & -T_F & 0
\end{pmatrix}
.\end{align}
The other combinations are computed analogously.

In general, one can easily see that for $i = j$
\begin{equation} \label{eq:titi}
\vT^{ a_1\dotsb\alpha_n}\,\hatt^c_i\, \hatt^c_i = C_i\, \vT^{ a_1\dotsb\alpha_n}
\,,\end{equation}
where $C_i = C_F$ for quarks and $C_i = C_A$ for gluons. By construction, the color basis $\vT^{ a_1\dotsb\alpha_n}$ conserves color, because each index corresponds to an external particle. Since $\hatt_i^c$ measures the color charge of the $i$th particle, color conservation implies
\begin{equation} \label{eq:sumti}
\vT^{a_1\dotsb\alpha_n} \Bigl(\sum_{i=1}^n \hatt_i^c\Bigr) = 0
\,.\end{equation}
As a simple example, consider $gq\bar{q}$ for which $\vT^{ a_1\alpha_2\balpha_3} \equiv T^{a_1}_{\alpha_2\balpha_3}$. In this case, \eq{sumti} gives
\begin{align}
&\img f^{a_1 c b_1} T^{b_1}_{\alpha_2\balpha_3} + T^c_{\alpha_2\bbeta_2} T^{a_1}_{\beta_2\balpha_3} - T^{a_1}_{\alpha_2\bbeta_3} T^c_{\beta_3\balpha_3}
\nn\\ & \quad
= \bigl(\img f^{a_1 c b_1} T^{b_1} + [T^c, T^{a_1}]\bigr)_{\alpha_2\balpha_3} = 0
\,.\end{align}

The total bare one-loop matrix element is given by summing \eq{Ic} for each external particle and \eq{Is} for each pair of distinct particles. The infrared logarithms $L_{{i\perp}}$ have to drop out in the sum of all UV-divergent contributions. To see that this is indeed the case, we can use \eq{titi} to rewrite the collinear contributions. Then, the sum of all $L_{{i\perp}}$ terms is proportional to
\begin{align} \label{eq:Lp2cancel}
&\vev{\Op^\dagger}^\tree \Bigr[\sum_i L_{{i\perp}} \hatt_i^c\,\hatt_i^c + \sum_{i < j}(L_{{i\perp}} + L_{{j\perp}})\, \hatt_i^c\, \hatt_j^c \Bigr]
\nn\\ & \quad
= \vev{\Op^\dagger}^\tree \Bigr(\sum_i L_{{i\perp}} \hatt_i^c\,\hatt_i^c + \sum_{i\neq j} L_{{i\perp}} \hatt_i^c\,\hatt_j^c \Bigr)
\nn\\ & \quad
= \vev{\Op^\dagger}^\tree \Bigl(\sum_i L_{{i\perp}} \hatt_i^c \Bigr)\Bigl(\sum_j\hatt_j^c\Bigr)
= 0
\,,\end{align}
where in the last step we used \eq{sumti}. For the same reason the $1/\eps^2$ poles in the soft diagrams cancel against half of the $1/\eps^2$ poles in the collinear diagrams. The remaining UV-divergent part of the matrix element is given by
\begin{align}\label{eq:Dmatrix}
\vev{\Op^\dagger}^\tree \hD
&= \vev{\Op^\dagger}^\tree\ \frac{\al_s}{4\pi} \biggl[n_g C_A \Bigl(\frac{1}{\eps^2} + \frac{1}{\eps}\Bigr)
\nn\\ &\quad
+ n_q C_F \Bigl(\frac{1}{\eps^2} + \frac{2}{\eps} \Bigr)
- \frac{2}{\eps} \hDe(\mu^2) \biggr]
\,,\end{align}
where the color mixing matrix is given by
\begin{equation} \label{eq:hDe_explicit}
\hDe(\mu^2) = -\sum_{i<j} \hatt_i^c\,\hatt_j^c\, L_{ij}
\,.\end{equation}
Combining this result with the identities in \eqs{titi}{sumti}, we can easily check that the $\mu$ dependence of $\hDe(\mu^2)$ is as in \eq{hDe_mudep}:
\begin{align}
\hDe(\mu^2) - \hDe(\mu_0^2)
&= - 2\sum_{i<j} \hatt_i^c\,\hatt_j^c\,\ln\Bigl(\frac{\mu_0}{\mu}\Bigr)
= \sum_i \hatt_i^c\,\hatt_i^c\,\ln\Bigl(\frac{\mu_0}{\mu}\Bigr)
 \nn\\
&= \id (n_g C_A + n_q C_F) \ln\Bigl(\frac{\mu_0}{\mu}\Bigr)
\,.\end{align}

We can now compute the anomalous dimension of the operators. From \eqs{Z_O}{Dmatrix}, we find at one loop
\begin{equation}\label{z_o}
\hZ_C = \id - \hD - \id \Bigl[\frac{n_g}{2}\,(Z_A-1) + \frac{n_q}{2}\,(Z_\xi-1) \Bigr]
\,,\end{equation}
which using \eq{ga_O} yields the one-loop anomalous dimension
\begin{equation} \label{eq:anomdim}
\hga_C(\mu)
= \frac{\al_s(\mu)}{4\pi} \bigl[4\hDe(\mu^2) - \id(n_g \beta_0 + n_q\,3 C_F) \bigr]
.\end{equation}
The coefficient of $4$ in front of $\hDe(\mu^2)$ is the one-loop cusp anomalous dimension coefficient [see \eq{cusp}]. The remaining terms determine the noncusp $\hga(\alpha_s)$ in \eq{gaO_general} at one loop,%
\begin{equation}
\hga(\alpha_s)
= - \frac{\al_s}{4\pi}\,(n_g \beta_0 + n_q\,3 C_F)\,\id
\,.\end{equation}

\subsection{Mixing Matrices}
\label{subsec:mixing}

In this section, we give explicit expressions for the mixing matrices for the color bases used in \secss{higgs}{vec}{pp}. For simplicity, we only give explicit expressions for up to four partons, but allow for additional colorless particles, such as a Higgs or vector boson. The matrices are straightforward to evaluate using the color relations in \app{color}, but become rather lengthy for more than four partons, due to the large number of allowed color structures, and are more easily evaluated in an automated way (see for example Ref.~\cite{Gerwick:2014gya}). For convenience, we introduce the following shorthand notation for sums and differences of logarithms $L_{ij}$,
\begin{align}
L_{ij \cdot kl \cdot \ldots} &= L_{ij} + L_{kl} + \ldots
\,, \nn \\
L_{ij \cdot \ldots/(kl\cdot \ldots)} &= (L_{ij \cdot \ldots}) -  (L_{kl\cdot\ldots})
\,,\end{align}
with $L_{ij} = \ln(-s_{ij}/\mu^2 -\img0)$ as defined in \eq{Lij_def}.

\subsubsection{Pure Gluon Mixing Matrices}

For $gg$ and $ggg$ in the bases used in \eq{H0_color} and \eqs{H1_color}{Z1g_color}, we have
\begin{align}\label{eq:mix23gluons}
\hDe_{gg}(\mu^2)
= C_A\, L_{12}
\,, \qquad
\hDe_{ggg}(\mu^2)
= \frac{1}{2}C_A\, L_{12 \cdot 13 \cdot 23}
\begin{pmatrix}
 1 & 0 \\
 0 & 1
\end{pmatrix}
.\end{align}
For $gggg$ in the basis used in \eqs{ggggH_color}{gggg_color}, we have
\begin{align} \label{eq:mix4gluons}
&\hDe_{gggg}(\mu^2)
\nn \\ & 
= \begin{pmatrix}
\frac{1}{2} C_A L_{12\cdot14\cdot23\cdot34} & 0 & 0 & 2T_F L_{14\cdot23/(13\cdot24)}  & 0 & 2T_F L_{12\cdot34/(13\cdot24)} \\
  0 & \frac{1}{2} C_A L_{12\cdot13\cdot24\cdot34} & 0 & 2T_F L_{13\cdot24/(14\cdot23)} & 2T_F L_{12\cdot34/(14\cdot23)} & 0 \\
  0 & 0 & \frac{1}{2} C_A L_{13\cdot14\cdot23\cdot24} & 0 & 2T_F L_{14\cdot23/(12\cdot34)} & 2T_F L_{13\cdot24/(12\cdot34)} \\
  T_F L_{12\cdot 34/(13 \cdot24)} & T_F L_{12\cdot34/(14\cdot23)} & 0 & C_A L_{12\cdot34} & 0 & 0 \\
  0 & T_F L_{13\cdot24/(14\cdot23)} & T_F L_{13\cdot 24/(12\cdot34)} & 0 & C_A L_{13\cdot 24} & 0 \\
  T_F L_{14\cdot23/(13\cdot24)} & 0 & T_F L_{14\cdot23/(12\cdot34)} & 0 & 0 & C_A L_{14\cdot23}
\end{pmatrix}
.\end{align}
For our color bases formed from multitrace color structures, the structure of the mixing matrices is simple. Since the mixing matrices are determined by single gluon exchange, cyclicity is maintained, and all that can occur in the mixing is that a single trace splits into two or two traces recombine into one. For example, the color structure $\tr[T^a T^b T^c T^d]$ can only mix with
\begin{align}
\tr[T^a T^b T^c T^d]
\,,\qquad
\tr[T^a T^b]\, \tr[T^c T^d]
\,,\qquad \text{and} \qquad
\tr[T^d T^a]\, \tr[T^b T^c]
\,.\end{align}
Therefore, although the mixing matrices quickly get large as the number of color structures grows, their structure remains relatively simple. (An alternative approach to the organization of the anomalous dimensions for a large number of partons has been given in Ref.~\cite{Platzer:2013fha}.) For the dijet case, i.e., in the absence of additional colorless particles, the kinematics simplifies to
\begin{equation} \label{eq:dijetkin}
s = s_{12} = s_{34}
\,,\qquad
t = s_{13} = s_{24}
\,,\qquad
u = s_{14} = s_{23}
\,,\end{equation}
and these matrices were given in Ref.~\cite{Kidonakis:1998nf}, which also gives their eigenvectors and eigenvalues.

\subsubsection{Mixing Matrices Involving $q \bar q$ Pairs}

For $q\bar q$ and $gq\bar q$ in the bases used in \eq{Z0_color} and \eqs{H1_color}{Z1q_color}, we have
\begin{align}\label{eq:mixquarks1}
\hDe_{q\bq}(\mu^2)
= C_F\, L_{12}
\,, \qquad
\hDe_{g\,q\bq}(\mu^2)
= \frac{1}{2}\Big[C_A L_{12 \cdot 13} + (2C_F -C_A) L_{23}\Big]\,.
\end{align}
For $q\bar q q'\bar q'$ in the basis used in \eqss{qqqqH_color}{qqqq_color}{qqqqV_color}, we have
\begin{align} \label{eq:mixing_qqqq}
\hDe_{q\bq\, q\bq}(\mu^2)
= \hDe_{q\bar{q}\, q'\bq'}(\mu^2)
= \begin{pmatrix}
  C_F\, L_{14 \cdot 23} + (C_F - \frac{1}{2}C_A)\, L_{12\cdot 34/(13\cdot24)} & T_F\, L_{14\cdot23/(13\cdot24)} \\
  T_F\, L_{12\cdot34/(13\cdot24)} & C_F\, L_{12\cdot34} + (C_F - \frac{1}{2}C_A)\, L_{14\cdot23/(13\cdot24)}
\end{pmatrix}
.\end{align}
For $gg q\bar q$ in the basis used in \eqss{ggqqH_color}{ggqq_color}{ggqqV_color}, we have 
\begin{align} \label{eq:mixing_ggqq}
\hDe_{gg\,q\bar q}(\mu^2)
&= \begin{pmatrix}
   \frac{1}{2} C_A\, L_{12\cdot13\cdot24} +(C_F - \frac{1}{2} C_A)\, L_{34} & 0 & T_F\, L_{13\cdot24/(14\cdot23)} \\
   0 & \frac{1}{2} C_A\, L_{12\cdot14\cdot23} + (C_F - \frac{1}{2} C_A)\, L_{34} & T_F\, L_{14\cdot23/(13\cdot24)}  \\
   T_F\, L_{12\cdot34/(14\cdot23)} & T_F\, L_{12\cdot34/(13\cdot24)} &  C_A\, L_{12} +  C_F\, L_{34}
   \end{pmatrix}
.\end{align}
Again, these simplify in the dijet case, for which they were given along with their eigenvectors and eigenvalues in Ref.~\cite{Kidonakis:1998nf}.

\subsection{Soft Function Evolution}\label{sec:soft_evolution}

In this section, we review the renormalization group evolution of the soft function, focusing on our use of the color basis notation of \subsec{color} for nonorthogonal bases.  We will consider the particular case of the $N$-jettiness event shape \cite{Stewart:2010tn}, which allows for a definition of exclusive $N$-jet production with a factorization theorem of the form of \eq{sigma}.

The color mixing matrices of the previous section are in general complex valued for physical kinematics. For a physical channel, some of the appearing $s_{ij}$ are positive, giving rise to imaginary terms from the logarithms, as in \eq{Lij_def}. Since the cross section is real, these imaginary terms generated by the renormalization group evolution must drop out of the final result. We start by describing the properties of the soft function that ensure that this is the case.

Recall that the hard function $\hH_\kappa$ for a particular partonic channel $\kappa$ has its color indices contracted with those of the soft function. Explicitly,
\begin{align}\label{eq:soft_hard_trace}
  \tr(\hH_\kappa \hS_\kappa)
  &= H_\kappa^{a_1 \cdots \al_n b_1 \cdots \bt_n} S_\kappa^{b_1 \cdots \bt_n a_1 \cdots \al_n}
  \\ \nn
  & =\sum_{\{ \lambda_i \}}
  \Bigl[C_{\lambda_1 \cdot\cdot(\cdot\cdot \lambda_n )}^{b_1 \cdots \bt_n}\Bigr]^*
  S_\kappa^{b_1 \cdots \bt_n a_1 \cdots \al_n} C_{\lambda_1 \cdot\cdot(\cdot\cdot \lambda_n )}^{a_1 \cdots \al_n} \,.
\end{align}
The soft function is defined as a vacuum matrix element of a product of soft Wilson lines $\widehat{Y}$ as
\begin{align}\label{eq:softfunction_def}
\hS_\kappa(M,\{n_i\})= \Big\langle 0 \Big |  {\rm\bar T}\,  \widehat{Y}^{ \dagger}(\{n_i\})\ \delta(M-\hat{M}\,)\ {\rm T}\, \widehat{Y} (\{n_i\}) \Big |  0 \Big \rangle ,
\end{align}
where $\widehat{Y}(\{n_i\})$ is a product of soft Wilson lines in the $n_i$ directions. It is a matrix in color space, and $\widehat{Y}^\dagger$ is its Hermitian conjugate. Here T and ${\rm \bar T}$ denote time ordering and antitime ordering respectively.  The matrices $\widehat{Y}$ and $\widehat{Y}^\dagger$ are multiplied with each other, i.e.\ one of the color indices of the corresponding Wilson lines are contracted, and the external indices correspond to $b_1 \cdots \bt_n$ and $a_1 \cdots \al_n$, respectively.  Thus, for example $\widehat{Y}^\dagger \widehat{Y} = \delta^{a_1 b_1} \cdots \delta^{\alpha_n\beta_n}$.
The dependence of the soft function on the particular measurement, as well as the details of the jet algorithm, are encoded in the measurement function $\hat{M}$, whose precise form is not relevant for the current discussion.

From the definition of the soft function in \eq{softfunction_def} we see that it is Hermitian, namely $(S_\kappa^{b_1 \cdots \bt_n a_1 \cdots \al_n})^* = S_\kappa^{a_1 \cdots \al_n b_1 \cdots \bt_n}$. In abstract notation, this means $\hS_\kappa^\dagger = \hS_\kappa$, which implies that the product $\vC^\dagger \widehat S_\kappa \vC$ appearing in the cross section is real, so imaginary terms that appear in the Wilson coefficients due to renormalization group evolution drop out in the final cross section.

While this argument is trivial in a basis independent form, it is important to emphasize that in a nonorthogonal basis it takes a slightly more complicated form. As discussed in \subsec{color}, in a specific nonorthogonal color basis, \eq{soft_hard_trace} takes the form
$\vC^\dagger\, \hS_\kappa \vC = \vC^{*T} \hT\, \hS_\kappa \vC$ as in \eq{trHS},
where the matrix $\hT$ is defined in \eq{hatT_def}.
Similarly, the matrix representation of $\hS_\kappa$ is not Hermitian with respect to the naive conjugate transpose of its components. Instead, the condition on the reality of the cross section is given by [see \eq{def_daggermatrix}]
\begin{align}
\hS_\kappa = \hS^\dagger_\kappa = \hT^{-1}\, \hS_\kappa^{*T}\, \hT
\,.\end{align}

The invariance of the cross section under the RGE
\begin{align}\label{eq:inv_xs}
\mu \frac{\df}{\df\mu} \sigma_N =0\,,
\end{align}
implies relations between the anomalous dimensions of the SCET functions appearing in the factorization theorem of \eq{sigma}. In particular, it allows the anomalous dimension of the soft function to be determined from the anomalous dimensions of the Wilson coefficients, along with the anomalous dimensions of the beam and jet functions. The anomalous dimensions of the jet and beam functions are proportional to the color-space identity. The anomalous dimensions of the beam and jet functions appearing in the $N$-jettiness factorization theorem are equal to all orders in perturbation theory \cite{Stewart:2010qs} allowing us to use only the jet function anomalous dimension in the following discussion.
Renormalization group consistency then implies that the contributions of the soft function anomalous dimension not proportional to the identity, including the color off-diagonal components, are completely determined by the anomalous dimensions of the Wilson coefficients.

The soft function for $N$-jettiness can be written in the general form of \eq{softfunction_def}, but with an explicit measurement function
\begin{align}
&\hS_\kappa(k_a, k_b, k_1,\ldots, k_N, \{ n_i \})
\\ & \qquad
=\Big\langle 0 \Big |  {\rm\bar T}\, \widehat{Y}^{ \dagger}( \left\{   n_i\right\})\, \prod\limits_i \delta(k_i-\mathcal{\hat T}_i\,)\  {\rm T}\, \widehat{Y} (\{n_i\}) \Big |  0 \Big \rangle\,. \nonumber
\end{align}
Here $\mathcal{\hat T}_i$ picks out the contribution to the $N$-jettiness observable from the momentum region $i$, whose precise definition can be found in Ref.~\cite{Jouttenus:2011wh}. The soft function for $N$-jettiness was first presented to NLO in Ref.~\cite{Jouttenus:2011wh}, and more recently analyzed to NNLO in Ref.~\cite{Boughezal:2015eha}.

The all-orders structure of the renormalization group evolution for the soft function can be derived from \eq{inv_xs}, and is given by \cite{Kelley:2010fn, Jouttenus:2011wh}
\begin{align}\label{eq:soft_evolution}
&\mu \frac{\df}{\df\mu} \hS_\kappa (\{k_i\},\mu)
\nn\\ & \qquad
= \int \biggl[  \prod\limits_i \df k'_i \biggr] \frac{1}{2} \Bigl[   \hga_S (\{ k_i - k'_i \}, \mu)\, \hS_\kappa (\{k_i'\},\mu)
\nn\\ & \qquad\qquad
+  \hS_\kappa (\{k_i'\},\mu)\, \hga_S^\dagger (\{ k_i - k'_i \}, \mu)  \Bigr]
\,.\end{align}
The soft anomalous dimension $\hga_S$, and its conjugate $\hga_S^\dagger$, are given in terms of the anomalous dimension $\gamma_J$ of the jet function and the anomalous dimension of the Wilson coefficients, $\hga_C$ defined in \eqs{RGE}{ga_O}, as
\begin{align}\label{eq:soft_anom}
\hga_S (\{k_i\},\mu)
&= -\id \sum\limits_i Q_i\, \gamma_J^i(Q_i k_i, \mu) \prod\limits_{j\neq i} \delta(k_j)
\nn \\ & \quad
-2 \hga_C^\dagger (\mu) \prod\limits_i \delta(k_i)
\,.\end{align}
(Here, the $Q_i$ are related to the precise $N$-jettiness definition, see Ref.~\cite{Jouttenus:2011wh}.)
The Hermitian conjugates of $\hga_C$ and $\hga_S$ above again refer to the abstract Hermitian conjugate in color space. In a nonorthogonal color basis, they are given in terms of the complex conjugate transpose components according to \eq{def_daggermatrix} as
\begin{equation}
\hga_C^\dagger = \hT^{-1}\, \hga_C^{*T}\, \hT
\,, \qquad
\hga_S^\dagger = \hT^{-1}\, \hga_S^{*T}\, \hT
\,.\end{equation}

%

\section{Conclusions}
\label{sec:conclusions}

In this chapter, we have presented a helicity operator approach to SCET. Helicities are naturally defined with respect to the external lightlike reference vectors specifying the jet directions in the effective theory, eliminating the need to consider complicated Lorentz and gamma matrix structures in the operator basis. The helicity operators correspond directly to physical states of definite helicity and color, which when combined with color organization techniques, greatly simplifies the construction of a minimal operator basis. Furthermore, the helicity operators are automatically crossing symmetric, and make manifest parity and charge conjugation symmetries, making it simple to determine relations amongst Wilson coefficients.

We demonstrated the utility of the helicity operator approach by explicitly constructing the basis valid to all orders in perturbation theory for a number of key processes at the LHC involving jets, and then determining the matching coefficients. In particular we considered $pp\to H + 0,1$ jets, $pp\to W/Z/\gamma + 0,1$ jets, and $pp\to 2$ jets at next-to-leading order, and $pp\to H + 2$ jets, $pp\to W/Z/\gamma + 2$ jets, and $pp\to 3$ jets at leading order.  We also discussed the dependence of this matching on the regularization scheme, considering schemes with helicities in $4$ and $d$ dimensions. An important and well-known simplification of the SCET approach is that when dimensional regularization is used for both IR and UV divergences, all loop graphs in the effective theory are scaleless, and thus vanish. As a result, the hard SCET Wilson coefficients in the $\overline{\rm MS}$ scheme, determined from matching QCD to SCET, are given directly by the IR-finite parts of color-ordered helicity amplitudes, defined using \eq{matching_general}. The use of our helicity operator basis therefore makes it simple to combine analytic resummation in SCET with fixed-order calculations of helicity amplitudes.

The all-orders structure for the renormalization group evolution of the helicity operator basis was discussed in detail. At leading power, distinct helicity structures do not mix, with renormalization group evolution causing mixing only in color space. This feature is made manifest at the level of the SCET Lagrangian due to the expansion in the soft and collinear limits. Subtleties associated with the use of nonorthogonal color bases were carefully treated, and expressions for the color sum matrix $\widehat T$ are given for the used color bases for all processes considered in the chapter. Explicit results are also given for the one-loop mixing matrices describing the renormalization group evolution in color space for the case of $pp\to \text{up to } 2$ jets with an arbitrary number of uncolored external particles and in a manifestly crossing symmetric form.

Combining the methods of this chapter with known expressions for jet, beam, and soft functions for particular exclusive jet cross sections, or jet shapes/observables, should facilitate analytic resummation for a large number of processes for which fixed-order amplitudes are known, or are soon to be calculated.

%% file: chap5.tex
\def\cA{\mathcal{A}}
\def\cB{\mathcal{B}}
\def\cC{\mathcal{C}} 
\def\cD{\mathcal{D}}
\def\cE{\mathcal{E}}
\def\cF{\mathcal{F}}
\def\cG{\mathcal{G}}
\def\cH{\mathcal{H}}
\def\cI{\mathcal{I}}
\def\cJ{\mathcal{J}}
\def\cK{\mathcal{K}}
\def\cL{\mathcal{L}}
\def\cM{\mathcal{M}}
\def\cN{\mathcal{N}}
\def\cO{\mathcal{O}}
\def\cS{\mathcal{S}}
\def\cT{\mathcal{T}}
\def\cP{\mathcal{P}}
\def\cJ{\mathcal{J}}
\def\cW{\mathcal{W}}
\def\cX{\mathcal{X}}
\def\cY{\mathcal{Y}}
\def\cZ{\mathcal{Z}}
\def\tr{{\rm tr}}

\def\bA{{\bf A}}
\def\bB{{\bf B}}
\def\bC{{\bf C}}
\def\bD{{\bf D}}
\def\bE{{\bf E}}
\def\bF{{\bf F}}
\def\bG{{\bf G}}
\def\bH{{\bf H}}
\def\bI{{\bf I}}
\def\bJ{{\bf J}}
\def\bK{{\bf K}}
\def\bR{{\bf R}}
\def\bS{{\bf S}}
\def\bV{{\bf V}}

\def\ba{{\bf a}}
\def\bb{{\bf b}}
\def\bc{{\bf c}}
\def\bg{{\bf g}}
\def\bk{{\bf k}}
\def\bp{{\bf p}}
\def\bq{{\bf q}}
\def\bn{{\bf n}}
\def\bs{{\bf s}}
\def\bt{{\bf t}}
\def\br{{\bf r}}
\def\bv{{\bf v}}

\def\vC{\vec C}
\def\hpo{\hat{p}_1}
\def\hpt{\hat{p}_2}
\def\hpos{\cancel{\hat{p}_1}}
\def\hpts{\cancel{\hat{p}_2}}
\def\hpob{\bar{\hat{p}}_1}
\def\hptb{\bar{\hat{p}}_2}
\def\hpobs{\cancel{\bar{\hat{p}}_1}}
\def\hptbs{\cancel{\bar{\hat{p}}_2}}
\def\imply{\quad\Longrightarrow\quad}
\def\And{\quad {\rm and} \quad}
\def\Or{\quad {\rm or} \quad}
\def\For{\quad {\rm for} \quad}
\def\Where{\quad {\rm where} \quad}
\def\Iff{\quad \iff \quad}
\def\Forall{\quad \text{for all }}
\def\dg{\dagger}
\def\Gc{\Gamma_{\rm cusp}}
\def\dxyzint{\int_0^1 dx\,dy\,dz\, 2\delta(x+y+z-1)\,}
\def\slash{\!\!\!/}
\newcommand{\dint}[1]{\int \frac{d^d #1}{(2\pi)^d}\,}
\newcommand{\dintmu}[1]{\mu^{2\epsilon}\int \frac{d^d #1}{(2\pi)^d}\,}
\newcommand{\fint}[1]{\int \frac{d^4 #1}{(2\pi)^4}\,}
\newcommand{\cn}[1]{\cancel{#1}}

\DeclareRobustCommand{\Sec}[1]{Sec.~\ref{#1}}
\DeclareRobustCommand{\Secs}[2]{Secs.~\ref{#1} and \ref{#2}}
\DeclareRobustCommand{\Secsdash}[2]{Secs.~\ref{#1} - \ref{#2}}
\DeclareRobustCommand{\App}[1]{App.~\ref{#1}}
\DeclareRobustCommand{\Tab}[1]{Table~\ref{#1}}
\DeclareRobustCommand{\Tabs}[2]{Tables~\ref{#1} and \ref{#2}}
\DeclareRobustCommand{\Fig}[1]{Fig.~\ref{#1}}
\DeclareRobustCommand{\Figs}[2]{Figs.~\ref{#1} and \ref{#2}}
\DeclareRobustCommand{\Eq}[1]{Eq.~(\ref{#1})}
\DeclareRobustCommand{\Eqs}[2]{Eqs.~(\ref{#1}) and (\ref{#2})}
\DeclareRobustCommand{\Ref}[1]{Ref.~\cite{#1}}
\DeclareRobustCommand{\Refs}[1]{Refs.~\cite{#1}}

\def\be{\begin{equation}}
\def\ee{\end{equation}}
\def\zhat{\hat{e}_z}
\def\xhat{\hat{e}_x}
\def\yhat{\hat{e}_y}
\def\phihat{\hat{e}_\phi}
\def\rhat{\hat{e}_r}
\def\thetahat{\hat{e}_\theta}
\def\rhohat{\hat{e}_\rho}
\def\musteq{\overset{!}{=}}

\def\l{\langle}
\def\r{\rangle}
\def\a{\alpha}
\def\ad{{\dot\alpha}}
\def\b{\beta}
\def\bd{{\dot\beta}}
\def\g{\gamma}
\def\gd{{\dot\gamma}}
\def\d{\delta}
\def\dd{{\dot\delta}}
\def\LO{\text{LO}}
\def\RPI{\text{RPI}}

\def\plr{\overleftrightarrow{\partial}}

\newcommand{\matrixxx}[9]{ \left(\begin{array}{ccc} #1&#2&#3 #4&#5&#6 #7&#8&#9 \end{array}\right) }
\newcommand{\vectorrr}[3]{ \left(\begin{array}{c} #1 #2 #3 \end{array}\right) }
\newcommand{\matrixx}[4]{ \left(\begin{array}{cc} #1&#2 #3&#4 \end{array}\right) }
\newcommand{\vectorr}[2]{ \left(\begin{array}{c} #1 #2 \end{array}\right) }

\newcommand{\disc}[1]{{\color{green}(#1)}}
\newcommand{\tcut}{\cT_{cut}^1}

\newcommand{\Pythia}{\textsc{Pythia}\xspace}
\newcommand{\Herwig}{\textsc{Herwig}\xspace}
\newcommand{\MCNLO}{\textsc{MC@NLO}\xspace}
\newcommand{\POWHEG}{\textsc{POWHEG}\xspace}
\newcommand{\Fehip}{\textsc{FEHiP}\xspace}
\newcommand{\FastJet}{\textsc{FastJet}\xspace}

\renewcommand{\arraystretch}{1.05}
\arraycolsep 2pt

\allowdisplaybreaks[3]

\renewcommand{\topfraction}{1}
\renewcommand{\textfraction}{0.0}
\setcounter{topnumber}{3}


%
%


%
%
%
%
%
%

\chapter{Jet Vetoes Interfering with $H\rightarrow WW$}\label{chap:WW}

Far off-shell Higgs production in $H \rightarrow WW,ZZ$, is a particularly powerful probe of Higgs properties, allowing one to disentangle Higgs width and coupling information unavailable in on-shell rate measurements. These measurements require an understanding of the cross section in the far off-shell region in the presence of realistic experimental cuts. We analytically study the effect of a $p_T$ jet veto on far off-shell cross sections, including signal-background interference, by utilizing hard functions in the soft collinear effective theory that are differential in the decay products of the $W/Z$.  Summing large logarithms of $\sqrt{\hat s}/\ptveto$, we find that the jet veto induces a strong dependence on the partonic centre of mass energy, $\sqrt{\hat s}$, and modifies distributions in $\sqrt{\hat s}$ or $M_T$. The example of $gg\rightarrow H \rightarrow WW$ is used to demonstrate these effects at next to leading log order. We also discuss the importance of jet vetoes and jet binning for the recent program to extract Higgs couplings and widths from far off-shell cross sections.




\section{Introduction}
\label{sec:intro}

With the recent discovery of a boson resembling a $126$ GeV Standard Model (SM) Higgs \cite{Aad:2012tfa,Chatrchyan:2012ufa}, a large program has begun to study in detail the properties of the observed particle \cite{Duhrssen:2004cv,Lafaye:2009vr,Bonnet:2011yx,Heinemeyer:2013tqa,Carmi:2012in,Carmi:2012yp,Djouadi:2012rh,Englert:2012wf,Klute:2012pu,Dobrescu:2012td,Plehn:2012iz,Corbett:2012ja,Belanger:2012gc,Farina:2012xp,Batell:2012ca,Espinosa:2012im,Espinosa:2012vu,Banerjee:2012xc,Barger:2012hv,Giardino:2012dp,Giardino:2012ww,Giardino:2013bma,Aad:2013wqa,Gainer:2013rxa,Belanger:2013xza,Belanger:2013kya,Ellis:2013lra,Cranmer:2013hia,Chen:2014pia}. Of fundamental interest are the couplings to SM particles and the total width of the observed boson, which is a sensitive probe of BSM physics  \cite{Djouadi:2005gj,Burgess:2000yq,Patt:2006fw,He:2011de,Raidal:2011xk,Englert:2011yb,Barbieri:2005ri}. Most studies have  focused on the extraction of Higgs properties from on-shell cross sections. In this case, the effect of jet vetoes and jet binning, which is required experimentally in many channels to reduce backgrounds, has been well studied theoretically \cite{Berger:2010xi,Banfi:2012yh,Banfi:2012jm,Becher:2012qa,Liu:2012sz,Tackmann:2012bt,Gangal:2013nxa,Becher:2013xia,Stewart:2013faa}. A jet veto, typically defined by requiring that there are no jets with $p_T \geq \ptveto$, introduces large logarithms, $\log(m_H/\ptveto)$, potentially invalidating the perturbative expansion, and requiring resummation for precise theoretical predictions. In this chapter, we analytically study the effect of an exclusive jet $p_T$-veto on off-shell particle production, resumming logarithms of $\sqrt{\hat s}/\ptveto$, where $\sqrt{\hat s}$ is the invariant mass of the off-shell particle. We use $gg\rightarrow H \rightarrow WW$ as an example to demonstrate these effects, although the formalism applies similarly to $gg\rightarrow H\rightarrow ZZ$ if a jet veto is imposed. We find that the off-shell cross section is significantly suppressed by a jet veto, and that the suppression has a strong dependence on $\sqrt{\hat s}$. This results in a modification of differential distributions in $\sqrt{\hat s}$, or any transverse mass variable, in the case that the invariant mass cannot be fully reconstructed. The jet veto also has an interesting interplay with signal-background interference effects, which typically contribute over a large range of $\sqrt{\hat s}$. We use two cases, $m_H=126$ GeV, and $m_H=600$ GeV, to demonstrate the effect of the jet veto on the signal-background interference in $gg \rightarrow H \rightarrow WW$.

There exist multiple motivations why it is important to have a thorough understanding of the far off-shell region in Higgs production, and the impact of a jet $p_T$ veto on this region. As has been emphasized in a number of recent chapters \cite{Dobrescu:2012td,Dixon:2013haa,Caola:2013yja,Campbell:2013una,Campbell:2013wga}, the separate extraction of the Higgs couplings and total width is not possible using only rate measurements for which the narrow width approximation (NWA) applies. In the NWA the cross section depends on the couplings and widths in the form
\be
\sigma^{\text{nwa}} \sim \frac{g_i^2 g_f^2}{\Gamma_H},
\ee
which is invariant under the rescaling
\be \label{eq:scaling_1}
 g_i \rightarrow \xi g_i,~~~~ \Gamma_H \rightarrow \xi^4 \Gamma_H,
\ee
  preventing their individual extraction from rate measurements alone.

The direct measurement of the width of the observed Higgs-like particle, expected to be close to its SM value of $\simeq 4$MeV, is difficult at the LHC, but is of fundamental interest as a window to new physics \cite{Djouadi:2005gj,Burgess:2000yq,Patt:2006fw,He:2011de,Raidal:2011xk,Englert:2011yb,Barbieri:2005ri}. It is also important for model independent measurements of the Higgs couplings. Proposals to measure the Higgs width include those that rely on assumptions on the nature of electroweak symmetry breaking \cite{Dobrescu:2012td}, direct searches for invisible Higgs decays \cite{CMS:2013bfa,CMS:2013yda,Bai:2011wz,Djouadi:2012zc,Englert:2012wf}, and a proposed measurement of the mass shift in $H\rightarrow \gamma \gamma$ relative to $H\rightarrow ZZ \rightarrow 4l$ caused by interference \cite{Dixon:2013haa}.

More recently, it has been proposed \cite{Caola:2013yja,Campbell:2013una,Campbell:2013wga} that the Higgs width can be bounded by considering the far off-shell production of the Higgs in decays to massive vector bosons. In this region there is a contribution from signal-background interference \cite{Campbell:2011cu,Kauer:2012ma,Passarino:2012ri,Kauer:2013qba}, and from far off-shell Higgs production \cite{Uhlemann:2008pm,Kauer:2013cga}. Far off-shell, the Higgs propagator is independent of $\Gamma_H$, giving rise to contributions to the total cross section that scale as
\be
\sigma^{\text{int}} \sim g_i g_f,~~~~ \sigma_H^{\text{off-shell}} \sim g_i^2 g_f^2,
\ee
for the signal-background interference and off-shell cross section respectively. The method proposed in \cite{Caola:2013yja} takes advantage of the fact that these components of the cross section scale differently than the NWA cross section. For example, in a scenario with large new physics contributions to the Higgs width, on-shell rate measurements at the LHC consistent with SM predictions enforce through \Eq{eq:scaling_1} that the Higgs couplings are also scaled as
$g_i\rightarrow g_i \left (    {\Gamma_H}/{\Gamma_H^{SM}} \right )^{1/4}$. The off-shell and interference contributions to the cross section are not invariant under this rescaling of the couplings, under which they are modified to
\begin{align}\label{Eq:Scaling}
 \sigma^{\text{int}} = \sqrt{\frac{\Gamma_H}{\Gamma^{SM}_H}}   \sigma_{SM}^{\text{int}},~~~~\sigma_H^{\text{off-shell}} =     \frac{\Gamma_H}{\Gamma^{SM}_H}      \sigma_{H, SM}^{\text{off-shell}}.
\end{align}
A measurement of the off-shell and interference cross section then allows for one to directly measure, or bound, the total Higgs width. 
This method is not completely model independent, indeed some of its limitations were recently discussed in \cite{Englert:2014aca}, along with a specific new physics model which decorrelated the on-shell and off-shell cross sections, evading the technique. However, interpreted correctly, this technique places restrictions on the Higgs width in many models of BSM physics. 
The study of the off-shell cross section as a means to bound the Higgs width was first discussed in the $H \rightarrow ZZ \rightarrow 4l$ channel \cite{Caola:2013yja,Campbell:2013una}, where the ability to fully reconstruct the invariant mass of the decay products allows for an easy separation of the on-shell and off-shell contributions. Recently, CMS has performed a measurement following this strategy and obtained a bound of $\Gamma_H \leq 4.2~ \Gamma_H^{SM}$ \cite{CMS:2014ala} .

The method was extended in \cite{Campbell:2013wga} to the $gg\rightarrow H\rightarrow WW \rightarrow l\nu l\nu$ channel. The WW channel has the advantage that the $2W$ threshold is closer than for $H\rightarrow ZZ$, as well as having a higher branching ratio to leptons, and a higher total cross section. It does however, also have the disadvantage of large backgrounds, which necessitate the use of jet vetoes, as well as final state neutrinos which prevent the reconstruction of the invariant mass. To get around the latter issue one can exploit the transverse mass variable
\be \label{eq:mt}
M_T^2=(E_T^{\text{miss}}+E_T^{\text{ll}})^2 -| \bf{p_T^{ll}+\bf{E}_T^{miss}}|^2,
\ee
which has a kinematic edge at $M_T=m_H$ for the signal. This variable was shown to be effective in separating the region where the off-shell and interference terms are sizeable, namely the high $M_T$ region, from the low $M_T$ region where on-shell production dominates, allowing for the extraction of a bound on the total Higgs width. Although the experimental uncertainties are currently large in the high $M_T$ region, the authors estimate that with a reduction in the background uncertainty to $\lesssim10\%$, the WW channel could be used to place a bound on the Higgs width competitive with, and complementary to the bound from the $H\rightarrow ZZ\rightarrow 4l$ channel. They therefore suggest a full experimental analysis focusing on the high-$M_T$ region of the $WW$channel. More generally, it was proposed in \cite{Coleppa:2014qja} that a similar method can also be used to probe couplings to heavy beyond the Standard Model states.

Independent of bounding the Higgs width, the study of the off-shell cross section opens up a new way to probe Higgs properties, which is particularly interesting as it probes particles coupling to the Higgs through loops over a large range of energies. Further benefits of the measurement of the off-shell cross section for constraining the parity properties of the Higgs, as well as for bounding higher dimensional operators were also discussed in \cite{Englert:2014aca}.

A full theoretical understanding of the far off-shell region, especially in the presence of realistic experimental cuts, is therefore well motivated to allow for a proper theoretical interpretation of the data, and of bounds on new physics. Indeed, the current limits on the Higgs width from the off-shell region are based on leading order calculations combined with a parton shower. There has recently been progress on the calculation of the perturbative amplitudes required for an NLO description of the off-shell cross section, including signal-background interference, with the calculation of the two loop master integrals with off-shell vector bosons \cite{Henn:2014lfa,Caola:2014lpa}. However, one aspect that has not yet been addressed theoretically is the effect of jet vetoes, and more generally jet binning, on far off-shell cross sections, and on the signal-background interference.

Jet vetoes and jet binning are used ubiquitously in LHC searches to reduce backgrounds. They are typically defined by constraining the $p_T$ of jets in the event. The $H\rightarrow WW$ channel is an example of such a search, where the exclusive zero jet bin, defined by enforcing that all jets in the event satisfy $p_T < \ptveto$, is used to reduce the large background from $t \bar t$ production. Indeed, the analysis of \cite{Campbell:2013wga} used the exclusive $N_{jet}=0$ bin in the large $M_T$ region to estimate the bound on the Higgs width achievable in the $H\rightarrow WW$ channel. Furthermore, the recent bound by CMS \cite{CMS:2014ala} of the Higgs width from the $H\rightarrow ZZ\rightarrow 2l2\nu$ channel used jet bins to optimize sensitivity, splitting data into exclusive $0$-jet and inclusive $1$-jet samples, which were each analyzed and then combined to give the limit. The proper interpretation of the off-shell cross section measurements requires understanding, preferably analytically, the impact of the jet veto and jet binning procedures.

As is well known, the jet veto introduces a low scale, typically $\ptveto\sim 25-30$GeV, into a problem which is otherwise characterized by the scale $Q$, of the hard collision. This causes large logarithms of the form $\alpha_s^n \log^m(Q/\ptveto),~m\leq 2n$, to appear in perturbation theory, forcing a reorganization of the perturbative expansion. Physically these logarithms arise due to constraints placed on the radiation in the event, which prevent a complete cancellation of real and virtual infrared contributions. A resummation to all orders in $\alpha_s$ is then required to make precise predictions. For the leading logarithms this resummation can be implemented by a parton shower. This approach is however difficult to systematically improve, and does not allow for higher order control of the logarithmic accuracy, or a systematic analysis of theoretical uncertainties in the correlations between jet bins. An alternative approach, which allows for the analytic resummation of large logarithms appearing in the cross section, is to match to the soft collinear effective theory (SCET) \cite{Bauer:2000ew,Bauer:2000yr,Bauer:2001ct,Bauer:2001yt,Bauer:2002nz}, which provides an effective field theory description of the soft and collinear limits of QCD. In SCET, large logarithms can be resummed through renormalization group evolution to desired accuracy, providing analytic control over the resummation. This framework also provides control over the theoretical uncertainties, including the proper treatment of correlations between jet bins~\cite{Berger:2010xi,Dittmaier:2012vm,Stewart:2011cf}.

The effect of jet vetoes on Higgs production in the on-shell region has attracted considerable theoretical interest \cite{Berger:2010xi,Banfi:2012yh,Banfi:2012jm,Becher:2012qa,Liu:2012sz,Tackmann:2012bt,Becher:2013xia,Stewart:2013faa}. For on-shell Higgs production, $Q\sim m_H$, and hence the resummation is of logarithms of the ratio $m_H/\ptveto$. The use of a jet clustering algorithm in the experimental analyses complicates resummation and factorization \cite{Ellis:2010rwa,Tackmann:2012bt}, and leads to logarithms of the jet radius parameter \cite{Tackmann:2012bt,Kelley:2012zs,Alioli:2013hba,Banfi:2012jm}. Current state of the art calculations achieve an NNLL$^\prime$+NNLO accuracy~\cite{Becher:2013xia,Stewart:2013faa}, along with the incorporation of the leading dependence on the jet radius, allowing for precise theoretical predictions in the presence of a jet veto. Such predictions are necessary for the reliable extractions of Higgs couplings from rate measurements. Indeed, the exclusive zero-jet Higgs cross section is found to decrease sharply as the $\ptveto$ scale is lowered.

In this chapter we use SCET to analytically study the effect of a jet veto on off-shell cross sections. In particular, we are interested in processes with contributions from a large range of $\hat s$, where $\sqrt{\hat s}$ is the partonic centre of mass energy. In \Sec{sec:2}, we present a factorization theorem allowing for the resummation of large logarithms of the form $\log \sqrt{\hat s}/\ptveto$, in the cross section for the production of a non-hadronic final state. Working to NLL order, and using canonical scales, for simplicity, gives \cite{Banfi:2012yh}
\begin{align} \label{eq:intro_fact}
\frac{d\sigma_0^{\rm NLL}(\ptveto)}{d\hat s\, d\Phi}=  & \big|\cM_{ij}(\mu=\sqrt{\hat s},\Phi)\big|^2  \int\!\! dx_a dx_b f_i(x_a,\mu=\ptveto)f_j(x_b,\mu=\ptveto)    \\ 
& \times  \delta(x_a x_b E_\text{cm}^2-\hat s)   e^{-2{\rm Re} K^i_{\rm NLL}(\sqrt{\hat s},\, \ptveto)} \,,
 \nonumber
\end{align}
where $\sigma_0(\ptveto)$ is the exclusive zero-jet cross section. In this formula, $f_i$, $f_j$ are the parton distribution functions (PDFs) for species $i,j$, $\cM_{ij}$ is the hard matrix element, $\Phi$ is the leptonic phase space, and $E_{\text{cm}}$ is the hadronic centre of mass energy. $K^i_{\text{NLL}}$ is a Sudakov factor, defined explicitly in \Sec{sec:2}, which depends only on the identity of the incoming partons. The form of \Eq{eq:intro_fact} shows that the effect of the jet veto can be captured independent of the hard underlying process, which enters into \Eq{eq:intro_fact} only through $\cM$. At higher logarithmic order a dependence on the jet algorithm is also introduced, but the ability to separate the effect of the jet veto from the particular hard matrix element using the techniques of factorization remains true, and allows one to make general statements about the effect of the jet veto.

The resummation of the large logarithms, $\log \sqrt{\hat s}/\ptveto$, introduced by the jet veto leads to a suppression of the exclusive zero-jet cross section, evident in \Eq{eq:intro_fact} through the Sudakov factor, and familiar from the case of on-shell production. The interesting feature in the case of off-shell effects is that this suppression depends on $\sqrt{\hat s}$. For example, when considering off-shell Higgs production, or signal-background interference, which contribute over a large range of $\sqrt{\hat s}$, the jet veto re-weights contributions from different $\sqrt{\hat s}$ regions in a strongly $\sqrt{\hat s}$ dependent manner. In particular, this modifies differential distributions in $\sqrt{\hat s}$, or any similar variable, such as $M_T$. This is of particular interest for the program to place bounds on the Higgs width using the off-shell cross section in channels which require a jet veto, as this procedure requires an accurate description of the shape of the differential cross section. Furthermore, the jet veto has an interesting effect on the signal-background interference, which often exhibits cancellations from regions widely separated in $\sqrt{\hat s}$. The study of these effects is the subject of this chapter.

Our outline is as follows. In \Sec{sec:2} we review the factorization theorem for the exclusive zero jet bin, with a jet veto on the $p_T$ of anti-$k_T$ jets, focussing on the dependence on $\sqrt{\hat s}$. \Sec{general} describes the generic effects of jet vetoes on off-shell production, including the dependence on the jet veto scale, the identity of the initial state partons and the hadronic centre of mass energy. In particular, we show that off-shell production in the exclusive zero-jet bin is suppressed by a strongly $\sqrt{\hat s}$ dependent Sudakov factor, and comment on the corresponding enhancement of the inclusive 1-jet cross section. In \Sec{sec:WW} we perform a case study for the $gg\rightarrow H\rightarrow WW\rightarrow l\nu l\nu$ process, resumming to NLL accuracy the off-shell cross section including the signal-background interference. For the signal-background interference, we consider two Higgs masses, $m_H=125$ GeV and $m_H=600$ GeV, whose interference depends differently on $\sqrt{\hat s}$, to demonstrate different possible effects of the jet veto on the signal-background interference. Since $\sqrt{\hat s}$ is not experimentally reconstructible for $H\to WW$, in \Sec{sec:mt_126}, we demonstrate the suppression as a function of $M_T$ caused by the jet-veto restriction. In \Sec{sec:bound} we discuss the effect of the jet veto and jet binning on the extraction of the Higgs width from the off-shell cross section in $H\rightarrow WW$ (commenting also on $H\to ZZ$). We conclude in \Sec{sec:conclusion}.

\section{Cross Sections with a Jet Veto: A Review}\label{sec:2}

In this section we review the factorization theorem, in the SCET formalism, for $pp \rightarrow L+0$-jets, where L is a non-hadronic final state. We consider a jet veto defined by clustering an event using an anti-$k_T$ algorithm with jet radius $R$ to define jets, $J$, and imposing the constraint that $p_T^J <\ptveto$ for all jets in the event. This is the definition of the jet veto currently used in experimental analyses, with the experimental value of $\ptveto$ typically $25-30$ GeV, and $R\simeq 0.4\text{-}0.5$. 

\subsection{Factorization Theorem}
Following the notation of \cite{Stewart:2013faa}, the factorization theorem for $pp \rightarrow L+0$-jets with a jet veto on $p_T$ can be computed in the framework of SCET. For a hard process where $L$ has invariant mass $\sqrt{\hat s}$ (on-shell or off-shell), we have 
\begin{align} \label{eq:fact}
&\frac{d\sigma_0(\ptveto)}{d \hat s}=\int d\Phi dx_a dx_b ~\delta(x_a x_bE_{cm}^2-\hat s) \sum_{i,j} H_{ij}(\sqrt{\hat s},\Phi,\mu) B_i(\sqrt{\hat s},\ptveto,R,x_a,\mu,\nu) \nonumber \\
&\times  B_j(\sqrt{\hat s},\ptveto,R,x_b,\mu,\nu)  S_{ij}(\ptveto,R,\mu,\nu)  
\nonumber \\
&
+\frac{d\sigma_0^{Rsub}(\ptveto,R)}{d \hat s}+\frac{d\sigma_0^{ns}(\ptveto,R,\mu_{ns})}{d\hat s}.
\end{align}
In this formula, $\Phi$ denotes the leptonic phase space, $i,j$ denote the initial partonic species, $H_{ij}$ is the hard function for a given partonic channel, $B_i$ are the beam functions which contain the PDFs, and $S_{ij}$ is the soft function, each of which will be reviewed shortly. Since this factorization theorem applies to the production of a color singlet final state, we either have $i=j=g$, or $i=q$, $j=\bar q$.   \Eq{eq:fact} is written as the sum of three terms. The first term in \Eq{eq:fact} contains the singular logarithmic terms, which dominate as $\ptveto \rightarrow 0$, or in the case of off-shell production that we are considering, as $\hat s \rightarrow \infty$, with $\ptveto$ fixed. The second term, $\sigma_0^{Rsub}$, contains corrections that are polynomial in the jet radius parameter $R$, and $\sigma_0^{ns}$ contains non-singular terms which vanish as $\ptveto \rightarrow 0$, and are suppressed relative to the singular terms when the ratio $\ptveto/\sqrt{\hat s}$ is small. 

The factorization theorem allows for each component of \Eq{eq:fact} to be calculated at its natural scale, and evolved via renormalization group evolution (RGE) to a common scale, resumming the large logarithms of $\ptveto/\sqrt{\hat s}$. For the case of a veto on the jet $p_T$, the factorization follows from SCET$_{\text{II}}$, where the RGE is in both the virtuality scale, $\mu$, and rapidity scale, $\nu$ \cite{Chiu:2011qc,Chiu:2012ir}. In this section, we will briefly summarize the components of the factorization theorem with a particular focus on their dependence on the underlying hard matrix element, the identity of the incoming partons, the jet algorithm, and the jet veto measurement. We will also review their RGE properties. Further details, including analytic expressions for the anomalous dimensions, can be found in \cite{Tackmann:2012bt,Stewart:2013faa,Berger:2010xi}, and references therein.

\subsection*{Soft Function}

The soft function $S_{ij}(\ptveto,R,\mu,\nu)$ describes the soft radiation from the incoming partons $i,j$ which are either both gluons or both quarks. It is defined as a matrix element of soft Wilson lines along the beam directions, with a measurement operator, $\cM^{jet}$, which enforces the jet veto condition:
\be
S_{ii}(\ptveto, R, \mu, \nu) = \bra{0} Y_{n_b} Y^\dagger_{n_a} \cM^{jet}(\ptveto,R)Y_{n_a} Y^\dagger_{n_b} \ket{0}.
\ee
The soft function depends only on the identity of the incoming partons, through the representation of the Wilson lines, which has not been made explicit.  It also depends on the definition of the jet veto through the measurement function.

The soft function is naturally evaluated at the soft scale $\mu_S \sim \ptveto$, and $\nu_S \sim \ptveto$, and satisfies a multiplicative renormalization in both $\mu$ and $\nu$. The solution is given by
\begin{align}
&S_{ii}(\ptveto,R,\mu,\nu)=S_{ii}(\ptveto,R,\mu_S,\nu_S) \exp \left[\log\frac{\nu}{\nu_S}\gamma_\nu^i(\ptveto,R,\mu_S)  \right] 
\exp \bigg[ \int\limits_{\mu_S}^{\mu} \frac{d\mu'}{\mu'} \gamma^i_S(\mu',\nu) \bigg] \,.
\end{align}
Further details including expressions for the anomalous dimensions are given in \cite{Stewart:2013faa}.

In the case of interest, where the jets are defined using a clustering algorithm with a finite R, the soft function also contains clustering logarithms from the clustering of correlated soft emissions, which first arise at NNLL. These appear in the cross section as logarithms of the jet radius, $\log(R)$, but are not resummed by the RGE. For experimentally used values of R, the first of these logarithms is large \cite{Banfi:2012jm}, while the leading $\cO(\alpha_s^3)$ term  was recently calculated and found to be small \cite{Alioli:2013hba}. We therefore treat these $\log(R)$ factors in fixed order perturbation theory. We discuss the impact of these logarithms on our results in \Sec{sec:conv}.

\subsection*{Beam Function}
The beam function \cite{Fleming:2006cd,Stewart:2009yx,Stewart:2010qs}, $B_i$, describes the collinear initial state-radiation from an incoming parton, $i$, as well as its extraction from the colliding protons through a parton distribution function. The beam function depends only on the identity of the incoming parton $i$, and the measurement function. 

In the case of a $\ptveto$, the beam function can be calculated perturbatively by matching onto the standard PDFs at the beam scale $\mu_B\sim \ptveto$, $\nu_B\sim \sqrt{\hat s}$:
\begin{align}
B_i(\sqrt{\hat s},\ptveto,R,x,\mu_B,\nu_B)=\sum_j \int\limits_x^1 \frac{dz}{z} 
  \cI_{ij}\big(\sqrt{\hat s},\ptveto,R,z,\mu_B,\nu_B\big) f_j\Big(\frac{x}{z},\mu_B\Big)
\end{align}
The lowest order matching coefficient is
\be
\cI_{ij}=\delta_{ij} \delta(1-z)
\ee
so that to leading order the beam function is simply the corresponding PDF, but evaluated at the beam scale $\mu_B\simeq \ptveto$. This was seen explicitly in the NLL expansion of \Eq{eq:intro_fact}. Higher order matching coefficients involve splitting functions, allowing for a mixing between quarks and gluons. This matching procedure corresponds to the measurement of the proton at the scale $\mu_B \sim \ptveto$ by the jet veto. Above the scale $\mu_B$, the beam function satisfies a multiplicative RGE in both virtuality, $\mu$ and rapidity, $\nu$, describing the evolution of an incoming jet for the off-shell parton of species $i$. Unlike the RGE for the PDFs, the RGE for the beam function leaves the identity and momentum fraction of the parton unchanged. The solution to the RGE is given by
\begin{align}
&B_{i}(\sqrt{\hat s},\ptveto,R,x,\mu,\nu)=B_{i}(\sqrt{\hat s},\ptveto,R,x,\mu_B,\nu_B) \exp \left[\frac{1}{2}\log\frac{\nu_B}{\nu}\gamma_\nu^i(\ptveto,R,\mu_B)  \right] \nonumber \\
&\times \exp \bigg[ ~\int\limits_{\mu_B}^{\mu} \frac{d\mu'}{\mu'} \gamma^i_B(\sqrt{\hat s},\mu',\nu) ~\bigg],
\end{align}
which resums the logarithmic series associated with the collinear radiation from the incoming parton. Further details and expressions for the anomalous dimensions are again given in \cite{Stewart:2013faa}.

As with the soft function, the beam function also contains logarithms and polynomial dependence  on the jet radius, $R$, from the clustering of collinear emissions. These logarithms can be numerically significant, but are not resummed by the RGE. We again treat these terms in fixed order perturbation theory.

\subsection*{Hard Function}
The hard function $H_{ij}$ encodes the dependence of the singular term of \Eq{eq:fact} on the underlying hard partonic matrix element of the $pp \rightarrow L+0$-jets process. It can be obtained by matching QCD onto an appropriate SCET operator at the scale $\sqrt{\hat s}$, giving a Wilson coefficient, $C_{ij}$. The Wilson coefficient satisfies a standard RGE in virtuality, allowing it to be evolved to the scale $\mu$. The hard function is then given by the square of the Wilson coefficient
\be
H_{ij}(Q,\mu)=|C_{ij}(Q,\mu)|^2 \,,
\ee
where $Q$ denotes dependence on all variables associated with the final leptons as well as parameters like the top-mass, and the Higgs and $W/Z$ masses and widths. The solution to the RGE equation for the hard function is
\begin{align}
  H_{ij}(Q,\mu) = H_{ij}(Q,\mu_H)  \Big| e^{-K^i(\sqrt{\hat s},\mu_H,\mu)} \Big|^2 
\,,
\end{align}
where the Sudakov form factor is
\begin{align} \label{eq:K}
  K^i(\sqrt{\hat s},\mu_H,\mu) &=\int_{\mu_H}^{\mu} \frac{d\mu'}{\mu'}  \gamma_H^i(\sqrt{\hat s},\mu') 
  \nonumber \\
  &= 2 K_{\Gamma_{\rm cusp}^i}(\mu_H,\mu) - K_{\gamma_H^i}(\mu_H,\mu) 
    - \ln\Big(\frac{-\hat s-i0}{\mu_H^2}\Big) \:  \eta_{\Gamma_{\rm cusp}^i}(\mu_H,\mu) \,.
\end{align}
Here the integrals involve the $\beta$-function and anomalous dimensions
\begin{align}  \label{eq:Ks}
K_{\Gamma^i_{\rm cusp}} &= \int_{\alpha_s(\mu_H)}^{\alpha_s(\mu)}  \frac{d\alpha_s}{\beta(\alpha_s)} \Gamma^i_{\rm cusp}(\alpha_s) \int_{\alpha_s(\mu_H)}^{\alpha_s} \frac{d\alpha_s'}{\beta(\alpha_s')} 
  \,,  
& \eta_{\Gamma_{\rm cusp}^i} &= \int_{\alpha_s(\mu_H)}^{\alpha_s(\mu)}  \frac{d\alpha_s}{\beta(\alpha_s)} \Gamma_{\rm cusp}^i(\alpha_s) \,,
\nonumber \\
  K_{\gamma_H^i} &= \int_{\alpha_s(\mu_H)}^{\alpha_s(\mu)}  \frac{d\alpha_s}{\beta(\alpha_s)} \gamma_H^i(\alpha_s)
 \,,
\end{align}
where  the channel $i$ is either for quarks or gluons. Here the cusp and regular anomalous dimensions are  $\Gamma_{\rm cusp}^i(\alpha_s) = \sum_{k=0}^\infty \Gamma^i_k (\alpha_s/4\pi)^{k+1}$,
$\gamma_H^i(\alpha_s) = \sum_{k=0}^\infty \gamma^i_k (\alpha_s/4\pi)^{k+1}$, respectively, and the $\beta$-function is $\beta(\alpha_s) = -2 \alpha_s \sum_{k=0}^\infty \beta_k (\alpha_s/4\pi)^{k+1}$ so $\beta_0=11 C_A/3-2n_f/3$.

Explicit results for the functions in Eq.~(\ref{eq:Ks}) can be found for example in Ref.~\cite{Berger:2010xi}.
Since we will be considering far off-shell production, and including signal-background interference effects, which have not been discussed in SCET factorization theorems before, we will discuss in more detail the definition of the hard function for the specific case of $gg\rightarrow l\nu l\nu$ in \Sec{sec:hardfunc}.

The beam and soft functions are universal, depending only on the given measurement and the identity of the incoming partons, it is the hard function that needs to be calculated separately for different processes. The beam and soft functions are known to NNLL for the case of a jet veto defined using a cut on $p_T$, and it is the hard coefficient that prevents resummation to NNLL for several cases of interest. In particular, since we are interested here in the case of off-shell production, one needs the full top mass dependence of loops, significantly complicating the computation. Indeed, for the case of signal-background interference for $gg \rightarrow H\rightarrow WW\rightarrow l\nu l\nu$, only the leading order hard function is known \cite{Campbell:2011cu}, while for direct gluon-fusion Higgs production, analytic results exist for the NLO virtual corrections including quark mass dependence \cite{Harlander:2005rq}. This restricts our predictions to NLL accuracy for signal-background interference for $gg \rightarrow H\rightarrow WW\rightarrow l\nu l\nu$.

\subsection*{Non-Singular Terms}
The non-singular term $\sigma_0^{ns}(\ptveto,R,\mu_{ns})$ is an additive correction to the factorization theorem, containing terms that vanish as $\ptveto \rightarrow 0$. This term scales as $\ptveto/\sqrt{\hat s}$. The non-singular piece is important when $\ptveto$ is of the same order as $\sqrt{\hat s}$, where both singular and non-singular pieces contribute significantly to the cross section.  In this chapter, we will be focusing on the effect of a jet veto on far off-shell effects, and we will therefore always be considering the case that $\ptveto \ll \sqrt{\hat s}$. We will therefore not discuss the non-singular pieces of the cross section, and focus on the singular contributions.

\subsection*{Uncorrelated Emissions}
Beginning with two emissions, the jet algorithm can cluster uncorrelated emissions from the soft and collinear sectors \cite{Tackmann:2012bt,Becher:2012qa,Banfi:2012jm}. This produces terms proportional to powers of $R^2$, which can formally be treated as power corrections for $R\ll 1$, and are included in $\sigma_0^{Rsub}$. For the jet radii of $0.4 \text{-}0.5$ used by the experimental collaborations, these effects are numerically very small, especially compared to the $\log R$ terms from correlated emissions. We make use of the expressions from \cite{Stewart:2013faa}.

\subsection{Expansion to NLL}\label{sec:NLL}

It is useful to consider the factorization theorem at NLL order with canonical scale choices, to see the main factors that control its behaviour. The result at NLL was first given in  \cite{Banfi:2012yh} for on-shell production with $\sqrt{\hat s}=m_H$. Allowing for off-shell production, and using canonical scales, the cross section with a $\ptveto$ cut is given at NLL by
\begin{align} \label{eq:eqNLL}
\frac{d\sigma_0^{\rm NLL}(\ptveto)}{d\hat s\, d\Phi}=  & \big|\cM_{ij}(\mu=\sqrt{\hat s},\Phi)\big|^2  \int\!\! dx_a dx_b f_i(x_a,\mu=\ptveto)f_j(x_b,\mu=\ptveto)    \\ 
& \times  \delta(x_a x_b E_\text{cm}^2-\hat s)   e^{-2{\rm Re} K^i_{\rm NLL}(\sqrt{\hat s},\, \ptveto)} \,,
 \nonumber
\end{align}
where $\Phi$ are phase space variables for the final state leptonic decay products.
In this equation, $f_i$ and $f_j$ are the appropriate PDFs, for example, they are both $f_g$ for the case of gluon-fusion since direct contributions from the quark PDFs do not enter until NNLL order. For a partonic center of mass energy $\sqrt{\hat s}$, \Eq{eq:eqNLL} resums to NLL accuracy the logarithms of $ \sqrt{\hat s}/ \ptveto$. \Eq{eq:eqNLL} does not include the non-singular contribution to the cross section. As discussed previously, in the far off-shell region, $\ptveto \ll \sqrt{\hat s}$, and the singular contributions to the cross section dominate. It should also be emphasized that at NLL one is not sensitive to the jet algorithm or jet radius, as at $\cO(\alpha_s)$ there is only a single soft or collinear emission. Although the R dependence is important for accurate numerical predictions, it does not effect the qualitative behaviour of the jet veto. The R dependence appears in the factorization theorem at NNLL.

The only dependence on the hard partonic process in \Eq{eq:eqNLL} is in the matrix element $\cM_{ij}(\hat s)$. The Sudakov form factor $K^i$ given in Eq.~(\ref{eq:K}) arises from restrictions on real radiation in QCD, and depends only on the identity of the incoming partons. At NLL the Sudakov factor is given by
\begin{align}  \label{KNLL}
K_{\rm NLL}^i(\sqrt{\hat s},\mu_H,\mu) &= - \frac{\Gamma_0^i}{2\beta_0^2} \bigg\{ 
   \frac{4\pi}{\alpha_s(\mu_H)} \Big( 1-\frac{1}{r} - \ln r\Big) + \bigg( \frac{\Gamma_1^i}{\Gamma_0^i}-\frac{\beta_1}{\beta_0}\bigg) (1-r+\ln r) + \frac{\beta_1}{2\beta_0} \ln^2r \bigg\} 
\nonumber \\
  & \quad + \frac{\gamma_0^i}{2\beta_0}\ln r
   + \ln\Big(\frac{-\hat s-i0}{\mu_H^2}\Big) \frac{\Gamma_0^i}{2\beta_0} \bigg\{ \ln r +
  \frac{\alpha_s(\mu_H)}{4\pi} \bigg( \frac{\Gamma_1^i}{\Gamma_0^i}-\frac{\beta_1}{\beta_0}\bigg)(r-1)\bigg\}
  \,,
\end{align}
where $r=\alpha_s(\mu)/\alpha_s(\mu_H)$. The form in Eq.~(\ref{KNLL}) allows for the use of complex scales, such as $\mu_H=-i \sqrt{\hat s}$, to minimize the appearance of large $\pi^2$ factors in the Hard function.  On the other hand, with canonical scales we would take $K_{\rm NLL}^i=K_{\rm NLL}^i(\sqrt{\hat s},\ptveto)$.  At LL order the terms with $\Gamma_1$, $\beta_1$, and $\gamma_0$ do not yet contribute and using the LL running coupling we can write ${\rm Re} K_{\rm LL}^i(\sqrt{\hat s},\ptveto)=-(4 C/\beta_0) \ln\sqrt{\hat s}/\ptveto\, [
1 + \ln(1-2\lambda)/(2\lambda)]$ where $\lambda= \alpha_s(\sqrt{\hat s})\frac{\beta_0}{4\pi} \ln\sqrt{\hat s}/\ptveto$.
For gluon-fusion, $C=C_A=3$, whereas for a quark-antiquark initial state, $C=C_F=4/3$.

There are two important features of the expression in Eq.~(\ref{eq:eqNLL}) compared with the case of no jet veto. First, the PDFs are evaluated at the scale $\mu=\ptveto$ instead of $\mu=\sqrt{\hat s}$. Secondly, the cross section is multiplied by a Sudakov factor, which depends on logs of the ratio $\sqrt{\hat s}/\ptveto$. These have a strong impact on the cross section, which will be the focus of \Sec{general}.

\section{Jet Vetoes and Off-Shell Effects}\label{general}

In this section we will discuss quite generally the effect of jet vetoes on off-shell cross sections.  We focus on the dependence on the identity of the initial state partons, and the relation between the exclusive $0$-jet and inclusive $1$-jet bins. We conclude with a discussion of the dependence on the hadronic centre of mass energy. For simplicity, in this section we will use the NLL expansion of \Eq{eq:eqNLL} with canonical scale choices. The NLL expansion demonstrates the essential features that persist at higher logarithmic order, and makes transparent how these effects depend on various parameters of interest. This serves for the purpose of demonstrating the generic effects of jet vetoes, and their dependencies. In \Sec{sec:WW}, we will perform a more detailed study for the specific case of $gg\rightarrow H\rightarrow WW$. 

Unlike on-shell effects, which contribute to the cross section over a small region in $\sqrt{\hat s}$, of order the width, off-shell effects, including signal-background interference and off-shell production, typically contribute over a large range of values of $\sqrt{\hat s}$. In this case the $\sqrt{\hat s}$ dependence of the jet veto suppression can produce interesting effects. In particular, it modifies differential distribution in $\sqrt{\hat s}$, or any substitute such as $M_T$ in cases where the full invariant mass cannot be reconstructed, such as $H\rightarrow WW$. Furthermore, for signal-background interference, the $\sqrt{\hat s}$ dependence of the jet veto suppression can cause an enhancement or suppression of the interference relative to the on-shell contribution to the cross section, or enhance/suppress interference contributions with different signs relative to one another. 

With this motivation, we now study the $\sqrt{\hat s}$ dependence of the jet veto suppression to the exclusive zero jet cross section using the NLL expression of \Sec{sec:NLL}. The benefit of the factorized expression is that this discussion can be carried out essentially independent of the matrix element prefactor $|{\cal M}_{ij}|^2$. From \Eq{eq:eqNLL}, the NLL cross section is modified compared with the LO cross section, only by the evaluation of the PDFs at the jet veto scale, and by the Sudakov factor, which is a function of $\sqrt{\hat s}$. To study the suppression due to the jet veto as a function of $\sqrt{\hat s}$, we will therefore consider
\begin{align} \label{eq:ratio}
E_0(\hat s)=\left .   \left (  \frac{  d\sigma_0^{\rm NLL}(\ptveto)  } {d\sqrt{\hat s}}   \right )   \middle /      \left (  \frac{d\sigma}{d\sqrt{\hat s}} \right)      \right . ,
\end{align}
where $\sigma_0^{NLL}(\ptveto)$ is the NLL exclusive zero jet cross section. In \Eq{eq:ratio}, the cross section in the denominator is evaluated to LO, namely to the same order as the matrix element that appears in \Eq{eq:eqNLL} for the NLL resummed cross section. When forming this combination, one could choose to evaluate the denominator at various orders, for example using the full NLO result calculated without the $\ptveto$. Since NLO corrections are typically large, especially for gluon initiated processes, this would typically decrease the above ratio. However, we have in mind an application to processes, such as signal-background interference in $gg\rightarrow l\nu l\nu$, for which the NLO corrections are not yet known, so that current calculations are restricted to LO results. In this case, we can incorporate the effect of the jet veto at NLL using \Eq{eq:eqNLL}, and the ratio of \Eq{eq:ratio} will characterize the effect of the resummation compared to previous calculations in the literature \cite{Campbell:2011cu,Campbell:2013una,Campbell:2013wga}. This approach also has the benefit that it can be done independent of the particular matrix element, as the NLO corrections are clearly process dependent. However, all of the general features described in this section persist to NNLL resummation, as will be demonstrated in \Sec{sec:WW}. As was mentioned previously, at NLL one doesn't have sensitivity to the jet radius R. While this dependence is important for precise predictions, it does not dominate the behaviour of the jet vetoed cross section as a function of $\hat s$, or modify in any way the conclusions of this section. 

For numerical calculations in this section we use the NLO PDF fit of Martin, Stirling, Thorne and Watt \cite{Martin:2009iq} with $\alpha_s(m_Z)=0.12018$. Unless otherwise stated, we use a hadronic center of mass energy of $E_\text{cm}=8$ TeV. In \Sec{sec:Ecm} we discuss the dependence on the $E_{\text{cm}}$, comparing behaviour at $8,13,$ and $100$ TeV.

\begin{figure}
\begin{center}
\subfloat[]{\label{fig:glumi_ratio}
\includegraphics[width=7.0cm]{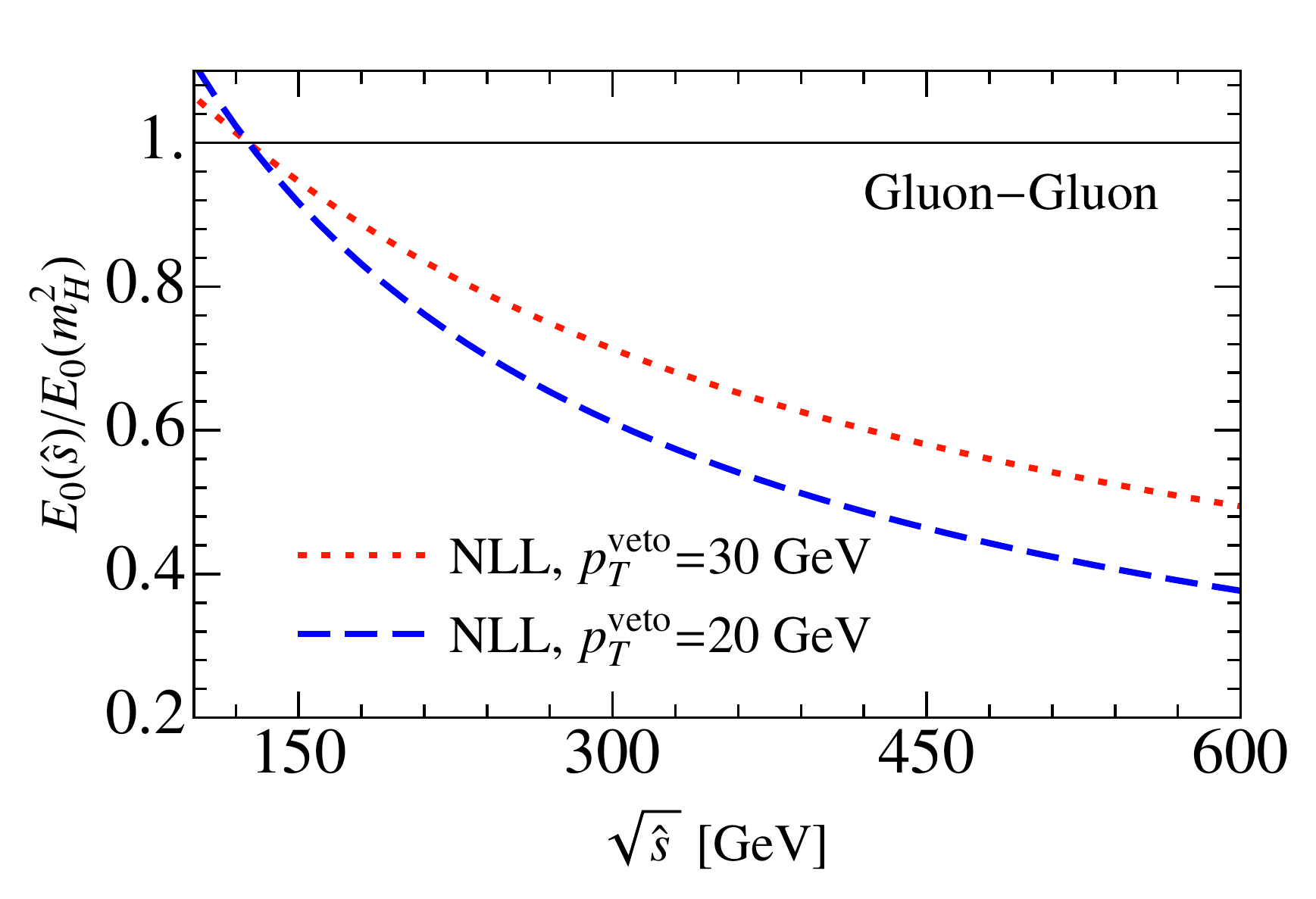}
}
$\qquad$
\subfloat[]{\label{fig:qlumi_ratio}
\includegraphics[width=7cm]{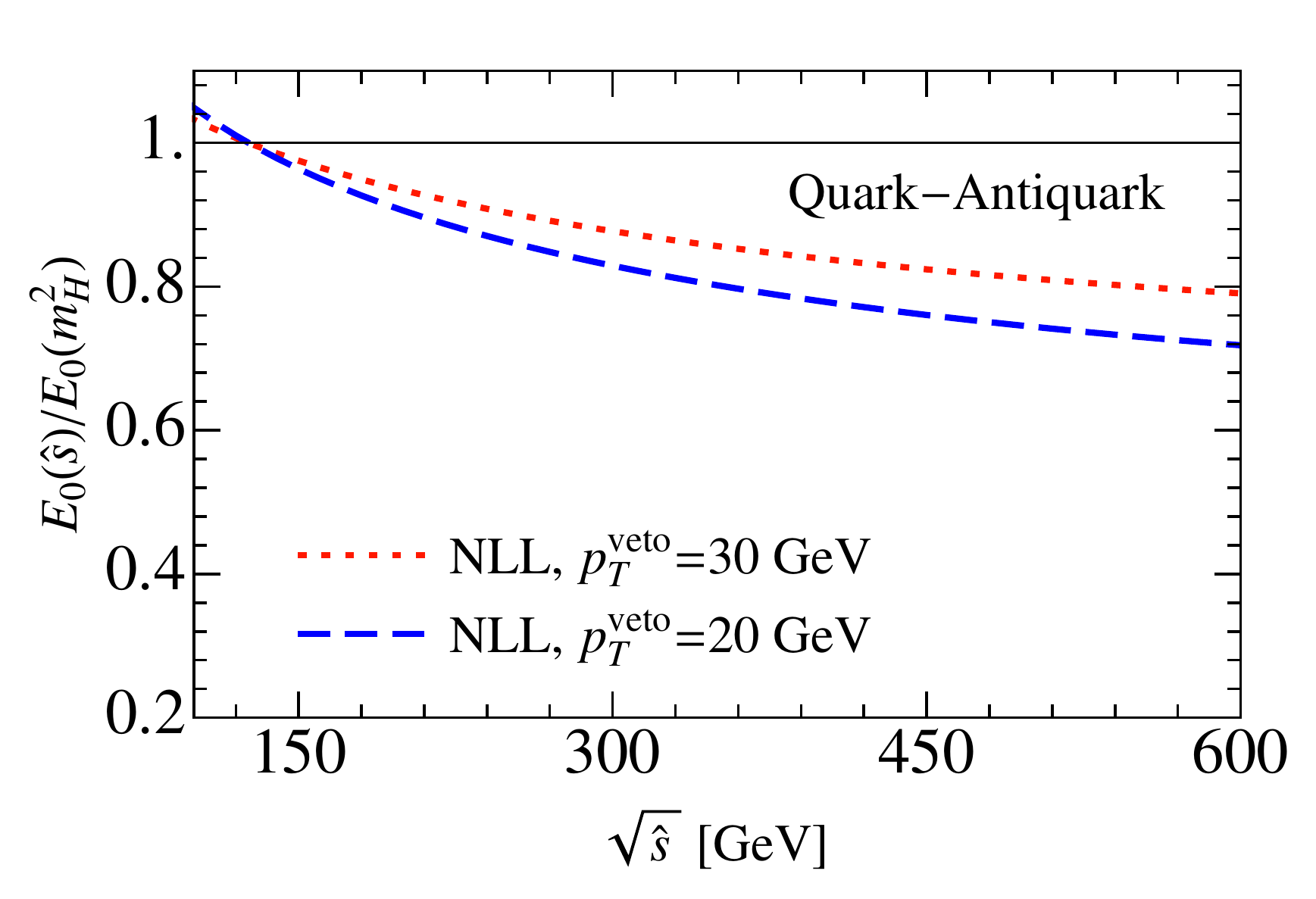}
}
\end{center}
\caption{The ratio  $E_0(\hat s)/E_0(m^2_H)$, for both a gluon-gluon initiated process in (a), and a quark-antiquark initiated process in (b). In both cases we consider $\ptveto=20,30$ GeV. The jet veto causes an $\hat s$ dependent suppression, which is significantly stronger for initial state gluons than initial state quarks, due to the larger colour factor appearing in the Sudakov.}
\label{fig:lumi_ratio}
\end{figure}

In \Fig{fig:glumi_ratio} we demonstrate the effect of the jet veto for a gluon-gluon initial state, as a function of $\sqrt{\hat s}$.\footnote{Note that a similar effect was considered in \cite{Berger:2010xi} which performed resummation for gluon fusion Higgs production with a veto on the global beam thrust event shape, as a function of the Higgs mass.} We plot the ratio $E_0( \hat s)/E_0(m^2_H)$, for $m_H=126$ GeV. We have chosen to plot this particular ratio to focus on the $\hat s$ dependence, rather than the impact that the jet veto has on the on-shell Higgs production cross section which is given by $E_0(m^2_H)$. The ratio $E_0( \hat s)/E_0(m^2_H)$ describes the impact of the jet-veto for  off-shell effects relative to its impact for on-shell production.   It will also be useful when discussing the impact on Higgs width bounds in \Sec{sec:bound}. \Fig{fig:glumi_ratio} shows that the suppression of the exclusive zero-jet cross section has a strong dependence on $\hat s$. The comparison between $\ptveto=20$ GeV, and $\ptveto=30$ GeV shows that a lower cut on the $p_T$ of emissions causes a more rapid suppression, as expected. We have chosen to use the values $\ptveto=20,~30$ GeV, because CMS currently uses $\ptveto=30$ GeV, and although the ATLAS collaboration uses $\ptveto=25$ GeV, the $\ptveto=20$ GeV cut demonstrates the effects of a fairly extreme jet veto. \Fig{fig:glumi_ratio} demonstrates that at scales of $\sqrt{\hat s} \simeq 500$ GeV, the suppression relative to that for on-shell production is of order $50\%$.

\subsection{Quarks vs. Gluons}\label{sec:qvsg}
We now consider the difference in the jet veto suppression for quark initiated and gluon initiated processes. This is relevant in the case where multiple partonic channels contribute to a given process, or if the signal and background processes are predominantly from different partonic channels. This is the case for both $gg\rightarrow H\rightarrow WW,~ZZ$, which have large $q\bar q$ initiated backgrounds. The factorization theorem in \Eq{eq:fact} allows one to easily study the dependence of the jet veto suppression on the identity of the incoming partons, which is carried by the hard, beam, and soft functions. The difference in the suppression arises from the differences in the anomalous dimensions, where for the 0-jet cross section, they involve $C_F$ for quarks, and $C_A$ for gluons. The clustering and correlation logarithms are also multiplied by the colour factors $C_F$ and $C_A$. This phenomenon is similar to quark vs. gluon discrimination for jets~\cite{Altheimer:2013yza}, where the same factors of $C_F$ and $C_A$ appear in the Sudakov and allow one to discriminate between quark and gluon jets. However, in this case, the discrimination is between incoming quarks and gluons. 

Comparing \Fig{fig:glumi_ratio} and \Fig{fig:qlumi_ratio}, we see a significant difference between a gluon-gluon and quark-antiquark initial state. The jet veto suppression increases more rapidly with $\hat s$ in the case of gluon-fusion induced processes than quark anti-quark induced processes. The suppression due to the jet veto being approximately twice as large for the case of gluon-fusion as for quark-antiquark fusion, for the values considered in \Fig{fig:lumi_ratio}. (Note that for the quark-antiquark initial state, we have used the up quark for concreteness, however, the result is approximately independent of flavour for the light quarks, with the suppression being dominated by the flavour independent Sudakov factor. A small dependence on flavour comes from the scale change in the PDF.) The effect of the jet veto is therefore of particular interest for gluon initiated processes, such as Higgs production through gluon-gluon fusion, to be discussed in \Sec{sec:WW}. This difference in the suppression is interesting for a proper analysis of the backgrounds for $H\rightarrow WW,~ZZ$ in the off-shell region, and deserves further study since one may wish to vary $\ptveto$ as a function of $\sqrt{\hat s}$ or $M_T$.

\subsection{Inclusive 1-Jet Cross Section}\label{sec:1jet}
We have up to this point focused on the exclusive zero jet cross section. However, since the total inclusive cross section is unaffected by the jet veto, the inclusive 1-jet cross section has the same logarithmic structure as the exclusive zero-jet cross section, and can be related to the exclusive zero jet cross section by 
\begin{equation}
\frac{d\sigma_{\geq 1}(\ptveto)}{d \sqrt{\hat s}}=\frac{d\sigma}{d \sqrt{\hat s}}-\frac{d\sigma_{0}(\ptveto)}{d \sqrt{\hat s}}.   \label{eq:1jet}
\end{equation}
In this equation, $\sigma_{\geq 1}(\ptveto)$ is the inclusive 1-jet cross section defined by requiring at least one jet with $p_T\geq \ptveto$, $\sigma_0$ is the exclusive zero-jet cross section and $\sigma$ is the inclusive cross section. This relation allows us to discuss the properties of the inclusive 1-jet bin as a function of $\hat s$ using the factorization theorem for the exclusive 0-jet cross section. Of particular interest is the split of the total cross section between the exclusive zero-jet bin and the inclusive 1-jet bin, and the migration between the two bins as a function of $\hat s$. This relation also implies a correlation between the theory uncertainties for the resummation for the two jet bins, which is important for experimental analyses using jet binning \cite{Stewart:2011cf}.

\begin{figure}
\begin{center}
\includegraphics[width=8cm]{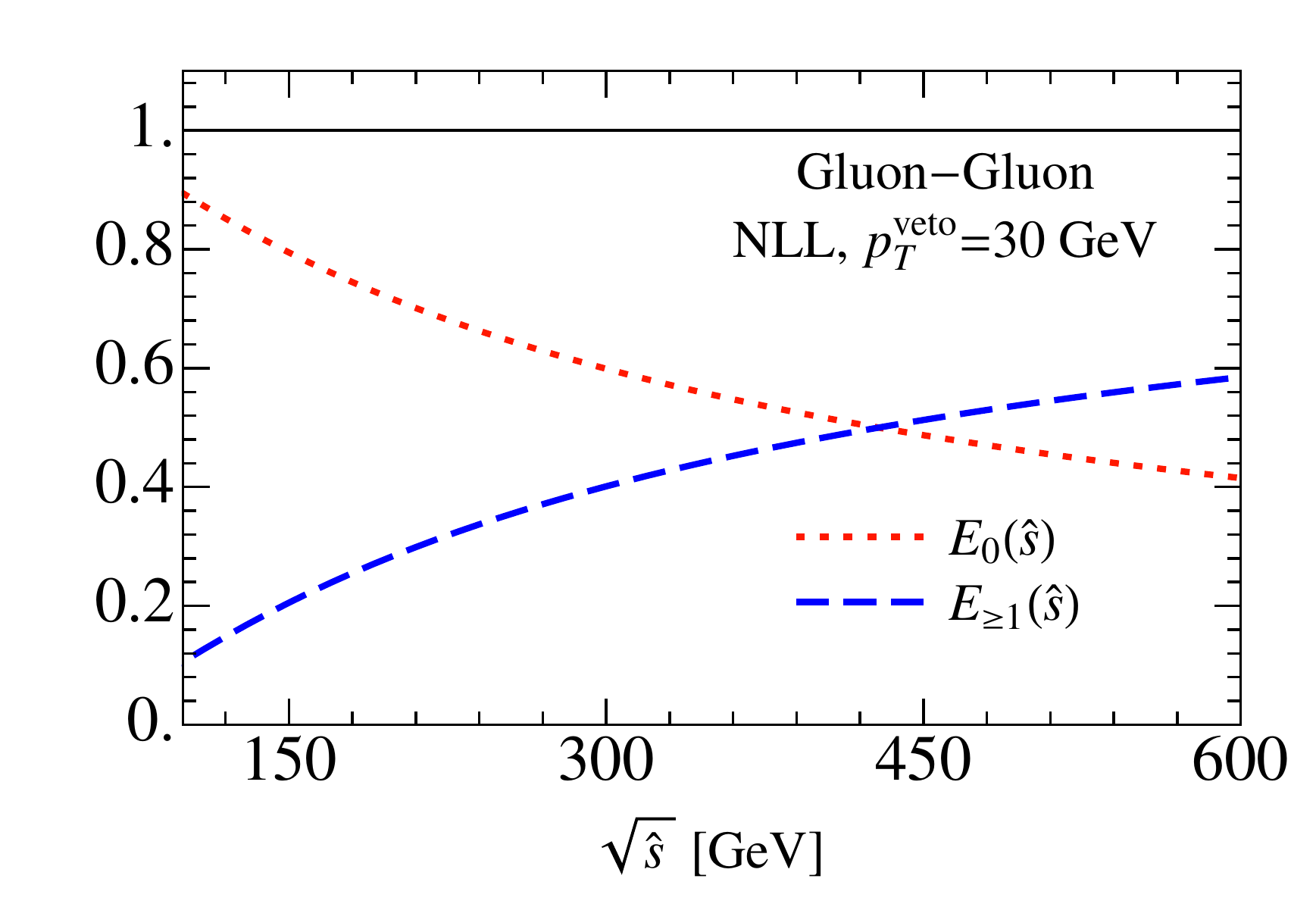}
\end{center}
\caption{The ratios $E_0(\hat s)$, $E_{\geq 1}(\hat s)$ for a gluon-gluon initial state, and $\ptveto=30$ GeV. There is a large migration from the exclusive 0-jet bin to the inclusive 1-jet bin as a function of $\hat s$. This phenomenon is important for understanding the impact of jet binning on off-shell cross sections.}
\label{fig:1jetincl}
\end{figure}

In \Fig{fig:1jetincl} we plot $E_0(\hat s)$, and
\begin{align} \label{eq:ratio_1jet}
E_{\geq 1}(\hat s)=\left .   \left (  \frac{  d\sigma_{\geq 1}^{NLL}(\ptveto)  } {d\sqrt{\hat s}}   \right )   \middle /      \left (  \frac{d\sigma}{d\sqrt{\hat s}} \right)      \right . ,
\end{align}
as a function of $\hat s $ for a gluon-gluon initial state with $\ptveto=30$ GeV. The behaviour in this plot is of course evident from \Fig{fig:lumi_ratio}, but it is interesting to interpret it in this fashion: as an $\hat s$ dependent migration between jet bins. Although our  calculation is only for the inclusive 1-jet bin, the dominant increase will be in the exclusive 1-jet bin.  

This migration between the jet bins as a function of $\hat s$ is important for the proper understanding of the off-shell cross section predictions in the presence of jet vetos. For CMS's recent off-shell $H\rightarrow ZZ\rightarrow 2l2\nu$ analysis, ignoring the VBF category, the events were categorized into exclusive zero jet, and inclusive one jet bins \cite{CMS:2014ala}, both of which have high sensitivity, due to the clean experimental signal. For the case of $H\rightarrow WW$, exclusive 0, 1, and 2 jet bins are used, although the experimental sensitivity is largest in the 0-jet bin, where the backgrounds are minimized. 

The effect of the migration is therefore different in the two cases. For $H\rightarrow ZZ$, since the backgrounds are easier to control, the jets that migrate from the exclusive 0-jet bin are captured in the inclusive 1-jet bin. Since both are used in the experiment, there is not a significant loss in analysis power. Accurate predictions for the two jet bins should still be used, and the correlations in the theory uncertainties due to resummation should still be treated properly.  For the case of $H\rightarrow WW$, where the jet veto plays a more essential role in removing backgrounds, the migration causes a loss in sensitivity. For example, the analysis of \cite{Campbell:2013wga} used the exclusive zero jet bin of $H\rightarrow WW$ to bound the Higgs width without a treatment of the $\hat s$ dependence induced by the jet veto. 
This will be discussed further in \Sec{sec:bound}. 
Calculations for the exclusive 1-jet and 2-jet bins are more difficult. Although NLL resummed results exist for the case $p_T^{jet} \sim \sqrt{\hat s}$ \cite{Liu:2013hba,Liu:2012sz}, the treatment of $p_T^{jet} \ll \sqrt{\hat s}$ is more involved~\cite{Boughezal:2013oha}.  The latter is the kinematic configuration of interest for far off-shell production.

\subsection{Variation with $E_\text{cm}$}\label{sec:Ecm}
Here we comment briefly on the dependence of the exclusive zero jet cross section on the hadronic centre of mass energy, $E_\text{cm}$. This is of course of interest as the LHC will resume at $E_\text{cm}=13$ TeV in the near future, and Higgs coupling and width measurements are important benchmarks for future colliders at higher energies. Here we only discuss the $\hat s$ dependence of the suppression due to the jet veto, the ratio of \Eq{eq:ratio}, on $E_\text{cm}$. Of course, with an increased $E_\text{cm}$, one can more easily achieve higher $\hat s$, allowing for off-shell production over a larger range, magnifying the importance of off-shell effects. We will discuss this for the specific case of $gg\rightarrow H \rightarrow WW$ in \Sec{sec:WW}.

\begin{figure}
\begin{center}
\includegraphics[width=8cm]{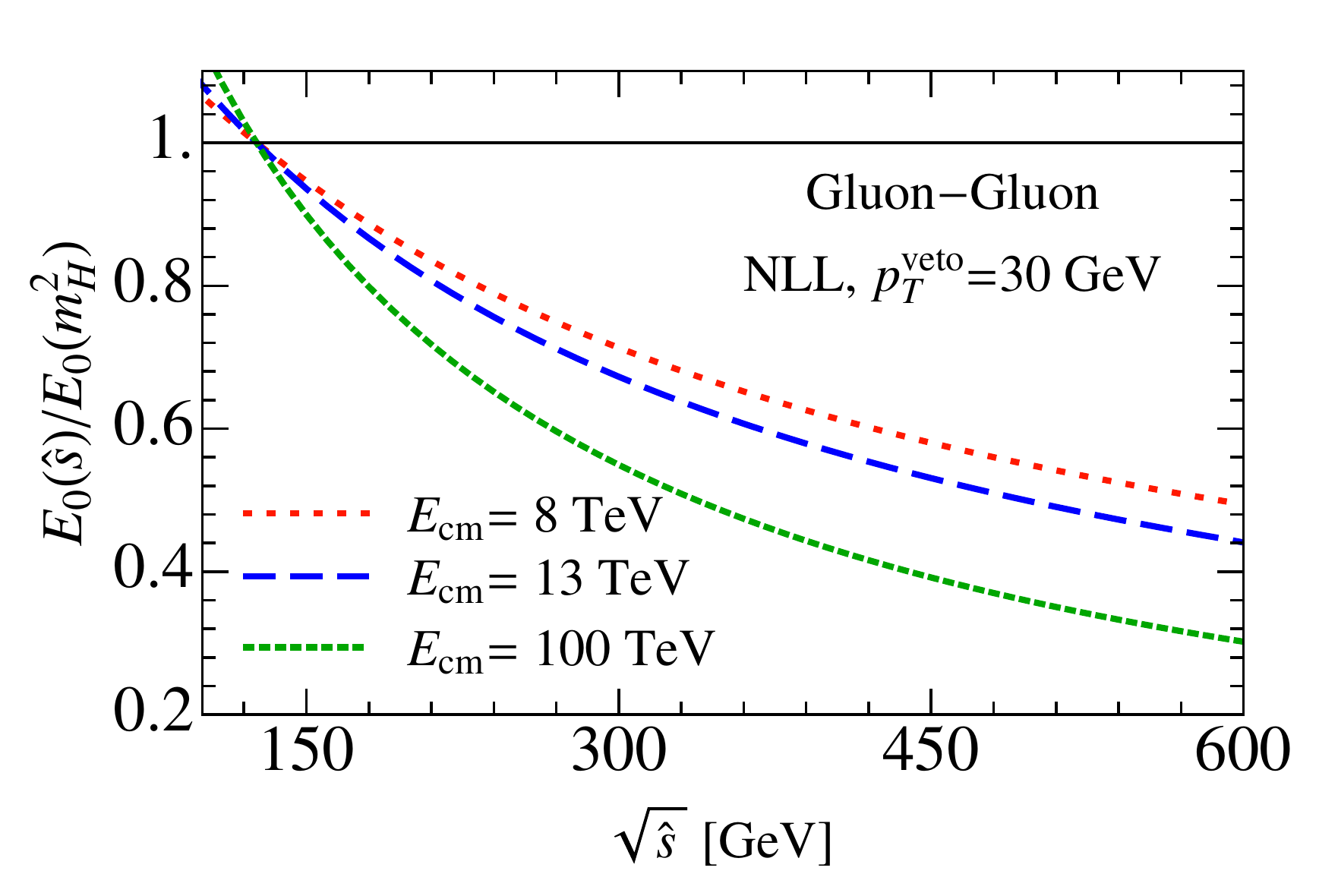}
\end{center}
\caption{A comparison of the effect of the jet veto at $E_\text{cm}=8,13,100$ TeV for a gluon-gluon initial state, and $\ptveto=30$ GeV. At higher $E_\text{cm}$ a larger suppression in the exclusive zero jet bin is observed, due to the larger range of Bjorken $x$ probed. }
\label{fig:Ecm}
\end{figure}

In \Fig{fig:Ecm} we compare the ratio $E_0(\hat s)/E_0(m^2_H)$ for $E_\text{cm}=8,13,100$ TeV. As the value of $E_{\text{cm}}$ is raised, the $\hat s $ dependence of the jet veto suppression systematically increases. Although the effect is relatively small between $8$ TeV and $13$ TeV,  it is significant at $100$ TeV. A similar effect was discussed in \cite{Campbell:2013qaa} where the exclusive zero jet fraction for on-shell Higgs production was observed to decrease with increasing $E_{\text{cm}}$. Since the Sudakov factor is independent of $E_\text{cm}$, this difference arises due to the fact that as the $E_{\text{cm}}$ is increased, the PDFs are probed over a larger range of Bjorken $x$, including smaller $x_{a,b}$ values. In the NLL factorization theorem of \Eq{eq:eqNLL} the PDFs are evaluated at the scale $\ptveto$ instead of at the scale $\hat s$. The impact of this change of scales in the PDFs depends on the $x$ values probed, and causes an increasing suppression as $E_{\text{cm}}$ is increased. 

For the majority of this chapter we will restrict ourselves to $E_{\text{cm}}=8$TeV, although in \Sec{sec:higherEcm} we will further discuss the effect of an increased $E_{\text{cm}}$.

\section{$gg\rightarrow H\rightarrow WW$: A Case Study}\label{sec:WW}

In this section we use $gg\rightarrow H\rightarrow WW$ to discuss the effect of an exclusive jet veto in more detail. $H\rightarrow WW$ is a particularly interesting example to demonstrate the $\sqrt{\hat s}$ dependence of the jet veto suppression since it has a sizeable contribution from far off-shell production \cite{Uhlemann:2008pm,Kauer:2013cga}, and furthermore has interference with continuum $gg\rightarrow WW\rightarrow l\nu l\nu$ production, which contributes over a large range of $\sqrt{\hat s}$ \cite{Campbell:2011cu,Kauer:2012ma,Kauer:2013qba}. A jet veto is also required experimentally for this channel due to large backgrounds. For the signal-background interference, we will consider two different Higgs masses, $m_H=126$ GeV and $m_H=600$ GeV, which have interference which depend differently on $\sqrt{\hat s}$ and therefore cover two interesting scenarios for the different effects that the jet veto can have.

In \Sec{sec:hardfunc} we discuss in detail the hard coefficients, and the matching to SCET.  Default parameters are given in \Sec{sec:nums}.  In \Sec{sec:conv} we use $gg\rightarrow H\rightarrow WW \rightarrow l\nu l\nu$, which can be calculated to NLL and NNLL, to study the convergence in the off-shell region. The extension to NNLL allows us to study the effect of the finite radius of the jet veto. In \Sec{sec:int_126} we show results for the NLL resummation for the signal-background interference. Although we are unable to go to NNLL without the NLO hard function for the interference, the results of \Sec{sec:conv} give us confidence that the NLL result is capturing the dominant effects imposed by the jet veto restriction. In \Sec{sec:mt_126} we consider jet veto suppression in the exclusive zero jet bin as a function of the experimental observable $M_T$.

\subsection{Hard Function and Matching to SCET}\label{sec:hardfunc}
In this section we discuss the hard function appearing in the SCET factorization theorem, which carries the dependence on the hard underlying process. This is discussed in some detail, as we will be considering signal-background interference, which has not previously been discussed in the language of SCET.

\begin{figure}
\begin{center}
\subfloat[]{
\includegraphics[width=6.0cm]{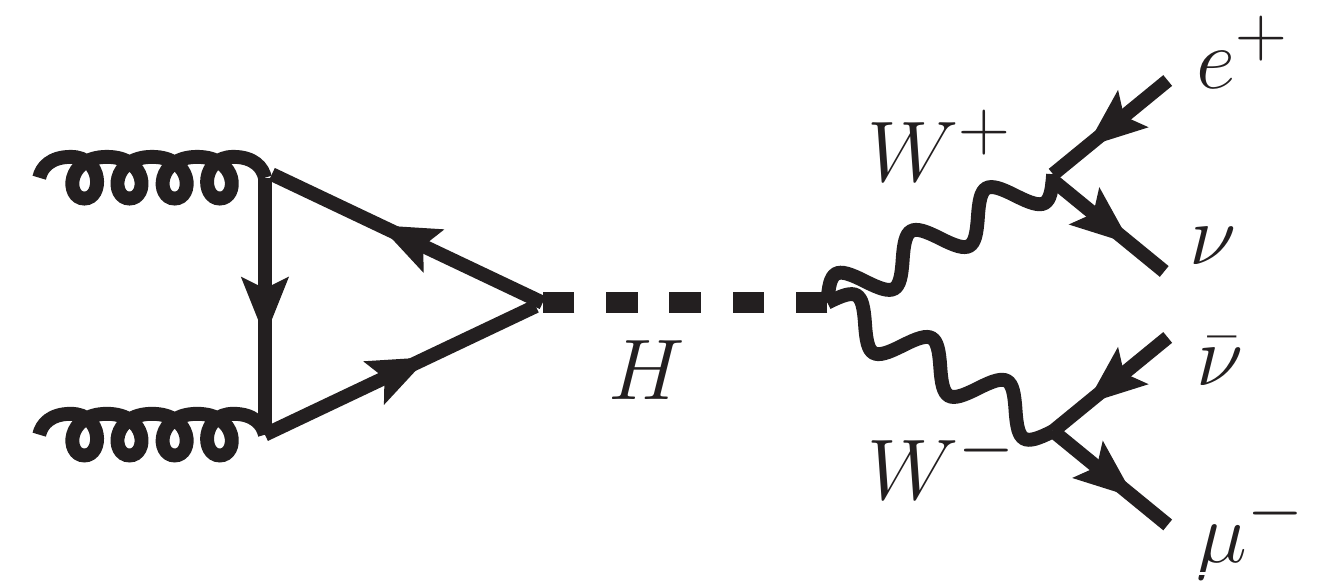}
}
$\qquad$
\subfloat[]{
\includegraphics[width=5.0cm]{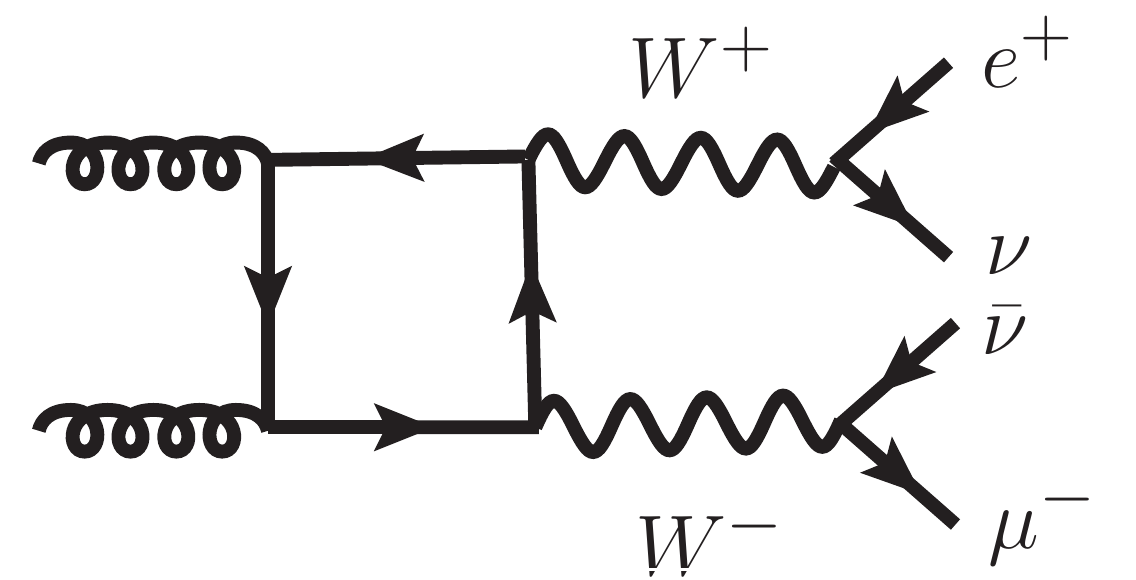}
}
\end{center}
\caption{LO Higgs mediated, (a), and continuum, (b), diagrams contributing to the process $gg\rightarrow l\nu l\nu$. These are matched onto the helicity basis of SCET operators given in \Eq{eq:operators}.}
 \label{fig:diagrams}
\end{figure}

It was shown in \cite{Campbell:2011cu} that only two Feynman diagram topologies contribute to the process $gg\rightarrow \nu_e e^+ \mu^- \bar \nu_\mu$ at LO, due to a cancellation between diagrams with an s-channel Z boson. The two diagrams that contribute are the gluon-fusion Higgs diagram, and a quark box diagram for the continuum production, both of which are shown in \Fig{fig:diagrams}. The $gg\rightarrow  \nu_e e^+ \mu^- \bar \nu_\mu$ cross section consists of Higgs production, the continuum production, and the interference between the two diagrams. Although the interference contribution is small when considering on-shell Higgs production, it becomes important in the off-shell region.  

In the effective field theory formalism, these two diagrams are matched onto effective operators in SCET. It is convenient both for understanding the interference, and for comparing with fixed order QCD calculations to work in a helicity and color operator basis in SCET \cite{Stewart:2012yh,Moult:2015aoa}. For this process the color structure is unique, as we are considering the production of a color singlet state from two gluons. We therefore focus on the helicity structure. The helicity of the outgoing leptons is fixed by the structure of the weak interactions, so we need only construct a helicity basis for the incoming gluons. We write the amplitudes for the above diagrams as
\be
\cA_H(1_g^{h_1},2_g^{h_2},3_{\nu_e}^-,4_{\bar e}^+,5_{\mu}^-,6_{\bar \nu_\mu}^+),~~\cA_\cC(1_g^{h_1},2_g^{h_2},3_{\nu_e}^-,4_{\bar e}^+,5_{\mu}^-,6_{\bar \nu_\mu}^+)
\ee  
where the subscripts $H$, $\cC$ denote the Higgs mediated, and continuum box mediated diagrams respectively, and the superscripts denote helicity. In the following we will mostly suppress the lepton arguments, as their helicities are fixed, and focus on the gluon helicities. 

Since the SM Higgs boson is a scalar, we have
\be
\cA_H(1_g^{-},2_g^{+})=\cA_H(1_g^{+},2_g^{-})=0.
\ee
In this chapter, our focus is on the Higgs production and the signal-background interference. Since there is no interference between distinct helicity configurations, we can therefore also ignore the continuum production diagrams with the $+-$, $-+$ helicity configuration. These do contribute to the background, however their contribution is small compared to the $q\bar q \rightarrow l\nu l\nu$ process. 

The above amplitudes are matched onto operators in the effective theory. The SCET operators at leading power are constructed from collinear gauge-invariant gluon fields \cite{Bauer:2000yr,Bauer:2001ct}
\be
\cB^\mu_{n,\omega \perp}= \frac{1}{g} \left[ \delta(\omega +\bar {\cP}_n) W_n^\dagger(x) i \cD^\mu_{n\perp} W_n(x)   \right]
\ee
where $n$, $\bar n$ are lightlike vectors along the beamline. The collinear covariant derivative is defined as
\be
i \cD^\mu_{n\perp} =\cP^\mu_{n\perp}+gA^\mu_{n\perp},
\ee
with $\cP$ a label operator which extracts the label component of the momentum in the effective theory, and $W_n$ is a Wilson line defined by
\be
W_n(x)= \left [   \sum\limits_{\text{perms}} \exp \left (  -\frac{g}{\bar \cP_n} \bar n \cdot A_n(x)  \right )    \right ].
\ee
A  helicity basis of SCET operators for the process of interest is given by
\begin{align}\label{eq:operators}
&\cO^{++}=\frac{1}{2}\cB^a_{n+}\cB^a_{\bar n+} J_{34-}J_{56-}\\
&\cO^{--}=\frac{1}{2}\cB^a_{n-}\cB^a_{\bar n-}   J_{34-}J_{56-},
\end{align}
where the $1/2$ is a bosonic symmetry factor to simplify matching to the effective theory. We have defined collinear gluon fields of definite helicity by
\be
\cB^a_{i \pm} =-\epsilon_{\mp \mu} \cB^{a,\mu}_{n_i,\omega_i \perp_i},
\ee
where $\epsilon_{\mp \mu}$ are polarization vectors, as well as leptonic currents of definite helicity
\be
J_{ij-}=\epsilon_{+}^\mu(p_i,p_j)\frac{\bar \psi_{i-}\psi_{j-}}{\sqrt{2}[ji]}.
\ee
In this expression, and in the expressions for the Wilson coefficients given below, we will use the standard spinor helicity notation, with $\langle ij \rangle= \bar u_-(p_i) u_+(p_j)$, and $[ ij ]= \bar u_+(p_i) u_-(p_j)$.

Note that we use Hard functions that are fully differential in the leptonic momenta. This allows for realistic experimental cuts on the leptonic phase space to be straightforwardly incorporated.

It is important to note that operators with distinct external helicities do not mix under the SCET RGE at leading power. The jets from the incoming partons, which are described by the beam functions, can only exchange soft gluons, described by the soft function. At leading power, the soft gluons cannot exchange spin, only color, and therefore the RGE can only mix Wilson coefficients in color space, which in this case is trivial.  This allows one to consistently neglect the operators $\cO^{+-}$, $\cO^{-+}$, which would arise from matching the $\cA_{\cC}(1_g^{-},2_g^{+})$, and $\cA_{\cC}(1_g^{+},2_g^{-})$ onto SCET. They do not contribute to the process of interest, and do not mix under the RGE with the operators that do contribute. 

We are interested in considering both the direct Higgs production and signal-background interference separately, so it is convenient to maintain this distinction in SCET. Although the SCET operators are the same in both cases, we can separate the Wilson coefficient into a component from the Higgs mediated diagram, and a component from the box mediated continuum diagram. We then have four Wilson coefficients
\be
C^H_{++},~C^H_{--},~C^C_{++},~C^C_{--}  \,.
\ee
Since the operators are in a helicity basis, these four Wilson coefficients are simply the finite part of the helicity amplitudes for the given processes (or more specifically for ${\overline{\rm MS}}$ Wilson coefficients in SCET are the finite part of the helicity amplitudes computed in pure dimensional regularization). These were computed in \cite{Campbell:2011cu}, and can be obtained from the MCFM code \cite{Campbell:2010ff}. The Wilson coefficients for the Higgs mediated process depend on the Higgs and W boson widths and masses, as well as the invariants $s_{12}, s_{34}, s_{56}$. The explicit leading order Wilson coefficients for the Higgs mediated process are given by
\begin{align}
&C^H_{--}(m_H,\Gamma_H, s_{12}, s_{34}, s_{56})= \left( \frac{g_w^4 g_s^2}{16\pi^2} \right) \cP_H(s_{12})   \cP_W(s_{34})    \cP_W(s_{56}) \frac{\langle 12 \rangle \langle 35 \rangle [64]}{[21] s_{34} s_{56}} F_H(s_{12}) , \\
&C^H_{++}(m_H,\Gamma_H, s_{12}, s_{34}, s_{56})= \left( \frac{g_w^4 g_s^2}{16\pi^2} \right) \cP_H(s_{12})   \cP_W(s_{34})    \cP_W(s_{56}) \frac{[ 12 ] \langle 35 \rangle [64]}{\langle 21\rangle s_{34} s_{56}} F_H(s_{12}),
\end{align}
where the function $\cP_i$ is the ratio of the propagator for the particle species $i$ to that of the photon,
\be
\cP_i(s)=\frac{s}{s-m_i^2+i\Gamma_i m_i}   \,.
\ee
We have also used $F_H (s_{12})$ for the usual loop function for gluon-fusion Higgs production
\be
F_H(s_{12})=\sum\limits_{q=t,b} \frac{m_q^2}{s_{12}}\bigg[   2+\Big(  \frac{4m_q^2}{s_{12}}-1  \Big) g\Big(\frac{m_q^2}{s_{12}} \Big)   \bigg] \,,
\ee
with
\be
g(x)= \begin{cases} 
      \frac{1}{2} \left[ \log \left(  \frac{1+\sqrt{1-4x}}{1-\sqrt{1-4x}}   \right ) -i\pi   \right]^2 & x< \frac{1}{4} \\
      -2 \left ( \sin^{-1} \left (  \frac{1}{2\sqrt{x}}\right)   \right )^2 & x\geq \frac{1}{4}. 
   \end{cases}
\ee

The Wilson coefficients $C_{++}^C$ and $C_{--}^C$ for the box diagram depend on the W mass and width, as well as the kinematic invariants formed by the external momenta. In the presence of massive quarks in the loops, they are extremely lengthy, so we do not reproduce them here. We refer interested readers to \cite{Campbell:2011cu}, and the MCFM code from which we have extracted the required results for our analysis. We have verified that our extracted expressions reproduce quoted numerical results and distributions in \cite{Campbell:2011cu}.

The Hard coefficient, $H$, appearing in the factorization theorem, \Eq{eq:fact} is given by the square of the Wilson coefficients:
\be
H=|C_{++}^H+C_{++}^C|^2 + |C_{--}^H+C_{--}^C|^2  + | C_{+-}^C |^2 + | C_{-+}^C |^2 
\ee
As is typically done in the case of squared matrix elements, we can separate the hard function into the sum of a hard function for the Higgs mediated process $H^H$, a hard function for the interference $H^{int}$, and a hard function for the background arising as the square of Wilson coefficient for the continuum process $H^C$ (which we will not use here). For the first two we have
\begin{align}
&H^H=|C_{++}^H|^2+|C_{--}^H|^2\\
&H^{int}=2\text{Re}\left [C_{++}^H (C_{++}^C)^\dagger\right ]   +   2\text{Re}\left [C_{--}^H (C_{--}^C)^\dagger\right ].
\end{align}
This decomposition allows us to discuss the resummation of the interference and the Higgs mediated processes separately in the effective theory, in a language that is identical to that used in Feynman diagram calculations. In \Secs{sec:conv}{sec:int_126} we will discuss the effect of resummation on both the Higgs mediated contribution and the signal-background interference.

\subsection{Parameters for Numerical Calculations} \label{sec:nums}

For the numerical results, we use the default set of electroweak parameters from MCFM, following \cite{Campbell:2013wga,Campbell:2011cu}:
\begin{align*}
&m_W=80.398\,\text{GeV},  &m_Z&=91.1876\,\text{GeV} \,, \\
&\Gamma_W=2.1054\,\text{GeV},  &\Gamma_Z& =2.4952\,\text{GeV}\,, \\
&m_t=172.5\,\text{GeV}, &m_b&=4.4\,\text{GeV}\,, \\
&G_F=1.16639 \times 10^{-5}\, \text{GeV}^{-2},  &\sin^2 \theta_W &=0.222646\,, \\
&\alpha_{e.m.}(m_Z)=\frac{1}{132.338} \,.
\end{align*}
We use the following two Higgs mass/width combinations to demonstrate the dependence on the Higgs mass: 
\begin{align*}
&m_H=126\,{\rm GeV},\qquad \Gamma_H=0.004307 \,\text{GeV} \,,\\
&m_H=600\,{\rm GeV},\qquad \Gamma_H=122.5\,\text{GeV}  \,,
\end{align*}
where the widths are determined from HDECAY~\cite{Djouadi:1997yw}.
We use the NLO PDF fit of Martin, Stirling, Thorne and Watt \cite{Martin:2009iq} with $\alpha_s(m_Z)=0.12018$. 

The results in this section were obtained using the analytic results for the partonic process documented in \cite{Campbell:2011cu}. Scalar loop integrals were evaluated using the LoopTools package \cite{Hahn:2000jm}, and phase space integrals were done using the Cuba integration package \cite{Hahn:2004fe}. For all the results presented in this section, we have integrated over the leptonic phase space, and allow for off-shell vector bosons.

\subsection{Off-Shell Higgs Production} \label{sec:conv}

We begin by studying the effect of the jet veto on far off-shell Higgs production in $gg\rightarrow H \rightarrow WW\rightarrow e^+\nu_e \mu \bar \nu_\mu$. While a full analysis of the off-shell region also requires the inclusion of signal-background interference, for which the hard function to NLO is not known, we use the off-shell Higgs mediated process to study the convergence of the resummed predictions. In particular, one is first sensitive to the jet radius at NNLL. The ability to perform the resummation to NNLL for the Higgs mediated signal enables us to assess the convergence of the resummed predictions in the off-shell region. It also allows us to check that the NLL result, which will be used when signal-background interference is included, accurately captures the effect of the jet veto reasonably well. In particular, we will focus on the shape of the differential distribution in $\hat s$. As will be discussed in more detail in \Sec{sec:bound}, the procedure for extracting a bound on the Higgs width from the off-shell cross section uses a rescaling procedure to the on-shell cross section. Because of this, the shape, but not the normalization of the distribution is important for an accurate application of this method. Therefore, as in \Sec{general} we will rescale the differential cross sections by $E_0(m_H^2)$, allowing us to focus just on the shape.
\begin{figure}
\begin{center}
\subfloat[]{\label{fig:conv_30}
\includegraphics[width=6cm]{figures/pureHiggs30R5}
}
$\qquad$
\subfloat[]{\label{fig:conv_20}
\includegraphics[width=6cm]{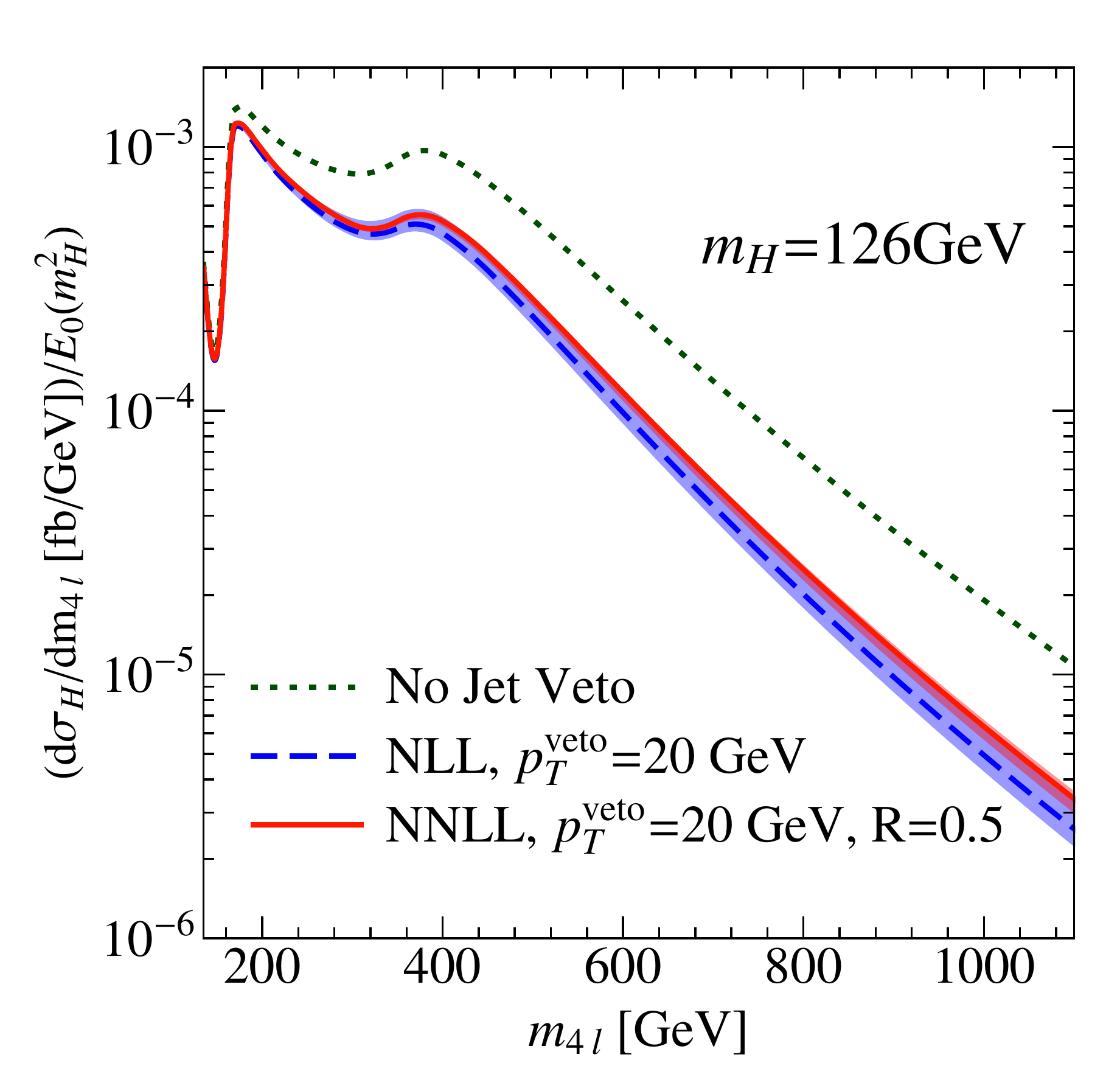}
}
\end{center}
\caption{The off-shell Higgs cross section in the exclusive zero jet bin for $\ptveto=30$ GeV in (a), and $\ptveto=20$ GeV in (b), with $R=0.5$ in both cases. Results are normalized by the jet veto suppression at the Higgs mass, such that the on-shell cross section is the same in all cases, allowing one to focus on the modification to the shape of the distribution. NLL and NNLL results are similar, with a small modification due to the finite jet radius, which is not present in the NLL calculation. }
\label{fig:conv}
\end{figure}

The NNLL calculation requires the NNLL beam and soft functions, which are known in the literature for a jet veto defined by a cut on $p_T$ \cite{Stewart:2013faa}, as well as the virtual part of the NLO gluon-fusion hard function. The NLO virtual contributions for gluon fusion Higgs production are known analytically with full dependence on the top and bottom quark mass \cite{Harlander:2005rq}, which is necessary, as in the off-shell region one transitions through the $\sqrt{\hat s} =2m_t$ threshold.\footnote{The analytic NLO virtual corrections were also used in \cite{Banfi:2013eda} to study the dependence of the jet veto on the b-quark mass for the case of on-shell Higgs production.} The NLO hard function is determined by matching onto the gluon-fusion operators in SCET, as discussed in \Sec{sec:hardfunc}. We do not include in our calculation the non-singular pieces, as we focus on the region $\ptveto \ll \sqrt{\hat s}$, where the singular contributions dominate, and are not interested in the transition to the region $\ptveto \sim \sqrt{\hat s}$.

In \Fig{fig:conv} we plot the resummed distribution, normalized to the jet veto suppression at the Higgs mass: $(d\sigma_0/dm_{4l})/E_0(m_H^2)$, for off-shell $gg \rightarrow H \rightarrow WW \rightarrow e^+\nu_e \mu \bar \nu_\mu$. Note that in the case without a jet veto, the jet veto suppression at the Higgs mass is defined to be $1$. We have integrated over the leptonic phase space. Here $m_{4l}=\sqrt{\hat s}$ is the invariant mass of the 4 lepton final state. In \Fig{fig:conv_30} we use $\ptveto=30$ GeV, and in \Fig{fig:conv_20} we use $\ptveto=20$ GeV. In both cases, we use a jet radius of $R=0.5$, as is currently used by the CMS collaboration. The uncertainty bands are rough uncertainty estimates from scale variations by a factor of 2. Note that in the calculation, we use a five flavour scheme, even above $m_t$ since the difference with using a six flavour coupling is well within our error band.

\Figs{fig:conv_30}{fig:conv_20} show a small modification to the differential distribution between NLL and NNLL. This arises primarily due to the clustering logarithms, which introduce dependence on the jet radius, which is not present at NLL. The $R$ dependence reproduces the expected physical dependence of the cross section on $R$: for a fixed $\ptveto$ cut, the restriction on radiation from the initial partons becomes weaker as the jet radius is decreased, causing a smaller suppression of the cross section. Despite this, the shape is well described by the NLL result.  In particular, the NLL result captures the dominant effect of the exclusive jet veto on the off-shell cross section. This is important for the resummation of the interference, considered in \Sec{sec:int_126}. In this case, higher order results are not available (for some approximate results, see \cite{Bonvini:2013jha}), and therefore one is restricted to an NLL resummation. However, the results of this section demonstrate that the NLL result accurately captures the effects of the jet veto on the shape of the distribution as a function of $\hat s$.

\subsection{Signal-Background Interference}\label{sec:int_126}
Signal-background interference for the process $gg\rightarrow  l\nu l\nu$ has been well studied in the literature\cite{Campbell:2011cu,Kauer:2012ma,Kauer:2013qba}. The interference comes almost exclusively from the $\sqrt{\hat s}>2M_W$ region. For a light Higgs, $m_H< 2M_W$, this means that the interference comes entirely from $\sqrt{\hat s} > m_H$. For a heavy Higgs, the Higgs width is sufficiently large that there are contributions to the interference from a wide range of $\sqrt{\hat s}$. The signal-background interference is therefore, in both cases, an interesting process on which to demonstrate the effect of the jet veto.

The NLO virtual corrections are not available for the interference process, restricting the resummation accuracy to NLL. However, as argued in \Sec{sec:conv} if one is interested in the shape of the distribution, and not the normalization, the NLL captures the effects of the jet veto. One thing that cannot be known without a full calculation of the NLO virtual contributions to the interference is if the NLO virtual contributions for the interference are different than for the signal. For the case of interference in $H\rightarrow \gamma \gamma$ where they are known, the virtual contributions for the interference were found to be smaller than for the signal \cite{Dixon:2013haa}. Due to the similar structure of the diagrams for $H\rightarrow WW$, the same could certainly be true. However, we expect this to be a minor correction compared to the effects of the jet veto. In particular, we do not expect the K-factor to have strong $\hat s$ dependence, which is the important effect captured by the resummation. In this section we use the LO result for $gg \rightarrow e\nu \mu \nu$, fully differential in the lepton momenta, which is available in the MCFM code, and is documented in \cite{Campbell:2011cu}.

We begin by reviewing the notation for the signal-background interference in $gg\rightarrow e \nu \mu \nu$ at LO following \cite{Campbell:2011cu}. It is convenient to pull out the dependence on $m_H$ and $\Gamma_H$ coming from the s-channel Higgs propagator. Defining $\widetilde C^H=(\hat s-m_H^2+im_H \Gamma_H)C^H$, we can separate the Hard function for the signal-background interference into its so called ``Imaginary'' and ``Real'' contributions:
\be \label{eq:int_decomp}
H^{int}=\frac{2(\hat s-m_H^2)}{(\hat s-m_H^2)^2+m_H^2 \Gamma_H^2} \text{Re}\left [ \widetilde C^H( C^C)^\dagger \right ]+\frac{2m_H \Gamma_H}{(\hat s-m_H^2)^2+m_H^2 \Gamma_H^2}\text{Im}\left [ \widetilde C^H (C^C)^\dagger \right ].
\ee
In \Eq{eq:int_decomp} there is a sum over helicities of the Wilson coefficients, which for notational convenience has not been made explicit. Note that the imaginary part of the interference is multiplied by an explicit factor of $\Gamma_H$, and is therefore negligible for a light Higgs. 

\begin{figure}
\begin{center}
\subfloat[]{\label{fig:int_126}
\includegraphics[width=7cm]{figures/int_126} 
}
$\qquad$
\subfloat[]{\label{fig:int_600}
\includegraphics[width=6.9cm]{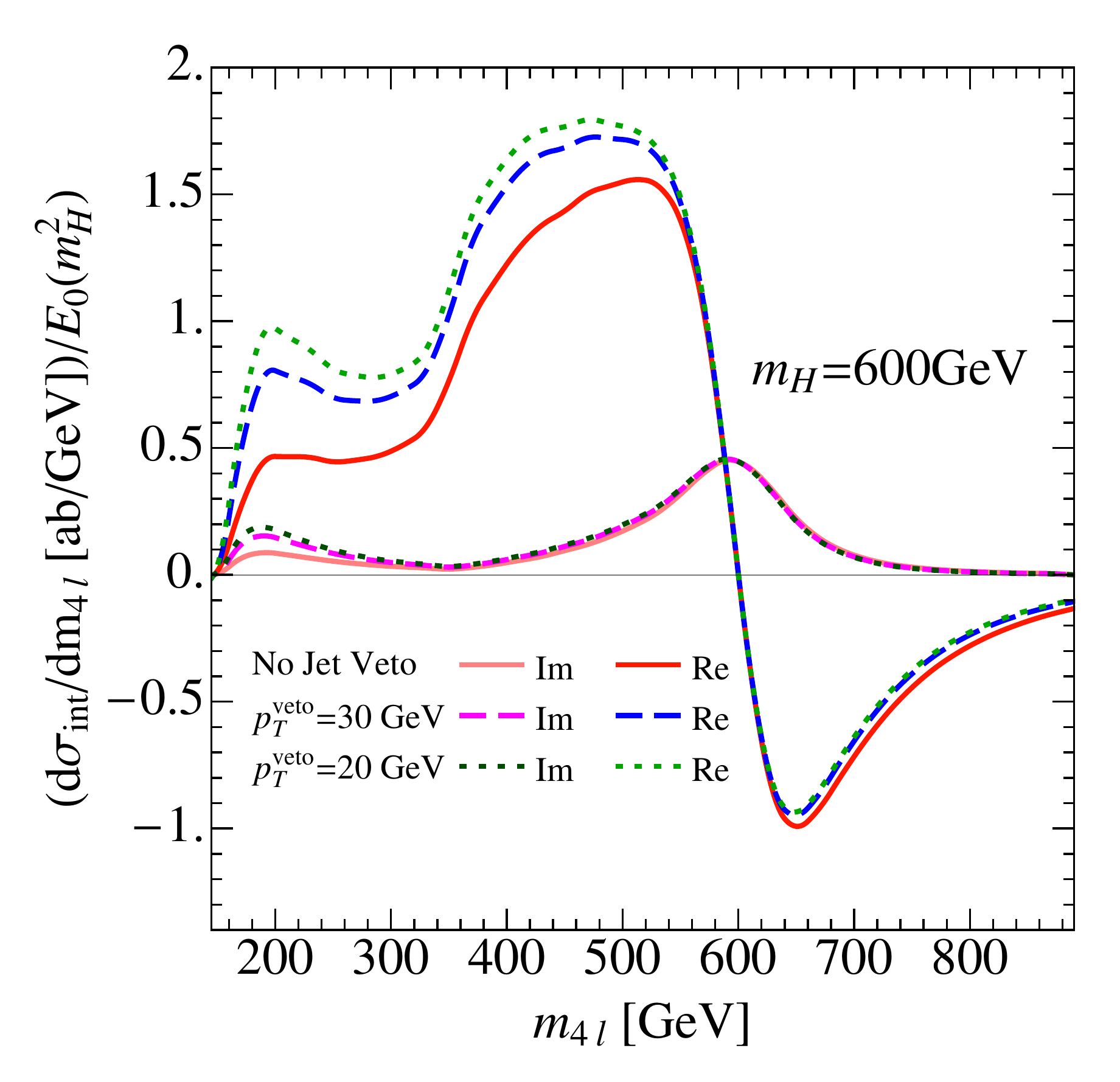}
}
\end{center}
\caption{Signal-background interference in $gg\rightarrow e^+ \nu_e \mu \bar \nu_\mu$ for (a) $m_H=126$ GeV, and (b) $m_H=600$ GeV. NLL predictions are shown for $\ptveto=20,30$ GeV, and have been rescaled by the jet veto efficiency at $m_H$. The size of the signal-background interference relative to the on-shell cross section is enhanced by the jet veto for a heavy Higgs, whereas it is suppressed for a light Higgs.  For $m_H=126$ GeV the jet veto causes a significant reduction of the cross section in the far off-shell region relative to the on-shell cross section.}
\label{fig:int}
\end{figure}

The interference without a jet veto is shown in \Fig{fig:int_126} for a $126$ GeV Higgs and \Fig{fig:int_600} for a $600$ GeV Higgs, as a function of $m_{4l}$. We have integrated over the phase space of the leptons, including allowing for off-shell vector bosons. The interference is negligible below the $\sqrt{\hat s}=2m_W$ threshold. For the case of $m_H=126$ GeV the only non-negligible contribution is the real part of the interference above the Higgs pole, which gives a negative contribution to the total cross section. In the case of $m_H=600$ GeV, there is significant interference both above and below the Higgs pole, and from both the real and imaginary parts. The interference below the pole dominates, leading to a net positive contribution to the total cross section. We have chosen these two Higgs masses, where the interference has a different $\sqrt{\hat s}$ dependence, so as to demonstrate the different effects that a jet veto can have on signal-background interference.

\Fig{fig:int} also shows as a function of $\sqrt{\hat s}$ the result for interference including a jet veto of $\ptveto=20,30$ GeV with NLL resummation, which can be compared with the interference without a jet veto. To make the interpretation of \Fig{fig:int} as simple as possible, we have rescaled the interference by $E_0(m^2_H)$, the jet veto efficiency at $m_H$. Therefore, enhancements and suppressions in the jet vetoed interference correspond to enhancements and suppressions of the interference relative to the on-shell Higgs contribution when a jet veto is applied. As expected from the discussion in \Sec{general}, we find a significant suppression of the interference at higher $\sqrt{\hat s}$, and this suppression increases with $\sqrt{\hat s}$.
 For $m_H=126$ GeV, shown in \Fig{fig:int_126}, the interference comes entirely from above $\sqrt{\hat s}=m_H$, and is therefore more highly suppressed by the jet veto relative to the on-shell Higgs cross section. 
 However, the situation is quite different for  the case of $m_H=600$ GeV, shown in \Fig{fig:int_600}.  Here the dominant contribution to the interference is from the real part in \eq{int_decomp}, which changes sign at $\sqrt{\hat s}=m_H$. The real part of the interference coming from below $\sqrt{\hat s}=m_H$ is positive and is partly cancelled by negative interference from above $\sqrt{\hat s}=m_H$ if we integrate over $\hat s$. The jet veto suppresses the on-shell cross section and the negative interference from above $\sqrt{\hat s}=m_H$ more than the contribution from the positive interference below $\sqrt{\hat s}=m_H$, and therefore the jet veto acts to enhance the interference contribution relative to the signal. This enhancement is significant in the case of $m_H=600$ GeV, as the interference has contributions starting at $m_{4l} \simeq 2m_W$, where the suppression due to the jet veto is smaller. To quantify this further we can consider the effect of the jet veto on the ratio
\be\label{eq:int_ratio}
R_I =\frac{\sigma_{H+I}}{\sigma_{H}} \,,
\ee
where $\sigma_{H+I}$ is the cross section including the signal-background interference, and $\sigma_H$ is the Higgs mediated cross section. The behaviour of this ratio is different for the two Higgs masses considered. Numerical values of $R_I$ are shown in \Tab{tab:ratio}. The effect of interference for $m_H=126$ GeV with or without the jet veto is fairly small, and would be made even smaller when cuts are made to eliminate interference.  However, for $m_H=126$ GeV the effect of the jet-veto can also be significantly amplified when cuts are used to maximize sensitivity to the Higgs width. For example, the analysis of \cite{Campbell:2013wga} considered the region $M_T>300$ GeV to bound the Higgs width. Since $m_{4l} \geq M_T$, we see from \Fig{fig:int_126} that in this region the effect of the exclusive jet veto is by no means a small effect, giving a suppression of $\sim 1.5- 2$. A representative error band from scale variation is also shown in \Fig{fig:int_126}. The effect on the derived bound will be discussed in \Sec{sec:bound}.

These two examples demonstrate that a jet veto can have an interesting interplay with signal-background interference, enhancing or suppressing its contribution relative to the Higgs mediated cross section, depending on the particular form of the interference. A detailed understanding of the interference is of phenomenological interest for both a light and heavy Higgs. In the case of $m_H=126$ GeV, the interference can be efficiently removed by cuts when studying the on-shell cross section \cite{Campbell:2011cu}, but is important for the understanding of the off-shell cross section. In the case of a heavy Higgs, the interference is important for heavy Higgs searches \cite{Campbell:2011cu,TheATLAScollaboration:2013zha,Chatrchyan:2013yoa}, where it is a large effect, and cannot be easily removed by cuts. The effect of the jet veto must therefore be incorporated in such searches.
\begin{table}[t]
\begin{center}
\begin{tabular}{c|c|c|c}
&No Veto & $\ptveto=30$ GeV & $\ptveto=20$ GeV \\ 
\hline
$m_H=126$ GeV & 0.92 & 0.94 &  0.95\\
$m_H=600$ GeV &1.38& 1.49 &  1.54 \\
\end{tabular}
\end{center}
\caption{Values of $R_I=\frac{\sigma_{H+I}}{\sigma_{H}}$ for $m_H=126$ GeV and $m_H=600$ GeV for two different values of $\ptveto$. As is clear from \Fig{fig:int}, the jet veto causes a suppression of the importance of the interference relative to the Higgs mediated process for a light Higgs, and an enhancement for a heavy Higgs. }
\label{tab:ratio}
\end{table}

\subsection{Suppression as a Function of $M_T$}\label{sec:mt_126}

We have so far discussed the effect of the jet veto on the cross section as a function of $\sqrt{\hat s}$, as the Sudakov factor is explicitly a function of $\sqrt{\hat s}$. However, in the case of $H\rightarrow WW\rightarrow l \nu l\nu$, the total invariant mass of the leptons cannot be reconstructed. A substitute for $\sqrt{\hat s}$, used in \cite{Campbell:2011cu}, and which is measured by the CMS and ATLAS collaborations \cite{ATLAS:2013wla,TheATLAScollaboration:2013zha,CMS:eya} is the transverse mass variable, $M_T$ defined in \Eq{eq:mt}.

\begin{figure}
\begin{center}
\includegraphics[width=6.8cm]{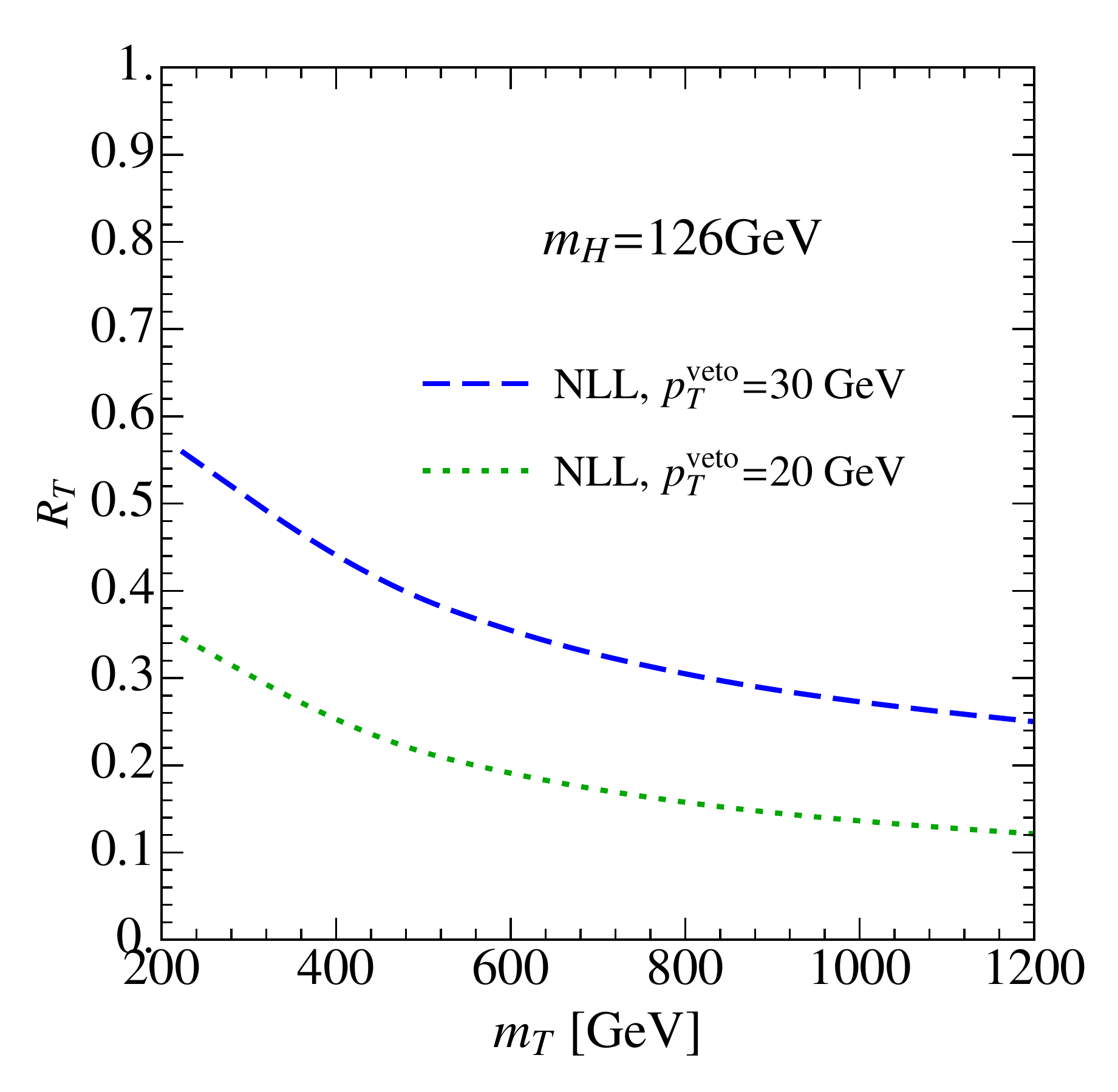}
\end{center}
\caption{Suppression of the exclusive zero jet cross section for off-shell Higgs production as a function of $M_T$. The Sudakov factor is controlled by $\sqrt{\hat s}$, but since $\sqrt{\hat s} \geq M_T$, a larger suppression is observed as a function of $M_T$.} \label{fig:MT}
\end{figure}

Similarly to the ratios considered in \Sec{general} of the exclusive zero jet cross section to the total cross section, as a function of $\hat s$, in \Fig{fig:MT}, we plot the variable
\be \label{eq:MTratio}
R_T=\left .   \left (  \frac{  d\sigma_0^{NLL}(\ptveto)  } {dM_T}   \right )   \middle /      \left (  \frac{d\sigma}{dM_T} \right)      \right . ,
\ee
for $gg\rightarrow H\rightarrow WW\rightarrow \bar \nu_\mu \mu e^+ \nu_e$ in the far off-shell region. Since $M_T$ is designed as a proxy for $\hat s$, the behaviour is as expected from the discussion of \Sec{general}, however, since $\sqrt{\hat s}\geq M_T$, the events contributing at a given $M_T$ all have a larger $\sqrt{\hat s}$. Since it is the $\sqrt{\hat s}$ that governs the Sudakov suppression due to the jet veto, the suppression due to the jet veto at a given $M_T$ is larger than at the same value of $\sqrt{\hat s}$.

We should note that while the values of $M_T$ at which the suppression due to the jet veto becomes significant are larger than is normally considered, or studied experimentally, the authors of \cite{Campbell:2013wga} show that with an improved understanding of the backgrounds in the ATLAS $N_{jet}=0$ bin of the $WW$ data, the $M_T>300$ GeV region, where the jet veto effects are indeed significant, can be used to place a competitive bound on the Higgs width. As will be discussed in \Sec{sec:bound}, their method relies heavily on having an accurate description of the shape of the shape of the $M_T$ distribution, which is modified by the jet veto. This section demonstrates that in the exclusive zero jet bin, there is a suppression by a factor of $\sim 2$ above $M_T>300$ GeV, which is a significant effect. This will cause a corresponding weakening of the bound on the Higgs width by a similar factor, which we discuss further in \Sec{sec:bound}.

\subsection{From $8$ TeV to $13$ TeV}\label{sec:higherEcm}
Since the focus will soon shift to the $13$ TeV LHC, in this section we briefly comment on how the effects discussed in the previous sections will be modified at higher $E_{\text{cm}}$. In \Sec{sec:Ecm} we noted that at higher $E_{\text{cm}}$ the jet veto suppression has an increased dependence on $\hat s$ due to the larger range of Bjorken $x$ that is probed in the PDFs. The larger range of available $x$ increases the gluon luminosity at high $\hat s$ allowing for an increased contribution to the cross section from far off-shell effects \cite{Campbell:2013una,Campbell:2013wga}, and increasing the range over which they contribute, potentially amplifying the effects of the jet veto discussed in the previous sections.

\begin{figure}
\begin{center}
\includegraphics[width=8cm]{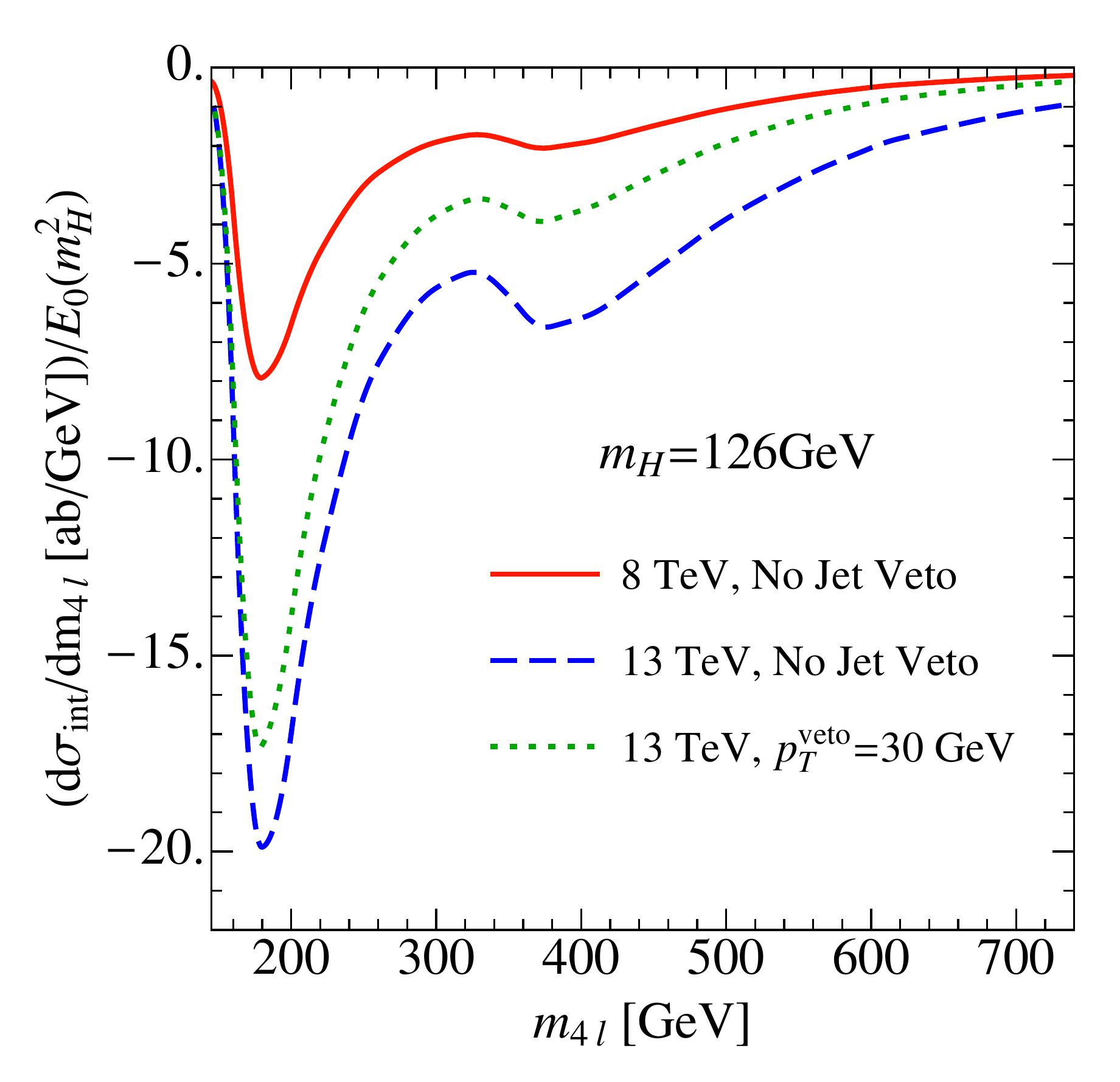}
\end{center}
\caption{A comparison of the signal-background interference, and impact of the jet veto, at $8$, $13$ TeV. At a higher centre of mass energy there is a larger contribution to the cross section from higher $m_{4l}$, where the effect of the exclusive jet veto is largest.}
\label{fig:int_13TeV}
\end{figure}
In this section we consider one example to demonstrate this point, the signal-background interference for $m_H=126$ GeV. The signal-background interference distribution for $E_\text{cm}=8$ and $13\,$TeV is shown in \Fig{fig:int_13TeV}, along with the signal-background interference in the exclusive zero-jet bin at $13$ TeV. As was done in \Fig{fig:int} for the NLL predictions, we have normalized the distribution by the jet veto suppression at $m_H$ so that the suppression is relative to the on-shell production. The most obvious modification compared with $E_\text{cm}=8$ TeV, is the significant enhancement of the signal-background interference cross section, due to the large enhancement of the gluon luminosity at larger $\hat s$. In particular, the contribution to the signal-background interference cross section from the peak at $m_{4l}=2m_t$ is enhanced at higher $E_\text{cm}$, relative to the contribution from $m_{4l}\sim 2m_W$. Since there is a larger relative contribution from higher invariant masses, where the suppression due to the jet veto is larger, the effect of an exclusive jet veto is larger at $13$ TeV. This is in addition to the fact that at $13$ TeV, the $\hat s$ dependence of the suppression due to the jet veto is slightly stronger, as was demonstrated in \Sec{sec:Ecm}.

We again emphasize that when cuts are applied to gain sensitivity to the off-shell region, the effect of the jet veto is not small. In particular, for $13$TeV, there is a significant region above $m_{4l}\sim350$GeV where the suppression due to the jet veto is $\gtrsim 2$, as is seen in \Fig{fig:int_13TeV}.

Although we have focused on the effect of an increased centre of mass energy on a particular observable, the conclusions apply generically, for example, for the $H\rightarrow ZZ\rightarrow 4l, 2l2\nu$ channel, which exhibits similar signal-background interference. Indeed as the centre of mass energy is increased one has the ability to probe phenomena over an increasingly large range of $\hat s$. This amplifies the effects of off-shell physics, as well as the effect of an exclusive jet veto. These effects will be important in any physics channel for which a jet veto is applied, and for which one is interested in the physics over a range of $\hat s$.

\section{Effect of Jet Vetoes on Higgs Width Bounds}\label{sec:bound}

In this section we discuss the impact of jet binning and jet vetoes for the recent program to use the off-shell cross section in $H\rightarrow WW,~ZZ$ to bound the Higgs width \cite{Caola:2013yja,Campbell:2013una,CMS:2014ala,Campbell:2013wga}. Although we have focussed on the case of $H\rightarrow WW$, we will review also the strategy for $H\rightarrow ZZ$ which is similar, also exhibiting a large contribution from the far off-shell cross section, as well as signal-background interference analogous to that in $H\rightarrow WW$. We first discuss the procedure used to bound the Higgs width, and then relate it to our discussion of the suppression of the off-shell cross section in the exclusive zero jet bin. Our focus will be on the  effect of the jet vetoes, rather than carrying out a complete numerical analysis. In particular the proper incorporation of backgrounds, and additional experimental cuts is beyond the scope of this chapter.

The method used to bound the Higgs width in Refs.~\cite{Caola:2013yja,Campbell:2013una,Campbell:2013wga} can be phrased in a common language for both $H\rightarrow ZZ, ~WW$. It is based on the different scalings of the on-shell, off-shell and interference contributions to the Higgs cross section, as discussed in \Sec{sec:intro}. Recalling the scaling from \Eq{Eq:Scaling}, the total cross section can be written as
\begin{align}\label{eq:scaling_bound}
\sigma_{H+I}=A+B \left ( \frac{\Gamma_H}{\Gamma_H^{SM}} \right)+C\sqrt{\frac{\Gamma_H}{\Gamma_H^{SM}}},
\end{align}
where the coefficients $A,B,C$ correspond to on-shell, off-shell Higgs mediated, and signal-background interference contributions, respectively. The coefficients depend strongly on the set of cuts that are applied.
To extract a bound on the Higgs width, the procedure is as follows. First, one determines a normalization factor between the experimental data and theoretical prediction, which is as independent as possible of the Higgs width. This can be done for $WW$ by using a strict $M_T$ cut, for example $0.75m_H\leq M_T \leq m_H$, and for $ZZ$ by using a strict cut on $\hat s$. In both cases, this essentially eliminates the coefficients $B,C$ corresponding to the off-shell production and interference. Once this normalization factor is determined, it is scaled out from the entire differential distribution, so that we now must consider the ratio of offshell and onshell production cross sections.  One can then compute the predicted number of events above some $M_T$, or $\hat s$ value, for example, $M_T,~\sqrt{\hat s} \geq 300$ GeV. In this region, the interference and off-shell production dominate the cross section, so that the coefficients $B,C$ are significant, and the expected number of events is sensitive to the Higgs width, as can be seen from the scalings in \Eq{eq:scaling_bound}. By comparing with the number of events observed by the experiment in this region, one can place a bound on the Higgs width. 

This method relies on the ability to normalize the theoretical prediction to data in the low $M_T$, or $\hat s$ region, which is insensitive to the dependence on the Higgs width, and then use the same normalization in the high $M_T$, or $\hat s$ region where there is a large sensitivity to the Higgs width through off-shell production and signal-background interference. However, to be able to do this, one needs to have an accurate theory prediction for the shape of the $M_T$, or $ \hat s $ distribution, particularly in the high $M_T$, or $\hat s$ region. 

As we have seen, the jet veto significantly modifies the shape of the $M_T$, or $\hat s$ distribution, causing it to fall off more rapidly at high $M_T$, or $\hat s$.   Often we presented our results by normalizing the offshell cross sections to the cross section at the Higgs mass. Given the agreement between theory and experiment at $m_H$, this normalization corresponds exactly to what is done if the theory prediction is normalized to the experimental data in the on-shell region, and therefore shows the extent to which a prediction without the inclusion of a jet veto overestimates the contribution to the cross section at high $M_T$  or $ \hat s$ compared with that in the exclusive zero jet bin. 

In both the $H\rightarrow ZZ$, and $H\rightarrow WW$ analyses, jet vetoes or jet binning are used, so it is interesting to consider how they will effect the width bounds. Their use in the two channels is quite different so we will discuss them separately. 

For the case of $H\rightarrow WW$ the jet veto plays an important role because the exclusive zero jet bin dominates the sensitivity, so the jet veto has a large impact on the potential Higgs width bound. This is because effectively it is more difficult to recover the jets which migrate out of the zero jet bin. The plots of the off-shell distributions in \Sec{sec:WW} show the extent to which a prediction without the inclusion of a jet veto overestimates the contribution to the cross section at high $M_T$. This will lead to a weakening on the bound of the Higgs mass, compared with a calculation that does not incorporate the effect of the jet veto. For example, in \cite{Campbell:2013wga}, which first proposed the use of the $H\rightarrow WW$ channel, the estimated sensitivity was derived by comparing an inclusive calculation for the off-shell cross section with data in the exclusive zero jet bin. Here the effect of the restriction to the zero jet bin is not small, and will worsen the bounds by a factor of $\sim 2$, as can be seen by the suppression of the far off-shell cross section in the exclusive zero jet bin, shown in \Sec{sec:WW}.  In an analogous experimental analysis this Sudakov suppression from the jet veto will be accounted for up to some level of precision by the use of a parton shower.  Because this is such a large effect, we believe that an experimental analysis of the high $M_T$ region performed to bound the Higgs width, would benefit from using an analytic calculation of the jet veto suppression in the exclusive zero jet bin, instead of relying on the parton shower. We have demonstrated that the resummation, including the signal-background interference, can be achieved to NLL. Once the NLO virtual corrections are calculated for the interference, these results can also easily be extended to NNLL.

In the $H\rightarrow ZZ\rightarrow 2l2\nu$ channel the situation is different, as the jet binning procedure is used to optimize sensitivity, splitting the data into exclusive 0-jet and inclusive 1-jet categories with comparable bounds coming from each category~\cite{CMS:2014ala}.  Because the inclusive 1-jet channel is still experimentally clean, the large migration to the inclusive 1-jet bin discussed in \Sec{sec:1jet} should have a small (or no) impact on the width bounds derived from the combination of the two channels in $H\to ZZ$. A proper treatment of the migration of events with changing $\hat s$ is still important when considering any improvement that is obtained by utilizing jet binning.  The analytic results for the Sudakov form factor discussed here for $H\to WW$ could be utilized for jet bins for $H\to ZZ$ in a straightforward manner.

\section{Conclusions}\label{sec:conclusion}

In this chapter a study of the effect of jet vetoes on off-shell cross sections was made. A factorization theorem in SCET allowed us to analytically treat the summation of Sudakov logarithms, and make a number of general statements about the effect of the jet veto. In particular, the restriction on radiation imposed by the jet veto causes a suppression to the exclusive 0-jet cross section, and correspondingly an enhancement of the inclusive 1-jet cross section, which depends strongly on $\hat s$. For gluon initiated processes the $\hat s$ dependence of the suppression is greater than for quark initiated processes, which is important for channels where the signal is dominated by one production channel, and the background by another. 

The fact that the jet veto suppression is $\hat s$ dependent has interesting effects on differential distributions in $\hat s$, as well as on signal-background interference. To demonstrate these effects, we considered the $gg\rightarrow H\rightarrow WW$ process, which has large off-shell contributions as well as signal-background interference. We performed an NLL resummation for the $gg\rightarrow H\rightarrow WW\rightarrow l\nu l\nu$ process, including a discussion of the resummation for the signal-background interference, for $m_H=126$ GeV, and $m_H=600$ GeV. These two examples demonstrated that depending on the structure of the interference, the jet veto can either enhance or suppress interference effects relative to the on-shell production. For a low mass Higgs a suppression is observed, while for a high mass Higgs there is a significant enhancement in the interference. These effects must be properly incorporated in high mass Higgs searches that use jet vetoes.

The modification of differential distributions in $\hat s$ or $M_T$ due to the $\hat s$ dependence of the jet veto suppression is particularly relevant to a recent program to bound the Higgs width using the off-shell cross section \cite{Caola:2013yja,Campbell:2013una,Campbell:2013wga,CMS:2014ala}. In particular, for the $H\rightarrow WW$ channel, where an exclusive 0-jet veto is imposed to mitigate large backgrounds, the jet veto weakens the bound on the Higgs width by a factor of $\sim 2$ compared to estimates without accounting for the jet veto. Furthermore, since the suppression in the exclusive 0-jet bin corresponds to an enhancement in the inclusive 1-jet bin, and the migration is significant as a function of $\sqrt{\hat s}$ a proper understanding of the effect of the jet veto is crucial for experimental analyses which use jet binning.  
This migration may for example play some role in $H\rightarrow ZZ\rightarrow 2l2\nu$, which was recently used by CMS to place a bound on the off-shell Higgs width, and which uses jet binning in exclusive 0-jet and inclusive 1-jet bins \cite{CMS:2014ala}. 

We presented a factorization theorem in SCET which allows for the resummation of large logarithms of $\sqrt{\hat s}/ \ptveto$, including for the signal-background interference, in a systematically improveable manner. This allows for the analytic study of the effect of the jet veto on the exclusive 0-jet and inclusive 1-jet bins, including the correlations in their theory uncertainties. A complete NNLL calculation would require the calculation of the NLO virtual corrections to the interference, but would allow for the analytic incorporation of jet radius effects, and would place the study of the off-shell cross section on a firmer theoretical footing. Furthermore, since our hard functions are fully differential in leptonic momenta, realistic experimental cuts on the leptonic phase space can be easily implemented. We leave a more detailed investigation, including the treatment of such cuts, and a calculation of the effect of the jet veto on the backgrounds, for future study.

With the LHC beginning its $13$ TeV run in the near future, the importance of the effects discussed in this chapter will be amplified. As the centre of mass energy is raised, the range of $\hat s$ which can be probed increases. This typically increases the importance of off-shell effects, as well as the impact of the jet veto, which is essential for an accurate description of the differential distribution in $\hat s$.  In general a proper theoretical understanding of jet vetoes and jet binning for large $\hat s$ can be achieved through resummation, and is important when theoretical cross sections are needed for the interpretation of experimental results.




%% file: chap_conc.tex

\DeclareRobustCommand{\Sec}[1]{Sec.~\ref{#1}}
\DeclareRobustCommand{\Secs}[2]{Secs.~\ref{#1} and \ref{#2}}
\DeclareRobustCommand{\App}[1]{App.~\ref{#1}}
\DeclareRobustCommand{\Tab}[1]{Table~\ref{#1}}
\DeclareRobustCommand{\Tabs}[2]{Tables~\ref{#1} and \ref{#2}}
\DeclareRobustCommand{\Fig}[1]{Fig.~\ref{#1}}
\DeclareRobustCommand{\Figs}[2]{Figs.~\ref{#1} and \ref{#2}}
\DeclareRobustCommand{\Eq}[1]{Eq.~(\ref{#1})}
\DeclareRobustCommand{\Eqs}[2]{Eqs.~(\ref{#1}) and (\ref{#2})}
\DeclareRobustCommand{\Ref}[1]{Ref.~\cite{#1}}
\DeclareRobustCommand{\Refs}[1]{Refs.~\cite{#1}}

\chapter{Conclusions}

In this thesis I have discussed applications of effective field theories, in particular SCET, to describing QCD at the LHC. In particular, I have focused on the development of extensions of SCET to describe additional hierarchical scales present in jet substructure, as well as formalisms in SCET to facilitate the description of more complicated hard scattering processes.

In the area of jet substructure I have introduced new observables to distinguish hadronically decaying boosted electroweak resonances from QCD background jets. These observables are theoretically clean, and offer advantages over previously proposed observables. Motivated by these observables, I have proposed effective field theory descriptions of subjet configurations allowing for a systematic resummation and factorization of jet substructure observables, and performed the first calculation of such an observable.  For the calculation of precision cross sections at the LHC, I have developed a helicity operator formalism in SCET to facilitate the matching of fixed order results onto the effective theory. This greatly simplifies the analysis of final states with multiple particles.

The effective theory tools developed in this thesis should aid in furthering our understanding of fundamental physics using the Large Hadron Collider over the coming decades.

%% file: appa.tex
\chapter{One-Loop Calculations for $D_2$}

\section{Conventions and SCET Notation}

In the body of the text we have presented the required factorization theorems for studying the two-prong substructure of jets using the $D_2$ observable. Although all the factorization theorems were presented, only heuristic descriptions of the functions appearing in the factorization theorems were presented in an attempt to appeal to a broader audience, and so as to not distract the reader with technical complications. In these appendices, we give the operator definitions of the functions appearing in the factorization theorems of \Sec{sec:Fact}, and calculate the functions to one-loop accuracy. 

In this appendix we begin by summarizing some notation and conventions. The factorization theorems presented in this chapter are formulated in the language of SCET \cite{Bauer:2000yr,Bauer:2001ct,Bauer:2001yt,Bauer:2002nz}. We assume that the reader has some familiarity with the subject, and will only define our particular notation, and review the definition for common SCET objects. We refer readers unfamiliar with SCET to the reviews \cite{iain_notes,Becher:2014oda}.

SCET is formulated as a multipole expansion in the momentum components along the jet directions. Since we take the jet directions to be lightlike, it is
convenient to work in terms of light-cone coordinates. We define two light-cone vectors
\begin{equation}
n^\mu = (1, \vec{n})
\,,\qquad
\bn^\mu = (1, -\vec{n})
\,,\end{equation}
with $\vec{n}$ a unit three-vector, which satisfy the relations $n^2 = \bn^2 = 0$ and  $n\cdot\bn = 2$.
We can then write any four-momentum $p$ as
\begin{equation} \label{eq:lightcone_dec}
p^\mu = \bn\sdt p\,\frac{n^\mu}{2} + n\sdt p\,\frac{\bn^\mu}{2} + p^\mu_{n\perp}
\,.\end{equation}
A particle in the $n$-collinear sector has momentum $p$ close to the $\vec{n}$ direction,
so that its momentum scales like
$(n\!\cdot\! p, \bn \!\cdot\! p, p_{n\perp}) \sim
\bn\!\cdot\! p$ $\,(\la^2,1,\la)$, with $\la \ll 1$ 
a small parameter. The parameter $\lambda$ is a generic substitute for the power counting parameters in the different factorization theorems presented in \Sec{sec:Fact}, and since our factorization theorems involve multiple scales, there are generically multiple distinct $\lambda$s.

In the effective field theory, the momentum of the particles in the $n$-collinear sector are multipole expanded, and written as
\begin{equation} \label{eq:label_dec}
p^\mu = \lp^\mu + k^\mu = \bn\sdt\lp\, \frac{n^\mu}{2} + \lp_{n\perp}^\mu + k^\mu\,,
\,\end{equation}
where $\bn\cdot\lp$ and $\lp_{n\perp}$ are large momentum
components, which label fields, while $k$ is a small residual momentum, suppressed by powers of $\lambda$. This gives rise to an effective theory expansion in powers of $\la$.

SCET fields for quarks and gluons in the $n$-collinear sector, $\xi_{n,\lp}(x)$ and
$A_{n,\lp}(x)$, are labeled by the lightlike vector of their collinear sector, $n$ and their large momentum $\lp$. We will write the fields in a mixed position space/momentum space notation, using position space for the residual momentum and momentum space for  the large momentum components. 
The residual momentum dependence can be extracted using the derivative operator $\img\partial^\mu \sim k$, while the large label momentum is obtained from the momentum label operator $\cP_n^\mu$.

Operators and matrix elements in SCET are constructed from collinearly gauge-invariant quark and gluon fields, defined as \cite{Bauer:2000yr,Bauer:2001ct}
\begin{align}
\chi_{n,\w}(x) &= \left[\delta(\w - \bnP_n)\, W_n^\dagger(x)\, \xi_n(x) \right]
\,,\\
\cB_{n,\w\perp}^\mu(x)
&= \frac{1}{g}\left[\delta(\w + \bnP_n)\, W_n^\dagger(x)\,\img D_{n\perp}^\mu W_n(x)\right]
\,.\end{align}
The $\perp$ derivative in the definition of the SCET fields is defined using the label momenta operator as
\begin{equation}
\img D_{n\perp}^\mu = \cP^\mu_{n\perp} + g A^\mu_{n\perp}\,,
\end{equation}
and
\begin{equation} 
W_n(x) = \left[\sum_\text{perms} \exp\left(-\frac{g}{\bnP_n}\,\bn\sdt A_n(x)\right)\right]\,,
\end{equation}
is a Wilson line of $n$-collinear gluons. 
We use the common convention that the label operators in the definition of the SCET fields only act inside the square brackets. Although the Wilson line $W_n(x)$ is a non-local operator, it is localized with respect to the residual position $x$, and we can therefore treat $\chi_{n,\w}(x)$ and $\cB_{n,\w}^\mu(x)$ as local quark and gluon fields when constructing operators. The operator definitions for jet functions in these appendices are given in terms of these collinear gauge invariant quark and gluon SCET fields.

Our operator definitions will also involve matrix elements of eikonal Wilson lines, which arise from the soft-collinear factorization through the BPS field redefinition at the Lagrangian level \cite{Bauer:2001yt}. The Wilson lines extend from the origin to infinity along the direction of a lightlike vector, $q$, specifying their directions. Explicitly
\begin{align}
S_q={\bf P} \exp \left( ig \int\limits_0^\infty ds\, q\cdot A(x+sq)    \right)\,.
\end{align}
Here $\bf P$ denotes path ordering, and $A$ is the appropriate gauge field for any sector which couples eikonally to a collinear sector with label $q$ (for example collinear-soft, soft, boundary soft), and the color representation has been suppressed. All Wilson lines are taken to be outgoing, since we consider the case of jet production from $e^+e^-$ collisions.

Throughout this chapter we have considered the production of two jets, one of which has a possible two-prong substructure, in an $e^+e^-$ collider. This implies the presence of at most three Wilson lines in the soft or collinear soft function. With only three Wilson lines, all possible color structures can be written as a sum of color-singlet traces. In the more general case, with more than three Wilson lines, the soft function is a color matrix which must be traced against the hard functions, which are also matrices in color space, appearing in the factorization theorem for the cross section (see e.g.~\Refs{Ellis:2010rwa,Jouttenus:2011wh} for more details).

In \App{sec:ninja_app} through \App{sec:signal_app} we will give operator definitions for the functions appearing in the factorization theorems in terms of matrix elements of the SCET operators, $\chi_{n,\w}(x)$ and $\cB_{n,\w}^\mu(x)$, as well as products of soft Wilson lines. These matrix elements can be calculated using the leading power SCET Lagrangian, which can be found in \Refs{Bauer:2000yr,Bauer:2001ct,Bauer:2001yt,Bauer:2002nz}, or by using eikonal Feynman rules in the soft functions, and known results for the splitting functions to calculate the jet functions \cite{Ritzmann:2014mka}. We will use the latter approach, as it considerably simplifies the calculations at one-loop.

\section{One Loop Calculations of Collinear Subjets Functions}\label{sec:ninja_app}

In this appendix we collect the calculations relevant to the calculation in the collinear subjets region of phase space, and explicitly show the cancellation of anomalous dimensions. The calculation follows closely that of \Ref{Bauer:2011uc}, with the exception of the form of the measurement function. Nevertheless, the calculation is presented in detail, as the SCET$_+$ effective theory has not been widely used.

\subsection*{Kinematics and Notation}
For our general kinematic setup, we will denote by $Q$ the center of mass energy of the $e^+e^-$ collisions, so that $Q/2$ is the energy deposited in a hemisphere. i.e. the four-momenta of the two hemispheres are
\begin{align}
p_{\text{hemisphere}_1}=\left( \frac{Q}{2},\vec p_1   \right)\,, \qquad p_{\text{hemisphere}_2}=\left( \frac{Q}{2},-\vec p_1   \right)
\end{align}
so
\begin{align}
s=Q^2\,.
\end{align}
We will also denote the energy in a jet at intermediate stages of the calculation by $E_J$, but we will write our final results in terms of $Q$.

We work in the region where one  hemispherical jet splits into two hard subjets, assume the power counting $z\sim \frac{1}{2}$, with $z$ being the energy fraction of one of the jets. We further assume the power counting relations between the energy correlation functions valid in the collinear subjets region, as discussed in \Sec{sec:power_counting}. We adopt the following notation to describe the kinematics of the subjets
\begin{align}
&\text{Subjet a,b momenta: }& &p_a,p_b\\
&\text{Subjet a,b spatial directions: }& &\hat{n}_{a}, \hat{n}_b\\
&\text{Thrust axis: }& &\hat{n}=\frac{\hat{n}_{a}+ \hat{n}_b}{|\hat{n}_{a}+ \hat{n}_b|}\\
&\text{Light-cone vectors: }& &n=(1,\hat{n}),\,\bar{n}=(1,-\hat{n}),\nonumber\\
& & &n_{a,b}=(1,\hat{n}_{a,b}),\,\bar{n}_{a,b}=(1,-\hat{n}_{a,b})\,.
\end{align}
In the collinear soft region of phase space, we have $n_a \cdot n_b \ll 1$. When performing expansions, we can work to leading order in $n_a \cdot n_b$, and must use a consistent power counting. It is therefore useful to collect some kinematic relations between vectors which are valid at leading power. These will be useful for later evaluations of the measurement function and integrand at leading power.
These kinematics satisfy the following useful relations
\begin{align} \label{eq:cs_kinematics1}
n\cdot n_a=n\cdot n_b= \frac{n_a \cdot n_b}{4}\,
\end{align}
\begin{align}
\bar{n}\cdot n_a=\bar{n}\cdot n_b=2\,,
\end{align}
\begin{align}
n_{\perp a,b}\cdot \bar{n}_{\perp a,b}=-n_{\perp a,b}\cdot n_{\perp a,b}=\hat{n}_{\perp a,b}\cdot\hat{n}_{\perp a,b}=\frac{n_a\cdot n_b}{2}\,. \label{eq:cs_kinematics2}
\end{align}
For a particle with the power counting of collinear sector $a$ or $b$, we have the following simplified relations
\begin{align}\label{energy_in_diverse_coor}
p_a\sim\frac{1}{2}(\bar{n}\cdot p_a) n_a,&\quad p_b\sim\frac{1}{2}(\bar{n}\cdot p_b) n_b\,,\\
p_a^0\sim\frac{1}{2}(\bar{n}\cdot p_a),&\quad p_b^0\sim\frac{1}{2}(\bar{n}\cdot p_b)\,,
\end{align}
which are true to leading order in the power counting.
Finally, we label the energy fractions carried in each subjet by
\begin{align}\label{energy_fractions}
z_{a,b}&=\frac{2p_{a,b}^0}{Q}=\frac{\bar n\cdot p_{a,b}}{Q}\,,
\end{align}
where the second relation is true to leading power.

The value of $\ecf{2}{\alpha}$ is given to leading power by the subjet splitting
\begin{align}\label{eq:lp_e2}
\ecf{2}{\alpha}&=\frac{1}{E_J^2} E_a E_b \left( \frac{2p_a\cdot p_b}{E_a E_b}  \right)^{\alpha/2}   \\
&=2^{\alpha/2}  z_az_b\left(n_a\cdot n_b\right)^{\alpha/2}\,.
\end{align}

In the collinear soft region of phase space, the 3-point energy correlation function is dominated by the correlation  between two particles in different subjets, with a third collinear, soft, or collinear-soft particle. Depending on the identity of the third particle, the power counting of the observable is different. We begin by collecting expressions for the $\ecf{3}{\alpha}$ observable for a single soft, collinear-soft, or collinear emission, which will be required for the one-loop calculations.

For three emissions, with momenta $k_1, k_2, k_3$, the general expression for the three point energy correlation function is
\begin{align}
\ecf{3}{\alpha}&=\frac{1}{E_J^3}k_1^0 k_2^0 k_3^0 \left(\frac{2 k_1 \cdot k_2}{k_1^0 k_2^0}  \right)^{\alpha/2}     \left(\frac{2k_1\cdot k_3}{k_1^0 k_3^0}  \right)^{\alpha/2}      \left(\frac{2k_2 \cdot k_3}{k_2^0 k_3^0}  \right)^{\alpha/2}\,.
\end{align}
For an emission collinear with one of the subjets, where we have the splitting $p_{a,b}\rightarrow k_1+k_2$, we can write $\ecf{3}{\alpha}$ entirely in terms of $k_1\cdot k_2$, $n_a\cdot n_b$, and $\bar n_a \cdot k_{1,2}$, because there is a hierarchy between the opening angle of the dipole, and the opening angle of the splitting. At leading power it is given by
\begin{align}\label{eq:collinear_limits_eec3}
\ecf{3}{\alpha}&=_{k_1,k_2\parallel n_a}   2^{5\alpha/2}z_b (n_a\cdot n_b)^{\alpha}\left(\frac{k_1\cdot k_2}{Q^2}\right)^{\frac{\alpha}{2}}\left(\frac{\bar n_a\cdot k_1}{Q}\right)^{1-\frac{\alpha}{2}}\left(\frac{\bar n_a\cdot k_2}{Q}\right)^{1-\frac{\alpha}{2}}\,,\\
\ecf{3}{\alpha}&=_{k_1,k_2\parallel n_b} 2^{5\alpha/2}z_a (n_a\cdot n_b)^{\alpha}\left(\frac{k_1\cdot k_2}{Q^2}\right)^{\frac{\alpha}{2}}\left(\frac{\bar n_b\cdot k_1}{Q}\right)^{1-\frac{\alpha}{2}}\left(\frac{\bar n_b\cdot k_2}{Q}\right)^{1-\frac{\alpha}{2}}\,.
\end{align}

For a soft emission off of the dipole, with momentum $k$, which cannot resolve the opening angle of the dipole, we have
\begin{align}
n_a\cdot k\rightarrow n\cdot k\,, \qquad n_b\cdot k \rightarrow n\cdot k\,,
\end{align}
at leading power. We then find
\begin{align}\label{eq:soft_limits_eec3}
\ecf{3}{\alpha}&=   2^{3\alpha/2+1} z_a z_b (n_a \cdot n_b)^{\alpha/2}    \left(\frac{\bar{n}\cdot k+n\cdot k}{2Q}\right)^{1-\alpha}\left(\frac{n\cdot k}{Q}\right)^{\alpha}\,,
\end{align}
where we have used the full expression for the energy of the soft particle, as it is not boosted.

For a third collinear-soft emission $k$ off of the $p_{a,b}$ partons, for which there is no hierarchy between the opening angle of the dipole and the opening angle of the emission (i.e. a collinear soft emission), $\ecf{3}{\alpha}$ is given by
\begin{align}\label{eq:collinear_soft_limits_eec3}
\ecf{3}{\alpha}&=2^{3\alpha/2+1}z_az_b\left(n_a\cdot n_b\right)^{\alpha/2}\left(\frac{\bar{n}\cdot k}{2Q}\right)^{1-\alpha}\left(\frac{n_a\cdot k}{Q}\right)^{\alpha/2}\left(\frac{n_b\cdot k}{Q}\right)^{\alpha/2}\,.
\end{align}

For the SCET operators involved in the matching calculation, we follow the notation of \Ref{Bauer:2011uc}, defining
\begin{align}
\mathcal{O}_2=\bar \chi_n Y^\dagger_n \Gamma Y_{\bar n} \chi_{\bar n}\,,
\end{align}
which is the usual SCET operator for $e^+e^-\to$ dijets, and
\begin{align}
\mathcal{O}_3=\bar \chi_{n_a} \mathcal{B}_{\perp n_b}^{A} \left [ X^\dagger_{n_a} X_{n_b} T^A X_{n_b} V_{\bar n}    \right]_{ij}  \left[ Y^\dagger_n Y_{\bar n} \right]_{jk}    \Gamma  \left[ \chi_{\bar n} \right]_k\,,
\end{align}
which is the SCET$_+$ operator describing the production of the collinear subjets. Throughout this section, we will not be careful with the Dirac structure of the operators, as it is largely irrelevant to our discussion. With this in mind, we have not made the Lorentz indices explicit on the operators.
Here we have chosen to write the Wilson line corresponding to the gluon in the fundamental representation. Note that the two stage matching onto SCET$_+$ makes it clear that the partonic configuration in which the two collinear subjets are both quarks is power suppressed. In the operators $\mathcal{O}_2, \mathcal{O}_3$, we have used $Y$ to denote soft Wilson lines, and $X,V$ to denote collinear-soft Wilson lines. In the definitions of the factorized functions below, we will refer to all Wilson lines as $S$, as after factorization, no confusion can arise.

\subsection*{Definitions of Factorized Functions}

The functions appearing in the collinear subjets factorization theorem of \Eq{eq:NINJA_fact} have the following SCET operator definitions:
\begin{itemize}
\item Hard Matching Coefficient for Dijet Production
\begin{equation}
H\left(Q^2,\mu \right)=\left |C\left(Q^2,\mu\right) \right|^2\,,
\end{equation}
where $C\left(Q^2,\mu\right)$ is the Wilson coefficient obtained from matching the full theory QCD current $\bar \psi \Gamma \psi$ onto the SCET dijet operator $\bar \chi_n \Gamma \chi_{\bar n}$
\begin{align}
\langle q\bar q | \bar \psi \Gamma \psi |0\rangle=C\left(Q^2,\mu\right) \langle q\bar q | \mathcal{O}_2 |0\rangle\,.
\end{align}
When accounting for the Lorentz structure, there is a contraction with the leptonic tensor, which we have dropped for simplicity. See \Ref{Ellis:2010rwa} for a detailed discussion.
\item Hard Splitting Function:
\begin{align}
&H_2 \left(\ecf{2}{\alpha},z_a, \mu\right)= \left|C_2\left(\ecf{2}{\alpha},z_a, \mu\right)\right|^2\,,
\end{align}
where $C_2\left(\ecf{2}{\alpha},z_a, \mu\right)$ is the Wilson coefficient in the matching from $\mathcal{O}_2$ to $\mathcal{O}_3$, namely the relation between the following matrix elements
\begin{align}
\langle q\bar q g| \mathcal{O}_2 |0\rangle=C_2\left(\ecf{2}{\alpha},z_a, \mu\right) \langle q\bar q g| \mathcal{O}_3 |0\rangle\,.
\end{align}
\item Jet Function:
\begin{align}
J_{n_{a,b}}\Big(\ecf{3}{\alpha}\Big)&=\\
&\hspace{-0.5cm}\frac{(2\pi)^3}{C_F}\text{tr}\langle 0|\frac{\bar{n}\!\!\!\slash _{a,b}}{2}\chi_{n_{a,b}}(0) \delta(Q-\bar{n}_{a,b}\cdot{\mathcal P})\delta^{(2)}(\vec{{\mathcal P}}_{\perp})\delta\Big(\ecf{3}{\alpha}-\ecfop{3}{\alpha}\Big)\bar{\chi}_{n_{a,b}}(0)|0\rangle \nonumber
\end{align}
For simplicity, we have given the definition of the quark jet function. The gluon jet function is defined identically but with the SCET collinear invariant gluon field, $\cB_{n_{a,b},\perp}$, instead of the collinear invariant quark field.
\item Soft Function:
\begin{align}\label{eq:def_soft_function}
S_{n \, \bar n }\Big(\ecf{3}{\alpha};R\Big)&=\frac{1}{C_{A}}\text{tr}\langle 0|T\{S_{n } S_{\bar n }\}  \delta\Big(\ecf{3}{\alpha}-\Theta_{R}\ecfop{3}{\alpha}\Big)\bar{T}\{S_{n } S_{\bar n }\} |0\rangle
\end{align}
\item Collinear-Soft Function:
\begin{align}
\hspace{-1cm}S_{n_a \,n_b\,\bar{n}}\Big(\ecf{3}{\alpha}\Big)&=\text{tr}\langle 0|T\{S_{n_a } S_{n_b} S_{\bar{n}}\}\delta\Big(\ecf{3}{\alpha}-\ecfop{3}{\alpha}\Big)\bar{T}\{S_{n_a } S_{n_b} S_{\bar{n}}\} |0\rangle
\end{align}
\end{itemize}
In each of these definitions, we have defined an operator, $\ecfop{3}{\alpha}$, which measures the contribution to $\ecf{3}{\alpha}$ from final states, and must be appropriately expanded following the power counting of the sector on which it acts, as was shown explicitly in \Eq{eq:collinear_limits_eec3}, \Eq{eq:soft_limits_eec3}, and \Eq{eq:collinear_soft_limits_eec3}. These operators can be written in terms of the energy-momentum tensor of the full or effective theory \cite{Sveshnikov:1995vi,Korchemsky:1997sy,Lee:2006nr,Bauer:2008dt}, but we can simply view them as returning the value of $\ecf{3}{\alpha}$ as measured on a particular perturbative state. The soft function is also sensitive to the jet function definition, which is included through the operator $\Theta_R$. To simplify the notation, we have strictly speaking only defined in the in-jet contribution to the soft function. Additionally, we assume that some IRC safe observable is also measured in the out-of-jet region, although this will play little role in our discussion, so we have not made it explicit.

\subsection*{Hard Matching Coefficient for Dijet Production}

The hard matching coefficient for dijet production, $H(Q^2,\mu)$, appears in the factorization theorems in each region of phase space. $H(Q^2,\mu)$ is the well known hard function for the production of a $q \bar q$ pair in $e^+e^-$ annihilation. It is defined by
\begin{equation}\label{eq:hard_matching_coeff}
H\left(Q^2, \mu\right)= \left|C\left(Q^2,\mu\right)\right|^2\,,
\end{equation}
where $C\left(Q^2,\mu\right)$ is the Wilson coefficient obtained from matching the full theory QCD current $\bar \psi \Gamma \psi$ onto the SCET dijet operator $\bar \chi_n \Gamma \chi_{\bar n}$. This Wilson coefficient is well known (see, e.g., \Refs{Bauer:2003di,Manohar:2003vb,Ellis:2010rwa,Bauer:2011uc} ), and is given at one-loop by
\begin{equation}
C(Q^2,\mu)=1+\frac{\alpha_s(\mu)\, C_F}{4\pi}\left ( -\log^2\left[ \frac{-Q^2}{\mu^2}  \right]+3\log \left[ \frac{-Q^2}{\mu^2} \right]-8+\frac{\pi^2}{6}   \right )\,.
\end{equation}
The branch cut in the logarithms must be taken as $-Q^2 \to -Q^2-i\epsilon$. The hard function satisfies a multiplicative RGE, given by
\begin{equation}
\mu \frac{d}{d\mu}\ln H\left(Q^2,\mu\right)=2\text{Re}\left[ \gamma_{C}\left(Q^2,\mu\right)\right]\,,
\end{equation}
where $\gamma_{C}(Q^2,\mu)$ is the anomalous dimension for the Wilson coefficient, which is given to one-loop by
\begin{equation}\label{eq:hard_anom_dim}
\gamma_C(Q^2,\mu)=\frac{\alpha_s C_F}{4\pi}\left(  4\log\left[  \frac{-Q^2}{\mu^2} \right] -6  \right )\,.
\end{equation}

\subsection*{Hard Splitting Function}

The hard splitting function can be calculated using known results for the one-loop splitting functions \cite{Kosower:1999rx} or from the result for $e^+e^- \to 3$ jets \cite{Ellis:1980wv}. However, since at leading power the measurement of the 2-point energy correlation functions define the energy fractions and splitting angle, it is simplest to change variables in the results of \Ref{Bauer:2011uc}, where the hard splitting function matching was performed for jet mass. Using the notation $t=s_{q g}$, and $x=s_{q \bar q}/Q^2$, \Ref{Bauer:2011uc} gave the matching coefficient to one-loop as
{\small \begin{align} \label{eq:hardsplit_t}
H_2^{q\to qg}(t,x,\mu)&=Q^2 \frac{\alpha_s(\mu) C_F}{2\pi} \frac{1}{t} \frac{1+x^2}{1-x} \left \{ 1+\frac{\alpha_s(\mu)}{2\pi} \left[ \left( \frac{C_A}{2} -C_F \right) \left( 2 \log \frac{t}{\mu^2} \log\, x +\log^2 x +2\text{Li}_2(1-x)  \right) \right. \right. \nonumber \\
&\hspace{-1.5cm}\left. \left. -\frac{C_A}{2} \left (  \log^2 \frac{t}{\mu^2}-\frac{7\pi^2}{6}+2\,\log \frac{t}{\mu^2}\log (1-x) +\log^2(1-x)+2\text{Li}_2 (x) \right)+(C_A-C_F)\frac{1-x}{1+x^2} \right] \right\}\,.
\end{align}}
We can now perform a change of variables to rewrite this in terms of $\ecf{2}{\alpha}$, and the subjet energy fractions, using the leading power relation of \Eq{eq:lp_e2}, and the kinematic relations valid in the collinear subjets region of phase space. We find
\begin{equation}\label{eq:relate_to_frank}
t=\frac{Q^2}{2}\frac{(z_a z_b)^{1-2/\alpha} \left( \ecf{2}{\alpha}  \right)^{2/\alpha}  }{2}\,, \qquad x=z_q\,,
\end{equation}
and
{\small \begin{align}
H_2^{q\to qg}(\ecf{2}{\alpha},z_q,\mu)&= \frac{\alpha_s(\mu) C_F}{\alpha \pi} \frac{1}{  \ecf{2}{\alpha}      } \frac{1+z_q^2}{1-z_q}   
\\
& \hspace{-2.75cm}\times\left \{ 1+\frac{\alpha_s(\mu)}{2\pi} \left[ \left( \frac{C_A}{2} -C_F \right) \left( 2 \log\left( \frac{Q^2}{\mu^2} \frac{(z_az_b)^{1-2/\alpha}    \left( \ecf{2}{\alpha}  \right)^{2/\alpha}      }{4} \right) \log\, z_q +\log^2 z_q +2\text{Li}_2(1-z_q)  \right) \right. \right. \nonumber \\
&\hspace{-1.5cm}\left. \left. -\frac{C_A}{2} \left (  \log^2 \left( \frac{Q^2}{\mu^2} \frac{(z_az_b)^{1-2/\alpha}    \left( \ecf{2}{\alpha}  \right)^{2/\alpha}      }{4} \right)-\frac{7\pi^2}{6} \right. \right. \right.  \nonumber \\
& \hspace{-3cm}\left. \left. \left.+2\,\log \left( \frac{Q^2}{\mu^2} \frac{(z_az_b)^{1-2/\alpha}    \left( \ecf{2}{\alpha}  \right)^{2/\alpha}      }{4} \right)\log (1-z_q) +\log^2(1-z_q)+2\text{Li}_2 (z_q) \right)+(C_A-C_F)\frac{1-z_q}{1+z_q^2} \right] \right\}\,. \nonumber
\end{align}}
Note that the hard splitting function depends on the partons involved in the split, which in our case we have taken to be $q\to qg$, and therefore singled out $z_q$, which is the energy fraction of the quark jet (defined identically to $z_a$, $z_b$). Throughout the rest of this appendix, we will, whenever possible, write results in terms of $z_a$, and $z_b$ for generic partons, using general Casimirs. Since we consider the case $q\to qg$, we will calculate the jet functions for both quark and gluon jets, and therefore the results in this appendix are sufficient to treat general two-prong substructure, where the prongs are associated with generic partons by using the hard splitting function for other partonic splittings. 

For completeness, we also present the one-loop results for $g\to gg$ and $g\to q\bar q$ splittings.  While one-loop, and even two-loop, splitting helicity amplitudes exist in the literature \cite{Kosower:1999rx,Bern:2004cz,Badger:2004uk}, to our knowledge, the one-loop unpolarized splitting functions have not not been explicitly written down before.  Using the results from \Refs{Kosower:1999rx,Badger:2004uk}, the one-loop function for the $g\to gg$ splitting is
\begin{align} \label{eq:hardsplit_gg}
H_2^{g\to gg}(s_{gg},z,\mu)&= \frac{\alpha_s(\mu) C_A}{2\pi} \frac{1}{s_{gg}} \left(
\frac{z}{1-z}+\frac{1-z}{z}+z(1-z)
\right) \left \{ 1+\frac{\alpha_s(\mu)}{2\pi}N\left[
\log\frac{\mu^2}{s_{gg}}\log\left(
z(1-z)
\right)\right.\right.\nonumber\\
&
\hspace{-1.5cm}\left.\left.
-\frac{1}{2}\log^2\frac{s_{gg}}{\mu^2}-\frac{1}{2}\log^2\frac{z}{1-z}+\frac{5\pi^2}{12}+\left(
\frac{1}{3}-\frac{n_F}{3N}
\right)\frac{z(1-z)}{1+z^4+(1-z)^4}
\right]
\right\} \,.
\end{align}
Here, we have expressed the result in terms of the numbers of colors, $N$, of the gauge theory and number of active quarks, $n_F$.  Note that $C_A = N$.  The virtuality of the splitting is
\begin{equation}
s_{gg} = 2z_az_bE_J^2(n_a\cdot n_b) \,,
\end{equation}
where $a$ and $b$ denote the final-state gluons in the splitting.
Its anomalous dimension to one-loop is
\begin{align}
\gamma_{g\to gg}=\frac{\alpha_s(\mu)}{\pi}\left[
N\,\log\frac{s_{gg}}{\mu^2}+N\,\log\,z(1-z)-\frac{\beta_0}{2}
\right]\,.
\end{align}
For the one-loop result of the $g\to q\bar q$ splitting, we have
\begin{align}
 \label{eq:hardsplit_qq}
H_2^{g\to q\bar q}(s_{q\bar q},z,\mu)&= \frac{\alpha_s(\mu) n_F}{2\pi} \frac{1}{s_{q\bar q}} \left(
z^2+(1-z)^2
\right) \left \{ 1+\frac{\alpha_s(\mu)}{2\pi}\left[
N\, \log\frac{\mu^2}{s_{q\bar q}}\log(z(1-z))\right.\right.\nonumber\\
&
\hspace{-1.5cm}
+\frac{3}{2}\frac{1}{N}\log\frac{\mu^2}{s_{q\bar q}}-\frac{2n_F}{3}\log\frac{\mu^2}{s_{q\bar q}}+\frac{13}{6}N\log\frac{\mu^2}{s_{q\bar q}}+\frac{1}{2N}\log^2\frac{\mu^2}{s_{q\bar q}}
\nonumber \\
&
\hspace{-1.5cm}
\left.\left.
-\frac{1}{N}\frac{7\pi^2}{12}-N\frac{\pi^2}{6}-\frac{N}{2}\log^2\frac{z}{1-z}+\frac{40}{9}N-\frac{10}{9}n_F
\right]
\right\}\,.
\end{align}
Note that, in terms of the number of colors, $$C_F = \frac{N^2-1}{2N}\,.$$
Its anomalous dimension is 
\begin{align}
\gamma_{g\to q\bar q} = \frac{\alpha_s(\mu)}{\pi}\left[
\frac{1}{N}\log\frac{\mu^2}{s_{q\bar q}}+N\,\log(z(1-z))+\frac{\beta_0}{2}-3C_F
\right]\,.
\end{align}

%

\subsection*{Global Soft Function}
In this section we calculate the global soft function. The global soft modes can resolve the boundaries of the jet, so the jet algorithm constraint cannot be expanded. However, the soft modes do not resolve the dipole of the collinear splitting. The global soft function therefore has two Wilson lines in the $n$ and $\bar n$ directions. A general one-loop soft function can be written as
\begin{align}
S^{(1)}_{G}\left(\ecf{3}{\alpha}\right)&=\frac{1}{2}\sum_{i\neq j}\mathbf{T}_i\cdot\mathbf{T}_j \,S_{G,\,ij}^{(1)}\left(\ecf{3}{\alpha}\right)\,,
\end{align}
where $\mathbf{T}_i$ is the color generator of leg $i$ in the notation of \Refs{Catani:1996jh,Catani:1996vz}, and the sum runs over all pairs of legs. Here we have only the contribution from $i,j=n, \bar n$, but we still perform this extraction of the color structure to keep the results generic.

The one-loop integrand for the soft function is given by
{\small \begin{align} \label{eq:soft_integrand_cs}
S_{G,\,n \bar{n}}^{(1)}(\ecf{3}{\alpha})&=\\
&
\hspace{-2cm}
-g^2 \left( \frac{\mu^{2}  e^{\gamma_E }  }{4\pi} \right) ^{\epsilon}         \int[d^dk]_+\frac{2n\cdot \bar{n}}{n\cdot k\,k\cdot \bar{n}}\Theta\left(\text{tan}^2\frac{R}{2}-\frac{n\cdot k}{\bar{n}\cdot k}\right)\delta\left(\ecf{3}{\alpha}-N_S\left(\frac{\bar{n}\cdot k+n\cdot k}{2Q}\right)^{1-\alpha}\left(\frac{n\cdot k}{Q}\right)^{\alpha}\right) \nonumber
\end{align}}
with $d=4-2\epsilon$, and where here we have extracted the normalization factor
 \begin{align}
N_S&=2^{3\alpha/2+1} z_az_b\left(n_a\cdot n_b\right)^{\alpha/2}\,,
\end{align}
following the expression for the three point energy correlation function in the soft power counting, given in \Eq{eq:soft_limits_eec3}. The first $\Theta$-function in \Eq{eq:soft_integrand_cs} implements the jet algorithm constraint, which is simple for a single emission. To simplify notation, we also use the following shorthand for the measure for a positive energy, on-shell, collinear particle
\begin{align}\label{eq:pos_energy}
[d^dk]_+&=\frac{d^dk}{(2\pi)^d}2\pi\Theta(\nbar\cdot k)\delta(k^2)\,.
\end{align}
To perform this integral, it is convenient to make the change of variables
\begin{align}
\bar{n}\cdot k=v\,, \qquad
n\cdot k=v\, u\,,
\end{align}
which factorizes the jet algorithm constraint and the measurement function. The integrals can then be evaluated using standard techniques.
Performing all the integrals but the $u$ integral, and transforming to Laplace space, $\ecf{3}{\alpha}\rightarrow\ecflp{3}{\alpha}$, gives
 \begin{align}
\tilde{S}_{G,\,n \bar{n}}^{(1)}(\ecflp{3}{\alpha})=-\frac{g^2 e^{-\epsilon\gamma_E}\Gamma(-2\epsilon)}{(2\pi)^2\Gamma(1-\epsilon)}\Bigg(\frac{e^{\gamma_E}\mu\ecflp{3}{\alpha}N_S}{2^{1-\alpha}Q}\Bigg)^{2\epsilon}   \int_{0}^{\text{tan}^2\frac{R}{2}} \frac{du}{u^{1+\epsilon(1-2\alpha)}}(1+u)^{2\epsilon(1-\alpha)}\,. 
\end{align}
This can be integrated exactly in terms of hypergeometric functions, 
\begin{align}
 \int_{0}^{\text{tan}^2\frac{R}{2}} \frac{du}{u^{1+\epsilon(1-2\alpha)}}(1+u)^{2\epsilon(1-\alpha)}=   \frac{\Gamma(-\epsilon(1-2\alpha))}{\Gamma(1-\epsilon(1-2\alpha))}    &\\
 &\hspace{-8cm} \times\left (  \tan^2 \frac{R}{2} \right)^\epsilon    \left(   \frac{\tan^2 \frac{R}{2}}{1+\tan^2 \frac{R}{2}} \right)^{-2(1-\alpha)\epsilon}     \!\,_2F_1\left[1,-2(1-\alpha)\epsilon;1-(1-2\alpha)\epsilon;\frac{\text{tan}^2\frac{R}{2}}{1+\text{tan}^2\frac{R}{2}}\right]\,,\nonumber
\end{align}
where we have used both a Pfaff and an Euler transformation to extract the singular behavior from the hypergeometric function.
We therefore have
\begin{align}
\tilde{S}_{G,\,n \bar{n}}^{(1)}(\ecflp{3}{\alpha})=-\frac{\alpha_s}{\pi}\frac{e^{-\epsilon\gamma_E}\Gamma(-2\epsilon)}{\Gamma(1-\epsilon)}\Bigg(\frac{e^{\gamma_E}\mu\ecflp{3}{\alpha}N_S}{2^{1-\alpha}Q}  \tan \frac{R}{2}    \Bigg)^{2\epsilon}   \frac{\Gamma(-\epsilon(1-2\alpha))}{\Gamma(1-\epsilon(1-2\alpha))}    &\\
 &
 \hspace{-10cm} 
 \times   \left(   \frac{\tan^2 \frac{R}{2}}{1+\tan^2 \frac{R}{2}} \right)^{-2(1-\alpha)\epsilon}     \!\,_2F_1\left[1,-2(1-\alpha)\epsilon;1-(1-2\alpha)\epsilon;\frac{\text{tan}^2\frac{R}{2}}{1+\text{tan}^2\frac{R}{2}}\right]\,.\nonumber
\end{align}

Expanding in $\epsilon$ (throughout these appendices we use the {\tt HypExp} package \cite{Huber:2005yg,Huber:2007dx} for expansions of hypergeometric functions) and separating in divergent and finite pieces, we find
\begin{align}\label{eq:gsoft_final}
\tilde{S}_{G,\,n \bar{n}}^{(1)\text{div}}(\ecflp{3}{\alpha})&=  \frac{\alpha_s}{2\pi}\frac{1}{(2 \alpha -1) \epsilon ^2}+  \frac{\alpha_s}{\pi} \frac{ \log \left[\frac{e^{\gamma_E}\mu\ecflp{3}{\alpha}N_S}{2^{1-\alpha}Q}\right]}{ (2 \alpha -1) \epsilon } 
+\frac{\alpha_s}{2\pi}\frac{\log\left[\tan^2\frac{R}{2}\right]}{\epsilon}
\,, 
\end{align}
\begin{align}
\tilde{S}_{G,\,n \bar{n}}^{(1)\text{fin}}(\ecflp{3}{\alpha})&=   \frac{\alpha_s}{\pi}\Bigg\{
\frac{\log^2 \left[\frac{e^{\gamma_E}\mu\ecflp{3}{\alpha}N_S}{2^{1-\alpha}Q}\right]}{2\alpha-1}
+\log \left[\frac{e^{\gamma_E}\mu\ecflp{3}{\alpha}N_S}{2^{1-\alpha}Q}\right]\,\log\left[\tan^2\frac{R}{2}\right] \\
&
+\frac{\pi^2}{8(2\alpha-1)}+\frac{2\alpha-1}{4}\log^2\left[\tan^2\frac{R}{2}\right]+(\alpha-1)\text{Li}_2\left[
-\tan^2\frac{R}{2}
\right]
\Bigg\} \,, \nonumber
\end{align}
where $\text{Li}_2$ is the dilogarithm function.

\subsection*{Jet Function}\label{sec:NINJA_jet_calc}

To calculate the jet function, we use the approach of \Ref{Ritzmann:2014mka} and integrate the appropriate splitting functions against our measurement function. In the power counting of the jet function, we can expand the jet algorithm constraint
\begin{align}\label{eq:jet_expand}
\Theta\left(\text{tan}^2\frac{R}{2}-\frac{n\cdot k}{\bar{n}\cdot k}\right) \to 1\,.
\end{align}

The one-loop jet function in the $n_a$ direction is then given by 
\begin{equation}
J_{i,n_{a}}^{(1)}(Q_J,\ecf{3}{\alpha})=\int d\Phi_2^c\, \sigma_2^c\, \delta\Bigg(\ecf{3}{\alpha}-N_J\left(\frac{\bar{n}_{a}\cdot k_1}{Q}\right)^{1-\alpha/2}\left(\frac{\bar{n}_{a}\cdot k_2}{Q}\right)^{1-\alpha/2}\left(\frac{k_1\cdot k_2}{Q^2}\right)^{\alpha/2}\Bigg)\,.
\end{equation}
The two particle collinear phase space is given by \cite{Giele:1991vf}
\begin{equation}
d\Phi_2^c= 2(2\pi)^{3-2\epsilon} Q_J \left[  d^d k_1  \right]_+   \left[  d^d k_2  \right]_+    \delta (Q_J-\bar{n}_{a}\cdot k_1-\bar{n}_{a}\cdot k_2) \delta^{d-2}(k_{1\perp} +k_{2\perp})\,,
\end{equation}
and
\begin{equation}
\sigma_2^c=\left(   \frac{\mu^2 e^{\gamma_E}}{4\pi} \right)^{\epsilon} \frac{2g^2}{s} P_{i}(z)\,,
\end{equation}
 where
 \begin{equation}
 P_{q}(z)=C_F \left[  \frac{1+z^2}{1-z} -\epsilon(1-z) \right]\,,
 \end{equation}
 and
 \begin{equation}
 P_g=C_A \left[ \frac{z}{1-z}+\frac{1-z}{z}+z(1-z) \right]+\frac{n_f}{2} \left[ 1-\frac{2z(1-z)}{1-\epsilon} \right]\,,
 \end{equation}
 which includes both the $g\to gg$ and $g\to q\bar q$ contributions.
Explicitly, the integrand is then given by
\begin{align}
J_{i,n_{a}}^{(1)}(Q_J,\ecf{3}{\alpha})&=\left( \frac{\mu^2 e^{\gamma_E}}{4\pi} \right)^\epsilon 2(2\pi)^{3-2\epsilon}  Q_J 2g^2\int[d^dk_1]_+\int[d^dk_2]_+\frac{ P_{i}\left(\frac{\bar{n}_{a}\cdot k_1}{Q_J}\right)}{2k_1\cdot k_2}\\
&
\hspace{1cm}
\times
\delta(Q_J-\bar{n}_{a}\cdot k_1-\bar{n}_{a}\cdot k_2)\delta^{d-2}(\vec{k}_{1\perp}+\vec{k}_{2\perp})\nonumber\\
&
\hspace{1cm}
\times
\delta\Bigg(\ecf{3}{\alpha}-N_J\left(\frac{\bar{n}_{a}\cdot k_1}{Q}\right)^{1-\alpha/2}\left(\frac{\bar{n}_{a}\cdot k_2}{Q}\right)^{1-\alpha/2}\left(\frac{k_1\cdot k_2}{Q^2}\right)^{\alpha/2}\Bigg)\,,\nonumber
\end{align}
where we have extracted the normalization factor
 \begin{align}
N_J&= 2^{5\alpha/2} (n_a\cdot n_b)^{\alpha} z_b\,,
\end{align}
for simplicity, following the expression of \Eq{eq:collinear_limits_eec3} for the three point energy correlation function in the power counting for the emission of a single collinear particle. Furthermore, note that we have used $Q_J= z_a Q$ in this expression.

The integrals can be performed using standard techniques, and we find, after transforming to Laplace space, $\ecf{3}{\alpha}\rightarrow\ecflp{3}{\alpha}$, for the jet function in the $n_a$ direction
\begin{align}\label{eq:gjet_final}
\tilde{J}_{g,n_{a}}^{(1)}(Q_J,\ecflp{3}{\alpha})&=\frac{\alpha_s}{2\pi}C_A\Bigg(\frac{\alpha}{(\alpha -1) \epsilon^2}   +\frac{2 L_\alpha^{J,a}\left(  \ecflp{3}{\alpha}  \right) }{(\alpha-1) \epsilon}   +\frac{1}{ \epsilon} \frac{11C_A-2n_f}{6C_A}  \\
   &
    \hspace{-2cm}
    -\frac{\alpha  \pi^2}{12 (\alpha -1)}+\frac{\pi^2}{3 (\alpha -1) \alpha }-\frac{67 }{9 \alpha }+\frac{2\pi^2}{3 \alpha }+\frac{2 L_\alpha^{J,a}\left(  \ecflp{3}{\alpha}  \right)^2}{ (\alpha -1) \alpha }+\frac{11     L_\alpha^{J,a}\left(  \ecflp{3}{\alpha}  \right)}{3 \alpha }\nonumber \\
   &
   \hspace{-2cm}
   +\frac{67}{9}-\frac{2\pi^2}{3}-\frac{2 n_f L_\alpha^{J,a}\left(  \ecflp{3}{\alpha}  \right)    }{3 C_A \alpha }+\frac{13 n_f}{9 C_A \alpha }-\frac{23 n_f}{18 C_A }\Bigg)\,,\nonumber
\end{align}
 for gluon jets, and 
\begin{align}\label{eq:qjet_final}
\tilde{J}_{q,n_{a}}^{(1)}(Q_J,\ecflp{3}{\alpha})=\frac{\alpha_s}{2\pi}C_F\Bigg(&-\frac{\alpha}{\epsilon^2(1-\alpha)}+\frac{3}{2\epsilon}-\frac{2}{\epsilon(1-\alpha)}   L_\alpha^{J,a}\left(  \ecflp{3}{\alpha}  \right)\\
&
\hspace{1cm}
-\frac{2}{\alpha(1-\alpha)}L_\alpha^{J,a}\left(  \ecflp{3}{\alpha}  \right)^2+\frac{3}{\alpha}L_\alpha^{J,a}\left(  \ecflp{3}{\alpha}  \right)-\frac{\pi^2}{6\alpha}\nonumber\\
&
\hspace{1cm}
-\frac{\pi^2}{4\alpha(1-\alpha)}+\frac{3\pi^2(1-\alpha)}{4\alpha}+\frac{1}{2\alpha}-\frac{13(1-\alpha)}{2\alpha}\Bigg) \,, \nonumber
\end{align}
for quark jets respectively.  The jet function for the $n_b$ direction can be trivially found from $a\rightarrow b$.

Here we have used $L_\alpha^{J,a}\left(  \ecflp{3}{\alpha}  \right)$to denote the logarithm appearing in the jet functions. The argument of this logarithm depends on the subjet energy fraction. We indicate the specific logarithm for the subjet via the notation
\begin{align}\label{eq:jet_log}
L_\alpha^{J,a}\left(  \ecflp{3}{\alpha}  \right)&=\log\left[N_J\ecflp{3}{\alpha}e^{\gamma_E }\left(\frac{\mu}{\sqrt{2}Q}\right)^\alpha z_a^{2-\alpha}\right]\,.
\end{align}

\subsection*{Collinear-Soft Function}\label{sec:csoft_calc}
We now calculate the collinear-soft function. The collinear-soft modes couple eikonally to the collinear sector, and so the collinear-soft function has the one-loop form
\begin{align}
S^{(1)}_{c}\left(\ecf{3}{\alpha}\right)&=\frac{1}{2}\sum_{i\neq j}\mathbf{T}_i\cdot\mathbf{T}_j \,S_{c,\,ij}^{(1)}\left(\ecf{3}{\alpha}\right)\,,
\end{align}
where $\mathbf{T}_i$ is the color generator of leg $i$ in the notation of \Refs{Catani:1996jh,Catani:1996vz}, and the sum runs over all pairs of legs. Since the collinear-soft modes resolves the dipole from the collinear splitting, there are three Wilson lines, $n_a, n_b, \bar n$ to which the collinear-soft modes couple. We calculate separately the contributions arising from the pair of legs $n_a, n_b$, and from the pairs $n_{a,b}, \bar n$. In both cases the integral involves the jet algorithm constraint. In the power counting of the collinear-soft modes, this constraint can be expanded as
\begin{align}\label{eq:jet_expand2}
\Theta\left(\text{tan}^2\frac{R}{2}-\frac{n\cdot k}{\bar{n}\cdot k}\right) \to 1\,.
\end{align}
If this expansion was not performed, the contribution of the collinear soft modes sensitive to the jet radius $R$, would be removed by a soft zero bin subtraction.

\subsubsection*{$n_a$, $n_b$ Contribution:}

We begin by calculating the contribution from the emission between the $n_a$, $n_b$ eikonal lines. The integrand is given by
\begin{align}
S_{c,\,n_a n_b}^{(1)}(\ecf{3}{\alpha})&= \\
&\hspace{-2cm} -g^2   \left( \frac{\mu^{2}  e^{\gamma_E }  }{4\pi} \right)^{\epsilon}       \int[d^dk]_+\frac{2n_a\cdot n_b}{n_{a}\cdot k\,k\cdot n_b}\delta\left(\ecf{3}{\alpha}-N_{CS}\left(\frac{\bar{n}\cdot k}{2Q}\right)^{1-\alpha}\left(\frac{n_a\cdot k}{Q}\right)^{\alpha/2}\left(\frac{n_b\cdot k}{Q}\right)^{\alpha/2}\right)\,, \nonumber
\end{align}
where we have extracted the normalization factor
 \begin{align}\label{eq:ncs}
N_{CS}&=2^{3\alpha/2+1}z_az_b\left(n_a\cdot n_b\right)^{\alpha/2}\,,
\end{align}
for simplicity, following the expression of \Eq{eq:collinear_soft_limits_eec3} for the three point energy correlation function in the power counting for the emission of a single collinear-soft particle.

To perform the calculation, we go to the light-cone basis defined by $n,\bar{n}$. We then have
\begin{align}
n_a\cdot k&=\frac{n\cdot n_a}{2}\bar{n}\cdot k+\frac{\bar{n}\cdot n_a}{2}n\cdot k+k_\perp\cdot n_{a\perp}\nonumber\\
&=\frac{n\cdot n_a}{2}\bar{n}\cdot k+\frac{\bar{n}\cdot n_a}{2}n\cdot k-(\bar{n}\cdot k n\cdot k)^{1/2} |\hat{n}_{a\perp}|\text{cos }\theta \,,\\
n_b\cdot k&=\frac{n\cdot n_b}{2}\bar{n}\cdot k+\frac{\bar{n}\cdot n_b}{2}n\cdot k+k_\perp\cdot n_{b\perp}\nonumber\\
&=\frac{n\cdot n_b}{2}\bar{n}\cdot k+\frac{\bar{n}\cdot n_b}{2}n\cdot k+(\bar{n}\cdot k n\cdot k)^{1/2}|\hat{n}_{a\perp}|\text{cos }\theta \,,
\end{align}
where $\theta$ denotes the angle between the particle $k$ and the $n$ axis. In the above kinematic relations, we have made use of the fact that since $\hat{n}\sim \hat{n}_a+\hat{n}_b$, $k_\perp\cdot n_{b\perp}=-k_\perp\cdot n_{a\perp}$. Rewriting the integrand for a positive energy gluon in terms of $\theta$, we find
\begin{align}
\int[d^dk]_+&=\frac{1}{2^{4-2\epsilon}\pi^{\frac{5}{2}-\epsilon}\Gamma(\frac{1}{2}-\epsilon)}\int_0^\infty\frac{dn\cdot k}{n\cdot k^{\epsilon}}\int_0^\infty\frac{d\bar{n}\cdot k}{\bar{n}\cdot k^{\epsilon}}\int_0^\pi d\theta\,\text{sin}^{-2\epsilon}\,\theta \,, \\
&= c_\epsilon \int_0^\infty\frac{dn\cdot k}{n\cdot k^{\epsilon}}\int_0^\infty\frac{d\bar{n}\cdot k}{\bar{n}\cdot k^{\epsilon}}\int_0^\pi d\theta\,\text{sin}^{-2\epsilon}\,\theta \,,
\end{align}
for $d=4-2\epsilon$.
To simplify our expressions, we have extracted the following constant
\begin{align}
c_{\epsilon}&=\frac{1}{2^{4-2\epsilon}\pi^{\frac{5}{2}-\epsilon}\Gamma(\frac{1}{2}-\epsilon)}\,.
\end{align}

In the collinear soft region of phase space, we power count $n_a \cdot n_b \ll 1$. We can therefore work to leading power in $n_a\cdot n_b$ in the integrand. Using the relations of \Eq{eq:cs_kinematics1}- \Eq{eq:cs_kinematics2}, and expanding to leading power in $n_a\cdot n_b$, we have
\begin{align}
n_a\cdot k\,n_b\cdot k&=\left(n\cdot k+\frac{n_a\cdot n_b}{8} \bar{n}\cdot k\right)^2-\frac{n_a \cdot n_b}{2}(n\cdot k \, \bar{n}\cdot k)\,\text{cos}^2\theta\,.
\end{align}
Note that in our power counting, $n \cdot k \sim n_a \cdot n_b$, so that this expression scales homogeneously. To perform the integral, we make the change of variables
\begin{align}
\bar n \cdot k = v, \qquad n \cdot k =v w \left(  \frac{n_a \cdot n_b}{8}  \right)\,.
\end{align}
We then have
\begin{align}
n_a\cdot k\,n_b\cdot k&=  v^2 \left(  \frac{n_a \cdot n_b}{8}  \right)^2    \left[  (1+w)^2-4w\,\cos^2 \theta  \right]\\
&= v^2 \left(  \frac{n_a \cdot n_b}{8}  \right)^2    \left[  (1-w)^2+4w\,\sin^2 \theta  \right]\,.
\end{align}
The one loop expression for the collinear soft function can then be written
\begin{align}
S_{c,\,n_a n_b}^{(1)}(\ecf{3}{\alpha})&= \\
&\hspace{-2cm} -g^2   \left( \frac{\mu^{2}  e^{\gamma_E}  }{4\pi} \right)^{\epsilon}    16c_\epsilon   \left(   \frac{ n_a\cdot n_b   }{ 8   }     \right)^{-\epsilon}      \int_0^\infty\frac{dw}{w^{\epsilon}}             \int_0^\infty\frac{dv}{v^{1+2\epsilon}}         \int_0^\pi d\theta\,\text{sin}^{-2\epsilon}\,\theta          \frac{1}{   (1-w)^2+4w\,\sin^2 \theta     }    \nonumber \\
&      \times\delta  \left(\ecf{3}{\alpha}-  \frac{N_{CS} }{2^{1-\alpha}}\frac{v}{Q}           \left(\frac{n_a\cdot n_b}{8} \right)^{\alpha}   \left[   (1-w)^2+4w\,\sin^2 \theta  \right]^{\alpha/2}            \right)\,, \nonumber
\end{align}
The $v$ integral is straightforward. Transforming to Laplace space,  $\ecf{3}{\alpha}\rightarrow\ecflp{3}{\alpha}$, we find
\begin{align}
S_{c,\,n_a n_b}^{(1)}(\ecflp{3}{\alpha})&=  -g^2 \Gamma(-2\epsilon)     \left(    \frac{  \mu^2 N_{CS}^2 e^{\gamma_E}  (\ecflp{3}{\alpha})^2      \left( \frac{n_a \cdot n_b}{8}   \right)^{-1+2\alpha}  }{4\pi 4^{1-\alpha}  Q^2    }  \right)^{\epsilon} \\
&
\hspace{2cm}
\times16 c_\epsilon          \int_0^\infty\frac{dw}{w^{\epsilon}}    \int_0^\pi d\theta\,\text{sin}^{-2\epsilon}\,\theta     \left[ (1-w)^2+4w\,\sin^2 \theta   \right]^{-1+\alpha \epsilon}\,.   \nonumber
\end{align}
The $\theta$ integral can be performed exactly in terms of hypergeometric functions using
\begin{align}
\int_0^\pi d\theta\,\text{sin}^{-2\epsilon}\,\theta     \left[ (1-w)^2+4w\,\sin^2 \theta   \right]^{-1+\alpha \epsilon} &= \\
&\hspace{-4cm}    \frac{\Gamma[1/2-\epsilon] \Gamma[1/2]}{\Gamma[1-\epsilon]} (1-w)^{2(-1+\alpha\epsilon)}   \,_2F_1\left[1-\alpha \epsilon, 1/2-\epsilon, 1-\epsilon, -\frac{4w}{(1-w)^2}\right] \,, \nonumber
\end{align}
which can be rewritten using a Pfaff transformation as
\begin{align}
&\int_0^\pi d\theta\,\text{sin}^{-2\epsilon}\,\theta     \left[ (1-w)^2+4w\,\sin^2 \theta   \right]^{-1+\alpha \epsilon} =    \frac{\Gamma[1/2-\epsilon] \Gamma[1/2]}{\Gamma[1-\epsilon]}   \\
& \hspace{1cm} \times (1+w)^{-1+2\epsilon}  \left( (1-w)^2\right)^{-1/2-(1-\alpha)\epsilon}  \,_2F_1\left[ 1/2-\epsilon,-\epsilon+\alpha\epsilon, 1-\epsilon, \frac{4w}{(1+w)^2}\right] \,. \nonumber
\end{align}
The remaining integral in $w$ is given by
\begin{align}
S_{c,\,n_a n_b}^{(1)}(\ecflp{3}{\alpha})&=  -g^2 \Gamma(-2\epsilon)     \left(    \frac{  \mu^2 N_{CS}^2 e^{\gamma_E}  (\ecflp{3}{\alpha})^2      \left( \frac{n_a \cdot n_b}{8}   \right)^{-1+2\alpha}  }{4\pi 4^{1-\alpha}  Q^2    }  \right)^{\epsilon}   16 c_\epsilon     \frac{\Gamma[1/2-\epsilon] \Gamma[1/2]}{\Gamma[1-\epsilon]}  \nonumber \\
& \hspace{-1.5cm}          \int_0^\infty\frac{dw}{w^{\epsilon}}   (1+w)^{-1+2\epsilon}  \left( (1-w)^2\right)^{-1/2-(1-\alpha)\epsilon}  \,_2F_1\left[ 1/2-\epsilon,-\epsilon+\alpha\epsilon, 1-\epsilon, \frac{4w}{(1+w)^2}\right] \,. 
\end{align}
Re-mapping the integral to the unit interval, we have
\begin{align}
S_{c,\,n_a n_b}^{(1)}(\ecflp{3}{\alpha})&=  -g^2 \Gamma(-2\epsilon)     \left(    \frac{  \mu^2 N_{CS}^2 e^{\gamma_E}  (\ecflp{3}{\alpha})^2      \left( \frac{n_a \cdot n_b}{8}   \right)^{-1+2\alpha}  }{4\pi 4^{1-\alpha}  Q^2    }  \right)^{\epsilon}    16 c_\epsilon      \frac{\Gamma[1/2-\epsilon] \Gamma[1/2]}{\Gamma[1-\epsilon]} \nonumber\\
& \hspace{-2.5cm}           \int_0^1dw \Big(w^{-\epsilon}+w^{(1-2\alpha)\epsilon}\Big)  (1+w)^{-1+2\epsilon}  (1-w)^{-1-2(1-\alpha)\epsilon}  \,_2F_1\left[ 1/2-\epsilon,-\epsilon+\alpha\epsilon, 1-\epsilon, \frac{4w}{(1+w)^2}\right] \,. \nonumber
\end{align}
We could not perform this integral exactly, but it can be done as a Laurent expansion in $\epsilon$ by expanding the hypergeometric function as 
 \begin{align}
\,_2F_1\left[\frac{1}{2}-\epsilon,-(1-\alpha)\epsilon;1-\epsilon;\frac{4w}{(1+w)^2}\right]&=1-2(1-\alpha)\epsilon\,\text{ln}(1+w)+{\mathcal O}(\epsilon^2)\,,
\end{align}
which is valid for $0\leq w \leq 1$, and we have truncated the expansion at ${\mathcal O}(\epsilon^2)$ as we are only interested in the terms up to ${\mathcal O}(\epsilon^0)$ in the one-loop result.
We then have
\begin{align}
S_{c,\,n_a n_b}^{(1)}(\ecflp{3}{\alpha})&=  -g^2 \Gamma(-2\epsilon)     \left(    \frac{  \mu^2 N_{CS}^2 e^{\gamma_E}  (\ecflp{3}{\alpha})^2      \left( \frac{n_a \cdot n_b}{8}   \right)^{-1+2\alpha}  }{4\pi 4^{1-\alpha}  Q^2    }  \right)^{\epsilon}   16 c_\epsilon     \frac{\Gamma[1/2-\epsilon] \Gamma[1/2]}{\Gamma[1-\epsilon]} \nonumber  \\
& \hspace{-1cm}          \int_0^1dw \Big(w^{-\epsilon}+w^{(1-2\alpha)\epsilon}\Big) (1+w)^{-1+2\epsilon}  (1-w)^{-1-2(1-\alpha)\epsilon} \left( 1-2(1-\alpha)\epsilon\,\text{ln}(1+w)   \right)\,. 
\end{align}
For the remaining integral in $w$, we have
 \begin{align}
& \int_0^1dw \Big(w^{-\epsilon}+w^{(1-2\alpha)\epsilon}\Big)    (1+w)^{-1+2\epsilon}  (1-w)^{-1-2(1-\alpha)\epsilon}    \left( 1-2(1-\alpha)\epsilon\,\text{ln}(1+w)   \right)= \nonumber \\
& \int_0^1dw \Big(w^{-\epsilon}+w^{(1-2\alpha)\epsilon}\Big)    (1+w)^{-1+2\epsilon}  (1-w)^{-1-2(1-\alpha)\epsilon}     \\
&\hspace{1cm}  -2(1-\alpha)\epsilon   \int_0^1dw \Big(w^{-\epsilon}+w^{(1-2\alpha)\epsilon}\Big)    (1+w)^{-1+2\epsilon}  (1-w)^{-1-2(1-\alpha)\epsilon}  \,   \log(1+w)  \,. \nonumber
 \end{align}
 The first integral can be done in terms of hypergeometric functions, while the second can be done using plus functions (for a detailed discussion of their properties, see e.g. \cite{Ligeti:2008ac}), and applying the identity
 \begin{align}
 \frac{1}{z^{1+a\epsilon}}=-\frac{1}{a\epsilon}\delta(z) +\sum \limits_{i=0}^{\infty}   \frac{(-a\epsilon)^i}{i!} \mathcal{D}_i (z)\,,
 \end{align}
 with
 \begin{align}
 \mathcal{D}_i (z)=\left[  \frac{\log^i z}{z}  \right]_+\,.
 \end{align}
 We find 
  \begin{align}
 & \int_0^1dw \Big(w^{-\epsilon}+w^{(1-2\alpha)\epsilon}\Big)  \Big((1+w)^2\Big)^{-\frac{1}{2}+\epsilon}\Big((1-w)^2\Big)^{-\frac{1}{2}-(1-\alpha)\epsilon} \left( 1-2(1-\alpha)\epsilon\,\text{ln}(1+w)   \right) \nonumber \\
&\hspace{0cm}= \frac{\Gamma[2(\alpha-1) \epsilon] \Gamma[1-\epsilon]}{ \Gamma[1-3\epsilon+2\alpha \epsilon]} \,_2F_1[1-2\epsilon,1-\epsilon; 1-3\epsilon+2\alpha \epsilon;-1]       
\\    &+       \frac{\Gamma[2(\alpha-1) \epsilon] \Gamma[1+\epsilon-2\alpha \epsilon]}{ \Gamma[1-\epsilon]} \,_2F_1[1-2\epsilon,1+\epsilon-2\alpha\epsilon;1-\epsilon;-1] \nonumber\\
&+ 2^{2\epsilon} \log 2 -2(1-\alpha)\epsilon \left( \log^2 2 -\frac{\pi^2}{12}   \right)  \nonumber \\
&=\frac{1}{(2 \alpha -2) \epsilon }+\frac{\alpha 
   \log (2)}{\alpha -1}+\log (2)+\frac{\epsilon  \left(-\pi ^2 \alpha ^2+36 \alpha ^2 \log ^2(2)+3 \pi ^2 \alpha -24 \alpha 
   \log ^2(2)-2 \pi ^2\right)}{12 (\alpha -1)}   \, . \nonumber
 \end{align}
Therefore, in total, we have
 \begin{align}
S_{c,\,n_a n_b}^{(1)}(\ecflp{3}{\alpha})&=  -g^2 \Gamma(-2\epsilon)     \left(    \frac{  \mu^2 N_{CS}^2 e^{\gamma_E}  (\ecflp{3}{\alpha})^2      \left( \frac{n_a \cdot n_b}{8}   \right)^{-1+2\alpha}  }{4\pi 4^{1-\alpha}  Q^2    }  \right)^{\epsilon}   16 c_\epsilon     \frac{\Gamma[1/2-\epsilon] \Gamma[1/2]}{\Gamma[1-\epsilon]} \nonumber  \\
& \hspace{-1.75cm}   \left(  \frac{1}{(2 \alpha -2) \epsilon }+\frac{\alpha 
   \log (2)}{\alpha -1}+\log (2)+\frac{\epsilon  \left(-\pi ^2 \alpha ^2+36 \alpha ^2 \log ^2(2)+3 \pi ^2 \alpha -24 \alpha 
   \log ^2(2)-2 \pi ^2\right)}{12 (\alpha -1)}      \right)     \,. 
\end{align}
Expanding in $\epsilon$, and keeping only the divergent piece, as relevant for the anomalous dimensions, we find 
 \begin{align} \label{eq:csoft_nanb_final}
\tilde{S}_{c,\,n_a n_b}^{(1)\text{div}}(\ecflp{3}{\alpha})&= \frac{\alpha_s}{\pi}\frac{1}{  (\alpha -1) \epsilon ^2}+2\frac{\alpha_s}{\pi}\frac{ \left(2 \alpha  \log (2)+ \log \left[   \frac{  \mu N_{CS} e^{\gamma_E}  (\ecflp{3}{\alpha})      \left( \frac{n_a \cdot n_b}{8}   \right)^{-1/2+\alpha}  }{2^{1-\alpha}  Q    }  \right]- \log (2)\right)}{   (\alpha -1) \epsilon }\nonumber \\
&=\frac{\alpha_s}{\pi}\frac{1}{  (\alpha -1) \epsilon ^2}+2\frac{\alpha_s}{\pi}\frac{L_\alpha^{cs}}{   (\alpha -1) \epsilon }\,,
\end{align}
where
\begin{align}\label{eq:csoft_log_def}
L_\alpha^{cs}= \log \left(    \frac{  \mu N_{CS} e^{\gamma_E}  (\ecflp{3}{\alpha})      \left( n_a \cdot n_b   \right)^{-1/2+\alpha}  }{\sqrt{2}  Q    }  \right)\,.
\end{align}

\subsubsection*{$n_a,\bar n$ and $n_b,\bar n$ Contributions:}

We now calculate the $n_a, \bar n$ contribution to the collinear-soft function. The $n_b, \bar n$ contribution will be identical. The one-loop integrand is given by
\begin{align}
S_{c,\,n_a \bar{n}}^{(1)}(\ecf{3}{\alpha})&= \\
&\hspace{-2cm}-g^2  \left( \frac{\mu^{2}  e^{\gamma_E }  }{4\pi} \right)^{\epsilon}    \int[d^dk]_+\frac{2n_a\cdot \bar{n}}{n_{a}\cdot k\,k\cdot \bar{n}}\delta\left(\ecf{3}{\alpha}-N_{CS}\left(\frac{\bar{n}\cdot k}{2Q}\right)^{1-\alpha}\left(\frac{n_a\cdot k}{Q}\right)^{\alpha/2}\left(\frac{n_b\cdot k}{Q}\right)^{\alpha/2}\right)\,,\nonumber
\end{align}
where we have again extracted the normalization factor
\begin{align}
N_{CS}&= 2^{3\alpha/2+1} z_az_b\left(n_a\cdot n_b\right)^{\frac{\alpha}{2}}\,.
\end{align}

As with the $n_a \cdot n_b$ contribution, we expand the integrand to leading power in $n_a \cdot n_b$ using
\begin{align}
n_a\cdot k\,n_b\cdot k&=\left(n\cdot k+\frac{n_a\cdot n_b}{8} \bar{n}\cdot k\right)^2-\frac{n_a \cdot n_b}{2}(n\cdot k \, \bar{n}\cdot k)\,\text{cos}^2\theta\,, \\
n_a\cdot \bar n&=2 \,,\\
n_a \cdot k&=\frac{n_a \cdot n_b}{8} \bar n\cdot k   +n\cdot k -\left( n\cdot k \bar n \cdot k    \right)^{1/2}    \sqrt{\frac{n_a \cdot n_b}{2}} \cos \theta\,.
\end{align}
To perform the integral, it is again convenient to make the change of variables
\begin{align}
\bar n \cdot k = v, \qquad n \cdot k =v w \left(  \frac{n_a \cdot n_b}{8}  \right)\,.
\end{align}
We then have
\begin{align}
n_a\cdot k\,n_b\cdot k&= v^2 \left(  \frac{n_a \cdot n_b}{8}  \right)^2    \left[  (1-w)^2+4w\,\sin^2 \theta  \right]\,, \\
n_a\cdot k&=\frac{n_a \cdot n_b}{8} v  +v w \left(  \frac{n_a \cdot n_b}{8}  \right) -\left(v^2 w \left(  \frac{n_a \cdot n_b}{8}  \right)   \right)^{1/2}    \sqrt{\frac{n_a \cdot n_b}{2}} \cos \theta \nonumber \\
&=v \left(  \frac{n_a \cdot n_b}{8}  \right) \left(   1+w-2\sqrt{w} \cos \theta \right)\,.
\end{align}

The one-loop expression for the contribution to the collinear soft function can then be written
\begin{align}
S_{c,\,n_a \bar{n}}^{(1)}(\ecf{3}{\alpha})&= \\
&\hspace{-1cm} -g^2   \left( \frac{\mu^{2}  e^{\gamma_E}  }{4\pi} \right)^{\epsilon}   4 c_\epsilon \left( \frac{n_a \cdot n_b}{8} \right)^{-\epsilon}  \int_0^\infty\frac{dw}{w^{\epsilon}}             \int_0^\infty\frac{dv}{v^{1+2\epsilon}}         \int_0^\pi d\theta\,\text{sin}^{-2\epsilon}\,\theta          \frac{1}{   1+w-2\sqrt{w} \cos \theta     }    \nonumber \\
&      \delta  \left(\ecf{3}{\alpha}-  \frac{N_{CS} }{2^{1-\alpha}}\frac{v}{Q}           \left(\frac{n_a\cdot n_b}{8} \right)^{\alpha}   \left[   (1-w)^2+4w\,\sin^2 \theta  \right]^{\alpha/2}            \right)\,. \nonumber
\end{align}
This integral can be performed in a similar manner to the $n_a\cdot n_b$ integral. The $v$ integral is straightforward, after transforming to Laplace space $\ecf{3}{\alpha}\to \ecflp{3}{\alpha}$, we find
 \begin{align}
 S_{c,\,n_a \bar{n}}^{(1)}(\ecflp{3}{\alpha})&=  -g^2    4 c_\epsilon    \Gamma(-2\epsilon)     \left(    \frac{  \mu^2 N_{CS}^2 e^{\gamma_E}  (\ecflp{3}{\alpha})^2      \left( \frac{n_a \cdot n_b}{8}   \right)^{-1+2\alpha}  }{4\pi 4^{1-\alpha}  Q^2    }  \right)^{\epsilon}  \nonumber \\
&    \hspace{2cm}     \int_0^\infty\frac{dw}{w^{\epsilon}}                  \int_0^\pi d\theta\,\text{sin}^{-2\epsilon}\,\theta          \frac{\left[  (1-w)^2+4w\,\sin^2 \theta  \right]^{\alpha \epsilon}}{   1+w-2\sqrt{w} \cos \theta     } \,.
\end{align}
We now focus on the integral
\begin{align}
 \int_0^\infty\frac{dw}{w^{\epsilon}}                  \int_0^\pi d\theta\,\text{sin}^{-2\epsilon}\,\theta          \frac{\left[  (1-w)^2+4w\,\sin^2 \theta  \right]^{\alpha \epsilon}}{   1+w-2\sqrt{w} \cos \theta     }\,.
\end{align}
Remapping to the unit interval,  we find
\begin{align}
& \int_0^1 du   \,  \left[  u^{-\epsilon}+u^{-1+(1-2\alpha)\epsilon} \right]               \int_0^\pi d\theta\,\text{sin}^{-2\epsilon}\,\theta      \frac{\left[  (1-u)^2+4u\,\sin^2 \theta  \right]^{\alpha \epsilon}}{   1+u-2\sqrt{u} \cos \theta     }     \nonumber\\
&= \int_0^1 du   \, \left[  u^{-\epsilon}+u^{-1+(1-2\alpha)\epsilon} \right]      \nonumber \\
& \hspace{2cm} \times        \int_0^\pi d\theta\,\text{sin}^{-2\epsilon}\,\theta         \left[  (1-u)^2+4u\,\sin^2 \theta  \right]^{\alpha \epsilon-1}    (1+u+2\sqrt{u} \cos \theta )     
\end{align} 
The $\theta$ integral can be performed in terms of hypergeometric functions using
\begin{align}
&\int_0^\pi d\theta\,\text{sin}^{-2\epsilon}\,\theta     \left[ (1-u)^2+4u\,\sin^2 \theta   \right]^{-1+\alpha \epsilon} =    \frac{\Gamma[1/2-\epsilon] \Gamma[1/2]}{\Gamma[1-\epsilon]}   \\
& \hspace{1cm} \times (1+u)^{-1+2\epsilon}  \left( (1-u)^2\right)^{-1/2-(1-\alpha)\epsilon}  \,_2F_1\left[ 1/2-\epsilon,-\epsilon+\alpha\epsilon, 1-\epsilon, \frac{4u}{(1+u)^2}\right] \,, \nonumber
\end{align}
and
\begin{align}
 \int_0^\pi d\theta\,\text{sin}^{-2\epsilon}\,\theta         \left[  (1-u)^2+4u\,\sin^2 \theta  \right]^{\alpha \epsilon-1}     \cos \theta = 0\,,
\end{align}
by symmetry.

The hypergeometric function has the expansion
 \begin{align}
\,_2F_1\left[\frac{1}{2}-\epsilon,-(1-\alpha)\epsilon;1-\epsilon;\frac{4u}{(1+u)^2}\right]&=1-2(1-\alpha)\epsilon\,\text{ln}(1+u)+{\mathcal O}(\epsilon^2)\,,
\end{align}
which is valid for $0\leq u \leq 1$, 

The final $u$ integral is then
\begin{align}
&    \frac{\Gamma[1/2-\epsilon] \Gamma[1/2]}{\Gamma[1-\epsilon]} \int_0^1 du   \, \left[  u^{-\epsilon}+u^{-1+(1-2\alpha)\epsilon} \right]          (1+u)^{2\epsilon}  \left( (1-u)^2\right)^{-1/2-(1-\alpha)\epsilon} \nonumber \\
&  \hspace{8cm} \times    \left(  1-2(1-\alpha)\epsilon\,\text{ln}(1+u)  \right)\,.
\end{align}

We expect this integral to contribute both $\frac{1}{(1-\alpha) \epsilon}$ and $\frac{1}{(1-2\alpha)\epsilon}$ poles, unlike the $n_a n_b$ contribution, which are evident in the $u\to 1$ and $u\to 0$ limits respectively. We need to do the integral to $\mathcal{O}(\epsilon)$ to get the finite pieces, but only $\mathcal{O}(\epsilon^0)$ to get the anomalous dimensions, which is sufficient for now. We have
\begin{align}
&   = \frac{\Gamma[1/2-\epsilon] \Gamma[1/2]}{\Gamma[1-\epsilon]} \int_0^1 du   \,  u^{-\epsilon}         (1+u)^{2\epsilon}  \left( (1-u)^2\right)^{-1/2-(1-\alpha)\epsilon}  \nonumber \\
&   -2(1-\alpha)\epsilon\, \frac{\Gamma[1/2-\epsilon] \Gamma[1/2]}{\Gamma[1-\epsilon]} \int_0^1 du   \,  u^{-\epsilon}         (1+u)^{2\epsilon}  \left( (1-u)^2\right)^{-1/2-(1-\alpha)\epsilon} \log(1+u) \nonumber \\
&+ \frac{\Gamma[1/2-\epsilon] \Gamma[1/2]}{\Gamma[1-\epsilon]} \int_0^1 du   \,  u^{-1+(1-2\alpha)\epsilon}           (1+u)^{2\epsilon}  \left( (1-u)^2\right)^{-1/2-(1-\alpha)\epsilon}  \nonumber \\
&-2(1-\alpha)\epsilon\, \frac{\Gamma[1/2-\epsilon] \Gamma[1/2]}{\Gamma[1-\epsilon]} \int_0^1 du   \,  u^{-1+(1-2\alpha)\epsilon}           (1+u)^{2\epsilon}  \left( (1-u)^2\right)^{-1/2-(1-\alpha)\epsilon}  \log(1+u)   
\end{align}
This integral can be done systematically using $+$-functions, but to the order we need the result, it is easier to use subtractions, evaluate the log at the value of the singularity, and then perform the integral in terms of hypergeometric functions. The integral can be written
\begin{align}
&   = \frac{\Gamma[1/2-\epsilon] \Gamma[1/2]}{\Gamma[1-\epsilon]} \int_0^1 du   \,  u^{-\epsilon}         (1+u)^{2\epsilon}  \left( (1-u)^2\right)^{-1/2-(1-\alpha)\epsilon}   \\
&   -2(1-\alpha)\epsilon\, \frac{\Gamma[1/2-\epsilon] \Gamma[1/2]}{\Gamma[1-\epsilon]} \int_0^1 du   \,  u^{-\epsilon}         (1+u)^{2\epsilon}  \left( (1-u)^2\right)^{-1/2-(1-\alpha)\epsilon} \log(2) \nonumber \\
&+ \frac{\Gamma[1/2-\epsilon] \Gamma[1/2]}{\Gamma[1-\epsilon]} \int_0^1 du   \,  u^{-1+(1-2\alpha)\epsilon}           (1+u)^{2\epsilon}  \left( (1-u)^2\right)^{-1/2-(1-\alpha)\epsilon}  \nonumber \\
&-2(1-\alpha)\epsilon\, \frac{\Gamma[1/2-\epsilon] \Gamma[1/2]}{\Gamma[1-\epsilon]} \int_0^1 du   \,  u^{-1+(1-2\alpha)\epsilon}         (1+u)^{2\epsilon}     \left[  \left( (1-u)^2\right)^{-1/2-(1-\alpha)\epsilon} -1\right] \log(2)   \,, \nonumber
\end{align}
which gives
\begin{align}
&   = \frac{\Gamma[1/2-\epsilon] \Gamma[1/2]}{\Gamma[1-\epsilon]}                    \frac{ \Gamma[1-\epsilon] \Gamma[-2(1-\alpha)\epsilon]   }{  \Gamma[1-\epsilon-2(1-\alpha) \epsilon]   }\, _2F_1[-2\epsilon,1-\epsilon; 1-\epsilon-2(1-\alpha)\epsilon;-1]   \nonumber \\
&   -2(1-\alpha)\epsilon\, \frac{\Gamma[1/2-\epsilon] \Gamma[1/2]}{\Gamma[1-\epsilon]}  \log(2)      \frac{ \Gamma[1-\epsilon] \Gamma[-2(1-\alpha)\epsilon]   }{  \Gamma[1-\epsilon-2(1-\alpha) \epsilon]   }\, _2F_1[-2\epsilon,1-\epsilon; 1-\epsilon-2(1-\alpha)\epsilon;-1]        \nonumber \\
&+ \frac{\Gamma[1/2-\epsilon] \Gamma[1/2]}{\Gamma[1-\epsilon]}  \frac{ \Gamma[ (1-2\alpha)\epsilon] \Gamma[-2(1-\alpha)\epsilon]   }{  \Gamma[  (1-2\alpha)\epsilon-2(1-\alpha)\epsilon       ]   }\, _2F_1[ -2\epsilon,(1-2\alpha)\epsilon;(1-2\alpha)\epsilon-2(1-\alpha)\epsilon;-1]  \nonumber \\
&-2(1-\alpha)\epsilon\, \frac{\Gamma[1/2-\epsilon] \Gamma[1/2]}{\Gamma[1-\epsilon]}  \log(2) \nonumber \\
& \times \left(        \frac{ \Gamma[ (1-2\alpha)\epsilon] \Gamma[-2(1-\alpha)\epsilon]   }{  \Gamma[  (1-2\alpha)\epsilon-2(1-\alpha)\epsilon       ]   }\, _2F_1[ -2\epsilon,(1-2\alpha)\epsilon;(1-2\alpha)\epsilon-2(1-\alpha)\epsilon;-1]  \right. \nonumber \\
&\hspace{5.5cm}\left.  -  \frac{\Gamma[(1-2\alpha)\epsilon] }{\Gamma[1+(1-2\alpha)\epsilon]} \,_2 F_1[-2\epsilon,1;1+(1-2\alpha)\epsilon;-1]   \right) \,.
\end{align}
Expanding this to  $\mathcal{O}(\epsilon^0)$ gives
\begin{align}
=\frac{\pi }{(\alpha -1) \epsilon }-\frac{\pi }{(2 \alpha -1) \epsilon } -\frac{2\pi  \log (2)}{2 \alpha -1}+\frac{4 \pi  \log (2)}{\alpha -1}+2\pi  \log (2)
\end{align}
We then have
 \begin{align}
 S_{c,\,n_a \bar{n}}^{(1)}(\ecflp{3}{\alpha})&=  -g^2     4 c_\epsilon    \Gamma(-2\epsilon)     \left(    \frac{  \mu^2 N_{CS}^2 e^{\gamma_E}  (\ecflp{3}{\alpha})^2      \left( \frac{n_a \cdot n_b}{8}   \right)^{-1+2\alpha}  }{4\pi 4^{1-\alpha}  Q^2    }  \right)^{\epsilon}  \nonumber \\
&    \hspace{2cm}    \,\times    \left(\frac{\pi }{(\alpha -1) \epsilon }-\frac{\pi }{(2 \alpha -1) \epsilon } -\frac{2\pi  \log (2)}{2 \alpha -1}+\frac{4 \pi  \log (2)}{\alpha -1}+2\pi  \log (2)   \right)\,.
\end{align}
Extracting just the divergent pieces so as to get the anomalous dimensions, we find
 \begin{align}\label{eq:csoft_nbarna_final}
 S_{c,\,n_a \bar{n}}^{(1)}(\ecflp{3}{\alpha})&= \frac{\alpha _s}{2\pi  (\alpha -1) \epsilon ^2}-\frac{\alpha _s}{2\pi  (2 \alpha
   -1) \epsilon ^2}+\frac{ \log\left[   \frac{  \mu N_{CS} e^{\gamma_E}  (\ecflp{3}{\alpha})      \left( \frac{n_a \cdot n_b}{8}   \right)^{-1/2+\alpha}  }{2^{1-\alpha}  Q    }     \right] \alpha _s}{\pi  (\alpha -1) \epsilon }    \nonumber \\
   &        -\frac{ \log \left[   \frac{  \mu N_{CS} e^{\gamma_E}  (\ecflp{3}{\alpha})      \left( \frac{n_a \cdot n_b}{8}   \right)^{-1/2+\alpha}  }{2^{1-\alpha}  Q    }   \right] \alpha _s}{\pi  (2 \alpha
   -1) \epsilon }    +\frac{ \log (2) \alpha _s}{\pi  (\alpha -1) \epsilon }+\frac{\log (2) \alpha _s}{\pi  \epsilon }\,,
\end{align}
which can be simplified to
\begin{align}
 S_{c,\,n_a \bar{n}}^{(1)}(\ecflp{3}{\alpha})&=\frac{\alpha _s}{2\pi  (\alpha -1) \epsilon ^2}-\frac{\alpha _s}{2\pi  (2 \alpha -1) \epsilon ^2}\nonumber\\
 &\hspace{2cm} +\frac{\alpha_s}{\pi (\alpha-1) \epsilon} L_\alpha^{cs} \left( \ecflp{3}{\alpha}  \right)-\frac{\alpha_s}{\pi (2\alpha-1) \epsilon} L_\alpha^{cs} \left( \ecflp{3}{\alpha}  \right)\,,
\end{align}
where, as for the logarithm in the $n_a\, n_b$ contribution, \Eq{eq:csoft_log_def}, the logarithm that appears is
\begin{align}
L_\alpha^{cs} \left( \ecflp{3}{\alpha}  \right)
= \log \left(    \frac{  \mu N_{CS} e^{\gamma_E}  (\ecflp{3}{\alpha})      \left( n_a \cdot n_b   \right)^{-1/2+\alpha}  }{\sqrt{2}  Q    }  \right)\,.
\end{align}

The contribution from an emission between the $n_b$ and $\bar n$ Wilson lines is identical, so we have
\begin{align}\label{eq:csoft_log}
S_{c,\,n_b \bar{n}}^{(1)}  \left(\ecf{3}{\alpha}\right)=S_{c,\,n_a \bar{n}}^{(1)}   \left(\ecf{3}{\alpha} \right)\,.
\end{align}

Note that for both the $\bar n\, n_a$ and $\bar n\, n_b$ contributions, and unlike for the $n_a \, n_b$ contribution, we have $1/\epsilon$ contributions both of the soft form $1/(1-2\alpha)$, and of the collinear form, $1/(1-\alpha)$. This will be crucial to achieve the cancellation of anomalous dimensions, as required for the consistency of the collinear subjets factorization theorem.

It is interesting to note that this structure is very different than that which appeared for the case of the $N$-subjettiness observable in \Ref{Bauer:2011uc}. In this case only a single angular exponent appears throughout the calculation, unlike both the $1/(1-2\alpha)$ and $1/(1-\alpha)$ that we find here, and the divergent pieces of the $\bar n\, n_a$ and $\bar n\, n_b$ contributions vanish.

\subsection*{Cancellation of Anomalous Dimensions}\label{sec:bkg_cancel_anomdim}

We now review the renormalization group evolution of each of the functions in the factorization theorem, and show that sum of the anomalous dimensions vanishes, as required for renormalization group consistency.  

The hard function satisfies a multiplicative RGE, given by
\begin{equation}
\mu \frac{d}{d\mu}\log\, H(Q^2,\mu)=\gamma_H (Q^2,\mu)=2\text{Re}\left[ \gamma_C(Q^2,\mu)\right]\,,
\end{equation}
where
\begin{equation}
\gamma_C(Q^2,\mu)=\frac{\alpha_s C_F}{4\pi}\left(  4\log\left[  \frac{-Q^2}{\mu^2} \right] -6  \right )\,,
\end{equation}
is the anomalous dimension of the dijet Wilson coefficient. Explicitly
\begin{align}
\gamma_H (Q^2,\mu)=\frac{\alpha_s C_F}{2\pi}\left(  4\log\left[  \frac{Q^2}{\mu^2} \right] -6  \right )\,.
\end{align}

The anomalous dimension of the hard splitting function $H_2$ can be extracted from \Ref{Bauer:2011uc} by performing a change of variables. It satisfies a multiplicative RGE
 \begin{align}
\mu \frac{d}{d\mu}H_2(t,x,\mu)=\gamma_{H_2}(t,x,\mu) H_2(t,x,\mu)\,, 
\end{align}
with anomalous dimension
\begin{align}
\gamma_{H_2}(t,x,\mu)=\frac{\alpha_s(\mu)}{2\pi} \left [  2C_A \log \frac{t}{\mu^2} +4\left( C_F-\frac{C_A}{2}\right ) \log\, x +2C_A \log(1-x)-\beta_0 \right]\,.
\end{align}
Here $\beta_0$ is defined with the normalization
\begin{equation}
\beta_0=\frac{11 C_A}{3}-\frac{2n_f}{3 } \,.
\end{equation}
Converting to $\ecf{2}{\alpha}$ by performing the change of variables given in \Eq{eq:relate_to_frank}, we find
\begin{align}\label{eq:h2_anom}
\gamma_{H_2}\left(\ecf{2}{\alpha},z_q,\mu\right)&=\frac{\alpha_s(\mu)}{2\pi} \left [  2C_A \log\left( \frac{Q^2}{\mu^2} \frac{(z_az_b)^{1-2/\alpha}    \left( \ecf{2}{\alpha}  \right)^{2/\alpha}      }{4} \right) \right. \\
&\hspace{3cm}\left.+4\left( C_F-\frac{C_A}{2}\right ) \log\, z_q +2C_A \log(1-z_q)-\beta_0        \vphantom{\log\left( \frac{Q^2}{\mu^2} \frac{(z_az_b)^{1-2/\alpha}    \left( \ecf{2}{\alpha}  \right)^{2/\alpha}      }{4} \right)}        \right]\,. \nonumber
\end{align}
Since the anomalous dimensions of the jet, soft and collinear-soft functions are written in terms of $\ecflp{3}{\alpha}$, $z_a$, $z_b$, and $n_a\cdot n_b$, for demonstrating cancellation of anomalous dimensions, it is convenient to replace $\ecf{2}{\alpha}$ in \Eq{eq:h2_anom} with its leading power expression from \Eq{eq:lp_e2}. We then have
\begin{align}\label{eq:h2_anom2}
\gamma_{H_2}\left(\ecf{2}{\alpha},z_q,\mu\right)&=\frac{\alpha_s(\mu)}{2\pi} \left [  2C_A \log\left( \frac{Q^2}{\mu^2} \frac{ z_az_b\,    n_a \cdot n_b     }{2} \right) \right. \\
&\hspace{3cm}\left.+4\left( C_F-\frac{C_A}{2}\right ) \log\, z_q +2C_A \log(1-z_q)-\beta_0        \vphantom{ \log\left( \frac{Q^2}{\mu^2} \frac{ z_az_b\,    n_a \cdot n_b     }{2} \right)}        \right]\,. \nonumber
\end{align}
Note that $1-z_q=z_g$.

The jet functions satisfy multiplicative RGEs in Laplace space (they satisfy convolutional RGEs in $\ecf{3}{\alpha}$, see \Ref{Ellis:2010rwa} for a detailed discussion)
\begin{align}
\mu\frac{d}{d\mu}\log\,\tilde{J}_{g,q\,n}    \left(Q_J,\ecflp{3}{\alpha}\right)&=\gamma_{g,q}^{\alpha}     \left(Q_J,\ecflp{3}{\alpha} \right)\,,
\end{align}
where the one-loop anomalous dimension is determined from \Eqs{eq:gjet_final}{eq:qjet_final}, and is given by
\begin{align}\label{eq:jet_anoms}
\gamma_{g,q}^{\alpha}\left(Q_J,\ecflp{3}{\alpha} \right)&=-2\frac{\alpha_s}{\pi}\frac{C_{g,q}}{(1-\alpha)}L_\alpha^{J,a}\left(  \ecflp{3}{\alpha}  \right)+\gamma_{g,q}\,,
\end{align}
where the logarithm $L_\alpha^{J,a}\left(  \ecflp{3}{\alpha}  \right)$ was defined in \Eq{eq:jet_log}, and is given by
\begin{align} 
L_\alpha^{J,a}\left(  \ecflp{3}{\alpha}  \right)&=\log\left[N_J\ecflp{3}{\alpha}e^{\gamma_E }\left(\frac{\mu}{\sqrt{2}Q}\right)^\alpha z_a^{2-\alpha}\right]\,.
\end{align}
Here $C_{g,q}$ is the appropriate Casimir ($C_A$ for gluon jets and $C_F$ for quark jets), and with $\gamma_{g,q}$ the standard functions
 \begin{align}
\gamma_{q}= \frac{3\alpha_s C_F}{2\pi}\,, \qquad
\gamma_{g}=\frac{\alpha_s}{\pi}\frac{11C_A- 2n_f}{6}\,.
\end{align}
For subjet $b$, we simply have $a\to b$.

Similarly, the soft function satisfies a multiplicative RGE in Laplace space 
\begin{align}
\mu\frac{d}{d\mu}\log\,\tilde{S}_{G}    \left(\ecflp{3}{\alpha}\right)&=\gamma_G     \left(\ecflp{3}{\alpha}    \right)\,,
\end{align}
with one-loop anomalous dimension determined by \Eq{eq:gsoft_final}, and given by
\begin{align}
\gamma_G    \left(\ecflp{3}{\alpha}\right)&=\frac{-2\alpha_s}{\pi(1-2\alpha)}{\mathbf T}_n\cdot{\mathbf T}_{\bar{n}} \,L_\alpha^{G} \left(\ecflp{3}{\alpha}\right)  \,.
\end{align}
Here the logarithm is given by
\begin{align}
L_\alpha^{G}\left(\ecflp{3}{\alpha}\right)&=\log\left[\frac{e^{\gamma_E}\mu\ecflp{3}{\alpha}N_S}{2^{1-\alpha}Q}\right]-\frac{(1-2\alpha)}{2}\log\left[\text{tan}^2\frac{R}{2}\right]\,.
\end{align}
Finally, the collinear soft function satisfies a multiplicative RGE in Laplace space
\begin{align}
\mu\frac{d}{d\mu}\log\,S^{}_{c}     \left(\ecflp{3}{\alpha}\right)&=\gamma_{cs}     \left(\ecflp{3}{\alpha}\right)\,,
\end{align}
with the one-loop anomalous dimension determined by \Eqs{eq:csoft_nanb_final}{eq:csoft_nbarna_final}
\begin{align}
\gamma_{cs}   \left(\ecflp{3}{\alpha}\right)&={\mathbf T}_a\cdot{\mathbf T}_b\gamma_{ab}    \left(\ecflp{3}{\alpha}\right)+{\mathbf T}_a\cdot{\mathbf T}_{\bar{n}}\gamma_{a\bar{n}}\left(\ecflp{3}{\alpha}\right)+{\mathbf T}_{\bar{n}}\cdot{\mathbf T}_b\gamma_{\bar{n}b}  \left(\ecflp{3}{\alpha}\right)\,,
\end{align}
where 
\begin{align}
\gamma_{ab}   \left (\ecflp{3}{\alpha}\right )&=     \frac{-4\alpha_s}{\pi(1-\alpha)}L_\alpha^{cs}\left(\ecflp{3}{\alpha}\right)\,,\\
\gamma_{a\bar{n}}    \left(\ecflp{3}{\alpha}\right)&=\gamma_{b\bar{n}} \left(\ecflp{3}{\alpha}\right)=\frac{-2\alpha_s}{\pi(1-\alpha)}L_\alpha^{cs}\left(\ecflp{3}{\alpha}\right)+\frac{2\alpha_s}{\pi(1-2\alpha)}L_\alpha^{cs}\left(\ecflp{3}{\alpha}\right)\,.
\end{align}
The argument of the logarithm appearing in the collinear soft function, was defined in \Eq{eq:csoft_log_def}, and is given by
\begin{align}
L_\alpha^{cs}= \log \left(    \frac{  \mu N_{CS} e^{\gamma_E}  (\ecflp{3}{\alpha})      \left( n_a \cdot n_b   \right)^{-1/2+\alpha}  }{\sqrt{2}  Q    }  \right)\,.
\end{align}

We can now explicitly check the cancellation of anomalous dimensions. We consider the particular partonic subprocess $e^+e^- \to \bar q q \to \bar q q g$ for which we have explicitly given the hard splitting function, in which case the color algebra can be simplified and written entirely in terms of Casimirs using the color conservation relations
\begin{align}
{\mathbf T}_n={\mathbf T}_q+{\mathbf T}_g\,,\\
{\mathbf T}_n+{\mathbf T}_{\bar n}=0\,.
\end{align}
We then have
\begin{align}
{\mathbf T}_n\cdot{\mathbf T}_{\bar{n}}&=-C_F\,,\\
{\mathbf T}_q\cdot{\mathbf T}_{g}&=-\frac{C_A}{2}\,,\\
{\mathbf T}_q\cdot{\mathbf T}_{\bar{n}}&=\frac{C_A}{2}-C_F\,,\\
{\mathbf T}_g\cdot{\mathbf T}_{\bar{n}}&=-\frac{C_A}{2}\,, \\
{\mathbf T}_g\cdot{\mathbf T}_{g}&=C_A\,, \\
{\mathbf T}_n\cdot{\mathbf T}_{n}&=C_F\,.
\end{align}
However, for most of the cancellation of the anomalous dimensions, it will be convenient to work in the abstract color notation, so as not to need to use relations between the color Casimirs.

The independence of the total cross section under renormalization group evolution implies the following relation between anomalous dimensions
 \begin{align}\label{eq:sum_anom_zero}
\gamma_H \left(Q^2,\mu  \right)+\gamma_{H_2}\left(\ecf{2}{\alpha},z_q,\mu\right)+\gamma_{g}^{\alpha}\left(\ecflp{3}{\alpha}\right)+\gamma_{q}^{\alpha}\left(\ecflp{3}{\alpha}\right)+\gamma_G\left(\ecflp{3}{\alpha}\right)+\gamma_{cs}\left(\ecflp{3}{\alpha}\right)\sim0\,,
\end{align}
where the $\sim$ means up to a term corresponding to the measurement of the jet in the $\bar n$ direction, and the out-of-jet contribution to the soft function, which is independent of the $\ecflp{3}{\alpha}$ measurement, and the kinematics of the substructure, namely $n_a\cdot n_b$, $z_a$, and $z_b$. We will make this relation precise shortly.

We now show explicitly that this cancellation occurs, and how it arises, which provides a non-trivial cross-check on the collinear-subjets factorization theorem. Substituting in the expressions above, we find
\begin{align}
&\sum_\gamma =\gamma_H \left(Q^2,\mu  \right)+\gamma_{H_2}\left(\ecf{2}{\alpha},z_q,\mu\right) \nonumber \\
&+  \left[   -    {\mathbf T}_a \cdot{\mathbf T}_{b}  \frac{4\alpha_s L_\alpha^{cs}\left(\ecflp{3}{\alpha}\right)}{\pi (1-\alpha)}     \right. \nonumber \\
&\left. \hspace{0.0cm}-2 {\mathbf T}_a \cdot{\mathbf T}_{\bar n}  \left( \frac{\alpha_s L_\alpha^{cs}\left(\ecflp{3}{\alpha}\right)}{\pi (1-\alpha)}-\frac{\alpha_s L_\alpha^{cs}\left(\ecflp{3}{\alpha}\right)}{\pi (1-2\alpha)}   \right)         -2 {\mathbf T}_b \cdot{\mathbf T}_{\bar n}  \left( \frac{\alpha_s L_\alpha^{cs}\left(\ecflp{3}{\alpha}\right)}{\pi (1-\alpha)}-\frac{\alpha_s L_\alpha^{cs}\left(\ecflp{3}{\alpha}\right)}{\pi (1-2\alpha)}   \right)  \right] \nonumber \\
&- \left[      {\mathbf T}_n \cdot{\mathbf T}_{\bar n}   \frac{2\alpha_s L_\alpha^{G}\left(\ecflp{3}{\alpha}\right)}{\pi (1-2\alpha)}   \right]        -   C_A   \frac{2\alpha_s L_\alpha^{g}\left(  \ecflp{3}{\alpha}  \right)}{\pi (1-\alpha)} +\gamma_g        - C_F   \frac{2\alpha_s L_\alpha^{q}\left(  \ecflp{3}{\alpha}  \right)}{\pi (1-\alpha)} +\gamma_q \,.
\end{align}
To make manifest the separate cancellations, we use the color conservation relation ${\mathbf T}_n={\mathbf T}_a+{\mathbf T}_b$ in the soft anomalous dimension, and ${\mathbf T}_{\bar n}=-{\mathbf T}_a-{\mathbf T}_b$ in the $1/(1-\alpha)$ pieces of the collinear soft anomalous dimensions. Grouping together collinear like terms ($1/(1-\alpha)$) and soft like terms ($1/(1-2\alpha)$), we then have
\begin{align}
&\sum_\gamma =\gamma_H \left(Q^2,\mu  \right)+\gamma_{H_2}\left(\ecf{2}{\alpha},z_q,\mu\right) \nonumber \\
&-  \left[      ({\mathbf T}_a+{\mathbf T}_b )\cdot{\mathbf T}_{\bar n}   \frac{2\alpha_s L_\alpha^{G}\left(\ecflp{3}{\alpha}\right)}{\pi (1-2\alpha)}   \right]+ \left[ {\mathbf T}_a \cdot{\mathbf T}_{\bar n}  \frac{2\alpha_s L_\alpha^{cs}\left(\ecflp{3}{\alpha}\right)}{\pi (1-2\alpha)}           + {\mathbf T}_b \cdot{\mathbf T}_{\bar n} \frac{2\alpha_s L_\alpha^{cs}\left(\ecflp{3}{\alpha}\right)}{\pi (1-2\alpha)}   \right] \nonumber \\
&-       {\mathbf T}_a \cdot{\mathbf T}_{b}  \frac{4\alpha_s L_\alpha^{cs}\left(\ecflp{3}{\alpha}\right)}{\pi (1-\alpha)} - \left[  {\mathbf T}_a \cdot   (-{\mathbf T}_a-{\mathbf T}_b)  \frac{2\alpha_s  L_\alpha^{cs}\left(\ecflp{3}{\alpha}\right) }{\pi (1-\alpha)}           + {\mathbf T}_b \cdot(-{\mathbf T}_a-{\mathbf T}_b)  \frac{2\alpha_s  L_\alpha^{cs}\left(\ecflp{3}{\alpha}\right)}{\pi (1-\alpha)}   \right] \nonumber \\
&-      C_A   \frac{2\alpha_s L_\alpha^{g}\left(  \ecflp{3}{\alpha}  \right)}{\pi (1-\alpha)} +\gamma_g  -     C_F   \frac{2\alpha_s L_\alpha^{q}\left(  \ecflp{3}{\alpha}  \right)}{\pi (1-\alpha)} +\gamma_q \,.
\end{align}
Since all the logs are linear in the $\ecflp{3}{\alpha}$, we immediately see that the color conservation relations have led to the cancellation of the $\ecflp{3}{\alpha}$ dependence in the soft like pieces between the $\bar nn_b$ and $\bar nn_a$ contributions to the collinear soft function with the global soft contribution, and the cancellation between the collinear like pieces involve all three contributions to the collinear soft function, as well as the jet functions. This nontrivial cancellation supports the validity of the collinear subjets factorization theorem.

It is also straightforward to check that the dependence on $\ecf{2}{\alpha}$ as well as on the jet energy fractions also cancels, although this is more tedious to perform step by step. We therefore simply quote the summed result of the anomalous dimensions, to make clear the meaning of the equivalence relation in \Eq{eq:sum_anom_zero}. We have
\begin{align}
&\gamma_H \left(Q^2,\mu  \right)+\gamma_{H_2}\left(\ecf{2}{\alpha},z_q,\mu\right)+\gamma_{g}^{\alpha}\left(\ecflp{3}{\alpha}\right)+\gamma_{q}^{\alpha}\left(\ecflp{3}{\alpha}\right)+\gamma_G\left(\ecflp{3}{\alpha}\right)+\gamma_{cs}\left(\ecflp{3}{\alpha}\right)= \nonumber \\
&\hspace{2cm}-\frac{3\alpha_s C_F}{2\pi}-\frac{  \alpha_s  C_F    \log \left[ \tan^2 \frac{R}{2}   \right] }{\pi} -\frac{  \alpha_s  C_F   \log \frac{\mu^2}{Q^2}}{\pi}\,.
\end{align}
These remaining terms are exactly those expected to cancel against the out-of-jet contribution; see, e.g., \Ref{Ellis:2010rwa} for a detailed discussion. 

The out-of-jet jet function is then given by the unmeasured jet function of \Ref{Ellis:2010rwa}
\begin{align}
\mu\frac{d}{d\mu}\ln J_{oj }(R_B) &= \frac{2\alpha_s C_F}{\pi} \log\left[    \frac{\mu}{ Q \tan \frac{ R_B}{2} }   \right]+\frac{3\alpha_s C_F}{2\pi}\,,
\end{align}
where here $R_B$ is the radius of the recoiling jet. For simplicity, throughout this chapter, we have taken $R_B=R$.

The out-of-jet contribution to the soft function has a pure cusp anomalous dimension  \cite{Ellis:2010rwa}
\begin{align}
\mu\frac{d}{d\mu}\ln S_{oj }(R_B) &= \frac{2\alpha_s C_F}{\pi} \log\left[  \tan^2 \frac{ R}{2}  \right]   \,.
\end{align}

\section{One Loop Calculations of Soft Subjet Functions}\label{sec:softjet_app}
In this appendix we give the operator definitions and one-loop results for the functions appearing in the factorization theorem of \Eq{fact_inclusive_form_1} for the soft subjet region of phase space. The factorization theorem in the soft subjet region of phase space was first presented in \Ref{Larkoski:2015zka}, where all functions were calculated to one-loop, and a detailed discussion of the structure of the required zero bin subtractions was given. This calculation was performed with a broadening axis cone algorithm, however it was argued in \Sec{sec:soft_jet} that to leading power, the factorization theorem is identical in the case of an anti-$k_T$ algorithm. Because of this, in this appendix we give only the final results for the one-loop anomalous dimensions, and the tree level matching for the soft subjet production, as are required for the resummation considered in this chapter. The interested reader is referred to \Ref{Larkoski:2015zka} for the detailed calculation, as well as a discussion of the intricate zero bin structure of the factorization theorem, which is only briefly mentioned in this appendix.         

\subsection*{Definitions of Factorized Functions}
The functions appearing in the soft subjet factorization theorem of \Eq{fact_inclusive_form_1} have the following SCET operator definitions:
\begin{itemize}
\item Hard Matching Coefficient for Dijet Production
\begin{equation}
H(Q^2,\mu)=|C(Q^2,\mu)|^2\,,
\end{equation}
where $C\left(Q^2,\mu\right)$ is the Wilson coefficient obtained from matching the full theory QCD current $\bar \psi \Gamma \psi$ onto the SCET dijet operator $\bar \chi_n \Gamma \chi_{\bar n}$
\begin{align}
\langle q\bar q | \bar \psi \Gamma \psi |0\rangle=C\left(Q^2,\mu\right) \langle q\bar q | \mathcal{O}_2 |0\rangle\,.
\end{align}
As before, we have neglected the contraction with the Leptonic tensor.
\item Soft Subjet Jet Function:
{\small\begin{align}
&J_{\sja }\Big(\ecf{3}{\alpha}\Big)=\\
& \hspace{.25cm}
\frac{(2\pi)^3}{C_A}\text{tr}\langle 0|\mathcal{B}_{\perp_{\sja}}^{\mu}(0)\Theta_{O}(B)\delta(Q_{SJ}-\sjabar \cdot{\mathcal P})\delta^{(2)}(\vec{{\mathcal P}}_{\perp_{SJ}})\delta\Big(\ecf{3}{\alpha}-\Theta_{FJ}\ecfop{3}{\alpha}\big|_{SJ}\Big)\,\mathcal{B}_{\perp_{\sja}\mu}(0)|0\rangle \nonumber
\end{align}}
\item Jet Function:
{\small\begin{align}
\hspace{-1cm}
J_{n}\Big(\ecf{3}{\alpha}\Big)&=\frac{(2\pi)^3}{C_F}\text{tr}\langle 0|\frac{\bar{n}\!\!\!\slash}{2}\chi_{n}(0) \Theta_{O}(B)\delta(Q-\bar{n}\cdot{\mathcal P})\delta^{(2)}(\vec{{\mathcal P}}_{\perp})\delta\Big(\ecf{3}{\alpha}-\Theta_{FJ}\ecfop{3}{\alpha}\big|_{HJ}\Big)\bar{\chi}_n(0)|0\rangle
\end{align}}
\item Boundary Soft Function:
\begin{align}
S_{\sja \,\sjabar }\Big(\ecf{3}{\alpha};R\Big)&=\frac{1}{C_{A}}\text{tr}\langle 0|T\{S_{\sja } S_{\sjabar }\} \Theta_{O}(B) \delta\Big(\ecf{3}{\alpha}-\Theta_{FJ}\ecfop{3}{\alpha}\big|_{BS}\Big)\bar{T}\{S_{\sja } S_{\sjabar }\} |0\rangle
\end{align}
\item Soft Subjet Soft Function:
\begin{align}
\hspace{-1cm}S_{\sja \,n\,\bar{n}}\Big(\ecf{3}{\alpha},B;R\Big)&=\text{tr}\langle 0|T\{S_{\sja } S_{n} S_{\bar{n}}\}\Theta_{O}(B)\delta\Big(\ecf{3}{\alpha}-\Theta_{FJ}\ecfop{3}{\alpha}\big|_{S}\Big)\bar{T}\{S_{\sja } S_{n} S_{\bar{n}}\} |0\rangle
\end{align}
\end{itemize}

The definitions of these functions include measurement operators, which when acting on the final state, return the value of a given observable. The operator $\ecfop{3}{\alpha}$ measures the contribution to $\ecf{3}{\alpha}$ from final states, and must be appropriately expanded following the power counting of the sector on which it acts. Expressions for the expansions in the power counting of the different sectors will be given shortly, after kinematic notation has been set up. The operators $\Theta_{FJ}$, and $\Theta_{O}$ constrain the measured radiation to be in the jet or out of the jet, respectively, and will be defined shortly. 

\subsection*{Kinematics and Notation}

For our general kinematic setup, we will denote by $Q$ the center of mass energy of the $e^+e^-$ collisions, so that $Q/2$ is the energy deposited in a hemisphere. i.e. the four-momenta of the two hemispheres are
\begin{align}
p_{\text{hemisphere}_1}=\left( \frac{Q}{2},\vec p_1   \right)\,, \qquad p_{\text{hemisphere}_2}=\left( \frac{Q}{2},-\vec p_1   \right)
\end{align}
so
\begin{align}
s=Q^2\,.
\end{align}

We are now interested in the regime where there is a wide angle soft subjet carrying a small energy fraction, and an energetic subjet, carrying the majority of the energy fraction. We will label the lightcone directions of the energetic subjet by $n,\bar n$, and the lightcone directions of the soft subjet as $n_{sj}, \bar n_{sj}$. We will use the variable $z_{sj}$ to label the energy fraction of the soft subjet, namely
\begin{align}
E_{sj}=z_{sj} \frac{Q}{2}\,, \qquad z_{sj} \ll 1\,.
\end{align}

In this region of phase space, to leading power the value of the two point energy correlation function is set by the two subjets, and is given by
\begin{align}
\ecf{2}{\alpha}= 2^{\alpha/2} z_{sj} \left(  n\cdot n_{sj} \right)^{\alpha/2}   \,.
\end{align}

The action of the measurement function $\ecfop{3}{\alpha}$ on a arbitrary state for each of the factorized sectors contributing to the 3-point energy correlation function measurement is given by
\begin{align}\label{eq:action_measurements}
\ecfop{3}{\alpha}\big|_{SJ}\Big|X_{sj}\Big\rangle&=\sum_{k_i,k_j\in X_{sj}} N_{SJ} \frac{\sjabar\cdot k_i}{Q}\frac{\sjabar\cdot k_j}{Q}\left(\frac{k_i\cdot k_j}{\sjabar\cdot k_i\sjabar\cdot k_j}\right)^{\frac{\alpha}{2}}\Big|X_{sj}\Big\rangle\,,\\
\ecfop{3}{\alpha}\big|_{HJ}\Big|X_{hj}\Big\rangle&=\sum_{k_i,k_j\in X_{hj}} N_{HJ}\frac{\nbar\cdot k_i}{Q}\frac{\nbar\cdot k_j}{Q}\left(\frac{k_i\cdot k_j}{\nbar\cdot k_i\nbar\cdot k_j}\right)^{\frac{\alpha}{2}}\Big|X_{hj}\Big\rangle\,,\\
\ecfop{3}{\alpha}\big|_{BS}\Big|X_{bs}\Big\rangle&=\sum_{k\in X_{bs}}N_{BS} \frac{\sjabar\cdot k}{Q}\left(\frac{\sja\cdot k}{\sjabar\cdot k}\right)^{\frac{\alpha}{2}}\Big|X_{bs}\Big\rangle\,,\\
\ecfop{3}{\alpha}\big|_{S}\Big|X_{s}\Big\rangle&=\sum_{k\in X_{s}}N_S\frac{k^0}{Q}\left(\frac{\sja\cdot k}{k^0}\frac{n\cdot k}{k^0}\right)^{\frac{\alpha}{2}}\Big|X_{s}\Big\rangle\,,
\end{align}
where, for simplicity, we have extracted the normalization factors
\begin{alignat}{2}\label{eq:norm_factors}
N_{SJ}&=2^{5\alpha/2}(n\cdot\sja)^{\alpha}\,, & \qquad
N_{HJ}&=2^{5\alpha/2}z_{sj}(n\cdot\sja)^{\alpha}\,,\\
N_{BS}&=2^{2\alpha}z_{sj}(n\cdot\sja)^{\alpha}\,,& \qquad
N_{S}&=2^{1+3\alpha/2} z_{sj}(n\cdot\sja)^{\alpha/2}\,.
\end{alignat}%

These expressions follow from properly expanding the definition of the energy correlation function measurements in the power counting of each of the sectors. Note that on the jet sectors, the 3-point correlation measurement becomes an effective 2-point correlation measurement, since the 2-point energy correlation function is set by the initial splitting of the subjet.

The in-jet restriction, $\Theta_{FJ}$, is given by
\begin{align}
\Theta_{FJ}(k)&=\Theta\left(\tan ^2\frac{R}{2}-\frac{n\cdot k}{\bar{n}\cdot k}\right)\,.
\end{align}
The jet restriction must also be expanded following the power counting of the given sector. We will see that this is actually quite subtle for the soft subjet modes, since the angle between the soft subjet axis and the boundary of the jet has a non-trivial power counting. In particular, the expansion of $\Theta_{FJ}(k)$ is different for the soft subjet jet and boundary soft modes, and will demonstrate the necessity of performing the complete factorization of the soft subjet dynamics into jet and boundary soft modes. Finally, since we are considering the case where the out-of-jet scale $B$ is much less than the in-jet scale, the operator
$$
\Theta_{O}(B)
$$
must also be included in the definition of the soft subjet functions. This operators vetoes out-of-jet radiation above the scale $B$. The explicit expression for $\Theta_{O}(B)$ expanded in the power counting of each of the factorized sectors can be found in \Ref{Larkoski:2015zka}.

\subsection*{Hard Matching Coefficient for Dijet Production}

The hard matching coefficient for dijet production, $H(Q^2,\mu)$, is identical to that for the collinear subjets factorization theorem by hard-collinear-soft factorization,  and is given in
\Eq{eq:hard_matching_coeff}.


\subsection*{Hard Matching for Soft Subjet Production}
The hard matching coefficient $H^{sj}(\sje,\sjtheta)$ is determined by the finite parts of the logarithm of the soft matrix element for a single soft state
\begin{align}
H^{sj}(\sje,\sja)&=\text{tr}\langle 0|T\{S_nS_{\bar{n}}\}|sj \rangle\langle sj|\bar{T}\{S_nS_{\bar{n}}\}|0\rangle_{\text{fin}}\,.
\end{align}
The virtual corrections of the effective theory cancel the IR divergences of this matrix element, giving a finite matching coefficient. This matrix element can be calculated from the square of the soft gluon current \cite{Berends:1988zn,Catani:2000pi}, which is known to two loop order \cite{Duhr:2013msa,Li:2013lsa}. The tree level hard matching coefficient for the soft subjet production is given by
\begin{align}
H^{sj(\text{tree})}_{n\bar{n}}(\sje,\sja)&=\frac{\alpha_s C_F}{\pi\sje}\frac{n\cdot \bar{n}}{n\cdot\sja\,\sja\cdot\bar{n}}\,.
\end{align}
The results of \Ref{Catani:2000pi} can be used to determine the soft subjet production matching from an arbitrary number of hard jets at one loop.

\subsection*{Anomalous Dimensions}\label{app:Anom_Dim}

In this section we collect the one-loop anomalous dimensions for all the functions calculated in this appendix. The two hard functions satisfy multiplicative renormalization group equations. For the dijet production hard function, we have
\begin{equation}
\mu \frac{d}{d\mu}\ln H(Q^2,\mu)=\gamma_H (Q^2,\mu)=2\text{Re}\left[ \gamma_C(Q^2,\mu)\right]\,.
\end{equation}
Explicitly
\begin{align}
\gamma_H (Q^2,\mu)=\frac{\alpha_s C_F}{2\pi}\left(  4\log\left[  \frac{Q^2}{\mu^2} \right] -6  \right )\,.
\end{align}

For the soft subjet production hard function, we have
\begin{equation}
\mu\frac{d}{d\mu}\ln H^{sj}_{n\bar{n}}(\sje,\sja,\mu) =-\frac{\alpha_s C_A}{\pi}\ln \Bigg[\frac{2\mu^2\bar{n}\cdot n}{Q^2 z_{sj}^2 n\cdot\sja\,\sja\cdot\bar{n}}\Bigg]-\frac{\alpha_s}{2\pi}\beta_0\,.
\end{equation}

The jet, boundary soft, and global soft functions satisfy multiplicative renormalization group equations in Laplace space, where the Laplace conjugate variable to $\ecf{3}{\alpha}$ will be denoted $\eeclp{3}{\alpha}$

The jet function for the soft subjet satisfies the RGE
\begin{align}
\mu\frac{d}{d\mu}\ln J_{\sja }\Big(\eeclp{3}{\alpha}\Big) &=-4 \frac{\alpha_s C_A}{2\pi (1-\alpha)} \log\left[  2^{-\alpha/2} \eeclp{3}{\alpha}   e^{\gamma_E}   z_{sj}^2 \left( \frac{\mu}{z_{sj} Q} \right)^\alpha N_{SJ}  \right]+\frac{\alpha_s}{2\pi}\beta_0\,,
\end{align}
where the normalization factor $N_{SJ}$ was defined in  \Eq{eq:norm_factors}. We have assumed that the soft subjet is a gluon jet, as it is this case that exhibits the soft singularity of QCD.

The jet function for the hard subjet, which we have assumed to be a quark jet, satisfies the RGE
\begin{align}
\mu\frac{d}{d\mu}\ln J_{hj }\Big(\eeclp{3}{\alpha}\Big) &=-4 \frac{\alpha_s C_F}{2\pi (1-\alpha)} \log\left[  2^{-\alpha/2} \eeclp{3}{\alpha}   e^{\gamma_E}   \left( \frac{\mu}{ Q} \right)^\alpha N_{HJ}  \right]+\frac{3\alpha_s C_F}{2\pi}\,.
\end{align}
where the normalization factor $N_{HJ}$ was defined in  \Eq{eq:norm_factors}.

Since the soft subjet factorization theorem is sensitive to the boundary of the jet, it is also necessary to include out-of-jet contributions. We assume that nothing is measured on the recoiling jet. The out-of-jet jet function is then given by the unmeasured jet function of \Ref{Ellis:2010rwa}
\begin{align}
\mu\frac{d}{d\mu}\ln J_{oj }(R_B) &= \frac{2\alpha_s C_F}{\pi} \log\left[    \frac{\mu}{ Q \tan \frac{ R_B}{2} }   \right]+\frac{3\alpha_s C_F}{2\pi}\,,
\end{align}
where here $R_B$ is the radius of the recoiling jet. For simplicity, throughout this chapter, we have taken $R_B=R$.

The boundary soft function, satisfies the RGE
\begin{align}
\mu\frac{d}{d\mu}\ln  S_{\sja \,\sjabar }\Big(\eeclp{3}{\alpha};R\Big)&=\frac{\alpha_s C_A}{\pi (1-\alpha)} \log \left[  \frac{\mu}{Q}   \eeclp{3}{\alpha} e^{\gamma_E} 2^{1-\alpha}N_{BS}     \left(  \frac{\bar n \cdot \sja}{n \cdot \sja}   \tan^4 \frac{R}{2}    \right)^{\frac{-(1-\alpha)}{2}}   \right. \nonumber\\
&\hspace{4cm}    \left.     \left(   1-   \frac{n \cdot \sja}{  \bar n \cdot \sja  \tan^2 \frac{R}{2}  }     \right)^{-(1-\alpha)}      \right]\,.
\end{align}
where the normalization factor $N_{BS}$ was defined in  \Eq{eq:norm_factors}.

For the soft function, it is necessary to perform a refactorization into in-jet and out-of-jet contributions along the lines of \Ref{Ellis:2010rwa}. This is particularly important in the present case, since as was discussed in detail in \Ref{Larkoski:2015zka}, the out-of-jet contribution to the soft function is sensitive to the large logarithm, $\log\left[  \tan^2 \frac{R}{2} -   \tan^2 \frac{\theta_{sj} }{2}   \right]$, but due to zero bin subtractions, the in-jet contribution to the soft function does not exhibit such a sensitivity.

The in-jet anomalous dimension has both $C_A$ and $C_F$ contributions. It is given by
\begin{align}
\gamma_{GS}^{(\text{in})}=& -\left(   \frac{C_A}{2}-C_F  \right)       \left(  \frac{2 \alpha_s}{\pi (1-\alpha)}  \log[T]  -\frac{2\alpha_s}{\pi}  \log\left[       \frac{2\tan \frac{R}{2}}{  \tan \frac{\theta_{sj} }{2}  } \right]           \right) \nonumber\\
&-\left(   \frac{-C_A}{2} \right)       \left(  \frac{3 \alpha_s}{\pi (1-\alpha)}  \log[T]  +\frac{2\alpha_s}{\pi}  \log\left[    \frac{1}{2\bar n \cdot \sja}   \frac{\tan \frac{R}{2}}{  \tan \frac{\theta_{sj} }{2}  } \right]     \right)\nonumber \\
&-\left(   \frac{-C_A}{2} \right)       \left(  \frac{ \alpha_s}{\pi (1-\alpha)}  \log[T]  -\frac{2\alpha_s}{\pi}  \log\left[       \bar n \cdot \sja \right]   \right) \\
&=(C_A+2C_F)\frac{\alpha_s}{\pi (1-\alpha)} \log[T]+\frac{C_A \alpha_s}{\pi} \log \left[    \frac{   \tan^2 \frac{R}{2}     }{  (\bar n \cdot \sja)^2 \tan^2 \frac{\theta_{sj} }{2}        }     \right] \nonumber\\
&\hspace{4cm}- \frac{2C_F \alpha_s}{\pi}   \log \left[  \frac{2\tan\frac{R}{2}}{   \tan \frac{\theta_{sj} }{2}  }   \right]\,,
\end{align}
where in the first equality we have separated the contributions from a gluon between the three different Wilson lines, and to simplify the expression we have extracted the argument of the logs
\begin{align}
T&= e^{\gamma_E} N_S\frac{ \eeclp{3}{\alpha}\, \mu }{Q \,\text{tan}^{1-\alpha }\frac{\sjtheta}{2}}\Big(\frac{n\cdot \sja}{2}\Big)^{\alpha/2}\,.
\end{align}
We choose the canonical scale for the in-jet soft function by minimizing the arguments of the $C_A$ log. Namely, we rewrite the anomalous dimension as
\begin{align}
\gamma_{GS}^{(\text{in})}=&  \frac{   (C_A +2C_F) \alpha_s    }{\pi (1-\alpha)   }      \log \left[   T  \left( \frac{    \tan \frac{R}{2}    }{      (\bar n \cdot \sja) \tan \frac{\theta_{sj} }{2}    }  \right)^{2(1-\alpha)} \right]    \nonumber \\
&\hspace{2cm} + \frac{  2C_F \alpha_s     }{ \pi(1-\alpha)        }     \log \left[   \left(  \frac{    \tan \frac{\theta_{sj} }{2}   } { 2 \tan \frac{R}{2}    }   \right)^{1-\alpha}       \left(  \frac{  \tan \frac{R}{2}    } {  (\bar n \cdot \sja) \tan \frac{\theta_{sj} }{2}    }   \right)^{-2(1-\alpha)}   \right]\,.
\end{align}
The argument of the second logarithm is formally an $\mathcal{O}(1)$ number in the soft subjet region of phase space, and is treated as the non-cusp anomalous dimension. The argument of the first logarithm is used to set the scale.

The out-of-jet anomalous dimension is purely non-cusp, and is given by
\begin{align}
\gamma_{GS}^{(\text{out})}=&  -\left(   \frac{C_A}{2}-C_F  \right)\frac{2\alpha_s}{\pi} \log  \left[  \tan \frac{R}{2} \tan \frac{R_B }{2}         \right]    \\
&\hspace{0cm} -\frac{\alpha_s C_A}{2\pi}  \log \left[ \frac{ \tan^2 \frac{R}{2}        }{  \tan^2 \frac{R}{2}  -    \tan^2 \frac{\theta_{sj} }{2}      }    \right]         -\frac{\alpha_s C_A}{2\pi}  \log \left[   \frac{ 1            }{    \tan^2 \frac{R_B}{2}    \left(   \tan^2 \frac{R}{2} -   \tan^2 \frac{\theta_{sj} }{2}     \right)           }     \right]\,.   \nonumber
\end{align}
The natural scale for the out-of-jet soft function is
\begin{align}
\mu_{\text{out}}=\frac{2 n\cdot \sja B}{\tan \frac{\theta_{sj} }{2}}\,,
\end{align}
where $B$ is the out-of-jet scale. We set $B=Q\left( \ecf{2}{\alpha}\right)^2$ as discussed in \Sec{sec:soft_jet}.

For consistency of our soft subjet factorization theorem, the sum of the anomalous dimensions listed above should cancel.  Indeed, one can explicitly check that the anomalous dimensions satisfy the consistency condition
\begin{align}
&\mu \frac{d}{d\mu}\ln H(Q^2,\mu)+\mu\frac{d}{d\mu}\ln H^{sj}_{n\bar{n}}(\sje,\sja,\mu)+\mu\frac{d}{d\mu}\ln J_{\sja }\Big(\eeclp{3}{\alpha}\Big)  +    \mu\frac{d}{d\mu}\ln J_{hj }\Big(\eeclp{3}{\alpha}\Big)    \nonumber \\
&+\mu\frac{d}{d\mu}\ln J_{oj }(R_B)    +\mu\frac{d}{d\mu}\ln  S_{\sja \,\sjabar }\Big(\eeclp{3}{\alpha};R\Big)+\mu\frac{d}{d\mu}\ln  S_{\sja \,n\,\bar{n}}\Big(\eeclp{3}{\alpha},B;R, R_B\Big) = 0 \,.
\end{align}
This cancellation is highly non-trivial, involving intricate cancellations between a large number of scales, providing support for the structure of our factorization theorem at the one-loop level. Some further details on the structure of the cancellations, particularly on the dependence of the angle between the soft subjet axis and the boundary, are discussed in \Ref{Larkoski:2015zka}.


\section{Soft Subjet Collinear Zero Bin}\label{sec:soft_subjet_cbin}

In this appendix we summarize the one-loop anomalous dimensions, and required tree level matrix elements for the calculation of the collinear zero bin of the soft subjet factorization theorem, which is required to interpolate between the collinear subjets and soft subjets factorization theorem. Although all the ingredients in this appendix can be obtained straightforwardly from \App{app:Anom_Dim} using the standard zero bin procedure \cite{Manohar:2006nz}, we explicitly summarize the results here for completeness. 

To perform the zero-bin, all anomalous dimensions and matrix elements of the soft subjet factorization theorem are written in terms of $\ecf{2}{\alpha}$ and $z_{sj}$, and then the limit
\begin{align}
\frac{\ecf{2}{\alpha}}{z_{sj}} \to 0
\end{align}
is taken. We will therefore write the anomalous dimensions and matrix elements in this section in terms of $\ecf{2}{\alpha}$, $z_{sj}$, and $\eeclp{3}{\alpha}$. To keep the notation as simple as possible, we will use only a tilde to denote a collinear zero binned matrix element or anomalous dimension, e.g. $\gamma_{GS}^{(\text{in})} \to \tilde \gamma_{GS}^{(\text{in})}$.

\subsection*{Hard Matching for Soft Subjet Production}

The collinear binned hard matching coefficient for soft subjet production is given at tree level by
\begin{align}
\tilde H^{sj(\text{tree})}_{n\bar{n}}(\sje, \ecf{2}{\alpha})&=    \frac{\alpha_s   C_F}{\pi} \frac{2}{\alpha}       \frac{1   }{  \sje \ecf{2}{\alpha}   }\,.
\end{align}

\subsection*{Anomalous Dimensions}
Since the renormalization group evolution of all functions in the zero bin is identical to in the soft subjet factorization theorem, here we simply list the results for the zero binned one-loop anomalous dimensions:
\begin{align}
 \gamma_{ \tilde H}     &=\frac{\alpha_s C_F}{2\pi}\left(  4\log\left[  \frac{Q^2}{\mu^2} \right] -6  \right )   \,, \\
 \gamma_{\tilde H^{sj}_{n\bar{n}}   } &= -\frac{  2C_A \alpha_s  }{   \pi } \log \left[ \frac{   2\mu      } {       Q      z_{sj}^{1-1/\alpha}  \left(      \ecf{2}{\alpha}  \right)^{1/\alpha}    }     \right]   -\frac{\alpha_s}{2\pi} \beta_0   \,, \\
 \gamma_{\tilde J_{oj}}&=   \frac{2\alpha_s C_F}{\pi} \log\left[    \frac{\mu}{ Q \tan \frac{ R_B}{2} }   \right]+\frac{3\alpha_s C_F}{2\pi}  \,, \\
\tilde \gamma_{GS}^{(\text{out})}&=  \frac{C_A \alpha_s}{2 \pi}  \log\left[  \tan^2 \frac{R}{2} \tan^2 \frac{R_B}{2}  \right]  -\frac{\left(   \frac{C_A}{2}-C_F \right)   \alpha_s}{\pi} \log\left[  \tan^2\frac{R}{2} \tan^2 \frac{R_B}{2}  \right]  \,, \\
 \gamma_{\tilde J_{hj}}&=  -4 \frac{\alpha_s C_F}{2\pi (1-\alpha)} \log\left[  2^{-\alpha/2} \eeclp{3}{\alpha}   e^{\gamma_E}   \left( \frac{\mu}{ Q} \right)^\alpha N_{HJ}  \right]+\frac{3\alpha_s C_F}{2\pi} \,, \\
 \gamma_{\tilde J_{sj}}&= -4 \frac{\alpha_s C_A}{2\pi (1-\alpha)} \log\left[  2^{-\alpha/2} \eeclp{3}{\alpha}   e^{\gamma_E}   z_{sj}^2 \left( \frac{\mu}{z_{sj} Q} \right)^\alpha N_{SJ}  \right]+\frac{\alpha_s}{2\pi}\beta_0  \,, \\
 \tilde \gamma_{GS}^{(\text{in})}&= \frac{C_F \alpha_s}{\pi (1-\alpha)} \log \left[ 2^{1-\alpha} \tan^{-3(1-\alpha)} \frac{R}{2} \left(  \frac{2^{-\alpha } \ecf{2}{\alpha}  }{z_{sj}}  \right)^{-3+3/\alpha}  \right]  \nonumber \\
 & +     \frac{(C_A+2C_F) \alpha_s}{\pi (1-\alpha)} \log \left[  \frac{2^{-1+4\alpha}   \mu \eeclp{3}{\alpha}   e^{\gamma_E}  z_{sj}   }{  Q }  \tan^{2(1-\alpha)} \frac{R}{2}   \left(  \frac{2^{-\alpha } \ecf{2}{\alpha}  }{z_{sj}}  \right)^{5-3/\alpha}        \right]   \,, \\
 \gamma_{ \tilde S_{\sja \,\sjabar }}&=  \frac{C_A \alpha_s}{\pi(1-\alpha)}     \log  \left[ \frac{\mu \eeclp{3}{\alpha}   e^{\gamma_E} 2^{1+2\alpha}  \tan^{2(\alpha-1)} \frac{R}{2}    }{Q} \left(  \frac{2^{-\alpha } \ecf{2}{\alpha}  }{z_{sj}}  \right)^{1+1/\alpha}       \right] \,.
\end{align}

As for the soft subjet anomalous dimensions, one can check that the zero binned anomalous dimensions satisfy the consistency relation
\begin{align}
&\mu \frac{d}{d\mu}\ln \tilde H(Q^2,\mu)+\mu\frac{d}{d\mu}\ln  \tilde H^{sj}_{n\bar{n}}(\sje,\sja,\mu)+\mu\frac{d}{d\mu}\ln  \tilde J_{\sja }\Big(\eeclp{3}{\alpha}\Big)  +    \mu\frac{d}{d\mu}\ln  \tilde J_{hj }\Big(\eeclp{3}{\alpha}\Big)    \nonumber \\
&+\mu\frac{d}{d\mu}\ln  \tilde J_{oj }(R_B)    +\mu\frac{d}{d\mu}\ln   \tilde S_{\sja \,\sjabar }\Big(\eeclp{3}{\alpha};R\Big)+\mu\frac{d}{d\mu}\ln   \tilde S_{\sja \,n\,\bar{n}}\Big(\eeclp{3}{\alpha},B;R, R_B\Big) = 0 \,,
\end{align}
as required for the consistency of the factorization theorem.


\section{One Loop Calculations of Signal Factorization Theorem}\label{sec:signal_app}

In this section we give the operator definitions, and one-loop results for the functions appearing in the factorization theorem of \Eq{eq:signal_fact} for the signal contribution from $Z\to q\bar q$. These are formulated in the SCET$_+$ effective theory of \Ref{Bauer:2011uc}, in an attempt to have a consistent approach to factorization for both the signal and background distributions. In the collinear subjets region of phase space the two are identical (including identical power counting for the modes) up to the absence of global soft modes for the signal distribution. Alternatively, the factorization theorem for the signal region can be formulated by boosting the factorization theorems for appropriately chosen $e^+e^-$ event shapes, as was considered in \Ref{Feige:2012vc}. While this approach is less in the spirit of developing effective field theory descriptions for jet substructure that was pursued in this chapter, it has the potential advantage of being easily able to relate to higher order known results for event shapes.

\subsection*{Definitions of Factorized Functions}

The functions appearing in the collinear subjets factorization theorem of \Eq{eq:NINJA_fact} have the following SCET operator definitions:
\begin{itemize}
\item Hard Matching Coefficient:
\begin{align}
H_Z\left(Q^2\right)=\left |C_Z\left(Q^2\right) \right|^2\,,
\end{align}
where $C_Z\left(Q^2\right)$ is the matrix element for the process $e^+e^- \to ZZ$, and also includes the leptonic decay of one of the $Z$ bosons. Since we use the narrow width approximation, flat polarization distributions for the $Z$, and normalize our distributions to unity, it will play no role in our calculation.
\item Jet Functions:
\begin{align}
J_{n_{a,b}}\Big(\ecf{3}{\alpha}\Big)&=\\
&\hspace{-1cm}\frac{(2\pi)^3}{C_F}\text{tr}\langle 0|\frac{\bar{n}\!\!\!\slash _{a,b}}{2}\chi_{n_{a,b}}(0) \delta(Q-\bar{n}_{a,b}\cdot{\mathcal P})\delta^{(2)}(\vec{{\mathcal P}}_{\perp})\delta\Big(\ecf{3}{\alpha}-  \ecfop{3}{\alpha} \Big)\bar{\chi}_{n_{a,b}}(0)|0\rangle \nonumber
\end{align}
\item Collinear-Soft Function:
\begin{align}
\hspace{-1cm} S_{c, \,n_a n_b} \Big(\ecf{3}{\alpha}\Big)&=\text{tr}\langle 0|T\{ S_{n_a } S_{n_b} \}\delta\Big(\ecf{3}{\alpha}-\ecfop{3}{\alpha}\Big)\bar{T}\{ S_{n_a } S_{n_b} \} |0\rangle
\end{align}
\end{itemize}
As in \App{sec:ninja_app} and \App{sec:softjet_app}, the operator, $\ecfop{3}{\alpha}$, measures the contribution to $\ecf{3}{\alpha}$ from final states, and must be appropriately expanded following the power counting of the sector on which it acts. Since the power counting is identical as for the collinear subjets factorization theorem, the expansions are given in \Eq{eq:collinear_limits_eec3}, and \Eq{eq:collinear_soft_limits_eec3}. In the collinear subjets region that we consider for the signal, all modes are boosted, and so there is no dependence on the jet algorithm at leading power.

\subsection*{Hard Matching Coefficient}

The hard matching coefficient for the process $e^+e^-\to ZZ$, with one $Z$ decaying leptonically, $H_Z(Q^2)$, does not carry an SCET anomalous dimension (hence we have dropped the $\mu$ dependence), as it is colorless. Because we work in the narrow width approximation, at a fixed $Q^2$, and consider only normalized distributions, it is therefore irrelevant to our discussion.

\subsection*{Matrix Element for $Z \to q\bar q$ Decay}
The anomalous dimension for the $Z\to q\bar q$ splitting function appearing in the factorization theorem of \Eq{eq:NINJA_fact} is the same as that for the SCET quark bilinear operator, which was given in \Eq{eq:hard_anom_dim}, but evaluated at the appropriately boosted scale. 

For simplicity, in this chapter we do not account for spin correlations, and assume a flat profile in the polarization of the $Z$ boson. The tree level $Z\to q\bar q$ matrix element is well known and first calculated in \Ref{Altarelli:1979ub}. The full matrix element is known to two loops \cite{Matsuura:1988sm}.

The anomalous dimension depends only on the color structure, and is therefore the same as the anomalous dimension for the hard matrix element for $e^+e^-\to q\bar q$, namely
\begin{align}
\gamma_{H_Z}=1+\frac{\alpha_s C_F}{2\pi} \left (  -8 +\frac{\pi^2}{6} -\log^2 \left[  \frac{\mu_H^2}{\mu^2}  \right]     +3 \log \left[  \frac{\mu_H^2}{\mu^2}  \right]  \right)\,.
\end{align}
Here $\mu_H$ is the scale of the splitting. It is essential for the cancellation of anomalous dimensions that the scale $\mu_H$ is equal to the invariant mass of the jet. In terms of the energy correlation functions, this is given by
\begin{align}
m_J^2&=\frac{Q^2 \left[ z (1-z) \right]^{1-2/\alpha}  \left(   \ecf{2}{\alpha} \right)^{2/\alpha}    }{4} \nonumber \\
&=\frac{Q^2 z (1-z) n_a \cdot n_b   }{2}\,.
\end{align}
The necessity for the appearance of the jet mass as the scale in the anomalous dimension is due to the fact that it is a Lorentz invariant quantity, and as has been discussed in \Ref{Feige:2012vc}, the factorization theorem for the case of the boosted boson can be obtained by boosting an $e^+e^-$ event shape, where it is of course known that the scale $Q^2$ of the off-shell $Z$, or $\gamma$ is the scale appearing in the hard anomalous dimension.

\subsection*{Jet Functions}
The jet functions appearing in the signal factorization theorem are identical to the quark (and antiquark) jet functions calculated in \App{sec:NINJA_jet_calc} for the collinear subjets region of phase space.  This is because the power counting is identical in the two cases and the jet functions are only sensitive to the color of the jet that they describe. Therefore we do not repeat them here.

\subsection*{Collinear-Soft Function}
The power counting for the signal is identical to the power counting for the collinear subjet region for the QCD background. However, the collinear-soft function contains only Wilson lines along the collinear subjet directions. The collinear-soft function for the QCD background was calculated in pairs of dipoles in \App{sec:csoft_calc}, and therefore the contribution from a collinear-soft exchange between the $n_a$ and $n_b$ Wilson lines can simply be extracted from that calculation. The result for this contribution is given by
 \begin{align}
S_{c,\,n_a n_b}^{(1)}(\ecflp{3}{\alpha})&=  -g^2 \Gamma(-2\epsilon)     \left(    \frac{  \mu^2 N_{CS}^2 e^{\gamma_E}  (\ecflp{3}{\alpha})^2      \left( \frac{n_a \cdot n_b}{8}   \right)^{-1+2\alpha}  }{4\pi 4^{1-\alpha}  Q^2    }  \right)^{\epsilon}   16 c_\epsilon     \frac{\Gamma[1/2-\epsilon] \Gamma[1/2]}{\Gamma[1-\epsilon]} \nonumber  \\
& \hspace{-1cm}   \left(  \frac{1}{(2 \alpha -2) \epsilon }+\frac{\alpha 
   \log (2)}{\alpha -1}+\log (2)+\frac{\epsilon  \left(-\pi ^2 \alpha ^2+36 \alpha ^2 \log ^2(2)+3 \pi ^2 \alpha -24 \alpha 
   \log ^2(2)-2 \pi ^2\right)}{12 (\alpha -1)}      \right)     \,,
\end{align}
where we recall that the normalization factor is given by
 \begin{align}
N_{CS}&=2^{3\alpha/2+1}z_az_b\left(n_a\cdot n_b\right)^{\alpha/2}\,,
\end{align}
as defined in \Eq{eq:ncs}. Also note that we have factored out the color generators, so that the collinear-soft function is defined as
\begin{align}
S^{(1)}_{c}\left(\ecf{3}{\alpha}\right)&=\frac{1}{2}\sum_{i\neq j}\mathbf{T}_i\cdot\mathbf{T}_j S_{c,\,ij}^{(1)} \left(\ecf{3}{\alpha}\right)\,,
\end{align}
which is the generic form of the collinear-soft (or soft) function to one-loop.

Expanding in $\epsilon$, and keeping only the divergent piece, as relevant for the anomalous dimensions, we find 
 \begin{align} 
\tilde{S}_{c,\,n_a n_b}^{(1)\text{div}}(\ecflp{3}{\alpha})&= \frac{\alpha_s}{\pi}\frac{1}{  (\alpha -1) \epsilon ^2}+2\frac{\alpha_s}{\pi}\frac{ \left(2 \alpha  \log (2)+ \log \left[   \frac{  \mu N_{CS} e^{\gamma_E}  (\ecflp{3}{\alpha})      \left( \frac{n_a \cdot n_b}{8}   \right)^{-1/2+\alpha}  }{2^{1-\alpha}  Q    }  \right]- \log (2)\right)}{   (\alpha -1) \epsilon }\nonumber \\
&=\frac{\alpha_s}{\pi}\frac{1}{  (\alpha -1) \epsilon ^2}+2\frac{\alpha_s}{\pi}\frac{L_\alpha^{cs}}{   (\alpha -1) \epsilon }\,,
\end{align}
where
\begin{align}
L_\alpha^{cs}= \log \left(    \frac{  \mu N_{CS} e^{\gamma_E}  (\ecflp{3}{\alpha})      \left( n_a \cdot n_b   \right)^{-1/2+\alpha}  }{\sqrt{2}  Q    }  \right)\,.
\end{align}
Since there is no global-soft function the cancellation of anomalous dimensions, to be discussed shortly, requires that only $1/(1-\alpha)$ contributions appear in the collinear soft function, as is observed.

\subsection*{Cancellation of Anomalous Dimensions}\label{sec:signal_cancel_anom}

It is also interesting to explicitly check the cancellation of anomalous dimensions for the signal factorization theorem as formulated in SCET$_+$ to further confirm the cancellation mechanism which took place for the background distribution. The functions appearing in the signal factorization theorem obey identical evolution equations to those for the background distribution, which were explicitly given in \App{sec:bkg_cancel_anomdim}, so we do not repeat them here.

The independence of the total cross section under renormalization group evolution implies the following relation between anomalous dimensions
 \begin{align} \label{eq:consistency_signal}
\gamma_{H_Z}  +\gamma_{q}^{\alpha}\left(\ecflp{3}{\alpha}\right)+\gamma_{\bar q}^{\alpha}\left(\ecflp{3}{\alpha}\right)+\gamma_{cs}\left(\ecflp{3}{\alpha}\right)= 0\,.
\end{align}
Here $\gamma_{H_Z} $ is the anomalous dimension of the $Z\to q\bar q$ matrix element,  $\gamma_{q}^{\alpha}\left(\ecflp{3}{\alpha}\right)$ and $\gamma_{\bar q}^{\alpha}\left(\ecflp{3}{\alpha}\right)$ are the anomalous dimensions of the quark and antiquark jet functions and  $\gamma_{cs}\left(\ecflp{3}{\alpha}\right)$ is the anomalous dimension of the collinear soft function.

For the case of $Z\to q \bar q$, we have the color conservation relation
\begin{align}\label{eq:color_cons_signal}
{\mathbf T}_q+{\mathbf T}_{\bar q}&=0\,.
\end{align}
The explicit values of the relevant Casimirs are
\begin{align}
{\mathbf T}_q\cdot{\mathbf T}_{q}=C_F\,, \qquad {\mathbf T}_{\bar q}\cdot{\mathbf T}_{\bar q}=C_F\,, \qquad {\mathbf T}_q\cdot{\mathbf T}_{\bar q}=-C_F\,,
\end{align}
however, for most of the cancellation of the anomalous dimensions, it will be convenient to work in the abstract color notation.

Substituting the explicit expressions for the anomalous dimensions into the consistency relation of \Eq{eq:consistency_signal}, we find
\begin{align}\label{eq:casimirs_signal}
&\sum_\gamma =\gamma_{H_Z} +  \left[   -    {\mathbf T}_q \cdot{\mathbf T}_{\bar q}  \frac{4\alpha_s L_\alpha^{cs}\left(\ecflp{3}{\alpha}\right)}{\pi (1-\alpha)}     \right] \nonumber \\
&  \hspace{4cm}     -   C_F   \frac{2\alpha_s L_\alpha^{g}\left(  \ecflp{3}{\alpha}  \right)}{\pi (1-\alpha)} +\gamma_q        - C_F   \frac{2\alpha_s L_\alpha^{q}\left(  \ecflp{3}{\alpha}  \right)}{\pi (1-\alpha)} +\gamma_q \,,
\end{align}
where $L_\alpha^{g}\left(  \ecflp{3}{\alpha}  \right)$ and $L_\alpha^{q}\left(  \ecflp{3}{\alpha}  \right)$, were defined in \Eq{eq:jet_anoms}.

As expected, all contributions are collinear in nature, having a $1/(1-\alpha)$ dependence, and using the color conservation relation of \Eq{eq:color_cons_signal} along with the explicit expressions for the Casimirs of \Eq{eq:casimirs_signal}, we immediately see the cancellation of the $\ecflp{3}{\alpha}$ dependence. It is also straightforward to check the cancellation of the remaining dependencies. It is a nice consistency check on the calculation that the cancellation occurs in exactly the same way as for the background cancellation, namely between the $ {\mathbf T}_q \cdot{\mathbf T}_{\bar q}$ contribution and the jet functions. It is important to emphasize that the cancellation only occurs if the scale of the splitting is given by the invariant mass of the jet, as expected from boosting $e^+e^-$ event shapes.

\section{Soft Haze Factorization Theorem}\label{app:softhaze}

For completeness, we list the operator definitions of the functions appearing in the soft haze factorization theorems. We also give the explicit forms of the measurement operators expanded in the appropriate kinematics. 

The quark jet functions are given as:
\begin{align}
J_{n}\Big(\ecf{2}{\alpha}\Big)&= \nonumber\\
&\hspace{-0.5cm}\frac{(2\pi)^3}{C_F}\text{tr}\langle 0|\frac{\bar{n}\!\!\!\slash _{a,b}}{2}\chi_{n_{a,b}}(0) \delta(Q-\bar{n}_{a,b}\cdot{\mathcal P})\delta^{(2)}(\vec{{\mathcal P}}_{\perp})\delta\Big(\ecf{2}{\alpha}-\ecfop{2}{\alpha}\Big)\bar{\chi}_{n_{a,b}}(0)|0\rangle\,.
\end{align}
The gluon jet functions are similarly defined. The soft functions appearing in the factorization theorems \eqref{eq:fact_soft_haze} and \eqref{eq:fact_soft_haze2} are: 
\begin{align}\label{eq:def_soft_function_SH_1}
S_{n \, \bar n }\Big(\ecf{2}{\beta},\ecf{2}{\alpha},\ecf{3}{\alpha};R\Big)&=\frac{1}{C_{A}}\text{tr}\langle 0|T\{S_{n } S_{\bar n }\}  \delta\Big(\ecf{2}{\beta}-\Theta_{R}\ecfop{2}{\beta}\Big)\delta\Big(\ecf{2}{\alpha}-\Theta_{R}\ecfop{2}{\alpha}\Big)\nonumber\\
&\qquad\qquad\delta\Big(\ecf{3}{\alpha}-\Theta_{R}\ecfop{3}{\alpha}\Big)\bar{T}\{S_{n } S_{\bar n }\} |0\rangle\,,\\
\label{eq:def_soft_function_SH_2}
S_{n \, \bar n }\Big(\ecf{2}{\alpha},\ecf{3}{\alpha};R\Big)&=\frac{1}{C_{A}}\text{tr}\langle 0|T\{S_{n } S_{\bar n }\}  \delta\Big(\ecf{2}{\alpha}-\Theta_{R}\ecfop{2}{\alpha}\Big)\delta\Big(\ecf{3}{\alpha}-\Theta_{R}\ecfop{3}{\alpha}\Big)\bar{T}\{S_{n } S_{\bar n }\} |0\rangle\,.
\end{align}

The action of the energy correlation functions on the collinear and soft haze states are given as:
\begin{align}
\ecfop{2}{\alpha}\big|_{C}\Big|X_{n}\Big\rangle&=\sum_{k,p\in X_{n}}\frac{\nbar\cdot k}{Q}\frac{\nbar\cdot p}{Q}\left(\frac{8\,p\cdot k}{\nbar\cdot p\,\nbar\cdot k}\right)^{\frac{\alpha}{2}}\Big|X_{n}\Big\rangle\,,\\
\ecfop{2}{\alpha}\big|_{SH}\Big|X_{s}\Big\rangle&=\sum_{k\in X_{s}}2\frac{k^0}{Q}\left(\frac{2 n\cdot k}{k^0}\right)^{\frac{\alpha}{2}}\Big|X_{s}\Big\rangle\,,\\
\ecfop{3}{\alpha}\big|_{SH}\Big|X_{s}\Big\rangle&=\sum_{k,p\in X_{s}}4\frac{k^0}{Q}\frac{p^0}{Q}\left(\frac{2 n\cdot k}{k^0}\frac{2 n\cdot p}{p^0}\frac{2 p\cdot k}{p^0k^0}\right)^{\frac{\alpha}{2}}\Big|X_{s}\Big\rangle\,.
\end{align}

\section{Summary of Canonical Scales}\label{sec:canonical_merging_scales}

As many of our factorization theorems involve a large number of scales, in this section we summarize for convenience the scales used in the resummation. Unless otherwise indicated, all scales are taken to be the canonical scales of the logarithms appearing in the factorization theorems.

When performing the numerical resummation, we perform the renormalization group evolution in Laplace space, and compute the cumulative distribution. We then perform the scale setting at the level of the cumulative distribution and numerically differentiate to derive the differential $D_2$ spectrum. While this is formally equivalent to scale setting in the differential distribution when working to all orders in perturbation theory, differences between scale setting in the differential and cumulative distribution arise when working to fixed order in perturbation theory \cite{Almeida:2014uva}. We have not investigated the size of the effect that this has on our $D_2$ distributions. We utilized only two loop running of $\alpha_s$, to be consistent with the Monte Carlos, and avoided the Landau pole by freezing out the running coupling at a specific $\mu_{Landau}\sim 1$ GeV.

Throughout this appendix we will use $z_q$ and $z_g$ to denote the energy fractions of the quark and gluon subjets, respectively. For simplicity, we restrict to the case $\alpha=\beta$. Finally, we estimate the soft out-of-jet radiation scale to be:
\begin{align}
B\approx Q\Big(\ecf{2}{\alpha}\Big)^2
\end{align}
This is consistent with the jet algorithm constraint given by \Eq{eq:out_jet_estimate}.

\subsection*{Collinear Subjets}

We take the canonical scales for the functions appearing in the collinear subjets factorization theorem as
\begin{align}
\mu_H&=Q\,, \\
\mu_{H_2}&=\frac{Q \left( \ecf{2}{\alpha} \right)^{1/\alpha}     z_q^{\frac{1}{2}-  \frac{1}{\alpha}}    z_g^{\frac{1}{2}-  \frac{1}{\alpha}}    }{2}\,,\\
\mu_{J_g}&=\frac{ e^{-\gamma_E/\alpha} Q \left( \ecf{2}{\alpha} \right)^{-2/\alpha}    \left( \ecflp{3}{\alpha} \right)^{1/\alpha}           z_q^{ \frac{1}{\alpha}}    z_g   }{2}\,,\\
\mu_{J_q}&=\frac{ e^{-\gamma_E/\alpha} Q \left( \ecf{2}{\alpha} \right)^{-2/\alpha}    \left( \ecflp{3}{\alpha} \right)^{1/\alpha}           z_q    z_g^{ \frac{1}{\alpha}}   }{2}\,,\\
\mu_{CS}&=\frac{ e^{-\gamma_E} Q \left( \ecf{2}{\alpha} \right)^{-3+1/\alpha}     \ecflp{3}{\alpha}           z_q^{ 2-\frac{1}{\alpha}}     z_g^{2- \frac{1}{\alpha}}   }{2} \,,\\
\mu_{S}^{(\text{in})}&= \frac{ 4^{-\alpha} e^{-\gamma_E} Q \left( \ecf{2}{\alpha} \right)^{-1}     \ecflp{3}{\alpha}       \tan^2\frac{R}{2}       }{2} \,, \\
\mu_{S}^{(\text{out})}&=B
\end{align}
where the scales are indexed by the name of the associated function in the factorization theorem.

\subsection*{Soft Subjets}
We take the canonical scales for the functions appearing in the soft subjets factorization theorem as
\begin{align}
\mu_H&=Q\,, \\
\mu_{H_{sj}}&=\frac{   Q  \left(   \ecf{2}{\alpha} \right)^{1/\alpha}  z_{sj}^{(\alpha-1)/\alpha}   \sqrt{4- \left(\ecf{2}{\alpha} \right)^{2/\alpha}  z_{sj}^{-2/\alpha}  }      }{         4           }\,, \\
\mu_{S_{n_{sj} \bar n_{sj}  }}&=   2^{-\alpha}  e^{-\gamma_E}  Q   \tan^{2(1-\alpha)}\frac{R}{2}    \left(\ecf{2}{\alpha}     \right)^{-(1+\alpha)/\alpha}   \eeclp{3}{\alpha}   \left(  z_{sj}  \right)^{1/\alpha}  \nonumber \\
&  \hspace{0.5cm}     \left( 1-\frac{1}{4}  \left(\ecf{2}{\alpha}     \right)^{2/\alpha}  z_{sj}^{-2/\alpha}    \right)^{(1-\alpha)/2} 
 \left(  \frac{ \left( 1+\tan^{2}\frac{R}{2} \right) \left(   \ecf{2}{\alpha} \right)^{2/\alpha}   -4 \tan^{2}\frac{R}{2} z_{sj}^{2/\alpha}    }{  \tan^{2}\frac{R}{2}   \left(  \left(   \ecf{2}{\alpha} \right)^{2/\alpha}   -4z_{sj}^{2/\alpha}   \right)       }   \right)^{1-\alpha}  \,, \\
\mu^{(\text{in})}_{S_{n_{sj} n \bar n  }}&= 2^{-2+\alpha}  e^{-\gamma_E}  Q   \tan^{2(\alpha-1)}\frac{R}{2}    \left(\ecf{2}{\alpha}     \right)^{-5+3/\alpha}   \eeclp{3}{\alpha}   \left(  z_{sj}  \right)^{4-3/\alpha}  \nonumber \\
&\hspace{4cm}     \left( 1-\frac{1}{4}  \left(\ecf{2}{\alpha}     \right)^{2/\alpha}  z_{sj}^{-2/\alpha}    \right)^{(1-\alpha)/2}   \,, \\
\mu^{(\text{out})}_{S_{n_{sj} n \bar n  }}&= \frac{2 n\cdot \sja B}{\tan \frac{\theta_{sj} }{2}}   \,, \\
\mu_{J_{hj}}&=   \frac{ Q e^{-\gamma_E/\alpha}    \left(   \ecf{2}{\alpha} \right)^{-2/\alpha}     \left( \eeclp{3}{\alpha}   \right)^{1/\alpha}   z_{sj}^{1/\alpha}  }{   2     }    \,, \\
\mu_{J_{\bar n}}&= Q \tan \frac{ R_B}{2}   \,, \\
\mu_{J_{n_{sj}}}&=\frac{ Q e^{-\gamma_E/\alpha}    \left(   \ecf{2}{\alpha} \right)^{-2/\alpha}     \left( \eeclp{3}{\alpha}   \right)^{1/\alpha}   z_{sj}  }{   2     }  \,.
\end{align}

\subsection*{Soft Subjet Collinear Zero Bin}

We take the canonical scales for the functions appearing in the collinear zero bin of the soft subjets factorization theorem as
\begin{align}
\mu_{\tilde H}&=Q\,, \\
\mu_{\tilde H_{sj}}&=   \frac{ Q  \left(  \ecf{2}{\alpha} \right)^{1/\alpha}  z_{sj}^{(\alpha-1)/\alpha} }{2}\,, \\
\mu_{\tilde S_{n_{sj} \bar n_{sj}  }}&=  \frac{   2^{-\alpha}Q e^{-\gamma_E} \tan^{2(1-\alpha)} \frac{R}{2} \eeclp{3}{\alpha} z_{sj}^{1/\alpha}       }{    \left(  \ecf{2}{\alpha} \right)^{(1+\alpha)/\alpha}         }   \,, \\
\mu^{(\text{in})}_{S_{n_{sj} n \bar n  }}&=    \frac{   2^{-2+\alpha} Q e^{-\gamma_E} \tan^{2(\alpha-1)} \frac{R}{2} \eeclp{3}{\alpha} z_{sj}^{4-3/\alpha}       }{    \left(  \ecf{2}{\alpha} \right)^{5-3/\alpha}         }   \,, \\
\mu^{(\text{out})}_{\tilde S_{n_{sj} n \bar n  }}&= \frac{2 n\cdot \sja B}{\tan \frac{\theta_{sj} }{2}}    \,, \\
\mu_{\tilde J_{hj}}&= \frac{ e^{-\gamma_E/\alpha}  Q \left(\eeclp{3}{\alpha} \right)^{1/\alpha}  z_{sj}^{1/\alpha}  }{  2   \left( \ecf{2}{\alpha} \right)^{2/\alpha}    } \,, \\
\mu_{\tilde J_{oj}}&= Q \tan \frac{ R_B}{2}   \,, \\
\mu_{\tilde J_{n_{sj}}}&= \frac{ e^{-\gamma_E/\alpha}  Q \left(\eeclp{3}{\alpha} \right)^{1/\alpha}  z_{sj}  }{   2    \left( \ecf{2}{\alpha} \right)^{2/\alpha}    }\,.
\end{align}

\subsection*{Scale Variation}
Here we list all the variations that went into the scale uncertainties of the QCD background calculations. Any common scale between the soft subjet factorization and its collinear bin are always varied together. Hence we will only discuss variations of the soft subjet and collinear subjets. It is important to note that $\mu^{(\text{out})}_{S_{n_{sj} n \bar n  }}$ of the soft subjet is not exactly the same as the $\mu^{(\text{out})}_{S}$ of the collinear factorization. The extra angular factor improves cancellation with the soft subjet collinear zero bin in the collinear region of the phase space. In the soft subjet region, the angular factor becomes an $O(1)$ number. Given the arbitrariness of the out-of-jet scale setting, we included several different schemes.
\begin{itemize}
\item Splitting scales $\mu_{H_2}$ and $\mu_{H_{sj}}$ from half to twice canonical.
\item $\mu_{Landau}$ where the running of the coupling is frozen from $0.5$ GeV to $1.5$ GeV, canonical is $1$ GeV.
\item All in-jet soft scales $\mu^{(\text{in})}_{S_{n_{sj} n \bar n  }}, \mu_{\tilde S_{n_{sj} \bar n_{sj}  }}, \mu_{CS}$, and $\mu_{S}$ from half to twice canonical. This included the scales in the collinear factorization and soft subjet factorization being varied together, and independently.
\item All out-of-jet soft scales $\mu^{(\text{out})}_{S_{n_{sj} n \bar n  }}, \mu_{S}^{(out)}$ from half to twice canonical. This included the scales in the collinear factorization and soft subjet factorization being varied together, and independently.
\item Soft subjet out-of-jet soft scale $\mu^{(\text{out})}_{S_{n_{sj} n \bar n  }}= Q z_{sj}^2$ from half to twice canonical. Also in this scheme the splitting scales were varied from half to twice canonical, and $\mu_{Landau}$ from $0.5$ GeV to $1.5$ GeV.
\item Soft subjet out-of-jet soft scale $\mu^{(\text{out})}_{S_{n_{sj} n \bar n  }}= \mu_{S}^{(out)}$ from half to twice canonical. Also in this scheme the splitting scales were varied from half to twice canonical, and $\mu_{Landau}$ from $0.5$ GeV to $1.5$ GeV.
\end{itemize}

The final uncertainty bands were taken as the envolope of these variations. Though these variations do not cover all perturbative functions that can be varied, we believe that they are representative of NLL uncertainties.


\section{Renormalization Group Evolution of the Shape Function}\label{app:shape_RGE}

In this appendix we briefly summarize some of the properties of the non-perturbative shape function used in the analysis of the $D_2$ observable, including hadron mass effects, so as to ensure that the level of renormalization group evolution of the parameter $\Omega_D$ is consistent with our results at both $1$ TeV and $91$ GeV, as discussed in \Secs{sec:Hadronization}{sec:LEP}, respectively. There we found that the value of $\Omega_D$ was approximately equal at the two energies, to within our uncertainties. As in the text, we assume that the dominant non-perturbative corrections arise from the global soft modes of the collinear subjets factorization theorem, so that we are working with a soft function with Wilson lines only along the $n$ and $\bar n$ directions. We follow closely the formalism originally developed in \Ref{Mateu:2012nk}.

In \Ref{Mateu:2012nk} it was shown that for dijet observables which can be written in terms of the rapidity $y$ and the transverse velocity $r$, defined as
\begin{align}
r=\frac{p_\perp}{\sqrt{p_\perp^2 +m_H^2  }}\,,
\end{align}
where $m_H$ is a light hadron mass, have a leading power correction that is universal, for event shapes with the same $r$ dependence. Furthermore, the leading power corrections can be written as an integral over an $r$ dependent power correction,
\begin{align}\label{eq:omega_int}
\Omega_D=\int\limits_0^1 dr g(r)\, \Omega_D(r),
\end{align}
where $g(r)$ is a function of $r$ which depends only on the definition of the event shape (see \Ref{Mateu:2012nk}), and $\Omega_D(r)$ exhibits a multiplicative renormalization group evolution in $r$, which is independent of $y$. In particular, for $\Omega_D$, we have
\begin{align}
\mu \frac{d}{d\mu} \Omega_D (r,\mu) =\gamma_{\Omega_D}(r,\mu)  \Omega_D (r,\mu)=\left( -\frac{\alpha_s C_A}{\pi} \log(1-r^2)\Omega_D (r,\mu) \right)\,,
\end{align}
to one loop accuracy \cite{Mateu:2012nk}. This renormalization group equation can be solved exactly for each $r$, however, the computation of $\Omega_D$ using \Eq{eq:omega_int} requires knowledge of the exact $r$ dependence of $\Omega_D (r,\mu)$ at a particular scale. However, it was shown that to order $\alpha_s$, only a single non-perturbative parameter is required to described the evolution, so that one can write
\begin{align}
\Omega_D (\mu) =\Omega_D(\mu_0)   +   \frac{\alpha_s (\mu_0) C_A}{\pi}   \log \left(    \frac{\mu}{\mu_0} \right)  \Omega_D^{\text{ln}}(\mu_0)\,,
\end{align}
where apart from the non-perturbative parameter $\Omega_D(\mu_0)$ evaluated at a particular scale, we have also had to introduce the non-perturbative parameter $\Omega_D^{\text{ln}}(\mu_0)$, which captures the logarithmic running (hence the notation).

The additional non-perturbative parameter $\Omega_D^{\text{ln}}(\mu_0)$ is not well constrained in the literature, and therefore as a simple estimate to make sure that the values used for $\Omega_D$ at both LEP energies and at $1$ TeV are consistent, we consider the estimate $\Omega_D^{\text{ln}}(\mu_0)=\Omega_D(\mu_0)$. Making this approximation, we find the difference between the values of $\Omega_D$ as relevant for LEP and our $1$ TeV analysis to differ by $\lesssim 0.1$, with the value at LEP being lower. This is small compared to our uncertainties, and compared to the scaling in the shift of the first moment with $E_J$ and $m_J$. However, it is an important check that the values of $\Omega_D$ that we use are consistent with each other in our different analyses, and could be important in analyses for which jets are probed over large energy ranges.

\section{Comparison of MC Generators for Single Emission Observables}\label{app:twoemissionMC}

Throughout this chapter, we have extensively compared different Monte Carlo generators both at parton and hadron level for the observable $D_2$, which is set by two emissions off the initiating quark. We found significant differences between different Monte Carlo generators, and as compared with our analytic calculation, particularly at parton level. After hadronization, differences remained but these were quantitative differences, not differences in the shapes of distributions. For reference, in this appendix we compare the Monte Carlo generators used in this chapter, at both parton and hadron level for an observable set by a single emission off of the initiating parton, namely the jet mass. Observables set by a single emission have been extensively studied in the literature, and are well understood. There exist automated codes for their resummation to NNLL \cite{Banfi:2004yd,Banfi:2014sua}, and they have been extensively used to tune Monte Carlo generators. We therefore expect to see much better agreement than for the $D_2$ observable, demonstrating that $D_2$ is a more differential probe of the perturbative shower structure.\footnote{Differences between Monte Carlo generators for single emission observable can also be accentuated by departing from jet mass, and considering angularities, or energy correlation functions, or differences between quark and gluon jets, for which limited data from LEP can be used for tuning \cite{Larkoski:2014pca,Sakaki:2015iya}.}

\begin{figure}
\begin{center}
\subfloat[]{\label{fig:D2_mass_a}
\includegraphics[width= 7.2cm]{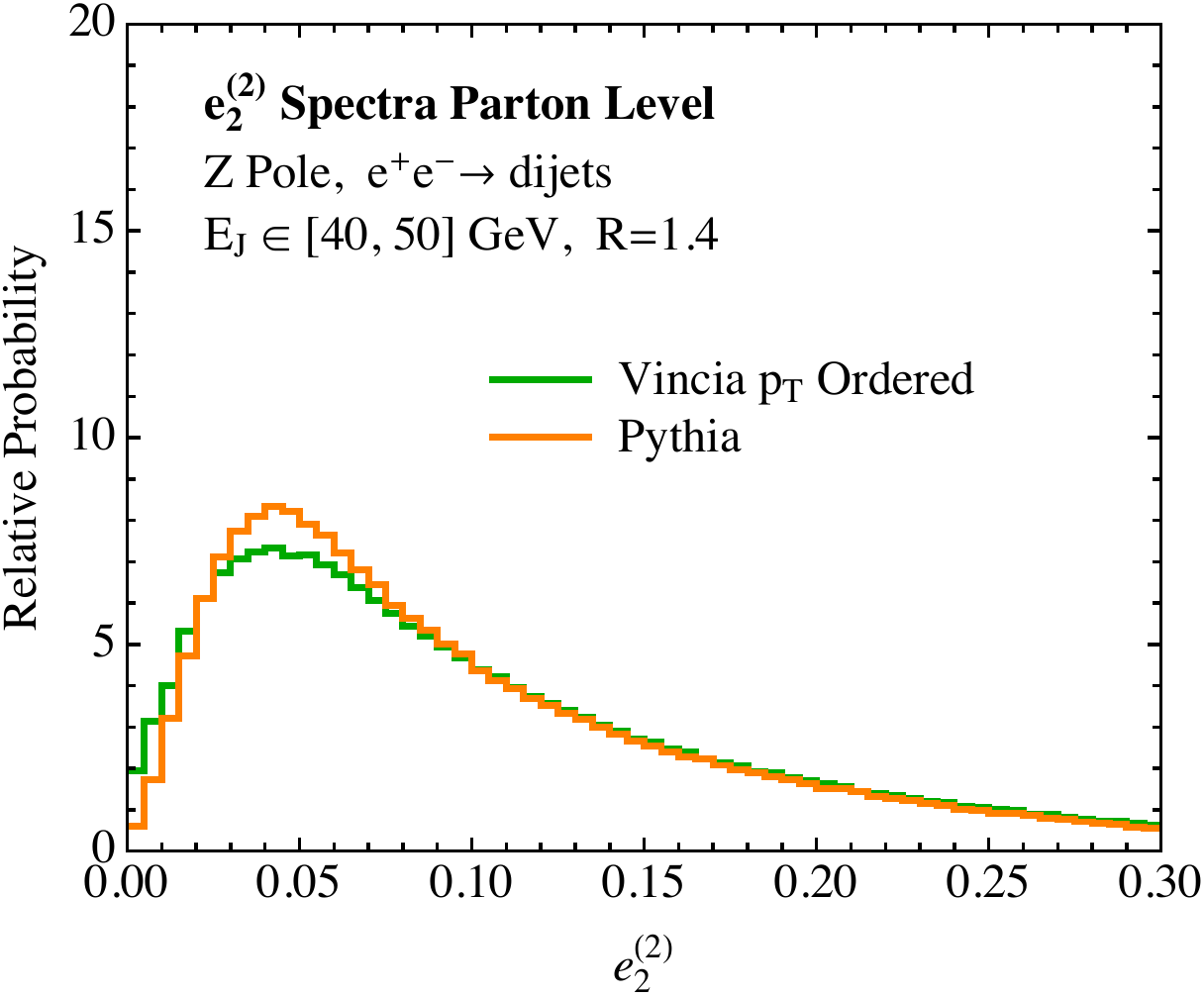}
}
\ 
\subfloat[]{\label{fig:D2_mass_b}
\includegraphics[width = 7.2cm]{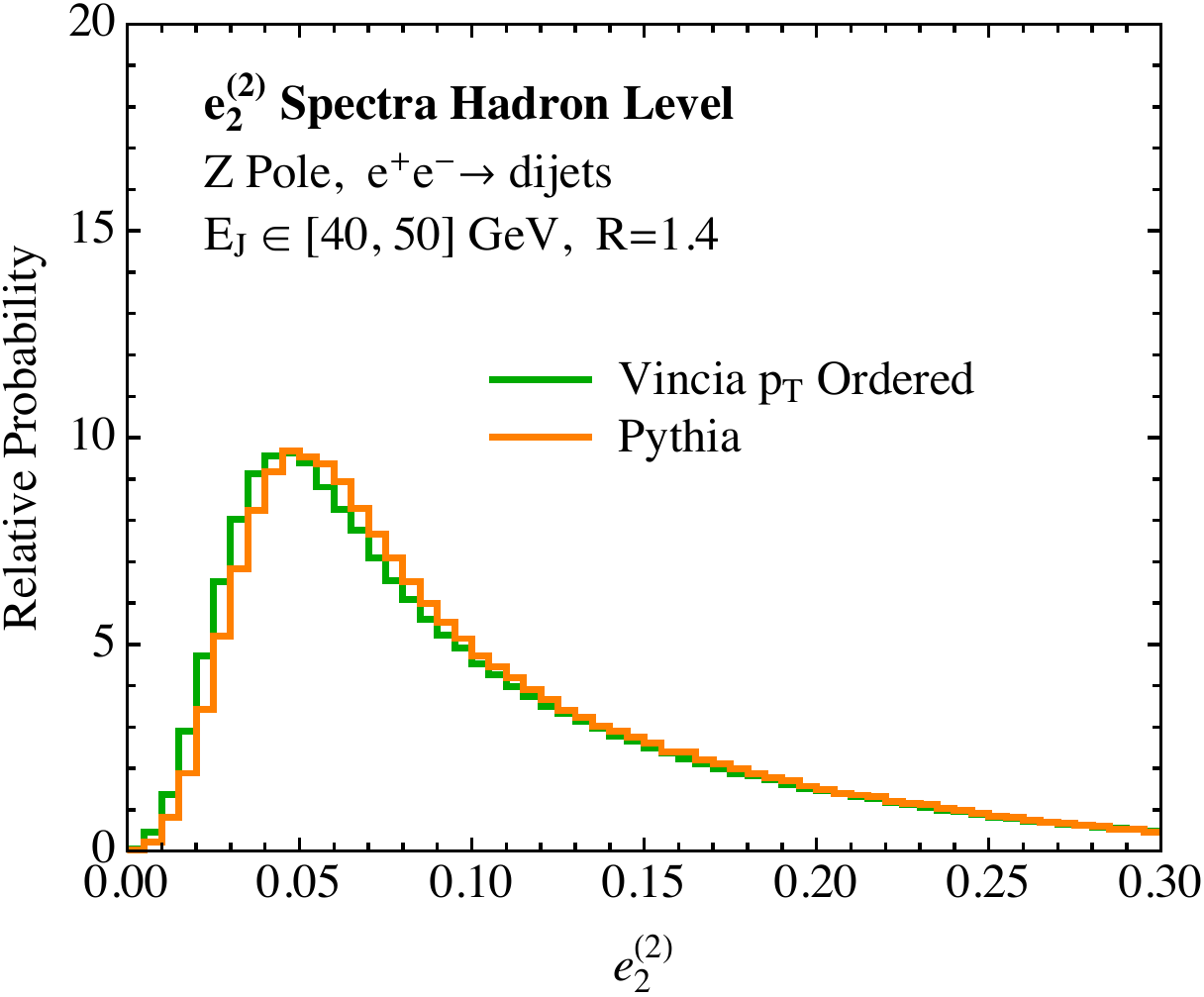}
}
\end{center}
\vspace{-0.2cm}
\caption{A comparison of the $\ecf{2}{2}$ spectrum as measured on quark initiated jets at the $Z$ pole from the \pythia{} and $p_T$-ordered \vincia{} Monte Carlo generators. Results are shown both for parton level Monte Carlo in a), and for hadron level Monte Carlo in b). 
}
\label{fig:D2_mass}
\end{figure}

In \Fig{fig:D2_mass} we compare the $\ecf{2}{2}$ spectra both at parton and hadron level for the \pythia{} and \vincia{} event generators at the $Z$ pole. We choose to the use $\ecf{2}{2}$ instead of the jet mass, as it is dimensionless. The level of agreement should be contrasted with \Fig{fig:D2_LEP} for the $D_2$ observable at the $Z$ pole, with and without hadronization. In particular, for the $\ecf{2}{2}$ observable, there is excellent agreement in the distributions at parton level, which is not true for $D_2$. For $D_2$, the disagreement is largely remedied by hadronization, while for $\ecf{2}{2}$, the level of disagreement before and after hadronization is much more similar. This supports our claim that the $D_2$ observable provides a more differential probe of the perturbative shower in particular, and could be used to improve its description.

\begin{figure}
\begin{center}
\subfloat[]{\label{fig:D2_mass_1TeV_a}
\includegraphics[width= 7.2cm]{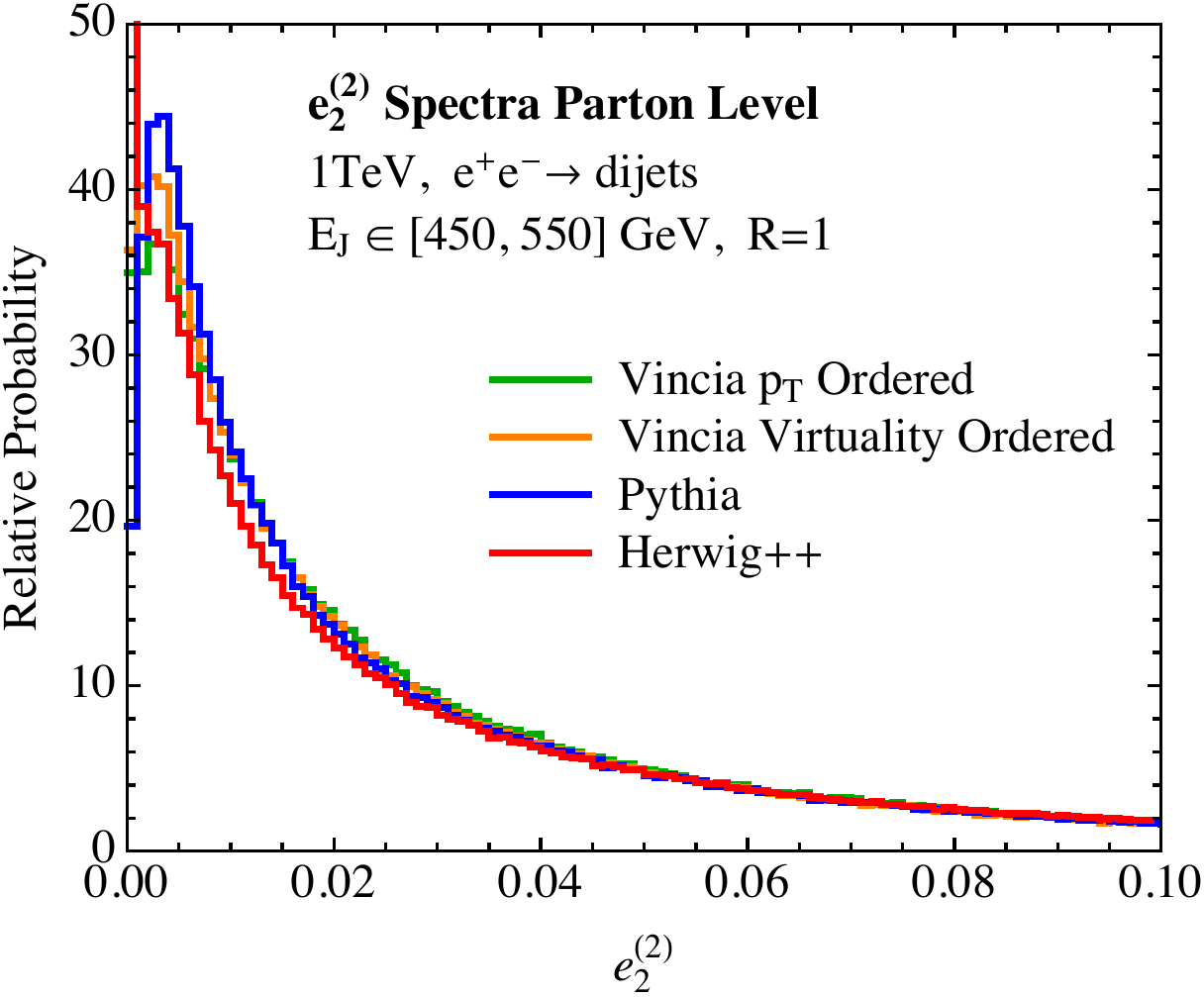}
}
\ 
\subfloat[]{\label{fig:D2_mass_1TeV_b}
\includegraphics[width = 7.2cm]{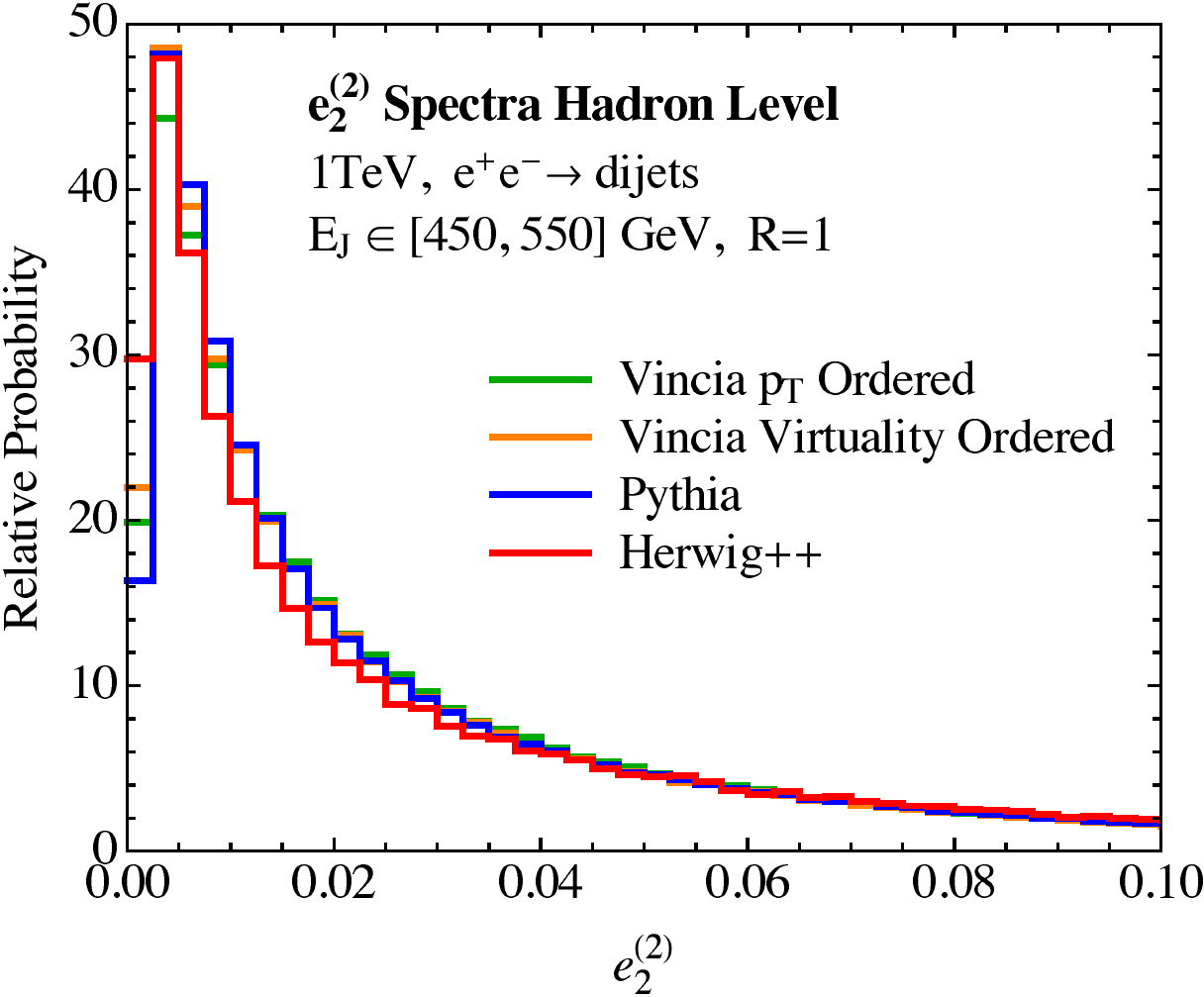}
}
\end{center}
\vspace{-0.2cm}
\caption{A comparison of the $\ecf{2}{2}$ spectrum as measured on quark initiated jets at a center of mass energy of $1$ TeV from the \pythia{}, $p_T$-ordered \vincia{}, virtuality ordered \vincia{}, and \herwigpp{} Monte Carlo generators. Results are shown both for parton level Monte Carlo in a), and for hadron level Monte Carlo in b).  
}
\label{fig:D2_mass_1TeV}
\end{figure}

In \Fig{fig:D2_mass_1TeV} we compare the $\ecf{2}{2}$ spectra both at parton and hadron level for the \pythia{}, $p_T$-ordered \vincia{}, virtuality ordered \vincia{}, and \herwigpp{} event generators at a center of mass energy of $1$ TeV and jet radius $R=1$, as was used for the majority of numerical comparisons with analytic calculations throughout the chapter. The level of agreement in \Fig{fig:D2_mass_1TeV} should be compared with that for the $D_2$ spectra throughout \Sec{sec:results}. In particular, it is interesting to compare the level of agreement observed for the partonic  $\ecf{2}{2}$ spectra as compared with the partonic $D_2$ spectra in \Fig{fig:MC_compare}. There is still some difference between the \herwigpp{} spectrum at parton level and those of \vincia{} and \pythia{}, however, this is to be expected, as these Monte Carlos have different hadronization models and the comparison at parton level should be taken with caution. At hadron level, all Monte Carlos also agree well for the $\ecf{2}{2}$ spectra.

\begin{figure}
\begin{center}
\subfloat[]{\label{fig:D2_mass_500GeV_cut}
\includegraphics[width= 7.2cm]{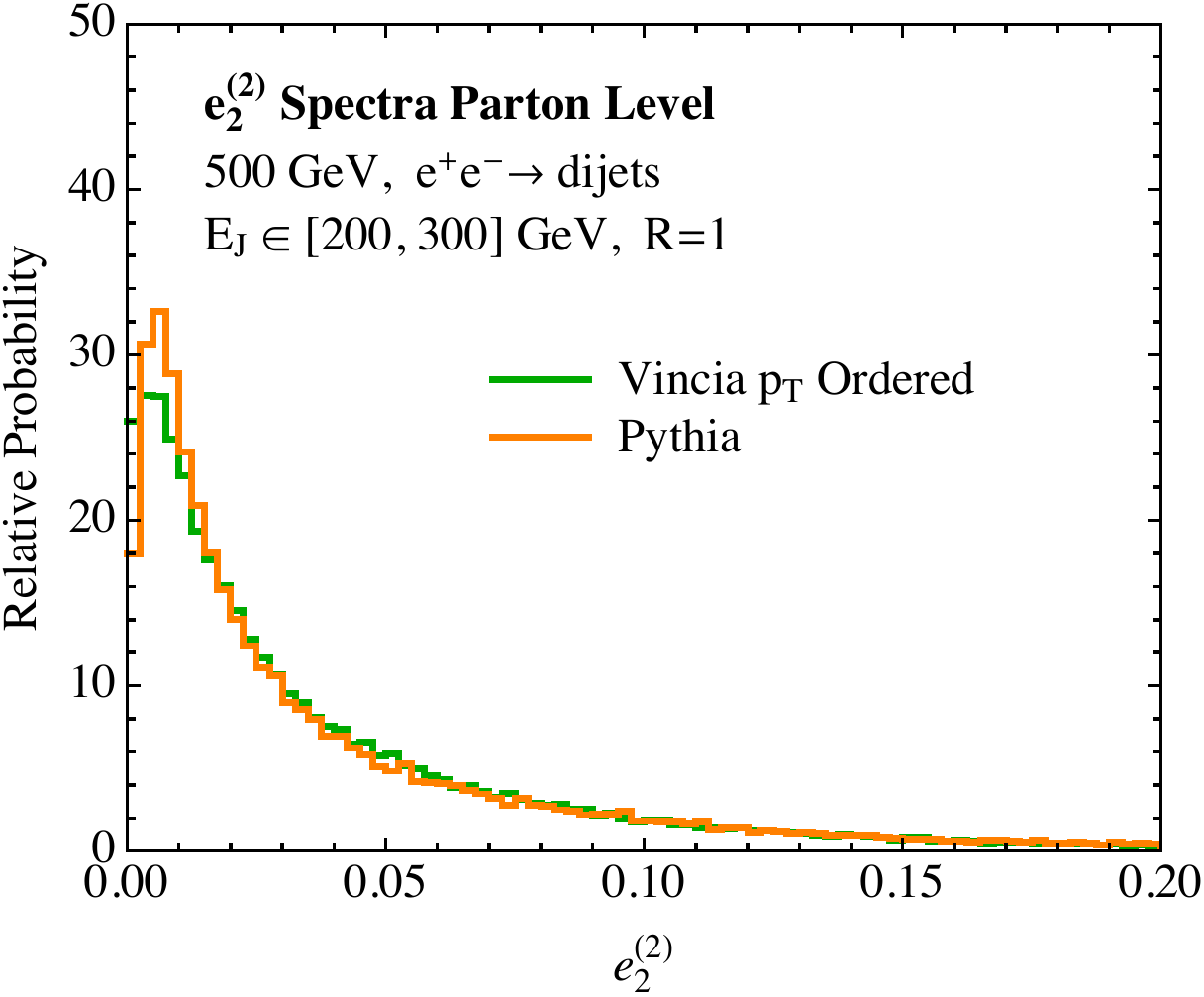}
}
\ 
\subfloat[]{\label{fig:D2_mass_2TeV_cut}
\includegraphics[width = 7.2cm]{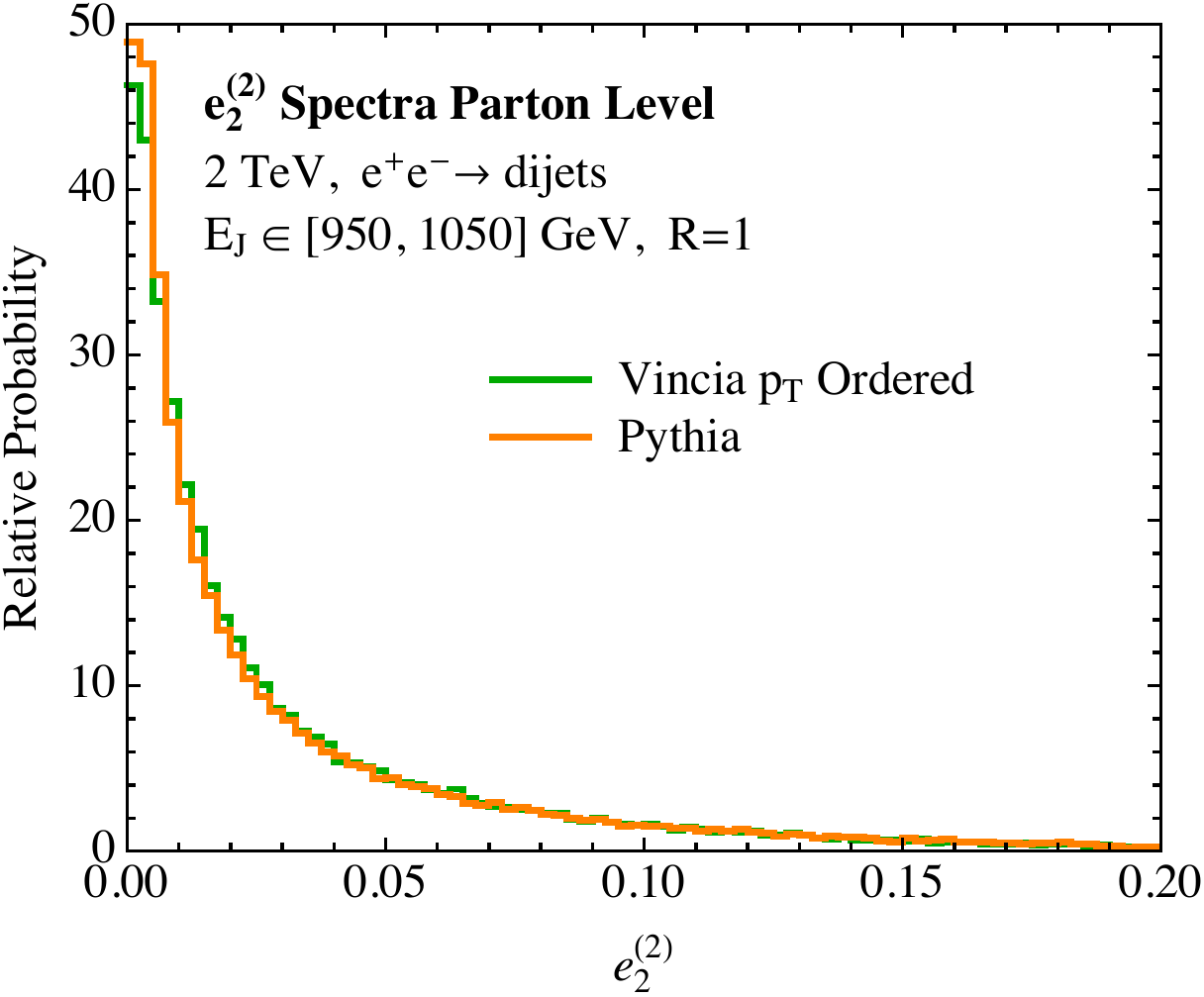}
}
\end{center}
\vspace{-0.2cm}
\caption{A comparison of the $\ecf{2}{2}$ spectrum as measured on quark initiated jets at a center of mass energy of $500$ GeV in a). and $2$ TeV in b). Results are shown for both the \pythia{}, and $p_T$-ordered \vincia{} Monte Carlo generators at parton level.
}
\label{fig:D2_mass_cut_500and2000}
\end{figure}

For completeness, in this appendix we will also include parton level plots of the $\ecf{2}{2}$ distributions for the other parameter ranges that were explored in detail in the text. In \Fig{fig:D2_mass_cut_500and2000} we show the $\ecf{2}{2}$ distributions at a center of mass energy of $500$ GeV and $2$ TeV, the two energies considered in the text. Only the \pythia{} and $p_T$-ordered \vincia{} generators are considered. The level of agreement between the different generators for $\ecf{2}{2}$ should be compared with the level of agreement for the $D_2$ spectra at these two energies, shown in \Fig{fig:E_dependence}. While for the $D_2$ observable, there was a significant discrepancy between the two generators at $2$ TeV, even in the general shape of the distribution, for $\ecf{2}{2}$, the distributions from the two generators agree quite well both at $500$ GeV and $2$ TeV. In particular, they exhibit a similar peak position and shape of the distributions.

\begin{figure}
\begin{center}
\subfloat[]{\label{fig:dep_app_1}
\includegraphics[width= 7.2cm]{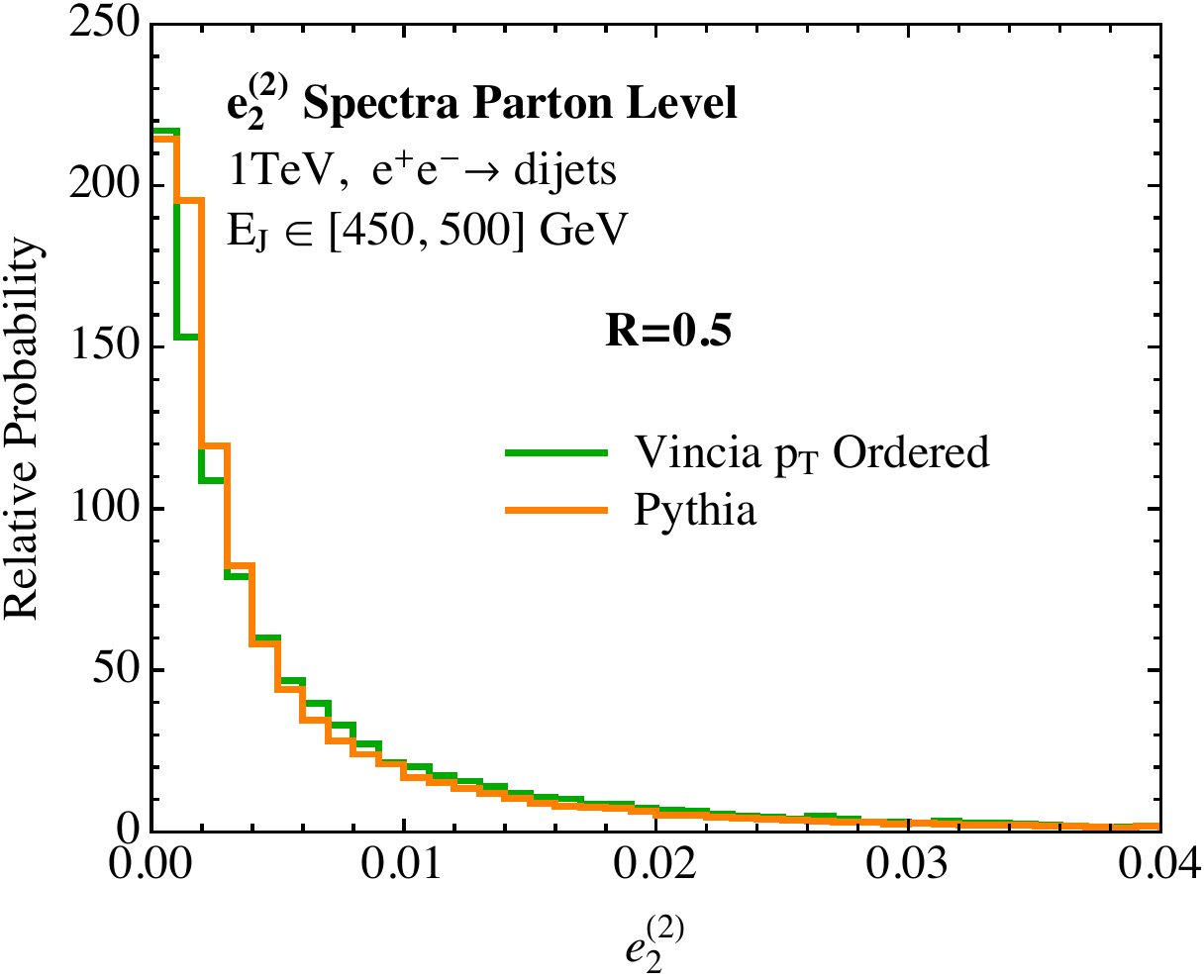}
}
\ 
\subfloat[]{\label{fig:dep_app_2}
\includegraphics[width = 7.2cm]{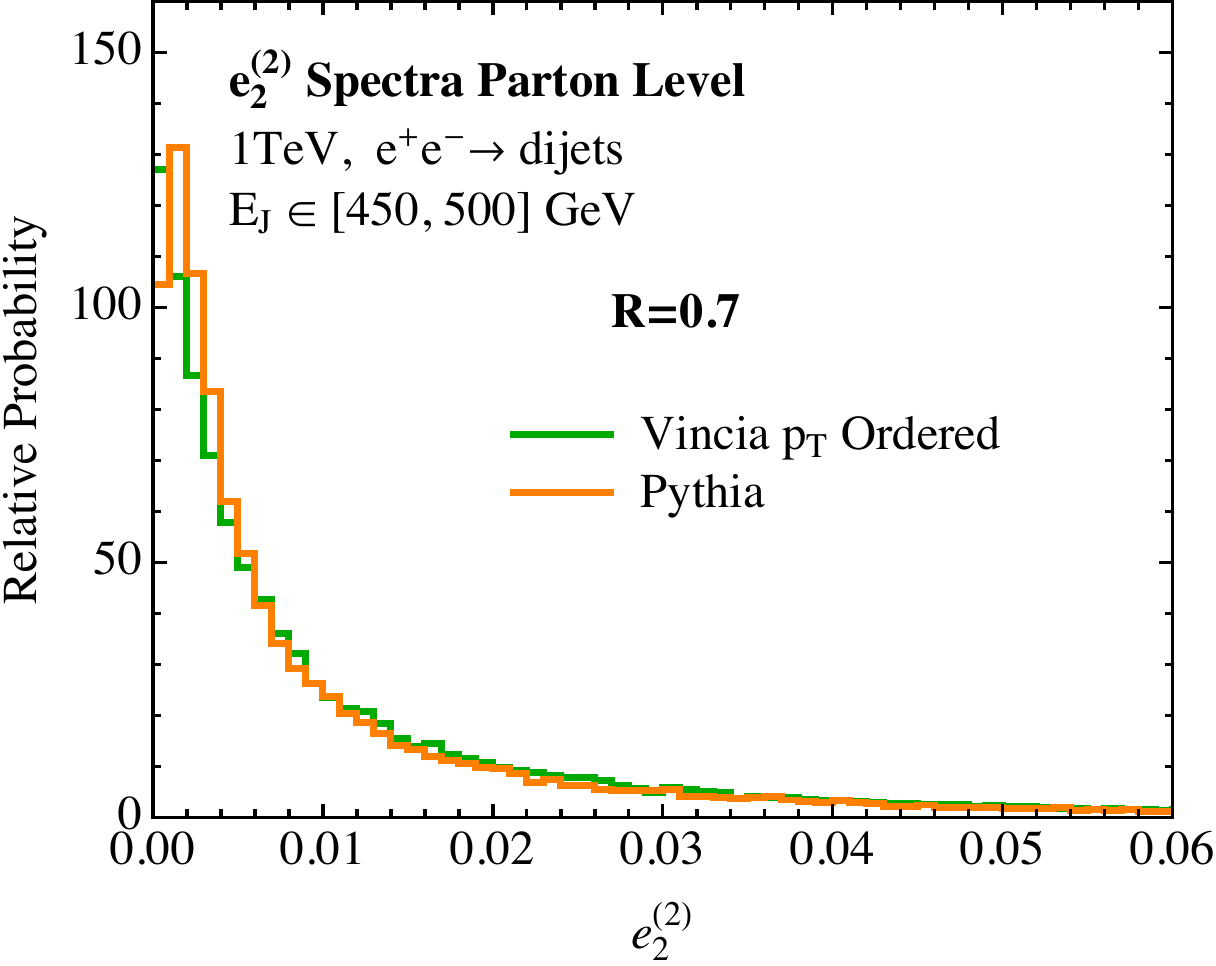}
}\\
\subfloat[]{\label{fig:dep_app_3}
\includegraphics[width= 7.2cm]{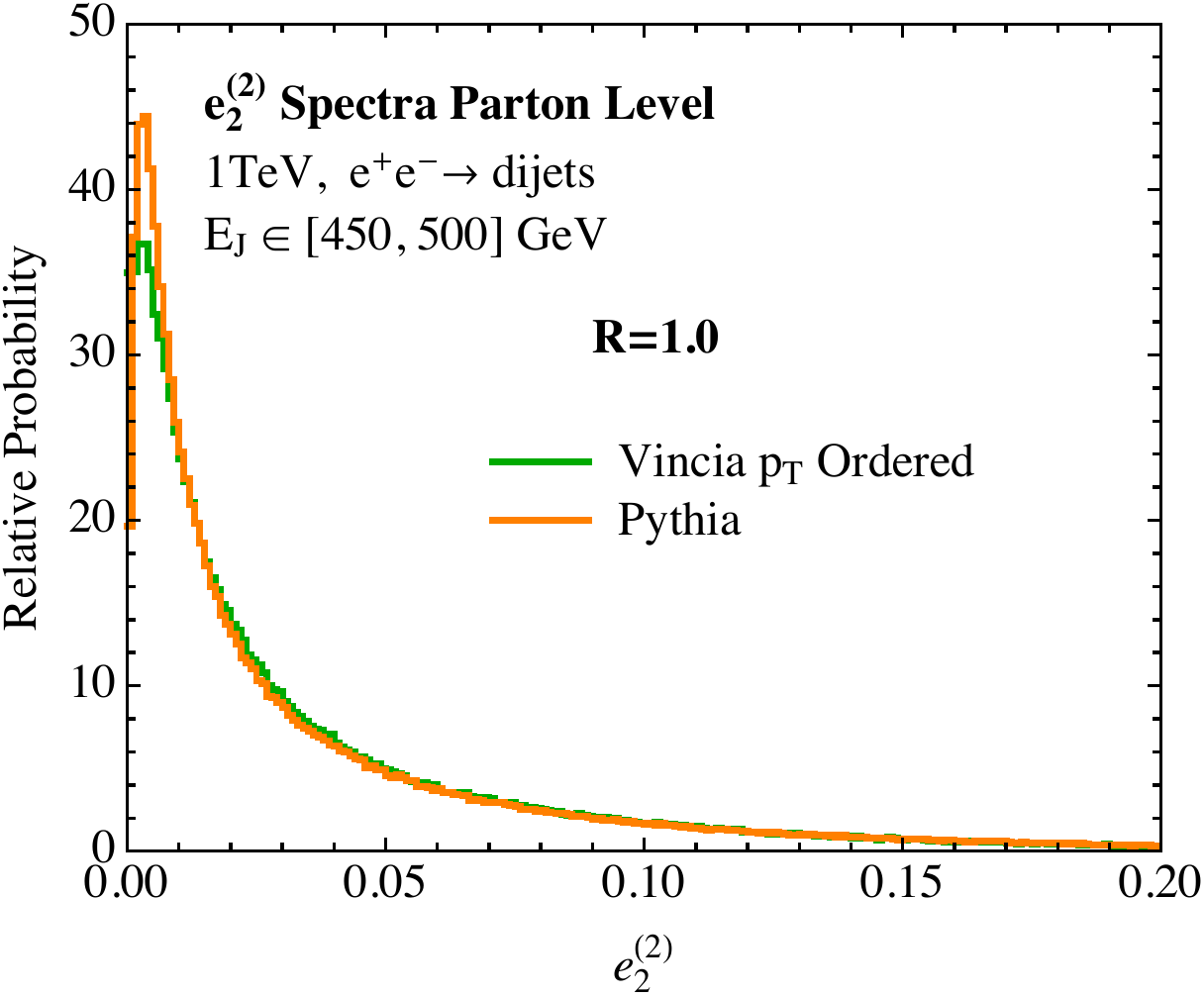}
}
\ 
\subfloat[]{\label{fig:Rdep_app_4}
\includegraphics[width = 7.2cm]{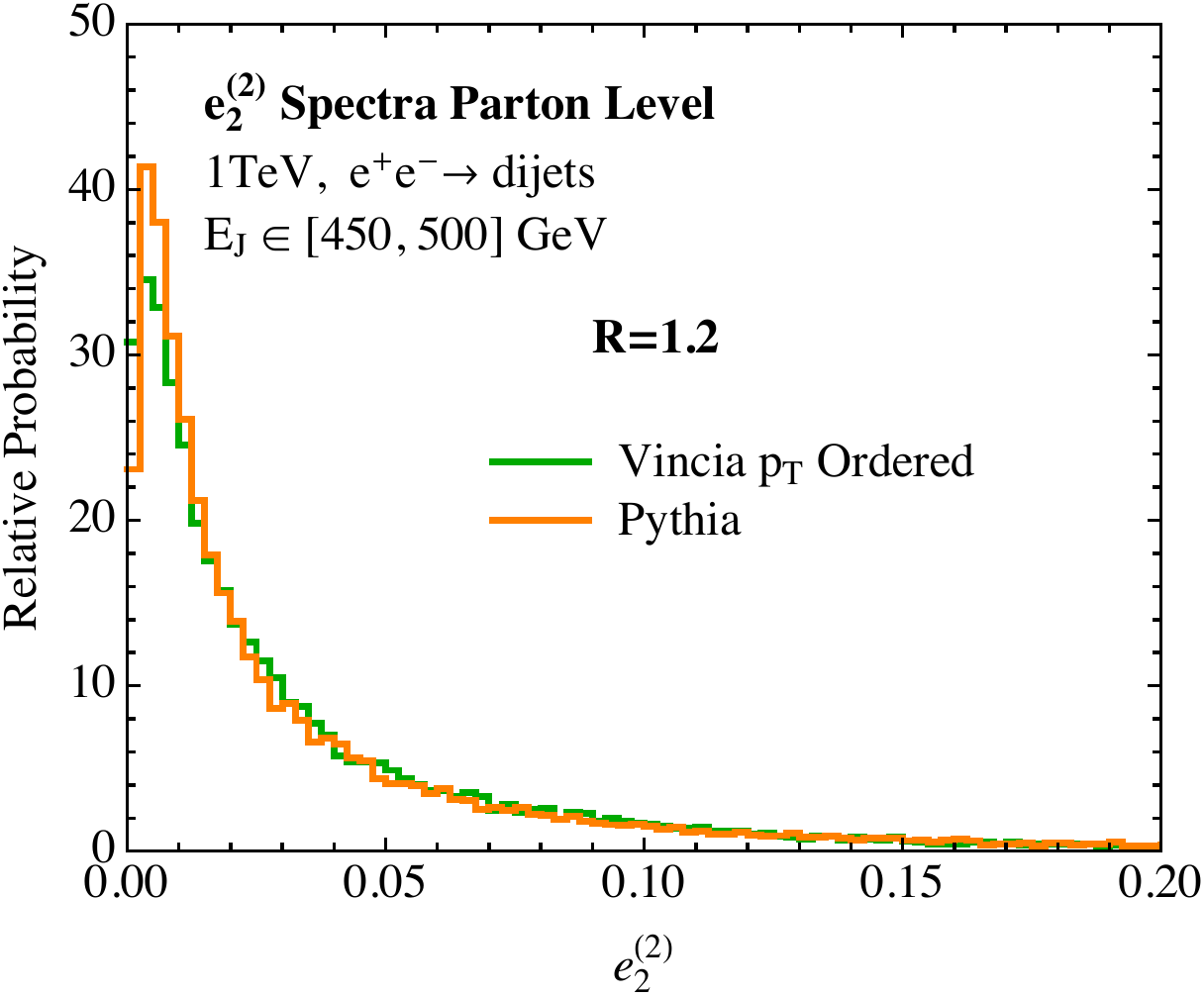}
}
\end{center}
\vspace{-0.2cm}
\caption{A comparison of the $\ecf{2}{2}$ spectrum as measured on quark initiated jets for different $R$ values at a center of mass energy of $1$ TeV from the \pythia{}, and $p_T$-ordered \vincia{} Monte Carlo generators at parton level. Results are shown $R=0.5,0.7,1.0, 1.2$ in a).-d). respectively.}
\label{fig:appendix_Rdependences}
\end{figure}

\begin{figure}
\begin{center}
\subfloat[]{
\includegraphics[width= 7.2cm]{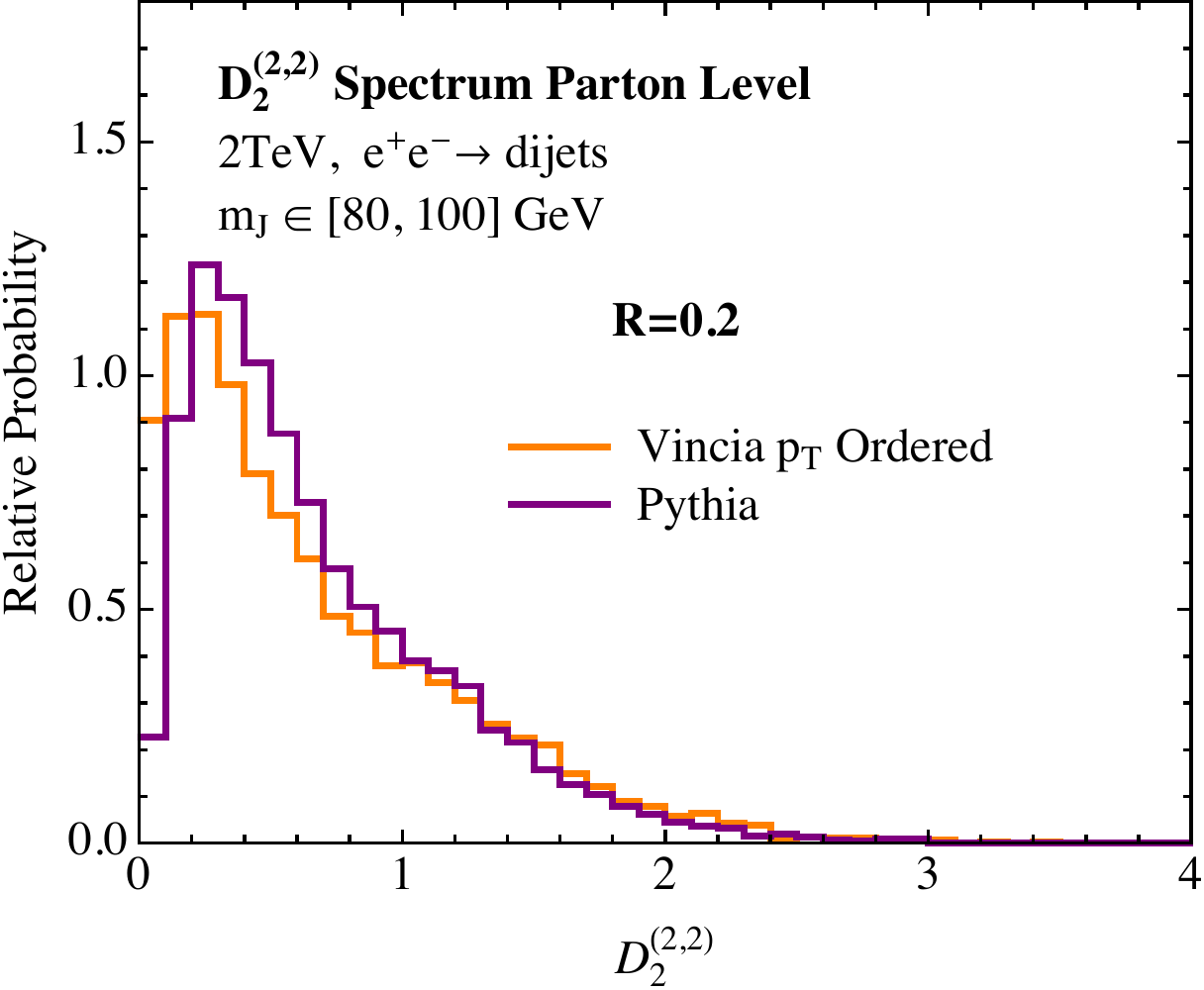}
}
\ 
\subfloat[]{
\includegraphics[width = 7.2cm]{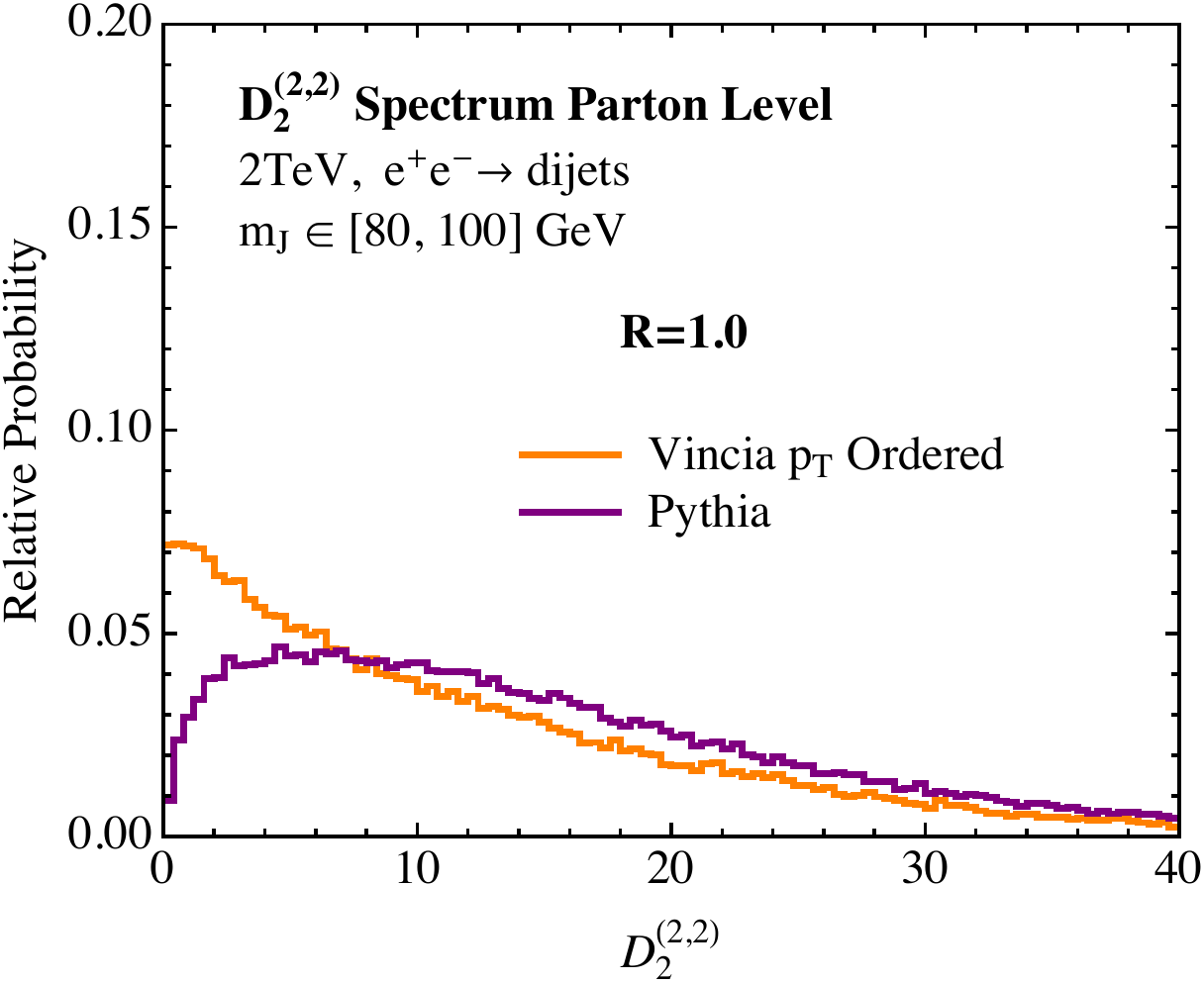}
}
\end{center}
\vspace{-0.2cm}
\caption{A comparison of the $D_2$ spectrum as measured on quark initiated jets at a center of mass energy of $2$ TeV from the \pythia{}, $p_T$-ordered \vincia{} Monte Carlo generators at parton level. A jet radius of $R=0.2$ is used in a) and $R=1.0$ is used in b).  
}
\label{fig:D2_app_R021}
\end{figure}

In \Fig{fig:appendix_Rdependences}, we consider the $R$ dependence of the parton level $\ecf{2}{2}$ distributions as measured in \pythia{} and $p_T$-ordered \vincia{}, as was considered in \Fig{fig:R_dependence} in the text for the $D_2$ observable. Unlike for the $D_2$ distributions, we see good agreement at parton level over the entire range of $R$. To conclude our discussion of $R$ dependence at parton level, we also include in \Fig{fig:D2_app_R021} a comparison of the parton level $D_2$ spectra as measured in in \pythia{} and $p_T$-ordered \vincia{} at $2$ TeV, with $R=0.2$ and $R=1.0$. As was referenced in \Sec{sec:jet_energy}, while poor agreement between the two generators is seen for $R=1$, comparably good agreement is seen at $R=0.2$. This further supports our claim that the discrepancy between the two generators arises from the soft subjet region of phase space, which is poorly described by the \pythia{} generator, but which can be removed by considering small radius jets. We view the ability to perform analytic calculations of observables which are sensitive to the substructure of the jet in this manner as an opportunity to improve the perturbative description of the QCD shower as implemented in Monte Carlo generators.

%% file: appb.tex
\chapter{Explicit Results for Helicity Amplitudes}

\section{Spinor and Color Identities}
\label{app:useful}

\subsection{Spinor Algebra}
\label{app:helicity}

The overall phase of the spinors $\ket{p\pm}$ is not determined by the Dirac equation, $\Sl p\, \ket{p\pm} = 0$, and so can be chosen freely. 
In the Dirac representation,
\begin{equation}
\ga^0 = \begin{pmatrix} 1 & 0 \\ 0 & -1 \end{pmatrix}
\,,\quad
\ga^i = \begin{pmatrix} 0 & \sigma^i \\ -\sigma^i & 0 \end{pmatrix}
\,,\quad
\ga_5 = \begin{pmatrix} 0 & 1 \\ 1 & 0 \end{pmatrix}
\,,\end{equation}
and taking $n_i^\mu = (1,0,0,1)$, we have the standard solutions~\cite{Dixon:1996wi}
\begin{equation} \label{eq:ket_explicit}
\ket{p+} = \frac{1}{\sqrt{2}}
\begin{pmatrix}
   \sqrt{p^-} \\
   \sqrt{p^+} e^{\img \phi_p} \\
   \sqrt{p^-} \\
   \sqrt{p^+} e^{\img \phi_p}
\end{pmatrix}
,\quad
\ket{p-} = \frac{1}{\sqrt{2}}
\begin{pmatrix}
    \sqrt{p^+} e^{-\img\phi_p} \\
    -\sqrt{p^-} \\
    -\sqrt{p^+} e^{-\img\phi_p} \\
    \sqrt{p^-}
\end{pmatrix}
,\end{equation}
where
\begin{equation}
p^\pm = p^0 \mp p^3
\,,\qquad
\exp(\pm \img \phi_p) = \frac{p^1 \pm \img p^2}{\sqrt{p^+ p^-}}
\,.\end{equation}
For negative $p^0$ and $p^\pm$ we use the usual branch of the square root, such that for $p^0 > 0$
\begin{equation}
\ket{(-p)\pm} = \img \ket{p\pm}
\,.\end{equation}
The conjugate spinors, $\bra{p\pm}$, are defined as
\begin{equation}
\bra{p\pm} = \mathrm{sgn}(p^0)\, \overline{\ket{p\pm}}
\,.\end{equation}
The additional minus sign for negative $p^0$ is included to use the same branch of the square root for both types of spinors, i.e., for $p^0 > 0$
\begin{equation}
\bra{(-p)\pm} = - \overline{\ket{(-p)\pm}} = -(-\img) \bra{p\pm} = \img\bra{p\pm}
\,.\end{equation}
In this way all spinor identities are automatically valid for both positive and negative momenta, which makes it easy to use crossing symmetry. The additional signs only appear in relations which involve explicit complex conjugation. The most relevant is
\begin{equation} \label{eq:spin_conj}
\braket{p-}{q+}^* = \mathrm{sgn}(p^0 q^0)\, \braket{q+}{p-}
\,.\end{equation}

The spinor products are denoted by
\begin{equation}
\langle p q \rangle = \braket{p-}{q+}
\,,\qquad
[p q] = \braket{p+}{q-}
\,.\end{equation}
Similarly, for products involving additional gamma matrices, we write
\begin{align}
\bra{p}\gamma^\mu |q] &= \bra{p-} \gamma^\mu \ket{q-}
\,,\quad 
[p|\gamma^\mu\ket{q} = \bra{p+} \gamma^\mu \ket{q+}
\,, \\
\bra{p}k |q] &= \bra{p-} {\Sl k} \ket {q-}
\,,\quad
[p | k \ket{q} = \bra{p+} {\Sl k} \ket {q+}
\,,\\
\bra{p}qk \ket{l} &= \bra{p-} {\Sl q} {\Sl k} \ket {l+}
\,,\quad
[p|qk|l] = \bra{p+} {\Sl q} {\Sl k} \ket {l-}
\,,\end{align}
etc.

Some useful identities, that follow directly from the definition of the spinors, are
\begin{align}
&\ang{pq} = - \ang{qp}
\,,\qquad
[pq] = - [qp]
\,,\\
 &[p|\ga^\mu \ket{p}=\bra{p} \ga^\mu |p] = 2p^\mu
\,.\end{align}
From the completeness relations
\begin{align} \label{eq:comp}
 \ket{p\pm}\bra{p\pm} = \frac{1 \pm \ga_5}{2}\,{\Sl p}
\,, \\
{\Sl p}=|p]\bra{p}+\ket{p}[p|,
\end{align}
one finds
\begin{equation}
\langle pq \rangle [qp] = \frac{1}{2}\,\tr\bigl\{(1 - \ga_5) {\Sl p} {\Sl q} \bigr\} = 2 p\cdot q
\,.\end{equation}
Combining this with \eq{spin_conj}, it follows that
\begin{equation}
 |\ang{pq}| = |[pq]| = \sqrt{|2p \cdot q|}
\,.\end{equation}
The completeness relation is also useful to reduce typical expressions like
\begin{equation}
[p|q\ket{k}= [pq]\langle qk\rangle\,,
\end{equation}
to spinor products.

Charge conjugation invariance of the current, the Fierz identity and the Schouten identity are
\begin{align} \label{eq:Fierzetc}
\bra{p}\gamma^\mu |q]  &= [q|\gamma^\mu \ket{p}
\,,\\
[p|\gamma_\mu \ket{q} [k|\gamma^\mu \ket{l}&= 2[pk]\ang{lq}
\,, \\
\ang{pq} \ang{kl} &= \ang{pk} \ang{ql} + \ang{pl} \ang{kq}
\,.\end{align}
Finally, momentum conservation $\sum_{i=1}^n p_i = 0$ implies
\begin{equation} \label{eq:spinormomcons}
\sum_{i=1}^n [ji] \ang{ik} = 0
\,.\end{equation}

From \eq{ket_explicit}, we see that under parity the spinors transform as
\begin{equation} \label{eq:spinorparity}
\ket{p^\P\pm} = \pm e^{\pm \img\phi_p}\gamma^0\,\ket{p\mp}
\,,\end{equation}
and therefore
\begin{align}
\ang{p^\P q^\P} &= -e^{\img(\phi_p + \phi_q)} [pq] \label{eq:app_parity1}
\,,\\
[p^\P q^\P] &= -e^{-\img(\phi_p + \phi_q)} \ang{pq} \label{eq:app_parity2}
\,.\end{align}

When applying the above result to a helicity amplitude, the phases which appear are determined by the little group scaling (see e.g. Refs.~\cite{Dixon:1996wi, Dixon:2013uaa,Elvang:2013cua} for a review). The little group is the subgroup of the Lorentz transformations that fixes a particular momentum. In terms of the spinor helicity variables, the action of the little group, which preserves the momentum vector $p$, is given by
\begin{equation}
|p\rangle \to z |p\rangle, \qquad [p| \to \frac{1}{z}[p|\,.
\end{equation}
In the case that the particle with momentum $p$ has helicity $h$, the corresponding helicity amplitude scales as $z^{-2h}$ under the little group scaling. This property of the helicity amplitudes then predicts the phases that appear in the amplitude under a parity transformation.

The following completeness relation for the polarization vectors is also useful
\begin{align}
\sum\limits_{\lambda=\pm} \epsilon^\lambda_\mu(p,q) \left (\epsilon^\lambda_\nu(p,q) \right )^*=-g_{\mu \nu}+\frac{p_\mu q_\nu+p_\nu q_\mu}{p\cdot q}.
\end{align}

In SCET the collinear quark fields produce projected spinors 
\begin{align}
\ket{p\pm}_{n}=\frac{{\Sl n}{\Sl {\bar n}}}{4}\ket{p\pm}\,.
\end{align}
The projected spinor trivially satisfies the relation
\begin{equation}
{\Sl n}\,\Bigl(\frac{{\Sl n}{\Sl {\bar n}}}{4}\ket{p\pm}\Bigr) = 0
\,,\end{equation}
so it is proportional to $\ket{n\pm}$. Working in the basis in \eq{ket_explicit}, we have 
\begin{align}\label{eq:ketn}
\frac{{\Sl n}{\Sl {\bar n}}}{4} \ket{p}=& \sqrt{p^0}\left[ \cos\left (\frac{\theta_n}{2}\right )\cos \left(\frac{\theta_p}{2}\right )\right. \nonumber \\
& \left.+e^{i(\phi_p-\phi_n)} \sin \left(\frac{\theta_n}{2}\right )   \sin \left(\frac{\theta_p}{2} \right )   \right ] \ket{n}\,, \nonumber \\
\frac{{\Sl n}{\Sl {\bar n}}}{4} |p]=& \sqrt{p^0}\left [ e^{i \left(\phi _p-\phi _n\right)} \cos \left(\frac{\theta _n}{2}\right)  \cos \left(\frac{\theta _p}{2}\right) \right. \nonumber \\
&\left. +\sin \left(\frac{\theta _n}{2}\right) \sin \left(\frac{\theta _p}{2}\right)   \right ] |n]\,.
\end{align}
Here $\theta_n, \phi_n$, and $\theta_p, \phi_p$, are the polar and azimuthal angle of the $n$ and $p$ vectors, respectively.
In particular, we see that choosing $n^\mu=p^\mu/p^0$, which can always be done at leading power since there is a single particle per collinear sector, we have $\phi_p=\phi_n$, $\theta_p=\theta_n$, and the simple relation
\begin{align}
\frac{{\Sl n}{\Sl {\bar n}}}{4} \ket{p\pm}=\sqrt{\frac{\bn\cdot p}{2}}\: \ket{n\pm}\,.
\end{align}

\subsection{Color Algebra}
\label{app:color}

The generators $t^a_r$ of a general irreducible representation $r$ of $\mathrm{SU(N)}$ satisfy
\begin{equation}
[t^a_r, t^b_r] = \img f^{abc} \, t_r^c
\,,\quad
t_r^a\, t_r^a = C_r\mathbf{1}
\,,\quad
\tr[t^a_r\,t^b_r] = T_r\,\delta^{ab}
\,,\end{equation}
where $f^{abc}$ are completely antisymmetric, and $C_r$ is the quadratic Casimir of the representation $r$.
The normalization $T_r$ is given by $T_r = C_r d_r/d$, where $d_r$ is the dimension of the representation
and $d$ the dimension of the Lie algebra.

We denote the generators in the fundamental representation by $t_F^a = T^a$, and the overall
normalization is fixed by choosing a specific value for $T_F$. 
The adjoint representation is given by $(t_A^a)_{bc} = -\img f^{abc}$, which implies
\begin{equation}
f^{acd} f^{bcd} = C_A\,\delta^{ab}
\,.\end{equation}
We also define the symmetric structure constants as 
\begin{equation}
d^{abc}=\frac{1}{T_F} \tr[T^a\{ T^b,T^c \}]
\,.
\end{equation}
For the fundamental and adjoint representations we have $d_F = N$, $d_A = d = N^2 - 1$, and so
\begin{equation}
C_F = \frac{N^2 - 1}{2N}
\,,\qquad
C_A = N
\,,
\end{equation}
where we have chosen the standard normalization
\begin{equation}
T_F = \frac{1}{2}
\,.\end{equation}
Throughout the text, and for the amplitudes in the appendices, we have kept $T_F$ arbitrary. This can be done using $C_F = T_F (N^2 - 1)/N$, $C_A = 2T_F N$. The strong coupling constant, $g_s$, can be kept convention independent, by using $g_s \to g_s/\sqrt{2T_F}$.

Some additional useful color identities are
\begin{align}
t^a_r t^b_r t^a_r &= \Bigl(C_r - \frac{C_A}{2}\Bigr)\, t^b_r
\,,\\
T^a T^b T^c T^a &= T_F^2\, \delta^{bc} \mathbf{1} + \Bigl(C_F - \frac{C_A}{2} \Bigr)\, T^b T^c
\,,\end{align}
where the second relation is equivalent to the completeness relation
\begin{equation}
T^a_{\alpha \bbeta}\,T^a_{\gamma \bdelta} = T_F \Bigl(\de_{\alpha \bdelta}\,\de_{\gamma \bbeta} - \frac{1}{N} \de_{\alpha \bbeta}\, \de_{\gamma \bdelta} \Bigr)
\,.\end{equation}
We also have
\begin{align}
T^b\, \img f^{bac}\, T^c &= \frac{C_A}{2}\, T^a
\,,\\
T^c\, \img f^{cad} \img f^{dbe} T^e
&= T_F^2\,\delta^{ab}\,\mathbf{1} + \frac{C_A}{2}\,T^a T^b
\,.\end{align}

\subsection{QCD Color Decompositions}
\label{app:color_decomp}

Here we briefly review a common color decomposition for QCD NLO amplitudes \cite{Berends:1987me, Mangano:1987xk, Mangano:1988kk, Bern:1990ux}. The color bases used for the processes discussed in the text are specific examples of the decompositions given below, and were chosen to facilitate the extraction of the matching coefficients from the amplitudes literature. For a pedagogical introduction to color decompositions in QCD amplitudes see for example Refs.~\cite{Dixon:1996wi, Dixon:2013uaa}.

For an $n$ gluon process, a one-loop color decomposition in terms of fundamental generators $T^a$ is given by
\begin{align} \label{eq:gluon_decomp}
\cA_n(g_1\cdots g_n)
&= g_s^{n-2}  \sum_{\si \in S_n/Z_n}\! \tr[T^{a_{\si(1)}} \cdot \cdot \cdot T^{a_{\si(n)}}]
\Big[ A_{n}^{\text{tree}}\bigl(\si(1), \cdot \cdot \cdot, \si(n)\bigr)
+ g_s^2\, C_A A_{n;1}\bigl(\si(1), \cdot \cdot \cdot, \si(n)\bigr) \Big]
\nn\\ &
+  g_s^n  \sum\limits_{c=3}^{\left \lfloor{n/2}\right \rfloor+1}     \!\!\sum_{\si \in S_n/S_{c-1,n-c+1}}\!\!\! \tr[ T^{a_{\si(1)}}  \cdot \cdot \cdot    T^{a_{\si(c-1)}}] \tr[ T^{a_{\si(c)}}  \cdot \cdot \cdot    T^{a_{\si(n)}}]
A_{n;c}\bigl(\si(1),\cdot \cdot \cdot,\si(n)\bigr)
\,,\end{align}
where $A_{n;1}$, $A_{n}^{\text{tree}}$ are primitive amplitudes, which can be efficiently calculated using unitarity methods, and the $A_{n;c}$ are partial amplitudes which can be written as sums of permutations of the primitive amplitudes. The amplitudes appearing in this decomposition are separately gauge invariant. In this formula, $S_n$ is the permutation group on $n$ elements, and $S_{i,j}$ is the subgroup of $S_{i+j}$ which leaves the given trace structure invariant. At tree level, only the single trace color structure appears.

In the case that additional noncolored particles are also present, an identical decomposition exists, since the color structure is unaffected. For example, for a process involving $n$ gluons and a Higgs particle, the amplitude satisfies the same decomposition as in \eq{gluon_decomp}, but with the partial and primitive amplitudes in \eq{gluon_decomp} simply replaced by $A\bigl(\phi, \si(1),\cdot \cdot \cdot,\si(n)\bigr)$, where $\phi$ denotes the Higgs particle  \cite{Berger:2006sh}.

A similar decomposition exists for processes involving $q \bar q$ pairs. For example, the one-loop decomposition for a process with a $q\bar q$ pair and $n-2$ gluons is given by \cite{Bern:1994fz}
\begin{align} \label{eq:qq_decomp}
&\cA_n \bigl( \bar q_1 q_2 g_3\dots g_n  \bigr) \nn \\
&=g_s^{n-2}\sum_{\si \in S_{n-2}}\!    \bigl( T^{a_{\si(3)}} \cdot \cdot \cdot T^{a_{\si(n)}}   \bigr)_{\alpha \bbeta}
\Big[A_{n}^{\text{tree}}\bigl( 1_{\bar q}, 2_q;\si(3), \cdot \cdot \cdot, \si(n)\bigr) 
+ g_s^2\, C_A A_{n;1}\bigl(1_{\bar q}, 2_q;\si(3), \cdot \cdot \cdot, \si(n)\bigr)\Big]
\nn\\ &
+g_s^n  \sum\limits_{c=3}^{n-3} ~    \!\!\sum_{\si \in S_{n-2}/Z_{c-1}}\!\!\! \tr[ T^{a_{\si(3)}}  \cdot \cdot \cdot    T^{a_{\si(c+1)}}] \bigl ( T^{a_{\si(c+2)}}  \cdot \cdot \cdot    T^{a_{\si(n)}} \bigr )_{\alpha \bbeta}
A_{n;c}\bigl(1_{\bar q}, 2_q;\si(3),\cdot \cdot \cdot,\si(n)\bigr) 
\nn\\ &
+g_s^n \!\! \!\!\!\! \!\!\sum_{\si \in S_{n-2}/Z_{n-3}}\!\!\! \tr[ T^{a_{\si(3)}}  \cdot \cdot \cdot    T^{a_{\si(n-1)}}] \bigl (  T^{a_{\si(n)}} \bigr )_{\alpha \bbeta}
A_{n;n-2}\bigl(1_{\bar q}, 2_q;\si(3),\cdot \cdot \cdot,\si(n)\bigr) 
\nn\\ &
+g_s^n \!\! \!\!\!\! \!\!  \sum_{\si \in S_{n-2}/Z_{n-2}}\!\!\! \tr[ T^{a_{\si(3)}}  \cdot \cdot \cdot    T^{a_{\si(n)}}]\delta_{\alpha \bbeta}
A_{n;n-1}\bigl(1_{\bar q}, 2_q;\si(3),\cdot \cdot \cdot,\si(n)\bigr) 
\,.\end{align}
This decomposition is easily extended to the case of additional $q\bar q$ pairs.
As with the gluon case, the same color decomposition also applies if additional uncolored particles are included in the amplitude.

For more than five particles, the one-loop color decompositions given above do not give a complete basis of color structures beyond one loop, since color structures with more than two traces can appear. A complete basis of color structures is required for the SCET basis to guarantee a consistent RGE. A convenient basis of color structures for one-loop matching is then given by extending the one-loop decomposition to involve all higher trace structures.

\section{Helicity Amplitudes for Higgs $+$ Jets}
\label{app:Hamplitudes}

In this appendix we give explicit results for the hard matching coefficients for $H+0,1,2$ jets. We only explicitly consider gluon-fusion processes, where the Higgs couples to two gluons through a top-quark loop, and additional jets correspond to additional gluons, or quark antiquark pairs. When matching onto SCET we perform a one-step matching and directly match full QCD onto SCET, as was done for $H+0$ jets in Ref.~\cite{Berger:2010xi}. Most QCD results are obtained in the limit of infinite top quark mass, by first integrating out the top quark and matching onto an effective $ggH$ interaction,
\begin{equation} \label{eq:HiggsLeff}
\cL_\hard = \frac{C_1}{12\pi v} H G_{\mu\nu}^a G^{\mu\nu\,a}
\,,\end{equation}
which is then used to compute the QCD amplitudes. Here $v = (\sqrt{2}G_F)^{-1/2} = 246\GeV$. From the point of view of the one-step matching from QCD onto SCET, using \eq{HiggsLeff} is just a convenient way to compute the full QCD amplitude in the $m_t\to \infty$ limit. In particular, the $\alpha_s$ corrections to $C_1$ in \eq{HiggsLeff} are included in the amplitudes below, and therefore also in the SCET Wilson coefficients. In this way, if higher-order corrections in $1/m_t$ or the exact $m_t$ dependence for a specific amplitude are known, they can easily be included in the QCD amplitudes and the corresponding SCET Wilson coefficients. We illustrate this for the case of $H+0$ jets below.

We separate the QCD amplitudes into their IR-divergent and IR-finite parts
\begin{align}
A &= A_\div + A_\fin\,,
\nonumber \\
B &= B_\div + B_\fin\,,
\end{align}
where $A_\fin$, $B_\fin$ enter the matching coefficients in \sec{higgs}. For simplicity, we drop the subscript ``$\fin$'' for those amplitudes that have no divergent parts, i.e. for $A_\div = 0$ we have $A_\fin \equiv A$. For the logarithms we use the notation
\begin{align}
L_{ij} &= \ln\Bigl(-\frac{s_{ij}}{\mu^2} - \img 0 \Bigr)
\,,\nn\qquad
L_{ij/H} = \ln\Bigl(-\frac{s_{ij}}{\mu^2} - \img 0 \Bigr) - \ln\Bigl(-\frac{m_H^2}{\mu^2} - \img 0 \Bigr)
\,.\end{align}

\subsection{\boldmath $H+0$ Jets}
\label{app:H0amplitudes}

We expand the amplitudes in powers of $\alpha_s(\mu)$ as
\begin{equation} \label{eq:H_expand}
A = \frac{2T_F\alpha_s(\mu)}{3\pi v}\, \sum_{n=0}^\infty A^{(n)} \Bigl(\frac{\alpha_s(\mu)}{4\pi}\Bigr)^n
\,.\end{equation}
The amplitudes with opposite helicity gluons vanish to all orders because of angular momentum conservation,
\begin{align}
A(1^\pm, 2^\mp; 3_H)
&= 0,
\end{align}
corresponding to the fact that the helicity operators for these helicity configurations were not included in the basis of \eq{ggH_basis}.
The lowest order helicity amplitudes including the full $m_t$ dependence are given by
\begin{align}
A^\zero(1^+, 2^+; 3_H)
&= \frac{s_{12}}{2}\frac{[12]}{\ang{12}}\, F^\zero\Bigl(\frac{s_{12}}{4m_t^2}\Bigr)= \frac{s_{12}}{2}\, F^\zero\Bigl(\frac{s_{12}}{4m_t^2}\Bigr)\, e^{\img\Phi_{++ H}}
\,,\nn\\
A^\zero(1^-, 2^-; 3_H)
&=\frac{s_{12}}{2}\frac{\ang{12}}{[12]}\, F^\zero\Bigl(\frac{s_{12}}{4m_t^2}\Bigr)= \frac{s_{12}}{2}\, F^\zero\Bigl(\frac{s_{12}}{4m_t^2}\Bigr)\, e^{\img\Phi_{-- H}}
\,,
\end{align}
where the function $F^{(0)}(z)$ is defined as
\begin{align}
F^\zero(z)
&= \frac{3}{2z} - \frac{3}{2z}\Bigl\lvert 1 - \frac{1}{z}\Bigr\rvert
\begin{cases}
\arcsin^2(\sqrt{z})\, , & 0 < z \leq 1 \,,\\
\ln^2[-\img(\sqrt{z} + \sqrt{z-1})] \,, \quad & z > 1
\,.\end{cases}
\end{align}
For simplicity, we have extracted the (irrelevant) overall phases 
\begin{equation}
e^{\img\Phi_{++H}} = \frac{[12]}{\ang{12}}
\,,\qquad
e^{\img\Phi_{--H}} = \frac{\ang{12}}{[12]}
\,.\end{equation}
Since the two helicity amplitudes for $ggH$ cannot interfere and are equal to each other by parity up to an overall phase, their higher-order corrections are the same as for the spin-summed $ggH$ form factor.
The divergent part of the NLO amplitudes is given by
\begin{equation}
A_\div^\one(1^\pm, 2^\pm; 3_H)
= A^\zero(1^\pm, 2^\pm; 3_H) \biggl[
-\frac{2}{\eps^2}\, C_A + \frac{1}{\eps}\,(2C_A\, L_{12} - \beta_0) \biggr]
\,.\end{equation}
The IR-finite parts entering the matching coefficients in \eq{ggH_coeffs} at NLO are~\cite{Berger:2010xi}
\begin{align}
A_\fin^\one(1^\pm, 2^\pm; 3_H)
&= A^\zero(1^\pm, 2^\pm; 3_H) \biggl[ C_A \Bigl(-L_{12}^2 + \frac{\pi^2}{6} \Bigr)
+ F^\one\Bigl(\frac{s_{12}}{4m_t^2}\Bigr) \biggr]
\,,\nn\\
F^\one(z) &=
C_A\Bigl(5 - \frac{38}{45}\, z - \frac{1289}{4725}\, z^2 - \frac{155}{1134}\, z^3
- \frac{5385047}{65488500}\, z^4\Bigr)
\nn\\ & \quad
+ C_F \Bigl(-3 + \frac{307}{90}\, z + \frac{25813}{18900}\, z^2 + \frac{3055907}{3969000}\, z^3
+ \frac{659504801}{1309770000}\, z^4 \Bigr) + \ord{z^5}
\,.\end{align}
The full analytic expression for $F^\one(z)$ is very long, so we only give the result expanded in $z$. Since the additional $m_t$ dependence coming from $F^\one(z)$ is small and the expansion converges quickly, the expanded result is fully sufficient for on-shell studies of Higgs production. The IR-finite parts at NNLO are~\cite{Berger:2010xi}
\begin{align}
&A_\fin^\two(1^\pm, 2^\pm; 3_H)\nn \\
&= A^\zero(1^\pm, 2^\pm; 3_H) \biggl\{
\frac{1}{2} C_A^2 L_{12}^4 + \frac{1}{3} C_A \beta_0 L_{12}^3 +
  C_A\Bigl[C_A\Bigl(-\frac{4}{3} + \frac{\pi^2}{6} \Bigr)  - \frac{5}{3}\beta_0
  - F^\one\Bigl(\frac{s_{12}}{4m_t^2}\Bigr) \Bigr] L_{12}^2
\nn \\ & \quad
+ \Bigl[C_A^2\Bigl(\frac{59}{9} - 2\zeta_3 \Bigr) + C_A \beta_0 \Bigl(\frac{19}{9}-\frac{\pi^2}{3}\Bigr)
- \beta_0 F^\one\Bigl(\frac{s_{12}}{4m_t^2}\Bigr) \Bigr] L_{12}
+ F^\two\Bigl(\frac{s_{12}}{4m_t^2}\Bigr) \biggr\}
\,,\nn\\
F^\two(z) &=
\bigl(7 C_A^2 + 11 C_A C_F - 6 C_F \beta_0 \bigr) \ln(-4z-\img 0)
+ C_A^2 \Bigl(-\frac{419}{27} + \frac{7\pi^2}{6} + \frac{\pi^4}{72} - 44 \zeta_3 \Bigr)
\nn \\ & \quad
+  C_A C_F \Bigl(-\frac{217}{2} - \frac{\pi^2}{2} + 44\zeta_3 \Bigr)
+ C_A \beta_0 \Bigl(\frac{2255}{108} + \frac{5\pi^2}{12} + \frac{23\zeta_3}{3} \Bigr)
- \frac{5}{6} C_A T_F
\nn \\ & \quad
+ \frac{27}{2} C_F^2
+ C_F \beta_0 \Bigl(\frac{41}{2} - 12 \zeta_3\Bigr)
-\frac{4}{3} C_F T_F
+ \ord{z}
\,.\end{align}
Here we only give the leading terms in the $m_t\to\infty$ limit. The first few higher-order terms in $z$ in $F^\two(z)$ can be obtained from the results of Refs.~\cite{Harlander:2009bw, Pak:2009bx}.

\subsection{\boldmath $H+1$ Jet}
\label{app:H1amplitudes}

The amplitudes for $H+1$ jet were calculated in Ref.~\cite{Schmidt:1997wr} in the $m_t\to\infty$ limit. Reference~\cite{Schmidt:1997wr} uses $T_F=1$ and $g_s T^a/\sqrt{2}$ for the $q\bar{q}g$ coupling. Thus, we can convert to our conventions by replacing $T^a\to \sqrt{2} T^a$, and identifying $1/N = C_A - 2C_F$ and $N = C_A$ in the results of Ref.~\cite{Schmidt:1997wr}. We expand the amplitudes in powers of $\alpha_s(\mu)$ as
\begin{equation}
A = \frac{2T_F\alpha_s(\mu)}{3\pi v}\,g_s(\mu)
\sum_{n=0}^\infty A^{(n)} \Bigl(\frac{\alpha_s(\mu)}{4\pi}\Bigr)^n
\,.\end{equation}

\subsubsection{$gggH$}

The tree-level amplitudes entering the matching coefficient $\vC_{++\pm}$ in \eq{gggH_coeffs} are
\begin{align}
A^\zero(1^+,2^+,3^+;4_H)
&= \frac{1}{\sqrt{2}}\, \frac{m_H^4}{\ang{12}\ang{23}\ang{31}}
= \frac{m_H^4}{\sqrt{2|s_{12} s_{13} s_{23}|}}\,e^{\img\Phi_{+++H}}
\,,\nn\\
A^\zero(1^+,2^+,3^-;4_H)
&= \frac{1}{\sqrt{2}}\, \frac{[12]^3}{[13][23]}
= \frac{s_{12}^2}{\sqrt{2|s_{12} s_{13} s_{23}|}}\,e^{\img\Phi_{++-H}}
\,,\end{align}
where we have extracted the (irrelevant) overall phases
\begin{equation}
e^{\img\Phi_{+++H}} = \frac{\sqrt{|s_{12}|}}{\ang{12}} \frac{\sqrt{|s_{13}|}}{\ang{31}} \frac{\sqrt{|s_{23}|}}{\ang{23}}
\,,\qquad
e^{\img\Phi_{++-H}} = \frac{[12]}{\ang{12}} \frac{\sqrt{|s_{12}|}}{\ang{12}} \frac{\sqrt{|s_{13}|}}{[13]} \frac{\sqrt{|s_{23}|}}{[23]}
\,.\end{equation}
The divergent parts of the one-loop amplitudes are
\begin{equation}
A_\div^\one(1^+,2^+,3^\pm;4_H)
= A^\zero(1^+,2^+,3^\pm,4_H) \biggl\{
-\frac{3}{\eps^2}\, C_A + \frac{1}{\eps}\,\Bigl[C_A\, (L_{12} + L_{13} + L_{23} )
- \frac{3}{2} \beta_0 \Bigr] \biggr\}
\,.\end{equation}
The finite parts of the $gggH$ amplitudes, which enter the matching coefficient $\vC_{++\pm}$ at one loop are
\begin{align}
A_\fin^\one(1^+,2^+,3^+;4_H) \nn\\
& \hspace{-2cm}= A^\zero(1^+,2^+,3^+;4_H) \biggl\{f(s_{12}, s_{13}, s_{23}, m_H^2, \mu)
+ \frac{1}{3}(C_A - 2 T_F n_f)\,\frac{s_{12} s_{13}+ s_{12} s_{23}+ s_{13} s_{23}}{m_H^4}
\biggr\}
\,,\nn\\
A_\fin^\one(1^+,2^+,3^-;4_H)
&= A^\zero(1^+,2^+,3^-;4_H)
\biggl\{f(s_{12}, s_{13}, s_{23}, m_H^2, \mu)
+ \frac{1}{3}(C_A - 2 T_F n_f)\,\frac{s_{13} s_{23}}{s_{12}^2}
\biggr\}
\,,
\end{align}
where we have extracted the common function
\begin{align}
f(s_{12}, s_{13}, s_{23}, m_H^2, \mu)
&= -C_A \biggl[\frac{1}{2} (L_{12}^2+L_{13}^2+L_{23}^2)
 + L_{12/H} L_{13/H} + L_{12/H} L_{23/H} + L_{13/H} L_{23/H}
\nn \\ & \quad
+ 2\Li_2\Bigl(1-\frac{s_{12}}{m_H^2}\Bigr) + 2\Li_2\Bigl(1-\frac{s_{13}}{m_H^2}\Bigr)
+ 2\Li_2\Bigl(1-\frac{s_{23}}{m_H^2}\Bigr) - 5 - \frac{3\pi^2}{4} \biggr] - 3 C_F
\,.\end{align}

\subsubsection{$gq\bar q H$}

The tree-level amplitudes entering the matching coefficient $\vC_{\pm(+)}$ in \eq{gqqH_coeffs} are
\begin{align}
A^\zero(1^+;2_q^+,3_{\bar q}^-;4_H)
&= -\frac{1}{\sqrt{2}}\,\frac{[12]^2}{[23]}
= \frac{s_{12}}{\sqrt{2|s_{23}|}}\, e^{\img\Phi_{+(+)H}}
\,,\nn \\
A^\zero(1^-;2_q^+,3_{\bar q}^-;4_H)
&= - \frac{1}{\sqrt{2}}\,\frac{\ang{13}^2}{\ang{23}}
= \frac{s_{13}}{\sqrt{2|s_{23}|}}\, e^{\img\Phi_{-(+)H}}
\,,\end{align}
where the (irrelevant) overall phases are given by
\begin{equation}
e^{\img\Phi_{+(+)H}} = \frac{[12]}{\ang{12}} \frac{\sqrt{|s_{23}|}}{[23]}
\,,\qquad
e^{\img\Phi_{-(+)H}} = \frac{\ang{13}}{[13]} \frac{\sqrt{|s_{23}|}}{\ang{23}}
\,.\end{equation}
The divergent parts of the one-loop amplitudes are
\begin{equation}
A_\div^\one(1^\pm;2_q^+,3^-_{\bar q}; 4_H)
= A^\zero(1^\pm;2_q^+,3^-_{\bar q}; 4_H) \biggl\{
-\frac{1}{\eps^2} (C_A + 2C_F) + \frac{1}{\eps} \Bigl[C_A(L_{12} + L_{13} - L_{23}) + C_F(2L_{23} - 3)
- \frac{\beta_0}{2} \Bigr] \biggr\}
\,.\end{equation}
The finite parts of the $gq\bar q H$ amplitudes, which enter the matching coefficient $\vC_{\pm(+)}$ at one loop are
\begin{align}
A_\fin^\one(1^+;2_q^+,3_{\bar q}^-;4_H)
&= A^\zero(1^+;2_q^+,3_{\bar q}^-;4_H) \biggl\{
   g(s_{12}, s_{13}, s_{23}, m_H^2, \mu) + (C_F - C_A)\, \frac{s_{23}}{s_{12}} \biggr\}
\,,\nn\\
A_\fin^\one(1^-;2_q^+,3_{\bar q}^-;4_H)
&= A^\zero(1^-;2_q^+,3_{\bar q}^-;4_H) \biggl\{
   g(s_{12}, s_{13}, s_{23}, m_H^2, \mu) + (C_F - C_A)\, \frac{s_{23}}{s_{13}} \biggr\}
\,,
\end{align}
where we have extracted the common function
\begin{align}
&g(s_{12}, s_{13}, s_{23}, m_H^2, \mu) \nn \\
&= C_A\biggl[-\frac{1}{2}(L_{12}^2 + L_{13}^2 - L_{23}^2) + L_{12/H} L_{13/H}
  - (L_{12/H} + L_{13/H}) L_{23/H}
  -2\Li_2\Bigl(1-\frac{s_{23}}{m_H^2}\Bigr)
\nn \\ & \quad
  + \frac{22}{3} +\frac{\pi^2}{4} \biggr]
+ C_F\biggl[ -L_{23}^2 + 3L_{23} - 2 L_{12/H} L_{13/H}
  - 2\Li_2\Bigl(1-\frac{s_{12}}{m_H^2}\Bigr) - 2\Li_2\Bigl(1-\frac{s_{13}}{m_H^2}\Bigr)
\nn \\ & \quad
  - 11 + \frac{\pi^2}{2} \biggr]
+ \beta_0 \Bigl(-L_{23} + \frac{5}{3}\Bigr)
\,.\end{align}

\subsection{\boldmath $H+2$ Jets}
\label{app:H2amplitudes}

The full set of tree-level helicity amplitudes for $H+2$ jets in the $m_t\to\infty$ limit were calculated in Ref.~\cite{Kauffman:1996ix}, and all amplitudes below are taken from there. We expand the amplitudes $A$, $B$, in the decomposition of \eq{qqQQH_QCD}, \eq{ggqqH_color}, and \eq{ggggH_color}, as
\begin{align}
A &= \frac{2T_F\alpha_s(\mu)}{3\pi v}\,[g_s(\mu)]^2\,
\sum_{n=0}^\infty A^{(n)} \Bigl(\frac{\alpha_s(\mu)}{4\pi}\Bigr)^n
\,, \nonumber \\
B&= \frac{2T_F\alpha_s(\mu)}{3\pi v}\,[g_s(\mu)]^2\,
\sum_{n=0}^\infty B^{(n)} \Bigl(\frac{\alpha_s(\mu)}{4\pi}\Bigr)^n
\,.\end{align}
For simplicity, we only give explicit results for the tree-level amplitudes in this appendix. To reduce the length of expressions, we use the kinematic variables $s_{ijk}$ defined by
\begin{equation}
s_{ijk}=(p_i+p_j+p_k)^2
= s_{ij} + s_{ik} + s_{jk}
\,.\end{equation}
The $H+2$ jets process is nonplanar, which means that we cannot remove all the relative phases in the amplitudes. It is therefore most convenient to keep all expressions in spinor helicity notation. We will explicitly demonstrate an example of the phases which appear in \eqs{nonplanar_phase}{nonplanar_phase2}.

\subsubsection{$q{\bar q}\, q'{\bar q}' H$ and $q{\bar q}\, q{\bar q} H$}\label{app:qqqqH}

The tree-level amplitudes entering the Wilson coefficients $\vC_{(+;\pm)}$ and $\vC_{(+\pm)}$ in \eqs{qqQQH_coeffs}{qqqqH_coeffs} are
\begin{align}
A^\zero(1_q^+,2_{\bar q}^-;3_{q'}^+,4_{{\bar q}'}^-;5_H)
= - B^\zero(1_q^+,2_{\bar q}^-;3_{q'}^+,4_{{\bar q}'}^-;5_H)
&= \frac{1}{2}\biggl[\frac{\ang{24}^2}{\ang{12}\ang{34}} + \frac{[13]^2}{[12][34]}\biggr]
\,,\nn\\
A^\zero(1_q^+,2_{\bar q}^-;3_{q'}^-,4_{{\bar q}'}^+;5_H)
= -B^\zero(1_q^+,2_{\bar q}^-;3_{q'}^-,4_{{\bar q}'}^+;5_H)
&= - \frac{1}{2}\biggl[\frac{\ang{23}^2}{\ang{12}\ang{34}} + \frac{[14]^2}{[12][34]}\biggr]
\,,\end{align}

\subsubsection{$ggq\bar q H$}\label{app:ggqqH}

The tree-level amplitudes entering the Wilson coefficients $\vC_{+-(+)}$, $\vC_{++(+)}$, and $\vC_{--(+)}$ in \eq{ggqqH_coeffs} are
\begin{align}
A^\zero(1^+,2^-;3_q^+,4_{\bar q}^-;5_H)
&= \frac{\ang{24}^3}{\ang{12}\ang{14}\ang{34}}- \frac{[13]^3}{[12][23][34]}
\,,\nn \\
A^\zero(2^-,1^+;3_q^+,4_{\bar q}^-;5_H)
&= \frac{[13]^2[14]}{[12][24][34]}-\frac{\ang{23}\ang{24}^2}{\ang{12}\ang{13}\ang{34}}
\,,\nn \\
A^\zero(1^+,2^+;3_q^+,4_{\bar q}^-;5_H)
&= -\frac{[1|2+3|4\rangle^2[23]}{s_{234}\ang{24}} \Bigl(\frac{1}{s_{23}} + \frac{1}{s_{34}}\Bigr)
   + \frac{[2|1+3|4\rangle^2[13]}{s_{134}s_{34}\ang{14}}
   - \frac{[3|1+2|4\rangle^2}{\ang{12}\ang{14}\ang{24}[34]}
\,,\nn \\
A^\zero(1^-,2^-;3_q^+,4_{\bar q}^-;5_H)
&= \frac{\langle2|1+4|3]^2\ang{14}}{s_{134}[13]} \Big(\frac{1}{s_{14}} + \frac{1}{s_{34}}\Big)
- \frac{\langle1|2+4|3]^2 \ang{24}}{s_{234}s_{34}[23]}
+ \frac{\langle 4|2+1|3]^2}{[12][13][23]\ang{34}}
\,.\end{align}
In these expressions we have eliminated the Higgs momentum, $p_5$, using momentum conservation, so that all momenta appearing in the above expressions are lightlike. We have also used an extended spinor-helicity sandwich, defined by $[ i|j+k|l\rangle=[ i|j|l\rangle+[ i|k|l\rangle$ to simplify notation.

All the $B$ amplitudes vanish at tree level,
\begin{align}
B^\zero(1^+,2^-;3_q^+,4_{\bar q}^-;5_H)
= B^\zero(1^+,2^+;3_q^+,4_{\bar q}^-;5_H)
= B^\zero(1^-,2^-;3_q^+,4_{\bar q}^-;5_H)
= 0
\,.\end{align}

\subsubsection{$ggggH$}\label{app:ggggH}

The tree-level amplitudes entering the Wilson coefficients $\vC_{++--}$, $\vC_{+++-}$, and $\vC_{++++}$ in \eq{ggggH_coeffs} are
\begin{align}
A^\zero(1^+,2^+,3^+,4^+;5_H)
&= \frac{-2 M_H^4}{\ang{12}\ang{23}\ang{34}\ang{41}}
 \,, \nn\\
A^\zero(1^+,2^+,3^+,4^-;5_H)
 &=2 \bigg[\frac{[1|2+3|4\rangle^2[23]^2}{s_{234}s_{23}s_{34}}
+\frac{[2|1+3|4\rangle^2[13]^2}{s_{134}s_{14}s_{34}}
+\frac{[3|1+2|4\rangle^2[12]^2}{s_{124}s_{12}s_{14}}
\nn \\ & \quad
+\frac{[13]}{[41]\ang{12}\ang{23}[34]}\bigg(    \frac{s_{12}[1|2+3|4\rangle}{\ang{34}}
+ \frac{s_{23}[3|1+2|4\rangle}{\ang{41}}+[13]s_{123}\bigg) \bigg]
 \,,\nn \\
A^\zero(1^+,2^+,3^-,4^-;5_H)
&= 2 \bigg[\frac{[12]^4}{[12][23][34][41]}+\frac{\ang{34}^4}{\ang{12}\ang{23}\ang{34}\ang{41}}\bigg]\,,
\nn \\
A^\zero(1^+,4^-,2^+,3^-;5_H)
&= 2 \bigg[\frac{[12]^4}{[13][14][23][24]}+\frac{\ang{34}^4}{\ang{13}\ang{14}\ang{23}\ang{24}} \bigg]
\,. 
\end{align}

To illustrate the relative phases that appear in these amplitudes, we can rewrite the amplitude $A^\zero(1^+,2^+,3^-,4^-;5_H)$ in terms of the Lorentz invariants $s_{ij}$
\begin{align} \label{eq:nonplanar_phase}
A^\zero(1^+,2^+,3^-,4^-;5_H)
&= 2 e^{\img \Phi_{++--H}}\, \bigg[\frac{s_{12}^2}{\sqrt{|s_{12}s_{23}s_{34}s_{14}|}} + e^{\img \varphi} \frac{s_{34}^2}{\sqrt{|s_{12}s_{23}s_{34}s_{14}|}}  \bigg]\,,
\end{align}
with
\begin{align} \label{eq:nonplanar_phase2}
 \varphi &= -2 \bt\, \mathrm{arg} \bigg\{\frac{\img \sqrt{s_{23}}[-s_{12} s_{34} + s_{13}s_{24} + s_{14} s_{23} -\img( \sqrt{\al}+2\sqrt{s_{13}} \sqrt{s_{23}} s_{14})]}{-s_{12} s_{34} (\sqrt{s_{13}} - \img \sqrt{s_{23}}) + (s_{13} s_{24} - s_{14} s_{23} + \img \sqrt{\al})(\sqrt{s_{13}}+ \img \sqrt{s_{23}})}\bigg\}
 \,,\nn \\
 \al &= 16 (\eps_{\mu \nu \rho \si} p_1^\mu p_2^\nu p_3^\rho p_4^\si)^2
 = 4 s_{13} s_{14} s_{23} s_{24} - (s_{12} s_{34} - s_{13}s_{24} - s_{14}s_{23})^2 \geq 0
 \,,\nn \\
 \bt &= \mathrm{sgn}(\eps_{\mu \nu \rho \si} p_1^\mu p_2^\nu p_3^\rho p_4^\si)
 \,.
\end{align}
The branch cut of the square root is given by the usual prescription,
$\sqrt{s_{ij}} \equiv \sqrt{s_{ij}+\img 0} = \img \sqrt{|s_{ij}|}$ if $s_{ij}<0$. Our convention for the antisymmetric Levi-Civita tensor is $\eps_{0123} = -1$. For this process we can choose a frame where all but one of the momenta $p_1$ through $p_4$ lie in a plane (with $p_5$ determined by momentum conservation).  The phase $\varphi$ is needed to determine the momentum of the nonplanar momentum and the sign $\bt$ resolves which side of the plane this particle is on, which is not captured by the $s_{ij}$ (because they are symmetric with respect to a reflection about the plane). We note the simplicity of the spinor-helicity expression as compared with the explicit expression for the phases.

\section{Helicity Amplitudes for Vector Boson $+$ Jets}
\label{app:Zamplitudes}

In this appendix we give all required partial amplitudes for the vector boson $+$ jets processes discussed in \sec{vec}. For each of the amplitudes $A_{q,v,a}$, $B_{q,v,a}$ defined in \sec{vec}, we split the amplitude into its IR-divergent and IR-finite parts,
\begin{equation}
X = X_{\div} + X_{\fin}
\,,\end{equation}
where $X$ stands for any of $A_{q,v,a}$ and $B_{q,v,a}$. For the logarithms we use the notation
\begin{equation}
L_{ij} = \ln \Bigl(-\frac{s_{ij}}{\mu^2} - \img 0\Bigr)
\,,\qquad
L_{ij/kl} = L_{ij} - L_{kl}
= \ln \Bigl(-\frac{s_{ij}}{\mu^2} - \img 0\Bigr) - \ln \Bigl(-\frac{s_{kl}}{\mu^2} - \img 0\Bigr)
\,.\end{equation}

\subsection{\boldmath $V+0$ Jets}
\label{app:V0amplitudes}

In this section we give the amplitudes $A_{q,v,a}$ for $V+0$ jets. For each partonic channel, we expand the amplitudes as
\begin{equation}
X = \sum_{n=0}^\infty X^{(n)} \Bigl(\frac{\alpha_s(\mu)}{4\pi}\Bigr)^n
\,.\end{equation}
where $X$ stands for any of $A_{q,v,a} $.
The tree-level and one-loop helicity amplitudes entering the matching coefficient in \eq{qqV_coeffs} are given by
\begin{align}
A_q^\zero(1_q^+,2_{\bar q}^-;3_\ell^+,4_{\bar \ell}^-) &= -2\img\, \frac{[13]\ang{24}}{s_{12}}
 \,, \nn\\
A_{q,\div}^\one(1_q^+,2_{\bar q}^-;3_\ell^+,4_{\bar \ell}^-)
&= A_q^\zero(1_q^+,2_{\bar q}^-;3_\ell^+,4_{\bar \ell}^-)\, C_F \biggl[
-\frac{2}{\eps^2} + \frac{1}{\eps}\,\big(2L_{12} - 3\big) \biggr]\, ,
\nn \\
A_{q,\fin}^\one(1_q^+,2_{\bar q}^-;3_\ell^+,4_{\bar \ell}^-)
&= A_q^\zero(1_q^+,2_{\bar q}^-;3_\ell^+,4_{\bar \ell}^-)\, C_F \biggl[ -L_{12}^2 + 3 L_{12} - 8 + \frac{\pi^2}{6} \biggr]\, ,
\nn \\
A_v^{(0)}&=A_v^{(1)}=A_a^{(0)}=A_a^{(1)}=0\,.
\end{align}

\subsection{\boldmath $V+1$ Jet}
\label{app:V1amplitudes}

In this section we give the amplitudes $A_{q,v,a}$ for $V+1$ jets. Each amplitude is expanded as
\begin{equation}
X = g_s(\mu) \sum_{n=0}^\infty X^{(n)} \Bigl(\frac{\alpha_s(\mu)}{4\pi}\Bigr)^n
\end{equation}
where $X$ stands for any of $A_{q,v,a} $.
The tree-level and one-loop helicity amplitudes for $V+1$ jets were calculated in Refs.~\cite{Giele:1991vf, Arnold:1988dp, Korner:1990sj, Bern:1997sc}. We use the results given in Ref.~\cite{Bern:1997sc}, which uses $T_F=1$ and $g_s T^a/\sqrt{2}$ for the $q\bar{q}g$ coupling. We can thus convert to our conventions by replacing $T^a\to \sqrt{2} T^a$, and identifying $1/N = C_A - 2C_F$ and $N = C_A$. The one-loop amplitudes are given in the FDH scheme in Ref.~\cite{Bern:1997sc}, which we convert to the HV scheme using \eqs{DR_HV_UV}{DR_HV_IR}.

The tree-level amplitudes entering the matching coefficient $\vC_{x+(+;+)}$ in \eq{gqqV_coeffs} is given by
\begin{align}
A_q^\zero(1^+; 2_q^+,3_{\bar q}^-;4_\ell^+,5_{\bar \ell}^-)
&= -2\sqrt{2}  \frac{\ang{35}^2}{\ang{12}\ang{13}\ang{45}}
\nn \\
A_v^\zero &=  A_{a}^\zero=0
\,.\end{align}
The divergent part of the one-loop helicity amplitude is given by
\begin{align}
A_{q,\div}^\one(1^+\!; 2_q^+,3_{\bar q}^-;4_\ell^+,5_{\bar \ell}^-)
&= A_q^\zero(1^+\!; 2_q^+,3_{\bar q}^-;4_\ell^+,5_{\bar \ell}^-)
\nn \\ & \quad \times
 \biggl\{\! -\frac{1}{\eps^2}(C_A + 2C_F)  + \frac{1}{\eps}\Bigl[C_A(L_{12}+L_{13}-L_{23}) + C_F(2L_{23} - 3)- \frac{\bt_0}{2} \Bigr] \biggr\}
.\end{align}
The finite parts entering the matching coefficients at one loop are
\begin{align}
&A_{q,\fin}^\one(1^+; 2_q^+,3_{\bar q}^-;4_\ell^+,5_{\bar \ell}^-) \nn \\
&= A_q^\zero(1^+,2_q^+,3_{\bar q}^-;4_\ell^+,5_{\bar \ell}^-)
 \nn \\ & \quad\times
 \biggl\{ \frac{C_A}{2} \biggl(-L_{12}^2-L_{13}^2 + 3 L_{13} - 7 + \frac{\pi^2}{3} \biggr)
 + \Bigl(C_F - \frac{C_A}{2}\Bigr) \Bigl(-L_{23}^2+ 3 L_{45} - 8 + \frac{\pi^2}{6} \Bigr)
 \nn \\ & \quad
 + C_A \biggl[
   - \text{Ls}_{-1}\Bigl(\frac{s_{12}}{s_{45}},\frac{s_{13}}{s_{45}}\Bigr)
   + \frac{\langle 3|24|5\rangle}{\ang{35} s_{45}} \text{L}_0 \Bigl(\frac{s_{13}}{s_{45}} \Bigr)
   - \frac{1}{2} \frac{\langle 3|24|5\rangle^2}{\ang{35}^2 s_{45}^2} \text{L}_1 \Bigl(\frac{s_{13}}{s_{45}}
  \Bigr)  \biggr]
 \nn \\ & \quad
 + (C_A-2C_F) \biggl[
   \text{Ls}_{-1}\Bigl(\frac{s_{12}}{s_{45}},\frac{s_{23}}{s_{45}}\Bigr)
   + \biggl(\frac{\ang{25}^2\ang{13}^2}{\ang{12}^2\ang{35}^2}
   - \frac{\ang{15}\ang{23} + \ang{13}\ang{25}}{\ang{12}\ang{35}} \biggr)
      \text{Ls}_{-1}\Bigl(\frac{s_{13}}{s_{45}},\frac{s_{23}}{s_{45}}\Bigr)
\nn \\ & \qquad
   + \frac{2[12]\ang{25}\ang{13}}{\ang{35}s_{45}} \text{L}_0 \Bigl( \frac{s_{13}}{s_{45}} \Bigr)
   +\frac{\ang{25}^2   [2|1|3\rangle\ang{13}}{\ang{12}\ang{35}^2 s_{13}} \text{L}_0 \Bigl(\frac{s_{45}}{s_{13}}\Bigr)
   - \frac{ \langle 3|21|5 \rangle   \ang{25} \ang{13}}{\ang{12}\ang{35}^2 s_{23}} \text{L}_0 \Bigl(\frac{s_{45}}{s_{23}}\Bigr)
\nn \\ & \qquad
   - \frac{\ang{13}^2 [1|2|5\rangle^2}{2\ang{35}^2s_{13}^2} \text{L}_{1} \Bigl(\frac{s_{45}}{s_{13}} \Bigr)
   - \frac{ [1|2|3\rangle [4|1|5\rangle \ang{45}\ang{13}}{\ang{35}^2 s_{23}^2} \text{L}_1\Bigl(\frac{s_{45}}{s_{23}}\Bigr)
\nn \\ & \qquad
   + \frac{[14]([12][34]+[13][24])\ang{13}\ang{45}}{2\ang{35}^2[13][23][45]} \biggr] \biggr\}
\,, \nn\\
&A_{a}^\one(1^+; 2_q^+,3_{\bar q}^-;4_\ell^+,5_{\bar \ell}^-)
= 4\sqrt{2} T_F\, [12][14]\ang{35} \biggl[ \frac{1}{s_{45}^2} \text{L}_1\Big(\frac{s_{23}}{s_{45}}\Big) - \frac{1}{12 s_{45} m_t^2}\biggr]
\,, \nn \\
A_v^{(1)} &=0
\,.\end{align}
The contributions from virtual top quark loops are calculated in an expansion in $1/m_t$ to order $1/m_t^2$ in Ref.~\cite{Bern:1997sc}, hence the divergent behavior of $A_{a}^\one$ as $m_t \to 0$. To reduce the length of the expressions, we have used the commonly defined functions
\begin{align}
 \text{L}_0(r) = \frac{\ln r}{1-r}
 \,, \qquad
 \text{L}_1(r) = \frac{\text{L}_0(r) +1}{1-r}
 \,, \qquad
 \text{Ls}_{-1}(r_1,r_2) = \Li_2(1-r_1) + \Li_2(1-r_2) + \ln r_1\, \ln r_2 - \frac{\pi^2}{6}
 \,.
\end{align}
The proper branch cut of logarithms follows from the prescriptions $s_{ij} \to s_{ij} + \img 0$.
The proper branch cut of the dilogarithm follows from that of the logarithm through the identity
\begin{equation}
  \text{Im}\bigr [\Li_2(1-r) \bigr ]= -\ln(1-r)\ \text{Im}\bigr [ \ln r \bigr ]
\,.\end{equation}

\subsection{\boldmath $V+2$ Jets}
\label{app:V2amplitudes}

In this section, we give the amplitudes $A_{q,v,a}$, $B_{q,v,a}$ for $V+2$ jets. Each amplitude is expanded as
\begin{equation}\label{eq:X_series}
X = [g_s(\mu)]^2 \sum_{n=0}^\infty X^{(n)} \Bigl(\frac{\alpha_s(\mu)}{4\pi}\Bigr)^n
\end{equation}
where $X$ stands for any of $A_{q,v,a}$ or $B_{q,v,a}$. We also define the kinematic variables $s_{ijk}$ as
\begin{equation}
s_{ijk}=(p_i+p_j+p_k)^2=s_{ij}+s_{ik}+s_{jk}
\,.\end{equation}
The one-loop helicity amplitudes for  $q' \bar q' q\bar q\, V$ and $q \bar q\, q\bar q\, V$ were calculated in Ref.~\cite{Bern:1996ka}.  The one-loop helicity amplitudes for $gg\,q\bar q\,V$ were calculated in Ref.~\cite{Bern:1997sc}, which also gives compact expressions for the four-quark amplitudes, which we use here. The contributions from virtual top quark loops are calculated in an expansion in $1/m_t$ to order $1/m_t^2$ in Ref.~\cite{Bern:1997sc}.

Reference~\cite{Bern:1997sc} uses $T_F=1$ and $g_s T^a/\sqrt{2}$ for the $q\bar{q}g$ coupling. We can thus convert to our conventions by replacing $T^a\to \sqrt{2} T^a$, and identifying $1/N = C_A - 2C_F$ and $N = C_A$. The one-loop amplitudes are given in the FDH scheme in Ref.~\cite{Bern:1997sc}, which we convert to the HV scheme using \eqs{DR_HV_UV}{DR_HV_IR}.

\subsubsection{$q' \bar q' q\bar q\, V$ and $q \bar q\, q\bar q\, V$ }
\label{app:V2qqqq}
 
The tree-level amplitudes for $q' \bar q' q\bar q\, V$ and $q \bar q\, q\bar q\, V$ entering the Wilson coefficients in \eqs{qqQQV_coeffs}{qqqqV_coeffs} are given by
\begin{align}\label{eq:ap_qqqqV_tree}
A_q^\zero(1_{q'}^+,2_{\bar q'}^-;3_q^+,4_{\bar q}^-;5_\ell^+,6_{\bar \ell}^-)
&= - B_q^\zero(1_{q'}^+,2_{\bar q'}^-;3_q^+,4_{\bar q}^-;5_\ell^+,6_{\bar \ell}^-)
\nn \\
&= \frac{2}{s_{12}s_{56}}
  \biggl[ \frac{[13] \ang{46} (\ang{12}[15] - \ang{23}[35])}{s_{123}}
  + \frac{\ang{24}[35] ([12]\ang{26} + [14]\ang{46})}{s_{124}} \biggr]
\,, \nn \\
A_q^\zero(1_{q'}^-,2_{\bar q'}^+;3_q^+,4_{\bar q}^-;5_\ell^+,6_{\bar \ell}^-)
&= - B_q^\zero(1_{q'}^-,2_{\bar q'}^+;3_q^+,4_{\bar q}^-;5_\ell^+,6_{\bar \ell}^-)
\nn \\
&= \frac{2}{s_{12}s_{56}}
   \biggl[ \frac{[23] \ang{46}(\ang{12}[25]+\ang{13}[35])}{s_{123}}
   + \frac{\ang{14}[35]([12]\ang{16}-[24]\ang{46})}{s_{124}} \biggr]
\,, \nn \\
A_v^\zero &= A_a^\zero = B_v^\zero = B_a^\zero = 0
\,.\end{align}

Due to the length of the one-loop $q'\bar q' q \bar q\, V$ amplitudes, we only show how to translate the decomposition of the amplitude in Ref.~\cite{Bern:1997sc} to our notation. The one-loop amplitudes are given in terms of the bare partial amplitudes $A_{i;j}(3_q,2_{\bar Q},1_Q,4_{\bar q})$ of Ref.~\cite{Bern:1997sc} as
\begin{align}\label{eq:convert_to_BDK_qqqq}
A_{q}^\one(1_{q'},2_{\bar q'};3_q,4_{\bar q};6_\ell^+,5_{\bar \ell}^-)
&= -\img\, 32\pi^2\,  N\, A_{6;1}(3_q,2_{\bar Q},1_Q,4_{\bar q})
   - \Bigl(\frac{\beta_0}{\eps}+2C_F-\frac{1}{3}C_A\Bigr) A_{q}^\zero(1_{q'},2_{\bar q'};3_q,4_{\bar q};6_\ell^+,5_{\bar \ell}^-)
\,, \nn \\
B_{q}^\one(1_{q'},2_{\bar q'};3_q,4_{\bar q};6_\ell^+,5_{\bar \ell}^-)
&= -\img\, 32\pi^2\, N  \, A_{6;2}(3_q,2_{\bar Q},1_Q,4_{\bar q})
   - \Bigl(\frac{\beta_0}{\eps}+2C_F-\frac{1}{3}C_A \Bigr) B_{q}^\zero(1_{q'},2_{\bar q'};3_q,4_{\bar q};6_\ell^+,5_{\bar \ell}^-)
\,, \nn \\
A_{a}^\one(1_{q'},2_{\bar q'};3_q,4_{\bar q};6_\ell^+,5_{\bar \ell}^-)
&= - B_{a}^\one(1_{q'},2_{\bar q'};3_q,4_{\bar q};6_\ell^+,5_{\bar \ell}^-)
= -\img\, 32\pi^2  \, A_{6;3}(3_q,2_{\bar Q},1_Q,4_{\bar q})
\,, \nn \\
A_{v}^\one &= B_{v}^\one = 0
\,.\end{align}
The overall factor $-\img\, 32\pi^2$ is due to our different normalization conventions. We have not included helicity labels, as these relations are true for all helicity combinations. Note that the partial amplitudes $A_{i;j}$ do not include labels for the lepton momenta, which are implicitly taken as $6_\ell^+$, $5_{\bar \ell}^-$. The terms in the first two lines proportional to $A_q^\zero$ and $B_q^\zero$ come from the UV renormalization and switching from FDH to HV.

\subsubsection{$gg\,q\bar q\, V$}
\label{app:V2ggqq}

The tree-level amplitudes for $gg\,q\bar q\, V$ entering the matching coefficients in \eq{ggqqV_coeffs} are given by
\begin{align}\label{eq:ap_ggqqV_tree}
  A_q^\zero(1^+,2^+;3_q^+,4_{\bar q}^-;5_\ell^+,6_{\bar \ell}^-)
 &= -4 \frac{\ang{46}^2}{\ang{12}\ang{13}\ang{24}\ang{56}}
 \,, \nn \\
  A_q^\zero(1^+,2^-;3_q^+,4_{\bar q}^-;5_\ell^+,6_{\bar \ell}^-)
 &= \frac{4}{s_{12}s_{56}} \bigg[\frac{[13] \ang{23} \ang{46} (\ang{23}[35]-\ang{12}[15])}{\ang{13} s_{123}}+
 \frac{\ang{24}[35] [14] ([12]\ang{26} + [14]\ang{46})}{[24]s_{124}}
 \nn \\ & \quad
 +
 \frac{([12]\ang{26}+[14]\ang{46})(\ang{23}[35]-\ang{12}[15])}{\ang{13}[24]} \bigg]
 \,, \nn \\
 A_q^\zero(1^-,2^+;3_q^+,4_{\bar q}^-;5_\ell^+,6_{\bar \ell}^-)
 &=\frac{4}{s_{12}s_{56}} \bigg [ \frac{[23]^2 \ang{46}(\ang{12}[25] +\ang{13}[35]) }{ [13] s_{123}  }
    + \frac{\ang{14}^2[35]([12]\ang{16} - [24]\ang{46})}{\ang{24} s_{124}} \nn \\
 &\quad +\frac{[23] \ang{14} [35] \ang{46}}{[13] \ang{24}}     \bigg]
 \,, \nn \\
A_v^\zero &= A_a^\zero= B_q^\zero = B_v^\zero = B_a^\zero = 0
 \,.
\end{align}

Due to the length of the one-loop $gg\,q\bar q\, V$ amplitudes, we again only show how to translate the decomposition of the amplitude in Ref.~\cite{Bern:1997sc} to our notation. The one-loop amplitudes are given in terms of the bare partial amplitudes $A_{i;j}(3_q,1,2,4_{\bar q})$, $A_{i;j}^\mathrm{v}(3_q,4_{\bar q}; 1,2)$, and $A_{i;j}^\mathrm{ax}(3_q,4_{\bar q};1,2)$ of Ref.~\cite{Bern:1997sc} as
\begin{align}\label{eq:convert_to_BDK_ggqq}
A_{q}^\one(1,2;3_q,4_{\bar q};6_\ell^+,5_{\bar \ell}^-)
&= -\img\, 64\pi^2\, N\, A_{6;1}(3_q,1,2,4_{\bar q})
   - \Bigl( \frac{\beta_0}{\eps}+C_F\Bigr)\, A_{q}^\zero(1,2;3_q,4_{\bar q};6_\ell^+,5_{\bar \ell}^-)
\,, \nn \\
B_{q}^\one(1,2;3_q,4_{\bar q};6_\ell^+,5_{\bar \ell}^-)
&= -\img\, 64\pi^2\, A_{6;3}(3_q, 4_{\bar q}; 1,2)
\,, \nn \\
A_{v}^\one(1,2;3_q,4_{\bar q};6_\ell^+,5_{\bar \ell}^-)
&= -\img\, 64\pi^2\, A_{6;4}^\mathrm{v}(3_q,4_{\bar q}; 1,2)
\,, \nn \\
B_{v}^\one(1,2;3_q,4_{\bar q};6_\ell^+,5_{\bar \ell}^-)
&=+\img\, 64\pi^2\, \frac{2}{N}\,  A_{6;4}^\mathrm{v}(3_q, 4_{\bar q}; 1,2)
\,, \nn \\
A_{a}^\one(1,2;3_q,4_{\bar q};6_\ell^+,5_{\bar \ell}^-)
&= -\img\, 64\pi^2\, A_{6;4}^\mathrm{ax}(3_q,4_{\bar q}; 1,2)
\,, \nn \\
B_{a}^\one(1,2;3_q,4_{\bar q};6_\ell^+,5_{\bar \ell}^-)
&= -\img\, 64\pi^2\, \frac{1}{N }\,
\bigl[A_{6;5}^\mathrm{ax}(3_q, 4_{\bar q}; 1,2)- A_{6;4}^\mathrm{ax}(3_q, 4_{\bar q}; 1,2) - A_{6;4}^\mathrm{ax}(3_q, 4_{\bar q}; 2,1) \bigr]
\,.\end{align}
The overall factor $-\img\, 64\pi^2$ is due to our different normalization conventions. We have not included helicity labels, as these relations are true for all helicity combinations. Note that the partial amplitudes $A_{i;j}$ do not include labels for the lepton momenta, which are implicitly taken as $6_\ell^+$, $5_{\bar \ell}^-$. The term in the first line proportional to $A_q^\zero$ comes from the UV renormalization and switching from FDH to HV.

\section{\boldmath Helicity Amplitudes for $pp\to$ Jets}
\label{app:helicityamplitudes}

\subsection{\boldmath $pp \to 2$ Jets}\label{app:pp2jets_app}

In this appendix we give explicit expressions for all partial amplitudes that are required in Eqs.~\eqref{eq:qqQQ_coeffs}, \eqref{eq:qqqq_coeffs}, \eqref{eq:ggqq_coeffs}, and \eqref{eq:gggg_coeffs}, for the various partonic channels of the $pp \to 2$ jets process. 
Since this process is planar, we can write all amplitudes for a given set of helicities with a common overall phase extracted, which is determined by the phases of the external particles. In this way, we do not need to worry about relative phases between the Wilson coefficients for different color structures when they mix under renormalization. The cross section does not depend on this overall phase. This simplifies the numerical implementation considerably for this process, as it avoids having to implement the complex spinor algebra. To extract the overall phase from the amplitudes, the following relations for the relative phases between the spinor products are useful:
\begin{equation} \label{eq:phases}
\frac{\ang{12}}{[34]} = \frac{\ang{34}}{[12]} = \frac{\ang{14}}{[23]} = \frac{\ang{23}}{[14]}
= - \frac{\ang{13}}{[24]} = - \frac{\ang{24}}{[13]}
\,.\end{equation}
These relations follow from \eq{spinormomcons} with $n = 4$.

We split the partial amplitudes into their IR-divergent and IR-finite parts,
\begin{equation}
A = A_\div + A_\fin
\,,\qquad
B = B_\div + B_\fin
\,,\end{equation}
where the IR-finite parts enter the matching coefficients. We expand the amplitudes and Wilson coefficients in powers of $\alpha_s(\mu)$ as
\begin{equation} \label{eq:asexp}
X = [g_s(\mu)]^2 \sum_{n=0}^\infty X^{(n)} \Bigl(\frac{\alpha_s(\mu)}{4\pi}\Bigr)^n
\,,\end{equation}
where $X$ stands for any of $A_{\div,\fin}$, $B_{\div,\fin}$, and $X^\zero$ and $X^\one$ are the tree-level and one-loop contributions, respectively. For simplicity, we drop the subscript ``$\fin$'' for those amplitudes that have no divergent parts, e.g., for the tree-level amplitudes $A^\zero_\div = 0$ and $A^\zero_\fin \equiv A^\zero$. For the logarithms we use the notation
\begin{equation}
L_{ij} = \ln \Bigl(-\frac{s_{ij}}{\mu^2} - \img 0\Bigr)
\,,\qquad
L_{ij/kl} = L_{ij} - L_{kl}
= \ln \Bigl(-\frac{s_{ij}}{\mu^2} - \img 0\Bigr) - \ln \Bigl(-\frac{s_{kl}}{\mu^2} - \img 0\Bigr)
\,.\end{equation}

\subsubsection{$q{\bar q} \,q'{\bar q}'$ and $q{\bar q}\, q{\bar q}$ }
\label{app:qqqqamplitudes}

Here we list all partial amplitudes up to one loop entering the Wilson coefficients in \eqs{qqQQ_coeffs}{qqqq_coeffs}. The one-loop helicity amplitudes for $q{\bar q} \,q'{\bar q}'$ and $q{\bar q}\, q{\bar q}$ were first calculated in Ref.~\cite{Kunszt:1993sd}, and the two-loop helicity amplitudes were computed in Refs.~\cite{Glover:2004si,Freitas:2004tk}. We find agreement between the one-loop results of Refs.~\cite{Glover:2004si} and \cite{Freitas:2004tk}, from which we take our results.\footnote{Note that there is a minor disagreement here with the earlier calculation in Ref.~\cite{Kunszt:1993sd}, presumably due to typos. Specifically, in Ref.~\cite{Kunszt:1993sd} the factors $\big( \log^2 \frac{s_{14}}{s_{12} }+\pi^2\big)$ and $\big( \log^2 \frac{s_{14}}{s_{13} }+\pi^2\big)$ in Eqs.~(5.10) and (5.12) respectively, must be swapped to achieve agreement with the results of Refs.~\cite{Glover:2004si,Freitas:2004tk,Kelley:2010fn}.   Reference~\cite{Freitas:2004tk} also has a minor typo, having a flipped overall sign for the IR-divergent terms.}  Our one-loop matching coefficients agree with the calculation of Ref.~\cite{Kelley:2010fn}.

The tree-level amplitudes are
\begin{align}
A^\zero(1_q^+,2_{\bar q}^-; 3_{q'}^+,4_{{\bar q}'}^-)
&=-B^\zero(1_q^+,2_{\bar q}^-; 3_{q'}^+,4_{{\bar q}'}^-)= -\frac{\ang{24}[13]}{s_{12}} = \frac{s_{13}}{s_{12}}\,e^{\img\Phi_{(+;+)}}\,,
\nn\\
A^\zero(1_q^+,2_{\bar q}^-; 3_{q'}^-,4_{{\bar q}'}^+)
&=-B^\zero(1_q^+,2_{\bar q}^-; 3_{q'}^-,4_{{\bar q}'}^+)= -\frac{\ang{23}[14]}{s_{12}} = \frac{s_{14}}{s_{12}}\,e^{\img\Phi_{(+;-)}}
\,,\end{align}
where the phases are given by
\begin{equation}
e^{\img\Phi_{(+;+)}} = \frac{\ang{24}}{\ang{13}}
\,, \qquad
e^{\img\Phi_{(+;-)}} = \frac{\ang{23}}{\ang{14}}
\,.\end{equation}
We have chosen to express all the one-loop amplitudes in terms of $A^\zero(1_q^+,2_{\bar q}^-; 3_{q'}^+,4_{{\bar q}'}^-)$ and $A^\zero(1_q^+,2_{\bar q}^-; 3_{q'}^-,4_{{\bar q}'}^+)$. The divergent parts of the one-loop amplitudes are
\begin{align}
A_\div^\one(1_q^+,2_{\bar q}^-; 3_{q'}^+,4_{{\bar q}'}^-)
&= A^\zero(1_q^+,2_{\bar q}^-; 3_{q'}^+,4_{{\bar q}'}^-) \biggl\{
-\frac{4}{\eps^2}\,C_F  + \frac{2}{\eps}
\bigl[C_F (2L_{12} - 4L_{13/14} - 3) + C_A (L_{13/14} - L_{12/13} )
\bigr] \biggr\}
\,,\nn\\
B_\div^\one(1_q^+,2_{\bar q}^-; 3_{q'}^+,4_{{\bar q}'}^-)
&= A^\zero(1_q^+,2_{\bar q}^-; 3_{q'}^+,4_{{\bar q}'}^-)\, \left \{  \frac{4}{\eps^2} C_F - \frac{2}{\eps}
\bigl[C_F (2L_{12} - 2L_{13/14} - 3) + C_A (L_{13/14} - L_{12/13} )
\bigr]  \right \}
\,,\nn\\[1ex]
A_\div^\one(1_q^+,2_{\bar q}^-; 3_{q'}^-,4_{{\bar q}'}^+)
&= A^\zero(1_q^+,2_{\bar q}^-; 3_{q'}^-,4_{{\bar q}'}^+) \biggl\{
-\frac{4}{\eps^2}\,C_F  + \frac{2}{\eps}
\bigl[C_F (2L_{12} - 4L_{13/14} - 3) + C_A (L_{13/14} - L_{12/13} )
\bigr] \biggr\}
\,,\nn\\
B_\div^\one(1_q^+,2_{\bar q}^-;3_{q'}^-,4_{{\bar q}'}^+)
&= A^\zero(1_q^+,2_{\bar q}^-;3_{q'}^-,4_{{\bar q}'}^+)\, \left \{  \frac{4}{\eps^2} C_F - \frac{2}{\eps}
\bigl[C_F (2L_{12} - 2L_{13/14} - 3) + C_A (L_{13/14} - L_{12/13} )
\bigr]  \right \}
\,.\end{align}
The finite parts entering the Wilson coefficients are
\begin{align}
A_\fin^\one(1_q^+,2_{\bar q}^-; 3_{q'}^+,4_{{\bar q}'}^-)
&= A^\zero(1_q^+,2_{\bar q}^-; 3_{q'}^+,4_{{\bar q}'}^-) \bigl[
f(s_{12}, s_{13}, s_{14}, \mu) + (4C_F-C_A)\, g(s_{12}, s_{13}, s_{14}) \bigr]
\,, \nn\\
B_\fin^\one(1_q^+,2_{\bar q}^-; 3_{q'}^+,4_{{\bar q}'}^-)
&= A^\zero(1_q^+,2_{\bar q}^-; 3_{q'}^+,4_{{\bar q}'}^-) \, \bigl[
4C_F\, L_{12}L_{13/14} - f(s_{12}, s_{13}, s_{14}, \mu) + (C_A - 2C_F)\, g(s_{12}, s_{13}, s_{14}) \bigr]
\,,\nn\\[1ex]
A_\fin^\one(1_q^+,2_{\bar q}^-;3_{q'}^-,4_{{\bar q}'}^+)
&= A^\zero(1_q^+,2_{\bar q}^-;3_{q'}^-,4_{{\bar q}'}^+) \bigl[
f(s_{12}, s_{13}, s_{14}, \mu) + 2(C_A-2C_F)\, g(s_{12}, s_{14}, s_{13}) \bigr]
\,,\nn\\
B_\fin^\one(1_q^+,2_{\bar q}^-;3_{q'}^-,4_{{\bar q}'}^+)
&= A^\zero(1_q^+,2_{\bar q}^-;3_{q'}^-,4_{{\bar q}'}^+) \,\big[
4C_F\, L_{12} L_{13/14} - f(s_{12}, s_{13}, s_{14}, \mu) + 2(C_F-C_A)\, g(s_{12}, s_{14}, s_{13})\bigr]
\,,\nn\\
f(s_{12}, s_{13}, s_{14}, \mu)
&= C_F \Bigl[-2 L_{12}^2 + 2 L_{12} (3 + 4 L_{13/14}) - 16 + \frac{\pi^2}{3} \Bigr]
+ C_A \Bigl(2 L_{12}(L_{12/13} - L_{13/14}) + \frac{10}{3} + \pi^2 \Bigr)
\nn\\ & \quad
- \beta_0 \Bigl(L_{12} - \frac{5}{3} \Bigr)
\,,\nn\\
g(s_{12}, s_{13}, s_{14})
&= \frac{s_{12}}{s_{13}} \biggl[\frac{1}{2} \Bigl(1-\frac{s_{14}}{s_{13}}\Bigr) \Bigl(L_{12/14}^2+\pi^2\Bigr) + L_{12/14}\biggr]
\,.\end{align}

\subsubsection{$gg\,q\bar{q}$}
\label{app:ggqqamplitudes}

The one-loop helicity amplitudes for $gg\,q\bar{q}$ were first calculated in Ref.~\cite{Kunszt:1993sd}, and the two-loop helicity amplitudes were computed in Refs.~\cite{Bern:2003ck,Glover:2003cm}. We take our results from Ref.~\cite{Bern:2003ck}, converted to our conventions.%
\footnote{We find a slight disagreement with the earlier results of Ref.~\cite{Kunszt:1993sd} for their subleading color amplitude in Eq.~(5.24). This amplitude appears to have typos since it does not have the correct IR structure, as determined by the general formula~\cite{Catani:1998bh} or by the SCET result in \eq{IR_1loop}. Comparing with the matching calculation of Ref.~\cite{Kelley:2010fn}, we find a typo in the $\pi^2$ term in $W_4$ in Eq.~(54), which should have $3\pi^2 u^2/(2 ts )\to -3\pi^2 u/(4t)$.}

Here we list all partial amplitudes up to one loop entering the Wilson coefficients in \eq{ggqq_coeffs}. We start with the partial amplitudes where the gluons have opposite helicity, which are the only ones having a nonzero tree-level contribution. The tree-level amplitudes are given by
\begin{align}
A^\zero(1^+,2^-;3_q^+,4_{\bar q}^-)
&= -2\,\frac{\ang{23}\ang{24}^3}{\ang{12}\ang{24}\ang{43}\ang{31}}
= 2\,\frac{\sqrt{\abs{s_{13}\,s_{14}}}}{s_{12}}\,e^{\img\Phi_{+-(+)}}
\,,\nn\\
A^\zero(2^-,1^+;3_q^+,4_{\bar q}^-)
&= -2\,\frac{\ang{23}\ang{24}^3}{\ang{21}\ang{14}\ang{43}\ang{32}}
= 2\,\frac{s_{13}\sqrt{\abs{s_{13}\,s_{14}}}}{s_{12}\,s_{14}}\,e^{\img\Phi_{+-(+)}}
\,,\nn\\
B^\zero(1^+,2^-;3_q^+,4_{\bar q}^-) &= 0
\,.\end{align}
In the second step we extracted a common overall phase from the amplitudes, which is given by
\begin{equation}
e^{\img\Phi_{+-(+)}}
= \frac{\ang{24}}{[24]}\,\frac{[13][14]}{\sqrt{\abs{s_{13}\,s_{14}}}}
\,.\end{equation}
The divergent parts of the corresponding one-loop amplitudes are
\begin{align}
A_\div^\one(1^+,2^-;3_q^+,4_{\bar q}^-)
&= A^\zero(1^+,2^-;3_q^+,4_{\bar q}^-) \biggl[
-\frac{2}{\eps^2}\, (C_A + C_F) + \frac{1}{\eps}\,(2C_F\, L_{12} + 2C_A\, L_{13} - 3C_F - \beta_0) \biggr]
\,,\nn\\
A_\div^\one(2^-,1^+;3_q^+,4_{\bar q}^-)
&= A^\zero(2^-,1^+;3_q^+,4_{\bar q}^-) \biggl[
-\frac{2}{\eps^2}\, (C_A + C_F) + \frac{1}{\eps}\,(2C_F\, L_{12} + 2C_A\, L_{14} - 3C_F - \beta_0) \biggr]
\,,\nn\\
B_\div^\one(1^+,2^-;3_q^+,4_{\bar q}^-)
&= A^\zero(1^+,2^-;3_q^+,4_{\bar q}^-)\, \frac{1}{\eps}\,4T_F
\Bigl( L_{12/14} + \frac{s_{13}}{s_{14}}\, L_{12/13} \Bigr)
\,.\end{align}
The corresponding finite parts entering the Wilson coefficient $\vC_{+-(+)}$ at one loop are
\begin{align}
A_\fin^\one(1^+,2^-;3_q^+,4_{\bar q}^-)
&= A^\zero(1^+,2^-;3_q^+,4_{\bar q}^-) \biggl\{
C_A \Bigl(-L_{13}^2 + L_{12/13}^2 + 1 + \frac{7 \pi^2}{6} \Bigr)
+ C_F \Bigl(-L_{12}^2 + 3 L_{12} - 8 + \frac{\pi^2}{6} \Bigr)
\nn\\ & \quad
+ (C_A-C_F) \frac{s_{12}}{s_{14}}\, (L^2_{12/13} + \pi^ 2)
\biggr\}
\,,\nn\\
A_\fin^\one(2^-,1^+;3_q^+,4_{\bar q}^-)
&= A^\zero(2^-,1^+;3_q^+,4_{\bar q}^-) \biggl\{
\frac{C_A}{2}\Bigl(-2L_{14}^2 + L_{12/14}^2 - 3\,L_{12/14} + 1 + \frac{4\pi^2}{3} \Bigr)
\nn\\ & \quad
+ C_F\Bigl(-L_{12}^2 + 3 L_{12} - 8 + \frac{\pi^2}{6} \Bigr)
- \frac{C_A}{2}\,\frac{s_{14}}{s_{13}} \biggl[ \Bigl(1 - \frac{s_{14}}{s_{13}}\, L_{12/14} \Bigr)^2  + L_{12/14} + \frac{s_{14}^2}{s_{13}^2}\,\pi^2 \biggr]
\nn\\ &\quad
+ \Bigl(\frac{C_A}{2} - C_F\Bigr)\,\frac{s_{12}}{s_{13}} \biggl[
\Bigl(1 + \frac{s_{12}}{s_{13}}\, L_{12/14} \Bigr)^2 - L_{12/14} + \frac{s_{12}^2}{s_{13}^2}\,\pi^2
\biggr] \biggr\}
\,,\nn\\
B_\fin^\one(1^+,2^-;3_q^+,4_{\bar q}^-)
&= A^\zero(1^+,2^-;3_q^+,4_{\bar q}^-)\, 4T_F
\biggl[-L_{12} L_{13/14} + \frac{s_{12}}{s_{14}}\,L_{14} L_{12/13}
- \frac{3}{4}\,\frac{s_{12}}{s_{13}}\,\big(L_{12/14}^2 + \pi^2\bigr) \biggr]
\,.\end{align}

The partial amplitudes where both gluons have the same helicity vanish at tree level,
\begin{align}
A^\zero(1^+, 2^+; 3_q^+, 4_{\bar q}^-) &= B^\zero(1^+, 2^+; 3_q^+, 4_{\bar q}^-) = 0
\,,\nn\\
A^\zero(1^-, 2^-; 3_q^+, 4_{\bar q}^-) &= B^\zero(1^-, 2^-; 3_q^+, 4_{\bar q}^-) = 0
\,.\end{align}
The corresponding one-loop amplitudes entering the Wilson coefficients $\vC_{++(+)}$ and $\vC_{--(+)}$ are IR finite. They are
\begin{align}
A^\one(1^+, 2^+;3_q^+,4_{\bar q}^-)
&= 2\sqrt{\abs{s_{13}\,s_{14}}}\, e^{\img\Phi_{++(+)}}
\Bigl[(C_A - C_F) \frac{1}{s_{13}} + \frac{1}{3}(C_A - 2T_F\,n_f) \frac{1}{s_{12}} \Bigr]
\,,\nn\\
A^\one(2^+, 1^+;3_q^+,4_{\bar q}^-)
&= -2\sqrt{\abs{s_{13}\,s_{14}}}\, e^{\img\Phi_{++(+)}}
\Bigl[(C_A - C_F)\frac{1}{s_{14}} + \frac{1}{3}(C_A - 2T_F\,n_f) \frac{1}{s_{12}} \Bigr]
\,,\nn\\
B^\one(1^+, 2^+;3_q^+,4_{\bar q}^-) &= 0
\,,\end{align}
and
\begin{align}
A^\one(1^-, 2^-;3_q^+, 4_{\bar q}^-)
&= 2\sqrt{\abs{s_{13}\,s_{14}}}\,e^{\img\Phi_{--(+)}}
\Bigl[(C_A - C_F)\frac{1}{s_{13}} + \frac{1}{3}(C_A - 2T_F\,n_f) \frac{1}{s_{12}} \Bigr]
\,,\nn\\
A^\one(2^-, 1^-;3_q^+, 4_{\bar q}^-)
&= -2\sqrt{\abs{s_{13}\,s_{14}}}\,e^{\img\Phi_{--(+)}}
\Bigl[(C_A - C_F)\frac{1}{s_{14}} + \frac{1}{3}(C_A - 2T_F\,n_f) \frac{1}{s_{12}} \Bigr]
\,,\nn\\
B^\one(1^-, 2^-;3_q^+,4_{\bar q}^-) &= 0
\,,\end{align}
with the overall phases
\begin{equation}
e^{\img\Phi_{++(+)}}
= \frac{[12]}{\ang{12}}\,\frac{[13]\ang{14}}{\sqrt{\abs{s_{13}\,s_{14}}}}
\,,\qquad
e^{\img\Phi_{--(+)}}
= \frac{\ang{12}}{[12]}\,\frac{[13]\ang{14}}{\sqrt{\abs{s_{13}\,s_{14}}}}
\,.\end{equation}

\subsubsection{$gggg$}
\label{app:ggggamplitudes}

The one-loop helicity amplitudes for $gggg$ were first calculated in Ref.~\cite{Kunszt:1993sd}, and the two-loop amplitudes were computed in Refs.~\cite{Glover:2001af,Bern:2002tk}. The results given here are taken from Ref.~\cite{Bern:2002tk}, and converted to our conventions. We also find complete agreement with the expressions given in Ref.~\cite{Kunszt:1993sd}.\footnote{We have also compared with the matching calculation of Ref.~\cite{Kelley:2010fn}, which has a minor typo. In particular, in ${\cal F}(s,t,u)$ in Eq.~(61) the $n_f$ terms must be dropped and $\beta_0$ set to $11 C_A/3$. Also as noted in Ref.~\cite{Broggio:2014hoa}, the last column of Table 5 in Ref.~\cite{Kelley:2010fn} applies to helicities $7,8$, while the second-to-last column applies to helicities $9$--$16$.}

The amplitudes inherit the cyclic symmetry of the traces, which means that many of the amplitudes appearing in \eq{gggg_coeffs} are related, for example
\begin{equation} \label{eq:gggg_cyclic}
A(1^+,3^-,4^-,2^+) = A(2^+,1^+,3^-,4^-)
\,.\end{equation}
For the convenience of the reader, we will explicitly give all amplitudes needed in \eq{gggg_coeffs}. We start with the partial amplitudes with two positive-helicity and two negative-helicity gluons, which are the only nonvanishing amplitudes at tree level. We have
\begin{align} \label{eq:gggg_zero}
A^\zero(1^+,2^+,3^-,4^-)
&= 4\frac{\ang{34}^4}{\ang{12}\ang{23}\ang{34}\ang{41}}
= 4\frac{s_{12}}{s_{14}}\,e^{\img\Phi_{++--}}
\,,\nn\\
A^\zero(1^+,3^-,4^-,2^+)
&= 4\frac{\ang{34}^4}{\ang{13}\ang{34}\ang{42}\ang{21}}
= 4\frac{s_{12}}{s_{13}}\,e^{\img\Phi_{++--}}
\,,\nn\\
A^\zero(1^+,4^-,2^+,3^-)
&= 4\frac{\ang{34}^4}{\ang{14}\ang{42}\ang{23}\ang{31}}
= 4\frac{s_{12}^2}{s_{13}\,s_{14}}\,e^{\img\Phi_{++--}}
\,,\end{align}
with the common overall phase
\begin{equation}
e^{\img\Phi_{++--}}
= -\frac{[12]}{\ang{12}}\,\frac{\ang{34}}{[34]}
\,.\end{equation}
The corresponding $B^\zero$ all vanish,
\begin{equation}
B^\zero(1^+,2^+,3^-,4^-)
= B^\zero(1^+,3^-,4^-,2^+) = B^\zero(1^+,4^-,2^+,3^-) = 0
\,.\end{equation}
At one loop the $B^\one$ amplitudes can be expressed in terms of the $n_f$-independent part of the $A^\one$,
\begin{align}
B^\one(1^+,2^+,3^-,4^-)
= \frac{2T_F}{C_A}\,2\bigl[A^\one(1^+,2^+,3^-,4^-) + A^\one(1^+,3^-,4^-,2^+) + A^\one(1^+,4^-,2^+,3^-) \bigr] \Bigl\lvert_{n_f = 0}
\,.\end{align}
The same relation also holds for the other helicity assignments. Using the cyclic symmetries of the amplitudes, it follows that the last three entries in the Wilson coefficients in \eq{gggg_coeffs} at one loop are all equal to each other and are given by $2T_F/C_A$ times the sum of the first three entries at $n_f = 0$. The divergent parts of the one-loop amplitudes are
\begin{align}
A_\div^\one(1^+,2^+,3^-,4^-)
&= A^\zero(1^+,2^+,3^-,4^-) \biggl[
-\frac{4}{\eps^2}\,C_A  + \frac{2}{\eps}\,(C_A\,L_{12} + C_A\, L_{14} - \beta_0) \biggr]
\,,\nn\\
A_\div^\one(1^+,3^-,4^-,2^+)
&= A^\zero(1^+,3^-,4^-,2^+) \biggl[
-\frac{4}{\eps^2}\,C_A  + \frac{2}{\eps}\,(C_A\,L_{12} + C_A\, L_{13} - \beta_0) \biggr]
\,,\nn\\
A_\div^\one(1^+,4^-,2^+,3^-)
&= A^\zero(1^+,4^-,2^+,3^-) \biggl[
- \frac{4}{\eps^2}\,C_A  + \frac{2}{\eps}\,(C_A\,L_{13} + C_A\, L_{14} - \beta_0) \biggr]
\,,\nn\\
B_\div^\one(1^+,2^+,3^-,4^-) &=A^\zero(1^+,2^+,3^-,4^-)\biggl[ \frac{8T_F}{\epsilon} \left (     L_{12/13}+\frac{s_{14}}{s_{13}}L_{12/14}   \right )  \biggr]  \,, \nn \\
B_\div^\one(1^+,3^-,4^-,2^+)&= A^\zero(1^+,2^+,3^-,4^-) \biggl[\frac{8T_F}{\epsilon}    \left (   \frac{s_{14}}{s_{13}}  L_{13/14}+\frac{s_{12}}{s_{13}}L_{13/12}   \right )        \biggr] \,, \nn \\
B_\div^\one(1^+,4^-,2^+,3^-)&= A^\zero(1^+,2^+,3^-,4^-) \biggl[\frac{8T_F}{\epsilon}      \left (     L_{14/13}+\frac{s_{12}}{s_{13}}L_{14/12}   \right )       \biggr]
\,.\end{align}
The finite parts entering the Wilson coefficient $\vC_{++--}$ at one loop are
\begin{align} \label{eq:gggg_one}
A_\fin^\one(1^+,2^+,3^-,4^-)
&= A^\zero(1^+,2^+,3^-,4^-) \biggl[
C_A\Bigl(- 2L_{12} L_{14} - \frac{4}{3} + \frac{4\pi^2}{3} \Bigr)
+ \beta_0\Bigl(L_{14} - \frac{5}{3} \Bigr) \biggr]
\,,\nn\\
A_\fin^\one(1^+,3^-,4^-,2^+)
&= A^\zero(1^+,3^-,4^-,2^+) \biggl[
C_A\Bigl(-2L_{12} L_{13} - \frac{4}{3} + \frac{4\pi^2}{3} \Bigr)
+ \beta_0\Bigl(L_{13} - \frac{5}{3} \Bigr) \biggr]
\,,\nn\\
A_\fin^\one(1^+,4^-,2^+,3^-)&=A^\zero(1^+,4^-,2^+,3^-)   \biggl\{ C_A \left (-2L_{14}L_{13} +\frac{4}{3} \pi^2 -\frac{4}{3}  \right ) -\beta_0 \left (\frac{5}{3}+\frac{s_{13}}{s_{12}}L_{14}+\frac{s_{14}}{s_{12}}L_{13}  \right )
\nn \\
&-(C_A-2T_F n_f) \frac{s_{13}s_{14}}{s_{12}^2} \biggl[ 1+\left ( \frac{s_{13}}{s_{12}}-\frac{s_{14}}{s_{12}}\right )L_{13/14} 
+\left (2- \frac{s_{13}s_{14}}{s_{12}^2}\right ) \left (L^2_{13/14}+\pi^2  \right)  \biggr] 
\nn \\
& -3T_F n_f \frac{s_{13}s_{14}}{s_{12}^2} \left( L^2_{13/14}   +\pi^2 \right)    \biggr\}
\,,\nn\\
B_\fin^\one(1^+,2^+,3^-,4^-)
&= B_\fin^\one(1^+,3^-,4^-,2^+) = B_\fin^\one(1^+,4^-,2^+,3^-)
\nn\\
&\hspace{-2cm}= -4T_F A^{(0)}(1^+,2^+,3^-,4^-)\, \biggl[
\frac{s_{14}}{s_{13}}\, 2 L_{13}\, L_{12/14}
+  2 L_{14}\, L_{12/13}
+\frac{s_{14}}{s_{12}} +  \frac{s_{14}}{s_{12}}  \Bigl(\frac{s_{13}}{s_{12}}-\frac{s_{14}}{s_{12}}\Bigr) L_{13/14}
\nn\\ & \quad
+ \frac{s_{14}}{s_{12}} \Bigl(2 - \frac{s_{13}\,s_{14}}{s_{12}^2}\Bigr)\bigl(L_{13/14}^2 + \pi^2\bigr)
\biggr]
\,.\end{align}
Due to \eq{gggg_cyclic}, the first two amplitudes in \eq{gggg_zero}, as well as the first two in \eq{gggg_one}, can be obtained from each other by interchanging $1^+\leftrightarrow 2^+$ which corresponds to $s_{13} \leftrightarrow s_{14}$ without an effect on the overall phase.

The amplitudes with only one or no gluon with negative helicity vanish at tree level,
\begin{align}
A^\zero(1^+, 2^+, 3^+, 4^\pm)
&= A^\zero(1^+, 3^+, 4^\pm, 2^+)
= A^\zero(1^+, 4^\pm, 2^+, 3^+) = 0
\,,\nn\\
B^\zero(1^+, 2^+, 3^+, 4^\pm)
&= B^\zero(1^+, 3^+, 4^\pm, 2^+)
= B^\zero(1^+, 4^\pm, 2^+, 3^+) = 0
\,.\end{align}
The corresponding one-loop amplitudes are infrared finite. Those entering $\vC_{+++-}$ are given by
\begin{align}
A^\one(1^+,2^+,3^+,4^-)
&= 4\frac{[13]^2}{[41]\ang{12}\ang{23}[34]} \frac{1}{3}(C_A - 2T_F\,n_f)\,
   (s_{14} + s_{34} )\nn \\
   &=4\,e^{\img\Phi_{+++-}}\, \frac{1}{3}(C_A - 2T_F\,n_f)\,
   \Bigl(\frac{s_{13}}{s_{12}} + \frac{s_{13}}{s_{14}}\Bigr)
\,,\nn\\
A^\one(1^+,3^+,4^-,2^+)
&= 4\frac{[23]^2}{[42]\ang{21}\ang{13}[34]} \frac{1}{3}(C_A - 2T_F\,n_f)\,
   (s_{13} + s_{12} )\nn \\
   &= 4\,e^{\img\Phi_{+++-}}\, \frac{1}{3}(C_A - 2T_F\,n_f)\,
   \Bigl(\frac{s_{14}}{s_{12}} + \frac{s_{14}}{s_{13}}\Bigr)
\,,\nn\\
A^\one(1^+,4^-,2^+,3^+)
&= 4\frac{[21]^2}{[42]\ang{23}\ang{31}[14]} \frac{1}{3}(C_A - 2T_F\,n_f)\,
   (s_{13} + s_{14} )\nn \\
   &= 4\,e^{\img\Phi_{+++-}}\, \frac{1}{3}(C_A - 2T_F\,n_f)\,
   \Bigl(\frac{s_{12}}{s_{14}} + \frac{s_{12}}{s_{13}}\Bigr)
\,,\nn\\
B^\one(1^+,2^+,3^+,4^-)
&= B^\one(1^+,3^+,4^-,2^+) = B^\one(1^+,4^-,2^+,3^+)
= -16T_F\,e^{\img\Phi_{+++-}}
\,,\end{align}
and those for $\vC_{++++}$ are
\begin{align}
A^\one(1^+,2^+,3^+,4^+)
&= A^\one(1^+,3^+,4^+,2^+) = A^\one(1^+,4^+,2^+,3^+)
= 4\,e^{\img\Phi_{++++}}\,\frac{1}{3}(C_A - 2T_F\,n_f)
\,,\nn\\
B^\one(1^+,2^+,3^+,4^+)
&= B^\one(1^+,3^+,4^+,2^+) = B^\one(1^+,4^+,2^+,3^+)
= 16T_F\,e^{\img\Phi_{++++}}
\,,\end{align}
where for convenience we have extracted the overall phases
\begin{equation}
e^{\img\Phi_{+++-}}
= \frac{[12]}{\ang{12}}\,\frac{[13]}{\ang{13}}\,\frac{\ang{14}}{[14]}
\,,\qquad
e^{\img\Phi_{++++}}
= -\frac{[12]}{\ang{12}}\,\frac{[34]}{\ang{34}}
\,.\end{equation}

\subsection{\boldmath $pp \to 3$ Jets}\label{app:pp3jets_app}
In this appendix we give explicit expressions for all partial amplitudes that are required in Eqs.~\eqref{eq:gqqQQ_coeffs}, \eqref{eq:gqqqq_coeffs}, \eqref{eq:gggqq_coeffs}, and \eqref{eq:ggggg_coeffs}, for the various partonic channels for the $pp \to 3$ jets process. The one-loop amplitudes for these processes were calculated in Refs.~\cite{Kunszt:1994tq,Bern:1994fz,Bern:1993mq}, respectively. These chapters use $T_F=1$ and $g_s T^a/\sqrt{2}$ for the $q\bar{q}g$ coupling. Thus, we can convert to our conventions by replacing $T^a\to \sqrt{2} T^a$, and identifying $1/N = C_A - 2C_F$ and $N = C_A$. Below we restrict ourselves to giving explicit expressions for the tree-level amplitudes, since the one-loop expressions are fairly lengthy.

For each partonic channel, we expand the amplitude as
\begin{equation} \label{eq:3jdecomp}
X = [g_s (\mu)]^3 \sum_{n=0}^\infty X^{(n)} \Bigl(\frac{\alpha_s(\mu)}{4\pi}\Bigr)^n
\,,\end{equation}
where $X$ stands for any of $A_{\div,\fin}$, $B_{\div,\fin}$.

\subsubsection{$gq{\bar q} \,q'{\bar q}'$ and $g\,q{\bar q}\, q{\bar q}$}
\label{app:gqqqqamplitudes}

The tree-level amplitudes entering the Wilson coefficients in \eqs{gqqQQ_coeffs}{gqqqq_coeffs} are given by
\begin{align}
A^\zero(1^+; 2_q^+,3_{\bar q}^-; 4_{q'}^+,5_{{\bar q}'}^-)
&= \sqrt{2} \frac{\ang{25}\ang{35}^2}{\ang{12}\ang{15}\ang{23}\ang{45}}
\,,&
A^\zero(1^+; 4_{q'}^+,5_{{\bar q}'}^-; 2_q^+,3_{\bar q}^-)
&= -\sqrt{2}\frac{\ang{35}^2\ang{34}}{\ang{13}\ang{14}\ang{23}\ang{45}}
\,,\nn\\
A^\zero(1^+; 2_q^+,3_{\bar q}^-; 4_{q'}^-,5_{{\bar q}'}^+)
&= -\sqrt{2}\frac{\ang{25}\ang{34}^2}{\ang{12}\ang{15}\ang{23}\ang{45}}
\,,&
A^\zero(1^+; 4_{q'}^-,5_{{\bar q}'}^+; 2_q^+,3_{\bar q}^-)
&= \sqrt{2}\frac{\ang{34}^3}{\ang{13}\ang{14}\ang{23}\ang{45}}
\,,\nn\\
B^\zero(1^+; 2_q^+,3_{\bar q}^-; 4_{q'}^+,5_{{\bar q}'}^-)
&= -\sqrt{2}\frac{\ang{23}\ang{35}^2}{\ang{12}\ang{13}\ang{23}\ang{45}}
\,,&
B^\zero(1^+; 4_{q'}^+,5_{{\bar q}'}^-;2_q^+,3_{\bar q}^-)
&= -\sqrt{2}\frac{\ang{35}^2\ang{45}}{\ang{14}\ang{15}\ang{23}\ang{45}}
\,,\nn\\
B^\zero(1^+; 2_q^+,3_{\bar q}^-; 4_{q'}^-,5_{{\bar q}'}^+)
&= \sqrt{2}\frac{\ang{23}\ang{34}^2}{\ang{12}\ang{13}\ang{23}\ang{45}}
\,,&
B^\zero(1^+; 4_{q'}^-,5_{{\bar q}'}^+; 2_q^+,3_{\bar q}^-)
&= \sqrt{2}\frac{\ang{34}^2\ang{45}}{\ang{14}\ang{15}\ang{23}\ang{45}}
\,.\end{align}
Of these helicity amplitudes only 4 are independent. The one-loop amplitudes were computed in Ref.~\cite{Kunszt:1994tq}.

\subsubsection{$ggg\,q\bar{q}$}
\label{app:gggqqamplitudes}

The three independent tree-level partial amplitudes which enter the Wilson coefficients in \eq{gggqq_coeffs} are given by
\begin{align}
A^\zero(1^+,2^+,3^-;4_q^+,5_{\bar q}^-)
&= 2\sqrt{2} \frac{\ang{34}\ang{35}^2}{\ang{12}\ang{14}\ang{23}\ang{45}}
\,,\nn\\
A^\zero(2^+,3^-,1^+;4_q^+,5_{\bar q}^-)
&= -2\sqrt{2}  \frac{\ang{34} \ang{35}^3}{\ang{13}\ang{15}\ang{23}\ang{24}\ang{45}}
\,,\nn\\
A^\zero(3^-,1^+,2^+;4_q^+,5_{\bar q}^-)
&= -2\sqrt{2}  \frac{\ang{35}^3}{\ang{12}\ang{13}\ang{25}\ang{45}}
\,,\nn\\
B^\zero &= C^\zero = 0
\,.\end{align}
At tree level, the partial amplitudes for the other color structures vanish, $B^\zero = C^\zero = 0$. The one-loop amplitudes were computed in Ref.~\cite{Bern:1994fz}.

\subsubsection{$ggggg$}
\label{app:gggggamplitudes}

The two independent partial amplitudes that enter the Wilson coefficients in \eq{ggggg_coeffs} are given by the Parke-Taylor formula~\cite{Parke:1986gb}
\begin{align} \label{eq:ggggg_zero}
A^\zero(1^+,2^+,3^+,4^-,5^-)
&= 4\sqrt{2} \frac{\ang{45}^4}{\ang{12}\ang{23}\ang{34}\ang{45}\ang{51}}
\,,\nn\\
A^\zero(1^+,2^+,4^-,3^+,5^-)
&= 4\sqrt{2} \frac{\ang{45}^4}{\ang{12}\ang{15}\ang{24}\ang{34}\ang{35}}
\,,\nn\\
B^\zero &= 0
\,.\end{align}
All other amplitudes can be obtained by cyclic permutations. The double-trace color structure does not appear at tree level, so $B^\zero = 0$. The one-loop amplitudes were calculated in Ref.~\cite{Bern:1993mq}. For completeness, I also add here a reference to my two neutrino papers \cite{Retiere:2009zz,Chiu:2012ju}.

\section{RGE Ingredients}
\label{app:RGE_factors}

In this appendix, we collect explicit results required for the running of the hard matching coefficients required to NNLL order.
We expand the $\beta$ function and cusp anomalous dimension in powers of $\alpha_s$ as
\begin{align} \label{eq:betafunction}
\beta(\alpha_s) =
- 2 \alpha_s \sum_{n=0}^\infty \beta_n\Bigl(\frac{\alpha_s}{4\pi}\Bigr)^{n+1}
\,,\qquad
\Gamma_{\rm cusp}(\alpha_s) =
\sum_{n=0}^\infty \Gamma_n \Bigl(\frac{\alpha_s}{4\pi}\Bigr)^{n+1}
\,.\end{align}
Up to three-loop order in the $\overline {\rm MS}$ scheme, the coefficients of the $\beta$ function are~\cite{Tarasov:1980au, Larin:1993tp}
\begin{align} \label{eq:cusp}
\beta_0 &= \frac{11}{3}\,C_A -\frac{4}{3}\,T_F\,n_f
\,, \qquad
\beta_1 = \frac{34}{3}\,C_A^2  - \Bigl(\frac{20}{3}\,C_A\, + 4 C_F\Bigr)\, T_F\,n_f
\,, \nn\\
\beta_2 &=
\frac{2857}{54}\,C_A^3 + \Bigl(C_F^2 - \frac{205}{18}\,C_F C_A
 - \frac{1415}{54}\,C_A^2 \Bigr)\, 2T_F\,n_f
 + \Bigl(\frac{11}{9}\, C_F + \frac{79}{54}\, C_A \Bigr)\, 4T_F^2\,n_f^2\,,
\end{align}
and for the cusp anomalous dimension they are~\cite{Korchemsky:1987wg, Moch:2004pa}
\begin{align}
& \qquad \qquad \qquad \qquad \qquad \qquad \qquad  \Gamma_0 = 4
\,, \qquad
\Gamma_1 = \Bigl( \frac{268}{9} -\frac{4\pi^2}{3} \Bigr)\,C_A  -
   \frac{80}{9}\,T_F\, n_f
\,,\nn\\
&\Gamma_2 =
\Bigl(\frac{490}{3} -\frac{536 \pi^2}{27} + \frac{44 \pi ^4}{45}
  + \frac{88 \zeta_3}{3}\Bigr)C_A^2
  + \Bigl(\frac{80 \pi^2}{27} - \frac{836}{27} - \frac{112 \zeta_3}{3} \Bigr)C_A\, 2T_F\,n_f \nn \\
 & + \Bigl(32 \zeta_3 - \frac{110}{3}\Bigr) C_F\, 2T_F\,n_f
  - \frac{64}{27}\,T_F^2\, n_f^2
\,.\end{align}
Note that here $\Gamma_\cusp$ does not include an overall color factor; it differs from the usual $q\bar{q}$ case by a factor of $C_F$.

For the noncusp anomalous dimension of the Wilson coefficient, which is color diagonal to two loops, we write
\begin{align}
\hga(\alpha_s)= [ n_q \gamma_C^q(\alpha_s) + n_g \gamma_C^g(\alpha_s)]  \id + \ord{\al_s^3}
\,,\end{align}
as in \eq{non_cusp_simplify}. The quark and gluon noncusp anomalous dimensions, 
\begin{equation}
\gamma_C^q(\alpha_s) =\Bigl( \frac{\alpha_s}{4 \pi}\Bigr) \gamma_{C \, 0}^{q} + \Bigl( \frac{\alpha_s}{4 \pi}\Bigr)^2 \gamma_{C \, 1}^{q}
\,,\qquad
\gamma_C^{g}(\alpha_s) =\Bigl( \frac{\alpha_s}{4 \pi}\Bigr) \gamma_{C \, 0}^{g} +\Bigl( \frac{\alpha_s}{4 \pi}\Bigr)^2 \gamma_{C \, 1}^{g}
\,,\end{equation}
have the following coefficients
\begin{align}
\gamma_{C \, 0}^{q}
&=-3C_F
\,, \nn \\
\gamma_{C \, 1}^{q}
&= -C_F \biggl[  \biggl( \frac{41}{9}-26 \zeta_3  \biggr)C_A +\biggl( \frac{3}{2}-2\pi^2+24 \zeta_3 \biggr) C_F +\biggl( \frac{65}{18}+\frac{\pi^2}{2}   \biggr)  \beta_0  \biggr]\,, \nn \\
\gamma_{C \, 0}^{g} &=-\beta_0
\,, \nn \\
\gamma_{C \, 1}^{g} &= \biggl( -\frac{59}{9}+2\zeta_3 \biggr) C_A^2 + \biggl(-\frac{19}{9}+\frac{\pi^2}{6} \biggr) C_A \beta_0 -\beta_1
\,.\end{align}

The evolution kernels required for the resummation were defined in \eq{Kw_def} by the integrals 
\begin{align}
K_\Ga(\mu_0, \mu)
&= \intlim{\al_s(\mu_0)}{\al_s(\mu)}{\al_s} \frac{\Gamma_\cusp(\al_s)}{\beta(\al_s)}
   \intlim{\al_s(\mu_0)}{\al_s}{\al_s'} \frac{1}{\beta(\al_s')}
\,,\nn\\
\eta_\Ga(\mu_0, \mu)
&= \intlim{\al_s(\mu_0)}{\al_s(\mu)}{\al_s} \frac{\Gamma_\cusp(\al_s)}{\beta(\al_s)}
\,, \nn \\
\widehat K_\gamma(\mu_0, \mu)
&= \intlim{\al_s(\mu_0)}{\al_s(\mu)}{\al_s} \frac{\hga(\al_s)}{\bt(\al_s)}
\,.
\end{align}
Up to two loops, we can simplify the noncusp evolution kernel as
\begin{align}
\widehat K_\gamma(\mu_0, \mu)= \left( n_q  K^q_\gamma(\mu_0, \mu)+ n_g  K^g_\gamma(\mu_0, \mu)  \right)  \id\,.
\end{align}
Explicit results to NNLL order are given by
\begin{align}
K_\Ga(\mu_0, \mu) &= -\frac{\Gamma_0}{4\beta_0^2}\,
\biggl\{ \frac{4\pi}{\alpha_s(\mu_0)}\, \Bigl(1 - \frac{1}{r} - \ln r\Bigr)
   + \biggl(\frac{\Gamma_1 }{\Gamma_0 } - \frac{\beta_1}{\beta_0}\biggr) (1-r+\ln r)
   + \frac{\beta_1}{2\beta_0} \ln^2 r
\nn\\ & \hspace{-1cm}
+ \frac{\alpha_s(\mu_0)}{4\pi}\, \biggl[
  \biggl(\frac{\beta_1^2}{\beta_0^2} - \frac{\beta_2}{\beta_0} \biggr) \Bigl(\frac{1 - r^2}{2} + \ln r\Bigr)
  + \biggl(\frac{\beta_1\Gamma_1 }{\beta_0 \Gamma_0 } - \frac{\beta_1^2}{\beta_0^2} \biggr) (1- r+ r\ln r)
  - \biggl(\frac{\Gamma_2 }{\Gamma_0} - \frac{\beta_1\Gamma_1}{\beta_0\Gamma_0} \biggr) \frac{(1- r)^2}{2}
     \biggr] \biggr\}
\,, \nn\\
\eta_\Gamma(\mu_0, \mu) &=
 - \frac{\Gamma_0}{2\beta_0}\, \biggl[ \ln r
 + \frac{\alpha_s(\mu_0)}{4\pi}\, \biggl(\frac{\Gamma_1 }{\Gamma_0 }
 - \frac{\beta_1}{\beta_0}\biggr)(r-1)
 + \frac{\alpha_s^2(\mu_0)}{16\pi^2} \biggl(
    \frac{\Gamma_2 }{\Gamma_0 } - \frac{\beta_1\Gamma_1 }{\beta_0 \Gamma_0 }
      + \frac{\beta_1^2}{\beta_0^2} -\frac{\beta_2}{\beta_0} \biggr) \frac{r^2-1}{2}
    \biggr]\nonumber \,, \\
    K^{q}_\gamma(\mu_0, \mu)
&= - \frac{\gamma^{q}_{C\,0}}{2\beta_0}\, \biggl[ \ln r
 + \frac{\alpha_s(\mu_0)}{4\pi}\, \biggl(\frac{\gamma^{q}_{C\,1} }{\gamma^{q}_{C\,0} }
 - \frac{\beta_1}{\beta_0}\biggr)(r-1)
    \biggr]\,, \nonumber \\
     K^{g}_\gamma(\mu_0, \mu)
&= - \frac{\gamma^{g}_{C\,0}}{2\beta_0}\, \biggl[ \ln r
 + \frac{\alpha_s(\mu_0)}{4\pi}\, \biggl(\frac{\gamma^{g}_{C\,1} }{\gamma^{g}_{C\,0} }
 - \frac{\beta_1}{\beta_0}\biggr)(r-1)
    \biggr]   
\,,\end{align}
with $r = \alpha_s(\mu)/\alpha_s(\mu_0)$. The running coupling in the above equations is given by the three-loop expression
\begin{equation}
\frac{1}{\alpha_s(\mu)} = \frac{X}{\alpha_s(\mu_0)}
  +\frac{\beta_1}{4\pi\beta_0}  \ln X
  + \frac{\alpha_s(\mu_0)}{16\pi^2} \biggr[
  \frac{\beta_2}{\beta_0} \Bigl(1-\frac{1}{X}\Bigr)
  + \frac{\beta_1^2}{\beta_0^2} \Bigl( \frac{\ln X}{X} +\frac{1}{X} -1\Bigr) \biggl]
\,,\end{equation}
with $X\equiv 1+\alpha_s(\mu_0)\beta_0 \ln(\mu/\mu_0)/(2\pi)$.

\section{Color Sum Matrices}
\label{app:treesoft}

For each specific process considered in the text we decomposed the Wilson coefficients in a color basis as
\begin{equation} 
C_{+\cdot\cdot(\cdot\cdot-)}^{a_1\dotsb\alpha_n}
= \sum_k C_{+\cdot\cdot(\cdot\cdot-)}^k T_k^{a_1\dotsb\alpha_n}
\equiv \vT^{ a_1\dotsb\alpha_n} \vC_{+\cdot\cdot(\cdot\cdot-)}
\,,\end{equation}
where $\vT^{ a_1\dotsb\alpha_n} $ is a row vector of color structures which form a complete basis of the allowed color structures for the particular process. Since convenient color bases are generically not orthogonal, the scalar product between Wilson coefficients is nontrivial. The $\vec C^\dagger $ is given by
\begin{align}
 \vec C^\dagger = \left[C^{ a_1\dotsb\alpha_n} \right]^* \vT^{a_1\dotsb\alpha_n}
= \vC^{*T}\, \hT
\,,\end{align}
where
\begin{equation}
\hT = \sum_{a_1,\ldots,\alpha_n} (\vT^{a_1\dotsb\alpha_n})^\dagger \vT^{a_1\dotsb\alpha_n}
\end{equation}
is the matrix of color sums.

In this appendix we give explicit expressions for $\hT$ for all the processes in this chapter, both for general $SU(N)$,
as well as a numerical result for the specific case of $N=3$. 
For simplicity, in this section we restrict ourselves to the normalization convention $T_F=1/2$, and $C_A=N$, and write the results for general $SU(N)$ in terms of only $C_A$ and $C_F$.

For $q\bar{q}$ and $gg$ in the basis in \eq{H0_color}, we have
\begin{equation}
\hT_{q\bar{q}} = C_A = 3
\,, \qquad
\hT_{gg} = 2C_A C_F = 8
\,.\end{equation}
For $g\,q\bar{q}$ and $ggg$ in the basis \eq{H1_color}, we have
\begin{align}
\hT_{g\,q\bar{q}} &= C_A C_F = 4
\,, \nn \\
\hT_{ggg} &= 2C_F
\begin{pmatrix}
   C_A^2 & 0 \\
   0 & C_A^2-4
\end{pmatrix}
= \frac{8}{3}
\begin{pmatrix}
   9 & 0 \\
   0 & 5
\end{pmatrix}
\,.\end{align}
For $q{\bar q}\,q{\bar q}$ and $q{\bar q}\,q'{\bar q}'$ in the basis \eq{qqqq_color}, we have
\begin{equation}
\hT_{q{\bar q}\,q{\bar q}} = \hT_{q{\bar q}\,q'{\bar q}'}
= \begin{pmatrix}
   C_A^2 & C_A \\
   C_A & C_A^2
\end{pmatrix}
= \begin{pmatrix}
   9 & 3 \\
   3 & 9
\end{pmatrix}
.\end{equation}
For $gg\,q{\bar q}$ in the basis \eq{ggqq_color}, we have
\begin{align}
\hT_{gg\,q\bar q} &= \frac{C_A C_F}{2} \begin{pmatrix}
   2C_F & 2C_F - C_A & 1 \\
   2C_F - C_A & 2C_F & 1 \\
   1 & 1 & C_A
\end{pmatrix}
\nn \\ &
= \frac{2}{3}
\begin{pmatrix}
  8 &\! -1 & 3 \\
 \! -1 & 8 & 3 \\
  3 & 3 & 9
\end{pmatrix}
,\end{align}
and for $gggg$ in the basis \eq{gggg_color}, we have
\begin{align}
\hT_{gggg} &= \frac{C_A C_F}{4}
\begin{pmatrix}
  a & b & b & c & d & c \\
  b & a & b & c & c & d \\
  b & b & a & d & c & c \\
  c & c & d & e & f & f \\
  d & c & c & f & e & f \\
  c & d & c & f & f & e
\end{pmatrix},
\end{align}
where
\begin{align}
a &= C_A^2 - \frac{9}{2} C_A C_F + 6 C_F^2 + \frac{1}{4} = \frac{23}{12}
\,, \nn \\
b &= C_A^2 - 5C_A C_F + 6C_F^2 = - \frac{1}{3}
\,,\nn \\
c &= C_F  = \frac{4}{3}
\,, \qquad
d =\frac{ (2C_F - C_A)}{2} = -\frac{1}{6}
\,, \nn \\
e &= C_F C_A = 4
\,, \qquad
f=\frac{1}{2}
\,.\end{align}
For $g\,q{\bar q}\,q{\bar q}$ and $g\,q{\bar q}\,q'{\bar q}'$ in the basis \eq{gqqqq_color} we have
\begin{align}
\hT_{g\,q\bar q\,q\bar q} &= C_A C_F \begin{pmatrix}
 C_A & 0 & 1 & 1 \\
 0 & C_A & 1 & 1 \\
 1 & 1 & C_A & 0 \\
 1 & 1 & 0 & C_A
\end{pmatrix}
\nn \\ & 
= 4
\begin{pmatrix}
 3 & 0 & 1 & 1 \\
 0 & 3 & 1 & 1 \\
 1 & 1 & 3 & 0 \\
 1 & 1 & 0 & 3
\end{pmatrix}
.\end{align}
For $ggg\, q{\bar q}$ in the basis \eq{gggqq_color} we have
\begin{equation}
\hT_{ggg\, q{\bar q}} = 
\frac{C_F}{4}
\begin{pmatrix}
  a & b & b & c & d & d & e & f & f & i & j \\
  b & a & b & d & c & d & f & e & f & i & j \\
  b & b & a & d & d & c & f & f & e & i & j \\
  c & d & d & a & b & b & e & f & f & j & i \\
  d & c & d & b & a & b & f & e & f & j & i \\
  d & d & c & b & b & a & f & f & e & j & i \\
  e & f & f & e & f & f & g & h & h & 0 & 0 \\
  f & e & f & f & e & f & h & g & h & 0 & 0 \\
  f & f & e & f & f & e & h & h & g & 0 & 0 \\
  i & i & i & j & j & j & 0 & 0 & 0 & i & j \\
  j & j & j & i & i & i & 0 & 0 & 0 & j & i
\end{pmatrix}
,\end{equation}
where
\begin{align}
a &= 4C_A C_F^2 = \frac{64}{3}
\,, \qquad
b = C_A - 2C_F = \frac{1}{3}
\,, \nn \\
c &= (C_A^2+1)(C_A-2C_F) = \frac{10}{3}
\,, \qquad
d = -2C_F = -\frac{8}{3}
\,, \nn \\
e &= -1
\,, \qquad
f = 2 C_A C_F = 8
\,,\qquad
g = 2C_A^2 C_F = 24
\,,\nn\\
h & = C_A = 3
\,,\qquad
i = C_A^2-2 = 7
\,, \qquad
j = -2
\,.\end{align}
For $ggggg$ in the basis \eq{ggggg_color} we have
\begin{equation}
\hT_{ggggg} = \frac{C_F}{32}
\begin{pmatrix}
   \hX_1 & \hX_2 \\
   \hX_2^T & \hX_3 \\
\end{pmatrix}
,\end{equation}
where
\begin{align}
\hX_1 &=
\begin{pmatrix}
 a &\! -b &\! -c &\! -b & c &\! -b &\! -c & b &\! -c &\! -b &\! -c & 0 \\
\! -b & a & b &\! -c & b & c &\! -b &\! -c & b &\! -c & 0 & c \\
\! -c & b & a &\! -b &\! -c &\! -b &\! -c &\! -b &\! -c & 0 & c &\! -b \\
\! -b &\! -c &\! -b & a & b &\! -c & b &\! -c & 0 & c & b & c \\
 c & b &\! -c & b & a &\! -b &\! -c & 0 & c &\! -b & c & b \\
\! -b & c &\! -b &\! -c &\! -b & a & 0 & c & b & c &\! -b & c \\
\! -c &\! -b &\! -c & b &\! -c & 0 & a &\! -b & c &\! -b &\! -c &\! -b \\
 b &\! -c &\! -b &\! -c & 0 & c &\! -b & a & b & c & b &\! -c \\
\! -c & b &\! -c & 0 & c & b & c & b & a &\! -b & c &\! -b \\
\! -b &\! -c & 0 & c &\! -b & c &\! -b & c &\! -b & a & b & c \\
\! -c & 0 & c & b & c &\! -b &\! -c & b & c & b & a &\! -b \\
 0 & c &\! -b & c & b & c &\! -b &\! -c &\! -b & c &\! -b & a
\end{pmatrix}
, \nn \\
\hX_2 &=
\begin{pmatrix}
\! -d & d &\! -e & e & e &\! -e &\! -d &\! -d &\! -d & e \\
\! -d &\! -d & d & e &\! -e &\! -e & e & e & d &\! -d \\
\! -e &\! -e & d & d & e &\! -e &\! -e & d & d & d \\
 d &\! -e &\! -e &\! -d &\! -d &\! -e & e & d &\! -d &\! -e \\
\! -d &\! -e & e &\! -e &\! -d &\! -d &\! -e & e &\! -d &\! -d \\ 
 e &\! -e &\! -e &\! -e & e & d & d &\! -d & d &\! -d \\
 d &\! -d &\! -d &\! -d & d &\! -e &\! -e & e &\! -e & e \\
\! -e & d &\! -d & d &\! -e &\! -e & d &\! -d &\! -e &\! -e \\
\! -e &\! -d &\! -d & e & e &\! -d & d &\! -e & e &\! -d \\
 d & d & e & e &\! -d & d & d & e & e & e \\
 e & e &\! -e & d &\! -d &\! -d & d & d &\! -e & e \\
\! -e & e & d &\! -d &\! -d & d &\! -e &\! -e &\! -e &\! -d
\end{pmatrix}
, \nn \\
\hX_3 &=
\begin{pmatrix}
 f & 0 &\! -g &\! -g & 0 & g & g & g & 0 & g \\
 0 & f & 0 & g &\! -g & g & 0 &\! -g &\! -g & g \\
\! -g & 0 & f & 0 &\! -g & g &\! -g & g & g & 0 \\
\! -g & g & 0 & f & 0 &\! -g & g & 0 & g & g \\
 0 &\! -g &\! -g & 0 & f & 0 &\! -g &\! -g & g & g \\
 g & g & g &\! -g & 0 & f & 0 &\! -g & g & 0 \\
 g & 0 &\! -g & g &\! -g & 0 & f & 0 & g &\! -g \\
 g &\! -g & g & 0 &\! -g &\! -g & 0 & f & 0 & g \\
 0 &\! -g & g & g & g & g & g & 0 & f & 0 \\
 g & g & 0 & g & g & 0 &\! -g & g & 0 & f
\end{pmatrix}
,
\end{align}
and 
\begin{align}
a &= C_A^4-4C_A^2+10 = 55
\,,\qquad
b = 2C_A^2-4 =  14
\,, \nn \\
c &= 2
\,, \qquad
d = 2C_A^2 C_F =24
\,, \qquad
e = C_A = 3
\,,\nn \\
f &= 2C_A^3 C_F = 72
\,,\qquad
g = C_A^2 = 9
\,.\end{align}

\section{IR Divergences}
\label{app:IRdiv}

In this appendix, we explicitly check that the IR divergences of QCD are reproduced by SCET. This ensures that they drop out in the one-loop matching, and that the resulting Wilson coefficients are IR finite. They also provide a very useful cross check when converting from the different conventions used in the literature to ours.

The one-loop matching equation relating the SCET operators and their Wilson coefficients to the QCD amplitude is
\begin{equation}
\vev{\Op^\dagger}^\zero \vec{C}^\one  + \vev{\Op^\dagger}^\one \vec{C}^\zero  = -\img \cA^\one
\,.
\end{equation}
First we determine the residues of the propagators entering the LSZ reduction formula. Regulating both UV and IR divergences in dimensional regularization, all bare loop integrals in SCET are scaleless and vanish, i.e.\ the UV and IR divergences cancel. In particular, for the self-energy diagrams, we have
\begin{equation}
    \Sigma = \Sigma_\UV + \Sigma_\IR = 0
\,.\end{equation}
The UV divergences $\Sigma_\UV$ plus possible additional UV finite terms $\Sigma_x$ (as dictated by the renormalization scheme) determine the wave function renormalization $Z_\xi$. The remainder $\Sigma_\IR-\Sigma_x$ enters the residue $R_\xi$
\begin{align}
Z_\xi^{-1} &= 1 - \frac{\df(\Sigma_\UV+\Sigma_x)}{\df{\Sl p}}\biggr|_{{\Sl p}=0}
\,,\nn \\
R_\xi^{-1} &= 1 - \frac{\df(\Sigma_\IR-\Sigma_x)}{\df{\Sl p}}\biggr|_{{\Sl p}=0}
\,.\end{align}
At one loop in pure dimensional regularization, we then have $R_\xi = Z_\xi^{-1}$, and similarly for gluons $R_A = Z_A^{-1}$. In the on-shell scheme $\Sigma_x=\Sigma_\IR$, so with pure dimensional regularization $Z_\xi = R_\xi = Z_A = R_A = 1$.

Since all loop diagrams contributing to $\vev{\Op^\dagger}^\one$ vanish, the only nonzero contributions come from the counterterm in \eq{Z_O} and the one-loop residues. At one loop we find
\vspace{9mm}

\begin{align}\label{eq:IR_1loop}
 \vev{\Op^\dagger}^\one  \vC^{(0)}
    &= \vev{\Op^\dagger}^{(0)} \Bigl[ \bigl(Z_\xi^{n_q/2}\, Z_A^{n_g/2}\, \hZ_C - 1 \bigr)
+ \big(R_\xi^{n_q/2} R_A^{n_g/2} - 1\big)  \Bigr] \vC^{(0)}
    = \vev{\Op^\dagger}^{(0)} (\hZ_C-1) \vC^{(0)}
\nn \\
    &= \vev{\Op^\dagger}^{(0)}\, \frac{\al_s}{4\pi} \biggl[-\frac{1}{\eps^2}\,(n_g C_A + n_q C_F)
+ \frac{1}{\eps}\Bigl( -\frac{1}{2}n_g \beta_0 - \frac{3}{2} n_q C_F + 2\hDe(\mu^2)\Bigr) \biggr] 
\vC^{(0)}
\,,\end{align}
where we used the explicit expression for $\hZ_C$ derived in \subsec{loops}. One can easily check that this exactly reproduces the IR-divergent parts of the QCD amplitudes. For example, for $gg\,q\bar{q}$, we have
\begin{equation}
\biggl[-\frac{1}{\eps^2}(2 C_A + 2 C_F) + \frac{1}{\eps}\Bigl( -\beta_0 -3 C_F + 2\hDe_{gg\,q\bar{q}}(\mu^2)\Bigr) \biggr]
\vC_{+-(+)}^{(0)} (p_1,p_2;p_3,p_4)
    = \begin{pmatrix}
    A_\div^\one(1^+,2^-;3_q^+,4_{\bar q}^-) \\
    A_\div^\one(2^-,1^+;3_q^+,4_{\bar q}^-) \\
    B_\div^\one(1^+,2^-;3_q^+,4_{\bar q}^-) \\
  \end{pmatrix}
\,.\end{equation}
Hence, the IR divergences in  $\vev{\Op^\dagger}^\one \vC^{(0)}$ and $\cA^\one$ cancel each other and do not enter in $\vC^\one$, as must be the case.

%% file: biblio.tex
\begin{singlespace}
\bibliography{thesis}
\bibliographystyle{jhep}
\end{singlespace}

%% file: main.bbl
\providecommand{\href}[2]{#2}\begingroup\raggedright\begin{thebibliography}{100}

\bibitem{Hoang:2014wka}
A.~H. Hoang, D.~W. Kolodrubetz, V.~Mateu, and I.~W. Stewart, {\it {C-parameter
  Distribution at N${}^3$LL$^\prime$ including Power Corrections}},
  \href{http://arxiv.org/abs/1411.6633}{{\tt arXiv:1411.6633}}.

\bibitem{Abbate:2010xh}
R.~Abbate, M.~Fickinger, A.~H. Hoang, V.~Mateu, and I.~W. Stewart, {\it {Thrust
  at $N^3LL$ with Power Corrections and a Precision Global Fit for
  alphas(mZ)}},  {\em Phys. Rev. D} {\bf 83} (2011) 074021,
  [\href{http://arxiv.org/abs/1006.3080}{{\tt arXiv:1006.3080}}].

\bibitem{collaboration:2015aa}
T.~A. collaboration, {\it {Search for resonances with boson-tagged jets in 3.2
  fb−1 of p p collisions at √ s = 13 TeV collected with the ATLAS
  detector}}, .

\bibitem{collaboration:2016aa}
T.~A. collaboration, {\it {Studies of $b$-tagging performance and jet
  substructure in a high ${p_T}$ $g\rightarrow b\bar{b}$ rich sample of
  large-$R$ jets from $pp$ collisions at $\sqrt{s}=8$ TeV with the ATLAS
  detector}}, .

\bibitem{Larkoski:2014gra}
A.~J. Larkoski, I.~Moult, and D.~Neill, {\it {Power Counting to Better Jet
  Observables}},  {\em JHEP} {\bf 1412} (2014) 009,
  [\href{http://arxiv.org/abs/1409.6298}{{\tt arXiv:1409.6298}}].

\bibitem{Aad:2015rpa}
{\bf ATLAS} Collaboration, G.~Aad et~al., {\it {Identification of Boosted,
  Hadronically Decaying W Bosons and Comparisons with ATLAS Data Taken at
  $\sqrt{s} = 8$ TeV}},  \href{http://arxiv.org/abs/1510.05821}{{\tt
  arXiv:1510.05821}}.

\bibitem{atlas_recent:2015}
G.~Aad et~al., {\it {Performance of jet substructure techniques in early
  $\sqrt{s}=13$ TeV $pp$ collisions with the ATLAS detector}},  {\em
  ATLAS-CONF-2015-035} (2015), no.~ATLAS-CONF-2015-035.

\bibitem{Larkoski:2014zma}
A.~J. Larkoski, I.~Moult, and D.~Neill, {\it {Building a Better Boosted Top
  Tagger}},  \href{http://arxiv.org/abs/1411.0665}{{\tt arXiv:1411.0665}}.

\bibitem{Larkoski:2015uaa}
A.~J. Larkoski and I.~Moult, {\it {The Singular Behavior of Jet Substructure
  Observables}},  {\em Phys. Rev.} {\bf D93} (2016) 014017,
  [\href{http://arxiv.org/abs/1510.08459}{{\tt arXiv:1510.08459}}].

\bibitem{Larkoski:2014tva}
A.~J. Larkoski, I.~Moult, and D.~Neill, {\it {Toward Multi-Differential Cross
  Sections: Measuring Two Angularities on a Single Jet}},
  \href{http://arxiv.org/abs/1401.4458}{{\tt arXiv:1401.4458}}.

\bibitem{Larkoski:2015zka}
A.~J. Larkoski, I.~Moult, and D.~Neill, {\it {Non-Global Logarithms,
  Factorization, and the Soft Substructure of Jets}},
  \href{http://arxiv.org/abs/1501.04596}{{\tt arXiv:1501.04596}}.

\bibitem{Larkoski:2015kga}
A.~J. Larkoski, I.~Moult, and D.~Neill, {\it {Analytic Boosted Boson
  Discrimination}},  \href{http://arxiv.org/abs/1507.03018}{{\tt
  arXiv:1507.03018}}.

\bibitem{Stewart:2013faa}
I.~W. Stewart, F.~J. Tackmann, J.~R. Walsh, and S.~Zuberi, {\it {Jet $p_T$
  resummation in Higgs production at $NNLL'+NNLO$}},  {\em Phys. Rev.} {\bf
  D89} (2014), no.~5 054001, [\href{http://arxiv.org/abs/1307.1808}{{\tt
  arXiv:1307.1808}}].

\bibitem{Moult:2014pja}
I.~Moult and I.~W. Stewart, {\it {Jet Vetoes interfering with $H \to WW$}},
  {\em JHEP} {\bf 1409} (2014) 129, [\href{http://arxiv.org/abs/1405.5534}{{\tt
  arXiv:1405.5534}}].

\bibitem{Khachatryan:2014iha}
{\bf CMS} Collaboration, V.~Khachatryan et~al., {\it {Constraints on the Higgs
  boson width from off-shell production and decay to Z-boson pairs}},  {\em
  Phys. Lett.} {\bf B736} (2014) 64--85,
  [\href{http://arxiv.org/abs/1405.3455}{{\tt arXiv:1405.3455}}].

\bibitem{Collins:1988ig}
J.~C. Collins, D.~E. Soper, and G.~Sterman, {\it Soft gluons and
  factorization},  {\em Nucl. Phys. B} {\bf 308} (1988) 833.

\bibitem{Gaunt:2014ska}
J.~R. Gaunt, {\it {Glauber Gluons and Multiple Parton Interactions}},  {\em
  JHEP} {\bf 07} (2014) 110, [\href{http://arxiv.org/abs/1405.2080}{{\tt
  arXiv:1405.2080}}].

\bibitem{Zeng:2015iba}
M.~Zeng, {\it {Drell-Yan process with jet vetoes: breaking of generalized
  factorization}},  \href{http://arxiv.org/abs/1507.01652}{{\tt
  arXiv:1507.01652}}.

\bibitem{Rothstein:2016bsq}
I.~Z. Rothstein and I.~W. Stewart, {\it {An Effective Field Theory for Forward
  Scattering and Factorization Violation}},
  \href{http://arxiv.org/abs/1601.04695}{{\tt arXiv:1601.04695}}.

\bibitem{Stewart:2009yx}
I.~W. Stewart, F.~J. Tackmann, and W.~J. Waalewijn, {\it {Factorization at the
  LHC: From PDFs to Initial State Jets}},  {\em Phys. Rev. D} {\bf 81} (2010)
  094035, [\href{http://arxiv.org/abs/0910.0467}{{\tt arXiv:0910.0467}}].

\bibitem{Bauer:2000ew}
C.~W. Bauer, S.~Fleming, and M.~E. Luke, {\it {Summing Sudakov logarithms in B
  -> X(s gamma) in effective field theory}},  {\em Phys.Rev.} {\bf D63} (2000)
  014006, [\href{http://arxiv.org/abs/hep-ph/0005275}{{\tt hep-ph/0005275}}].

\bibitem{Bauer:2000yr}
C.~W. Bauer, S.~Fleming, D.~Pirjol, and I.~W. Stewart, {\it An effective field
  theory for collinear and soft gluons: Heavy to light decays},  {\em Phys.
  Rev. D} {\bf 63} (2001) 114020,
  [\href{http://arxiv.org/abs/hep-ph/0011336}{{\tt hep-ph/0011336}}].

\bibitem{Bauer:2001ct}
C.~W. Bauer and I.~W. Stewart, {\it {Invariant operators in collinear effective
  theory}},  {\em Phys.Lett.} {\bf B516} (2001) 134--142,
  [\href{http://arxiv.org/abs/hep-ph/0107001}{{\tt hep-ph/0107001}}].

\bibitem{Bauer:2001yt}
C.~W. Bauer, D.~Pirjol, and I.~W. Stewart, {\it Soft-collinear factorization in
  effective field theory},  {\em Phys. Rev. D} {\bf 65} (2002) 054022,
  [\href{http://arxiv.org/abs/hep-ph/0109045}{{\tt hep-ph/0109045}}].

\bibitem{Bauer:2002nz}
C.~W. Bauer, S.~Fleming, D.~Pirjol, I.~Z. Rothstein, and I.~W. Stewart, {\it
  Hard scattering factorization from effective field theory},  {\em Phys. Rev.
  D} {\bf 66} (2002) 014017, [\href{http://arxiv.org/abs/hep-ph/0202088}{{\tt
  hep-ph/0202088}}].

\bibitem{Manohar:2002fd}
A.~V. Manohar, T.~Mehen, D.~Pirjol, and I.~W. Stewart, {\it Reparameterization
  invariance for collinear operators},  {\em Phys. Lett. B} {\bf 539} (2002)
  59--66, [\href{http://arxiv.org/abs/hep-ph/0204229}{{\tt hep-ph/0204229}}].

\bibitem{Chay:2002vy}
J.~Chay and C.~Kim, {\it {Collinear effective theory at subleading order and
  its application to heavy-light currents}},  {\em Phys. Rev. D} {\bf 65}
  (2002) 114016, [\href{http://arxiv.org/abs/hep-ph/0201197}{{\tt
  hep-ph/0201197}}].

\bibitem{Bauer:2002aj}
C.~W. Bauer, D.~Pirjol, and I.~W. Stewart, {\it {Factorization and endpoint
  singularities in heavy to light decays}},  {\em Phys.Rev.} {\bf D67} (2003)
  071502, [\href{http://arxiv.org/abs/hep-ph/0211069}{{\tt hep-ph/0211069}}].

\bibitem{Marcantonini:2008qn}
C.~Marcantonini and I.~W. Stewart, {\it {Reparameterization Invariant Collinear
  Operators}},  {\em Phys. Rev.} {\bf D79} (2009) 065028,
  [\href{http://arxiv.org/abs/0809.1093}{{\tt arXiv:0809.1093}}].

\bibitem{Larkoski:2013eya}
A.~J. Larkoski, G.~P. Salam, and J.~Thaler, {\it {Energy Correlation Functions
  for Jet Substructure}},  {\em JHEP} {\bf 1306} (2013) 108,
  [\href{http://arxiv.org/abs/1305.0007}{{\tt arXiv:1305.0007}}].

\bibitem{Thaler:2011gf}
J.~Thaler and K.~Van~Tilburg, {\it {Maximizing Boosted Top Identification by
  Minimizing N-subjettiness}},  {\em JHEP} {\bf 1202} (2012) 093,
  [\href{http://arxiv.org/abs/1108.2701}{{\tt arXiv:1108.2701}}].

\bibitem{Thaler:2010tr}
J.~Thaler and K.~Van~Tilburg, {\it {Identifying Boosted Objects with
  N-subjettiness}},  {\em JHEP} {\bf 1103} (2011) 015,
  [\href{http://arxiv.org/abs/1011.2268}{{\tt arXiv:1011.2268}}].

\bibitem{Larkoski:2014uqa}
A.~J. Larkoski, D.~Neill, and J.~Thaler, {\it {Jet Shapes with the Broadening
  Axis}},  {\em JHEP} {\bf 1404} (2014) 017,
  [\href{http://arxiv.org/abs/1401.2158}{{\tt arXiv:1401.2158}}].

\bibitem{Procura:2014cba}
M.~Procura, W.~J. Waalewijn, and L.~Zeune, {\it {Resummation of
  Double-Differential Cross Sections and Fully-Unintegrated Parton Distribution
  Functions}},  {\em JHEP} {\bf 1502} (2015) 117,
  [\href{http://arxiv.org/abs/1410.6483}{{\tt arXiv:1410.6483}}].

\bibitem{Chien:2015cka}
Y.-T. Chien, A.~Hornig, and C.~Lee, {\it {Soft-collinear mode for jet cross
  sections in soft collinear effective theory}},  {\em Phys. Rev.} {\bf D93}
  (2016), no.~1 014033, [\href{http://arxiv.org/abs/1509.04287}{{\tt
  arXiv:1509.04287}}].

\bibitem{Pietrulewicz:2016nwo}
P.~Pietrulewicz, F.~J. Tackmann, and W.~J. Waalewijn, {\it {Factorization and
  Resummation for Generic Hierarchies between Jets}},
  \href{http://arxiv.org/abs/1601.05088}{{\tt arXiv:1601.05088}}.

\bibitem{Larkoski:2014pca}
A.~J. Larkoski, J.~Thaler, and W.~J. Waalewijn, {\it {Gaining (Mutual)
  Information about Quark/Gluon Discrimination}},
  \href{http://arxiv.org/abs/1408.3122}{{\tt arXiv:1408.3122}}.

\bibitem{Dasgupta:2001sh}
M.~Dasgupta and G.~Salam, {\it {Resummation of nonglobal QCD observables}},
  {\em Phys.Lett.} {\bf B512} (2001) 323--330,
  [\href{http://arxiv.org/abs/hep-ph/0104277}{{\tt hep-ph/0104277}}].

\bibitem{Larkoski:2015npa}
A.~J. Larkoski and I.~Moult, {\it {Nonglobal correlations in collider
  physics}},  {\em Phys. Rev.} {\bf D93} (2016), no.~1 014012,
  [\href{http://arxiv.org/abs/1510.05657}{{\tt arXiv:1510.05657}}].

\bibitem{Larkoski:2014wba}
A.~J. Larkoski, S.~Marzani, G.~Soyez, and J.~Thaler, {\it {Soft Drop}},  {\em
  JHEP} {\bf 1405} (2014) 146, [\href{http://arxiv.org/abs/1402.2657}{{\tt
  arXiv:1402.2657}}].

\bibitem{Frye:2016okc}
C.~Frye, A.~J. Larkoski, M.~D. Schwartz, and K.~Yan, {\it {Precision physics
  with pile-up insensitive observables}},
  \href{http://arxiv.org/abs/1603.06375}{{\tt arXiv:1603.06375}}.

\bibitem{Frye:2016aiz}
C.~Frye, A.~J. Larkoski, M.~D. Schwartz, and K.~Yan, {\it {Factorization for
  groomed jet substructure beyond the next-to-leading logarithm}},
  \href{http://arxiv.org/abs/1603.09338}{{\tt arXiv:1603.09338}}.

\bibitem{Moult:2015aoa}
I.~Moult, I.~W. Stewart, F.~J. Tackmann, and W.~J. Waalewijn, {\it {Employing
  Helicity Amplitudes for Resummation}},
  \href{http://arxiv.org/abs/1508.02397}{{\tt arXiv:1508.02397}}.

\bibitem{Kolodrubetz:2016uim}
D.~W. Kolodrubetz, I.~Moult, and I.~W. Stewart, {\it {Building Blocks for
  Subleading Helicity Operators}},  \href{http://arxiv.org/abs/1601.02607}{{\tt
  arXiv:1601.02607}}.

\bibitem{Caola:2013yja}
F.~Caola and K.~Melnikov, {\it {Constraining the Higgs boson width with ZZ
  production at the LHC}},  {\em Phys.Rev.} {\bf D88} (2013) 054024,
  [\href{http://arxiv.org/abs/1307.4935}{{\tt arXiv:1307.4935}}].

\bibitem{Abdesselam:2010pt}
A.~Abdesselam, E.~B. Kuutmann, U.~Bitenc, G.~Brooijmans, J.~Butterworth,
  et~al., {\it {Boosted objects: A Probe of beyond the Standard Model
  physics}},  {\em Eur.Phys.J.} {\bf C71} (2011) 1661,
  [\href{http://arxiv.org/abs/1012.5412}{{\tt arXiv:1012.5412}}].

\bibitem{Altheimer:2012mn}
A.~Altheimer, S.~Arora, L.~Asquith, G.~Brooijmans, J.~Butterworth, et~al., {\it
  {Jet Substructure at the Tevatron and LHC: New results, new tools, new
  benchmarks}},  {\em J.Phys.} {\bf G39} (2012) 063001,
  [\href{http://arxiv.org/abs/1201.0008}{{\tt arXiv:1201.0008}}].

\bibitem{Altheimer:2013yza}
A.~Altheimer, A.~Arce, L.~Asquith, J.~Backus~Mayes, E.~Bergeaas~Kuutmann,
  et~al., {\it {Boosted objects and jet substructure at the LHC. Report of
  BOOST2012, held at IFIC Valencia, 23rd-27th of July 2012}},  {\em
  Eur.Phys.J.} {\bf C74} (2014) 2792,
  [\href{http://arxiv.org/abs/1311.2708}{{\tt arXiv:1311.2708}}].

\bibitem{CMS:2011xsa}
{\bf CMS Collaboration} Collaboration, {\it {Jet Substructure Algorithms}}, .

\bibitem{Miller:2011qg}
{\bf ATLAS Collaboration} Collaboration, D.~W. Miller, {\it {Jet substructure
  in ATLAS}},  Tech. Rep. ATL-PHYS-PROC-2011-142, 2011.

\bibitem{ATLAS-CONF-2012-066}
{\bf ATLAS Collaboration} Collaboration, {\it Studies of the impact and
  mitigation of pile-up on large-$r$ and groomed jets in atlas at $\sqrt{s}=7$
  tev},  Tech. Rep. ATLAS-CONF-2012-066, CERN, Geneva, Jul, 2012.

\bibitem{Chatrchyan:2012mec}
{\bf CMS Collaboration} Collaboration, S.~Chatrchyan et~al., {\it {Shape,
  transverse size, and charged hadron multiplicity of jets in pp collisions at
  7 TeV}},  {\em JHEP} {\bf 1206} (2012) 160,
  [\href{http://arxiv.org/abs/1204.3170}{{\tt arXiv:1204.3170}}].

\bibitem{ATLAS:2012jla}
{\bf ATLAS Collaboration} Collaboration, {\it {Studies of the impact and
  mitigation of pile-up on large-$R$ and groomed jets in ATLAS at $\sqrt{s}=7$
  TeV}}, .

\bibitem{Aad:2012meb}
{\bf ATLAS Collaboration} Collaboration, G.~Aad et~al., {\it {ATLAS
  measurements of the properties of jets for boosted particle searches}},  {\em
  Phys.Rev.} {\bf D86} (2012) 072006,
  [\href{http://arxiv.org/abs/1206.5369}{{\tt arXiv:1206.5369}}].

\bibitem{ATLAS:2012kla}
{\bf ATLAS Collaboration} Collaboration, {\it {Performance of large-R jets and
  jet substructure reconstruction with the ATLAS detector}}, .

\bibitem{ATLAS:2012am}
{\bf ATLAS Collaboration} Collaboration, G.~Aad et~al., {\it {Jet mass and
  substructure of inclusive jets in $\sqrt{s}=7$ TeV $pp$ collisions with the
  ATLAS experiment}},  {\em JHEP} {\bf 1205} (2012) 128,
  [\href{http://arxiv.org/abs/1203.4606}{{\tt arXiv:1203.4606}}].

\bibitem{Aad:2013gja}
{\bf ATLAS Collaboration} Collaboration, G.~Aad et~al., {\it {Performance of
  jet substructure techniques for large-$R$ jets in proton-proton collisions at
  $\sqrt{s}$ = 7 TeV using the ATLAS detector}},  {\em JHEP} {\bf 1309} (2013)
  076, [\href{http://arxiv.org/abs/1306.4945}{{\tt arXiv:1306.4945}}].

\bibitem{Aad:2013fba}
{\bf ATLAS Collaboration} Collaboration, G.~Aad et~al., {\it {Measurement of
  jet shapes in top pair events at sqrt(s) = 7 TeV using the ATLAS detector}},
  \href{http://arxiv.org/abs/1307.5749}{{\tt arXiv:1307.5749}}.

\bibitem{TheATLAScollaboration:2013tia}
{\bf ATLAS Collaboration} Collaboration, {\it {Performance and Validation of
  Q-Jets at the ATLAS Detector in pp Collisions at $\sqrt{s}$=8 TeV in 2012}},
  Tech. Rep. ATLAS-CONF-2013-087, ATLAS-COM-CONF-2013-099, 2013.

\bibitem{TheATLAScollaboration:2013sia}
{\bf ATLAS Collaboration} Collaboration, {\it {Jet Charge Studies with the
  ATLAS Detector Using $\sqrt{s} = 8$ TeV Proton-Proton Collision Data}},
  Tech. Rep. ATLAS-CONF-2013-086, ATLAS-COM-CONF-2013-101, 2013.

\bibitem{TheATLAScollaboration:2013ria}
{\bf ATLAS Collaboration} Collaboration, {\it {Performance of pile-up
  subtraction for jet shapes}},  Tech. Rep. ATLAS-CONF-2013-085,
  ATLAS-COM-CONF-2013-100, 2013.

\bibitem{TheATLAScollaboration:2013pia}
{\bf ATLAS Collaboration} Collaboration, {\it {Pile-up subtraction and
  suppression for jets in ATLAS}},  Tech. Rep. ATLAS-CONF-2013-083,
  ATLAS-COM-CONF-2013-097, 2013.

\bibitem{CMS:2013uea}
{\bf CMS Collaboration} Collaboration, C.~Collaboration, {\it {Identifying
  Hadronically Decaying Vector Bosons Merged into a Single Jet}}, .

\bibitem{CMS:2013kfa}
{\bf CMS Collaboration} Collaboration, C.~Collaboration, {\it {Performance of
  quark/gluon discrimination in 8 TeV pp data}},  Tech. Rep.
  CMS-PAS-JME-13-002, 2013.

\bibitem{CMS:2013wea}
{\bf CMS Collaboration} Collaboration, C.~Collaboration, {\it {Pileup Jet
  Identification}},  Tech. Rep. CMS-PAS-JME-13-005, 2013.

\bibitem{CMS-PAS-JME-10-013}
{\bf CMS Collaboration} Collaboration, {\it Jet substructure algorithms},
  Tech. Rep. CMS-PAS-JME-10-013, CERN, Geneva, 2011.

\bibitem{CMS-PAS-QCD-10-041}
{\bf CMS Collaboration} Collaboration, {\it Measurement of the subjet
  multiplicity in dijet events from proton-proton collisions at sqrt(s) = 7
  tev},  Tech. Rep. CMS-PAS-QCD-10-041, CERN, Geneva, 2010.

\bibitem{Aad:2014gea}
{\bf ATLAS Collaboration} Collaboration, G.~Aad et~al., {\it {Light-quark and
  gluon jet discrimination in pp collisions at $\sqrt{s}$ = 7 TeV with the
  ATLAS detector}},  \href{http://arxiv.org/abs/1405.6583}{{\tt
  arXiv:1405.6583}}.

\bibitem{LOCH:2014lla}
{\bf ATLAS} Collaboration, P.~Loch, {\it {Studies of jet shapes and jet
  substructure in proton-proton collisions at $\sqrt{s} =$ 7 TeV with ATLAS}},
  {\em PoS} {\bf EPS-HEP2013} (2013) 442.

\bibitem{CMS:2014fya}
{\bf CMS Collaboration} Collaboration, C.~Collaboration, {\it {Boosted Top Jet
  Tagging at CMS}}, .

\bibitem{CMS:2011bqa}
{\bf CMS Collaboration} Collaboration, {\it {Search for BSM ttbar Production in
  the Boosted All-Hadronic Final State}},  Tech. Rep. CMS-PAS-EXO-11-006, 2011.

\bibitem{Fleischmann:2013woa}
{\bf ATLAS, CMS Collaboration} Collaboration, S.~Fleischmann, {\it {Boosted top
  quark techniques and searches for $t\bar{t}$ resonances at the LHC}},  {\em
  J.Phys.Conf.Ser.} {\bf 452} (2013), no.~1 012034.

\bibitem{Pilot:2013bla}
{\bf ATLAS, CMS Collaboration} Collaboration, J.~Pilot, {\it {Boosted Top
  Quarks, Top Pair Resonances, and Top Partner Searches at the LHC}},  {\em EPJ
  Web Conf.} {\bf 60} (2013) 09003.

\bibitem{TheATLAScollaboration:2013qia}
{\bf ATLAS Collaboration} Collaboration, {\it {Performance of boosted top quark
  identification in 2012 ATLAS data}},  Tech. Rep. ATLAS-CONF-2013-084,
  ATLAS-COM-CONF-2013-074, 2013.

\bibitem{Chatrchyan:2012ku}
{\bf CMS Collaboration} Collaboration, S.~Chatrchyan et~al., {\it {Search for
  Anomalous $t\bar{t}$ Production in the Highly-Boosted All-Hadronic Final
  State}},  {\em JHEP} {\bf 1209} (2012) 029,
  [\href{http://arxiv.org/abs/1204.2488}{{\tt arXiv:1204.2488}}].

\bibitem{Chatrchyan:2012sn}
{\bf CMS Collaboration} Collaboration, S.~Chatrchyan et~al., {\it {Search for a
  Higgs boson in the decay channel $H$ to ZZ(*) to $q$ qbar $\ell^-$ l+ in $pp$
  collisions at $\sqrt{s}=7$ TeV}},  {\em JHEP} {\bf 1204} (2012) 036,
  [\href{http://arxiv.org/abs/1202.1416}{{\tt arXiv:1202.1416}}].

\bibitem{CMS:2013cda}
{\bf CMS Collaboration} Collaboration, C.~Collaboration, {\it {Search for a
  Standard Model-like Higgs boson decaying into WW to l nu qqbar in pp
  collisions at sqrt s = 8 TeV}}, .

\bibitem{CMS:2014afa}
{\bf CMS Collaboration} Collaboration, C.~Collaboration, {\it {Search for
  pair-produced vector-like quarks of charge -1/3 decaying to bH using boosted
  Higgs jet-tagging in pp collisions at sqrt(s) = 8 TeV}}, .

\bibitem{CMS:2014aka}
{\bf CMS Collaboration} Collaboration, C.~Collaboration, {\it {Search for
  top-Higgs resonances in all-hadronic final states using jet substructure
  methods}}, .

\bibitem{Catani:1991bd}
S.~Catani, G.~Turnock, and B.~Webber, {\it {Heavy jet mass distribution in e+
  e- annihilation}},  {\em Phys.Lett.} {\bf B272} (1991) 368--372.

\bibitem{Chien:2010kc}
Y.-T. Chien and M.~D. Schwartz, {\it {Resummation of heavy jet mass and
  comparison to LEP data}},  {\em JHEP} {\bf 1008} (2010) 058,
  [\href{http://arxiv.org/abs/1005.1644}{{\tt arXiv:1005.1644}}].

\bibitem{Chien:2012ur}
Y.-T. Chien, R.~Kelley, M.~D. Schwartz, and H.~X. Zhu, {\it {Resummation of Jet
  Mass at Hadron Colliders}},  {\em Phys.Rev.} {\bf D87} (2013) 014010,
  [\href{http://arxiv.org/abs/1208.0010}{{\tt arXiv:1208.0010}}].

\bibitem{Dasgupta:2012hg}
M.~Dasgupta, K.~Khelifa-Kerfa, S.~Marzani, and M.~Spannowsky, {\it {On jet mass
  distributions in Z+jet and dijet processes at the LHC}},  {\em JHEP} {\bf
  1210} (2012) 126, [\href{http://arxiv.org/abs/1207.1640}{{\tt
  arXiv:1207.1640}}].

\bibitem{Jouttenus:2013hs}
T.~T. Jouttenus, I.~W. Stewart, F.~J. Tackmann, and W.~J. Waalewijn, {\it {Jet
  Mass Spectra in Higgs $+$ One Jet at NNLL}},  {\em Phys.~Rev.~D} {\bf 88}
  (2013) 054031, [\href{http://arxiv.org/abs/1302.0846}{{\tt
  arXiv:1302.0846}}].

\bibitem{Feige:2012vc}
I.~Feige, M.~D. Schwartz, I.~W. Stewart, and J.~Thaler, {\it {Precision Jet
  Substructure from Boosted Event Shapes}},  {\em Phys.Rev.Lett.} {\bf 109}
  (2012) 092001, [\href{http://arxiv.org/abs/1204.3898}{{\tt
  arXiv:1204.3898}}].

\bibitem{Dasgupta:2013ihk}
M.~Dasgupta, A.~Fregoso, S.~Marzani, and G.~P. Salam, {\it {Towards an
  understanding of jet substructure}},  {\em JHEP} {\bf 1309} (2013) 029,
  [\href{http://arxiv.org/abs/1307.0007}{{\tt arXiv:1307.0007}}].

\bibitem{Dasgupta:2013via}
M.~Dasgupta, A.~Fregoso, S.~Marzani, and A.~Powling, {\it {Jet substructure
  with analytical methods}},  {\em Eur.Phys.J.} {\bf C73} (2013), no.~11 2623,
  [\href{http://arxiv.org/abs/1307.0013}{{\tt arXiv:1307.0013}}].

\bibitem{Larkoski:2013paa}
A.~J. Larkoski and J.~Thaler, {\it {Unsafe but Calculable: Ratios of
  Angularities in Perturbative QCD}},  {\em JHEP} {\bf 1309} (2013) 137,
  [\href{http://arxiv.org/abs/1307.1699}{{\tt arXiv:1307.1699}}].

\bibitem{Larkoski:2014bia}
A.~J. Larkoski and J.~Thaler, {\it {Aspects of Jets at 100 TeV}},
  \href{http://arxiv.org/abs/1406.7011}{{\tt arXiv:1406.7011}}.

\bibitem{Walsh:2011fz}
J.~R. Walsh and S.~Zuberi, {\it {Factorization Constraints on Jet
  Substructure}},  \href{http://arxiv.org/abs/1110.5333}{{\tt
  arXiv:1110.5333}}.

\bibitem{Catani:1992jc}
S.~Catani, G.~Turnock, and B.~Webber, {\it {Jet broadening measures in $e^{+}
  e^{-}$ annihilation}},  {\em Phys.Lett.} {\bf B295} (1992) 269--276.

\bibitem{Dokshitzer:1998kz}
Y.~L. Dokshitzer, A.~Lucenti, G.~Marchesini, and G.~Salam, {\it {On the QCD
  analysis of jet broadening}},  {\em JHEP} {\bf 9801} (1998) 011,
  [\href{http://arxiv.org/abs/hep-ph/9801324}{{\tt hep-ph/9801324}}].

\bibitem{Banfi:2004yd}
A.~Banfi, G.~P. Salam, and G.~Zanderighi, {\it {Principles of general
  final-state resummation and automated implementation}},  {\em JHEP} {\bf
  0503} (2005) 073, [\href{http://arxiv.org/abs/hep-ph/0407286}{{\tt
  hep-ph/0407286}}].

\bibitem{Cacciari:2008gp}
M.~Cacciari, G.~P. Salam, and G.~Soyez, {\it {The anti-$k_t$ jet clustering
  algorithm}},  {\em JHEP} {\bf 04} (2008) 063,
  [\href{http://arxiv.org/abs/0802.1189}{{\tt arXiv:0802.1189}}].

\bibitem{Bertolini:2013iqa}
D.~Bertolini, T.~Chan, and J.~Thaler, {\it {Jet Observables Without Jet
  Algorithms}},  {\em JHEP} {\bf 1404} (2014) 013,
  [\href{http://arxiv.org/abs/1310.7584}{{\tt arXiv:1310.7584}}].

\bibitem{Salambroadening}
G.~Salam, {\it {Unpublished}},  {\em Unpublished}.

\bibitem{Blazey:2000qt}
G.~C. Blazey, J.~R. Dittmann, S.~D. Ellis, V.~D. Elvira, K.~Frame, et~al., {\it
  {Run II jet physics}},  \href{http://arxiv.org/abs/hep-ex/0005012}{{\tt
  hep-ex/0005012}}.

\bibitem{Rakow:1981qn}
P.~E. Rakow and B.~Webber, {\it {Transverse Momentum Moments of Hadron
  Distributions in QCD Jets}},  {\em Nucl.Phys.} {\bf B191} (1981) 63.

\bibitem{Ellis:1986ig}
R.~K. Ellis and B.~Webber, {\it {QCD Jet Broadening in Hadron Hadron
  Collisions}},  {\em Conf.Proc.} {\bf C860623} (1986) 74.

\bibitem{Cacciari:2011ma}
M.~Cacciari, G.~P. Salam, and G.~Soyez, {\it {FastJet User Manual}},  {\em Eur.
  Phys. J. C} {\bf 72} (2012) 1896, [\href{http://arxiv.org/abs/1111.6097}{{\tt
  arXiv:1111.6097}}].

\bibitem{fjcontrib}
``Fastjet contrib.'' http://fastjet.hepforge.org/contrib/.

\bibitem{Beneke:1997zp}
M.~Beneke and V.~A. Smirnov, {\it {Asymptotic expansion of Feynman integrals
  near threshold}},  {\em Nucl.Phys.} {\bf B522} (1998) 321--344,
  [\href{http://arxiv.org/abs/hep-ph/9711391}{{\tt hep-ph/9711391}}].

\bibitem{Smirnov:2001in}
V.~A. Smirnov, {\it {`Strategy of regions': Expansions of Feynman diagrams both
  in Euclidean and pseudo-Euclidean regimes}},
  \href{http://arxiv.org/abs/hep-ph/0101152}{{\tt hep-ph/0101152}}.

\bibitem{Stewart:2010tn}
I.~W. Stewart, F.~J. Tackmann, and W.~J. Waalewijn, {\it {N-Jettiness: An
  Inclusive Event Shape to Veto Jets}},  {\em Phys.Rev.Lett.} {\bf 105} (2010)
  092002, [\href{http://arxiv.org/abs/1004.2489}{{\tt arXiv:1004.2489}}].

\bibitem{Bauer:2011uc}
C.~W. Bauer, F.~J. Tackmann, J.~R. Walsh, and S.~Zuberi, {\it {Factorization
  and Resummation for Dijet Invariant Mass Spectra}},  {\em Phys.Rev.} {\bf
  D85} (2012) 074006, [\href{http://arxiv.org/abs/1106.6047}{{\tt
  arXiv:1106.6047}}].

\bibitem{Soyez:2012hv}
G.~Soyez, G.~P. Salam, J.~Kim, S.~Dutta, and M.~Cacciari, {\it {Pileup
  subtraction for jet shapes}},  {\em Phys.Rev.Lett.} {\bf 110} (2013), no.~16
  162001, [\href{http://arxiv.org/abs/1211.2811}{{\tt arXiv:1211.2811}}].

\bibitem{Alwall:2014hca}
J.~Alwall, R.~Frederix, S.~Frixione, V.~Hirschi, F.~Maltoni, et~al., {\it {The
  automated computation of tree-level and next-to-leading order differential
  cross sections, and their matching to parton shower simulations}},
  \href{http://arxiv.org/abs/1405.0301}{{\tt arXiv:1405.0301}}.

\bibitem{Sjostrand:2006za}
T.~Sjostrand, S.~Mrenna, and P.~Z. Skands, {\it {PYTHIA 6.4 Physics and
  Manual}},  {\em JHEP} {\bf 0605} (2006) 026,
  [\href{http://arxiv.org/abs/hep-ph/0603175}{{\tt hep-ph/0603175}}].

\bibitem{Sjostrand:2007gs}
T.~Sjostrand, S.~Mrenna, and P.~Z. Skands, {\it {A Brief Introduction to PYTHIA
  8.1}},  {\em Comput.Phys.Commun.} {\bf 178} (2008) 852--867,
  [\href{http://arxiv.org/abs/0710.3820}{{\tt arXiv:0710.3820}}].

\bibitem{Marchesini:1991ch}
G.~Marchesini, B.~Webber, G.~Abbiendi, I.~Knowles, M.~Seymour, et~al., {\it
  {HERWIG: A Monte Carlo event generator for simulating hadron emission
  reactions with interfering gluons. Version 5.1 - April 1991}},  {\em
  Comput.Phys.Commun.} {\bf 67} (1992) 465--508.

\bibitem{Corcella:2000bw}
G.~Corcella, I.~Knowles, G.~Marchesini, S.~Moretti, K.~Odagiri, et~al., {\it
  {HERWIG 6: An Event generator for hadron emission reactions with interfering
  gluons (including supersymmetric processes)}},  {\em JHEP} {\bf 0101} (2001)
  010, [\href{http://arxiv.org/abs/hep-ph/0011363}{{\tt hep-ph/0011363}}].

\bibitem{Corcella:2002jc}
G.~Corcella, I.~Knowles, G.~Marchesini, S.~Moretti, K.~Odagiri, et~al., {\it
  {HERWIG 6.5 release note}},  \href{http://arxiv.org/abs/hep-ph/0210213}{{\tt
  hep-ph/0210213}}.

\bibitem{Bahr:2008pv}
M.~Bahr, S.~Gieseke, M.~Gigg, D.~Grellscheid, K.~Hamilton, et~al., {\it
  {Herwig++ Physics and Manual}},  {\em Eur.Phys.J.} {\bf C58} (2008) 639--707,
  [\href{http://arxiv.org/abs/0803.0883}{{\tt arXiv:0803.0883}}].

\bibitem{Butterworth:2008iy}
J.~M. Butterworth, A.~R. Davison, M.~Rubin, and G.~P. Salam, {\it {Jet
  substructure as a new Higgs search channel at the LHC}},  {\em Phys. Rev.
  Lett.} {\bf 100} (2008) 242001, [\href{http://arxiv.org/abs/0802.2470}{{\tt
  arXiv:0802.2470}}].

\bibitem{Ellis:2009su}
S.~D. Ellis, C.~K. Vermilion, and J.~R. Walsh, {\it {Techniques for improved
  heavy particle searches with jet substructure}},  {\em Phys.Rev.} {\bf D80}
  (2009) 051501, [\href{http://arxiv.org/abs/0903.5081}{{\tt
  arXiv:0903.5081}}].

\bibitem{Ellis:2009me}
S.~D. Ellis, C.~K. Vermilion, and J.~R. Walsh, {\it {Recombination Algorithms
  and Jet Substructure: Pruning as a Tool for Heavy Particle Searches}},  {\em
  Phys.Rev.} {\bf D81} (2010) 094023,
  [\href{http://arxiv.org/abs/0912.0033}{{\tt arXiv:0912.0033}}].

\bibitem{Krohn:2009th}
D.~Krohn, J.~Thaler, and L.-T. Wang, {\it {Jet Trimming}},  {\em JHEP} {\bf
  1002} (2010) 084, [\href{http://arxiv.org/abs/0912.1342}{{\tt
  arXiv:0912.1342}}].

\bibitem{Krohn:2013lba}
D.~Krohn, M.~D. Schwartz, M.~Low, and L.-T. Wang, {\it {Jet Cleansing: Pileup
  Removal at High Luminosity}},  \href{http://arxiv.org/abs/1309.4777}{{\tt
  arXiv:1309.4777}}.

\bibitem{Berta:2014eza}
P.~Berta, M.~Spousta, D.~W. Miller, and R.~Leitner, {\it {Particle-level pileup
  subtraction for jets and jet shapes}},  {\em JHEP} {\bf 1406} (2014) 092,
  [\href{http://arxiv.org/abs/1403.3108}{{\tt arXiv:1403.3108}}].

\bibitem{Cacciari:2014gra}
M.~Cacciari, G.~P. Salam, and G.~Soyez, {\it {SoftKiller, a particle-level
  pileup removal method}},  \href{http://arxiv.org/abs/1407.0408}{{\tt
  arXiv:1407.0408}}.

\bibitem{Bertolini:2014bba}
D.~Bertolini, P.~Harris, M.~Low, and N.~Tran, {\it {Pileup Per Particle
  Identification}},  \href{http://arxiv.org/abs/1407.6013}{{\tt
  arXiv:1407.6013}}.

\bibitem{Gallicchio:2011xq}
J.~Gallicchio and M.~D. Schwartz, {\it {Quark and Gluon Tagging at the LHC}},
  {\em Phys.Rev.Lett.} {\bf 107} (2011) 172001,
  [\href{http://arxiv.org/abs/1106.3076}{{\tt arXiv:1106.3076}}].

\bibitem{Gallicchio:2012ez}
J.~Gallicchio and M.~D. Schwartz, {\it {Quark and Gluon Jet Substructure}},
  {\em JHEP} {\bf 1304} (2013) 090, [\href{http://arxiv.org/abs/1211.7038}{{\tt
  arXiv:1211.7038}}].

\bibitem{Gallicchio:2011xc}
J.~Gallicchio and M.~D. Schwartz, {\it {Pure Samples of Quark and Gluon Jets at
  the LHC}},  {\em JHEP} {\bf 1110} (2011) 103,
  [\href{http://arxiv.org/abs/1104.1175}{{\tt arXiv:1104.1175}}].

\bibitem{Adams:2015hiv}
D.~Adams, A.~Arce, L.~Asquith, M.~Backovic, T.~Barillari, et~al., {\it {Towards
  an Understanding of the Correlations in Jet Substructure}},
  \href{http://arxiv.org/abs/1504.00679}{{\tt arXiv:1504.00679}}.

\bibitem{CMS:2014joa}
{\bf CMS} Collaboration, {\it {V Tagging Observables and Correlations}},  Tech.
  Rep. CMS-PAS-JME-14-002, 2014.

\bibitem{Aad:2014haa}
{\bf ATLAS} Collaboration, G.~Aad et~al., {\it {Measurement of the
  cross-section of high transverse momentum vector bosons reconstructed as
  single jets and studies of jet substructure in $pp$ collisions at
  ${\sqrt{s}}$ = 7 TeV with the ATLAS detector}},  {\em New J.Phys.} {\bf 16}
  (2014), no.~11 113013, [\href{http://arxiv.org/abs/1407.0800}{{\tt
  arXiv:1407.0800}}].

\bibitem{Khachatryan:2015axa}
{\bf CMS} Collaboration, V.~Khachatryan et~al., {\it {Search for vector-like T
  quarks decaying to top quarks and Higgs bosons in the all-hadronic channel
  using jet substructure}},  \href{http://arxiv.org/abs/1503.01952}{{\tt
  arXiv:1503.01952}}.

\bibitem{CMS:1900uua}
{\bf CMS} Collaboration, C.~Collaboration, {\it {Search for pair-produced
  vector-like top quark partners decaying to bW in the fully hadronic channel
  using jet substructure at 8 TeV}}, .

\bibitem{Khachatryan:2015bma}
{\bf CMS} Collaboration, V.~Khachatryan et~al., {\it {Search for a massive
  resonance decaying into a Higgs boson and a W or Z boson in hadronic final
  states in proton-proton collisions at sqrt(s) = 8 TeV}},
  \href{http://arxiv.org/abs/1506.01443}{{\tt arXiv:1506.01443}}.

\bibitem{Aad:2015owa}
{\bf ATLAS} Collaboration, G.~Aad et~al., {\it {Search for high-mass diboson
  resonances with boson-tagged jets in proton-proton collisions at $\sqrt{s}$ =
  8 TeV with the ATLAS detector}},  \href{http://arxiv.org/abs/1506.00962}{{\tt
  arXiv:1506.00962}}.

\bibitem{Field:2012rw}
M.~Field, G.~Gur-Ari, D.~A. Kosower, L.~Mannelli, and G.~Perez, {\it
  {Three-Prong Distribution of Massive Narrow QCD Jets}},  {\em Phys.Rev.} {\bf
  D87} (2013), no.~9 094013, [\href{http://arxiv.org/abs/1212.2106}{{\tt
  arXiv:1212.2106}}].

\bibitem{Dasgupta:2015yua}
M.~Dasgupta, A.~Powling, and A.~Siodmok, {\it {On jet substructure methods for
  signal jets}},  {\em JHEP} {\bf 08} (2015) 079,
  [\href{http://arxiv.org/abs/1503.01088}{{\tt arXiv:1503.01088}}].

\bibitem{Seymour:1997kj}
M.~Seymour, {\it {Jet shapes in hadron collisions: Higher orders, resummation
  and hadronization}},  {\em Nucl.Phys.} {\bf B513} (1998) 269--300,
  [\href{http://arxiv.org/abs/hep-ph/9707338}{{\tt hep-ph/9707338}}].

\bibitem{Li:2011hy}
H.-n. Li, Z.~Li, and C.-P. Yuan, {\it {QCD resummation for jet substructures}},
   {\em Phys. Rev. Lett.} {\bf 107} (2011) 152001,
  [\href{http://arxiv.org/abs/1107.4535}{{\tt arXiv:1107.4535}}].

\bibitem{Larkoski:2012eh}
A.~J. Larkoski, {\it {QCD Analysis of the Scale-Invariance of Jets}},  {\em
  Phys. Rev. D} {\bf 86} (2012) 054004,
  [\href{http://arxiv.org/abs/1207.1437}{{\tt arXiv:1207.1437}}].

\bibitem{Jankowiak:2012na}
M.~Jankowiak and A.~J. Larkoski, {\it {Angular Scaling in Jets}},  {\em JHEP}
  {\bf 1204} (2012) 039, [\href{http://arxiv.org/abs/1201.2688}{{\tt
  arXiv:1201.2688}}].

\bibitem{Chien:2014nsa}
Y.-T. Chien and I.~Vitev, {\it {Jet Shape Resummation Using Soft-Collinear
  Effective Theory}},  \href{http://arxiv.org/abs/1405.4293}{{\tt
  arXiv:1405.4293}}.

\bibitem{Chien:2014zna}
Y.-T. Chien, {\it {Resummation of Jet Shapes and Extracting Properties of the
  Quark-Gluon Plasma}},  {\em Int.J.Mod.Phys.Conf.Ser.} {\bf 37} (2015)
  1560047, [\href{http://arxiv.org/abs/1411.0741}{{\tt arXiv:1411.0741}}].

\bibitem{Isaacson:2015fra}
J.~Isaacson, H.-n. Li, Z.~Li, and C.~P. Yuan, {\it {Factorization for
  substructures of boosted Higgs jets}},
  \href{http://arxiv.org/abs/1505.06368}{{\tt arXiv:1505.06368}}.

\bibitem{Krohn:2012fg}
D.~Krohn, M.~D. Schwartz, T.~Lin, and W.~J. Waalewijn, {\it {Jet Charge at the
  LHC}},  {\em Phys. Rev. Lett.} {\bf 110} (2013) 212001,
  [\href{http://arxiv.org/abs/1209.2421}{{\tt arXiv:1209.2421}}].

\bibitem{Waalewijn:2012sv}
W.~J. Waalewijn, {\it {Calculating the Charge of a Jet}},  {\em Phys.Rev.} {\bf
  D86} (2012) 094030, [\href{http://arxiv.org/abs/1209.3019}{{\tt
  arXiv:1209.3019}}].

\bibitem{Bertolini:2015pka}
D.~Bertolini, J.~Thaler, and J.~R. Walsh, {\it {The First Calculation of
  Fractional Jets}},  {\em JHEP} {\bf 05} (2015) 008,
  [\href{http://arxiv.org/abs/1501.01965}{{\tt arXiv:1501.01965}}].

\bibitem{Bhattacherjee:2015psa}
B.~Bhattacherjee, S.~Mukhopadhyay, M.~M. Nojiri, Y.~Sakaki, and B.~R. Webber,
  {\it {Associated jet and subjet rates in light-quark and gluon jet
  discrimination}},  {\em JHEP} {\bf 1504} (2015) 131,
  [\href{http://arxiv.org/abs/1501.04794}{{\tt arXiv:1501.04794}}].

\bibitem{Almeida:2008yp}
L.~G. Almeida, S.~J. Lee, G.~Perez, G.~F. Sterman, I.~Sung, et~al., {\it
  {Substructure of high-$p_T$ Jets at the LHC}},  {\em Phys.Rev.} {\bf D79}
  (2009) 074017, [\href{http://arxiv.org/abs/0807.0234}{{\tt
  arXiv:0807.0234}}].

\bibitem{Larkoski:2015lea}
A.~J. Larkoski, S.~Marzani, and J.~Thaler, {\it {Sudakov Safety in Perturbative
  QCD}},  {\em Phys.Rev.} {\bf D91} (2015), no.~11 111501,
  [\href{http://arxiv.org/abs/1502.01719}{{\tt arXiv:1502.01719}}].

\bibitem{Korchemsky:1999kt}
G.~P. Korchemsky and G.~Sterman, {\it {Power corrections to event shapes and
  factorization}},  {\em Nucl. Phys. B} {\bf 555} (1999) 335--351,
  [\href{http://arxiv.org/abs/hep-ph/9902341}{{\tt hep-ph/9902341}}].

\bibitem{Korchemsky:2000kp}
G.~P. Korchemsky and S.~Tafat, {\it {On power corrections to the event shape
  distributions in QCD}},  {\em JHEP} {\bf 10} (2000) 010,
  [\href{http://arxiv.org/abs/hep-ph/0007005}{{\tt hep-ph/0007005}}].

\bibitem{Giele:2007di}
W.~T. Giele, D.~A. Kosower, and P.~Z. Skands, {\it {A simple shower and
  matching algorithm}},  {\em Phys. Rev.} {\bf D78} (2008) 014026,
  [\href{http://arxiv.org/abs/0707.3652}{{\tt arXiv:0707.3652}}].

\bibitem{Giele:2011cb}
W.~T. Giele, D.~A. Kosower, and P.~Z. Skands, {\it {Higher-Order Corrections to
  Timelike Jets}},  {\em Phys. Rev.} {\bf D84} (2011) 054003,
  [\href{http://arxiv.org/abs/1102.2126}{{\tt arXiv:1102.2126}}].

\bibitem{GehrmannDeRidder:2011dm}
A.~Gehrmann-De~Ridder, M.~Ritzmann, and P.~Z. Skands, {\it {Timelike
  Dipole-Antenna Showers with Massive Fermions}},  {\em Phys. Rev.} {\bf D85}
  (2012) 014013, [\href{http://arxiv.org/abs/1108.6172}{{\tt
  arXiv:1108.6172}}].

\bibitem{Ritzmann:2012ca}
M.~Ritzmann, D.~A. Kosower, and P.~Skands, {\it {Antenna Showers with Hadronic
  Initial States}},  {\em Phys. Lett.} {\bf B718} (2013) 1345--1350,
  [\href{http://arxiv.org/abs/1210.6345}{{\tt arXiv:1210.6345}}].

\bibitem{Hartgring:2013jma}
L.~Hartgring, E.~Laenen, and P.~Skands, {\it {Antenna Showers with One-Loop
  Matrix Elements}},  {\em JHEP} {\bf 10} (2013) 127,
  [\href{http://arxiv.org/abs/1303.4974}{{\tt arXiv:1303.4974}}].

\bibitem{Larkoski:2013yi}
A.~J. Larkoski, J.~J. Lopez-Villarejo, and P.~Skands, {\it {Helicity-Dependent
  Showers and Matching with VINCIA}},  {\em Phys. Rev.} {\bf D87} (2013), no.~5
  054033, [\href{http://arxiv.org/abs/1301.0933}{{\tt arXiv:1301.0933}}].

\bibitem{Salam:2001bd}
G.~Salam and D.~Wicke, {\it {Hadron masses and power corrections to event
  shapes}},  {\em JHEP} {\bf 0105} (2001) 061,
  [\href{http://arxiv.org/abs/hep-ph/0102343}{{\tt hep-ph/0102343}}].

\bibitem{Mateu:2012nk}
V.~Mateu, I.~W. Stewart, and J.~Thaler, {\it {Power Corrections to Event Shapes
  with Mass-Dependent Operators}},  {\em Phys.Rev.} {\bf D87} (2013), no.~1
  014025, [\href{http://arxiv.org/abs/1209.3781}{{\tt arXiv:1209.3781}}].

\bibitem{Brandt:1978zm}
S.~Brandt and H.~Dahmen, {\it {Axes and Scalar Measures of Two-Jet and
  Three-Jet Events}},  {\em Z.Phys.} {\bf C1} (1979) 61.

\bibitem{Catani:1993hr}
S.~Catani, Y.~L. Dokshitzer, M.~Seymour, and B.~Webber, {\it {Longitudinally
  invariant $K_t$ clustering algorithms for hadron hadron collisions}},  {\em
  Nucl.Phys.} {\bf B406} (1993) 187--224.

\bibitem{Ellis:1993tq}
S.~D. Ellis and D.~E. Soper, {\it {Successive combination jet algorithm for
  hadron collisions}},  {\em Phys.Rev.} {\bf D48} (1993) 3160--3166,
  [\href{http://arxiv.org/abs/hep-ph/9305266}{{\tt hep-ph/9305266}}].

\bibitem{Dokshitzer:1997in}
Y.~L. Dokshitzer, G.~Leder, S.~Moretti, and B.~Webber, {\it {Better jet
  clustering algorithms}},  {\em JHEP} {\bf 9708} (1997) 001,
  [\href{http://arxiv.org/abs/hep-ph/9707323}{{\tt hep-ph/9707323}}].

\bibitem{Wobisch:1998wt}
M.~Wobisch and T.~Wengler, {\it {Hadronization corrections to jet
  cross-sections in deep inelastic scattering}},
  \href{http://arxiv.org/abs/hep-ph/9907280}{{\tt hep-ph/9907280}}.

\bibitem{Wobisch:2000dk}
M.~Wobisch, {\it {Measurement and QCD analysis of jet cross-sections in deep
  inelastic positron proton collisions at $\sqrt{s} = 300$~GeV}},  2000.

\bibitem{Appleby:2002ke}
R.~Appleby and M.~Seymour, {\it {Nonglobal logarithms in interjet energy flow
  with kt clustering requirement}},  {\em JHEP} {\bf 0212} (2002) 063,
  [\href{http://arxiv.org/abs/hep-ph/0211426}{{\tt hep-ph/0211426}}].

\bibitem{Banfi:2005gj}
A.~Banfi and M.~Dasgupta, {\it {Problems in resumming interjet energy flows
  with $k_t$ clustering}},  {\em Phys. Lett. B} {\bf 628} (2005) 49--56,
  [\href{http://arxiv.org/abs/hep-ph/0508159}{{\tt hep-ph/0508159}}].

\bibitem{Banfi:2010pa}
A.~Banfi, M.~Dasgupta, K.~Khelifa-Kerfa, and S.~Marzani, {\it {Non-global
  logarithms and jet algorithms in high-pT jet shapes}},  {\em JHEP} {\bf 08}
  (2010) 064, [\href{http://arxiv.org/abs/1004.3483}{{\tt arXiv:1004.3483}}].

\bibitem{Kelley:2012kj}
R.~Kelley, J.~R. Walsh, and S.~Zuberi, {\it {Abelian Non-Global Logarithms from
  Soft Gluon Clustering}},  {\em JHEP} {\bf 1209} (2012) 117,
  [\href{http://arxiv.org/abs/1202.2361}{{\tt arXiv:1202.2361}}].

\bibitem{Ellis:2010rwa}
S.~D. Ellis, C.~K. Vermilion, J.~R. Walsh, A.~Hornig, and C.~Lee, {\it {Jet
  Shapes and Jet Algorithms in SCET}},  {\em JHEP} {\bf 1011} (2010) 101,
  [\href{http://arxiv.org/abs/1001.0014}{{\tt arXiv:1001.0014}}].

\bibitem{Fleming:2007xt}
S.~Fleming, A.~H. Hoang, S.~Mantry, and I.~W. Stewart, {\it {Top Jets in the
  Peak Region: Factorization Analysis with NLL Resummation}},  {\em Phys. Rev.
  D} {\bf 77} (2008) 114003, [\href{http://arxiv.org/abs/0711.2079}{{\tt
  arXiv:0711.2079}}].

\bibitem{Kelley:2011aa}
R.~Kelley, M.~D. Schwartz, R.~M. Schabinger, and H.~X. Zhu, {\it {Jet Mass with
  a Jet Veto at Two Loops and the Universality of Non-Global Structure}},  {\em
  Phys.Rev.} {\bf D86} (2012) 054017,
  [\href{http://arxiv.org/abs/1112.3343}{{\tt arXiv:1112.3343}}].

\bibitem{Hornig:2011tg}
A.~Hornig, C.~Lee, J.~R. Walsh, and S.~Zuberi, {\it {Double Non-Global
  Logarithms In-N-Out of Jets}},  {\em JHEP} {\bf 1201} (2012) 149,
  [\href{http://arxiv.org/abs/1110.0004}{{\tt arXiv:1110.0004}}].

\bibitem{Nagy:1997yn}
Z.~Nagy and Z.~Trocsanyi, {\it {Next-to-leading order calculation of four jet
  shape variables}},  {\em Phys.Rev.Lett.} {\bf 79} (1997) 3604--3607,
  [\href{http://arxiv.org/abs/hep-ph/9707309}{{\tt hep-ph/9707309}}].

\bibitem{Nagy:1998bb}
Z.~Nagy and Z.~Trocsanyi, {\it {Next-to-leading order calculation of four jet
  observables in electron positron annihilation}},  {\em Phys.Rev.} {\bf D59}
  (1999) 014020, [\href{http://arxiv.org/abs/hep-ph/9806317}{{\tt
  hep-ph/9806317}}].

\bibitem{Nagy:2001fj}
Z.~Nagy, {\it {Three jet cross-sections in hadron hadron collisions at
  next-to-leading order}},  {\em Phys.Rev.Lett.} {\bf 88} (2002) 122003,
  [\href{http://arxiv.org/abs/hep-ph/0110315}{{\tt hep-ph/0110315}}].

\bibitem{Nagy:2001xb}
Z.~Nagy and Z.~Trocsanyi, {\it {Multijet cross-sections in deep inelastic
  scattering at next-to-leading order}},  {\em Phys.Rev.Lett.} {\bf 87} (2001)
  082001, [\href{http://arxiv.org/abs/hep-ph/0104315}{{\tt hep-ph/0104315}}].

\bibitem{Nagy:2003tz}
Z.~Nagy, {\it {Next-to-leading order calculation of three jet observables in
  hadron hadron collision}},  {\em Phys. Rev. D} {\bf 68} (2003) 094002,
  [\href{http://arxiv.org/abs/hep-ph/0307268}{{\tt hep-ph/0307268}}].

\bibitem{Manohar:2006nz}
A.~V. Manohar and I.~W. Stewart, {\it {The Zero-Bin and Mode Factorization in
  Quantum Field Theory}},  {\em Phys.Rev.} {\bf D76} (2007) 074002,
  [\href{http://arxiv.org/abs/hep-ph/0605001}{{\tt hep-ph/0605001}}].

\bibitem{Fischer:2014bja}
N.~Fischer, S.~Gieseke, S.~Pl{\"a}tzer, and P.~Skands, {\it {Revisiting
  radiation patterns in $e^+e^-$ collisions}},  {\em Eur.Phys.J.} {\bf C74}
  (2014), no.~4 2831, [\href{http://arxiv.org/abs/1402.3186}{{\tt
  arXiv:1402.3186}}].

\bibitem{Fischer:2015pqa}
{\bf OPAL} Collaboration, N.~Fischer, S.~Gieseke, S.~Kluth, S.~Pl{\"a}tzer, and
  P.~Skands, {\it {Measurement of observables sensitive to coherence effects in
  hadronic Z decays with the OPAL detector at LEP}},
  \href{http://arxiv.org/abs/1505.01636}{{\tt arXiv:1505.01636}}.

\bibitem{Buckley:2011ms}
A.~Buckley, J.~Butterworth, S.~Gieseke, D.~Grellscheid, S.~Hoche, et~al., {\it
  {General-purpose event generators for LHC physics}},  {\em Phys.~Rept.} {\bf
  504} (2011) 145--233, [\href{http://arxiv.org/abs/1101.2599}{{\tt
  arXiv:1101.2599}}].

\bibitem{Skands:2012ts}
P.~Skands, {\it {Introduction to QCD}},
  \href{http://arxiv.org/abs/1207.2389}{{\tt arXiv:1207.2389}}.

\bibitem{Seymour:2013ega}
M.~H. Seymour and M.~Marx, {\it {Monte Carlo Event Generators}},
  \href{http://arxiv.org/abs/1304.6677}{{\tt arXiv:1304.6677}}.

\bibitem{Gieseke:2013eva}
S.~Gieseke, {\it {Simulation of jets at colliders}},  {\em
  Prog.Part.Nucl.Phys.} {\bf 72} (2013) 155--205.

\bibitem{Hoche:2014rga}
S.~H{\"o}che, {\it {Introduction to parton-shower event generators}},
  \href{http://arxiv.org/abs/1411.4085}{{\tt arXiv:1411.4085}}.

\bibitem{Gleisberg:2003xi}
T.~Gleisberg, S.~Hoeche, F.~Krauss, A.~Schalicke, S.~Schumann, et~al., {\it
  {SHERPA 1. alpha: A Proof of concept version}},  {\em JHEP} {\bf 0402} (2004)
  056, [\href{http://arxiv.org/abs/hep-ph/0311263}{{\tt hep-ph/0311263}}].

\bibitem{Gleisberg:2008ta}
T.~Gleisberg, S.~Hoeche, F.~Krauss, M.~Schonherr, S.~Schumann, et~al., {\it
  {Event generation with SHERPA 1.1}},  {\em JHEP} {\bf 0902} (2009) 007,
  [\href{http://arxiv.org/abs/0811.4622}{{\tt arXiv:0811.4622}}].

\bibitem{Lonnblad:1992tz}
L.~Lonnblad, {\it {ARIADNE version 4: A Program for simulation of QCD cascades
  implementing the color dipole model}},  {\em Comput.Phys.Commun.} {\bf 71}
  (1992) 15--31.

\bibitem{Hoche:2015sya}
S.~H{\"o}che and S.~Prestel, {\it {The midpoint between dipole and parton
  showers}},  \href{http://arxiv.org/abs/1506.05057}{{\tt arXiv:1506.05057}}.

\bibitem{Platzer:2011bc}
S.~Platzer and S.~Gieseke, {\it {Dipole Showers and Automated NLO Matching in
  Herwig++}},  {\em Eur.Phys.J.} {\bf C72} (2012) 2187,
  [\href{http://arxiv.org/abs/1109.6256}{{\tt arXiv:1109.6256}}].

\bibitem{Platzer:2009jq}
S.~Platzer and S.~Gieseke, {\it {Coherent Parton Showers with Local Recoils}},
  {\em JHEP} {\bf 1101} (2011) 024, [\href{http://arxiv.org/abs/0909.5593}{{\tt
  arXiv:0909.5593}}].

\bibitem{Dasgupta:2002bw}
M.~Dasgupta and G.~P. Salam, {\it {Accounting for coherence in interjet E(t)
  flow: A Case study}},  {\em JHEP} {\bf 0203} (2002) 017,
  [\href{http://arxiv.org/abs/hep-ph/0203009}{{\tt hep-ph/0203009}}].

\bibitem{Tackmann:2012bt}
F.~J. Tackmann, J.~R. Walsh, and S.~Zuberi, {\it {Resummation Properties of Jet
  Vetoes at the LHC}},  {\em Phys.Rev.} {\bf D86} (2012) 053011,
  [\href{http://arxiv.org/abs/1206.4312}{{\tt arXiv:1206.4312}}].

\bibitem{Dasgupta:2014yra}
M.~Dasgupta, F.~Dreyer, G.~P. Salam, and G.~Soyez, {\it {Small-radius jets to
  all orders in QCD}},  {\em JHEP} {\bf 1504} (2015) 039,
  [\href{http://arxiv.org/abs/1411.5182}{{\tt arXiv:1411.5182}}].

\bibitem{Bosch:2004th}
S.~Bosch, B.~Lange, M.~Neubert, and G.~Paz, {\it {Factorization and shape
  function effects in inclusive B meson decays}},  {\em Nucl.Phys.} {\bf B699}
  (2004) 335--386, [\href{http://arxiv.org/abs/hep-ph/0402094}{{\tt
  hep-ph/0402094}}].

\bibitem{Hoang:2007vb}
A.~H. Hoang and I.~W. Stewart, {\it {Designing Gapped Soft Functions for Jet
  Production}},  {\em Phys. Lett. B} {\bf 660} (2008) 483--493,
  [\href{http://arxiv.org/abs/0709.3519}{{\tt arXiv:0709.3519}}].

\bibitem{Ligeti:2008ac}
Z.~Ligeti, I.~W. Stewart, and F.~J. Tackmann, {\it {Treating the b quark
  distribution function with reliable uncertainties}},  {\em Phys. Rev. D} {\bf
  78} (2008) 114014, [\href{http://arxiv.org/abs/0807.1926}{{\tt
  arXiv:0807.1926}}].

\bibitem{Akhoury:1995sp}
R.~Akhoury and V.~I. Zakharov, {\it {On the universality of the leading, 1/Q
  power corrections in QCD}},  {\em Phys.Lett.} {\bf B357} (1995) 646--652,
  [\href{http://arxiv.org/abs/hep-ph/9504248}{{\tt hep-ph/9504248}}].

\bibitem{Dokshitzer:1995zt}
Y.~L. Dokshitzer and B.~R. Webber, {\it {Calculation of power corrections to
  hadronic event shapes}},  {\em Phys. Lett. B} {\bf 352} (1995) 451--455,
  [\href{http://arxiv.org/abs/hep-ph/9504219}{{\tt hep-ph/9504219}}].

\bibitem{Lee:2006fn}
C.~Lee and G.~F. Sterman, {\it {Universality of nonperturbative effects in
  event shapes}},  {\em eConf} {\bf C0601121} (2006) A001,
  [\href{http://arxiv.org/abs/hep-ph/0603066}{{\tt hep-ph/0603066}}].

\bibitem{Lee:2007jr}
C.~Lee, {\it {Universal nonperturbative effects in event shapes from
  soft-collinear effective theory}},  {\em Mod.Phys.Lett.} {\bf A22} (2007)
  835--851, [\href{http://arxiv.org/abs/hep-ph/0703030}{{\tt hep-ph/0703030}}].

\bibitem{Stewart:2014nna}
I.~W. Stewart, F.~J. Tackmann, and W.~J. Waalewijn, {\it {Dissecting Soft
  Radiation with Factorization}},  {\em Phys.Rev.Lett.} {\bf 114} (2015), no.~9
  092001, [\href{http://arxiv.org/abs/1405.6722}{{\tt arXiv:1405.6722}}].

\bibitem{Beneke:1998ui}
M.~Beneke, {\it {Renormalons}},  {\em Phys.Rept.} {\bf 317} (1999) 1--142,
  [\href{http://arxiv.org/abs/hep-ph/9807443}{{\tt hep-ph/9807443}}].

\bibitem{Gardi:2000yh}
E.~Gardi, {\it {Perturbative and nonperturbative aspects of moments of the
  thrust distribution in e+ e- annihilation}},  {\em JHEP} {\bf 0004} (2000)
  030, [\href{http://arxiv.org/abs/hep-ph/0003179}{{\tt hep-ph/0003179}}].

\bibitem{Hornig:2009vb}
A.~Hornig, C.~Lee, and G.~Ovanesyan, {\it {Effective Predictions of Event
  Shapes: Factorized, Resummed, and Gapped Angularity Distributions}},  {\em
  JHEP} {\bf 0905} (2009) 122, [\href{http://arxiv.org/abs/0901.3780}{{\tt
  arXiv:0901.3780}}].

\bibitem{Achard:2004sv}
{\bf L3} Collaboration, P.~Achard et~al., {\it {Studies of hadronic event
  structure in $e^{+} e^{-}$ annihilation from 30-GeV to 209-GeV with the L3
  detector}},  {\em Phys.Rept.} {\bf 399} (2004) 71--174,
  [\href{http://arxiv.org/abs/hep-ex/0406049}{{\tt hep-ex/0406049}}].

\bibitem{Gehrmann:2009eh}
T.~Gehrmann, M.~Jaquier, and G.~Luisoni, {\it {Hadronization effects in event
  shape moments}},  {\em Eur.Phys.J.} {\bf C67} (2010) 57--72,
  [\href{http://arxiv.org/abs/0911.2422}{{\tt arXiv:0911.2422}}].

\bibitem{Abbate:2012jh}
R.~Abbate, M.~Fickinger, A.~H. Hoang, V.~Mateu, and I.~W. Stewart, {\it
  {Precision Thrust Cumulant Moments at $N^3$LL}},  {\em Phys.~Rev.~D} {\bf 86}
  (2012) 094002, [\href{http://arxiv.org/abs/1204.5746}{{\tt
  arXiv:1204.5746}}].

\bibitem{Hoang:2015hka}
A.~H. Hoang, D.~W. Kolodrubetz, V.~Mateu, and I.~W. Stewart, {\it {A Precise
  Determination of $\alpha_s$ from the C-parameter Distribution}},
  \href{http://arxiv.org/abs/1501.04111}{{\tt arXiv:1501.04111}}.

\bibitem{Liu:2012sz}
X.~Liu and F.~Petriello, {\it {Resummation of jet-veto logarithms in hadronic
  processes containing jets}},  {\em Phys.~Rev.~D} {\bf 87} (2013) 014018,
  [\href{http://arxiv.org/abs/1210.1906}{{\tt arXiv:1210.1906}}].

\bibitem{Liu:2013hba}
X.~Liu and F.~Petriello, {\it {Reducing theoretical uncertainties for exclusive
  Higgs-boson plus one-jet production at the LHC}},  {\em Phys.Rev.} {\bf D87}
  (2013), no.~9 094027, [\href{http://arxiv.org/abs/1303.4405}{{\tt
  arXiv:1303.4405}}].

\bibitem{guido_talk}
J.~Talbert, {\it {Automated Calculations of Dijet Soft Functions}},  {\em SCET
  Conference} (2015).

\bibitem{Boughezal:2015eha}
R.~Boughezal, X.~Liu, and F.~Petriello, {\it {$N$-jettiness soft function at
  next-to-next-to-leading order}},  {\em Phys.Rev.} {\bf D91} (2015), no.~9
  094035, [\href{http://arxiv.org/abs/1504.02540}{{\tt arXiv:1504.02540}}].

\bibitem{Banfi:2002hw}
A.~Banfi, G.~Marchesini, and G.~Smye, {\it {Away-from-jet energy flow}},  {\em
  JHEP} {\bf 08} (2002) 006, [\href{http://arxiv.org/abs/hep-ph/0206076}{{\tt
  hep-ph/0206076}}].

\bibitem{Weigert:2003mm}
H.~Weigert, {\it {Nonglobal jet evolution at finite N(c)}},  {\em Nucl.Phys.}
  {\bf B685} (2004) 321--350, [\href{http://arxiv.org/abs/hep-ph/0312050}{{\tt
  hep-ph/0312050}}].

\bibitem{Hatta:2013iba}
Y.~Hatta and T.~Ueda, {\it {Resummation of non-global logarithms at finite
  $N_c$}},  {\em Nucl. Phys.} {\bf B874} (2013) 808--820,
  [\href{http://arxiv.org/abs/1304.6930}{{\tt arXiv:1304.6930}}].

\bibitem{Caron-Huot:2015bja}
S.~Caron-Huot, {\it {Resummation of non-global logarithms and the BFKL
  equation}},  \href{http://arxiv.org/abs/1501.03754}{{\tt arXiv:1501.03754}}.

\bibitem{Boughezal:2015aha}
R.~Boughezal, C.~Focke, W.~Giele, X.~Liu, and F.~Petriello, {\it {Higgs boson
  production in association with a jet using jettiness subtraction}},  {\em
  Phys.Lett.} {\bf B748} (2015) 5--8,
  [\href{http://arxiv.org/abs/1505.03893}{{\tt arXiv:1505.03893}}].

\bibitem{Boughezal:2015dva}
R.~Boughezal, C.~Focke, X.~Liu, and F.~Petriello, {\it {$W$-boson production in
  association with a jet at next-to-next-to-leading order in perturbative
  QCD}},  \href{http://arxiv.org/abs/1504.02131}{{\tt arXiv:1504.02131}}.

\bibitem{Gaunt:2015pea}
J.~Gaunt, M.~Stahlhofen, F.~J. Tackmann, and J.~R. Walsh, {\it {N-jettiness
  Subtractions for NNLO QCD Calculations}},
  \href{http://arxiv.org/abs/1505.04794}{{\tt arXiv:1505.04794}}.

\bibitem{Boughezal:2015dra}
R.~Boughezal, F.~Caola, K.~Melnikov, F.~Petriello, and M.~Schulze, {\it {Higgs
  Boson Production in Association with a Jet at Next-to-Next-to-Leading
  Order}},  \href{http://arxiv.org/abs/1504.07922}{{\tt arXiv:1504.07922}}.

\bibitem{Larkoski:2015yqa}
A.~J. Larkoski, F.~Maltoni, and M.~Selvaggi, {\it {Tracking down hyper-boosted
  top quarks}},  {\em JHEP} {\bf 1506} (2015) 032,
  [\href{http://arxiv.org/abs/1503.03347}{{\tt arXiv:1503.03347}}].

\bibitem{Fleming:2007qr}
S.~Fleming, A.~H. Hoang, S.~Mantry, and I.~W. Stewart, {\it {Jets from massive
  unstable particles: Top-mass determination}},  {\em Phys. Rev. D} {\bf 77}
  (2008) 074010, [\href{http://arxiv.org/abs/hep-ph/0703207}{{\tt
  hep-ph/0703207}}].

\bibitem{DeCausmaecker:1981bg}
P.~De~Causmaecker, R.~Gastmans, W.~Troost, and T.~T. Wu, {\it {Multiple
  Bremsstrahlung in Gauge Theories at High-Energies. 1. General Formalism for
  Quantum Electrodynamics}},  {\em Nucl. Phys. B} {\bf 206} (1982) 53.

\bibitem{Berends:1981uq}
F.~A. Berends, R.~Kleiss, P.~De~Causmaecker, R.~Gastmans, W.~Troost, et~al.,
  {\it {Multiple Bremsstrahlung in Gauge Theories at High-Energies. 2. Single
  Bremsstrahlung}},  {\em Nucl. Phys. B} {\bf 206} (1982) 61.

\bibitem{Gunion:1985vca}
J.~Gunion and Z.~Kunszt, {\it {Improved Analytic Techniques for Tree Graph
  Calculations and the G g q anti-q Lepton anti-Lepton Subprocess}},  {\em
  Phys. Lett. B} {\bf 161} (1985) 333.

\bibitem{Xu:1986xb}
Z.~Xu, D.-H. Zhang, and L.~Chang, {\it {Helicity Amplitudes for Multiple
  Bremsstrahlung in Massless Nonabelian Gauge Theories}},  {\em Nucl. Phys. B}
  {\bf 291} (1987) 392.

\bibitem{Berends:1987me}
F.~A. Berends and W.~Giele, {\it {Recursive Calculations for Processes with n
  Gluons}},  {\em Nucl.Phys.} {\bf B306} (1988) 759.

\bibitem{Mangano:1987xk}
M.~L. Mangano, S.~J. Parke, and Z.~Xu, {\it {Duality and Multi - Gluon
  Scattering}},  {\em Nucl.Phys.} {\bf B298} (1988) 653.

\bibitem{Mangano:1988kk}
M.~L. Mangano, {\it {The Color Structure of Gluon Emission}},  {\em Nucl.Phys.}
  {\bf B309} (1988) 461.

\bibitem{Bern:1990ux}
Z.~Bern and D.~A. Kosower, {\it {Color decomposition of one loop amplitudes in
  gauge theories}},  {\em Nucl.Phys.} {\bf B362} (1991) 389--448.

\bibitem{Bern:1994zx}
Z.~Bern, L.~J. Dixon, D.~C. Dunbar, and D.~A. Kosower, {\it {One-Loop n-Point
  Gauge Theory Amplitudes, Unitarity and Collinear Limits}},  {\em Nucl. Phys.
  B} {\bf 425} (1994) 217--260,
  [\href{http://arxiv.org/abs/hep-ph/9403226}{{\tt hep-ph/9403226}}].

\bibitem{Bern:1994cg}
Z.~Bern, L.~J. Dixon, D.~C. Dunbar, and D.~A. Kosower, {\it {Fusing gauge
  theory tree amplitudes into loop amplitudes}},  {\em Nucl.Phys.} {\bf B435}
  (1995) 59--101, [\href{http://arxiv.org/abs/hep-ph/9409265}{{\tt
  hep-ph/9409265}}].

\bibitem{Giele:1991vf}
W.~Giele and E.~Glover, {\it {Higher order corrections to jet cross-sections in
  e+ e- annihilation}},  {\em Phys. Rev. D} {\bf 46} (1992) 1980--2010.

\bibitem{Arnold:1988dp}
P.~B. Arnold and M.~Reno, {\it {The Complete Computation of High p(t) W and Z
  Production in 2nd Order QCD}},  {\em Nucl. Phys. B} {\bf 319} (1989) 37.

\bibitem{Giele:1993dj}
W.~Giele, E.~Glover, and D.~A. Kosower, {\it {Higher order corrections to jet
  cross-sections in hadron colliders}},  {\em Nucl. Phys. B} {\bf 403} (1993)
  633--670, [\href{http://arxiv.org/abs/hep-ph/9302225}{{\tt hep-ph/9302225}}].

\bibitem{Bern:1997sc}
Z.~Bern, L.~J. Dixon, and D.~A. Kosower, {\it {One-loop amplitudes for e+ e- to
  four partons}},  {\em Nucl. Phys. B} {\bf 513} (1998) 3--86,
  [\href{http://arxiv.org/abs/hep-ph/9708239}{{\tt hep-ph/9708239}}].

\bibitem{Ellis:2008qc}
R.~Ellis, W.~Giele, Z.~Kunszt, K.~Melnikov, and G.~Zanderighi, {\it {One-loop
  amplitudes for W+3 jet production in hadron collisions}},  {\em JHEP} {\bf
  01} (2009) 012, [\href{http://arxiv.org/abs/0810.2762}{{\tt
  arXiv:0810.2762}}].

\bibitem{Berger:2009zg}
C.~Berger, Z.~Bern, L.~J. Dixon, F.~Febres~Cordero, D.~Forde, et~al., {\it
  {Precise Predictions for $W$ + 3 Jet Production at Hadron Colliders}},  {\em
  Phys.Rev.Lett.} {\bf 102} (2009) 222001,
  [\href{http://arxiv.org/abs/0902.2760}{{\tt arXiv:0902.2760}}].

\bibitem{Berger:2009ep}
C.~Berger, Z.~Bern, L.~J. Dixon, F.~Febres~Cordero, D.~Forde, et~al., {\it
  {Next-to-Leading Order QCD Predictions for W+3-Jet Distributions at Hadron
  Colliders}},  {\em Phys.Rev.} {\bf D80} (2009) 074036,
  [\href{http://arxiv.org/abs/0907.1984}{{\tt arXiv:0907.1984}}].

\bibitem{Berger:2010vm}
C.~Berger, Z.~Bern, L.~J. Dixon, F.~Febres~Cordero, D.~Forde, et~al., {\it
  {Next-to-Leading Order QCD Predictions for Z, $gamma^*$+3-Jet Distributions
  at the Tevatron}},  {\em Phys.~Rev.~D} {\bf 82} (2010) 074002,
  [\href{http://arxiv.org/abs/1004.1659}{{\tt arXiv:1004.1659}}].

\bibitem{Berger:2010zx}
C.~Berger, Z.~Bern, L.~J. Dixon, F.~Febres~Cordero, D.~Forde, et~al., {\it
  {Precise Predictions for W + 4 Jet Production at the Large Hadron Collider}},
   {\em Phys.Rev.Lett.} {\bf 106} (2011) 092001,
  [\href{http://arxiv.org/abs/1009.2338}{{\tt arXiv:1009.2338}}].

\bibitem{Ita:2011wn}
H.~Ita, Z.~Bern, L.~Dixon, F.~Febres~Cordero, D.~Kosower, et~al., {\it {Precise
  Predictions for Z + 4 Jets at Hadron Colliders}},  {\em Phys.Rev.} {\bf D85}
  (2012) 031501, [\href{http://arxiv.org/abs/1108.2229}{{\tt
  arXiv:1108.2229}}].

\bibitem{Bern:2013gka}
Z.~Bern, L.~Dixon, F.~Febres~Cordero, S.~H{\"o}che, H.~Ita, et~al., {\it
  {Next-to-Leading Order $W + 5$-Jet Production at the LHC}},  {\em Phys.Rev.}
  {\bf D88} (2013), no.~1 014025, [\href{http://arxiv.org/abs/1304.1253}{{\tt
  arXiv:1304.1253}}].

\bibitem{Bern:1993mq}
Z.~Bern, L.~J. Dixon, and D.~A. Kosower, {\it {One loop corrections to five
  gluon amplitudes}},  {\em Phys. Rev. Lett.} {\bf 70} (1993) 2677--2680,
  [\href{http://arxiv.org/abs/hep-ph/9302280}{{\tt hep-ph/9302280}}].

\bibitem{Kunszt:1993sd}
Z.~Kunszt, A.~Signer, and Z.~Trocsanyi, {\it {One loop helicity amplitudes for
  all 2 $\to$ 2 processes in QCD and N=1 supersymmetric Yang-Mills theory}},
  {\em Nucl. Phys. B} {\bf 411} (1994) 397--442,
  [\href{http://arxiv.org/abs/hep-ph/9305239}{{\tt hep-ph/9305239}}].

\bibitem{Kunszt:1994tq}
Z.~Kunszt, A.~Signer, and Z.~Trocsanyi, {\it {One loop radiative corrections to
  the helicity amplitudes of QCD processes involving four quarks and one
  gluon}},  {\em Phys. Lett. B} {\bf 336} (1994) 529--536,
  [\href{http://arxiv.org/abs/hep-ph/9405386}{{\tt hep-ph/9405386}}].

\bibitem{Bern:1994fz}
Z.~Bern, L.~J. Dixon, and D.~A. Kosower, {\it {One loop corrections to two
  quark three gluon amplitudes}},  {\em Nucl. Phys. B} {\bf 437} (1995)
  259--304, [\href{http://arxiv.org/abs/hep-ph/9409393}{{\tt hep-ph/9409393}}].

\bibitem{Bern:2011ep}
Z.~Bern, G.~Diana, L.~Dixon, F.~Febres~Cordero, S.~Hoeche, et~al., {\it
  {Four-Jet Production at the Large Hadron Collider at Next-to-Leading Order in
  QCD}},  {\em Phys.Rev.Lett.} {\bf 109} (2012) 042001,
  [\href{http://arxiv.org/abs/1112.3940}{{\tt arXiv:1112.3940}}].

\bibitem{Badger:2012pf}
S.~Badger, B.~Biedermann, P.~Uwer, and V.~Yundin, {\it {NLO QCD corrections to
  multi-jet production at the LHC with a centre-of-mass energy of $\sqrt{s}=8$
  TeV}},  {\em Phys. Lett.} {\bf B718} (2013) 965--978,
  [\href{http://arxiv.org/abs/1209.0098}{{\tt arXiv:1209.0098}}].

\bibitem{Badger:2013yda}
S.~Badger, B.~Biedermann, P.~Uwer, and V.~Yundin, {\it {Next-to-leading order
  QCD corrections to five jet production at the LHC}},  {\em Phys.~Rev.~D} {\bf
  89} (2014) 034019, [\href{http://arxiv.org/abs/1309.6585}{{\tt
  arXiv:1309.6585}}].

\bibitem{Campbell:2006xx}
J.~M. Campbell, R.~K. Ellis, and G.~Zanderighi, {\it {Next-to-Leading order
  Higgs + 2 jet production via gluon fusion}},  {\em JHEP} {\bf 10} (2006) 028,
  [\href{http://arxiv.org/abs/hep-ph/0608194}{{\tt hep-ph/0608194}}].

\bibitem{Kauffman:1996ix}
R.~P. Kauffman, S.~V. Desai, and D.~Risal, {\it {Production of a Higgs boson
  plus two jets in hadronic collisions}},  {\em Phys. Rev. D} {\bf 55} (1997)
  4005--4015, [\href{http://arxiv.org/abs/hep-ph/9610541}{{\tt
  hep-ph/9610541}}].

\bibitem{Schmidt:1997wr}
C.~R. Schmidt, {\it {H --> g g g (g q anti-q) at two loops in the large-M(t)
  limit}},  {\em Phys. Lett. B} {\bf 413} (1997) 391--395,
  [\href{http://arxiv.org/abs/hep-ph/9707448}{{\tt hep-ph/9707448}}].

\bibitem{Dixon:2009uk}
L.~J. Dixon and Y.~Sofianatos, {\it {Analytic one-loop amplitudes for a Higgs
  boson plus four partons}},  {\em JHEP} {\bf 0908} (2009) 058,
  [\href{http://arxiv.org/abs/0906.0008}{{\tt arXiv:0906.0008}}].

\bibitem{Badger:2009hw}
S.~Badger, E.~Nigel~Glover, P.~Mastrolia, and C.~Williams, {\it {One-loop Higgs
  plus four gluon amplitudes: Full analytic results}},  {\em JHEP} {\bf 1001}
  (2010) 036, [\href{http://arxiv.org/abs/0909.4475}{{\tt arXiv:0909.4475}}].

\bibitem{Campbell:2010cz}
J.~M. Campbell, R.~K. Ellis, and C.~Williams, {\it {Hadronic production of a
  Higgs boson and two jets at next-to-leading order}},  {\em Phys. Rev. D} {\bf
  81} (2010) 074023, [\href{http://arxiv.org/abs/1001.4495}{{\tt
  arXiv:1001.4495}}].

\bibitem{vanDeurzen:2013rv}
H.~van Deurzen, N.~Greiner, G.~Luisoni, P.~Mastrolia, E.~Mirabella, et~al.,
  {\it {NLO QCD corrections to the production of Higgs plus two jets at the
  LHC}},  {\em Phys. Lett. B} {\bf 721} (2013) 74--81,
  [\href{http://arxiv.org/abs/1301.0493}{{\tt arXiv:1301.0493}}].

\bibitem{Cullen:2013saa}
G.~Cullen, H.~van Deurzen, N.~Greiner, G.~Luisoni, P.~Mastrolia, et~al., {\it
  {Next-to-Leading-Order QCD Corrections to Higgs Boson Production Plus Three
  Jets in Gluon Fusion}},  {\em Phys.~Rev.~Lett.} {\bf 111} (2013), no.~13
  131801, [\href{http://arxiv.org/abs/1307.4737}{{\tt arXiv:1307.4737}}].

\bibitem{Campanario:2013fsa}
F.~Campanario, T.~M. Figy, S.~Pl{\"a}tzer, and M.~Sj{\"o}dahl, {\it
  {Electroweak Higgs Boson Plus Three Jet Production at Next-to-Leading-Order
  QCD}},  {\em Phys. Rev. Lett.} {\bf 111} (2013), no.~21 211802,
  [\href{http://arxiv.org/abs/1308.2932}{{\tt arXiv:1308.2932}}].

\bibitem{Binoth:2010ra}
{\bf SM and NLO Multileg Working Group} Collaboration, J.~R. Andersen et~al.,
  {\it {The SM and NLO multileg working group: Summary report}},
  \href{http://arxiv.org/abs/1003.1241}{{\tt arXiv:1003.1241}}.

\bibitem{AlcarazMaestre:2012vp}
{\bf SM and NLO MULTILEG Working Group, SM MC Working Group} Collaboration,
  J.~Alcaraz~Maestre et~al., {\it {The SM and NLO Multileg and SM MC Working
  Groups: Summary Report}},  \href{http://arxiv.org/abs/1203.6803}{{\tt
  arXiv:1203.6803}}.

\bibitem{Ossola:2007ax}
G.~Ossola, C.~G. Papadopoulos, and R.~Pittau, {\it {CutTools: a program
  implementing the OPP reduction method to compute one-loop amplitudes}},  {\em
  JHEP} {\bf 03} (2008) 042, [\href{http://arxiv.org/abs/0711.3596}{{\tt
  arXiv:0711.3596}}].

\bibitem{Berger:2008sj}
C.~Berger, Z.~Bern, L.~J. Dixon, F.~Febres~Cordero, D.~Forde, et~al., {\it {An
  Automated Implementation of On-Shell Methods for One-Loop Amplitudes}},  {\em
  Phys.Rev. D} {\bf 78} (2008) 036003,
  [\href{http://arxiv.org/abs/0803.4180}{{\tt arXiv:0803.4180}}].

\bibitem{Binoth:2008uq}
T.~Binoth, J.-P. Guillet, G.~Heinrich, E.~Pilon, and T.~Reiter, {\it {Golem95:
  A Numerical program to calculate one-loop tensor integrals with up to six
  external legs}},  {\em Comput. Phys. Commun.} {\bf 180} (2009) 2317--2330,
  [\href{http://arxiv.org/abs/0810.0992}{{\tt arXiv:0810.0992}}].

\bibitem{Mastrolia:2010nb}
P.~Mastrolia, G.~Ossola, T.~Reiter, and F.~Tramontano, {\it {Scattering
  AMplitudes from Unitarity-based Reduction Algorithm at the Integrand-level}},
   {\em JHEP} {\bf 08} (2010) 080, [\href{http://arxiv.org/abs/1006.0710}{{\tt
  arXiv:1006.0710}}].

\bibitem{Badger:2010nx}
S.~Badger, B.~Biedermann, and P.~Uwer, {\it {NGluon: A Package to Calculate
  One-loop Multi-gluon Amplitudes}},  {\em Comput. Phys. Commun.} {\bf 182}
  (2011) 1674--1692, [\href{http://arxiv.org/abs/1011.2900}{{\tt
  arXiv:1011.2900}}].

\bibitem{Fleischer:2010sq}
J.~Fleischer and T.~Riemann, {\it {A Complete algebraic reduction of one-loop
  tensor Feynman integrals}},  {\em Phys.~Rev.~D} {\bf 83} (2011) 073004,
  [\href{http://arxiv.org/abs/1009.4436}{{\tt arXiv:1009.4436}}].

\bibitem{Cullen:2011kv}
G.~Cullen, J.-P. Guillet, G.~Heinrich, T.~Kleinschmidt, E.~Pilon, et~al., {\it
  {Golem95C: A library for one-loop integrals with complex masses}},
  \href{http://arxiv.org/abs/1101.5595}{{\tt arXiv:1101.5595}}.

\bibitem{Hirschi:2011pa}
V.~Hirschi et~al., {\it {Automation of one-loop QCD corrections}},  {\em JHEP}
  {\bf 05} (2011) 044, [\href{http://arxiv.org/abs/1103.0621}{{\tt
  arXiv:1103.0621}}].

\bibitem{Bevilacqua:2011xh}
G.~Bevilacqua, M.~Czakon, M.~V. Garzelli, A.~van Hameren, A.~Kardos, C.~G.
  Papadopoulos, R.~Pittau, and M.~Worek, {\it {HELAC-NLO}},  {\em Comput. Phys.
  Commun.} {\bf 184} (2013) 986--997,
  [\href{http://arxiv.org/abs/1110.1499}{{\tt arXiv:1110.1499}}].

\bibitem{Cullen:2011ac}
G.~Cullen, N.~Greiner, G.~Heinrich, G.~Luisoni, P.~Mastrolia, et~al., {\it
  {Automated One-Loop Calculations with GoSam}},  {\em Eur. Phys. J.} {\bf C72}
  (2012) 1889, [\href{http://arxiv.org/abs/1111.2034}{{\tt arXiv:1111.2034}}].

\bibitem{Cascioli:2011va}
F.~Cascioli, P.~Maierhofer, and S.~Pozzorini, {\it {Scattering Amplitudes with
  Open Loops}},  {\em Phys. Rev. Lett.} {\bf 108} (2012) 111601,
  [\href{http://arxiv.org/abs/1111.5206}{{\tt arXiv:1111.5206}}].

\bibitem{Badger:2012pg}
S.~Badger, B.~Biedermann, P.~Uwer, and V.~Yundin, {\it {Numerical evaluation of
  virtual corrections to multi-jet production in massless QCD}},  {\em Comput.
  Phys. Commun.} {\bf 184} (2013) 1981--1998,
  [\href{http://arxiv.org/abs/1209.0100}{{\tt arXiv:1209.0100}}].

\bibitem{Actis:2012qn}
S.~Actis, A.~Denner, L.~Hofer, A.~Scharf, and S.~Uccirati, {\it {Recursive
  generation of one-loop amplitudes in the Standard Model}},  {\em JHEP} {\bf
  1304} (2013) 037, [\href{http://arxiv.org/abs/1211.6316}{{\tt
  arXiv:1211.6316}}].

\bibitem{Peraro:2014cba}
T.~Peraro, {\it {Ninja: Automated Integrand Reduction via Laurent Expansion for
  One-Loop Amplitudes}},  {\em Comput. Phys. Commun.} {\bf 185} (2014)
  2771--2797, [\href{http://arxiv.org/abs/1403.1229}{{\tt arXiv:1403.1229}}].

\bibitem{Cullen:2014yla}
G.~Cullen, H.~van Deurzen, N.~Greiner, G.~Heinrich, G.~Luisoni, et~al., {\it
  {G$\scriptsize{O}$S$\scriptsize{AM}$-2.0: a tool for automated one-loop
  calculations within the Standard Model and beyond}},  {\em Eur. Phys. J.}
  {\bf C74} (2014), no.~8 3001, [\href{http://arxiv.org/abs/1404.7096}{{\tt
  arXiv:1404.7096}}].

\bibitem{Becher:2008cf}
T.~Becher and M.~D. Schwartz, {\it {A Precise determination of $\alpha_s$ from
  LEP thrust data using effective field theory}},  {\em JHEP} {\bf 07} (2008)
  034, [\href{http://arxiv.org/abs/0803.0342}{{\tt arXiv:0803.0342}}].

\bibitem{Alioli:2012fc}
S.~Alioli, C.~W. Bauer, C.~J. Berggren, A.~Hornig, F.~J. Tackmann, et~al., {\it
  {Combining Higher-Order Resummation with Multiple NLO Calculations and Parton
  Showers in GENEVA}},  {\em JHEP} {\bf 1309} (2013) 120,
  [\href{http://arxiv.org/abs/1211.7049}{{\tt arXiv:1211.7049}}].

\bibitem{Alioli:2013hqa}
S.~Alioli, C.~W. Bauer, C.~Berggren, F.~J. Tackmann, J.~R. Walsh, and
  S.~Zuberi, {\it {Matching Fully Differential NNLO Calculations and Parton
  Showers}},  {\em JHEP} {\bf 06} (2014) 089,
  [\href{http://arxiv.org/abs/1311.0286}{{\tt arXiv:1311.0286}}].

\bibitem{Alioli:2015toa}
S.~Alioli, C.~W. Bauer, C.~Berggren, F.~J. Tackmann, and J.~R. Walsh, {\it
  {Drell-Yan production at NNLL$'+$NNLO matched to parton showers}},  {\em
  Phys.~Rev.~D} {\bf 92} (2015), no.~9 094020,
  [\href{http://arxiv.org/abs/1508.01475}{{\tt arXiv:1508.01475}}].

\bibitem{Baikov:2009bg}
P.~A. Baikov, K.~G. Chetyrkin, A.~V. Smirnov, V.~A. Smirnov, and
  M.~Steinhauser, {\it {Quark and gluon form factors to three loops}},  {\em
  Phys. Rev. Lett.} {\bf 102} (2009) 212002,
  [\href{http://arxiv.org/abs/0902.3519}{{\tt arXiv:0902.3519}}].

\bibitem{Gehrmann:2010ue}
T.~Gehrmann, E.~W.~N. Glover, T.~Huber, N.~Ikizlerli, and C.~Studerus, {\it
  {Calculation of the quark and gluon form factors to three loops in QCD}},
  \href{http://arxiv.org/abs/1004.3653}{{\tt arXiv:1004.3653}}.

\bibitem{Becher:2009th}
T.~Becher and M.~D. Schwartz, {\it {Direct photon production with effective
  field theory}},  {\em JHEP} {\bf 02} (2010) 040,
  [\href{http://arxiv.org/abs/0911.0681}{{\tt arXiv:0911.0681}}].

\bibitem{Becher:2011fc}
T.~Becher, C.~Lorentzen, and M.~D. Schwartz, {\it {Resummation for W and Z
  production at large pT}},  {\em Phys. Rev. Lett.} {\bf 108} (2012) 012001,
  [\href{http://arxiv.org/abs/1106.4310}{{\tt arXiv:1106.4310}}].

\bibitem{Becher:2012xr}
T.~Becher, C.~Lorentzen, and M.~D. Schwartz, {\it {Precision Direct Photon and
  W-Boson Spectra at High $p_T$ and Comparison to LHC Data}},  {\em
  Phys.~Rev.~D} {\bf 86} (2012) 054026,
  [\href{http://arxiv.org/abs/1206.6115}{{\tt arXiv:1206.6115}}].

\bibitem{Becher:2013vva}
T.~Becher, G.~Bell, C.~Lorentzen, and S.~Marti, {\it {Transverse-momentum
  spectra of electroweak bosons near threshold at NNLO}},  {\em JHEP} {\bf
  1402} (2014) 004, [\href{http://arxiv.org/abs/1309.3245}{{\tt
  arXiv:1309.3245}}].

\bibitem{Huang:2014mca}
F.~P. Huang, C.~S. Li, H.~T. Li, and J.~Wang, {\it {Renormalization-group
  improved predictions for Higgs boson production at large $p_T$}},  {\em
  Phys.~Rev.~D} {\bf 90} (2014), no.~9 094024,
  [\href{http://arxiv.org/abs/1406.2591}{{\tt arXiv:1406.2591}}].

\bibitem{Becher:2014tsa}
T.~Becher, G.~Bell, C.~Lorentzen, and S.~Marti, {\it {The transverse-momentum
  spectrum of Higgs bosons near threshold at NNLO}},  {\em JHEP} {\bf 1411}
  (2014) 026, [\href{http://arxiv.org/abs/1407.4111}{{\tt arXiv:1407.4111}}].

\bibitem{Chiu:2008vv}
J.-y. Chiu, R.~Kelley, and A.~V. Manohar, {\it {Electroweak Corrections using
  Effective Field Theory: Applications to the LHC}},  {\em Phys. Rev. D} {\bf
  78} (2008) 073006, [\href{http://arxiv.org/abs/0806.1240}{{\tt
  arXiv:0806.1240}}].

\bibitem{Zhu:2010mr}
H.~X. Zhu, C.~S. Li, J.~Wang, and J.~J. Zhang, {\it {Factorization and
  resummation of s-channel single top quark production}},  {\em JHEP} {\bf
  1102} (2011) 099, [\href{http://arxiv.org/abs/1006.0681}{{\tt
  arXiv:1006.0681}}].

\bibitem{Kelley:2010qs}
R.~Kelley and M.~D. Schwartz, {\it {Threshold Hadronic Event Shapes with
  Effective Field Theory}},  {\em Phys. Rev. D} {\bf 83} (2011) 033001,
  [\href{http://arxiv.org/abs/1008.4355}{{\tt arXiv:1008.4355}}].

\bibitem{Wang:2010ue}
J.~Wang, C.~S. Li, H.~X. Zhu, and J.~J. Zhang, {\it {Factorization and
  resummation of t-channel single top quark production}},
  \href{http://arxiv.org/abs/1010.4509}{{\tt arXiv:1010.4509}}.

\bibitem{Ahrens:2010zv}
V.~Ahrens, A.~Ferroglia, M.~Neubert, B.~D. Pecjak, and L.~L. Yang, {\it
  {Renormalization-Group Improved Predictions for Top-Quark Pair Production at
  Hadron Colliders}},  {\em JHEP} {\bf 09} (2010) 097,
  [\href{http://arxiv.org/abs/1003.5827}{{\tt arXiv:1003.5827}}].

\bibitem{Ahrens:2011mw}
V.~Ahrens, A.~Ferroglia, M.~Neubert, B.~D. Pecjak, and L.-L. Yang, {\it
  {RG-improved single-particle inclusive cross sections and forward-backward
  asymmetry in $t\bar t$ production at hadron colliders}},  {\em JHEP} {\bf
  1109} (2011) 070, [\href{http://arxiv.org/abs/1103.0550}{{\tt
  arXiv:1103.0550}}].

\bibitem{Li:2014ula}
H.~T. Li, C.~S. Li, and S.~A. Li, {\it {Renormalization group improved
  predictions for $t\bar{t}W^\pm$ production at hadron colliders}},  {\em Phys.
  Rev. D} {\bf 90} (2014), no.~9 094009,
  [\href{http://arxiv.org/abs/1409.1460}{{\tt arXiv:1409.1460}}].

\bibitem{Liu:2014oog}
Z.~L. Liu, C.~S. Li, J.~Wang, and Y.~Wang, {\it {Resummation prediction on the
  jet mass spectrum in one-jet inclusive production at the LHC}},  {\em JHEP}
  {\bf 04} (2015) 005, [\href{http://arxiv.org/abs/1412.1337}{{\tt
  arXiv:1412.1337}}].

\bibitem{Kelley:2010fn}
R.~Kelley and M.~D. Schwartz, {\it {1-loop matching and NNLL resummation for
  all partonic 2 to 2 processes in QCD}},  {\em Phys. Rev. D} {\bf 83} (2011)
  045022, [\href{http://arxiv.org/abs/1008.2759}{{\tt arXiv:1008.2759}}].

\bibitem{Broggio:2014hoa}
A.~Broggio, A.~Ferroglia, B.~D. Pecjak, and Z.~Zhang, {\it {NNLO hard functions
  in massless QCD}},  {\em JHEP} {\bf 12} (2014) 005,
  [\href{http://arxiv.org/abs/1409.5294}{{\tt arXiv:1409.5294}}].

\bibitem{Stewart:2012yh}
I.~W. Stewart, F.~J. Tackmann, and W.~J. Waalewijn, {\it {Combining Fixed-Order
  Helicity Amplitudes With Resummation Using SCET}},  {\em PoS} {\bf LL2012}
  (2012) 058, [\href{http://arxiv.org/abs/1211.2305}{{\tt arXiv:1211.2305}}].

\bibitem{Stewart:2010pd}
I.~W. Stewart, F.~J. Tackmann, and W.~J. Waalewijn, {\it {The Beam Thrust Cross
  Section for Drell-Yan at NNLL Order}},  {\em Phys. Rev. Lett.} {\bf 106}
  (2011) 032001, [\href{http://arxiv.org/abs/1005.4060}{{\tt
  arXiv:1005.4060}}].

\bibitem{Berger:2010xi}
C.~F. Berger, C.~Marcantonini, I.~W. Stewart, F.~J. Tackmann, and W.~J.
  Waalewijn, {\it {Higgs Production with a Central Jet Veto at NNLL+NNLO}},
  {\em JHEP} {\bf 1104} (2011) 092, [\href{http://arxiv.org/abs/1012.4480}{{\tt
  arXiv:1012.4480}}].

\bibitem{Jouttenus:2011wh}
T.~T. Jouttenus, I.~W. Stewart, F.~J. Tackmann, and W.~J. Waalewijn, {\it {The
  Soft Function for Exclusive N-Jet Production at Hadron Colliders}},  {\em
  Phys.Rev.} {\bf D83} (2011) 114030,
  [\href{http://arxiv.org/abs/1102.4344}{{\tt arXiv:1102.4344}}].

\bibitem{Kang:2013nha}
D.~Kang, C.~Lee, and I.~W. Stewart, {\it {Using 1-Jettiness to Measure 2 Jets
  in DIS 3 Ways}},  {\em Phys. Rev.} {\bf D88} (2013) 054004,
  [\href{http://arxiv.org/abs/1303.6952}{{\tt arXiv:1303.6952}}].

\bibitem{Kang:2013lga}
Z.-B. Kang, X.~Liu, and S.~Mantry, {\it {1-jettiness DIS event shape: NNLL+NLO
  results}},  {\em Phys. Rev. D} {\bf 90} (2014), no.~1 014041,
  [\href{http://arxiv.org/abs/1312.0301}{{\tt arXiv:1312.0301}}].

\bibitem{Stewart:2015waa}
I.~W. Stewart, F.~J. Tackmann, J.~Thaler, C.~K. Vermilion, and T.~F. Wilkason,
  {\it {XCone: N-jettiness as an Exclusive Cone Jet Algorithm}},  {\em JHEP}
  {\bf 11} (2015) 072, [\href{http://arxiv.org/abs/1508.01516}{{\tt
  arXiv:1508.01516}}].

\bibitem{Catani:2000vq}
S.~Catani, D.~de~Florian, and M.~Grazzini, {\it {Universality of nonleading
  logarithmic contributions in transverse momentum distributions}},  {\em Nucl.
  Phys.} {\bf B596} (2001) 299--312,
  [\href{http://arxiv.org/abs/hep-ph/0008184}{{\tt hep-ph/0008184}}].

\bibitem{Becher:2010tm}
T.~Becher and M.~Neubert, {\it {Drell-Yan production at small $q_T$, transverse
  parton distributions and the collinear anomaly}},  {\em Eur.Phys.J.} {\bf
  C71} (2011) 1665, [\href{http://arxiv.org/abs/1007.4005}{{\tt
  arXiv:1007.4005}}].

\bibitem{GarciaEchevarria:2011rb}
M.~G. Echevarria, A.~Idilbi, and I.~Scimemi, {\it {Factorization Theorem For
  Drell-Yan At Low $q_T$ And Transverse Momentum Distributions
  On-The-Light-Cone}},  {\em JHEP} {\bf 1207} (2012) 002,
  [\href{http://arxiv.org/abs/1111.4996}{{\tt arXiv:1111.4996}}].

\bibitem{Chiu:2012ir}
J.-Y. Chiu, A.~Jain, D.~Neill, and I.~Z. Rothstein, {\it {A Formalism for the
  Systematic Treatment of Rapidity Logarithms in Quantum Field Theory}},  {\em
  JHEP} {\bf 05} (2012) 084, [\href{http://arxiv.org/abs/1202.0814}{{\tt
  arXiv:1202.0814}}].

\bibitem{Catani:2013tia}
S.~Catani, L.~Cieri, D.~de~Florian, G.~Ferrera, and M.~Grazzini, {\it
  {Universality of transverse-momentum resummation and hard factors at the
  NNLO}},  {\em Nucl. Phys.} {\bf B881} (2014) 414--443,
  [\href{http://arxiv.org/abs/1311.1654}{{\tt arXiv:1311.1654}}].

\bibitem{Delenda:2006nf}
Y.~Delenda, R.~Appleby, M.~Dasgupta, and A.~Banfi, {\it {On QCD resummation
  with k(t) clustering}},  {\em JHEP} {\bf 0612} (2006) 044,
  [\href{http://arxiv.org/abs/hep-ph/0610242}{{\tt hep-ph/0610242}}].

\bibitem{Bauer:2008jx}
C.~W. Bauer, A.~Hornig, and F.~J. Tackmann, {\it {Factorization for generic jet
  production}},  {\em Phys. Rev. D} {\bf 79} (2009) 114013,
  [\href{http://arxiv.org/abs/0808.2191}{{\tt arXiv:0808.2191}}].

\bibitem{Kelley:2012zs}
R.~Kelley, J.~R. Walsh, and S.~Zuberi, {\it {Disentangling Clustering Effects
  in Jet Algorithms}},  \href{http://arxiv.org/abs/1203.2923}{{\tt
  arXiv:1203.2923}}.

\bibitem{Banfi:2012yh}
A.~Banfi, G.~P. Salam, and G.~Zanderighi, {\it {NLL+NNLO predictions for
  jet-veto efficiencies in Higgs-boson and Drell-Yan production}},  {\em JHEP}
  {\bf 1206} (2012) 159, [\href{http://arxiv.org/abs/1203.5773}{{\tt
  arXiv:1203.5773}}].

\bibitem{Becher:2012qa}
T.~Becher and M.~Neubert, {\it {Factorization and NNLL Resummation for Higgs
  Production with a Jet Veto}},  {\em JHEP} {\bf 1207} (2012) 108,
  [\href{http://arxiv.org/abs/1205.3806}{{\tt arXiv:1205.3806}}].

\bibitem{Banfi:2012jm}
A.~Banfi, P.~F. Monni, G.~P. Salam, and G.~Zanderighi, {\it {Higgs and Z-boson
  production with a jet veto}},  {\em Phys. Rev. Lett.} {\bf 109} (2012)
  202001, [\href{http://arxiv.org/abs/1206.4998}{{\tt arXiv:1206.4998}}].

\bibitem{Becher:2013xia}
T.~Becher, M.~Neubert, and L.~Rothen, {\it {Factorization and
  $N^{3}LL_{p}$+NNLO predictions for the Higgs cross section with a jet veto}},
   {\em JHEP} {\bf 1310} (2013) 125,
  [\href{http://arxiv.org/abs/1307.0025}{{\tt arXiv:1307.0025}}].

\bibitem{Shao:2013uba}
D.~Y. Shao, C.~S. Li, and H.~T. Li, {\it {Resummation Prediction on Higgs and
  Vector Boson Associated Production with a Jet Veto at the LHC}},  {\em JHEP}
  {\bf 02} (2014) 117, [\href{http://arxiv.org/abs/1309.5015}{{\tt
  arXiv:1309.5015}}].

\bibitem{Li:2014ria}
Y.~Li and X.~Liu, {\it {High precision predictions for exclusive $VH$
  production at the LHC}},  {\em JHEP} {\bf 06} (2014) 028,
  [\href{http://arxiv.org/abs/1401.2149}{{\tt arXiv:1401.2149}}].

\bibitem{Jaiswal:2014yba}
P.~Jaiswal and T.~Okui, {\it {Explanation of the $WW$ excess at the LHC by
  jet-veto resummation}},  {\em Phys.~Rev.~D} {\bf 90} (2014), no.~7 073009,
  [\href{http://arxiv.org/abs/1407.4537}{{\tt arXiv:1407.4537}}].

\bibitem{Gangal:2014qda}
S.~Gangal, M.~Stahlhofen, and F.~J. Tackmann, {\it {Rapidity-Dependent Jet
  Vetoes}},  {\em Phys. Rev. D} {\bf 91} (2015), no.~5 054023,
  [\href{http://arxiv.org/abs/1412.4792}{{\tt arXiv:1412.4792}}].

\bibitem{Becher:2014aya}
T.~Becher, R.~Frederix, M.~Neubert, and L.~Rothen, {\it {Automated NNLL $+$ NLO
  resummation for jet-veto cross sections}},  {\em Eur. Phys. J. C} {\bf 75}
  (2015), no.~4 154, [\href{http://arxiv.org/abs/1412.8408}{{\tt
  arXiv:1412.8408}}].

\bibitem{Ellis:2009wj}
S.~D. Ellis, A.~Hornig, C.~Lee, C.~K. Vermilion, and J.~R. Walsh, {\it
  {Consistent Factorization of Jet Observables in Exclusive Multijet
  Cross-Sections}},  {\em Phys. Lett. B} {\bf 689} (2010) 82--89,
  [\href{http://arxiv.org/abs/0912.0262}{{\tt arXiv:0912.0262}}].

\bibitem{Becher:2015gsa}
T.~Becher and X.~Garcia~i Tormo, {\it {Factorization and resummation for
  transverse thrust}},  {\em JHEP} {\bf 06} (2015) 071,
  [\href{http://arxiv.org/abs/1502.04136}{{\tt arXiv:1502.04136}}].

\bibitem{Procura:2009vm}
M.~Procura and I.~W. Stewart, {\it {Quark Fragmentation within an Identified
  Jet}},  {\em Phys. Rev.} {\bf D81} (2010) 074009,
  [\href{http://arxiv.org/abs/0911.4980}{{\tt arXiv:0911.4980}}]. [Erratum:
  Phys. Rev.D83,039902(2011)].

\bibitem{Liu:2010ng}
X.~Liu, {\it {SCET approach to top quark decay}},  {\em Phys. Lett.} {\bf B699}
  (2011) 87--92, [\href{http://arxiv.org/abs/1011.3872}{{\tt
  arXiv:1011.3872}}].

\bibitem{Jain:2011xz}
A.~Jain, M.~Procura, and W.~J. Waalewijn, {\it {Parton Fragmentation within an
  Identified Jet at NNLL}},  {\em JHEP} {\bf 05} (2011) 035,
  [\href{http://arxiv.org/abs/1101.4953}{{\tt arXiv:1101.4953}}].

\bibitem{Procura:2011aq}
M.~Procura and W.~J. Waalewijn, {\it {Fragmentation in Jets: Cone and Threshold
  Effects}},  {\em Phys.~Rev.~D} {\bf 85} (2012) 114041,
  [\href{http://arxiv.org/abs/1111.6605}{{\tt arXiv:1111.6605}}].

\bibitem{Chang:2013rca}
H.-M. Chang, M.~Procura, J.~Thaler, and W.~J. Waalewijn, {\it {Calculating
  Track-Based Observables for the LHC}},  {\em Phys. Rev. Lett.} {\bf 111}
  (2013) 102002, [\href{http://arxiv.org/abs/1303.6637}{{\tt
  arXiv:1303.6637}}].

\bibitem{Bauer:2013bza}
C.~W. Bauer and E.~Mereghetti, {\it {Heavy Quark Fragmenting Jet Functions}},
  {\em JHEP} {\bf 1404} (2014) 051, [\href{http://arxiv.org/abs/1312.5605}{{\tt
  arXiv:1312.5605}}].

\bibitem{Ritzmann:2014mka}
M.~Ritzmann and W.~J. Waalewijn, {\it {Fragmentation in Jets at NNLO}},  {\em
  Phys.Rev.} {\bf D90} (2014) 054029,
  [\href{http://arxiv.org/abs/1407.3272}{{\tt arXiv:1407.3272}}].

\bibitem{Dixon:1996wi}
L.~J. Dixon, {\it {Calculating scattering amplitudes efficiently}},
  \href{http://arxiv.org/abs/hep-ph/9601359}{{\tt hep-ph/9601359}}.

\bibitem{Dixon:2013uaa}
L.~J. Dixon, {\it {A brief introduction to modern amplitude methods}},
  \href{http://arxiv.org/abs/1310.5353}{{\tt arXiv:1310.5353}}.

\bibitem{Keppeler:2012ih}
S.~Keppeler and M.~Sjodahl, {\it {Orthogonal multiplet bases in SU(Nc) color
  space}},  {\em JHEP} {\bf 1209} (2012) 124,
  [\href{http://arxiv.org/abs/1207.0609}{{\tt arXiv:1207.0609}}].

\bibitem{Sjodahl:2015qoa}
M.~Sjodahl and J.~Thor{\'e}n, {\it {Decomposing color structure into multiplet
  bases}},  \href{http://arxiv.org/abs/1507.03814}{{\tt arXiv:1507.03814}}.

\bibitem{Kidonakis:1998nf}
N.~Kidonakis, G.~Oderda, and G.~Sterman, {\it {Evolution of color exchange in
  {QCD} hard scattering}},  {\em Nucl. Phys. B} {\bf 531} (1998) 365--402,
  [\href{http://arxiv.org/abs/hep-ph/9803241}{{\tt hep-ph/9803241}}].

\bibitem{Maltoni:2002mq}
F.~Maltoni, K.~Paul, T.~Stelzer, and S.~Willenbrock, {\it {Color flow
  decomposition of QCD amplitudes}},  {\em Phys. Rev. D} {\bf 67} (2003)
  014026, [\href{http://arxiv.org/abs/hep-ph/0209271}{{\tt hep-ph/0209271}}].

\bibitem{Catani:1996pk}
S.~Catani, M.~H. Seymour, and Z.~Trocsanyi, {\it {Regularization scheme
  independence and unitarity in QCD cross sections}},  {\em Phys. Rev. D} {\bf
  55} (1997) 6819--6829, [\href{http://arxiv.org/abs/hep-ph/9610553}{{\tt
  hep-ph/9610553}}].

\bibitem{tHooft:1972fi}
G.~'t~Hooft and M.~Veltman, {\it {Regularization and Renormalization of Gauge
  Fields}},  {\em Nucl. Phys. B} {\bf 44} (1972) 189--213.

\bibitem{Bern:1991aq}
Z.~Bern and D.~A. Kosower, {\it {The Computation of loop amplitudes in gauge
  theories}},  {\em Nucl. Phys. B} {\bf 379} (1992) 451--561.

\bibitem{Bern:2002zk}
Z.~Bern, A.~De~Freitas, L.~J. Dixon, and H.~L. Wong, {\it {Supersymmetric
  regularization, two-loop QCD amplitudes and coupling shifts}},  {\em Phys.
  Rev. D} {\bf 66} (2002) 085002,
  [\href{http://arxiv.org/abs/hep-ph/0202271}{{\tt hep-ph/0202271}}].

\bibitem{Siegel:1979wq}
W.~Siegel, {\it {Supersymmetric Dimensional Regularization via Dimensional
  Reduction}},  {\em Phys. Lett. B} {\bf 84} (1979) 193.

\bibitem{Broggio:2015dga}
A.~Broggio, C.~Gnendiger, A.~Signer, D.~St{\"o}ckinger, and A.~Visconti, {\it
  {SCET approach to regularization-scheme dependence of QCD amplitudes}},
  \href{http://arxiv.org/abs/1506.05301}{{\tt arXiv:1506.05301}}.

\bibitem{Kilgore:2011ta}
W.~B. Kilgore, {\it {Regularization Schemes and Higher Order Corrections}},
  {\em Phys. Rev. D} {\bf 83} (2011) 114005,
  [\href{http://arxiv.org/abs/1102.5353}{{\tt arXiv:1102.5353}}].

\bibitem{Boughezal:2011br}
R.~Boughezal, K.~Melnikov, and F.~Petriello, {\it {The four-dimensional
  helicity scheme and dimensional reconstruction}},
  \href{http://arxiv.org/abs/1106.5520}{{\tt arXiv:1106.5520}}.

\bibitem{Altarelli:1980fi}
G.~Altarelli, G.~Curci, G.~Martinelli, and S.~Petrarca, {\it {QCD Nonleading
  Corrections to Weak Decays as an Application of Regularization by Dimensional
  Reduction}},  {\em Nucl. Phys. B} {\bf 187} (1981) 461.

\bibitem{Dawson:1990zj}
S.~Dawson, {\it {Radiative corrections to Higgs boson production}},  {\em Nucl.
  Phys. B} {\bf 359} (1991) 283--300.

\bibitem{Djouadi:1991tka}
A.~Djouadi, M.~Spira, and P.~M. Zerwas, {\it {Production of Higgs bosons in
  proton colliders: QCD corrections}},  {\em Phys. Lett. B} {\bf 264} (1991)
  440--446.

\bibitem{Spira:1995rr}
M.~Spira, A.~Djouadi, D.~Graudenz, and P.~M. Zerwas, {\it {Higgs boson
  production at the LHC}},  {\em Nucl. Phys. B} {\bf 453} (1995) 17--82,
  [\href{http://arxiv.org/abs/hep-ph/9504378}{{\tt hep-ph/9504378}}].

\bibitem{Harlander:2005rq}
R.~Harlander and P.~Kant, {\it {Higgs production and decay: Analytic results at
  next-to-leading order QCD}},  {\em JHEP} {\bf 12} (2005) 015,
  [\href{http://arxiv.org/abs/hep-ph/0509189}{{\tt hep-ph/0509189}}].

\bibitem{Anastasiou:2006hc}
C.~Anastasiou, S.~Beerli, S.~Bucherer, A.~Daleo, and Z.~Kunszt, {\it {Two-loop
  amplitudes and master integrals for the production of a Higgs boson via a
  massive quark and a scalar-quark loop}},  {\em JHEP} {\bf 01} (2007) 082,
  [\href{http://arxiv.org/abs/hep-ph/0611236}{{\tt hep-ph/0611236}}].

\bibitem{Harlander:2000mg}
R.~V. Harlander, {\it {Virtual corrections to g g --> H to two loops in the
  heavy top limit}},  {\em Phys. Lett. B} {\bf 492} (2000) 74--80,
  [\href{http://arxiv.org/abs/hep-ph/0007289}{{\tt hep-ph/0007289}}].

\bibitem{Harlander:2009bw}
R.~V. Harlander and K.~J. Ozeren, {\it {Top mass effects in Higgs production at
  next-to-next-to-leading order QCD: virtual corrections}},  {\em Phys. Lett.
  B} {\bf 679} (2009) 467--472, [\href{http://arxiv.org/abs/0907.2997}{{\tt
  arXiv:0907.2997}}].

\bibitem{Pak:2009bx}
A.~Pak, M.~Rogal, and M.~Steinhauser, {\it {Virtual three-loop corrections to
  Higgs boson production in gluon fusion for finite top quark mass}},  {\em
  Phys. Lett. B} {\bf 679} (2009) 473--477,
  [\href{http://arxiv.org/abs/0907.2998}{{\tt arXiv:0907.2998}}].

\bibitem{Gehrmann:2011aa}
T.~Gehrmann, M.~Jaquier, E.~W.~N. Glover, and A.~Koukoutsakis, {\it {Two-Loop
  QCD Corrections to the Helicity Amplitudes for $H \to$ 3 partons}},  {\em
  JHEP} {\bf 02} (2012) 056, [\href{http://arxiv.org/abs/1112.3554}{{\tt
  arXiv:1112.3554}}].

\bibitem{Neill:2009mz}
D.~Neill, {\it {Analytic Virtual Corrections for Higgs Transverse Momentum
  Spectrum at $O(\alpha_s^2/m_t^3)$ via Unitarity Methods}},
  \href{http://arxiv.org/abs/0911.2707}{{\tt arXiv:0911.2707}}.

\bibitem{Dawson:1991au}
S.~Dawson and R.~Kauffman, {\it {Higgs boson plus multi - jet rates at the
  SSC}},  {\em Phys.Rev.Lett.} {\bf 68} (1992) 2273--2276.

\bibitem{DelDuca:2001fn}
V.~Del~Duca, W.~Kilgore, C.~Oleari, C.~Schmidt, and D.~Zeppenfeld, {\it
  {Gluon-fusion contributions to H + 2 jet production}},  {\em Nucl. Phys. B}
  {\bf 616} (2001) 367--399, [\href{http://arxiv.org/abs/hep-ph/0108030}{{\tt
  hep-ph/0108030}}].

\bibitem{Berger:2006sh}
C.~F. Berger, V.~Del~Duca, and L.~J. Dixon, {\it {Recursive Construction of
  Higgs-Plus-Multiparton Loop Amplitudes: The Last of the Phi-nite Loop
  Amplitudes}},  {\em Phys.Rev. D} {\bf 74} (2006) 094021,
  [\href{http://arxiv.org/abs/hep-ph/0608180}{{\tt hep-ph/0608180}}].

\bibitem{Badger:2007si}
S.~D. Badger, E.~W.~N. Glover, and K.~Risager, {\it {One-loop phi-MHV
  amplitudes using the unitarity bootstrap}},  {\em JHEP} {\bf 07} (2007) 066,
  [\href{http://arxiv.org/abs/0704.3914}{{\tt arXiv:0704.3914}}].

\bibitem{Glover:2008ffa}
E.~W.~N. Glover, P.~Mastrolia, and C.~Williams, {\it {One-loop phi-MHV
  amplitudes using the unitarity bootstrap: The General helicity case}},  {\em
  JHEP} {\bf 08} (2008) 017, [\href{http://arxiv.org/abs/0804.4149}{{\tt
  arXiv:0804.4149}}].

\bibitem{Badger:2009vh}
S.~Badger, J.~M. Campbell, R.~Ellis, and C.~Williams, {\it {Analytic results
  for the one-loop NMHV Hqqgg amplitude}},  {\em JHEP} {\bf 0912} (2009) 035,
  [\href{http://arxiv.org/abs/0910.4481}{{\tt arXiv:0910.4481}}].

\bibitem{Korner:1990sj}
J.~Korner and P.~Sieben, {\it {Use of helicity methods in evaluating loop
  integrals: A QCD example}},  {\em Nucl. Phys. B} {\bf 363} (1991) 65--82.

\bibitem{Garland:2001tf}
L.~Garland, T.~Gehrmann, E.~N. Glover, A.~Koukoutsakis, and E.~Remiddi, {\it
  {The Two loop QCD matrix element for $e+ e- \to$ 3 jets}},  {\em Nucl.Phys.}
  {\bf B627} (2002) 107--188, [\href{http://arxiv.org/abs/hep-ph/0112081}{{\tt
  hep-ph/0112081}}].

\bibitem{Garland:2002ak}
L.~Garland, T.~Gehrmann, E.~N. Glover, A.~Koukoutsakis, and E.~Remiddi, {\it
  {Two loop QCD helicity amplitudes for $e+ e\to $ three jets}},  {\em
  Nucl.Phys.} {\bf B642} (2002) 227--262,
  [\href{http://arxiv.org/abs/hep-ph/0206067}{{\tt hep-ph/0206067}}].

\bibitem{vanderBij:1988ac}
J.~van~der Bij and E.~Glover, {\it {Z BOSON PRODUCTION AND DECAY VIA GLUONS}},
  {\em Nucl. Phys. B} {\bf 313} (1989) 237.

\bibitem{Gehrmann:2013vga}
T.~Gehrmann, L.~Tancredi, and E.~Weihs, {\it {Two-loop QCD helicity amplitudes
  for $g\,g \to Z\,g$ and $g\,g \to Z\,\gamma $}},  {\em JHEP} {\bf 1304}
  (2013) 101, [\href{http://arxiv.org/abs/1302.2630}{{\tt arXiv:1302.2630}}].

\bibitem{Bern:1996ka}
Z.~Bern, L.~J. Dixon, D.~A. Kosower, and S.~Weinzierl, {\it {One-loop
  amplitudes for e+ e- --> anti-q q anti-Q Q}},  {\em Nucl. Phys. B} {\bf 489}
  (1997) 3--23, [\href{http://arxiv.org/abs/hep-ph/9610370}{{\tt
  hep-ph/9610370}}].

\bibitem{Bern:2003ck}
Z.~Bern, A.~De~Freitas, and L.~J. Dixon, {\it {Two loop helicity amplitudes for
  quark gluon scattering in QCD and gluino gluon scattering in supersymmetric
  Yang-Mills theory}},  {\em JHEP} {\bf 0306} (2003) 028,
  [\href{http://arxiv.org/abs/hep-ph/0304168}{{\tt hep-ph/0304168}}].

\bibitem{Glover:2003cm}
E.~W.~N. Glover and M.~E. Tejeda-Yeomans, {\it {Two-loop QCD helicity
  amplitudes for massless quark- massless gauge boson scattering}},  {\em JHEP}
  {\bf 06} (2003) 033, [\href{http://arxiv.org/abs/hep-ph/0304169}{{\tt
  hep-ph/0304169}}].

\bibitem{Anastasiou:2000kg}
C.~Anastasiou, E.~N. Glover, C.~Oleari, and M.~Tejeda-Yeomans, {\it {Two-loop
  QCD corrections to the scattering of massless distinct quarks}},  {\em
  Nucl.~Phys.} {\bf B601} (2001) 318--340,
  [\href{http://arxiv.org/abs/hep-ph/0010212}{{\tt hep-ph/0010212}}].

\bibitem{Anastasiou:2000ue}
C.~Anastasiou, E.~N. Glover, C.~Oleari, and M.~Tejeda-Yeomans, {\it {Two loop
  QCD corrections to massless identical quark scattering}},  {\em Nucl.~Phys.}
  {\bf B601} (2001) 341--360, [\href{http://arxiv.org/abs/hep-ph/0011094}{{\tt
  hep-ph/0011094}}].

\bibitem{Glover:2004si}
E.~W.~N. Glover, {\it {Two-loop QCD helicity amplitudes for massless
  quark-quark scattering}},  {\em JHEP} {\bf 04} (2004) 021,
  [\href{http://arxiv.org/abs/hep-ph/0401119}{{\tt hep-ph/0401119}}].

\bibitem{Freitas:2004tk}
A.~De~Freitas and Z.~Bern, {\it {Two-loop helicity amplitudes for quark-quark
  scattering in QCD and gluino-gluino scattering in supersymmetric Yang-Mills
  theory}},  {\em JHEP} {\bf 09} (2004) 039,
  [\href{http://arxiv.org/abs/hep-ph/0409007}{{\tt hep-ph/0409007}}].

\bibitem{Glover:2001af}
E.~N. Glover, C.~Oleari, and M.~Tejeda-Yeomans, {\it {Two loop QCD corrections
  to gluon-gluon scattering}},  {\em Nucl.~Phys.} {\bf B605} (2001) 467--485,
  [\href{http://arxiv.org/abs/hep-ph/0102201}{{\tt hep-ph/0102201}}].

\bibitem{Bern:2002tk}
Z.~Bern, A.~De~Freitas, and L.~J. Dixon, {\it {Two loop helicity amplitudes for
  gluon-gluon scattering in QCD and supersymmetric Yang-Mills theory}},  {\em
  JHEP} {\bf 0203} (2002) 018, [\href{http://arxiv.org/abs/hep-ph/0201161}{{\tt
  hep-ph/0201161}}].

\bibitem{MertAybat:2006mz}
S.~{Mert Aybat}, L.~J. Dixon, and G.~Sterman, {\it {The two-loop soft anomalous
  dimension matrix and resummation at next-to-next-to leading pole}},  {\em
  Phys. Rev. D} {\bf 74} (2006) 074004,
  [\href{http://arxiv.org/abs/hep-ph/0607309}{{\tt hep-ph/0607309}}].

\bibitem{Aybat:2006wq}
S.~M. Aybat, L.~J. Dixon, and G.~F. Sterman, {\it {The Two-loop anomalous
  dimension matrix for soft gluon exchange}},  {\em Phys. Rev. Lett.} {\bf 97}
  (2006) 072001, [\href{http://arxiv.org/abs/hep-ph/0606254}{{\tt
  hep-ph/0606254}}].

\bibitem{Almelid:2015jia}
{\O}.~Almelid, C.~Duhr, and E.~Gardi, {\it {Three-loop corrections to the soft
  anomalous dimension in multi-leg scattering}},
  \href{http://arxiv.org/abs/1507.00047}{{\tt arXiv:1507.00047}}.

\bibitem{Gardi:2009zv}
E.~Gardi and L.~Magnea, {\it {Infrared singularities in QCD amplitudes}},  {\em
  Nuovo Cim.} {\bf C32N5-6} (2009) 137--157,
  [\href{http://arxiv.org/abs/0908.3273}{{\tt arXiv:0908.3273}}].

\bibitem{Gardi:2009qi}
E.~Gardi and L.~Magnea, {\it {Factorization constraints for soft anomalous
  dimensions in QCD scattering amplitudes}},  {\em JHEP} {\bf 03} (2009) 079,
  [\href{http://arxiv.org/abs/0901.1091}{{\tt arXiv:0901.1091}}].

\bibitem{Dixon:2009ur}
L.~J. Dixon, E.~Gardi, and L.~Magnea, {\it {On soft singularities at three
  loops and beyond}},  {\em JHEP} {\bf 02} (2010) 081,
  [\href{http://arxiv.org/abs/0910.3653}{{\tt arXiv:0910.3653}}].

\bibitem{Becher:2009cu}
T.~Becher and M.~Neubert, {\it {Infrared singularities of scattering amplitudes
  in perturbative QCD}},  {\em Phys. Rev. Lett.} {\bf 102} (2009) 162001,
  [\href{http://arxiv.org/abs/0901.0722}{{\tt arXiv:0901.0722}}].

\bibitem{Becher:2009qa}
T.~Becher and M.~Neubert, {\it {On the Structure of Infrared Singularities of
  Gauge-Theory Amplitudes}},  {\em JHEP} {\bf 06} (2009) 081,
  [\href{http://arxiv.org/abs/0903.1126}{{\tt arXiv:0903.1126}}].

\bibitem{DelDuca:2011ae}
V.~Del~Duca, C.~Duhr, E.~Gardi, L.~Magnea, and C.~D. White, {\it {The Infrared
  structure of gauge theory amplitudes in the high-energy limit}},  {\em JHEP}
  {\bf 1112} (2011) 021, [\href{http://arxiv.org/abs/1109.3581}{{\tt
  arXiv:1109.3581}}].

\bibitem{Bret:2011xm}
V.~Del~Duca, C.~Duhr, E.~Gardi, L.~Magnea, and C.~D. White, {\it {An infrared
  approach to Reggeization}},  {\em Phys.~Rev.~D} {\bf 85} (2012) 071104,
  [\href{http://arxiv.org/abs/1108.5947}{{\tt arXiv:1108.5947}}].

\bibitem{DelDuca:2013ara}
V.~Del~Duca, G.~Falcioni, L.~Magnea, and L.~Vernazza, {\it {High-energy QCD
  amplitudes at two loops and beyond}},  {\em Phys. Lett.} {\bf B732} (2014)
  233--240, [\href{http://arxiv.org/abs/1311.0304}{{\tt arXiv:1311.0304}}].

\bibitem{Caron-Huot:2013fea}
S.~Caron-Huot, {\it {When does the gluon reggeize?}},
  \href{http://arxiv.org/abs/1309.6521}{{\tt arXiv:1309.6521}}.

\bibitem{Catani:1996jh}
S.~Catani and M.~Seymour, {\it {The Dipole formalism for the calculation of QCD
  jet cross-sections at next-to-leading order}},  {\em Phys.Lett.} {\bf B378}
  (1996) 287--301, [\href{http://arxiv.org/abs/hep-ph/9602277}{{\tt
  hep-ph/9602277}}].

\bibitem{Catani:1996vz}
S.~Catani and M.~Seymour, {\it {A General algorithm for calculating jet
  cross-sections in NLO QCD}},  {\em Nucl.Phys.} {\bf B485} (1997) 291--419,
  [\href{http://arxiv.org/abs/hep-ph/9605323}{{\tt hep-ph/9605323}}].

\bibitem{Gerwick:2014gya}
E.~Gerwick, S.~Hoeche, S.~Marzani, and S.~Schumann, {\it {Soft evolution of
  multi-jet final states}},  {\em JHEP} {\bf 1502} (2015) 106,
  [\href{http://arxiv.org/abs/1411.7325}{{\tt arXiv:1411.7325}}].

\bibitem{Platzer:2013fha}
S.~Pl{\"a}tzer, {\it {Summing Large-$N$ Towers in Colour Flow Evolution}},
  {\em Eur. Phys. J.} {\bf C74} (2014), no.~6 2907,
  [\href{http://arxiv.org/abs/1312.2448}{{\tt arXiv:1312.2448}}].

\bibitem{Stewart:2010qs}
I.~W. Stewart, F.~J. Tackmann, and W.~J. Waalewijn, {\it {The Quark Beam
  Function at NNLL}},  {\em JHEP} {\bf 1009} (2010) 005,
  [\href{http://arxiv.org/abs/1002.2213}{{\tt arXiv:1002.2213}}].

\bibitem{Aad:2012tfa}
{\bf ATLAS Collaboration} Collaboration, G.~Aad et~al., {\it {Observation of a
  new particle in the search for the Standard Model Higgs boson with the ATLAS
  detector at the LHC}},  {\em Phys.Lett.} {\bf B716} (2012) 1--29,
  [\href{http://arxiv.org/abs/1207.7214}{{\tt arXiv:1207.7214}}].

\bibitem{Chatrchyan:2012ufa}
{\bf CMS Collaboration} Collaboration, S.~Chatrchyan et~al., {\it {Observation
  of a new boson at a mass of 125 GeV with the CMS experiment at the LHC}},
  {\em Phys.Lett.} {\bf B716} (2012) 30--61,
  [\href{http://arxiv.org/abs/1207.7235}{{\tt arXiv:1207.7235}}].

\bibitem{Duhrssen:2004cv}
M.~Duhrssen, S.~Heinemeyer, H.~Logan, D.~Rainwater, G.~Weiglein, et~al., {\it
  {Extracting Higgs boson couplings from CERN LHC data}},  {\em Phys.Rev.} {\bf
  D70} (2004) 113009, [\href{http://arxiv.org/abs/hep-ph/0406323}{{\tt
  hep-ph/0406323}}].

\bibitem{Lafaye:2009vr}
R.~Lafaye, T.~Plehn, M.~Rauch, D.~Zerwas, and M.~Duhrssen, {\it {Measuring the
  Higgs Sector}},  {\em JHEP} {\bf 0908} (2009) 009,
  [\href{http://arxiv.org/abs/0904.3866}{{\tt arXiv:0904.3866}}].

\bibitem{Bonnet:2011yx}
F.~Bonnet, M.~Gavela, T.~Ota, and W.~Winter, {\it {Anomalous Higgs couplings at
  the LHC, and their theoretical interpretation}},  {\em Phys.Rev.} {\bf D85}
  (2012) 035016, [\href{http://arxiv.org/abs/1105.5140}{{\tt
  arXiv:1105.5140}}].

\bibitem{Heinemeyer:2013tqa}
{\bf LHC Higgs Cross Section Working Group} Collaboration, S.~Heinemeyer
  et~al., {\it {Handbook of LHC Higgs Cross Sections: 3. Higgs Properties}},
  \href{http://arxiv.org/abs/1307.1347}{{\tt arXiv:1307.1347}}.

\bibitem{Carmi:2012in}
D.~Carmi, A.~Falkowski, E.~Kuflik, T.~Volansky, and J.~Zupan, {\it {Higgs After
  the Discovery: A Status Report}},  {\em JHEP} {\bf 1210} (2012) 196,
  [\href{http://arxiv.org/abs/1207.1718}{{\tt arXiv:1207.1718}}].

\bibitem{Carmi:2012yp}
D.~Carmi, A.~Falkowski, E.~Kuflik, and T.~Volansky, {\it {Interpreting LHC
  Higgs Results from Natural New Physics Perspective}},  {\em JHEP} {\bf 1207}
  (2012) 136, [\href{http://arxiv.org/abs/1202.3144}{{\tt arXiv:1202.3144}}].

\bibitem{Djouadi:2012rh}
A.~Djouadi, {\it {Precision Higgs coupling measurements at the LHC through
  ratios of production cross sections}},  {\em Eur.Phys.J.} {\bf C73} (2013)
  2498, [\href{http://arxiv.org/abs/1208.3436}{{\tt arXiv:1208.3436}}].

\bibitem{Englert:2012wf}
C.~Englert, M.~Spannowsky, and C.~Wymant, {\it {Partially (in)visible Higgs
  decays at the LHC}},  {\em Phys.Lett.} {\bf B718} (2012) 538--544,
  [\href{http://arxiv.org/abs/1209.0494}{{\tt arXiv:1209.0494}}].

\bibitem{Klute:2012pu}
M.~Klute, R.~Lafaye, T.~Plehn, M.~Rauch, and D.~Zerwas, {\it {Measuring Higgs
  Couplings from LHC Data}},  {\em Phys.Rev.Lett.} {\bf 109} (2012) 101801,
  [\href{http://arxiv.org/abs/1205.2699}{{\tt arXiv:1205.2699}}].

\bibitem{Dobrescu:2012td}
B.~A. Dobrescu and J.~D. Lykken, {\it {Coupling spans of the Higgs-like
  boson}},  {\em JHEP} {\bf 1302} (2013) 073,
  [\href{http://arxiv.org/abs/1210.3342}{{\tt arXiv:1210.3342}}].

\bibitem{Plehn:2012iz}
T.~Plehn and M.~Rauch, {\it {Higgs Couplings after the Discovery}},  {\em
  Europhys.Lett.} {\bf 100} (2012) 11002,
  [\href{http://arxiv.org/abs/1207.6108}{{\tt arXiv:1207.6108}}].

\bibitem{Corbett:2012ja}
T.~Corbett, O.~Eboli, J.~Gonzalez-Fraile, and M.~Gonzalez-Garcia, {\it {Robust
  Determination of the Higgs Couplings: Power to the Data}},  {\em Phys.Rev.}
  {\bf D87} (2013) 015022, [\href{http://arxiv.org/abs/1211.4580}{{\tt
  arXiv:1211.4580}}].

\bibitem{Belanger:2012gc}
G.~Belanger, B.~Dumont, U.~Ellwanger, J.~Gunion, and S.~Kraml, {\it {Higgs
  Couplings at the End of 2012}},  {\em JHEP} {\bf 1302} (2013) 053,
  [\href{http://arxiv.org/abs/1212.5244}{{\tt arXiv:1212.5244}}].

\bibitem{Farina:2012xp}
M.~Farina, C.~Grojean, F.~Maltoni, E.~Salvioni, and A.~Thamm, {\it {Lifting
  degeneracies in Higgs couplings using single top production in association
  with a Higgs boson}},  {\em JHEP} {\bf 1305} (2013) 022,
  [\href{http://arxiv.org/abs/1211.3736}{{\tt arXiv:1211.3736}}].

\bibitem{Batell:2012ca}
B.~Batell, S.~Gori, and L.-T. Wang, {\it {Higgs Couplings and Precision
  Electroweak Data}},  {\em JHEP} {\bf 1301} (2013) 139,
  [\href{http://arxiv.org/abs/1209.6382}{{\tt arXiv:1209.6382}}].

\bibitem{Espinosa:2012im}
J.~Espinosa, C.~Grojean, M.~Muhlleitner, and M.~Trott, {\it {First Glimpses at
  Higgs' face}},  {\em JHEP} {\bf 1212} (2012) 045,
  [\href{http://arxiv.org/abs/1207.1717}{{\tt arXiv:1207.1717}}].

\bibitem{Espinosa:2012vu}
J.~R. Espinosa, M.~Muhlleitner, C.~Grojean, and M.~Trott, {\it {Probing for
  Invisible Higgs Decays with Global Fits}},  {\em JHEP} {\bf 1209} (2012) 126,
  [\href{http://arxiv.org/abs/1205.6790}{{\tt arXiv:1205.6790}}].

\bibitem{Banerjee:2012xc}
S.~Banerjee, S.~Mukhopadhyay, and B.~Mukhopadhyaya, {\it {New Higgs
  interactions and recent data from the LHC and the Tevatron}},  {\em JHEP}
  {\bf 1210} (2012) 062, [\href{http://arxiv.org/abs/1207.3588}{{\tt
  arXiv:1207.3588}}].

\bibitem{Barger:2012hv}
V.~Barger, M.~Ishida, and W.-Y. Keung, {\it {Total Width of 125 GeV Higgs
  Boson}},  {\em Phys.Rev.Lett.} {\bf 108} (2012) 261801,
  [\href{http://arxiv.org/abs/1203.3456}{{\tt arXiv:1203.3456}}].

\bibitem{Giardino:2012dp}
P.~P. Giardino, K.~Kannike, M.~Raidal, and A.~Strumia, {\it {Is the resonance
  at 125 GeV the Higgs boson?}},  {\em Phys.Lett.} {\bf B718} (2012) 469--474,
  [\href{http://arxiv.org/abs/1207.1347}{{\tt arXiv:1207.1347}}].

\bibitem{Giardino:2012ww}
P.~P. Giardino, K.~Kannike, M.~Raidal, and A.~Strumia, {\it {Reconstructing
  Higgs boson properties from the LHC and Tevatron data}},  {\em JHEP} {\bf
  1206} (2012) 117, [\href{http://arxiv.org/abs/1203.4254}{{\tt
  arXiv:1203.4254}}].

\bibitem{Giardino:2013bma}
P.~P. Giardino, K.~Kannike, I.~Masina, M.~Raidal, and A.~Strumia, {\it {The
  universal Higgs fit}},  \href{http://arxiv.org/abs/1303.3570}{{\tt
  arXiv:1303.3570}}.

\bibitem{Aad:2013wqa}
{\bf ATLAS Collaboration} Collaboration, G.~Aad et~al., {\it {Measurements of
  Higgs boson production and couplings in diboson final states with the ATLAS
  detector at the LHC}},  {\em Phys.Lett.} {\bf B726} (2013) 88--119,
  [\href{http://arxiv.org/abs/1307.1427}{{\tt arXiv:1307.1427}}].

\bibitem{Gainer:2013rxa}
J.~S. Gainer, J.~Lykken, K.~T. Matchev, S.~Mrenna, and M.~Park, {\it
  {Geolocating the Higgs Boson Candidate at the LHC}},  {\em Phys.Rev.Lett.}
  {\bf 111} (2013) 041801, [\href{http://arxiv.org/abs/1304.4936}{{\tt
  arXiv:1304.4936}}].

\bibitem{Belanger:2013xza}
G.~Belanger, B.~Dumont, U.~Ellwanger, J.~Gunion, and S.~Kraml, {\it {Global fit
  to Higgs signal strengths and couplings and implications for extended Higgs
  sectors}},  {\em Phys.Rev.} {\bf D88} (2013) 075008,
  [\href{http://arxiv.org/abs/1306.2941}{{\tt arXiv:1306.2941}}].

\bibitem{Belanger:2013kya}
G.~Belanger, B.~Dumont, U.~Ellwanger, J.~Gunion, and S.~Kraml, {\it {Status of
  invisible Higgs decays}},  {\em Phys.Lett.} {\bf B723} (2013) 340--347,
  [\href{http://arxiv.org/abs/1302.5694}{{\tt arXiv:1302.5694}}].

\bibitem{Ellis:2013lra}
J.~Ellis and T.~You, {\it {Updated Global Analysis of Higgs Couplings}},  {\em
  JHEP} {\bf 1306} (2013) 103, [\href{http://arxiv.org/abs/1303.3879}{{\tt
  arXiv:1303.3879}}].

\bibitem{Cranmer:2013hia}
K.~Cranmer, S.~Kreiss, D.~Lopez-Val, and T.~Plehn, {\it {A Novel Approach to
  Higgs Coupling Measurements}},  \href{http://arxiv.org/abs/1401.0080}{{\tt
  arXiv:1401.0080}}.

\bibitem{Chen:2014pia}
Y.~Chen, E.~Di~Marco, J.~Lykken, M.~Spiropulu, R.~Vega-Morales, et~al., {\it
  {8D Likelihood Effective Higgs Couplings Extraction Framework in the Golden
  Channel}},  \href{http://arxiv.org/abs/1401.2077}{{\tt arXiv:1401.2077}}.

\bibitem{Djouadi:2005gj}
A.~Djouadi, {\it {The Anatomy of electro-weak symmetry breaking. II. The Higgs
  bosons in the minimal supersymmetric model}},  {\em Phys.Rept.} {\bf 459}
  (2008) 1--241, [\href{http://arxiv.org/abs/hep-ph/0503173}{{\tt
  hep-ph/0503173}}].

\bibitem{Burgess:2000yq}
C.~Burgess, M.~Pospelov, and T.~ter Veldhuis, {\it {The Minimal model of
  nonbaryonic dark matter: A Singlet scalar}},  {\em Nucl.Phys.} {\bf B619}
  (2001) 709--728, [\href{http://arxiv.org/abs/hep-ph/0011335}{{\tt
  hep-ph/0011335}}].

\bibitem{Patt:2006fw}
B.~Patt and F.~Wilczek, {\it {Higgs-field portal into hidden sectors}},
  \href{http://arxiv.org/abs/hep-ph/0605188}{{\tt hep-ph/0605188}}.

\bibitem{He:2011de}
X.-G. He and J.~Tandean, {\it {Hidden Higgs Boson at the LHC and Light Dark
  Matter Searches}},  {\em Phys.Rev.} {\bf D84} (2011) 075018,
  [\href{http://arxiv.org/abs/1109.1277}{{\tt arXiv:1109.1277}}].

\bibitem{Raidal:2011xk}
M.~Raidal and A.~Strumia, {\it {Hints for a non-standard Higgs boson from the
  LHC}},  {\em Phys.Rev.} {\bf D84} (2011) 077701,
  [\href{http://arxiv.org/abs/1108.4903}{{\tt arXiv:1108.4903}}].

\bibitem{Englert:2011yb}
C.~Englert, T.~Plehn, D.~Zerwas, and P.~M. Zerwas, {\it {Exploring the Higgs
  portal}},  {\em Phys.Lett.} {\bf B703} (2011) 298--305,
  [\href{http://arxiv.org/abs/1106.3097}{{\tt arXiv:1106.3097}}].

\bibitem{Barbieri:2005ri}
R.~Barbieri, T.~Gregoire, and L.~J. Hall, {\it {Mirror world at the large
  hadron collider}},  \href{http://arxiv.org/abs/hep-ph/0509242}{{\tt
  hep-ph/0509242}}.

\bibitem{Gangal:2013nxa}
S.~Gangal and F.~J. Tackmann, {\it {Next-to-leading-order uncertainties in
  Higgs+2 jets from gluon fusion}},  {\em Phys.Rev.} {\bf D87} (2013), no.~9
  093008, [\href{http://arxiv.org/abs/1302.5437}{{\tt arXiv:1302.5437}}].

\bibitem{Dixon:2013haa}
L.~J. Dixon and Y.~Li, {\it {Bounding the Higgs Boson Width Through
  Interferometry}},  {\em Phys.Rev.Lett.} {\bf 111} (2013) 111802,
  [\href{http://arxiv.org/abs/1305.3854}{{\tt arXiv:1305.3854}}].

\bibitem{Campbell:2013una}
J.~M. Campbell, R.~K. Ellis, and C.~Williams, {\it {Bounding the Higgs width at
  the LHC using full analytic results for $gg \rightarrow 2 \text{e} 2\mu$}},
  \href{http://arxiv.org/abs/1311.3589}{{\tt arXiv:1311.3589}}.

\bibitem{Campbell:2013wga}
J.~M. Campbell, R.~K. Ellis, and C.~Williams, {\it {Bounding the Higgs width at
  the LHC: complementary results from $H \to WW$}},
  \href{http://arxiv.org/abs/1312.1628}{{\tt arXiv:1312.1628}}.

\bibitem{CMS:2013bfa}
{\bf CMS Collaboration} Collaboration, C.~Collaboration, {\it {Search for an
  Invisible Higgs Boson}}, .

\bibitem{CMS:2013yda}
{\bf CMS Collaboration} Collaboration, C.~Collaboration, {\it {Search for
  invisible Higgs produced in association with a Z boson}}, .

\bibitem{Bai:2011wz}
Y.~Bai, P.~Draper, and J.~Shelton, {\it {Measuring the Invisible Higgs Width at
  the 7 and 8 TeV LHC}},  {\em JHEP} {\bf 1207} (2012) 192,
  [\href{http://arxiv.org/abs/1112.4496}{{\tt arXiv:1112.4496}}].

\bibitem{Djouadi:2012zc}
A.~Djouadi, A.~Falkowski, Y.~Mambrini, and J.~Quevillon, {\it {Direct Detection
  of Higgs-Portal Dark Matter at the LHC}},  {\em Eur.Phys.J.} {\bf C73} (2013)
  2455, [\href{http://arxiv.org/abs/1205.3169}{{\tt arXiv:1205.3169}}].

\bibitem{Campbell:2011cu}
J.~M. Campbell, R.~K. Ellis, and C.~Williams, {\it {Gluon-Gluon Contributions
  to W+ W- Production and Higgs Interference Effects}},  {\em JHEP} {\bf 1110}
  (2011) 005, [\href{http://arxiv.org/abs/1107.5569}{{\tt arXiv:1107.5569}}].

\bibitem{Kauer:2012ma}
N.~Kauer, {\it {Signal-background interference in gg
  $\rightarrow$H$\rightarrow$VV}},  {\em PoS} {\bf RADCOR2011} (2011) 027,
  [\href{http://arxiv.org/abs/1201.1667}{{\tt arXiv:1201.1667}}].

\bibitem{Passarino:2012ri}
G.~Passarino, {\it {Higgs Interference Effects in $gg \to ZZ$ and their
  Uncertainty}},  {\em JHEP} {\bf 1208} (2012) 146,
  [\href{http://arxiv.org/abs/1206.3824}{{\tt arXiv:1206.3824}}].

\bibitem{Kauer:2013qba}
N.~Kauer, {\it {Interference effects for H $\to$ WW/ZZ $\to
  \ell\bar{\nu}_\ell\bar{\ell}\nu_\ell$ searches in gluon fusion at the LHC}},
  {\em JHEP} {\bf 1312} (2013) 082, [\href{http://arxiv.org/abs/1310.7011}{{\tt
  arXiv:1310.7011}}].

\bibitem{Uhlemann:2008pm}
C.~Uhlemann and N.~Kauer, {\it {Narrow-width approximation accuracy}},  {\em
  Nucl.Phys.} {\bf B814} (2009) 195--211,
  [\href{http://arxiv.org/abs/0807.4112}{{\tt arXiv:0807.4112}}].

\bibitem{Kauer:2013cga}
N.~Kauer, {\it {Inadequacy of zero-width approximation for a light Higgs boson
  signal}},  {\em Mod.Phys.Lett.} {\bf A28} (2013) 1330015,
  [\href{http://arxiv.org/abs/1305.2092}{{\tt arXiv:1305.2092}}].

\bibitem{Englert:2014aca}
C.~Englert and M.~Spannowsky, {\it {Limitations and Opportunities of Off-Shell
  Coupling Measurements}},  \href{http://arxiv.org/abs/1405.0285}{{\tt
  arXiv:1405.0285}}.

\bibitem{CMS:2014ala}
{\bf CMS Collaboration} Collaboration, C.~Collaboration, {\it {Constraints on
  the Higgs boson width from off-shell production and decay to ZZ to llll and
  llvv}}, .

\bibitem{Coleppa:2014qja}
B.~Coleppa, T.~Mandal, and S.~Mitra, {\it {Coupling Extraction From Off-Shell
  Cross-sections}},  \href{http://arxiv.org/abs/1401.4039}{{\tt
  arXiv:1401.4039}}.

\bibitem{Henn:2014lfa}
J.~M. Henn, K.~Melnikov, and V.~A. Smirnov, {\it {Two-loop planar master
  integrals for the production of off-shell vector bosons in hadron
  collisions}},  \href{http://arxiv.org/abs/1402.7078}{{\tt arXiv:1402.7078}}.

\bibitem{Caola:2014lpa}
F.~Caola, J.~M. Henn, K.~Melnikov, and V.~A. Smirnov, {\it {Non-planar master
  integrals for the production of two off-shell vector bosons in collisions of
  massless partons}},  \href{http://arxiv.org/abs/1404.5590}{{\tt
  arXiv:1404.5590}}.

\bibitem{Dittmaier:2012vm}
S.~Dittmaier, S.~Dittmaier, C.~Mariotti, G.~Passarino, R.~Tanaka, et~al., {\it
  {Handbook of LHC Higgs Cross Sections: 2. Differential Distributions}},
  \href{http://arxiv.org/abs/1201.3084}{{\tt arXiv:1201.3084}}.

\bibitem{Stewart:2011cf}
I.~W. Stewart and F.~J. Tackmann, {\it {Theory Uncertainties for Higgs and
  Other Searches Using Jet Bins}},  {\em Phys.~Rev.~D} {\bf 85} (2012) 034011,
  [\href{http://arxiv.org/abs/1107.2117}{{\tt arXiv:1107.2117}}].

\bibitem{Alioli:2013hba}
S.~Alioli and J.~R. Walsh, {\it {Jet Veto Clustering Logarithms Beyond Leading
  Order}},  {\em JHEP} {\bf 1403} (2014) 119,
  [\href{http://arxiv.org/abs/1311.5234}{{\tt arXiv:1311.5234}}].

\bibitem{Chiu:2011qc}
J.-y. Chiu, A.~Jain, D.~Neill, and I.~Z. Rothstein, {\it {The Rapidity
  Renormalization Group}},  {\em Phys.Rev.Lett.} {\bf 108} (2012) 151601,
  [\href{http://arxiv.org/abs/1104.0881}{{\tt arXiv:1104.0881}}].

\bibitem{Fleming:2006cd}
S.~Fleming, A.~K. Leibovich, and T.~Mehen, {\it {Resummation of Large Endpoint
  Corrections to Color-Octet $J/\psi$ Photoproduction}},  {\em Phys. Rev. D}
  {\bf 74} (2006) 114004, [\href{http://arxiv.org/abs/hep-ph/0607121}{{\tt
  hep-ph/0607121}}].

\bibitem{Martin:2009iq}
A.~D. Martin, W.~J. Stirling, R.~S. Thorne, and G.~Watt, {\it {Parton
  distributions for the LHC}},  {\em Eur. Phys. J. C} {\bf 63} (2009) 189--285,
  [\href{http://arxiv.org/abs/0901.0002}{{\tt arXiv:0901.0002}}].

\bibitem{Boughezal:2013oha}
R.~Boughezal, X.~Liu, F.~Petriello, F.~J. Tackmann, and J.~R. Walsh, {\it
  {Combining Resummed Higgs Predictions Across Jet Bins}},  {\em Phys.Rev.}
  {\bf D89} (2014) 074044, [\href{http://arxiv.org/abs/1312.4535}{{\tt
  arXiv:1312.4535}}].

\bibitem{Campbell:2013qaa}
J.~Campbell, K.~Hatakeyama, J.~Huston, F.~Petriello, J.~R. Andersen, et~al.,
  {\it {Report of the Snowmass 2013 energy frontier QCD working group}},
  \href{http://arxiv.org/abs/1310.5189}{{\tt arXiv:1310.5189}}.

\bibitem{Campbell:2010ff}
J.~M. Campbell and R.~Ellis, {\it {MCFM for the Tevatron and the LHC}},  {\em
  Nucl.Phys.Proc.Suppl.} {\bf 205-206} (2010) 10--15,
  [\href{http://arxiv.org/abs/1007.3492}{{\tt arXiv:1007.3492}}].

\bibitem{Djouadi:1997yw}
A.~Djouadi, J.~Kalinowski, and M.~Spira, {\it {HDECAY: A Program for Higgs
  boson decays in the standard model and its supersymmetric extension}},  {\em
  Comput.Phys.Commun.} {\bf 108} (1998) 56--74,
  [\href{http://arxiv.org/abs/hep-ph/9704448}{{\tt hep-ph/9704448}}].

\bibitem{Hahn:2000jm}
T.~Hahn, {\it {Automatic loop calculations with FeynArts, FormCalc, and
  LoopTools}},  {\em Nucl.Phys.Proc.Suppl.} {\bf 89} (2000) 231--236,
  [\href{http://arxiv.org/abs/hep-ph/0005029}{{\tt hep-ph/0005029}}].

\bibitem{Hahn:2004fe}
T.~Hahn, {\it {CUBA: A Library for multidimensional numerical integration}},
  {\em Comput. Phys. Commun.} {\bf 168} (2005) 78--95,
  [\href{http://arxiv.org/abs/hep-ph/0404043}{{\tt hep-ph/0404043}}].

\bibitem{Banfi:2013eda}
A.~Banfi, P.~F. Monni, and G.~Zanderighi, {\it {Quark masses in Higgs
  production with a jet veto}},  {\em JHEP} {\bf 1401} (2014) 097,
  [\href{http://arxiv.org/abs/1308.4634}{{\tt arXiv:1308.4634}}].

\bibitem{Bonvini:2013jha}
M.~Bonvini, F.~Caola, S.~Forte, K.~Melnikov, and G.~Ridolfi, {\it
  {Signal-background interference effects for $gg \to H \to W^+ W^-$ beyond
  leading order}},  {\em Phys.Rev.} {\bf D88} (2013) 034032,
  [\href{http://arxiv.org/abs/1304.3053}{{\tt arXiv:1304.3053}}].

\bibitem{TheATLAScollaboration:2013zha}
T.~A. collaboration, {\it {Search for a high-mass Higgs boson in the $H
  \rightarrow WW \rightarrow l\nu l\nu$ decay channel with the ATLAS detector
  using 21 fb$^{-1}$ of proton-proton collision data}}, .

\bibitem{Chatrchyan:2013yoa}
{\bf CMS Collaboration} Collaboration, S.~Chatrchyan et~al., {\it {Search for a
  standard-model-like Higgs boson with a mass in the range 145 to 1000 GeV at
  the LHC}},  {\em Eur.Phys.J.} {\bf C73} (2013) 2469,
  [\href{http://arxiv.org/abs/1304.0213}{{\tt arXiv:1304.0213}}].

\bibitem{ATLAS:2013wla}
{\bf ATLAS Collaboration} Collaboration, {\it {Measurements of the properties
  of the Higgs-like boson in the $WW^{(\ast)} \to \ell \nu \ell \nu$ decay
  channel with the ATLAS detector using 25 fb$^{-1}$ of proton-proton collision
  data}}, .

\bibitem{CMS:eya}
{\bf CMS Collaboration} Collaboration, {\it {Evidence for a particle decaying
  to W+W- in the fully leptonic final state in a standard model Higgs boson
  search in pp collisions at the LHC}}, .

\bibitem{iain_notes}
I.~W. Stewart and C.~W. Bauer, ``Lectures on the soft-collinear effective
  theory.''

\bibitem{Becher:2014oda}
T.~Becher, A.~Broggio, and A.~Ferroglia, {\it {Introduction to Soft-Collinear
  Effective Theory}},  \href{http://arxiv.org/abs/1410.1892}{{\tt
  arXiv:1410.1892}}.

\bibitem{Sveshnikov:1995vi}
N.~A. Sveshnikov and F.~V. Tkachov, {\it Jets and quantum field theory},  {\em
  Phys. Lett. B} {\bf 382} (1996) 403--408,
  [\href{http://arxiv.org/abs/hep-ph/9512370}{{\tt hep-ph/9512370}}].

\bibitem{Korchemsky:1997sy}
G.~P. Korchemsky, G.~Oderda, and G.~Sterman, {\it Power corrections and
  nonlocal operators},  \href{http://arxiv.org/abs/hep-ph/9708346}{{\tt
  hep-ph/9708346}}.

\bibitem{Lee:2006nr}
C.~Lee and G.~F. Sterman, {\it {Momentum Flow Correlations from Event Shapes:
  Factorized Soft Gluons and Soft-Collinear Effective Theory}},  {\em
  Phys.~Rev.~D} {\bf 75} (2007) 014022,
  [\href{http://arxiv.org/abs/hep-ph/0611061}{{\tt hep-ph/0611061}}].

\bibitem{Bauer:2008dt}
C.~W. Bauer, S.~P. Fleming, C.~Lee, and G.~Sterman, {\it {Factorization of $e^+
  e^-$ Event Shape Distributions with Hadronic Final States in Soft Collinear
  Effective Theory}},  {\em Phys. Rev. D} {\bf 78} (2008) 034027,
  [\href{http://arxiv.org/abs/0801.4569}{{\tt arXiv:0801.4569}}].

\bibitem{Bauer:2003di}
C.~W. Bauer, C.~Lee, A.~V. Manohar, and M.~B. Wise, {\it {Enhanced
  nonperturbative effects in Z decays to hadrons}},  {\em Phys. Rev. D} {\bf
  70} (2004) 034014, [\href{http://arxiv.org/abs/hep-ph/0309278}{{\tt
  hep-ph/0309278}}].

\bibitem{Manohar:2003vb}
A.~V. Manohar, {\it {Deep inelastic scattering as $x \to 1$ using
  soft-collinear effective theory}},  {\em Phys. Rev. D} {\bf 68} (2003)
  114019, [\href{http://arxiv.org/abs/hep-ph/0309176}{{\tt hep-ph/0309176}}].

\bibitem{Kosower:1999rx}
D.~A. Kosower and P.~Uwer, {\it {One loop splitting amplitudes in gauge
  theory}},  {\em Nucl.Phys.} {\bf B563} (1999) 477--505,
  [\href{http://arxiv.org/abs/hep-ph/9903515}{{\tt hep-ph/9903515}}].

\bibitem{Ellis:1980wv}
R.~K. Ellis, D.~A. Ross, and A.~E. Terrano, {\it {The Perturbative Calculation
  of Jet Structure in e+ e- Annihilation}},  {\em Nucl. Phys.} {\bf B178}
  (1981) 421.

\bibitem{Bern:2004cz}
Z.~Bern, L.~J. Dixon, and D.~A. Kosower, {\it {Two-loop g ---> gg splitting
  amplitudes in QCD}},  {\em JHEP} {\bf 08} (2004) 012,
  [\href{http://arxiv.org/abs/hep-ph/0404293}{{\tt hep-ph/0404293}}].

\bibitem{Badger:2004uk}
S.~Badger and E.~N. Glover, {\it {Two loop splitting functions in QCD}},  {\em
  JHEP} {\bf 0407} (2004) 040, [\href{http://arxiv.org/abs/hep-ph/0405236}{{\tt
  hep-ph/0405236}}].

\bibitem{Huber:2005yg}
T.~Huber and D.~Maitre, {\it {HypExp: A Mathematica package for expanding
  hypergeometric functions around integer-valued parameters}},  {\em
  Comput.Phys.Commun.} {\bf 175} (2006) 122--144,
  [\href{http://arxiv.org/abs/hep-ph/0507094}{{\tt hep-ph/0507094}}].

\bibitem{Huber:2007dx}
T.~Huber and D.~Maitre, {\it {HypExp 2, Expanding Hypergeometric Functions
  about Half-Integer Parameters}},  {\em Comput.Phys.Commun.} {\bf 178} (2008)
  755--776, [\href{http://arxiv.org/abs/0708.2443}{{\tt arXiv:0708.2443}}].

\bibitem{Berends:1988zn}
F.~A. Berends and W.~Giele, {\it {Multiple Soft Gluon Radiation in Parton
  Processes}},  {\em Nucl.Phys.} {\bf B313} (1989) 595.

\bibitem{Catani:2000pi}
S.~Catani and M.~Grazzini, {\it {The soft gluon current at one loop order}},
  {\em Nucl.Phys.} {\bf B591} (2000) 435--454,
  [\href{http://arxiv.org/abs/hep-ph/0007142}{{\tt hep-ph/0007142}}].

\bibitem{Duhr:2013msa}
C.~Duhr and T.~Gehrmann, {\it {The two-loop soft current in dimensional
  regularization}},  {\em Phys.Lett.} {\bf B727} (2013) 452--455,
  [\href{http://arxiv.org/abs/1309.4393}{{\tt arXiv:1309.4393}}].

\bibitem{Li:2013lsa}
Y.~Li and H.~X. Zhu, {\it {Single soft gluon emission at two loops}},  {\em
  JHEP} {\bf 1311} (2013) 080, [\href{http://arxiv.org/abs/1309.4391}{{\tt
  arXiv:1309.4391}}].

\bibitem{Altarelli:1979ub}
G.~Altarelli, R.~K. Ellis, and G.~Martinelli, {\it {Large Perturbative
  Corrections to the Drell-Yan Process in QCD}},  {\em Nucl. Phys. B} {\bf 157}
  (1979) 461.

\bibitem{Matsuura:1988sm}
T.~Matsuura, S.~C. {van der Marck}, and W.~L. {van Neerven}, {\it {The
  Calculation of the Second Order Soft and Virtual Contributions to the
  Drell-Yan Cross-Section}},  {\em Nucl. Phys. B} {\bf 319} (1989) 570.

\bibitem{Almeida:2014uva}
L.~G. Almeida, S.~D. Ellis, C.~Lee, G.~Sterman, I.~Sung, et~al., {\it
  {Comparing and counting logs in direct and effective methods of QCD
  resummation}},  {\em JHEP} {\bf 1404} (2014) 174,
  [\href{http://arxiv.org/abs/1401.4460}{{\tt arXiv:1401.4460}}].

\bibitem{Banfi:2014sua}
A.~Banfi, H.~McAslan, P.~F. Monni, and G.~Zanderighi, {\it {A general method
  for the resummation of event-shape distributions in $e^{+} e^{?}$
  annihilation}},  {\em JHEP} {\bf 05} (2015) 102,
  [\href{http://arxiv.org/abs/1412.2126}{{\tt arXiv:1412.2126}}].

\bibitem{Sakaki:2015iya}
Y.~Sakaki, {\it {Evolution variable dependence of jet substructure}},
  \href{http://arxiv.org/abs/1506.04811}{{\tt arXiv:1506.04811}}.

\bibitem{Elvang:2013cua}
H.~Elvang and Y.-t. Huang, {\it {Scattering Amplitudes}},
  \href{http://arxiv.org/abs/1308.1697}{{\tt arXiv:1308.1697}}.

\bibitem{Catani:1998bh}
S.~Catani, {\it {The Singular behavior of QCD amplitudes at two loop order}},
  {\em Phys. Lett.} {\bf B427} (1998) 161--171,
  [\href{http://arxiv.org/abs/hep-ph/9802439}{{\tt hep-ph/9802439}}].

\bibitem{Parke:1986gb}
S.~J. Parke and T.~R. Taylor, {\it {An Amplitude for $n$ Gluon Scattering}},
  {\em Phys. Rev. Lett.} {\bf 56} (1986) 2459.

\bibitem{Retiere:2009zz}
F.~Retiere et~al., {\it {Characterization of Multi Pixel Photon Counters for
  T2K Near Detector}},  {\em Nucl. Instrum. Meth.} {\bf A610} (2009) 378--380.

\bibitem{Chiu:2012ju}
C.~S. Chiu, C.~Ignarra, L.~Bugel, H.~Chen, J.~M. Conrad, B.~J.~P. Jones,
  T.~Katori, and I.~Moult, {\it {Environmental Effects on TPB
  Wavelength-Shifting Coatings}},  {\em JINST} {\bf 7} (2012) P07007,
  [\href{http://arxiv.org/abs/1204.5762}{{\tt arXiv:1204.5762}}].

\bibitem{Tarasov:1980au}
O.~V. Tarasov, A.~A. Vladimirov, and A.~Y. Zharkov, {\it {The Gell-Mann-Low
  Function of QCD in the Three Loop Approximation}},  {\em Phys. Lett. B} {\bf
  93} (1980) 429--432.

\bibitem{Larin:1993tp}
S.~A. Larin and J.~A.~M. Vermaseren, {\it {The three-loop QCD $\beta$ function
  and anomalous dimensions}},  {\em Phys. Lett. B} {\bf 303} (1993) 334--336,
  [\href{http://arxiv.org/abs/hep-ph/9302208}{{\tt hep-ph/9302208}}].

\bibitem{Korchemsky:1987wg}
G.~P. Korchemsky and A.~V. Radyushkin, {\it {Renormalization of the Wilson
  Loops Beyond the Leading Order}},  {\em Nucl. Phys. B} {\bf 283} (1987)
  342--364.

\bibitem{Moch:2004pa}
S.~Moch, J.~A.~M. Vermaseren, and A.~Vogt, {\it {The three-loop splitting
  functions in QCD: The non-singlet case}},  {\em Nucl. Phys. B} {\bf 688}
  (2004) 101--134, [\href{http://arxiv.org/abs/hep-ph/0403192}{{\tt
  hep-ph/0403192}}].

\end{thebibliography}\endgroup
